\documentclass[a4paper,12pt]{book}
\usepackage{amsthm,amsfonts,amssymb,euscript}
\usepackage{latexsym, multicol, fancybox}
\usepackage{graphicx}
\usepackage{color}
\usepackage{amsmath, amsthm, amssymb, bm}
\usepackage{epstopdf}
\usepackage{caption}
\usepackage{psfrag}
\usepackage{verbatim}
\usepackage{accents}

\newtheorem{theorem}{Theorem}[section]
\newtheorem{lemma}[theorem]{Lemma}
\newtheorem{proposition}[theorem]{Proposition}
\newtheorem{corollary}[theorem]{Corollary}
\newtheorem{definition}[theorem]{Definition}
\newtheorem{remark}[theorem]{Remark}

\newtheorem*{thmmain}{Main Theorem}
\newtheorem*{thmM0}{Theorem M0}
\newtheorem*{thmM1}{Theorem M1}
\newtheorem*{thmM2}{Theorem M2}
\newtheorem*{thmM3}{Theorem M3}
\newtheorem*{thmM4}{Theorem M4}
\newtheorem*{thmM5}{Theorem M5}
\newtheorem*{thmM6}{Theorem M6}
\newtheorem*{thmM7}{Theorem M7}
\newtheorem*{thmM8}{Theorem M8}

\numberwithin{equation}{section}

\newcommand{\bea}{\begin{eqnarray}}
\newcommand{\eea}{\end{eqnarray}}
\def\beaa{\begin{eqnarray*}}
\def\eeaa{\end{eqnarray*}}
\def\ba{\begin{array}}
\def\ea{\end{array}}
\def\be#1{\begin{equation} \label{#1}}
\def \eeq{\end{equation}}
\def\bsplit{\begin{split}}
\newcommand{\nn}{\nonumber}

\def\lab{\label}
\def\les{\lesssim}

\def\c{\cdot}
\def\lot{\mbox{l.o.t.}}
\def\dual{{\,^\star \mkern-2mu}}
\def\tr{\mbox{tr}}
\def\Gab{{\bf \Gamma}}
\def\rh{r_\TT}
\def\rhot{\tilde{\rho}}
\newcommand{\nabb}{\nab\mkern-13mu /\,}

\newcommand{\near}{\!\!\nearrow\,}

\newcommand{\rr}{r_\TT}

\newcommand{\mr}{\textrm{Match}}
\newcommand{\Axis}{\mathfrak{A}}
\def\Good{\mbox{Good}}

\def\ov{\overline}

\def\ovu{\overset{\circ}{ u}}
\def\ovd{\overset{\circ}{ d}}
\def\ovs{\overset{\circ}{ s}}

\def\ovr{\overset{\circ}{ r}}
\def\ovS{\overset{\circ}{ \S}}

\def\dddov{ \,  \ovd \hspace{-2.4pt}    \mkern-6mu /}

\newcommand{\Mext}{\,{}^{(ext)}\mathcal{M}}
\def\MMext{\,{}^{(ext)}\mathcal{M}}
\newcommand{\Mint}{\,{}^{(int)}\mathcal{M}}

\newcommand{\sint}{\,{}^{(int)}s}
\newcommand{\sext}{\,{}^{(ext)}s}
\newcommand{\rint}{\,{}^{(int)}r}
\newcommand{\rext}{\,{}^{(ext)}r}
\newcommand{\mint}{\,{}^{(int)}m}
\newcommand{\mext}{\,{}^{(ext)}m}
\newcommand{\Lext}{\,{}^{(ext)}\mathcal{L}_0}
\newcommand{\Lint}{\,{}^{(int)}\mathcal{L}_0}

\def\pid{\dot{\pi}}
\def\pidX{\,^{(X)}\pid}

\renewcommand{\div}{\mbox{div }}

\newcommand{\divv}{\mbox{div}\mkern-19mu /\,\,\,\,}
\newcommand{\lapp}{\mbox{$\bigtriangleup  \mkern-13mu / \,$}}
\newcommand{\curll}{\mbox{curl}\mkern-19mu /\,\,\,\,}

\def\nab{\nabla}

\def\pr{\partial}
\def\Div {\mbox{\bf Div}}

\def\dkb{ \, \mathfrak{d}     \mkern-9mu /}
\def\dk{\mathfrak{d}}

\def\dddS{\ddd^{\,\S}}
\def\ddsS{\ddd^{\,\S, \star}}

\def\dtb{\widetilde{\mathfrak{d}}}

\def\dkout{\dk_{\near}}

 \def\nabbc{\,^{(c)}\nabb }
 \def\lappc{\,^{(c)}\mkern-2.5mu\lapp}

\def\DDs{ \, \DD \hspace{-2.4pt}\dual    \mkern-16mu /}
\def\DDd{ \, \DD \hspace{-2.4pt}    \mkern-8mu /}

\def\dds{ \, d  \hspace{-2.4pt}\dual    \mkern-14mu /}
\def\ddd{ \,  d \hspace{-2.4pt}    \mkern-6mu /}

\def\rhod{ \,^\star  \hspace{-2.2pt} \rho}

\def\epg{\overset{\circ}{\ep}}  
\def\ug{\overset{\circ}{u}}   
\def\sg{\overset{\circ}{s}}               
\def\rg{\overset{\circ}{r}}

\def\ovu{\overset{\circ}{ u}}
\def\ovd{\overset{\circ}{ d}}
\def\ovs{\overset{\circ}{ s}}

\def\ovr{{\overset{\circ}{ r}\,}}
\def\ovm{{\overset{\circ}{ m}\,}}
\def\ovS{\overset{\circ}{ S}}
\def\ovga{\overset{\circ}{\ga}\,}
\def\ovgS{\overset{\circ}{\gS}\,}
\def\dg{\overset{\circ}{\de}}
\def\ovka{\overset{\circ}{ \ka}}
\def\ovkab{\overset{\circ}{ \kab}}
\def\ovI{{\overset{\circ}{ I}}}

  \def\gazero{\ga^{(0)}}

  \def\muc{\check{\mu}}

       \def\hnn{h^{(n+1)}}
         \def\hbnn{\underline{h}^{(n+1)}}
              \def\hn{h^{(n)}}
  \def\hbn{\underline{h}^{(n)}}    
                   \def\wn{w^{(n)}}   
                \def\wnn{w^{(n+1)}}

                            \def\en{e^{(n)}}   
                \def\enn{e^{(n+1)}}     
             \def\hb{\underline{h}}

\def\dkbS{{\dkb^\S}}
\def\vthS{\vth^\S}
\def\vthbS{\vthb^\S}
\def\etaS{\eta^\S}
\def\etabS{\etab^\S}

\def\kabS{\kab^\S}
\def\omS{\om^\S}
\def\ombS{\omb^\S}
\def\zeS{\ze^\S}
\def\aS{\a^\S}
\def\aaS{\aa^\S}
\def\bS{\b^\S}
\def\bbS{\bb^\S}
\def\rhoS{\rho^\S}
\def\gaS{{\ga^\S}\,}
\def\vsiS{\vsi^\S}

\def\GaS{\Ga^\S}

\def\ombcS{\ombc^\S}

\def\nuS{\nu^\S}

\def\a{\alpha}
\def\b{\beta}
\def\ga{\gamma}
\def\Ga{\Gamma}
\def\de{\delta}
\def\De{\Delta}
\def\ep{\epsilon}
\def\la{\lambda}
\def\La{\Lambda}
\def\si{\sigma}
\def\vsi{\varsigma}
\def\Si{\Sigma}
\def\om{\omega}
\def\Om{\Omega}

\def\th{{\theta}}
\def\Th{{\Theta}}
\def\ka{\kappa}
\def\ze{\zeta}
\def\Up{\Upsilon}

\def\vphi{{\varphi}}
\def\vth{\vartheta}

\def\thb{\underline{\th}}
\def\vthb{\underline{\vth}}
\renewcommand{\aa}{\protect\underline{\a}}
\newcommand{\bb}{\protect\underline{\b}}
\def\omb{{\underline{\om}}}
\def\Lb{{\underline{L}}}

\def\lb{{\underline  l}}

\def\Omb{\underline{\Omega}}
\def\ub{{\underline{u}} }
\newcommand{\chib}{\underline{\chi}}
\newcommand{\xib}{\underline{\xi}}
\newcommand{\etab}{\underline{\eta}}

\def\kab{\underline{\kappa}}
\def\vsib{\underline{\varsigma}}
\def\Lab{{\underline{\La}}}
\def\Bb{\underline{B}}
\def\Nb{\underline{N}}

\def\AA{{\mathcal A}}
\def\BB{{\mathcal B}}
\def\CC{{\mathcal C}}
\def\DD{{\mathcal D}}
\def\EE{{\mathcal E}}
\def\FF{{\mathcal F}}

\def\HH{{\mathcal H}}
\def\II{{\mathcal I}}
\def\JJ{{\mathcal J}}
\def\KK{{\mathcal K}}
\def\Lie{{\mathcal L}}
\def\LL{{\mathcal L}}
\def\MM{{\mathcal M}}
\def\NN{{\mathcal N}}

\def\PP{{\mathcal P}}
\def\QQ{{\mathcal Q}}
\def\RR{{\mathcal R}}
\def\SS{{\mathcal S}}
\def\TT{{\mathcal T}}
\def\UU{{\mathcal U}}

\def\C{{\bf C}}
\def\D{{\bf D}}

\def\G{{\bf G}}

\def\M{{\bf M}}
\def\N{{\bf N}}

\def\R{{\bf R}}

\def\S{{\bf S}}
\def\T{{\bf T}}
\def\U{{\bf U}}

\def\Z{{\bf Z}}

\def\g{{\bf g}}

\def\Ab{\underline{A}}
\def\Bb{\underline{B}}
\def\Eb{\underline{E}}

\def\Vb{\underline{V}\,}
\def\lb{{\underline{l}}}
\def\mub{{\underline{\mu}}}
\def\fb{\underline{f}}
\def\CCb{\underline{\CC}}

\def\NNN{{\Bbb N}}

\def\RRR{{\Bbb R}}
\def\SSS{{\Bbb S}}

\def\qf{\frak{q}}

\def\Dk{\mathfrak{D}}
\def\Nk{\mathfrak{N}}

\def\Rk{\mathfrak{R}}
\def\Gk{\mathfrak{G}}
\def\Ik{\mathfrak{I}}

\def\hk{\mathfrak{h}}
\def\af{\mathfrak{a}}
\def\sk{\mathfrak{s}}

\def\EEt{\tilde{\EE}}

\def\Rc{\check R}

\def\rhoc{\check \rho}
\def\Gac{\check \Gamma}

\def\kac{\check \ka}
\def\kabc{\check{\underline{\ka}}}
\def\Omc{\check{\Omega}}
\def\Oc{\check{\Omega}}
\def\Ombc{\check{\underline{\Omega}}}
\def\Obc{\check{\underline{\Omega}}}
\def\omc{\check \omega}
\def\ombc{\underline{\check \omega}}
\def\vsic{\check{\vsi}}

\def\piT{{\,^{(T)} \pi }}
\def\piR{{\,^{(R)} \pi }}

\def\LaX{\,^{(X)}\La}

\newcommand{\piX}{\,^{(X)}\pi}
\newcommand{\piY}{\,^{(Y)}\pi}

\def\chih{\widehat{\chi}}
\def\chibh{\widehat{\chib}}
\def\trch{\tr \chi}
\def\trchb{\tr \chib}

\def\aS{\,^{(1+3)} \hspace{-2.2pt}\a} 
\def\bS{\,^{(1+3)} \hspace{-2.2pt}\b} 
\def\rhoS{\,^{(1+3)} \hspace{-2.2pt}\rho} 
\def\rhodS{\,^{(1+3)} \hspace{-2.2pt}\rhod} 
\def\bbS{\,^{(1+3)} \hspace{-2.2pt}\bb} 
\def\aaS{\,^{(1+3)} \hspace{-2.2pt}\aa} 

\def\chiS{\,^{(1+3)} \hspace{-2.2pt}\chi} 
\def\chibS{\,^{(1+3)} \hspace{-2.2pt}\chib} 
\def\chihS{\,^{(1+3)} \hspace{-2.2pt}\chih} 
\def\chibhS{\,^{(1+3)} \hspace{-2.2pt}\chibh}
\def\trchS{\,^{(1+3)} \hspace{-2.2pt}\trch} 
\def\trchbS{\,^{(1+3)} \hspace{-2.2pt}\trchb}
\def\xiS{\,^{(1+3)} \hspace{-2.2pt}\xi} 
\def\xibS{\,^{(1+3)} \hspace{-2.2pt}\xib} 
\def\etaS{\,^{(1+3)} \hspace{-2.2pt}\eta} 
\def\etabS{\,^{(1+3)} \hspace{-2.2pt}\etab} 
\def\zeS{\,^{(1+3)} \hspace{-2.2pt}\ze} 
\def\omS{\,^{(1+3)} \hspace{-2.2pt}\om} 
\def\ombS{\,^{(1+3)} \hspace{-2.2pt}\omb} 
\def\fS{\,^{(1+3)} \hspace{-2.2pt} f} 
\def\divvS{\,^{(1+3)} \hspace{-2.2pt}\divv} 

\def\nabbS{\,^{(1+3)} \hspace{-2.2pt}\nabb}

\def\hot{\widehat{\otimes}}
\def\c{\cdot}
\def \f12{\frac 1 2 }
\def\ov{\overline}
\def\Db{\dot{\D}}
\def\squared{\dot{\square}}
\def\QQh{\widehat{\QQ}}
\def\err{\mbox{Err}}

\def\psic{\,\check{\psi}}

\def\Mor{\mbox{Mor}}
\def\Morr{\mbox{Morr}}

\def\ec{\check{e}}

\def\Edot{\dot{E}}

\def\gS{ g \mkern-8.5mu/\,}

\def\com{\mbox{Com}}
\def\comb{\underline{Com}}
\def\err{\mbox{Err}}

\newcommand{\deh}{\delta_{\mathcal{H}}}
\newcommand{\dec}{\delta_{dec}}

\newcommand{\dee}{\delta_{extra}}
\newcommand{\dt}{\delta_B}

\def\ntrap{trap\mkern-18 mu\big/\,}
\def\Sitrap{\,^{(trap)}\Si}
\def\Mtrap{\,^{(trap)}\MM}
\def\Mntrap{\,^{(\ntrap)}\MM}
\def\Sext{\,{}^{(ext)}\Si}
\def\Sint{\,{}^{(int)}\Si}

 \def\Rbrev{\breve{R}}
        \def\Tbrev{\breve{T}}
        \def\thbrev{\breve{\th}}

\newcommand{\Ng}{N_g}
\newcommand{\Nm}{N_m[\qf]}

\def\EEd{\dot{\EE}}
        \def\Ed{\dot{E}}
\def\Fd{\dot{F}}

\def\Bdot{\dot{B}}

\def\psic{\,\check{\psi}}
\def\qfc{\check{\qf}}

\def\qfk{\qf^{(k)}}

\def\Jc{\check{J}}

\def\pidX{\,^{(X)}\pid}

\def\pidd{\ddot{\pi}}
\def\piddX{\,^{(X)} \pidd}

\def\pitX{\,^{(X)} \widetilde{\pi}}

\newcommand{\idl}{\LL_0}
\newcommand{\sintl}{\,{}^{(int)}s_{\idl}}
\newcommand{\sextl}{\,{}^{(ext)}s_{\idl}}
\newcommand{\rintl}{\,{}^{(int)}r_{\idl}}
\newcommand{\rextl}{\,{}^{(ext)}r_{\idl}}

\newcommand{\Blue}{\textcolor{blue}}

 \def\ddsSn{\ddd^{\, \S(n), \star}\,}

\def\MMextend{\MM^{(extend)} }

\begin{document}

\title{Global Nonlinear Stability  of Schwarzschild Spacetime under Polarized Perturbations}  

\author{Sergiu Klainerman and  J\'er\'emie Szeftel}

\maketitle

\centerline{Abstract}
We   prove the nonlinear  stability of the Schwarzschild  spacetime under axially symmetric  polarized perturbations, i.e. solutions of the  Einstein vacuum equations
 for asymptotically flat  $1+3$ dimensional Lorentzian metrics  which admit  a hypersurface orthogonal spacelike Killing vectorfield   with closed orbits.
  While   building  on the  remarkable  advances made       in last   15 years on    establishing  quantitative  linear stability, the paper     introduces   a series of new ideas among which  we emphasize   the \textit{general covariant modulation} (GCM)  procedure  which allows  us  to construct, dynamically, the center of mass frame of the  final state. The mass of the final state itself   is tracked using  the well known   Hawking mass relative to a well adapted foliation itself connected to the center of mass frame. 
  
  Our work here is the first to prove the nonlinear stability of Schwarzschild  in   a  restricted    class of  nontrivial  perturbations. To a large extent, the restriction to this class of perturbations is only needed to ensure  that  the final state of evolution is   another 
Schwarzschild space. We are thus confident that our  procedure  may  apply in a more general setting.

\tableofcontents


\chapter{INTRODUCTION}



\section{Bare-bones version of the main theorem}


The goal of the book  is to    prove the nonlinear  stability of the Schwarzschild  spacetime under axially symmetric  polarized perturbations, i.e. solutions of the  Einstein vacuum equations \eqref{EVE}
 for asymptotically flat  $1+3$ dimensional Lorentzian metrics  which admit  a hypersurface orthogonal spacelike Killing vectorfield  $\Z$  with closed orbits.
  Recall that the Schwarzschild  metric $\g_m$ of  mass $m>0$   is a stationary, spherically symmetric solution of \eqref{EVE},  which takes the form, in standard coordinates,
 \beaa
\g_m =  -\left(1-\frac{2m}{r}\right) dt^2 + \left(1-\frac{2m}{r}\right)^{-1} dr^2 + r^2\left ( d\th^2 +\sin^2\th  d\vphi^2\right).
 \eeaa
  This class of    perturbations  allows us    to restrict our  analysis to the  case when the final state of evolution is    itself a Schwarzschild spacetime.   This is not the case in general, as a typical perturbation  of Schwarzschild may  approach a  member  of the Kerr family with small angular momentum.

The simplest  version of  our main theorem can be stated as  follows.
\begin{theorem}[Main Theorem (first version)] The future globally hyperbolic   development  of  an   \emph{axially symmetric, polarized\footnote{See section  \ref{subs:axial-polarized} for a precise definition of axial symmetry and polarization.  This property  is preserved by the Einstein equations, i.e.  if the data is axially symmetric, polarized, so is  its development.}},  asymptotically  flat   initial data set, sufficiently close  (in a specified  topology)  to a Schwarzschild  initial data set   of  mass $m_0>0$,  has a complete    future null infinity  $\II^+$ and converges 
in  its causal past  $\JJ^{-1}(\II^{+})$  to another  nearby  Schwarzschild solution of mass $m_{\infty}$ close to $m_0$. 
 \end{theorem}
 
Our  theorem  is  an important     step in   the  long standing effort   to prove  the full   nonlinear stability of  Kerr spacetimes $\KK(a,m)$, in the sub-extremal regime $|a|<m$.      We give a succinct   review below of  some  of       the most important results   which have been obtained so far in this direction.


\section{The Kerr family}


Consider  solutions to  the Einstein vacuum  equations (EVE),
\bea
\label{EVE}
\R_{\a\b}=0
\eea
with   $\R_{\a\b}$ the Ricci curvature of a four dimensional, 
Lorentzian  manifold $(\MM, \g)$.  Solutions of the equations are invariant under general  diffeomorphisms of $\Phi:\MM\longrightarrow \MM$, i.e. if $\g$ verifies EVE  so does  its pull back $\Phi^\# \g$.
We recall that an initial
data set $(\Si_{(0)}, g_{(0)}, k_{(0)})$  consists of a $3$ dimensional manifold $\Si_{(0)}$
 together with  a  complete Riemannian metric $g_{(0)}$ and  a symmetric 2-tensor $k_{(0)}$ which verify    a well known  set of constraint equations (see for instance the introduction in \cite{Ch-Kl}).  A Cauchy development of an initial data set  is a globally hyperbolic  space-time   $(\MM,\g)$, verifying EVE together with 
  an  embedding $i:\Si_{(0)}\longrightarrow \MM$ such that 
$i_*(g_{(0)}), i_*(k_{(0)})$  are the first and second fundamental forms
of $i(\Si_{(0)})$ in $\MM$. A well known  foundational result  in GR
 associates a unique   maximal, global hyperbolic,  future development 
    to all sufficiently regular  initial data sets, see \cite{Bruhat}, \cite{Br-Ger}\footnote{See also \cite{Sb} for a modern treatment.}.   
We  further  restrict  the discussion  to asymptotically flat initial data sets,
i.e.  assume that outside a sufficiently large compact set $K$,
 $\Si_{(0)}\setminus K$ is diffeomorphic  to the complement of the unit
 ball  in $\RRR^3$ and admits a system of coordinates in which $ g_{(0)}$ is asymptotically euclidean and $k_{(0)}$ vanishes at an appropriate order at infinity.
 
 EVE admits a remarkable   two parameter family of explicit     solutions, the  Kerr spacetimes  $\KK(a,m)$, $0\le a\le m$,  which are  stationary,  axisymmetric and asymptotically flat. In the usual  Boyer-Lindquist coordinates  they take the form,  
   \begin{equation}\label{zq1}
\g_{a,m}=-\frac{q^2\Delta}{\Sigma^2}(dt)^2+\frac{\Sigma^2(\sin\theta)^2}{q^2}\Big(d\varphi-\frac{2amr}{\Sigma^2}dt\Big)^2 +\frac{q^2}{\Delta}(dr)^2+q^2(d\theta)^2,
\end{equation}
where
\begin{equation}\label{zq2}
\begin{cases}
&\Delta=r^2+a^2-2mr, \\
&  q^2=r^2+a^2(\cos\theta)^2,\\
&\Sigma^2=(r^2+a^2)q^2+2mra^2(\sin\theta)^2=(r^2+a^2)^2-a^2(\sin\theta)^2\Delta.
\end{cases}
\end{equation}
 Among  them 
  one distinguishes the Schwarzschild family of spherically symmetric   solutions,  of mass $m>0$,
\bea
\g_m=-\left(1-\frac{2m}{r}\right) dt^2+\left(1-\frac{2m}{r}\right)^{-1} dr^2 +r^2 d\si_{\SSS^2}. 
\label{eq:Schw}
\eea 
Though the metric seems singular at $r=2m$   ($r=r_+$, the largest  root   of $\De(r)=0$,  in the case of Kerr)
it turns out that one can glue together two regions $r>2m$ and two regions $r<2m$ of the Schwarzschild metric to obtain a metric which is smooth along the null hypersurface   $\EE=\{r=2m\}$
 called the Schwarzschild  event  horizon.   The portion of  $r<2m$  to the future
 of the hypersurface  $t=0$ is a black hole whose future boundary $r=0$ is 
 singular.    The region $r>2m$, free of  singularities,  is called the domain of outer communication.  
  The more general family of Kerr solutions, which   are both stationary and axially symmetric,  possesses (in addition to   well defined
  event horizons, black holes and domains of outer communication)   Cauchy horizons  ($r=r_-$ for   the smallest root $r=r_-$ of $\De(r)=0$) inside the black hole region   across which predictability seems to fail\footnote{I.e. various  smooth extensions are possible.}. Again, one
  can easily check, from the precise nature of the Kerr metric, that the region
  outside  the event horizon,  i.e. outside the  Kerr black hole,  is free of singularities\footnote{The generalization of this    observation   to all  but an exceptional set of     initial data   is   the celebrated  \textit{weak cosmic censorship conjecture} of Penrose.}. 
   Note that    the Kerr spacetimes 
  $\KK(a,m)$ possess two Killing vectorfields; the stationary  vectorfield    $\T=\pr_t$,  which is timelike   in the 
  asymptotic region, away  from the horizon, and  the  axial symmetric Killing vectorfield $\Z=\pr_\varphi$. In the particular case of Schwarzschild,  $\T$ is also  orthogonal    to the hypersurfaces  $t=$const.
 
Here are some other important properties of the Kerr family.
\begin{itemize}
\item The Kerr solution   has a   remarkable algebraic feature, encoded        in the so called Petrov type $D$ property, according to which  it
  admits, at every point a pair of    null vectors $(l, \lb)$, normalized  by the condition $\g(l,\lb)=-2$, called principal null vectors,   such that all components  of the Riemann curvature tensor vanish identically   except for  the two  independent components  
   \beaa
   \R(l, \lb, l, \lb), \quad \dual \R(l, \lb, l, \lb)
   \eeaa
    with $\dual \R$ the Hodge dual of $\R$.
    
\item   In addition to the symmetries provided  by   the Killing vectorfields   $\T$ and       $\Z$,  the Kerr solution possesses a nontrivial Killing tensor   i.e. a symmetric $2$-covariant tensor $\C$ (the Carter tensor)  verifying,
 \beaa
  \D_{(\a } \C_{\b\ga)}=0.
 \eeaa
   
\item The Kerr family  is distinguished   among all stationary solutions of  EVE by the vanishing of a  four tensor  called the Mars-Simon tensor, see \cite{Mars}. 
\end{itemize}


\section{Stability of Kerr}



\subsection{Stability of Kerr conjecture}


The  nonlinear stability of the Kerr    family is one of the most pressing  issues   in mathematical GR  today.   Roughly, the problem is to show that all spacetime developments of  initial data sets,    sufficiently close to the initial data set of a Kerr spacetime, behave    in the large like  a (typically another)  Kerr solution.  This is not only a deep mathematical question but one with serious astrophysical  implications. Indeed, if  the Kerr family would be  unstable  under  perturbations, black holes  would be nothing more than mathematical artifacts.  Here is a more precise formulation of the conjecture.  

   {\bf Conjecture} (Stability of Kerr conjecture).\,\,{\it  Vacuum initial data sets, 
   sufficiently close to Kerr initial data, have a maximal development with complete
   future null infinity\footnote{This means, roughly, that observers  which are  far away  
    from the black hole  may live forever.   } and with   domain of outer communication which
   approaches  (globally)  a nearby Kerr solution.}
   
    So far,  the only   space-time for which  full nonlinear  stability has been established is the Minkowski space, corresponding to the particular case $a=m=0$. The result was  first proved in  \cite{Ch-Kl},   see also   \cite{KlNi}, \cite{L-Rodn}, \cite{Bi} and \cite{HiVa:Minkstab}.   
\begin{theorem}[Global stability of   Minkowski]    Any
asymptotically flat initial data set, which is sufficiently close to the trivial one, 
has a  regular, complete,    maximal development\footnote{  The  complete result    in  \cite{Ch-Kl}     also provides
very precise information about the decay of the curvature tensor along
null and timelike directions as well as many   other geometric information 
concerning the causal structure of the corresponding  spacetime. 
Of particular interest are \textit{peeling properties} i.e. the precise decay rates 
of  various components of the curvature tensor along future  null  geodesics.
}. \label{thm:stability.mink}
\end{theorem}

Here are, very schematically, the main ideas 
 in the proof of stability of Minkowski space.
 
  \begin{itemize}
\item[ (I) ] Perturbations radiate and decay 
{ \textit{sufficiently fast}}             (just fast enough!)  to insure  convergence.

\item[ (II)]
  Interpret  the   Bianchi identities as   a Maxwell like system. 
 This  is an effective,  \textit{invariant},  way to       treat the hyperbolic
  character  of the equations.

\item[ (III) ]Rely on four    important  PDE  advances    of late last century:
\end{itemize}

\begin{enumerate}
\item[(i)]   Vectorfield  approach  to  get  decay  based on   {\textit{  approximate}}  Killing and conformal Killing  symmetries  of the  equations, see \cite{Kl-vectorfield}, \cite{Kl-null},  \cite{Kl-vect2}, \cite{Ch-Kl0}.  
\item[(ii)]   Generalized energy estimates   using
  both the  Bianchi identities and  the approximate Killing  and conformal Killing vector fields.
  \item[(iii)]   The  \textit{null condition} identifies the deep  mechanism for     nonlinear stability, 
i.e.  the specific structure of the nonlinear terms    enables stability  despite the slow
  decay  rate of the  perturbations, see \cite{Kl-ICM},  \cite{Kl-null}, \cite{Chr}.
\item[(iv)]    Complex boot-strap argument according to which  one makes  educated assumptions 
about the behavior of the space-time and then proceeds to       show that   they are in fact 
 satisfied. This  amounts to a \textit{conceptual linearization}, i.e. a method  by which the equations become, essentially, linear  without   actually linearizing them.
 \end{enumerate}
 
  There are three, related,  major obstacles in passing from the stability of Minkowski  to that of the Kerr family.
  
 \begin{enumerate}
 \item The first   can be understood in the general framework of nonlinear hyperbolic or dispersive equations.
 Given a nonlinear equation $\NN[\phi]=0$ and a stationary solution $\phi_0$   we have two notions of stability,
 \textit{orbital stability}, according to which  small perturbations    of $\phi_0$    lead to solutions $\phi$ which remain close, in some  norm (typically $L^2$ based )  for all time,   and  \textit{asymptotical stability}, according to which
 the perturbed solutions converge, as $t\to \infty$,   to  a nearby  stationary solution.    Note that the second notion is  far stronger, and much more precise,  than the first  and that orbital stability  can only be  established  (without   appealing to the  the stronger version) only   for    equations with  very weak  nonlinearities. For quasilinear equations, such as the Einstein field equations,   a proof of stability requires, necessarily, a proof of    asymptotic stability.      This    must then be based on     a    detailed  understanding of the decay properties of the linearized\footnote{ It is irrelevant  whether a specific linearization procedure needs to be implemented; what is important here is to identify the linear mechanism for decay, such as the  Maxwell system in the case of the stability of Minkowski space mentioned above.} equations.
  
 One is thus led to  study  the linearized equations $\NN'[\phi_0]\psi =0$,  with $\NN'[\phi_0]$ the Fr\'echet  derivative of $\NN$ at $\phi_0$, which, in  many important cases,  are hyperbolic\footnote{ In the case of EVE the linearized  equations are  linear hyperbolic only   after  we mod out  the linearized version of general  coordinate transformations.}   systems  with variable coefficients  that  typically present instabilities. 
  In the  exceptional  situation,  when  nonlinear  stability can ultimately  be established,   one  can  tie  all   the  instability   modes of the linearized system  to  two  properties of the nonlinear equation:  
\begin{enumerate}
\item  The   presence of a  continuous\footnote{In the  case of the stability of   Kerr  we have a $2$ parameter family of solutions $\KK(a,m)$.}${}^{,}$\footnote{This is responsible of the fact   that a small perturbation of  the   fixed stationary solution $\phi_0$  may not converge to  $\phi_0$ but to  another nearby  stationary solution.}  family  of other  stationary solutions  of   $\NN[\phi]=0$      near $\phi_0$.  
   
\item The presence of  a continuous family of diffeomorphisms\footnote{In the case of EVE, any diffeomorphism  has that property.   }  of the background manifold 
   which  map, by pull back, solutions to solutions.
\end{enumerate}   
For  a typical  stationary solution $\phi_0$,  both properties exist and  generate  nontrivial solutions of the linearized equation  $\NN'[\phi_0]\psi=0$.
    In the case of   relatively simple scalar     nonlinear equations, where the symmetry  group of the equation is  small,   an effective   strategy    of dealing with this problem  (known  under the name of modulation theory)  has been developed,  see   for example  \cite{Ma-Me}, \cite{Me-R}.  In the case   of the Einstein equations this problem is compounded by    the    large invariance  group of  the equations, i.e. all diffeomorphisms of the spacetime manifold.  To deal with  both  problems and establish stability   one has to
   \begin{itemize}
   \item Track the  parameters $(a_f, m_f)$ of the  final Kerr spacetime. 
   \item  Track the   coordinate  system (gauge condition)  relative to which  we have decay for  all  linearized quantities. Such a coordinate  system cannot be imposed a-priori, it has to emerge dynamically  in the construction of  the spacetime.  
   \end{itemize}

\item As described earlier, the fundamental insight   in the stability of the Minkowski space  was that we can treat the Bianchi identities as a Maxwell system in a slightly perturbed Minkowski space by using  the vectorfield method.  This  cannot work for  perturbations of Kerr due to the fact that some of the null components of the  curvature tensor\footnote{With respect to  the so called  principal null directions.}  are non-trivial in Kerr.

\item  Even if  we   can establish   a  useful version of   linearization (i.e. one which addresses   the above mentioned problems), there are still major obstacles  in understanding their  decay properties.  Indeed,  when one  considers  the simplest,    relevant,
  linear equation  on a fixed Kerr background,  i.e. the  wave equation $\square_\g \psi=0$ (often referred to as the\textit{ poor's man linearization} of EVE),  one 
  encounters serious difficulties  even to  prove the boundedness of solutions  for  the most reasonable,
   smooth, compactly supported,  data.   Below is a  very short description of these.
 \begin{itemize}
       
 \item \textit{The problem of   trapped null geodesics.}   This  concerns the existence of null geodesics\footnote{In the Schwarzschild case, these geodesics are located on the so-called photon sphere $r=3m$.} neither crossing the event horizon nor escaping to null infinity, along which solutions   can concentrate for arbitrary long times.  This   leads  to degenerate        energy estimates  which require a very delicate analysis.

\item \textit{The   trapping   properties of the     horizon.}       
    The horizon  itself  is ruled  by null  geodesics, which do not     communicate with  null infinity and can thus concentrate energy.         This problem was solved   by understanding      the  so called     red-shift effect associated to  the event horizon,  which more than    counteracts this type of  trapping.

\item \textit{The problem of superradiance.} This  is essentially the failure of the stationary Killing field $\T=\partial_t$
to be everywhere timelike in the domain of outer communications and, thus, the failure of the associated  conserved energy to be positive. Note that this problem is absent in Schwarzschild and, in general,   for axially symmetric solutions.
\item   \textit{Superposition problem.}       This is the problem  of combining the estimates  in the near  region, close to the horizon, (including the  ergoregion  and trapping) with estimates in the asymptotic region, where the spacetime looks Minkowskian.
\end{itemize}
 
\item The full linearized system of EVE around Kerr, usually referred to as the linearized gravity system (LGS), whatever its formulation, presents far more difficulties beyond those mentioned above concerning the poor man's linear scalar wave equation on Kerr, see the discussion below.  
  
\end{enumerate}

   Historically,    two versions of LGS  have been   considered.
   \begin{itemize}
   \item[(a)] At the level of the metric  itself, i.e. if   $\G$  denotes the Einstein  tensor,  $\G_{\a\b}=\R_{\a\b}-\frac 1 2 \R \g_{\a\b}$, 
   \bea
\label{Lin. Grav}
\G'(\g_0) \, \de \g =0.
\eea

   \item[(b)]  Via  the Newman-Penrose (NP) formalism,  based  on null frames. 
   \end{itemize}
   In what  follows we  review  the main known  results  concerning  solutions to the linearized equations on a Kerr background.


\subsection{Formal Mode Analysis}


The first important  results  concerning both items  (3) and (4)  above were  obtained by physicists  based on the classical  method 
of  separation of variables  and   formal mode analysis.
  In the particular case where $\g_0$ is the Schwarzschild metric, the  linearized equations \eqref{Lin. Grav}   
can be formally decomposed into modes,  by associating t-derivatives with multiplication by $ i \om  $ and angular derivatives with multiplication by $ l$,  i.e. the eigenvalues  of the spherical laplacian.
A similar  decomposition, using oblate  spheroidal harmonics, can be done in Kerr.
 The formal study of fixed modes from the point of view of \textit{metric perturbations}  as in \eqref{Lin. Grav}   was initiated by  Regge-Wheeler \cite{Re-W} who discovered the master Regge-Wheeler equation for   odd-parity perturbations. This study was completed by Vishveshwara \cite{Vishev} and Zerilli  \cite{Ze}.  A gauge-invariant formulation of \textit{metric perturbations}  was then given by Moncrief  \cite{Moncr}.  An alternative approach via the Newman-Penrose  (NP) formalism      was  first  undertaken  by Bardeen-Press \cite{Bar-Press}. This latter type of analysis was later extended to the Kerr family by Teukolsky  \cite{Teuk} who made the  important  discovery that  the extreme  curvature components, relative   to  a principal null frame, satisfy  decoupled, separable,  wave equations.   These extreme curvature components    also turn out to be \textit{gauge invariant}  in the sense that small perturbations of the frame lead  to quadratic errors in  their  expression.     The  full  extent   of what could be done  by  mode analysis, in both approaches,    can be found in    Chandrasekhar's   book  \cite{Chand}. Chandrasekhar  also introduced (see \cite{Chand2}) a  transformation theory relating  the two approaches.  More precisely, he exhibits a transformation which connects  the Teukolsky  equations  to the       Regge-Wheeler  one.  This  transformation    was further  elucidated and extended  by R. Wald \cite{Wald} and recently by  Aksteiner and al \cite{Akst}.
   The full  mode stability, i.e. lack of exponentially growing modes,  for the Teukolsky equation on Kerr is  due  to  Whiting \cite{Whit} (see also \cite{Yacov} for a stronger quantitive version).

 
\subsection{Vectorfield Method} 
\lab{subsection:vectorfieldmethod}

Note that mode stability     is far from establishing  even boundedness of solutions to the linearized equations. To achieve that and, in addition, to   derive realistic decay estimates one needs  an entirely different approach based on  a far reaching  extension of the classical  vectorfield  method\footnote{Method based on the symmetries of Minkowski space   to derive uniform, robust,   decay  for nonlinear wave equations, see  \cite{Kl-vectorfield},  \cite{Kl-null}, \cite{Kl-vect2},  \cite{Ch-Kl0}.} used in the proof of  the nonlinear stability of Minkowski \cite{Ch-Kl}. The new vectorfield method compensates  for   the lack of  enough Killing and conformal Killing vectorfields   on a Schwarzschild  or  Kerr  background  by   introducing  new vectorfields  whose deformation tensors have    coercive properties    in different regions of spacetime, not necessarily causal.    The new   method   has emerged  in the last 15 years   in connection to the study of boundedness and decay  for the    scalar wave equation in $Kerr(a,m)$,
\bea
\square_{\g_{a,m}} \psi=0.
\eea 
 The starting and most  demanding   part of the new method is  the derivation of  a global,  simultaneous,  \textit{Energy--Spacetime Morawetz}   estimate  which degenerates 
  in the trapping region. This task is  somewhat  easier in Schwarzschild,  or    for axially symmetric solutions in Kerr,  where the trapping region  is  restricted to a    smooth hypersurface.  The first  such estimates, in Schwarzschild,    were proved
  by Blue and Soffer in \cite{B-S1}, \cite{B-S2} followed by  a long sequence of further  improvements in \cite{B-St}, \cite{DaRo1}, \cite{MaMeTaTo} etc. See also  \cite{I-Kl1} and \cite{St}    for    a vectorfield     method   treatment  of  the  axially symmetric  case in Kerr with applications to nonlinear equations.
      In the absence of axial symmetry   the derivation of    an    Energy-Morawetz estimate   in    $Kerr(a,m)$, $|a/m|\ll 1 $ requires  a more refined  analysis involving either      Fourier decompositions,   see   \cite{DaRo2}, \cite{TaTo}, or  a systematic use of the second order Carter operator, see \cite{A-Blue1}.    The   derivation of    such an estimate  in the full sub-extremal case $|a|<m$  is even more subtle  and   was recently achieved  by Dafermos, Rodnianski and Shlapentokh-Rothman \cite{mDiRySR2014}  by  combining mode decomposition with   the vectorfield method.
      
      Once an Energy-Morawetz estimate is established  one  can  commute  with the  time translation   vectorfield  and  the so called \textit{Red Shift} vectorfield, first introduced in  \cite{DaRo1},      to derive uniform bounds  for  solutions.  The most efficient  way  to also  get decay, and  solve the\textit{ superposition  problem},  is  due to  Dafermos and Rodnianski,  see \cite{Da-Ro3},    based on  the presence of  a family of \textit{$r^p$-weighted},  quasi-conformal  vectorfields  defined in the  far $r$ region of spacetime\footnote{These replace the scaling  and inverted time translation  vectorfields used in \cite{Kl-vectorfield}  or their  corresponding  deformations used in \cite{Ch-Kl}.  A recent improvement of the method, relevant to our work here, allowing one to derive higher order decay  can be found in \cite{AnArGa}. }.

       
\subsection{Linear Stability of the Schwarzschild space-time}    


A  first quantitative (i.e.  which provides precise decay estimates)   proof   of the  linear stability of Schwarzschild  spacetime has recently been established\footnote{ A  somewhat weaker version of  linear stability of Schwarzschild    was   subsequently   proved   in  \cite{HKW} by using the original, direct,   Regge-Wheeler, Zerilli  approach combined with the vectorfield method and    adapted gauge  choices.    } by Dafermos, Holzegel and Rodnianksi  in \cite{D-H-R}, via the  NP formalism (expressed in a double null foliation\footnote{This is possible in Schwarzschild where the principal null directions are integrable.}). It is important to note    that while the Teukolsky equation   (in the NP formalism)   is separable, and thus amenable to mode analysis, it  is not Lagrangian and thus  cannot be treated   by  direct    energy type estimates. To overcome this difficulty    \cite{D-H-R}  relies  on  a  new physical space version of   the Chandrasekhar transformation \cite{Chand2}, which takes solutions  of the Teukolsky equations to  solutions of  Regge-Wheeler, which  is manifestly both  Lagrangian and coercive.  After  quantitative decay   has been established for this latter equation, based on the new vectorfield method,   the  physical space form of the transformation allows    one to     derive   quantitative decay for solutions of the original Teukolsky equation.   Once  decay  estimates  for the  Teukolsky  equation have been established, the remaining work in \cite{D-H-R}   is to bound all other curvature  and Ricci coefficients associated to the double null foliation.  This   last step  requires carefully chosen gauge conditions along the event horizon  of  the fixed Schwarzschild background. This final gauge is itself then quantitatively bounded in terms of the initial data, giving thus  a comprehensive statement of linear  stability.

        
 \subsection{Nonlinear Stability  of  Schwarzschild  under  axially symmetric, polarized,  perturbations}
 
 
In the passage   from linear to nonlinear stability  of Schwarzschild   one has to overcome   major new difficulties. Some are  similar to those  encountered in the stability of Minkowski \cite{Ch-Kl} such as,   
      \begin{enumerate}
      \item Need of an appropriate geometric   setting  which takes into account  the  decay and peeling properties of  the curvature. In \cite{Ch-Kl} this was   achieved with the help of the   foliation  of the perturbed spacetime given by  two optical functions $^{(int)}u $  and $^{(ext)} u$  and a maximal time function $t$. The  exterior  optical function  $^{(ext)} u$, which   was initialized at infinity, was essential  to  derive the   decay and peeling properties  along null directions while   $^{(int)}u $, initialized  on a timelike axis,   was responsible   for covering the   interior, non-radiative, back scattering,  decay.
      
      \item  The peeling and decay  estimates  have to be derived by    some version of  the geometric    vectorfield  method  which relates decay    to   generalized energy type estimates.  
       
      \item  The peeling and decay estimates mentioned above  should be sufficiently strong  to be able  to deal with the error terms  generated by 
      the vectorfield method.  For this to happen, the error terms need to exhibit  an appropriate   null structure. 
      \end{enumerate}
  
     The new main difficulties are as follows:
   \begin{enumerate}
  \item  One needs a procedure which allows  to  take into account  the  change of mass and detect its final value.  Note also that we need to   restrict the nature of  the perturbations to  insure  that the final state  of  a perturbation of           Schwarzschild is still Schwarzschild.   
   
   \item  While in the stability of Minkowski space all  components of the curvature  tensor   where expected  to approach  zero, this is no longer true. Indeed, the middle  curvature component (relative to   an adapted null frame), ought to converge   to its  respective value in the final   Schwarzschild    spacetime. This statement is unfortunately hard to quantify since  that  value    depends both on the final mass and  on the corresponding    Schwarzschild coordinates. Moreover, some of  the other curvature components,  which are expected to converge to zero,    are also  ill defined since
      a small change of  the null  frame can produce   small linear distortion to the basic equation which these curvature components verify. Note that this   difficulty   was absent in the stability of the Minkowski space where small changes in the frame produce only  quadratic errors.
      
      \item The  classical   vector-field   method  used in the  nonlinear stability of Minkowski space  was based on the construction, together  with the spacetime, of an adequate  family of  approximate Killing and        conformal Killing vectorfields which mimic  the role played by  the corresponding  vectorfields in Minkowski space  in establishing uniform decay estimates. The Schwarzschild space however  has   a much more limited  set of   Killing vectorfields and no useful conformal Killing ones. As mentioned above, this problem appears already   in  the analysis of  the standard scalar linear wave equation in Schwarzschild. 
      
    \item  As in the stability of the Minkowski space, one needs  to make   gauge conditions  to insure that we are measuring decay relative to an appropriate  center  of mass frame.  Yet,  as we  saw above,  it is no longer true that  small perturbations of  the null   frame produce only quadratic errors for the curvature, as   was the case  in the stability of Minkowski space.  In fact, the center of mass frame  of the perturbed   black hole      continuously  changes in response  to  incoming radiation. This, the so called \textit{recoil problem},   does not  occur   in linear theory.
\end{enumerate}
    
Here is a  very short  summary of how we solve these new challenges in our work.

\begin{enumerate}
\item We   resolve the first  difficulty  by restricting  our  analysis  to axially symmetric, polarized perturbations and by tracking the mass  using  a quantity,  called  the quasi-local    Hawking  mass,       for which we  derive simple propagation equations which establish 
monotonicity  of the mass up to errors which are quadratic  with respect to the perturbations.

\item We resolve the second difficulty 
 by  making use of the  fact   that the extreme components  of the curvature  are, up to quadratic terms,  invariant  under   null  frame transformations.  As in \cite{D-H-R},   we  also  make use   of    a transformation,  similar to that of Chandrasekhar mentioned above,
 which maps    the extreme  components  of the curvature  to  a new   quantity  $\qf$, defined up to quadratic errors,   that verifies   a  Regge-Wheeler type  equation.   Once we     manage to   control  $\qf$, i.e. to derive quantitative decay estimates for it,  we can also control, in principle\footnote{  Provided that one can deal with the nonlinear terms.},   the two extreme  curvature invariants $\a$  and $\aa$,  the first by   inverting the Chandrasekhar   transformation  and the second  by using a variant of the Teukolsky- Starobinski identities.
   One is then left with the arduous task of recovering\footnote{In the linear setting this was  partially achieved in \cite{DHR2}.}  all other null  components of the curvature   tensor and all connection  coefficients.

 \item The third difficulties manifests itself  in the most  sensitive part of the entire  argument, i.e.  in the  task   of deriving   quantitative decay  estimates for  $\qf$   by  making use of the Regge-Wheeler type equation
  it verifies.  To do this we rely on  the new vectorfield method as outlined  in  subsection\ref{subsection:vectorfieldmethod} above. The main new difficulties   are:
   \begin{enumerate}
   \item[(i)] The vectorfield method introduces new  error terms, not present in  linear theory. To estimate these terms we need   precise decay informations, off   the final Schwarzschild space,  for  all   connection  coefficients and curvature  of the perturbation.
   \item[(ii)] The most  difficult  terms are those  due to the quadratic errors  made in  the derivation of the  Regge-Wheeler equation for $\qf$.    As in the proof  the stability of the Minkowski space  the precise  rates of decay for various   curvature and connection coefficients, i.e. the  peeling properties of the perturbation,   and the   the  precise structure  of these  error terms  is of  fundamental importance. 
   \end{enumerate}

\item We  solve the  fourth   and most important new difficulty
 by   a procedure  we   call  \textit{General Covariant Modulation} (GCM). This procedure, which takes advantage   of  the full covariance  of the Einstein equations, allows us  to construct  the perturbed spacetime  by a  continuity argument involving   finite   GCM admissible spacetimes  $\MM $  as   represented  in Figure \ref{fig1-introd}. 
\begin{figure}[h!]
\centering
\includegraphics[scale=0.5]{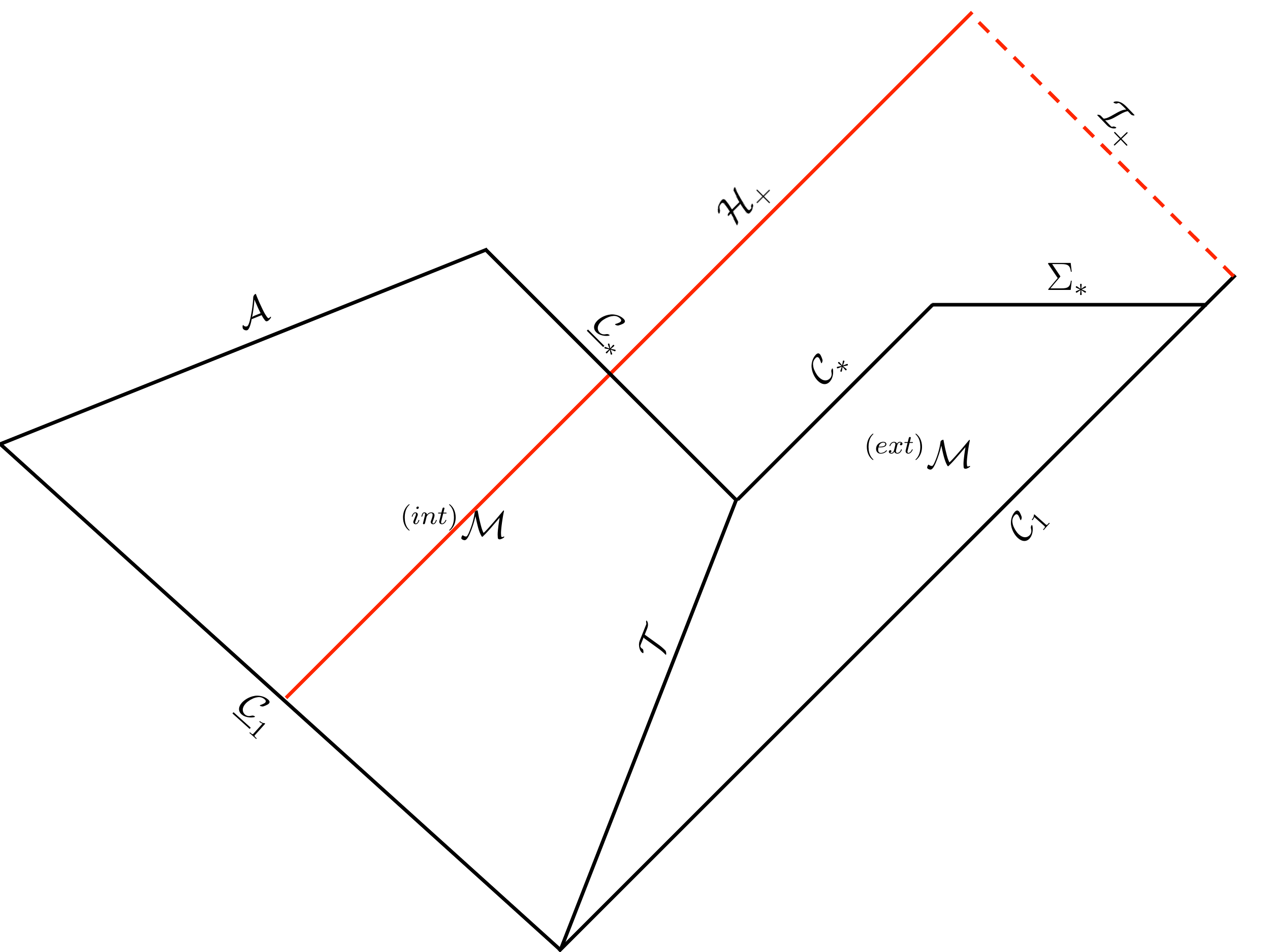}
\caption{The GCM admissible space-time $\mathcal{M}$}
\lab{fig1-introd}
\end{figure}
The past boundaries  $\CCb_1\cup\CC_1$  are incoming and outgoing  null hypersurfaces   on which the initial perturbation is prescribed.  The future boundaries   consists of the union $\AA\cup \CCb_* \cup \CC_* \cup \Si_*$ 
 where $\AA$  and  $ \Si_*$  are spacelike,  $   \CCb_*$ is incoming null, $ \CC_*$ outgoing null.   The boundary $\AA$ is chosen so  that,  in the limit when   $\MM$ converges to the final state,   is included in  the  perturbed    black hole.    The spacelike  boundary $\Si_*$  plays a fundamental role in our construction as seen below.   The spacetime $\MM$ also contains a timelike hypersurface  $\TT$  which divides  $\MM$ into  an exterior region we call $\Mext  $ and an  interior one $\Mint$.  We say that   $\MM$ is    a GCM admissible spacetime 
 if  it verifies the following properties.
 \begin{enumerate}
 \item[(i)]  The  far region  $\Mext  $ is foliated by a geodesic foliation induced by   an outgoing  optical function $u$  initialized on  $\Si_*$
  
\item[(ii)]  The  near region $\Mint $ is foliated by a geodesic foliation induced by   an incoming   optical function $\ub$ 
initialized at $\TT$ such that its level sets on $\TT$ coincide with  those of $u$.  

\item[(iii)] The foliation induced on $\Si_* $  is  such that    specific  geometric quantities  take Schwarzschildian values.  
We refer to these  as GCM conditions. These conditions are dynamically reset in the   continuation  process on which our proof is based.

\item[(iv)]  The area radius  $r(u)$ of  the  spheres of constant $u$ along $\Si_*$   is far greater than the corresponding value of $u$.  This condition allows us to  simplify  somewhat the null structure and Bianchi      equations  induced 
on $\Si_*$ and corresponds to the expectation that    the  spacelike hypersurfaces $\Si_*$ converges to the  null infinity of the  final state of the perturbation.

 \end{enumerate}

\item    The GCM conditions  together with the control  derived on $\qf$, $\a$  and $\aa$ mentioned earlier allows us to   control  all null connection  and  curvature coefficients along  on $\Si_* $, i.e. to derive  appropriated   decay estimates for them.    These  estimates    can then be  transported  to  $\Mext$  using the   the full scope of the   null structure and null Bianchi identities associated to the  outgoing geodesic  foliation. 

\item The decay estimates  in $\Mext $ can then be used as initial condition along  the timelike   hypersurface
 $\TT$  for the incoming foliation of $\Mint$.  These  allows us to also  derive  appropriate decay estimates 
  for   all null connection  and  curvature coefficients  of the foliation induced by $\ub$.
  
  \item  The precise decay estimates   derived in  5   are  sufficiently  strong  to allow us to control   all error terms  generated  in the process of estimating $\qf$,  as  mentioned in  3.  
 
\end{enumerate}
 Note that  in Figure   \ref{fig1-introd}, one   starts with initial conditions on  the union of null hypersurfaces $\CC_1\cup\CCb_1$
 rather than an initial  spacelike hypersurface $\Si_{(0)}$. One can justify this simplification based on   the results of  \cite{KlNi}, \cite{KlNi2}, see Remark \ref{rem:expaningwhywecanstratinthecontextofthecaracteristicCauchyproblem}. The full red line $\HH_+$ represents the  future event horizon of the perturbed  Schwarzschild. The line  $\TT$  represents   the timelike  hypersurface  separating   $\Mint $ from $\Mext$. In deriving  decay estimates  the precise choice  of $\TT$  is irrelevant. A choice, however,  needs  to be made in order to  avoid a  derivative loss for  our  top energy estimates\footnote{See \cite{Ch-Kl} for a similar situation.}.
 
  The spacetime is constructed by a continuity argument, i.e. we assume   that the spacetime    terminating at $\CC_*\cup\CCb_*$   saturates   a given bootstrap assumption  ({\bf BA}) and show,  by a long sequence of  a-priori estimates which take advantage   of the smallness of  the  initial perturbation,   that  $({\bf BA})$ can be improved
  and the spacetime extended past $\CC_*\cup\CCb_*\cup \Si_*$.

 Our work here is the first to prove the nonlinear stability of Schwarzschild  in   a  restricted    class of  nontrivial  perturbations, i.e. perturbations for which new  ideas, such  as  our GCM   procedure  are  needed. To a large extent, the restriction to this class of perturbations is only needed to ensure  that  the final state of evolution is   another 
Schwarzschild space. We are thus confident that our  procedure  may  apply in a more general setting.
  We would like to single  out two other recent important contributions to nonlinear stability of black holes.
In the context of  asymptotically flat  Einstein vacuum   equations  the result of Dafermos-Holzegel-Rodnianski \cite{DHR2} constructs a   class of Kerr   black hole solutions  starting from  future infinity while  Hintz-Vasy \cite{HVas}\footnote{See also \cite{Hintz} for the stability of Kerr-Newman de Sitter.}   prove   the nonlinear  stability of   Kerr-de Sitter, for small angular momentum, in the context of the Einstein vacuum equations with a nontrivial positive cosmological constant. Though the two results are very different they  share in common   the  fact that the perturbations  they treat  decay exponentially. This makes the analysis significantly easier  than in our case when the decay is  barely  enough to control the nonlinear terms.

 
\section{Organization} 


The paper  is organized as follow.   In Chapter \ref{chap:preliminaries}  we  introduce the main quantities,  equations and basic tools needed later.  It  is our main reference kit providing  all  main null structure and null  Bianchi equations,  in general  null frames, in the   context of  axially symmetric polarized   spacetimes.       Though   we work with the reduced equations, i.e the equations  reduced  by the symmetries, most of the  work in the paper  does not  really depend of the reduction.   Besides insuring that the final state  is   a Schwarzschild space  the reduction only plays a significant role in the GCM construction.

   Chapter \ref{chapter:maintheoremstatementbootstrapandinterresults}, the heart of the paper,   contains   the precise version of our main theorem, its main conclusions  as well as a full strategy of its proof,  divided in nine supporting intermediate results, Theorems {\bf M0}--{\bf M8}.  
We also give a short description of the proof of each theorem. 

 In the  other chapters of  this  paper we give complete proofs of      Theorems, {\bf M0}--{\bf M8}  and a full description of our GCM procedure.   
 
  The reader   versed in the formalism of null structure and Bianchi equations, as discussed in \cite{Ch-Kl}, is encouraged to  glance  fast over  Chapter \ref{chap:preliminaries}, to get familiarized with the notation,  and then move directly to Chapter \ref{chapter:maintheoremstatementbootstrapandinterresults}.


 \section{Acknowledgements}
 
 
 This work would be inconceivable without the  remarkable advances made in the last sixty years on black holes. The  works of  Regge-Weeler, Carter,  Teukolsky,   Chandrasekhar, Wald etc.,  made  during  the so called   \textit{golden age} of   black hole physics in the sixties and seventies,
 have greatly influenced our understanding of invariant quantities  and the wave equations they satisfy.  The advances made in the last fifteen years, quoted earlier,  which   have led to  the development of new mathematical  methods to    derive  the decay of waves   on  black holes spacetimes, are even more  immediately   relevant  to  our work.    In particular   we would like to single out    the  direct  influence  of  Dafermos-Holzegel-Rodnianski  \cite{D-H-R} in  the gestation of  our own ideas in this paper.  Finally, the  work on the nonlinear stability of the Minkowski space\footnote{See also \cite{KlNi}.} in \cite{Ch-Kl},   a milestone in the mathematical GR,   has significantly instructed our work here.  We would like  to  thank  E. Giorgi for   
   her careful   proofreading of   various sections of the manuscript.     Various discussions we had with  S. Aksteiner  were very useful.
       Finally  we thank our wives Anca and  Emilie  for their  incredible patience, understanding and support during our  many years of work  on  this project.

The first author   has been supported  by the  NSF grant  DMS 1362872.  He would like to thank the mathematics departments  of Paris 6, Cergy-Pontoise  and   IHES   for their  hospitality during  his many visits in the last  six years.  The second author is supported by the ERC grant  ERC-2016 
CoG 725589 EPGR.


\chapter{PRELIMINARIES}\lab{chap:preliminaries}



\section{Axially symmetric polarized spacetimes}



\subsection{Axial symmetry}\label{subs:axial-polarized}


   We consider  vacuum,    four dimensional,   simply connected,    axially symmetric  spacetimes $(\MM,\g, \Z) $      with $\g$ 
Lorentzian  and $\Z$ an  axial  Killing vectorfield   on $\MM$.
We denote by $\Axis$  the axis of   symmetry, i.e. the points   on $\MM$ for which 
$X:=\g(\Z, \Z)=0$.     In the case of interest for us    we   assume  $ dX\neq 0$  and that  $\Axis$ is a smooth   
 manifold of codimension 2.  The Ernst potential of the spacetime    is given by,
 \beaa
 \si_\mu&:=&\D_\mu(-\Z^\a\Z_\a)- i\in_{\mu\b\ga\de}\Z^\b \D^\ga\Z^\de.
 \eeaa
 The $1$-form   $\si_\mu dx^\mu$ is closed   and thus  there exists a function $\si:\MM \to\mathbb{C}$, called the $\Z$- Ernst potential,
  such that  $ \si_\mu =\D_\mu\si.$
   Note also  that    $\D_\mu\g(\Z, \Z)=2 \G_{\mu\la} \Z^\la = - \Re(\si_\mu) $   where $G_{\mu\nu}=D_{\mu}Z_{\nu}$.   Hence 
 we can choose  the potential $\si$  such that $ \Re\si=- X$.
 By a      standard calculation one can show that,
 \beaa
 \square \si=-X^{-1} \D_\mu \si \D^\mu \si.
 \eeaa

  \begin{definition}
 An axially symmetric   Lorentzian manifold $(\MM, \g, \Z)$       is said to be polarized  if   the Ernst potential  $\sigma$ is real, i.e.       $\si=-X$.  In that case the metric $\g$ can be written in the form,
\bea
\g= Xd\vphi^2+g_{ab} dx^a dx^b
\eea
where $X$ and $g$ are independent of $\vphi$.   We refer to  the orbit space $\MM/\Z$ as the reduced space   and the metric $g=g_{ab} dx^a dx^b $ as the reduced metric.   Note that the reduced space  $(\MM/\Z, g)$ is smooth away from the axis $\Axis$. Moreover the scalar $X$ verifies the wave equation,
\bea
\label{equation:waveforX}
\square_\g  X=X^{-1} \D_\mu X\D^\mu X.
\eea
 \end{definition}
 
    We denote  by $\R$,  resp. $R$  the curvature  tensor  of  the spacetime  metric  $\g$, respectively  $g$,  and  by $\square_\g$, resp     $\square_g$     the   d'Alembertian with respect  to $\g$  and resp.   the reduced metric  $g$.  We also denote by $\Gab$ the Christoffel symbols of $\g$ and by $\Ga$ the ones of $g$.
Note that the only non-vanishing Christoffel symbols are:
\bea
\label{christoffel}
  \Gab^\vphi_{\vphi b}=\frac 1 2 X^{-1}\pr_b X,\qquad
  \Gab^a_{\vphi\vphi}=-\frac 1 2 g^{as}\pr_s X,\qquad 
\Gab^a_{bc}=\Ga^a_{bc}.
\eea

One can      easily  prove the following.
\begin{proposition} The scalar  curvature $R$ of the reduced metric $g$  of   an axially symmetric polarized  Einstein vacuum spacetime   vanishes identically\footnote{This is an easy consequence of the  equation \eqref{equation:waveforX}.}. Moreover,  
setting    $\Phi:=\frac 1 2 \log X$
we find,
\bea
\label{eq: R-Phi}
R_{ab}=D_a D_b \Phi+D_a \Phi D_b \Phi,\qquad 
\square_g \Phi=-D^a\Phi D_a \Phi.
\eea
Also,
\bea
\R_{a\vphi b}\,^\vphi &=& -\frac 1 2X^{-1} D_a D_b X+\frac 1 4  X^{-2} D_a XD_b X=-R_{ab},  \nn\\
\R_{ac b}\,^\vphi &=& 0,\label{eq:Curv1}\\
\R_{abc}\,^d&=&R_{abc}\,^d, \nn
\eea
and,
\bea
\label{Ricci}
R_{abcd}&=& g_{ac}R_{bd}+g_{bd}R_{ac}-g_{ad} R_{bc}-g_{bc}R_{ad}.
\eea

Finally,  when applied to  $\Z$-invariant   functions,
 \bea
  \square_\g&=&\square_g +g^{ab} \pr_a \Phi \pr_b.              \label{red-square}
 \eea
\end{proposition}

\begin{remark}
The wave equation in  \eqref{eq: R-Phi} is equivalent to 
\bea
\square_\g \Phi&=&0   \label{eq: R- square-Phi}.
\eea
\end{remark}
 
\begin{remark}
 Schwarzschild spacetime is     axially symmetric polarized  with,
\beaa
X=r^2(\sin\th)^2,\qquad  \Phi=\log(r)+\log(\sin\th).
\eeaa
\end{remark}


\subsection{$\Z$-frames}


We consider  orthonormal   frames $e_0, e_1, e_\th=e_2,  e_\vphi=X^{-1/2}  \Z$,    with  $X:=\g(\Z, \Z)$,  which  are   $\Z$- equivariant , i.e.
$[\Z, e_\a]=0$. From now on, the index $\vphi$ is referring to the frame rather than the coordinates

\begin{lemma}
Setting  $ (\La_\a )_{ \b\ga  }:=       \g(D_ \a e_\ga ,e_\b)$ we have,
\bea
\label{axial-pol:frame}
(\La_\vphi)_{a\vphi}=-D_a \Phi,        \quad (\La_\vphi)_{ab}=(\La_a)_{b\vphi}=0,              \qquad   \forall a=0,1,2,
\eea
and,
 \bea
 \begin{split}
 \label{frame-red}
 \D_a e_b &=D_a e_b, \\
 \D_a e_\vphi &=0,\\
 \D_\vphi   e_a&=(\La_\vphi)_{\vphi a } e_\vphi =  (D_a \Phi ) e_\vphi, \\
 \D_\vphi e_\vphi&=(\La_\vphi)_{a\vphi} e_a =-D^a \Phi  e_a.
 \end{split}
 \eea
\end{lemma} 

\begin{proof}
Straightforward verification.  
\end{proof}

 \begin{lemma}
 We have,
 \beaa
  \D_s \R_{abcd}&=& D_s R_{abcd},\\
 \D_s \R_{\vphi bcd} &=& 0,\\
  \D_s \R_{\vphi b \vphi d}   &=& - D_s R_{bd},\\
   \D_\vphi \R_{abcd}  &=&0,\\
     \D_\vphi \R_{\vphi bcd}    &=&D^s \Phi R_{s bcd}+D_c \Phi  R_{bd}- D_d\Phi  R_{bc},\\
     \D_ \vphi  \R_{\vphi b\vphi d }&=& 0.
     \eeaa
     \end{lemma}
     
 \begin{proof}
 Straightforward verification.  
 \end{proof}
 
 \begin{definition} We say  that a spacetime  tensor  $\U$  is  $\Z$-invariant   if $\Lie_\Z U =0$
and $\Z$- invariant polarized  if its contractions to  an odd number   of   $e_\vphi=  X^{-1/2} \Z$  vanish identically.
\end{definition}

\begin{proposition}    All higher covariant derivatives of the Riemann  curvature tensor  $\R$ of an axially symmetric 
polarized spacetime $(\M, \g, \Z)$  are   $\Z$-invariant, polarized.
\end{proposition}

\begin{proof}
The statement   has been already verified  above for  both $\R $ and $\D\R$.  It suffices  to show that,  given an arbitrary $\Z$-invariant, polarized  tensor  $\U$, its covariant   derivative   $\D\U$ is also  $\Z$-invariant, polarized.
The invariance   is immediate.  To show polarization  we   consider  all   frame components  of $\D\U$ with 
 respect  to  our adapted equivariant frame    $e_1, e_2, e_3, e_\vphi $.    Assume   first   that    the  components  of $\D\U$   contain  only one $e_\vphi$. These are,
 \beaa
 \D_\vphi \U_a, \qquad \D_a \U_{b\vphi c}
 \eeaa
 with various combinations of  horizontal    indices  $a, b, c$. 
 Now, in view of the polarization property  of  $\U$ and the  
 relations $\D_a e_b= D_a e_b, \, \D_a  e_\vphi=0$ we  easily deduce,
 \beaa
  \D_a \U_{b\vphi c}=e_a  \U_{b\vphi c}-  \U_{\D_a b\vphi c}-\U_{ b \D_a \vphi c}-\U_{b\vphi  c}=0.
 \eeaa
 Similarly, since  $ e_\vphi( U_a)=  X^{-1/2}  \Z( U_a)=X^{-1/2} \Lie_\Z U_a=0$ and $\D_\vphi  e_a$ is proportional to $e_\vphi$, 
  \beaa
  \D_\vphi \U_a=e_\vphi  ( \U_a)-  \U_{\D_\vphi } e_a=0.
 \eeaa
 Similarly  we can check   that the contraction of  $\D\U$  with  any odd number of $e_\vphi$    must be   zero.
\end{proof}

In what follows we shall refer to $\Z$-invariant, polarized  tensors as  simply $\Z$-polarized.


\subsection{Axis of symmetry}


  We denote by $\Axis$ the axis of  symmetry   of $\Z$,  i.e. the  set  of  zeroes of $X=\g(\Z, \Z)$.  Since   we  assume  $dX\neq 0$, $\Axis $ is  a smooth  timelike   submanifold of dimension $2$. In view of the definition   of axial symmetry   every    trajectory   of $\Z$ is closed  and   intersects     $\Axis$ at one point.   The following regularity result at $\Axis$ holds true.
  \begin{lemma}
At the axis of symmetry $\Axis$ we have,
\bea
\label{regularityofPhi}
\frac{\g^{\mu\nu} \pr_\mu  X \pr_\nu  X}{4X} =e^{2\Phi} \g^{\mu\nu} \pr_\mu  \Phi \pr_\nu  \Phi    \longrightarrow 1.
\eea
\end{lemma} 
  
\begin{proof}
This is a classical result, see for example \cite{Ma-Seg}. We provide a proof for the convenience of the reader. We introduce a coordinates system $(x^0, x^1, x^2, x^3)$ centered at a point $q=(0,0,0,0)$ on the axis such that the Christoffel symbols of the metric vanish at $q$ and ${\pr_{x^0}}_{|_q}$ and ${\pr_{x^1}}_{|_q}$ are tangent to the axis at $q$. In particular, in this coordinates system, the matrix $\pr_\a\Z^\mu(q)$ is given by
\beaa
\pr_\a\Z^\mu(q) &=& \left(\ba{cc}
0 & 0\\
0 & A
\ea\right),
\eeaa
where $A$ is an antisymmetric matrix. Note that we used the fact that $\Z$ vanishes on the axis, that $q$ belongs to the axis, and that $\pr_\a\Z^\mu(q)$ is antisymmetric since $\Z$ is Killing. Now, if $x(\vphi)$ denotes an orbit of $\Z$ close to $q$, and $y=(x^2, x^3)$, we have in particular from Taylor formula
\beaa
\frac{dy}{d\vphi} =Ay+O(y^2).
\eeaa
Hence
\beaa
\exp(-\vphi A)y(\vphi) &=& y(0)+O(\vphi y^2)
\eeaa
and since $y(2\pi)=y(0)$ in view of the $2\pi$-periodicity of the orbits of $\Z$, we infer
\beaa
\exp(-2\pi A)y(0) &=& y(0)+O(y^2).
\eeaa
As $y(0)$ can be taken arbitrarily small, we infer that $\exp(2\pi A)$ is the $2\times 2$ identity matrix. Since $A$ is antisymmetric and non zero, its eigenvalues necessarily are $i$ and $-i$, and hence $A^TA=I$. This yields
\beaa
A^\a\,_\mu A_\ga\,^\nu A_{\a\nu} = A^\a\,_\mu (A^TA)_{\ga\a} = A_{\ga\mu} 
\eeaa
and hence
\beaa
\pr^\a\Z_\mu(q) \pr_\ga\Z^\nu(q)\pr_\a\Z_\nu(q) = \pr_\ga\Z_\mu(q).
\eeaa
Finally, since $\Z$ vanishes on the axis, and since the coordinates system we use in this lemma has vanishing Christoffel symbols at $q$, we have as $|x|$ goes to 0
\beaa
\frac{\g^{\mu\nu} \pr_\mu  X \pr_\nu  X}{4X} &=& \frac{\Z^\mu\D^\a\Z_\mu \Z^\nu\D_\a\Z_\nu}{\Z^\mu\Z_\mu} = \frac{\pr_\b\Z^\mu(q) x^\b\pr^\a\Z_\mu(q) \pr_\ga\Z^\nu(q) x^\ga\pr_\a\Z_\nu(q)}{\pr_\b\Z^\mu(q) x^\b\pr_\ga\Z_\mu(q) x^\ga}+O(x).
\eeaa
Together with the previous identity, we infer near any point $q$ on the axis
\beaa
\frac{\g^{\mu\nu} \pr_\mu  X \pr_\nu  X}{4X} \longrightarrow  1.
\eeaa
This concludes the proof of the lemma.
\end{proof}

 We note that  $\Z$-polarized, smooth,  vectorfields  are automatically tangent to $\Axis$.  This is the content of the following.
  \begin{lemma}
  \label{Lemma:tangentonaxis}
Any,  regular (i.e. smooth) $\Z$-polarized   vectorfield $\U$     is tangent to the axis $\Axis$.
\end{lemma}

\begin{proof}
Let $\U$ a polarized $\Z$-invariant regular vectorfield. Since it is $\Z$-invariant, we have
\beaa
0=[\Z, \U]=\Z^\a \D_\a \U- \U^\a\D_\a\Z.
\eeaa
Since $\Z=0$ on the axis and $\U$ is regular (hence bounded on the axis)  we infer that
$\U^\a\D_\a\Z = 0\textrm{ on }\Axis.$
In view of \eqref{frame-red},
\beaa
\U^\a\D_\a\Z=\U(e^\phi)e_\vphi,
\eeaa
and since $e_\vphi$ is unitary, we infer that
\beaa
 \U(X^{1/2})=\U(e^\phi)=0\textrm{ on }\Axis
\eeaa
and hence $\U(X)=0$         when $X=0$. 
\end{proof}

\begin{corollary}
Let  $u$ be a smooth regular   optical  function, i.e. $\g^{\a\b}\D_\a u\D_\b u=0$, which is $\Z$ -invariant, i.e. $\Z(u)=0$.
Then  its associated null geodesic generator $L=-\g^{\a\b}\pr_\a u \pr_\b $ is $\Z$-invariant, polarized, tangent to the axis of symmetry $\Axis$. 
\end{corollary}

\begin{proof}
It is easy to check that $L$ is $\Z$-invariant, polarized.  It must therefore be tangent to $\Axis$ in view of Lemma \ref{Lemma:tangentonaxis}.
\end{proof}


\subsection{$\Z$-polarized $S$-  surfaces}


 Throughout our work   we   shall  deal     various   $\Z$-  polarized,  $S$- foliations   i.e. foliations   given by compact  $2$- surfaces  $S$ with induced metrics of the form,
  \bea
 \label{eq:Smetric}
  \gS=  \ga d\th^2 +  X d\vphi^2, \qquad  \ga=\ga(\th)>0, \qquad \th\in[0, \pi].
 \eea
 Here $\ga$ and $X$ are independent of $\vphi$ and  $\Phi=\frac 1 2 \log X$ vanishes   on the  poles $\th=0$ and $\th=\pi$.
 The regularity condition \eqref{regularityofPhi}  takes the form,
 \bea
 \label{regularityofPhiS}
\lim_{\sin\th \to 0} \Big(e_\th(e^\Phi)\Big)^2 = 1
 \eea
 where $e_\th$ is the unit vector,
 \beaa
  e_\th :=\ga^{-1/2} \pr_\th.
  \eeaa

  We denote  the  induced covariant derivative $\nabb$ and   define    the  volume radius  of $S$   by the formula
\beaa
|S|=4\pi r^2
\eeaa
where $|S|$ is the volume of the surface using the volume form of the   metric $\gS$. Note also that the area element on $S$ is given by
\beaa
\sqrt{\ga}e^\Phi d\th d\vphi.
\eeaa

 In this subsection we   record    some basic  general   formulas  concerning these surfaces.
 We  consider adapted   orthonormal frames   
 \beaa
 e_\th, e_\vphi= X^{-1/2} \Z= X^{-1/2} \pr_\vphi.
 \eeaa
    Note that in view of  \eqref{frame-red} we  have,   
\bea
\label{frame-red:inducedS}
\nabb_\vphi   e_\vphi=- (e_\th \Phi) e_\th, \qquad  \nabb_\vphi   e_\th=(e_\th \Phi ) e_\vphi, \qquad \nabb_\th   e_\th=\nabb_\th   e_\vphi=0.
\eea
In what follows, we consider    $\Z$-invariant   polarized tensors tangent to $S$ or simply polarized $k$-tensors on $S$.

In view of  Lemma \ref{Lemma:tangentonaxis}, a   regular $\Z$-polarized  tensor on $S$ must  vanish on the axis
 of symmetry i.e. at  $\th=0$ and $\th=\pi$.  More precisely we have,
 \begin{lemma} 
 \label{Lemma:axisofsymmetryS}
 The following facts  hold true  for $\Z$-polarized tensors on $S$.
 \begin{enumerate}
 \item  If $U$ is a $1$-form then, on the axis of symmetry\footnote{Note that the component $U_\vphi$ must automatically vanish on $S$.} of $\Z$,  (i.e.  for $\th=0$ and $\th=\pi$), 
 \beaa
 U_\th:=U(e_\th)=0
 \eeaa
 \item  For a  covariant $2$-tensor, then, on the axis of symmetry\footnote{Note that the components $U_{\th\vphi}, U_{\vphi\th}$  must automatically vanish on $S$.} of $\Z$,  (i.e.  for $\th=0$ and $\th=\pi$), 
 \beaa
 U_{\th\th}=U_{\vphi\vphi}=0.
 \eeaa
 \end{enumerate}
 Similar  statements can be deduced for higher order  tensors.
 \end{lemma} 
 
 \begin{proof}
 Immediate consequence of Lemma \ref{Lemma:tangentonaxis}.
 \end{proof}

\begin{lemma}
\label{lemma:Gauss-curvatureS1}
The       Gauss curvature $K$  of the metric  \eqref{eq:Smetric}   can be expressed in terms of  the          polar function   $\Phi:=\frac 1 2 \log X$   by the formula,
\bea
\label{eq:lap-Phi=K}
\lapp \Phi=-K.
\eea
\end{lemma} 

\begin{proof}
Direct  calculation using the form of the $\gS$ metric in \eqref{eq:Smetric}. 
\end{proof}


\subsubsection{Basic operators on $S$}  


We recall  (see \cite{Ch-Kl} chapter 2)  the following operations which  preserve  the space of   fully symmetric traceless  tensors:

\begin{definition} We denote by $\SS_k$ the set of  $k$-covariant  polarized tensors which are fully symmetric  and traceless, i.e. which verify,
\beaa
f_{A_1\ldots A_k}=f_{(A_1\ldots A_k)}, \qquad  \gS^{A_1 A_2} f_{A_1 A_2\ldots A_k}=0.
\eeaa
We  define   the following  operators on  $\SS_k$-tensors.
\begin{enumerate}
\item   The   operator $\DDd_k$ which  takes $\SS_k$ into $\SS_{k-1}$ is the divergence operator,
\beaa
(\DDd_k f)_{A_2,\ldots A_k}:&=&(\divv f)_{A_2,\ldots A_k}:= \gS^{AB} \nabb_B f_{A A_2,\ldots A_k}.
\eeaa

\item  The   operator  $  \DDs_k$ which       takes $\SS_{k-1}$  into $\SS_{k}$ is  the fully  symmetrized, traceless, covariant derivative operator\footnote{ For an arbitrary $k$-tensor,  $f_{(A_1\ldots A_k)}= \sum_{\si\in\Pi}  f_{A_{\si(1)} \ldots A_\si(k) } $. In the particular  case       when $k=1$  we get  $(\DDs_1 f)_A =  -\nabb_A f $ and when $k=2 $ we get $\DDs_2 f_{AB}= -\frac 1 2 ( \nabb_A f_B +\nabb _B f_A -  \gS_{AB} \divv f) $.},
\beaa
( \DDs_k f) _{A_1\ldots A_k}:&=&\begin{cases} -\nabb_{A_1} f,\qquad \qquad \qquad  \qquad \qquad \qquad \qquad \qquad \qquad \,k=1,\\
-\frac{1}{k} \nabb_{(A_1 }f_{A_2\ldots A_k)}    +\frac{1}{ k(k-1)}  \gS_{(A_1 A_2}( \divv  f  )_{A_3\ldots A_k)}, \qquad \quad  k\ge 2. 
\end{cases}
\eeaa

\item The operator $\lapp_k$ takes $\SS_k $ to $\SS_k$,
\beaa
(\lapp_k f)_{A_1\ldots A_k}:=\gS^{BC} \nabb_B \nabb_C f_{A_1\ldots A_k}.
\eeaa
\end{enumerate}
\end{definition}

\begin{remark}
Note that  if $f\in \SS_k$  then $\curll   f:= \in^{BC} \nabb_B f_{CA_1\ldots A_k}=0$. 
\end{remark} 

\begin{lemma}
\label{lemma:Nabbf-DDs}
Given $f\in \SS_k$, $k\ge 1 $, we have the identity,
\bea
\nabb_B f_{A_1\ldots A_k}= -( \DDs_{k+1}  f)_{BA_1\ldots A_k} +\frac{1}{k} \gS_{(B A_1 } ( \DDd_{k} f ) _{A_2\ldots A_k)}. 
\eea
In other words the covariant derivatives of  any tensor in $\SS_k$  can be expressed as a linear combination
 of  $\DDs_{k+1}f$ and  $\gS \otimes \DDd_{k-1} f$.
\end{lemma}

\begin{proof} The proof follows easily from definitions and the vanishing of the $\curll$.
For example, if  $k=2$, 
\beaa
 3 \nabb_B f_{A_1A_2} &=& \left( \nabb_B f_{A_1A_2}+ \nabb_{A_1}  f_{A_2 B} +\nabb_{A_2} f_{B A_1} \right)\\
 &+&\left( \nabb_B f_{A_1A_2}-\nabb_{A_1} f_{BA_2}\right) + \left( \nabb_B f_{A_1A_2}-\nabb_{A_2} f_{A_1 B}\right) \\
 &=&-3\left[(\DDs_3 f)_{BA_1 A_2} - \gS_{A_1 A_2}  (\DDd_2 f)_B- \gS_{ A_2 B}  (\DDd_2 f)_{A_1}-\gS_{A_1 A_2} (\DDd_2 f)_B\right]\\
 &=& -3\left[ ( \DDs_{3}  f)_{BA_1 A_2} -\frac 1 2 \gS_{(B A_1 } ( \DDd_{3} f ) _{A_2)} \right].
\eeaa
\end{proof}

We  can easily check that $  \DDs_k $ is the formal adjoint of $\DDd_k$,  i.e.,
\beaa
\int_S ( \DDd_k f) g\, =\int_S  f ( \DDs_k \, g ).
\eeaa
It is also  easy to check that the kernels of $\DDd_k$ are trivial for all $k\ge 1$ (see  also Chapter 2 in \cite{Ch-Kl}).
The kernel of $\DDs_1:\SS_0\longrightarrow \SS_1 $  consists   of  constants on $S$ while the kernel of  $\DDs_2$
 consists of constant multiple of   co-vectors  $f$ with $f_\th= C e^\Phi$. 
Moreover,
\bea
\begin{split}
\label{eq:dcalident}
\DDs_1\c\DDd_1&=-\lapp_1+K,\qquad \DDd_1 \c\DDs_1=-\lapp_0,\\
\DDs_2 \c\DDd_2 &=-\f12\lapp_2+K,\qquad \DDd_2\c\DDs_2=-\f12(\lapp_1+K).
\end{split}
\eea
Similar identities also hold for higher $k$. Using \eqref{eq:dcalident} one can  also prove the following (see  also Chapter 2 in \cite{Ch-Kl}).
\begin{proposition} 
\label{prop:2D-hodge}
Let $(S,\gS)$ be
a compact manifold with Gauss curvature $K$. We have,

{\bf i.)}\quad The following identity holds for vectorfields  $  f  \in \SS_1$,
\beaa
\int_S\big(|\nabb   f |^2+K|   f |^2\big)=\int_S|\DDd_1   f   |^2.
\eeaa

{\bf ii.)}\quad The following identity holds for symmetric, traceless tensors in $\SS_2$,
\beaa
\int_S\big(|\nabb    f  |^2+2K| f  |^2\big)=2\int_S |\DDd_2   f  |^2.
\eeaa

{\bf iii.)}\quad The following identity holds for  scalars $f\in \SS_0$,
\beaa
  \int_S|\nabb  f |^2=\int_S|\DDs_1\, f|^2.
\eeaa

{\bf iv.)}\quad  The following identity holds for vectors $   f  \in \SS_1 $,
\beaa
\int_S \big(|\nabb  f   |^2-K|  f  |^2\big)=2\int_S|\DDs_2   f   |^2.
\eeaa
\label{prop:hodgeident}
\end{proposition}

\begin{proof}
All statements  appear in \cite{Ch-Kl}. 
\end{proof}

\begin{proposition}
we have for $f\in \SS_0$,
\beaa
\int_S\left(|\nabb^2 f|^2+ K |\nabb f|^2 \right)     = \int_S|\lapp_0  f|^2.
\eeaa
 Moreover, under mild  assumptions on the curvature such as
  \beaa
 K= \frac{1}{r^2}+ O\left(\frac{\ep}{r^2}\right), \qquad  r e_\th ( K )  =O\left(\frac{\ep}{r^2}\right),
 \eeaa
 for any $f\in\SS_k$, $k\ge 1 $,
\beaa
\int_S\left(|\nabb^2 f|^2+r^{-2}  |\nabb f|^2 \right)           \les \int_S|\lapp_k  f|^2 + O(\ep) r^{-4}\int_S |f|^2. 
\eeaa
\end{proposition}

\begin{proof}
Follows from the standard Bochner identity on $S$.
\end{proof}


\subsubsection{Reduced Picture}


  \begin{lemma}
  \label{lemma:Reduceddds-ddd}
  The following relations hold true between  the spacetime picture and the reduced one.
  \begin{enumerate}
  \item 
  Let $ \fS\in \SS_k$  sucht that $\fS_{\th\ldots\th}= f$. Then,
  \bea
  \label{equation:DDsktoddsk}
  \begin{split}
 ( \DDd_k \fS)_{\th\ldots\th}&=  e_\th (f) + k  e_\th(\Phi)  f. \\
 \end{split}
  \eea
  
  \item If   $f\in \SS_0$ we have,
  \beaa
  \dds_1 f=-e_\th(f).
  \eeaa
  
  \item   If $ \fS\in \SS_{k-1} $, $k\ge 2$,   such that $\fS_{\th\ldots\th}= f$ we have,
  \bea
 2  (\DDs_k \,\fS)_{\th\ldots\th}&=-e_\th(f) +(k-1)  e_\th(\Phi) f .
  \eea
  
  \item 
    Let $ \fS\in \SS_k$  sucht that $\fS_{\th\ldots\th}= f$. Then,
  \beaa
  \lapp_k \fS_{\th_1\ldots\th_k} &=&  e_\th( e_\th f) + e_\th(\Phi)  e_\th f - k^2 \big (e_\th(\Phi)\big)^2 f.
  \eeaa
     \end{enumerate}
  \end{lemma}
  
  \begin{proof}
  The proof follows easily from  the definitions  of $\DDd_k, \DDs_k,\, \lapp_k$  and  the  formulae \eqref{frame-red:inducedS}. We check below the formula \eqref{equation:DDsktoddsk}.
  \beaa
 -  ( \DDs_k \fS)_{\th\ldots\th}&=&e_\th f - \frac 12  ( \DDd_{k-1} f)_{\th\ldots \th} = e_\th(f)-\frac 1 2 ( e_\th  f+(k-1) e_\th(\Phi) f )\\
 &=&\frac 1 2 \left(e_\th f-(k-1) e_\th(\Phi) f\right)
  \eeaa
  as desired.
  \end{proof}
  
  \begin{definition}
  \label{definition:reducedderivatives}
  We say that a scalar $f$  is a  reduced $k$-scalar  on $S$   if    there is a  $\Z$-invariant, polarized, $k$-covector  $\fS\in\SS_k$ such that,
  \beaa
  f= \fS_{\th\ldots\th}.
  \eeaa
     We   denote by $\mathfrak{s}_k$       the set of $k$ reduced scalars. 
     
  \begin{itemize} 
  \item Given a  $k$ reduced scalar $f$, reduced from $\fS$ we define,
  \beaa
  |\nabb f|^2 = |\nabb \fS|^2, \qquad |\nabb^l f|^2 =|\nabb^l \fS|^2.  
  \eeaa
   
  \item Given a $k$-reduced   scalar $f$ on $S$ we   define,
  \beaa
   \ddd_k f&:=   e_\th (f) + k  e_\th(\Phi)  f.
  \eeaa

 \item  Given a $(k-1)$-reduced  scalar $f\in S_{k-1}$ we   define,
  \beaa
   \dds_k f&:=- e_\th(f) +(k-1)  e_\th(\Phi) f.
  \eeaa

\item   Given a $k$-reduced   scalar $f\in \mathfrak{s}_k $ we   define,
   \beaa
    \lapp_k f&:= e_\th( e_\th f) + e_\th(\Phi)  e_\th f - k^2 \big (e_\th(\Phi)\big)^2 f. 
   \eeaa
   \end{itemize}
   In view of Lemma \ref{lemma:Reduceddds-ddd} we  have,
   \beaa
   \ddd_k f&=& (\DDd_k \fS)_{\th\ldots\th}
   \eeaa
   and, 
   \beaa
   \dds_k f =\begin{cases}  (\DDs_k \fS)_{\th\ldots\th},\,  \qquad\, k=1,\\
   2 (\DDs_k \fS)_{\th\ldots\th}, \qquad k\ge 2.
   \end{cases}
   \eeaa
   
  Clearly $\ddd_k $  takes $k$-reduced scalars into $(k-1)$-reduced scalars,  $\dds_k $ takes $(k-1)$-reduced ones into $k$-reduced and $\lapp_k$ takes $k$-reduces scalars into $k$-reduced scalars.    
  \end{definition}
  
       \begin{remark}\lab{rem:vanishingontheaxisofreducesscalars}
     Note that, in view of Lemma \ref{Lemma:axisofsymmetryS}, any  reduced scalar in  $\mathfrak{s}_k$, 
   for $k\ge 1 $,  must vanish on the axis of symmetry of $\Z$, i.e. at the two poles.
   \end{remark}
  
   \begin{remark}
   The operator  $\ddd_k$ and $\dds_k$    can only be applied to $k$-reduced, resp   $(k-1)$-reduced  scalars.  Thus  whenever we write a sequence  of operators   involving $\ddd_k$,  $\dds_k$ we understand from the context  to which
    type of $k$-reduced scalars they are applied, see for example  the proposition below. The same remark applies to $\lapp_k$.
   \end{remark}
   
 \begin{remark}
\label{Remark:$ethf$}
 Note that for  given reduced scalar $f\in \mathfrak{s}_k$    and  $h\in \mathfrak{s}_1$     we can write,
\beaa
 h e_\th(f) &=& \frac 1 2 h  \left(\ddd_k f-\dds_{k+1} f\right).
\eeaa
The term $ h  \ddd_k  f  $  is the reduced form of   a tensor product of $\,^{(1+3)}  h$ with $ \DDd_k \,^{(1+3)}  f $ while 
$h  \dds_{k+1}  f $ is the reduced form of  a  contraction between $\,^{(1+3)}  h$ and $\DDs_{k+1} \,^{(1+3)} \psi$
This can be formalized precisely  using Lemma \ref{lemma:Nabbf-DDs}.  The Remark   will be useful  in what follows, for example  in Lemma \ref{lemma:commTwithe3e4}.
\end{remark} 
   
  \begin{remark}\lab{rem:simpleintegrationon2surfaceswithareaelement}
     The duality between  the operators $\ddd_k$ and $\dds_k$    follows in view of  the duality of $\DDd_k$ and $\DDs_k$. It             can also be  interpreted directly   in terms of the    area element   $ \sqrt{\ga}  e^\Phi d\th d\vphi$,
   \beaa
    \int_S (\ddd_k  f g  - g \dds_k g) da_S   &=&\int_S  e_\th(fg) + e_\th(\Phi)  fg  =
      \int_0^\pi \int_0^{2\pi}  \left( e_\th(fg) + e_\th(\Phi)  fg \right)\sqrt{\ga}  e^\Phi d\th d\vphi\\
      &=& \int_0^\pi \int_0^{2\pi} \pr_\th ( e^\Phi fg)  d\th d\vphi=0.
   \eeaa   
  \end{remark}

\begin{proposition}
\label{prop:DDd--ddd}
 The following identities hold true,
 \bea
 \bsplit
 \dds_{k} \ddd_k &=-\lapp_k+ k K,\\
  \ddd_k \dds_{k} &=-\lapp_{k-1} - ({k-1} )K.
 \end{split}
 \eea
 In particular for $k=1, 2$
 \beaa
\dds_1\ddd_1&=&-\lapp_1+K,\quad \ddd_1 \dds_1=-\lapp_0,\quad \dds_2\ddd_2 = -\lapp_2 +2K,\quad  \ddd_2\dds_2=-\lapp_1 -K.
\eeaa

Moreover, note the following commutation formulas
\beaa
\ddd_k\dds_k - \dds_{k-1}\ddd_{k-1} &=& -2(k-1)K,\\
-\ddd_k\lapp_k +\lapp_{k-1}\ddd_k &=& K\ddd_k -ke_\th(K),\\
-\dds_k\lapp_{k-1} +\lapp_k\dds_k &=& (2k-1)K\dds_k +(k-1)e_\th(K).
\eeaa
\end{proposition}

\begin{proof}
We have, for a $k$ reduced scalar $f$,
\beaa
-\dds_{k}\ddd_k f&=&( e_\th - (k-1) e_\th(\Phi) )( e_\th(f)+k e_\th(\Phi)  f)\\
&=& e_\th( e_\th(f) )+ k  e_\th(\Phi)  e_\th f+  k ( e_\th  e_\th \Phi) f - (k-1) e_\th(\Phi) )( e_\th(f)+k e_\th(\Phi)  f)\\
&=& e_\th( e_\th(f) )+   e_\th(\Phi)  e_\th f +  k ( e_\th  e_\th \Phi) f- k(k-1)\big( e_\th(\Phi)\big)^2.
\eeaa
In view of Lemma \ref{lemma:Gauss-curvatureS1} we have, since $\Phi$ is a scalar 
\beaa
-K=\lapp \Phi&=&   e_\th  e_\th (\Phi)+\big( e_\th(\Phi)\big)^2.
\eeaa
Therefore,
\beaa
-\dds_{k}\ddd_{k} f&=&  e_\th( e_\th(f) )+  e_\th(\Phi)  e_\th f+ k\left(-K- \big( e_\th(\Phi)\big)^2 \right) f- k(k-1)\big( e_\th(\Phi)\big)^2\\
&=&  e_\th( e_\th(f) )+  e_\th(\Phi)  e_\th f -k K f-  k^2  \big( e_\th(\Phi)\big)^2\\
&=&\lapp_k f   -k K f.
\eeaa
Similarly, for a $(k-1)$-reduced $f$,
\beaa
-\ddd_k\dds_{k} f&=&  ( e_\th +k  e_\th(\Phi) )( e_\th(f)-(k-1)  e_\th(\Phi)  f)\\
&=& e_\th( e_\th(f) )+ k  e_\th(\Phi)  e_\th f-  (k-1)  ( e_\th  e_\th \Phi) f - (k-1) e_\th(\Phi)  e_\th(f)\\
&-& k (k-1)  \big( e_\th(\Phi)\big)^2 f\\
&=& e_\th( e_\th(f) )+   e_\th(\Phi)  e_\th f- (k-1) \left(  -K-    \big( e_\th(\Phi)\big)^2    \right) - k (k-1)  \big( e_\th(\Phi)\big)^2 f\\
&=& e_\th( e_\th(f) )+   e_\th(\Phi)  e_\th f+(k-1) K -(k-1)^2 \big( e_\th(\Phi)\big)^2 f\\
&=&\lapp_{k-1} f  +(k-1) K.
\eeaa

Next, we check the commutation formulas. We have
\beaa
\ddd_k\dds_k - \dds_{k-1}\ddd_{k-1} &=& -\lapp_{k-1} -(k-1)K-\Big(-\lapp_{k-1}+(k-1)K\Big)\\
&=&  -2(k-1)K
\eeaa
from which we infer
\beaa
\ddd_k(-\lapp_k) &=& \ddd_k(\dds_k\ddd_k-kK)\\
&=& \ddd_k\dds_k\ddd_k -kK\ddd_k -ke_\th(K)\\
&=&  \Big(\dds_{k-1}\ddd_{k-1} -2(k-1)K\Big)\ddd_k  -kK\ddd_k -ke_\th(K)\\
&=&  \Big(-\lapp_{k-1}+(k-1)K \Big)\ddd_k  -(k-2)K\ddd_k -ke_\th(K)
\eeaa
and hence
\beaa
-\ddd_k\lapp_k +\lapp_{k-1}\ddd_k &=& K\ddd_k -ke_\th(K).
\eeaa
Also, we have
\beaa
\dds_k(-\lapp_{k-1}) &=& \dds_k(\ddd_k\dds_k+(k-1)K)\\
&=& \dds_k\ddd_k\dds_k +(k-1)K\dds_k +(k-1)e_\th(K)\\
&=& \Big(-\lapp_k+kK\Big)\dds_k+(k-1)K\dds_k +(k-1)e_\th(K)
\eeaa
and hence
\beaa
-\dds_k\lapp_{k-1} +\lapp_k\dds_k &=& (2k-1)K\dds_k +(k-1)e_\th(K)
\eeaa
as desired.
\end{proof}


\subsubsection{A remarkable identity}


First, note the following observation which follows immediately from the form of $\dds_2$. 
\begin{lemma}
The kernel of $\dds_2$ is spanned by $e^\Phi$. 
\end{lemma}

The above lemma, in connection with a Poincar\'e inequality for $\dds_2$, see \eqref{eq:poincaredds2takingintoaccountitskernelell=1mode}, will result in the need of a specific treatment for the projection of some of the quantities on the kernel of $\dds_2$. This motivates the following definition.

\begin{definition}[The $\ell=1$ mode]
For a $1$-reduced scalar $f$, the $\ell=1$ mode denotes its projection on the kernel of $\dds_2$, i.e. 
\beaa
\int_Sf e^\Phi.
\eeaa

For a $0$-reduced scalar $f$, the $\ell=1$ mode denotes the projection of $e_\th(f)$ on the kernel of $\dds_2$, i.e. 
\beaa
\int_Se_\th(f) e^\Phi.
\eeaa
\end{definition}

\begin{remark}
The above definition is motivated by the fact that, in Schwarzschild, this corresponds to the projection on the $\ell=1$ spherical harmonic\footnote{In general, there are 3 spherical harmonics corresponding to $\ell=1$, but only one is axially symmetric. This is why we have only one projection instead of 3 in our case.}.
\end{remark}

We are now ready to state the following remarkable identity which  will play a crucial role later in the paper. 
\begin{lemma}[Vanishing of the $\ell=1$ mode of the Gauss curvature]\label{lemma:Gauss-curvatureS}
The $\ell=1$ mode of $K$ vanishes identically, i.e.  
\bea
\label{eq:badmode-K}
\int_S e_\th(K)e^\Phi=0.
\eea
\end{lemma}

\begin{proof}
       To prove \eqref{eq:badmode-K} we write,
                   \beaa
-\int_Se_\th(K)e^\Phi &=& \int_S \dds_1  (K)  e^\Phi=\int_S K \ddd_1 ( e^\Phi)= 2 \int_S  K e_\th(\Phi) e^\Phi.
      \eeaa 
       Thus, in view of    \eqref{eq:lap-Phi=K}, using in addition   $\lapp\Phi =e_\th(e_\th(\Phi))+e_\th(\Phi)^2$  
       \beaa
     -  \int_Se_\th(K)e^\Phi &=&  2 \int_S  \lapp \Phi  e_\th(\Phi) e^\Phi= 2\int_S \big(  e_\th(e_\th(\Phi))+e_\th(\Phi)^2\big)  e_\th(\Phi)e^\Phi\\
       &=&       \int_S   \ddd_2\big( (e_\th\Phi)^2 \big) e^\Phi=                -\int_S  (e_\th\Phi)^2 \dds_2(e^\Phi) =0
       \eeaa
       as desired\footnote{Note that the boundary term  which appears from the last integration by parts  has the form
       $(\pr_\th \Phi)^2 e^{2\Phi} (\pi)-   (\pr_\th \Phi)^2 e^{2\Phi} (0) $  and hence vanishes in view of  the regularity condition  \eqref{regularityofPhiS},  see also the computation in Remark \ref{rem:simpleintegrationon2surfaceswithareaelement}.}.
           \end{proof}


\subsubsection{Poincar\'e inequalities on 2-spheres}


Proposition \ref{prop:2D-hodge} takes the following reduced form,
\begin{proposition} 
\label{prop:2D-hodge-reduced}
 The following identities  hold true for                reduced    $k$-scalars   $f\in\mathfrak{s}_k$.     

{\bf i.)}\quad  If   $  f  \in \mathfrak{s}_1$,
\be{eq:hodgeident1}
\int_S\big(|\nabb   f |^2+K   f ^2\big)=\int_S|\ddd_1   f   |^2.
\end{equation} 

{\bf ii.)}\quad If $ f\in \mathfrak{s}_2$,
\be{eq:hodgeident2}
\int_S\big(|\nabb    f  |^2+4K  f ^2\big)=2\int_S |\ddd_2   f  |^2.
\end{equation} 

{\bf iii.)}\quad  If  $f\in \mathfrak{s}_0$,
\be{eq:hodgeident3}
  \int_S|\nabb  f |^2=\int_S|\dds_1\, f|^2.
\end{equation} 

{\bf iv.)}\quad   If  $   f  \in \mathfrak{s}_1 $,
\be{eq:hodgeident3*}
\int_S \big(|\nabb  f   |^2-K  f  ^2\big)= \int_S|\dds_2   f   |^2.
\end{equation}

{\bf v.)}\quad  If $f\in  \mathfrak{s}_0$,
\bea
\int_S |\nabb^2 f|^2 +\int_S K|\nabb f|^2 =\int_S|\lapp_0 f|^2.
\eea
Under mild assumptions on the Gauss  curvature  $K$, such as 
 \beaa
 K= \frac{1}{r^2}+ O\left(\frac{\ep}{r^2}\right), \qquad  r e_\th ( K )  =O\left(\frac{\ep}{r^2}\right). 
 \eeaa
We also have  for $f\in \mathfrak{s}_k$, $k \ge 1 $,
 \bea
 \label{eq:hodgeident8}
        \|\nabb^2 f\|_{L^2(S) }^2+ r^{-2}\|\nabb f\|^2_{L^2(S)}  &\les &  \|  \lapp_k   f  \| _{L^2(S)}^2  + \ep  r^{-4}\| f\|_{L^2(S)}^2. 
 \eea
\label{prop:hodgeident:secondtimethislabelappears}
\end{proposition}

\begin{proof}
The proof of the above statements can be either derived from their space-time version or  checked directly.   
\end{proof}

\begin{lemma}
\label{lemma:nabbfS}
The   following relations hold between $\Z$-polarized   $S$-tensors and reduced scalars\footnote{ Note that the  expressions on the left of the inequalities below  should be interpreted as applying  to the  spacetime tensor from which 
 $f$ is reduced.}.
\begin{itemize}

\item  If  $f\in \mathfrak{s}_0$
\beaa
|\nabb   f |^2 &=&|e_\th f|^2, \\
|\nabb^2 f |^2&=& |e_\th(e_\th f)|^2+|e_\th \Phi   e_\th f|^2  .
\eeaa

\item  If  $f\in \mathfrak{s}_1$,
\beaa
|\nabb  f |^2 =  |e_\th f |^2 +| e_\th(\Phi)|^2 |f|^2.
\eeaa

\item
If  $f\in\mathfrak{s}_2$, 
\beaa
|\nabb f  | ^2 =2 \left( |e_\th f|^2 +4| e_\th(\Phi)|^2 |f|^2\right).
\eeaa
\end{itemize}
\end{lemma} 

\begin{proof}
If $f\in \mathfrak{s}_0$, 
\beaa
|\nabb^2 f|^2 &=&\nab_A\nabb_B f \nabb^A\nabb^B f = |\nabb_\th \nabb_\th f|^2+ |\nabb_\vphi \nabb_\vphi  f|^2= |e_\th(e_\th f)|^2+|e_\th \Phi   e_\th f|^2.  
\eeaa
  If $f\in \mathfrak{s}_1$ is reduced from a $\Z$ invariant, polarized vector $F$,
  \beaa
  |\nabb  f  |^2 &=&\nab_A F_B \nabb^A  F^B=| \nabb_\th F_{\th} |^2 +| \nabb_\vphi F_{\vphi} |^2\\
  &=&|e_\th f|^2+|e_\th \Phi f |^2.
  \eeaa  
If    $f\in \mathfrak{s}_2$ is reduced from a     symmetric, traceless   $\Z$-invariant, polarized tensor $F=\fS$  we have,
\beaa
|\nabb  f |^2 &=&| \nabb_\th F_{\th\th} |^2+2| \nabb_\th F_{\vphi\th} |^2+ | \nabb_\th F_{\vphi\vphi} |^2+| \nabb_\vphi F_{\th\th} |^2+2| \nabb_\vphi F_{\vphi\th} |^2+ | \nabb_\vphi F_{\vphi\vphi} |^2\\
&=&| \nabb_\th F_{\th\th} |^2 +| \nabb_\th F_{\vphi\vphi} |^2+2| \nabb_\vphi F_{\vphi\th} |^2
\eeaa
and,
\beaa
 \nabb_\th F_{\th\th}&=& e_\th f=-  \nabb_\th F_{\vphi\vphi},  \\
\nabb_\vphi F_{\vphi\th}&=& e_\th \Phi F_{\th\th}-  e_\th \Phi F_{\vphi\vphi}= 2 e_\th \Phi  f. 
\eeaa
Thus,
\beaa
|\nabb  f|^2 =2|e_\th f|^2+8(e_\th \Phi f)^2 
\eeaa
as desired.
\end{proof}

\begin{proposition}[Poincar\'e]   
\label{proposition:Poincaree} 
The   following  inequalities  hold for $k$-reduced scalars.
\begin{enumerate}
\item If $f\in \mathfrak{s}_0$,
\bea
\int_S |\nabb^2 f|^2 &\ge &\int_S K  |\dds_1 f|^2.
\eea
\item 
 If  $f\in \mathfrak{s}_1$
 \bea
\int_S |\nabb f|^2 &\ge & \int_S K f^2.  
\eea
\item
If  $f\in \mathfrak{s}_2$,
\bea
\int_S |\nabb f|^2  &\ge &4 \int_S K f^2. 
\eea
\end{enumerate}
\end{proposition}

\begin{proof}
We  first prove the result for  $f\in \mathfrak{s}_2$. 
According to Lemma \ref{lemma:nabbfS},
\beaa
2^{-1}|\nabb f|^2&=&  |e_\th f|^2 +4| e_\th(\Phi)|^2 |f|^2=(e_\th f - 2 e_\th(\Phi) f)^2+4 f ( e_\th f ) e_\th(\Phi) \\
&=&(e_\th f - 2 e_\th(\Phi) f)^2+ 2 e_\th( f^2)  e_\th(\Phi). 
\eeaa
Hence,
\beaa
2^{-1}\int_S |\nabb F|^2 da_S &=&\int_S (e_\th f - 2 e_\th(\Phi) f)^2 da_S+ 
2\int_S e_\th( f^2)  e_\th(\Phi) \sqrt{\ga}  e^\Phi d\th d\vphi.
\eeaa
Now,
\beaa
\int_S e_\th( f^2)  e_\th(\Phi) \sqrt{\ga}  e^\Phi d\th d\vphi&=& \int_0^\pi \int_0^{2\pi} \pr_\th( f^2)  e_\th(\Phi)   e^\Phi d\th d\vphi=-
 \int_0^\pi \int_0^{2\pi}  f^2  e_\th( e^\Phi e_\th \Phi) \sqrt{\ga} d\th d\vphi\\
 &=& -\int_0^\pi \int_0^{2\pi}\left( e_\th e_\th \Phi +( e_\th\Phi)^2 \right)  f^2 e^\Phi \sqrt{\ga}d\th d\vphi= \int_S K  f^2  da_S.
\eeaa
Hence,
\beaa
\int_S |\nabb f|^2&\ge & 4 \int_S K   f^2 
\eeaa
as desired.

The result for $f\in \mathfrak{s}_1$ is   proved in the same way. 

If $f\in \mathfrak{s}_0$  we write, according to Lemma \ref{lemma:nabbfS},
 \beaa
 |\nabb^2 f|^2 &=& |e_\th(e_\th f)|^2+|e_\th \Phi   e_\th f|^2 =|e_\th h|^2 + |e_\th \Phi|^2|e_\th f|^2    \\
 &=&(e_\th  e_\th f  -  e_\th(\Phi) e_\th f )^2+  e_\th[ ( e_\th   f )^2 ]\,   e_\th(\Phi). 
 \eeaa
 Integrating by parts as before,
 \beaa
 \int_S e_\th[ ( e_\th   f )^2 ]\,   e_\th(\Phi) da_S &=&-
 \int_0^\pi \int_0^{2\pi}   (e_\th f)^2  e_\th( e^\Phi e_\th \Phi) \sqrt{\ga} d\th d\vphi=  \int_S K  (e_\th f)^2 da_S.
 \eeaa
 Thus,
  \beaa
\int_S  |\nabb^2 f|^2 &\ge &  \int_S K  (e_\th f)^2.
\eeaa
\end{proof}

As a corollary we deduce the following,
\begin{corollary}
The following hold true for reduced scalars,
\begin{enumerate}
\item
If $f\in \mathfrak{s}_1$,
 \bea
\int_S |\ddd_1 f |^2  &\ge & \int_S 2K f^2. 
\eea
\item
If $f\in\mathfrak{s}_2$,
\bea
\int_S |\ddd_2 f |^2 &\ge &4 \int_S K f^2.  
\eea
\end{enumerate}
\end{corollary}

\begin{proof}
According to \eqref{eq:hodgeident1},
\beaa
\int_S\big(|\nabb   f |^2+K   f ^2\big)=\int_S|\ddd_1   f   |^2.
\eeaa
We deduce,
\beaa
\int_S|\ddd_1   f   |^2&\ge & 2  \int_S K   f^2. 
\eeaa

According to \eqref{eq:hodgeident2}
\beaa
\int_S\big(|\nabb    f  |^2+4 K f ^2\big)=2\int_S |\ddd_2   f  |^2.
\eeaa
Hence,
\beaa
2\int_S |\ddd_2   f  |^2 &\ge& 8 \int_SK    f ^2 
\eeaa
as desired.
\end{proof}

\begin{corollary}
\label{Corollary:proposition-Poincaree}
Under the following mild assumptions on the Gauss curvature
 \beaa
 K= \frac{1}{r^2}+ O\left(\frac{\ep}{r^2}\right), \qquad  r e_\th ( K )  =O\left(\frac{\ep}{r^2}\right), 
 \eeaa
the following holds.
\begin{enumerate}
\item If $f\in\mathfrak{s}_0$  is orthogonal to the kernel of $\dds_1$, i.e.  $\int_Sf=0$,
then, we have
 \bea
\int_S |\dds_1 f |^2  &\ge & 2\int_S (1+O(\ep))K f^2. 
\eea

\item
If $f\in \mathfrak{s}_1$  is orthogonal to the kernel of $\dds_2$, i.e $\int_Sfe^\Phi=0$,
then, we have
 \bea\lab{eq:Poincareinequalityforddd1anddds2whenorthgonaltoexpPhi}
\int_S |\ddd_1 f |^2  \ge  6\int_S(1+O(\ep))K f^2\textrm{ and } \int_S |\dds_2 f |^2  \ge  4\int_S(1+O(\ep))K f^2.
\eea
\end{enumerate} 
\end{corollary}

\begin{proof}
We start with the first assertion. If $f\in\mathfrak{s}_0$ satisfies $\int_Sf=0$
then, $f$ is orthogonal to 1 which generates the kernel of $\dds_1$, and hence, $f$ is in the image of $\ddd_1$, i.e. there exists $h\in\mathfrak{s}_1$ such that
\beaa
f=\ddd_1h.
\eeaa
We deduce
\beaa
\int_S(\dds_1(f))^2 &=& \int_S(\dds_1\ddd_1 h)^2= \int_S(\dds_1\ddd_1)^2h h.
\eeaa
Now, the above Poincar\'e inequality for $\ddd_1$ and the assumption on $K$ implies a lower bound for the spectrum of the selfadjoint operator $\dds_1\ddd_1$ by $2K(1+O(\ep))$, and hence
\beaa
\int_S(\dds_1(f))^2 &\geq&  2\int_SK(1+O(\ep))(\dds_1\ddd_1)h h\\
&\geq& 2\int_S(1+O(\ep))(\ddd_1h)^2\\
&\geq& 2\int_S(1+O(\ep))f^2
\eeaa
which yields the first assertion.

Assume now that $f\in\mathfrak{s}_1$ satisfies  $\int_Sfe^\Phi=0$
i.e. ,$f$ is orthogonal to $e^\Phi$ which generates the kernel of $\dds_2$, and hence, $f$ is in the image of $\ddd_2$, i.e. there exists $h\in\mathfrak{s}_2$ such that
\beaa
f=\ddd_2h.
\eeaa
We deduce
\beaa
\int_S(\ddd_1f)^2 &=& \int_S(\ddd_1\ddd_2h)^2 \int_S\dds_1\ddd_1\ddd_2 h \ddd_2h= \int_S(\ddd_2\dds_2+2K)\ddd_2 h \ddd_2h\\
&=& \int_S(\dds_2\ddd_2)^2h h+\int 2K(\ddd_2h)^2.
\eeaa
Now, the above Poincar\'e inequality for $\ddd_2$ and the assumption on $K$ implies a lower bound for the spectrum of the selfadjoint operator $\dds_2\ddd_2$ by $4K(1+O(\ep))$, and hence
\beaa
\int_S(\ddd_1f)^2 &\geq& 4\int_SK(1+O(\ep))(\dds_2\ddd_2)h h+\int 2K(\ddd_2h)^2\\
&\geq& 6\int_S(1+O(\ep))(\ddd_2h)^2\\
&\geq& 6\int_S(1+O(\ep))f^2.
\eeaa
Together with the fact that
\beaa
\int_S(\dds_2f)^2 =\int_S\ddd_2\dds_2f f = \int_S(\dds_1\ddd_1-2K)f f=\int_S(\ddd_1f)^2-2\int_SKf^2, 
\eeaa
this yields the second assertion and concludes the proof of the corollary.
\end{proof}
 \begin{lemma}
\label{Lemma:poincarefor-dds2}
Assume that
 \beaa
 K= \frac{1}{r^2}+ O\left(\frac{\ep}{r^2}\right), \qquad  r e_\th ( K )  =O\left(\frac{\ep}{r^2}\right), \qquad \int_S e^{2\Phi} =r^4\left(\frac{8\pi}{3} +O(\ep)\right).
 \eeaa
 Then,  If $f\in \mathfrak{s}_1$,  we have the estimate,
\bea\lab{eq:poincaredds2takingintoaccountitskernelell=1mode}
 \int_S\big| f|^2 &\les&  r^2  \int_S |\dds_2  f   |^2+r^{-4}\left|\int_S e^\Phi f\right|^2.
\eea
More precisely,
\bea
f=  \frac{\int_S f e^\Phi} {\int_S e^{2\Phi} }  e^{\Phi}+ f^\perp
\eea
with 
\beaa
 \int_S\big| f^\perp |^2 &\les&  r^2  \int_S |\dds_2  f   |^2.
\eeaa
\end{lemma}

\begin{proof}
According to  Corollary \ref{Corollary:proposition-Poincaree}, see \ref{eq:Poincareinequalityforddd1anddds2whenorthgonaltoexpPhi},
if $f\in \mathfrak{s}_1$  is orthogonal to the kernel of $\dds_2$, i.e $\int_Sfe^\Phi=0$,
then, we have
 \beaa
 \int_S |\dds_2 f |^2  \ge  4\int_S(1+O(\ep))K f^2.
\eeaa
As a consequence $ f^\perp= f-   \left( \frac{\int_S f e^\Phi} {\int_S e^{2\Phi} }\right)  e^{\Phi}$      verifies,
\beaa
r^{-2} \int_S | f^\perp|^2 &\les&  \int_S |\dds_2(  f^\perp) |^2= \int_S |\dds_2  f   |^2 
\eeaa
from which we derive,
\beaa
 \int_S\big| f-   \left( \frac{\int_S f e^\Phi} {\int_S e^{2\Phi} }\right)  e^{\Phi}\big| ^2 &\les &r^2  \int_S |\dds_2  f   |^2 
\eeaa
or,
\beaa
 \int_S\big| f|^2 &=&  r^2  \int_S |\dds_2  f   |^2 +\left|\int_S e^\Phi f\right|^2 \frac{1}{ \int_\S e^{2\Phi}}\\
 &\le&  r^2  \int_S |\dds_2  f   |^2+  r^{-4} \left |\int_S e^\Phi f\right|^2
\eeaa
as desired.
\end{proof}


\subsubsection{Higher derivative operators and spaces}


\begin{definition}\lab{definitionhkspacesS}
Given $f$ a $k$-reduced  scalar  and    $s$ a positive integer   we define,
\bea\lab{def:angularderivativesonreducedksclars}
\dkb^s f=
\begin{cases}
 r^{2p}\lapp_k ^p, \qquad\qquad  \mbox{if} \quad  s=2p,\\
 r^{2p+1} \ddd_k  \lapp_k^p ,\qquad \mbox{if} \quad  s=2p+1.
\end{cases}
\eea
We also define the norms,
\bea
\| f\|_{\frak h_s(S)}:&=&\sum_{i=0}^s\| \dkb^i f\|_{L^2(S)}.
\eea
\end{definition}

\begin{lemma}
Assume  the  Gauss curvature $K$ of $S$ verifies  the condition, 
\beaa
K= \frac{1}{r^2}+ O(\ep), \qquad  | r^i \nabb^i K |=O(\ep), \qquad  1\le i\le [s/2]+1.
\eeaa
Then, the following holds.
\begin{enumerate}
\item If  $f$  is  a $k$-scalar,  reduced from $\fS$, we have,
\bea
\label{eq:similaritynormsonS}
  \| f\|_{\hk_s(S)}    \sim \sum_{j=0}^ s r^j\|\nabb^jf\|_{L^2(S)}
\eea
where $\nabb$ denotes the usual covariant derivative operator on $S$.
\item Equivalently, the norm  $ r^{-s}  \| f\|_{\hk_s(S)}$ of a reduced scalar $f\in\sk_s(S)$ can be defined   as the sum of $L^2$ norms 
of any  allowable sequence of Hodge operators $\ddd_a$, $\dds_a$ applied to $f$.
\end{enumerate}
\end{lemma} 

\begin{proof}
For $s=1, 2 $  the proof  of the  first part  follows  immediately from Proposition  \ref{prop:2D-hodge-reduced}. For higher $s$ the proof follows, step by step,   by a simple commutation argument between  covariant derivatives and $\lapp_k$   and applications of Proposition   \ref{prop:2D-hodge-reduced}. The proof of  the second  part follows from our reduced  elliptic estimates and  definition of the reduced  Hodge operators.
\end{proof} 

As a consequence of the lemma we can   derive the reduce form of the standard Sobolev and product Sobolev 
inequalities. Before stating the result we pause to define the product of two  reduced  scalars.
\begin{definition}
Let $f\in \frak{s}_a$  be reduced from an   $\SS_a$ tensor and $ g\in \frak{s}_b $ reduced from an $\SS_b$  tensor.
We define the product $f\c g$  to be the reduction of any product between  the corresponding  tensors  on $S$, i.e. any contraction of the tensor  product   between  them.
Thus $f\c g\in \sk_{a+b-2c}$  where $c$ denotes the number of indices affected by the contraction.
\end{definition}

{\bf Examples. } Here are the  most relevant examples  for us.
\begin{itemize}
\item $f\in \frak{s}_0, \, g\in     \frak{s}_k  $         in which case $f\c g\in \frak{s}_k $ and equals $f g$.
 
\item  $f\in \frak{s}_1, \, g\in \frak{s}_k$ in which case $f\c g\in \frak{s}_{k-1}$ or $f\c g\in \frak{s}_{k+1}$
 and in both cases  $f\c g=fg$ as  simple product of  the reduced scalars.
 
\item $f\in \frak{s}_2,\,   g\in \frak{s}_k$  in which case $f\c g\in \frak{s}_{k-2}$ or $f\c g\in \frak{s}_{k}$ or 
$f\c g\in \frak{s}_{k+2}$. In the first case  $f\c g=2 fg$.    In the second case  and third cases $f\c g= fg$
 as  simple product of  the reduced scalars.
\end{itemize}

\begin{lemma}
\label{lemma:Hodge-contractions}
 Let $f\in \frak{s}_a(S)$,  $g\in \frak{s}_b(S)$, $a\ge b$, $a>0$,   and $f\c g\in \sk_{a+b-2c}$  where $0\le c\le \frac 1 2 (a-b)$ denotes the order of contraction.
 Then, 
 \bea
 \bsplit
\ddd_{a+b-2c} (f g)&=  f  \ddd_b g+ g \left(   (1-\frac{c}{a})  \ddd_ a  f- \frac{c}{a}  \dds_{a+1} f\right),\\
 \dds_{a+b-2c+1} (f g)&= f \dds_{b+1} g+\left(  (-1+\frac{c}{a} )    \ddd_a f +\frac{c}{a} \dds_{a+1} f\right).
 \end{split}
 \eea
\end{lemma}

\begin{proof}
Assume $ a\ge b$ and $c\le \frac{a-b}{2}$. We write,
\beaa
\ddd_{a+b-2c} (f g)&=&  f  \ddd_b g+  g \left( e_\th f  +(a-2 c) e_\th(\Phi)   f  \right).
\eeaa
We look for  reals  $A, B$ wit $A+B=1$ such that
\beaa
 e_\th f  +(a-2 c) e_\th(\Phi)   f&=& A  \ddd_ a  f- B  \dds_{a+1} f =    e_\th f+  a (A-B) e_\th \Phi f.
\eeaa
Therefore,
\beaa
 a (1-2B)= a- 2 c 
\eeaa
i.e. $B=\frac{c}{a}$, $A=1-\frac{c}{a}$ and we derive,
\beaa
\ddd_{a+b-2c} (f g)&=&  f  \ddd_b g+ g \left(   (1-\frac{c}{a})  \ddd_ a  f- \frac{c}{a}  \dds_{a+1} f\right).
\eeaa
Also,
\beaa
\dds_{a+b-2c+1} (f g)&=& f \dds_{b+1} g+ g\left( - e_\th (f)   + (a-2 c) e_\th(\Phi)   f  \right).
\eeaa
As before we write, with $A+B=-1$
\beaa
 - e_\th (f)   + (a-2 c) e_\th(\Phi)   f&=&  A  \ddd_ a  f- B  \dds_{a+1} f =   - e_\th f+  a (A-B) e_\th \Phi f. 
\eeaa
Hence,
\beaa
- a (-1 -2B) = (a-2 c)
\eeaa
i.e. $B=-\frac{c}{a}$, $A=-1 +\frac{c}{a}$.
Hence,
\beaa
\dds_{a+b-2c+1} (f g)&=& f \dds_{b+1} g+\left(  (-1+\frac{c}{a} )    \ddd_a f +\frac{c}{a} \dds_{a+1} f\right)
\eeaa
as desired.
\end{proof}

\begin{proposition}\lab{prop:sobolevandproductesitmatesonS}
The following results hold true for $k$-reduced scalars on $S$,
\begin{enumerate}
\item 
 If $f \in \frak{s}_k$  we have,
 \beaa
 \|f\|_{L^\infty(S)} &\les& r^{-1} \| f\|_{\hk_2(S)}.
 \eeaa
 \item  Given two reduced scalars $f , g$ we have,
 \beaa
 \| f\c g \|_{\hk_s(S)} &\les& r^{-1}\left( \| f\|_{\hk_{[s/2]+2}  (S)} \|g\|_{\hk_s(S)} + \| g\|_{\hk_{[s/2]+2}  (S)} \|f\|_{\hk_s(S)} \right)
 \eeaa
 where $[s/2]$  denotes   the largest integer smaller than $s/2$.
 \end{enumerate}
 \end{proposition}

\begin{proof}
Both  statements are   classical for $\SS_k(S)$ tensors  with respect to the norm on the right hand side of \eqref{eq:similaritynormsonS}.  A direct proof can also be derived 
using Lemma \ref{lemma:Hodge-contractions} and the equivalence  definition of the  $\hk_s(S)$ norms.
\end{proof}


\subsubsection{$S$-averages}


\begin{definition}
Given any $f\in\mathfrak{s}_0$ we denote its average   by,
\beaa
\bar{f}:&=&\frac{1}{|S|} \int_S f, \qquad \check{f}:= f-\bar{f}.
\eeaa
\end{definition}

The following follows immediately from the definition.
\begin{lemma}
\label{lemma:averagesorproducts}
For any  two scalar reduced scalars $f$ and $g$ in $\mathfrak{s}_0$
we have
\beaa
\overline{fg} &=& \overline{f}\,\overline{g}+ \overline{\check{f}\check{g}},
\eeaa
and,
\beaa
fg-\overline{fg}= \check{f}\,\overline{g} +\overline{f}\check{g}+ \big(\check{f} \check{g}-\overline{\check{f}\check{g}}\big).
\eeaa
\end{lemma}

\begin{remark}
In view of the notations above, we may rewrite the Poincar\'e inequality for $\dds_1$ as follows. Under mild assumptions on the Gauss curvature ($K= r^{-2}+ O(\ep r^{-2}), \,\,  r e_\th ( K )  =O(\ep r^{-2})$), we have for any $f\in\mathfrak{s}_0$
 \beaa
\int_S |\dds_1 f |^2  &\ge & 2\int_S (1+O(\ep))K (\check{f})^2. 
\eeaa
\end{remark}


\subsection{Invariant $S$-foliations}


In this section we     record the main equations    associated to   general,  $\Z$-invariant  Einstein vacuum spacetimes $(\MM, \g)$. 
We start by recalling  the   spacetime framework of \cite{Ch-Kl} and then we    show  how the   null structure and Bianchi identities 
 simplify in the reduced picture. Throughout   this section we consider   given an invariant   $S$-foliation\footnote{From now on, an  invariant  $S$ foliation is       automatically assumed to be a $\Z$ invariant polarized foliation.}  and a fixed 
         adapted null   pair   $e_3, e_4$, i.e.   future directed  $\Z$- invariant, polarized,    null vectors orthogonal to  the leaves $S$  of the foliation   such as $\g(e_3, e_4)=-2$. 
         \begin{definition}
         We denote by $\SS_k(\MM)$ the set of $k$-covariant polarized   tensors on $\MM$  tangent to the $S$-foliation and which  restrict to $\SS_k(S)$ on any
         $S$-surface of the foliation and by  $\sk_k(\MM)$  their  corresponding  reductions.
         \end{definition}


\subsubsection{Spacetime   null   decompositions}


   Following  \cite{Ch-Kl}  we   define the spacetime Ricci coefficients,
   \bea
   \begin{split}
   \chiS_{AB}:&=\g(\D_A  e_4, e_B), \qquad \xiS_A:=\frac 1 2 \g(\D_4 e_4, e_A), \qquad \etaS_A:=\frac 1 2 \g(\D_3 e_4, e_A),\\
   \zeS_A:&=\frac 1 2 \g( \D_A e_4, e_3), \qquad \omS:=\frac 1 4 \g(D_4 e_4, e_3),
   \end{split}
   \eea
   and interchanging $e_3, e_4$,
    \bea
   \begin{split}
   \chibS_{AB}:&=\g(\D_A  e_3, e_B), \qquad \quad \xibS_A:=\frac 1 2 \g(\D_3 e_3, e_A), \qquad \etabS_A:=\frac 1 2 \g(\D_4 e_3, e_A),\\
   \zeS_A:&=-\frac 1 2 \g( \D_A e_3, e_4), \qquad \ombS:=\frac 1 4 \g(D_3 e_3, e_4).
   \end{split}
   \eea
We also define the  spacetime  null  curvature components,
\bea
\begin{split}
\aS_{AB}:&=\R_{A4B4}, \qquad\quad \bS_A
:=\frac 1 2 \R_{A434}, \qquad\,\, \rhoS:=\frac 14 \R_{3434},\\
\aaS_{AB}:&=\R_{A3B3}, \qquad  \bbS_{A} :=\frac 1 2 \R_{A334}, \qquad\quad  \rhodS:=\frac 1 4  \dual \R _{3434}.
\end{split}
\eea


\subsubsection{Reduced  null  decompositions}
   
   
   We define the spacetime   Ricci coefficients as follows 

\begin{definition}[\textit{Ricci coefficients}]
Let  $e_3, e_4, e_\th$ be a  reduced   null frame.   The following scalars 
\bea
\label{eq-red:Ricci-coeff}
\begin{split}
\chi&=g(D_\th e_4,e_\th),  \qquad \, \,
\chib =g(D_\th e_3,e_\th),\\
\eta &=\frac{1}{2} g(D_{3} e_4,e_\th),\qquad 
 \etab =\frac{1}{2} g(D_{4} e_3,e_\th),\\
\xi &=\frac{1}{2} g(D_4 e_4,e_\th), \qquad 
\xib =\frac{1}{2} g(D_3 e_3,e_\th),\\
\om&=\frac{1}{4} g(D_{4} e_4,e_3), \qquad 
\omb=\frac{1}{4}g(D_3 e_3,e_4),\\
\ze &=\frac{1}{2}g(D_\th e_4,e_3),
\end{split}
\eea
are called the Ricci coefficients associated to our canonical null pair.
\end{definition}

 \begin{lemma}
The following lemma follows   easily   from the definitions,
\bea
\label{eq-red:Ricci form}
\begin{split}
D_4 e_4 &=-  2\om e_4 +2\xi e_\th , \qquad  \qquad \, D_3 e_3 =-2\omb e_3 +2\xib e_\th,\\
D_4 e_3 &= 2\om e_3 +2\etab e_\th, \qquad  \qquad \quad D_3 e_4=2\omb  e_4+2\eta e_\th,\\
D_4 e_\th &= \etab  e_4+\xi e_3, \qquad  \qquad \qquad  D_3 e_\th =\xib e_4+ \eta e_3,\\
D_{\th}  e_4 &= -\ze e_4 +\chi e_\th, \qquad  \qquad\quad  D_{\th} e_3=\ze e_3 +\chib e_\th,\\
D_\th e_\th &=  \frac{1}{2}\chib e_4+\frac{1}{2}\chi e_3.
\end{split}
\eea
\end{lemma}

\begin{definition}\label{def:nullcomponents}
The null components of the  Ricci curvature   tensor\footnote{Recall    that  the scalar  curvature  of the reduced metric $g$    vanishes,  $R=0$, and hence $R_{34}=R_{\th\th}$.}  of the metric $g$ are  denoted  by
 \beaa
 R_{33}=\aa,\quad R_{44}=\a, \quad R_{3\th}=\bb,\quad  R_{4\th}=\b, \quad R_{\th\th}=R_{34}=\rho, \quad R_{34}=\rho.
 \eeaa 
 \end{definition}

 
\subsubsection{Comparison to  the space-time  frame}\label{sect:comparisonreduced-spacetime}


Let $e_3, e_4, e_\th$  be a null frame     for the  reduced metric  $g$   and       $e_3, e_4, e_\th,     e_\vphi= X^{-1/2} \pr_\vphi$  the 
augmented   adapted  $3+1$ frame for $\g$.  Recall   that we have   denoted,
   \beaa
\chiS, \xiS, \etaS, \etabS, \zeS,  \omS, \chibS,  \xibS, \ombS,
\eeaa
the  standard  (as  defined in \cite{Ch-Kl} )  space-time  Ricci coefficients  and by
 \beaa
\aS, \bS, \rhoS, \rhodS, \bbS, \aaS,
\eeaa
 the   null decomposition of  the curvature  tensor 
$\R$.    
\begin{proposition} 
\label{prop:reduced-spacetime-RandGa}
The  following relations  between the  spacetime and reduced  Ricci and curvature  null components hold true,
\begin{itemize}
\item 
We have,
\beaa
\begin{split}
&\chiS_{\th\th}=\chi, \quad \chiS_{\th\vphi}=0, \quad \chiS_{\vphi\vphi}= e_4(\Phi),\\
&\chibS_{\th\th}=\chi, \quad \chiS_{\th\vphi}=0, \quad \chiS_{\vphi\vphi}= e_3(\Phi),\\
&\aS_{\th\th}=\a, \quad \aS_{\th\vphi}=0, \quad \aS_{\vphi\vphi}=-\a,\\
&\aaS_{\th\th}=\aa, \quad \aaS_{\th\vphi}=0, \quad \aaS_{\vphi\vphi}=-\a.
\end{split}
\eeaa

\item
All $e_\vphi$ components   of $\etaS, \etabS, \zeS, \xiS,\xibS, \bS,\bbS$ vanish and,
\beaa
\etaS_\th=\eta, \,  \etabS_\th=\etab,\,  \zeS_\th=\ze, \,  \xiS_\th=\xi, \, 
 \xibS_\th=\xib,\, 
 \eeaa
 and\footnote{Note the change of sign for the  $\bb$ component.}
 \beaa
   \bS_\th=\b, \, \bbS_\th=-\bb. 
   \eeaa
Also,
\beaa
\omS=\om, \quad \ombS=\omb, \quad \rhoS=\rho, \quad \rhodS=0.
\eeaa

\item  We have, 
\beaa
\trchS =\chi+e_4(\Phi), \qquad \trchbS=\chib+e_3(\Phi).
\eeaa
Recalling, see  definition in  \cite{Ch-Kl}, 
\beaa
\chihS_{AB}=\chiS_{AB}-\frac 1 2 (\trchS)  \gS_{AB}, \qquad \chibhS_{AB}=\chibS_{AB}-\frac 1 2 (\trchbS)  \gS_{AB},
\eeaa
we have
\beaa
\chihS_{\th\th}&=&\frac 1 2 \left( \chi-e_4(\Phi) \right), \qquad  \chibhS_{\th\th}=\frac 1  2 \left( \chib-e_3(\Phi)\right).
\eeaa
\end{itemize}
\end{proposition}

\begin{proof}
We check only  the   less obvious relations  such as those        involving the null components of curvature. Using 
\eqref{eq:Curv1} and \eqref{Ricci}  we deduce,
\beaa
\aS_{\th\th}&=&\R_{\th 4\th 4} =R_{\th4\th4}=g_{\th\th} R_{44}=\a,\\
\aS_{\th\vphi}&=&\R_{\th 4\vphi 4}=0,\\
\aS_{\vphi\vphi}&=&\R_{\vphi 4\vphi 4}=-R_{44}=-\a,\\
2 \bS_{\th}&=&\R_{\th 343}=R_{\th 343}=-g_{34} R_{\th3}= 2 \b,\\
2 \bS_{\vphi}&=&\R_{\vphi 343}=0,\\
4\rhoS&=&\R_{ 34 34 }=R_{ 34 34 } =-2 g_{34}R_{34}= 4 \rho,\\
4\rhodS&=&\dual \R_{3434}=0,\\
2\bbS_\th&=& \R_{\th 334} =R_{\th334}= g_{34} R_{\th 3 } =-2 \bb.
\eeaa
\end{proof}

\begin{definition}
We introduce the notation,
\beaa
\vth:&=&\chi-e_4( \Phi),\qquad   \ka: =\trchS =\chi+e_4 (\Phi ),\\
\vthb:&=&\chib-e_3 (\Phi), \qquad   \kab: =\trchbS =\chib+e_3( \Phi ).
\eeaa
Thus,
\beaa
\chihS_{\th\th}= \chihS_{\vphi\vphi}  &=&\frac 1 2 \vth, \qquad \chibhS_{\th\th}   =\chibhS_{\vphi\vphi}          =\frac 1 2 \vthb.
\eeaa
In particular, $\chi=\frac 1 2 (\vth+\ka)$ and $\chib=\frac 1 2 (\vthb+\kab)$.
\end{definition}

\begin{remark}
In view of Proposition \ref{prop:reduced-spacetime-RandGa}   we have,
\begin{enumerate}
\item   The quantities $\ka, \kab, \om, \omb , \rho$  are reduced  scalars in $\mathfrak{s}_0$.
\item The quantities $\eta, \etab, \ze, \xi, \xib, \b, \bb$ are reduced scalars in $\mathfrak{s}_1$.
\item  The quantities $\vth, \vthb, \a, \aa$ are  reduced scalars in $\mathfrak{s}_2$.
\end{enumerate}
\end{remark}


\subsubsection{Commutation identities}


 We record first  the  commutation relations between the elements of the frame,
 \beaa
\,[e_\th, e_3]&=&                  \frac 1 2 (  \kab  +\vthb)  e_\th     +(\ze-\eta) e_3-\xib e_4, \\
\,[e_\th, e_4]&=&\frac  1 2 (\ka +\vth)e_\th -(\ze+\etab) e_4-\xi e_3,\\
\,[e_3, e_4]&=& 2 \omb e_4-2\om e_3+2(\eta-\etab) e_\th.
\eeaa
\begin{lemma}
\label{Le:comme3e4}
\begin{enumerate} The following commutation formulae hold true for reduced scalars.

 \item If $f\in \mathfrak{s}_k$,
\bea
\bsplit
\,[\ddd_k, e_3] f&=\frac 1 2 \kab  \ddd_k f+\comb_k(f),\\
\comb_k(f)&=- \frac 1 2 \vthb  \dds_{k+1} f + (\ze-\eta) e_3 f - k \eta  e_3\Phi f -\xib( e_4  f  +k e_4(\Phi)  f )- k \bb f,\\
\,[\ddd_k, e_4] &=\frac 1 2 \ka  \ddd_k f+\com_k(f),\\
\com_k(f)&=- \frac 1 2 \vth  \dds_{k+1} f - (\ze+\etab) e_4 f - k \etab  e_4\Phi f -\xi( e_3  f  +k e_3(\Phi)  f )- k \b f.
\end{split}
\eea
\item If  $f\in \mathfrak{s}_{k-1}$
\bea
\bsplit
\,[\dds_k, e_3]f &=\frac 1 2 \kab  \dds_k f+\comb^*_k(f),\\
\comb^*_k(f)&=- \frac 1 2 \vthb  \ddd_{k-1} f - (\ze-\eta) e_3 f  -(k-1)  \eta  e_3\Phi f +\xib(e_4  f  -(k-1)  e_4(\Phi)  f )\\
&- (k-1) \bb f,\\
\,[\dds_k, e_4]f &=\frac 1 2 \ka  \dds_k f+\com^*_k(f),\\
\com^*_k(f)&=- \frac 1 2 \vth  \ddd_{k-1} f + (\ze+\etab) e_4 f - (k-1)  \etab  e_4\Phi f +\xi( e_3  f  -(k-1) e_3(\Phi)  f )\\
&- (k-1)  \b f.
\end{split}
\eea
\end{enumerate}
\end{lemma}

\begin{proof}
We write,
\beaa
\,[e_\th+k e_\th(\Phi),  e_3] f &=&[e_\th, e_3]f- k ( e_3 e_\th \Phi) f.
\eeaa
Recall that (see \eqref{eq: R-Phi}),  $D_a D_b \Phi=  R_{ab}- D_a \Phi D_b \Phi $. Hence,
\beaa
 e_3 e_\th \Phi-\xib e_4 \Phi -\eta e_3\Phi= D_3 D_\th  \Phi&=&  R_{3\th}- D_3 \Phi D_\th\Phi  = \bb- e_3\Phi e_\th \Phi.
\eeaa
Thus,
\beaa
e_3 e_\th \Phi&=&\bb- e_3\Phi e_\th \Phi+\eta e_3(\Phi) +\xib e_4 \Phi.
\eeaa
We deduce, since $e_3\Phi=\frac1 2 (\kab-\vthb)$,
\beaa
\,[e_\th+k e_\th(\Phi),  e_3] f &=&[e_\th, e_3]f- k\left(  \bb- e_3\Phi e_\th \Phi+\eta e_3(\Phi) +\xib e_4 \Phi     \right)f\\
&=&\frac 1 2(\kab+\vthb)  e_\th f +(\ze-\eta) e_3 f-\xib e_4f- k\left(  \bb- e_3\Phi e_\th \Phi+\eta e_3(\Phi) +\xib e_4 \Phi     \right)f\\
&=&\frac 1 2(\kab+\vthb)  e_\th f + k \frac 1 2  (\kab-\vthb)e_\th\Phi f \\
&+&  (\ze-\eta) e_3 f - k \eta  e_3\Phi f -\xib( e_4  f  +k e_4(\Phi)  f )- k \bb f \\
&=&\frac 1 2 \kab  \ddd_k f +\frac 1 2 \vthb  ( e_\th f - k  e_\th \Phi f)\\
&+&  (\ze-\eta) e_3 f - k \eta  e_3\Phi f -\xib( e_4  f  +k e_4(\Phi)  f )- k \bb f
\eeaa
i.e., recalling the definition of $\dds_{k+1}$,
\beaa
\,[e_\th+k e_\th(\Phi),  e_3] f &=&\frac 1 2 \kab  \ddd_k f+\comb_k(f),\\
\comb(f)_k &=&- \frac 1 2 \vthb  \dds_{k+1} f + (\ze-\eta) e_3 f - k \eta  e_3\Phi f -\xib( e_4  f  +k e_4(\Phi)  f )- k \bb f.
\eeaa
The other commutation formulae are proved in the same manner.
\end{proof}


\subsection{Schwarzschild spacetime} 


In standard coordinates  the   Schwarzschild metric has the form,
 \bea
 ds^2=-\Up  dt^2 + \Up ^{-1} dr^2 + r^2 d\th^2+X^2  d\vphi^2, \qquad
 \eea
 where,
 \beaa
 \Up:= 1-\frac{2m}{r}, \qquad   X= r^2 \sin^2 \th.
 \eeaa

 We denote   by $\T$ the  stationary Killing vectorfield  $\T=\pr_t$  and by $\Z=\pr_\vphi$  the   axial symmetric  one. Recall the regular,  $\Z$-invariant  optical functions  in the exterior region $r\ge 2m$ of Schwarzschild
 \bea
u=t-r_*, \qquad  \ub=t+r_*, \qquad  \frac{dr_*}{dr}= \Up^{-1}
\eea
with $r^*=r+2 m \log(\frac{r}{2m}-1)$. The corresponding   null geodesic generators are, 
\bea
\Lb:=-g^{ab}\pr_a v  \pr_b=\Up^{-1}\pr_t -\pr_r,         \qquad  L:=-g^{ab}\pr_a u  \pr_b=\Up^{-1}\pr_t +\pr_r.
\eea
Clearly,
\beaa
g(L, L)=g(\Lb,\Lb)=0, \quad  g(L, \Lb)=-2\Up^{-1},\qquad D_L L=D_\Lb \Lb=0.
\eeaa

\begin{definition}
We  can use the null geodesic generators $L, \Lb$  to define the following          canonical null pairs. In all cases  all curvature components 
 vanish identically  except, 
 \bea
 \rhoS=-\frac{2m}{r^3}.
 \eea
 
\begin{enumerate}
\item The          null frame $(e_3, e_4) $        for which $e_3$ is geodesic  (which is regular towards the future for all $r>0$) is given by
\bea
\label{eq:regular-nullpair}
e_3=\Lb=\Up^{-1}\pr_t-\pr_r,\,\, \qquad e_4=\Up L=\pr_t+\Up\pr_r,\,\,\qquad  \Up=1-\frac{2m}{r}.
\eea
All  Ricci  coefficients  vanish  except,
\beaa
\chi=\frac{\Up}{r},\,\,\quad  \chib=-\frac{1}{r},\,\,\quad  \om=-\frac{m}{r^2},\,\, \quad \omb=0.
\eeaa
\item The null frame $(e_3, e_4)$  for which $e_4$ is geodesic.    
\beaa
e_4 =L= \Up^{-1}\pr_t+\pr_r,\qquad  e_3 =\Up \Lb =\pr_t-\Up \pr_r. 
\eeaa
All  Ricci  coefficients  vanish except,
\beaa
\chi=\frac{1}{r},\qquad \chib=-\frac{\Up}{r}, \qquad \om=0, \qquad \omb =\frac{m}{r^2}. 
\eeaa
\end{enumerate}
\end{definition}

Note  that   the null pair  \eqref{eq:regular-nullpair} is regular along the future event horizon  as  can be   easily seen by studying 
the behavior\footnote{i.e.        the null  geodesics  in the    direction of $\Lb$  reach    the horizon in   finite proper time. Note that, on the other hand, the past null  geodesics in the  direction of $L$ still meet the horizon in infinite proper time. }.  of  future directed ingoing null geodesics near  $r=2m$.


\section{Main Equations}


In this section we  translate   the       null structure  and null Bianchi identities associated to   an $S$-foliation 
in the reduced picture.  We  start     with general,   $\Z$-invariant,   $S$ foliation . We then consider  the special case 
of geodesic foliations.


\subsection{Main equations for general $S$-foliations}\label{subsect:generalmainequations}


We   consider a fixed $\Z$-invariant $S$-foliation  with a fixed $\Z$-invariant null frame $e_3, e_4$.


\subsubsection{Null structure equations} 


We simply  translate   the  well known spacetime    null structure equations (see\footnote{Note however that   the notation  in \cite{Ch-Kl}  are     different, see       section 7.3  for the definitions. }   proposition  7.4.1 in \cite{Ch-Kl})  in the reduced picture.  Thus   the spacetime equation\footnote{For convenience  we  drop  the    $^{(1+3)} $ labels in what follows.}, 
\beaa
\nabb_3 \chibh+\trchb\, \chibh &=&\nabb \hot\xib        -2\omb \chibh +(\eta+\etab-2\ze)\hot \xib-\aa
\eeaa
becomes\footnote{recall that  $\chihS_{\th\th} = \frac 12 \vth$},
\bea
e_3(\vthb)+\kab \, \vthb =2( e_\th(\xib) - e_\th(\Phi)  \xib) - 2\omb \, \vthb +2(\eta+\etab-2\ze)\, \xib-2\aa.
\eea
The spacetime equation,
\beaa
e_3(\trchb)+\frac 1 2 \trchb^2=2 \divv\xib  -2\omb \trchb+2\xib\c( \eta+\etab-2\ze) -\chibh\c\chibh
\eeaa
becomes,
\bea
e_3(\kab)+\frac 12 \kab^2 +2 \omb \,\kab &=&2( e_\th\xib+ e_\th(\Phi) \xib)+2(\eta+\etab-2\ze)  \xib- \frac 1 2 \vthb\, \vthb.
\eea
The spacetime equation,
\beaa
\nabb_4\chibh+\frac 12  \trch \,\chibh  &=&  \nabb\hot \etab+2 \om \chibh -\frac 1 2 \trchb \chih  +\xi\hot\xib + \etab\hot \etab
\eeaa
becomes,
\beaa
e_4\vthb +\frac 12 \ka\, \vthb - 2\om \vthb &=&2( e_\th\etab  - e_\th(\Phi) \etab )-\frac 12 \trchb \,\vth+2(\xi\, \xib+\etab^2).
\eeaa
The spacetime equation,
\beaa
\nabb_4 \trchb +\frac 12  \trch\, \trchb =2 \divv \etab+ 2\rho +2\om\,\trchb -\chih\c \chibh +2(\xi\c \xib+\etab\c \etab)
\eeaa
becomes,
\beaa
e_4(\kab)+\frac 1 2 \ka\, \kab -2\om \kab &=& 2( e_\th \etab +   e_\th(\Phi) \etab) + 2\rho -\frac 1 2  \vth\, \vthb +2(\xi\, \xib+\etab\, \etab).
\eeaa
The spacetime equation,
\beaa
\nabb_3 \ze&=&-\bb -2 \nabb \omb -\chibh\c(\ze+\eta)-\frac 1  2 \trchb (\ze+\eta)+2\omb(\ze-\eta)+(\chih +\frac 12 \trch)\xib+2\om \xib
\eeaa
becomes   (note that  $\bbS=-\bb$ !),
\beaa
e_3 \ze +\frac 1  2 \kab (\ze+\eta) -2\omb(\ze-\eta)&=&\bb -2 e_\th(\omb)+2\om \xib+\frac 12 \ka \,\xib  -\frac  1 2 \vthb(\ze+\eta)+\frac 1 2 \vth \,\xib.
\eeaa

The spacetime equation,
\beaa
\nabb_4 \xib -\nabb_3 \etab =-\bb+ 4\om \xib+\chibh\c(\etab-\eta)+\frac 12 \trchb (\etab-\eta) 
\eeaa
becomes\footnote{Note that  $\bbS=-\bb$      and   $ \dds_1 f =-e_\th(f) $.},
\beaa
e_4(\xib) - e_3(\etab)=\bb + 4\om \xib+\frac 12 \kab  (\etab-\eta)+\frac 1 2 \vthb (\etab-\eta).
\eeaa

The spacetime equation,
\beaa
\nabb_4 \omb+\nabb_3\om&=&\rho+ 4\om\omb +\xi\c\xib+\ze\c(\eta-\etab) -\eta\c\etab
\eeaa
becomes
\beaa
 e_4 \omb+e_3\om&=&\rho+ 4\om\omb +\xi\,\xib+\ze(\eta-\etab) -\eta\, \etab.
\eeaa

The spacetime  Codazzi  equation,
\beaa
\divvS \chibhS&=&\bbS+\frac 1 2 (\nabbS \trchbS -\trchbS \zeS)+\chibhS\c \zeS
\eeaa
becomes\footnote{Note that         $\bbS_\th=-\bb$, $\chibhS=\frac 1 2\vthb$.},
\beaa
\frac 1 2 (e_\th(\vthb)+2 e_\th(\Phi)\vthb) &=&-\bb+\frac 1 2 (e_\th(  \kab) -\kab  \ze)+\frac 1 2 \vthb  \ze.
\eeaa
The Gauss  equation,
\beaa
K&=&-\frac 1 4 \trchS \trchbS +\frac 1 2 \chihS \chibhS -\rhoS
\eeaa
becomes,
\beaa
K&=& -\frac 1 4 \ka \kab +\frac 1 4 \vth\vthb -\rho.
\eeaa

We summarize the  results in the following  proposition.
\begin{proposition}
\label{prop:null.structure-general}
\label{eq:general-NS}
\bea
\begin{split}
e_3(\vthb)+\kab \, \vthb  + 2\omb\,  \vthb &=-2\aa-2\dds_2\,\xib  +2(\eta+\etab-2\ze)\,  \xib,\\
e_3(\kab)+\frac 12 \kab^2 +2 \omb \,\kab &=2\ddd_1\xib+2(\eta+\etab-2\ze)  \xib -\frac 1 2 \vthb^2,\\
e_4\vthb +\frac 12 \ka\, \vthb - 2\om \vthb &=-2\dds_2\,\etab -\frac 12 \kab  \,\vth+2(\xi\, \xib+\etab^2),          
\\
e_4(\kab)+\frac 1 2 \ka\, \kab -2\om \kab &= 2\ddd_1\etab + 2\rho -\frac 1 2  \vth\, \vthb +2(\xi\,  \xib+\etab\,\etab),\\
e_3 \ze +\frac 1  2 \kab (\ze+\eta) -2\omb(\ze-\eta)&=\bb +2\dds_1\,\omb+2\om \xib+\frac 12 \ka \,\xib  -\frac  1 2 \vthb(\ze+\eta)+\frac 1 2 \vth \,\xib,\\
e_4(\xib) - 4\om \xib- e_3(\etab)&=\bb +\frac 12 \kab (\etab-\eta)+\frac 1 2 \vthb (\etab-\eta),\\
e_4 \omb+e_3\om&=\rho+ 4\om\omb +\xi\,\xib+\ze(\eta-\etab) -\eta\, \etab.\\
\\
\end{split}
\eea
In view of the symmetry $e_3-e_4$, we also derive,
\bea
\begin{split}
e_4(\vth)+\ka \, \vth +2\om \vth &=-2\a-2\dds_2\, \xi +2(\etab+\eta+2\ze)  \xi,\\
e_4(\ka)+\frac 12 \ka^2 +2 \om \,\ka &=2\ddd_1\xi+2(\etab+\eta + 2\ze)  \xi -\frac 1 2 \vth^2,\\
e_3\vth +\frac 12 \kab\, \vth - 2\omb \vth &=-2\dds_2\,\eta -\frac 12 \ka \,\vthb+2(\xib\, \xi+\eta^2), 
\\
e_3(\ka)+\frac 1 2 \kab\, \ka -2\omb \ka &= 2\ddd_1\eta + 2\rho -\frac 1 2  \vthb\, \vth +2(\xib\,  \xi+\eta\, \eta),\\
-e_4 \ze +\frac 1  2 \ka (-\ze+\etab) +2\om(\ze+\etab)&=\b +2\dds_1\om+2\omb \xi+\frac 12 \kab \,\xi  -\frac  1 2 \vth(-\ze+\etab)-\frac 1 2 \vthb \,\xi,\\
e_3(\xi) - e_4(\eta)&=\b + 4\omb \xi+\frac 12 \ka  (\eta-\etab)+\frac 1 2 \vth (\eta-\etab),\\
e_4 \omb +e_3\om&=\rho+ 4\om\omb +\xi\,\xib+\ze(\eta-\etab) -\eta\, \etab.\\
\end{split}
\eea
We also have the Codazzi equations,
\beaa
\ddd_2\vthb &=-2\bb -\dds_1\,\kab -\ze \kab  + \vthb \,  \ze,\\
\ddd_2\vth &=-2\b -\dds_1\,\ka + \ze\ka -   \vth \, \ze,
\eeaa
and the Gauss  equation,
\beaa
K=-\rho -\frac 1 4 \ka\,\kab +\frac 1  4 \vth\, \vthb.
\eeaa
\end{proposition}


\subsection{Null Bianchi identities}


We now translate the spacetime null  Bianchi identities of  \cite{Ch-Kl} (see proposition 7.3.2.) in the reduced  picture.
The spacetime equation (note that    $\DDs_2 \b:=-\frac 12 \nabbS\otimes \b$),
\beaa
\nabb_3\a+\frac 1 2 \trchb\,\a&=&-2 \DDs_2\, \b+4\omb \a -3(\chih \rho+\dual\, \chih \rhod)+(\ze+4\eta)\otimes \b
\eeaa
becomes (note that    $\rhod=0$),
\bea
e_3 \a+\frac 1 2 \kab \a&=&(e_\th(\b)- (e_\th \Phi)\b)+4\omb \a - \frac  3 2 \vth \rho  +(\ze+4\eta) \b.
\eea

The spacetime equation,
\beaa
\nabb_4 \b+ 2\trch  \b&=&\divv\a -2\om \b +(2\ze+\etab)\c \a +3(\xi \rho +\dual\xi \, \rhod)
\eeaa
becomes,
\bea
e_4\b +2\ka \b &=&( e_\th \a+2(e_\th \Phi)\a)- 2\om \b+ (2\ze+\etab) \a+ 3\xi \rho.
\eea

The spacetime equation,
\beaa
\nabb_3 \b+\trchb \b &=&\DDs_1(-\rho, \rhod)+2\chih \c\bb +2\omb\, \b +\xib\c \a+3(\eta \rho +\dual\eta\,   \rhod)
\eeaa
becomes (recall $\bbS_\th=-\bb$),
\bea
e_3 \b+ \kab \b &=&e_\th(\rho) +2\omb \b   + 3\eta \rho- \vth \bb +\xib \a. 
\eea

The spacetime equation,
\beaa
e_4 \rho+\frac 3 2 \trch \rho&=&\divv\b -\frac 1 2 \chibh \c \a +\ze\c\b +2(\etab \c\b-\xi\c\bb)
\eeaa
becomes,
\bea
e_4 \rho+\frac 3 2 \ka \rho&=&(e_\th(\b)  +(e_\th \Phi)\b)  -\frac 1 2 \vthb \, \a +\ze\, \b +2(\etab \,\b+ \xi\,\bb).
\eea
Indeed note that,
\beaa
\chibhS \c \aS=2 \chibhS_{\th\th} \aS_{\th\th}= \vthb \a.
\eeaa

All other  equations in the  proposition below are derived  using the $e_3-e_4$ symmetry. We summarize the results  in the following proposition.
\begin{proposition}
\label{prop:reduced-Bianchi}
\bea
\begin{split}
e_3 \a+\frac 1 2 \kab \a&=-\dds_2\,\b+4\omb \a -\frac 3 2  \vth \rho  +(\ze+4\eta) \b,\\
e_4\b +2\ka \b &=\ddd_2\a- 2\om \b+ (2\ze+\etab) \a+ 3\xi \rho,\\
e_3 \b+ \kab \b &=-\dds_1\rho +2\omb \b   + 3\eta \rho- \vth \bb +\xib \a, \\
e_4 \rho+\frac 3 2 \ka \rho&=\ddd_1 \b  -\frac 1 2 \vthb \, \a +\ze\, \b +2(\etab \,\b+ \xi\,\bb),
\\
\\
e_3 \rho+\frac 3 2 \kab \rho&=\ddd_1\bb  -\frac 1 2 \vth \, \aa - \ze\, \bb +2(\eta \,\bb+ \xib\,\b),\\
e_4 \bb+ \ka \bb &= -\dds_1\rho +2\om \bb   + 3\etab  \rho- \vthb  \b +\xi \aa,\\
e_3\bb +2\kab \,\bb &= \ddd_2\aa - 2\omb\, \bb+ (-2\ze+\eta) \aa+ 3\xib \rho,\\
e_4 \aa+\frac 1 2 \ka\,  \aa&= -\dds_2\,\bb +4\om \aa -\frac 3 2  \vthb  \rho  +(-\ze+4\etab) \bb.
\end{split}
\eea
\end{proposition}


\subsubsection{Mass aspect functions}


We define     the mass aspect functions,
\bea
\label{def:massaspectfunctions.general}
\bsplit
\mu:&=- \ddd_1 \ze -\rho+\frac 1 4  \vth \vthb,\\
\mub:&=\ddd_1\ze -\rho+\frac 1 4  \vth \vthb.
\end{split}
\eea
One can  derive  useful propagation equations,   in the $e_4$ direction for  $\mu$ and  in the $e_3$  direction for $\mub$   by using the null structure 
 and  null  Bianchi equations, see \cite{Ch-Kl} and \cite{KlNi}.
In the next section  we   will do this    in the context of null-geodesic foliations.


\subsection{Hawking mass}


\begin{definition}
\label{def:Hawking-mass}
The Hawking  mass  $m=m(S)$  of $S$  is    defined by the formula, 
  \bea
\frac{2m}{r}=1+\frac{1}{16\pi}\int_{S_{}}\ka \kab.
\eea
\end{definition}

\begin{proposition}\label{prop:identityaverage-forwgeodfoliation}
The following identities hold true.
\begin{enumerate}
\item 
The average of  $\rho$      is given by the formulas,
\bea
\label{identity:overlinerho}
\overline{\rho}  &=& -\frac{2m}{r^3} + \frac{1}{16\pi r^2}\int_S\vth\vthb.
\eea

\item The average of the mass aspect function is,
\bea
\overline{\mu}&=&\overline{\mub}=\frac{2m}{r^3}.
\eea

\item The average of  $\ka$ and  $\kab$  are related by,
\bea\lab{identity:overlinekaandoverlinekab}
\overline{\ka}\,\overline{\kab} &=& -\frac{4\Up}{r^2}-\overline{\check{\ka}\check{\kab}}
\eea
where $\Up=1-\frac{2m}{r}$.
\end{enumerate}
\end{proposition}

\begin{proof}
We have from the Gauss equation
\beaa
K &=& -\frac{1}{4}\ka\kab + \frac{1}{4}\vth\vthb -\rho.
\eeaa
Integrating on $S$ and using the Gauss Bonnet formula, we infer
\beaa
4\pi &=& -\frac{1}{4}\int_S\ka\kab + \frac{1}{4}\int_S\vth\vthb -\int_S\rho.
\eeaa
Together with the definition of the Hawking mass, we infer
\beaa
\int_S\rho  &=& -4\pi\left(1+ \frac{1}{16\pi}\int_S\ka\kab\right) + \frac{1}{4}\int_S\vth\vthb\\
&=& -\frac{8\pi m}{r} + \frac{1}{4}\int_S\vth\vthb
\eeaa
and hence
\beaa
\overline{\rho}  &=& -\frac{2m}{r^3} + \frac{1}{16\pi r^2}\int_S\vth\vthb.
\eeaa
which proves  our first identity.
The second identity follows easily  from the definition of  $\mu, \mub$  and the  first formula.
Thus, for example,
\beaa
  \overline{\mu} =\frac{1}{|S|}    \int_S \mu&=& \frac{1}{|S|} \int_S\left( -\ddd_1 \ze-\rho+ \frac{1}{4}\vth\vthb\right)=-\overline{\rho}  +\frac{1}{4|S|}  \int_S \vth\vthb =\frac{2m}{r^3}.
\eeaa
To prove the last identity we  remark that, in view of the definition of the Hawking mass,
\beaa
-\Up= \frac{2m}{r} -1&=&\frac{1}{16\pi}\int_S\ka\kab= \frac{1}{16\pi}\left(|S|\overline{\ka}\,\overline{\kab}+\int_S\check{\ka}\check{\kab}\right)
\eeaa
and hence
\beaa
\overline{\ka}\,\overline{\kab} &=& -\frac{16\pi\Up}{|S|}-\frac{1}{|S|}\int_S\check{\ka}\check{\kab}\\
&=& -\frac{4\Up}{r^2}-\overline{\check{\ka}\check{\kab}}.
\eeaa
This concludes the proof of the proposition.
\end{proof}


\subsection{Outgoing Geodesic   foliations}\lab{sec:mainequationsforoutgoiggeodesicfoliations}


We     restrict our attention to geodesic foliations, i.e. geodesic foliations  by    $\Z$ invariant optical functions.


\subsubsection{Basic definitions}


Assume  given an outgoing  optical function $u$, i.e.   $\Z$-invariant  solution of the equation,
\beaa
\g^{\a\b}\pr_\a u\pr_\b u =g^{ab}\pr_a u \pr_b u =0
\eeaa
and $L=-g^{ab} \pr_b u \pr_a  $ its  null geodesic generator. We choose $e_4$ such that,
\bea
e_4=\vsi L, \qquad  L(\vsi)=0.
\eea

\begin{remark}
In our definition of  a GCM  admissible spacetime, see  section  \ref{section:GCMspacetime},   we initialize   $\vsi$ on the spacelike hypersurface $\Si_*$. 
\end{remark}

 We then choose $s$    such that 
 \bea
 e_4 (s)=1.
 \eea
The functions $u, s$ generate  what is called an  outgoing  geodesic foliation.  Let $S_{u,s}$    be       the    $2$-surfaces  of intersection    between the level surfaces of $u$ and $s$. We  choose  $e_3 $  the unique  $\Z$-invariant  null vectorfield orthogonal to  $S_{u,s}$   and such that $g(e_3, e_4)=-2$. We  then  let $e_\th$  to be unit   tangent to $S_{u, s}$, $\Z$-invariant and  orthogonal to $\Z$.   We also  introduce
 \bea 
 \underline{\Omega} := e_3(s).
 \eea
 
 \begin{lemma}
 \label{le:outgoinggeodesicfoliation}
We have
\bea
\om=\xi=0,  \qquad  \etab = -\ze,
\eea
\bea\lab{geodesic-foliation-M4}
\bsplit
 \vsi&=\frac{2}{e_3(u)},\\
  e_4(\vsi) &=0,\\
  e_\th(\log\vsi) &= \eta-\ze,\\
 e_\th(\Omb)&=-\xib- (\eta-\ze)\Omb,\\
  e_4(\underline{\Omega})&=-2\omb.
 \end{split}
\eea
\end{lemma}

\begin{proof}
Recall that $L$ is geodesic, $e_4=\vsi L$ and $L(\vsi)=0$. This immediately implies that $e_4$  is geodesic, and hence we have 
\beaa
\om=\xi=0.
\eeaa
Applying the vectorfield  
\beaa
[e_4, e_\th]=(\etab+\ze) e_4+\xi e_3 -\chi e_\th
\eeaa
 to $s$, and since $e_4(s)=1$ and $e_\th(s)=0$, we derive,
 \beaa
 \etab+\ze=0.
 \eeaa 
 
Next, note that
\beaa
e_3(u)=g(e_3, -L)=-\vsi^{-1}  g(e_3,e_4)=\frac{2}{\vsi}
\eeaa
and hence
\beaa
\vsi &=& \frac{2}{e_3(u)}.
\eeaa
Applying the vectorfield  
\beaa
[e_3, e_\th]=\xib e_4+(\eta-\ze)e_3-\chib e_\th
\eeaa
to $u$  and making use of the relation  $e_4(u)=e_\th(u)=0$  we deduce,
\beaa
(\eta-\ze)e_3(u)= e_3(e_\th u)- e_\th e_3(u)=- e_\th e_3(u)
\eeaa
which together with the identity $\vsi=2/e_3(u)$ yields
\beaa
\eta-\ze&=&-e_\th \log (e_3 u)= -e_\th \log\left(\frac{2}{\vsi} \right)=e_\th(\log\vsi)
\eeaa
and hence
\beaa
e_\th(\log\vsi) &=& \eta-\ze.
\eeaa

Applying the vectorfield  
\beaa
[e_3, e_\th]=\xib e_4+(\eta-\ze)e_3-\chib e_\th
\eeaa
 to $s$ we deduce, since $e_4(s)=1$, $e_\th(s)=0$ and $e_3(s)=\Omb$,
\beaa
e_\th (\Omb)&=& -\xib-(\eta-\ze) \Omb. 
\eeaa

Finally applying 
\beaa
[e_4,e_3]=   -2\omb  e_4- 2 (\eta-\etab) e_\th + 2\om e_3 
\eeaa
to $s$, and using $e_4(s)=1$ and $e_\th(s)=0$,  we infer   $e_4(e_3(s))=-2\omb$,  i.e. $e_4(\underline{\Omega})=-2\omb$ as desired. 
\end{proof}

\begin{remark}
In the particular case when $\vsi$ is constant we   have $\eta=\ze=-\etab$.  In   Schwarzschild, relative  to     the standard outgoing geodesic frame,  we  have
\beaa
\vsi=1, \qquad  \underline{\Omega}=-\Up=-\left(1-\frac{2m}{r}\right).
\eeaa
\end{remark}


\subsubsection{Basic equations}


 \begin{proposition}
 \label{propos:basiceqts-geod}
  Relative to an outgoing geodesic foliation   we have
 \begin{enumerate}
 \item The  reduced null structure equations   take the form,
  \beaa
\begin{split}
e_4(\vth)+\ka \, \vth  &=-2\a, \\
e_4(\ka)+\frac 12 \ka^2  &= -\frac 1 2 \vth^2,\\
e_4 \ze + \ka\ze &=-\b -   \vth\ze,\\
e_4(\eta-\ze)+\frac{1}{2}\ka(\eta-\ze) &=   -\frac 1 2 \vth (\eta-\ze),\\
e_4\vthb +\frac 12 \ka\, \vthb  &=2\dds_2\,\ze -\frac 12 \kab  \,\vth+2\ze^2,\\
e_4(\kab)+\frac 1 2 \ka\, \kab  &= -2\ddd_1\ze + 2\rho -\frac 1 2  \vth\, \vthb +2\ze^2,\\
e_4 \omb &=\rho +\ze(2\eta+\ze),\\
e_4(\xib) &= - e_3(\ze) +\bb -\frac 12 \kab (\ze+\eta)-\frac 1 2 \vthb (\ze+\eta),\\
\end{split}
\eeaa
and
\beaa
\bsplit
e_3(\vthb)+\kab \, \vthb  + 2\omb\,  \vthb &=-2\aa-2\dds_2\,\xib  +2(\eta-3\ze)\,  \xib,\\
e_3(\kab)+\frac 12 \kab^2 +2 \omb \,\kab &=2\ddd_1\xib+2(\eta-3\ze)  \xib -\frac 1 2 \vthb^2,\\
e_3 \ze +\frac 1  2 \kab (\ze+\eta) -2\omb(\ze-\eta)&=\bb +2\dds_1\,\omb+\frac 12 \ka \,\xib  -\frac  1 2 \vthb(\ze+\eta)+\frac 1 2 \vth \,\xib,\\
e_3\vth +\frac 12 \kab\, \vth - 2\omb \vth &=-2\dds_2\,\eta -\frac 12 \ka \,\vthb+2\eta^2,\\
e_3(\ka)+\frac 1 2 \kab\, \ka -2\omb \ka &= 2\ddd_1\eta + 2\rho -\frac 1 2  \vthb\, \vth +2\eta^2,
\end{split}
\eeaa
and
\beaa
\ddd_2\vthb &=& -2\bb -\dds_1\,\kab -\ze \kab  + \vthb \,  \ze,\\
\ddd_2\vth &=& -2\b -\dds_1\,\ka + \ze\ka -   \vth \, \ze,\\
K &=& -\rho -\frac 1 4 \ka\,\kab +\frac 1  4 \vth\, \vthb.
\eeaa
\item
The  null    Bianchi identities are given in this case by
\beaa
\begin{split}
e_3 \a+\frac 1 2 \kab \a&=-\dds_2\,\b+4\omb \a -\frac 3 2  \vth \rho  +(\ze+4\eta) \b,\\
e_4\b +2\ka \b &=\ddd_2\a+ \ze\a,\\
e_3 \b+ \kab \b &=-\dds_1\rho +2\omb \b   + 3\eta \rho- \vth \bb +\xib \a, \\
e_4 \rho+\frac 3 2 \ka \rho&=\ddd_1 \b  -\frac 1 2 \vthb \, \a -\ze \b,
\end{split}
\eeaa
\beaa
\bsplit
e_3 \rho+\frac 3 2 \kab \rho&=\ddd_1\bb  -\frac 1 2 \vth \, \aa - \ze\, \bb +2(\eta \,\bb+ \xib\,\b),\\
e_4 \bb+ \ka \bb &= -\dds_1\rho    - 3\ze \rho- \vthb  \b,\\
e_3\bb +2\kab \,\bb &= \ddd_2\aa - 2\omb\, \bb+ (-2\ze+\eta) \aa+ 3\xib \rho,\\
e_4 \aa+\frac 1 2 \ka\,  \aa&= -\dds_2\,\bb  -\frac 3 2  \vthb  \rho  -5\ze \bb.
\end{split}
\eeaa
\item The mass aspect function  $\mu =-\ddd_1\ze-\rho+\frac 1 4 \vth\vthb $,  defined           in \eqref{def:massaspectfunctions.general} verifies the transport equation,
\beaa
e_4(\mu) +\frac{3}{2}\ka\mu &=&\err[e_4\mu],\\
\err[e_4\mu]:&=&  \frac{1}{2}\ka\ze^2 +e_\th(\ka)\ze+\ddd_1(\vth\ze) -\frac{1}{8}\kab\vth^2.
\eeaa
\end{enumerate}
 \end{proposition}

 \begin{proof}
 Concerning  the  null structure equations 
we only need to derive the equation for $\eta-\ze$. According  to Proposition \eqref{prop:null.structure-general}
 we have,
 \beaa
 e_3(\xi) - e_4(\eta)&=\b + 4\omb \xi+\frac 12 \ka  (\eta-\etab)+\frac 1 2 \vth (\eta-\etab)
 \eeaa
 which becomes
 \beaa
 e_4 \eta &=&-\b -\frac 12 \ka (\eta-\etab)-\frac 1 2 \vth(\eta-\etab)
 \eeaa
 and,
 \beaa
 -e_4 \ze +\frac 1  2 \ka (-\ze+\etab) +2\om(\ze+\etab)&=\b +2\dds_1\om+2\omb \xi+\frac 12 \kab \,\xi  -\frac  1 2 \vth(-\ze+\etab)-\frac 1 2 \vthb \,\xi
 \eeaa
 which becomes,
 \beaa
  e_4 \ze &=& - \ka\ze -\b - \vth\ze.
 \eeaa
 Hence,
 \beaa
 e_4(\ze-\eta)&=&-\ka\ze  -\vth \ze+\frac 1 2 \ka(\eta-\etab) +\frac 1 2 \vth(\eta-\etab)\\
 &=&\ka\left(-\ze+\frac 1 2 (\eta-\etab) \right)+\vth\left(-\ze+\frac 1 2 (\eta-\etab) \right).
 \eeaa
 Since $\ze=-\etab$ we deduce $-\ze+\frac 1 2 (\eta-\etab) =\frac 1 2 (-\ze+\eta)$ and  thus,
 \beaa
 e_4(\ze-\eta)&=&-\ka(\ze-\eta)-\vth(\ze-\eta)
 \eeaa
 as desired.

The Bianchi equations  equations follow immediately from the general equations derived in the previous section.
It only remains to check   the equation verified by the mass aspect function $\mu$.  We have
\beaa
e_4(\mu) &=& -[e_4, \ddd_1]\ze -\ddd_1e_4(\ze)-e_4(\rho)+\frac{1}{4}e_4(\vth\vthb)\\
&=& \frac{1}{2}\ka\ddd_1\ze -\frac{1}{2}\vth\dds_2\ze+e_4(\Phi)\ze^2-\b\ze -\ddd_1(- \ka\ze -\b -\vth\ze)\\
&& -\left(-\frac 3 2 \ka \rho+\ddd_1 \b  -\frac 1 2 \vthb \, \a -\ze \b\right)\\
&&+\frac{1}{4}\vth \left(-\frac 12 \ka\, \vthb +2\dds_2\,\ze -\frac 12 \kab  \,\vth+2\ze^2\right)+\frac{1}{4}\vthb\left(-\ka \, \vth -2\a\right)\\
&=& \frac{3}{2}\ka\left(\ddd_1\ze+\rho-\frac{1}{4}\vth\vthb\right) -\frac{1}{2}\vth\dds_2\ze+\frac{1}{2}(\ka-\vth)\ze^2 +e_\th(\ka)\ze +\ddd_1(\vth\ze)\\
&& +\frac{1}{4}\vth \left(2\dds_2\,\ze - \frac 12 \kab\vth+2\ze^2\right)\\
\eeaa
and hence
\beaa
e_4(\mu) +\frac{3}{2}\ka\mu &=&  \frac{1}{2}\ka\ze^2 +e_\th(\ka)\ze+\ddd_1(\vth\ze) -\frac{1}{8}\kab\vth^2
\eeaa
as desired. This concludes the proof of the proposition. 
\end{proof}


 \subsubsection{Transport equations for $S$-averages}


\begin{proposition}
\label{prop:outgoinggeod-e3e4averages}\lab{prop:outgoinggeod-e3e4averages-M4}
For any scalar function $f$, we have,
\bea
\bsplit
e_4\left(\int_Sf\right) &= \int_S(e_4(f)+\ka f),\\
e_3\left(\int_Sf\right) &= \int_S(e_3(f)+\kab f)+\err\left[e_3\left(\int_S f\right)\right],
\end{split}
\eea
where the error term is given by the formula
\beaa
\err\left[e_3\left(\int_S f\right)\right]:&=&-\vsi^{-1} \check{\vsi} \int_S(e_3(f)+\kab f) +\vsi^{-1}\int_S \check{\vsi}(e_3(f)+\kab f) \\
 &+&\left(\Obc +\varsigma^{-1}\ov{\Omb}\check{\varsigma}\right)\int_S(e_4f +\ka f) 
- \varsigma^{-1}\ov{\Omb}\int_S\check{\varsigma}(e_4f +\ka f) \\
&-&\varsigma^{-1}\int_S\Obc\, \varsigma(e_4f +\ka f).
\eeaa

In particular, we have 
\bea
\bsplit
e_4(r) &= \frac{r}{2}\overline{\ka},\qquad
e_3(r) = \frac{r}{2}\left( \ov{\kab}+\Ab\right)
\end{split}
\eea
where
\bea\lab{eq:defininitionofAb}
\Ab:&=&-\vsi^{-1}\ov{\kab}\check{\vsi} +\ov{\ka}\left(\Obc +\vsi^{-1}\ov{\Omb}\check{\vsi}\right) + \vsi^{-1}\ov{\check{\vsi}\kabc}  -\vsi^{-1}\ov{\Omb}\,\ov{\check{\vsi}\kac}  -\vsi^{-1}\ov{\Obc\vsi\ka}.
\eea
\end{proposition}

\begin{proof}
See section \ref{sec:proofofprop:outgoinggeod-e3e4averages}.
\end{proof}

\begin{corollary}
\label{corr:transportintSfePhi}
For a reduced scalar $f$, we have
\beaa
e_4\left(\int_Sfe^\Phi\right) &=& \int_S\left(e_4(f)+\left(\frac{3}{2}\ka-\frac{1}{2}\vth\right)f\right)e^\Phi
\eeaa
and
\beaa
e_3\left(\int_Sfe^\Phi\right) &=& \int_S\left(e_3(f)+\left(\frac{3}{2}\kab-\frac{1}{2}\vthb\right)f\right)e^\Phi +\err\left[e_3\left(\int_S fe^\Phi\right)\right].
\eeaa
\end{corollary}

\begin{proof}
In view of Proposition \ref{prop:outgoinggeod-e3e4averages}, we have
\beaa
e_4\left(\int_Sfe^\Phi\right) &=& \int_S\Big(e_4(fe^\Phi)+\ka fe^\Phi\Big)\\
&=& \int_S\Big(e_4(f)+(\ka+e_4\Phi) f\Big)e^\Phi\\
&=& \int_S\left(e_4(f)+\left(\frac{3}{2}\ka-\frac{1}{2}\vth\right)f\right)e^\Phi
\eeaa
as desired.

Also, using again Proposition \ref{prop:outgoinggeod-e3e4averages}, we have
\beaa
e_3\left(\int_Sfe^\Phi\right) &=& \int_S\Big(e_3(fe^\Phi)+\kab fe^\Phi\Big)+\err\left[e_3\left(\int_S fe^\Phi\right)\right]\\
&=&  \int_S\Big(e_3(f)+(\kab+e_3\Phi) f\Big)e^\Phi +\err\left[e_3\left(\int_S fe^\Phi\right)\right]\\
&=& \int_S\left(e_3(f)+\left(\frac{3}{2}\kab-\frac{1}{2}\vthb\right)f\right)e^\Phi +\err\left[e_3\left(\int_S fe^\Phi\right)\right]
\eeaa
as desired.
\end{proof}

\begin{corollary}
\label{corr:transportcheckf}
Given a scalar function $f$ we have,
\bea
\bsplit
e_4(\overline{f}) &= \overline{e_4(f)}+\overline{\check{\ka\,}\check{f}},\\
e_4(\check{f}) &=e_4(f) -\overline{e_4(f)} - \overline{\check{\ka\,}\check{f}},
\end{split}
\eea
and
\bea
\bsplit
e_3\left(\overline{f}\right) &= \overline{e_3(f)}+\err[e_3\ov{f}],\\
e_3(\check{f}) &=e_3(f) -\overline{e_3(f)}-\err[e_3(\ov{f})],
\end{split}
\eea
where,
\bea
\bsplit
\err[e_3( \ov{f})]&=-\vsi^{-1} \check{\vsi} \, \left( \ov{e_3f+\kab f} -\ov{\kab} \ov{f} \right)+ \vsi^{-1} \,\left(  \ov{\check{\vsi}(e_3f+\kab f)}- \ov{\check{\vsi}\kabc} \, \ov{f}\right)\\
&+\left(\Obc +\vsi^{-1}\ov{\Omb}\check{\vsi}\right) \,\left( \ov{e_4f +\ka f)}-\ov{\ka}\, \ov{f}\right) -\vsi^{-1}\ov{\Omb}\, \left(\ov{ \check{\vsi}(e_4f +\ka f)}- \ov{\check{\vsi} \kac}     \,\ov{f} \right)\\
&-\vsi^{-1}\left(  \ov{\Obc\vsi(e_4f +\ka f)}  -  \ov{\Obc\vsi \,\ka}\ov{f}   \right)+\ov{\kabc \check{f}}.
\end{split}
\eea
\end{corollary}

\begin{proof}
We have, recalling Lemma \ref{lemma:averagesorproducts} and $|S|=4\pi r^2$,
\beaa
e_4(\overline{f}) &=& e_4\left(\frac{\int_Sf}{|S|}\right)=  \frac{1}{|S|}\int_S(e_4(f)+\ka f)-\frac{e_4(|S|)}{|S|} \, \ov{f}   =   
  \ov{e_4(f)+\ka f}    -2\frac{e_4 r}{r}\ov{f}\\
&=&\overline{e_4(f)}+\overline{\ka\,  f}-\overline{\ka\,}\overline{f}= \overline{e_4(f)}+\overline{\check{\ka\,}\check{f}}.
\eeaa
This also yields
\beaa
e_4(\check{f}) &=& e_4(f) - e_4(\overline{f})= e_4(f) -\overline{e_4(f)} - \overline{\check{\ka}\,\check{f}}
\eeaa
as desired. 

Similarly,
\beaa
e_3(\overline{f}) &=& e_3\left(\frac{\int_Sf}{|S|}\right)=  \frac{1}{|S|} e_3\left( \int_\S f\right) - \frac{2 e_3(r)}{r} \ov{f}\\
&=&  \frac{1}{|S|} \int_S( e_3 f +\kab f ) +\frac{1}{|S|}\err\left[e_3\left(\int_S f\right)\right] - (\ov{\kab} +\Ab)\ov{f}\\
&=&\ov{e_3(f)} +\ov{\kab f}-\ov{\kab} \ov{f} + \frac{1}{|S|}\err\left[e_3\left(\int_S f\right)\right] -\Ab \ov{f}\\
&=&\ov{e_3(f)} +\ov{\kabc \check{f}}  +\frac{1}{|S|}\err\left[e_3\left(\int_S f\right)\right] -\Ab \ov{f}.
\eeaa
 We deduce,
 \beaa
 e_3(\overline{f}) &=& \ov{ e_3 (f)   }+\err[e_3( \ov{f})]
 \eeaa
where, recalling the definitions of  the error terms $\err\left[e_3\left(\int_S f\right)\right]$ and $\Ab$,
\beaa
\err[e_3( \ov{f})]&=& \ov{\kabc \check{f}}+\frac{1}{|S|}\err\left[e_3\left(\int_S f\right)\right] -\Ab \,\ov{f}\\
&=& \ov{\kabc \check{f}} -\vsi^{-1} \check{\vsi} \, \ov{e_3f+\kab f} + \vsi^{-1} \, \ov{\check{\vsi}(e_3f+\kab f)} +\left(\Obc +\vsi^{-1}\ov{\Omb}\check{\vsi}\right) \, \ov{e_4f +\ka f}\\
&-&\vsi^{-1}\ov{\Omb}\, \ov{ \check{\vsi}(e_4f +\ka f)} -\vsi^{-1}\, \ov{\Obc\vsi(e_4f +\ka f)}\\
&-&\ov{f}\left(-\vsi^{-1}\ov{\kab}\check{\vsi}+ \vsi^{-1}\ov{\check{\vsi}\kabc}   +\ov{\ka}\left(\Obc +\vsi^{-1}\ov{\Omb}\check{\vsi}\right)-\vsi^{-1}\ov{\Omb}\,\ov{\check{\vsi}\kac} -\vsi^{-1}\ov{\Obc \vsi\ka}\right),\\
\eeaa
i.e.,
\beaa
\err[e_3( \ov{f})]&=&  \ov{\kabc \check{f}} -\vsi^{-1} \check{\vsi} \, \left( \ov{e_3f+\kab f} -\ov{\kab} \ov{f} \right)+ \vsi^{-1} \,\left(  \ov{\check{\vsi}(e_3f+\kab f)}- \ov{\check{\vsi}\kabc} \, \ov{f}\right)\\
&+&\left(\Obc +\vsi^{-1}\ov{\Omb}\check{\vsi}\right) \,\left( \ov{e_4f +\ka f)}-\ov{\ka}\, \ov{f}\right) -\vsi^{-1}\ov{\Omb}\, \left(\ov{ \check{\vsi}(e_4f +\ka f)}- \ov{\check{\vsi} \kac}     \,\ov{f} \right)\\
&-&\vsi^{-1}\left(  \ov{\Obc\vsi(e_4f +\ka f)}  -  \ov{\Obc\vsi\ka }\ov{f}  \right)
\eeaa
as stated. Finally
\beaa
e_3 (\check f)&=& e_3 f-e_3(\ov{f})= e_3 f- \ov{ e_3 (f)   }- \err[e_3\ov{f}]
\eeaa 
which ends the proof of the corollary.
\end{proof}

The following is  also an immediate application of  Proposition \ref{prop:outgoinggeod-e3e4averages}.
 \begin{corollary}
 \label{le:transportrp-f}   If   $f$ verifies the scalar equation
 \beaa
 e_4 (f) + \frac p 2  \ov{\ka}   f=F,
 \eeaa
 then,
 \beaa
 e_4(r^p f)&=&         r^p F. 
 \eeaa
 \end{corollary}


\subsubsection{Commutation identities revisited}


We  revisit  the general commutation identities of Lemma \ref{Le:comme3e4}
in   an outgoing  geodesic foliation.   
\begin{lemma}
\label{Le:comme3e4-outgeodesic}
The following commutation formulae holds true,
\begin{enumerate}     
 \item If $f\in \mathfrak{s}_k$,
 \bea
\bsplit
 \,[r\ddd_k, e_4]f &=r\left[ \com_k(f) + \frac 1 2 \kac \ddd_k  f \right],\\
 \,[r\ddd_k, e_3]f&= r\left[ \comb_k(f) + \frac 1 2(- \Ab+\kabc) \ddd_k  f \right].
 \end{split}
 \eea
 
\item If  $f\in \mathfrak{s}_{k-1}$

\bea
\bsplit
\,[r\dds_k, e_4]f &=r\left[\com^*_k(f)  + \frac 1 2 \kac  \dds_k f   \right],   \\
\,[r \dds_k, e_3]f &= r\left[\comb^*_k(f)  + \frac 1 2(- \Ab+\kabc)  \dds_k f   \right].       
\end{split}
\eea
\end{enumerate}
Also, we have
\beaa
\comb_k(f)&=&- \frac 1 2 \vthb  \dds_{k+1} f + (\ze-\eta) e_3 f - k \eta  e_3\Phi f -\xib( e_4  f  +k e_4(\Phi)  f )- k \bb f,\\
\com_k(f)&=&- \frac 1 2 \vth  \dds_{k+1} f  + k \ze  e_4\Phi f - k \b f,\\
\comb^*_k(f)&=&- \frac 1 2 \vthb  \ddd_{k-1} f - (\ze-\eta) e_3 f  -(k-1)  \eta  e_3\Phi f +\xib(e_4  f  -(k-1)  e_4(\Phi)  f )\\
&- &(k-1) \bb f,\\
\com^*_k(f)&=&- \frac 1 2 \vth  \ddd_{k-1} f  + (k-1)  \ze  e_4\Phi f - (k-1)  \b f.
\eeaa
\end{lemma}

\begin{proof}
We make use of  the commutation   Lemma  \ref{Le:comme3e4} and the definition of  $ \Ab$, see Proposition \ref{prop:outgoinggeod-e3e4averages},  to write, for $f\in \mathfrak{s}_k$,
\beaa
\,[r\ddd_k, e_4]f&=&r[\ddd_k, e_4]f - e_4(r)\ddd_k  f \\
&=&\frac 1 2r\ka \ddd_k f +r\com_k(f)- \frac r 2 \ov{\ka} \ddd_k f\\
&=& r\left[ \com_k(f)+\frac 1 2 \kac \ddd_k  f \right]\\
\,[r\ddd_k, e_3]f&=&r[\ddd_k, e_4]f - e_3(r) \ddd_k f \\
&=&\frac 1 2r\kab  \ddd_k f +r\comb_k(f) -  \frac{r}{2}(\Ab+\ov{\kab}) \ddd_k f\\
&=& r\left[ \comb_k(f) + \frac 1 2(- \Ab + \kabc)\ddd_k  f \right].
\eeaa
The remaining formulae are proved in the same manner. Also, the form of $\comb_k(f)$, $\com_k(f)$, $\comb^*_k(f)$ and $\com^*_k(f)$ follows from 
Lemma  \ref{Le:comme3e4} and the fact that we have $\xi=\etab+\ze=0$ in an outgoing geodesic foliation.
\end{proof}

We also record here  for future use the following lemma.
\begin{lemma}
\label{lemma:commTwithe3e4}
Let $\T=\frac 12 \left(e_3 +\Up e_4\right)$, with $\Up=1-\frac{2m}{r} $.  We have,
\bea
\begin{split}
\,[\T, e_4]&=\left(\left( \omb-\frac{m}{r^2} \right)     -\frac{m}{2r} \left( \ov{\ka} -\frac{2}{r}\right)+\frac{e_4 (m)}{r}  \right)e_4+(\eta+\ze) e_\th,\\
\,[\T, e_3]&=\left(-\Up\left(\omb-\frac{m}{r^2}\right)      -\frac{m}{2r}   \left(\ov{\kab} +\frac{2\Up}{r}\right) -\frac{m}{2r} \Ab  +\frac{e_3(m)}{r}\right) e_4 -(\eta+\ze)\Up e_\th.
\end{split}
\eea
\end{lemma}

\begin{proof}
Recall that $[e_3, e_4]= 2\omb e_4+ 2(\eta+\ze) e_\th$. Thus,
\beaa
\,[\T, e_4]&=&\frac 12 [e_3 +\Up e_4, e_4]=\frac 1 2\left(  2\omb e_4+ 2(\eta+\ze) e_\th   -e_4\left(1-\frac{2m}{r} \right) e_4\right)\\
&=&\left( \omb-\frac{m}{r^2} e_4(r) +\frac{e_4 (m)}{r}  \right)e_4+(\eta+\ze )e_\th\\
&=&\left(\left( \omb-\frac{m}{r^2} \right)     -\frac{m}{r^2} ( e_4(r)-1) +\frac{e_4 (m)}{r}  \right)e_4+(\eta+\ze) e_\th\\
&=&\left(\left( \omb-\frac{m}{r^2} \right)     -\frac{m}{r^2} \left(\frac{r}{2} \ov{\ka} -1\right)+\frac{e_4 (m)}{r}  \right)e_4+(\eta+\ze ) e_\th\\
&=&\left(\left( \omb-\frac{m}{r^2} \right)     -\frac{m}{2r } \left( \ov{\ka} -\frac{2}{r}\right)+\frac{e_4 (m)}{r}  \right)e_4+(\eta+\ze ) e_\th
\eeaa
and,
\beaa
\,[\T, e_3]&=&\frac 12 [e_3 +\Up e_4, e_3]= \frac 12\left(\Up \left( -2\omb e_4-2(\eta+\ze) e_\th \right)- e_3\left(1-\frac{2m}{r}\right) e_4\right)\\
&=&\left(-\Up\omb -\frac{m}{r^2} e_3(r) +\frac{e_3(m)}{r}\right) e_4 -\Up(\eta+ \ze )e_\th\\
&=&\left(-\Up\left(\omb-\frac{m}{r^2}\right)     -\Up\frac{m}{r^2}  -\frac{m}{r^2}  \frac r 2 \left(\ov{\kab} +\Ab   \right) +\frac{e_3(m)}{r}\right) e_4  -\Up(\eta+ \ze )e_\th\\
&=&\left(-\Up\left(\omb-\frac{m}{r^2}\right)      -\frac{m}{2r}   \left(\ov{\kab} +\frac{2\Up}{r}\right) -\frac{m}{2r} \Ab +\frac{e_3(m)}{r}\right) e_4  -\Up(\eta+ \ze )e_\th
\eeaa
which concludes the proof of the lemma.
\end{proof}

\begin{remark}
\label{Remark:ze-th- f}
When applying the formulas of Lemma \ref{lemma:commTwithe3e4} to a $k$ reduced  scalar $f\in \mathfrak{s}_k$,  the term $ (\eta+ \ze)e_\th (f)$  should correspond to a  reduced scalar. In fact, recalling  Remark \ref{Remark:$ethf$},
  we can write, 
  \beaa
  \ze e_\th(f) &=& \frac 1 2\ze  \left(\ddd_k f-\dds_{k+1} f\right)
  \eeaa
  which can indeed be shown to be a $k$-reduced scalar in $\mathfrak{s}_k$.
\end{remark}


\subsubsection{Derivatives of the Hawking mass}


\begin{proposition}[Derivatives of the Hawking mass]\label{prop:derivativesHawkingmass}
We have the following identities for the Hawking mass,
\bea
e_4(m) &=&  \frac{r}{32\pi}\int_S\err_1,
\eea
and
\bea
\nn e_3(m)&=&   \left(1- \vsi^{-1} \check{\vsi}\right)\frac {r}{32\pi}\int_S\underline{\err}_1+\left(\Obc +\vsi^{-1}\ov{\Omb}\check{\vsi}\right)\frac {r}{32\pi}\int_S\err_1\\
\nn&&+ \vsi^{-1}\frac {r}{32\pi} \int_S\check{\vsi}\,\left(2 \ov{\rho}\kabc+2 \rhoc\ov{\kab}+2 \kab\ddd_1 \eta +2\ka \ddd_1 \xib+\underline{\err}_2\right)\\
\nn&&-\vsi^{-1}\frac{r}{32\pi} \int_S(\ov{\Omb}\check{\vsi}+\Obc\vsi)\left(2 \ov{\rho}\kac+2 \rhoc\ov{\ka}-2\ka\ddd_1\ze+\err_2\right)\\
&&-\frac{m}{r}\vsi^{-1}\left[-\ov{\check{\vsi}\kabc}+  \ov{\Omb}\,\ov{\check{\vsi}\kac } +  \ov{\Obc \vsi \ka}\right],
\eea
where 
\beaa
\err_1 &:=& 2\kac\rhoc+2e_\th(\ka)\ze-\frac{1}{2}\kab\vth^2 -\frac{1}{2}\kac\vth\vthb +2\ka\ze^2,\\ 
\underline{\err}_1 &:=& 2 \rhoc\kabc-2 e_\th(\kab)\eta -2e_\th(\ka)\xib-\frac{1}{2}\kabc\vth\vthb +2\kab\eta^2+2\ka \big(\eta-3\ze\big)\xib-\frac 1 2 \ka \vthb^2,\\
\err_2 &:=& 2\rhoc\kac-\frac{1}{2}\kab\vth^2 -\frac{1}{2}\ka\vth\vthb +2\ka\ze^2,\\
\underline{\err}_2 &:=& 2\rhoc\kabc +\kab\left(2\eta^2-\frac 1 2 \vth \vthb \right)+2\ka \big(\eta-3\ze\big)\xib-\frac 1 2 \ka \vthb^2.
\eeaa
\end{proposition}

\begin{proof}
The proof relies on the definition of the Hawking mass $m$ given by the formula $\frac{2m}{r}=1+\frac{1}{16\pi}\int_S\ka\kab$, Proposition \ref{prop:outgoinggeod-e3e4averages}, and the the null structure equations for $e_4(\ka)$, $e_4(\kab)$, $e_3(\ka)$ and $e_3(\kab)$ provided by Proposition \ref{propos:basiceqts-geod}. We refer to section \ref{sec:proofofprop:derivativesHawkingmass} for the details.
\end{proof}


\subsubsection{Transport equations for  main averaged quantities}


\begin{lemma}\lab{lemma:transportequationforoverlinekaintheoutgoinggeodesicfoliation}
The following equations hold true
\bea
\bsplit
e_4\left(\ov{\ka}-\frac{2}{r}\right)  + \frac{1}{2}\ov{\ka}\left(\ov{\ka}-\frac{2}{r} \right) &=   -\frac{1}{4}\ov{\vth^2}  + \frac{1}{2}\ov{\check{\ka}^2},\\
e_4\left(\ov{\omb}-\frac{m}{r^2}\right) &= \ov{\rho}+\frac{2m}{r^3} +\frac{m}{r^2}\left(\ov{\ka}-\frac{2}{r}\right) -\frac{e_4(m)}{r^2} +3\ov{\ze(2\eta+\ze)}+\ov{\check{\ka}\check{\omb}},
\end{split}
\eea
and
\bea
\nn&& e_3\left(\ov{\ka}-\frac 2 r \right) +\frac 1 2  \ov{ \kab} \left(\ov{\ka}-\frac 2 r \right)\\
\nn &=&  2 \ov{\omb}\,\left( \ov{\ka}-\frac 2 r \right)+\frac 4 r\left(\ov{ \omb}-\frac{m}{r^2}\right)+ 2\left(  \ov{\rho} +\frac{2m}{r^3}\right) -\vsi^{-1}\left( -\frac{1}{2}\ov{\kab}\, \ov{\ka}+2\ov{\omb}\,\ov{\ka}+2\ov{\rho} \right)\check{\vsi}\\
 && -\frac{1}{2}\ov{\ka}^2\left(\Obc +\vsi^{-1}\ov{\Omb}\check{\vsi}\right) -\frac{1}{r}\vsi^{-1}\ov{\kab}\check{\vsi} +\frac{1}{r}\ov{\ka}\left(\Obc +\vsi^{-1}\ov{\Omb}\check{\vsi}\right)+\err\left[e_3\left(\ov{\ka}-\frac 2 r \right)\right],
 \eea
 where,
 \bea
\nn\err\left[e_3\left(\ov{\ka}-\frac 2 r \right)\right] &:=&2\ov{\eta^2}+ 2\ov{\check{\omb}\, \check{\ka}}  -  \frac 1 2\ov{ \kac\,\kabc}    -\frac{1}{2}\ov{\vth\vthb} + \frac{1}{r}\vsi^{-1}\ov{\check{\vsi}\kabc}  -\frac{1}{r}\vsi^{-1}\ov{\Omb}\,\ov{\check{\vsi}\kac}  -\frac{1}{r}\vsi^{-1}\ov{\Obc\vsi\ka}  \\
\nn&&-\vsi^{-1} \check{\vsi} \, \left( \ov{\frac{1}{2}\kac\kabc+2\ombc\kac    -\frac{1}{2}\vth\vthb +2\eta^2}  \right)\\
\nn&&+ \vsi^{-1} \,\left(  \ov{\check{\vsi}\left(\frac{1}{2}\ka\kab+2\omb\ka +2\rhoc +2\ddd_1\eta  -\frac{1}{2}\vth\vthb +2\eta^2\right)}- \ov{\check{\vsi}\kabc} \, \ov{\ka}\right)\\
\nn&&+\left(\Obc +\vsi^{-1}\ov{\Omb}\check{\vsi}\right)\left(\ov{\frac{1}{2}\kac^2 -\frac{1}{4}\vth^2}\right) -\vsi^{-1}\ov{\Omb}\, \left(\ov{ \check{\vsi}\left(\frac{1}{2}\ka^2 -\frac{1}{4}\vth^2\right)}- \ov{\check{\vsi} \kac}     \,\ov{\ka} \right)\\
&&-\vsi^{-1}\left(  \ov{\Obc\vsi\left(\frac{1}{2}\ka^2 -\frac{1}{4}\vth^2\right)}  -  \ov{\Obc\vsi \,\ka} \,\ov{\ka}  \right) +\ov{\kabc\kac}.
\eea
\end{lemma}

\begin{proof}
The proof relies on Corollary \ref{corr:transportcheckf} and the null structure equations for $e_4(\ka)$ and $e_3(\ka)$ provided by Proposition \ref{propos:basiceqts-geod}. We refer to section \ref{sec:proofoflemma:transportequationforoverlinekaintheoutgoinggeodesicfoliation} for the details.
\end{proof}


\subsubsection{Transport equations for  main checked  quantities}


\begin{proposition}[Transport equations for checked quantities]
\label{propos:transportaverages}
We have the following transport equations in the  $e_4$ direction,
\bea
\begin{split}
e_4\check{\ka}+\ov{\ka}\, \check{\ka}&=\err[e_4\check{\ka}],\\
\err[e_4\check{\ka}]:&=-\frac 1 2 \check{\ka}^2 -\frac 1 2 \ov{\check{\ka}^2} -\frac 1 2 (\vth^2-\ov{\vth^2}),\\
e_4\check{\kab} +\frac1 2 \ov{\ka} \check{\kab}+\frac 1 2 \check{\ka} \ov{\kab} &=-2 \ddd_1\ze+2\check{\rho}+\err[e_4\check{\kab}],\\
\err[e_4\check{\kab}]:&=-\frac 1 2 \check{\ka}\check{\kab}-\frac 1 2 \ov{\check{\ka}\check{\kab}} +\left( -\frac 1 2  \vth\vthb +2\ze^2\right)        -\ov{\left( -\frac 1 2  \vth\vthb +2\ze^2\right) },\\
e_4\check{\omb}&=\check{\rho}+\err[e_4\check{\omb}],\\
\err[e_4\check{\omb}]:&=  -\ov{\check{\ka} \check{\omb}}+  (\ze(2\eta+\ze) -\ov{\ze(2\eta+\ze)}),
\end{split}
\eea
\bea
\bsplit
e_4\check{\rho}+\frac  3 2 \ov{\ka} \check{\rho}+\frac 3 2 \ov{\rho}\check{\ka}&=\ddd_1\b+\err[e_4\check{\rho}],\\
\err[e_4\check{\rho}]:&=-\frac 3 2 \check{\ka}\check{\rho} +\frac 12 \ov{ \check{\ka}\check{\rho}} - \left(\frac{1}{2}\vthb\a +\ze\b\right)+\ov{\left(\frac{1}{2}\vthb\a +\ze\b\right)},\\
 e_4\check{\mu} +\frac 3 2\ov{ \ka}\check{ \mu} +\frac  3 2 \ov{\mu}\check{\ka}& =\err[e_4\check{\mu}],\\
\err[e_4\check{\mu}]\ :&=-\frac 3 2 \check{\ka}\check{\mu}+\frac 1 2 
 \ov{ \check{\ka}\check{\mu}}+\err[e_4\mu]-\ov{\err[e_4\mu]},\\
  e_4(\Obc) &= -2 \check{\omb} + \ov{\check{\ka}\Obc}.
\end{split}
\eea
Also in the $e_3$ direction,
\bea
\bsplit
e_3(\kac) &= 2\ddd_1 \eta +2 \rhoc-\frac 1 2\left (\kab \kac+\ka \kabc\right)+  2 \left(\omb \kac+\ka \ombc\right) \\
& +\vsi^{-1}\left( -\frac{1}{2}\ov{\kab}\, \ov{\ka}+2\ov{\omb}\,\ov{\ka}+2\ov{\rho} \right)\check{\vsi} +\frac{1}{2}\ov{\ka}^2\left(\Obc +\vsi^{-1}\ov{\Omb}\check{\vsi}\right) +\err[e_3\kac],\\
e_3(\kabc) +\kab \, \kabc &= 2\ddd_1\xib- 2\left(\ombc\,\kab+\omb\, \kabc\right) +\vsi^{-1} \check{\vsi} \, \left( -\frac 12 \ov{\kab}^2 -2 \ov{\omb} \,\ov{\kab} \right)\\
&-\left(\Obc +\vsi^{-1}\ov{\Omb}\check{\vsi}\right) \,\left( -\frac 1 2 \ov{\ka}\, \ov{\kab}  + 2\ov{\rho}\right) +\err[e_3(\kabc)],\\
e_3\check{\rho} +\frac 3 2  \ov{\kab} \check{\rho}&= - \frac 3 2 \ov{\rho}\check{\kab}+ \ddd_1\bb -\frac{3}{2}\ov{\kab}\, \ov{\rho}\vsi^{-1} \check{\vsi} +\frac{3}{2}\ov{\ka}\, \ov{\rho}\left(\Obc +\vsi^{-1}\ov{\Omb}\check{\vsi}\right) +\err[e_3\check{\rho}],
\end{split}
\eea
 with error terms given by,
\bea
\bsplit
\err[e_3\kac]&:= 2\left( \eta^2-\ov{\eta^2}\right)-\frac 12 \ov{\kac \kabc}+ 2\ov{\ombc\kac}-\frac 12 \left(\vth\vthb- \ov{\vth\vthb} \right)\\
&+\vsi^{-1} \check{\vsi} \, \left( \ov{\frac{1}{2}\kac\kabc+2\ombc\kac    -\frac{1}{2}\vth\vthb +2\eta^2}  \right)\\
&- \vsi^{-1} \,\left(  \ov{\check{\vsi}\left(\frac{1}{2}\ka\kab+2\omb\ka +2\rhoc +2\ddd_1\eta  -\frac{1}{2}\vth\vthb +2\eta^2\right)}- \ov{\check{\vsi}\kabc} \, \ov{\ka}\right)\\
&-\left(\Obc +\vsi^{-1}\ov{\Omb}\check{\vsi}\right)\left(\ov{\frac{1}{2}\kac^2 -\frac{1}{4}\vth^2}\right) +\vsi^{-1}\ov{\Omb}\, \left(\ov{ \check{\vsi}\left(\frac{1}{2}\ka^2 -\frac{1}{4}\vth^2\right)}- \ov{\check{\vsi} \kac}     \,\ov{\ka} \right)\\
&+\vsi^{-1}\left(  \ov{\Obc\vsi\left(\frac{1}{2}\ka^2 -\frac{1}{4}\vth^2\right)}  -  \ov{\Obc\vsi \,\ka} \,\ov{\ka}  \right) -\ov{\kabc\kac},
\end{split}
\eea
\bea
\bsplit
\err[e_3(\kabc)] &:= - \frac 1 2 \ov{\kabc^2}-2 \ov{\ombc\, \kabc}+ 2(\eta-3 \ze) \xib-\ov{2(\eta-3 \ze) \xib} -\frac{1}{2}\left(\vthb^2-\ov{\vthb^2}\right)\\
&- \vsi^{-1} \,\left(  \ov{\check{\vsi}\left(\frac 12 \kab^2 -2 \omb \,\kab+2\ddd_1\xib+2(\eta-3\ze)  \xib -\frac 1 2 \vthb^2\right)}- \ov{\check{\vsi}\kabc} \, \ov{\kab}\right)\\
&+\vsi^{-1}\ov{\Omb}\, \left(\ov{ \check{\vsi}\left(\frac 1 2 \ka\, \kab -2\ddd_1\ze + 2\rho -\frac 1 2  \vth\, \vthb +2\ze^2\right)}- \ov{\check{\vsi} \kac}     \,\ov{\kab} \right)\\
&+\vsi^{-1}\left(  \ov{\Obc\vsi\left(\frac 1 2 \ka\, \kab -2\ddd_1\ze + 2\rho -\frac 1 2  \vth\, \vthb +2\ze^2\right)}  -  \ov{\Obc\vsi \,\ka} \,\ov{\kab}  \right) -\ov{\kabc^2},
\end{split}
\eea
and
\bea
\bsplit
\err[e_3\check{\rho}] &:=  - \left(\frac{1}{2}\vth\aa + \ze\bb -2\eta \bb -2\xib\b\right)+\ov{\left(\frac{1}{2}\vth\aa + \ze\bb -2\eta \bb -2\xib\b\right)}-\frac{3}{2}\kabc\rhoc \\
& +\vsi^{-1} \check{\vsi} \, \left( \ov{-\frac 1 2 \kabc \rhoc  -\frac 1 2 \vth \, \aa - \ze\, \bb +2(\eta \,\bb+ \xib\,\b)} \right)\\
&- \vsi^{-1} \,\left(  \ov{\check{\vsi}\left(-\frac 1 2 \kab\rho+ \ddd_1\bb  -\frac 1 2 \vth \, \aa - \ze\, \bb +2(\eta \,\bb+ \xib\,\b)\right)}- \ov{\check{\vsi}\kabc} \, \ov{\rho}\right)\\
& -\left(\Obc +\vsi^{-1}\ov{\Omb}\check{\vsi}\right) \,\left( \ov{-\frac 1 2 \kac \rhoc  -\frac 1 2 \vthb \, \a -\ze \b}\right)\\
& +\vsi^{-1}\ov{\Omb}\, \left(\ov{ \check{\vsi}\left(-\frac 1 2 \ka \rho+\ddd_1 \b  -\frac 1 2 \vthb \, \a -\ze \b\right)}- \ov{\check{\vsi} \kac}     \,\ov{\rho} \right)\\
&+\vsi^{-1}\left(  \ov{\Obc\vsi\left(-\frac 1 2 \ka \rho+\ddd_1 \b  -\frac 1 2 \vthb \, \a -\ze \b\right)}  -  \ov{\Obc\vsi \,\ka}   \right) -\ov{\kabc\rhoc}.
\end{split}
\eea
\end{proposition}

\begin{proof}
The proof relies on  Corollary \ref{corr:transportcheckf}  and the null structure equations of   Proposition    \ref{propos:basiceqts-geod}. We refer to section \ref{sec:proofofpropos:transportaverages} for the details.
\end{proof}


\subsection{Additional equations}


We derive below additional equations  for $\omb, \eta, \xib$.
 \begin{proposition}
 \label{prop:eqtsfor-ometaxib}
The following identities hold true for a general  forward  geodesic foliation.
\begin{itemize}
\item The scalar   $\omb$ verifies
\beaa
  2\dds_1\omb &=& -\frac{1}{2}\ka\xib+\left(\frac{1}{2}\kab  +2\omb +\frac{1}{2}\vthb\right)\eta + e_3(\ze) -\bb \\
&& +\frac{1}{2}\kab\ze  -2\omb\ze +\frac{1}{2}\vthb\ze  -\frac{1}{2}\vth\xib.
 \eeaa

\item  The  reduced $1$-form  $\eta$ verifies
\beaa
 2\ddd_2\dds_2\eta&=& \ka\left( -e_3(\ze) +\bb\right) -e_3(e_\th(\ka)) - \ka\left(\frac{1}{2}\kab\ze  -2\omb\ze \right)+6\rho\eta-\kab e_\th \ka\\
 &-&\frac 1 2 \ka e_\th(\kab) +2\omb e_\th(\ka) + 2e_\th(\rho)+\err[\ddd_2\dds_2\eta],\\
\err[\ddd_2\dds_2\eta]&=& \left(2\ddd_1\eta -\frac{1}{2}\ka\vthb +2\eta^2    \right)\eta+2e_\th(\eta^2)-\ka\left(\frac{1}{2}\vthb\ze  -\frac{1}{2}\vth\xib\right) -\frac{1}{2}\vthb e_\th(\ka)\\
& -&\left(2\ddd_1\eta-\frac{1}{2}\vth\vthb+2\eta^2\right)\ze  -\frac 1 2  e_\th(\vthb\, \vth) -\frac{1}{2}\vth^2\xib-\frac 3 2 \vth \vthb \eta.
 \eeaa
 
\item  The reduced $1$-form $\xib$ verifies
\beaa
2\ddd_2\dds_2\xib&=&-e_3(e_\th(\kab))+ \kab\left(e_3(\ze) -\bb\right) +\kab^2\ze-\frac 3 2 \kab e_\th \kab + 6\rho\xib -2\omb e_\th(\kab)\\
&+&\err[\ddd_2\dds_2\xib],\\
\err[\ddd_2\dds_2\xib]&=&\left(2\ddd_1\xib+\frac{1}{2}\kab\,\vthb+2\eta\xib-\frac{1}{2}\vthb^2\right)\eta + 2e_\th(\eta\xib)  -\frac{1}{2}e_\th(\vthb^2)\\
&+&\kab\left(\frac{1}{2}\vthb\ze  -\frac{1}{2}\vth\xib\right)-\frac 1 2 \vthb e_\th \kab  -\frac{1}{2}\vth\vthb\xib
-\ze\left(2\ddd_1\xib+2(\eta-3\ze)\xib-\frac{1}{2}\vthb^2\right)\\
&+&\xib\Big(-\vth\vthb -2\ddd_1\ze+2\ze^2\Big)-6\eta\ze\xib -6e_\th(\ze\xib).
\eeaa
\end{itemize}
\end{proposition}

\begin{proof}
The proof relies on the null structure equations of Proposition    \ref{propos:basiceqts-geod}, in particular the ones for $e_3(\ze)$, $e_3(\ka)$ and $e_3(\kab)$. We refer to section \ref{sec:proofofprop:eqtsfor-ometaxib} for the details.
\end{proof}


\subsection{Ingoing geodesic foliation}\lab{sec:mainequationsforingoiggeodesicfoliations}


All the equations of section \ref{sec:mainequationsforoutgoiggeodesicfoliations} for outgoing geodesic foliations have a counterpart for ingoing geodesic foliations. The corresponding equations can be easily deduced from the ones in section \ref{sec:mainequationsforoutgoiggeodesicfoliations} by performing the following substitutions
\beaa
&& u\to \ub, \quad s\to s, \quad \CC_u\to \CC_{\ub},\quad S_{u,s}\to S_{\ub, s},\quad r\to r,\quad m\to m,\\
&& e_4\to e_3,  \quad e_3\to e_4, \quad e_\th\to e_\th, \quad e_4(s)=1\to e_3(s)=-1,\\
&&\a\to \aa,\quad \b\to, \bb,\quad \rho\to \rho,\quad \mu\to \mub, \quad \bb\to\b, \quad\aa\to\a, \\
&&\xi\to\xib, \quad\om\to \omb,\quad \ka\to \kab,\quad \vth\to \vthb, \quad\eta\to \etab, \quad\etab\to \eta, \quad\ze\to -\ze,\quad  \kab\to \ka,\\
&&\vthb\to \vth,\quad \omb\to \om,  \quad\xib\to \xi, \quad\Omb=e_3(s)\to \Om=e_4(s), \quad \vsi=\frac{2}{e_3(u)}\to \vsib=\frac{2}{e_4(\ub)}, \\
&& \overline{\ka}-\frac{2}{r}\to \overline{\kab}+\frac{2}{r}, \quad\overline{\kab}+\frac{2\Up}{r}\to \overline{\ka}-\frac{2\Up}{r}, \quad\overline{\omb}-\frac{m}{r^2}\to \overline{\om}+\frac{m}{r^2}, \quad \overline{\rho}+\frac{2m}{r^3}\to \overline{\rho}+\frac{2m}{r^3},\\
&&\overline{\mu}-\frac{2m}{r^3}\to \overline{\mub}-\frac{2m}{r^3}, \quad \overline{\Omb}+\Up\to \overline{\Om}-\Up,  \quad  \overline{\vsi}-1\to \overline{\vsib}-1, \\
&& \Ab=\frac{2}{r}e_3(r)-\ov{\kab}\to A=\frac{2}{r}e_4(r)-\ov{\ka}.
\eeaa


\subsection{Adapted coordinates systems}



\subsubsection{$(u,s, \th, \vphi)$ coordinates}


 \begin{proposition}
 \label{prop:outgoinggeodesiccoordinates}
Consider, in addition   to the  functions   $u, s, \vphi$  an additional    $\Z$ invariant      function  $\th$.   Then, relative to the  coordinates system
$(u, s,\th, \vphi)$,  the following hold true,
\begin{enumerate}
\item
 The spacetime  metric takes the form,
 \bea
 \label{eq:propoutgoinggeodesiccoordinates1}
  g=-2\vsi du ds +\vsi^2\Omb du^2+ \ga\left( d\th- \frac 1 2 \vsi(\underline{b}-\Omb b) du- b ds \right)^2+e^{2\Phi}(d\vphi)^2
 \eea
where,
\bea
\Omb= e_3(s), \qquad b=e_4(\th), \qquad \underline{b}=e_3(\th), \qquad \ga^{-1} =  e_\th(\th)^2. 
\eea

 \item In these coordinates the reduced frame takes the form,
  \bea
   \label{eq:propoutgoinggeodesiccoordinates2}
  \pr_s=e_4- b\sqrt{\ga}e_\th , \,\,  \pr_u =\vsi\left(\frac{1}{2}e_3-\frac{1}{2}\Omb e_4-\frac{1}{2}\sqrt{\ga}(\underline{b}-b\Omb)e_\th\right),\,\, \pr_\th=\sqrt{\ga}e_\th. 
 \eea
 
  \item In the particular case when $b=e_4(\th)=0$ we have,
 \bea
  \label{eq:propoutgoinggeodesiccoordinates3}
 e_4(\ga)=2\chi \ga, \qquad e_4(\underline{b})=-2(\ze+\eta)\ga^{-1/2}.
 \eea
\end{enumerate}
\end{proposition}

\begin{proof}
First, from the fact that $(e_3, e_4, e_\th)$ forms a null frame, we easily verify that \eqref{eq:propoutgoinggeodesiccoordinates2} holds. Then, \eqref{eq:propoutgoinggeodesiccoordinates1} immediately follows from \eqref{eq:propoutgoinggeodesiccoordinates2} and the  fact that $(e_3, e_4, e_\th)$ forms a null frame.

To prove the last statement, when $ b=e_4(\th)=0$, 
we start with, 
\beaa
 [e_4 ,   e_3 ]= 2\om e_3-2\omb e_4 +2 (\etab-\eta) e_\th=               -2(\ze+\eta) e_\th -2\omb e_4.
 \eeaa
Applying this to $\th$ we derive,
 \beaa
 [e_4 ,   e_3 ](\th)=(-2(\ze+\eta) e_\th -2\omb e_4 )(\th)=-2(\ze+\eta)  e_\th(\th)=-2(\ze+\eta) \ga^{-1/2}.
 \eeaa
 We deduce,
 \beaa
e_4(\underline{b}) =  e_4 (e_3 (\th))=-2(\ze+\eta) \ga^{-1/2}.
 \eeaa 
 
 To prove the equation for $\ga$ we make use of, 
 \beaa
  [e_4, e_\th]=(\etab+\ze) e_4 +\xi e_3 -\chi e_\th =-\chi e_\th 
  \eeaa
 so that
 \beaa
 \, e_4 e_\th(\th) =  [e_4, e_\th](\th) &=&-\chi e_\th(\th) =-\chi \ga^{-1/2}.
 \eeaa
 Thus 
 \beaa
 e_4 (\ga^{-1/2})=-\chi \ga^{-1/2}
 \eeaa 
 from which
 \beaa
 e_4(\ga)=2 \chi \ga.
 \eeaa
This concludes the proof of the lemma.
\end{proof}

\begin{remark} 
In  Schwarzschild, relative  to  the above coordinate system,  we  have 
\beaa
\vsi=1, \qquad\underline{\Omega}=-\Up,\qquad b=\underline{b}=0, \qquad \ga=r^2,\qquad e^\Phi=r\sin\th,
\eeaa
so that we obtain outgoing Eddington-Finkelstein coordinates. 
\end{remark}

\begin{remark}
The $(u,s, \th, \vphi)$ coordinates system, with the choice $b=0$ (i.e. $\th$ is transported by $e_4(\th)=0$), will be used in section \ref{sec:mainstatementsonGCMSprocedure} and Chapter \ref{chap:proofofGCMprocedure} in connection with our GCM procedure. 
\end{remark}


\subsubsection{$(u,r, \th, \vphi)$ coordinates}


  \begin{proposition}\label{Prop:coordin(u,r,th)}
Consider, in addition   to the  functions   $u, r, \vphi$  an additional    $\Z$ invariant      function  $\th$.    Relative to the coordinates $(u, r, \th, \vphi)$ the following hold true,
    \begin{enumerate}
    \item The spacetime metric takes the form,
  \bea\lab{eq:thiistheformofthemetricintheurthetacoordinatessystem}
  \bsplit
g&=-\frac{4\vsi}{r\ov{\ka}} du dr+\frac{\vsi^2(\ov{\kab}+\Ab)}{\ov{\kab}} du^2 +\ga\left(  d\th-\frac 1 2\vsi\underline{b} du -\frac b 2  \Th\right)^2 
\end{split}
\eea
where,
\bea
b=e_4(\th), \qquad \underline{b}=e_3(\th),\qquad \ga=\frac{1}{(e_\th(\th))^2}
\eea
and,
\beaa
\Th:=\frac{4}{r\overline{\ka}} d r - \vsi\left(\frac{\overline{\kab}+\Ab}{\overline{\ka}}\right) du.
\eeaa

\item The reduced coordinates derivatives take the form,
\bea
\label{eq:coord-vfields-urth:fisttimeeverthisappearsinthepaper}
\bsplit
\pr_r&=\frac{2}{r\overline{\ka}}e_4- \frac{2\sqrt{\ga} }{ r\ov{\ka}} b e_\th, \\
\pr_\th &= \sqrt{\ga}e_\th,\\
\pr_u&=\vsi\left[\frac{1}{2}e_3-\frac{1}{2}\frac{\overline{\kab}+\Ab}{\overline{\ka}}e_4-\frac{1}{2}\sqrt{\ga}\left( \underline{b}  - \left(\frac{\overline{\kab}+\Ab}{\overline{\ka}}\right) b  \right)e_\th\right]. 
\end{split}
\eea

\item To control $e^\Phi$, we will rely on the following transport equation 
\bea\label{equations:transportequatione4diectionexponetialPhi}
e_4\left(\frac{e^\Phi}{r\sin\th}-1\right) &=& \frac{e^\Phi}{2r\sin\th}\left(\check{\ka}-\vth\right).
\eea
\end{enumerate}
\end{proposition}

  \begin{proof}
First, from the fact that $(e_3, e_4, e_\th)$ forms a null frame, we easily verify that \eqref{eq:coord-vfields-urth:fisttimeeverthisappearsinthepaper} holds. Then, \eqref{eq:thiistheformofthemetricintheurthetacoordinatessystem} immediately follows from \eqref{eq:coord-vfields-urth:fisttimeeverthisappearsinthepaper} and the  fact that $(e_3, e_4, e_\th)$ forms a null frame.
      
  It remains to prove  \eqref{equations:transportequatione4diectionexponetialPhi}. It follows from
    \beaa
e_4\left(\frac{e^\Phi}{r\sin\th}-1\right) &=&  \frac{e^\Phi}{r\sin\th}\left(e_4(\Phi) -\frac{e_4(r)}{r}\right)= \frac{e^\Phi}{r\sin\th}\left(\frac{1}{2}(\ka-\vth) -\frac{\overline{\ka}}{2}\right)\\
&=& \frac{e^\Phi}{2r\sin\th}\left(\check{\ka}-\vth\right)
\eeaa
which concludes the proof of the lemma.
  \end{proof}
  
\begin{remark} 
In  Schwarzschild, relative  to  the above coordinate system,  we  have 
\beaa
\overline{\ka}=\frac{2}{r},\qquad \overline{\kab}=-\frac{2\Up}{r}, \qquad \vsi=1,\qquad \Ab=0, \qquad b= \underline{b}=0, \qquad \ga=r^2,\qquad e^\Phi=r\sin\th,
\eeaa
so that we obtain outgoing Eddington-Finkelstein coordinates. 
\end{remark}

\begin{remark}\lab{rem:whereistheurthetacoordinatesused}
The $(u,r, \th, \vphi)$ coordinates system, with the choice \eqref{def:definitionofthetausedatnullinfinity:originalplaceappearing} for $\th$ introduced below, will be used in Proposition  \ref{prop:finalcoordinatessystem} to prove the convergence to the outgoing Eddington-Finkelstein coordinates of Schwarzschild. 
\end{remark}


\subsubsection{$(\ub,r, \th, \vphi)$ coordinates}


We easily deduce an analog statement  relative to $(\ub, r, \th, \vphi)$ coordinates.

  \begin{proposition}\label{Prop:coordin(ub,r,th)}
Consider, in addition   to the  functions   $\ub, r, \vphi$  an additional    $\Z$ invariant      function  $\th$.    Relative to the coordinates $(\ub, r, \th, \vphi)$ the following hold true,
    \begin{enumerate}
    \item The spacetime metric takes the form,
  \bea\lab{eq:thiistheformofthemetricintheubarrthetacoordinatessystem}
  \bsplit
g&=-\frac{4\vsib}{r\ov{\kab}} d\ub dr+\frac{\vsib^2(\ov{\ka}+A)}{\ov{\ka}} d\ub^2 +\ga\left(  d\th-\frac 1 2\vsib b d\ub -\frac{\underline{b}}{2}  \underline{\Th}\right)^2 
\end{split}
\eea
where,
\bea
b=e_4(\th), \qquad \underline{b}=e_3(\th),\qquad \ga=\frac{1}{(e_\th(\th))^2}
\eea
and,
\beaa
\underline{\Th}:=\frac{4}{r\overline{\kab}} d r - \vsib\left(\frac{\overline{\ka}+A}{\overline{\kab}}\right) d\ub.
\eeaa

\item The reduced coordinates derivatives take the form,
\bea
\label{eq:coord-vfields-ubarrth:fisttimeeverthisappearsinthepaper}
\bsplit
\pr_r&=\frac{2}{r\overline{\kab}}e_3- \frac{2\sqrt{\ga} }{ r\ov{\kab}} \underline{b} e_\th, \\
\pr_\th &= \sqrt{\ga}e_\th,\\
\pr_\ub &=\vsib\left[\frac{1}{2}e_4-\frac{1}{2}\frac{\overline{\ka}+A}{\overline{\kab}}e_3-\frac{1}{2}\sqrt{\ga}\left( b  - \left(\frac{\overline{\ka}+A}{\overline{\kab}}\right) \underline{b}  \right)e_\th\right]. 
\end{split}
\eea

\item To control $e^\Phi$, we will rely on the following transport equation 
\bea\label{equations:transportequatione3directionexponetialPhi}
e_3\left(\frac{e^\Phi}{r\sin\th}-1\right) &=& \frac{e^\Phi}{2r\sin\th}\left(\check{\kab}-\vthb\right).
\eea
\end{enumerate}
\end{proposition}
  
\begin{remark} 
In  Schwarzschild, relative  to  the above coordinate system,  we  have 
\beaa
\overline{\ka}=\frac{2}{r},\qquad \overline{\kab}=-\frac{2\Up}{r}, \qquad \vsib=1,\qquad A=0,\qquad b= \underline{b}=0, \qquad \ga=r^2,\qquad e^\Phi=r\sin\th,
\eeaa
so that we obtain ingoing Eddington-Finkelstein coordinates. 
\end{remark}

\begin{remark}\lab{rem:whereistheubrthetacoordinatesused}
The $(\ub,r, \th, \vphi)$ coordinates system, with the choice \eqref{def:definitionofthetausedatnullinfinity:originalplaceappearing} for $\th$ introduced below, will be used in Proposition  \ref{prop:finalcoordinatessystem:bis} to prove the convergence to the ingoing Eddington-Finkelstein coordinates of Schwarzschild. 
\end{remark}

  
\subsubsection{Initialization of $\th$}


We now introduce the coordinate function $\th$ that will be used for the $(u,r,\th,\vphi)$ coordinates system and for the $(\ub,r,\th,\vphi)$ coordinates system, see Remarks \ref{rem:whereistheurthetacoordinatesused} and \ref{rem:whereistheubrthetacoordinatesused}.     
\begin{lemma}\label{Lemma:choiceof-theta}
Let $\th\in[0, \pi]$ be the $\Z$-invariant scalar on $\MM$  defined by,
\bea
\lab{def:definitionofthetausedatnullinfinity:originalplaceappearing}
\th:= \cot^{-1}\left(re_\th(\Phi)\right).
\eea
 Then,        
\bea\lab{eq:computationfoexponentialPhiintermsofafrak}
\frac{e^{\Phi}}{r\sin\th} &=&   \sqrt{1+\af}.
\eea
where,
\bea
\lab{eq:definitionofaf:firsttimeeveritisintroduced}
\af:= \frac{e^{2\Phi}}{r^2}+(e_\th(e^\Phi))^2-1. 
\eea
Moreover, we have in an outgoing geodesic foliation
\beaa
 re_\th(\th)  &=& 1+\frac{r^2(K-\frac{1}{r^2})}{1+(re_\th(\Phi))^2},\\
  e_3(\th)   &=& -\frac{r\bb +\frac{r}{2}\left(-\check{\kab}+\Ab +\vthb\right)e_\th(\Phi)+r\xib e_4(\Phi) +r\eta e_3(\Phi)}{1+(re_\th(\Phi))^2},\\
   e_4(\th)     &=& -\frac{r\b +\frac{r}{2}\left(-\check{\ka} +\vth\right)e_\th(\Phi) -r\ze e_3(\Phi)}{1+(re_\th(\Phi))^2},
  \eeaa 
  and analog identities hold for an ingoing geodesic foliation.
 \end{lemma}

\begin{proof}
In view of the definition of $\th$, we have $\th\in[0,\pi]$, $\sin\th\geq 0$ and 
\beaa
\sin\th &=& \frac{1}{\sqrt{1+\cot\th^2}}= \frac{1}{\sqrt{1+(re_\th(\Phi))^2}}= \frac{e^{\Phi}}{\sqrt{e^{2\Phi}+(re_\th(e^\Phi)))^2}}\\
&=& \frac{e^{\Phi}}{r\sqrt{\frac{e^{2\Phi}}{r^2}+(e_\th(e^\Phi))^2}}=\frac{e^{\Phi}}{r\sqrt{ 1+\af}}.
\eeaa
Hence
\beaa
\frac{e^{\Phi}}{r\sin\th} &=&  \sqrt{\frac{e^{2\Phi}}{r^2}+(e_\th(e^\Phi))^2}=\sqrt{1+\af}.
\eeaa

 Also, we compute
 \beaa
 re_\th(\th) &=& -\frac{r^2e_\th e_\th(\Phi)}{1+(re_\th(\Phi))^2}.
 \eeaa 
  Next, recall that we have
 \beaa
 e_\th e_\th(\Phi) &=& -K-(e_\th(\Phi))^2.
 \eeaa
 We infer
\beaa
 re_\th(\th) &=& \frac{r^2(K+(e_\th(\Phi))^2)}{1+(re_\th(\Phi))^2}= 1+\frac{r^2(K-\frac{1}{r^2})}{1+(re_\th(\Phi))^2}.
 \eeaa 
 as  desired.
  
  Also, we have in an outgoing geodesic foliation
  \beaa
   e_4(\th) &=& -\frac{re_4 e_\th(\Phi)+e_4(r)e_\th(\Phi)}{1+(re_\th(\Phi))^2}\\
   &=& -\frac{r(D_4D_\th\Phi+D_{D_4e_\th}\Phi)+e_4(r)e_\th(\Phi)}{1+(re_\th(\Phi))^2}\\
    &=& -\frac{r\b +r\left(\frac{e_4(r)}{r}-e_4(\Phi)\right)e_\th(\Phi) -r\ze e_4(\Phi)}{1+(re_\th(\Phi))^2}\\
    &=& -\frac{r\b +\frac{r}{2}\left(-\check{\ka} +\vth\right)e_\th(\Phi) -r\ze e_4(\Phi)}{1+(re_\th(\Phi))^2}. 
  \eeaa
   Finally, we compute in an outgoing geodesic foliation
  \beaa
 e_3(\th) &=& -\frac{re_3 e_\th(\Phi)+e_3(r)e_\th(\Phi)}{1+(re_\th(\Phi))^2}\\
 &=& -\frac{r(D_3D_\th\Phi+D_{D_3e_\th}\Phi)+e_3(r)e_\th(\Phi)}{1+(re_\th(\Phi))^2}\\
 &=& -\frac{r\bb +r\left(\frac{e_3(r)}{r}-e_3(\Phi)\right)e_\th(\Phi)+r\xib e_4(\Phi) +r\eta e_3(\Phi)}{1+(re_\th(\Phi))^2}\\
&=& -\frac{r\bb +\frac{r}{2}\left(-\check{\kab}+\Ab +\vthb\right)e_\th(\Phi)+r\xib e_4(\Phi) +r\eta e_3(\Phi)}{1+(re_\th(\Phi))^2}. 
  \eeaa 
   This concludes the proof of the lemma.
 \end{proof}

In view of \eqref{eq:computationfoexponentialPhiintermsofafrak}, we will need to control the quantity $\af$ defined in \eqref{eq:definitionofaf:firsttimeeveritisintroduced}. To this end, we will need the following lemma.  
\begin{lemma}\lab{lemma:transportequationforPhiinthee4direction}
 The quantity $\af$ defined in \eqref{eq:definitionofaf:firsttimeeveritisintroduced} vanishes on the axis of symmetry  and 
verifies the following identities  in an outgoing geodesic foliation,
\beaa
e_4(\af) &=& \frac{(\check{\ka}-\vth)e^{2\Phi}}{r^2} +2e_\th(e^\Phi) \Big(\b-e_4(\Phi)\ze\Big)e^\Phi,\\
e_\th(\af) &=& 2e_\th(\Phi)e^{2\Phi}\left(\left(\rho+\frac{2m}{r^3}\right)+\frac{1}{4}\left(\ka\kab+\frac{4\Up}{r^2}\right) -\frac{1}{4}\vth\vthb\right),\\
e_3(\af) &=& \frac{\Big(\check{\kab}-\Ab-\vthb\Big)e^{2\Phi}}{r^2} +2e_\th(e^\Phi) \Big(\bb+e_3(\Phi)\eta+\xib e_4(\Phi)\Big)e^\Phi,
\eeaa
 and analog identities hold in an outgoing geodesic foliation.
\end{lemma}

 \begin{proof}
 The vanishing on the axis  follow easily   from the fact that both  $e^{2\Phi}$ 
and   $e_\th(e^\Phi))^2-1$ vanish on the axis (see \eqref{regularityofPhiS}).  
To prove the second part of the lemma we  recall that, with respect to the reduced metric (see equation  \eqref{eq: R-Phi}),
\beaa
R_{ab}=D_a D_b \Phi+D_a \Phi D_b \Phi, 
\eeaa
and (see Definition \ref{def:nullcomponents})  
\beaa
 R_{3\th}=\bb,\quad  R_{4\th}=\b, \quad R_{\th\th}=R_{34}=\rho, \quad R_{34}=\rho.
 \eeaa
Starting with  the definition  $\af= \frac{e^{2\Phi}}{r^2}+(e_\th(e^\Phi))^2-1$,  
we compute  in an outgoing geodesic foliation
\beaa
e_4(\af) &=& \frac{2e_4(\Phi)e^{2\Phi}}{r^2} - \frac{2e_4(r)e^{2\Phi}}{r^3}+2e_\th(e^\Phi) e_4(e_\th(e^\Phi))\\
&=& \frac{(\ka-\vth)e^{2\Phi}}{r^2} - \frac{\overline{\ka}e^{2\Phi}}{r^2}+2e_\th(e^\Phi) \Big(e_4(e_\th(\Phi))+e_\th(\Phi)e_4(\Phi)\Big)e^\Phi\\
&=& \frac{(\check{\ka}-\vth)e^{2\Phi}}{r^2} +2e_\th(e^\Phi) \Big(\b-e_4(\Phi)\ze\Big)e^\Phi.
\eeaa
Also
\beaa
e_\th(\af) &=& \frac{2e_\th(\Phi)e^{2\Phi}}{r^2} +2e_\th(e^\Phi) e_\th(e_\th(e^\Phi))\\
&=& \frac{2e_\th(\Phi)e^{2\Phi}}{r^2} +2e_\th(e^\Phi) \Big(e_\th(e_\th(\Phi))+e_\th(\Phi)^2\Big)e^\Phi\\
&=& \frac{2e_\th(\Phi)e^{2\Phi}}{r^2} +2e_\th(e^\Phi) \Big(\rho+D_{D_\th e_\th}\Phi\Big)e^\Phi\\
&=& \frac{2e_\th(\Phi)e^{2\Phi}}{r^2} +2e_\th(e^\Phi) \Big(\rho+\frac{1}{2}\chi e_3\Phi+\frac{1}{2}\chib e_4\Phi\Big)e^\Phi\\
&=& \frac{2e_\th(\Phi)e^{2\Phi}}{r^2} +2e_\th(e^\Phi) \Big(\rho+\frac{1}{4}\ka\kab -\frac{1}{4}\vth\vthb\Big)e^\Phi\\
&=& 2e_\th(\Phi)e^{2\Phi}\left(\left(\rho+\frac{2m}{r^3}\right)+\frac{1}{4}\left(\ka\kab+\frac{4\Up}{r^2}\right) -\frac{1}{4}\vth\vthb\right).
\eeaa
Finally, we have in an outgoing geodesic foliation
\beaa
e_3(\af) &=& \frac{2e_3(\Phi)e^{2\Phi}}{r^2} - \frac{2e_3(r)e^{2\Phi}}{r^3}+2e_\th(e^\Phi) e_3(e_\th(e^\Phi))\\
&=& \frac{(\kab-\vthb)e^{2\Phi}}{r^2} - \frac{\Big(\overline{\kab}+\Ab\Big)e^{2\Phi}}{r^2}+2e_\th(e^\Phi) \Big(e_3(e_\th(\Phi))+e_\th(\Phi)e_3(\Phi)\Big)e^\Phi\\
&=& \frac{\Big(\check{\kab}-\Ab-\vthb\Big)e^{2\Phi}}{r^2} +2e_\th(e^\Phi) \Big(\bb+e_3(\Phi)\eta+\xib e_4(\Phi)\Big)e^\Phi.
\eeaa
This concludes the proof of the lemma.
\end{proof}

\begin{remark}
The  function $\th$  defined by  \eqref{def:definitionofthetausedatnullinfinity}  defines
\begin{itemize}
\item together with  the functions $(u, r, \vphi)$,  a  regular coordinates system   with the axis of symmetry corresponding  to  $\th=0, \pi$,

\item together with  the functions $(u, r, \vphi)$, a  regular coordinates system   with the axis of symmetry corresponding  to  $\th=0, \pi$.
\end{itemize}
\end{remark}


\section{Perturbations of   Schwarzschild   and invariant quantities}


Recall that in Schwarzschild   all Ricci  coefficients  $\xi, \xib, \vth, \vthb, \eta, \etab, \ze$  and curvature
 components $\a,\aa,\b, \bb$  vanish identically.        In addition  the check quantities   $\check{\ka}, \check{\kab}, \check{\om}, \check{\omb}  $ and $ \check{\rho}$ also vanish.       Thus,   roughly, we expect  that  in perturbations of Schwarzschild     these  quantities stay   small,  i.e. of oder $O(\ep)$ for a sufficiently  small $\ep$. More precisely we say that    a       smooth,  vacuum,  $\Z$-invariant, polarized   spacetime  is an $O(\ep)$-perturbation of Schwarzschild, or simply $O(\ep)$-Schwarzschild,   if   the following are true relative to a $\Z$-invariant null frame $e_3, e_4, e_\th$,
 \bea
\xi, \, \xib,\,  \vth, \, \vthb,\,  \eta,\,  \etab,\,  \ze,\, \check{\ka}, \, \check{\kab},\,  \check{\om},\,  \check{\omb}    \qquad \a,\, \aa,\, \b,\,  \bb\, , \check{\rho} =O(\ep)
\eea
Moreover,
\bea
e_3(r)-\frac{r}{2} \ov{\kab}=O(\ep), \qquad  e_4(r)-\frac{r}{2} \ov{\ka}=O(\ep),
\eea
where $r$ is the area radius of the $2$-spheres generated by $e_\th, e_\vphi$, see \eqref{eq:Smetric}.

   In reality, of course, we expect  that small perturbations of Schwarzschild,  remain       not  only      close to the original Schwarzschild   but also    converge to a    nearby Schwarzschild solution but for the discussion below   this will suffice.

 
\subsection{Null frame transformations}


Our definition of  $O(\ep)$-Schwarzschild perturbations   does not specify a particular  frame. In what follows we investigate  how 
  the main Ricci and  curvature    quantities   change    relative to  frame transformations, i.e linear transformations 
   which take  null frames  into null frames.   

 \begin{lemma}
 \label{lemma:SSMe:general.composite}
 A general null transformation can be written in   the form,
  \bea
\label{SSMe:general.composite}
\begin{split}
e_4'&=\la\left(e_4 + f e_\th +\frac 1 4 f^2  e_3\right),\\
e_\th'&=\left(1+\frac 1 2   f \fb\right) e_\th  + \frac 1 2  \fb e_4+\frac 1 2 f\left(1+ \frac 1 4 f \fb\right) e_3,\\
e_3'&= \la^{-1} \left( \left(1+\frac 12  f \fb +\frac{1}{16} f^2 \fb^2\right)  e_3 +\fb\left(1+\frac 1 4 f\fb\right) e_\th + \frac 1 4 \fb^2  e_4\right).
\end{split}
\eea
\end{lemma}

\begin{proof}
It is straightforward  to check  that  the transformation \eqref{SSMe:general.composite} takes null frames into null frames. One can also check  that it can be  written in the form  $\mbox{ type}(3)\circ \mbox{ type}(1)\circ\mbox{ type}(2)$ where the type 1   transformations  fix $e_3$, $ i.e. (\la=1, \fb=0$), type 2  transformations  fix $e_4$, i.e. $ (\la=1, f=0)$ and  type 3 transformations   keep the directions of $e_3, e_4$ i.e.  $ (f=\fb=0)$.
\end{proof}

 \begin{remark}
   Note that   $f, \fb$ are     reduced   from    spacetime $1$ forms    while $\la$   is reduced from a scalar.
\end{remark}

\begin{remark}  
A transformation  consistent  with $O(\ep)$- Schwarzschild spacetimes must have $f, \fb =O(\ep)$ and $a:=\log \la =O(\ep)$.   
\end{remark}

\begin{proposition}[Transformation formulas]
\label{prop:transformations1}
Under a general transformation of  type \eqref{SSMe:general.composite},  the   Ricci coefficients and curvature components  transform as follows:
\bea
\begin{split}
\xi' &= \la^2\left(\xi+\frac{1}{2}\la^{-1}e_4'(f)+\om f + \frac{1}{4}f\ka\right)+\la^2\err(\xi, \xi'),\\
\err(\xi, \xi')&= \frac{1}{4}f\vth+\lot,\\ 
\xib' &= \la^{-2}\left(\xib+\frac{1}{2}\la e_3'(\fb)+\omb\,\fb + \frac{1}{4}\fb\,\kab\right)+\la^{-2}\err(\xib, \xib'),\\
\err(\xib, \xib')&=  -\frac 18\la\fb^2 e_3'(f)+ \frac{1}{4}\fb\,\vthb+\lot,
\end{split}
\eea
\bea
\bsplit
\ze' &= \ze - e_\th'(\log(\la))  +\frac 14 (- f\kab +\fb \ka ) + \fb \om- f  \omb+\err(\ze,\ze'), \\
\err(\ze,\ze')&= \frac{1}{2}\fb e_\th'(f)  +\frac 1 4( - f \vthb +\fb \vth)+\lot,\\
     \eta'&= \eta +\frac{1}{2}\la e_3'(f)     +\frac 1 4 \ka \fb   -f\omb +\err(\eta, \eta'),\\
     \err(\eta, \eta')&= \frac{1}{4}\fb\vth+\lot, \\
     \etab'&= \etab + \frac{1}{2} \la^{-1}e_4'(\fb)     +\frac 1 4 \kab f   -\fb\om +\err(\etab, \etab'),\\
     \err(\etab, \etab')&=   - \frac{1}{8}\fb^2\la^{-1}e_4'(f) +\frac{1}{4}f\vthb+\lot,
\end{split}
\eea
\bea
\bsplit
\ka'&=\la\left( \ka+ \ddd_1\,\!'(f)    \right) +\la\err(\ka,\ka'),\\
   \err(\ka,\ka')&=  f(\ze+\eta)    +\fb\xi       -\frac 1 4 f^2\kab +f\fb\om -f^2\omb+\lot,\\
\kab'&=\la^{-1}\left( \kab+ \ddd_1\,\!'(\fb)    \right) +\la^{-1}\err(\kab,\kab'),\\
   \err(\kab,\kab')&= -\frac{1}{4}\fb^2e_\th'(f) +\fb(-\ze+\etab)    +f\xib       -\frac 1 4 \fb^2\ka +f\fb\omb -\fb^2\om+\lot,
\end{split}
\eea
\bea
\bsplit
\vth' &= \la\left(\vth- \dds_2\,\!'(f)   \right) + \la\err(\vth,\vth'),\\
\err(\vth,\vth') &= f(\ze+\eta)    +\fb\xi            +\frac 1 4  f\fb \ka +f\fb\om -f^2\omb+\lot\\
\vthb' &= \la^{-1}\left(\vthb- \dds_2\,\!'(\fb)   \right) + \la^{-1}\err(\vthb,\vthb'),\\
\err(\vthb,\vthb') &=  -\frac{1}{4}\fb^2 e_\th'(f)+\fb(-\ze+\etab)    +f\xib            +\frac 1 4  f\fb \kab +f\fb\omb -\fb^2\om+\lot, 
\end{split}
\eea 
\bea
\begin{split}
\om' &= \la\left(\om -\frac{1}{2}\la^{-1}e_4'(\log(\la))\right)+\la\err(\om, \om'),\\
\err(\om, \om')&= \frac{1}{4}\fb e_4'(f)  +\frac{1}{2}\om f\fb - \frac{1}{2}f\etab +\frac{1}{2}\fb\xi +\frac{1}{2}f\ze -\frac{1}{8}\kab f^2+\frac{1}{8}f\fb \ka-\frac{1}{4}\omb f^2+\lot,\\
\omb' &= \la^{-1}\left(\omb +\frac{1}{2}\la e_3'(\log(\la))\right)+\la^{-1}\err(\omb, \omb'),\\
\err(\omb, \omb')&=  - \frac{1}{4}\fb e_3'(f)  +\omb f\fb - \frac{1}{2}\fb\eta +\frac{1}{2}f\xib -\frac{1}{2}\fb\ze -\frac{1}{8}\ka\fb^2+\frac{1}{8}f\fb \kab-\frac{1}{4}\om\fb^2+\lot 
\end{split}
\eea 
  The  lower order terms we denote by $\lot$  are linear 
  with respect       $\Ga=\{\xi,\xib,\vth, \ka, \eta,\etab,\ze,\kab,  \vthb\}$  and quadratic or  higher order in $f,\fb$, and do not contain derivatives of these latter.

Also,
\bea
\bsplit
\a' &= \la^2\a+\la^2\err(\a, \a'),\\
\err(\a, \a') &= 2f\b+\frac{3}{2}f^2\rho+\lot,\\
\b' &= \la\left(\b+\frac{3}{2}\rho f\right)+\la\err(\b,\b'),\\
\err(\b,\b') &= \frac{1}{2}\fb\a+\lot,\\
\rho' &= \rho+\err(\rho,\rho'),\\
\err(\rho,\rho') &= \frac{3}{2}\rho f\fb+\fb\b  +f\bb+\lot,\\
 \bb' &= \la^{-1}\left(\bb+\frac{3}{2}\rho\fb\right)+\la^{-1}\err(\bb,\bb'),\\
\err(\bb,\bb') &= \frac{1}{2}f\aa+\lot,\\
\aa' &= \la^{-2}\aa+\la^{-2}\err(\aa, \aa'),\\
\err(\aa, \aa') &= 2\fb\,\bb+\frac{3}{2}\fb^2\rho+\lot
\end{split}
\eea
    The  lower order terms we denote by $\lot$ are linear   with respect to the curvature quantities    $\a,\b, \rho, \bb, \,\aa$  
    and quadratic  or higher order  in  $f,\fb$, and do not contain derivatives of these latter.
    \end{proposition}

\begin{proof}
See Appendix \ref{sec:proofofprop:transformations1}.
\end{proof}

\begin{lemma}
\label{remark:lotxixi'}
In the particular case when $\la=1, \fb=0$, we have  
  \beaa
\begin{split}
e_4'&= e_4 + f e_\th +\frac 1 4 f^2  e_3,\\
e_\th'&= e_\th +\frac 1 2 f e_3,\\
e_3'&=  e_3,
\end{split}
\eeaa
and 
\beaa
\xi' &=& \xi+\frac{1}{2}e_4'f+\frac{1}{4}\ka f+f\om+\frac{1}{4}f\vth +\frac{1}{4}f^2\eta-\frac 1 4 f^2\etab+\frac{1}{2}f^2\ze-\frac{1}{16}f^3\kab\\
&&-\frac{1}{4}f^3\omb-\frac{1}{16}f^3\vthb -\frac{1}{16}f^4\xib,\\
\om' &=& \om+\frac{1}{2}f\ze -\frac{1}{2}\etab f - \frac{1}{4}f^2\omb  -\frac{1}{8}f^2\kab -\frac{1}{8}f^2\vthb -\frac{1}{8}f^3\xib,\\
\ze' &=& \ze     -\left(\frac{1}{4}\kab + \omb\right)f   -f\left(\frac{1}{4}\vthb +\frac{1}{2}f\xib\right),\\
\eta' &=& \eta+ \frac{1}{2}e_3'(f) -f\omb -\frac{1}{4}f^2\xib.
\eeaa
\end{lemma}

\begin{proof}
The proof follows from Proposition \ref{prop:transformations1} by setting $\la=1, \fb=0$. Since we need precise formulas for the error terms, we provide a proof in section \ref{sec:remark:lotxixi'}.
\end{proof}

\begin{lemma}[Transport equations for $(f, \fb, \la)$]\lab{lemma:transportequationsforffbandlambda}
Assume that we have in the new null frame $(e_3', e_4', e_\th')$ of type \eqref{SSMe:general.composite}
\beaa
\xi'=0,\quad \om'=0,\quad \ze'+\etab'=0.
\eeaa
Then, $(\fb, f, \log(\la))$ satisfy the following transport equations 
\beaa
\la^{-1}e_4'(f)  +\left(\frac{\ka}{2}+2\om\right) f &=& -2\xi+E_1(f, \Ga),\\
\la^{-1}e_4'(\log(\la)) &=&  2\om+E_2(f, \Ga),\\
\la^{-1}e_4'(\fb) + \frac{\ka}{2}\fb &=& -2(\ze+ \etab)  + 2e_\th'(\log(\la))      + 2f  \omb+E_3(f, \fb, \Ga),
\eeaa
where $E_1$, $E_2$ and $E_3$ are given by
\beaa
E_1(f, \Ga) &=& -\frac{1}{2}\vth f+\lot,\\
E_2(f, \Ga) &=&  f\ze- \frac{1}{2}f^2\omb -\etab f - \frac{1}{4}f^2\kab +\lot,\\
E_3(f, \fb, \Ga) &=& -\fb e_\th'(f)  -\frac{1}{2}\fb \vth     +\lot,
\eeaa
Here, $\lot$ denote terms which are cubic or higher order in $f, \fb$ (or in $f$ only in the case of $E_1$ and $E_2$) and $\Gac$ and do not contain derivatives of these quantities, where $\Ga$ and $\Gac$ denotes the Ricci coefficients and renormalized Ricci coefficients w.r.t. the original null frame $(e_3, e_4, e_\th)$.
\end{lemma}

\begin{proof}
See section \ref{sec:proofoflemma:transportequationsforffbandlambda}.
\end{proof}

To avoid a potential log loss for the third equation in Lemma \ref{lemma:transportequationsforffbandlambda}, i.e. the transport equation for $\fb$, we state the following renormalized version of the lemma.

\begin{corollary}\lab{cor:transportequationsforffbandlambda}
Assume given a null frame $(e_3, e_4, e_\th)$ associated to an outgoing geodesic foliation as in section \ref{sec:mainequationsforoutgoiggeodesicfoliations}, and let $r$ denote the corresponding area radius.  
Assume that we have in the new null frame $(e_3', e_4', e_\th')$ of type \eqref{SSMe:general.composite}
\beaa
\xi'=0,\quad \om'=0,\quad \ze'+\etab'=0.
\eeaa
 Then, $(\fb, f, \log(\la))$ satisfy the following transport equations 
\beaa
\la^{-1}e_4'(rf) &=& E_1'(f, \Ga),\\
\la^{-1}e_4'(\log(\la)) &=&  E_2'(f, \Ga),\\
\la^{-1}e_4'\Big(r\fb-2r^2e_\th'(\log(\la))+rf\Omb\Big) &=& E_3'(f, \fb, \la, \Ga),
\eeaa
where
\beaa
E_1'(f, \Ga) &=& -\frac{r}{2}\kac f -\frac{r}{2}\vth f+\lot,\\
E_2'(f, \Ga) &=&  f\ze- \frac{1}{2}f^2\omb -\etab f - \frac{1}{4}f^2\kab +\lot,\\
E_3'(f, \fb, \la, \Ga)  &=& -\frac{r}{2}\kac\fb+ r^2\left(\kac-\left(\ov{\ka}-\frac{2}{r}\right)\right)e_\th'(\log(\la)) +r^2\Big(\ddd_1'(f)+\la^{-1}\vth'\Big)e_\th'(\log(\la))\\
&&   -\frac{r}{2}\kac\Omb f    +rE_3(f, \fb, \Ga)-2r^2e_\th'(E_2(f, \Ga))+r\Omb E_1(f, \Ga),
\eeaa
and where $E_1$, $E_2$ and $E_3$ are given in Lemma \ref{lemma:transportequationsforffbandlambda}.
\end{corollary}

\begin{proof}
See section \ref{sec:proofofcor:transportequationsforffbandlambda}.
\end{proof}

    
\subsection{Schematic notation $\Ga_g$ and $\Ga_b$}\lab{sec:schematicnotaionsGagGab}   
  

Many of the identities which we present   below,   contain a huge number of  $O(\ep^2)$ terms.  In what follows we introduce schematic  notation meant to keep track  of  
  the most important error   terms. Note that the decomposition below between  the terms  $\Ga_g$ and $\Ga_b$  is consistent with our main bootstrap assumptions {\bf BA-E} on energy and {\bf BA-D} on decay, see section \ref{section:bootstrap}.

\begin{definition}
\label{definition-errortermsforsquareqf}
We divide the   small  connection coefficient terms  (relative to an arbitrary null frame) into\footnote{In the frames we are using, we have in fact $\xi=0$ for $r\geq 4m_0$ so that it behaves in fact better than the other components of $\Ga_g^{(0)}$.}
\beaa
\Ga_{g}^{(0)} = \left\{r\xi, \, \vth, \, \ze, \, \etab, \,\frac{2}{r}e_4(r)-\ka, \,\frac{1}{r}e_\th(r)\right\}, \qquad \qquad  \Ga_{b}^{(0)}=\left\{\eta, \,\vthb, \,\xib, \,\frac{2}{r}e_3(r)-\kab\right\}.
\eeaa
For higher derivatives we  introduce,
\beaa
\Ga_{g}^{(1)} = \Big\{\dk\Ga_g^{(0)},\, r^2e_\th(\om), \, re_\th(\ka), \, re_\th(\kab)\Big\}, \qquad \qquad  \Ga_{b}^{(1)}=\Big\{\dk\Ga_b^{(0)}, \, re_\th(\omb)\Big\},
\eeaa
and for $s\geq 2$,
\beaa
\Ga_g^{(s)} = \dk^{\leq s}\Ga_g, \qquad  \Ga_b^{(s)} = \dk^{\leq s}\Ga_b,
\eeaa
where we have introduced the notations
$$\dk=\{e_3, re_4, \dkb\},$$
with angular derivatives $\dkb$ of reduced scalars in $\mathfrak{s}_k$ defined by \eqref{def:angularderivativesonreducedksclars}. 
\end{definition}

\begin{remark}
\label{remark-notation-qf} According to  the main bootstrap   assumptions   {\bf BA-E}, {\bf BA-D} (see         section \ref{section:bootstrap}), 
  the terms $\Ga_b$ behave worse in powers of $r$ than  the terms in $\Ga_g$. 
  Thus, in the calculations below, we  replace   the terms of  the form $\Ga^{(s)}_g+\Ga^{(s)}_b$   by $\Ga^{(s)}_b$.   Given the form of the bootstrap assumptions, we may also replace $r^{-1}\Ga^{(s)}_b$ by $\Ga^{(s)}_g$.
  We will denote $l.o.t.$ the cubic and higher error terms in $\Gac, \Rc$.    We also include in $\lot $ terms which decay faster in powers of $r$  than    the main  quadratic terms.
  \end{remark}


\subsection{The invariant quantity $\qf$}\lab{sec:discussionanddefintionofinvariantquantities}


Note  from the transformation formulas of Proposition \ref{prop:transformations1} that the only quantities which   remain invariant up to quadratic  or higher order  error terms   are $\a$,  $\aa$ and $\rho$. Among these only $\a, \aa$   vanish in Schwarzschild. We call  such quantities   $O(\ep^2)$ invariant. In what follows we show that, in addition to these two invariants,  there exist other  important invariants.

\begin{lemma}
\label{lemma:def: gen-qf}
The expression,
\beaa
e_3(e_3(\a))+(2\kab -6\omb)e_3(\a)+\left(-4e_3(\omb)+8\omb^2-8\omb\,\kab+\frac{1}{2}\kab^2\right)\a
\eeaa
is an $O(\ep^2)$ invariant. It is also a conformal invariant, i.e. invariant  under transformations \eqref{SSMe:general.composite}
 with $f=\fb=0$.
\end{lemma}

\begin{proof}
Clearly  the quantity  vanishes in Schwarzschild and is an  $O(\ep^2)$ invariant.         For a conformal transformation, the result follows by  a straightforward application of the transformation properties of Proposition \ref{prop:transformations1} in the particular case where $f=\fb=0$.
\end{proof}

\begin{remark}
Alternatively one can also   define  the corresponding quantity   obtained  by interchanging $e_3, e_4$, i.e.
\beaa
e_4(e_4(\aa))+(2\ka -6\om)e_4(\aa)+\left(-4e_4(\om)+8\om^2-8\om\ka+\frac{1}{2}\ka^2\right)\aa.
\eeaa
Note that it differs by $O(\ep^2)$ from the previous one.
\end{remark}

\begin{definition}\lab{definition:materquantity-qf}
Given a general null frame $(e_4, e_3, e_\th)$, and given a scalar function $r$ satisfying the assumptions for section \ref{sec:schematicnotaionsGagGab}, i.e. 
\beaa
\frac{2}{r}e_4(r)-\ka\in\Ga_g,\quad \frac{1}{r}e_\th(r)\in\Ga_g, \quad \frac{2}{r}e_3(r)-\kab\in\Ga_b,
\eeaa
we defined our main quantity $\qf$ as
\bea
\label{def: gen-qf}
\qf := r^4\left[e_3(e_3(\a))+(2\kab -6\omb)e_3(\a)+\left(-4e_3(\omb)+8\omb^2-8\omb\,\kab+\frac{1}{2}\kab^2\right)\a\right].
\eea
\end{definition}

    
\subsection{Several identities for $\qf$}   
  
    
In this section, we state three identities involving the quantity $\qf$ defined by \eqref{def: gen-qf}. All calculations are made   in  a general frame.
\begin{proposition}
\label{prop:alternateformulaforqfinvolvingtwoangularderrivativesofrho}
We have
\bea\lab{eq:alternateformulaforqfinvolvingtwoangularderrivativesofrho}
\qf &=& r^4\left( \dds_2\dds_1\rho+\frac{3}{4}\kab\rho\vth +\frac{3}{4}\ka\rho\vthb\right) +\err[\qf] 
\eea
with error term written schematically in the form
\bea
\lab{eq:alternateformulaforqfinvolvingtwoangularderrivativesofrho-err}
\err[\qf]&=&r^4 e_3 \eta \c \b+ r^ 2 \dk^{\le 1 }\big( \Ga_b\c\Ga_g).
\eea
\end{proposition}

\begin{proof}
See section \ref{sec:proofofprop:alternateformulaforqfinvolvingtwoangularderrivativesofrho}
\end{proof}

 The following consequence of  Proposition \ref{prop:alternateformulaforqfinvolvingtwoangularderrivativesofrho} 
 will prove to be very useful in the sequel.
  
\begin{proposition}
\label{Le:Teuk-Star1}
We have
\bea
\label{eq:Le-Teuk-Star1}
\nn e_3(r\qf) &=& r^5\left\{ \dds_2\dds_1\ddd_1\bb   -\frac{3}{2}\rho\dds_2\dds_1\kab -\frac{3}{2}\kab\rho\dds_2\ze -\frac{3}{2}\ka\rho\aa + \frac{3}{4}(2\rho^2-\ka\kab\rho)\vthb\right\}\\
&&+\err[e_3(r\qf)],
 \eea
 where the error term $\err[e_3(r\qf)]$ is given schematically by
 \bea
\err[e_3(r\qf)] &=& r\Ga_b \qf +r^5  \dk^{\le 1} \big(e_3 \eta \c \b\big) + r^3  \dk^{\le 2 }\big( \Ga_b\c\Ga_g\big).
\eea
\end{proposition}

\begin{proof}
See section \ref{sec:appendix:proofpropLe:Teuk-Star1}.
\end{proof}

We deduce from Proposition \ref{Le:Teuk-Star1} the following nonlinear version of the Teukolsky-Starobinski identity.

\begin{proposition}
\lab{Prop:Teuk-Star-main(seeApp)}\lab{Prop:Teuk-Star-main}
The following identity holds true in $\Mint$,
\begin{equation}\lab{eq:Teuk-Star-main(seeApp)}
e_3(r^2e_3(r\qf))+2\omb r^2e_3(r\qf) = r^7\left\{  \dds_2\dds_1\ddd_1\ddd_2\aa  +\frac{3}{2}\rho\Big(\kab e_4 -\ka e_3\Big)\aa\right\}+\err[TS],
\end{equation}
where the error term $\err[TS]$ is given schematically by
\beaa
\err[TS]&=&r^4\big( \dkb \Ga_b+  r \Ga_b\c \Ga_b) \c   \aa+r^2\big( \Ga_b e_3(r\qf)+( \dk^{\le 1} \Ga_b) r\qf \Big)\\
&+& r^7 \dk^{\le 2} \big(e_3 \eta \c \b\big)+r^5  \dk^{\le 3 }\big( \Ga_b\c\Ga_g\big).
\eeaa
\end{proposition}

\begin{proof}
See section \ref{appendix:Teuk-Star}.
\end{proof}


\section{Invariant wave equations}
\label{section:invariantwaveequations}

 
In this section,  we  write wave equations for the invariant quantities $\a$, $\aa$  and $\qf$.


\subsection{Preliminaries} 
\lab{subsection:invariantwaveequations}


\begin{lemma}
\label{le:square(psi)-null-frame} With respect to a general $S$-foliation 
we have, for a  reduced scalar $\psi\in \mathfrak{s}_0$,
\bea
\begin{split}
\square_\g \psi&=-\frac 1 2 \left(e_3 e_4  + e_4 e_3\right)       \psi+\lapp\psi  +\left(\omb -\frac 1 2\kab\right) e_4\psi+\left(\om -\frac 1 2 \ka\right) e_3\psi \\
&+(\eta+\etab) e_\th \psi.
\end{split}
\eea
Also,
\beaa
\square_\g \psi&=&-e_3 e_4 \psi +\lapp\psi+\left(2\omb -\frac 1 2 \kab\right) e_4\psi- \frac 1 2 \ka e_3\psi+2 \eta e_\th \psi,\\
\square_\g \psi&=&-e_4 e_3 \psi +\lapp\psi+\left(2\om -\frac 1 2 \ka\right) e_3\psi- \frac 1 2 \kab e_4\psi+2 \etab e_\th \psi.
\eeaa
\end{lemma}

\begin{proof}
We calculate, in spacetime,
\beaa
\square_\g\psi&=&  \g ^{34} \D_3\D_4\psi +  \g^ {43} \D_4\D_3 \psi +\de^{AB}\D_A\D_B\psi=-\frac 12 (\D_3\D_4 +\D_4\D_3)\psi+\g^{AB}\D_A\D_B\psi.
\eeaa
Now,
\beaa
\de^{AB}\D_A\D_B\psi&=&\lapp \psi -\frac 1 2 \trchS e_3\psi-\frac 1 2 \trchbS e_4 \psi, \\
\D_3\D_4\psi&=& e_3 e_4 \psi- 2\omb e_4\psi-2\eta e_\th \psi,\\
\D_4\D_3\psi &=& e_4 e_3 \psi- 2\om e_3\psi-2\etab e_\th \psi.
\eeaa
Hence,
\beaa
\square_\g\psi&=& -\frac 12 (e_3e_4 +e_4e_3)\psi+\lapp\psi-\frac 1 2 \trchS e_3\psi-\frac 1 2 \trchbS e_4 \psi\\
&+&\omb e_4\psi+ \eta e_\th \psi+\om e_3\psi+ \etab e_\th \psi\\
&=&-\frac 12 (e_3e_4 +e_4e_3)\psi+\lapp\psi +\left(\omb -\frac 1 2 \trchbS\right) e_4\psi+\left(\om -\frac 1 2 \trchS\right) e_3\psi\\
&+&(\eta+\etab) e_\th \psi.
\eeaa
Since,
\beaa
\frac 1 2 e_4 e_3 \psi=\frac 1 2 e_3 e_4 \psi+\om e_3 \psi-\omb e_4\psi+(\etab-\eta) e_\th \psi
\eeaa
we also have,
\beaa
\square_\g \psi&=&-e_3 e_4 \psi+\lapp\psi+\left(2\omb -\frac 1 2 \trchbS\right) e_4\psi- \frac 1 2 \trchS e_3\psi+2 \eta e_\th \psi.
\eeaa
Since $\ka=\trchS, \,\kab=\trchbS$, this concludes the proof of the lemma.
\end{proof}

\begin{definition}
\label{Definition:square-k}
Given a reduced $k$-scalar $\psi\in\mathfrak{s}_k$ we define,
\bea
\begin{split}
\square_k \psi&=-\frac 1 2 \left(e_3 e_4  + e_4 e_3\right)       \psi+\lapp_k\psi  +(\omb -\frac 1 2 \trchb) e_4\psi+(\om -\frac 1 2 \trch) e_3\psi \\
&- (\eta+\etab)  e_\th \psi.
\end{split}
\eea
Equivalently, we have
\beaa
\square_k \psi&=&-e_3 e_4 \psi +\lapp_k\psi+\left(2\omb -\frac 1 2 \kab\right) e_4\psi- \frac 1 2 \ka e_3\psi+2 \eta e_\th \psi,\\
\square_k \psi&=&-e_4 e_3 \psi +\lapp_k\psi+\left(2\om -\frac 1 2 \ka\right) e_3\psi- \frac 1 2 \kab e_4\psi+2 \etab e_\th \psi.
\eeaa
\end{definition}

\begin{remark}
Not that the terms $\eta e_\th \psi, \etab e_\th \psi $ have to be interpreted  as in Remark 
\ref{Remark:$ethf$}, i.e.
\beaa
 \eta e_\th \psi  &=& \frac 1 2 \eta  \left(\ddd_k \psi -\dds_{k+1} \psi\right).
\eeaa
The term $\eta \ddd_k\psi $  is the reduced form of   a tensor product of $\etaS$ with $ \DDd_k \,^{(1+3)} \psi$ while 
$\eta \dds_{k+1} \psi$ is the reduced form of  a  contraction between  the $1$ form $\etaS$ and  $k+1$ tensor    $\DDs_{k+1} \,^{(1+3)} \psi$.
\end{remark} 

\begin{remark}
Recall  that (recall Definition \ref{definition:reducedderivatives}),
 \beaa
    \lapp_k f&:= e_\th( e_\th f) + e_\th(\Phi)  e_\th f - k^2 \big (e_\th(\Phi)\big)^2 f.
   \eeaa
   Thus, for a $\psi\in \mathfrak{s}_k$, we  have,
   \beaa
   \lapp_k \psi= \lapp \psi -  k^2 \big (e_\th(\Phi)\big)^2   \psi.
   \eeaa
 \end{remark}


\subsubsection{Spacetime  interpretation of   Definition \ref{Definition:square-k}}
\lab{subsubsection-spacetimeinterpfprqf}


The linearized  equation verified by our main   quantity  $ \qf$, which   will  be derived  in the next subsection, has the form,
\bea
\label{equation:square2psi+Vpsi}
\square_2 \psi+V\psi = 0.
\eea
with $V$ a scalar potential.
In what follows we give  simple spacetime interpretation of the equation.

Given a mixed spacetime   tensor in   $\T^k \M\otimes   \T_S^l \M  $  of the form $U_{\mu_1\ldots \mu_k,  A_1\ldots A_L}$ where $e_\mu$ is an orthonormal frame  on $\MM$ with $(e_A)_{A=1,2}$  tangent to $S$.
We define,
\beaa
\Db_\mu U_{\nu_1\ldots \nu_k,  A_1\ldots A_L}&=& e_\mu U_{\nu_1\ldots \nu_k,  A_1\ldots A_l}-U_{\D_\mu\nu_1\ldots \nu_k,  A_1\ldots A_l}-\ldots- U_{\nu_1\ldots  \D_\mu\nu_k,  A_1\ldots A_l}\\
&-& U_{\nu_1\ldots \nu_k,   \Db_\mu A_1\ldots A_l}-  U_{\nu_1\ldots \nu_k,   A_1 \ldots \Db_\mu A_l}
\eeaa
with $\Db_\mu A$  denoting  the projection of  $\D_{e_\mu}  e_A$ on  $S$.
One can easily  check   the commutator  formulae,
 \beaa
( \Db _\mu\Db_\nu -\Db_\nu\Db _\mu)\Psi_A&=&\R_{A}\, ^   B\,_{   \mu\nu}\Psi_B,
 \\
( \Db _\mu\Db_\nu -\Db_\nu\Db _\mu)\Psi_{\la A}&=&   \R_{\la }\, ^   \si \,_{   \mu\nu}\Psi_{\si A}+    
            \R_{A}\, ^   B\,_{   \mu\nu}\Psi_{\la B}.
 \eeaa
 Define,
 \beaa
\squared_\g \Psi := \g^{\mu\nu} \Db_\mu\Db_ \nu \Psi.
\eeaa
         Consider the following  Lagrangian for   $\Psi=\Psi_{AB} \in \SS_2$.
 \beaa
 \LL[\Psi]&=& \gS^{A_1B_1} \gS^{A_2B_2}  \left(   \g^{\mu\nu}         \Db_\mu  \Psi_ {A_1 A_2} \Db_\mu  \Psi_{B_1 B_2}  +
 V \Psi_{A_1 A_2} \Psi_{B_1B_2} \right).
 \eeaa

\begin{proposition}
\label{prop:spacetimeversioofwaveforqf}
The Euler- Lagrange equations       for  the Lagrangian $\LL[\Psi]$ above are given by:
\bea
\squared\Psi= V \Psi
\eea
  and its reduced for $\psi =\Psi_{\th\th} $  is precisely \eqref{equation:square2psi+Vpsi}.
 \end{proposition}

\begin{proof}
Straightforward  verification.
\end{proof}


\subsection{Wave equations for $\a$, $\aa$, and $\qf$}\lab{sec:waveeqationforalphaalphabarandqfrak}


We start with the wave equations for $\a$ and $\aa$, which are derived in a general null frame.  
\begin{proposition}      
\label{proposition:wave-a-aa-qf}  
The following identities  hold true.
\begin{enumerate}
\item The invariant  quantity $\a\in \sk_2$ verifies the Teukolsky  wave  equation,
\bea\lab{eq:Teukolskiequationforalpha} 
\bsplit
 \square_2\a &= -4\omb e_4(\a)+ (4 \om+2\ka) e_3(\a) + V \a +\err[\square_\g \a],\\
V&=-4\rho   - 4 e_4(\omb)  -8\om\omb  +2\om\, \kab    -10 \ka\, \omb+\frac 1 2 \ka\,\kab,
\end{split}
\eea
where $\err[\square_\g \a]$ is given schematically by
\beaa
\err(\square_\g\a) &=& \Ga_g e_3(\a) + r^{-1}\dk^{\leq 1}\Big((\eta, \Ga_g)(\a, \b)\Big)    +\xi (e_3(\b), r^{-1}\dk\rhoc)+\lot    
\eeaa
where $\lot$ denote terms which are quadratic and enjoy better decay properties or are higher order and decay at least as good.

\item The invariant  quantity $\aa\in \sk_2$ verifies the Teukolsky  wave  equation,
\bea\lab{eq:Teukolskiequationforalphabar}
\bsplit
 \square_2\aa&= -4\om e_3(\aa)+ (4 \omb+2\kab) e_4(\aa) +\Vb \aa+\err[\square_\g \aa],\\
\Vb&=-4\rho   - 4 e_3(\om)  -8\om\omb  +2\omb\ka    -10 \kab\, \om+\frac 1 2 \ka\,\kab,
\end{split}
\eea
where 
\beaa
\err(\square_\g\aa) &=& r^{-1}\dk(\Ga_b\aa)+\dk(\Ga_b\bb) +\lot
\eeaa
\end{enumerate}
\end{proposition}

\begin{proof}
See appendix \ref{appendix-Prop.lemma:waveeqalphawithquadraticterms}
\end{proof}

We may now state the wave equation satisfied by $\qf$. 
\begin{theorem}
\label{thm:wave-qf}
The invariant  scalar quantity $\qf$ defined in \eqref{def: gen-qf} 
 verifies  the  equation,
\bea
\label{thmwaveqf:improperfrom}
\square_2 \qf  +\ka \kab\, \qf=\err[\square_2\qf]
\eea
where $\err[\square_2\qf]$ is $O(\ep^2)$.

 If $\qf$ is defined relative to a null frame satisfying, in addition to the assumptions of section \ref{sec:schematicnotaionsGagGab}, that $\eta\in\Ga_g$ and $\xi=0$ for $r\geq 4m_0$, the error term is then given schematically by
\bea\lab{thmwaveqf:schematicformerrorterm}
\err[\square_2\qf]&= r^2 \dk^{\le 2}(\Ga_g \c (\a, \b) )         + e_3 \Big( r^3 \dk^{\le 2}(\Ga_g \c (\a, \b) )   \Big) +  \dk^{\le 1 } (\Ga_g \c \qf)+\lot
\eea
\end{theorem}

\begin{proof}
See appendix \ref{appendix-Proof-thmwaveqf}.
\end{proof}

\begin{remark}\lab{rmk:whyweneedaglobalframewithbettereta}
Note that the main frame used in this paper is an outgoing geodesic null frame in $r\geq 4m_0$ so that $\xi=0$, but unfortunately, as it turns out,  $\eta\in \Ga_b$. This would not allow us to control the error term appearing in \eqref{thmwaveqf:improperfrom}. To overcome this problem, we are forced to define $\qf$ relative to a different frame where $\xi=0$ still holds for $r\geq 4m_0$ and for which we have in addition $\eta\in\Ga_g$, see Proposition \ref{prop:existenceandestimatesfortheglobalframe:bis} for the existence of such a frame. See also the discussion at the beginning of section \ref{subsection:constructionsecondframeinMext}. 
\end{remark}

The remark above leads us to the following.
\begin{remark}\lab{def:wherewementionthatwealwaysexpressqfinthesecondglobalframe}
The quantity $\qf$   we will be working with for the rest of the paper  is defined, according to equation \eqref{def: gen-qf},  relative to the global frame of Proposition \ref{prop:existenceandestimatesfortheglobalframe:bis} for which  $\eta\in \Ga_g$. It is only in such a frame that $\qf$ verifies the correct decay estimates.
\end{remark}


\chapter{MAIN THEOREM}\lab{chapter:maintheoremstatementbootstrapandinterresults}



\section{General covariant modulated admissible spacetimes}\lab{section:GCMspacetime}


Note that all definitions below are consistent with the framework of $\Z$-invariant polarized spacetimes.


\subsection{Initial data layer}\lab{sec:defintionoftheinitialdatalayer}


Recall that $m_0>0$ is given as the mass of the Schwarzschild solution to which the initial data is $\ep_0$ close.  Let $\deh>0$ be a sufficiently small constant which will be specified later.

\begin{definition}[Initial data layer]
\label{definition:initialdatalayer}
We consider a spacetime region $(\LL_0, g)$, sketched  below in figure  \ref{fig0}, where
\begin{itemize}
\item The metric $g$ is a reduced metric from a Lorentzian spacetime metric $\g$ close to Schwarzschild in a suitable topology\footnote{This topology will be specified in our initial data layer assumptions, see \eqref{def:initialdatalayerassumptions} as well as section \ref{sec:initialdatalayernorm}.}. 

\item $\LL_0=\Lext\cup\Lint$. 

\item The intersection   $\Lext\cap\Lint$ is non trivial.
\end{itemize}
Furthermore, our initial data layer $(\LL_0, g)$ satisfies
\begin{enumerate}
\item {\bf Boundaries.} The future and past  boundaries  of $\LL_0$ are  given by
\beaa
\pr^+\LL_0&=& \AA_0 \cup \CCb_{(2,\idl)}\cup \CC_{(2,\idl)},\\
\pr^- \LL_0 &=& \CC_{(0,\idl)}\cup \CCb_{(0,\idl)},
\eeaa
where
\begin{enumerate}
  \item  The  past outgoing    null boundary of the far region $\Lext$ is denoted by $\CC_{(0,\idl)}$. 
 
    \item  The  past incoming    null boundary of the near region  $\Lint$ is denoted by $\CCb_{(0,\idl)}$.  
 
 \item  $\Lext$ is unbounded in the future outgoing null directions.
   
  \item  The  future outgoing    null boundary of the far region $\Lext$ is denoted by $\CC_{(2,\idl)}$. 
   
   \item  The  future incoming    null boundary of the near region  $\Lint$ is denoted by $\CCb_{(2,\idl)}$.  
 
  \item  The  future spacelike boundary of the near region $\Lint$ is denoted by $\AA_0$.
  
 \end{enumerate}

\item {\bf Foliations of $\LL_0$ and adapted null frames.} The spacetime $\LL_0=\Lext\cup \Lint$ is foliated as follows
\begin{enumerate}

\item  The  far region $\Lext$ is foliated by two functions $(u_{\idl}, \sextl)$ such that
\begin{itemize}
\item  $u_{\idl}$ is an outgoing  optical function on $\Lext$ whose leaves are denoted by $\CC_{(u_{\idl},\idl)}$. 

\item  $\sextl$ is an affine parameter  along the level hypersurfaces  of $u_{\idl}$, i.e. 
\beaa
\,{}^{(ext)}L_0( \sextl)=1\textrm{  where }\,{}^{(ext)}L_0:=-g^{ab} \pr_b (u_{\idl}) \pr_a.
\eeaa

\item We denote by $(\,{}^{(ext)}(e_0)_3, \,{}^{(ext)}(e_0)_4, \,{}^{(ext)}(e_0)_\th)$ the null frame adapted to the outgoing geodesic foliation $(u_{\idl}, \sextl)$ on $\Lext$. 

\item Let $\rextl$  denote the area radius of the 2-spheres $S(u_{\idl}, \sextl)$ of this foliation. 


\item The    outgoing  future null boundary $\CC_{(2,\idl)}$ corresponds precisely  to $u_{\idl}=1$ and   the outgoing   past null  boundary $\CC_{(0,\idl)}$  corresponds to  $u_{\idl}=-1$.

\item The foliation  by $u_{\idl}$ of $\Lext$ terminates at the time like boundary 
$$\left\{\rextl= 2m_0\left(1+\frac{\deh}{4}\right)\right\}.$$
\end{itemize}

\item  The near region $\Lint$ is foliated by two functions $(\ub_{\idl}, \sintl)$ such that  
\begin{itemize}
\item  $\ub_{\idl}$ is an ingoing  optical function on $\Lint$ whose leaves are denoted by $\CC_{(\ub_{\idl},\idl)}$. 

\item  $\sintl$ is an affine parameter  along the level hypersurfaces  of $\ub_{\idl}$, i.e. 
\beaa
\,{}^{(int)}\Lb_0( \sintl)=-1\textrm{  where }\,{}^{(ext)}\Lb_0:=-g^{ab} \pr_b (\ub_{\idl}) \pr_a.
\eeaa

\item We denote by $(\,{}^{(int)}(e_0)_3, \,{}^{(int)}(e_0)_4, \,{}^{(int)}(e_0)_\th)$ the null frame adapted to the outgoing geodesic foliation $(u_{\idl}, \sintl)$ on $\Lint$. 

\item Let $\rintl$  denote the area radius of the 2-spheres $S(\ub_{\idl}, \sintl)$ of this foliation. 

\item The $(\ub_{\idl},\sint)$ foliation is initialized on $\rextl=2m_0(1+\frac{\deh}{2})$ as it will be made precise below. 

\item The foliation  by $\ub_{\idl}$, of $\Lint$ terminates at the space like boundary 
$$\AA_0=\left\{\rintl= 2m_0(1-2\deh)\right\}.$$
  where $m_0$ and $\deh$ have been defined above.

\item The    ingoing  future null boundary $\CCb_{(2,\idl)}$ corresponds precisely  to $\ub_{\idl}=2$ and   the ingoing   past null  boundary $\CCb_{(0,\idl)}$  corresponds to  $\ub_{\idl}=0$.

\item The foliation  by $\ub_{\idl}$ of $\Lint$ terminates at the time like boundary 
$$\Big\{\rintl= 2m_0\left(1+2\deh\right)\Big\}.$$
\end{itemize}
\end{enumerate}

\item {\bf Initializations  of the $(\ub_{\idl},\sint)$ foliation.} 
 
 The $(\ub_{\idl},\sintl)$ foliation is initialized on $\rextl=2m_0(1+\deh)$ 
by setting,
\beaa
\ub_{\idl}&=&u_{\idl}, \qquad \sintl=\sextl
\eeaa
and, with $\la_0= {}^{(ext)}\la_0=1-\frac{2m_0}{\rextl}$,
\beaa
\,{}^{(int)}(e_0)_4=\la_0\,{}^{(ext)}(e_0)_4,\quad  {}^{(int)}(e_0)_3=\la_0^{-1}\,{}^{(ext)}(e_0)_3,\quad 
\,{}^{(int)}(e_0)_\th=\,{}^{(ext)}(e_0)_\th.
\eeaa

\item {\bf Coordinates system on $\Lext(\rextl\geq 4m_0)$.} In $\Lext(\rextl\geq 4m_0)$, there exists adapted coordinates $(u_{\idl},\sext_{\idl},\th_{\idl}, \vphi)$ with $b=0$, see   Proposition \ref{prop:outgoinggeodesiccoordinates}, such that the  spacetime  metric $\g$ takes the form,
\begin{equation}
 \label{reducedmetric-geodesicfoliation:intheinitialdatalayer}
  \g = -2 du_{\idl}\, d\left(\sext_{\idl}\right) +\Omb_{\idl} (du_{\idl})^2+ \ga_{\idl}\left( d\th_{\idl} - \frac 1 2 \underline{b}_{\idl} du_{\idl}\right)^2 + e^{2\Phi} d\vphi^2. 
 \end{equation}
\end{enumerate} 
\end{definition}

\begin{figure}[h!]
\centering
\includegraphics[scale=0.5]{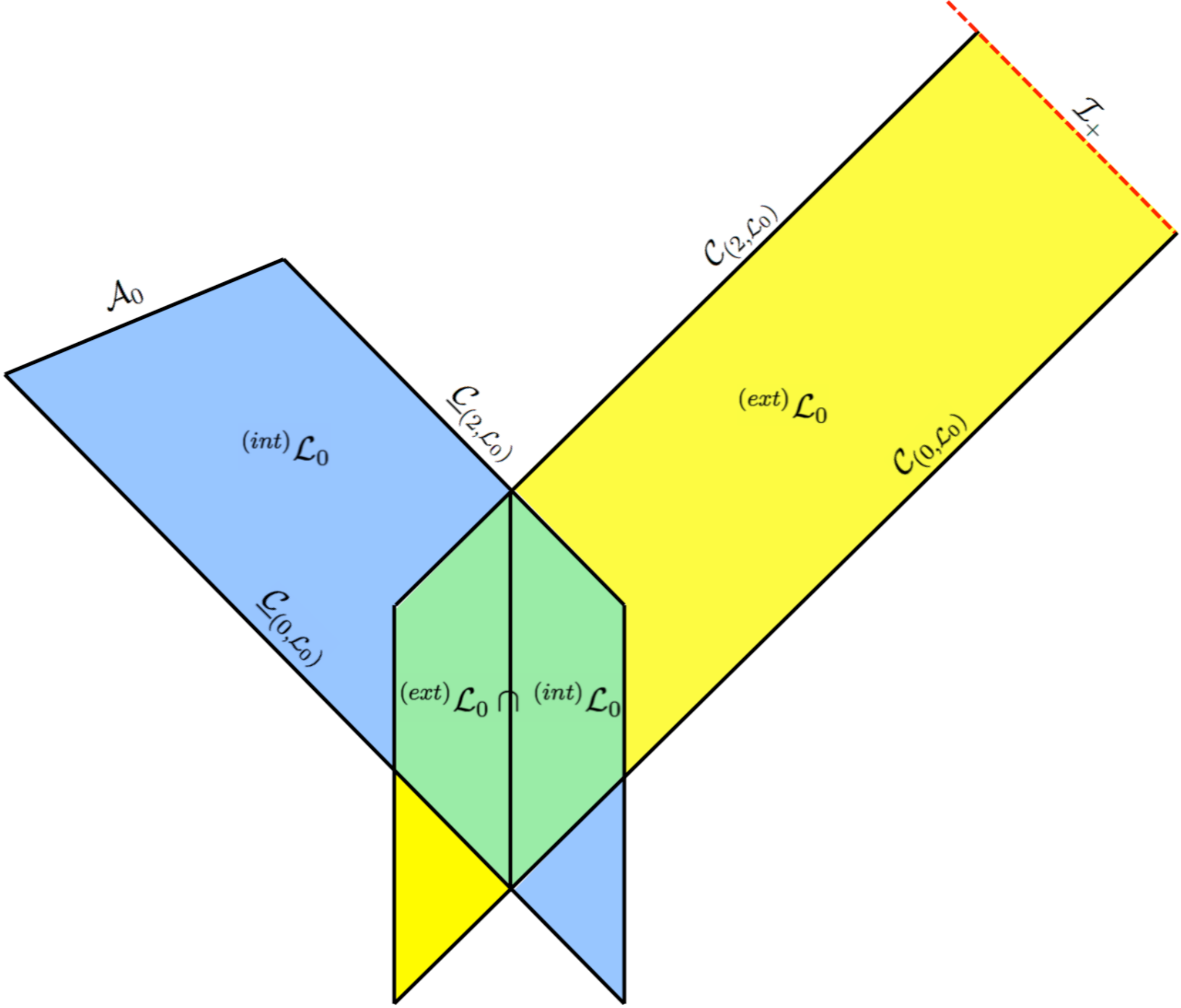}
\caption{The initial data layer $\LL_0$}
\label{fig0}
\end{figure}


\subsection{Main Definition}\lab{sec:defintioncanonicalspacetime}


Recall that $m_0>0$ is given as the mass of the Schwarzschild solution to which the initial data is $\ep_0$ close, and that $\deh>0$ is a sufficiently small constant which will be specified later. 

\begin{definition}[GCM-admissible spacetime]
\label{definition:canonical-spacetime}
We consider a spacetime $(\MM, g)$, sketched  below in figure  \ref{fig1}, where
\begin{itemize}
\item The metric $g$ is a reduced metric from a Lorentzian spacetime metric $\g$ close to Schwarzschild in a suitable topology\footnote{This topology will be specified in our bootstrap assumptions, see \eqref{def:bootstrapasumptionsglobalnorms} as well as section \ref{section:main-norms}.}. 

\item $\MM=\Mext\cup\Mint$ 

\item $\TT=\Mext\cap\Mint$ is a time-like hyper-surface.
\end{itemize}
$(\MM, g)$ is called a general covariant modulated admissible (or shortly GCM-admissible) spacetime  if it is defined as follows
\begin{enumerate}
\item {\bf Boundaries.} The future and past  boundaries  of $\MM$ are  given by
\beaa
\pr^+\MM&=& \AA \cup \CCb_*\cup \CC_*\cup\Sigma_*,\\
\pr^- \MM&=& \CC_1\cup \CCb_1,
\eeaa
where
\begin{enumerate}
\item The past boundary  $ \CC_1\cup \CCb_1$ is included in  the  initial data   layer $\mathcal{L}_0$, defined in section \ref{sec:defintionoftheinitialdatalayer}, in which the metric on $\MM$ is specified to be  
 a small perturbation of the Schwarzschild data. 
 
 \item  The  future spacelike boundary of the far region $\Mext$ is denoted by $\Sigma_*$.
  
  \item  The  future outgoing    null boundary of the far region $\Mext$ is denoted by $\CC_*$.  
  
   \item  The  future incoming    null boundary of the near region  $\Mint$ is denoted by $\CCb_*$.  
 
  \item  The  future spacelike boundary of the near region $\Mint$ is denoted by $\AA$.
  
  \item The time-like boundary  $\TT$, separating  $\Mext$ from $\Mint$,  starts at $\CCb_1\cap\CC_1$ and terminates at $\CCb_*\cap \CC_*$.
 \end{enumerate}

\item {\bf Foliations of $\MM$ and adapted null frames.} The spacetime $\MM=\Mext\cup \Mint$ is foliated as follows
\begin{enumerate}
\item  The  far region $\Mext$ is foliated by two functions $(u, \sext)$ such that
\begin{itemize}
\item  $u$ is an outgoing  optical function on $\Mext$, initialized on $ \Sigma_*$,  whose leaves are denoted by $\CC(u)$. 

\item  $\sext$ is an affine parameter  along the level hypersurfaces  of $u$, i.e. 
\beaa
L( \sext)=1\textrm{  where }L:=-g^{ab} \pr_b u \pr_a.
\eeaa
\item The $(u,\sext)$ foliation is initialized on $ \Sigma_*$  as it will be made precise below. 

\item We denote by $(\,{}^{(ext)}e_3, \,{}^{(ext)}e_4, \,{}^{(ext)}e_\th)$ the null frame adapted to the outgoing geodesic foliation $(u, \sext)$ on $\Mext$  where ${}^{(ext)}e_4=L$. 

\item Let $\rext$ and $\,{}^{(ext)}m$ respectively the area radius and the Hawking mass of the 2-spheres $S(u, \sext)$ of this foliation.

\item The    outgoing  future null boundary $\CC_*$ corresponds precisely  to $u=u_*$ and   the outgoing   past null  boundary $\CC_1$  corresponds to  $u=1$.

\item The foliation  by $u$ of $\Mext$ terminates at the time like boundary 
$$\TT=\left\{\rext= \rh\right\}$$
where $\rh$ satisfies\footnote{A specific choice of $\rh$ will be made in section \ref{subsection:discTheoremM8}, see \eqref{eq:choiceofRTTismadebythisinfimum},  in the context of a Lebesgue point argument needed to recover the top order derivatives.}
\beaa
2m_0\left(1+\frac{\deh}{2}\right)\leq \rh\leq 2m_0\left(1+\frac{3\deh}{2}\right).
\eeaa
\end{itemize}

\item  The near region $\Mint$ is foliated by two functions $(\ub, \sint)$ such that  
\begin{itemize}
\item  $\ub$ is an ingoing  optical function on $\Mint$, initialized on $\TT$, whose leaves are denoted by $\CCb(\ub)$. 

\item  $\sint$ is an affine parameter  along the level hypersurfaces  of $\ub$, i.e. 
\beaa
\Lb( \sint)=-1\textrm{  where }\Lb:=-g^{ab} \pr_b \ub \pr_a.
\eeaa
\item The $(\ub,\sint)$ foliation is initialized on $\TT$ as it will be made precise below. 

\item We denote by $(\,{}^{(int)}e_3, \,{}^{(int)}e_4, \,{}^{(int)}e_\th)$ the null frame adapted to the outgoing geodesic foliation $(u, \sint)$ on $\Mint$  where ${}^{(int)}e_3=\Lb$.  

\item Let $\rint$ and $\,{}^{(int)}m$ respectively the area radius and the Hawking mass of the 2-spheres $S(\ub, \sint)$ of this foliation.

\item The foliation  by $\ub$ of $\Mint$ terminates at the space like boundary 
$$\AA=\left\{\rint= 2m_0(1-\deh)\right\}$$
where $m_0 $ and $\deh$ have been defined above.

\item The    ingoing  future null boundary $\CCb_*$ corresponds precisely  to $\ub=u_*$ and   the ingoing   past null  boundary $\CCb_1$  corresponds to  $\ub=1$.
\end{itemize}
\end{enumerate}

\item {\bf GCM foliation of $\Sigma_*$.} The $(u, \sext)$-foliation of $\Mext$ restricted to the spacelike hypersurface $\Sigma_*$ has the following properties  
\begin{enumerate}
\item There exists a constant $c_{\Sigma_*}$ such that 
\beaa
\Sigma_*:=\{u+\rext=c_{\Sigma_*}\}.
\eeaa

\item We have\footnote{See \eqref{eq:behaviorofronSigmastar} for the precise condition.}
\bea\lab{eq:behaviorofronSigmastarrough}
r\gg u_*^4\textrm{ on }\Sigma_*.
\eea

\item $\sext$ satisfies\footnote{Recall that $\sext$ satisfies on $\Mext$ the transport equation $L(\sext)=1$ and thus needs to be initialized on a hypersurface transversal to $L$, chosen here to be $\Si_*$.}
\beaa
\sext=\rext\,\,\textrm{ on }\Si_*.
\eeaa

\item We say that $\Sigma_*$ is a general covariant modulated hypersurface\footnote{More generally, a  GCM hypersurface  is one  with the property  that  we can specify, using the   full covariance of the Einstein equations,  a number  of  vanishing conditions (equal to the number of degrees of freedom 
 of the  diffeomorphism  group)  for  well-chosen components  of $\Gac$.} (or shortly GCM hypersurface) if relative to the above defined null frame of $\Mext$, the following conditions  hold\footnote{The existence of such hypersurfaces is an essential part of our construction.} along  $ \Sigma_*$
\bea
\begin{split}
\ka=\frac{2}{r},\,\,\, \,\dds_2\dds_1\kab=0,\,\,\,  \,\dds_2\dds_1\mu=0,\\
\int_S\eta e^\Phi=0,\,\,\, \int_S\xib e^\Phi=0,\,\,\, a\big|_{SP}=1, 
\end{split}
\eea
where $a$ is the unique scalar function such that $\nu=e_3+ae_4$ is tangent to $\Sigma_*$, and $SP$ denotes the south poles of the spheres on $\Si_*$. Moreover we also assume
\bea
\int_{S_*}\b e^\Phi=0,\,\,\, \int_{S_*}e_\th(\kab) e^\Phi=0,\textrm{ with }S_*:=\Sigma_*\cap\CC_*.
\eea
Note that the role of the GCM foliation of $\Sigma_*$ is to initialize the $(u, \sext)$-foliation of $\Mext$. 

\item In view of the definition of $\nu$ and $\vsi$, we have $\nu(u)=e_3(u)+ae_4(u)=2/\vsi$. $\nu$ being tangent to $\Si_*$, $u$ is thus transported along $\Si_*$, and hence defined up to a constant. To calibrate $u$ on $\Si_*$, we fix the value $u=1$ as follows
\bea\lab{eq:calibrationofubythechoiceofS1intersectingSPconeIDL}
S_1=\Si_*\cap\{u=1\}\textrm{ is such that }S_1\cap\CC_{(1,\LL_0)}\cap SP\neq \emptyset,  
\eea
i.e. $S_1$  is the unique sphere of $\Si_*$ such that its south pole intersects the south pole of one of the sphere of the outgoing null cone $\CC_{(1,\LL_0)}$ of the initial data layer. 
\end{enumerate}

\item {\bf Initialization the $(\ub, \sint)$-foliation on $\TT$.} The $(\ub,\sint)$ foliation is initialized on $\TT$  such that,
\beaa
\ub=u, \qquad  \sint=\sext
\eeaa
In particular, the 2-spheres $S(u, \sint)$ coincide on $\TT$ with $S(u, \sext)$ and      $\rint=\rext$.
Moreover, the null frame $(\,{}^{(int)}e_3, \,{}^{(int)}e_4, \,{}^{(int)}e_\th)$ is defined on $\TT$ by the following  renormalization,
\beaa
\,{}^{(int)}e_4=  \la\,{}^{(ext)}\, e_4,\qquad {}^{(int)}e_3= \la^{-1}\, \,{}^{(ext)}e_3, \qquad {}^{(int)}e_\th=\,{}^{(ext)}e_\th\,\,\textrm{ on }\TT
\eeaa
where
\beaa
\la=\,{}^{(ext)}\la=1-\frac{2\,{}^{(ext)}m}{\rext}.
\eeaa
\end{enumerate} 
\end{definition}

\begin{figure}[h!]
\centering
\includegraphics[scale=0.5]{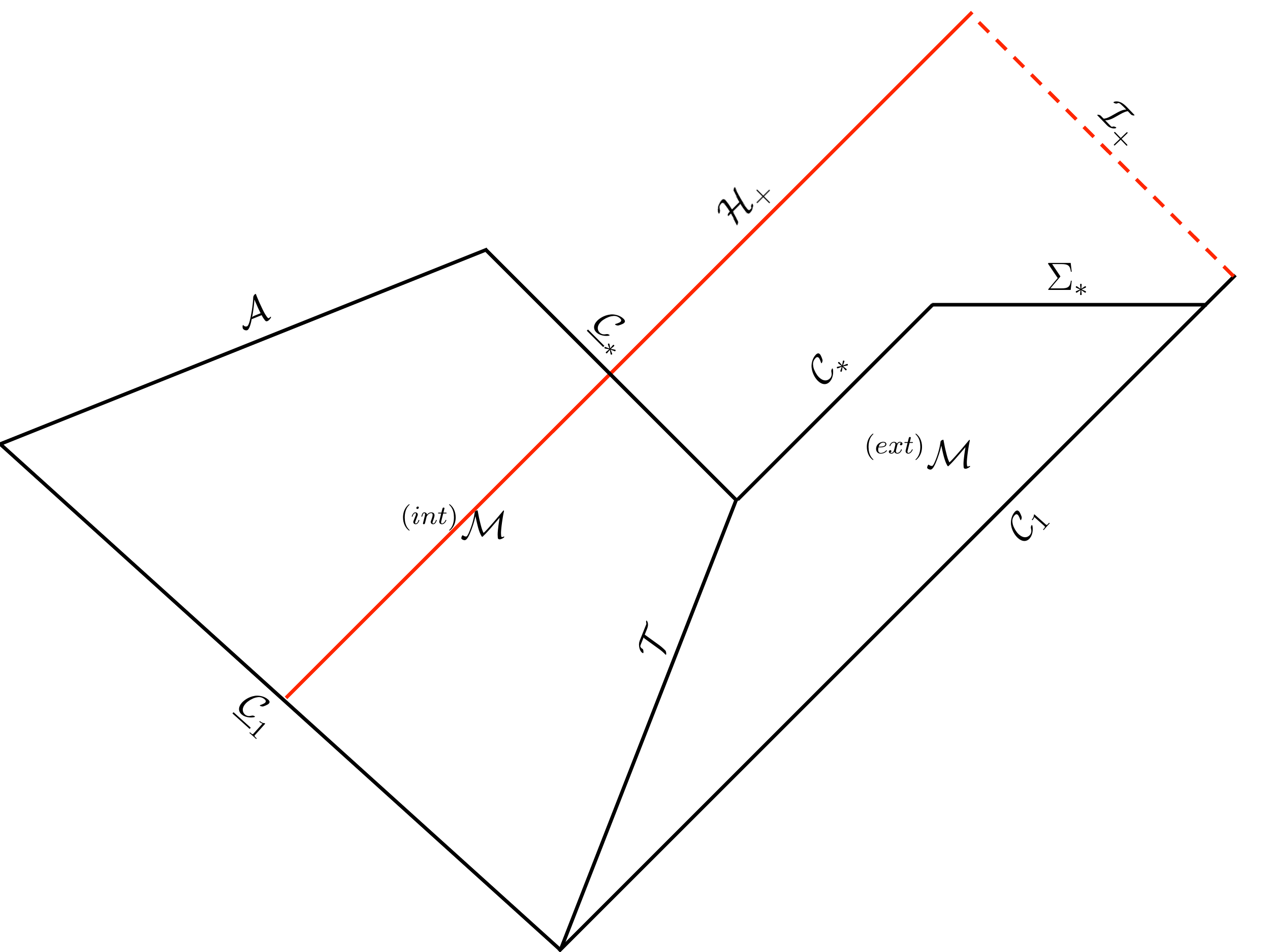}
\caption{The GCM admissible space-time $\mathcal{M}$}
\label{fig1}
\end{figure}

\begin{remark}
In Schwarzschild, $u= t-r_*$, $\ub= t+r_*$,  with $ \frac{dr_*}{dr}= \Up^{-1} $, and
\beaa
{}^{(ext)}e_4&=&\Up^{-1} \pr_t +\pr_r , \qquad  {}^{(ext)}e_3= \pr_t -\Up \pr_r,\\
{}^{(int)} e_4&=& \pr_t +\Up \pr_r , \qquad\quad   {}^{(int)}e_3=\Up^{-1} \pr_t- \pr_r.
\eeaa
\end{remark}


\subsection{Renormalized curvature components and Ricci coefficients}


For convenience, we introduce in this section a notation for renormalized curvature components and Ricci coefficients.

\begin{definition}[Renormalized curvature components and Ricci coefficients in $\Mext$]
We introduce the following notations in $\Mext$
\beaa
\,{}^{(ext)}\Rc &=& \Big\{ \a,\, \b, \, \rhoc, \, \check{\mu},\, \bb, \, \aa\Big\},\qquad   \,{}^{(ext)}\Gac = \Big\{ \kac, \, \vth, \, \ze, \, \eta, \,\kabc,\, \vthb,\, \ombc,\, \xib\Big\}, 
\eeaa
where, recall,
\beaa
 \rhoc= \rho-\overline{ \rho},\,\,\,\,  \check{\mu}= \mu-\overline{\mu},\,\,\,\,  \kac= \ka-\overline{ \ka},\,\,\,\,  \kabc= \kab-\overline{ \kab},\,\,\,\, \ombc= \omb-\overline{\omb},
\eeaa
and 
$$\xi=\om=0,\,\,\,\,  \etab=-\ze.$$
Note that all the above quantities are defined with respect to the outgoing geodesic foliation of $\Mext$ (see section \ref{sec:mainequationsforoutgoiggeodesicfoliations}), and that the averages are taken with respect to that corresponding 2-spheres.
\end{definition}

\begin{definition}[Renormalized curvature components and Ricci coefficients in $\Mint$]
We introduce the following notations in $\Mint$
\beaa
\,{}^{(int)}\Rc &=& \Big\{ \a,\, \b, \, \rhoc, \, \check{\mub},\, \bb, \, \aa\Big\},\\
 \,{}^{(int)}\Gac &=& \Big\{\xi,\, \omc,\,  \,\kac, \, \vth, \, \ze, \, \etab, \, \kabc,\, \vthb\Big\}, 
\eeaa
where we have defined
\beaa
 \rhoc= \rho-\overline{ \rho},\,\,\,\,  \check{\mub}= \mub-\overline{\mub},\,\,\,\,  \kac= \ka-\overline{ \ka},\,\,\,\,  \kabc= \kab-\overline{ \kab},\,\,\,\, \omc= \om-\overline{ \om},
\eeaa
and we recall that 
$$\xib=\omb=0,\,\,\,\, \eta=\ze,\,\,\,\, \overline{\mub}-\frac{2m}{r^3}=0.$$
Note that all the above quantities are defined with respect to the ingoing geodesic foliation of $\Mint$ (see section \ref{sec:mainequationsforingoiggeodesicfoliations}), and that the averages are taken with respect to that corresponding 2-spheres.
\end{definition}

\begin{remark}
In Schwarzschild, we have
$$\,{}^{(ext)}\Rc=0,\,\,\,\,\,{}^{(int)}\Rc=0,\,\,\,\,\,{}^{(ext)}\Gac=0,\,\,\,\, \,{}^{(int)}\Gac=0.$$
\end{remark}


\section{Main norms}\label{section:main-norms}



\subsection{Main norms in $\Mext$}\label{section:main-normsextregion}


All quantities appearing in this section are defined relative to the $\Mext$ frame adapted to the $(u, \sext)$ foliation. In particular, recall that with respect to this frame, we have
$$\xi=\om=0,\qquad \etab=-\ze.$$

Recall the definition \eqref{def:angularderivativesonreducedksclars} of higher order angular derivatives $\dkb^s$ of reduced scalars in $\mathfrak{s}_k$. We introduce the notations
$$\dk=\{e_3, re_4, \dkb\}.$$

\begin{definition}
We introduce the vectorfield $\T$ defined on $\Mext$ as 
\bea
\T:=\frac{1}{2}\left( \left(1-\frac{2 m}{ r}\right)e_4+e_3\right).
\eea
We also introduce the vectorfield $\N$ is defined on $\Mext$ by
\bea\label{def:prrvectorfield}
\N: = \frac{1}{2}\left( \left(1-\frac{2  m}{  r}\right)e_4 - e_3\right).
\eea
\end{definition}

\begin{remark}
In Schwarzschild, we have
$$\T=\pr_t,\,\,\,\, \N=\left(1-\frac{2m_0}{r}\right)\pr_r$$
in the standard $(t,r,\th, \varphi)$ coordinates.
\end{remark}

We are ready to  introduce our norms in $\Mext$.


\subsubsection{$L^2$ curvature norms in $\Mext$}


Let $\dt>0$ a small constant to be specified later. We introduce the weighted curvature norms,
\beaa
\Big(\,{}^{(ext)}\mathfrak{R}_0^{\geq 4m_0}[\Rc]\Big)^2 &:=&     \sup_{1\leq u\leq u_*}   \int_{\CC_u(r\geq 4m_0)} \Big(  r^{4+\dt} \a^2+r^4\b^2    \Big)\\
&+&\int_{\Sigma_*}\Big(r^{4+\dt}(\a^2+\b^2)+r^4(\check{\rho})^2+r^2\bb^2+\aa^2\Big)\\
&+&\int_{\Mext(r\geq 4m_0)}\Big(r^{3+\dt} (\a^2+\b^2) +r^{3-\dt}(\rhoc)^2 +r^{1-\dt}\bb^2 +r^{-1-\dt}\aa^2        \Big),
\eeaa
\beaa
\Big(\,{}^{(ext)}\mathfrak{R}_0^{\leq 4m_0}[\Rc]\Big)^2 &:=&   \int_{\Mext(r\leq 4m_0)}\left(1-\frac{3m}{r}\right)^2|\Rc|^2,
\eeaa
and
\beaa
\,{}^{(ext)}\mathfrak{R}_0[\Rc]: = \,{}^{(ext)}\mathfrak{R}_0^{\geq 4m_0}[\Rc] + \,{}^{(ext)}\mathfrak{R}_0^{\leq 4m_0}[\Rc].
\eeaa

For any nonzero integer $k$, we introduce the following higher derivatives norms
\beaa
\Big(\,{}^{(ext)}\mathfrak{R}_k[\Rc]\Big)^2 &:=&       \Big(\,{}^{(ext)}\mathfrak{R}_0[\dk^{\leq k}\Rc]\Big)^2 +\int_{\Mext(r\leq 4m_0)}\Big(|\dk^{\leq k-1}\N\Rc|^2+|\dk^{\leq k-1}\Rc|^2\Big).
\eeaa

\begin{remark}
Note that  the derivative in the $\N$ direction,  unlike all  other first derivatives of $\Rc$,  appear in the spacetime integral  $\int_{\Mext(r\leq 4m_0)}$ with top number of derivatives. This reflects the fact the $\N$- derivatives  do not degenerate at $r=3m$ in the Morawetz estimate. 
\end{remark}


\subsubsection{$L^2$ Ricci coefficients norms in $\Mext$}


 For any $k\geq 2$, we introduce the following  norms
\beaa
\Big(\,{}^{(ext)}\mathfrak{G}_k^{\geq 4m_0}\left[\Gac\right]\Big)^2 &:=& \int_{\Sigma_*}\Bigg[r^2\Big((\dk^{\leq k}\vth)^2+(\dk^{\leq k}\check{\ka})^2+(\dk^{\leq k}\ze)^2+(\dk^{\leq k}\check{\kab})^2\Big)+(\dk^{\leq k}\vthb)^2\\
\nn&+&(\dk^{\leq k}\eta)^2+(\dk^{\leq k}\check{\omb})^2+(\dk^{\leq k}\xib)^2\Bigg]\\
&+& \sup_{\la\geq 4m_0}\Bigg(\int_{\{r=\la\}}\Bigg[\la^2\Big((\dk^{\leq k}\vth)^2+(\dk^{\leq k}\check{\ka})^2+(\dk^{\leq k}\ze)^2\Big)\\
\nn&+&\la^{2-\dt}(\dk^{\leq k}\check{\kab})^2+(\dk^{\leq k}\vthb)^2+(\dk^{\leq k}\eta)^2+(\dk^{\leq k}\check{\omb})^2+\la^{-\dt}(\dk^{\leq k}\xib)^2\Bigg]\Bigg),
\eeaa
\beaa
\Big(\,{}^{(ext)}\mathfrak{G}_k^{\leq 4m_0}\left[\Gac\right]\Big)^2 &:=& \int_{\Mext(\leq 4m_0)}\left|\dk^{\leq k}\left(\Gac\right)\right|^2,
\eeaa
and 
\beaa
\,{}^{(ext)}\mathfrak{G}_k\left[\Gac\right]: = \,{}^{(ext)}\mathfrak{G}_k^{\leq 4m_0}\left[\Gac\right] + \,{}^{(ext)}\mathfrak{G}_k^{\geq 4m_0}\left[\Gac\right].
\eeaa


\subsubsection{Decay norms in $\Mext$}


Let $\dec>0$ a small constant to be specified later. We define
\beaa
\,{}^{(ext)}\mathfrak{D}_0[\a] &:=&\sup_{\Mext}\left(r^2(2r+u)^{1+\dec}+r^3(2r+u)^{\frac{1}{2}+\dec}\right)|\a|,\\
\,{}^{(ext)}\mathfrak{D}_0[\b] &:=&\sup_{\Mext}\Big(r^2(2r+u)^{1+\dec}+r^3(2r+u)^{\frac{1}{2}+\dec}\Big)|\b|,\\
\,{}^{(ext)}\mathfrak{D}_0[\check{\rho}] &:=&\sup_{\Mext}\Big(r^2u^{1+\dec}+r^3u^{\frac{1}{2}+\dec}\Big)|\check{\rho}|,\\
\,{}^{(ext)}\mathfrak{D}_0[\check{\mu}] &:=&\sup_{\Mext}r^3u^{1+\dec}|\check{\mu}|,\\
\,{}^{(ext)}\mathfrak{D}_0[\bb] &:=&\sup_{\Mext}r^2u^{1+\dec}|\bb|,\\
\,{}^{(ext)}\mathfrak{D}_0[\aa] &:=&\sup_{\Mext}ru^{1+\dec}|\aa|,
\eeaa
and 
\beaa
\,{}^{(ext)}\mathfrak{D}_0[\Rc] := \,{}^{(ext)}\mathfrak{D}_0[\a]+\,{}^{(ext)}\mathfrak{D}_0[\b] +\,{}^{(ext)}\mathfrak{D}_0[\check{\rho}] +\,{}^{(ext)}\mathfrak{D}_0[\check{\mub}] +\,{}^{(ext)}\mathfrak{D}_0[\bb] +\,{}^{(ext)}\mathfrak{D}_0[\aa].
\eeaa

Also, we introduce the following higher derivatives norms
\beaa
\,{}^{(ext)}\mathfrak{D}_1[\Rc] &:=&       \,{}^{(ext)}\mathfrak{D}_0[\Rc]+\,{}^{(ext)}\mathfrak{D}_0[\dk\Rc]\\
&+&\sup_{\Mext}\left(r^3(2r+u)^{1+\dec}+r^4(2r+u)^{\frac{1}{2}+\dec}\right)|e_3(\a)|\\
&+&\sup_{\Mext}\Big(r^3u^{1+\dec}+r^4u^{\frac{1}{2}+\dec}\Big)|e_3(\b)|+\sup_{\Mext}r^3u^{1+\dec}|e_3(\check{\rho})|,
\eeaa
and for any integer $k\geq 2$
\beaa
\,{}^{(ext)}\mathfrak{D}_k[\Rc]:=       \,{}^{(ext)}\mathfrak{D}_1[\dk^{\leq k-1}\Rc].
\eeaa

Also, we define 
\beaa
\,{}^{(ext)}\mathfrak{D}_0[\check{\ka}] &:=&\sup_{\Mext}r^2u^{1+\dec}|\check{\ka}|,\\
\,{}^{(ext)}\mathfrak{D}_0[\vth] &:=&\sup_{\Mext}\Big(ru^{1+\dec}+r^2u^{\frac{1}{2}+\dec}\Big)|\vth|,\\
\,{}^{(ext)}\mathfrak{D}_0[\ze] &:=&\sup_{\Mext}\Big(ru^{1+\dec}+r^2u^{\frac{1}{2}+\dec}\Big)|\ze|,\\
\,{}^{(ext)}\mathfrak{D}_0[\check{\kab}] &:=&\sup_{\Mext}\Big(ru^{1+\dec}+r^2u^{\frac{1}{2}+\dec}\Big)|\check{\kab}|,\\
\,{}^{(ext)}\mathfrak{D}_0[\vthb] &:=&\sup_{\Mext}ru^{1+\dec}|\vthb|,\\
\,{}^{(ext)}\mathfrak{D}_0[\eta] &:=&\sup_{\Mext}ru^{1+\dec}|\eta| +\left(\int_{\Sigma_*}u^{2+2\dec}\eta^2\right)^{\frac{1}{2}},\\
\,{}^{(ext)}\mathfrak{D}_0[\check{\omb}] &:=&\sup_{\Mext}ru^{1+\dec}|\check{\omb}|,\\
\,{}^{(ext)}\mathfrak{D}_0[\xib] &:=&\sup_{\Mext}ru^{1+\dec}|\xib|,
\eeaa
and
\beaa
\,{}^{(ext)}\mathfrak{D}_0[\Gac] &:=& \,{}^{(ext)}\mathfrak{D}_0[\check{\ka}] + \,{}^{(ext)}\mathfrak{D}_0[\vth] +\,{}^{(ext)}\mathfrak{D}_0[\ze]+\,{}^{(ext)}\mathfrak{D}_0[\check{\kab}]  +\,{}^{(ext)}\mathfrak{D}_0[\vthb] +\,{}^{(ext)}\mathfrak{D}_0[\eta]\\
&+&\,{}^{(ext)}\mathfrak{D}_0[\check{\omb}]+\,{}^{(ext)}\mathfrak{D}_0[\xib].
\eeaa
Also, we introduce the following higher derivatives norms
\beaa
\,{}^{(ext)}\mathfrak{D}_1[\Gac]:= \,{}^{(ext)}\mathfrak{D}_0[\dk\Gac]  + \sup_{\Mext}r^2u^{1+\dec}|e_3(\vth, \ze, \check{\kab})| 
\eeaa
and for any integer $k\geq 2$
\beaa
\,{}^{(ext)}\mathfrak{D}_k[\Gac]:=       \,{}^{(ext)}\mathfrak{D}_1[\dk^{\leq k-1}\Gac].
\eeaa

\begin{remark}
The integral bootstrap assumption on $\Sigma_*$ for $\eta$ will only be needed in the proof of Proposition \ref{prop:constructionsecondframeinMext} and recovered in Proposition  \ref{Prop.Flux-bb-vthb-eta-xib}. In fact, other components satisfy an analog integral estimate  on $\Sigma_*$: this is the case of $\vthb$, $\xib$ and $r\bb$, see  Proposition  \ref{Prop.Flux-bb-vthb-eta-xib}. But $\eta$ is the only component for which we need to make this type of  bootstrap assumption. 
\end{remark}


\subsection{Main norms in $\Mint$}\label{section:main-normsintregion}


All quantities appearing in this section are defined relative to the $\Mint$ frame adapted to the $(\ub, \sint)$ foliation.


\subsubsection{$L^2$ based norms in $\Mint$}


We introduce the curvature norms,
\beaa
\Big(\,{}^{(int)}\mathfrak{R}_0[\Rc]\Big)^2 &:=&   \int_{\Mint}|\Rc|^2.
\eeaa
For any nonzero integer $k$, we introduce the following higher derivatives norms
\beaa
\,{}^{(int)}\mathfrak{R}_k[\Rc]:=       \,{}^{(int)}\mathfrak{R}_0[\dk^{\leq k}\Rc].
\eeaa

For any $k\geq 0$, we introduce the following  norms
\beaa
\Big(\,{}^{(int)}\mathfrak{G}_k[\Gac]\Big)^2 &:=& \int_{\Mint}|\dk^{\leq k}\Gac|^2.
\eeaa


\subsubsection{Decay norms in $\Mint$}


We define
\beaa
\,{}^{(int)}\mathfrak{D}_0[\Rc] := \sup_{\Mint}\ub^{1+\dec}|\Rc|,\,\,\,\, \,{}^{(int)}\mathfrak{D}_0[\Gac] := \sup_{\Mint}\ub^{1+\dec}|\Gac|.
\eeaa

Also, we introduce the following higher derivatives norms for any integer $k\geq 1$
\beaa
\,{}^{(int)}\mathfrak{D}_k[\Rc]:=       \,{}^{(int)}\mathfrak{D}_0[\dk^{\leq k}\Rc],\,\,\,\, \,{}^{(int)}\mathfrak{D}_k[\Gac]:=       \,{}^{(int)}\mathfrak{D}_0[\dk^{\leq k}\Gac].
\eeaa


\subsection{Combined norms}\lab{sec:definitionofconcatenatednorm}


We define the following norms $\MM$ by combining our above norms on $\Mext$ and $\Mint$
\beaa
\Nk^{(En)}_k & :=& \,{}^{(ext)}\Rk_k[\Rc]+\,{}^{(ext)}\Gk_k[\Gac]+ \,{}^{(int)}\Rk_k[\Rc]+\,{}^{(int)}\Gk_k[\Gac],\\
\Nk^{(Dec)}_k & :=& \,{}^{(ext)}\Dk_k[\Rc] +\,{}^{(int)}\Dk_k[\Gac]+\,{}^{(ext)}\Dk_k[\Rc] +\,{}^{(int)}\Dk_k[\Gac]. 
\eeaa


\subsection{Initial layer norm}\lab{sec:initialdatalayernorm}


Recall the notations of section \ref{sec:defintionoftheinitialdatalayer} concerning the initial data layer $\LL_0$. Recall that the constant $m_0>0$ is the mass of the initial Schwarzschild spacetime relative to which our initial perturbation is measured. We define   the initial layer norm to be\footnote{Recall that the initial data layer foliations satisfy $\etab+\ze=0$, as well as $\xi=\om=0$ on $\Lext$ and $\eta=\ze$ as well as $\xib=\omb=0$ on $\Lint$.},
\beaa
\Ik_k &:=&\, ^{(ext)} \Ik_k+\, ^{(int)} \Ik_k+\Ik_k'
\eeaa 
where
\beaa
\, ^{(ext)} \Ik_0&:=&  \sup_{\Lext}  \left[ r^{\frac{7}{2}  +\dt}\left( |\a| + |\b|\right)+r^3\ \left|\rho+\frac{2m_0}{r^3}\right| +   r^2 |\bb|+r|\aa|    \right]\\
&+& \sup_{\Lext}  r^2\left(|\vth|+\left|\ka-\frac{2}{r}\right|+|\ze|+ \left|\kab+\frac{2\left(1-\frac{2m_0}{r}\right)}{r}\right|\right)\\
&+& \sup_{\Lext} r \left(|\vthb|+\left|\omb-\frac{m_0}{r^2}\right|+|\xib|\right)\\
&+&\sup_{\Lext\left(\rext_0\geq 4m_0\right)}\left(r\left|\frac{\ga}{r^2}-1\right|+r|\underline{b}|+|\Omb+\Up| +|\vsi-1|+r\left|\frac{e^\Phi}{r\sin\th}-1\right| \right),
\eeaa
\beaa
\, ^{(int)} \Ik_0&:=&  \sup_{\Lint} \left(  |\aa|+|\bb| +\left|\rho+\frac{2m_0}{r^3}\right| +|\b|  +|\a|    \right)\\
&+& \sup_{\Lint}   \left(|\vth|+\left|\ka-\frac{2\left(1-\frac{2m_0}{r}\right)}{r}\right|+|\ze| + \left|\kab+\frac{2}{r}\right|+|\vthb|+\left|\om+\frac{m_0}{r^2}\right|+|\xi| \right),
\eeaa
\beaa
\Ik_0'&:=&  \sup_{\Lint\cap\Lext} \left(  |f|+|\fb| +|\log(\la_0^{-1}\la)|   \right),\qquad \la_0= {}^{(ext)}\la_0=1-\frac{2m_0}{\rextl},
\eeaa
with $\Ik_k$ the corresponding  higher derivative norms obtained by replacing   each   component by $\dk^{\le k}$ of it. In the definition of $\Ik_0'$ above, $(f, \fb, \la)$ denote the transition functions of Lemma \ref{lemma:SSMe:general.composite}  from the frame of the outgoing part $\Lext$ of the initial data layer to the frame of the ingoing part $\Lint$ of the initial data layer in the region $\Lint\cap\Lext$. 

\begin{remark}  
 Note that  In the definition of   $\, ^{(ext)} \Ik_k$ we allow a higher power  of $r$   in front $\a$, $\b$ and their derivatives than what    it is consistent with the results  of  \cite{Ch-Kl} and \cite{KlNi}. The additional $r^{\dt}$ power, for $\dt$ small,   is consistent instead with the result  of   \cite{KlNi2}.
\end{remark}


\section{Main Theorem}



\subsection{Smallness constants}\lab{sec:discussionofsmallnessconstantforthemaintheorem}


Before stating our main theorem, we first introduce the following constants that will be involved in its statement.
\begin{itemize}
\item The constant $m_0>0$ is the mass of the initial Schwarzschild spacetime relative to which our initial perturbation is measured. 

\item The integer $k_{large}$ which corresponds to the maximum number of derivatives of the solution.

\item The size of the initial data layer norm is measured by $\ep_0>0$. 

\item The size of the bootstrap assumption norms are measured by $\ep>0$.

\item $\deh>0$ measures the width of the region $|r-2m_0|\leq 2m_0\deh$ where the redshift estimate holds and which includes in particular the region $\Mint$.

\item $\dec$ is tied to decay estimates  in $u$, $\ub$  for $\check{\Gamma}$ and $\Rc$.

\item $\dt$ is involved in the $r$-power of the $r^p$ weighted estimates for curvature.
\end{itemize}

  In what follows $m_0$ is a fixed constant, $\deh$, $\dt$, and $\dec$ are fixed, sufficiently small, universal constants, and $k_{large}$ is a  fixed, sufficiently large, universal constant, chosen such that
\bea\lab{eq:constraintsonthemainsmallconstantsepanddelta}
0<\deh,\,\, \dec, \,\,\dt \ll \min\{m_0, 1\},\qquad \dt> 2\dec, \qquad k_{large}\gg \frac{1}{\dec}.
\eea
Then, $\ep$ and $\ep_0$ are chosen such that
\bea\lab{eq:constraintsonthemainsmallconstantsepanddelta:bis}
\ep_0, \ep\ll \min\left\{\deh, \dec, \dt, \frac{1}{k_{large}}, m_0, 1\right\}
\eea
and
\bea\lab{eq:constraintbetweenepep*ep0}
\ep=\ep_0^{\frac{2}{3}}.
\eea

Using the definition of $\ep_0$, we may now precise the behavior \eqref{eq:behaviorofronSigmastarrough} of $r$ on $\Sigma_*$
\bea\lab{eq:behaviorofronSigmastar}
\inf_{\Sigma_*}r\geq \ep_0^{-\frac{2}{3}}u_*^4.
\eea

From now on, in the rest of the paper, $\lesssim$ means bounded by a constant depending only on geometric universal constants (such as Sobolev embeddings, elliptic estimates,...) as well as the constants 
$$m_0,\, \deh,\, \dec, \,\dt, \, k_{large}$$
\textit{but not on} $\ep$ and $\ep_0$.


\subsection{Statement of the main theorem}\lab{sec:statementofthemaintheorempreciseversion}


We are now ready to give the following precise version of our main theorem. 

\begin{thmmain}[Main theorem, version 2]
There exists a sufficiently large integer $k_{large}$ and a sufficiently small constant $\ep_0>0$ such that given an initial layer defined as in section \ref{sec:defintionoftheinitialdatalayer} and satisfying the bound
\bea\lab{def:initialdatalayerassumptions}
\Ik_{k_{large}+5}\leq \ep_0,
\eea
there exists a globally hyperbolic development with a complete future null infinity $\II_+$ and a complete future horizon $\HH_+$ together with  foliations and adapted null frames verifying the admissibility conditions of section \ref{sec:defintioncanonicalspacetime}  such that following bound is satisfied 
\bea\lab{def:bootstrapasumptionsglobalnorms}
\Nk^{(En)}_{k_{large}}+\Nk^{(Dec)}_{k_{small}}\le C\ep_0
\eea
where $C$ is a large enough universal constant and where $k_{small}$ is given by 
\bea\lab{eq:choiceksmallmaintheorem}
k_{small}=\left \lfloor\frac 1 2 k_{large}\right \rfloor +1.
\eea
In particular, 
\begin{itemize}
\item On $\Mext$, we have
\beaa
 |\a|, |\b|&\les& \min\left\{ \frac{\ep_0}{r^3(u+2r)^{\frac{1}{2}+\dec}},\,   \frac{\ep_0}{r^2(u+2r)^{1+\dec}}\right\},\\
 |\rhoc| &\les & \min\left\{ \frac{\ep_0}{r^3 u ^{\frac{1}{2}+\dec}},\,   \frac{\ep_0}{r^2 u^{1+\dec}}\right\},\\
  |\bb| &\les& \frac{\ep_0}{r^2u^{1+\dec}},\\
   |\aa| &\les&\frac{\ep_0}{ru^{1+\dec}},
\eeaa
and 
\beaa
|\kac|&\les& \frac{\ep_0}{r^2u^{1+\dec}},\\
|\vth|, |\ze|, |\kabc| &\les& \min\left\{ \frac{\ep_0}{r^2u^{\frac{1}{2}+\dec}}\frac{\ep_0}{ru^{1+\dec}}\right\},\\
|\eta|, |\vthb|, |\ombc|, |\xib| & \les& \frac{\ep_0}{ru^{1+\dec}}.
\eeaa

\item On $\Mint$ we have, with $\Gac=\{\check{\kab}, \vthb, \ze, \etab, \kac, \vth, \omc, \xi\},  \Rc=\{ \a,\b,\rhoc, \bb, \aa\}$,
\beaa
|\Gac, \Rc| \les \frac{\ep_0}{\ub^{1+\dec}}.
\eeaa

\item The Bondi mass converges as $u\to +\infty$ along $\II_+$ to the final Bondi mass which we denote by $m_\infty$.  The final Bondi mass verifies the estimate
$$\left|\frac{m_\infty}{m_0}-1\right|\les\ep_0.$$
In particular $m_\infty>0 $.

\item The Hawking mass $m$ satisfies
\beaa
\frac{|m-m_\infty|}{m_0}\les \begin{cases}
 \displaystyle\frac{\ep_0}{u^{1+\dec}}\quad   \textrm{ on }\Mext,\\[2mm]
 \displaystyle  \frac{\ep_0}{\ub^{1+\dec}}\quad \textrm{ on }\Mint.
 \end{cases}
\eeaa

\item The location of the future Horizon $\HH_+$ satisfies 
\beaa
r=2m_\infty+O\left(\frac{\sqrt{\ep_0}}{\ub^{1+\frac{\dec}{2}}}\right)\textrm{ on }\HH_+.
\eeaa

\item On $\Mext$, we have
\beaa
\left|\rho+\frac{2m_\infty}{r^3}\right| &\les& \min\left\{ \frac{\ep_0}{r^3u^{\frac{1}{2}+\dec}}, \frac{\ep_0}{r^2u^{1+\dec}}\right\},\\
\left|\ka-\frac{2}{r}\right| &\les &\frac{\ep_0}{r^2u^{1+\dec}},\\
 \left|\kab+\frac{2\left(1-\frac{2m_\infty}{r}\right)}{r}\right| &\les& \min\left\{ \frac{\ep_0}{r^2u^{\frac{1}{2}+\dec}}\frac{\ep_0}{ru^{1+\dec}}\right\},\\
  \left|\omb-\frac{m_\infty}{r^2}\right| & \les& \frac{\ep_0}{ru^{1+\dec}}.
\eeaa

\item On $\Mint$, we have.
\beaa
\left|\rho+\frac{2m_\infty}{r^3}\right|, \left|\kab+\frac{2}{r}\right|, \left|\ka-\frac{2\left(1-\frac{2m_\infty}{r}\right)}{r}\right|, \left|\om+\frac{m_\infty}{r^2}\right| \les \frac{\ep_0}{\ub^{1+\dec}}.
\eeaa

\item On $\Mext$, the space-time metric $\g$ is given in the $(u, r, \th, \vphi)$ coordinates system by
\beaa
\g &=& \g_{m_\infty, \Mext} +O\left(\frac{\ep_0}{u^{1+\dec}}\right)\Big((dr, du, rd\th)^2, r^2(\sin\th)^2(d\vphi)^2\Big)
\eeaa
where $\g_{m_\infty, \Mext}$ denotes the Schwarzschild metric of mass $m_\infty>0$ in outgoing Eddington-Finkelstein coordinates, i.e.
\beaa
\g_{m_\infty, \Mext} &:=& -2du dr -\left(1-\frac{2m_{\infty}}{r}\right) (du)^2+ r^2\Big((d\th)^2+(\sin\th)^2(d\vphi)^2\Big).
\eeaa

\item On $\Mint$, the space-time metric $\g$ is given in the $(\ub, r, \th, \vphi)$ coordinates system by
\beaa
\g &=& \g_{m_\infty, \Mint} +O\left(\frac{\ep_0}{\ub^{1+\dec}}\right)\Big((dr, d\ub, rd\th)^2, r^2(\sin\th)^2(d\vphi)^2\Big)
\eeaa
where $\g_{m_\infty, \Mext}$ denotes the Schwarzschild metric of mass $m_\infty>0$ in ingoing Eddington-Finkelstein coordinates, i.e.
\beaa
\g_{m_\infty, \Mint} &:=& 2d\ub dr -\left(1-\frac{2m_{\infty}}{r}\right) (d\ub)^2+ r^2\Big((d\th)^2+(\sin\th)^2(d\vphi)^2\Big).
\eeaa
\end{itemize}
Note that analog statements of the above estimates also hold for $\dk^k$ derivatives with $k\leq k_{small}$.
\end{thmmain}

\begin{remark}\lab{rem:expaningwhywecanstratinthecontextofthecaracteristicCauchyproblem}
In this paper, we choose to specify the closeness to Schwarzschild of our initial data in the context of the Characteristic Cauchy problem. Note that the conclusions of our main theorem can be immediately extended to the case where  the data are specified to be close to Schwarzschild on a spacelike hypersurface $\Sigma$. Indeed, one can reduce this latter case to our situation by invoking
\begin{itemize}
\item The results in \cite{KlNi} \cite{KlNi2} which allow us to control the causal region between $\Sigma$ and the outgoing part of the initial data layer\footnote{Note that the results of \cite{KlNi2} are consistent with our initial data layer assumptions.}.

\item A standard local existence result which controls the finite causal region between $\Sigma$ and the ingoing part of the initial data layer.
\end{itemize}
\end{remark}

\begin{remark}
In the context of the previous remark, we note that the constant $m_0>0$ appearing in the initial data layer norm of the assumption \eqref{def:initialdatalayerassumptions} of our main theorem does not necessarily coincide with the ADM mass of the corresponding   initial data set on  the spacelike  hypersurface $\Si$. With respect to this ADM mass, we would recover the well known inequality stating that the final Bondi mass is smaller than the ADM mass.
\end{remark}


\section{Bootstrap assumptions and first consequences}



\subsection{Main bootstrap assumptions}\label{section:bootstrap}


We assume that the combined  norms $\Nk^{(En)}_k$ and $\Nk^{(Dec)}_k$ defined    in   section \ref{section:main-norms} verifies   the following bounds

{\it{\bf BA-E} (Bootstrap Assumptions on energies and weighted energies)
\bea\lab{def:bootstrapasumptionsglobalnormonenergie}
\Nk^{(En)}_{k_{large}}\le \ep,
\eea

{\bf BA-D} (Bootstrap Assumptions on decay)
\bea\lab{def:bootstrapasumptionsglobalnormondecay}
\Nk^{(Dec)}_{k_{small}}\le \ep.
\eea}

In the remaining of section \ref{section:bootstrap}, we state several simple consequences of the bootstrap assumptions which will be proved in Chapter \ref{chap:proofofconsequencesbootass}.


\subsection{Control of the initial data}


While  the smallness constant  involved in the bootstrap assumptions is $\ep>0$, we need the smallness constant  involved in the control of the initial data to be $\ep_0>0$. This is achieved in the theorem below.
\begin{thmM0}
Assume that the initial data layer  $\LL_0$, as   defined  in section \ref{sec:defintionoftheinitialdatalayer},  satisfies
\beaa
\Ik_{k_{large}+5}\leq \ep_0.
\eeaa
Then under the bootstrap assumptions {\bf BA-D} on decay, the following holds true on the initial data hypersurface $\CC_1\cup\CCb_1$,
\beaa
\max_{0\leq k\leq k_{large}}&&\Bigg\{ \sup_{\CC_1}  \left[r^{\frac{7}{2}  +\de_B}\left( |\dk^k\,{}^{(ext)}\a| + |\dk^k\,{}^{(ext)}\b|\right)+r^{\frac{9}{2}  +\de_B}|\dk^{k-1}e_3(\,{}^{(ext)}\a)|   \right]\\
\nn&&+ \sup_{\CC_1} \left[r^3\ \left|\dk^k\left(\,{}^{(ext)}\rho+\frac{2m_0}{r^3}\right)\right|+r^2|\dk^k\,{}^{(ext)}\bb|+r|\dk^k\,{}^{(ext)}\aa|\right]\Bigg\} \les \ep_0,
\eeaa
\beaa
\nn\max_{0\leq k\leq k_{large}}\sup_{\CCb_1}\Bigg[ |\dk^k\,{}^{(int)}\a| + |\dk^k\,{}^{(int)}\b|+ \left|\dk^k\left(\,{}^{(int)}\rho+\frac{2m_0}{r^3}\right)\right| &&\\
 +|\dk^k\,{}^{(int)}\bb|+|\dk^k\,{}^{(int)}\aa|\Bigg] &\les& \ep_0,
\eeaa
and
\beaa
\sup_{\CC_1\cup\CCb_1}\left|\frac{m}{m_0}-1\right|\les \ep_0.
\eeaa
\end{thmM0}


\subsection{Control of averages and of the Hawking mass}


The following two lemma are simple consequence of the bootstrap assumptions and will be proved in section \ref{section:proofoflemma:estimatesonceandforallforaverages}.
\begin{lemma}[Control of averages]\lab{lemma:estimatesonceandforallforaverages}
Assume given a GCM admissible spacetime $\MM$  as defined in section \ref{sec:defintioncanonicalspacetime} 
verifying the bootstrap   assumption  for  some sufficiently small   $\ep>0$. Then, we have
\beaa
\sup_{\Mext}u^{1+\dec}\left(r^3\left|\dk^{\leq k_{small}}\left(\overline{\ka}-\frac{2}{r}\right)\right|+r^3\left|\dk^{\leq k_{small}}\left(\overline{\rho}+\frac{2m}{r^3}\right)\right|\right)&\les&\ep_0,\\
\sup_{\Mext}u^{1+\dec}\left(r^2\left|\dk^{\leq k_{small}}\left(\overline{\kab}+\frac{2\Up}{r}\right)\right|+r^2\left|\dk^{\leq k_{small}}\left(\overline{\omb}-\frac{m}{r^2}\right)\right|\right)&\les&\ep_0,
\eeaa
\beaa
\sup_{\Mext}u^{\frac{1}{2}+\dec}\left(r^3\left|\dk^{\leq k_{large}}\left(\overline{\ka}-\frac{2}{r}\right)\right|+r^3\left|\dk^{\leq k_{large}}\left(\overline{\rho}+\frac{2m}{r^3}\right)\right|\right)&\les&\ep_0,\\
\sup_{\Mext}u^{\frac{1}{2}+\dec}\left(r^2\left|\dk^{\leq k_{large}}\left(\overline{\kab}+\frac{2\Up}{r}\right)\right|+r^2\left|\dk^{\leq k_{large}}\left(\overline{\omb}-\frac{m}{r^2}\right)\right|\right)&\les&\ep_0,
\eeaa
\beaa
\sup_{\Mint}u^{1+\dec}\left(\left|\dk^{\leq k_{small}}\left(\overline{\ka}-\frac{2\Up}{r}\right)\right|+\left|\dk^{\leq k_{small}}\left(\overline{\rho}+\frac{2m}{r^3}\right)\right|\right)&\les&\ep_0,\\
\sup_{\Mint}u^{1+\dec}\left(\left|\dk^{\leq k_{small}}\left(\overline{\kab}+\frac{2}{r}\right)\right|+\left|\dk^{\leq k_{small}}\left(\overline{\om}+\frac{m}{r^2}\right)\right|\right)&\les& \ep_0,
\eeaa
\beaa
\sup_{\Mint}u^{\frac{1}{2}+\dec}\left(\left|\dk^{\leq k_{large}}\left(\overline{\ka}-\frac{2\Up}{r}\right)\right|+\left|\dk^{\leq k_{large}}\left(\overline{\rho}+\frac{2m}{r^3}\right)\right|\right) &\les&\ep_0,\\
\sup_{\Mint}u^{\frac{1}{2}+\dec}\left(\left|\dk^{\leq k_{large}}\left(\overline{\kab}+\frac{2}{r}\right)\right|+\left|\dk^{\leq k_{large}}\left(\overline{\om}+\frac{m}{r^2}\right)\right|\right)&\les& \ep_0.
\eeaa
Also, we have
\beaa
\sup_{\Mext}\Big(u^{1+\dec}r\left|\dk^{\leq k_{small}}\left(\overline{\Omb}+\Up\right)\right|+u^{\frac{1}{2}+\dec}r\left|\dk^{\leq k_{large}}\left(\overline{\Omb}+\Up\right)\right|\Big) \les\ep_0,
\eeaa
\beaa
\sup_{\Mint}\Big(u^{1+\dec}\left|\dk^{\leq k_{small}}\left(\overline{\Om}-\Up\right)\right|+u^{\frac{1}{2}+\dec}\left|\dk^{\leq k_{large}}\left(\overline{\Om}-\Up\right)\right|\Big) \les\ep_0.
\eeaa
Finally, recall that $\overline{\mu}$ and $\overline{\mub}$ are given by the following formula 
\beaa
\overline{\mu}=\frac{2m}{r^3}\quad \textrm{ \, on\,  }\Mext , \qquad \overline{\mub}=\frac{2m}{r^3}\textrm{ \, on\,  }\Mint.
\eeaa
\end{lemma}

\begin{lemma}[Control of the Hawking mass]\lab{lemma:estimatesonceandforallforHawkingmass}
Assume given a GCM admissible spacetime $\MM$  as defined in section \ref{sec:defintioncanonicalspacetime} 
verifying the bootstrap   assumption  for  some sufficiently small   $\ep>0$. Then, we have 
\beaa
\max_{0\leq k\leq k_{large}}\sup_{\Mext}u^{1+\dec}\Big(|\dk^ke_3(m)| +r|\dk^ke_4(m)|\Big) &\les& \ep_0,\\
\max_{0\leq k\leq k_{large}}\sup_{\Mint}\ub^{1+\dec}\Big(|\dk^ke_3(m)|+|\dk^ke_4(m)|\Big) &\les& \ep_0.
\eeaa
The $e_4 $ derivatives behave better in powers of $r$,
\beaa
\max_{0\leq k\leq k_{small}}\sup_{\Mext}r^2u^{1+\dec}|\dk^ke_4(m)| & \les &\ep_0,\\
\max_{0\leq k\leq k_{large}}\sup_{\Mext}r^2u^{\frac{1}{2}+\dec}|\dk^ke_4(m)| & \les &\ep_0.
\eeaa
Moreover,
\beaa
\sup_{\MM}\left|\frac{m}{m_0}-1\right| &\les& \ep_0.
\eeaa
\end{lemma}


\subsection{Control of coordinates system}


The following two propositions on the existence of a suitable coordinates system both in $\Mext$ and in $\Mint$ are also consequences of the bootstrap assumptions and will be proved in section \ref{section:proofofprop:finalcoordinatessystem}.

\begin{proposition}[Control of a coordinates system  on $\Mext$]\lab{prop:finalcoordinatessystem}
Let $\th\in[0, \pi]$ be the $\Z$-invariant scalar on $\MM$  defined by \eqref{def:definitionofthetausedatnullinfinity:originalplaceappearing}, i.e.
\bea
\lab{def:definitionofthetausedatnullinfinity}
\th= \cot^{-1}\left(re_\th(\Phi)\right).
\eea
Consider the $(u,r,\th, \vphi)$ coordinates system introduced in Proposition \ref{Prop:coordin(u,r,th)}. Then, relative to these $(u,r,\th,\vphi)$ coordinates,  
\begin{enumerate}
\item The spacetime metric takes the form,
  \bea
  \bsplit
 g&=-\frac{4\vsi}{r\ov{\ka}} du dr+\frac{\vsi^2(\ov{\kab}+\Ab)}{\ov{\kab}} du^2 +\ga\left(  d\th-\frac 1 2\vsi\underline{b} du -\frac b 2  \Th\right)^2 
  \end{split}
\eea
where,
\bea
b=e_4(\th), \qquad \underline{b}=e_3(\th),\qquad \ga=\frac{1}{(e_\th(\th))^2}
\eea
and,
\beaa
\Th=\frac{4}{r\overline{\ka}} d r - \vsi\left(\frac{\overline{\kab}+\Ab}{\overline{\ka}}\right) du.
\eeaa

\item The reduced coordinates derivatives take the form,
\bea
\bsplit
\pr_r&=\frac{2}{r\overline{\ka}}e_4- \frac{2\sqrt{\ga} }{ r\ov{\ka}} b e_\th, \\
\pr_\th &= \sqrt{\ga}e_\th,\\
\pr_u&=\vsi\left[\frac{1}{2}e_3-\frac{1}{2}\frac{\overline{\kab}+\Ab}{\overline{\ka}}e_4-\frac{1}{2}\sqrt{\ga}\left( \underline{b}  - \left(\frac{\overline{\kab}+\Ab}{\overline{\ka}}\right) b  \right)e_\th\right]. 
\end{split}
\eea

 \item The following estimates hold true:
 \beaa
 \max_{0\leq k\leq k_{small}}\sup_{\Mext}\Big(ru^{\frac{1}{2}+\dec}+u^{1+\dec}\Big)\left(\left|\dk^k\left(\frac{\ga}{r^2}-1\right)\right|+r\left|\dk^kb\right|\right) &\les& \ep,\\
 \max_{0\leq k\leq k_{small}}\sup_{\Mext}u^{1+\dec}\left(\left|\dk^k\Obc\right|+\left|\dk^k(\vsi-1)\right|+r\left|\dk^k\underline{b}\right|\right) &\les& \ep.
 \eeaa
  Also, $e^\Phi$ satisfies 
 \beaa
 \max_{0\leq k\leq k_{small}}\sup_{\Mext}\Big(ru^{\frac{1}{2}+\dec}+u^{1+\dec}\Big)\left|\dk^k\left(\frac{e^\Phi}{r\sin\th}-1\right)\right| &\les& \ep.
 \eeaa
\end{enumerate}
\end{proposition}
 
  \begin{proposition}[Control of a coordinates system  on $\Mint$]\lab{prop:finalcoordinatessystem:bis}
Let $\th\in[0, \pi]$ be the $\Z$-invariant scalar on $\MM$  defined by \eqref{def:definitionofthetausedatnullinfinity}. Consider the $(\ub,r,\th, \vphi)$ coordinates system introduced in Proposition \ref{Prop:coordin(ub,r,th)}. Then, relative to these $(\ub,r,\th,\vphi)$ coordinates,
\begin{enumerate}
\item The spacetime metric takes the form,  
  \bea\lab{reducedmetric-geodesicfoliation:bis}
  \bsplit
g&=-\frac{4\vsib}{r\ov{\kab}} d\ub dr+\frac{\vsib^2(\ov{\ka}+A)}{\ov{\ka}} d\ub^2 +\ga\left(  d\th-\frac 1 2\vsib b d\ub -\frac{\underline{b}}{2}  \underline{\Th}\right)^2 
\end{split}
\eea
where,
\bea
b=e_4(\th), \qquad \underline{b}=e_3(\th),\qquad \ga=\frac{1}{(e_\th(\th))^2}
\eea
and,
\beaa
\underline{\Th}:=\frac{4}{r\overline{\kab}} d r - \vsib\left(\frac{\overline{\ka}+A}{\overline{\kab}}\right) d\ub.
\eeaa

\item The reduced coordinates derivatives take the form,
\bea
\label{transf:coords-frame:bis}
\bsplit
\pr_r&=\frac{2}{r\overline{\kab}}e_3- \frac{2\sqrt{\ga} }{ r\ov{\kab}} \underline{b} e_\th, \\
\pr_\th &= \sqrt{\ga}e_\th,\\
\pr_\ub &=\vsib\left[\frac{1}{2}e_4-\frac{1}{2}\frac{\overline{\ka}+A}{\overline{\kab}}e_3-\frac{1}{2}\sqrt{\ga}\left( b  - \left(\frac{\overline{\ka}+A}{\overline{\kab}}\right) \underline{b}  \right)e_\th\right]. 
\end{split}
\eea

 \item  The following estimates hold true:
 \beaa
 \max_{0\leq k\leq k_{small}}\sup_{\Mint}\ub^{1+\dec}\left(\left|\dk^k\Omc\right|+\left|\dk^k(\vsib-1)\right|+\left|\dk^k\left(\frac{\ga}{r^2}-1\right)\right|+\left|\dk^kb\right|+\left|\dk^k\underline{b}\right|\right) &\les& \ep.
 \eeaa
  Also, $e^\Phi$ satisfies 
 \beaa
 \max_{0\leq k\leq k_{small}}\sup_{\Mint}\ub^{1+\dec}\left|\dk^k\left(\frac{e^\Phi}{r\sin\th}-1\right)\right| &\les& \ep.
 \eeaa
 \end{enumerate}
 \end{proposition}


\subsection{Pointwise bounds for high order derivatives}


We will need later to interpolate between the estimates provided by the bootstrap assumptions on decay and the bootstrap assumptions on energy. To this end, we will need the following consequence of the bootstrap assumptions on weighted energies. 
\begin{proposition}\lab{prop:pointwiseboundsforhighorderderivatives}
The Ricci coefficients and curvature components satisfy the following pointwise estimates on $\MM$
\beaa
\max_{k\leq k_{large}-5}\sup_{\MM}&&\Big\{r^{\frac{7}{2}+\frac{\dt}{2}}\big(|\dk^k\a|+|\dk^k\b|\big)+ r^3\big(|\dk^k\mu|+|\dk^k\rhoc|\big)\\
&&+r^2\big(|\dk^k\check{\ka}|+|\dk^k\ze|+|\dk^k\vth|+|\dk^k\check{\kab}|+|\dk^k\bb|\big)\\
&&+r\big(|\dk^k\vthb|+|\dk^k\vthb|+|\dk^k\check{\omb}|+|\dk\xib|+|\dk^k\aa|\big)\Big\} \les \ep.
\eeaa
\end{proposition}


\subsection{Construction of a second frame in $\Mext$}
\label{subsection:constructionsecondframeinMext}


Recall that the quantity $\qf$ satisfies the following wave equation, see \eqref{thmwaveqf:improperfrom}, 
\beaa
\square_2\qf+\ka\kab\qf &=& \err[\square_2\qf]
\eeaa
where the nonlinear term $\err[\square_2\qf]$ has the schematic structure exhibited in \eqref{thmwaveqf:schematicformerrorterm}. Also, 
recall that according to our bootstrap assumption on decay and Proposition \ref{prop:pointwiseboundsforhighorderderivatives}, $\eta$ satisfies on $\Mext$
\beaa
|\dk^{\leq k_{small}}\eta| \leq \frac{\ep}{ru^{1+\dec}}, \qquad |\dk^{\leq k_{large}-5}\eta| \les \frac{\ep}{r}.
\eeaa
As discuss in Remark \ref{rmk:whyweneedaglobalframewithbettereta}, this decay in $r^{-1}$ is too weak to derive suitable decay for $\qf$. We thus need to provide another frame for $\Mext$. This is the aim of the following proposition.
\begin{proposition}\lab{prop:constructionsecondframeinMext}
Let an integer $k_{loss}$ and a small constant $\de_0>0$ satisfying\footnote{Recall from \eqref{eq:constraintsonthemainsmallconstantsepanddelta} and \eqref{eq:choiceksmallmaintheorem} that we have
\beaa
0<\dec\ll 1, \qquad \dec\, k_{large}\gg 1, \qquad k_{small}=\left \lfloor\frac 1 2 k_{large}\right \rfloor +1.
\eeaa
In particular, we have $\dec(k_{large}-k_{small})\gg 1$ 
and hence the exists an integer $k_{loss}$ satisfying the required constraints.}
\bea\lab{eq:constraintsonklossandde0forsecondframeofMext}
16\leq k_{loss} \leq   \frac{\dec}{3}(k_{large}-k_{small}), \qquad \de_0:=\frac{k_{loss}}{k_{large}-k_{small}}.
\eea
Let $(e_4, e_3, e_\th)$ the outgoing geodesic null frame of $\Mext$. There exists another frame $(e_4', e_3', e_\th')$ of $\Mext$  provided by
\beaa
\begin{split}
e_4'&= e_4 + f e_\th +\frac 1 4 f^2  e_3,\\
e_\th'&= e_\th +\frac 1 2 f e_3,\\
e_3'&=  e_3,
\end{split}
\eeaa
such that the Ricci coefficients and curvature components with respect to that frame satisfy 
\beaa
\xi'=0,
\eeaa
\beaa
\nn\max_{0\leq k\leq k_{small}+k_{loss}}\sup_{\Mext}&&\Bigg\{\Big(r^2u^{\frac{1}{2}+\dec-2\de_0}+ru^{1+\dec-2\de_0}\Big)|\dk^k\Ga_g'|+ru^{1+\dec-2\de_0}|\dk^k\Ga_b'|\\
\nn&&+r^2u^{1+\dec-2\de_0}\left|\dk^{k-1}e_3'\left(\ka'-\frac{2}{r}, \kab'+\frac{2\Up}{r}, \vth', \ze', \etab', \eta'\right)\right|\\
\nn&&+\Big(r^{\frac{7}{2}+\frac{\dt}{2}}+r^3u^{\frac{1}{2}+\dec-2\de_0}+r^2u^{1+\dec-2\de_0}\Big)\Big(|\dk^k\a'|+|\dk^k\b'|\Big)\\
\nn&& +\left(r^{\frac{9}{2}+\frac{\dt}{2}}+r^3u^{1+\dec}+r^4u^{\frac{1}{2}+\dec-2\de_0}\right)|\dk^{k-1}e_3'(\a')|\\
\nn&&+\Big(r^3u^{1+\dec}+r^4u^{\frac{1}{2}+\dec-2\de_0}\Big)|\dk^{k-1}e_3'(\b')|\\
\nn&&+\Big(r^3u^{\frac{1}{2}+\dec-2\de_0}+r^2ru^{1+\dec-2\de_0}\Big)|\dk^k\rhoc'|\\
&&+u^{1+\dec-2\de_0}\Big(r^2|\dk^k\bb'|+r|\dk^k\aa'|\Big)\Bigg\} \les \ep,
\eeaa
where we have used the notation\footnote{Here, $r$ and $m$ denote respectively the area radius and the Hawking mass of the outgoing geodesic foliation of $\Mext$, i.e. $r=\rext$ and $m=\mext$. In particular, while $e_\th(r)=e_\th(m)=0$, we have in general $e_\th'(r)\neq 0$ and $e_\th'(m)\neq 0$.} 
\beaa
\Ga_g' &=& \left\{r\om', \,\ka'-\frac{2}{r}, \, \vth', \, \ze',  \, \eta', \, \etab', \, \kab'+\frac{2\Up}{r}, \, r^{-1}(e_4'(r)-1), r^{-1}e_\th'(r), \, e_4'(m) \right\},\\
 \Ga_b' &=& \left\{\vthb', \omb'-\frac{m}{r^2}, \xib', \, r^{-1}(e_3'(r)+\Up),  \, r^{-1}e_3'(m)\right\}.
\eeaa
Furthermore, $f$ satisfies 
\bea\lab{eq:estimateforfinconstructionsecondframeinMext}
\begin{split}
|\dk^kf| &\les \frac{\ep}{ru^{\frac{1}{2}+\dec-2\de_0}+u^{1+\dec-2\de_0}}, \,\,\,\textrm{ for }k\leq k_{small}+k_{loss}+2\textrm{ on }\Mext,\\
 |\dk^{k-1}e_3'f| &\les \frac{\ep}{ru^{1+\dec-2\de_0}}\,\,\,\textrm{ for }k\leq k_{small}+k_{loss}+2\textrm{ on }\Mext.
\end{split}
\eea
\end{proposition}

\begin{remark}
The crucial point of Proposition \ref{prop:constructionsecondframeinMext} is that in the new frame $(e_4', e_3', e_\th')$ of $\Mext$, $\eta'$ belongs to $\Ga_g'$ and thus displays a better decay in $r^{-1}$ than $\eta$ corresponding to the outgoing geodesic frame $(e_4, e_3, e_\th)$ of $\Mext$.
\end{remark}


\section{Global null frames}\lab{section-globaleframe}


In this section, we  construct 2 smooth global   frames on $\MM$ by matching the frame of $\Mint$ on the one hand with a renormalization of the frame on $\Mext$, and on the other hand, with a renormalization of the second frame of $\Mext$ given by Proposition \ref{subsection:constructionsecondframeinMext}.


\subsection{Extension of frames}\lab{sec:extensionofframes}


To construct the first global frame, we need to extend the frame $({}^{(int)}e_4, {}^{(int)}e_3, {}^{(int)}e_\th)$ of $\Mint$ slightly into $\Mext$, and the 
frame $({}^{(ext)}e_4, {}^{(ext)}e_3, {}^{(ext)}e_\th)$ of $\Mext$ slightly into $\Mint$. We keep the same labels for the extended frame, 
i.e. $({}^{(int)}e_4, {}^{(int)}e_3, {}^{(int)}e_\th)$ represents the extended frame of $\Mint$ in $\Mext$ and vice versa. This convention also applies to the Ricci coefficients, curvature components, area radius and Hawking mass of the extended frames.

Note that these extensions require, in addition to the initialization of the frames on $\TT$, to initialize 
\begin{enumerate}
\item $({}^{(ext)}e_4, {}^{(ext)}e_3, {}^{(ext)}e_\th)$ on $\CCb_*$ by
\beaa
({}^{(ext)}e_4, {}^{(ext)}e_3, {}^{(ext)}e_\th)=(({}^{(int)}\Up)^{-1}{}^{(int)}e_4, {}^{(int)}\Up{}^{(int)}e_3, {}^{(int)}e_\th).
\eeaa

\item $({}^{(int)}e_4, {}^{(int)}e_3, {}^{(int)}e_\th)$ on $\CC_*$ by
\beaa
({}^{(int)}e_4, {}^{(int)}e_3, {}^{(int)}e_\th)=({}^{(ext)}\Up {}^{(ext)}e_4, ({}^{(ext)}\Up)^{-1}{}^{(ext)}e_3, {}^{(ext)}e_\th).
\eeaa
\end{enumerate}


\subsection{Construction of the first global frame}
\label{subsection:constructionglobalframe}


We start with the definition of the region where the frame of $\Mint$ and a conformal renormalization of the frame of $\Mext$ will be matched.
\begin{definition}\lab{def:cutofffunctionforthematchingregion}
We define the matching region as the spacetime region
\beaa
\mr:=\left(\Mext\cap\left\{\rint\leq 2m_0\left(1+\frac{3}{2}\deh\right)\right\}\right)\cup\left(\Mint\cap\left\{\rint\geq 2m_0\left(1+\frac{1}{2}\deh\right)\right\}\right),
\eeaa
where, as explained in the previous section, $\rint$ denotes the area radius of the ingoing geodesic foliation of $\Mint$ and its extension to $\Mext$. 
\end{definition}

Here is our main proposition concerning our first global frame.
\begin{proposition}\lab{prop:existenceandestimatesfortheglobalframe}
There exists a global null frame defined on $\Mint\cup\Mext$ and denoted by $({}^{(glo)}e_4, {}^{(glo)}e_3, {}^{(glo)}e_\th)$ such that
\begin{itemize}
\item[(a)] In $\Mext\setminus\mr$, we have
\beaa
({}^{(glo)}e_4, {}^{(glo)}e_3, {}^{(glo)}e_\th)= \left({}^{(ext)}\Up\,{}^{(ext)}e_4, {}^{(ext)}\Up^{-1}{}^{(ext)}e_3, {}^{(ext)}e_\th\right).
\eeaa

\item[(b)] In $\Mint\setminus\mr$, we have
\beaa
({}^{(glo)}e_4, {}^{(glo)}e_3, {}^{(glo)}e_\th) = \left({}^{(int)}e_4, {}^{(int)}e_3, {}^{(int)}e_\th\right).
\eeaa

\item[(c)] In the matching region, we have
\beaa
\max_{0\leq k\leq k_{small}-2}\sup_{\mr\cap\Mint}\ub^{1+\dec}\left|\dk^k({}^{(glo)}\Gac, {}^{(glo)}\Rc)\right| &\les&\ep,\\
\max_{0\leq k\leq k_{small}-2}\sup_{\mr\cap\Mext}u^{1+\dec}\left|\dk^k({}^{(glo)}\Gac, {}^{(glo)}\Rc)\right| &\les& \ep,\\
\max_{0\leq k\leq k_{large}-1}\left(\int_{\mr}\left|\dk^k({}^{(glo)}\Gac, {}^{(glo)}\Rc)\right|^2\right)^{\frac{1}{2}} &\les& \ep,
\eeaa
where ${}^{(glo)}\Rc$ and ${}^{(glo)}\Gac$ are given by
\beaa
{}^{(glo)}\Rc &=& \left\{\a, \b, \rho+\frac{2m}{r^3}, \bb, \aa\right\},\\
{}^{(glo)}\Gac &=& \left\{\xi, \om+\frac{m}{r^2}, \ka-\frac{2\Up}{r}, \vth, \ze, \eta, \etab, \kab+\frac{2}{r}, \vthb, \omb, \xib\right\}.
\eeaa

\item[(d)] Furthermore, we may also choose the global frame such that, in addition, one of the following  two possibilities hold,
\begin{itemize}
\item[i.] We have on  all $\Mext$ 
\beaa
({}^{(glo)}e_4, {}^{(glo)}e_3, {}^{(glo)}e_\th)= \left({}^{(ext)}\Up\,{}^{(ext)}e_4, {}^{(ext)}\Up^{-1}{}^{(ext)}e_3, {}^{(ext)}e_\th\right).
\eeaa

\item[ii.] We have on all  $\Mint$ 
\beaa
({}^{(glo)}e_4, {}^{(glo)}e_3, {}^{(glo)}e_\th) = \left({}^{(int)}e_4, {}^{(int)}e_3, {}^{(int)}e_\th\right).
\eeaa
\end{itemize}
\end{itemize}
\end{proposition}

\begin{remark}
\label{remark-globalframeforM1}
The global frame on $\MM$ of Proposition \ref{prop:existenceandestimatesfortheglobalframe} will be used to construct the second global frame in the next section, see Proposition \ref{prop:existenceandestimatesfortheglobalframe:bis}. It will also be used to recover high order derivatives in Theorem M8 (stated in section \ref{sec:endoftheproofofmaintheorem}), see section \ref{sec:presentationoftheglobalframeusedintheproofofThmM8}.   
\end{remark}


\subsection{Construction of the second global frame}
\label{subsection:constructionsecondglobalframe}


We start with the definition of the region where first global frame of $\MM$ (i.e. the one of Proposition \ref{prop:existenceandestimatesfortheglobalframe}) and a conformal renormalization of the frame second frame of $\Mext$ (i.e. the one of Proposition \ref{prop:constructionsecondframeinMext}) will be matched.
\begin{definition}\lab{def:cutofffunctionforthematchingregion:bis}
We define the matching region as the spacetime region
\beaa
\mr':=\Mext\cap\left\{\frac{7m_0}{2}\leq \rext\leq 4m_0\right\},
\eeaa 
where $\rext$ denotes the area radius of the outgoing geodesic foliation of $\Mext$.
\end{definition}

Here is our main proposition concerning our second global frame.
\begin{proposition}\lab{prop:existenceandestimatesfortheglobalframe:bis}
Let an integer $k_{loss}$ and a small constant $\de_0>0$ satisfying \eqref{eq:constraintsonklossandde0forsecondframeofMext}. There exists a global null frame $({}^{(glo')}e_4, {}^{(glo')}e_3, {}^{(glo')}e_\th)$ defined on $\Mint\cup\Mext$ such that
\begin{itemize}
\item[(a)] In $\Mext\cap\{\rext\geq 4m_0\}$, we have
\beaa
({}^{(glo')}e_4, {}^{(glo')}e_3, {}^{(glo')}e_\th)= \left({}^{(ext)}\Up\,{}^{(ext)}e_4', {}^{(ext)}\Up^{-1}{}^{(ext)}e_3', {}^{(ext)}e_\th'\right),
\eeaa
where $({}^{(ext)}e_4', {}^{(ext)}e_3', {}^{(ext)}e_\th')$ denotes the second frame of $\Mext$, i.e. the fame of Proposition \ref{prop:constructionsecondframeinMext}.

\item[(b)] In $\Mint\cup(\Mext\cap\{\rext\leq \frac{7m_0}{2}\})$, we have
\beaa
({}^{(glo')}e_4, {}^{(glo')}e_3, {}^{(glo')}e_\th) = ({}^{(glo)}e_4, {}^{(glo)}e_3, {}^{(glo)}e_\th),
\eeaa
where $({}^{(glo)}e_4, {}^{(glo)}e_3, {}^{(glo)}e_\th)$ denotes the first global frame of $\MM$, i.e. the frame of Proposition \ref{prop:existenceandestimatesfortheglobalframe}.
 
\item[(c)] In the matching region, we have
\beaa
\max_{0\leq k\leq k_{small}+k_{loss}}\sup_{\mr'}u^{1+\dec-2\de_0}\left|\dk^k({}^{(glo')}\Gac, {}^{(glo')}\Rc)\right| &\les& \ep,
\eeaa
where ${}^{(glo')}\Rc$ and ${}^{(glo')}\Gac$ are given by
\beaa
{}^{(glo')}\Rc &=& \left\{\a, \b, \rho+\frac{2m}{r^3}, \bb, \aa\right\},\\
{}^{(glo')}\Gac &=& \left\{\xi, \om+\frac{m}{r^2}, \ka-\frac{2\Up}{r}, \vth, \ze, \eta, \etab, \kab+\frac{2}{r}, \vthb, \omb, \xib\right\}.
\eeaa
with the Ricci coefficients and curvature components being the one associated to the frame $({}^{(glo')}e_4, {}^{(glo')}e_3, {}^{(glo')}e_\th)$. 

\item[(d)] Furthermore, we may also choose the global frame such that, in addition, one of the following  two possibilities hold,
\begin{itemize}
\item[i.] We have on  $\Mext\cap\{\rext\geq \frac{15m_0}{4}\}$ 
\beaa
({}^{(glo')}e_4, {}^{(glo')}e_3, {}^{(glo')}e_\th)= \left({}^{(ext)}\Up\,{}^{(ext)}e_4', {}^{(ext)}\Up^{-1}{}^{(ext)}e_3', {}^{(ext)}e_\th'\right).
\eeaa

\item[ii.] We have on $\Mint\cup(\Mext\cap\{\rext\leq \frac{15m_0}{4}\})$ 
\beaa
({}^{(glo')}e_4, {}^{(glo')}e_3, {}^{(glo')}e_\th) = ({}^{(glo)}e_4, {}^{(glo)}e_3, {}^{(glo)}e_\th).
\eeaa
\end{itemize}
\end{itemize}
\end{proposition}

\begin{remark}
\label{remark-globalframeforM1:bis}
The global frame on $\MM$ of Proposition \ref{prop:existenceandestimatesfortheglobalframe:bis} will be needed to derive decay estimates for the quantity $\qf$ in Theorem M1 (stated in section \ref{sect:main-intermediateresults}). 
\end{remark}


\section{Proof of the main theorem}



\subsection{Main intermediate results}
\lab{sect:main-intermediateresults}


We are ready to   state our main intermediary results.
\begin{thmM1}
Assume given a GCM admissible spacetime $\MM$  as defined in section \ref{sec:defintioncanonicalspacetime} 
verifying the bootstrap   assumptions\footnote{Recall in particular that the conclusions of Theorem M0 hold under the bootstrap assumptions {\bf BA-E} and {\bf BA-D}.} {\bf BA-E} and {\bf BA-D}  for  some sufficiently small   $\ep>0$. Then, if $\ep_0>0$ is sufficiently small, there exists $\dee>\dec$ such that we have the following   estimates in  $\MM$,
\beaa
 \max_{0\leq k\leq k_{small}+20}\sup_{\Mext}\left\{\Big(ru^{\frac{1}{2}+\dee}+u^{1+\dee}\Big)|\dk^k\qf|+ru^{1+\dee}|\dk^ke_3\qf|\right\}\\
+\max_{0\leq k\leq k_{small}+20}\sup_{\Mint}\ub^{1+\dee}|\dk^k\qf|    &\les&  \ep_0.
\eeaa
Moreover, $\qf$ also satisfies the following estimate
\beaa
 \max_{0\leq k\leq k_{small}+21}\ub^{2+2\dee}\int_{\Mint(\geq \ub)}|\dk^k\qf|^2 +\max_{0\leq k\leq k_{small}+20}u^{2+2\dee}\int_{\Sigma_*(\geq u)}|\dk^ke_3\qf|^2 &\les& \ep_0^2.
\eeaa
\end{thmM1}

\begin{thmM2}
Under  the  same assumptions as above we  have the following decay estimates for $ \,{}^{(ext)}\a$
\beaa
\max_{0\leq k\leq k_{small}+20}\sup_{\Mext}\Big(\frac{r^2(2r+u)^{1+\dee}}{\log(1+u)}+r^3(2r+u)^{\frac{1}{2}+\dee}\Big)\Big(|\dk^k\,{}^{(ext)}\a|+r|\dk^ke_3\,{}^{(ext)}\a|\Big) &\les& \ep_0.
\eeaa
\end{thmM2}

\begin{thmM3}
Under  the  same assumptions as above 
 we  have the following decay estimates for $\aa$
\beaa
\,{}^{(int)}\Dk_{k_{small}+16 }[\aa]      \les  \ep_0, \qquad \max_{0\leq k\leq k_{small}+18}\int_{\Sigma_*}u^{2+2\dee}|\dk^k\aa|^2 \les \ep_0^2.
\eeaa
\end{thmM3}

\begin{thmM4}
Under  the  same assumptions as above we also have the following decay estimates  in $\Mext$
\beaa
 \,{}^{(ext)}\Dk_{k_{small}+8} [\Rc] + \,{}^{(ext)}\Dk_{k_{small}+8} [\Gac]    \les  \ep_0.
\eeaa
\end{thmM4}

\begin{thmM5}
Under  the  same assumptions as above we also have the following decay estimates  for $\Rc$ and $\Gac$    in    $\Mint$
\beaa
 \,{}^{(int)}\Dk_{k_{small}+5} [\Rc] + \,{}^{(int)}\Dk_{k_{small}+5} [\Gac]  \les  \ep_0.
\eeaa
\end{thmM5}

Note that, as an immediate consequence of Theorem M2 to Theorem M5 we have obtained, under  the  same assumptions as above, the following improvement of our bootstrap assumptions on decay
\bea\lab{eq:finalimprovmentofabsolutlyallbootstrapasumptionseverywhereonMMforallderivatives}
\Nk^{(Dec)}_{k_{small}+5}\les \ep_0.
\eea


\subsection{End of the proof of the main theorem}\lab{sec:endoftheproofofmaintheorem}


\begin{definition}[Definition of $\aleph(u_*)$]
Let $\ep_0>0$ and $\ep>0$ be given small constants satisfying the constraint \eqref{eq:constraintbetweenepep*ep0}. Let $\aleph(u_*)$ be the set of all GCM admissible spacetimes $\MM$ defined in section \ref{sec:defintioncanonicalspacetime} such that
\begin{itemize}
\item $u_*$ is the value of $u$ on the last  outgoing slice $\CC_*$, 

\item $u_*$ satisfies \eqref{eq:behaviorofronSigmastar},

\item the bootstrap assumptions \eqref{def:bootstrapasumptionsglobalnormonenergie} \eqref{def:bootstrapasumptionsglobalnormondecay} hold true, i.e., relative to the combined norms defined in   section \ref{sec:definitionofconcatenatednorm}, we have
\beaa
\Nk^{(En)}_{k_{large}}\leq \ep,\,\,\,\,\Nk^{(Dec)}_{k_{small}}\le \ep.
\eeaa
\end{itemize}
\end{definition}

\begin{definition}
Let $\UU$ be the set of all values of $u_*\geq 0$ such that the spacetime $\aleph(u_*)$ exists. 
\end{definition}

The following theorem shows that $\UU$ is not empty. 
\begin{thmM6}\lab{th:M6}
There exists $\de_0>0$ small enough such that for sufficiently small constants $\ep_0>0$ and $\ep>0$ satisfying the constraints \eqref{eq:constraintbetweenepep*ep0} \eqref{eq:behaviorofronSigmastar}, we have $[1,1+\de_0]\subset\UU$. 
\end{thmM6}

In view of Theorem M6, we may define $U_*$ as the supremum over all value of $u_*$ that belongs to $\UU$. 
\beaa
U_* :=\sup_{u_*\in\UU}u_*.
\eeaa

Assume by contradiction that 
$$U_*<+\infty.$$
Then, by the continuity of the flow, $U_*\in \UU$. Furthermore, according to the consequence \eqref{eq:finalimprovmentofabsolutlyallbootstrapasumptionseverywhereonMMforallderivatives} of Theorem M2 to Theorem M5, the bootstrap assumptions on decay \eqref{def:bootstrapasumptionsglobalnormondecay} on any spacetime of $\aleph(U_*)$ are improved by
\beaa
\Nk^{(Dec)}_{k_{small}+5}\les \ep_0.
\eeaa
To reach a contradiction, we still need an extension procedure for spacetimes in $\aleph(u_*)$ to larger values of $u$, as well as to improve our bootstrap assumptions on weighted energies  \eqref{def:bootstrapasumptionsglobalnormonenergie}. This is done in two steps.
\begin{thmM7}
Any GCM admissible spacetime in $\aleph(u_*)$ for some $0<u_*<+\infty$ such that 
\beaa
\Nk^{(Dec)}_{k_{small}+5}\les \ep_0,
\eeaa
has a GCM admissible extension (satisfying \eqref{eq:behaviorofronSigmastar}), i.e. $u_*'>u_*$, initialized by Theorem M0, which verifies  
\beaa
\Nk^{(Dec)}_{k_{small}}\les \ep_0.
\eeaa
\end{thmM7}

\begin{remark}\lab{rmk:infactimproveddecayofThM7holdsforrTininterval}
Recall that the definition of a GCM admissible spacetime in section \ref{sec:defintioncanonicalspacetime} is such that $\TT=\{r=\rh\}$ for some $\rh$ satisfying
\bea\lab{eq:rangeforthechoiceofrh}
2m_0\left(1+\frac{\deh}{2}\right)\leq \rh\leq 2m_0\left(1+\frac{3\deh}{2}\right).
\eea
All results obtained so far, in particular Theorems M0--M7, hold for any choice of $\rh$ satisfying \eqref{eq:rangeforthechoiceofrh}, see Remark \ref{remark:whydotheresultsofThemM0-M7holdforanyrhproofThM8} for a more precise statement. It is at this stage, in Theorem M8 below, that we need to make a specific choice of $\rh$ in the context of a Lebesgue point argument required for the control of top order derivatives. This choice will be made in \eqref{eq:choiceofRTTismadebythisinfimum:bis}.
\end{remark}

\begin{thmM8}
There exists a choice of $\rh$ satisfying \eqref{eq:rangeforthechoiceofrh} such that the GCM admissible spacetime exhibited in Theorem M7 satisfies in addition 
\beaa
\Nk^{(En)}_{k_{large}}\les \ep_0
\eeaa
and therefore belongs to $\aleph(u_*')$. In particular $u_*'$ belongs to $\UU$. 
\end{thmM8}

In view of Theorem M8, we have reached a contradiction, and hence
$$U_*=+\infty$$
so that the spacetime may be continued forever. This concludes the proof of the main theorem.\\


\subsection{Conclusions}



\subsubsection{The Penrose diagram of $\MM$}


{\bf Complete future null infinity.} We first deduce from our estimate that our spacetime $\MM$ has a complete future null infinity $\II_+$. The portion of null infinity of $\MM$ corresponds to the limit $r\to+\infty$ along the leaves $\CC_u$ of the outgoing geodesic foliation of $\Mext$. As $\CC_u$ exists for all $u\geq 0$ with suitable estimates, it suffices to prove that $u$ is an affine parameter of $\II_+$. To this end, recall from our main theorem that the estimates $\Nk^{(Dec)}_{k_{small}}\les \ep_0$ hold which implies in particular\footnote{Using also Proposition  \ref{prop:finalcoordinatessystem} for the control of $\vsi$.}
\bea\lab{eq:estimate1intheconclusionsforxibombandvsi}
\sup_{\Mext}ru^{1+\dec}\left(|\xib|+\left|\omb-\frac{m}{r^2}\right|+r^{-1}|\vsi-1|\right) &\les&\ep_0.
\eea
As $|m-m_0|\les\ep_0 m_0$, see Lemma  \ref{lemma:estimatesonceandforallforHawkingmass}, $m$ is bounded. We infer that 
\beaa
\lim_{\CC_u, r\to +\infty}\xib, \omb =0\textrm{ for all }1\leq u<\infty.
\eeaa
In view of the identity 
\beaa
D_3e_3 &=& -2\omb e_3 + 2\xib e_\th,
\eeaa
we infer that $e_3$ is a null geodesic generator of $\II_+$. Since we have $e_3(u)=\frac{2}{\vsi}$ with $|\vsi-1|\les\ep_0$ in view of \eqref{eq:estimate1intheconclusionsforxibombandvsi}, $u$ is an  affine parameter of $\II_+$ so that $\II_+$ is indeed complete.

{\bf Existence of a future event horizon.} Next, note that the estimates $\Nk^{(Dec)}_{k_{small}}\les \ep_0$ also imply
\beaa
\sup_{\Mint}\ub^{1+\dec}\left(\left|\kab+\frac{2}{r}\right|+\left|\ka-\frac{2\left(1-\frac{2m}{r}\right)}{r}\right|\right) &\les&\ep_0.
\eeaa
In particular, considering the spacetime region $r\leq 2m_0(1-\deh/2)$ of $\Mint$, and in view of the estimate $|m-m_0|\les\ep_0 m_0$, we infer, for all $r\le 2m_0(1-\deh/2)$, that 
\beaa
\ka\le 2 \frac{r-2m}{r^2}+O(\ep_0)\les \frac{2}{r^2} (r-2m_0+ 2m_0-2m)+O(\ep_0) \les  \frac{2m_0}{r^2} (- \deh+ \ep_0  ) +O(\ep_0).
\eeaa
Thus, since  $0<\ep_0\ll\deh\ll 1$, we deduce,
\beaa
\sup_{\Mint\left(r\leq 2m_0\left(1-\frac{\deh}{2}\right)\right)}\ka &\leq& -\frac{\deh}{2m_0\left(1-\frac{\deh}{2}\right)^2} +O(\ep_0)\\
&\leq& -\frac{\deh}{4m_0}. 
\eeaa
 Thus, all 2-spheres $S(\ub, s)$ of the ingoing geodesic foliation of $\Mint$ which are located in the spacetime region $r\leq 2m_0(1-\deh/2)$ of $\Mint$ are trapped. This implies that the past of $\II_+$ in $\MM$ does not contain this region, and hence $\MM$ contains the event horizon $\HH_+$ of a black hole in its interior. Moreover, since the timelike hyper surface $\TT$ is foliated by the outgoing null cones $\CC_u$ of $\Mext$, it is in the past of $\II_+$. Hence, since $\TT$ is one of the boundaries of $\Mint$, $\HH_+$ is actually located in the interior of the region $\Mint$.  

{\bf Asymptotic stationarity of $\MM$.} Recall that we have introduced a vectorfield $\T$ in $\Mext$ as well as one in $\Mint$ by
\beaa
\T=e_3+\Up e_4\textrm{ in }\Mext,\quad \T=e_4+\Up e_3\textrm{ in }\Mint.
\eeaa
We can easily express all components of $\piT$ in terms of  $\Gac$, $ e_3(m), e_4 m$. Thus, making  us of  the estimate $\Nk^{(Dec)}_{k_{small}}\les \ep_0$ 
of our main theorem, we deduce,
\beaa
|{}^{(\T)}\pi|\les \frac{\ep_0}{ru^{1+\dec}}\textrm{ in }\Mext\textrm{ and }|{}^{(\T)}\pi|\les \frac{\ep_0}{\ub^{1+\dec}}\textrm{ in }\Mint.
\eeaa
In particular, $\T$ is an asymptotically Killing vectorfield and hence our spacetime $\MM$ is asymptotically stationary.\\

\begin{figure}[h!]
\centering
\includegraphics[scale=0.5]{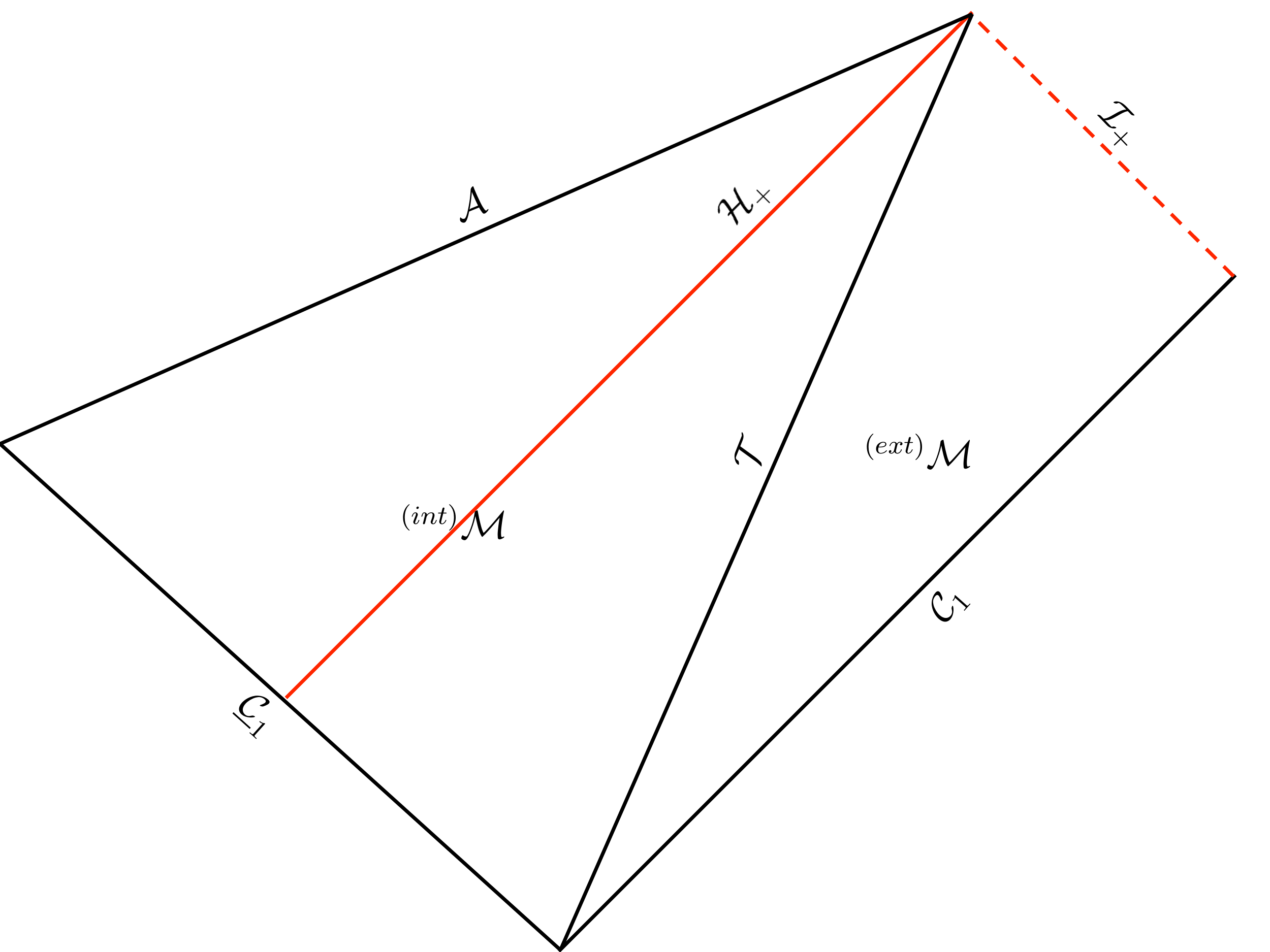}
\caption{The Penrose diagram of the space-time $\mathcal{M}$}
\label{fig:penrosediagramconclusionmaintheorem}
\end{figure}

The above conclusions regarding $\II_+$ and $\HH_+$ allow us to draw the Penrose diagram of $\MM$, see Figure \ref{fig:penrosediagramconclusionmaintheorem}.

\subsubsection{Limits  at null infinity and Bondi mass}


Recall the following formula for the derivative of the Hawking mass in $\Mext$, see Proposition \ref{prop:derivativesHawkingmass}
\beaa
e_4(m) &=&  \frac{r}{32\pi}\int_S\left(-\frac{1}{2}\kab\vth^2  -\frac{1}{2}\check{\ka}\vth\vthb+2\check{\ka}\check{\rho}+2e_\th(\ka)\ze  +2\ka\ze^2\right).
\eeaa
As a simple corollary of  the decay estimates of our main theorem, i.e.,  $\Nk^{(Dec)}_{k_{small}}\les \ep_0$, we deduce,
\bea\lab{eq:estimatefore4derivativeofHawkingmassinMext}
|e_4(m)| &\les& \frac{\ep^2_0}{r^2u^{1+2\dec}}.
\eea
Since $r^{-2}$ is integrable, we infer the existence of a limit to $m$ as $r\to +\infty$ along $\CC_u$
\beaa
M_B(u) &=& \lim_{r\to +\infty} m(u, r)\textrm{ for all }1\leq u<+\infty
\eeaa
where $M_B(u)$ is the so-called Bondi mass.
 
Next, we recall the following formula in $\Mext$, see Proposition \ref{propos:basiceqts-geod}
\beaa
e_4(\vthb) + \frac{1}{2}\ka\vthb &=& 2\dds_2\ze -\frac{1}{2}\kab\vth+2\ze^2.
\eeaa 
In view of $\Nk^{(Dec)}_{k_{small}}\les \ep_0$, we  deduce
\beaa
|e_4(r\vthb)| &\les& \frac{\ep_0}{r^2u^{\frac{1}{2}+\dec}}.
\eeaa
Since $r^{-2}$ is integrable, we infer the existence of a limit to $r\vthb$ as $r\to +\infty$ along $\CC_u$
\beaa
\underline{\Theta}(u,\cdot) &=& \lim_{r\to +\infty}r\vthb(r, u, \cdot)\textrm{ for all }1\leq u<+\infty.
\eeaa
  On the other hand, in view of     $\Nk^{(Dec)}_{k_{small}}\les \ep_0$  again,   
 \beaa
 r|\vthb| &\les& \frac{\ep_0}{u^{1+\dec}}, \qquad \mbox{on} \quad \Mext.
 \eeaa
 We infer that
 \beaa
 |\underline{\Theta}(u,\cdot)| &\les& \frac{\ep_0}{u^{1+\dec}}\textrm{ for all }1\leq u<+\infty.
 \eeaa


\subsubsection{The spheres at null infinity are round}


The Gauss curvature is given by the formula,
\beaa
K &=& -\rho-\frac{1}{4}\ka\kab +\frac{1}{4}\vth\vthb.
\eeaa
Thus, in view of our estimates in $\Mext$,
\beaa
\left|K -\frac{1}{r^2}\right| &\les& \frac{\ep_0}{r^3u^{\frac{1}{2}+\dec}}
\eeaa
so that 
\beaa
\lim_{r\to +\infty}r^2K =1.
\eeaa
In particular the spheres at null infinity are round.

 
\subsubsection{A Bondi mass formula} 


Using the formula for $e_3(m)$ in $\Mext$, see Proposition \ref{prop:derivativesHawkingmass}, together with the estimates $\Nk^{(Dec)}_{k_{small}}\les \ep_0$, we deduce
 \beaa
\left|e_3(m) + \frac{r}{64\pi}\int_S\ka\vthb^2\right| &\les& \frac{\ep_0^2}{ru^{\frac{3}{2}+2\dec}}
 \eeaa
 and hence
 \beaa
\left|e_3(m) + \frac{1}{8|S|}\int_S(r\vthb)^2\right| &\les& \frac{\ep_0^2}{ru^{\frac{3}{2}+2\dec}}.
 \eeaa 
Letting $r\to +\infty$ along $\CC_u$, and using that the spheres at null infinity are round, we infer in view of the definition of $M_B$ and $\underline{\Theta}$
\beaa
e_3(M_B)(u) &=& -\frac{1}{8}\int_{\mathbb{S}^2}\underline{\Theta}^2(u,\cdot)\textrm{ for all }1\leq u<+\infty.
\eeaa
Since $e_3(u)=\frac{2}{\vsi}$ and $e_3$ is orthogonal to the spheres foliating $\II_+$, we infer $e_3=\frac{2}{\vsi}\pr_u$. Thus, we obtain the following Bondi mass type formula 
\beaa
\pr_uM_B(u) &=& -\frac{\vsi}{16}\int_{\mathbb{S}^2}\underline{\Theta}^2(u,\cdot)\textrm{ for all }1\leq u<+\infty,
\eeaa
with $\vsi$ satisfying \eqref{eq:estimate1intheconclusionsforxibombandvsi}.


\subsubsection{Final Bondi mass} 


In view of the estimate
 \beaa
 |\underline{\Theta}(u,\cdot)| &\les& \frac{\ep_0}{u^{1+\dec}}\textrm{ for all }1\leq u<+\infty,
 \eeaa
and the control for $\vsi$ in \eqref{eq:estimate1intheconclusionsforxibombandvsi}, we infer that
\beaa
|\pr_uM_B(u)| &\les& \frac{\ep_0^2}{u^{2+2\dec}}\textrm{ for all }1\leq u<+\infty.
\eeaa 
In particular, since $u^{-2-2\dec}$ is integrable, the limit along $\II_+$ exists
\beaa
M_B(+\infty)=\lim_{u\to +\infty}M_B(u)
\eeaa
and is the so-called final Bondi mass. We denote it as $m_\infty$, i.e. $m_\infty=M_B(+\infty)$.

{\bf Control of $m-m_\infty$.} We have as a consequence of the above estimate for $\pr_uM_B$ and the definition of $m_\infty$
\beaa
|M_B(u)-m_\infty| &\les& \frac{\ep_0^2}{u^{1+2\dec}}\textrm{ for all }1\leq u<+\infty. 
\eeaa
 Also, recall from \eqref{eq:estimatefore4derivativeofHawkingmassinMext} that we have  obtained in $\Mext$
 \beaa
|e_4(m)| &\les& \frac{\ep^2_0}{r^2u^{1+2\dec}}
\eeaa
which yields, together with the definition of $M_B(u)$, by integration in $r$ at fixed $u$
\beaa
|m(r,u)-M_B(u)| &\les& \frac{\ep^2_0}{ru^{1+2\dec}}\textrm{ in }\Mext.
\eeaa
We infer
\bea\lab{eq:estimateformminusfinalBondimassonMext}
\sup_{\Mext}u^{1+2\dec}|m-m_{\infty}| &\les& \ep^2_0.
\eea
Also, recall the following formula for the derivative of the Hawking mass in $\Mint$, see Proposition \ref{prop:derivativesHawkingmass} in the context of an outgoing geodesic foliation,
\beaa
e_3(m) &=&  \frac{r}{32\pi}\int_S\left(-\frac{1}{2}\ka\vthb^2  -\frac{1}{2}\check{\kab}\vth\vthb+2\check{\kab}\check{\rho}-2e_\th(\kab)\ze  +2\kab\ze^2\right).
\eeaa
 Together with the estimates $\Nk^{(Dec)}_{k_{small}}\les \ep_0$, we deduce
 \beaa
 |e_3(m)| &\les& \frac{\ep^2_0}{\ub^{2+2\dec}}\textrm{ on }\Mint
 \eeaa
and hence by integration in $r$ at fixed $\ub$,   for $r\in[2m_0(1-\deh), \rr]$,
\beaa
\left|m(r,\ub)-m\Big(\rr , \ub\Big)\right| &\les& \frac{\ep^2_0  }{\ub^{2+2\dec}}  m_0\deh  \qquad     \textrm{ on }\Mint.
\eeaa
According to \eqref{eq:estimateformminusfinalBondimassonMext}, since $\{r=\rr\}=\TT=\Mext\cap\Mint\subset \Mext$, 
and since $\ub=u$ in $\TT$ by the initialization of $\ub$, 
\beaa
\ub^{1+2\dec}\left|m\Big(\rr , \ub\Big)-m_{\infty}\right| &\les& \ep^2_0.
\eeaa
We deduce
\bea\lab{eq:estimateformminusfinalBondimassonMint}
\sup_{\Mint}\ub^{1+2\dec}|m-m_{\infty}| &\les& \ep_0^2.
\eea
Combining \eqref{eq:estimateformminusfinalBondimassonMext} and \eqref{eq:estimateformminusfinalBondimassonMint} with the estimate
\beaa
\sup_{\MM}|m-m_0| &\les& \ep_0 m_0,
\eeaa
in the statement of our  main theorem (see also Lemma \ref{lemma:estimatesonceandforallforHawkingmass}), we infer that
\beaa
|m_\infty -m_0|\les \ep_0 m_0.
\eeaa
In particular       we  deduce  that   $m_\infty>0$ since  $\ep_0$ can be made arbitrarily small.


\subsubsection{Coordinates systems on $\Mext$ and $\Mint$}


In view of Proposition \ref{prop:finalcoordinatessystem}, and together with the control of the averages $\overline{\ka}$, $\overline{\kab}$ provided by Lemma \ref{lemma:estimatesonceandforallforaverages}, the control of $\check{\ka}$ provided by the estimates $\Nk^{(Dec)}_{k_{small}}\les \ep_0$, and the control of $m-m_\infty$ obtained in \eqref{eq:estimateformminusfinalBondimassonMext}, we infer for the space-time metric $\g$ on $\Mext$ in the $(u, r, \th, \vphi)$ coordinates system
\beaa
\g &=& \g_{m_\infty, \Mext} +O\left(\frac{\ep_0}{u^{1+\dec}}\right)\Big((dr, du, rd\th)^2, r^2(\sin\th)^2(d\vphi)^2\Big)
\eeaa
where $\g_{m_\infty, \Mext}$ denotes the Schwarzschild metric of mass $m_\infty>0$ in outgoing Eddington-Finkelstein coordinates, i.e.
\beaa
\g_{m_\infty, \Mext} &=& -2du dr -\left(1-\frac{2m_{\infty}}{r}\right) (du)^2+ r^2\Big((d\th)^2+(\sin\th)^2(d\vphi)^2\Big).
\eeaa

Also, in view of Proposition \ref{prop:finalcoordinatessystem:bis}, and together with the control of the averages $\overline{\ka}$, $\overline{\kab}$ provided by Lemma \ref{lemma:estimatesonceandforallforaverages}, the control of $\check{\kab}$ provided by the estimates $\Nk^{(Dec)}_{k_{small}}\les \ep_0$, and the control of $m-m_\infty$ obtained in \eqref{eq:estimateformminusfinalBondimassonMint}, we infer for the space-time metric $\g$ on $\Mint$ in the $(\ub, r, \th, \vphi)$ coordinates system
\beaa
\g &=& \g_{m_\infty, \Mint} +O\left(\frac{\ep_0}{\ub^{1+\dec}}\right)\Big((dr, d\ub, rd\th)^2, r^2(\sin\th)^2(d\vphi)^2\Big)
\eeaa
where $\g_{m_\infty, \Mext}$ denotes the Schwarzschild metric of mass $m_\infty>0$ in ingoing Eddington-Finkelstein coordinates, i.e.
\beaa
\g_{m_\infty, \Mint} &=& 2d\ub dr -\left(1-\frac{2m_{\infty}}{r}\right) (d\ub)^2+ r^2\Big((d\th)^2+(\sin\th)^2(d\vphi)^2\Big).
\eeaa

{\bf Asymptotic of the future event horizon.} We show below that $\HH_+$ is located in the following region of $\Mint$
\bea\lab{eq:asymptoticlocationofthehorizonintheconlusionofthemaintheorem}
2m\left(1 -\frac{\sqrt{\ep_0}}{\ub^{1+\dec}}\right)\leq r\leq 2m\left(1 +\frac{\sqrt{\ep_0}}{\ub^{1+\frac{\dec}{2}}}\right)\textrm{ on }\HH_+\textrm{ for any }1\leq\ub<+\infty.
\eea
Note first that the lower bound follows from the fact that 
\beaa
\sup_{\Mint\left(r\leq 2m\left(1 -\frac{\sqrt{\ep_0}}{\ub^{1+\dec}}\right)\right)}\ka &\leq& -\frac{\frac{\sqrt{\ep_0}}{\ub^{1+\dec}}}{m\left(1-\frac{\sqrt{\ep_0}}{\ub^{1+\dec}}\right)^2} +O\left(\frac{\ep_0}{\ub^{1+\dec}}\right)\\
&\leq& -\frac{\sqrt{\ep_0}}{2m_0\ub^{1+\dec}} <0.
\eeaa

Concerning the upper bound, we need to show that any 2-sphere 
\bea\lab{eq:asymptoticspheretothehorizon}
S(\ub_1):=S\left(\ub_1, r=2m\left(1 +\frac{\sqrt{\ep_0}}{\ub_1^{1+\frac{\dec}{2}}}\right)\right),\quad 1\leq \ub_1<+\infty
\eea
is in the past of $\II_+$. Since $\Mext$ is in the past of $\II_+$, it suffices to show that the forward outgoing null cone emanating from any 2-sphere \eqref{eq:asymptoticspheretothehorizon} reaches $\Mext$ in finite time. 

Assume, by contradiction, that there exists an outgoing null geodesic, denoted by $\ga$, perpendicular to $S(\ub_1)$, that does not reach $\Mext$ in finite time. Let $e_4'$ be the geodesic generator of $\ga$. In view of Lemma \ref{lemma:SSMe:general.composite} on general null frame transformation, and denoting by $(e_4, e_3, e_\th)$ the null frame\footnote{Recall that we assume by contradiction that $\ga$ does not reach $\Mext$ and hence stays in $\Mint$.} of $\Mint$, we look for $e_4'$ under the form 
\beaa
e_4' &=& \la\left(e_4 + f e_\th +\frac 1 4 f^2  e_3\right),
\eeaa
and the fact that $e_4'$ is geodesic implies the following transport equations along $\ga$ for $f$ and $\la$ in view of Lemma \ref{lemma:transportequationsforffbandlambda} (applied\footnote{i.e. we keep the direction of  $e_3$  fixed.} with $\fb=0$)
\beaa
\la^{-1}e_4'(f)  +\left(\frac{\ka}{2}+2\om\right) f &=& -2\xi+E_1(f, \Ga),\\
\la^{-1}e_4'(\log(\la)) &=&  2\om+E_2(f, \Ga),
\eeaa
where $E_1$ and $E_2$ are given schematically by
\beaa
E_1(f, \Ga) &=& -\frac{1}{2}\vth f+\lot,\\
E_2(f, \Ga) &=&  f\ze- \frac{1}{2}f^2\omb -\etab f - \frac{1}{4}f^2\kab +\lot
\eeaa
Here, $\lot$ denote terms which are cubic or higher order in $f$ and $\Ga$ denotes the Ricci coefficients w.r.t. the original null frame $(e_3, e_4, e_\th)$ of $\Mint$.  

We then proceed as follows
\begin{enumerate}
\item First, we initialize $f$ and $\la$ as follows on the $\ga\cap S(\ub_1)$
\beaa
f=0,\quad \la=1\textrm{ on }\ga\cap S(\ub_1).
\eeaa

\item Then, we initiate a continuity argument by assuming for some 
$$\ub_1<\ub_2<\ub_1+\left(\frac{\ub_1}{\ep_0}\right)^{\frac{\dec}{2}}$$ 
that we have
\bea\lab{eq:bootstrapasumptiononfandUpfortheasymptoticlocationofthehorizon}
|f|\leq \frac{\sqrt{\ep_0}}{\ub_1^{\frac{1}{2}+\dec}},\quad \Up\geq \frac{\sqrt{\ep_0}}{2\ub_1^{1+\frac{\dec}{2}}},\quad 0<\la<+\infty\textrm{ on }\ga(\ub_1, \ub_2)\cap\Mint
\eea
where $\ga(\ub_1, \ub_2)$ denotes the portion of $\ga$ in $\ub_1\leq \ub\leq\ub_2$.

\item We have
$$\la^{-1}e_4'(\ub)=e_4(\ub)+\frac{1}{4}f^2e_3(\ub)=\frac{2}{\vsib}.$$
Relying on our control of the ingoing geodesic foliation of $\Mint$, the above assumption for $f$ and the transport equation for $f$, we obtain on $\ga(\ub_1, \ub_2)\cap\Mint$
\beaa
\sup_{\ga(\ub_1, \ub_2)\cap \Mint}|f| &\les& \frac{\ep_0}{\ub_1^{1+\dec}}(\ub_2-\ub_1)\\
&\les& \frac{\ep_0^{1-\frac{\dec}{2}}}{\ub_1^{1+\frac{\dec}{2}}}
\eeaa
which improves our assumption in \eqref{eq:bootstrapasumptiononfandUpfortheasymptoticlocationofthehorizon} on $f$.

\item We have in view of the control of $f$
$$\la^{-1}e_4'(r)=e_4(r)+\frac{1}{4}f^2e_3(r)=\Up+O\left(\frac{\ep_0}{\ub_1^{1+\dec}}\right).$$
This yields
\beaa
\la^{-1}e_4'(\log(\Up)) &=& \frac{\frac{2m}{r^2}e_4(r)-\frac{2}{r}\la^{-1}e_4'(m)}{\Up}\\
&=& \frac{\frac{2m}{r^2}\Up+O\left(\frac{\ep_0}{\ub_1^{1+\dec}}\right)}{\Up}.
\eeaa
Thanks to our assumption on the lower bound of $\Up$, we infer
\beaa
\la^{-1}e_4'(\log(\Up)) &=& \frac{2m}{r^2}(1+O(\sqrt{\ep_0}))
\eeaa
and since we are in $\Mint$
\beaa
\la^{-1}e_4'(\log(\Up)) &\geq& \frac{1}{3m_0}.
\eeaa
Integrating from $\ub=\ub_1$, we deduce
\beaa
\Up &\geq& \frac{\sqrt{\ep_0}}{(1+\sqrt{\ep_0})\ub_1^{1+\frac{\dec}{2}}}\exp\left(\frac{\ub-\ub_1}{3m_0}\right)
\eeaa
which is an improvement of our assumption in \eqref{eq:bootstrapasumptiononfandUpfortheasymptoticlocationofthehorizon} on $\Up$. 

\item In view of the control of $f$ and of the ingoing geodesic foliation of $\Mint$, we rewrite the transport equation for $\la$ as
\beaa
\la^{-1}e_4'(\log(\la)) &=&  2\om+E_2(f, \Ga)\\
&=& -\frac{2m}{r^2}+O\left(\frac{\ep_0}{\ub_1^{1+\dec}}\right).
\eeaa
Since we have obtained above the other hand 
\beaa
\la^{-1}e_4'(\log(\Up)) &=& \frac{2m}{r^2}(1+O(\sqrt{\ep_0}))
\eeaa
we immediately infer
\beaa
\la^{-1}e_4'(\log(\la)\Up^2)>0,\quad \la^{-1}e_4'(\log(\la)\sqrt{\Up})<0.
\eeaa
Integrating from $\ub=\ub_1$, this yields
\beaa
 \left(\frac{\sqrt{\ep_0}}{(1+\sqrt{\ep_0})\ub^{1+\frac{\dec}{2}}}\right)^2\Up^{-2}\leq\la\leq \left(\frac{\sqrt{\ep_0}}{(1+\sqrt{\ep_0})\ub^{1+\frac{\dec}{2}}}\right)^\frac{1}{2}\Up^{-\frac{1}{2}}.
\eeaa
Since $\Up$ has an explicit lower bounded in view of our previous estimate, as well as an explicit upper bound since we are in $\Mint$, this yields an improvement of our assumptions in \eqref{eq:bootstrapasumptiononfandUpfortheasymptoticlocationofthehorizon} for $\la$.

\item Since we have improved all our bootstrap assumptions \eqref{eq:bootstrapasumptiononfandUpfortheasymptoticlocationofthehorizon}, we infer by a continuity argument the following bound
\beaa
\Up &\geq& \frac{\sqrt{\ep_0}}{(1+\sqrt{\ep_0})\ub_1^{1+\frac{\dec}{2}}}\exp\left(\frac{\ub-\ub_1}{3m_0}\right)\textrm{ on }\ga\left(\ub_1, \ub_1+\left(\frac{\ub_1}{\ep_0}\right)^{\frac{\dec}{2}}\right)\cap\Mint.
\eeaa
Now, in this $\ub$ interval, we may choose
\beaa
\ub_3:=\ub_1+3m_0\left(1+\frac{\dec}{2}\right)\log\left(\frac{\ub_1}{\ep_0}\right)
\eeaa
for which we have $\Up\geq 1$. This is a contradiction since $\Up=O(\deh)$ in $\Mint$. Thus, we deduce that $\ga$ reaches $\Mext$ before $\ub=\ub_3$, a contradiction to our assumption on $\ga$. This concludes the proof of  \eqref{eq:asymptoticlocationofthehorizonintheconlusionofthemaintheorem}. 
\end{enumerate}


\section{The General Covariant Modulation procedure}\lab{sec:mainstatementsonGCMSprocedure}


 The role of this section is to give a short description of the results concerning our General Covariant Modulation (GCM) procedure, which is at the heart of our proof. We will apply it  in   $^{(ext)}\MM$ under  our main bootstrap assumptions {\bf BA-E} {\bf BA-D}.  The proof of the results stated in this section will be proved in Chapter \ref{chap:proofofGCMprocedure}.

 
 \subsection{Spacetime assumptions for the GCM procedure}  
 
 
 To state our results, which are local in nature  it is  convenient to consider   axially symmetric  polarized   spacetime regions  $\RR$ 
  foliated by two functions $(u, s)$ such that
\begin{itemize}
\item On $\RR$, $(u, s)$ defines an outgoing  geodesic foliation as in section \ref{sec:mainequationsforoutgoiggeodesicfoliations}.

\item We denote by $(e_3, e_4, e_\th)$ the null frame adapted to the outgoing geodesic foliation $(u, s)$ on $\RR$. 

\item We denote by $\ovS$ a fixed sphere of $\RR$ 
\bea
\ovS &:=& S(\ug, \sg)
\eea
and  by $\rg$ the area radius of $\ovS$, where $S(u, s)$ denote the 2-spheres  of the outgoing geodesic foliation $(u, s)$ on $\RR$.

\item In adapted coordinates $(u,s,\th, \vphi)$ with $b=0$, see   Proposition \ref{prop:outgoinggeodesiccoordinates}, the  spacetime  metric $\g$ in $\RR $  takes the form,  with $\Omb=e_3(s), \, \,  \underline{b}=e_3 (\th)$,
\bea
 \label{reducedmetric-geodesicfoliation-GCM:00}
  \g=-2\vsi du ds +\vsi^2 \Omb du^2+ \ga\left( d\th- \frac 1 2 \vsi\underline{b} du\right)^2 + e^{2\Phi} d\vphi^2 , \, 
 \eea
 where $\th$ is chosen such that $b=e_4(\th)=0$.
   
\item
 The  spacetime  metric induced on $S(u,s)$ is given by,
 \bea
\gS= \ga d\th^2 + e^{2\Phi}   d\vphi^2. 
 \eea
 \item The relation between the null frame and coordinate system is given by
 \bea
 \label{transf:coords-frame:00}
 e_4 =\pr_s, \qquad e_3 =\frac{2}{\vsi} \pr_u+\Omb  \pr_s+\underline{b}  \pr_\th,\qquad e_\th=\ga^{-1/2} \pr_\th. 
 \eea 
 \item
We denote the induced metric on $\ovS$ by 
 \beaa
 \ovgS=\ovga d\th^2 + e^{2\Phi} d\vphi^2.
 \eeaa
\end{itemize}

\begin{definition} 
\label{defintion:regionRRovr:00}
Let $0<\dg\leq \epg $   two sufficiently   small   constants.
 Let  $(\ug, \sg)$ real numbers so that
\bea\lab{eq:rangeofugandsg:00}
1  \leq \ug<+\infty,\quad  4m_0\leq\sg < +\infty.
\eea

We define  $\RR=\RR(\dg, \epg)$  to be the region
\bea
\lab{definition:RR(dg,epg):00}
\RR:=\left\{|u-\ug|\leq\de_{\RR} ,\quad |s-\sg|\leq  \de_\RR \right\}, \qquad \de_{\RR}:=\dg \big(\epg\big)^{-\frac{1}{2}},
\eea
such that  assumption {\bf A1-A3} below  with constant $\epg$  on  the background foliation of $\RR$,   are verified. The  smaller constant $\dg$          controls the size of    the GCMS quantities  as it will be made precise   below.  
\end{definition}

Consider  the  renormalized  Ricci  and  curvature  components associated to the $(u,s)$ geodesic foliation of $\RR$
  \beaa
\bsplit
\Gac :&=\left\{ \check{\ka}, \, \vth,\, \ze, \, \eta,\,  \ov{\ka}-\frac 2 r,\,  \ov{\kab}   +\frac{2\Up}{r}, \check{\kab},\, \vthb, \, \xib,\,  \check{\omb}, \,   \ov{ \omb}  -\frac{m}{r^2},  \, \check{\Omb}, \, \big(\ov{\Omb}+\Up\big),\, \big(\ov{\vsi}+1\big)\right \},\\
\Rc:&=\left\{ \a,\,  \b,\, \rhoc,\, \ov{\rho}+\frac{2m}{r^3},\, \bb, \, \aa\right\}.
\end{split}
\eeaa
Since our foliation is outgoing geodesic we also have, 
\bea
\xi=\om=0, \quad \etab+\ze=0.
\eea 
We decompose  $\Gac=\Ga_g\cup \Ga_b$ where,
\bea
\lab{definition:Gag-Gab-GCM:00}
\bsplit
\Ga_g&=\left\{ \check{\ka}, \, \vth,\, \ze, \, \kabc, \,\ov{\ka}-\frac 2 r,\,  \ov{\kab}   +\frac{2\Up}{r}   \right\},\\
\Ga_b&=\left\{\eta,\,   \vthb, \, \xib,\,  \check{\omb}, \,    \ov{ \omb}  -\frac{m}{r^2},  \, r^{-1}\Ombc,\, r^{-1} \vsic ,  r^{-1}   \big(\ov{\Omb}+\Up\big),\,  r^{-1}\big(\ov{\vsi}-1\big)\right\}.\\
\end{split}
\eea

Given an integer $s_{max}$, we assume the following\footnote{In applications, $s_{max}=k_{small}+4$  in Theorem  M7, and $s_{max}=k_{large }+5$  in Theorem M0 and Theorem M6.} 
 
{\bf A1.} For  $k\le s_{max}$, we have on $\RR$
\bea
\bsplit
\| \Ga_g\|_{k, \infty}&\les  \epg  r^{-2},\\
\| \Ga_b\|_{k, \infty}&\les  \epg  r^{-1},
\end{split}
\eea
and,
\bea
\bsplit
\|\a, \b, \rhoc, \muc\|_{k, \infty}&\les \epg r^{-3},\\
\|e_3( \a, \b)\|_{k-1, \infty}&\les \epg r^{-4},\\
\|\bb\|_{k,\infty}&\les \epg r^{-2},\\
\|\aa\|_{k,\infty}&\les \epg r^{-1}.
\end{split}
\eea

{\bf A2.} We have, with $m_0$ denoting the mass of the unperturbed spacetime,
\bea\lab{eq:assumtionsonthegivenusfoliationforGCMprocedure:Hawkingmass:00} 
\sup_{\RR}\left|\frac{m}{m_0}-1\right| &\les& \epg.
\eea

{\bf A3.}  The metric coefficients are assumed to satisfy the following assumptions  in $\RR$, for all $k\le s_{max}$
\bea
\lab{eq:assumtionsonthegivenusfoliationforGCMprocedure:bis:00} 
 r \left\|\left ( \frac{\ga}{r^2}-1 ,\,\,  \underline{b},  \, \, \frac{e^\Phi}{r\sin\th}-1\right)\right\|_{\infty, k}+
\left \|   \Omb+\Up\right\|_{\infty, k} +\left \|  \vsi-1\right\|_{\infty, k}\les \epg
\eea

We will  assume, in addition,  that the following small GCM conditions   hold true on $\RR$,
\begin{equation}
\lab{Main:TheoremsGCM-assumptions2:00}
\left|\ov{\ka}-\frac{2}{r}\right|+\left|\dk^k\kac\right|+r^2\left|\dk^{k-2}\dds_2\dds_1\kab\right| + r^3\left|\dk^{k-2}\dds_2\dds_1\mu\right|  \les \dg r^{-2} \,\textrm{ for all }k\le s_{max},
\end{equation}
\bea
\lab{Main:TheoremsGCM-assumptions3:00}
r^{-2}  \left|\int_S \eta e^\Phi \right|  &\les&    \dg, \quad 
 r^{-2}    \left|\int_S \xib e^\Phi \right|  \les   \dg.
 \eea
 Also, 
  \bea
 \lab{Main:TheoremsGCM-assumptions4:00}
    \left| \left(\frac{2}{\vsi}+\Omb\right)\Big|_{SP}- 1-\frac{2m}{r} \right| &\les  \dg.
 \eea

Additionally we may assume on $\RR$
 \bea
\lab{Main:TheoremsGCM-assumptions5:00}
 \bsplit
  r   \left| \int_S \b e^\Phi\right| &\les   \dg, \qquad    \left| \int_S e_\th (\kab)  e^\Phi\right| &\les   \dg.
  \end{split}
 \eea

  
 \subsection{Deformations of surfaces}


  \begin{definition}
 \label{definition:Deformations:00}
 We say that    $\S$ is an $O(\epg)$\,  $\Z$-polarized deformation of $ \ovS$ if there exists    a map $\Psi:\ovS\longrightarrow \S $ of the form,
 \bea
 \Psi(\ovu, \ovs, \th, \vphi)=\left( \ovu+ U(\th), \ovs+S(\th), \th, \vphi \right)
 \eea
 where   $U, S$ are smooth  functions   defined on the interval  $[0,\pi]$
 of amplitude at most $\epg$.  
 We denote by $\psi$ the reduce map defined on the interval $ [0,\pi]$,
 \bea
 \psi(\th)= ( \ovu+ U(\th), \ovs+S(\th), \th).
 \eea
 We restrict ourselves to deformations which fix the South Pole, i.e. 
 \bea
 U(0)=S(0)=0.
 \eea
  \end{definition}

  
   \subsection{Adapted frame transformations}
 
  
 We consider    general null transformations introduced in Lemma \ref{lemma:SSMe:general.composite},
   \bea
\lab{SSMe:general.composite-GCMdescription:00}
\begin{split}
e_4'&=\la\left(e_4 + f e_\th +\frac 1 4 f^2  e_3\right),\\
e_\th'&=\left(1+\frac 1 2   f \fb\right) e_\th  + \frac 1 2  \fb e_4+\frac 1 2 f\left(1+ \frac 1 4 f \fb\right) e_3,\\
e_3'&= \la^{-1} \left( \left(1+\frac 12  f \fb +\frac{1}{16} f^2 \fb^2\right)  e_3 +\fb\left(1+\frac 1 4 f\fb\right) e_\th + \frac 1 4 \fb^2  e_4\right).
\end{split}
\eea

\begin{definition}
 Given a deformation $\Psi:\ovS\longrightarrow \S$ we  say that  a new frame   $(e_3', e_4',  e_\th')$, obtained from the standard frame $(e_3, e_4, e_\th)$  via the transformation \eqref{SSMe:general.composite-GCMdescription:00},  is $\S$-adapted  if  we have,
 \bea
 e_\th'= e_\th^\S=\frac{1}{(\ga^\S)^{1/2} } \psi_\# (\pr_\th)
 \eea
 where $\psi_\#  (\pr_\th)$   is the push-forward   defined  by the deformation map $\psi$.
\end{definition}

The condition translates into the following relations between the functions  $U, S$ defining the deformation  and  the transition functions  $(f, \fb)$.
 \bea
 \label{eq:DeformS-66-GCMdescription:00}
  \bsplit
   \vsi^\# \pr_\th  U&= \left((\gaS)^\#  \right)^{1/2}   f^\#\left( 1 +\frac 1 4 (f\fb)^\#\right),\\
     \pr_\th S-\frac{\vsi^\#}{2} \Omb^\#  \pr_\th U&= \frac 1 2   \left((\gaS)^\#  \right)^{1/2}\fb^\#,
  \\
     (\ga^ \S)^\# &=\ga^\#+(\vsi^\#)^2 \left(\Omb+\frac{1}{4} \underline{b}^2  \ga  \right)^\#\, ( \pr_\th U )^2 -2 \vsi^\# \pr_\th U \pr_\th  S-(\ga \vsi  \underline{b})^\# \pr_\th  U,\\
     U(0)&= S(0)=0.
     \end{split}
  \eea

  
 \subsection{GCM results}\lab{sec:statementofGCMresultsinChapter3}
 

\begin{theorem}[GCMS-I]
\lab{thm:introd:GCMI:00}
Consider the region $\RR$ as above,     verifying  the assumptions {\bf A1}--{\bf A3}  and the  small GCM conditions\footnote{Here,   the other assumptions  \eqref{Main:TheoremsGCM-assumptions3:00} \eqref{Main:TheoremsGCM-assumptions4:00}  are not needed.} \eqref{Main:TheoremsGCM-assumptions2:00}.
Let $\ovS$ denote the  sphere $\ovS=S(\ovu, \ovs)$. 
For any fix   $\La, \Lab \in \RRR$  verifying,
\bea
|\La|,\,  |\Lab|  &\les & \dg \big(\ovr\big)^{2},
\eea
\begin{enumerate}
\item  There 
exists a unique  GCM sphere $\S=\S^{(\La, \Lab)}$, which is a deformation\footnote{In the sense of Definition \ref{definition:Deformations:00}.} of $\ovS$,   and an adapted  null frame   $e_3^\S, e_\th^\S, e_4^\S$,  such that  the  following  GCMS conditions are verified\footnote{ $\Ga^\S$, $R^\S$  denote the Ricci and curvature components with respect to the  adapted frame on $\S$.}   components.
 \bea
 \lab{Results:ThmGCMI-1Descr:00}
\ddsS_2\ddsS_1\kab^\S=\ddsS_2\ddsS_1\mu^\S=0, \qquad \ka^\S=\frac{2}{r^\S}.
\eea
In addition
\bea
\lab{Results:ThmGCMI-2Descr:00}
\int_{\S}  f  e^\Phi=\La, \qquad 
\int_{\S} \fb e^\Phi=\Lab, 
\eea
where $(f, \fb)$ belong to  the triplet $(f, \fb, \la=e^a)$ which denote  the change of frame coefficients from the frame of $\ovS$ to the one of $\S$.  
\item The transition functions  $(f, \fb, \log\la)$ verify, 
\bea
\lab{Results:ThmGCMI-1-F:00}
\big\| (f, \fb, \log\la) \big\|_{\hk_k(\S)} &\les \dg, \qquad k\le s_{max}+1. 
\eea
\item The area radius  $r^\S$  and Hawking mass $m^\S$of $\,\S$ verify,
\bea\lab{eq:comparisionareadraiusandHwakingmassonnewGCMsphere:00}
\big| r^\S-\ovr\big|\les \dg, \qquad \big| m^\S-\ovm\big|\les \dg.
\eea
\end{enumerate}
\end{theorem} 

The precise version of Theorem \ref{thm:introd:GCMI:00} and its proof   are given in  section \ref{section-ExistenceGCMspheres}.

\begin{theorem}[GCMS-II] 
\lab{thm:introd:GCMII:00}
In addition to  the assumptions    of Theorem \ref{thm:introd:GCMI:00} we  also  assume that \eqref{Main:TheoremsGCM-assumptions5:00} holds true. 
Then,
\begin{enumerate}
\item There exists a unique GCM sphere  $\S$, which is a deformation of $\ovS$,  such that  in addition to  \eqref{Results:ThmGCMI-1Descr:00}    the following  GCMS conditions  also  hold true on $\S$.
 \bea
 \label{Results:ThmGCMI-2}
\int_{\S}  \b^\S e^\Phi=0, \qquad 
\int_{\S}  e^\S_\th(\kabS) e^\Phi=0.
\eea
\item The transition functions  $(f, \fb, \log\la)$ verify the estimates \eqref{Results:ThmGCMI-1-F:00}.
\item The area radius  $r^\S$  and Hawking mass $m^\S$of $\,\S$ verify \eqref{eq:comparisionareadraiusandHwakingmassonnewGCMsphere:00}.
\end{enumerate}
\end{theorem}
 The precise version of Theorem \ref{thm:introd:GCMII:00} and its proof   are given in \ref{section:Corollary-ExistenceGCMS}.

\begin{theorem}[GCMH]\lab{th:existenceofGCMH:00}
Consider the region $\RR$ as above,     verifying  the assumptions {\bf A1}--{\bf A3}  and the  small GCM conditions  \eqref{Main:TheoremsGCM-assumptions2:00}--\eqref{Main:TheoremsGCM-assumptions4:00}. 
Let
 $\S_0=\S_0[\ovu, \ovs, \La_0, \Lab_0]$   the deformation of $\ovS$ constructed  in Theorem GCMS-I above. 
 
 There exists a smooth  spacelike hypersurface  $\Si_0\subset\RR$  passing through $\S_0$,       a  scalar function $u^\S$ defined  on $ \Si_0$,  whose level surfaces  are topological spheres denoted by $\S$,  and a smooth collection of   constants $\La^\S, \Lab^\S$ verifying,
\beaa
\La^{\S_0}=\La_0, \qquad \Lab^{\S_0} =\Lab_0,
\eeaa
    such that the following conditions are verified:
 \begin{enumerate}
\item   The surfaces    $\S$   of constant $u^\S$   verifies  all the properties   stated in 
Theorem  GCMS-I  for the prescribed  constants  $\La^\S, \Lab^\S$. 
  In particular  they come endowed  with   null frames  $(e_4^\S,  e_\th^\S, e_3^\S)$ such that
  \begin{itemize}
  \item[i.]
For each $\S$ the    GCM conditions \eqref{Results:ThmGCMI-1Descr:00},  \eqref{Results:ThmGCMI-2Descr:00}
hold   with $\La=\La^\S, \Lab=\Lab^\S$. 
  \item[iii.] The   transversality conditions  hold true on each $\S$.
  \bea
\lab{Transversality-Si*-intr.:00}
\xi^\S=0, \qquad \om^\S=0, \qquad \etab^\S+\ze^\S=0.
\eea
  \end{itemize}

\item We have, for some constant $c_{\Si_0}$,  
\bea
u^\S+r^\S=c_{\Si_0} , \qquad \textit{along} \quad \Si_0.
\eea
\item
Let $\nu^\S$  be the unique vectorfield tangent to the hypersurface $\Si_0$, normal to $\S$, and normalized by $g(\nu^\S, e_4^\S)=-2$. There exists a unique scalar function $a^\S$ on $\Si_0$ such that $\nu^\S$ is given by
 \beaa
 \nu^\S =e_3^\S+ a^\S e_4^\S.
 \eeaa
 The following normalization condition  holds true  at the  South Pole $SP$  of every sphere $\S$, i.e. at $\th=0$,
 \bea
 \lab{eq:ThmGCMH-Spole-a^S:00}
 a^\S\Big|_{SP}=-1 -\frac{2m^\S}{r^\S}.
 \eea
 
 \item Under the additional transversality condition\footnote{Here  the average of $\ka^\S$ is taken on $\S$. In view of the GCM conditions \eqref{Results:ThmGCMI-1Descr:00} we deduce $e_4^\S(r^\S)=1$.} on $\Si_0$
\bea
e_4^\S(u^\S)=0,\qquad e_4(r^s) = \frac{r^\S}{2}\ov{\ka^\S}= 1.
\eea
the Ricci coefficients $\eta^\S, \xib^\S$ are well defined and verify,
\bea
\lab{eq:ThmGCMH-badmodesetaxib:00}
\int_\S \eta^\S e^\Phi= \int_\S \xib^S e^\Phi=0.
\eea

\item The transition functions  $ (f, \fb, \log \la)$ verify the estimates   \eqref{Results:ThmGCMI-1-F:00}.

\item The area radius  $r^\S$  and Hawking mass $m^\S$of $\,\S$ verify \eqref{eq:comparisionareadraiusandHwakingmassonnewGCMsphere:00}.
\end{enumerate}
\end{theorem}

 The precise version of Theorem \ref{th:existenceofGCMH:00} and its proof   are given in section \ref{section:GCM-hypersurface}.
 
 \bigskip
 
 In addition to the three theorems  stated above we also prove  three rigidity  statements which   can be regarded  as  corollaries.
 \begin{corollary}[Rigidity I]
Assume $\S$ is a sphere in $\RR$ which verifies the the GCM conditions
\bea
\ka^\S=\frac{2}{r^\S}, \quad \dds_2\,^\S\dds_1\,^\S\kab^\S=\dds_2\,^\S\dds_1\,^\S\mu^\S=0.
\eea
Assume also that the background foliation  verify on $\RR$  the  conditions {\bf A1-A3}.  
Then the transition functions  $(f, \fb, \log\la)$ from the background frame  of $\RR$ to    to that of $\S$ 
 verifies the  estimates
 \beaa
\|(f, \fb, \log(\la))\|_{\hk_{s_{max}+1}(\S) } &\les& \dg+\left|\int_{\S}fe^\Phi\right|+\left|\int_{\S}\fb e^\Phi\right|.
\eeaa
Moreover, The area radius  $r^\S$  and Hawking mass $m^\S$of $\,\S$ verify \eqref{eq:comparisionareadraiusandHwakingmassonnewGCMsphere:00}.
\end{corollary}
 
\begin{corollary}[Rigidity II]
\lab{corr:GCM-rigidity2:00}
Assume given a sphere $\S\subset \RR$ endowed with a compatible  frame $e_3^\S, e_\th^\S, e_4^\S $   which verifies  the GCM conditions
\bea
\ka^\S=\frac{2}{r^\S}, \quad \dds_2\,^\S\dds_1\,^\S\kab^\S=\dds_2\,^\S\dds_1\,^\S\mu^\S=0.
\eea
In addition, we assume
\begin{enumerate}
\item  We have,
\bea
 r \left|\int_{\S} \b^\S e^\Phi\right|+\left|\int_{\S}  e^\S_\th (\kab^\S) e^\Phi\right| &\les& \dg.   
\eea

\item The background foliation  verifies  on $\RR$  the  conditions {\bf A1-A3}. 
\end{enumerate}
 Then the transition functions $(f, \fb, \log\la) $  from the background frame of $\RR$ to that of $\S$ 
 verifies the estimates
\beaa
\|(f, \fb, \log(\la))\|_{\hk_{s_{max}+1}(\S)} &\les& \dg.
\eeaa
Moreover, the area radius  $r^\S$  and Hawking mass $m^\S$of $\,\S$ verify \eqref{eq:comparisionareadraiusandHwakingmassonnewGCMsphere:00}.
\end{corollary}

 \begin{corollary}[ Rigidity III]
Assume given a GCM hypersurface $\Si_0\subset \RR$  foliated by surfaces $\S$ such that 
\beaa
\ka^\S=\frac{2}{r^\S}, \quad \dds_2\,^\S\dds_1\,^\S\kab^\S=\dds_2\,^\S\dds_1\,^\S\mu^\S=0, \quad \int_{\S}\eta^\S e^\Phi=0, \quad \int_{\S}\xib^\S e^\Phi=0.
\eeaa
Assume in addition that   for a  specific  sphere  $\S_0$  on $\Sigma_0$ we have the following additional  information:
\begin{enumerate}
\item  The transition functions $f, \fb $ from the background foliation  to $\S_0$  verify
\bea
\int_{\S}f e^\Phi=O(\dg), \quad \int_{\S}\fb e^\Phi=O(\dg).
\eea
\item  In $\RR$,  the Ricci  and curvature  coefficients  of the background foliation 
 verify  the assumptions  {\bf A1--A3}.
\end{enumerate}

 Then,  for all derivatives of  the transition functions along $\S$,
\beaa
 \|\dk^{\le s_{max}+1} (f, \fb, \log \la) \|_{L^2(\S)}\les \dg. 
\eeaa
Moreover, The area radius  $r^\S$  and Hawking mass $m^\S$of $\,\S$ verify \eqref{eq:comparisionareadraiusandHwakingmassonnewGCMsphere:00}.
\end{corollary}

 
 \subsection{Main ideas }
 
 
   Both theorems   GCMS-II and GCMH  are based on  Theorem GSMS-I.  They are   heavily based on 
  the transformation formulas     for the Ricci and curvature  coefficients     
     recorded in Proposition \ref{prop:transformations1}.

     
     \subsubsection{Sketch of the proof of  Theorems   GSMS-I and  GSMS-II}   
     
      
           A given deformation  $\Psi :\ovS\longrightarrow \S$  is fixed by the  parameters
      $U,  S$ and transition functions $F=(f, \fb, \la)$ connected by the system \eqref{eq:DeformS-66-GCMdescription:00}.       Making use  of the transformation formulas  one can show that
      the  GCMS conditions \eqref{Results:ThmGCMI-1Descr:00}-\eqref{Results:ThmGCMI-2Descr:00} are  if and only  if  the transition functions $ F $   verify a coercive  nonlinear   elliptic Hodge  system  of the form  $\DD_{\Psi} F=B(\Psi)$,  where  the operator $\DD_\Psi$ depends on the deformation $\Psi $  and  the right hand side $B$, depends on on both $\Psi$  and  the   background foliation     (see Proposition \ref{proposition-GCMS-system} for the precise form of the system).  To find    a desired GSMS deformation we   have to solve  a   coupled system  between the   transport type equations in  \eqref{eq:DeformS-66-GCMdescription:00}  and the   elliptic  coercive system   $\DD_{\Psi} F=0$  of Proposition   \ref{proposition-GCMS-system}. 
      
      The actual proof is thus  based on an iteration procedure for a sequence of deformation spheres $\S(n)$  of $\ovS$ 
       given by the maps $\Psi^{(n)}=(U^{(n)}, S^{(n)})  :\ovS\longrightarrow \S(n)$
  and the corresponding transition functions  $ \log \la ^{(n)},  f^{(n)},  \fb^{(n)}$.
 The iteration procedure for  the quintets $  Q^{(n)}= (U^{(n)}, S^{(n)}, \log \la ^{(n)},  f^{(n)},  \fb^{(n)})$, starting with the trivial quintet   $Q^{(0)}$  corresponding to the   zero  deformation,  is described  in section
 \ref{section:iterationprocedureGCM}. The main steps   in the proof are as follows. 
  \begin{enumerate}
  \item   Given  the triplet  $ \log\la ^{(n)}, f^{(n)}, \fb^{(n)}) $      the pair  $(U^{(n)}, S^{(n)})$ defines the deformation  sphere $\S(n)$ and the corresponding pull back map  $\#_n:\ovS\longrightarrow  \S(n)$ according to the equation \eqref{eq:DeformS-66-GCMdescription:00}.
  
  \item
 Given   the pair  $\Psi ^{(n)}=(U^{(n)}, S^{(n)})$ and the deformation sphere $\S(n)$ 
   we   define the triplet    $(\log\la  ^{(n+1)}, f^{(n+1)},  \fb ^{(n+1)}) $  by solving the corresponding  elliptic system  
   \beaa
   \DD_{\Psi(n)} F^{(n+1)} = B(\Psi^{(n)})
   \eeaa
   This step is based on the crucial  apriori estimates of section \ref{subsection:apriori-GCM}.

\item  Given the  new pair     $( f^{(n+1)}, \fb^{(n+1)})$   we make use  of   the equations  \eqref{eq:DeformS-66-GCMdescription:00}
  to find a unique new map $( U^{(n+1)}, S^{(n+1)})$ and thus the new deformation sphere $\S(n+1)$.

  \item The convergence  of the iterates  $Q^{(n)} $,  described in subsection  \ref{section:convergenceofiterates-GCM} in  the boundedness Proposition \ref{Prop:BondsforQn}   and   the contraction   Proposition \ref{Proposition-contraction}.     The latter   requires us to  carefully  compare  the iterates  $Q^{(n)} $,  $Q^{(n+1)} $ by pulling them back to $\ovS$.  One has to be particularly careful with the behavior of the iterates on the axis of symmetry. 
  \end{enumerate}

Theorem    GSMS-II, which is  an easy consequence of Theorem    GSMS-I is proved in  section \ref{section:Corollary-ExistenceGCMS} and the transformation formulas which  relate  $ \int_{\S}  \b^\S e^\Phi$ to  $ \La=\int_{\S}   f  e^\Phi$
 and $\int_{\S}  e^\S_\th(\kabS) e^\Phi$ to $\Lab =\int_{\S}   \fb e^\Phi$.  One can show that there exist choices of $\La, \Lab $
  such that $ \int_{\S}  \b^\S e^\Phi= \int_{\S}  e^\S_\th(\kabS) e^\Phi=0.$

  
  \subsubsection{Sketch of the proof of  Theorem  GCMH}  
  
  
     The proof of Theorem  GCMH makes use  of  Theorem  GCMS-I to construct  $\Si_0$ as a  union of GCMS     spheres. 
  
  {\bf Step 1.}  Theorem  GCMS-I  allows to construct,
       for every  value of the parameters $(u, s)$  in
 $\RR$ (i.e. such that the background  spheres  $S(u,s)\subset \RR$) and every  real numbers  $(\La, \Lab) $,  a unique  GCM sphere   $\S[u, s, \La, \Lab]$, as a   $\Z$-polarized deformation of $S(u,s)$.   In particular  \eqref{Results:ThmGCMI-1Descr:00} and \eqref{Results:ThmGCMI-2Descr:00}  are verified and  $\S_0=\S_0[\ovu, \ovs, \La_0, \Lab_0]$.
  
  {\bf Step 2.}  We  look for  functions $\Psi(s), \La(s), \Lab(s)$  such that  
  \begin{enumerate}
  \item  We have,
  \beaa
  \Psi(\ovs)=\ovu, \qquad   \La(\ovs)=\La_0, \qquad  \Lab(\ovs)=\Lab_0.
  \eeaa
  \item  The  resulting hypersurface $\Si_0=\cup_{s} \S[\Psi(s), s, \La(s), \Lab(s)] $  verifies 
  \beaa
  u^\S+r^\S =c_{\Si_0} , \qquad \textit{along} \quad \Si_0.
  \eeaa
  \item   The additional GCM conditions 
    \eqref{eq:ThmGCMH-Spole-a^S:00} and \eqref {eq:ThmGCMH-badmodesetaxib:00} of Theorem GCMH are verified.
  \end{enumerate}
   These conditions  lead to a first  order  differential  system for $\Psi(s), \La(s), \Lab(s) $, with prescribed  initial conditions at $\ovs$  which allows us to   determine the desired surface. 
   The proof is given in detail in section \ref{section:GCM-hypersurface}.

   
  \subsubsection{Sketch of the proof of  the Rigidity  Corollaries}  
  
  
  The Rigidity I  corollary,  see Corollary \ref{corr:GCM-rigidity2},   is proved   in exactly the same way as Proposition \ref{prop.GCMSequations-fixedS} which is one of the main steps  in the proof of  Theorem  GCMS-I.  The rigidity II   corollary, see Corollary \ref{corr:GCM-rigidity2},  is based on    Rigidity  I and a simple variation of  Lemma \ref{Lemma:ExistenceGCMS2}.  The  proof of the  rigidity  III corollary, see Corollary \ref{corr:GCM-rigidity3}, is essentially part of the proof of Theorem  GCMH, once the  existence part of the theorem has been established.


\section{Overview of the proof of Theorem M0-M8}


In this section, we provide a brief overview of the proof of Theorem M0-M8. In addition to the null frame adapted to the outgoing foliation of $\Mext$ and to the null frame adapted to the ingoing foliation of $\Mint$, we have also introduced 2 global frames on $\MM=\Mint\cup\Mext$ as well as associated scalars $r$ and $m$ in section \ref{section-globaleframe}. Unless otherwise specified, when we discuss a particular spacetime region, i.e. $\Mext$, $\Mint$ or $\MM$, it should be understood that the frame as well as $r$ and $m$ are the ones corresponding to that region.


\subsection{Discussion of Theorem M0}
\label{subsection:disc-ThmM0}


{\bf Step 1.} Recall our GCM conditions on $S_*=\Si_*\cap\CC_*$
\beaa
\int_{S_*}e_\th(\kab)e^\Phi =0, \qquad \int_{S_*}\b e^\Phi =0.
\eeaa
Recall that $\nu=e_3+a_*e_4$ is the unique tangent vectorfield to $\Si_*$ which is orthogonal to $e_\th$ and normalized by $\g(\nu, e_4)=-2$. Using the null structure equation for $e_3(\kab)$ and $e_3(\b)$, as well as  $e_4(\kab)$ and $e_4(\b)$, we obtain transport equations along $\Si_*$ in the $\nu$ direction for 
\beaa
\int_{S}e_\th(\kab)e^\Phi \textrm{ and } \int_{S}\b e^\Phi =0.
\eeaa
Integrating these transport equations in $\nu$, we propagate the control on $S_*$ to $\Si_*$. In particular, we obtain the following estimates on $S_1=\Si_*\cap\CC_1$,
\bea\lab{eq:usefulesitateinthediscussionofThmM0}
\left|\int_{S_1}e_\th(\kab)e^\Phi\right|+\left|\int_{S_1}\b e^\Phi\right| &\les& \ep^2+\frac{\ep}{r}\les\ep_0,
\eea
where we used in the last inequality the dominance condition of $r$ on $\Si_*$, see \eqref{eq:behaviorofronSigmastar}.  

{\bf Step 2.} We consider the transition functions $(f, \fb, \la)$ from the frame of the initial data layer to the frame of $\Mext$. Since
\begin{itemize}
\item $S_1$ is a sphere of $\Mext$ in the initial data layer,

\item $S_1$ is a sphere of the GCM hypersurface $\Si_*$,

\item the estimate \eqref{eq:usefulesitateinthediscussionofThmM0} holds on $S_1$,
\end{itemize}
we can invoke the GCM Corollary Rigidity II and III of section \ref{sec:statementofGCMresultsinChapter3} which yields, together with a Sobolev embedding on $S_1$
\bea\lab{eq:usefulesitateinthediscussionofThmM0:bis}
 \sup_{S_1}\Big(r|\dk^{\le k_{large}+1} (f, \fb, \log \la) |+|m-m_0|\Big)\les \ep_0. 
\eea

{\bf Step 3.} Relying on the transport equations in $e_4$ for $(f, \fb, \la)$, see Corollary \ref{cor:transportequationsforffbandlambda},  and Proposition \ref{prop:derivativesHawkingmass} for $m$, we propagate \eqref{eq:usefulesitateinthediscussionofThmM0:bis} to $\CC_1$, and then, proceeding similarly, in the $e_3$ direction to $\CCb_1$ which yields
\beaa
 \sup_{\CC_1\cup\CCb_1}\Big(r|\dk^{\le k_{large}+1} (f, \fb, \log \la) |+|m-m_0|\Big) \les \ep_0. 
\eeaa
Together with the control of the  initial data layer foliation and the transformation formulas of Proposition \ref{prop:transformations1}, we then obtain the desired estimates on $\CCb_1\cup\CCb_1$ for the curvature components.


\subsection{Discussion of Theorem M1}
\label{subsection:disc-ThmM1}


Here are the main steps in the proof  of Theorem M1.

{\bf Step 1.} Consider the global frame on $\MM$ constructed in Proposition \ref{prop:existenceandestimatesfortheglobalframe:bis} and the definition of    $\qf$ on $\MM$ with respect to that frame, see section  \ref{sec:discussionanddefintionofinvariantquantities} for the definition of $\qf$ with respect to any null frame. According to Theorem \ref{thm:wave-qf} we have,
\bea
\label{eq:equationforqf-description}
\square_2 \qf + V \qf = N,\quad V=\ka\kab
\eea
where the  nonlinear term  $N=\err[\square_\g \qf]$  is   a long  expression  of  terms  quadratic, or higher  order, in $\Gac, \Rc$ involving  various  powers of $ r$.  
Making use of the symbolic notation introduced in  definition \ref{definition-errortermsforsquareqf} we have, see \eqref{thmwaveqf:schematicformerrorterm}, 
\beaa
\err[\square_2\qf]&=& r^2 \dk^{\le 2}(\Ga_g \c (\a, \b) )         + e_3 \Big( r^3 \dk^{\le 2}(\Ga_g \c (\a, \b) )   \Big) +  \dk^{\le 1 } (\Ga_g \c \qf)+\lot
\eeaa
where the terms   denoted  by $\lot $  are higher order in $(\Gac, \Rc)$.

\begin{remark}\lab{rmk:whyweneedaglobalframewithbettereta:bisnew}
Recall from Remark \ref{rmk:whyweneedaglobalframewithbettereta} that the above good structure of the error term $\err[\square_2\qf]$ only holds in a frame for which 
 $\xi=0$ for $r\geq 4m_0$ and  $\eta\in \Ga_g$. This is why, in Theorem M1, $\qf$ is defined relative to the global  frame of Proposition \ref{prop:existenceandestimatesfortheglobalframe:bis}, see also Remark \ref{def:wherewementionthatwealwaysexpressqfinthesecondglobalframe}.
\end{remark}

{\bf Step 2.}  We   follow the Dafermos-Rodnianski version of the vector-field method  to derive desired decay estimates.  We recall that, in the context of a wave equation of the  form $\square_{ (Sch) } \psi=0$ on Schwarzschild 
spacetime,    their strategy  consists  in the  following:
\begin{itemize}
\item Start by deriving  Morawetz-energy  type  estimates  for   $ \psi$ with nondegenerate flux energies    and     the usual degeneracy  of bulk integrals at $r= 3m$. 
\item Derive $r^p$ weighted  estimates for $0<p<2$  and use them, in conjunction to the Morawetz estimates, to derive 
decay estimates.   
\item The decay estimates obtained by using the standard $r^p$ weighted approach are too weak to be useful in our nonlinear approach.  We  improve them by making use of a recent   variation of the Dafermos-Rodnianski approach due to 
Angelopoulos, Aretakis and Gajic \cite{AnArGa} which is based on first  commuting   the   wave equation is $\square_{(Sch)} \psi=0$
  with  $r^2(e_4+ r^{-1} )$ and then  repeating the process  described  for the  resulting new equation.  This procedure allows   to derive   the improved decay estimates  consistent  with our decay norms.
  \item  Derive estimates for higher derivatives by  commuting with  $\T$, $r\dkb$, the red-shift 
vectorfield, and $re_4$.
\end{itemize}

{\bf Step 3.}   The estimates mentioned in step 2  have  to be adapted to  the case  of  our   equation \eqref{eq:equationforqf-description}.   There are three main differences to take into account
  \begin{itemize}
  \item The application of the vectorfield method  to   our context produces   various nontrivial  commutator  terms 
  which   have to  be absorbed.   This is  taken care by our bootstrap assumption for $\Gac, \Rc$, as well as, in some cases, by integration by parts.
   
  \item The presence of the potential $V$  is mostly  advantageous  but various  modifications have to be nevertheless 
   made, especially  near the trapping region\footnote{At the linear level, on a Schwarzschild   spacetime, this step was also treated (minus the improved decay) in the paper \cite{D-H-R}.}.
 
   \item  The presence of the nonlinear term $N$ is  the most important   complication. The  precise  null structure of $N$
   is essential and various  integrations by parts  are  needed. 
   
 \item The quadratic terms involving $\eta$  in $N$ can only be treated provided the definition of $\qf$ is done with respect to the  global frame on $\MM$ constructed in Proposition \ref{prop:existenceandestimatesfortheglobalframe:bis}, for with $\eta$ behaves better in powers of $r^{-1}$. 
  \end{itemize}


\subsection{Discussion of Theorem M2} 


Recall from section \ref{sec:discussionanddefintionofinvariantquantities} that $\qf$ is defined with respect to a general null frame as follows 
\beaa
\qf= r^4\left(e_3(e_3(\a))+(2\kab -6\omb)e_3(\a)+\left(-4e_3(\omb)+8\omb^2-8\omb\,\kab+\frac{1}{2}\kab^2\right)\a\right)
\eeaa
which yields the following transport equation for $\a$
\beaa
e_3(e_3(\a))+(2\kab -6\omb)e_3(\a)+\left(-4e_3(\omb)+8\omb^2-8\omb\,\kab+\frac{1}{2}\kab^2\right)\a &=& \frac{\qf}{r^4}.
\eeaa
Recall also that $\qf$, controlled in Theorem M1, is defined w.r.t. the global frame of Proposition \ref{prop:existenceandestimatesfortheglobalframe:bis} whose normalization is such that, in particular, $\omb$ is a small quantity. Also, since we have 
$$e_3(r)=\frac{r}{2}\kab+\lot$$
we infer
\beaa
e_3(e_3(r^2\a)) &=& \frac{\qf}{r^2}+\lot
\eeaa
Integrating twice this transport equation from $\CC_1$ where we control the initial data - and in particular $\a$ - in view of Theorem M0, and using the decay for $\qf$ provided by Theorem M1, we deduce\footnote{Recall that $\dee$ has been introduced in Theorem M1 and satisfies $\dee>\dec$.} 
\beaa
\sup_{\Mext}\left(\frac{r^2(2r+u)^{1+\dee}}{\log(1+u)}+r^3(2r+u)^{\frac{1}{2}+\dee}\right)|\dk^{\leq k_{small}+20}\a| &\les&\ep_0,\\
\sup_{\Mext}\left(\frac{r^3(2r+u)^{1+\dee}}{\log(1+u)}+r^4(2r+u)^{\frac{1}{2}+\dee}\right)|\dk^{\leq k_{small}+19}e_3(\a)| &\les& \ep_0.
\eeaa

Now that we control $\a$ in the global frame of Proposition \ref{prop:existenceandestimatesfortheglobalframe:bis}, we need to go back to the frame of $\Mext$. By invoking the relationships between our various frame of $\Mext$, see Proposition \ref{prop:existenceandestimatesfortheglobalframe:bis} and Proposition \ref{prop:constructionsecondframeinMext}, and the transformation formula for $\a$, we infer 
\beaa
\,{}^{(ext)}\Dk_{k_{small}+20 }\left[\,{}^{(ext)}\a\right]    & \les & \ep_0
\eeaa
and hence the conclusion of Theorem M2.


\subsection{Discussion of Theorem M3} 


Here are the main steps in the proof  of Theorem M3.

{\bf Step 1.} To derive decay estimates for $\aa$ in $\MM$, we first recall the following Teukolsky-Starobinski identity, see \eqref{eq:Teuk-Star-main(seeApp)},
\beaa
e_3(r^2e_3(r\qf))+2\omb r^2e_3(r\qf) &=& r^7\left\{  \dds_2\dds_1\ddd_1\ddd_2\aa  +\frac{3}{2}\kab\rho e_4\aa -\frac{3}{2}\ka\rho e_3(\aa) \right\} +\lot
 \eeaa
where $\lot$ denotes terms which are quadratic of higher, and where all quantities are defined w.r.t. the global frame of Proposition \ref{prop:existenceandestimatesfortheglobalframe:bis}. Then, introducing the vectorfield
\beaa
\widetilde{T}= e_4-\frac{1}{\overline{\kab}}\Big(\overline{\ka}+\overline{\kab}\Oc-\overline{\check{\kab}\Oc}\Big)e_3,
\eeaa
we rewrite the identity  as
\bea\lab{eq:forwardparabolicequationforaainschematicformforhtediscussionofTheoremM3}
 6m\widetilde{T}\aa+r^4\dds_2\dds_1\ddd_1\ddd_2\aa  &=& \frac{1}{r^3}\Big(e_3(r^2e_3(r\qf))+2\omb r^2e_3(r\qf)\Big) +\lot
\eea
As it turns out, see Remark \ref{remark:basicpropertiesofvectorfieldwidetildeT}, this is a forward parabolic equation on each hyper surface of contant $r$ in $\Mint$.

{\bf Step 2.} Thanks to
\begin{itemize}
\item the control in $\Mint$  of the RHS of \eqref{eq:forwardparabolicequationforaainschematicformforhtediscussionofTheoremM3} which follows from the decay estimates of Theorem M1 for $\qf$, as well as the bootstrap assumptions for the quadratic and higher order terms,

\item the control of $\aa$ on $\CCb_1$ - i.e. of the initial data of \eqref{eq:forwardparabolicequationforaainschematicformforhtediscussionofTheoremM3} - provided by Theorem M0,

\item parabolic estimates for the forward parabolic equation \eqref{eq:forwardparabolicequationforaainschematicformforhtediscussionofTheoremM3},
\end{itemize}
we obtain the desired  decay estimates for $\aa$ in $\Mint$. 

{\bf Step 3.} It remains to control $\aa$ on $\Sigma_*$. Recall that $\nu$ denotes the unique tangent vectorfield to $\Sigma_*$ which can be written as $\nu=e_3+ae_4$. The Teukolsky-Starobinski identity of Step 1 can then be written as
\bea
 6m\nu\aa+r^4\dds_2\dds_1\ddd_1\ddd_2\aa  &=& \frac{1}{r^3}\Big(e_3(r^2e_3(r\qf))+2\omb r^2e_3(r\qf)\Big) +\lot
\eea
where $\lot$ denotes terms which are quadratic of higher, as well as terms which are linear but display additional decay in $r$. This is a forward parabolic equation along $\Si_*$. To obtain the desired decay for $\aa$ along $\Si_*$, one then proceeds as in Step 2, using in addition, for the linear term with extra decay in $r$, the behavior \eqref{eq:behaviorofronSigmastar} of $r$ on $\Si_*$.


\subsection{Discussion of Theorem M4}


Here are the main steps in the proof  of Theorem M4.

{\bf Step 1.} We derive decay estimates for the spacelike GCM hypersurface $\Sigma_*$. More precisely, thanks to 
\begin{itemize}
\item  the GCM conditions on $ \Sigma_*$
\beaa
\ka=\frac{2}{r},\,\,\, \,\dds_2\dds_1\kab=0,\,\,\,  \,\dds_2\dds_1\mu=0,\,\,\, \int_S\eta e^\Phi=0,\,\,\, \int_S\xib e^\Phi=0,
\eeaa

\item the control of $\qf$ in $\Mext$, established in Theorem M1, and hence in particular on $\Sigma_*$,

\item the control of $\a$ of the outgoing geodesic foliation in  $\Mext$, established in Theorem M2, and hence in particular on $\Sigma_*$,

\item the control of $\aa$ on  $\Sigma_*$, established in Theorem M3,

\item the fact that the following condition holds on $\Sigma_*$
\beaa
r\big|_{\Sigma_*}\geq \ep_0^{-\frac{2}{3}}u^4,
\eeaa

\item the identity \eqref{eq:alternateformulaforqfinvolvingtwoangularderrivativesofrho} relating $\qf$ to derivatives of $\rho$, i.e.
\beaa
\qf &=& r^4\left(\dds_2\dds_1\rho+\frac{3}{4}\rho\kab\vth+\frac{3}{4}\rho\ka\vthb+\cdots\right),
\eeaa

\item elliptic estimates for Hodge operators on the 2-spheres foliating $\Sigma_*$,
\end{itemize}
we infer the control with improved decay of all Ricci and curvature components on the spacelike hypersurface $\Sigma_*$.

{\bf Step 2.} We derive decay estimates for the outgoing geodesic foliation of $\Mext$. More precisely: 
\begin{itemize}
\item First, we propagate the estimates involving only $u^{-\frac{1}{2}-\dec}$ decay in $u$ from $\Si_*$ to $\Mext$. 

\item We then focus on the harder to recover estimates, i.e. the ones involving $u^{-1-\dec}$ decay in $u$. We proceed as follows.
\begin{itemize}
\item We first propagate  the main GCM quantities $\check{\ka}$, $\check{\mu}$, and a renormalized quantity involving $\check{\kab}$ (see the quantity $\Xi$ in Lemma \ref{Lemma:EstimateXi})  from $\Sigma_*$ to $\Mext$. 
\item We then recover the estimates involving $u^{-1-\dec}$ decay in $u$ on $\TT$. To this end, we use that we control the main GCM quantities, $\aa$ from Theorem M3 (since $\TT$ belongs both to $\Mext$ and $\Mint$), $\qf$ and $\a$ from Theorem M1--M2, and the estimates are then derived somewhat in the spirit of the ones on $\Sigma_*$, in particular by relying on elliptic estimates for Hodge operators on the 2-spheres foliating $\TT$. 

\item To recover the remaining estimates in $\Mext$ involving $u^{-1-\dec}$ decay in $u$, we integrate the transport equations in $e_4$ forward from $\TT$, which concludes the proof of Theorem M4.
\end{itemize}
\end{itemize}


\subsection{Discussion of Theorem M5}


Here are the main steps in the proof  of Theorem M5.

{\bf Step 1.} We first derive decay estimates for the ingoing geodesic foliation of $\Mint$ on the timelike hyper surface $\TT$. More precisely, thanks to
\begin{itemize}
\item the fact that the null frame of $\Mint$ is defined on $\TT$ as a simple conformal renormalization of the null frame of $\Mext$ in view of its initialization, see section \ref{sec:defintioncanonicalspacetime},

\item the control of the outgoing geodesic foliation of $\Mext$ on $\TT$ obtained in Theorem M4,
\end{itemize}
this allows us to transfer the decay estimates for $({}^{(ext)}\Rc, {}^{(ext)}\Gac)$ to $({}^{(int)}\Rc, {}^{(int)}\Gac)$ on $\TT$.
  
{\bf Step 2.} We derive on $\Mint$ decay estimates for the ingoing geodesic foliation of $\Mint$. More precisely, thanks to
\begin{itemize}
\item the improve decay estimates for $\aa$ in  $\Mint$ derived in Theorem M3,

\item the improved decay estimates for $\Gac$ and $\Rc$ on $\TT$ derived in the Step 1,

\item the null structure equations and Bianchi identities,
\end{itemize}
we infer  $O(\ep_0 u^{-1-\dec})$  decay estimates for $\Gac$ and $\Rc$ corresponding to the ingoing geodesic foliation of $\Mint$ which concludes the proof of Theorem M5.


\subsection{Discussion of Theorem M6}


{\bf Step 1.} Using
\begin{itemize}
\item[(a)] The control of the initial data layer,

\item[(b)] Theorem GCMS-II  of section \ref{sec:statementofGCMresultsinChapter3},

\item[(c)] Theorem GCMH  of section \ref{sec:statementofGCMresultsinChapter3}, 
\end{itemize}
we produce a smooth hypersurface $\Si_*$ in the initial data layer starting from a GCM sphere $S_*$, and satisfying all the required properties for the future spacelike boundary of a GCM admissible spacetime, according to item 3 of definition \ref{definition:canonical-spacetime}. 

{\bf Step 2.} We then consider the outgoing geodesic foliation  initialized on $\Si_*$ which foliates the region we denote $\Mext$, to the past of $\Si_*$, and included in the outgoing part $\Lext$ of the initial data layer. In order to control it, we consider the transition functions $(f,\fb, \la)$ from the background frame of the initial data  layer to the frame of $\Mext$. These functions satisfy transport equations in $e_4$ with right-hand side depending on  $(f,\fb, \la)$ and the Ricci coefficients of the background foliation. Integrating the transport equations from $\Si_*$, where $(f,\fb, \la)$ are under control as a by product of the use of Theorem GCMH in Step 1, we obtain the control of  $(f,\fb, \la)$ in $\Mext$. Using the transformation formulas of Proposition \ref{prop:transformations1}, and using the control of the initial data layer, we then infer the desired control (i.e. with $\ep_0$ smallness constant and suitable $r$-weights) for the Ricci coefficients and curvature components of the foliation of $\Mext$. 

{\bf Step 3.}  $\Mext$ terminates on a timelike hypersurface $\TT$ of constant area radius\footnote{With respect to the foliation of $\Mext$.}. We then consider the ingoing geodesic foliation initialized on $\TT$ according to item 4 of definition \ref{definition:canonical-spacetime}, which foliates the region we denote $\Mint$, included in the ingoing part $\Lint$ of the initial data layer. Proceeding as in Step 2, relying on transport equations in $e_3$ instead of $e_4$, we then derive the desired control (i.e. with $\ep_0$ smallness constant) for the Ricci coefficients and curvature components of the foliation of $\Mint$, thus concluding the proof of Theorem M6.


\subsection{Discussion of Theorem M7}


From the assumptions of Theorem M7 we are given a GCM admissible spacetime $\MM=\MM(u_*) \in\aleph(u_*)$  verifying   the following improved  bounds,  for a universal constant $C>0$,  
\beaa
 \Nk^{(Dec)}_{k_{small}+5}(\MM)\le  C \ep_0 
 \eeaa
  provided by Theorems M1-M5. We then proceed as follows.
  
  {\bf Step 1.} We extend $\MM$ by a local existence argument, to a strictly   larger  spacetime  $\MM^{(extend)}$,  with a naturally  extended foliation and the following slightly increased bounds
\beaa
 \Nk^{(Dec)}_{k_{small}+5}(\MMextend) \le 2C  \ep_0.
\eeaa
but which may  not verify our admissibility criteria. 

{\bf Step 2.} Using
\begin{itemize}
\item[(a)] The control of the extended spacetime $\MMextend$,

\item[(b)] Theorem GCMS-II  of section \ref{sec:statementofGCMresultsinChapter3},

\item[(c)] Theorem GCMH  of section \ref{sec:statementofGCMresultsinChapter3}, 
\end{itemize}
we produce a small piece of smooth GCM hypersurface $\widetilde{\Si}_*$ in $\MMextend\setminus\MM$ starting from a GCM sphere $\widetilde{S}_*$. 

{\bf Step 3.} By a continuity argument based on a priori estimates, we extend $\widetilde{\Si}_*$ all the way to the initial data layer, while ensuring that it remains in $\MMextend\setminus\MM$ and satisfying all the required properties for the future spacelike boundary of a GCM admissible spacetime, according to item 3 of definition \ref{definition:canonical-spacetime}.

{\bf Step 4.} We then consider the outgoing geodesic foliation  initialized on $\widetilde{\Si}_*$ which foliates the region we denote $\,^{(ext)}\widetilde{\MM}$,  included in the outgoing part of $\MMextend$. In order to control it, we consider the transition functions $(f,\fb, \la)$ from the background frame of the initial data  layer to the frame of $\,^{(ext)}\widetilde{\MM}$. These functions satisfy transport equations in $e_4$ with right-hand side depending on  $(f,\fb, \la)$ and the Ricci coefficients of the background foliation. Integrating the transport equations from $\widetilde{\Si}_*$, where $(f,\fb, \la)$ are under control as a by product of the use of Theorem GCMH in Step 2, we obtain the control of  $(f,\fb, \la)$ in $\,^{(ext)}\widetilde{\MM}$. Using the transformation formulas of Proposition \ref{prop:transformations1}, and using the control of the initial data layer, we then derive the desired control (i.e. with $\ep_0$ smallness constant and suitable $u$ and $r$ weights) for the Ricci coefficients and curvature components of the foliation of $\,^{(ext)}\widetilde{\MM}$.

{\bf Step 5.}  $\,^{(ext)}\widetilde{\MM}$ terminates on a timelike hypersurface $\widetilde{\TT}$ of constant area radius\footnote{With respect to the foliation of $\,^{(ext)}\widetilde{\MM}$.}. We then consider the ingoing geodesic foliation initialized on $\widetilde{\TT}$ according to item 4 of definition \ref{definition:canonical-spacetime}, which foliates the region we denote $\,^{(int)}\widetilde{\MM}$, included in the ingoing part of $\MMextend$. Proceeding as in Step 4, relying on transport equations in $e_3$ instead of $e_4$, we then derive the desired control (i.e. with $\ep_0$ smallness constant and suitable $\ub$-weights) for the Ricci coefficients and curvature components of the foliation of $\,^{(int)}\widetilde{\MM}$, thus concluding the proof of Theorem M7.


\subsection{Discussion of Theorem M8}\lab{subsection:discTheoremM8}


So far, we have only improved our bootstrap assumptions on decay estimates. We now improve our bootstrap assumptions on energies and weighted energies for $\check{R}$ and $\check{\Ga}$ relying on an iterative procedure recovering derivatives one by one\footnote{See also \cite{Holz} for a related strategy to recover higher order derivatives from the control of lower order ones.}.

{\bf Step 0.} Let $I_{m_0, \deh}$ the interval of $\mathbb{R}$ defined by
\bea
I_{m_0, \deh} &:=& \left[2m_0\left(1+\frac{\deh}{2}\right), 2m_0\left(1+\frac{3\deh}{2}\right)\right].
\eea
Recall that $\TT=\{r=\rh\}$, where $\rh\in I_{m_0, \deh}$, and note, see also Remark  \ref{rmk:infactimproveddecayofThM7holdsforrTininterval}, that the results of Theorems M0--M7 hold for any $\rh\in I_{m_0, \deh}$.

It is at this stage that we need to make a specific choice of $\rh$ in the context of a Lebesgue point argument. More precisely, we choose $\rh$ such that we have
\bea\lab{eq:choiceofRTTismadebythisinfimum}
\int_{\{r=\rh\}}|\dk^{\leq k_{large}}\Rc|^2 &=& \inf_{r_0\in I_{m_0, \deh}}\int_{\{r=r_0\}}|\dk^{\leq k_{large}}\Rc|^2.
\eea
In view of this definition, and since $\TT=\{r=\rh\}$, we infer that
\bea\lab{eq:consequenceofthechoiceofrhwhichisuseful}
\int_{\TT}|\dk^{\leq k_{large}}\Rc|^2  &\lesssim & \int_{\Mext\Big(r\in I_{m_0, \deh}\Big)}|\dk^{\leq k_{large}}\Rc|^2. 
\eea

\begin{remark}\lab{rmk:whatcanweassumetostarttheproofofTheoemM8}
From now on, we may thus assume that the spacetime $\MM$ satisfies the conclusions of Theorem M0 and Theorem M7, as well as \eqref{eq:consequenceofthechoiceofrhwhichisuseful}, and our goal is to prove Theorem M8, i.e. to prove that $\Nk^{(En)}_{k_{large}} \les\ep_0$ holds.
\end{remark}

{\bf Step 1.} The $O(\ep_0)$ decay estimates derived in Theorem M7 imply in particular the following (non sharp) consequence
\beaa
\Nk^{(En)}_{k_{small}} &\les&\ep_0,
 \eeaa
 where we recall\footnote{See sections \ref{section:main-normsextregion} and \ref{section:main-normsintregion} for the definition of our norms measuring energies for  curvature components and Ricci coefficients.} 
\beaa
\Nk^{(En)}_k & =& \,{}^{(ext)}\Rk_k[\Rc]+\,{}^{(ext)}\Gk_k[\Gac]+ \,{}^{(int)}\Rk_k[\Rc]+\,{}^{(int)}\Gk_k[\Gac].
\eeaa
This allows us to initialize our iteration scheme in the next step.

{\bf Step 2.} Next, for $J$ such that $k_{small}\leq J\leq k_{large}-1$, consider the iteration assumption 
\bea\lab{eq:iterationassumptiondiscussionThM8}
\Nk^{(En)}_{J}  &\les& \ep_\BB[J],
\eea
where
\bea\lab{eq:defofepJforiteration}
\ep_\BB[J] &:=& \sum_{j=k_{small}-2}^J (\ep_0)^{\ell(j)}\,\BB^{1-\ell(j)}+\ep_0^{\ell(J)} \BB, \qquad     \ell(j):=2^{k_{small}-2-j},\\
\BB &:=& \left(\int_{\Mext\Big(r\in I_{m_0, \deh}\Big)}|\dk^{\leq k_{large}}\Rc|^2\right)^{\frac{1}{2}}.
 \eea
In view of Step 1, \eqref{eq:iterationassumptiondiscussionThM8} holds for $J=k_{small}$. From now on, we assume that \eqref{eq:iterationassumptiondiscussionThM8} holds for $J$ such that $k_{small}\leq J\leq k_{large}-2$, and our goal is to show that this also holds for $J+1$ derivatives. 

{\bf Step 3.} Using the Teukolsky wave equations for $\a$ and $\aa$, as well as a wave equation for $\rhoc$, see Proposition \ref{prop:waveeqfor-rt}, we derive Morawetz type estimates for $J+1$ derivatives of these quantities in terms of $O(\ep_\BB[J]+\ep_0\Nk^{(En)}_{J+1})$.

{\bf Step 4.} Relying on Bianchi identities,  we also derive Morawetz type estimates for $J+1$ derivatives for $\b$ and $\bb$. As a consequence,  we obtain Morawetz type estimates for $J+1$ derivatives of all curvature components in terms of $O(\ep_\BB[J]+\ep_0\Nk^{(En)}_{J+1})$.

{\bf Step 5.} As a consequence of Step 4, we immediately obtain, for any $r_0\geq 4m_0$,
\beaa
\,{}^{(int)}\mathfrak{R}_{J+1}[\Rc]+\,{}^{(ext)}\mathfrak{R}_{J+1}[\Rc] &\leq& \,{}^{(ext)}\mathfrak{R}^{\geq r_0}_{J+1}[\Rc]+O(r_0^{10}(\ep_\BB[J]+\ep_0\Nk^{(En)}_{J+1})).
\eeaa

{\bf Step 6.} Relying on the Bianchi identities, we derive $r^p$-weighted estimates for $J+1$ derivatives of curvature on $r\geq r_0$ with $r_0\geq 4m_0$. We obtain 
\beaa
 \,{}^{(ext)}\mathfrak{R}^{\geq r_0}_{J+1}[\Rc] &\les& \frac{1}{r_0^{\dt}}\,{}^{(ext)}\mathfrak{G}^{\geq r_0}_k[\Gac]+r_0^{10}(\ep_\BB[J]+\ep_0\Nk^{(En)}_{J+1}).
 \eeaa 
 
 {\bf Step 7.} Next, we estimate  the Ricci coefficients of $\Mext$. To control them, we rely on the null structure equations in $\Mext$. Using the null structure equations in $\Mext$ and the GCM conditions on $\Sigma_*$, we derive the following weighted estimates for $J+1$ derivatives of the Ricci coefficients 
 \beaa
 \,{}^{(ext)}\mathfrak{G}_{J+1}[\Gac] &\les& \,{}^{(ext)}\mathfrak{R}_{J+1}[\Rc]+\ep_\BB[J]+\ep_0\Nk^{(En)}_{J+1}.
 \eeaa
Together with the estimates of Step 5 and Step 6, we infer for a large enough choice of $r_0$ 
\beaa
 \,{}^{(ext)}\mathfrak{G}_{J+1}[\Gac] + \,{}^{(int)}\mathfrak{R}_{J+1}[\Rc]+ \,{}^{(ext)}\mathfrak{R}_{J+1}[\Rc] &\les&\ep_\BB[J]+\ep_0\Nk^{(En)}_{J+1}.
 \eeaa
 
 {\bf Step 8.} Next, we estimate the Ricci coefficients of $\Mint$. Using the information on $\TT$ induced by Step 7 and the null structure equations in $\Mint$, we derive 
 \beaa
 \,{}^{(int)}\mathfrak{G}_{J+1}[\Gac] &\les& \,{}^{(int)}\mathfrak{R}_{J+1}[\Rc]+\ep_\BB[J]+\ep_0\Nk^{(En)}_{J+1}+\left(\int_{\TT}|\dk^{J+1}({}^{(ext)}\Rc)|^2\right)^{\frac{1}{2}}.
 \eeaa
 
We need to deal with the last term. Relying on a trace theorem in the spacetime region $\Mext(r\in I_{m_0, \deh})$, and the fact that $J+2\leq k_{large}$, we obtain
\beaa
\left(\int_{\TT}|\dk^{J+1}({}^{(ext)}\Rc)|^2\right)^{\frac{1}{2}} &\les& \left(\int_{\Mext\Big(r\in I_{m_0, \deh}\Big)}|\dk^{k_{large}}\Rc|^2\right)^{\frac{1}{4}} (\,{}^{(ext)}\mathfrak{R}_{J+1}[\Rc])^{\frac{1}{2}}\\
&&+\,{}^{(ext)}\mathfrak{R}_{J+1}[\Rc].
\eeaa

 {\bf Step 9.} The last estimate of Step 7 and the 2 estimates of Step 8 yield, for $\ep_0>0$ small enough,
\beaa
\Nk^{(En)}_{J+1} &\les& \ep_\BB[J]+\left(\int_{\Mext\Big(r\in I_{m_0, \deh}\Big)}|\dk^{k_{large}}\Rc|^2\right)^{\frac{1}{4}} \Big(\ep_\BB[J]+\ep_0\Nk^{(En)}_{J+1}\Big)^{\frac{1}{2}}.
\eeaa
In view of the definition \eqref{eq:defofepJforiteration} of $\ep_\BB[J]$, we infer that
\beaa
\Nk^{(En)}_{J+1} &\les& \ep_\BB[J+1]
\eeaa
 which is the iteration assumption \eqref{eq:iterationassumptiondiscussionThM8} for $J+1$ derivatives. We deduce that \eqref{eq:iterationassumptiondiscussionThM8} holds for all $J\leq k_{large}-1$, and hence
\beaa
\Nk^{(En)}_{k_{large}-1} &\les& \ep_\BB[k_{large}-1].
\eeaa 
 
{\bf Step 10.} Relying on the conclusion of Step 9, and arguing as in Step 3 to Step 7, we obtain the conclusion of Step 7 for $J=k_{large}-1$, i.e.
\beaa
 \,{}^{(ext)}\mathfrak{G}_{k_{large}}[\Gac] + \,{}^{(int)}\mathfrak{R}_{k_{large}}[\Rc]+ \,{}^{(ext)}\mathfrak{R}_{k_{large}}[\Rc] &\les&\ep_\BB[k_{large}-1]+\ep_0\Nk^{(En)}_{k_{large}}.
 \eeaa
We then infer that
\beaa
\ep_\BB[k_{large}-1]\les \ep_0+\ep_0\Nk^{(En)}_{k_{large}}
\eeaa 
which yields, together with the last estimate of Step 9,
\beaa
 \,{}^{(ext)}\mathfrak{G}_{k_{large}}[\Gac] + \,{}^{(int)}\mathfrak{R}_{k_{large}}[\Rc]+ \,{}^{(ext)}\mathfrak{R}_{k_{large}}[\Rc] &\les& \ep_0+\ep_0\Nk^{(En)}_{k_{large}}.
 \eeaa

{\bf Step 11.} It remains to recover $ \,{}^{(int)}\mathfrak{G}_{k_{large}}[\Gac]$. Arguing as for the first estimate of Step 8 with $J=k_{large}-1$, we have
 \beaa
 \,{}^{(int)}\mathfrak{G}_{k_{large}}[\Gac] &\les& \,{}^{(int)}\mathfrak{R}_{k_{large}}[\Rc]+\ep_\BB[k_{large}-1]+\ep_0\Nk^{(En)}_{k_{large}}+\left(\int_{\TT}|\dk^{k_{large}}({}^{(ext)}\Rc)|^2\right)^{\frac{1}{2}}.
 \eeaa
 Thanks to the outcome of Step 10, we deduce that 
 \beaa
 \,{}^{(int)}\mathfrak{G}_{k_{large}}[\Gac] &\les& \ep_0+\ep_0\Nk^{(En)}_{k_{large}}+\left(\int_{\TT}|\dk^{k_{large}}({}^{(ext)}\Rc)|^2\right)^{\frac{1}{2}}
 \eeaa
 and hence, for $\ep_0>0$ small enough, using again the last estimate of Step 10, 
 \beaa
 \Nk^{(En)}_{k_{large}} &\les& \ep_0+\left(\int_{\TT}|\dk^{k_{large}}({}^{(ext)}\Rc)|^2\right)^{\frac{1}{2}}.
 \eeaa
 
 It remains to estimate the last term of the RHS of the previous inequality. It is at this stage that we use the choice of $\rh$, or rather its consequence \eqref{eq:consequenceofthechoiceofrhwhichisuseful}, which implies
  \beaa
\left(\int_{\TT}|\dk^{k_{large}}({}^{(ext)}\Rc)|^2\right)^{\frac{1}{2}}  &\les& \ep_0+\ep_0\Nk^{(En)}_{k_{large}}
\eeaa
so that we finally obtain, for $\ep_0>0$ small enough,
\beaa
 \Nk^{(En)}_{k_{large}} &\les& \ep_0
 \eeaa
which concludes the proof of Theorem M8.


\section{Structure of the rest of the paper}


The rest of this paper is devoted to the proof of Theorem M0-M8, as well as our GCM procedure. More precisely, 
\begin{enumerate}
\item Theorem M0, together with other first consequences of the bootstrap assumptions, is proved in Chapter \ref{chap:proofofconsequencesbootass}.

\item Theorem M1 is proved in Chapter \ref{chapter:Thm:mainwavetheorem}.

\item Theorems M2 and M3 are proved in Chapter \ref{chap:proofoftheoremM2M3}.

\item Theorems M4 and M5 are proved in Chapter \ref{chap:proofoftheoremM4M5}.

\item  Theorems M6, M7 and M8 are proved in Chapter \ref{chap:proofoftheoremM0M7M8}.

\item Our GCM procedure is described in details in Chapter \ref{chap:proofofGCMprocedure}.

\item Chapter \ref{chapter:waveeqtionestimates} contains estimates for Regge-Wheeler type wave equations used in Theorem M1.

\item Many of the long calculations are to be found in the appendix.
\end{enumerate}


\chapter{CONSEQUENCES OF THE BOOTSTRAP ASSUMPTIONS}\lab{chap:proofofconsequencesbootass}



\section{Proof of Theorem M0}


According to the statement of Theorem M0 we consider   given  the initial layer  $\LL_0=\Lext\cup\Lint$  as defined in Definition \ref{definition:initialdatalayer}.
We also assume that  the initial layer norm verifies 
\bea
\lab{def:initialdatalayerassumptions-stronger:bisbisbis}
\sup_{k\le k_{large}+5} \Ik_k \les \ep_0
\eea
where $\Ik_k = ^{(ext)} \Ik_k+\, ^{(int)} \Ik_k+\Ik_k'$ 
and,
\beaa
\, ^{(ext)} \Ik_0&=&  \sup_{\Lext}  \left[ r^{\frac{7}{2}  +\de_B}\left( |\a| + |\b|\right)+r^3\ \left|\rho+\frac{2m_0}{r^3}\right| +   r^2 |\bb|+r|\aa|    \right]\\
&+& \sup_{\Lext}  r^2\left(|\vth|+\left|\ka-\frac{2}{r}\right|+|\ze|+ \left|\kab+\frac{2\left(1-\frac{2m_0}{r}\right)}{r}\right|\right)\\
&+& \sup_{\Lext} r \left(|\vthb|+\left|\omb-\frac{m_0}{r^2}\right|+|\xib|\right)\\
&+&\sup_{\Lext\left(\rext_0\geq 4m_0\right)}\left(r\left|\frac{\ga}{r^2}-1\right|+r|\underline{b}|+|\Omb+\Up| +|\vsi-1|+r\left|\frac{e^\Phi}{r\sin\th}-1\right| \right),
\eeaa
\beaa
\, ^{(int)} \Ik_0&=&  \sup_{\Lint} \left(  |\aa|+|\bb| +\left|\rho+\frac{2m_0}{r^3}\right| +|\b|  +|\a|    \right)\\
&+& \sup_{\Lint}   \left(|\vth|+\left|\ka-\frac{2\left(1-\frac{2m_0}{r}\right)}{r}\right|+|\ze| + \left|\kab+\frac{2}{r}\right|+|\vthb|+\left|\om+\frac{m_0}{r^2}\right|+|\xi| \right),
\eeaa
\beaa
\Ik_0'&=&  \sup_{\Lint\cap\Lext} \left(  |f|+|\fb| +|\log(\la_0^{-1}\la)|   \right),\qquad \la_0= {}^{(ext)}\la_0=1-\frac{2m_0}{\rextl},
\eeaa
with $\Ik_k$ the corresponding  higher derivative norms obtained by replacing   each   component by $\dk^{\le k}$ of it. In the definition of $\Ik_0'$ above, $(f, \fb, \la)$ denote the transition functions of Lemma \ref{lemma:SSMe:general.composite}  from the frame of the outgoing part $\Lext$ of the initial data layer to the frame of the ingoing part $\Lint$ of the initial data layer in the region $\Lint\cap\Lext$. 

We divide the proof  of  Theorem  M0 in the following steps.

{\bf Step 1.} We have the following lemma.
\begin{lemma}\lab{lemma:identitiesfortransportequatione3e4badmodeofbetaandethkab}
We have on $\Mext$
\beaa
e_4\left(\int_Se_\th(\kab)e^\Phi\right) &=& \int_S\Bigg(\frac{1}{2}\ka e_\th(\kab)  -\frac 1 2\kab e_\th(\ka)  + 4K\ze + 2e_\th(\rho) -\frac 1 2  e_\th(\vth\, \vthb)\\
&& - \vth e_\th(\kab)+2e_\th(\ze^2)\Bigg)e^\Phi,\\
e_4\left(\int_S\b e^\Phi\right) &=& \int_S\left(-\frac{1}{2}\ka \b + \ze\a -\frac{1}{2}\vth\b \right)e^\Phi,\\
e_3\left(\int_S\b e^\Phi\right) &=& -\frac{1}{4}\int_Se_\th(\ka\kab)e^\Phi +3\ov{\rho}\int_S\eta e^\Phi + \frac{1}{4}\int_Se_\th(\vth\vthb)e^\Phi \\
&&+\int_S\Bigg(\frac{1}{2}\kab \b   +2\omb \b   + 3\eta \rhoc- \vth \bb +\xib \a -\frac{1}{2}\vthb\b \Bigg)e^\Phi+\err\left[e_3\left(\int_S \b e^\Phi\right)\right],
\eeaa
and
\beaa
e_3\left(\int_Se_\th(\kab)e^\Phi\right) &=& \kab e_3\left(\int_S\ze e^\Phi\right)  -\ov{\kab}\int_S\bb e^\Phi \\
&&+ \int_S\Bigg(  - \kabc\bb  -\frac{1}{2}\kab^2\ze+6\rho\xib -2\omb e_\th(\kab) -\frac{1}{2}\vthb(e_\th(\kab)-\kab \ze)+\err[\ddd_2\dds_2\xib]\Bigg)e^\Phi\\
  &&+\err\left[e_3\left(\int_S e_\th(\kab) e^\Phi\right)\right] +\int_S\kabc\left(e_3(\ze)+\left(\frac{3}{2}\kab-\frac{1}{2}\vthb\right)\ze\right)e^\Phi \\
  &&-\kabc\int_S\left(e_3(\ze)+\left(\frac{3}{2}\kab-\frac{1}{2}\vthb\right)\ze\right)e^\Phi -\kab\err\left[e_3\left(\int_S\ze e^\Phi\right)\right].
\eeaa
\end{lemma}

\begin{proof}
We have in $\Mext$, see Proposition \ref{propos:basiceqts-geod}, 
\beaa
e_4(\kab)+\frac 1 2 \ka\, \kab  &= -2\ddd_1\ze + 2\rho -\frac 1 2  \vth\, \vthb +2\ze^2.
\eeaa
Together with the following commutation relation
 \beaa
\,[e_\th, e_4]&=&\frac  1 2 (\ka +\vth)e_\th,
\eeaa
we infer
\beaa
e_4(e_\th(\kab))+ \ka e_\th(\kab) +\frac{1}{2} \vth e_\th(\kab) +\frac 1 2\kab e_\th(\ka)  &=& 2\dds_1\ddd_1\ze + 2e_\th(\rho) -\frac 1 2  e_\th(\vth\, \vthb) +2e_\th(\ze^2).
\eeaa
Also, we have in view of  Proposition \ref{prop:eqtsfor-ometaxib} the following identity
 \beaa
e_3(e_\th(\kab)) -  \kab e_3(\ze)  &=&- 2\ddd_2\dds_2\xib - \kab\bb +\kab^2\ze -\frac 3 2 \kab e_\th \kab+6\rho\xib -2\omb e_\th(\kab)+\err[\ddd_2\dds_2\xib].
\eeaa

Next, in view of Corollary \ref{corr:transportintSfePhi}, we have in $\Mext$
\beaa
e_4\left(\int_Se_\th(\kab)e^\Phi\right) &=& \int_S\left(e_4(e_\th(\kab))+\left(\frac{3}{2}\ka-\frac{1}{2}\vth\right)e_\th(\kab)\right)e^\Phi\\
e_4\left(\int_S\b e^\Phi\right) &=& \int_S\left(e_4(\b)+\left(\frac{3}{2}\ka-\frac{1}{2}\vth\right)\b \right)e^\Phi,\\
e_3\left(\int_S\b e^\Phi\right) &=& \int_S\left(e_3(\b)+\left(\frac{3}{2}\kab-\frac{1}{2}\vthb\right)\b \right)e^\Phi+\err\left[e_3\left(\int_S \b e^\Phi\right)\right],
\eeaa
and
\beaa
&& e_3\left(\int_Se_\th(\kab)e^\Phi\right) - \kab e_3\left(\int_S\ze e^\Phi\right)\\
 &=& \int_S\left(e_3(e_\th(\kab))+\left(\frac{3}{2}\kab-\frac{1}{2}\vthb\right)e_\th(\kab)\right)e^\Phi+\err\left[e_3\left(\int_S e_\th(\kab) e^\Phi\right)\right]\\
 && -\kab\int_S\left(e_3(\ze)+\left(\frac{3}{2}\kab-\frac{1}{2}\vthb\right)\ze \right)e^\Phi -\kab\err\left[e_3\left(\int_S\ze e^\Phi\right)\right]\\
  &=& \int_S\left(e_3(e_\th(\kab))-\kab e_3(\ze)+\left(\frac{3}{2}\kab-\frac{1}{2}\vthb\right)(e_\th(\kab)-\kab \ze)\right)e^\Phi+\err\left[e_3\left(\int_S e_\th(\kab) e^\Phi\right)\right]\\
 && +\int_S\kabc\left(e_3(\ze)+\left(\frac{3}{2}\kab-\frac{1}{2}\vthb\right)\ze\right)e^\Phi -\kabc\int_S\left(e_3(\ze)+\left(\frac{3}{2}\kab-\frac{1}{2}\vthb\right)\ze\right)e^\Phi -\kab\err\left[e_3\left(\int_S\ze e^\Phi\right)\right].
\eeaa
Together with the above identities for $e_4(e_\th(\kab))$ and $e_3(e_\th(\kab))$, as well as the Bianchi identities of Proposition \ref{propos:basiceqts-geod} for $e_4(\b)$ and $e_3(\b)$, we infer
\beaa
e_4\left(\int_Se_\th(\kab)e^\Phi\right) &=& \int_S\Bigg(\frac{1}{2}\ka e_\th(\kab)  -\frac 1 2\kab e_\th(\ka)  + 2\dds_1\ddd_1\ze + 2e_\th(\rho) -\frac 1 2  e_\th(\vth\, \vthb)\\
&& - \vth e_\th(\kab)+2e_\th(\ze^2)\Bigg)e^\Phi,\\
e_4\left(\int_S\b e^\Phi\right) &=& \int_S\left(-\frac{1}{2}\ka \b +\ddd_2\a+ \ze\a -\frac{1}{2}\vth\b \right)e^\Phi,\\
e_3\left(\int_S\b e^\Phi\right) &=& \int_S\Bigg(\frac{1}{2}\kab \b  +e_\th(\rho) +2\omb \b   + 3\eta \rho- \vth \bb +\xib \a -\frac{1}{2}\vthb\b \Bigg)e^\Phi\\
&&+\err\left[e_3\left(\int_S \b e^\Phi\right)\right],
\eeaa
and
\beaa
&& e_3\left(\int_Se_\th(\kab)e^\Phi\right) - \kab e_3\left(\int_S\ze e^\Phi\right)\\
  &=& \int_S\Bigg( - 2\ddd_2\dds_2\xib - \kab\bb  -\frac{1}{2}\kab^2\ze+6\rho\xib -2\omb e_\th(\kab) -\frac{1}{2}\vthb(e_\th(\kab)-\kab \ze)+\err[\ddd_2\dds_2\xib]\Bigg)e^\Phi\\
  &&+\err\left[e_3\left(\int_S e_\th(\kab) e^\Phi\right)\right] +\int_S\kabc\left(e_3(\ze)+\left(\frac{3}{2}\kab-\frac{1}{2}\vthb\right)\ze\right)e^\Phi \\
  &&-\kabc\int_S\left(e_3(\ze)+\left(\frac{3}{2}\kab-\frac{1}{2}\vthb\right)\ze\right)e^\Phi -\kab\err\left[e_3\left(\int_S\ze e^\Phi\right)\right].
\eeaa
Using in particular the fact that $\dds_2(e^\Phi)=0$, that $\dds_2$ is the adjoint of $\ddd_2$, and the identity $\dds_1\ddd_1=\ddd_2\dds_2+2K$, we deduce
\beaa
e_4\left(\int_Se_\th(\kab)e^\Phi\right) &=& \int_S\Bigg(\frac{1}{2}\ka e_\th(\kab)  -\frac 1 2\kab e_\th(\ka)  + 4K\ze + 2e_\th(\rho) -\frac 1 2  e_\th(\vth\, \vthb)\\
&& - \vth e_\th(\kab)+2e_\th(\ze^2)\Bigg)e^\Phi,\\
e_4\left(\int_S\b e^\Phi\right) &=& \int_S\left(-\frac{1}{2}\ka \b + \ze\a -\frac{1}{2}\vth\b \right)e^\Phi,\\
e_3\left(\int_S\b e^\Phi\right) &=& \int_Se_\th(\rho) e^\Phi +3\ov{\rho}\int_S\eta e^\Phi+\int_S\Bigg(\frac{1}{2}\kab \b   +2\omb \b   + 3\eta \rhoc- \vth \bb +\xib \a -\frac{1}{2}\vthb\b \Bigg)e^\Phi\\
&&+\err\left[e_3\left(\int_S \b e^\Phi\right)\right],
\eeaa
and
\beaa
e_3\left(\int_Se_\th(\kab)e^\Phi\right) &=& \kab e_3\left(\int_S\ze e^\Phi\right)  -\ov{\kab}\int_S\bb e^\Phi \\
&&+ \int_S\Bigg(  - \kabc\bb  -\frac{1}{2}\kab^2\ze+6\rho\xib -2\omb e_\th(\kab) -\frac{1}{2}\vthb(e_\th(\kab)-\kab \ze)+\err[\ddd_2\dds_2\xib]\Bigg)e^\Phi\\
  &&+\err\left[e_3\left(\int_S e_\th(\kab) e^\Phi\right)\right] +\int_S\kabc\left(e_3(\ze)+\left(\frac{3}{2}\kab-\frac{1}{2}\vthb\right)\ze\right)e^\Phi \\
  &&-\kabc\int_S\left(e_3(\ze)+\left(\frac{3}{2}\kab-\frac{1}{2}\vthb\right)\ze\right)e^\Phi -\kab\err\left[e_3\left(\int_S\ze e^\Phi\right)\right].
\eeaa

Finally, from the identity \eqref{eq:badmode-K} for $e_\th(K)$ and the formula for $K$, we have
\beaa
\int_S e_\th(\rho)e^\Phi=-\frac{1}{4}\int_Se_\th(\ka\kab)e^\Phi + \frac{1}{4}\int_Se_\th(\vth\vthb)e^\Phi.
\eeaa
We deduce
\beaa
e_3\left(\int_S\b e^\Phi\right) &=& -\frac{1}{4}\int_Se_\th(\ka\kab)e^\Phi +3\ov{\rho}\int_S\eta e^\Phi + \frac{1}{4}\int_Se_\th(\vth\vthb)e^\Phi \\
&&+\int_S\Bigg(\frac{1}{2}\kab \b   +2\omb \b   + 3\eta \rhoc- \vth \bb +\xib \a -\frac{1}{2}\vthb\b \Bigg)e^\Phi+\err\left[e_3\left(\int_S \b e^\Phi\right)\right]
\eeaa
which concludes the proof of Lemma \ref{lemma:identitiesfortransportequatione3e4badmodeofbetaandethkab}.
\end{proof}

{\bf Step 2.} Using  the transport equations of Lemma \ref{lemma:identitiesfortransportequatione3e4badmodeofbetaandethkab} and the bootstrap assumptions on decay for $k=0, 1$ derivatives in $\Mext$, we infer in that region, and in particular on $\Si_*$
\beaa
\left|e_4\left(\int_Se_\th(\kab)e^\Phi\right)\right| &\les& \frac{\ep}{ru^{\frac{1}{2}+\dec}},\\
\left|e_4\left(\int_S\b e^\Phi\right)\right| &\les& \frac{\ep}{r^{\frac{3}{2}+\dec}},\\
\left|e_3\left(\int_S\b e^\Phi\right)\right| &\les& \left|\int_Se_\th(\ka\kab)e^\Phi\right| +r^{-3}\left|\int_S\eta e^\Phi\right| +\frac{\ep}{r^{\frac{3}{2}+\dec}}+\frac{\ep^2}{ru^{1+\dec}},\\
\left|e_3\left(\int_Se_\th(\kab)e^\Phi\right)\right| &\les& r^{-1}\left|e_3\left(\int_S\ze e^\Phi\right)\right|  +r^{-1}\left|\int_S\bb e^\Phi\right| +\frac{\ep}{ru^{\frac{1}{2}+\dec}}+\frac{\ep^2}{u^{2+2\dec}}.
\eeaa

Next, we focus on the two last estimates. Recall the following GCM conditions 
\beaa
\ka=\frac{2}{r}, \qquad \int_S\eta e^\Phi=0\,\,\textrm{ on }\Si_*.
\eeaa
We deduce on $\Si_*$
\beaa
\left|e_3\left(\int_S\b e^\Phi\right)\right| &\les& r^{-1}\left|\int_Se_\th(\kab)e^\Phi\right| +\frac{\ep}{r^{\frac{3}{2}+\dec}}+\frac{\ep^2}{ru^{1+\dec}}.
\eeaa
Also, projecting both Codazzi on $e^\Phi$, using $\dds_2(e^\Phi)=0$ and the fact that $\dds_2$ is the adjoint of $\ddd_2$, and using also the GCM condition for $\ka$ on $\Si_*$, we have on $\Si_*$
\beaa
\int_S\bb e^\Phi &=&  -\frac{1}{2}\int_S\dds_1\,\kab e^\Phi -\frac{1}{2}\int_S\ze \kab e^\Phi + \frac{1}{2}\int_S\vthb \,  \ze e^\Phi,\\
\int_S\ze e^\Phi  &=& r\int_S\b e^\Phi + \frac{r}{2}\int_S\vth \, \ze e^\Phi.
\eeaa
Together with the bootstrap assumptions on decay for $k=0, 1$ derivatives in $\Mext$, we infer on $\Si_*$
\beaa
\left|e_3\left(\int_S\ze e^\Phi\right)\right|  +\left|\int_S\bb e^\Phi\right| &\les& \frac{\ep}{u^{\frac{1}{2}+\dec}}.
\eeaa
We have thus on $\Si_*$
\beaa
\left|e_4\left(\int_Se_\th(\kab)e^\Phi\right)\right| &\les& \frac{\ep}{ru^{\frac{1}{2}+\dec}},\\
\left|e_4\left(\int_S\b e^\Phi\right)\right| &\les& \frac{\ep}{r^{\frac{3}{2}+\dec}},\\
\left|e_3\left(\int_S\b e^\Phi\right)\right| &\les& r^{-1}\left|\int_Se_\th(\kab)e^\Phi\right| +\frac{\ep}{r^{\frac{3}{2}+\dec}}+\frac{\ep^2}{ru^{1+\dec}},\\
\left|e_3\left(\int_Se_\th(\kab)e^\Phi\right)\right| &\les& \frac{\ep}{ru^{\frac{1}{2}+\dec}}+\frac{\ep^2}{u^{2+2\dec}}.
\eeaa
In view of  the behavior \eqref{eq:behaviorofronSigmastar} of $r$ on $\Sigma_*$, we infer on $\Si_*$
\beaa
\left|e_4\left(\int_Se_\th(\kab)e^\Phi\right)\right| &\les& \frac{\ep_0}{u^{2+2\dec}},\\
\left|e_4\left(\int_S\b e^\Phi\right)\right| &\les& \frac{\ep_0}{ru^{1+\dec}},\\
\left|e_3\left(\int_S\b e^\Phi\right)\right| &\les& r^{-1}\left|\int_Se_\th(\kab)e^\Phi\right| +\frac{\ep_0}{ru^{1+\dec}},\\
\left|e_3\left(\int_Se_\th(\kab)e^\Phi\right)\right| &\les& \frac{\ep_0}{u^{2+2\dec}}.
\eeaa

{\bf Step 3.} Let $\nu_*$ the unique tangent vector to $\Si_*$ which can be written as
\beaa
\nu_*= e_3+ae_4
\eeaa
where $a$ is a scalar function on $\Si_*$. Recall that there exists a constant $c_*$ such that $\Si_*=\{u+r=c_*\}$. We infer $\nu_*(u+r)=0$ and hence
\beaa
0 = e_3(u+r)+a e_4(u+r)=\frac{2}{\vsi}+\frac{r}{2}(\ov{\kab}+\Ab)+a\frac{r}{2}\ov{\ka}
\eeaa
which yields
\beaa
a &=& -\frac{\frac{2}{\vsi}+\frac{r}{2}(\ov{\kab}+\Ab)}{\frac{r}{2}\ov{\ka}}.
\eeaa
In view of the GCM condition on $\ka$ and the definition of the Hawking mass $m$, we have on $\Si_*$
\beaa
\ov{\ka}=\frac{2}{r}, \qquad \ov{\kab}=-\frac{2\Up}{r}
\eeaa
and hence, we have on $\Si_*$
\beaa
a &=& -\frac{2}{\vsi} +\Up -\frac{r}{2}\Ab.
\eeaa
The bootstrap assumptions on decay for $k=0$ derivatives in $\Mext$, the definition \eqref{eq:defininitionofAb} of $\Ab$, and the estimates for $\vsi$ and $\Omb$ yield the rough estimate\footnote{The estimates for $\Omb$ and $\vsi$ are proved later in Proposition \ref{prop:finalcoordinatessystem}. Since the proof does not rely on Theorem M0, we may use it here.}
\beaa
|a| &\les& 1.
\eeaa
Together with the fact that $\nu_*=e_3+ae_4$ and the estimates of Step 2, we infer
\beaa
\left|\nu_*\left(\int_Se_\th(\kab)e^\Phi\right)\right| &\les& \frac{\ep_0}{u^{2+2\dec}},\\
\left|\nu_*\left(\int_S\b e^\Phi\right)\right| &\les& r^{-1}\left|\int_Se_\th(\kab)e^\Phi\right| +\frac{\ep_0}{ru^{1+\dec}}.
\eeaa

{\bf Step 4.} Now, recall that we have the following GCM on the last sphere $S_*=\Sigma_*\cap\CC_*$ of $\Si_*$
\beaa
\int_{S_*}e_\th(\kab) e^\Phi = \int_{S_*}\b e^\Phi =0.
\eeaa
Integrating backward from $S_*$ the estimate for $e_\th(\kab)$ of Step 3 yields on $\Si_*$
\beaa
\left|\int_Se_\th(\kab)e^\Phi\right| &\les& \frac{\ep_0}{u^{1+\dec}}.
\eeaa
Plugging in the estimate for $\b$ of Step 3, we infer on $\Si_*$
\beaa
\left|\nu_*\left(\int_S\b e^\Phi\right)\right| &\les& \frac{\ep_0}{ru^{1+\dec}}.
\eeaa
Integrating backward from $S_*$ yields on $\Si_*$
\beaa
\left|\int_S\b e^\Phi\right| &\les& \frac{\ep_0}{ru^{\dec}}.
\eeaa
Thus, at the first sphere $S_1=\Si_*\cap\CC_1$ of $\Si_*$, we have obtained 
\bea\lab{eq:theintermadiaryestimateofTheoremM0onS1toinvoketherigidityresult}
\left|\int_{S_1}e_\th(\kab)e^\Phi\right|+r\left|\int_{S_1}\b e^\Phi\right| &\les& \ep_0. 
\eea

\begin{remark}\lab{rem:whycanweuseTheoremM0forTheoremM8}
Note that the only bootstrap assumptions used in the proof of Theorem M0  are  the bootstrap assumption {\bf BA-D} on decay for $k=0, 1$ derivatives. Indeed, to obtain \eqref{eq:theintermadiaryestimateofTheoremM0onS1toinvoketherigidityresult}, we have only used, in Steps 1--4, the bootstrap assumption {\bf BA-D} on decay for $k=0, 1$ derivatives, while, from now on, we will only rely on \eqref{eq:theintermadiaryestimateofTheoremM0onS1toinvoketherigidityresult} and the assumptions \eqref{def:initialdatalayerassumptions-stronger:bisbisbis} on the initial data layer. This observation will allow us to use the conclusions of Theorem M0, not only for the bootstrap spacetime $\MM$ in Theorem M1--M5, but also for the extended spacetime in the proof of Theorem M8, where the only assumptions are the one on decay (which are established for the extended spacetime in Theorem M7).
\end{remark}

{\bf Step 5.} On the sphere $S_1=\Si_*\cap\CC_1$ of $\Si_*$, we have in view of the GCM conditions of $\Si_*$ and \eqref{eq:theintermadiaryestimateofTheoremM0onS1toinvoketherigidityresult}
\bea\lab{eq:usefulesitateinthediscussionofThmM0:00}
\ka=\frac{2}{r}, \quad \dds_2\dds_1\kab=0, \quad \dds_2\dds_1\mu=0, \quad \left|\int_{S_1}e_\th(\kab)e^\Phi\right|+r\left|\int_{S_1}\b e^\Phi\right| \les \ep_0.
\eea
We consider the transition functions $(f, \fb, \la)$ from the frame of the outgoing part $\Lext$ of the initial data layer to the frame of $\Mext$. Since
\begin{itemize}
\item $S_1$ is a sphere of $\Mext$ in the initial data layer,

\item $S_1$ is a sphere of the GCM hypersurface $\Si_*$,

\item the estimate \eqref{eq:usefulesitateinthediscussionofThmM0:00} holds on $S_1$,
\end{itemize}
we can invoke the GCM corollary Rigidity II and III of section \ref{sec:statementofGCMresultsinChapter3} with the choice $\epg=\dg=\ep_0$, $s_{max}=k_{large}+5$, and with the background foliation being the one of the outgoing part $\Lext$  of the initial data layer. We obtain 
\beaa
\|\dk^{\le k_{large}+6} (f, \fb, \log \la) \|_{L^2(S_1)} +\sup_{S_1}\left|\frac{m}{m_0}-1\right| &\les& \ep_0
\eeaa
Together with the Sobolev embedding on $S_1$, we deduce
\bea\lab{eq:usefulesitateinthediscussionofThmM0:00:first}
 \sup_{S_1}\left(r|\dk^{\le k_{large}+4} (f, \fb, \log \la) |+\left|\frac{m}{m_0}-1\right|+\left|r-\rextl\right| \right)\les \ep_0. 
\eea

{\bf Step 6.} Recall from Corollary \ref{cor:transportequationsforffbandlambda} that $(\fb, f, \log(\la))$ satisfy the following transport equations along $\CC_1$
\beaa
\la^{-1}e_4'(rf) &=& E_1'(f, \Ga),\\
\la^{-1}e_4'(\log(\la)) &=&  E_2'(f, \Ga),\\
\la^{-1}e_4'\Big(r\fb-2r^2e_\th'(\log(\la))+rf\Omb\Big) &=& E_3'(f, \fb, \la, \Ga),
\eeaa
where, in view of the form of $E_1'$, $E_2'$, $E_3'$ in Corollary \ref{cor:transportequationsforffbandlambda} and the estimates \eqref{def:initialdatalayerassumptions-stronger:bisbisbis} for the Ricci coefficients of the outgoing part $\Lext$ of the initial data layer, we have
\beaa
|\dk^kE_1'(f, \Ga)|+|\dk^kE_2'(f, \Ga)| &\les& \frac{\ep_0^2}{r^2}+|\dk^{\leq k}f|^2\textrm{ for }k\leq k_{large}+5\textrm{ on }\CC_1
\eeaa 
and 
\beaa
|\dk^k E_3'(f, \fb, \la, \Ga)| &\les& \frac{\ep_0^2}{r^2}+|\dk^{\leq k+1}f|^2+|\dk^{\leq k+1}\log(\la)|^2\textrm{ for }k\leq k_{large}+5\textrm{ on }\CC_1.
\eeaa

Next, recall from Lemma \ref{lemma:commTwithe3e4} the following commutator identity
\beaa
\,[\T, e_4]&=&\left(\left( \omb-\frac{m}{r^2} \right)     -\frac{m}{2r} \left( \ov{\ka} -\frac{2}{r}\right)+\frac{e_4 (m)}{r}  \right)e_4+(\eta+\ze) e_\th
\eeaa
while from Lemma \ref{Le:comme3e4-outgeodesic}, we have schematically 
\beaa
[\dkb, e_4] &=& \big(\kac, \vth\big)\dkb+\big(\ze, r\b\big)
\eeaa
Together with the  fact that 
\beaa
\la^{-1}e_4' &=& e_4+fe_\th+\frac{f^2}{4}e_3,
\eeaa
the commutator above identities for $[T, e_4]$ and $[\dkb, e_4]$, as well as the estimates \eqref{def:initialdatalayerassumptions-stronger:bisbisbis} for the Ricci coefficients and curvature components of the outgoing part $\Lext$  of the initial data layer, we infer, for $k\leq k_{large}+5$,
\beaa
&&|\dk^k[T, \la^{-1}e_4']h|+|\dk^k[\dkb, \la^{-1}e_4']h|\\
 &\les& \frac{\ep_0}{r^2}|\dk^{\leq k+1}h|+\frac{1}{r}|\dk^{\leq k}(f\dk h)|+\frac{1}{r}|\dk^{\leq k}(h\dk f)|+|\dk^{\leq k}(f^2\dk h)|+|\dk^{\leq k}(hf\dk f)|.
\eeaa
By commuting first the transport equations in the direction $\la^{-1}e_4'$ with $(T, \dkb)^k$, and by using these transport equations to recover  the $e_4$ derivatives, we deduce 
\beaa
\la^{-1}e_4'(r\dk^kf) &=& E_{1,k}'(f, \Ga),\\
\la^{-1}e_4'(\dk^k\log(\la)) &=&  E_{2,k}'(f, \Ga),\\
\la^{-1}e_4'\Big(r\dk^k\big(\fb-2re_\th'(\log(\la))+f\Omb\big)\Big) &=& E_{3,k}'(f, \fb, \la, \Ga),
\eeaa
where we have
\beaa
|E_{1,k}'(f, \Ga)|+|E_{2,k}'(f, \Ga)| &\les& \frac{\ep_0^2}{r^2}+|\dk^{\leq k}f|^2\textrm{ for }k\leq k_{large}+5\textrm{ on }\CC_1
\eeaa 
and 
\beaa
|E_{3,k}'(f, \fb, \la, \Ga)| &\les& \frac{\ep_0^2}{r^2}+|\dk^{\leq k+1}f|^2+|\dk^{\leq k+1}\log(\la)|^2+|\dk^{\leq k}\fb|^2\textrm{ for }k\leq k_{large}+5\textrm{ on }\CC_1.
\eeaa
This allows us to propagate the estimates for $(f,\fb, \la)$ in \eqref{eq:usefulesitateinthediscussionofThmM0:00:first} on $S_1$ to any sphere on $\CC_1$, and hence
\bea\lab{eq:usefulesitateinthediscussionofThmM0:00:second}
 \sup_{\CC_1}\Big(r|\dk^{\le k_{large}+4} (f, \log \la) |\Big)+\sup_{\CC_1}\Big(r|\dk^{\le k_{large}+3}\fb |\Big) \les \ep_0. 
\eea

{\bf Step 7.} In view of \eqref{eq:usefulesitateinthediscussionofThmM0:00:second}, the change of frame formulas of Proposition \ref{prop:transformations1}, and  the estimates \eqref{def:initialdatalayerassumptions-stronger:bisbisbis} for the Ricci coefficients and curvature components of the outgoing part $\Lext$  of the initial data layer, we obtain
\bea\lab{eq:gatheringtheconclusionsyieldingtheproofofThmM0:1}
\max_{0\leq k\leq k_{large}}&&\Bigg\{ \sup_{\CC_1}  \left[r^{\frac{7}{2}  +\de_B}\left( |\dk^k\,{}^{(ext)}\a| + |\dk^k\,{}^{(ext)}\b|\right)+r^{\frac{9}{2}  +\de_B}|\dk^{k-1}e_3(\,{}^{(ext)}\a)|   \right]\\
\nn&&+ \sup_{\CC_1} \left[r^3\ \left|\dk^k\left(\,{}^{(ext)}\rho+\frac{2m_0}{r^3}\right)\right|+r^2|\dk^k\,{}^{(ext)}\bb|+r|\dk^k\,{}^{(ext)}\aa|\right]\Bigg\} \les \ep_0,
\eea
as well as 
\beaa
\max_{0\leq k\leq k_{large}}&&\Bigg\{ \sup_{\CC_1}  \Big[r^2\left( |\dk^k\,{}^{(ext)}(\kac, \vth, \ze, \kabc)| +r |\dk^k\,{}^{(ext)}\vthb|\right)\Big]\Bigg\} \les \ep_0.
\eeaa
Now, according to  Proposition \ref{prop:derivativesHawkingmass},  we have in $\Mext$
\beaa
\,{}^{(ext)}e_4(\,{}^{(ext)}m) &=&  \frac{\,{}^{(ext)}r}{32\pi}\int_S\Bigg(2\,{}^{(ext)}\kac\,{}^{(ext)}\rhoc+2\,{}^{(ext)}e_\th(\,{}^{(ext)}\ka)\,{}^{(ext)}\ze-\frac{1}{2}\,{}^{(ext)}\kab(\,{}^{(ext)}\vth)^2\\
&& -\frac{1}{2}\,{}^{(ext)}\kac\,{}^{(ext)}\vth\,{}^{(ext)}\vthb +2\,{}^{(ext)}\ka(\,{}^{(ext)}\ze)^2\Bigg),
\eeaa
which together with the above estimates yields
\beaa
\sup_{\CC_1}r^2\Big|\,{}^{(ext)}e_4(\,{}^{(ext)}m)\Big| &\les& \ep_0^2.
\eeaa
This allows us to propagate the estimates for ${}^{(ext)}m$ in \eqref{eq:usefulesitateinthediscussionofThmM0:00:first} on $S_1$ to any sphere on $\CC_1$, and hence
\bea\lab{eq:gatheringtheconclusionsyieldingtheproofofThmM0:2}
\sup_{\CC_1}\left|\frac{{}^{(ext)}m}{m_0}-1\right|\les \ep_0.
\eea
Also, we have
\beaa
\la^{-1}e_4'\left(\log\left(\frac{\rext}{\rextl}\right)\right) &=& \frac{\la^{-1}{}^{(ext)}e_4(\rext)}{\rext} - \frac{\,{}^{(ext)}(e_0)_4(\rextl)}{\rextl} - \frac{f^2}{4}\,{}^{(ext)}\frac{(e_0)_3(\rextl)}{\rext}\\
&=& \frac{1}{2}\Big(\la^{-1}\ka'-\ka\Big)+\frac{\la^{-1}}{2}\kac' - \frac{1}{2}\kac -\frac{rf^2}{8}(\ov{\kab}+\Ab)\\
&=& \frac{1}{2}\Big(\ddd_1'(f)+\err(\ka, \ka')\Big)+\frac{\la^{-1}}{2}\kac' - \frac{1}{2}\kac -\frac{rf^2}{8}(\ov{\kab}+\Ab) 
\eeaa
where we have denoted with primes the quantities with respect to the frame of $\Mext$ and without primes  the quantities with respect to the frame of $\Lext$. Together with the above estimates, this yields
\beaa
\sup_{\CC_1}r^2\left|\left(\log\left(\frac{\rext}{\rextl}\right)\right)\right| &\les& \ep_0.
\eeaa
This allows us to propagate the estimates for $\rext-\rextl$ in \eqref{eq:usefulesitateinthediscussionofThmM0:00:first} on $S_1$ to any sphere on $\CC_1$, and hence
\beaa
\sup_{\CC_1}\left|\rext-\rextl\right|\les \ep_0.
\eeaa

{\bf Step 8.} Recall that
\begin{itemize}
\item $({}^{(ext)}e_4, {}^{(ext)}e_3, {}^{(ext)}e_\th)$ denotes the null frame of $\Mext$,

\item $({}^{(int)}e_4, {}^{(int)}e_3, {}^{(ext)}e_\th)$ denotes the null frame of $\Mint$,

\item $(\,{}^{(ext)}(e_0)_3, \,{}^{(ext)}(e_0)_4, \,{}^{(ext)}(e_0)_\th)$ denotes the null frame of $\Lext$,

\item $(\,{}^{(int)}(e_0)_3, \,{}^{(int)}(e_0)_4, \,{}^{(int)}(e_0)_\th)$ denotes the null frame of $\Lint$.
\end{itemize}
Also, recall that the timelike hyper surface $\TT$ is given by 
\beaa
\TT=\{\rext=\rh\}\textrm{ where }2m_0\left(1+\frac{\deh}{2}\right)\leq \rh\leq 2m_0\left(1+\frac{3\deh}{2}\right)
\eeaa
to that $\TT\subset \Lint\cap\Lext$, and recall that the frame of $\Mint$ is initialed on $\TT$ as follows
\beaa
\,{}^{(int)}e_4=  \la\,{}^{(ext)}\, e_4,\qquad {}^{(int)}e_3= \la^{-1}\, \,{}^{(ext)}e_3, \qquad {}^{(int)}e_\th=\,{}^{(ext)}e_\th\,\,\textrm{ on }\TT
\eeaa
where
\beaa
\la=\,{}^{(ext)}\la=1-\frac{2\,{}^{(ext)}m}{\rext}.
\eeaa
Denoting 
\begin{itemize}
\item by $(f, \fb, \la)$ the transition functions  from the frame of the outgoing part $\Lext$ of the initial data layer to the frame of $\Mext$ as in Steps 5 to 7,

\item by $(f', \fb', \la')$ the transition functions  from the frame of the ingoing part $\Lint$ of the initial data layer to the frame of $\Mint$,

\item by $(\tilde{f}, \tilde{\fb}, \tilde{\la})$ the transition functions on $\Lint\cap \Lext$ from the frame outgoing part $\Lext$ of the initial data layer to the frame of the ingoing part $\Lint$ of the initial data layer, 
\end{itemize}
we obtain, using also that $\CC_1\cap\CCb_1\subset\TT$,  
\beaa
\sup_{\CC_1\cap\CCb_1}\Big(|\dk^{\le k_{large}+3} (f', \fb', \log(\la')) |\Big) &\les& \sup_{\CC_1\cap\CCb_1}\Big(|\dk^{\le k_{large}+3} (f, \fb, \log(\la)) |\Big)\\
&&+\sup_{\CC_1\cap\CCb_1}\Big(|\dk^{\le k_{large}+3} (\tilde{f}, \tilde{\fb}, \log(\Up_0^{-1}\tilde{\la})) |\Big)\\
&&+\sup_{\CC_1\cap\CCb_1}\Big(|\dk^{\le k_{large}+3}\log(\Up_0^{-1}\Up) |\Big)
\eeaa
where we have denoted 
\beaa
\Up_0=1-\frac{2m_0}{\rextl}, \qquad \Up=1-\frac{2\mext}{\rext}.
\eeaa
Together with the control of $(\tilde{f}, \tilde{\fb}, \log(\Up_0^{-1}\tilde{\la}))$ provided on $\Lint\cap\Lext$ by the estimates \eqref{def:initialdatalayerassumptions-stronger:bisbisbis},  the estimates \eqref{eq:usefulesitateinthediscussionofThmM0:00:second} for $(f, \fb, \la)$, and the estimates $\mext-m_0$ and $\rext-\rextl$ obtained in Step 7, we infer 
\beaa
\sup_{\CC_1\cap\CCb_1}\Big(|\dk^{\le k_{large}+3} (f', \fb', \log(\la')) |\Big) &\les& \ep_0.
\eeaa
 
{\bf Step 9.} Similarly to Step 6, we propagate the estimate for $(f', \fb', \log(\la')$ on $\CC_1\cap\CCb_1$ provided by Step 8 to $\CCb_1$ using the analog of Corollary \ref{cor:transportequationsforffbandlambda} in the ingoing direction $e_3$. We obtain the following analog of \eqref{eq:usefulesitateinthediscussionofThmM0:00:second}
\beaa
 \sup_{\CCb_1}\Big(|\dk^{\le k_{large}+3} (\fb, \log \la) |\Big)+\sup_{\CC_1}\Big(r|\dk^{\le k_{large}+2}f |\Big) \les \ep_0. 
\eeaa
Together with the change of frame formulas of Proposition \ref{prop:transformations1}, and  the estimates \eqref{def:initialdatalayerassumptions-stronger:bisbisbis} for the Ricci coefficients and curvature components of the ingoing part $\Lint$  of the initial data layer, we obtain
\bea\lab{eq:gatheringtheconclusionsyieldingtheproofofThmM0:3}
\nn\max_{0\leq k\leq k_{large}}\sup_{\CCb_1}\Bigg[ |\dk^k\,{}^{(int)}\a| + |\dk^k\,{}^{(int)}\b|+ \left|\dk^k\left(\,{}^{(int)}\rho+\frac{2m_0}{r^3}\right)\right| &&\\
 +|\dk^k\,{}^{(int)}\bb|+|\dk^k\,{}^{(int)}\aa|\Bigg] &\les& \ep_0,
\eea
as well as 
\beaa
\max_{0\leq k\leq k_{large}}&&\Bigg\{ \sup_{\CCb_1} |\dk^k\,{}^{(int)}(\kabc, \vthb, \ze, \kac, \vth)|\Bigg\} \les \ep_0.
\eeaa
Also, since we have as a consequence of   the initialization on $\TT$ of the ingoing geodesic foliation of $\Mint$
\beaa
\mint=\mext \,\,\textrm{ on }\,\,\CC_1\cap\CCb_1
\eeaa
we infer from the control of $\mext$ provided by Step 7
\beaa
|\mint-m_0| &\les& \ep_0\textrm{ on }\CC_1\cap\CCb_1.
\eeaa
We then propagate, similarly to Step 7, this bound to $\CCb_1$ and obtain 
\beaa
\sup_{\CCb_1}\left|\mint-m_0\right|\les \ep_0.
\eeaa
Together with \eqref{eq:gatheringtheconclusionsyieldingtheproofofThmM0:1}, \eqref{eq:gatheringtheconclusionsyieldingtheproofofThmM0:2} and \eqref{eq:gatheringtheconclusionsyieldingtheproofofThmM0:3}, this concludes the proof of Theorem M0.


\section{Control of averages and of the Hawking mass}\lab{section:proofoflemma:estimatesonceandforallforaverages}


In this section, we prove Lemma \ref{lemma:estimatesonceandforallforaverages} and Lemma \ref{lemma:estimatesonceandforallforHawkingmass}.


\subsection{Proof of Lemma \ref{lemma:estimatesonceandforallforaverages}}


{\bf Step 1.} We start with the control of $\overline{\rho}$ on $\MM$. Recall the identity \eqref{identity:overlinerho}
\beaa
\overline{\rho}  +\frac{2m}{r^3} &=&  \frac{1}{4}\overline{\vth\vthb}.
\eeaa
Thus, in view  of the bootstrap assumptions  {\bf BA-D}, {\bf BA-E}, we have,
\beaa
\Big|\overline{\rho}  +\frac{2m}{r^3} \Big|&\les&\ep^2 \min \{ r^{-3} u^{-\frac 3 2 -\dec}, r^{-2} u^{-2-2\dec}\} \qquad \mbox{in} \,  \Mext,\\
\Big|\overline{\rho}  +\frac{2m}{r^3} \Big|&\les&\ep^2 \ub^{-2-2\dec}  \qquad \mbox{in} \,  \Mint.
\eeaa
 Differentiating  the  equation with respect to $e_3, e_4$   we derive,
 \beaa
e_4\left(\overline{\rho}  +\frac{2m}{r^3} \right)&=&  \frac{1}{4}\overline{ e_4(\vth) \vthb      +\vth e_4(\vthb)}+\lot,\\
e_3\left(\overline{\rho}  +\frac{2m}{r^3} \right)&=&  \frac{1}{4}\overline{ e_3(\vth) \vthb      +\vth e_3(\vthb)}+\lot,\\
e_\th \left(\overline{\rho}  +\frac{2m}{r^3} \right)&=&0.
\eeaa
Taking higher derivatives in $e_3, e_4$ and making use of the bootstrap assumptions  {\bf BA-D}, {\bf BA-E}, we derive   in $\Mext$,
\beaa
\left|\dk^{\leq k_{small}}\left(\overline{\rho}+\frac{2m}{r^3}\right)\right|&\les &\ep^2 \min \{ r^{-3} u^{-\frac 3 2 -\dec}, r^{-2} u^{-2-2\dec}\}, \\
\left|\dk^{\leq k_{large}}\left(\overline{\rho}+\frac{2m}{r^3}\right)\right|  &\les&r^{-3} u^{-1/2-\dec},
\eeaa
and in $ \Mint$,
\beaa
\left|\dk^{\leq k_{small}}\left(\overline{\rho}+\frac{2m}{r^3}\right)\right|&\les &\ep^2 \ub^{-2-2\dec},\\
\left|\dk^{\leq k_{large}}\left(\overline{\rho}+\frac{2m}{r^3}\right)\right|&\les &\ep^2\ub^{-1 -\dec}.
\eeaa
In particular,
\beaa
\sup_{\Mext}u^{\frac{3}{2}+\dec}r^3\left|\dk^{\leq k_{small}}\left(\overline{\rho}+\frac{2m}{r^3}\right)\right|+\sup_{\Mext}u^{\frac{1}{2}+\dec}r^3\left|\dk^{\leq k_{large}}\left(\overline{\rho}+\frac{2m}{r^3}\right)\right| &\les& \ep_0,\\
\sup_{\Mint}\ub^{\frac{3}{2}+\dec}\left|\dk^{\leq k_{small}}\left(\overline{\rho}+\frac{2m}{r^3}\right)\right|+\sup_{\Mint}\ub^{\frac{1}{2}+\dec}\left|\dk^{\leq k_{large}}\left(\overline{\rho}+\frac{2m}{r^3}\right)\right| &\les& \ep_0
\eeaa

{\bf Step 2.} Next, we proceed with the control of $\overline{\ka}$ in $\Mext$.
Recalling  Lemma \ref{lemma:transportequationforoverlinekaintheoutgoinggeodesicfoliation}, we start with
\bea
\label{transportfor(ovk-2r)}
\bsplit
e_4\left(\overline{\ka}-\frac{2}{r}\right)  + \frac{1}{2}\overline{\ka}\left(\overline{\ka}-\frac{2}{r} \right) &=   -\frac{1}{4}\overline{\vth^2}  + \frac{1}{2}\overline{\check{\ka}^2}.
\end{split}
\eea
In view of Corollary \ref{le:transportrp-f}  we deduce, from the first equation,
\bea
\label{transportfor-r(ovk-2r)}
e_4\left( r\left(\ov{\ka}-\frac{2}{r}\right)\right)&= - r\left(\frac{1}{4}\overline{\vth^2}  + \frac{1}{2}\overline{\check{\ka}^2}\right).
\eea
 Making use of the GCM condition
\beaa
\ka &=& \frac{2}{r}\textrm{ on }\Sigma_*,
\eeaa
which yields
\beaa
\ov{\ka} &=& \frac{2}{r}\textrm{ on }\Sigma_*,
\eeaa
we deduce, integrating  \eqref{transportfor-r(ovk-2r)}  with respect to $r$ along $C_u$  from $\Sigma_*$, 
\beaa
\sup_{\Mext}u^{1+\dec}r^3\left|\overline{\ka}-\frac{2}{r}\right| &\les& \ep^2\les\ep_0.
\eeaa
Also, making use of  the bootstrap  assumptions {\bf BA-D}, {\bf BA-E} we easily deduce, 
\beaa
\sup_{\Mext}u^{1+\dec}r^3\left|\dk_{\near}^{\leq k_{small}+1}\left(\overline{\ka}-\frac{2}{r}\right)\right|  &\les& \ep^2\les \ep_0,\\
\sup_{\Mext}u^{\frac{1}{2}+\dec}r^3\left|\dk_{\near}^{\leq k_{large}+1}\left(\overline{\ka}-\frac{2}{r}\right)\right| &\les& \ep^2\les \ep_0.
\eeaa
We next commute \eqref{transportfor-r(ovk-2r)} with $e_3$  and derive,
\beaa
e_4 e_3 \left( r\left(\ov{\ka}-\frac{2}{r}\right)\right)&=& e_3\left( r\left(\frac{1}{4}\overline{\vth^2}  + \frac{1}{2}\overline{\check{\ka}^2}\right)\right)
-[e_3, e_4] \left( r\left(\ov{\ka}-\frac{2}{r}\right)\right)\\
&=& e_3\left( r\left(\frac{1}{4}\overline{\vth^2}  + \frac{1}{2}\overline{\check{\ka}^2}\right)\right)-2\omb  \left( r\left(\ov{\ka}-\frac{2}{r}\right)\right)- 2\ze 
 \left( r\left(\ov{\ka}-\frac{2}{r}\right)\right).
\eeaa
It is thus  easy to see that  we can   prove estimates of the type
\beaa
\sup_{\Mext}u^{1+\dec}r^3\left|\dk^{\leq k_{small}+1}\left(\overline{\ka}-\frac{2}{r}\right)\right|  &\les& \ep^2\les \ep_0,\\
\sup_{\Mext}u^{\frac{1}{2}+\dec}r^3\left|\dk^{\leq k_{large}+1}\left(\overline{\ka}-\frac{2}{r}\right)\right| &\les& \ep^2\les \ep_0,
\eeaa
provided that we can check that,
\beaa
\sup_{\Mext}u^{1+\dec}r^3\left|  e_3^{\leq k_{small}+1}\left(\overline{\ka}-\frac{2}{r}\right)\right|  &\les& \ep^2\les \ep_0,\\
\sup_{\Mext}u^{\frac{1}{2}+\dec}r^3\left|e_3^{\leq k_{large}+1}\left(\overline{\ka}-\frac{2}{r}\right)\right| &\les& \ep^2\les \ep_0.
\eeaa
The  difficulty in this case  is to  make sure that  we can control  terms of the type,
\beaa
e_3^{k+1}\left( r\left(\frac{1}{4}   e_3^{k+1} ( \overline{\vth^2} ) + \frac{1}{2} e_3^{k+1}   ( \overline{\check{\ka}^2})\right)\right)
\eeaa
using only  at most $k$ derivatives of  $\Gac, \Rc$. To see this we note that,
\beaa
 e_3( \overline{\vth^2})&=&\ov{e_3\vth^2}-(\Ombc\, \ov{\ka}-\ov{\Ombc \, \kac} )\ov{\vth^2}+\ov{\kabc  \check{\vth^2}},\\
 e_3 (\overline{\kac^2} )&=&\ov{e_3\kac^2}-(\Ombc\, \ov{\ka}-\ov{\Ombc \, \kac} )\ov{\kac^2}+\ov{\kabc  \check{\ka^2}},
\eeaa
and,
\bea
\label{eq:e3vth,kac}
\bsplit
e_3(\vth) + \frac{1}{2}\kab\vth -2\omb\vth &= -2\dds_2\ze -\frac{1}{2}\ka\vthb+2\ze^2,\\
 e_3\kac  +\frac 1 2  \overline{\kab}\,      \check{\ka}&=-2\check{\mu}-
\frac 1 2 \overline{\ka} \check{\kab} +2 (\check{\omb}\overline{\ka}+\overline{\omb}\check{\ka}) +\Obc \overline{\ka} \, \overline{\kab}+\err[e_3\check{\ka}],\\
\err[e_3\check{\ka}]:&=2(\ze^2-\overline{\ze^2}) + 2(\check{\omb}\check{\ka}-\overline{\check{\omb}\check{\ka}})-\frac 1 2 \check{\kab} \,\check{\ka} -\frac 1 2 \overline{ \check{\kab} \,\check{\ka}} -\overline{\Obc \check{\ka} }\, \overline{\kab}.
\end{split}
\eea
 We  thus derive,
 \beaa
\sup_{\Mext}u^{1+\dec}r^3\left|\dk^{\leq k_{small}+1}\left(\overline{\ka}-\frac{2}{r}\right)\right|+\sup_{\Mext}u^{\frac{1}{2}+\dec}r^3\left|\dk^{\leq k_{large}+1}\left(\overline{\ka}-\frac{2}{r}\right)\right| &\les& \ep_0.
\eeaa

{\bf Step 3.}  We  next estimate  $\overline{\kab}$ in $ \Mext$ making use  of the identity  \eqref{identity:overlinekaandoverlinekab} 
 derived in connection to the   Hawking mass
\beaa
\overline{\kab} +\frac{2\Up}{r} &=&   \frac{2\Up}{r\overline{\ka}}\left(\overline{\ka}-\frac{2}{r}\right) - \frac{1}{\overline{\ka}}\overline{\check{\ka}\check{\kab}}.
\eeaa
Thus, in view of the estimates for   $\overline{\ka}$ derived in step 2 we easily infer that,
\beaa
\sup_{\Mext}u^{\frac{3}{2}+\dec}r^2\left|\dk^{\leq k_{small}}\left(\overline{\kab}+\frac{2\Up}{r}\right)\right| +\sup_{\Mext}u^{\frac{1}{2}+\dec}r^2\left|\dk^{\leq k_{large}}\left(\overline{\kab}+\frac{2\Up}{r}\right)\right|  &\les& \ep_0.
\eeaa
as  desired.

{\bf Step 4.} We  estimate  $\overline{\omb}$ in $\Mext$  based on the following identity in Lemma \ref{lemma:transportequationforoverlinekaintheoutgoinggeodesicfoliation}
\beaa
e_3\left(\ov{\ka}-\frac 2 r \right) +\frac 1 2  \ov{ \kab} \left(\ov{\ka}-\frac 2 r \right) &=&  2 \ov{\omb}\,\left( \ov{\ka}-\frac 2 r \right)+\frac 4 r\left(\ov{ \omb}-\frac{m}{r^2}\right)+ 2\left(  \ov{\rho} +\frac{2m}{r^3}\right)\\
&& -\frac{1}{2}\overline{\ka}\left(\ov{\ka}-\frac 2 r \right)\check{\Omb} + 2\overline{\check{\omb}\check{\ka}} -\frac{1}{2}\overline{\vth\vthb} +2\overline{\ze^2}\\
&&+\frac{1}{2}\check{\Omb}\Big(-\overline{\vth^2} +\overline{\check{\ka}^2}\Big)
- \overline{\check{\Omb}(e_4(\check{\ka}) +\ka\check{\ka})}+\frac{1}{2}\overline{\check{\kab}\check{\ka}} -\frac{1}{r}\overline{\Obc\check{\ka}},
\eeaa
which we rewrite as
\beaa
 \ov{ \omb}-\frac{m}{r^2} &=& \frac{r}{4}\Bigg\{e_3\left(\ov{\ka}-\frac 2 r \right) +\frac 1 2  \ov{ \kab} \left(\ov{\ka}-\frac 2 r \right) - 2 \ov{\omb}\,\left( \ov{\ka}-\frac 2 r \right)  - 2\left(  \ov{\rho} +\frac{2m}{r^3}\right)\\
&& +\frac{1}{2}\overline{\ka}\left(\ov{\ka}-\frac 2 r \right)\check{\Omb} - 2\overline{\check{\omb}\check{\ka}} +\frac{1}{2}\overline{\vth\vthb} -2\overline{\ze^2} -\frac{1}{2}\check{\Omb}\Big(-\overline{\vth^2} +\overline{\check{\ka}^2}\Big)\\
&&+ \overline{\check{\Omb}(e_4(\check{\ka}) +\ka\check{\ka})}-\frac{1}{2}\overline{\check{\kab}\check{\ka}} +\frac{1}{r}\overline{\Obc\check{\ka}}\Bigg\}.
\eeaa
Using the estimates of $\overline{\rho}$ in Step 1, the estimates for $\overline{\ka}$ in Step 2, as well as  our bootstrap assumptions on decay and energy, we easily derive
\beaa
\sup_{\Mext}u^{1+\dec}r^2\left|\dk^{\leq k_{small}}\left(\overline{\omb}-\frac{m}{r^2}\right)\right|+\sup_{\Mext}u^{\frac{1}{2}+\dec}r^2\left|\dk^{\leq k_{large}}\left(\overline{\omb}-\frac{m}{r^2}\right)\right| &\les& \ep_0.
\eeaa

\begin{remark}
It is to estimate $k_{large}$ derivatives of $\overline{\omb}-mr^{-2}$ that we had   to control $k_{large}+1$ derivatives of $\overline{\ka}-2/r$ is Step 2.
\end{remark}

{\bf Step 5.} We  estimate  $\overline{\Omb}$ in $\Mext$. First we need the control of $\ov{\Omb}$ on $\Sigma_*$. To this end, we recall that $s$ is initialized on $\Sigma_*$ by $s=r$ so that 
\beaa
\nu(s-r) &=& 0\textrm{ on }\Si_*, \qquad \nu=e_3+ae_4,
\eeaa
where the scalar function $a$ is such that the vectorfield $\nu$ is tangent to $\Sigma_*$. On the other hand, we have $e_4(s)=1$ and 
\beaa
e_4(r)=\frac{r}{2}\ov{\ka}=1\textrm{ on }\Sigma_*
\eeaa
where we used the GCM condition $\ka=2/r$ on $\Sigma_*$. We infer $e_3(s)=e_3(r)$ on $\Sigma_*$ and hence
\beaa
\Omb &=& e_3(r)\textrm{ on }\Si_*.
\eeaa
This yields
\beaa
\ov{\Omb} &=& \ov{e_3(r)}=\frac{r\ov{\kab}}{2}+\ov{\Ab},
\eeaa
and hence, in view of the estimate for $\ov{\kab}$ of step 3, the fact that $\ov{\Ab}$ contains only quadratic terms in view of the formula for $\Ab$, and in view of the bootstrap assumptions on decay and energy, we infer
\beaa
\sup_{\Si_*}u^{1+\dec}r\left|\dk^{\leq k_{small}}\left(\overline{\Omb}-\frac{m}{r^2}\right)\right|+\sup_{\Si_*}u^{\frac{1}{2}+\dec}r\left|\dk^{\leq k_{large}}\left(\overline{\Omb}-\frac{m}{r^2}\right)\right| &\les& \ep_0.
\eeaa
Then, we use $e_4(\Omb)=-2\omb$ and  Corollary \ref{corr:transportcheckf} to obtain
\beaa
e_4(\ov{\Omb}) &=& -2\ov{\omb}+\ov{\check{\ka}\,\check{\Omb}}
\eeaa
and hence
\beaa
e_4(\ov{\Omb}+\Up) &=& -2\left(\ov{\omb}-\frac{m}{r^2}\right)+\frac{m}{r}\left(\ov{\ka}-\frac{2}{r}\right)+\ov{\check{\ka}\,\check{\Omb}} -\frac{2e_4(m)}{r}.
\eeaa
Commuting with $\dk$, integrating from $\Si_*$ where we have controlled $\ov{\Omb}$ above, and using the estimates of Step 2 for $\ov{\ka}$, Step 4 for $\ov{\omb}$, the bootstrap assumptions, and the estimates for $e_4(m)$ of Lemma \ref{lemma:estimatesonceandforallforHawkingmass} (which do not depend on the control of $\ov{\Omb}$), we infer
\beaa
\sup_{\Mext}u^{1+\dec}r\left|\dk^{\leq k_{small}}\left(\overline{\Omb}-\frac{m}{r^2}\right)\right|+\sup_{\Mext}u^{\frac{1}{2}+\dec}r\left|\dk^{\leq k_{large}}\left(\overline{\Omb}-\frac{m}{r^2}\right)\right| &\les& \ep_0.
\eeaa

{\bf Step 6.} Next, we control $\,{}^{(int)}\overline{\kab}$ on  the cylinder  $\TT$. From the initialization of the frame of $\Mint$ on $\TT$, we have
\beaa
\rint=\rext,\,\,\,\, \,{}^{(int)}\overline{\ka} = \Up\,{}^{(ext)}\overline{\ka},\,\,\,\, \,{}^{(int)}\overline{\kab} = \Up^{-1}\,{}^{(ext)}\overline{\kab}\textrm{ on }\TT.
\eeaa
Also, making use  of the identity  \eqref{identity:overlinekaandoverlinekab}  derived in connection to the   Hawking mass, we have
\beaa
\,{}^{(ext)}\overline{\kab} +\frac{2\Up}{r} &=&  \frac{2\Up}{r\,{}^{(ext)}\overline{\ka}}\left(\,{}^{(ext)}\overline{\ka}-\frac{2}{r}\right) - \frac{1}{\,{}^{(ext)}\overline{\ka}}\overline{\,{}^{(ext)}\check{\ka}\,{}^{(ext)}\check{\kab}}.
\eeaa
We deduce
\beaa
{}^{(int)}\overline{\kab} +\frac{2}{r} &=& \Up^{-1}\left(\,{}^{(ext)}\overline{\kab}+\frac{2\Up}{r}\right)= \frac{2}{r\,{}^{(ext)}\overline{\ka}}\left(\,{}^{(ext)}\overline{\ka}-\frac{2}{r}\right) - \frac{\Up^{-1}}{\,{}^{(ext)}\overline{\ka}}\overline{\,{}^{(ext)}\check{\ka}\,{}^{(ext)}\check{\kab}}\textrm{ on }\TT.
\eeaa
To derive higher tangential derivatives along $\TT$ we  remark   that the vectorfield 
\beaa
T_{\TT} = e_4-\frac{e_4(r)}{e_3(r)}e_3=e_4 -\frac{\ov{\ka}+A}{\ov{\kab}} e_3,
\eeaa
 together with $e_\th$, spans the tangent space to $\TT$.  The transversal derivatives, on the other hand, can be determined 
  with help of the equation,
  \beaa
e_3\left(    \overline{\kab}+\frac{2}{r}\right)  + \frac{1}{2}\overline{\kab}\left(\overline{\kab}+\frac{2}{r} \right) &=&   -\frac{1}{4}\overline{\vthb^2}  + \frac{1}{2}\overline{\check{\kab}^2}.
\eeaa
adapted to the $\Mint$ foliation.
  Making use of the estimates for $\,{}^{(ext)}\overline{\ka}$ in $\Mext$ derived in Step 2 and the bootstrap assumptions,  
we infer that,
\beaa
&&\sup_{\TT}u^{\frac{3}{2}+\dec}\left|\dk^{\leq k_{small}+1}\left({}^{(int)}\overline{\kab} +\frac{2}{r}\right)\right|+\sup_{\TT}u^{\frac{1}{2}+\dec}\left|\dk^{\leq k_{large}+1}\left({}^{(int)}\overline{\kab} +\frac{2}{r}\right)\right| \\
&\les& \ep_0+\sup_{\TT}u^{\frac{1}{2}+\dec}\left|\dk^{k_{large}+1}\left(\overline{\,{}^{(ext)}\check{\ka}\,{}^{(ext)}\check{\kab}}\right)\right|
\eeaa
Now, in view of the transport equations for $\,{}^{(ext)}e_4(\,{}^{(ext)}\check{\ka})$, $\,{}^{(ext)}e_3(\,{}^{(ext)}\check{\ka})$, $\,{}^{(ext)}e_4(\,{}^{(ext)}\check{\kab})$ and $\,{}^{(ext)}e_3(\,{}^{(ext)}\check{\ka})$, as well as the bootstrap assumptions, we have
\beaa
&& \sup_{\TT}u^{\frac{1}{2}+\dec}\left|\dk^{k_{large}+1}\left(\overline{\,{}^{(ext)}\check{\ka}\,{}^{(ext)}\check{\kab}}\right)\right|\\
&\les& \ep_0+\sup_{\TT}u^{\frac{1}{2}+\dec}\left|\overline{\,{}^{(ext)}\check{\kab}\dk^{k_{large}}\ddd_1(\,{}^{(ext)}\ze)}\right|+\sup_{\TT}u^{\frac{1}{2}+\dec}\left|\overline{\,{}^{(ext)}\check{\ka}\dk^{k_{large}}\ddd_1(\,{}^{(ext)}\ze)}\right| \\
&&+\sup_{\TT}u^{\frac{1}{2}+\dec}\left|\overline{\,{}^{(ext)}\check{\ka}\dk^{k_{large}}\ddd_1(\,{}^{(ext)}\xib)}\right| \\
&\les& \ep_0+\sup_{\TT}u^{\frac{1}{2}+\dec}\left|\overline{\dds_1\,{}^{(ext)}\check{\kab}\dk^{k_{large}}\,{}^{(ext)}\ze}\right|+\sup_{\TT}u^{\frac{1}{2}+\dec}\left|\overline{\dds_1\,{}^{(ext)}\check{\ka}\dk^{k_{large}}\,{}^{(ext)}\ze}\right| \\
&&+\sup_{\TT}u^{\frac{1}{2}+\dec}\left|\overline{\dds_1\,{}^{(ext)}\check{\ka}\dk^{k_{large}}\,{}^{(ext)}\xib}\right| \\
&\les&\ep_0
\eeaa
where we have integrated $\ddd_1$ by parts and used that $\dds_1$ is its adjoint. We infer
\beaa
\sup_{\TT}\ub^{\frac{3}{2}+\dec}\left|\dk^{\leq k_{small}+1}\left({}^{(int)}\overline{\kab} +\frac{2}{r}\right)\right|+\sup_{\TT}\ub^{\frac{1}{2}+\dec}\left|\dk^{\leq k_{large}+1}\left({}^{(int)}\overline{\kab} +\frac{2}{r}\right)\right| &\les& \ep_0.
\eeaa

{\bf Step 7.} From now on, we only work with the frame of $\Mint$.     Starting with the equation,
\beaa
e_3\left(\overline{\kab}+\frac{2}{r}\right)  + \frac{1}{2}\overline{\kab}\left(\overline{\kab}+\frac{2}{r} \right) &=&   -\frac{1}{4}\overline{\vthb^2}  + \frac{1}{2}\overline{\check{\kab}^2}.
\eeaa
Using the estimates of step 5 we can then proceed precisely as in Step 2 ( using the $\Mint$  counterpart of the equations \eqref{eq:e3vth,kac})
to derive,
\beaa
\sup_{\Mint}\ub^{1+\dec}\left|\dk^{\leq k_{small}+1}\left(\overline{\kab}+\frac{2}{r}\right)\right|+\sup_{\Mint}\ub^{\frac{1}{2}+\dec}\left|\dk^{\leq k_{large}+1}\left(\overline{\kab}+\frac{2}{r}\right)\right| &\les& \ep_0.
\eeaa

{\bf Step 8.} Finally, we estimate the remaining averages in $\Mint$, i.e. $\overline{\ka}$ and $\overline{\om}$.  To estimate $\ov{\ka}$ 
we make use once more of the identity,
\beaa
\overline{\ka} -\frac{2\Up}{r} &=&   -\frac{2\Up}{r\overline{\kab}}\left(\overline{\kab}+\frac{2}{r}\right) - \frac{1}{\overline{\kab}}\overline{\check{\ka}\check{\kab}}.
\eeaa
Making use of  the estimates of $\overline{\kab}$ in Step 5 as well as the bootstrap assumptions for $\check{\ka}$ and $\check{\kab}$  we easily derive,
\beaa
\sup_{\Mint}\ub^{1+\dec}\left|\dk^{\leq k_{small}}\left(\overline{\ka}-\frac{2\Up}{r}\right)\right|+\sup_{\Mint}\ub^{\frac{1}{2}+\dec}\left|\dk^{\leq k_{large}}\left(\overline{\ka}-\frac{2\Up}{r}\right)\right| &\les& \ep_0.
\eeaa
{\bf Step 9.}  To estimate $\ov{\om}$ we proceed as in Step 4 by making use of   the identity
\beaa
 \ov{ \om}+\frac{m}{r^2} &=& \frac{r}{4}\Bigg\{e_4\left(\ov{\kab}+\frac 2 r \right) +\frac 1 2  \ov{ \ka} \left(\ov{\kab}+\frac 2 r \right) - 2 \ov{\om}\,\left( \ov{\kab}+\frac 2 r \right)  - 2\left(  \ov{\rho} +\frac{2m}{r^3}\right)\\
&& +\frac{1}{2}\overline{\kab}\left(\ov{\kab}+\frac 2 r \right)\check{\Om} - 2\overline{\check{\om}\check{\kab}} +\frac{1}{2}\overline{\vth\vthb} -2\overline{\ze^2} -\frac{1}{2}\check{\Om}\Big(-\overline{\vthb^2} +\overline{\check{\kab}^2}\Big)\\
&&+ \overline{\check{\Om}(e_4(\check{\kab}) +\kab\check{\kab})}-\frac{1}{2}\overline{\check{\ka}\check{\kab}} +\frac{1}{r}\overline{\Omc\check{\kab}}\Bigg\}.
\eeaa
Thus, in  view of  the estimates of $\overline{\rho}$ in Step 1, the estimates for $\overline{\kab}$ in Step 5, the estimates of $\overline{\ka}$ above\footnote{It is to estimate $k_{large}$ derivatives of $\overline{\om}+m/r^2$ that we made sure   to control $k_{large}+1$ derivatives of $\overline{\kab}+2/r$.}, as well as the bootstrap assumptions {\bf BA-D}  and {\bf BA-E},   we deduce, 
\beaa
\sup_{\Mint}\ub^{1+\dec}\left|\dk^{\leq k_{small}}\left(\overline{\om}+\frac{m}{r^2}\right)\right|+\sup_{\Mint}\ub^{\frac{1}{2}+\dec}\left|\dk^{\leq k_{large}}\left(\overline{\om}+\frac{m}{r^2}\right)\right| &\les& \ep_0.
\eeaa

{\bf Step 10.} It remains to estimate $\ov{\Om}$ in $\Mint$. First we need the control of $\ov{\Om}$ on $\TT$. To this end, we recall that $s$ is initialized on $\TT$ by $s=r$ so that 
\beaa
T_\TT(s-r) &=& 0\textrm{ on }\TT, \qquad T_{\TT} = e_4 -\frac{\ov{\ka}+A}{\ov{\kab}} e_3,
\eeaa,
where the vectorfield has been introduced above and is tangent to $\TT$. On the other hand, we have $e_3(s)=-1$ and $e_3(r)=r\ov{\kab}/2$, and hence
\beaa
\Om &=& e_4(r)+\frac{\ov{\ka}+A}{\ov{\kab}}(-1-e_3(r))= \frac{r}{2}(\ov{\ka}+A)\left(1+\frac{\ov{\kab}-\frac{2}{r}}{\ov{\kab}}\right)\textrm{ on }\TT
\eeaa
This yields
\beaa
\ov{\Om} &=& \frac{r}{2}(\ov{\ka}+\ov{A})\left(1+\frac{\ov{\kab}-\frac{2}{r}}{\ov{\kab}}\right)\textrm{ on }\TT,
\eeaa
and hence, in view of the estimate for $\ov{\kab}$ of step 7, the estimate for $\ov{\ka}$ of step 8, the fact that $\ov{A}$ contains only quadratic terms in view of the formula for $A$, and in view of the bootstrap assumptions on decay and energy, we infer
\beaa
\sup_{\TT}\ub^{1+\dec}\left|\dk^{\leq k_{small}}\left(\overline{\Om}-\Up\right)\right|+\sup_{\TT}\ub^{\frac{1}{2}+\dec}\left|\dk^{\leq k_{large}}\left(\overline{\Om}-\Up\right)\right| &\les& \ep_0.
\eeaa
Then, we use the analog of the transport equation used to estimate $\ov{\Omb}$ in $\Mext$, i.e.
\beaa
e_3(\ov{\Om}-\Up) &=& 2\left(\ov{\om}+\frac{m}{r^2}\right)-\frac{m}{r}\left(\ov{\kab}+\frac{2}{r}\right)+\ov{\check{\kab}\,\check{\Om}} +\frac{2e_3(m)}{r}.
\eeaa
Commuting with $\dk$, integrating from $\TT$ where we have controlled $\ov{\Om}$ above, and using the estimates of Step 2 for $\ov{\ka}$, Step 4 for $\ov{\omb}$, the bootstrap assumptions, and the estimates for $e_4(m)$ of Lemma \ref{lemma:estimatesonceandforallforHawkingmass} (which do not depend on the control of $\ov{\Omb}$), we infer
\beaa
\sup_{\Mint}\ub^{1+\dec}\left|\dk^{\leq k_{small}}\left(\overline{\Om}-\Up\right)\right|+\sup_{\Mint}\ub^{\frac{1}{2}+\dec}\left|\dk^{\leq k_{large}}\left(\overline{\Om}-\Up\right)\right| &\les& \ep_0.
\eeaa
This concludes the proof of Lemma \ref{lemma:estimatesonceandforallforaverages}.


\subsection{Proof of Lemma \ref{lemma:estimatesonceandforallforHawkingmass}}


{\bf Step 1.} We start with the control of $e_3(m)$ and $e_4(m)$ in $\Mext$. According to  Proposition \ref{prop:derivativesHawkingmass}  we have in $\Mext$
\bea
e_4(m) &=&  \frac{r}{32\pi}\int_S\err_1,
\eea
and
\bea
\nn e_3(m)&=&   \left(1- \vsi^{-1} \check{\vsi}\right)\frac {r}{32\pi}\int_S\underline{\err}_1+\left(\Obc +\vsi^{-1}\ov{\Omb}\check{\vsi}\right)\frac {r}{32\pi}\int_S\err_1\\
\nn&&+ \vsi^{-1}\frac {r}{32\pi} \int_S\check{\vsi}\,\left(2 \ov{\rho}\kabc+2 \rhoc\ov{\kab}+2 \kab\ddd_1 \eta +2\ka \ddd_1 \xib+\underline{\err}_2\right)\\
\nn&&-\vsi^{-1}\frac{r}{32\pi} \int_S(\ov{\Omb}\check{\vsi}+\Obc\vsi)\left(2 \ov{\rho}\kac+2 \rhoc\ov{\ka}-2\ka\ddd_1\ze+\err_2\right)\\
&&-\frac{m}{r}\vsi^{-1}\left[-\ov{\check{\vsi}\kabc}+  \ov{\Omb}\,\ov{\check{\vsi}\kac } +  \ov{\Obc \vsi \ka}\right],
\eea
where 
\beaa
\err_1 &:=& 2\kac\rhoc+2e_\th(\ka)\ze-\frac{1}{2}\kab\vth^2 -\frac{1}{2}\kac\vth\vthb +2\ka\ze^2,\\ 
\underline{\err}_1 &:=& 2 \rhoc\kabc-2 e_\th(\kab)\eta -2e_\th(\ka)\xib-\frac{1}{2}\kabc\vth\vthb +2\kab\eta^2+2\ka \big(\eta-3\ze\big)\xib-\frac 1 2 \ka \vthb^2,\\
\err_2 &:=& 2\rhoc\kac-\frac{1}{2}\kab\vth^2 -\frac{1}{2}\ka\vth\vthb +2\ka\ze^2,\\
\underline{\err}_2 &:=& 2\rhoc\kabc +\kab\left(2\eta^2-\frac 1 2 \vth \vthb \right)+2\ka \big(\eta-3\ze\big)\xib-\frac 1 2 \ka \vthb^2.
\eeaa
Thus, according to the bootstrap assumption {\bf BA-D} on decay, we   deduce,
\beaa
|e_4(m)|&\les&\ep^2  r^{-2} u^{-1-\dec},\\
|e_3(m)|&\les&\ep^2   u^{-2-2 \dec}.
\eeaa
Moreover,  differentiating   the equations with respect to  $e_3, e_4$      and making use of both  bootstrap assumptions {\bf BA-D} {\bf BA-E} on decay and energy,  and integrating by part once the $e_\th$ derivative for the terms involving $e_\th(\ka)$ and $e_\th(\kab)$ when they contain top order derivatives,          we infer that,
\beaa
\max_{0\leq k\leq k_{small}}\sup_{\Mext}r^2u^{1+\dec}|\dk^ke_4(m)| & \les &\ep_0,\\
\max_{0\leq k\leq k_{large}}\sup_{\Mext}\Big(r^2u^{\frac{1}{2}+\dec}+ru^{1+\dec}\Big)|\dk^ke_4(m)| & \les &\ep_0,
\eeaa
as well as 
\beaa
\max_{0\leq k\leq k_{small}}\sup_{\Mext}u^{2+2\dec}|\dk^ke_3(m)| & \les &\ep_0,\\
\max_{0\leq k\leq k_{large}}\sup_{\Mext}u^{1+\dec}|\dk^ke_3(m)| & \les &\ep_0,
\eeaa
consistent with the statement of the lemma.

{\bf Step 2.} We  derive  the estimates on $\Mint$. According to the  analogue of Proposition \ref{prop:derivativesHawkingmass}  in the  situation of
  the incoming  geodesic foliations of $\Mint$, and proceeding as in Step 1, we easily derive,
\bea\lab{eq:estimatefore4mande3minMintforlemmaHawkingmass}
\max_{0\leq k\leq k_{large}}\sup_{\Mint}\ub^{1+\dec}\Big(|\dk^ke_3(m)|+|\dk^ke_4(m)|\Big) \les\ep^2 \les\ep_0.
\eea

{\bf Step 3.} We  estimate $m-m_0$ in $\Mext$.

 First, recall from Theorem M0 that we have
\bea\lab{eq:consequenceofTheoremM0forproofoflemmaHawkingmass}
\sup_{\CC_1\cup\CCb_1}|m-m_0|\les \ep_0 m_0.
\eea
We start with the control in $\Mext$. Note that  $\Mext$ is covered by integral curves of $e_3$ starting from $\CC_1$. Thus, integrating the $e_3 m$ equation 
and making use of the estimate $\sup_{\CC_1}|m-m_0| \les \ep_0 m_0$  as well as  the fact that $e_3(u)=2$, we  easily deduce  that,
\beaa
\sup_{\Mext}|m-m_0| &\les&\ep_0m_0+\ep^2 \les \ep_0 m_0.
\eeaa 

{\bf Step 4.} 
We   estimate   $|m-m_0|$ on $\TT$.

 In view of our initialization of the  ingoing geodesic foliation of $\Mint$ on $\TT$,
\beaa
\,{}^{(int)}\ka\,{}^{(int)}\kab = \,{}^{(ext)}\ka\,{}^{(ext)}\kab\textrm{ on }\TT. 
\eeaa
Since the spheres of both foliations agree on $\TT$, we infer from the definition of the Hawking mass,
\beaa
\,{}^{(int)}m=\,{}^{(ext)}m\textrm{ on }\TT.
\eeaa
 Using  the estimate for  $\,{}^{(ext)}m$   we infer that
\beaa
\sup_{\TT}|\,{}^{(int)}m-m_0| &\les&\ep_0 m_0.
\eeaa

{\bf Step 5.}  We estimate  $|m-m_0|$ on $\Mint$.

Note first that  in $\Mint$,
\beaa
e_3(r) + 1 &=& \frac{r}{2}\overline{\kab}+1= \frac{r}{2}\left(\overline{\kab}+\frac{2}{r}\right).
\eeaa
Thus, in view  of the estimate for  $\overline{\kab}+\frac{2}{r} $ derived in Lemma \ref{lemma:estimatesonceandforallforaverages}
\beaa
\sup_{\Mint}\left| e_3(r) +1\right| \les \ep^2.
\eeaa
Thus integrating  the estimate \eqref{eq:estimatefore4mande3minMintforlemmaHawkingmass} 
  in $r\in[ 2m_0(1-\deh), \rr]$, where we recall that $\rr\leq 2m_0(1+2\deh)$, we derive,
\beaa
\sup_{\Mint}|m-m_0| &\les&\ep_0m_0.
\eeaa
Since $\MM=\Mext\cup\Mint$ we infer that,
\beaa
\sup_{\MM}|m-m_0| &\les&\ep_0m_0.
\eeaa 
This concludes the proof of Lemma \ref{lemma:estimatesonceandforallforHawkingmass}.


\section{Control of coordinates systems}\lab{section:proofofprop:finalcoordinatessystem}


The goal of this section is to prove Propositions \ref{prop:finalcoordinatessystem} and \ref{prop:finalcoordinatessystem:bis}. In both cases, the first two claims, on the form of the spacetime metric in the corresponding coordinates system as well as on the expression of the coordinates vectorfield with respect to the null frame $(e_4, e_3, e_\th)$, is already proved in Propositions \ref{Prop:coordin(u,r,th)} and \ref{Prop:coordin(ub,r,th)}. So we only focus on the third claim, i.e. on estimating $\Obc$, $\Omc$, $\vsi$, $\vsib$, $\ga$, $b$, $\underline{b}$ and $e^\Phi$. The proof of Propositions \ref{prop:finalcoordinatessystem} and \ref{prop:finalcoordinatessystem:bis} thus reduces to the proof of the following lemma.

\begin{lemma}
Let $\th\in[0, \pi]$ be the $\Z$-invariant scalar on $\MM$  defined by \eqref{def:definitionofthetausedatnullinfinity:originalplaceappearing}, i.e.
\bea\lab{def:definitionofthetausedatnullinfinity:bisrepetita}
\th= \cot^{-1}\left(re_\th(\Phi)\right).
\eea
Let
\bea
b=e_4(\th), \qquad \underline{b}=e_3(\th),\qquad \ga=\frac{1}{(e_\th(\th))^2}.
\eea
Then, we have
\beaa
 \max_{0\leq k\leq k_{small}}\sup_{\Mext}\Big(ru^{\frac{1}{2}+\dec}+u^{1+\dec}\Big)\left(\left|\dk^k\left(\frac{\ga}{r^2}-1\right)\right|+r\left|\dk^kb\right|\right) &\les& \ep,\\
 \max_{0\leq k\leq k_{small}}\sup_{\Mext}u^{1+\dec}\left(\left|\dk^k\Obc\right|+\left|\dk^k(\vsi-1)\right|+r\left|\dk^k\underline{b}\right|\right) &\les& \ep,\\
 \max_{0\leq k\leq k_{small}}\sup_{\Mint}\ub^{1+\dec}\left(\left|\dk^k\Omc\right|+\left|\dk^k(\vsib-1)\right|+\left|\dk^k\left(\frac{\ga}{r^2}-1\right)\right|+\left|\dk^kb\right|+\left|\dk^k\underline{b}\right|\right) &\les& \ep.
 \eeaa
  Also, $e^\Phi$ satisfies 
 \beaa
 \max_{0\leq k\leq k_{small}}\sup_{\Mext}\Big(ru^{\frac{1}{2}+\dec}+u^{1+\dec}\Big)\left|\dk^k\left(\frac{e^\Phi}{r\sin\th}-1\right)\right| &\les& \ep,\\
 \max_{0\leq k\leq k_{small}}\sup_{\Mint}\ub^{1+\dec}\left|\dk^k\left(\frac{e^\Phi}{r\sin\th}-1\right)\right| &\les& \ep.
 \eeaa 
\end{lemma}

\begin{proof}
We prove the estimates in $\Mext$. The proof in $\Mint$ is similar and left to the reader. 

{\bf Step 1.} We start with the estimate for $\Obc$. Recall that 
\beaa
\dds_1\Obc = \xib
\eeaa
so that the bootstrap assumptions for $\xib$ imply on any 2-sphere of the foliation of $\Mext$ and for any $k\leq k_{small}$
\beaa
r^{\frac{1}{2}}\|\dk^kr\dds_1\Obc\|_{L^4(S)}+\|\dk^kr\dds_1\Obc\|_{L^2(S)} \les  r^2\sup_S|\dk^k\xib| \les \ep ru^{-1-\dec}.
\eeaa
In view of the commutation formulas of Lemma \ref{Le:comme3e4-outgeodesic} and of Proposition \ref{prop:DDd--ddd}, together with the bootstrap assumptions, we infer any $k\leq k_{small}$, schematically, 
\beaa
[\dk^k, r\dds_1] &=& O(\ep)\dk^{\leq k}+O(1)\dk^{\leq k-1},
\eeaa
and hence, 
\beaa
 r^{\frac{1}{2}}\|r\dds_1\dk^k\Obc\|_{L^4(S)}+\|r\dds_1\dk^k\Obc\|_{L^2(S)} &\les&  \ep ru^{-1-\dec}+\ep\|\dk^{\leq k}\Obc\|_{L^2(S)}+\ep r^{\frac{1}{2}}\|\dk^{\leq k}\Obc\|_{L^4(S)}\\
 &&+\|\dk^{\leq k-1}\Obc\|_{L^2(S)}+r^{\frac{1}{2}}\|\dk^{\leq k-1}\Obc\|_{L^4(S)}\\
 &\les&  \ep ru^{-1-\dec}+\ep\|r\dds_1\dk^{\leq k}\Obc\|_{L^2(S)}+\ep\|\dk^{\leq k}\Obc\|_{L^2(S)}\\
 &&+\|\dk^{\leq k-1}\Obc\|_{L^2(S)}+\|\dk^{\leq k}\Obc\|_{L^2(S)}^{\frac{1}{2}}\|\dk^{\leq k-1}\Obc\|_{L^2(S)}^{\frac{1}{2}},
\eeaa
where we used Gagliardo-Nirenberg on $S$. Together with the Poincar\'e inequality of Corollary \ref{Corollary:proposition-Poincaree} for $\dds_1$, we deduce
\beaa
 r^{\frac{1}{2}}\|r\dds_1\dk^k\Obc\|_{L^4(S)}+\|r\dds_1\dk^k\Obc\|_{L^2(S)}+\|\dk^k\Obc\|_{L^2(S)}  &\les& \ep ru^{-1-\dec}+ \|\dk^{\leq k-1}\Obc\|_{L^2(S)}.
\eeaa
By iteration, and using again Gagliardo-Nirenberg on $S$, we infer on any 2-sphere of the foliation of $\Mext$ and for any $k\leq k_{small}$
\beaa
 \|r\dds_1\dk^k\Obc\|_{L^4(S)}+\|\dk^k\Obc\|_{L^4(S)}  &\les& \ep r^{\frac{1}{2}}u^{-1-\dec},
\eeaa
and thus, by Sobolev embedding 
\beaa
\max_{0\leq k\leq k_{small}}\sup_{\Mext}u^{1+\dec}|\dk^k\Obc| &\les& \ep
\eeaa
which is the desired estimate for $\Obc$.

{\bf Step 2.} Next, we estimate $\vsi$. First, recall that we have
\beaa
  e_\th(\log\vsi) &= \eta-\ze.
\eeaa
Since the bootstrap assumptions for $\eta-\ze$ are at least as good as for $\xib$, we obtain, arguing as in Step 1 the following analog of the above estimate for $\Obc$
\beaa
\max_{0\leq k\leq k_{small}}\sup_{\Mext}u^{1+\dec}|\dk^k\check{\vsi}| &\les& \ep.
\eeaa

Now that we control $\check{\vsi}$, we turn to the estimate for $\vsi$. First, recall from the GCM on $\Sigma_*$ that we have
\beaa
u+r=c_{\Si_*}\textrm{ and }a\big|_{SP}=-1-\frac{2m}{r},\textrm{ where }\nu=e_3+ae_4\textrm{ and }\nu\textrm{ is tangent to }\Sigma_*,
\eeaa
with $c_{\Si_*}$ a constant, and $SP$ denoting the south pole of the spheres of $\Si_*$. We deduce on the south poles of $\Sigma_*$
\beaa
0 &=& \nu(u+r) = e_3(u)+e_3(r)+ae_4(r)=\frac{2}{\vsi}+e_3(r)-\left(1+\frac{2m}{r}\right)e_4(r)
\eeaa
and hence
\beaa
\frac{2}{\vsi} -2 &=& -\frac{r}{2}\left(\left(\ov{\kab}+\frac{2\Up}{r}\right)+\Ab-\left(1+\frac{2m}{r}\right)\left(\ov{\ka}-\frac{2}{r}\right)\right)\textrm{ on }SP\cap\Si_*.
\eeaa
Together with the fact that $\ov{\vsi}=\vsi - \check{\vsi}$, the above control of $\check{\vsi}$, the control of $\ov{\ka}$ and $\ov{\kab}$ provided by Lemma \ref{lemma:estimatesonceandforallforaverages}, the formula for $\Ab$, the control for $\Obc$ in Step 1, the bootstrap assumptions on decay, and the fact that $\ov{\vsi}$ is constant on the sphere,  we infer
\beaa
\max_{0\leq k\leq k_{small}}\sup_{\Si_*}u^{1+\dec}|\dk^k(\ov{\vsi}-1)| &\les& \ep.
\eeaa
Using $\vsi=\ov{\vsi}+\check{\vsi}$ and the above estimates for $\ov{\vsi}$ and $\check{\vsi}$, we obtain
\beaa
\max_{0\leq k\leq k_{small}}\sup_{\Si_*}u^{1+\dec}|\dk^k(\vsi-1)| &\les& \ep.
\eeaa
Finally, recall
\beaa
e_4(\vsi) &=& 0.
\eeaa 
Commuting with $\dk$, using the bootstrap assumptions on decay and the above control for $\vsi-1$ on $\Si_*$, we infer
\beaa
\max_{0\leq k\leq k_{small}}\sup_{\Mext}u^{1+\dec}|\dk^k(\vsi-1)| &\les& \ep.
\eeaa

\begin{remark}
In $\Mint$, we analogously transport $\vsib$ from the timelike hyper surface $\TT$. To estimate $\vsib$ on $\TT$, one uses the following identity (in the frame of $\Mint$)
\beaa
\frac{2}{\vsib} - 1 &=& -\frac{\ov{\ka}+A}{\Up\ov{\kab}}\left(\frac{2}{\vsi}-1\right) -\frac{A}{\Up\ov{\kab}} -\frac{\left(\ov{\ka}-\frac{2\Up}{r}\right)+\Up\left(\ov{\kab}+\frac{2}{r}\right)}{\Up\ov{\kab}} \textrm{ on }\TT.
\eeaa
This identity follows from the definition of $\vsi$ and $\vsib$, the identity for $e_3(r)$ and $e_4(r)$ in $\Mint$, the fact that $\ub=u$ on $\TT$, and that $\TT=\{r=\rh\}$ so that the vectorfield
\beaa
T_\TT = e_4 -\frac{e_4(r)}{e_3(r)}e_3 =e_4 -\frac{\ov{\ka}+A}{\ov{\kab}}e_3
\eeaa
is tangent to $\TT$.
\end{remark}

{\bf Step 3.} We make the auxiliary bootstrap assumption which will be recovered at the end of Step 5
\bea\lab{eq:localbootstrapassumptionintheproofofthecontrolofthecoordinatessystems}
\big| e^\Phi\big|&\leq&   2r, \qquad  \big|  e_\th (e^\Phi)\big| \leq 2. 
\eea

We start with the estimate for $e^\Phi$. Recall from \eqref{eq:computationfoexponentialPhiintermsofafrak} that the following identity holds
\bea\lab{eq:computationfoexponentialPhiintermsofafrak:weuseitnow} 
\frac{e^{\Phi}}{r\sin\th} &=&   \sqrt{1+\af}.
\eea
where $\af$ has been introduced in \eqref{eq:definitionofaf:firsttimeeveritisintroduced} by
\beaa
\af= \frac{e^{2\Phi}}{r^2}+(e_\th(e^\Phi))^2-1. 
\eeaa
In order to estimate $e^\Phi$, it thus suffices to estimate $\af$. 

{\bf Step 4.} Now, recall from Lemma \ref{lemma:transportequationforPhiinthee4direction}
 that $\af$ verifies the following identities on $\Mext$,
\beaa
e_4(\af) &=& \frac{(\check{\ka}-\vth)e^{2\Phi}}{r^2} +2e_\th(e^\Phi) \Big(\b-e_4(\Phi)\ze\Big)e^\Phi,\\
e_\th(\af) &=& 2e_\th(\Phi)e^{2\Phi}\left(\left(\rho+\frac{2m}{r^3}\right)+\frac{1}{4}\left(\ka\kab+\frac{4\Up}{r^2}\right) -\frac{1}{4}\vth\vthb\right),\\
e_3(\af) &=& \frac{\Big(\check{\kab}-\Ab-\vthb\Big)e^{2\Phi}}{r^2} +2e_\th(e^\Phi) \Big(\bb+e_3(\Phi)\ze+\xib e_4(\Phi)\Big)e^\Phi.
\eeaa
Together with our bootstrap assumptions on decay for in $\Mext$ for $\check{\ka}$, $\vth$, $\check{\kab}$, $\vthb$, $\b$, $\bb$, $\rho$, $\ze$, $\xib$ and $\Obc$ and the bootstrap assumption \eqref{eq:localbootstrapassumptionintheproofofthecontrolofthecoordinatessystems}, we infer
\beaa
\max_{1\leq k\leq k_{small}}\sup_{\Mext}\Big(ru^{\frac{1}{2}+\dec}+u^{1+\dec}\Big)\left|\dk^k\af\right| &\les& \ep.
\eeaa
In particular, we deduce,
\beaa
\sup_{\Mext}\Big(ru^{\frac{1}{2}+\dec}+u^{1+\dec}\Big)|\check{\af}| &\les& \ep.
\eeaa

{\bf Step 5.} To  estimate $\ov{\af}$ we make use of  equation  \eqref{regularityofPhiS} according to which
 \beaa
\Big(e_\th(e^\Phi)\Big)^2 = 1\qquad \textrm{ on the axis of symmetry}.
 \eeaa
 Since $e^{2\Phi}$ also vanishes there  we  infer that $\af=0$ on the axis.  Therefore,  on the axis, $ \check{\af}=-\ov{\af}$,   i.e.,
  \beaa
 \ov{\af}&=&- \check{\af}|_{\mbox{axis}}
  \eeaa
  and therefore,
  \beaa
  |\ov{\af}|&\les&|\check{\af}| \les \frac{\ep}{ru^{\frac{1}{2}+\dec}+u^{1+\dec}}.
  \eeaa
  We conclude that, 
\bea
\label{eq:derivatives-of-af}
\max_{0\leq k\leq k_{small}}\sup_{\Mext}\Big(ru^{\frac{1}{2}+\dec}+u^{1+\dec}\Big)\left|\dk^k\af\right| &\les& \ep.
\eea
In view of \eqref{eq:computationfoexponentialPhiintermsofafrak:weuseitnow} and \eqref{eq:derivatives-of-af}, we immediately infer
\beaa
\max_{0\leq k\leq k_{small}}\sup_{\Mext}\Big(ru^{\frac{1}{2}+\dec}+u^{1+\dec}\Big)\left|\dk^k\left(\frac{e^\Phi}{r\sin\th}-1\right)\right| &\les& \ep.
\eeaa
Together with  \eqref{eq:derivatives-of-af} and the definition of $\af$, this implies
\bea\lab{eq:localbootstrapassumptionintheproofofthecontrolofthecoordinatessystems:improved} 
\big| e^\Phi\big|=  (1+O(\ep))r\sin\th\leq\frac{3r}{2}, \quad  \big|  e_\th (e^\Phi)\big| = \sqrt{1 - \frac{e^{2\Phi}}{r^2}+\af} \leq |\cos\th|+O(\ep)\leq \frac{3}{2},
\eea
which is an improvement of the bootstrap assumption \eqref{eq:localbootstrapassumptionintheproofofthecontrolofthecoordinatessystems} which hence holds everywhere on $\Mext$. 

{\bf Step 6.} We now prove the estimates for $b$, $\underline{b}$ and $\ga$. Recall from Lemma \ref{Lemma:choiceof-theta} that $\th$ defined by \eqref{def:definitionofthetausedatnullinfinity:bisrepetita} satisfies 
\beaa
 re_\th(\th)  &=& 1+\frac{r^2(K-\frac{1}{r^2})}{1+(re_\th(\Phi))^2},\\
  e_3(\th)   &=& -\frac{r\bb +\frac{r}{2}\left(-\check{\kab}+\Ab+\vthb\right)e_\th(\Phi)+r\xib e_4(\Phi) +r\ze e_3(\Phi)}{1+(re_\th(\Phi))^2},\\
   e_4(\th)     &=& -\frac{r\b +\frac{r}{2}\left(-\check{\ka} +\vth\right)e_\th(\Phi) -r\ze e_3(\Phi)}{1+(re_\th(\Phi))^2}.  
  \eeaa 
In view of the definition of $b$, $\underline{b}$ and $\ga$, we infer 
  \beaa
  \frac{r}{\sqrt{\ga}}  &=& 1+\frac{r^2(K-\frac{1}{r^2})}{1+(re_\th(\Phi))^2},\\
  \underline{b}   &=& -\frac{r\bb +\frac{r}{2}\left(-\check{\kab}+\Ab +\vthb\right)e_\th(\Phi)+r\xib e_4(\Phi) +r\ze e_3(\Phi)}{1+(re_\th(\Phi))^2},\\
   b     &=& -\frac{r\b +\frac{r}{2}\left(-\check{\ka} +\vth\right)e_\th(\Phi) -r\ze e_3(\Phi)}{1+(re_\th(\Phi))^2}.  
  \eeaa 
  Also, we have in view of the definition of $\af$
  \beaa
  1+(re_\th(\Phi))^2 &=& 1+\frac{(e_\th(e^\Phi))^2}{\frac{e^{2\Phi}}{r^2}}=\frac{r^2}{e^{2\Phi}}(1+\af)
  \eeaa
 and hence
  \beaa
  \frac{r}{\sqrt{\ga}}  &=& 1+\frac{e^{2\Phi}}{r^2}\left(\frac{r^2(K-\frac{1}{r^2})}{1+\af}\right),\\
  \underline{b}   &=& -\frac{e^{2\Phi}}{r^2}\left(\frac{r\bb +\frac{r}{2}\left(-\check{\kab}+\Ab+\vthb\right)e_\th(\Phi)+r\xib e_4(\Phi) +r\ze e_3(\Phi)}{1+\af}\right),\\
   b     &=& -\frac{e^{2\Phi}}{r^2}\left(\frac{r\b +\frac{r}{2}\left(-\check{\ka} +\vth\right)e_\th(\Phi) -r\ze e_3(\Phi)}{1+\af}\right).  
  \eeaa
   
The bootstrap assumptions on decay in $\Mext$ for $\check{\ka}$, $\vth$, $\check{\kab}$, $\vthb$, $\b$, $\bb$, $\ze$, $\xib$ and $\Obc$, the estimate \eqref{eq:derivatives-of-af} for $\af$, the estimate  \eqref{eq:localbootstrapassumptionintheproofofthecontrolofthecoordinatessystems:improved}, and the identity
\beaa
K-\frac{1}{r^2} &=& -\frac{1}{4}\ka\kab+\frac{1}{4}\vth\vthb -\rho -\frac{1}{r^2}\\
&=& -\frac{1}{4}\left(\ka\kab+\frac{4\Up}{r^2}\right) -\left(\rho+\frac{2m}{r^3}\right)+\frac{1}{4}\vth\vthb
\eeaa
imply
\beaa
 \max_{0\leq k\leq k_{small}}\sup_{\Mext}\Big(ru^{\frac{1}{2}+\dec}+u^{1+\dec}\Big)\left(\left|\dk^k\left(\frac{r}{\sqrt{\ga}}-1\right)\right|+r\left|\dk^kb\right|\right) &\les& \ep,
\eeaa
and
\beaa
 \max_{0\leq k\leq k_{small}}\sup_{\Mext}ru^{1+\dec}\left|\dk^k\underline{b}\right| &\les& \ep.
\eeaa
In particular, we also have
\beaa
 \max_{0\leq k\leq k_{small}}\sup_{\Mext}\Big(ru^{\frac{1}{2}+\dec}+u^{1+\dec}\Big)\left|\dk^k\left(\frac{\ga}{r^2}-1\right)\right| &\les& \ep.
\eeaa
These are the desired estimate for $b$, $\underline{b}$ and $\ga$ in $\Mext$. This concludes the proof of the lemma.
\end{proof}

In this section, we also prove two useful lemmas concerning estimates on 2-spheres of $\Mext$ and $\Mint$. 
\begin{lemma}\lab{lemma:clearlyveryusefulforannoyingaxis1}
Let $\th\in[0, \pi]$ be the $\Z$-invariant scalar on $\MM$  defined by \eqref{def:definitionofthetausedatnullinfinity:originalplaceappearing}. Then, we have on $\MM$
\beaa
re_\th(\Phi) &=& \frac{\varpi}{\sin\th}
\eeaa
where $\varpi$ is a reduced 1-scalar satisfying 
\beaa
\sup_{\MM}|\varpi| &\leq& 2.
\eeaa
Also, we have
\beaa
\frac{1}{\sin\th} &\leq& 2|re_\th(\Phi)|+2\textrm{ on }\MM.
\eeaa
\end{lemma}

\begin{proof}
The proof is similar on $\Mext$ and $\Mint$ so we focus on $\Mext$. Recall from \eqref{eq:localbootstrapassumptionintheproofofthecontrolofthecoordinatessystems:improved}  that 
\beaa
\big|  e_\th (e^\Phi)\big| &\leq& \frac{3}{2}.
\eeaa
Furthermore, in view of Proposition \ref{prop:finalcoordinatessystem}, we have in particular
 \beaa
 \sup_{\Mext}\left|\frac{e^\Phi}{r\sin\th}-1\right| &\les& \ep.
 \eeaa
 Since we have
 \beaa
 \varpi &=&  r\sin\th e_\th(\Phi),
 \eeaa
 we deduce
 \beaa
 |\varpi| &=& \frac{r\sin\th}{e^\Phi}|e_\th(e^\Phi)|\leq \frac{3}{2}(1+O(\ep))\leq 2,
 \eeaa
 which is the desired estimate for $\varpi$.
 
 We now consider the upper bound for $(\sin\th)^{-1}$. Recall the definition \eqref{eq:definitionofaf:firsttimeeveritisintroduced} of $\af$ 
\beaa
\af= \frac{e^{2\Phi}}{r^2}+(e_\th(e^\Phi))^2-1.
\eeaa
 We infer
\beaa
r^2e_\th(\Phi)^2  &=& \frac{r^2(e_\th(e^\Phi))^2}{e^{2\Phi}} \\
&=& \frac{1+\af}{\frac{e^{2\Phi}}{r^2}} -1\\
&=& \frac{1+\af -(\sin\th)^2\left(1+\left(\frac{e^{2\Phi}}{r^2(\sin\th)^2}-1\right)\right)}{(\sin\th)^2\left(1+\left(\frac{e^{2\Phi}}{r^2(\sin\th)^2}-1\right)\right)}\\
&=& \frac{(\cos\th)^2+\af -(\sin\th)^2\left(\frac{e^{2\Phi}}{r^2(\sin\th)^2}-1\right)}{(\sin\th)^2\left(1+\left(\frac{e^{2\Phi}}{r^2(\sin\th)^2}-1\right)\right)}
\eeaa
and hence
\beaa
\sin\th|re_\th(\Phi)|  &=& \frac{\sqrt{(\cos\th)^2+\af -(\sin\th)^2\left(\frac{e^{2\Phi}}{r^2(\sin\th)^2}-1\right)}}{\sqrt{1+\left(\frac{e^{2\Phi}}{r^2(\sin\th)^2}-1\right)}}.
\eeaa
 Now, in view of \eqref{eq:derivatives-of-af}, $\af$ satisfies in particular 
\beaa
\sup_{\Mext}\left|\af\right| &\les& \ep.
\eeaa
Together with 
 \beaa
\sup_{\Mext}\left|\frac{e^\Phi}{r\sin\th}-1\right| &\les& \ep,
 \eeaa
we infer
 \beaa
\sin\th|re_\th(\Phi)|  &=& \frac{\sqrt{(\cos\th)^2+O(\ep)}}{\sqrt{1+O
(\ep)}}.
\eeaa
Thus, we deduce 
\beaa
\sin\th|re_\th(\Phi)| &\geq& \frac{\sqrt{2}}{2}(1+O(\ep))\geq \frac{1}{2}\textrm{ for }0\leq \th\leq \frac{\pi}{4}\textrm{ and }\frac{3\pi}{4}\leq\th\leq\pi.
\eeaa
On the other hand, we have
\beaa
\sin\th\geq \frac{\sqrt{2}}{2}\textrm{ on }\frac{\pi}{4}\leq\th\leq\frac{3\pi}{4}
\eeaa
and hence
\beaa
\frac{1}{\sin\th} &\leq& 2|re_\th(\Phi)|+2\textrm{ on }0\leq \th\leq \pi
\eeaa
which is the desired estimate. This concludes the proof of the lemma.
\end{proof}
 
\begin{lemma}\lab{lemma:clearlyveryusefulforannoyingaxis2}
Let $\th\in[0, \pi]$ be the $\Z$-invariant scalar on $\MM$  defined by \eqref{def:definitionofthetausedatnullinfinity:originalplaceappearing}. Then, for any reduced 1-scalar $h$, we have on any 2-sphere $S$ on $\Mext$ and of $\Mint$
\beaa
\sup_S\frac{|h|}{e^\Phi} \les r^{-1}\sup_S(|h|+|\dkb h|)\quad\textrm{ and }\quad\left\|\frac{h}{e^\Phi}\right\|_{L^2(S)}\les r^{-1}\|h\|_{\hk_1(S)}.
\eeaa
\end{lemma}

\begin{proof}
The proof is similar on $\Mext$ and $\Mint$ so we focus on $\Mext$. Recall that the 2-surface $S$ is parametrized by the coordinate $\th\in[0,\pi]$, and that the axis corresponds to the 2 poles $\th=0$ and $\th=\pi$. In view of 
 \beaa
\sup_{\Mext}\left|\frac{e^\Phi}{r\sin\th}-1\right| &\les& \ep,
 \eeaa 
we have
\beaa
\sup_{S\cap\{\frac{\pi}{4}\leq \th\leq \frac{3\pi}{4}\}}\frac{|h|}{e^\Phi} \les r^{-1}\sup_S|h|\quad\textrm{ and }\quad\left\|\frac{h}{e^\Phi}\right\|_{L^2(S\cap\{\frac{\pi}{4}\leq \th\leq \frac{3\pi}{4}\})}\les r^{-1}\|h\|_{L^2(S)}
\eeaa
which is the desired estimate for $\pi/4\leq\th\leq 3\pi/4$.

It remains to consider the portions $0\leq \th\leq \pi/4$ and $3\pi/4\leq\th\leq \pi$ of $S$. These regions can be treated analogously, so we focus on $0\leq\th\leq \pi/4$. Recall  from Remark \ref{rem:vanishingontheaxisofreducesscalars} that any  reduced scalar in  $\mathfrak{s}_k$, 
   for $k\ge 1 $,  must vanish on the axis of symmetry of $\Z$, i.e. at the two poles. In particular, $h$ must vanish at $\th=0$. We deduce
\beaa
\frac{h}{e^\Phi} = \frac{he^\Phi}{e^{2\Phi}}=\frac{\int_0^\th\pr_\th(e^\Phi h)}{e^{2\Phi}}=\frac{\int_0^\th\sqrt{\ga^\S}e_\th(e^\Phi h)}{e^{2\Phi}}=\frac{\int_0^\th\sqrt{\ga}e^\Phi\ddd_1h}{e^{2\Phi}}.
\eeaa
Since we have $|\ga|\les r$, we infer
\beaa
\frac{|h|}{e^\Phi} &\les& \frac{\int_0^\th e^\Phi|\dkb h|}{e^{2\Phi}}
\eeaa
and since
 \beaa
\sup_{\Mext}\left|\frac{e^\Phi}{r\sin\th}-1\right| &\les& \ep,
 \eeaa
we deduce
\beaa
\frac{|h|}{e^\Phi} &\les& r^{-1}\frac{\int_0^\th \sin(\th')|\dkb h|d\th'}{(\sin\th)^2}.
\eeaa
This yields
\beaa
\sup_{S\cap\{0\leq \th\leq\frac{\pi}{4}\}}\frac{|h|}{e^\Phi} \les r^{-1}\sup_S|\dkb h|
\eeaa
which is the desired sup norm estimate for $0\leq\th\leq\pi/4$.

It remains to control the $L^2$-norm on $0\leq\th\leq\pi/4$. We have in view of the above
\beaa
\left\|\frac{h}{e^\Phi}\right\|_{L^2(S\cap\{0\leq \th\leq\frac{\pi}{4}\})}^2&\les& r^{-2}\int_0^{\frac{\pi}{4}}\frac{\left(\int_0^\th \sin(\th')|\dkb h|d\th'\right)^2}{(\sin\th)^4}e^\Phi d\th\\
&\les& r^{-1}\int_0^{\frac{\pi}{4}}\left(\int_0^\th (\sin(\th'))^2|\dkb h|^2d\th'\right)\frac{d\th}{(\sin\th)^2}\\
&\les& r^{-1}\int_0^{\frac{\pi}{4}}(\sin\th)^2|\dkb h|^2\left(\int_\th^{\frac{\pi}{4}}\frac{d\th'}{(\sin(\th'))^2}\right)d\th\\
&\les& r^{-1}\int_0^{\frac{\pi}{4}}|\dkb h|^2\sin\th d\th\\
&\les& r^{-2}\int_0^{\frac{\pi}{4}}|\dkb h|^2e^\Phi d\th\\
&\les& r^{-2}\|\dkb h\|_{L^2(S)}^2
\eeaa
and hence 
\beaa
\left\|\frac{h}{e^\Phi}\right\|_{L^2(S\cap\{0\leq \th\leq\frac{\pi}{4}\})}\les r^{-1}\|\dkb h\|_{L^2(S)}
\eeaa
which is the desired $L^2(S)$ estimate for $0\leq\th\leq\pi/4$. This concludes the proof of the lemma.
\end{proof}


\section{Pointwise bounds for high order derivatives}


The goal of this section is to prove Proposition \ref{prop:pointwiseboundsforhighorderderivatives}. We deal first with the region $r\leq 4m_0$ as follows
\begin{enumerate}
\item The curvature components and Ricci coefficients satisfy in view of the bootstrap assumptions on energy
\beaa
\max_{k\leq k_{large}}\int_{\Mint}\Big(|\Rc|^2+|\Gac|^2\Big)+\max_{k\leq k_{large}-1}\int_{\Mext(r\leq 4m_0)}\Big(|\Rc|^2+|\Gac|^2\Big) &\leq&\ep^2.
\eeaa

\item We first take the trace on the ingoing null cones foliating $\Mint$ and the outgoing null cones foliating $\Mext(r\leq 4m_0)$ which looses one derivative. We thus obtain
\beaa
\max_{k\leq k_{large}-1}\sup_{1\leq \ub\leq u_*}\int_{\CC_{\ub}}\Big(|\Rc|^2+|\Gac|^2\Big)+\max_{k\leq k_{large}-2}\sup_{1\leq u\leq u_*}\int_{\CC_u(r\leq 4m_0)}\Big(|\Rc|^2+|\Gac|^2\Big) &\les&\ep^2.
\eeaa

\item We then take the trace on the 2-spheres $S$ foliation the null cones in $\Mint$ and $\Mext(r\leq 4m_0)$ to infer
\beaa
\max_{k\leq k_{large}-2}\sup_{\Mint}\Big(\|\Rc\|_{L^2(S)}+\|\Gac\|_{L^2(S)}\Big)+\max_{k\leq k_{large}-3}\sup_{\Mext(r\leq 4m_0)}\Big(|\Rc|^2+|\Gac|^2\Big) &\les&\ep.
\eeaa

\item Finally, using the Sobolev embedding on the 2-sphere $S$, which looses 2 derivatives, we deduce 
\beaa
\max_{k\leq k_{large}-4}\sup_{\Mint}\Big(|\Rc|+|\Gac|\Big)+\max_{k\leq k_{large}-5}\sup_{\Mext(r\leq 4m_0)}\Big(|\Rc|+|\Gac|\Big) &\les&\ep,
\eeaa
which is the desired estimate in the region $\Mint\cup\Mext(r\leq 4m_0)$.
\end{enumerate}

It remains to consider the region $\Mext(r\geq 4m_0)$. We proceed as follows

{\bf Step 1}. The Ricci coefficients satisfy in view of the bootstrap assumptions on energy
\beaa
&&\max_{k\leq k_{large}}\int_{\Sigma_*}\Bigg[r^2\Big((\dk^{\leq k}\vth)^2+(\dk^{\leq k}\check{\ka})^2+(\dk^{\leq k}\ze)^2+(\dk^{\leq k}\check{\kab})^2\Big)+(\dk^{\leq k}\vthb)^2\\
\nn&+&(\dk^{\leq k}\eta)^2+(\dk^{\leq k}\check{\omb})^2+(\dk^{\leq k}\xib)^2\Bigg]\\
&+& \sup_{\la\geq 4m_0}\Bigg(\int_{\{r=\la\}}\Bigg[\la^2\Big((\dk^{\leq k}\vth)^2+(\dk^{\leq k}\check{\ka})^2+(\dk^{\leq k}\ze)^2\Big)\\
\nn&+&\la^{2-\dt}(\dk^{\leq k}\check{\kab})^2+(\dk^{\leq k}\vthb)^2+(\dk^{\leq k}\eta)^2+(\dk^{\leq k}\check{\omb})^2+\la^{-\dt}(\dk^{\leq k}\xib)^2\Bigg]\Bigg) \leq\ep^2.
\eeaa
We take the trace on the 2-spheres $S$ foliating the timelike cylinders $\{r=r_0\}$, for $r_0\geq 4m_0$, which looses a derivative, and infer in particular
\beaa
 \max_{k\leq k_{large}-1}\sup_{\Mext(r\geq 4m_0)}&&\Big\{r\big(\|\dk^k\check{\ka}\|_{L^2(S)}+\|\dk^k\ze\|_{L^2(S)}+\|\dk^k\vth\|_{L^2(S)}\big)+r^{1-\frac{\dt}{2}}\|\dk^k\check{\kab}\|_{L^2(S)}\\
 &&+\|\dk^k\eta\|_{L^2(S)}+\|\dk^k\vthb\|_{L^2(S)}+\|\dk^k\check{\omb}\|_{L^2(S)}+r^{-\frac{\dt}{2}}\|\dk^k\xib\|_{L^2(S)}\Big\} \les\ep.
\eeaa
Also, we take the trace on the 2-spheres $S$ foliating the spacelike hyper surface $\Sigma_*$, which looses a derivative, and infer in particular
\beaa
 \max_{k\leq k_{large}-1}\sup_{\Sigma_*}r\|\dk^k\check{\kab}\|_{L^2(S)} &\les& \ep.
\eeaa

{\bf Step 2}. On can easily prove the following trace theorem
\beaa
\max_{k\leq k_{large}-1}\left(\sup_{r\geq 4m_0}r^{5+\dt}\int_S(\dk^k\a)^2 \right) &\les& \sup_{1\leq u\leq u_*}\int_{\CC_u}r^{4+\dt}(\dk^{\leq k_{large}}\a)^2,
\eeaa
which together with the bootstrap assumptions on energy for $\a$ in $\Mext(r\geq 4m_0)$ implies
\beaa
\max_{k\leq k_{large}-1}\left(\sup_{r\geq 4m_0}r^{5+\dt}\int_S(\dk^k\a)^2 \right) &\les&  \ep^2.
\eeaa

{\bf Step 3}. Using the trace theorem 
\beaa
\max_{k\leq k_{large}-1}\left(\sup_{r\geq 4m_0}r^{5}\int_S(\dk^k\b)^2 \right) &\les& \sup_{1\leq u\leq u_*}\int_{\CC_u}r^{4}(\dk^{\leq k_{large}}\b)^2,
\eeaa
we infer, together with the bootstrap assumptions on energy for $\b$ in $\Mext(r\geq 4m_0)$,
 \bea\lab{eq:firstnonsharppointwiseestimateforbetainMext}
\max_{k\leq k_{large}-1}\left(\sup_{r\geq 4m_0}r^{5}\int_S(\dk^k\b)^2 \right) &\les&  \ep^2.
\eea

The power of $r$ of the above estimate is not strong enough. To upgrade the estimate, recall that we have the Bianchi identity
\beaa
e_4(\b) +2\ka\b &=& \ddd_2\a+\ze\a.
\eeaa
This yields
\beaa
e_4\left(r^{5+\dt}\int_S\b^2\right) &=& \int_Sr^{5+\dt}\left(2\b e_4(\b)+\ka\b^2+b\frac{e_4(r)}{r}\b^2\right)\\
&=& \int_Sr^{5+\dt}\left(-\frac{1-\dt}{2}\ka\b^2+2\b(r^{-1}\dkb\a+\ze\a)-\frac{5+\dt}{2}\check{\ka}\b^2\right)
\eeaa
and hence
\beaa
e_4\left(r^{5+\dt}\int_S\b^2\right) +\frac{1-\dt}{2}\int_Sr^{5+\dt}\ka\b^2 &=& \int_Sr^{4+\dt}\left(2\b(\dkb\a+r\ze\a)-\frac{5+\dt}{2}\check{\ka}\b^2\right)\\
&\les& \left(\int_Sr^{4+\dt}(\dk^{\leq 1}\a)^2\right)^{\frac{1}{2}}\left(\int_Sr^{4+\dt}\b^2\right)^{\frac{1}{2}}+\ep\int_Sr^{4+\dt}\b^2
\eeaa
where we used the pointwise estimates of Step 1 for $\kac$ and $\ze$. We infer
\beaa
e_4\left(r^{5+\dt}\int_S\b^2\right) +\int_Sr^{4+\dt}\b^2 &\les& \int_Sr^{4+\dt}(\dk^{\leq 1}\a)^2.
\eeaa
Integrating, from $r\geq 6m_0$, we deduce
\beaa
\sup_{r\geq 6m_0}r^{5+\dt}\int_S\b^2+\sup_{1\leq u\leq u_*}\int_{\CC_u(r\geq 6m_0)}r^{4+\dt}\b^2 &\les& \sup_{1\leq u\leq u_*}\int_{\CC_u}r^{4+\dt}(\dk^{\leq 1}\a)^2+\int_{S_{r=6m_0}}\b^2\\
&\les& \ep^2,
\eeaa
where we used the bootstrap assumptions on energy for $\a$ in $\Mext(r\geq 4m_0)$ and the non sharp estimate \eqref{eq:firstnonsharppointwiseestimateforbetainMext} for $\b$. Using again \eqref{eq:firstnonsharppointwiseestimateforbetainMext}, we obtain
\beaa
\sup_{r\geq 4m_0}r^{5+\dt}\int_S\b^2 + \sup_{1\leq u\leq u_*}\int_{\CC_u(r\geq 4m_0)}r^{4+\dt}\b^2 &\les& \ep^2.
\eeaa

To discuss higher order derivatives, recall from Lemma \ref{Le:comme3e4-outgeodesic}  the following commutator, written in schematic form,
\beaa
 \,[\dkb, e_4] &= (\kac, \vth)\dkb +(\ze, r\b).
\eeaa
Also, recall from Lemma \ref{lemma:commTwithe3e4} the following commutator,
\beaa
\,[\T, e_4]&=\left(\left( \omb-\frac{m}{r^2} \right)     -\frac{m}{2r} \left( \ov{\ka} -\frac{2}{r}\right)+\frac{e_4 (m)}{r}  \right)e_4+(\eta+\ze) e_\th.
\eeaa
In view of the estimates of Step 1 for $k_{large}-1$ derivatives of $\kac$, $\vth$, $\ze$, $\eta$, $\ombc$, the pointwise estimates for $\b$ in \eqref{eq:firstnonsharppointwiseestimateforbetainMext}, the control of $\ov{\ka}$ in Lemma \ref{lemma:estimatesonceandforallforaverages}, and the control of $e_4(m)$ in Lemma \ref{lemma:estimatesonceandforallforHawkingmass}, we infer, schematically,
\beaa
\left\|\dk^k\Big(\,[\dkb, e_4]\b, \,\,[\T, e_4]\b\Big)\right\|_{L^2(S)} &\les& O(\ep r^{-2})\|\dk^{\leq k+1}\b\|_{L^2(S)}\textrm{ for }k\leq k_{large}-2.
\eeaa
Thus, commuting the Bianchi identity for $e_4(\b)$ with $\T$ and $\dkb$ together with the above commutator estimate, using the Bianchi identity to recover the $e_4$ derivatives, we obtain for higher order derivatives  
\beaa
&&\max_{k\leq k_{large}-1}\left(\sup_{r\geq 4m_0}r^{5+\dt}\int_S(\dk^k\b)^2 + \sup_{1\leq u\leq u_*}\int_{\CC_u(r\geq 4m_0)}r^{4+\dt}(\dk^k\b)^2\right)\\
 &\les& \sup_{1\leq u\leq u_*}\int_{\CC_u(r\geq 4m_0)}r^{4+\dt}(\dk^{\leq k_{large}}\a)^2\\
&\les& \ep^2.
\eeaa

{\bf Step 4}. Recall from Proposition \ref{propos:transportaverages} that we have
\beaa
e_4\check{\rho}+\frac  3 2 \ov{\ka} \check{\rho}+\frac 3 2 \ov{\rho}\check{\ka}&=&\ddd_1\b+\err[e_4\check{\rho}],\\
\err[e_4\check{\rho}]&=&-\frac 3 2 \check{\ka}\check{\rho} +\frac 12 \ov{ \check{\ka}\check{\rho}} - \left(\frac{1}{2}\vthb\a +\ze\b\right)+\ov{\left(\frac{1}{2}\vthb\a +\ze\b\right)}.
\eeaa
This yields
\beaa
e_4\left(r^{4}\int_S(\check{\rho})^2\right) &=& \int_Sr^{4}\left(2\rhoc e_4(\rhoc)+\ka\rhoc^2+4\frac{e_4(r)}{r}\rhoc^2\right)\\
&=& \int_Sr^{4}\Big(-3 \ov{\rho}\check{\ka}\rhoc+2\rhoc(r^{-1}\dkb\b+\err[e_4\check{\rho}])+\kac\rhoc^2\Big)
\eeaa
and hence
\beaa
e_4\left[\left(r^{4}\int_S(\check{\rho})^2\right)^{\frac{1}{2}}\right] &\les& \left[\int_Sr^4\Big((\ov{\rho}\check{\ka})^2+(r^{-1}\dkb\b)^2+(\err[e_4\check{\rho}])^2+\kac^2\rhoc^2\Big)\right]^\frac{1}{2}
\eeaa
Using the estimates of Step 1, 2 and 3 for $\kac$, $\ze$, $\vthb$, $\a$ and $\b$, and the control of $\ov{\rho}$ in Lemma  \ref{lemma:estimatesonceandforallforaverages}, we infer
\beaa
e_4\left[\left(r^{4}\int_S(\check{\rho})^2\right)^{\frac{1}{2}}\right] &\les& \frac{\ep}{r^{\frac{3}{2}+\frac{\dt}{2}}}+\frac{\ep}{r^2}\left(r^{4}\int_S(\check{\rho})^2\right)^{\frac{1}{2}}.
\eeaa
Integrating from $r=4m_0$, we control $\|\rhoc\|_{L^2(S)}$ from the control in $r\leq 4m_0$, we infer
\beaa
\sup_{r\geq 4m_0}r^{4}\int_S\rhoc^2  &\les& \ep^2.
\eeaa

Next, commuting the equation for $e_4(\rhoc)$ with $\T$ and $\dkb$ together with the commutator estimate of Step 3, using the equation for $e_4(\rhoc)$ to recover the $e_4$ derivatives, we obtain similarly for higher order derivatives  
\beaa
\max_{k\leq k_{large}-2}\sup_{r\geq 4m_0}r^{4}\int_S(\dk^k\rhoc)^2  &\les&  \ep^2.
\eeaa
 
{\bf Step 5}. Recall from Proposition \ref{propos:transportaverages} that we have the following transport equations in the  $e_4$ direction,
\beaa
e_4\check{\kab} +\frac1 2 \ov{\ka} \check{\kab}+\frac 1 2 \check{\ka} \ov{\kab} &=&-2 \ddd_1\ze+2\check{\rho}+\err[e_4\check{\kab}],\\
\err[e_4\check{\kab}]&=&-\frac 1 2 \check{\ka}\check{\kab}-\frac 1 2 \ov{\check{\ka}\check{\kab}} +\left( -\frac 1 2  \vth\vthb +2\ze^2\right)        -\ov{\left( -\frac 1 2  \vth\vthb +2\ze^2\right) }.
\eeaa
This yields
\beaa
e_4\left(r\int_S(\check{\kab})^2\right) &=& \int_Sr\left(2\kabc e_4(\kabc)+\ka\kabc^2+\frac{e_4(r)}{r}\kabc^2\right)\\
&=& \int_Sr\left(2\kabc\left(-\frac 1 2 \check{\ka} \ov{\kab}-2 \ddd_1\ze+2\check{\rho}+\err[e_4\check{\kab}]\right)+\kac\kabc^2\right)
\eeaa
and hence, using the the estimates of Step 1 and 4 for $\kac$, $\ze$, $\vthb$ and $\rhoc$, and the control of $\ov{\ka}$ and $\ov{\kab}$ in Lemma  \ref{lemma:estimatesonceandforallforaverages}, we infer
\beaa
e_4\left(r\int_S(\check{\kab})^2\right)  &\les& \frac{\ep}{r^2}\int_Sr\kabc^2+\ep r^{-\frac{3}{2}}\left(\int_Sr\kabc^2\right)^{\frac{1}{2}}
\eeaa
and hence
\beaa
e_4\left(\left(r\int_S(\check{\kab})^2\right)^{\frac{1}{2}}\right)  &\les& \frac{\ep}{r^2}\left(r\int_S(\check{\kab})^2\right)^{\frac{1}{2}}+\ep r^{-\frac{3}{2}}
\eeaa
Integrating backward from $\Sigma_*$, where $\kabc$ under control in view of Step 1, we infer
\beaa
\sup_{r\geq 4m_0}r^{2}\int_S\kabc^2  &\les& \ep^2.
\eeaa

Next, commuting the equation for $e_4(\kabc)$ with $\T$ and $\dkb$ together with the commutator estimate of Step 3, using the equation for $e_4(\kabc)$ to recover the $e_4$ derivatives, we obtain similarly for higher order derivatives  
\beaa
\max_{k\leq k_{large}-2}\sup_{r\geq 4m_0}r^{4}\int_S(\dk^k\kabc)^2  &\les&  \ep^2.
\eeaa

{\bf Step 6}. In view of Codazzi for $\vthb$, and the estimates of Step 1 on $\ze$, and $\vthb$ and of Step 3 on $\kabc$ in $\Mext(r\geq 4m_0)$, we infer
\beaa
\max_{k\leq k_{large}-2}\sup_{\Mext(r\geq 4m_0)}r\|\dk^k\bb\|_{L^2(S)} &\les& \ep.
\eeaa

{\bf Step 7}. In view of the null structure equation for $e_3(\kab)$, and the estimates of Step 1 on $\ombc$, $\ze$, $\eta$ and $\vthb$, and of Step 3 on $\kabc$ in $\Mext(r\geq 4m_0)$, we infer
\beaa
\max_{k\leq k_{large}-3}\sup_{\Mext(r\geq 4m_0)}\|\dk^k\xib\|_{L^2(S)} &\les& \ep.
\eeaa

{\bf Step 8}. In view of the Bianchi identity for $e_3(\bb)$, and the estimates of Step 1 on $\ombc$, $\ze$, and $\eta$, the estimates of Step 2 on $\rhoc$, of Step 3 on $\kabc$ and of Step 5 on $\xib$ in $\Mext(r\geq 4m_0)$, we infer
\beaa
\max_{k\leq k_{large}-3}\sup_{\Mext(r\geq 4m_0)}\|\dk^k\aa\|_{L^2(S)} &\les& \ep.
\eeaa

{\bf Step 9}. Gathering the estimates for Step 1 to Step 8, we have obtained
\beaa
&& \max_{k\leq k_{large}-1}\sup_{\Mext(r\geq 4m_0)}\Big\{r^{\frac{5}{2}+\frac{\dt}{2}}\big(\|\dk^k\a\|_{L^2(S)}+\|\dk^k\b\|_{L^2(S)}\big)\\
&+&r\big(\|\dk^k\check{\ka}\|_{L^2(S)}+\|\dk^k\ze\|_{L^2(S)}+\|\dk^k\vth\|_{L^2(S)}\big)+\|\dk^k\vthb\|_{L^2(S)}+\|\dk^k\vthb\|_{L^2(S)}+\|\dk^k\check{\omb}\|_{L^2(S)}\Big\}\\
&+& \max_{k\leq k_{large}-2}\sup_{\Mext(r\geq 4m_0)}\Big\{r^2\big(\|\dk^k\mu\|_{L^2(S)}+\|\dk^k\rhoc\|_{L^2(S)}\big)+r\big(\|\dk^k\check{\kab}\|_{L^2(S)}+\|\dk^k\bb\|_{L^2(S)}\big)\Big\}\\
&+& \max_{k\leq k_{large}-3}\sup_{\Mext(r\geq 4m_0)}\Big\{\|\dk\xib\|_{L^2(S)}+\|\dk^k\aa\|_{L^2(S)}\Big\} \les\ep.
\eeaa
Using the Sobolev embedding on the 2-sphere $S$ which looses 2 derivatives, and in view of the previous estimate on $\Mext(r\leq 4m_0)$, we infer
\beaa
\max_{k\leq k_{large}-5}\sup_{\MM}&&\Big\{r^{\frac{7}{2}+\frac{\dt}{2}}\big(|\dk^k\a|+|\dk^k\b|\big)+ r^3\big(|\dk^k\muc|+|\dk^k\rhoc|\big)\\
&&+r^2\big(|\dk^k\check{\ka}|+|\dk^k\ze|+|\dk^k\vth|+|\dk^k\check{\kab}|+|\dk^k\bb|\big)\\
&&+r\big(|\dk^k\vthb|+|\dk^k\vthb|+|\dk^k\check{\omb}|+|\dk\xib|+|\dk^k\aa|\big)\Big\} \les \ep
\eeaa
which is the desired estimate on $\Mext(r\geq 4m_0)$. This concludes the proof of Proposition \ref{prop:pointwiseboundsforhighorderderivatives}.


\section{Proof of Proposition \ref{prop:constructionsecondframeinMext}}


Let $(e_4, e_3, e_\th)$ the outgoing geodesic null frame of $\Mext$. We will exhibit another frame $(e_4', e_3', e_\th')$ of $\Mext$  provided by
\bea\lab{eq:formulachangeofframeforsecondframeMext}
\begin{split}
e_4'&= e_4 + f e_\th +\frac 1 4 f^2  e_3,\\
e_\th'&= e_\th +\frac 1 2 f e_3,\\
e_3'&=  e_3,
\end{split}
\eea
where $f$ is such that 
\bea\lab{eq:propertiesoffforsecondframeofMext}
f=0\textrm{ on }\Sigma_*\cap\CC_*, \quad \eta'=0\textrm{ on }\Sigma_*, \quad \xi'=0\textrm{ on }\Mext.
\eea
The desired estimates for the  Ricci coefficients and curvature components with respect to the new frame $(e_4', e_3', e_\th')$ of $\Mext$ will be obtained using
\begin{itemize}
\item the change of frame formulas of Proposition \ref{prop:transformations1}, applied to the change of frame from $(e_4, e_3, e_\th)$ to $(e_4', e_3', e_\th')$,

\item the estimates for $f$ on $\Mext$,

\item the estimates for the Ricci coefficients and curvature components with respect to the outgoing geodesic frame $(e_4, e_3, e_\th)$ of $\Mext$ provided by the bootstrap assumptions on decay and Proposition \ref{prop:pointwiseboundsforhighorderderivatives}.
\end{itemize}

{\bf Step 1}. We start by deriving an equation for $f$ on $\Mext$. In view of the condition $\xi'=0$ on $\Mext$,  
see \eqref{eq:propertiesoffforsecondframeofMext}, in view of $\xi=\om=0$ and $\etab=-\ze$ satisfied by the outgoing geodesic foliation of $\Mext$, and in view of Lemma \ref{remark:lotxixi'}, we have
\bea\lab{eq:equationfore4primefforsecondframeofMext}
\nn e_4'(f)+\frac{1}{2}\ka f &=& -\frac{1}{2}f\vth -\frac{1}{2}f^2\eta-\frac{3}{2}f^2\ze\\
&&+\frac{1}{8}f^3\kab+\frac{1}{2}f^3\omb+\frac{1}{8}f^3\vthb +\frac{1}{8}f^4\xib\,\,\textrm{ on }\Mext.
\eea

We also derive an equation for $f$ on $\Sigma_*$. In view of the condition $\eta'=0$ on $\Sigma_*$,  
see \eqref{eq:propertiesoffforsecondframeofMext}, and in view of Lemma \ref{remark:lotxixi'}, we have
\bea\lab{eq:equationfore3primefforsecondframeofMext}
e_3'(f) &=& -2\eta  +2f\omb +\frac{1}{2}f^2\xib\,\,\textrm{ on }\Sigma_*.
\eea
Now, since $u+r$ is constant on $\Sigma_*$, the following vectorfield 
\beaa
\nu_{\Si_*}' := e_3'+a'e_4', \qquad a' := -\frac{e_3'(u+r)}{e_4'(u+r)},
\eeaa
is tangent to $\Sigma_*$. We compute in view of the above
\beaa
\nu_{\Si_*}'(f) &=& e_3'(f)+a'e_4'(f)\\
&=& -2\eta  +2f\omb +\frac{1}{2}f^2\xib +a'\Bigg\{-\frac{1}{2}\ka f  -\frac{1}{2}f\vth -\frac{1}{2}f^2\eta-\frac{3}{2}f^2\ze\\
&&+\frac{1}{8}f^3\kab+\frac{1}{2}f^3\omb+\frac{1}{8}f^3\vthb +\frac{1}{8}f^4\xib\Bigg\}.
\eeaa
Using \eqref{eq:formulachangeofframeforsecondframeMext}, we have
\beaa
a' &=& -\frac{e_3'(u+r)}{e_4'(u+r)}\\
&=&  -\frac{e_3(u+r)}{\left(e_4 + f e_\th +\frac 1 4 f^2  e_3\right)(u+r)}\\
&=& -\frac{\frac{2}{\vsi}+\frac{r}{2}(\ov{\kab}+\Ab)}{\frac{r}{2}\ov{\ka}+\frac 1 4 f^2\left(\frac{2}{\vsi}+\frac{r}{2}(\ov{\kab}+\Ab)\right)}
\eeaa
and hence
\bea\lab{eq:equationsatsfiedbyfonSigma*forsecondframeMext}
\nn \nu_{\Si_*}'(f) &=& -2\eta  +2f\omb +\frac{1}{2}f^2\xib  -\frac{\frac{2}{\vsi}+\frac{r}{2}(\ov{\kab}+\Ab)}{\frac{r}{2}\ov{\ka}+\frac 1 4 f^2\left(\frac{2}{\vsi}+\frac{r}{2}(\ov{\kab}+\Ab)\right)}\Bigg\{-\frac{1}{2}\ka f  -\frac{1}{2}f\vth \\
&& -\frac{1}{2}f^2\eta -\frac{3}{2}f^2\ze +\frac{1}{8}f^3\kab+\frac{1}{2}f^3\omb+\frac{1}{8}f^3\vthb +\frac{1}{8}f^4\xib\Bigg\}\,\, \textrm{ on }\Mext. 
\eea

{\bf Step 2}. Next, we estimate $f$ on $\Sigma_*$.  Introducing an integer $k_{loss}$ and a small constant $\de_0>0$ satisfying 
\beaa
16\leq k_{loss} \leq  \frac{\dec}{3}(k_{large}-k_{small}), \qquad \de_0=\frac{k_{loss}}{k_{large}-k_{small}},
\eeaa
we assume the following local bootstrap assumption
\bea\lab{eq:localbootstraponSigma*forsecondframeMext}
|\dk^{\leq k_{small}+k_{loss}+2}f| \leq \frac{\sqrt{\ep}}{ru^{\frac{1}{2}+\dec-2\de_0}}\,\,\,\,\textrm{ on }u_1\leq u\leq u_*
\eea
where 
\beaa
1\leq u_1<u_*.
\eeaa
Since $f=0$ on $\Si_*\cap\CC_*$ in view of  \eqref{eq:propertiesoffforsecondframeofMext}, \eqref{eq:localbootstraponSigma*forsecondframeMext} holds for $u_1$ close enough to $u_*$, and our goal is to prove that we may in fact choose $u_1=1$ and replace $\sqrt{\ep}$ with $\ep$ in \eqref{eq:localbootstraponSigma*forsecondframeMext}. 

In view of  the estimates for the Ricci coefficients and curvature components with respect to the outgoing geodesic frame $(e_4, e_3, e_\th)$ of $\Mext$ provided by Proposition \ref{prop:pointwiseboundsforhighorderderivatives}, \eqref{eq:equationsatsfiedbyfonSigma*forsecondframeMext} yields
\beaa
\nu_{\Si_*}'(f) &=& -2\eta +h, \qquad |\dk^kh| \les r^{-1}(|\dk^{\leq k}f|+|\dk^{\leq k}f|^4)\textrm{ for }k\leq k_{large}-5.
\eeaa
Using commutator identities, using also \eqref{eq:equationfore4primefforsecondframeofMext} and \eqref{eq:equationfore3primefforsecondframeofMext}, and in view of \eqref{eq:localbootstraponSigma*forsecondframeMext}, we infer
\beaa
|\nu_{\Si_*}'(\dkb^kf)| &\les& |\dkb^{\leq k}\eta| + \frac{\sqrt{\ep}}{r^2u^{\frac{1}{2}+\dec-2\de_0}}\textrm{ for }k\leq k_{small}+k_{loss}+2, \,\, u_1\leq u\leq u_*.
\eeaa

Since $f=0$ on $\Si_*\cap\CC_*$ in view of  \eqref{eq:propertiesoffforsecondframeofMext}, and since $\nu_{\Si_*}'$ is tangent to $\Si_*$, we deduce on $\Si_*$, integrating along the integral curve of $\nu_{\Si_*}'$
\beaa
|\dkb^kf| &\les& \int_u^{u_*}|\dkb^{\leq k}\eta|+\frac{\sqrt{\ep}}{u^{\frac{1}{2}+\dec-2\de_0}}\int_u^{u_*}\frac{1}{\nu_{\Si_*}'(u')r^2}\textrm{ for }k\leq k_{small}+k_{loss}+2, \,\, u_1\leq u\leq u_*.
\eeaa
Since 
\beaa
\nu_{\Si_*}'(u) &=& e_3'(u)+a'e_4'(u)\\
&=& e_3(u)  -\frac{\frac{2}{\vsi}+\frac{r}{2}(\ov{\kab}+\Ab)}{\frac{r}{2}\ov{\ka}+\frac 1 4 f^2\left(\frac{2}{\vsi}+\frac{r}{2}(\ov{\kab}+\Ab)\right)}\left(e_4 + f e_\th +\frac 1 4 f^2  e_3\right)u\\
&=& \frac{2}{\vsi}  -\frac{f^2}{2\vsi}\frac{\frac{2}{\vsi}+\frac{r}{2}(\ov{\kab}+\Ab)}{\frac{r}{2}\ov{\ka}+\frac 1 4 f^2\left(\frac{2}{\vsi}+\frac{r}{2}(\ov{\kab}+\Ab)\right)}
\eeaa
we have
\beaa
\nu_{\Si_*}'(u) &=& 2+O(\ep)
\eeaa
and hence
\beaa
|\dkb^kf| &\les& \int_u^{u_*}|\dkb^{\leq k}\eta|+\frac{\sqrt{\ep}}{u^{\frac{1}{2}+\dec-2\de_0}}\int_u^{u_*}\frac{1}{r^2}\textrm{ for }k\leq k_{small}+k_{loss}+2, \,\,u_1\leq u\leq u_*.
\eeaa
Together with the behavior \eqref{eq:behaviorofronSigmastar} of $r$ on $\Sigma_*$, we infer
\beaa
|\dkb^kf| &\les& \int_u^{u_*}|\dkb^{\leq k}\eta|+\frac{\ep}{ru^{\frac{1}{2}+\dec-2\de_0}}\textrm{ for }k\leq k_{small}+k_{loss}+2,\,\, u_1\leq u\leq u_*.
\eeaa

Next, we estimate $\eta$. We have by interpolation, since $k_{loss}\leq k_{large}-k_{small}$,
\beaa
\|\dkb^{\leq k_{small}+k_{loss}+4}\eta\|_{L^2(S)} &\les& \|\dkb^{\leq k_{small}}\eta\|_{L^2(S)}^{1-\frac{k_{loss}+4}{k_{large}-k_{small}}}\|\dkb^{\leq k_{large}}\eta\|_{L^2(S)}^{\frac{k_{loss}+4}{k_{large}-k_{small}}},
\eeaa
and hence, using $\de_0>0$, we have
\beaa
&& \int_{\Si_*(\geq u)}|\dkb^{\leq k_{small}+k_{loss}+4}\eta| \\
&\les& \left(\int_{\Si_*(\geq u)}{u'}^{1+\de_0}|\dkb^{\leq  k_{small}+k_{loss}+4}\eta|^2\right)^{\frac{1}{2}}\\
&\les& \frac{1}{u^{\frac{1}{2}+\dec - 2\de_0}}\left(\int_{\Si_*}{u'}^{2+2\dec}|\dkb^{\leq k_{small}}\eta|^2\right)^{\frac{1}{2}-\frac{k_{loss}+4}{2(k_{large}-k_{small})}}\left(\int_{\Si_*}|\dkb^{\leq k_{large}}\eta|^2\right)^{\frac{k_{loss}+4}{2(k_{large}-k_{small})}}.
\eeaa
where we have used the fact that 
\beaa
\frac{k_{loss}+4}{k_{large}-k_{small}}(1+\dec)+\frac{\de_0}{2} & = & \left(\left(1+\frac{4}{k_{loss}}\right)(1+\dec)+\frac{1}{2}\right)\de_0\leq 2\de_0
\eeaa
and
\beaa
\frac{1}{2}+\dec - 2\de_0 = \frac{1}{2}+\dec -\frac{4k_{loss}}{k_{large}-k_{small}}\geq \dec>0
\eeaa
since $16\leq k_{loss} \leq   \frac{1}{8}(k_{large}-k_{small})$ and $\dec>0$ is small. Now, recall from the bootstrap assumptions on decay and energy for $\eta$ along $\Si_*$ that we have 
\beaa
\int_{\Si_*}u^{2+2\dec}|\dk^{\leq k_{small}}\eta|^2 + \int_{\Si_*}|\dk^{\leq k_{large}}\eta|^2 &\leq& \ep^2. 
\eeaa
We deduce
\beaa
\int_{\Si_*(\geq u)}|\dkb^{\leq k_{small}+k_{loss}+4}\eta| &\les& \frac{\ep}{u^{\frac{1}{2}+\dec - 2\de_0}}.
\eeaa
Together with the Sobolev embedding on the 2-spheres $S$ foliating $\Si_*$, as well as  the behavior \eqref{eq:behaviorofronSigmastar} of $r$ on $\Sigma_*$, we infer 
\beaa
\int_u^{u_*}|\dkb^{\leq k_{small}+k_{loss}+2}\eta| &\les& \frac{\ep}{u^{\frac{1}{2}+\dec-2\de_0}}.
\eeaa
Plugging in the above estimate  for $f$, we infer
\beaa
|\dkb^kf| &\les& \frac{\ep}{ru^{\frac{1}{2}+\dec-2\de_0}}\textrm{ for }k\leq k_{small}+k_{loss}+2,\,\, u_1\leq u\leq u_*.
\eeaa
Together with \eqref{eq:equationfore4primefforsecondframeofMext} and \eqref{eq:equationfore3primefforsecondframeofMext}, we recover $e_4$ and $e_3$ derivatives to deduce
\beaa
|\dk^kf| &\les& \frac{\ep}{ru^{\frac{1}{2}+\dec-2\de_0}}\textrm{ for }k\leq k_{small}+k_{loss}+2,\,\, u_1\leq u\leq u_*.
\eeaa
This is an improvement of the bootstrap assumption \eqref{eq:localbootstraponSigma*forsecondframeMext}. Thus, we may choose $u_1=1$, and $f$ satisfies the following estimate
\beaa
|\dk^kf| &\les& \frac{\ep}{ru^{\frac{1}{2}+\dec-2\de_0}}\textrm{ for }k\leq k_{small}+k_{loss}+2\textrm{ on }\Sigma_*.
\eeaa
Together with \eqref{eq:equationfore3primefforsecondframeofMext}, as well as  the behavior \eqref{eq:behaviorofronSigmastar} of $r$ on $\Sigma_*$, we infer 
\beaa
|\dk^{k-1}e_3'f| &\les& |\dk^{k-1}\eta|+\frac{\ep}{r^2}\\
&\les& \frac{\ep}{ru^{1+\dec-2\de_0}}\textrm{ for }k\leq k_{small}+k_{loss}+2\textrm{ on }\Sigma_*.
\eeaa
Collecting the two above estimates, we obtain
\begin{equation}\lab{eq:controlonSigma*forsecondframeMext}
|\dk^kf| \les \frac{\ep}{ru^{\frac{1}{2}+\dec-2\de_0}}, \quad |\dk^{k-1}e_3'f| \les \frac{\ep}{ru^{1+\dec-2\de_0}}\textrm{ for }k\leq k_{small}+k_{loss}+2\textrm{ on }\Sigma_*.
\end{equation}

{\bf Step 3}. Next, we estimate $f$ on $\Mext$. We assume the following local bootstrap assumption
\bea\lab{eq:localbootstraponMextforsecondframeMext}
|\dk^{\leq k_{small}+k_{loss}+2}f| \leq \frac{\sqrt{\ep}}{ru^{\frac{1}{2}+\dec-2\de_0}+u^{1+\dec-2\de_0}}\,\,\,\,\textrm{ on }r\geq r_1.
\eea
where $r_1\geq 4m_0$. In view of the control of $f$ on $\Si_*$ provided by \eqref{eq:controlonSigma*forsecondframeMext},  \eqref{eq:localbootstraponMextforsecondframeMext} holds for $r_1$ sufficiently large, and our goal is to prove that we may in fact choose $r_1=4m_0$ and replace $\sqrt{\ep}$ with $\ep$ in \eqref{eq:localbootstraponMextforsecondframeMext}. 

Recall \eqref{eq:equationfore4primefforsecondframeofMext}
\beaa
\nn e_4'(f)+\frac{1}{2}\ka f &=& -\frac{1}{2}f\vth -\frac{1}{2}f^2\eta-\frac{3}{2}f^2\ze\\
&&+\frac{1}{8}f^3\kab+\frac{1}{2}f^3\omb+\frac{1}{8}f^3\vthb +\frac{1}{8}f^4\xib\,\,\textrm{ on }\Mext.
\eeaa
In view of  the estimates for the Ricci coefficients and curvature components with respect to the outgoing geodesic frame $(e_4, e_3, e_\th)$ of $\Mext$ provided by Proposition \ref{prop:pointwiseboundsforhighorderderivatives}, 
\beaa
&&\left|\dk^k\left(-\frac{1}{2}f\vth -\frac{1}{2}f^2\eta-\frac{3}{2}f^2\ze+\frac{1}{8}f^3\kab+\frac{1}{2}f^3\omb+\frac{1}{8}f^3\vthb +\frac{1}{8}f^4\xib\right)\right|\\
&\les& \ep r^{-2}u^{-\frac{1}{2}}|\dk^{\leq k}f|+r^{-1}(|\dk^{\leq k}f|^2+|\dk^{\leq k}f|^4)\textrm{ for }k\leq k_{large}-5.
\eeaa
Using commutator identities, using also \eqref{eq:equationfore4primefforsecondframeofMext}, and in view of \eqref{eq:localbootstraponMextforsecondframeMext}, we infer\footnote{Note that 
\beaa
\dec-2\de_0 = \dec - \frac{2k_{loss}}{k_{large}-k_{small}} \geq \frac{\dec}{3}>0
\eeaa
where we have used the definition of $\de_0$ and the upper bound on $k_{loss}$.}
\beaa
e_4'\Big((\dkb, T)^kf\Big) +\frac{1}{2}\ka(\dkb, T)^kf &\leq& \frac{\ep}{r^3u^{1+\dec-2\de_0}}\textrm{ for }k\leq k_{small}+k_{loss}+2,\,\, r\geq r_1.  
\eeaa
Integrating backwards from $\Sigma_*$ where we have \eqref{eq:controlonSigma*forsecondframeMext}, we deduce\footnote{Note that \eqref{eq:controlonSigma*forsecondframeMext} yields
\beaa
|\dk^kf| &\les& \frac{\ep}{u^{1+\dec-2\de_0}}\,\,\,\textrm{ for }k\leq k_{small}+k_{loss}+2\textrm{ on }\Sigma_*.
\eeaa
in view of  the behavior \eqref{eq:behaviorofronSigmastar} of $r$ on $\Sigma_*$.}
\beaa
|(\dkb, T)^kf|  &\leq& \frac{\ep}{ru^{\frac{1}{2}+\dec-2\de_0}+u^{1+\dec-2\de_0}}\,\,\,\textrm{ for }k\leq k_{small}+k_{loss}+2,\,\, r\geq r_1.  
\eeaa
Together with \eqref{eq:equationfore4primefforsecondframeofMext}, we recover the $e_4$ derivatives and obtain
\beaa
|\dk^kf|  &\leq& \frac{\ep}{ru^{\frac{1}{2}+\dec-2\de_0}+u^{1+\dec-2\de_0}}\,\,\,\textrm{ for }k\leq k_{small}+k_{loss}+2,\,\, r\geq r_1.  
\eeaa
This is an improvement of the bootstrap assumption \eqref{eq:localbootstraponMextforsecondframeMext}. Thus, we may choose $r_1=4m_0$, and we have
\beaa
|\dk^kf| &\les& \frac{\ep}{ru^{\frac{1}{2}+\dec-2\de_0}+u^{1+\dec-2\de_0}}\,\,\,\textrm{ for }k\leq k_{small}+k_{loss}+2\textrm{ on }\Mext.
\eeaa
Also, commuting once \eqref{eq:equationfore4primefforsecondframeofMext} with $e_3'$, using the commutator identity $[e_3',e_4']=2\omb'e_4'-2\om' e_3'+(\eta'-\etab')e_\th'$, and proceeding as above to integrate backward from $\Sigma_*$ where $e_3'f$ is under control from \eqref{eq:controlonSigma*forsecondframeMext}, we also obtain 
\beaa
|\dk^{k-1}e_3'f| &\les& \frac{\ep}{ru^{1+\dec-2\de_0}}\,\,\,\textrm{ for }k\leq k_{small}+k_{loss}+2\textrm{ on }\Mext.
\eeaa
Collecting the two above estimates, we obtain
\bea\lab{eq:controlonMextforsecondframeMext}
\begin{split}
|\dk^kf| &\les \frac{\ep}{ru^{\frac{1}{2}+\dec-2\de_0}+u^{1+\dec-2\de_0}}, \,\,\,\textrm{ for }k\leq k_{small}+k_{loss}+2\textrm{ on }\Mext,\\
 |\dk^{k-1}e_3'f| &\les \frac{\ep}{ru^{1+\dec-2\de_0}}\,\,\,\textrm{ for }k\leq k_{small}+k_{loss}+2\textrm{ on }\Mext,
\end{split}
\eea
which is the desired estimate for $f$.

{\bf Step 4}. In view of Proposition \ref{prop:transformations1} applied to our particular case, i.e. a triplet $(f, ,\fb, \la)$ with $\fb=0$ and $\la=1$, and the fact that the frame $(e_4, e_3, e_\th)$ is outgoing geodesic, we have 
\beaa
\begin{split}
\xib' &=\xib,\\
\ze' &= \ze   -\frac 14 f\kab - f  \omb  -\frac 1 4 f \vthb +\lot,\\
     \eta'&= \eta +\frac{1}{2}e_3'(f)     -f\omb +\lot, \\
     \etab'&= -\ze      +\frac 1 4 \kab f  +\frac{1}{4}f\vthb+\lot,
\end{split}
\eeaa
\beaa
\bsplit
\ka'&= \ka+ \ddd_1\,\!'(f)   + f(\ze+\eta)     -\frac 1 4 f^2\kab  -f^2\omb+\lot,\\
\kab'&= \kab      +f\xib  +\lot,\\
\vth' &= \vth- \dds_2\,\!'(f)   +  f(\ze+\eta)    -f^2\omb+\lot\\
\vthb' &= \vthb    +f\xib  +\lot, 
\end{split}
\eeaa
\beaa
\begin{split}
\om' &=   f\ze -\frac{1}{8}\kab f^2-\frac{1}{4}\omb f^2+\lot,\\
\omb' &= \omb  +\frac{1}{2}f\xib,
\end{split}
\eeaa
and
\bea
\bsplit
\a' &= \a+ 2f\b+\frac{3}{2}f^2\rho+\lot,\\
\b' &= \b+\frac{3}{2}\rho f+\lot,\\
\rho' &= \rho  +f\bb+\lot,\\
 \bb' &= \bb+ \frac{1}{2}f\aa,\\
\aa' &= \aa.
\end{split}
\eea
where   the  lower order terms denoted by $\lot$  are linear 
  with respect   to    $\xi,\xib,\vth,\ka,  \eta,\etab,\ze,\kab,  \vthb$ and $\a,\b, \rho, \bb, \,\aa$,  and quadratic or  higher order in $f$, and do not contain derivatives of the latter. Together with  the estimates \eqref{eq:controlonMextforsecondframeMext} for $f$ on $\Mext$, and 
 the estimates for the Ricci coefficients and curvature components with respect to the outgoing geodesic frame $(e_4, e_3, e_\th)$ of $\Mext$ provided by the bootstrap assumptions on decay and Proposition \ref{prop:pointwiseboundsforhighorderderivatives}, we immediately infer
\bea\lab{eq:recoveryofalldesiredestimatesexceptforeta'forsecondframeMext}
\nn\max_{0\leq k\leq k_{small}+k_{loss}+1}\sup_{\Mext}&&\Bigg\{\Big(r^2u^{\frac{1}{2}+\dec-2\de_0}+ru^{1+\dec-2\de_0}\Big)|\dk^k(\Ga_g'\setminus\{\eta'\})|+ru^{1+\dec-2\de_0}|\dk^k\Ga_b'|\\
\nn&&+r^2u^{1+\dec-2\de_0}\left|\dk^{k-1}e_3'\left(\ka'-\frac{2}{r}, \kab'+\frac{2\Up}{r}, \vth', \ze', \etab'\right)\right|\\
\nn&&+\Big(r^3(u+2r)^{\frac{1}{2}+\dec-2\de_0}+r^2(u+2r)^{1+\dec-2\de_0}\Big)\Big(|\dk^k\a'|+|\dk^k\b'|\Big)\\
\nn&& +\left(r^3(2r+u)^{1+\dec}+r^4(2r+u)^{\frac{1}{2}+\dec-2\de_0}\right)|\dk^{k-1}e_3'(\a')|\\
\nn&&+\Big(r^3u^{1+\dec}+r^4u^{\frac{1}{2}+\dec-2\de_0}\Big)|\dk^{k-1}e_3'(\b')|\\
\nn&&+\Big(r^3u^{\frac{1}{2}+\dec-2\de_0}+r^2ru^{1+\dec-2\de_0}\Big)|\dk^k\rhoc'|\\
&&+u^{1+\dec-2\de_0}\Big(r^2|\dk^k\bb'|+r|\dk^k\aa'|\Big)\Bigg\} \les \ep
\eea
where we have introduced the notation 
\beaa
\Ga_g'\setminus\{\eta'\} &=& \left\{r\om', \,\ka'-\frac{2}{r}, \, \vth', \, \ze',  \, \etab', \, \kab'+\frac{2\Up}{r}, \, r^{-1}(e_4'(r)-1), r^{-1}e_\th'(r), \, e_4'(m) \right\}.
\eeaa
Note also, in view of the above transformation formula for $\om'$, i.e.
\beaa
\om' &=&   f\ze -\frac{1}{8}\kab f^2-\frac{1}{4}\omb f^2+\lot,
\eeaa
that we have in fact a gain of $r^{-1}$ for $\om'$ compared to \eqref{eq:recoveryofalldesiredestimatesexceptforeta'forsecondframeMext}, i.e.
\bea\lab{eq:evenbetterestimateforom'forsecondframeMext}
\max_{0\leq k\leq k_{small}+k_{loss}+1}\sup_{\Mext}\Big(r^3u^{\frac{1}{2}+\dec-2\de_0}+r^2u^{1+\dec-2\de_0}\Big)|\dk^k\om'| &\les& \ep.
\eea

We now focus on estimating $\eta'$. Proceeding as for the other Ricci coefficients would yield for $\eta'$ the same behavior than $\eta$ and hence a loss of $r^{-1}$ compared to the desired estimate. Instead, we rely on the following null structure equation which follow from Proposition \ref{prop:null.structure-general} and the fact that $\xi'=0$
\beaa
  e_4'(\eta' -\ze') + \frac 12 \ka'(\eta' -\ze') &=&  2\dds_1'\om' -\frac 1 2 \vth' (\eta'-\ze').
\eeaa
Next, 
\begin{itemize}
\item we commute with $\dkb'$ and $T'$, and we rely on the corresponding commutator identities,

\item we use the above equation for $e_4'(\eta')$ to recover the $e_4'$ derivatives, 

\item we rely on the estimates \eqref{eq:recoveryofalldesiredestimatesexceptforeta'forsecondframeMext}, as well as the estimate \eqref{eq:evenbetterestimateforom'forsecondframeMext} for $\om'$,
\end{itemize}
which allows us to derive
\beaa
  \left|e_4'(\dk^k(\eta'-\ze'))+\frac 12 \ka'\dk^k(\eta' -\ze')\right| &\les& \frac{\ep}{r^4u^{\frac{1}{2}+\dec-2\de_0}+r^3u^{1+\dec-2\de_0}}\\
  &&+\frac{\ep}{r^2}|\dk^{\leq k}(\eta' -\ze')|,\qquad k\leq k_{small}+k_{loss}.
\eeaa
Integrating backwards from $\Sigma_*$ where $\eta'=0$ in view of \eqref{eq:propertiesoffforsecondframeofMext}, and using the control $\ze'$ provided by \eqref{eq:recoveryofalldesiredestimatesexceptforeta'forsecondframeMext}, we infer
\beaa
&&\max_{0\leq k\leq k_{small}+k_{loss}}\sup_{\Mext}\Big(r^2u^{\frac{1}{2}+\dec-2\de_0}+ru^{1+\dec-2\de_0}\Big)|\dk^k\eta'| \\
&\les& \ep+\max_{0\leq k\leq k_{small}+k_{loss}}\sup_{\Mext}\Big(r^2u^{\frac{1}{2}+\dec-2\de_0}+ru^{1+\dec-2\de_0}\Big)|\dk^k\ze'|\\
&\les& \ep.
\eeaa
Also, commuting first the equation for $e_4'(\eta'-\ze')$ with $e_3'$, using the commutator identity $[e_3',e_4']=2\omb'e_4'-2\om' e_3'+(\eta'-\etab')e_\th'$, and proceeding as above to integrate backward from $\Sigma_*$, we also obtain 
\beaa
&&\max_{0\leq k\leq k_{small}+k_{loss}}\sup_{\Mext}r^2u^{1+\dec-2\de_0}|\dk^{k-1}e_3'\eta'| \\
&\les& \ep+\max_{0\leq k\leq k_{small}+k_{loss}}\sup_{\Mext}r^2u^{1+\dec-2\de_0}|\dk^{k-1}e_3'\ze'|\\
&\les& \ep.
\eeaa
Thus, together with \eqref{eq:recoveryofalldesiredestimatesexceptforeta'forsecondframeMext}, we infer
\beaa
\nn\max_{0\leq k\leq k_{small}+k_{loss}}\sup_{\Mext}&&\Bigg\{\Big(r^2u^{\frac{1}{2}+\dec-2\de_0}+ru^{1+\dec-2\de_0}\Big)|\dk^k\Ga_g'|+ru^{1+\dec-2\de_0}|\dk^k\Ga_b'|\\
\nn&&+r^2u^{1+\dec-2\de_0}\left|\dk^{k-1}e_3'\left(\ka'-\frac{2}{r}, \kab'+\frac{2\Up}{r}, \vth', \ze', \etab', \eta'\right)\right|\\
\nn&&+\Big(r^3(u+2r)^{\frac{1}{2}+\dec-2\de_0}+r^2(u+2r)^{1+\dec-2\de_0}\Big)\Big(|\dk^k\a'|+|\dk^k\b'|\Big)\\
\nn&& +\left(r^3(2r+u)^{1+\dec}+r^4(2r+u)^{\frac{1}{2}+\dec-2\de_0}\right)|\dk^{k-1}e_3'(\a')|\\
\nn&&+\Big(r^3u^{1+\dec}+r^4u^{\frac{1}{2}+\dec-2\de_0}\Big)|\dk^{k-1}e_3'(\b')|\\
\nn&&+\Big(r^3u^{\frac{1}{2}+\dec-2\de_0}+r^2ru^{1+\dec-2\de_0}\Big)|\dk^k\rhoc'|\\
&&+u^{1+\dec-2\de_0}\Big(r^2|\dk^k\bb'|+r|\dk^k\aa'|\Big)\Bigg\} \les \ep.
\eeaa
Together with the fact that $\xi'=0$ in view of \eqref{eq:propertiesoffforsecondframeofMext}, this concludes the proof of Proposition \ref{prop:constructionsecondframeinMext}.


\section{Existence and control of the global frames}



\subsection{Proof of Proposition \ref{prop:existenceandestimatesfortheglobalframe}}


To match the frame of $\Mint$ and a conformal renormalization of the frame of $\Mext$, we will need to introduce a cut-off function.
\begin{definition}
Let $\psi:\RRR\to\RRR$ a smooth cut-off function such that $0\leq\psi\leq 1$, $\psi=0$ on $(-\infty, 0]$ and $\psi=1$ on $[1, +\infty)$. We define $\psi_{m_0, \deh}$ as follows
\beaa
\psi_{m_0, \deh}(r)=\begin{cases} &1 \qquad  \mbox{if} \quad r \ge 2m_0\left(1+\frac 3 2 \deh\right),\\[1mm]
&0  \qquad  \mbox{if} \quad r \le 2m_0\left(1+\frac 1 2 \deh\right),
\end{cases}
\eeaa
and
\beaa
\psi_{m_0, \deh}(r)=\psi\left(\frac{r - 2m_0\left(1+\frac{1}{2}\deh\right)}{2m_0\deh}\right)\textrm{ on }2m_0\left(1+\frac 1 2 \deh\right)\leq r\leq 2m_0\left(1+\frac 3 2 \deh\right).
\eeaa
\end{definition} 

We are now ready to define the global frame of the statement of Proposition \ref{prop:existenceandestimatesfortheglobalframe}. 
\begin{definition}[Definition of the global frame]\lab{def:definitionoftheglobalframe}
We introduce a global null frame defined on $\Mext\cup\Mint$ and denoted by $({}^{(glo)}e_4, {}^{(glo)}e_3, {}^{(glo)}e_\th)$. The global frame is defined as follows
\begin{enumerate}
\item In $\Mext\setminus\mr$, we have
\beaa
({}^{(glo)}e_4, {}^{(glo)}e_3, {}^{(glo)}e_\th)= \left({}^{(ext)}\Up\,{}^{(ext)}e_4, {}^{(ext)}\Up^{-1}{}^{(ext)}e_3, {}^{(ext)}e_\th\right).
\eeaa

\item In $\Mint\setminus\mr$, we have
\beaa
({}^{(glo)}e_4, {}^{(glo)}e_3, {}^{(glo)}e_\th) = \left({}^{(int)}e_4, {}^{(int)}e_3, {}^{(int)}e_\th\right).
\eeaa

\item It remains to define the global frame on the matching region. We denote by $(f, \fb, \la)$ the reduced scalars such that we have in the matching region
\beaa
{}^{(ext)}e_4 &=&\la\left({}^{(int)}e_4 +  f {}^{(int)}e_\th +\frac 1 4 f^2  {}^{(int)}e_3\right),\\
{}^{(ext)}e_\th &=&  \left(1+\frac{1}{2}f\fb\right){}^{(int)}e_\th + \frac{\fb}{2}{}^{(int)}e_4+  \frac{f}{2}\left(1+\frac{f\fb}{4}\right){}^{(int)}e_3,\\
{}^{(ext)}e_3 &=& \la^{-1} \left(  \left(1+ \frac{1}{2}f\fb +\frac{1}{16}f^2\fb^2\right) {}^{(int)}e_3+ \fb\left(1+ \frac{f\fb}{4}\right) {}^{(int)}e_\th +\frac{\fb^2}{4}{}^{(int)}e_4\right),
\eeaa
where we recall that the frame of $\Mext$ has been extended to $\Mint$, see section \ref{sec:extensionofframes}. Then, in the matching region, the global frame  is given by 
\beaa
{}^{(glo)}e_4 &=&\la'\left({}^{(int)}e_4 +  f' {}^{(int)}e_\th +\frac 1 4 {f'}^2  {}^{(int)}e_3\right),\\
{}^{(glo)}e_\th &=&  \left(1+\frac{1}{2}f'\fb'\right){}^{(int)}e_\th + \frac{\fb'}{2}{}^{(int)}e_4+  \frac{f'}{2}\left(1+\frac{f'\fb'}{4}\right){}^{(int)}e_3,\\
{}^{(glo)}e_3 &=& {\la'}^{-1} \left(  \left(1+ \frac{1}{2}f'\fb' +\frac{1}{16}{f'}^2{\fb'}^2\right) {}^{(int)}e_3+ \fb'\left(1+ \frac{f'\fb'}{4}\right) {}^{(int)}e_\th +\frac{{\fb'}^2}{4}{}^{(int)}e_4\right),
\eeaa
where 
\bea\lab{eq:def:definitionoftheglobalframe}
\begin{split}
 f' &=\psi_{m_0, \deh}(\rint)f, \quad \fb'=\psi_{m_0, \deh}(\rint)\fb,\\ 
 \la' &=1-\psi_{m_0, \deh}(\rint)+\psi_{m_0, \deh}(\rint){}^{(ext)}\Up\la. 
\end{split}
\eea
\end{enumerate} 
\end{definition}

\begin{remark}\lab{rem:possibilitytoadaptglobalframetoMextorMint}
Recall that the smooth cut-off function $\psi$ in Definition \ref{def:cutofffunctionforthematchingregion}, allowing to define $\psi_{m_0, \deh}$, is such that we have in particular $\psi=0$ on $(-\infty, 0]$ and $\psi=1$ on $[1, +\infty)$. The following two special cases correspond to the properties (d) i. and (d) ii. of Proposition \ref{prop:existenceandestimatesfortheglobalframe}.
\begin{itemize}
\item If the cut-off $\psi$ in Definition \ref{def:cutofffunctionforthematchingregion} is such that $\psi=1$ on $[1/2, +\infty)$, then
\beaa
({}^{(glo)}e_4, {}^{(glo)}e_3, {}^{(glo)}e_\th)= \left({}^{(ext)}\Up\,{}^{(ext)}e_4, {}^{(ext)}\Up^{-1}{}^{(ext)}e_3, {}^{(ext)}e_\th\right)\textrm{ on }\Mext.
\eeaa

\item If the cut-off $\psi$ in Definition \ref{def:cutofffunctionforthematchingregion} is such that $\psi=0$ on $(-\infty, 1/2]$, then
\beaa
({}^{(glo)}e_4, {}^{(glo)}e_3, {}^{(glo)}e_\th) = \left({}^{(int)}e_4, {}^{(int)}e_3, {}^{(int)}e_\th\right)\textrm{ on }\Mint.
\eeaa
\end{itemize}
\end{remark}

\begin{definition}[Global area radius and Hawking mass]\lab{def:globalrandmfor1stglobalframe}
We definition an area radius and a Hawking mass on $\Mext\cup\Mint$ as follows
\begin{itemize}
\item On $\Mext\setminus\mr$, we have
\beaa
{}^{(glo)}r=\rext,\quad {}^{(glo)}m=\mext
\eeaa

\item On $\Mint\setminus\mr$, we have
\beaa
{}^{(glo)}r=\rint,\quad {}^{(glo)}m=\mint
\eeaa

\item On the matching region, we have
\beaa
{}^{(glo)}r &=& (1-\psi_{m_0, \deh}(\rint))\rint+\psi_{m_0, \deh}(\rint)\rext,\\
{}^{(glo)}m &=& (1-\psi_{m_0, \deh}(\rint))\mint+\psi_{m_0, \deh}(\rint)\mext.
\eeaa
\end{itemize}
\end{definition}

The following two lemmas provide the main properties of the global frame.  
\begin{lemma}\lab{lemma:identitiesrelatingtheglobalframetotheframeofMext}
We have in $\Mext\setminus\mr$ the following relations between the quantities in the respective frames
\beaa
&& {}^{(glo)}\a=\Up^2{}^{(ext)}\a,\quad {}^{(glo)}\b=\Up{}^{(ext)}\b,\quad {}^{(glo)}\rho+\frac{2m}{r^3}={}^{(ext)}\rho+\frac{2m}{r^3},\quad {}^{(glo)}\bb=\Up^{-1}{}^{(ext)}\bb,\\
&&{}^{(glo)}\aa=\Up^{-2}{}^{(ext)}\aa,\quad {}^{(glo)}\xi= 0, \,\,\,\, {}^{(glo)}\xib =   \Up^{-2}{}^{(ext)}\xib, \,\,\,\,{}^{(glo)}\ze=-{}^{(glo)}\etab={}^{(ext)}\ze,\\
&&  {}^{(glo)}\eta= {}^{(ext)}\eta,\,\,\,\, {}^{(glo)}\om +\frac{m}{r^2}=   -\frac{m}{2r}\left(\overline{{}^{(ext)}\ka}-\frac{2}{r}\right)+\frac{e_4(m)}{r},\\
&& {}^{(glo)}\omb =\Up^{-1}\left({}^{(ext)}\omb -\frac{m}{r^2}+\frac{m}{2\Up r}\left(\overline{{}^{(ext)}\kab}-\frac{2\Up}{r}\right)+\frac{m}{2\Up r}\Big({}^{(ext)}\Obc\overline{{}^{(ext)}\ka}-\overline{{}^{(ext)}\Obc{}^{(ext)}\check{\ka}}\Big) -\frac{e_3(m)}{\Up r}\right),\\
&&{}^{(glo)}\ka - \frac{2\Up}{r} = \Up\left({}^{(ext)}\ka-\frac{2}{r}\right), \,\,\,\, {}^{(glo)}\kab +\frac{2}{r}= \Up^{-1}\left({}^{(ext)}\kab+\frac{2\Up}{r}\right), \\
&& {}^{(glo)}\check{\ka}=\Up{}^{(ext)}\check{\ka},\,\,\,\, {}^{(glo)}\check{\kab}=\Up^{-1}{}^{(ext)}\check{\kab},\quad {}^{(glo)}\vth =\Up{}^{(ext)}\vth,  \,\,\,\,    {}^{(glo)}\vthb =\Up^{-1}  {}^{(ext)}\vthb.
\eeaa 
\end{lemma}

\begin{proof}
The proof follows immediately from the change of frame formula with the choice $(f=0, \fb=0, \la=\Up)$, the fact that $e_\th(\Up)=0$, and the fact that the frame of $\Mext$ is outgoing geodesic and thus satisfies in particular $\xi=\om=0$ and $\etab=-\ze$.
\end{proof}

\begin{lemma}[Control of the global frame in the matching region]\lab{lemma:estimatesfortheglobalframeinthematchingregion}
In the matching region, the following estimates holds for the global frame\footnote{We only need the first estimate for the proof of Proposition \ref{prop:existenceandestimatesfortheglobalframe}, but the second estimate will be needed in the proof of Theorem M8.}
\beaa
\max_{0\leq k\leq k_{small}-2}\left(\sup_{\mr\cap\Mint}\ub^{1+\dec}\left|\dk^k({}^{(glo)}\Gac, {}^{(glo)}\Rc)\right|+\sup_{\mr\cap\Mext}u^{1+\dec}\left|\dk^k({}^{(glo)}\Gac, {}^{(glo)}\Rc)\right|\right)\\
 +\max_{0\leq k\leq k_{large}-1}\left(\int_{\mr}\left|\dk^k({}^{(glo)}\Gac, {}^{(glo)}\Rc)\right|^2\right)^{\frac{1}{2}} &\les& \ep.
\eeaa
and
\beaa
\left(\int_{\mr}\left|\dk^{k_{large}}({}^{(glo)}\Gac, {}^{(glo)}\Rc)\right|^2\right)^{\frac{1}{2}} &\les& \ep+\left(\int_{\TT}\left|\dk^{k_{large}}({}^{(ext)}\Rc)\right|^2\right)^{\frac{1}{2}}.
\eeaa
\end{lemma}

\begin{remark}
The quantities associated to the global frame can be estimated as follows
\begin{itemize}
\item In $\Mint\setminus\mr$, the global frame coincides with the frame of $\Mint$, and hence, the quantities associated to the global frame satisfy the same estimates than the bootstrap assumptions for the frame of $\Mint$. 

\item In $\Mext\setminus\mr$, estimates for the quantities associated to the global frame follow from the identities of Lemma \ref{lemma:identitiesrelatingtheglobalframetotheframeofMext} together with  the bootstrap assumptions for the frame of $\Mext$. 

\item In $\mr$, the estimates  for the quantities associated to the global frame are provided by Lemma \ref{lemma:estimatesfortheglobalframeinthematchingregion}.
\end{itemize}
\end{remark}

The proof of Proposition \ref{prop:existenceandestimatesfortheglobalframe} easily follows from Definition \ref{def:definitionoftheglobalframe}, Remark \ref{rem:possibilitytoadaptglobalframetoMextorMint}, and Lemma \ref{lemma:estimatesfortheglobalframeinthematchingregion}. Thus, from now on, we focus on the proof of Lemma \ref{lemma:estimatesfortheglobalframeinthematchingregion} which is carried out in the next section.


\subsection{Proof of Lemma \ref{lemma:estimatesfortheglobalframeinthematchingregion}}\lab{sec:proofoflemma:estimatesfortheglobalframeinthematchingregion}


In this section, we prove Lemma \ref{lemma:estimatesfortheglobalframeinthematchingregion}. To ease the exposition, the quantities associated to the the frame of $\Mint$ are unprimed, the quantities associated to the  frame of $\Mext$ are primed,  and  the quantities associated to the the global frame are double-primed.
 
{\bf Step 1.} Let $(e_3, e_\th, e_4)$ denote the frame of $\Mint$ (and its extension) and $(e_3', e_\th', e_4')$ the frame of $\Mext$ (and its extension). We denote by $(f, \fb, \la)$ the reduced scalars such that 
\beaa
e_4'&=&\la\left(e_4 + f e_\th +\frac 1 4 f^2  e_3\right),\\
e_\th'&=&  \left(1+\frac{1}{2}f\fb\right)e_\th + \frac{\fb}{2}e_4+  \frac{f}{2}\left(1+\frac{f\fb}{4}\right)e_3,\\
e_3'&=& \la^{-1} \left(  \left(1+ \frac{1}{2}f\fb +\frac{1}{16}f^2\fb^2\right) e_3+\left(\fb+\frac 1 4 \fb^2 f\right) e_\th +\frac{\fb^2}{4}e_4\right).
\eeaa
Together with the initialization of the frame of $\Mext$ and $\Mint$ on $\TT$ in section \ref{sec:defintioncanonicalspacetime} (where the spheres coincide), we have in particular
\bea\lab{eq:initializationofffblambdaonTTforcontrolglobalframe}
f=\fb=0,\quad \la=\Up^{-1}\textrm{ on }\TT.
\eea
Also, recall from section \ref{sec:extensionofframes} that  in order for $(e_3', e_\th', e_4')$ to be defined everywhere on $\Mint\cap\mr$, we need - in addition to the above initialization of $(f, \fb, \la)$ on $\TT$, to initialize it also on $\CCb_*\cap\mr$ by
\bea\lab{eq:initializationofffblambdaonTTforcontrolglobalframe:bis}
f=\fb=0,\quad \la=\Up^{-1}\textrm{ on }\CCb_*\cap\mr.
\eea

{\bf Step 2.} Next, we control the change of frame $(f, \fb, \la)$ from $(e_3, e_\th, e_4)$ to $(e_3', e_\th', e_4')$ in the region $\Mint\cap\mr$. To this end, we rely on the transport equation of Lemma \ref{lemma:transportequationsforffbandlambda} together with the fact that $\om'=\xi'=\ze'+\etab'=0$.  Then, $(\fb, f, \log(\la))$ satisfy the following transport equations 
\beaa
\la^{-1}e_4'(f)  +\left(\frac{\ka}{2}+2\om\right) f &=& -2\xi+E_1(f, \Ga),\\
\la^{-1}e_4'(\log(\la)) &=&  2\om+E_2(f, \Ga),\\
\la^{-1}e_4'(\fb) + \frac{\ka}{2}\fb &=& -2(\ze+ \etab)  + 2e_\th'(\log(\la))      + 2f  \omb+E_3(f, \fb, \Ga),
\eeaa
where $E_1$, $E_2$ and $E_3$ are given by
\beaa
E_1(f, \Ga) &=& -\frac{1}{2}\vth f+\lot,\\
E_2(f, \Ga) &=&  f\ze- \frac{1}{2}f^2\omb -\etab f - \frac{1}{4}f^2\kab +\lot,\\
E_3(f, \fb, \Ga) &=& -\fb e_\th'(f)  -\frac{1}{2}\fb \vth     +\lot,
\eeaa
Here, $\lot$ denote terms which are cubic or higher order in $f, \fb$ (or in $f$ only in the case of $E_1$ and $E_2$) and $\Gac$ and do not contain derivatives of these quantities, where $\Ga$ and $\Gac$ denotes the Ricci coefficients and renormalized Ricci coefficients w.r.t. the original null frame $(e_3, e_4, e_\th)$. We rewrite the transport equation for $\log(\la)$ as
\beaa
&&\la^{-1}e_4'\left(\log\left(\Up\la\right)\right) \\
&=& \la^{-1}e_4'(\log(\la)) + \la^{-1}e_4'(\log(\Up))\\
&=&   2\om+E_2(f, \Ga) +\frac{1}{\Up}\left(e_4 + f e_\th +\frac 1 4 f^2  e_3\right)\Up\\
&=&   2\left(\om+\frac{m}{r^2}\right)+E_2(f, \Ga) +\frac{2}{\Up}\frac{m (e_4(r)-\Up)}{r^2} -\frac{2}{\Up}\frac{e_4(m)}{r} -\frac{1}{\Up}\left(f e_\th +\frac 1 4 f^2  e_3\right)\Up.
\eeaa

In view of the above transport equations for $f$, $\fb$ and $\la$, the initialization \eqref {eq:initializationofffblambdaonTTforcontrolglobalframe}  \eqref{eq:initializationofffblambdaonTTforcontrolglobalframe:bis} for $(f, \fb, \la)$ on $\TT\cup(\CCb_*\cap\mr)$, and the control of $\Ga$ induced by the bootstrap assumptions on $\Mint$, we easily deduce
\beaa
\max_{0\leq k\leq k_{small}}\sup_{\Mint\cap\mr}\ub^{1+\dec}\left|\dk^k(f, \log(\Up\la))\right|+\max_{0\leq k\leq k_{small}-1}\sup_{\Mint\cap\mr}\ub^{1+\dec}\left|\dk^k\fb\right| &\les& \ep,\\
\max_{0\leq k\leq k_{large}}\left(\int_{\Mint\cap\mr}\left|\dk^k(f, \log(\Up\la))\right|^2\right)^{\frac{1}{2}}+\max_{0\leq k\leq k_{large}-1}\left(\int_{\Mint\cap\mr}\left|\dk^k\fb\right|^2\right)^{\frac{1}{2}} &\les& \ep.
\eeaa

{\bf Step 3.} We need to improve the number of derivatives in the top order estimate for $(f, \fb, \log(\la))$. To this end, note first in view of the transformation formulas of Proposition \ref{prop:transformations1} and the control of $(f, \fb, \log(\la))$ provided by Step 2, we have in particular
\beaa
\max_{0\leq k\leq k_{large}-1}\left(\int_{\Mint\cap\mr}\left|\dk^k\Rc'\right|^2\right)^{\frac{1}{2}} &\les& \ep.
\eeaa
Relying on this estimate, the control of the Ricci coefficients associated to the outgoing null frame $(e_4', e_3', e_\th')$ on $\TT\cup(\Mint\cap\mr)$, and the null structure equations, we infer
\beaa
\max_{0\leq k\leq k_{large}-1}\left(\int_{\Mint\cap\mr}\left|\dk^k\Gac'\right|^2\right)^{\frac{1}{2}} &\les& \ep.
\eeaa
We refer to section \ref{sec:proofprop:improvementoftheiterationassupmtionThM8} for a completely analogous proof where the Ricci coefficients are recovered in $\Mint$ based on the control of the curvature components. 

In view of the transformation formulas of Proposition \ref{prop:transformations1}, which can be written schematically as
\beaa
\pr\Big(f, \fb, \log(\la)\Big) &=& F(f, \fb, \la, \Gac),
\eeaa
the control of $(f, \fb, \log(\la))$ provided by Step 2, and the above control of $\Ga'$, we infer
\beaa
\max_{0\leq k\leq k_{large}}\left(\int_{\Mint\cap\mr}\left|\dk^k(f, \fb, \log(\Up\la))\right|^2\right)^{\frac{1}{2}} &\les& \ep.
\eeaa 

{\bf Step 4.} We still need to control one more derivative of $(f, \fb, \log(\la))$. Repeating the process of Step 3, we use again the transformation formulas of Proposition \ref{prop:transformations1}  and then the final estimate of Step 3 for $(f, \fb, \log(\la))$ yields the following control for the curvature components
\beaa
\max_{0\leq k\leq k_{large}}\left(\int_{\Mint\cap\mr}\left|\dk^k\Rc'\right|^2\right)^{\frac{1}{2}} &\les& \ep.
\eeaa
Arguing as in Step 3, we infer\footnote{In Step 3, there is no term corresponding to the one integrated on $\TT$. This is due to the fact that for $k\leq k_{large}-1$, we have  thanks to the bootstrap assumptions on energy and a trace estimate
\beaa
\max_{0\leq k\leq k_{large}-1}\left(\int_{\TT}\left|\dk^k({}^{(ext)}\Rc)\right|^2\right)^{\frac{1}{2}}\les\ep.
\eeaa} 
\beaa
\max_{0\leq k\leq k_{large}}\left(\int_{\Mint\cap\mr}\left|\dk^k\Gac'\right|^2\right)^{\frac{1}{2}} &\les& \ep+\left(\int_{\TT}\left|\dk^{k_{large}}({}^{(ext)}\Rc)\right|^2\right)^{\frac{1}{2}}.
\eeaa
Using again the transformation formulas of Proposition \ref{prop:transformations1}, this yields the following control for $(f, \fb, \log(\la))$
\beaa
\max_{0\leq k\leq k_{large}+1}\left(\int_{\Mint\cap\mr}\left|\dk^k(f, \fb, \log(\Up\la))\right|^2\right)^{\frac{1}{2}} &\les& \ep.
\eeaa 
We have finally obtained for $(f, \fb, \la)$ in $\Mint\cap\mr$
\beaa
\max_{0\leq k\leq k_{small}-1}\sup_{\Mint\cap\mr}\ub^{1+\dec}\left|\dk^k(f, \fb, \log(\Up\la))\right| &\les& \ep,\\
\max_{0\leq k\leq k_{large}}\left(\int_{\Mint\cap\mr}\left|\dk^k(f, \fb, \log(\Up\la))\right|^2\right)^{\frac{1}{2}} &\les& \ep,\\
\left(\int_{\Mint\cap\mr}\left|\dk^{k_{large}+1}(f, \fb, \log(\Up\la))\right|^2\right)^{\frac{1}{2}} &\les& \ep+\left(\int_{\TT}\left|\dk^{k_{large}}({}^{(ext)}\Rc)\right|^2\right)^{\frac{1}{2}}.
\eeaa

{\bf Step 5.}  In addition to the estimate of $(f, \fb, \la)$ in $\Mint\cap\mr$ of Step 4, we need to estimate $(f, \fb, \la)$ in $\Mext\cap\mr$. To this end, we first control in $\Mext\cap\mr$ the reduced scalar $(f', \fb', \la')$ satisfying 
\beaa
e_3&=&\la'\left(e_3' + \fb' e_\th' +\frac 1 4 {\fb'}^2  e_4'\right),\\
e_\th&=&\left(1+\frac 1 2   f' \fb'\right) e_\th'  + \frac 1 2  f' e_3'+\frac 1 2 \left(\fb'+ \frac 1 4 f {\fb'}^2\right) e_4',\\
e_4&=& {\la'}^{-1} \left( \left(1+\frac 12  f' \fb' +\frac{1}{16} {f'}^2 {\fb'}^2\right)  e_4' +\left(f'+\frac 1 4 {f'}^2 \fb'\right) e_\th' + \frac 1 4 {f'}^2  e_3'\right).
\eeaa

Together with the initialization of the frame of $\Mext$ and $\Mint$ on $\TT$ in section \ref{sec:defintioncanonicalspacetime} (where the spheres coincide), we have in particular
\beaa
f'=\fb'=0,\quad \la'=\Up^{-1}\textrm{ on }\TT.
\eeaa
Also, recall from section \ref{sec:extensionofframes} that in order for $(e_3, e_\th, e_4)$ to be defined everywhere on $\Mext\cap\mr$, we need - in addition to the above initialization of $(f, \fb, \la)$ on $\TT$, to initialize it also on $\CC_*\cap\mr$ by
\bea\lab{eq:initializationofffblambdaonTTforcontrolglobalframe:ter}
f'=\fb'=0,\quad \la'=\Up^{-1}\textrm{ on }\CC_*\cap\mr.
\eea
Arguing similarly to Steps 1-4, we estimate $(f', \fb', \la')$ and $(\Gac, \Rc)$ in $\Mext\cap\mr$. We obtain 
\beaa
\max_{0\leq k\leq k_{small}-2}\sup_{\Mext\cap\mr}u^{1+\dec}\left|\dk^k(\Gac, \Rc)\right| +\max_{0\leq k\leq k_{large}-1}\left(\int_{\Mext\cap\mr}\left|\dk^k(\Gac, \Rc)\right|^2\right)^{\frac{1}{2}} &\les& \ep,\\
\left(\int_{\Mext\cap\mr}\left|\dk^{k_{large}}\Rc\right|^2\right)^{\frac{1}{2}} &\les& \ep,\\
\max_{0\leq k\leq k_{small}-1}\sup_{\Mext\cap\mr}u^{1+\dec}\left|\dk^k(f', \fb', \log(\Up'\la'))\right| &\les& \ep,\\
\max_{0\leq k\leq k_{large}}\left(\int_{\Mext\cap\mr}\left|\dk^k(f', \fb', \log(\Up'\la'))\right|^2\right)^{\frac{1}{2}} &\les& \ep,
\eeaa
and
\beaa
\left(\int_{\Mext\cap\mr}\left|\dk^{k_{large}}\Gac\right|^2\right)^{\frac{1}{2}} &\les& \ep+\left(\int_{\TT}\left|\dk^{k_{large}}({}^{(ext)}\Rc)\right|^2\right)^{\frac{1}{2}},\\
\left(\int_{\Mext\cap\mr}\left|\dk^{k_{large}+1}(f', \fb', \log(\Up'\la'))\right|^2\right)^{\frac{1}{2}} &\les& \ep+\left(\int_{\TT}\left|\dk^{k_{large}}({}^{(ext)}\Rc)\right|^2\right)^{\frac{1}{2}}.
\eeaa

{\bf Step 6.} As mentioned above, in addition to the estimate of $(f, \fb, \la)$ in $\Mint\cap\mr$ of Step 4, we need to estimate $(f, \fb, \la)$ in $\Mext\cap\mr$. To this end, we derive simple algebraic relations between $(f, \fb, \la)$ and $(f', \fb', \la')$ of Step 5. On the one hand, we have from the definition of $(f, \fb, \la)$
\beaa
g(e_4', e_3) &=& -2\la,\quad g(e_4', e_\th) = \la f,\quad g(e_\th', e_4) = -f\left(1+\frac{f\fb}{4}\right),\quad g(e_\th', e_3) = -\fb,\\ 
g(e_3', e_4) &=& -2\la^{-1}\left(1+\frac{f\fb}{2}+\frac{1}{16}f^2\fb^2\right),\quad g(e_3', e_\th) = \la^{-1}\fb\left(1+\frac{f\fb}{4}\right).
\eeaa
On the other hand, we have from the definition of $(f', \fb', \la')$ 
\beaa
g(e_3, e_4') &=& -2\la',\quad g(e_3, e_\th') = \la'\fb',\quad g(e_\th, e_4') = -f',\quad g(e_\th, e_3') = -\fb'\left(1+\frac{f'\fb'}{4}\right),\\
g(e_4, e_3') &=& -2{\la'}^{-1}\left(1+\frac{f'\fb'}{2}+\frac{1}{16}{f'}^2{\fb'}^2\right),\quad g(e_4, e_\th') = {\la'}^{-1}f'\left(1+\frac{f'\fb'}{4}\right).
\eeaa
We immediately infer
\beaa
\la'=\la,\quad f'=-\la f, \quad \fb'=-\la^{-1}\fb. 
\eeaa
In view of the estimates of Step 5, we infer
\beaa
\max_{0\leq k\leq k_{small}-1}\sup_{\Mext\cap\mr}u^{1+\dec}\left|\dk^k(f, \fb, \log(\Up\la))\right| &\les& \ep,\\
\max_{0\leq k\leq k_{large}}\left(\int_{\Mext\cap\mr}\left|\dk^k(f, \fb, \log(\Up\la))\right|^2\right)^{\frac{1}{2}} &\les& \ep,
\eeaa
and
\beaa
\left(\int_{\Mext\cap\mr}\left|\dk^{k_{large}+1}(f, \fb, \log(\Up\la))\right|^2\right)^{\frac{1}{2}} &\les& \ep.
\eeaa
Together with Step 4, this yields
\beaa
\max_{0\leq k\leq k_{small}-1}\sup_{\Mint\cap\mr}\ub^{1+\dec}\left|\dk^k(f, \fb, \log(\Up\la))\right| &\les& \ep,\\
\max_{0\leq k\leq k_{small}-1}\sup_{\Mext\cap\mr}u^{1+\dec}\left|\dk^k(f, \fb, \log(\Up\la))\right| &\les& \ep,\\
\max_{0\leq k\leq k_{large}}\left(\int_{\mr}\left|\dk^k(f, \fb, \log(\Up\la))\right|^2\right)^{\frac{1}{2}} &\les& \ep+\left(\int_{\TT}\left|\dk^{k_{large}}({}^{(ext)}\Rc)\right|^2\right)^{\frac{1}{2}},
\eeaa
and
\beaa
\max_{0\leq k\leq k_{large}+1}\left(\int_{\mr}\left|\dk^k(f, \fb, \log(\Up\la))\right|^2\right)^{\frac{1}{2}} &\les& \ep+\left(\int_{\TT}\left|\dk^{k_{large}}({}^{(ext)}\Rc)\right|^2\right)^{\frac{1}{2}}.
\eeaa

{\bf Step 7.} Next, we estimate $r'-r$ and $m'-m$. Note first the in view of the initialization of the foliations of $\Mext$ and $\Mint$ on $\TT$, as well as the initializations \eqref{eq:initializationofffblambdaonTTforcontrolglobalframe:bis} on $\CCb_*\cap\mr$ and \eqref{eq:initializationofffblambdaonTTforcontrolglobalframe:ter} on $\CC_*\cap\mr$, we have
\bea\lab{eq:initializationofffblambdaonTTforcontrolglobalframe:quatre}
r'=r,\quad m'=m\textrm{ on }\TT\cup\mr.
\eea
We start with the region $\Mint\cap\mr$. We have
\beaa
e_4'(r') = \frac{r'}{2}\overline{\ka}' = 1+\frac{r'}{2}\left(\overline{\ka}'-\frac{2}{r'}\right), \quad e_3'(r') = \frac{r'}{2}(\overline{\kab}'+\Ab') = -\Up'+\frac{r'}{2}\left(\overline{\kab}'+\frac{2\Up'}{r'}\right)+\frac{r'}{2}\Ab',
\eeaa
which together with the identities for $e_4'(m')$ and $e_3'(m')$ in the outgoing foliation of $\Mext$ and the control of the foliation of $\Mext$ in $\Mint\cap\mr$ established in Step 4 yields, using also $e_\th'(r')=e_\th'(m')=0$, 
\beaa
\max_{0\leq k\leq k_{small}-2}\sup_{\Mint\cap\mr}\ub^{1+\dec}\left|\dk^k(e_4'(r')-1, e_3'(r')+\Up', e_\th'(r'), e_4'(m'), e_3'(m'), e_\th'(m'))\right| &\les& \ep,\\
\max_{0\leq k\leq k_{large}-1}\left(\int_{\Mint\cap\mr}\left|\dk^k(e_4'(r')-1, e_3'(r')+\Up', e_\th'(r'), e_4'(m'), e_3'(m'), e_\th'(m'))\right|^2\right)^{\frac{1}{2}} &\les& \ep.
\eeaa
On the other hand, we have in view of the decomposition of $e_4'$, $e_3'$ and $e_\th'$ of Step 1
\beaa
e_4'(r) &=& \la\left(e_4+fe_\th+\frac{1}{4}f^2e_3\right)r \\
&=& \la\left(\frac{r}{2}(\overline{\ka}+A)+\frac{1}{4}f^2e_3(r)\right)\\
&=& 1+\Big(\la\Up-1\Big)+\la\left(\frac{r}{2}\left(\overline{\ka}-\frac{2\Up}{r}\right)+\frac{r}{2}A+\frac{1}{4}f^2e_3(r)\right),
\eeaa
\beaa
e_4'(m) &=& \la\left(e_4+fe_\th+\frac{1}{4}f^2e_3\right)m \\
&=& \la\left(e_4(m)+\frac{1}{4}f^2e_3(m)\right),
\eeaa
\beaa
e_3'(r) &=& \la^{-1} \left(  \left(1+ \frac{1}{2}f\fb +\frac{1}{16}f^2\fb^2\right) e_3+\left(\fb+\frac 1 4 \fb^2 f\right) e_\th +\frac{\fb^2}{4}e_4\right)r \\
&=& \la^{-1} \left( e_3(r)+ \left(\frac{1}{2}f\fb +\frac{1}{16}f^2\fb^2\right) e_3(r) +\frac{\fb^2}{4}e_4(r)\right)\\
&=& -\Up+\la^{-1}(\la\Up-1)+\la^{-1}\left(\frac{r}{2}\left(\overline{\kab}+\frac{2}{r}\right)+\left(\frac{1}{2}f\fb +\frac{1}{16}f^2\fb^2\right) e_3(r) +\frac{\fb^2}{4}e_4(r)\right),
\eeaa
\beaa
e_3'(m) &=& \la^{-1} \left(  \left(1+ \frac{1}{2}f\fb +\frac{1}{16}f^2\fb^2\right) e_3+\left(\fb+\frac 1 4 \fb^2 f\right) e_\th +\frac{\fb^2}{4}e_4\right)m \\
&=& \la^{-1} \left( \left(1+\frac{1}{2}f\fb +\frac{1}{16}f^2\fb^2\right) e_3(m) +\frac{\fb^2}{4}e_4(m)\right),
\eeaa
\beaa
e_\th'(r)&=&  \left(\left(1+\frac{1}{2}f\fb\right)e_\th + \frac{\fb}{2}e_4+  \frac{f}{2}\left(1+\frac{f\fb}{4}\right)e_3\right)r\\
&=&   \frac{\fb}{2}e_4(r)+  \frac{f}{2}\left(1+\frac{f\fb}{4}\right)e_3(r),
\eeaa
and
\beaa
e_\th'(r)&=&  \left(\left(1+\frac{1}{2}f\fb\right)e_\th + \frac{\fb}{2}e_4+  \frac{f}{2}\left(1+\frac{f\fb}{4}\right)e_3\right)m\\
&=&   \frac{\fb}{2}e_4(m)+  \frac{f}{2}\left(1+\frac{f\fb}{4}\right)e_3(m).
\eeaa
Together with the identities for $e_4(m)$ and $e_3(m)$ in the ingoing foliation of $\Mint$, the final estimates of Step 6 for $f$ and $\la$, and the bootstrap assumptions for the foliation of $\Mint$, we infer 
\beaa
\max_{0\leq k\leq k_{small}-2}\sup_{\Mint\cap\mr}\ub^{1+\dec}\left|\dk^k(e_4'(r)-1, e_3'(r)+\Up, e_\th'(r), e_4'(m), e_3'(m), e_\th'(m))\right| &\les& \ep,\\
\max_{0\leq k\leq k_{large}-1}\left(\int_{\Mint\cap\mr}\left|\dk^k(e_4'(r)-1, e_3'(r)+\Up, e_\th'(r), e_4'(m), e_3'(m), e_\th'(m))\right|^2\right)^{\frac{1}{2}} &\les& \ep.
\eeaa
We deduce 
\beaa
\max_{0\leq k\leq k_{small}-2}\sup_{\Mint\cap\mr}\ub^{1+\dec}\left|\dk^k(e_4'(r'-r), e_\th'(r-r'), \dk(m'-m))\right| &\les& \ep,\\
\max_{0\leq k\leq k_{large}-1}\left(\int_{\Mint\cap\mr}\left|\dk^k(e_4'(r'-r), e_\th'(r-r'), \dk(m'-m))\right|^2\right)^{\frac{1}{2}} &\les& \ep.
\eeaa
In particular, we have
\beaa
\sup_{\Mint\cap\mr}\ub^{1+\dec}\left|(e_4'(r'-r), e_4'(m'-m))\right| &\les& \ep,
\eeaa
and together with the initialization \eqref{eq:initializationofffblambdaonTTforcontrolglobalframe:quatre}, we integrate the transport equation from $\TT\cup(\Mint\cap\mr)$ and obtain
\beaa
\sup_{\Mint\cap\mr}\ub^{1+\dec}\left|(r'-r, m'-m)\right| &\les& \ep.
\eeaa
Together with the above estimates, and recovering the $e_3'(r'-r)$ using 
\beaa
e_3'(r'-r) &=& \Big(e_3'(r')+\Up'\Big) - \Big(e_3'(r)+\Up\Big) +2\left(\frac{m'}{r'}-\frac{m}{r}\right),
\eeaa
we infer
\beaa
\max_{0\leq k\leq k_{small}-1}\sup_{\Mint\cap\mr}\ub^{1+\dec}\left|\dk^k(r'-r, m'-m)\right| &\les& \ep,\\
\max_{0\leq k\leq k_{large}}\left(\int_{\Mint\cap\mr}\left|\dk^k(r'-r, m'-m)\right|^2\right)^{\frac{1}{2}} &\les& \ep.
\eeaa
Finally, arguing similarly in the region $\Mext\cap\mr$, we infer
\beaa
\max_{0\leq k\leq k_{small}-1}\sup_{\Mext\cap\mr}u^{1+\dec}\left|\dk^k(r'-r, m'-m)\right| &\les& \ep,\\
\max_{0\leq k\leq k_{large}}\left(\int_{\Mext\cap\mr}\left|\dk^k(r'-r, m'-m)\right|^2\right)^{\frac{1}{2}} &\les& \ep,
\eeaa
and hence
\beaa
\max_{0\leq k\leq k_{small}-1}\sup_{\Mint\cap\mr}\ub^{1+\dec}\left|\dk^k(r'-r, m'-m)\right| &\les& \ep,\\
\max_{0\leq k\leq k_{small}-1}\sup_{\Mext\cap\mr}u^{1+\dec}\left|\dk^k(r'-r, m'-m)\right| &\les& \ep,\\
\max_{0\leq k\leq k_{large}}\left(\int_{\mr}\left|\dk^k(r'-r, m'-m)\right|^2\right)^{\frac{1}{2}} &\les& \ep.
\eeaa

{\bf Step 8.}  Recall from Definition \ref{def:definitionoftheglobalframe} that we have defined the global null frame $(e_4'', e_3'', e_\th'')$ as 
\begin{itemize}
\item In $\Mint\setminus\mr$, $(e_4'', e_3'', e_\th'')=(e_4, e_3, e_\th)$.

\item In $\Mext\setminus\mr$, $(e_4'', e_3'', e_\th'')=(\Up e_4', \Up^{-1}e_3', e_\th')$.

\item In $\mr$, $(e_4'', e_3'', e_\th'')$ is given by the change of frame formula starting from $(e_4, e_3, e_\th)$ and with change of frame coefficients $(f'', \fb'', \la'')$ given by
\beaa
f''=\psi f, \quad \fb''=\psi \fb, \quad \la''=1-\psi+\psi\Up'\la,
\eeaa
see \eqref{eq:def:definitionoftheglobalframe}.
\end{itemize}

Also, recall that we have defined $r''$ and $m''$ as
\beaa
r''=(1-\psi)r+\psi r',\quad m''=(1-\psi)m+\psi m'.
\eeaa

{\bf Step 9.} In view of the transformation formulas of Proposition \ref{prop:transformations1}, we have schematically
\beaa
(\Gac'', \Rc'') &=& (\Gac, \Rc)+\dk(f'', \fb'', \la''-1)+f''+\fb''+(\la''-1)+(r''-r)+(m''-m).
\eeaa
In view of the definition of $(f'', \fb'', \la'')$ and $(r'', m'')$ in Step 8, we infer 
\beaa
(\Gac'', \Rc'') &=& (\Gac, \Rc)+\dk(f, \fb, \Up\la-1)+f+\fb+(\Up\la-1)+(r'-r)+(m'-m).
\eeaa
Together with the bootstrap assumptions in $\Mint$ for $(\Ga, \Rc)$, the estimates for $(\Ga, \Rc)$ in $\Mext$ provided by Step 5, the estimates for $(f, \fb, \la)$ provided by Step 6 in $\mr$, and the estimates for $r'-r$ and $m'-m$ provided by Step 7, we deduce
\beaa
\max_{0\leq k\leq k_{small}-2}\sup_{\Mint\cap\mr}\ub^{1+\dec}\left|\dk^k(\Gac'', \Rc'')\right| &\les& \ep,\\
\max_{0\leq k\leq k_{small}-2}\sup_{\Mext\cap\mr}u^{1+\dec}\left|\dk^k(\Gac'', \Rc'')\right| &\les& \ep,\\
\max_{0\leq k\leq k_{large}-1}\left(\int_{\mr}\left|\dk^k(\Gac'', \Rc'')\right|^2\right)^{\frac{1}{2}} &\les& \ep,\\
\left(\int_{\mr}\left|\dk^{k_{large}}(\Gac'', \Rc'')\right|^2\right)^{\frac{1}{2}} &\les& \ep+\left(\int_{\TT}\left|\dk^{k_{large}}({}^{(ext)}\Rc)\right|^2\right)^{\frac{1}{2}}.
\eeaa
Since the double-primed quantities correspond to the quantities associated to the the global frame, this concludes the proof of Lemma \ref{lemma:estimatesfortheglobalframeinthematchingregion}.


\subsection{Proof of Proposition \ref{prop:existenceandestimatesfortheglobalframe:bis}}


To match the first global frame of $\MM$ of Proposition \ref{prop:existenceandestimatesfortheglobalframe:bis} with a conformal renormalization of the second frame of $\Mext$ of Proposition \ref{prop:constructionsecondframeinMext}, we will need to introduce a cut-off function.
\begin{definition}
Let $\psi:\RRR\to\RRR$ a smooth cut-off function such that $0\leq\psi\leq 1$, $\psi=0$ on $(-\infty, 0]$ and $\psi=1$ on $[1, +\infty)$. We define $\psi_{m_0}$ as follows
\beaa
\psi_{m_0}(r)=\begin{cases} &1 \qquad  \mbox{if} \quad r \ge 4m_0,\\[1mm]
&0  \qquad  \mbox{if} \quad r \le \frac{7m_0}{2},
\end{cases}
\eeaa
and
\beaa
\psi_{m_0}(r)=\psi\left(\frac{2\left(r - \frac{7m_0}{2}\right)}{m_0}\right)\textrm{ on }\frac{7m_0}{2}\leq \rext\leq 4m_0.
\eeaa
\end{definition} 

We are now ready to define the second global frame, i.e. the global frame of the statement of Proposition \ref{prop:existenceandestimatesfortheglobalframe:bis}. 
\begin{definition}[Definition of the second global frame]\lab{def:definitionoftheglobalframe:bis}
We introduce a global null frame defined on $\Mext\cup\Mint$ and denoted by $({}^{(glo')}e_4, {}^{(glo')}e_3, {}^{(glo')}e_\th)$. The second global frame is defined as follows
\begin{enumerate}
\item In $\Mext\cap\{\rext\geq 4m_0\}$, we have
\beaa
({}^{(glo')}e_4, {}^{(glo')}e_3, {}^{(glo')}e_\th)= \left({}^{(ext)}\Up e_4', {}^{(ext)}\Up^{-1} e_3', e_\th'\right),
\eeaa
where $(e_4', e_3', e_\th')$ denotes the second frame of $\Mext$, i.e. the one constructed in of Proposition \ref{prop:constructionsecondframeinMext}.

\item In $\Mint\cup(\Mext\cap\{\rext\leq \frac{7m_0}{2}\})$, we have
\beaa
({}^{(glo')}e_4, {}^{(glo')}e_3, {}^{(glo')}e_\th) = \left({}^{(glo)}e_4, {}^{(glo)}e_3, {}^{(glo)}e_\th\right),
\eeaa
where $\left({}^{(glo)}e_4, {}^{(glo)}e_3, {}^{(glo)}e_\th\right)$ denotes the first global frame of $\MM$ of Proposition \ref{prop:existenceandestimatesfortheglobalframe:bis}. 

\item It remains to define the global frame on the matching region $\mr'$. We denote by $f$ the reduced scalar introduced in Proposition \ref{prop:constructionsecondframeinMext} such that we have in $\Mext$
\beaa
e_4' &=& {}^{(ext)}e_4 +  f {}^{(ext)}e_\th +\frac 1 4 f^2  {}^{(ext)}e_3,\\
e_\th' &=&  {}^{(ext)}e_\th +  \frac{f}{2}{}^{(ext)}e_3,\\
e_3' &=&  {}^{(ext)}e_3.
\eeaa
Then, in the matching region $\mr'$, the second global frame of $\MM$ is given by 
\beaa
{}^{(glo')}e_4 &=& \Up'\left({\Up'}^{-1}\,{}^{(glo)}e_4 +  f' {}^{(glo)}e_\th +\frac 1 4 {f'}^2 \Up'\, {}^{(glo)}e_3\right),\\
{}^{(glo')}e_\th &=&  {}^{(glo)}e_\th +  \frac{f'}{2}\Up'\,{}^{(glo)}e_3,\\
{}^{(glo')}e_3 &=&  {}^{(glo)}e_3,
\eeaa
where 
\bea\lab{eq:def:definitionoftheglobalframe:bis}
 f' =\psi_{m_0}(\rext)f, \qquad \Up'=1-\psi_{m_0}(\rext)+\psi_{m_0}(\rext)\,{}^{(ext)}\Up.
\eea
\end{enumerate} 
\end{definition}

\begin{remark}\lab{rem:possibilitytoadaptglobalframetoMextorMint:bis}
Recall that the smooth cut-off function $\psi$ in Definition \ref{def:cutofffunctionforthematchingregion:bis}, allowing to define $\psi_{m_0, \deh}$, is such that we have in particular $\psi=0$ on $(-\infty, 0]$ and $\psi=1$ on $[1, +\infty)$. The following two special cases correspond to the properties (d) i. and (d) ii. of Proposition \ref{prop:existenceandestimatesfortheglobalframe:bis}.
\begin{itemize}
\item If the cut-off $\psi$ in Definition \ref{def:cutofffunctionforthematchingregion:bis} is such that $\psi=1$ on $[1/2, +\infty)$, then
\beaa
({}^{(glo')}e_4, {}^{(glo')}e_3, {}^{(glo')}e_\th)= \left({}^{(ext)}\Up e_4', {}^{(ext)}\Up^{-1} e_3', e_\th'\right)\textrm{ on }\Mext\left(\rext\geq \frac{15m_0}{4}\right).
\eeaa

\item If the cut-off $\psi$ in Definition \ref{def:cutofffunctionforthematchingregion:bis} is such that $\psi=0$ on $(-\infty, 1/2]$, then
\beaa
({}^{(glo')}e_4, {}^{(glo')}e_3, {}^{(glo')}e_\th) = \left({}^{(glo)}e_4, {}^{(glo)}e_3, {}^{(glo)}e_\th\right)\textrm{ on }\Mint\cup\Mext\left(\rext\leq \frac{15m_0}{4}\right).
\eeaa
\end{itemize}
\end{remark}

\begin{remark}
When dealing with the second global frame $({}^{(glo')}e_4, {}^{(glo')}e_3, {}^{(glo')}e_\th)$, the area radius and Hawking mass that we use are the ones corresponding to the first global frame, i.e. ${}^{(glo)}r$ and ${}^{(glo)}m$. 
\end{remark}

The following two lemmas provide the main properties of the second global frame of $\MM$.  
\begin{lemma}\lab{lemma:identitiesrelatingtheglobalframetotheframeofMext:bis}
We have in $\Mext(r\geq 4m_0)$ the following relations between the quantities in the second global frame of $\MM$, i.e. $({}^{(glo')}e_4, {}^{(glo')}e_3, {}^{(glo')}e_\th)$, and the second frame of $\Mext$, i.e. $(e_4', e_3', e_\th')$,
\beaa
&& {}^{(glo')}\a=\Up^2\a',\quad {}^{(glo')}\b=\Up\b',\quad {}^{(glo')}\rho+\frac{2m}{r^3}=\rho'+\frac{2m}{r^3},\quad {}^{(glo')}\bb=\Up^{-1}\bb',\\
&&{}^{(glo')}\aa=\Up^{-2}\aa',\quad {}^{(glo')}\xi= 0, \,\,\,\, {}^{(glo')}\xib =   \Up^{-2}\xib', \,\,\,\,{}^{(glo')}\ze=-{}^{(glo')}\etab=\ze',\\
&& {}^{(glo')}\eta=\eta',\,\,\,\,{}^{(glo')}\om +\frac{m}{r^2}= \Up \om'+ \frac{m}{r^2}\left(1-e_4'(r)\right)+\frac{e_4'(m)}{r},\\
&& {}^{(glo')}\omb =\Up^{-1}\left(\omb' -\frac{m}{r^2}+\frac{m}{r^2}\left(1-\frac{e_3'(r)}{\Up}\right) -\frac{e_3'(m)}{\Up r}\right),\,\,\,\,{}^{(glo')}\ka - \frac{2\Up}{r} = \Up\left(\ka'-\frac{2}{r}\right),\\
&& {}^{(glo')}\kab +\frac{2}{r}= \Up^{-1}\left(\kab'+\frac{2\Up}{r}\right),\,\,\,\, {}^{(glo')}\vth =\Up\vth',  \,\,\,\,    {}^{(glo')}\vthb =\Up^{-1}\vthb'.
\eeaa 
\end{lemma}

\begin{proof}
The proof follows immediately from the change of frame formula with the choice $(f=0, \fb=0, \la=\Up)$, the fact that $e_\th(\Up)=0$, and the fact that the frame $(e_4', e_3', e_\th')$ is such that $\xi'=0$ and $\etab'=-\ze'$.
\end{proof}

\begin{lemma}[Control of the second global frame in the matching region]\lab{lemma:estimatesfortheglobalframeinthematchingregion:bis}
In the matching region, the following estimates holds for the second global frame
\beaa
\max_{0\leq k\leq k_{small}+k_{loss}}\sup_{\mr'}u^{1+\dec-2\de_0}\left|\dk^k({}^{(glo')}\Gac, {}^{(glo')}\Rc)\right| &\les& \ep.
\eeaa
\end{lemma}

\begin{remark}
The quantities associated to the second global frame can be estimated as follows
\begin{itemize}
\item In $\Mint\cup\Mext(\rext\leq \frac{7m_0}{2})$, the second global frame coincides with the first global frame, and hence, the quantities associated to the second global frame satisfy the same estimates than the corresponding quantities for the first global frame.

\item In $\Mext(\rext\geq 4m_0)$, estimates for the quantities associated to the second global frame follow from the identities of Lemma \ref{lemma:identitiesrelatingtheglobalframetotheframeofMext:bis} together with the estimates of Proposition \ref{prop:constructionsecondframeinMext} for the second frame of $\Mext$. 

\item In $\mr'$, the estimates  for the quantities associated to the global frame are provided by Lemma \ref{lemma:estimatesfortheglobalframeinthematchingregion:bis}.
\end{itemize}
\end{remark}

The proof of Proposition \ref{prop:existenceandestimatesfortheglobalframe:bis} easily follows from Definition \ref{def:definitionoftheglobalframe:bis}, Remark \ref{rem:possibilitytoadaptglobalframetoMextorMint:bis}, and Lemma \ref{lemma:estimatesfortheglobalframeinthematchingregion:bis}. Thus, from now on, we focus on the proof of Lemma \ref{lemma:estimatesfortheglobalframeinthematchingregion:bis} which is carried out below.

\begin{proof}[Proof of Lemma \ref{lemma:estimatesfortheglobalframeinthematchingregion:bis}]
Recall from definition \ref{def:definitionoftheglobalframe:bis} that we have in the matching region $\mr'$
\beaa
{}^{(glo')}e_4 &=& \Up'\left({\Up'}^{-1}\,{}^{(glo)}e_4 +  f' {}^{(glo)}e_\th +\frac 1 4 {f'}^2 \Up'\, {}^{(glo)}e_3\right),\\
{}^{(glo')}e_\th &=&  {}^{(glo)}e_\th +  \frac{f'}{2}\Up'\,{}^{(glo)}e_3,\\
{}^{(glo')}e_3 &=&  {}^{(glo)}e_3,
\eeaa
where 
\beaa
 f' =\psi_{m_0}(\rext)f, \qquad \Up'=1-\psi_{m_0}(\rext)+\psi_{m_0}(\rext)\,{}^{(ext)}\Up.
\eeaa
Now, since $\rext\geq \frac{7m_0}{2}$ on $\mr'$, we also have in that region
\beaa
({}^{(glo)}e_4, {}^{(glo)}e_3, {}^{(glo)}e_\th) = ({}^{(ext)}\Up\,{}^{(ext)}e_4, ({}^{(ext)}\Up)^{-1}\,{}^{(ext)}e_3, {}^{(ext)}e_\th).
\eeaa
We deduce on $\mr'$
\beaa
{}^{(glo')}e_4 &=& {}^{(ext)}\Up\left({}^{(ext)}e_4 +  f'' {}^{(ext)}e_\th +\frac 1 4 {f''}^2 {}^{(ext)}e_3\right),\\
{}^{(glo')}e_\th &=&  {}^{(ext)}e_\th +  \frac{f''}{2}\,{}^{(ext)}e_3,\\
{}^{(glo')}e_3 &=&  ({}^{(ext)}\Up)^{-1}\,{}^{(ext)}e_3,
\eeaa
where
\beaa
f'' &=& \Up'({}^{(ext)}\Up)^{-1}f' \\
&=& \Big(1-\psi_{m_0}(\rext)+\psi_{m_0}(\rext)\,{}^{(ext)}\Up\Big)({}^{(ext)}\Up)^{-1}\psi_{m_0}(\rext)f.
\eeaa
In view of the transformation formulas of Proposition \ref{prop:transformations1}, we deduce, schematically,
\beaa
\Big({}^{(glo')}\Gac, {}^{(glo')}\Rc\Big) &=& \Big({}^{(ext)}\Gac, {}^{(ext)}\Rc\Big)+\dk f+f.
\eeaa
Together with the bootstrap assumptions on decay and Proposition \ref{prop:pointwiseboundsforhighorderderivatives} for $({}^{(ext)}\Gac, {}^{(ext)}\Rc)$, and the estimate \eqref{eq:estimateforfinconstructionsecondframeinMext} for $f$, we infer
\beaa
\max_{0\leq k\leq k_{small}+k_{loss}}\sup_{\mr'}u^{1+\dec-2\de_0}\left|\dk^k({}^{(glo')}\Gac, {}^{(glo')}\Rc)\right| &\les& \ep
\eeaa
which concludes the proof of Lemma  \ref{lemma:estimatesfortheglobalframeinthematchingregion:bis}. 
\end{proof}


\chapter{DECAY ESTIMATES FOR $\qf$ (Theorem M1)}\label{chapter:Thm:mainwavetheorem}


The goal of the chapter is to prove Theorem M1, i.e. to derive decay estimates for    the quantity $\qf$  for $k\leq k_{small}+20$ derivatives. To this end, we will make use of the  wave equation satisfied by $\qf$ (see \eqref{thmwaveqf:improperfrom})
\bea
\lab{eq:Masterwaveequation-qf}
\square_2 \qf  +\ka \kab\, \qf&=& N,
\eea
where $N$ contains only quadratic or higher order terms. Now, in order to have a suitable right-hand side $N$, recall from the discussion in Remarks \ref{rmk:whyweneedaglobalframewithbettereta} and \ref{def:wherewementionthatwealwaysexpressqfinthesecondglobalframe} that $\qf$ is defined relative to the global null frame of Proposition \ref{prop:existenceandestimatesfortheglobalframe:bis} for which $\xi=0$ for $r\geq 4m_0$ and $\eta\in \Ga_g$. For such a global fame, $N$ is given schematically by, see \eqref{thmwaveqf:schematicformerrorterm}, 
\bea\lab{eq:Masterwaveequation-qf:theerrortermofit}
N &=& r^2 \dk^{\le 2}(\Ga_g \c (\a, \b) )         + e_3 \Big( r^3 \dk^{\le 2}(\Ga_g \c (\a, \b) )   \Big) +  \dk^{\le 1 } (\Ga_g \c \qf)+\lot
\eea


\section{Preliminaries}



\subsubsection{Smallness constants}


Recall from the beginning of section \ref{sec:statementofthemaintheorempreciseversion} the constant $m_0$ and the main small constants $\deh$, $\dt$, $\dec$, $\ep$ and $\ep_0$ such that 
\begin{itemize}
\item The constant $m_0>0$ is the mass of the initial Schwarzschild spacetime relative to which our initial perturbation is measured. 

\item The integer $k_{large}$ which corresponds to the maximum number of derivatives of the solution.

\item The size of the initial data layer norm is measured by $\ep_0>0$. 

\item The size of the bootstrap assumption norms are measured by $\ep>0$.

\item $\deh>0$ measures the width of the region $|r-2m_0|\leq 2m_0\deh$ where the redshift estimate holds and which includes in particular the region $\Mint$.

\item $\dec$ is tied to decay estimates  in $u$, $\ub$  for $\check{\Gamma}$ and $\Rc$.

\item $\dt$ is involved in the $r$-power of the $r^p$ weighted estimates for curvature.
\end{itemize}
Recall also that these constants satisfy in view of \eqref{eq:constraintsonthemainsmallconstantsepanddelta} \eqref{eq:constraintsonthemainsmallconstantsepanddelta:bis} \eqref{eq:constraintbetweenepep*ep0}
\beaa
\begin{split}
0<\deh,\,\, \dec, \,\,\dt \ll \min\{m_0, 1\},\qquad \dt> 2\dec, \qquad k_{large}\gg \frac{1}{\dec},
\end{split}
\eeaa
\beaa
\ep_0, \ep\ll \min\{\deh, \dec, \dt, m_0, 1\},
\eeaa
and
\beaa
\ep=\ep_0^{\frac{2}{3}}.
\eeaa

We will need the following additional small constants in this chapter
\begin{itemize}
\item $\dee>0$, tied to the decay of $\qf$, and is chosen such that $\delta_{extra}>\delta_{dec}$,

\item $\de>0$ for  various degeneracies, 

\item $\de_0>0$ which comes from interpolating between $k\le k_{small}$ derivatives
 of $(\Gac, \Rc)$ and  $k\le k_{large} $ derivatives of $(\Gac, \Rc)$, see Lemma \ref{le:interpolatedbootstrap},
 
\item $ q_0>0$ which will allow us to recover the fact that the decay for $\qf$ in Theorem M1 has an extra gain $u^{-(\dee-\dec)}$ compared from the expected behavior inferred from the bootstrap assumptions. 
\end{itemize}

We will choose $\dee$ such that
\beaa
 \dec<\dee<2\dec,\,\, \dt\geq 2\dee,
\eeaa
$\de$ and $\de_0$ such that 
\bea\lab{eq:deandde0muchsmallerthananyotherdeltaintheproof}
0<\ep, \ep_0\ll \de, \de_0\ll \dec, \dee, \deh, m_0, 1,
\eea
and $q_0$ such that\footnote{This will allow us to choose in the proof of Theorem M1, see \eqref{eq:wefinallymakethechoiceofq0striclylargethandeltadec},
\beaa
\dee=\frac{q_0-\de}{2}
\eeaa
which satisfies the desired estimate $\dee>\dec$ for $\de>0$ small enough.}  
\bea
2\dec<q_0<4\dec-4\de_0-4\de.
\eea


\subsection{The foliation of $\MM$ by $\tau$}\lab{sec:foliationofMMbytau}


Recall that the spacetime $\MM$ is decomposed as $\MM=\Mint\cup \Mext$ and that $u$ is an outgoing optical function on $\Mext$ while $\ub$ is an ingoing optical function. In this chapter, we rely on the global frame $(e_3, e_4, e_\th, e_\vphi)$ defined in    section  \ref{section-globaleframe}, and $r$ and $m$ denote the corresponding scalar functions associated to it. Also, we define the  trapping region region  $\MM_{trap} $ as,
\bea
\MM_{trap} &:=&     \left\{  \frac{5m_0}{2}\le r \le  \frac{7m_0}{2} \right\}.
\eea
Also, let $\Mntrap=\MM\setminus\Mtrap$ the complement of $\Mtrap$ in $\MM$.

                We foliate   our spacetime   domain $\MM$    by      $\Z$ invariant  hypersurfaces  $\Si(\tau) $ which are:
             \begin{itemize}
             \item  Incoming null    in $ \Mint$,        
              with $e_3$ as null incoming   generator. We denote  this portion 
              $\Sint(\tau)$. 
              
               \item   Strictly spacelike  in $\Mtrap$. We denote  this  portion   by $\Sitrap$. 
               
              \item  Outgoing null  in $\MM_{>4m_0}$.  We denote this portion by  $\Si_{>4m_0}(\tau)$.
              
              \item  The parameter $\tau$  of $\Si(\tau) $  can be chosen, smoothly, such that  
              \bea\lab{eq:defintionoftauininMtrapandnullregions}
              \tau:=\left\{\ba{ll}
              u & \text{in }\,\,\MM_{> 4m_0},  \\[1mm]
              u+r & \text{in }\,\,\MM_{trap},\\[1mm]
              \ub &  \text{in  }\,\,  \Mint.
              \ea\right.
              \eea
                          
            \item In particular, the   unit   normal in the region $\MM_{trap}$, i.e. the normal to $\Sitrap$, satisfies\footnote{$N_\Si$ is  given in view of its definition by 
            \beaa
            N_\Si &=& \frac{1}{\sqrt{2}\sqrt{e_4(r)(e_3(u)+e_3(r))}}\Big(e_4(r)e_3+(e_3(u)+e_3(r))e_4\Big)\\
               &=&   \frac{1}{\sqrt{2}\sqrt{2-\Up+O(\ep)}}\Big((1+O(\ep))e_3+(2-\Up+O(\ep))e_4\Big)
            \eeaa
            where we used the bootstrap assumptions.}
            \bea\lab{eq:upperandlowerboundforNSigmaonMMtrap}
            -2\leq g(N_\Si, e_4)\leq -1, \quad  -2\leq g(N_\Si, e_3)\leq -1\textrm{ on }\MM_{trap}.
            \eea
           \end{itemize}

              
\subsection{Assumptions for Ricci coefficients and curvature}\label{subsection:Main assumptions-wave}
\lab{sec:interpolatedboundsforGac}
              
              
Recall from  Remark \ref{def:wherewementionthatwealwaysexpressqfinthesecondglobalframe} that $\qf$ is defined, according to equation \eqref{def: gen-qf} in Lemma  \ref{lemma:def: gen-qf},  relative to the global frame of Proposition \ref{prop:existenceandestimatesfortheglobalframe:bis} for which  $\eta\in \Ga_g$ with the notation
 \beaa
\Ga_g  &=&      \Ga^{(0)}_{g} = \Big\{ \xi,\,  \vth, \,  \om+\frac{m}{r^2},\,  \ka-\frac{2\Up}{r}, \,   \eta,\,  \etab,\,  \ze, \,  A \Big\},\\
\Ga_b&=& \Ga^{(0)}_{b} =\Big\{\vthb,\,  \kab+\frac{2}{r}, \,  \underline{A}, \, \omb, \, \xib  \Big\},
\eeaa
where we recall that  
\beaa
\Up=1-\frac{2m}{r},  \quad  A=\frac{2}{r}e_4(r)-\ka, \quad \underline{A}=\frac{2}{r}e_3(r)-\kab.
\eeaa
Note also that  $\xi$ vanishes in $\Mext$ away from the matching region of Proposition \ref{prop:existenceandestimatesfortheglobalframe:bis}, and in particular for $r\geq 4m_0$.

For higher derivatives we write,
\beaa
\Ga_g^{(1)}&=& \Big\{ \dk \xi,\,  \dk\vth,\,  r e_\th \om, \,     r e_\th(\ka), \,        \dk \eta, \, \dk \etab,\dk  \ze,\,  \dk  A\Big\}\\
 \Ga^{(1)}_{b}&=&\Big\{\dk \vthb , re_\th(\kab), \dk \xib,\dk \underline{A}, \,  r e_\th \omb,  \dk \xib \Big\}
\eeaa
and for $s\ge 2$,
\beaa
 \Ga^{(s)}_{g}&=&\dk^{s-1} \Ga_g^{(1)},  \qquad   \Ga^{(s)}_{b}=\dk^{s-1} \Ga_b^{(1)}
\eeaa
Moreover we denote
\beaa
\Ga_g^{\le s} = \Big\{  \Ga^{(0)}_{g},    \Ga^{(1)}_{g}, \ldots    \Ga^{(s)}_{g}       \Big\}, \qquad \Ga_b^{\le s} = \Big\{  \Ga^{(0)}_{b},    \Ga^{(1)}_{b}, \ldots    \Ga^{(s)}_{b}  \Big\}.
\eeaa
  
With these notations, we may now state the estimates satisfied by the Ricci coefficients and curvature components.
\begin{lemma}
\label{le:interpolatedbootstrap}
Consider the global frame of  Proposition \ref{prop:existenceandestimatesfortheglobalframe:bis} and the above definition\footnote{Recall in particular that the global frame of  Proposition \ref{prop:existenceandestimatesfortheglobalframe:bis} is such that $\eta\in\Ga_g$.} of $\Ga_g$ and $\Ga_b$. 
Let an integer $k_{loss}$ and a small constant $\de_0>0$ satisfying\footnote{Recall that we have
\beaa
0<\dec\ll 1, \qquad \dec\, k_{large}\gg 1, \qquad k_{small}=\left \lfloor\frac 1 2 k_{large}\right \rfloor +1.
\eeaa
In particular, we have $\dec(k_{large}-k_{small})\gg 1$ 
and hence there exists an integer $k_{loss}$ satisfying the required constraints. We will in fact choose $k_{loss}=33$, see \eqref{eq:defofdelta0comingfromtheinterpolationlemma}.}
\bea\lab{eq:constraintsonklossandde0forsecondframeofMext:bis=ThmM1}
16\leq k_{loss} \leq   \frac{\dec}{3}(k_{large}-k_{small}), \qquad \de_0=\frac{k_{loss}}{k_{large}-k_{small}}.
\eea
Then, the Ricci coefficients and curvature components with respect to the global frame of  Proposition \ref{prop:existenceandestimatesfortheglobalframe:bis} satisfy 
\beaa
\xi=0\textrm{ on }r\geq 4m_0,
\eeaa
\beaa
\nn\max_{0\leq k\leq k_{small}+k_{loss}}\sup_{\MM}&&\Bigg\{\Big(r^2\tau^{\frac{1}{2}+\dec-2\de_0}+r\tau^{1+\dec-2\de_0}\Big)|\dk^k\Ga_g|+r\tau^{1+\dec-2\de_0}|\dk^k\Ga_b|\\
\nn&&+\Big(r^{\frac{7}{2}+\frac{\dt}{2}}+r^3\tau^{\frac{1}{2}+\dec-2\de_0}+r^2\tau^{1+\dec-2\de_0}\Big)\Big(|\dk^k\a|+|\dk^k\b|\Big)\\
\nn&&+\Big(r^3\tau^{\frac{1}{2}+\dec-2\de_0}+r^2\tau^{1+\dec-2\de_0}\Big)|\dk^k\rhoc|\\
&&+\tau^{1+\dec -2\de_0}\Big(r^2|\dk^k\bb|+r|\dk^k\aa|\Big)\Bigg\} \les \ep,
\eeaa
\beaa
\max_{0\leq k\leq k_{small}+k_{loss}}\sup_{\MM}&&\Bigg\{r^2\tau^{1+\dec-2\de_0}|\dk^{k-1}e_3(\Ga_g)|\\
&&+r^3(\tau+2r)^{1+\dec-2\de_0}\Big(|\dk^{k-1}e_3(\a)|+|\dk^ke_3(\b)|\Big)\Bigg\} \les \ep.
\eeaa
  \end{lemma}
  
 \begin{proof}
 In $r\geq 4m_0$, the global frame of  Proposition \ref{prop:existenceandestimatesfortheglobalframe:bis} coincides with a conformal renormalization of the second frame of $\Mext$, see Proposition \ref{prop:constructionsecondframeinMext}. The estimates there follow immediately from the ones of Proposition \ref{prop:constructionsecondframeinMext}. In the matching region $7/2m_0\leq r\leq 4m_0$, the estimates are stated in Proposition \ref{prop:existenceandestimatesfortheglobalframe:bis}. Finally, for $\Mext(r\leq 7/2m_0)$ and $\Mint$, the estimates follow directly from interpolation between the bootstrap assumptions on decay for $k\leq k_{small}$ and the pointwise estimates of Proposition \ref{prop:pointwiseboundsforhighorderderivatives} for $k\leq k_{large}-5$. 
  \end{proof}


\subsection{Structure of nonlinear terms}  


The following lemma will be important in what follows.
\begin{lemma}
\label{Lemma:structure-forN}
For the solution $\qf$ to the wave equation \eqref{eq:Masterwaveequation-qf}, the structure of the   error term $N$ can be written schematically as follows
\bea
\label{eq"effectivestructureforN}
N&=& \Ng+ e_3(r\Ng)+ \Nm
\eea
where,
\bea
\bsplit
\Ng &= r^2 \dk^{\le 2}(\Ga_g \c (\a, \b) ),\\
\Nm &=  \dk^{\le 1 } (\Ga_g \c \qf).
\end{split}
\eea
Moreover, for every $k\le k_{large}-3$ we have schematically,
 \bea
\label{eq"effectivestructureforN-s}
\dk ^k N&=& \dk^ {\le k}\Ng+ e_3(\dk^k(r\Ng))+ \dk^k \Nm.
\eea
\end{lemma} 

\begin{remark}
 In fact, \eqref{eq"effectivestructureforN} and \eqref{eq"effectivestructureforN-s} also contain lower order terms which are strictly better in powers of $r$ and  contain at most the same number of derivatives. For convenience, we drop them in the rest of the proof of Theorem M1.
\end{remark}

\begin{proof}
For $k=0$, this is an immediate consequence of \eqref{eq:Masterwaveequation-qf:theerrortermofit}. For the higher derivatives we write,
\beaa
\dk^k (e_3 (r\Ng))= e_3 (\dk^k(r\Ng))+ [\dk^k, e_3](r\Ng).
\eeaa
In view of the formula for $[e_3, \dkb]$ of Lemma \ref{Le:comme3e4-outgeodesic}, and the commutator formula for $[e_3, e_4]$, we have,  schematically,
\beaa
\,[e_3, e_3]=0, \quad \,[\dkb, e_3]=\Ga_b \dk  +\Ga_b, \quad [re_4, e_3]=\left(\frac{1}{r}+\Ga_b\right) \dk.
\eeaa
In view of our assumptions.
\beaa
\big| \dk^i(  \Ga_b)\big|\le r^{-1} \ep, \qquad i\le k_{large}-4,
\eeaa
$\Ga_b$ is at least as good as $r^{-1}$, and hence, we deduce,  schematically, 
\beaa
\,[\dk, e_3]&=& \frac{1}{r}\dk  +\frac{1}{r}.
\eeaa
On the other hand, we have, schematically,
\beaa
[\dk, r]=r
\eeaa
and hence, for $k\leq k_{large}-3$,
\beaa
\,[ \dk^k, e_3]  (r\Ng)&=&\sum_{i+j\le k-1 }\dk^i\left(\frac{1}{r}\dk+\frac{1}{r}\right) \dk^j(r\Ng)\\
&=& \dk^{\leq k}\Ng
\eeaa
as desired.
\end{proof}

  
\subsection{Main quantities}\lab{sec:mainquantitiesforcontrolofwaveequation}


 We restrict our attention to the region $\MM(\tau_1, \tau_2)=\MM\cap \{\tau_1\le \tau\le \tau_2\}$. For a given $\psi\in \sk_2(\MM)$  we introduce the following  quantities, for $0\le \tau_1<\tau_2\le \tau_*$.

                 
     \subsubsection{Morawetz bulk quantities} 
     
     
Consider the vectorfields,
 \bea\lab{def:TandR}
 T := \frac 1 2 \left(e_4+\Up e_3 \right), \qquad   R := \frac 1 2 \left(  e_4 - \Up  e_3\right).
 \eea
    
        Let $\th$  a smooth bump  function   equal $1$ on  $  |\Up|\le\de_\HH^{\frac{1}{10}} $  vanishing for  $|\Up|\ge2 \de_\HH^{\frac{1}{10}}$ and define the modified  vectorfields, 
      \bea
       \label{eq:Rc-Tc}
       \begin{split}
  \Rbrev&:=  \th  \frac 1 2 ( e_4-e_3) +(1-\th)        \Up^{-1} R=    \frac 1 2 \left[\thbrev e_4- e_3\right], \\
  \Tbrev&:=\th  \frac 1 2 ( e_4+e_3) +(1-\th)        \Up^{-1} T= \frac 1 2 \left[ \thbrev e_4+ e_3\right],
  \end{split}
      \eea 
      where $\thbrev=\th+\Up^{-1} (1-\th)$. Note that,     
        \bea
        \label{eq:cutoff-tildechi}
      \thbrev=\begin{cases}  1&\qquad \mbox{for} \quad |\Up|\le  \de_\HH^{\frac{1}{10}},\\
      \Up^{-1}&\qquad \mbox{for} \quad |\Up|\ge2 \de_\HH^{\frac{1}{10}}.
      \end{cases}
      \eea 
      
\begin{remark}  
Note that 
\beaa
\Rbrev+\Tbrev=e_4,\,\,  -\Rbrev+\Tbrev=e_3\textrm{ in }\Mint\textrm{ and }\Rbrev+\Tbrev=\Up^{-1} e_4, \,\, -\Rbrev+\Tbrev=e_3\textrm{ in }\MM_{>4m_0}.
\eeaa
\end{remark}

     We define the quantities
    \bea
    \label{def:Mor-bulk}
    \bsplit
\Mor[\psi](\tau_1, \tau_2):&=\int_{\MM(\tau_1, \tau_2)} \frac{1}{r^3}  |\Rbrev \psi|^2 +\frac{1}{r^4} |\psi|^2  +\left(1-\frac{3m}{r}\right)^2\frac{1}{r}\left(|\nabb\psi|^2 +\frac{1}{r^2}|\Tbrev\psi|^2 \right),\\
\\
\Morr[\psi](\tau_1, \tau_2):&=\Mor[\psi](\tau_1, \tau_2)+\int_{\MM_{> 4 m_0}(\tau_1, \tau_2)} r^{-1-\de}  |e_3(\psi)|^2 ,
\end{split}
\eea
with $m=m(\tau,  r)=m(u, r)$  the  Hawking mass in $\MM$. The  constant $\de>0$  is a sufficiently small quantity.  An equivalent   definition  for $\Morr[\psi](\tau_1, \tau_2)$  is given below,
\bea
\bsplit
\Morr[\psi](\tau_1, \tau_2)&=\int_{\Mtrap(\tau_1, \tau_2)}  | R\psi|^2 +r^{-2}  |\psi|^2     +\left(1-\frac{3m}{r}\right)^2\left(|\nabb\psi|^2 +\frac{1}{r^2}| T \psi|^2 \right)    \\
  &+ \int_{\Mntrap(\tau_1,\tau_2)}   r^{-3} \big(|e_4 \psi|^2      +r^{-1} |\psi|^2\big) + r^{-1} |\nabb\psi|^2        +    r^{-1-\de}|e_3\psi|^2 
  \end{split}
  \eea
  where $\Mntrap$ denotes the complement of $\Mtrap$.

  
  \subsubsection{Weighted bulk quantities}
 

  Define, for  $0<p<2$,
  \bea
  \label{notation:Mps}
\bsplit
\Bdot_{p\,; \,R}[\psi](\tau_1, \tau_2):&=\int_{\MM_{\ge  R}(\tau_1, \tau_2) }  r^{p-1}  \left( p | \ec_4(\psi) |^2 +(2-p)   |\nabb \psi|^2+   r^{-2}  |\psi|^2\right),\\
\\
B_{p}[\psi](\tau_1, \tau_2):&=\Morr[\psi](\tau_1, \tau_2)+\Bdot_{p\,; \,4m_0}[\psi](\tau_1, \tau_2).
\end{split}
\eea
The bulk quantity  $B_{p}[\psi](\tau_1, \tau_2) $ is equivalent to\footnote{This equivalence follows from the coarea formula and the fact that the lapse of the $\tau$-foliation is controlled uniformly from above and below.}
\beaa
 B_{p}[\psi](\tau_1, \tau_2) \simeq \int_{\tau_1}^{\tau_2}  M_{p-1}[\psi](\tau) d\tau
\eeaa
where,
\beaa
M_{p-1}[\psi](\tau)&=&\int_{\Si_{\le  4m_0} (\tau)}   |\Rbrev \psi|^2  +r^{-2} |\psi|^2  +\left(1-\frac{3m}{r}\right)^2\left(|\nabb\psi|^2 +\frac{m^2}{r^2}| \Tbrev\psi|^2 \right)\\
&+& \int_{\Si_{\ge   4m_0} (\tau)}   r^{p-1}  \left(p | e_4(\psi) |^2 + (2-p)   |\nabb \psi|^2+ r^{-2}|\psi|^2 \right)  +  \int_{\Si_{\ge   4m_0} (\tau)} r^{-1-\de} |e_3\psi|^2.
\eeaa
\begin{remark}
\label{rem:equivalentB-norms:0}
Note that, for $\de\leq p\leq 2-\de$,
\beaa
 B_{p}[\psi](\tau_1, \tau_2):&=&\Morr[\psi](\tau_1, \tau_2)+\Bdot_{p\,; \,4m_0}[\psi](\tau_1, \tau_2)
 \eeaa
 is equivalent to,
 \beaa
 B_{p}[\psi](\tau_1, \tau_2)&\simeq&\Morr[\psi](\tau_1, \tau_2)+\int_{\MM_{\ge  4m_0}(\tau_1, \tau_2) }  r^{p-1}  \left(  | \ec_4(\psi) |^2 +  |\nabb \psi|^2+  r^{-2} |e_3 \psi|^2 +  r^{-2}  |\psi|^2\right).
 \eeaa
 Indeed,
 \beaa
 \int_{\MM_{\ge  4m_0}(\tau_1, \tau_2) } r^{p-3 } |e_3\psi|^2 \les  
 \int_{\MM_{\ge  4m_0}(\tau_1, \tau_2) } r^{-1-\de}|e_3 \psi|^2. 
 \eeaa
 Therefore, since  $  r^2\left(| \ec_4(\psi) |^2 +  |\nabb \psi|^2\right) \les |\dk\psi|^2 $, we have,
 \bea
 B_{p}[\psi](\tau_1, \tau_2) &\simeq& \Morr[\psi](\tau_1, \tau_2)+\int_{\MM_{\ge  4m_0}(\tau_1, \tau_2) }  r^{p-3} 
  \left( |\dk \psi|^2 + |\psi|^2\right).
 \eea
 \end{remark}


  \subsubsection{Basic     energy-flux quantity}
  
  
The basic energy-flux quantity on a hypersurface $\Si(\tau)$  is defined by
\bea
\label{def:basic-energy}
E[\psi](\tau)= \int_{\Si(\tau)}\bigg( \frac 1 2  (N_\Si, e_3)^2  \,    |e_4 \psi|^2  +\frac 1 2 (N_\Si, e_4 )^2\,  |e_3\psi|^2 +|\nabb\psi|^2 + r^{-2}|\psi|^2 \bigg).
\eea
Here $N_\Si$ denotes a choice for the normal to $\Si$ so that in particular we have 
\bea
N_{\Si}=\begin{cases}
 N_{\Si}=e_3 \qquad \mbox{on} \quad \Sint,\\
 N_{\Si}=e_4 \qquad \mbox{on} \quad \Sext,
\end{cases}
\eea
and, in view of \eqref{eq:upperandlowerboundforNSigmaonMMtrap},
\bea
 (N_\Si, e_3)\leq -1\textrm{ and }(N_\Si, e_4)\leq -1\qquad  \mbox{on} \quad \Sitrap.
\eea


\subsubsection{Weighted Energy-Flux type quantities}


We have
\bea \label{notation:Eps:0}
  \Ed_{p\,; \,R}[\psi](\tau)&:=& \begin{cases}
\displaystyle \int_{\Si_{\ge  R}(\tau)}  r^p\Big(  |\ec_4\psi|^2+ r^{-2} |\psi|^2 \Big)\quad \textrm{ for }p\leq 1-\de,\\[5mm]
\displaystyle \int_{\Si_{\ge  R}(\tau)}  r^p\Big(|\ec_4\psi|^2+r^{-p-1-\de}|\psi|^2\Big)\quad\textrm{ for }p> 1-\de,
\end{cases}
\eea
and
 \bea\label{notation:Eps}
  E_{p}[\psi](\tau)&:=& E[\psi](\tau)+ \Ed_{p\,; \,4m_0}[\psi](\tau).
  \eea
    
Here  $\ec_4$ denotes the first order operator
\bea
\ec_4\psi =r^{-1} \Up^{-1}e_4(r\psi).
\eea

\begin{remark}
To control the weighted quantities \eqref{notation:Eps}, it will be convenient to introduce in $\Mext(r\geq 4m_0)$ the following renormalized frame
\beaa
e_4'=\Up^{-1}e_4,\quad e_3'=\Up e_3, \quad e_\th'=e_\th.
\eeaa
In particular, this yields
\beaa
\ec_4\psi =r^{-1}e_4'(r\psi).
\eeaa
Note also that we have the following alternate form
\beaa
\ec_4\psi  = e_4'\psi+ r^{-1} \psi +\frac{e_4'(r)-1}{r}\psi
\eeaa
where $e_4'(r)-1=\Up^{-1}e_4(r)-1=O(\ep r^{-1})$ in view of our assumption on $\Ga_g$. 
\end{remark}


\subsubsection{Flux quantities}


The boundary of $\MM(\tau_1, \tau_2)$ is given by
\beaa
\pr\MM(\tau_1, \tau_2) &=& \Sigma(\tau_1)\cup\Sigma(\tau_2)\cup\AA(\tau_1, \tau_2)\cup\Sigma_*(\tau_1, \tau_2).
\eeaa
Our basic flux quantity along the spacelike hypersurfaces $\AA$ and $\Sigma_*$ is given by
\bea
\nn F[\psi](\tau_1, \tau_2) &:=& \int_{\AA(\tau_1, \tau_2)}\Big( \deh^{-1}   |e_4 \Psi|^2  + \deh  |e_3\Psi|^2 + |\nabb \Psi|^2 + r^{-2} |\Psi|^2\Big)\\
&&  +\int_{\Sigma_*(\tau_1, \tau_2)}\Big(  |e_4 \Psi|^2  + |e_3\Psi|^2 + |\nabb \Psi|^2 + r^{-2} |\Psi|^2\Big),
\eea
with $\AA(\tau_1, \tau_2)=\AA\cap\MM(\tau_1, \tau_2)$ and $\Sigma_*(\tau_1, \tau_2)=\Sigma_*\cap\MM(\tau_1, \tau_2)$.


\subsubsection{Weighted flux quantities}

 
\bea
\begin{split}
\Fd_p[\psi](\tau_1, \tau_2) &:= \int_{\Sigma_*(\tau_1, \tau_2)}r^p\Big( |e_4\psi|^2+|\nabb\psi|^2 + r^{-2} |\psi|^2 \Big),\\
F_p[\psi](\tau_1, \tau_2) &:= F[\psi](\tau_1, \tau_2)+\Fd_p[\psi](\tau_1, \tau_2).
\end{split}
\eea

  
\subsubsection{Weighted  quantities  for the inhomogeneous term $N$}
    
    
Recall the decomposition \eqref{eq"effectivestructureforN} for the  inhomogeneous term $N$
\beaa
N&=& \Ng+ e_3(r\Ng)+ \Nm.
\eeaa    
We define, for $p\ge \de$, 
 \bea
\label{def:normI-Nde}
 \nn I_p[N_g]( \tau_1,\tau_2) &=& \bigg( \int_{\tau_1}^{\tau_2} d\tau \|N_g\|_{L^2(\Sitrap(\tau))}\bigg)^2+\int_{\Mntrap(\tau_1,\tau_2)} r^{1+p}|N_g|^2\\
\nn && +\int_{\Mntrap(\tau_1,\tau_2)} r^{2+p}|N_g||e_3(N_g)|+\sup_{\tau\in[\tau_1, \tau_2]}  \int_{\Si(\tau)} r^{p+2 } \big |N_g \big |^2\\
 &&+\int_{\Mntrap(\tau_1,\tau_2)} r^{3+\de}|e_3(N_g)|^2.
\eea

 \begin{remark}
 While $\Nm$ is present in the decomposition of  the  inhomogeneous term $N$, \eqref{def:normI-Nde} only contains a norm for $N_g$. In fact, $\Nm$ will always be absorbed by the left hand side wherever it appears.
\end{remark} 


  \subsubsection{Higher derivative  quantities}
  
  
  We define the higher order derivative quantities  $E^s[\psi],   \Mor^s[\psi],  \Morr^s[\psi],     E^s_{p}[\psi]$, $B^s_b[\psi]$,  $M ^s_{p}[\psi]$, $F^s[\psi]$, $F^s_p[\psi]$, $I_p^s[N_g]$
   by the obvious procedure,
   \beaa
   Q^s[\psi]= \sum_{k\le s} Q[\dk^k \psi].
   \eeaa
   
       \begin{remark}
    \label{rem:equivalentB-norms}
    Note that in view of Remark \ref{rem:equivalentB-norms:0} we  can also write, equivalently, for $p<2-\de$,
    \bea
    B_p^s[\psi](\tau_1, \tau_2)=Morr^s[\psi](\tau_1, \tau_2)+\int_{\MM_{>4m_0}(\tau_1,\tau_2)} r^{p-3}|\dk^{\le 1+s}\psi|^2.
    \eea
     \end{remark}
    


 

\subsubsection{Decay Norms}


We introduce,
\bea
\label{definition:normsdecay-psi}
\bsplit
\EE^s_{p,d}[\psi]:&= \sup_{0\le \tau\le \tau_*} (1+\tau)^d E^s_{p}[\psi](\tau),\\
\BB^s_{p, d}[\psi]:&= \sup_{0\le \tau\le \tau_*} (1+\tau)^d B^s_p[\psi](\tau, \tau_*), \\
&\simeq\sup_{0\le \tau\le \tau_*} (1+\tau)^d\int_{\tau}^{\tau_*} M^s_{p-1}[\psi](\tau') d\tau',\\
\FF^s_{p,d}[\psi]:&= \sup_{0\le \tau\le \tau_*} (1+\tau)^d F^s_{p}[\psi](\tau, \tau_*),\\
\II^s_{p,d}[N_g]:&= \sup_{0\le \tau\le \tau_*} (1+\tau)^d I^s_{p}[N_g](\tau, \tau_*).
\end{split}
\eea


\section{Proof of  Theorem M1}


Recall that we have to prove for $k\leq k_{small}+20$
\beaa
|\dk^k\qf|&\les& \ep_0 r^{-1} (1+\tau)^{-\frac{1}{2}-\dee},\\ 
  |\dk^k\qf|&\les& \ep_0 r^{-\frac{1}{2}}  (1+\tau) ^{-1-\dee}, \\
  |\dk^ke_3(\qf)| &\les& \ep_0 r^{-1} (1+\tau) ^{-1-\dee},
\eeaa
and 
\beaa
\int_{\Mint(\tau, \tau_*)}|\dk^ke_3\qf|^2 +\int_{\Sigma_*(\tau, \tau_*)}|\dk^ke_3\qf|^2 &\les& \ep_0^2(1+\tau) ^{-2-2\dee},
\eeaa
for some constant $\dee$ such that $\dec<\dee<2\dec$.

   
\subsection{Flux Decay Estimates for $\qf$}

    
The following result establishes decay of flux estimates for $\qf$.  
\begin{theorem}
\label{theorem:improveddecay-estimates-psi}
Let $0<q_0 <1$ be a fixed number and $s\le k_{small}+25$.  Then, for all $\de>0$ we have, with a constant $C$ depending only on $s$,  $\de$ and $q_0$ such that for all  $\de\le p\le 2-\de$, we have
\bea
\nn&&\EE^s_{p, 2+q_0-p}[\qf] +\BB^s_{p, 2+q_0-p}[\qf] +\FF^s_{p, 2+q_0-p}[\qf]  \\
  &\les&  \EE_{q_0}^{s+2}[\qfc] (0)+\EE_{2-\de}^{s+4} [\qf](0) +   \II^{s+5}_{q_0+2, 0}[\Ng]  + \II^{s+5}_{\de, 2+q_0 -\de}[\Ng],
\eea
where we recall that the  decay norms $\II_{p, d}^s[\Ng]$ are defined by,
\beaa
\II_{p,d}^s[\Ng] &=&\sup_{0\le \tau\le \tau_*} ( 1+\tau)^d I_{p}^s[\Ng] (\tau, \tau_*).
\eeaa
\end{theorem}
    
Theorem \ref{theorem:improveddecay-estimates-psi} will be proved in section \ref{sec:proofoftheorem:improveddecay-estimates-psi}. 

To prove Theorem M1   we  have to eliminate    the norms $\II^s_{p,d}[\Ng]$ on the right hand side of Theorem \ref{theorem:improveddecay-estimates-psi}.
\begin{proposition}
\label{Prop:Main-Thm1-1}
Let $s\le k_{small}+30$ and assume 
\bea
q_0<4\dec-4\de_0
\eea
where 
 \bea\lab{eq:defofdelta0comingfromtheinterpolationlemma}
 \de_0= \frac{33}{k_{large}-k_{small}}=\frac{33}{k_{large}-\lfloor \frac{k_{large}}{2}\rfloor-1}
 \eea
 is the  small constant  appearing in Lemma  \ref{le:interpolatedbootstrap}. Then, the following estimates hold true,
\beaa
  \II^s_{q_0+2, 0}[\Ng]  + \II^s_{\de, 2+q_0-\de }[\Ng]&\les& \ep^4. 
\eeaa
\end{proposition}

The proof of Proposition \ref{Prop:Main-Thm1-1} is postponed to section \ref{sec:proofofProp:Main-Thm1-1}. Together with Theorem  \ref{theorem:improveddecay-estimates-psi}, Proposition \ref{Prop:Main-Thm1-1}  immediately yields the proof of the following corollary. 
 
\begin{corollary}
\label{theorem:gen-decay-psic-improved}
In addition to the  assumptions  of  Theorem \ref{theorem:improveddecay-estimates-psi}  we assume
\bea
2\dec<q_0<4\dec-4\de_0
\eea
where $\de_0>0$ is given by \eqref{eq:defofdelta0comingfromtheinterpolationlemma}.   Then   for a sufficiently small  bootstrap constant $\ep>0$, for all $s\le k_{small}+25$ and for all $\de\le p\le 2-\de$, we have
\beaa
  \EE^s_{p, 2+q_0-p}[\qf]+ \BB^s_{p, 2+q_0-p}[\qf] +\FF^s_{p, 2+q_0-p}[\qf] &\les \EE^{s+2}_{q_0}[\qfc] (0)+\EE^{s+4}_{2-\de}[\qf](0)  +\ep^4. 
\eeaa
\end{corollary}


  \subsection{Proof of Theorem M1}\lab{sec:proofofTheoremM1atlast}
  

  Since   $\ep=\ep_0^{2/3}$, and in view of the control on $\qf$ at $\tau=0$ provided by Theorem M0,  we immediately deduce from Corollary \ref{theorem:gen-decay-psic-improved},
 For all        $0<q\le q_0$, $\de\le p\le 2-\de$,   and        $s\le k_{small}+25$,
\bea
\label{Equation:decaypd-psi''}
  \EE^s_{p, 2+q_0-p}[\qf]+ \BB^s_{p, 2+q_0-p}[\qf] +\FF^s_{p, 2+q_0-p}[\qf] &\les\ep_0^2. 
\eea

We will also need the following two propositions concerning $L^2$ estimates on spheres.
    \begin{proposition}
   \label{proposition:supnormdecay-psi}
On  any   $S=S(\tau, r)\subset\Si(\tau)$, for $s\le k_{small}+ 25$,
\bea
\label{eq:pointwisedecayS-1}
(1+\tau)^{1+q_0}\int_{S_r}|\qf^{(s)} |^2&\les& \left(\EE^s_{1+\de, 1+q_0-\de}[\qf]\right)^{\frac{1}{2}}\left(\EE^s_{1-\de, 1+q_0+\de}[\qf]\right)^{\frac{1}{2}}
\eea
and,
\bea
\label{eq:pointwisedecayS-1:bis}
r^{-1}(1+\tau)^{2+q_0-\de} \int_{S_r}|\qf^{(s)} |^2&\les& \EE^s_{\de, 2+q_0-\de}[\qf]. 
\eea
\end{proposition}

\begin{proposition}
\label{Prop:integrated decay-lastslice-repeat}
We have for $s\le k_{small}+ 25$
\bea\lab{eq:ffffffffflux:00}
(1+\tau)^{2+q_0-\de}\int_{\Sigma_*(\tau, \tau_*)}|e_3\dk^{\le s} \qf|^2 &\les& \FF^s_{\de, 2+q_0-\de}[\qf]. 
\eea
Also, on  any   $S=S(\tau, r)\subset\Si(\tau)$, for $s\le k_{small}+23$, we have
\bea\lab{eq:ffffffffflux:00:bis}
(1+\tau)^{2+q_0-\de}\int_{S_r}|e_3\dk^{\le s} \qf|^2 &\les& \ep_0^2+\FF^{s+1}_{\de, 2+q_0-\de}[\qf]+\EE^{s+2}_{\de, 2+q_0-\de}[\qf].
\eea
\end{proposition}

The proof of Proposition \ref{proposition:supnormdecay-psi} is postponed to section \ref{subs:pointwisedecay-psi}, and  
the proof of Proposition \ref{Prop:integrated decay-lastslice-repeat} is postponed to section \ref{sec:integrated decay-lastslice}.

We now conclude the proof of Theorem M1. Indeed, in view of \eqref{Equation:decaypd-psi''}, Proposition \ref{proposition:supnormdecay-psi} and Proposition \ref{Prop:integrated decay-lastslice-repeat}, we infer for  $s\le k_{small}+25$
\beaa
(1+\tau)^{2+q_0-\de}\int_{\Mint(\tau, \tau_*)}|\dk^{\le s+1} \qf|^2 &\les& \ep^2_0,\\
(1+\tau)^{1+q_0}\int_{S_r}|\qf^{(s)} |^2&\les& \ep_0^2,\\
r^{-1}(1+\tau)^{2+q_0-\de} \int_{S_r}|\qf^{(s)} |^2&\les& \ep_0^2,\\
(1+\tau)^{2+q_0-\de}\int_{\Sigma_*(\tau, \tau_*)}|e_3\dk^{\le s} \qf|^2 &\les& \ep^2_0,
\eeaa
and for  $s\le k_{small}+23$
\beaa
( 1+\tau)^{2+q_0-\de}\int_S|\dk^se_3\qf|^2 &\les&  \ep_0^2.
\eeaa
In view of the standard Sobolev inequality on  the $2$-surfaces $S$ i.e.,
\beaa
\|\psi\|_{L^\infty(S)}&\les & r^{-1} \| (r\nabb )^{\le 2} \psi\|_{L^2(S)},
\eeaa
we immediately infer for $s\le k_{small}+23$
\beaa
|\qf^{(s)} |&\les&  \ep_0 r^{-1} (1+\tau)^{-\frac{1}{2}-\frac{q_0}{2}},\\ 
  |\qf^{(s)} |&\les& \ep_0 r^{-\frac{1}{2}}  (1+\tau) ^{-1-\frac{q_0-\de}{2}}, 
\eeaa
and for $s\le k_{small}+21$
\beaa
  |\dk^se_3(\qf)| &\les& \ep_0 r^{-1}(1+\tau) ^{-1-\frac{q_0-\de}{2}}.
\eeaa
Recall that $q_0>2\dec$ and that $\de>0$ can be chosen arbitrarily small so that we have $q_0-\de>2\dec$. In particular, we may choose
\bea\lab{eq:wefinallymakethechoiceofq0striclylargethandeltadec}
\dee:=\frac{q_0-\de}{2},\quad \dee>\dec,
\eea
which together with the above estimates for $\qf$ implies for $s\le k_{small}+25$
\beaa
(1+\tau)^{2+q_0-\de}\int_{\Mint(\tau, \tau_*)}|\dk^{\le s+1} \qf|^2 &\les& \ep^2_0,\\
(1+\tau)^{2+2\dee}\int_{\Sigma_*(\tau, \tau_*)}|e_3\dk^{\le s} \qf|^2 &\les& \ep^2_0,
\eeaa
for $s\le k_{small}+23$
\beaa
|\qf^{(s)} |&\les& \ep_0 r^{-1} (1+\tau)^{-\frac{1}{2}-\dee},\\ 
  |\qf^{(s)} |&\les& \ep_0 r^{-\frac{1}{2}}  (1+\tau) ^{-1-\dee}, 
\eeaa
and for $s\le k_{small}+21$
\beaa
  |\dk^se_3(\qf)| &\les& \ep_0 r^{-1} (1+\tau) ^{-1-\dee}
\eeaa
as desired. This concludes the proof of Theorem M1.

 
 \subsection{Proof of Proposition \ref{Prop:Main-Thm1-1}}\lab{sec:proofofProp:Main-Thm1-1}


Recall that,
\beaa
 I_p[\Ng]( \tau_1,\tau_2) &=& \bigg( \int_{\tau_1}^{\tau_2} d\tau \|\Ng\|_{L^2(\Sitrap(\tau))}\bigg)^2+\int_{\Mntrap(\tau_1,\tau_2)} r^{1+p}|\Ng|^2\\
 && +\int_{\Mntrap(\tau_1,\tau_2)} r^{2+p}|\Ng||e_3(\Ng)|+\sup_{\tau\in[\tau_1, \tau_2]}  \int_{\Si(\tau)} r^{p+2 } \big |\Ng \big |^2\\
 && +\int_{\Mntrap(\tau_1,\tau_2)} r^{3+\de}|e_3(N_g)|^2
 \eeaa
and,
\beaa
\II_{p,  d}^s[\Ng] &=&\sup_{0\le \tau\le \tau_*} ( 1+\tau)^d I_{p}^s[\Ng] (\tau, \tau_*).
\eeaa
Since we have
\beaa
r^\de(1+\tau)^{2+q_0-\de} &\les& r^{2+q_0}+(1+\tau)^{2+q_0},
\eeaa
and
$$
\int_{\Mntrap(\tau,\tau_*)} r^2|\dk^{\leq s}e_3(\Ng)||\dk^{\leq s}\Ng|\les \int_{\Mntrap(\tau,\tau_*)} r|\dk^{\leq s}\Ng|^2+\int_{\Mntrap(\tau,\tau_*)} r^3|\dk^{\leq s}e_3(\Ng)|^2,
$$
we infer  
 \bea\lab{eq:afterinterpolationboundforIinstermsofhogestpowerofrandhighestpooweroftau}
 && \II^s_{q_0+2, 0}[\Ng]  + \II^s_{\de, 2+q_0 -\de }[\Ng]\\
 \nn&\les& \sup_{0\le \tau\le \tau_*}\Bigg[\int_{\Mntrap(\tau,\tau_*)} r^{4+q_0}  |\dk^{\leq s+1}\Ng|^2  +\sup_{\tau'\in[\tau, \tau_*]}  \int_{\Si(\tau')} r^{4+q_0} \big |\dk^{\leq s}\Ng \big |^2\\
\nn &&+(1+\tau)^{2+q_0}\Bigg(\int_{\Mntrap(\tau,\tau_*)} r|\dk^{\leq s}\Ng|^2+\int_{\Mntrap(\tau,\tau_*)} r^{3+\de}|\dk^{\leq s}e_3(\Ng)|^2\\
\nn&&+\sup_{\tau'\in[\tau, \tau_*]}  \int_{\Si(\tau')} r^2\big |\dk^{\leq s}\Ng \big |^2\Bigg)+ (1+\tau)^{2+q_0}\bigg( \int_\tau^{\tau_*} d\tau' \|\dk^{\leq s}\Ng\|_{L^2(\Sitrap(\tau'))}\bigg)^2\Bigg].
\eea

In order to prove Proposition \ref{Prop:Main-Thm1-1}, it suffices to estimate the right-hand side of \eqref{eq:afterinterpolationboundforIinstermsofhogestpowerofrandhighestpooweroftau}. To this end, we will estimate 
separately the terms  with highest power of $r$, i.e. the first two terms, and the terms with highest power the $\tau$, i.e. the four last terms.

 
\subsubsection{Terms  with highest power of $r$ in \eqref{eq:afterinterpolationboundforIinstermsofhogestpowerofrandhighestpooweroftau}}


We estimate the first two terms of \eqref{eq:afterinterpolationboundforIinstermsofhogestpowerofrandhighestpooweroftau}. Recall from Lemma \ref{Lemma:structure-forN} that 
\beaa
\Ng &=& r^2 \dk^{\le 2}(\Ga_g \c (\a, \b)).
\eeaa
Recall from Lemma \ref{le:interpolatedbootstrap}
we have
\beaa
\max_{0\leq k\leq k_{large}-3}\sup_{\Mext(r\geq 4m_0)}r^{\frac{7}{2}+\frac{\dt}{2}}\Big(|\dk^k\a|+|\dk^k\b|\Big) &\les& \ep.
\eeaa
We infer for $s\leq k_{large}-6$
\beaa
|\dk^{\leq s+1}\Ng| &\les& \ep r^{-\frac{7}{2}-\frac{\dt}{2}}|r^2\dk^{\leq s+3}\Ga_g|
\eeaa
and hence, for $s\leq k_{large}-6$, we deduce
\beaa
&& \int_{\Mntrap(\tau,\tau_*)} r^{4+q_0}  |\dk^{\leq s+1}\Ng|^2+\sup_{\tau'\in[\tau, \tau_*]}  \int_{\Si(\tau')} r^{4+q_0} \big |\dk^{\leq s}\Ng \big |^2\\
&\les& \ep^2\int_{\Mntrap(\tau,\tau_*)} r^{-3-\dt+q_0}(r^2\dk^{\leq s+3}\Ga_g)^2+\ep^2\sup_{\tau'\in[\tau, \tau_*]}  \int_{\Si(\tau')}r^{-3-\dt+q_0}(r^2\dk^{\leq s+2}\Ga_g)^2\Big).
\eeaa
Since we also have for $s\leq k_{large}-6$
\beaa
\sup_{r_0\geq 4m_0}\int_{\{r=r_0\}}(r^2\dk^{\leq s+3}\Ga_g)^2 \les\ep^2,\quad \int_{\MM_{r\leq 4m_0}}(\dk^{\leq s+3}\Ga_g)^2\les \ep^2,\quad \sup_{\MM}|r^2\dk^{\leq s+2}\Ga_g| &\les& \ep,
\eeaa
we deduce
\beaa
&& \int_{\Mntrap(\tau,\tau_*)} r^{4+q_0}  |\dk^{\leq s+1}\Ng|^2+\sup_{\tau'\in[\tau, \tau_*]}  \int_{\Si(\tau')} r^{4+q_0} \big |\dk^{\leq s}\Ng \big |^2\\
&\les& \ep^4\left(1+\int_{r\geq 4m_0}\frac{dr}{r^{1+\dt-q_0}}\right).
\eeaa
Since $q_0< 4\dec$ and $\dt\geq 4\dec$, we have $q_0<\dt$ and hence, we obtain for $s\leq k_{large}-6$
\beaa
\int_{\Mntrap(\tau,\tau_*)} r^{4+q_0}  |\dk^{\leq s+1}\Ng|^2+\sup_{\tau'\in[\tau, \tau_*]}  \int_{\Si(\tau')} r^{4+q_0} \big |\dk^{\leq s}\Ng \big |^2 &\les& \ep^4.
\eeaa
This is the desired control of the terms  with highest power of $r$ in \eqref{eq:afterinterpolationboundforIinstermsofhogestpowerofrandhighestpooweroftau}.

 
\subsubsection{Terms  with highest power of $\tau$ in \eqref{eq:afterinterpolationboundforIinstermsofhogestpowerofrandhighestpooweroftau}}

   
We estimate the four last terms of \eqref{eq:afterinterpolationboundforIinstermsofhogestpowerofrandhighestpooweroftau}. In view of Lemma \ref{le:interpolatedbootstrap} with $k_{loss}=33$, so that 
$$\de_0= \frac{33}{k_{large}-k_{small}-2}=\frac{33}{k_{large}-\lfloor \frac{k_{large}}{2}\rfloor-3},$$
we have
 \beaa
\Big|\dk^{\le  k_{small}+33}  \Ga_g \Big|&\les&  \ep r^{-2}\tau^{-1/2-\dec+2\de_0}, \\
\Big|\dk^{\le  k_{small}+33}  \Ga_g\Big|&\les& \ep r^{-1} \tau^{-1-\dec+2\de_0},\\
\Big|\dk^{\le  S+32}e_3\Ga_g\Big|&\les& \ep r^{-2} [\tau^{-1-\dec}]^{1- \de_0}\les \ep r^{-2} \tau^{\Blue{-1}-\dec+2\de_0},\\
\Big|\dk^{\le k_{small}+33}(  \a, \b)\Big |&\les& \ep  r^{-3}   (\tau+r) ^{-1/2-\dec+2\de_0},      \\
\Big|\dk^{\le k_{small}+33}(  \a, \b) \Big|&\les&  \ep r^{-2}   (\tau+r) ^{-1-\dec+2\de_0},\\
\Big|\dk^{\le  S+32}e_3(  \a, \b) \Big|&\les& \ep r^{-3-\frac{1}{2} \de_0}  [\tau^{-1-\dec} ] ^{1-\de_0}\les \ep r^{-3}\tau^{-1-\dec+2\de_0}.
\eeaa 
In particular, together with the bootstrap assumption for $k\leq k_{small}$, the pointwise bound 
\beaa
|\dk^{\leq k_{large}-5}\a|+|\dk^{\leq k_{large}-5}\b|\les \ep r^{-\frac{7}{2}-\frac{\dt}{2}}
\eeaa
and since $\Ng = r^2 \dk^{\le 2}(\Ga_g \c (\a, \b))$, we infer for $s\leq k_{small}+30$
\bea\lab{eq:usedlaterforthedecayofe3qfthankstoflux}
\begin{split}
|\dk^s\Ng| &\les \ep^2r^{-3}\tau^{-1-2\dec+2\de_0}\\
|\dk^s\Ng| &\les \ep^2r^{-1}\tau^{-2-2\dec+2\de_0},\\
|\dk^se_3(\Ng)| &\les \ep^2r^{-3}\tau^{-\frac{3}{2}-2\dec+2\de_0},\\
|\dk^se_3(\Ng)| &\les \ep^2r^{-\frac{7}{2}-\frac{\dt}{2}}\tau^{-1-\dec+2\de_0}.
\end{split}
\eea
Using these 4 bounds and interpolation, we infer for $\de>0$
\beaa
&&(1+\tau)^{2+q_0}\Bigg(\int_{\Mntrap(\tau,\tau_*)} r|\dk^{\leq s}\Ng|^2+\int_{\Mntrap(\tau,\tau_*)} r^{3+\de}|\dk^{\leq s}e_3(\Ng)|^2\\
&&+\sup_{\tau'\in[\tau, \tau_*]}  \int_{\Si(\tau')} r^2\big |\dk^{\leq s}\Ng \big |^2\Bigg) + (1+\tau)^{2+q_0}\bigg( \int_{\tau}^{\tau_*} d\tau' \|\dk^{\leq s}\Ng\|_{L^2(\Sitrap(\tau'))}\bigg)^2\\
 &\les& \ep^4(1+\tau)^{2+q_0}\int_{\Mntrap(\tau,\tau_*)} r(r^{-3}{\tau'}^{-1-2\dec+2\de_0})^{1+\de}(r^{-1}{\tau'}^{-2-2\dec+2\de_0})^{1-\de}\\
 &&+\ep^4(1+\tau)^{2+q_0}\int_{\Mntrap(\tau,\tau_*)} r^{3+\de}(r^{-3}{\tau'}^{-\frac{3}{2}-2\dec+2\de_0})^{2-2\de}(r^{-\frac{7}{2}-\frac{\dt}{2}}{\tau'}^{-1-\dec+2\de_0})^{2\de}\\
 &&+\ep^4(1+\tau)^{2+q_0}\sup_{\tau'\in[\tau, \tau_*]}  \int_{\Si(\tau')} r^2(r^{-3}{\tau'}^{-1-2\dec+2\de_0})^2\\
 && + \ep^4(1+\tau)^{2+q_0}\bigg( \int_\tau^{\tau_*} {\tau'}^{-2-2\dec+2\de_0}d\tau'\bigg)^2\\
 &\les& \ep^4(1+\tau)^{2+q_0}\int_{\Mntrap(\tau,\tau_*)} r^{-3-\de\dt}{\tau'}^{-3-4\dec+\de +4\de_0+2\de\dec} \\
&&+\ep^4(1+\tau)^{2+q_0}\sup_{\tau'\in[\tau, \tau_*]}  \int_{\Si(\tau')} r^{-4}{\tau'}^{-2-4\dec+4\de_0}+ \ep^4(1+\tau)^{2+q_0}\bigg( \int_\tau^{\tau_*} {\tau'}^{-2-2\dec+2\de_0}d\tau'\bigg)^2
\eeaa
and since $\de>0$, we obtain
\beaa
&&(1+\tau)^{2+q_0}\Bigg(\int_{\Mntrap(\tau,\tau_*)} r|\dk^{\leq s}\Ng|^2+\int_{\Mntrap(\tau,\tau_*)} r^{3+\de}|\dk^{\leq s}e_3(\Ng)|^2\\
&&+\sup_{\tau'\in[\tau, \tau_*]}  \int_{\Si(\tau')} r^2\big |\dk^{\leq s}\Ng \big |^2\Bigg) + (1+\tau)^{2+q_0}\bigg( \int_{\tau}^{\tau_*} d\tau' \|\dk^{\leq s}\Ng\|_{L^2(\Sitrap(\tau'))}\bigg)^2\\
  &\les& \ep^4(1+\tau)^{q_0 -4\dec+\de +4\de_0+2\de\dec}.
\eeaa
As we have $q_0<4\dec-4\de_0$, there exists $\de>0$ small enough such that 
$$q_0 -4\dec+\de +4\de_0+2\de\dec\leq 0,$$ 
and hence
\beaa
(1+\tau)^{2+q_0}\Bigg(\int_{\Mntrap(\tau,\tau_*)} r|\dk^{\leq s}\Ng|^2+\int_{\Mntrap(\tau,\tau_*)} r^{3+\de}|\dk^{\leq s}e_3(\Ng)|^2\\
+\sup_{\tau'\in[\tau, \tau_*]}  \int_{\Si(\tau')} r^2\big |\dk^{\leq s}\Ng \big |^2\Bigg) + (1+\tau)^{2+q_0}\bigg( \int_{\tau}^{\tau_*} d\tau' \|\dk^{\leq s}\Ng\|_{L^2(\Sitrap(\tau'))}\bigg)^2 &\les& \ep^4.
\eeaa
This is the desired control of the terms  with highest power of $\tau$ in \eqref{eq:afterinterpolationboundforIinstermsofhogestpowerofrandhighestpooweroftau}. Together with \eqref{eq:afterinterpolationboundforIinstermsofhogestpowerofrandhighestpooweroftau} and the above control of the terms  with highest power of $r$, we infer
 \beaa
 \nn&&  \II^s_{q_0+2, 0}[\Ng]  + \II^s_{\de, 2+q_0-\de }[\Ng]\\
 \nn&\les& \sup_{0\le \tau\le \tau_*}\Bigg[\int_{\Mntrap(\tau,\tau_*)} r^{4+q_0}  |\dk^{\leq s+1}\Ng|^2  +\sup_{\tau'\in[\tau, \tau_*]}  \int_{\Si(\tau')} r^{4+q_0} \big |\dk^{\leq s}\Ng \big |^2\\
\nn &&+(1+\tau)^{2+q_0}\Bigg(\int_{\Mntrap(\tau,\tau_*)} r|\dk^{\leq s}\Ng|^2+\int_{\Mntrap(\tau,\tau_*)} r^{3+\de}|\dk^{\leq s}e_3(\Ng)|^2\\
\nn&&+\sup_{\tau'\in[\tau, \tau_*]}  \int_{\Si(\tau')} r^2\big |\dk^{\leq s}\Ng \big |^2\Bigg)+ (1+\tau)^{2+q_0}\bigg( \int_\tau^{\tau_*} d\tau' \|\dk^{\leq s}\Ng\|_{L^2(\Sitrap(\tau'))}\bigg)^2\Bigg]\\
 &\les&\ep^4
\eeaa
which is the desired estimate. This concludes the proof of Proposition \ref{Prop:Main-Thm1-1}.


\section{Improved weighted estimates}\lab{section:inthissectionweprovethe2firstsupportingtheoremstoThM1} 


The goal of this section is to prove the two following theorems on improved weighted estimates.
\begin{theorem}
\label{theorem-combinedMor-r-weighted-improved}
Assume $\qf$ verifies following wave equation, see  \eqref{eq:Masterwaveequation-qf},
\beaa
\square_2 \qf  +\ka \kab\, \qf&=& N
\eeaa
with $N$ given, in view of Lemma \ref{Lemma:structure-forN}, by
\beaa
N=\Ng+e_3(r\Ng) + \Nm.
\eeaa
Then, for   any  $\de\le p\le 2-\de$, $ 0\le s\le k_{small}+30$, 
 \bea\lab{eq:Ineedalabeltodoaremarkbelow}
\sup_{\tau\in[\tau_1,\tau_2] }  E\,^s_{p}[\qf](\tau)+   B ^s_{p}[\qf](\tau_1, \tau_2) +   F ^s_{p}[\qf](\tau_1, \tau_2) 
\les  E\,^s _{p}[\qf](\tau_1)+   I ^{s+1}_p[ N_g]( \tau_1,\tau_2).
\eea
\end{theorem}

The next result deals with    weighted estimates for the quantity
\bea
\qfc&=& f_2  \ec_4 \qf,
\eea
where $f_2 $ is a fixed smooth function of $r$  defined as follows,
\bea
f_2(r)=
\begin{cases}
r^2 \qquad \mbox{for}\quad r\ge 6m_0,\\
0\, \,\qquad \mbox{for}\quad r\le  4m_0.
\end{cases}
\eea

 \begin{theorem}
\label{theorem-combinedMor-r-weighted-improved:bis}
Assume $\qf$ verifies equation, see  \eqref{eq:Masterwaveequation-qf},
\beaa
\square_2 \qf  +\ka \kab\, \qf&=& N
\eeaa
with,
\beaa
N=\Ng+e_3(r\Ng) + \Nm
\eeaa
as in Lemma \ref{Lemma:structure-forN}. Then,  for   any  $-1+\de<  q\le 1-\de$, $ 0\le s\le k_{small}+29$, 
\bea\lab{eq:Ineedalabeltodoaremarkbelow:bis}
 \sup_{\tau\in[\tau_1,\tau_2] }  E\,^s_{q}[\qfc](\tau)+   B^s _q[\qfc](\tau_1, \tau_2)      \les 
            E\,^s_{q} [\qfc](\tau_1) +    E\,^{s+1}_{q+1} [\qf](\tau_1)     + I_{q+2}^{s+2}[\Ng](\tau_1, \tau_2).
\eea
\end{theorem}

\begin{remark}
Note that in \eqref{eq:Ineedalabeltodoaremarkbelow} and \eqref{eq:Ineedalabeltodoaremarkbelow:bis}, the term $\Nm$ does not appear in the right-hand side since it      turns out that it can be absorbed by the left hand side. 
\end{remark}  

The proof of Theorem \ref{theorem-combinedMor-r-weighted-improved} is postponed to section \ref{sec:proofoftheorem-combinedMor-r-weighted-improved}, and the proof of Theorem \ref{theorem-combinedMor-r-weighted-improved:bis} is postponed to section \ref{sec:proofoftheorem-combinedMor-r-weighted-improved:bis}. These proofs will rely on weighted energy flux estimates introduced in the next section.


\subsection{Proof of Theorem \ref{theorem:Daf-Rodn1-psi-s}}\lab{sec:proofoftheorem:Daf-Rodn1-psi-s:0}


Assume given a spacetime $\MM$ verifying  the  bootstrap assumptions   with small  constant $\ep>0$. The proof of Theorem  \ref{theorem-combinedMor-r-weighted-improved} and Theorem \ref{theorem-combinedMor-r-weighted-improved:bis} will rely on estimates stated below for solutions $\psi\in\sk_2(\MM)$ of the equation,
\bea
\label{eq:masterwavepsi}
\square_2 \psi+V\psi =N, \qquad V=\ka\kab. 
\eea


\subsubsection{Basic weighted estimates}


 \begin{theorem}
 \label{theorem-combinedMor-r-weighted}
 Recall the definitions in  \eqref{notation:Eps},  \eqref{notation:Mps}. The  following holds  for   any $0\leq s\leq k_{small}+30$. For all $\de\le p\le 2-\de$, we have,
  \bea  
  \sup_{\tau\in[\tau_1,\tau_2] }   E^s_{p} [\psi](\tau)+   B^s_{p}[\psi](\tau_1, \tau_2)  + F^s_{p}[\psi](\tau_1, \tau_2) \les 
          E^s_{p}[\psi](\tau_1)+   J^s_p[\psi, N]( \tau_1,\tau_2),
                \label{eq:theorem-combinedMor-r-weighted}  
\eea
where, for $p\ge \de$, we have introduced the notation
\bea
\lab{def:normI-JnormforN}
\bsplit
J_{p, R} [\psi, N](\tau_1,\tau_2):&= \bigg| \int_{\MM_{\ge R}(\tau_1,\tau_2)  } r^ p \ec_4 \psi N\bigg|,\\
J_p[\psi, N](\tau_1,\tau_2):&=\bigg( \int_{\tau_1}^{\tau_2} d\tau \|N\|_{L^2(\Sitrap(\tau))}\bigg)^2+\int_{\Mntrap(\tau_1,\tau_2)}r^{1+\de}|N|^2\\
& +J_{p, 4m_0} [\psi, N](\tau_1,\tau_2), 
\end{split}
\eea
 and
   \beaa
    J^s_p[\psi, N](\tau_1,\tau_2)&:=&\sum_{k\le s}  J_p[\dk^k \psi, \dk^k N](\tau_1,\tau_2). 
   \eeaa
\end{theorem}

The proof of Theorem \ref{theorem-combinedMor-r-weighted} is postponed to section \ref{sec:proofoftheorem-combinedMor-r-weighted}.


\subsubsection{Higher weighted estimates}


The next result deals with    weighted estimates for the quantity
\bea
\psic&=& f_2  \ec_4 \psi,
\eea
where $f_2 $ is a fixed smooth function of $r$  defined as follows,
\bea
f_2(r)=
\begin{cases}
r^2 \qquad \mbox{for}\quad r\ge 6m_0,\\
0\, \,\qquad \mbox{for}\quad r\le  4m_0.
\end{cases}
\eea

\begin{theorem}\label{theorem:Daf-Rodn-estim2-psic}
 The  following holds  for   any $-1+\de<  q\le 1-\de$, $ 0\le s\le k_{small}+29$,  
    \bea
     \label{eq:Daf-Rodn-estim2-psic}
   \bsplit
  \sup_{\tau\in[\tau_1,\tau_2] }  E\,^s_q[\psic](\tau)+   B^s _q[\psic](\tau_1, \tau_2)      &\les 
           E\,^s_q [\psic](\tau_1)      +  \Jc^s_q[\psic, N] (\tau_1,\tau_2)  \\
&+    E\, ^{s+1}_{\max(q, \de)}[\psi](\tau_1)  + J^{s+1}_{\max(q, \de)}[\psi, N],
      &               \end{split}
\eea
where we have introduced the notation
 \beaa
  \Jc_q[\psic, N] (\tau_1,\tau_2)&:=&J_{q,4m_0}\left[ \psic,   r^2\left(e_4 N+\frac 3 r N\right) \right] (\tau_1,\tau_2)\\
  &=&  \int_{\MM_{\ge 4m_0}(\tau_1,\tau_2)  } r^ {q+2} \big(  \ec_4  \psic\big) \c\,  \left(e_4    N+\frac 3 r   N\right),
 \eeaa
and
 \beaa
  \Jc^s_q[\psic, N] (\tau_1,\tau_2) :=\sum_{k\le s}  \Jc_q[ \dk^k \psic, \dk^k N] (\tau_1,\tau_2).
 \eeaa
   \end{theorem}

The proof of Theorem \ref{theorem:Daf-Rodn-estim2-psic} is postponed to section \ref{sec:proofoftheorem:Daf-Rodn-estim2-psic}.

We now proceed to the proof of Theorem \ref{theorem-combinedMor-r-weighted-improved} and Theorem \ref{theorem-combinedMor-r-weighted-improved:bis} in the next 2 sections. The proofs will follow from the structure of the nonlinear term $N$ of $\qf$ provided by Lemma \ref{Lemma:structure-forN} and the use of Theorem  \ref{theorem-combinedMor-r-weighted} and Theorem  \ref{theorem:Daf-Rodn-estim2-psic}.

    
\subsection{Proof of Theorem    \ref{theorem-combinedMor-r-weighted-improved}}\lab{sec:proofoftheorem-combinedMor-r-weighted-improved}


     Applying Theorem  \ref{theorem-combinedMor-r-weighted} to   the equation for $\qf$,
  with $N$ given by Lemma \ref{Lemma:structure-forN},  we   derive  corresponding   estimates    with the norm
 $ J^s_p[\qf, N]( \tau_1,\tau_2)$ on the right hand side, i.e.  for $ 0\le s\le k_{small}+30$, and for $\de\leq  p\le 2-\de$,
\bea
    \label{eq-beforeimproved} 
\sup_{\tau\in[\tau_1,\tau_2] }  E\,^s_{p}[\qf](\tau)+   B ^s_{p}[\qf](\tau_1, \tau_2) +   F ^s_{p}[\qf](\tau_1, \tau_2) 
\les  E\,^s _{p}[\qf](\tau_1)+   J^s_p[\qf, N]( \tau_1,\tau_2).
\eea

To prove Theorem    \ref{theorem-combinedMor-r-weighted-improved}, it suffices, in view of \eqref{eq-beforeimproved},  to estimate $J^s_p[\qf, N]( \tau_1,\tau_2)$.  Recall that, see \eqref{def:normI-JnormforN} and \eqref{def:normI-Nde}
\beaa
 I_p[N]( \tau_1,\tau_2) &=& \bigg( \int_{\tau_1}^{\tau_2} d\tau \|N\|_{L^2(\Sitrap(\tau))}\bigg)^2+\int_{\Mntrap(\tau_1,\tau_2)} r^{1+p}|N|^2\\
 && +\int_{\Mntrap(\tau_1,\tau_2)} r^{2+p}|\Ng||e_3(\Ng)|+\sup_{\tau\in[\tau_1, \tau_2]}  \int_{\Si(\tau)} r^{p+2 } \big |N \big |^2\\
 && +\int_{\Mntrap(\tau_1,\tau_2)} r^{3+\de}|e_3(\Ng)|^2
 \eeaa
and,
\beaa
 J_{p, R}[\qf, N]&=& \bigg| \int_{\MM_{\ge R}(\tau_1,\tau_2)  } r^ p \ec_4(\qf) N\bigg|,\\
J_p[\qf, N](\tau_1,\tau_2) &=&\bigg( \int_{\tau_1}^{\tau_2} d\tau \|N\|_{L^2(\Sitrap(\tau))}\bigg)^2+\int_{\Mntrap(\tau_1,\tau_2)} r^{1+\de}|N|^2\\
&& +J^s_{p, 4m_0} [\qf, N](\tau_1,\tau_2),\\
J^s_p [\qf, N](\tau_1,\tau_2)&=&\sum_{k\leq s} J_p[\dk^k \qf, \dk^k N],\\
\eeaa

Recall also from \eqref{eq"effectivestructureforN-s}
\bea
\dk ^k N&=& \dk^ {\le k}\Ng+ e_3(\dk^k(r\Ng))+ \dk^k \Nm
\eea
 and  consider separately the three terms.
 
 {\bf Case of  $\Nm$.}  Recall that $\Nm= \dk^{\le 1 }(\Ga_g \c \qf)$. We have, schematically,
 \beaa
\dk^k  \Nm&=& \dk^{1+k}   (\Ga_g \c \qf) =\sum_{i+j=k+1}   \dk^{\leq i} \Ga_g    \dk^{\le j}  \qf.
 \eeaa
We make use of the following consequence of the bootstrap assumptions for $k\le k_{large}-5$
\beaa
\big|\dk^{\le k}\Ga_g\big|\le \ep r^{-2}
\eeaa
to deduce,
\bea
\label{eq:combined-improved0}
\big|\dk^k \Nm\big|&\les& \ep r^{-2}   \big| \dk^{\le k+1} \qf\big|. 
\eea
We deduce, 
\beaa
J^s_{p, 4m_0} [\qf, \Nm](\tau_1,\tau_2)&\les & \sum_{k\le s}  \int_{\MM_{\ge 4m_0}(\tau_1,\tau_2)  } r^ p\big| \ec_4 \qfk \big|\,  \big|\dk^k \Nm \big|\\
&\les&\ep  \sum_{k\le s}  \int_{\MM_{\ge 4m_0}(\tau_1,\tau_2)  } r^{p-3}  \big| \dk^{1+k} \qf\big|^2.
\eeaa
Thus, recalling Remark \ref{rem:equivalentB-norms}, we infer
\bea
\label{eq:combined-improved1}
J^s_{p, 4m_0} [\qf, \Nm](\tau_1,\tau_2)&\les &  \ep  B_p^s[\qf](\tau_1, \tau_2).
\eea

Next, we estimate in view of \eqref{eq:combined-improved0}
\beaa
 \int_{\Mntrap(\tau_1,\tau_2)} r^{1+\de} |\dk^k \Nm|^2 &\les& \ep \int_{\Mntrap(\tau_1,\tau_2)}r^{\de-3}  |\dk^{\leq k+1}\qf|^2
\eeaa
which yields, using again Remark \ref{rem:equivalentB-norms},
\bea
\label{eq:combined-improved2}
 \int_{\Mntrap(\tau_1,\tau_2)}r^{1+\de}  |\dk^k \Nm|^2&\les & \ep B^s_\de[\qf](\tau_1, \tau_2).
\eea

We next estimate  the integral
\beaa
\int_{\tau_1}^{\tau_2} d\tau \|\dk^k\Nm \|_{L^2(\Sitrap(\tau))}.
\eeaa
In view of the definition of $\Nm=\dk^{\le 1} (\Ga_g \c \qf) $,
\beaa
\dk^k  \Nm&=&
\dk^{\le k+1}   (\Ga_g \c \qf) =\sum_{i+j=k+1}   \dk^{\le i} \Ga_g \,   \dk^{\le j}  \qf\\
 &=&   \dk^{j\le (k+1)/2 } \Ga_g  \,  \dk^{\le k+1}  \qf + \dk^{j\le (k+1)/2 } \qf  \,  \dk^{\le k+1}  \Ga_g  =J_1+J_2.
\eeaa
Now, since $\frac{k+1}{2} \le k_{small}$ we have
\beaa
\Big| \dk^{j\le (k+1)/2 } \Ga_g \Big| &\les& \ep (1+\tau)^{-1-\dec}
\eeaa
Hence,
\beaa
 \| J_1 \|^2_{L^2(\Sitrap(\tau))}&=&\int_{\Sitrap (\tau)}\Big| \dk^{j\le (k+1)/2 } \Ga_g \Big|^2 \Big| \dk^{\le k+1}  \qf \Big|^2\\
 &\les&  \ep^2 (1+\tau)^{-2-2\dec}  E^s[\qf](\tau)
\eeaa
i.e.,
\beaa
 \| J_1 \|_{L^2(\Sitrap(\tau))}&\les& \ep (1+\tau)^{-1-\dec}  \left(E^s[\qf](\tau)\right)^{1/2}.
\eeaa

For $J_2$ we write,
\beaa
\| J_2 \|^2_{L^2(\Sitrap(\tau))}&=&\int_{\Sitrap (\tau)}\Big| \dk^{j\le (k+1)/2 } \qf \Big|^2 \Big| \dk^{\le k+1}  \Ga_g \Big|^2\\
&\les&\left(\sup_{\Sitrap(\tau)} \Big| \dk^{j\le (k+1)/2 } \qf \Big|\right)^2 \int_{\Sitrap (\tau)}  \Big| \dk^{\le k+1}  \Ga_g \Big|^2\\
&\les& \int_{\Sitrap (\tau)}  \Big| \dk^{\le (k+1)/2+2}\qf \Big|^2 \int_{\Sitrap (\tau)}  \Big| \dk^{\le k+1}  \Ga_g \Big|^2
\eeaa
or, since  $(k+1)/2+2\le s$,
\beaa
\| J_2 \|_{L^2(\Sitrap(\tau))}\les\left[\int_{\Sitrap (\tau)}  \Big| \dk^{\le s}\qf\Big|^2\right]^{1/2}  \left[\int_{\Sitrap (\tau)}  \Big| \dk^{\le k+1}  \Ga_g \Big|^2\right]^{1/2}.
\eeaa
In view of the above estimates for $J_1$ and $J_2$, we deduce, for all $k\le s \le k_{large}-5$
\beaa
&&\int_{\tau_1}^{\tau_2} d\tau \|\dk^k\Nm \|_{L^2(\Sitrap(\tau))}\\
 &\les& \ep \sup_{\tau_1\le \tau\le \tau_2} \left( E^s[\qf](\tau) \right)^{1/2}+  \int_{\tau_1}^{\tau_2}  d\tau \left[\int_{\Sitrap (\tau)}  \Big| \dk^{\le s}\qf \Big|^2\right]^{1/2}\left[\int_{\Sitrap (\tau)}  \Big| \dk^{\le s }  \Ga_g \Big|^2\right]^{1/2} \\
&\les& \ep \sup_{\tau_1\le \tau\le \tau_2} \left( E^s[\qf](\tau) \right)^{1/2}+\left(\int_{\Mntrap(\tau_1,\tau_2)}\Big| \dk^{\le s}\qf \Big|^2\right)^{\frac{1}{2}}
\left(\int_{ \MM_{r\le 4m_0}}  \Big| \dk^{\le s}  \Ga_g \Big|^2\right)^{1/2}
\eeaa
Making use of the following consequence of the bootstrap assumptions
\beaa
\left(\int_{ \MM_{r\le 4m_0}}  \Big| \dk^{\le s}  \Ga_g \Big|^2\right)^{1/2} &\les \ep, 
\eeaa
as well as the fact that 
\beaa
\int_{\Mntrap(\tau_1,\tau_2)}\Big| \dk^{\le s}\qf \Big|^2 &\les& \Morr^s[\qf](\tau_1, \tau_2),
\eeaa
we deduce,
\bea
\label{eq:combined-improved3}
\bigg( \int_{\tau_1}^{\tau_2} d\tau \|\dk^k\Nm\|_{L^2(\Sitrap(\tau))}\bigg)^2 \les \ep^2   \sup_{\tau_1\le \tau\le \tau_2}  E^s[\qf](\tau) + \ep^2\Morr^s[\qf](\tau_1, \tau_2)
\eea
which together  with  \eqref{eq:combined-improved2} and \eqref{eq:combined-improved1} yields for any $p\geq \de$
\bea
\label{eq:combined-improved5}
J^s_p[\qf, \Nm](\tau_1,\tau_2 )&\les&\ep^2   \sup_{\tau_1\le \tau\le \tau_2}  E^s[\qf](\tau) +\ep B_p^s[\qf](\tau_1, \tau_2).
\eea

{\bf Case of  $N_g$.} We write, as before,
\beaa
J^s_{p, 4m_0} [\qf, N_g](\tau_1,\tau_2)&\les & \sum_{k\le s}  \int_{\MM_{\ge 4m_0}(\tau_1,\tau_2)  } r^ p\big| \ec_4 \qfk  \dk^k N_g \big|\\
&\les&  \sum_{k\le s} \Big(  \int_{\MM_{\ge 4m_0}(\tau_1,\tau_2)  }  r^{p-1} \big|  \ec_4 \qfk    \big|^2\Big)^{1/2}
 \Big(  \int_{\MM_{\ge 4m_0}(\tau_1,\tau_2)  }  r^{p+1} \big| \dk^k N_g \big|^2\Big)^{1/2}.
\eeaa
Therefore,
\beaa
J^s_{p, 4m_0} [\qf, N_g](\tau_1,\tau_2)&\les & \left( B^s_p[\qf](\tau_1, \tau_2)\right)^{1/2} \left( I^s_p[\Ng](\tau_1, \tau_2) \right)^{1/2} \\
&\les& \de_1  B^s_p[\qf](\tau_1, \tau_2)+\de_1^{-1}  I^s_p[\Ng](\tau_1, \tau_2)
\eeaa
where $\de_1>0$ is   chosen sufficiently small so that we can  later absorb the term $\de_1  B^s_p[\qf](\tau_1, \tau_2)$ by the left hand side of our main estimate.

Also, we have in view of the definition of  $I^s_p[N]( \tau_1,\tau_2)$ and the fact that $p\geq \de$
\beaa
 \bigg( \int_{\tau_1}^{\tau_2} d\tau \|\dk^{\leq s}\Ng\|_{L^2(\Sitrap(\tau))}\bigg)^2+\int_{\Mntrap(\tau_1, \tau_2)}r^{1+\de}|\dk^{\leq s}\Ng|^2 &\les&  I^s_p[\Ng]( \tau_1,\tau_2).
\eeaa

Therefore, 
\beaa
J^s_p[\qf, N_g](\tau_1,\tau_2)&=& \bigg( \int_{\tau_1}^{\tau_2} d\tau \|\dk^{\leq s}\Ng\|_{L^2(\Sitrap(\tau))}\bigg)^2+\int_{\Mntrap(\tau_1, \tau_2)}r^{1+\de}|\dk^{\leq s}\Ng|^2\\
&& +J^s_{p, 4m_0} [\qf, \Ng](\tau_1,\tau_2)\\
&\les& I^s_\de[\Ng](\tau_1,\tau_2)  +\de_1^{-1}  I^s_p[\Ng](\tau_1, \tau_2)+\de_1  B^s_p[\qf](\tau_1, \tau_2),
\eeaa
i.e.,
\bea
\label{eq:combined-improved6}
J^s_p[\qf, N_g](\tau_1,\tau_2)&\les& \de_1^{-1}  I_p^s[\Ng](\tau_1,\tau_2) +\de_1  B^s_p[\qf](\tau_1, \tau_2).
\eea

{\bf Case of $e_3(r\Ng)$.} First, note that we have 
\beaa
\nn &&\bigg( \int_{\tau_1}^{\tau_2} d\tau \|\dk^{\leq s}e_3(r\Ng)\|_{L^2(\Sitrap(\tau))}\bigg)^2+\int_{\Mntrap(\tau_1,\tau_2)}r^{1+\de}|\dk^{\leq s}e_3(r\Ng)|^2\\
 &\les& \bigg( \int_{\tau_1}^{\tau_2} d\tau \|\dk^{\leq s+1}\Ng\|_{L^2(\Sitrap(\tau))}\bigg)^2+\int_{\Mntrap(\tau_1,\tau_2)}r^{1+\de}|\dk^{\leq s}\Ng|^2\\
&& +\int_{\Mntrap(\tau_1,\tau_2)}r^{3+\de}|\dk^{\leq s}e_3(\Ng)|^2
\eeaa
where we used the fact that $|\dk^{\leq s}e_3(r)|\les 1$ and $|\dk^{\leq s}r|\les r$. Hence, we infer in view of the definition of  $I^s_p[N]( \tau_1,\tau_2)$ and the fact that $p\geq \de$
\bea\lab{eq:combined-improved6:fore3Ng}
\nn &&\bigg( \int_{\tau_1}^{\tau_2} d\tau \|\dk^{\leq s}e_3(r\Ng)\|_{L^2(\Sitrap(\tau))}\bigg)^2+\int_{\Mntrap(\tau_1,\tau_2)}r^{1+\de}|\dk^{\leq s}e_3(r\Ng)|^2\\
 &\les&  I^{s+1}_p[\Ng]( \tau_1,\tau_2).
\eea

We then estimate
\beaa
J_{p, 4m_0}[\qfk,  e_3 (\dk^k(r\Ng))](\tau_1,\tau_2), \qquad k\le s.
\eeaa
To this end, we introduce a smooth cut-off function $\phi_{0}$     vanishing for $r\le  4m_0$ and  equal to $1$ for $r\ge  8m_0$. Then, we have 
\bea
\nn J_{p, 4m_0}[\qfk,  \dk^k(r\Ng)](\tau_1, \tau_2) &=& \left|\int_{\MM(\tau_1, \tau_2)} r^{p} \ec_4\qfk  \, e_3\dk^k(r\Ng)\right|\\
\nn&\les& J_{p,4m_0}[\qfk, \phi_0\dk^k(r\Ng)](\tau_1, \tau_2)\\
&&+J_{p,4m_0}[\qfk,(1- \phi_0)r\Ng](\tau_1, \tau_2). \label{eq:combined-improved6a}
\eea
In view of the fact that $1-\phi_0$ is supported in $r\leq 8m_0$, we easily obtain
\beaa
J_{p, 4m_0}[\qfk,(1- \phi_0)r\Ng](\tau_1, \tau_2) &\les&   \left(\sup_{\tau_1\le \tau\le \tau_2}  E^s[\qf](\tau)+ B^s_p[\qf](\tau_1, \tau_2)\right)^{1/2}\left(I^{s+1}_p[\Ng]( \tau_1,\tau_2)\right)^{\frac{1}{2}}
\eeaa
and hence 
\bea\lab{eq:eq:combined-improved11:before}
\nn&& J_{p, 4m_0}[\qfk,(1- \phi_0)r\Ng](\tau_1, \tau_2)\\
 &\les& \de_1\Big(\sup_{\tau_1\le \tau\le \tau_2}  E^s[\qf](\tau)+  B^s_p[\qf](\tau_1, \tau_2)\Big)+\de_1^{-1}  I^{s+1}_p[\Ng](\tau_1, \tau_2)
\eea
where $\de_1>0$ is   chosen sufficiently small so that we can  later absorb the terms $\de_1\sup_{\tau_1\le \tau\le \tau_2}  E^s[\qf](\tau)$ and $\de_1  B^s_p[\qf](\tau_1, \tau_2)$ by the left hand side of our main estimate.

It  remains to estimate the terms
\beaa
J_{p, 4m_0}[\qfk,  \phi_0 e_3 (\dk^k(r\Ng))](\tau_1,\tau_2), \qquad k\le s
\eeaa
which is supported for $r\geq 4m_0$. Note that $e_3(r\Ng)$  behaves like $r N_g$ and therefore  the same  sequence of estimates as for 
$N_g$ would lead to a loss of $r^{-1}$. For this reason we need  to integrate by parts  by parts in $e_3$.

\begin{proposition}
\label{Proposition:combinedimproved-crucial}
The following estimate holds true, for all $k\le s\le k_{large}-5$,
 \bea \lab{eq:combined-improved11}
  \sum_{k\leq s}J_{p,4m_0}[\qfk,  \phi_0e_3 (\dk^k(r\Ng))](\tau_1,\tau_2)\les  \de_1  B_p^s[\qf](\tau_1,\tau_2)+\de_1^{-1}  I_p^{s+1}[\Ng](\tau_1,\tau_2) 
  \eea
  for a sufficiently small $\de_1>0$.
\end{proposition}

We postponed the proof of Proposition \ref{Proposition:combinedimproved-crucial} to the end of the section. We are now in position to conclude the proof of Theorem \ref{theorem-combinedMor-r-weighted-improved}.

\begin{proof}[Proof of Theorem \ref{theorem-combinedMor-r-weighted-improved}] 
\eqref{eq:eq:combined-improved11:before} and \eqref{eq:combined-improved11} yield
 \beaa
  \bsplit
 \sum_{k\leq s}J_{p,4m_0}[\qfk,  e_3 (\dk^k(r\Ng))](\tau_1,\tau_2)&\les  \de_1  B_p^s[\qf](\tau_1,\tau_2)+\de_1^{-1}  I_p^{s+1}[\Ng](\tau_1,\tau_2).
  \end{split}
  \eeaa
  Together with \eqref{eq:combined-improved5}, \eqref{eq:combined-improved6} and \eqref{eq:combined-improved6:fore3Ng}, we infer
\beaa
J^s_p[\qf, N]( \tau_1,\tau_2) &\les&  (\de_1+\ep)  B_p^s[\qf](\tau_1,\tau_2)+\de_1^{-1}  I_p^{s+1}[\Ng](\tau_1,\tau_2)+\ep^2   \sup_{\tau_1\le \tau\le \tau_2}  E^s[\qf](\tau).
\eeaa
In view of \eqref{eq-beforeimproved}, this concludes the proof of Theorem \ref{theorem-combinedMor-r-weighted-improved}.
\end{proof}

The proof of Proposition \ref{Proposition:combinedimproved-crucial} will rely in particular on the following identity.
   \begin{lemma}
\label{lemma:formula-wave-rpsi}
The following hold true for any $\psi\in \sk_2$
\begin{itemize}
\item  We have, schematically,
 \bea\lab{eq:higherorderderivativesintheidentityformula-wave-rpsi:0}
e_3  e_4 (r\psi) &=& -r\square_2 \psi + r\lapp_2  \psi + r^{-1}\dk\psi. 
\eea

\item The following identity holds true, schematically,
  \bea\lab{eq:higherorderderivativesintheidentityformula-wave-rpsi}
   e_3  e_4 (r\dk^k\psi) =-\dk^{\leq k}(r\square_2 \psi )+ r\lapp_2(\dk^{\leq k}\psi)+r^{-1}\dk^{\leq k+1}\psi.
 \eea
 \end{itemize}
\end{lemma}

\begin{proof}
We start with the following identity for $\psi\in \sk_2$, see Definition \ref{Definition:square-k},
 \beaa
\square_2 \psi&=&-e_3 e_4\psi+\lapp_2\psi +\left(2\omb -\frac 1 2 \kab\right) e_4\psi  -\frac 1 2 \ka   e_3\psi +2\eta e_\th \psi
\eeaa
from which we deduce,
\beaa
r\square_2 \psi&=& -re_3 e_4\psi +r\left(\lapp_2\psi +\left(2\omb -\frac 1 2 \kab\right) e_4\psi  -\frac 1 2 \ka   e_3\psi +2\eta e_\th \psi\right).
\eeaa
On the other  hand,
\beaa
re_3 e_4\psi&=&  e_3( r e_4\psi)- (e_3 r) e_4\psi = e_3( e_4( r\psi)- e_4(r) \psi) - (e_3 r) e_4\psi\\
&=& e_3  e_4 (r\psi)  - e_4(r)e_3 \psi-  (e_3 r) e_4\psi-( e_3 e_4 r) \psi.
\eeaa
Hence,
\beaa
r\square_2 \psi&=& -e_3  e_4 (r\psi)  +e_4(r)e_3 \psi+  (e_3 r) e_4\psi+( e_3 e_4 r) \psi
+ r\lapp_2 \psi\\
&+&r\left(2\omb -\frac 1 2 \kab\right) e_4\psi- \frac 1 2  r\ka   e_3\psi +2r\eta e_\th \psi\\
&=&  -e_3  e_4 (r\psi) + r\lapp \psi+\left( e_4 r -\frac 1 2  r \ka\right) e_3\psi + \left( e_3 r -\frac 1 2  r\kab +2 r\omb    \right) e_4\psi +2r\eta e_\th \psi\\
&=&  -e_3  e_4 (r\psi) + r\lapp \psi+ \frac r 2 A e_3\psi+\frac r 2 \left(\Ab+ 4 \omb \right) e_4 \psi +2r\eta e_\th \psi
\eeaa
i.e.,
\beaa
e_3  e_4 (r\psi) &=&-r\square_2 \psi + r\lapp_2 \psi+ \frac r 2 A e_3\psi+\frac r 2 \left(\Ab+ 4 \omb \right) e_4 \psi +2r\eta e_\th \psi.
\eeaa
or, schematically,  in view of the definition of $\dk \psi$ and the estimate $|\omb|+r|\Ga_g|+|\Ga_b|\les r^{-1}$,
\beaa
e_3  e_4 (r\psi) &=&-r\square_2 \psi + r\lapp_2 \psi+\Big(r\Ga_g+\Ga_b+r^{-1}\Big) e_3\psi\\
&=&-r\square_2 \psi + r\lapp_2 \psi + r^{-1}\dk \psi
\eeaa
which is \eqref{eq:higherorderderivativesintheidentityformula-wave-rpsi:0}.

To  derive the identity for higher  derivatives we write, schematically,
 \beaa
\dk^k  e_3  e_4 (r\psi) &=&-\dk^k (r\square_2 \psi )+\dk^k ( r\lapp_2 \psi)+\dk^k(r \Ga_g \dk \psi).
 \eeaa
 We write,
 \beaa
\dk^k  e_3  e_4 (r  \psi)& =& e_3 e_4 ( r \dk^k \psi) +[ \dk^k, e_3 e_4 r]\psi = e_3 e_4 ( r \dk^k \psi) +[ \dk^k, e_3]\dk\psi +e_3[ \dk^k, e_4 r]\psi,\\
\dk^k ( r\lapp_2 \psi)&=&r\lapp_2 \dk^k \psi +[\dk^k, r\lapp]\psi=r\lapp_2 \dk^k \psi +[\dk^k, r^{-1}]\dk^2\psi +r^{-1}[\dk^k, r^2\lapp]\psi.
 \eeaa
In view of the identites for $[e_3, \dkb]$ and $[e_4, \dkb]$ of Lemma \ref{Le:comme3e4-outgeodesic}, the identities of Proposition \ref{prop:DDd--ddd} for commutation formulas involving $\ddd_k$ and $\dds_k$ derivatives, and the commutator formula for $[e_3, e_4]$, we have schematically
\beaa
&& [e_3, e_3]=0, \quad [\dkb, r^2\lapp]=\dkb+1,\quad [e_3, e_4r]=(r^{-1}+\Ga_g)\dk \\
&& [e_3, \dkb]=\Ga_b\dk+\Ga_b, \quad [e_4r, \dkb]=(r^2\xi+r\Ga_g)\dk+r\Ga_g
\eeaa 
In view of the estimates for $\Ga_g$, $\Ga_b$, and the fact that $\xi=0$ for $r\geq 4m_0$, we infer
\beaa
[ \dk^k, e_3] =r^{-1}\dk^{\leq k}, \quad [\dk^k, r^2\lapp]=\dk^{\leq k+1}, \quad [\dk^k, r^{-1}]=r^{-1}\dk^{\leq k-1}
\eeaa
and hence
 \beaa
\dk^k  e_3  e_4 (r  \psi)& =& e_3 e_4 ( r \dk^k \psi) +e_3[ \dk^k, e_4 r]\psi+r^{-1}\dk^{\leq k+1}\psi,\\
\dk^k ( r\lapp_2 \psi)&=& r\lapp_2 \dk^k \psi +r^{-1}\dk^{\leq k+1}\psi.
 \eeaa
 Also, we have
 \beaa
 [re_4, e_4r] = [re_4,e_4]r+e_4[re_4,r]=-e_4(r)e_4r-e_4re_4(r)=-2e_4r+r^{-1}\dk
 \eeaa
 and we infer by induction, schematically,
 \beaa
 [(re_4)^j, e_4r] = e_4r(re_4)^{\leq j-1}+r^{-1}\dk^{\leq j}
 \eeaa 
 so that, together with 
 \beaa
 [\dk_{\searrow}^{k-j},e_4r] = r^{-1}\dk^{\leq k-j},
 \eeaa
 we infer
 \beaa
[\dk^k,e_4r] &=& [(re_4)^j\dk_{\searrow}^{k-j},e_4r]= e_4r(re_4)^{\leq j-1}\dk_{\searrow}^{k-j}+r^{-1}\dk^{\leq k}.
\eeaa 
We deduce
  \beaa
   e_3  e_4 (r(re_4)^j\dk_{\searrow}^{k-j} \psi) &=&-(re_4)^j\dk_{\searrow}^{k-j}(r\square_2 \psi )+ r\lapp_2(\dk^k\psi)+r^{-1}\dk^{\leq k+1}\psi+   e_4r(re_4)^{\leq j-1}\dk_{\searrow}^{k-j}\psi.
 \eeaa
 We infer by induction on $j$
  \beaa
   e_3  e_4 (r(re_4)^j\dk_{\searrow}^{k-j} \psi) &=&-(re_4)^{\leq j}\dk_{\searrow}^{k-j}(r\square_2 \psi )+ r\lapp_2(\dk^{\leq k}\psi)+r^{-1}\dk^{\leq k+1}\psi
 \eeaa
 and hence
   \beaa
   e_3  e_4 (r\dk^k\psi) =-\dk^{\leq k}(r\square_2 \psi )+ r\lapp_2(\dk^{\leq k}\psi)+r^{-1}\dk^{\leq k+1}\psi
 \eeaa
 which is  \eqref{eq:higherorderderivativesintheidentityformula-wave-rpsi}. This concludes the proof of Lemma \ref{lemma:formula-wave-rpsi}.
\end{proof}

We now are in position to prove Proposition \ref{Proposition:combinedimproved-crucial}. 

\begin{proof}[Proof of Proposition \ref{Proposition:combinedimproved-crucial}]
We integrate by parts,
\bea\lab{eq:combined-improved007}
\nn J_{p,4m_0}[\qfk, \phi_0\,\dk^k(r\Ng)](\tau_1, \tau_2)&\les& \left|\int_{\MM(\tau_1, \tau_2)}  e_3 \left(  \phi_0(r)  r^{p}       \ec_4\qfk    \right)\dk^k(r\Ng)\right|+       |B^k_p(\tau_1)|+|B^k_p(\tau_2)|\\
&& +  \left|\int_{\MM(\tau_1, \tau_2)}  {\bf Div}(e_3) \phi_0(r)  r^{p}       \ec_4\qfk \dk^k(r\Ng)\right|
\eea
 where ${\bf Div}(e_3)$ denotes the spacetime divergence of $e_3$, and where the  with boundary terms are bounded by
  \beaa
  |B^k_p (\tau_1)|&\les& \int_{\Si (\tau_1)} r^{p}  |\ec_4\qfk |\,         |\dk^k(r\Ng)|,\\
   |B^k_p(\tau_2)|&\les& \int_{\Si (\tau_2)} r^{p}  |\ec_4\qfk|\,  |\dk^k(r\Ng)|.
   \eeaa   
   
  We estimate,
   \beaa
   |B^k_p(\tau)|&\les& \int_{\Si (\tau)} r^{p}  |\ec_4\qfk|\,  |\dk^k(r\Ng)|\les \Big( \int_{\Si (\tau)}  r^{p}|\ec_4\qfk|^2 \Big)^{1/2}  \Big(\int_{\Si (\tau)}  r^{p}|\dk^k(r\Ng)|^2 \Big)^{1/2}
   \\
   &\les&\Big(E^k_{p}[\qf](\tau)  \Big)^{1/2} \Big(\int_{\Si (\tau)}  r^{p+2}|\dk^k\Ng|^2 \Big)^{1/2}.
  \eeaa
We deduce, with $\de_1>0$  a sufficiently small constant, for any $\tau\in[\tau_1, \tau_2]$,
\bea
\label{eq:combined-improved7}
\bsplit
 \big|B^k_p(\tau_1)\big| &\les\de_1 \sup_{\tau_1\le \tau\le \tau_2}  E^k_{p}[\qf](\tau)  +\de_1^{-1}
   \sup_{\tau_1\le \tau\le \tau_2} \int_{\Si(\tau)} r^{p+2} |N_g^{\le k}|^2,\\
 \big|B^k_p(\tau_2) \big| &\les\de_1 \sup_{\tau_1\le \tau\le \tau_2}  E^k_{p}[\qf](\tau)  +\de_1^{-1}
   \sup_{\tau_1\le \tau\le \tau_2}  \int_{\Si(\tau)} r^{p+2} |N_g^{\le k}|^2.
   \end{split}
\eea

Next, notice that ${\bf Div}(e_3)=\kab-2\omb$ so that 
\beaa
|{\bf Div}(e_3)| &\les& r^{-1}.
\eeaa
Together with the fact that $e_3(\Phi_0(r))$ is supported in $4m_0\leq r\leq 8m_0$, the fact that $|e_3(r)|\les 1$ and
\beaa
r\ec_4\qfk = e_4(r\qfk)+O(r^{-1})e_4(\qfk),
\eeaa
we infer
\beaa
&& \left| \int_{\MM(\tau_1, \tau_2)}  e_3 \left(  \phi_0(r)  r^{p}       \ec_4\qfk    \right)\dk^k(r\Ng)\right| + \left|\int_{\MM(\tau_1, \tau_2)}  {\bf Div}(e_3) \phi_0(r)  r^{p}       \ec_4\qfk \dk^k(r\Ng)\right|\\
&\les &  \left| \int_{\MM(\tau_1, \tau_2)}  \phi_0(r)  r^{p-1}e_3e_4(r\qfk) \dk^k(r\Ng)\right|+ \int_{\MM_{\geq 4m_0}(\tau_1, \tau_2)}r^{p-1}|\ec_4(\qfk)||\dk^k(r\Ng)|\\
&&+ \int_{\MM_{4m_0\leq r\leq 8m_0}(\tau_1, \tau_2)}|\ec_4(\qfk)||\dk^k(r\Ng)|\\
&\les &  \left| \int_{\MM(\tau_1, \tau_2)}  \phi_0(r)  r^{p-1}e_3e_4(r\qfk) \dk^k(r\Ng)\right|\\
&&+ \left(\int_{\MM_{\geq 4m_0}(\tau_1, \tau_2)}r^{p-1}|\ec_4(\qfk)|^2\right)^{\frac{1}{2}}\left(\int_{\MM_{\geq 4m_0}(\tau_1, \tau_2)}r^{p+1}|\dk^{\leq k}\Ng|^2\right)^\frac{1}{2}\\
\eeaa
and hence
\beaa
&& \left| \int_{\MM(\tau_1, \tau_2)}  e_3 \left(  \phi_0(r)  r^{p}       \ec_4\qfk    \right)\dk^k(r\Ng)\right| + \left|\int_{\MM(\tau_1, \tau_2)}  {\bf Div}(e_3) \phi_0(r)  r^{p}       \ec_4\qfk \dk^k(r\Ng)\right|\\
&\les &  \left| \int_{\MM(\tau_1, \tau_2)}  \phi_0(r)  r^{p-1}e_3e_4(r\qfk) \dk^k(r\Ng)\right| +\left(B^s_p[\qf](\tau_1, \tau_2)\right)^{1/2}\left(I^s_p[\Ng]( \tau_1,\tau_2)\right)^{\frac{1}{2}}
\eeaa
which yields
\bea\lab{eq:IputalabelasIwillneeditlaterforsimilarproofforqfc}
\nn&& \left| \int_{\MM(\tau_1, \tau_2)}  e_3 \left(  \phi_0(r)  r^{p}       \ec_4\qfk    \right)\dk^k(r\Ng)\right| + \left|\int_{\MM(\tau_1, \tau_2)}  {\bf Div}(e_3) \phi_0(r)  r^{p}       \ec_4\qfk \dk^k(r\Ng)\right|\\
 &\les& \left| L^k\right|+\de_1 B^s_p[\qf](\tau_1, \tau_2)+\de_1^{-1}  I^s_p[\Ng](\tau_1, \tau_2)
\eea
where $\de_1>0$ is   chosen sufficiently small so that we can  later absorb the term $\de_1  B^s_p[\qf](\tau_1, \tau_2)$ by the left hand side of our main estimate, and where we have introduced the notation 
 \bea
    \label{eq:combined-improved-Lk}
     L^k:&=& \int_{\MM(\tau_1, \tau_2)}    \phi_0(r)      r^{p-1}    e_3  e_4\big(r\qfk \big) \dk^k(r\Ng).
    \eea
   
    It   remains to estimate the term $L^k$. Making use of Lemma \ref{lemma:formula-wave-rpsi},  we deduce 
  \beaa
     L^k&=& \int_{\MM(\tau_1, \tau_2)}    \phi_0(r)  r^{p-1}        e_3 e_4(r\qfk ) \dk^k(r\Ng)\\
     &=&- \int_{\MM(\tau_1, \tau_2)}    \phi_0(r)  r^{p-1}  \dk^{\leq k} (r\square_2 \qf )   \dk^k(r\Ng)\\
     &+& \int_{\MM(\tau_1, \tau_2)}\phi_0(r)  r^{p} \lapp_2( \dk^{\leq k} \qf)\dk^k(r\Ng)\\
     &+& \int_{\MM(\tau_1, \tau_2)}\phi_0(r) r^{p-2} \dk^{\leq k+1}\qf\, \dk^k(r\Ng)\\
   &=&  L^k_1+L^k_2+L^k_3.
    \eeaa

 We first estimate $L^k_3$ as follows 
\beaa
\big| L_3^k\big|&\les& \int_{\MM_{\geq 4m_0}(\tau_1, \tau_2)} r^{p-2}| \dk^{\le k+1} \qf| \, |\dk^k(r\Ng)|\\
&\les& \Big(\int_{\MM_{\geq 4m_0}(\tau_1, \tau_2)} r^{p-3} \big|\dk^{\le k+1} \qf\big|^2\Big)^{1/2} \Big(\int_{\MM_{\geq 4m_0}(\tau_1, \tau_2)} r^{p+1 }|\dk^{\leq k}\Ng |^2\Big)^{1/2}\
\eeaa 
 In view of Remark \ref{rem:equivalentB-norms}          we thus deduce,
 \beaa
 \big| L_3^k\big|&\les&\left( B_p^{k}[\qf]\right)^{1/2} \Big(\int_{\MM_{\geq 4m_0}(\tau_1, \tau_2)} r^{p+1 }|\dk^{\leq k}\Ng|^2\Big)^{1/2}\\
 &\les&  \left( B_p^{k}[\qf](\tau_1,\tau_2) \right)^{1/2} \left(  I_p^k[\Ng](\tau_1,\tau_2) \right)^{1/2}
\eeaa
and hence
 \bea
 \label{eq:combined-improved8}
 \big| L_3^k\big|&\les& \de_1 B^s_p[\qf](\tau_1, \tau_2)+\de_1^{-1}  I^s_p[\Ng](\tau_1, \tau_2)
\eea
where $\de_1>0$ is   chosen sufficiently small so that we can  later absorb the term $\de_1  B^s_p[\qf](\tau_1, \tau_2)$ by the left hand side of our main estimate.

 We now estimate  the term
 \beaa
 L^k_2&=& \int_{\MM(\tau_1, \tau_2)}\phi_0(r)  r^{p} \lapp_2( \dk^k \qf)\dk^k(r\Ng)
 \eeaa
 by first performing another integration by parts in the angular directions
 \beaa
 \big|L^k_2\big | &\les& \int_{\MM_{\geq 4m_0}(\tau_1, \tau_2)} r^{p-2} \big| \dk^{k+1}  \qf\big|  \big|  \dk^{k+1}(r\Ng)\big|\\
 &\les&\left( \int_{\MM_{\geq 4m_0}(\tau_1, \tau_2)} r^{p-3} \big| \dk^{k+1}  \qf\big|^2  \right)^{1/2}  \left( \int_{\MM_{\geq 4m_0}(\tau_1, \tau_2)} r^{p+1} \big| \dk^{\le k+1}\Ng\big|^2  \right)^{1/2} \\
 &\les&\left( B_p^k[\qf] (\tau_1, \tau_2)\right)^{1/2}  \left(  I_p^{k+1} [\Ng](\tau_1,\tau_2)\right)^{1/2}.
 \eeaa
 Hence,
  \bea
 \label{eq:combined-improved9}
 \big| L_2^k\big|&\les&  \de_1 B^s_p[\qf](\tau_1, \tau_2)+\de_1^{-1}  I^{s+1}_p[\Ng](\tau_1, \tau_2)
\eea
where $\de_1>0$ is   chosen sufficiently small so that we can  later absorb the term $\de_1  B^s_p[\qf](\tau_1, \tau_2)$ by the left hand side of our main estimate.
 
 It remains to estimate the term,
 \beaa
 L^k_1 &=&- \int_{\MM(\tau_1, \tau_2)}    \phi_0(r)  r^{p-1}   \dk^{\leq k} (r\square_2 \qf )   \dk^k(r\Ng).
 \eeaa
  Making use of the equation verified by $\qf$, i.e., $\square_2 \qf= -\ka \kab  \qf+ N$,  we deduce,
 \beaa
  \dk^k (r\square_2 \qf )&=&-\dk^k( r \ka \kab  \qf)+\dk^k  (rN).
 \eeaa
 Recall \eqref{eq"effectivestructureforN-s} 
  \beaa
\dk ^k N&=& \dk^ {\le k}\Ng+ e_3(\dk^k(r\Ng))+ \dk^k \Nm.
\eeaa
We infer
\beaa
\dk^{\leq k}(rN)&=& r\dk^{\leq k}N+\dk^{\leq k-1}N\\
&=& r\dk^ {\le k}\Ng+ re_3(\dk^{\leq k}(r\Ng))+ r\dk^{\leq k}\Nm
\eeaa 
and hence
\bea
  \label{eq:combined-improved10}
\nn |\dk^k (r\square_2 \qf )| &\les&  r^{-1}\big |\dk^{\le k} \qf \big|+ r \big|\dk^{\le k } N_g|   + r^2\big|\dk^{\le k }e_3(N_g)| +    r \big|\dk^k  \Nm|\\
 &\les& r^{-1}\big |\dk^{\le k+1} \qf \big|+ r \big|\dk^{\le k } N_g|+ r^2\big|\dk^{\le k }e_3(N_g)|.
 \eea
Note that we have used in the last inequality the form of $\Nm=\dk^{\leq 1}(\Ga_g\qf)$ and the fact that $|\Ga_g|\leq \ep r^{-2}$.  
 We deduce, using \eqref{eq:combined-improved10},
 \beaa
 \big|L^k_1\big| &\les& \int_{\MM_{\geq 4m_0}(\tau_1, \tau_2)}r^{p-1}|\dk^{\le k+1} \qf \big||\dk^{\leq k}\Ng| +\int_{\MM_{\geq 4m_0}(\tau_1, \tau_2)}r^{p+1}|\dk^{\leq k}\Ng|^2\\
 && +\int_{\MM_{\geq 4m_0}(\tau_1, \tau_2)}r^{p+2}|\dk^{\leq k}e_3(\Ng)||\dk^{\leq k}\Ng| \\
 &\les& \left( B_p^k[\qf](\tau_1,\tau_2) \right)^{1/2} \left(  I_p^k[\Ng](\tau_1,\tau_2) \right)^{1/2}+I_p^k[\Ng](\tau_1,\tau_2).
 \eeaa
We deduce
 \bea
  \label{eq:combined-improved15}
 \big| L^k_1\big|&\les& \de_1 B^s_p[\qf](\tau_1, \tau_2)+\de_1^{-1}  I^s_p[\Ng](\tau_1, \tau_2)
\eea
where $\de_1>0$ is   chosen sufficiently small so that we can  later absorb the term $\de_1  B^s_p[\qf](\tau_1, \tau_2)$ by the left hand side of our main estimate.
 
  Together with \eqref{eq:combined-improved8} and \eqref{eq:combined-improved9} we deduce,
  \bea
\big| L^k \big|&\les&  \de_1  B_p^{k}[\qf](\tau_1,\tau_2)+\de_1^{-1}  I_p^k[\Ng](\tau_1,\tau_2).
  \eea
  Together with \eqref{eq:combined-improved007}, \eqref{eq:combined-improved7} and \eqref{eq:combined-improved-Lk},  we infer,
  \beaa
  \bsplit
 \sum_{k\leq s}J_{p,4m_0}[\qfk, \phi_0\,\dk^k(r\Ng)](\tau_1, \tau_2) &\les  \de_1  B_p^s[\qf](\tau_1,\tau_2)+\de_1^{-1}  I_p^{s+1}[\Ng](\tau_1,\tau_2)
  \end{split}
  \eeaa
  which concludes the proof of  Proposition \ref{Proposition:combinedimproved-crucial}.
    \end{proof}


    \subsection{Proof of Theorem \ref{theorem-combinedMor-r-weighted-improved:bis}}\lab{sec:proofoftheorem-combinedMor-r-weighted-improved:bis}
  

   We apply  Theorem \ref{theorem:Daf-Rodn-estim2-psic}   to  the case when $\psi=\qf$.
   Hence,
     \bea
     \label{eq:Daf-Rodn-estim2-qfcc}
   \bsplit
    E\,^s_q[\qfc](\tau_2)+   B^s _q[\qfc](\tau_1, \tau_2)      &\les 
           E\,^s_q [\qfc](\tau_1)      +  \Jc^s_q[\qfc, N] (\tau_1,\tau_2)  \\
&+    E\, ^{s+1}_{\max(q,\de)}[\qf](\tau_1)  + J^{s+1}_{\max(q, \de)}[\qf, N](\tau_1, \tau_2).
                 \end{split}
\eea
Also, recall that 
 \beaa
\qfc&=& f_2  \ec_4 \qf,
\eeaa
where $f_2 $ is a fixed smooth function of $r$  defined as follows,
\bea
f_2(r)=
\begin{cases}
r^2 \qquad \mbox{for}\quad r\ge 6m_0,\\
0\, \,\qquad \mbox{for}\quad r\le  4m_0.
\end{cases}
\eea
In particular, $\qfc$ is supported in $r\geq 4m_0$, and hence, in view of Remark \ref{rem:equivalentB-norms:0}, 
      \bea\lab{eq-combinedimproved-qfc1}
  B_q[\qfc](\tau_1, \tau_2)  &\simeq &  \int_{\MM_{\ge 4m_0}(\tau_1, \tau_2)} r^{q-3}|\dk\qfc|^2,
 \eea
 where we have used the fact that $-1+\de\leq q\leq 1-\de$.
 
 First, notice that the proof of Theorem    \ref{theorem-combinedMor-r-weighted-improved} yields
 \beaa
 J^{s+1}_{\max(q, \de)}[\qf, N](\tau_1, \tau_2) &\les& \sup_{\tau_1\le \tau\le \tau_2}  E^{s+1}[\qf](\tau)+B_{\max(q, \de)}^{s+1}[\qf](\tau_1,\tau_2)+I_{\max(q, \de)}^{s+2}[\Ng](\tau_1,\tau_2).
 \eeaa
 Hence, using Theorem    \ref{theorem-combinedMor-r-weighted-improved}, together with the fact that $\max(q, \de)\leq 1-\de$, we infer
 \beaa
 J^{s+1}_{\max(q, \de)}[\qf, N](\tau_1, \tau_2) &\les& {E}_{\max(q, \de)}^{s+1}[\qf](\tau_1)+I_{\max(q, \de)}^{s+2}[\Ng](\tau_1,\tau_2).
 \eeaa
 Since $q\geq -1+\de$, we have $\max(q, \de)\leq \de\leq q+1$ and thus
 \bea\lab{eq:Iaddedthistermwhichshouldbeestimatedbutwasmissing-qfc}
 J^{s+1}_{\max(q, \de)}[\qf, N](\tau_1, \tau_2) &\les& {E}_{q+1}^{s+1}[\qf](\tau_1)+I_{q+1}^{s+2}[\Ng](\tau_1,\tau_2).
 \eea 
 
 It only remains to estimate the term
 \beaa
  \Jc^s_q[\qfc, N] (\tau_1,\tau_2) =\sum_{k\le s}  \Jc_q[ \dk^k \qfc, \dk^k N] (\tau_1,\tau_2)
 \eeaa
  with,
 \beaa
  \Jc_q[  \qfc, N] (\tau_1,\tau_2)&=&J_{q,4m_0}\left[ \qfc,   r^2\left(e_4 N+\frac 3 r N\right) \right] (\tau_1,\tau_2)\\
  &=&  \int_{\MM_{\ge 4m_0}(\tau_1,\tau_2)  } r^ {q+2} \big(  \ec_4  \qfc\big) \c\,  \left(e_4    N+\frac 3 r    N\right). 
 \eeaa
 We rewrite in the equivalent  form,
 \bea
  \lab{eq-combinedimproved-qfc2}
   \Jc_q[ \dk^k \qfc, \dk^k N] (\tau_1,\tau_2)&=&  \left|\int_{\MM_{\ge 4m_0}(\tau_1,\tau_2)  } r^q   \big(  r  \ec_4 \dk^k \qfc\big)\,\big( \dk^{k+1} N\big)\right|.
 \eea
 Using the identity \eqref{eq"effectivestructureforN-s},  we have 
 \beaa
 \dk^{k+1} N=\dk^{\le k+1}\Ng+ e_3(\dk^{\leq k+1}r\Ng)+ \dk^{k+1}\Nm.
 \eeaa
 
  The  integral due to  $ \dk^{\le k+1} N_g$ is treated as follows
   \beaa
   \Jc_q[ \dk^k \qfc, \dk^k N_g] (\tau_1,\tau_2) &\les& \int_{\MM_{\ge 4m_0}(\tau_1,\tau_2)  } r^q   \big| r  \ec_4 \dk^k \qfc\big|\,\big| \dk^{\leq k+1}
   \Ng\big|\\
   &\les& \Big(  \int_{\MM_{\ge 4m_0}(\tau_1,\tau_2)  }  r^{q-3} \big|  r\ec_4 \dk^k\qfc    \big|^2\Big)^{1/2}
 \Big(  \int_{\MM_{\ge 4m_0}(\tau_1,\tau_2)  }  r^{q+3} \big| \dk^{\leq k+1} N_g \big|^2\Big)^{1/2}
   \eeaa
 Therefore,
\bea\lab{eq-combinedimproved-qfc3}
\nn\Jc^s_q[\qfc, \Ng] (\tau_1,\tau_2) &\les & \left( B^s_q[\qfc](\tau_1, \tau_2)\right)^{1/2} \left( I^{s+1}_{q+2}[\Ng](\tau_1, \tau_2) \right)^{1/2} \\
&\les& \de_1B^s_q[\qfc](\tau_1, \tau_2)+\de_1^{-1}  I^{s+1}_{q+2}[\Ng](\tau_1, \tau_2)
\eea
where $\de_1>0$ is   chosen sufficiently small so that we can  later absorb the term $\de_1 B^s_q[\qf](\tau_1, \tau_2)$ by the left hand side of our main estimate.
   
  The  integral due to  $ \dk^{k+1}\Nm$ is treated as follows
  \beaa
   &&\Jc_q[ \dk^k \qfc, \dk^k\Nm] (\tau_1,\tau_2) \\
   &\les&  \int_{\MM_{\ge 4m_0}(\tau_1,\tau_2)  } r^q   \big|  r  \ec_4 \dk^k \qfc\big|\,\big| \dk^{k+1}\Nm\big|\\
   &\les& \int_{\MM_{\ge 4m_0}(\tau_1,\tau_2)  } r^{q+1}   \big|    \ec_4 \dk^k \qfc\big|\,\big| \dk^{\leq k+2}\qf\big|\,\big| \dk^{\leq k+2}\Ga_g\big|\\
    &\les& \ep\int_{\MM_{\ge 4m_0}(\tau_1,\tau_2)  } r^{q-1}\tau^{-\frac{1}{2}-\dec+2\de_0}   \big|    \ec_4 \dk^k \qfc\big|\,\big| \dk^{\leq k+2}\qf\big|\\
   &\les& \ep\left(\int_{\MM_{\ge 4m_0}(\tau_1,\tau_2)  } r^q \tau^{-1-2\dec+4\de_0} \big| \ec_4 \dk^k \qfc\big|^2\right)^{\frac{1}{2}}\left(\int_{\MM_{\ge 4m_0}\tau_1,\tau_2) }r^{q-2}  \big| \dk^{\leq k+2}\qf\big|^2\right)^{\frac{1}{2}}\\
    &\les& \ep\left(\sup_{\tau_1\leq\tau\leq\tau_2}E^s_q[\qfc](\tau)\right)^{\frac{1}{2}}\Big(B_{q+1}^{s+1}[\qf](\tau_1,\tau_2)\Big)^{\frac{1}{2}}\\
 \eeaa
 where we have used $|\Ga_g|\les \ep r^{-2}\tau^{-1/2-\dec+2\de_0}$ and $2\de_0<\dec$. Since $\de\leq q+1\leq 2-\de$ and $s\leq k_{small}+29$, we have in view of Theorem    \ref{theorem-combinedMor-r-weighted-improved} 
 \beaa
 B_{q+1}^{s+1}[\qf](\tau_1,\tau_2) &\les& E^{s+1}_{q+1}[\qf](\tau_1)+I^{s+2}_{q+1}[\Ng](\tau_1, \tau_2).
 \eeaa
 We infer
     \begin{equation}
     \lab{eq-combinedimproved-qfc3:bis}
   \sum_{k\leq s}\Jc_q[ \dk^k \qfc, \dk^k \Nm] (\tau_1,\tau_2) \les \ep^2\sup_{\tau_1\leq\tau\leq\tau_2}E^s_q[\qfc](\tau)+  E^{s+1}_{q+1}[\qf](\tau_1)+I^{s+2}_{q+1}[\Ng](\tau_1, \tau_2).
   \end{equation}
   
   It remains to estimate  the  integral due to  $e_3(\dk^{\leq k+1}r\Ng)$. We proceed as in Proposition \ref{Proposition:combinedimproved-crucial} by integration by parts, and obtain in particular the following analog of \eqref{eq:IputalabelasIwillneeditlaterforsimilarproofforqfc} 
\bea\lab{eq:thisistherighttimetoconcludetheproof:bis}
\Jc_q[ \dk^k \qfc, \dk^ke_3(r\Ng)] (\tau_1,\tau_2) &\les& \left| P^k\right|+\de_1 B^s_q[\qfc](\tau_1, \tau_2)+\de_1^{-1}  I^{s+1}_{q+2}[\Ng](\tau_1, \tau_2)
\eea
where $\de_1>0$ is   chosen sufficiently small so that we can  later absorb the term $\de_1  B^s_q[\qfc](\tau_1, \tau_2)$ by the left hand side of our main estimate, and where we have introduced the notation $P^k$ for the analog of $L^k$ in  \eqref{eq:combined-improved-Lk}, i.e.\footnote{    Recall that $\qfc$ is localized in $r\geq 4m_0$ so that we don't need in \eqref{eq-combinedimproved-qfc4} the cutoff function $\phi_0(r)$ introduced in Proposition \ref{Proposition:combinedimproved-crucial}.}
   \bea
     \lab{eq-combinedimproved-qfc4}
   P^k:=   \int_{\MM(\tau_1, \tau_2)}r^{q}    e_3  e_4\big(r\dk^k \qfc \big) \dk^{\leq k+1}(r\Ng).
    \eea

As in Lemma  \ref{lemma:formula-wave-rpsi},
\bea
  \lab{eq-combinedimproved-qfc5}
  e_3  e_4 (r\dk^k \qfc ) &=&-\dk^{\leq k} (r\square_2 \qfc  )+ r\lapp_2( \dk^{\leq k}\qfc)+r^{-1}\dk^{\leq k+1}\qfc.
 \eea
 We infer
   \beaa
   P^k &=&   \int_{\MM(\tau_1, \tau_2)}r^{q}    e_3  e_4\big(r\dk^k \qfc \big) \dk^{\leq k+1}(r\Ng)\\
   &=& -\int_{\MM(\tau_1, \tau_2)}r^{q}\dk^{\leq k} (r\square_2 \qfc  ) \dk^{\leq k+1}(r\Ng)\\
   &+& \int_{\MM(\tau_1, \tau_2)}r^{q+1}\lapp_2( \dk^{\leq k}\qfc) \dk^{\leq k+1}(r\Ng)\\
   &+& \int_{\MM(\tau_1, \tau_2)}r^{q-1}\dk^{\leq k+1}\qfc\,\, \dk^{\leq k+1}(r\Ng)\\
   &=& P_1^k+P_2^k+P_3^k.
    \eeaa
 The last two terms on the right can be treated  exactly as  the  the corresponding terms  in the treatment of $L^k$. This yields to the following analog of 
 \eqref{eq:combined-improved8} and \eqref{eq:combined-improved9}
 \bea\lab{eq:thisistherighttimetoconcludetheproof:0}
 \begin{split}
 \big| P_3^k\big|&\les \de_1 B^s_q[\qfc](\tau_1, \tau_2)+\de_1^{-1}  I^{s+1}_{q+2}[\Ng](\tau_1, \tau_2),\\[1mm]
 \big| P_2^k\big|&\les  \de_1 B^s_q[\qfc](\tau_1, \tau_2)+\de_1^{-1}  I^{s+2}_{q+2}[\Ng](\tau_1, \tau_2),
 \end{split}
\eea
where $\de_1>0$ is   chosen sufficiently small so that we can  later absorb the term $\de_1  B^s_p[\qfc](\tau_1, \tau_2)$ by the left hand side of our main estimate.
 
 It thus only remains to consider the analogous of the term $L^k_1$, i.e.
 \beaa
 P_1^k=   \int_{\MM(\tau_1, \tau_2)}          r^{q}   \dk^k (r\square_2 \qfc  )\dk^{\leq k+1}(r\Ng).
 \eeaa
Now, in view of Proposition  \ref{square-psic-modified}, $\qf$ verifies, schematically,
 \beaa
 \square_2 \qfc&=&   r^{-2}\dk^{\leq 1}\qfc +r^{-2}\dk^{\leq 2}\qf +r\dk^{\leq 1}N
 \eeaa
 so that
 \beaa
 \dk^k (r\square_2 \qfc  ) &=&   r^{-1}\dk^{\leq k+1}\qfc +r^{-1}\dk^{\leq k+2}\qf +r^2\dk^{\leq k+1}N\\
 &=&   r^{-1}\dk^{\leq k+1}\qfc +r^{-1}\dk^{\leq k+2}\qf +r^2\dk^{\leq k+1}\Ng+r^2\dk^{\leq k+1}\Nm+r^2\dk^{\leq k+1}e_3(r\Ng).
 \eeaa 
We infer the following decomposition of $P_1^k$
\beaa
P_1^k &=&  \int_{\MM(\tau_1, \tau_2)}          r^{q-1} \Big(\dk^{\leq k+1}\qfc +\dk^{\leq k+2}\qf\Big)\dk^{\leq k+1}(r\Ng)\\
&&+\int_{\MM(\tau_1, \tau_2)} r^{q+2}\dk^{\leq k+1}\Nm\dk^{\leq k+1}(r\Ng)\\
&&+\int_{\MM(\tau_1, \tau_2)} r^{q+2}\Big(\dk^{\leq k+1}\Ng+\dk^{\leq k+1}e_3(r\Ng)\Big)\dk^{\leq k+1}(r\Ng)\\
&=& P_{11}^k+P_{12}^k+P_{13}^k.
\eeaa

$P_{11}^k$ is estimated as $\Jc^s_q[\qfc, \Ng] (\tau_1,\tau_2)$, see \eqref{eq-combinedimproved-qfc3}, and hence
\beaa
|P_{11}^k| &\les & \left( B^s_q[\qfc](\tau_1, \tau_2)\right)^{1/2} \left( I^{s+1}_{q+2}[\Ng](\tau_1, \tau_2) \right)^{1/2} \\
&\les& \de_1B^s_q[\qfc](\tau_1, \tau_2)+B^s_{\max(q,\de)}[\qf](\tau_1, \tau_2)+\de_1^{-1}  I^{s+1}_{q+2}[\Ng](\tau_1, \tau_2)
\eeaa
which in view of Theorem    \ref{theorem-combinedMor-r-weighted-improved} yields
\bea\lab{eq:thisistherighttimetoconcludetheproof}
|P_{11}^k| &\les& \de_1B^s_q[\qfc](\tau_1, \tau_2)+E^{s+1}_{\max(q,\de)}[\qf](\tau_1)+\de_1^{-1}  I^{s+1}_{q+2}[\Ng](\tau_1, \tau_2).
\eea

Next, $P_{12}^k$ is estimated as follows
\beaa
|P_{12}^k| &\les & \int_{\MM_{\ge 4m_0}(\tau_1, \tau_2)} r^{q+3}|\dk^{\leq k+1}\Nm||\dk^{\leq k+1}\Ng|\\
&\les & \int_{\MM_{\ge 4m_0}(\tau_1, \tau_2)} r^{q+3}|\dk^{\leq k+2}\Ga_g||\dk^{\leq k+2}\qf||\dk^{\leq k+1}\Ng|\\
&\les & \ep\int_{\MM_{\ge 4m_0}(\tau_1, \tau_2)} r^{q+1}\tau^{-\frac{1}{2}-\dec+2\de_0}|\dk^{\leq k+2}\qf||\dk^{\leq k+1}\Ng|\\
&\les& \ep\left(\int_{\MM_{\ge 4m_0}(\tau_1, \tau_2)} r^{q-2}|\dk^{\leq k+2}\qf|^2\right)^{\frac{1}{2}}\left(\int_{\MM_{\ge 4m_0}(\tau_1, \tau_2)} r^{q+4}\tau^{-1-2\dec+4\de_0}|\dk^{\leq k+1}\Ng|^2\right)^{\frac{1}{2}}\\
&\les& \ep\Big(B_{q+1}^{s+1}[\qf](\tau_1,\tau_2)\Big)^{\frac{1}{2}}\left(\sup_{\tau\in[\tau_1,\tau_2]}\int_{\Sigma(\tau)}r^{q+4}|\dk^{\leq k+1}\Ng|^2\right)^{\frac{1}{2}}
\eeaa
where we have used $|\Ga_g|\les \ep r^{-2}\tau^{-1/2-\dec+2\de_0}$ and $2\de_0<\dec$. We infer
\bea\lab{eq:thisistherighttimetoconcludetheproof1}
|P_{12}^k| &\les& B_{q+1}^{s+1}[\qf](\tau_1,\tau_2)+I^{s+1}_{q+2}[\Ng](\tau_1, \tau_2).
\eea

Finally, $P_{13}^k$ is estimated as follows
\beaa
|P_{13}^k| &\les & \int_{\MM_{\ge 4m_0}(\tau_1, \tau_2)} r^{q+3}\Big(|\dk^{\leq k+1}\Ng|+|\dk^{\leq k+1}e_3(r\Ng)|\Big)|\dk^{\leq k+1}\Ng|\\
&\les & \int_{\MM_{\ge 4m_0}(\tau_1, \tau_2)} r^{q+3}|\dk^{\leq k+1}\Ng|^2 +\int_{\MM_{\ge 4m_0}(\tau_1, \tau_2)} r^{q+4}|\dk^{\leq k+1}e_3(\Ng)||\dk^{\leq k+1}\Ng|\\
&\les& I^{s+1}_{q+2}[\Ng](\tau_1, \tau_2).
\eeaa
Together with \eqref{eq:thisistherighttimetoconcludetheproof} and \eqref{eq:thisistherighttimetoconcludetheproof1}, we infer
\beaa
|P_1^k| &\leq& |P_{11}^k|+|P_{12}^k|+|P_{13}^k|\\
&\les& \de_1B^s_q[\qfc](\tau_1, \tau_2)+E^{s+1}_{\max(q,\de)}[\qf](\tau_1)+\de_1^{-1}  I^{s+1}_{q+2}[\Ng](\tau_1, \tau_2)+B_{q+1}^{s+1}[\qf](\tau_1,\tau_2).
\eeaa
Together with \eqref{eq:thisistherighttimetoconcludetheproof:0}, we deduce
\beaa
|P^k| &\leq& |P_1^k|+|P_2^k|+|P_3^k|\\
&\les& \de_1B^s_q[\qfc](\tau_1, \tau_2)+E^{s+1}_{\max(q,\de)}[\qf](\tau_1)+\de_1^{-1}  I^{s+2}_{q+2}[\Ng](\tau_1, \tau_2)+B_{q+1}^{s+1}[\qf](\tau_1,\tau_2).
\eeaa
Together with \eqref{eq:Daf-Rodn-estim2-qfcc}, \eqref{eq:Iaddedthistermwhichshouldbeestimatedbutwasmissing-qfc}, \eqref{eq-combinedimproved-qfc3}, \eqref{eq-combinedimproved-qfc3:bis} and \eqref{eq:thisistherighttimetoconcludetheproof:bis}, this concludes the proof of Theorem \ref{theorem-combinedMor-r-weighted-improved:bis}.


\section{Decay Estimates}\lab{sec:proofdecayestimatesfortheoremM1supp}


In this section we  prove the decay estimates. In particular
\begin{itemize}
\item In section \ref{sec:proofoftheorem:decay-psi-general}, we prove first flux decay estimates for $\qf$.

\item In section \ref{sec:proofoftheorem:gen-decay-psic}, we prove flux decay estimates for $\qfc$. 

\item In section \ref{sec:proofoftheorem:improveddecay-estimates-psi}, we prove Theorem \ref{theorem:improveddecay-estimates-psi}. 

\item In section \ref{subs:pointwisedecay-psi}, we prove Proposition \ref{proposition:supnormdecay-psi} on pointwise decay estimates for $\qf$.

\item In section \ref{sec:integrated decay-lastslice}, we prove Proposition \ref{Prop:integrated decay-lastslice-repeat} on flux estimates on $\Sigma_*$ and on improved pointwise estimates for $e_3(\qf)$.
\end{itemize}

The decay estimates rely on the norms \eqref{definition:normsdecay-psi} which we recall below.
\beaa
\bsplit
\EE^s_{p,d}[\psi] &= \sup_{0\le \tau\le \tau_*} (1+\tau)^d E^s_{p}[\psi](\tau),\\
\BB^s_{p, d}[\psi] &= \sup_{0\le \tau\le \tau_*} (1+\tau)^d\int_{\tau}^{\tau_*} M^s_{p-1}[\psi](\tau) d\tau,\\
\FF^s_{p,d}[\psi] &= \sup_{0\le \tau\le \tau_*} (1+\tau)^d F^s_{p}[\psi](\tau),\\
\II_{p,d}^s[\Ng] &=\sup_{0\le \tau\le \tau_*} ( 1+\tau)^d I_{p}^s[\Ng] (\tau, \tau_*).
\end{split}
\eeaa


\subsection{First Flux Decay Estimates}\lab{sec:proofoftheorem:decay-psi-general}


The goal of this section is to prove the following flux decay estimates for $\qf$.
\begin{theorem}
\label{theorem:decay-psi-general}
Assume $\qf$  verifies all the estimates of Theorem \ref{theorem-combinedMor-r-weighted-improved}.
Then the  following estimates   hold true   for all $s\le k_{small}+30$ and for all    $\de \le p\le 2-\de$ 
\begin{equation}\label{equation:decaypd-psi}
\EE^{s-[2-\de-p]}_{p,2-\de-p}[ \qf]  +  \BB^{s-[2-\de-p]}_{p, 2-\de-p}[\qf] +  \FF^{s-[2-\de-p]}_{p, 2-\de-p}[\qf]  \les \EE^s_{2-\de}[\qf](0)  +\II_{2-\de, 0}^{s+1}[\Ng]+\II_{\de, 2-2\de}^{s+1}[\Ng].
\end{equation}
Here $[x]$ denotes the least integer  greater or equal to $x$.
\end{theorem}
    
\begin{proof}
We make use of Theorem \ref{theorem-combinedMor-r-weighted-improved}    according to which we have, for $\de\le p\le 2-\de$, and $0\leq k\leq k_{small}+30$,
\beaa  
     E^s_{p}[\qf](\tau_2)+   B^s_{p}[\qf](\tau_1, \tau_2) +F^s_p[\qf](\tau_1,\tau_2)   &\les &
          E^s_{p}[\qf](\tau_1)+   I^{s+1}_p[\Ng]( \tau_1,\tau_2)    
             \eeaa
which we write in the form,
\bea
\label{eq:Daf-Rdn-1-s}
 E^s_{p} (\tau_2)+   \int_{\tau_1}^{\tau_2}      M_{p-1} ^s (\tau) d\tau    &\les E^s_{p} (\tau_1)+    I^{s+1}_p[\Ng](\tau_1, \tau_2), \quad \de\le  p\le 2-\de.
   \eea
In particular,
\beaa
 E^s_{2-\de} (\tau)+   \int_{\tau/2}^{\tau}      M_{1-\de } ^s (\la ) d\la     &\les E^s_{2-\de} (\tau/2)+\II_{2-\de, \,0} ^{s+1}[\Ng].
\eeaa
By the mean value theorem there exists $\tau_0\in[\tau/2, \tau]$  such that,
\beaa
  M^{s}_{1-\de } (\tau_0)        &\les&        \frac{1}{\tau}      \left( E^s_{2-\de}(\tau/2 ) +\II_{2-\de, \,0} ^{s+1}[\Ng] \right).
\eeaa
Since\footnote{Note that the loss of   derivative is due to the   degeneracy of   the bulk integral in the trapping region.}
\beaa
 E^{s-1}_{1-\de } (\tau)        &\les&       M^{s}_{1-\de } (\tau), 
\eeaa
we deduce,
\beaa
E^{s-1}_{1-\de } (\tau_0)   &\les&        \frac{1}{\tau}      \left( E^s_{2-\de}(\tau/2 ) +\II_{2-\de, \,0} ^{s+1}[\Ng]\right).
\eeaa
Moreover, applying  \eqref{eq:Daf-Rdn-1-s}   again  for $p=1-\de$, we deduce,
      \beaa
 E^{s-1}_{1-\de } (\tau)             +   \int_{\tau_0}^{\tau}      M_{-\de } ^{s-1} (\la ) d\la      &\les&   E^{s-1}_{1-\de}(\tau_0 ) +   (1+\tau)^{-1}  \II^s_{1-\de, 1}[\Ng]      \\
 &\les& (1+\tau)^{-1} \left(  E^s_{2-\de}(\tau/2 ) +\II_{2-\de, \,0} ^{s+1}[\Ng]+  \II^s_{1-\de, 1}[\Ng]     \right).
      \eeaa
      In particular,
      \bea\lab{eq:referenceforthisequationasitisusedagainlaterforinterpolationwithenergy}
       E^{s-1}_{1-\de } (\tau)     &\les& (1+\tau)^{-1} \left(  E^s_{2-\de}(\tau/2 ) +\II_{2-\de, \,0} ^{s+1}[\Ng]+  \II^s_{1-\de, 1}[\Ng]     \right).
       \eea
      Interpolating with
       \beaa
       E^{s}_{2-\de } (\tau)\les  E^s_{2-\de} (\tau/2)+\II_{2-\de, \,0} ^{s+1}[\Ng]
       \eeaa
      by using,
      \beaa
  E_p^s&\les& ( E^s_{p_1})^{\frac{p_2-p}{p_2-p_1} }(E^s_{p_2})^{\frac{p-p_1}{p_2-p_1}}, \quad p_1\leq p\leq p_2, 
 \eeaa
      we deduce
      \beaa
       E^{s-1}_1(\tau)\les( E^{s-1}_{1-\de}(\tau))^{1-\de}( E^{s-1}_{2-\de}(\tau))^\de \les (1+\tau)^{-1+\de} \left(  E^s_{2-\de}(\tau/2 ) +\II_{2-\de, \,0} ^{s+1}[\Ng]+  \II^s_{1-\de, 1}[\Ng]     \right).
      \eeaa
      The same inequality    hods  for $\tau$ replaced by $\tau/2$ i.e.,
      \bea
      \label{eq:enhanceddecay1}
       E^{s-1}_1(\tau/2 )\les (1+\tau)^{-1+\de} \left(  E^s_{2-\de}(\tau/4 ) +\II_{2-\de, \,0} ^{s+1}[\Ng]+  \II^s_{1-\de, 1}[\Ng]     \right).
      \eea

      We now repeat the procedure       starting  this time with  the  inequality   \eqref{eq:Daf-Rdn-1-s} for  $p=1$,
\beaa
E^{s-1}_{1}  (\tau )   +   \int_{\tau/2}^{\tau}      M_{0 } ^{s-1}  (\la ) d\la      &\les&    E^{s-1}_{1}  (\tau/2)+ 
   I^s_{1}[\Ng](\tau/2 ,\tau )  \\
   &\les   &  E^{s-1}_{1}  (\tau/2 )+   (1+\tau)^{-1+\de}    \II^s_{1, 1-\de}[\Ng]. 
\eeaa
Thus, in view of \eqref{eq:enhanceddecay1},
\beaa
 \int_{\tau/2}^{\tau}      M_{0 } ^{s-1}  (\la ) d\la      &\les& (1+\tau)^{-1+\de} \left(  E^s_{2-\de}(\tau/4 ) +\II_{2-\de, \,0} ^{s+1}[\Ng]+  \II^s_{1-\de, 1}[\Ng]  +  \II^s_{ 1,1-\de}[\Ng]   \right)
\eeaa
or,
since
\beaa
E^s_{2-\de}(\tau/4 )&\les& \EE^s_{2-\de}(0)+ \II^{s+1}_{2-\de,0}[\Ng],
\eeaa
we infer that,
\beaa
 \int_{\tau/2}^{\tau}      M_{0 } ^{s-1}  (\la ) d\la      &\les&B (1+\tau)^{-1+\de} 
\eeaa
where,
\bea
B:&=& \EE^s_{2-\de}(0 ) +\II_{2-\de, \,0} ^{s+1}[\Ng]+  \II^s_{1-\de, 1}[\Ng]  +  \II^s_{ 1,1-\de}[\Ng].
\eea
Repeating  the mean value    argument, we  can find $\tau_1\in[\tau/2, \tau]$  such that,
\beaa
M_0^{s-1}(\tau_1) &\les& \frac{1}{\tau} \int_{\tau/2}^{\tau}      M_{0 } ^{s-1}  (\la ) d\la \les  B(1+\tau)^{-2+\de}.
\eeaa   
We now  make    use  of  the fact   that   the energy norm $E^{s-1}$   is comparable    with  $M_0^{s-1}$   everywhere except  in the trapping region
where  we    lose a derivative. Thus
   \beaa
   E^{s-2}(\tau_1) \les M_0^{s-1}(\tau_1)
   \eeaa
   and therefore,
\bea
\label{eq:enhanceddecay2}
E^{s-2} (\tau_1)&\les B (1+\tau)^{-2+\de}. 
\eea
We would like now to compare $E^{s-2}(\tau)$  with $E^{s-2}(\tau_1)$   using the usual version of the energy inequality  and thus derive a similar estimate for  the former.
Unfortunately\footnote{The  loss of $\de$ is due to the fact that we are on a perturbation of Schwarzschild rather than on Schwarzschild.}, we don't have a closed energy inequality for $E$  and we therefore have instead to rely on  $E_\de$ for 
 which we have the inequality,
 \bea
 \label{eq:enhanceddecay3}
 E^{s-2}_\de(\tau)\les E^{s-2}_\de(\tau_1)+I_\de^{s-1}[\Ng](\tau_1,\tau).
 \eea
 We  also have in view of \eqref{eq:referenceforthisequationasitisusedagainlaterforinterpolationwithenergy}
 \beaa
 E^{s-2}_{1-\de}(\tau_1) &\les& (1+\tau)^{-1} \left(  E^s_{2-\de}(0) +\II_{2-\de, \,0} ^{s+1}[\Ng]+  \II^s_{1-\de, 1}[\Ng]     \right).
 \eeaa
 Interpolating  this last inequality  with \eqref{eq:enhanceddecay2} we deduce, for $\de>0$ sufficiently small
 \beaa
 E^{s-2}_\de(\tau_1) &\les&\big( E^{s-2} (\tau_1)\big )^{\frac{1-2\de}{1-\de}}  \big(E^{s-2}_{1-\de}(\tau_1 )\big)^{\frac{\de}{1-\de}}\\
 &\les& (1+\tau)^{-2+2\de}(B+ \EE_{1-\de}^{s-2}(0)+\II^{s-1}_{1-\de,0}[\Ng])\\
 &\les& (1+\tau)^{-2+2\de}B.
 \eeaa
 Thus, in view of \eqref{eq:enhanceddecay3},
\beaa
E^{s-2}_\de(\tau)\les E^{s-2}_\de(\tau_1)+I_\de^{s-1}[\Ng](\tau_1,\tau)\les  (1+\tau)^{-2+2\de}(B  +\II^{s-1}_{\de, 2-2\de}[\Ng])
\eeaa
i.e.,
\beaa
E^{s-2}_\de(\tau)\les  (1+\tau)^{-2+2\de}\big(\EE^s_{2-\de}(0 )+\II_{2-\de, \,0}^{s+1}[\Ng]+  \II^s_{1-\de, 1}[\Ng]  +  \II^s_{ 1,1-\de}[\Ng] +\II^{s-1}_{\de, 2-2\de}[\Ng]\big).
\eeaa
We infer
\beaa
\EE^{s-2}_{\de,  2-2\de} &\les& \big(\EE^s_{2-\de}(0 ) +\II_{2-\de, \,0}^{s+1}[\Ng]+  \II^s_{1-\de, 1}[\Ng]  +  \II^s_{ 1,1-\de}[\Ng] +\II^{s-1}_{\de, 2-2\de}[\Ng]\big)
\eeaa
which can be written in the shorter form (by  interpolation of the middle terms),
\bea\lab{eq:letsfinishthisproofatlastasiamgettingtiredofit}
\EE^{s-2}_{\de,  2-2\de} &\les& \EE^s_{2-\de}(0 ) +\II_{2-\de, \,0}^{s+1}[\Ng]+\II^{s+1}_{\de, 2-2\de}[\Ng].
\eea
Also, \eqref{eq:referenceforthisequationasitisusedagainlaterforinterpolationwithenergy} yields
\bea\lab{eq:letsfinishthisproofatlastasiamgettingtiredofit:1}
\nn\EE^{s-1}_{1-\de,1} &\les& \EE^s_{2-\de}(0 ) +\II_{2-\de, \,0}^{s+1}[\Ng]+\II^s_{1-\de, 1}[\Ng]\\
&\les& \EE^s_{2-\de}(0 ) +\II_{2-\de, \,0}^{s+1}[\Ng]+\II^{s+1}_{\de, 2-2\de}[\Ng].
\eea
while from  Theorem \ref{theorem-combinedMor-r-weighted-improved}, we have
\bea\lab{eq:letsfinishthisproofatlastasiamgettingtiredofit:2}
\EE^s_{2-\de,0} &\les&  \EE^s_{2-\de}(0 ) +\II_{2-\de, \,0}^{s+1}[\Ng].
\eea
Interpolating \eqref{eq:letsfinishthisproofatlastasiamgettingtiredofit} and \eqref{eq:letsfinishthisproofatlastasiamgettingtiredofit:1}, as well as \eqref{eq:letsfinishthisproofatlastasiamgettingtiredofit:1} and \eqref{eq:letsfinishthisproofatlastasiamgettingtiredofit:2}, we infer  for all $s\le k_{small}+30$ and for all    $\de \le p\le 2-\de$ 
\bea\lab{eq:letsfinishthisproofatlastasiamgettingtiredofit:3}
\EE^{s-[2-\de-p]}_{p,2-\de-p}[ \qf]  \les \EE^s_{2-\de}[\qf](0)  +\II_{2-\de, 0}^{s+1}[\Ng]+\II_{\de, 2-2\de}^{s+1}[\Ng].
\eea

Finally, making use of Theorem \ref{theorem-combinedMor-r-weighted-improved}  between $\tau$ and $\tau_*$, we have in particular 
\beaa  
B^{s-[2-\de-p]}_{p}[\qf](\tau, \tau_*)   +F^{s-[2-\de-p]}_p[\qf](\tau,\tau_*) &\les & E^{s-[2-\de-p]}_{p}[\qf](\tau)+   I^{s+1-[2-\de-p]}_p[\Ng]( \tau,\tau_*)\\
&\les&  (1+\tau)^{-(2-\de-p)}\Big(\EE^{s-[2-\de-p]}_{p,2-\de-p}[ \qf]+\II_{p, 2-\de-p}^{s+1}[\Ng]\Big)
\eeaa
and hence, we infer  for all $s\le k_{small}+30$ and for all    $\de \le p\le 2-\de$
\beaa  
\BB^{s-[2-\de-p]}_{p,2-\de-p}[\qf] +\FF^{s-[2-\de-p]}_{p,2-\de-p}[\qf]   &\les&  \EE^{s-[2-\de-p]}_{p,2-\de-p}[ \qf]+\II_{2-\de, 0}^{s+1}[\Ng]+\II_{\de, 2-2\de}^{s+1}[\Ng].
\eeaa
Together with \eqref{eq:letsfinishthisproofatlastasiamgettingtiredofit:3}, this concludes the proof of Theorem \ref{theorem:decay-psi-general}.
\end{proof}


\subsection{Flux Decay Estimates for $\qfc$}\lab{sec:proofoftheorem:gen-decay-psic}


The goal of this section is to prove the following flux decay estimates for $\qfc$.      

\begin{theorem}\label{theorem:gen-decay-psic}
The  following estimates hold for all  $q_0-1\le q\le q_0$,  where $q_0$ is a fixed number $\de<q_0\leq 1-\de$, and $s\le k_{small}+28$
\bea
\label{equation::gen-decay-psic}
\bsplit
\nn\EE^s_{q, q_0-q}[\qfc] + \BB^s_{q,q_0-q}[\qfc] &\les   \EE^s_{q_0}[\qfc] (0) +\EE^{s+2}_{2-\de}[\qf](0) +\II^{s+3}_{q_0+2, 0}[\Ng] + \II^{s+3}_{\de, 2+q_0-\de }[\Ng].
\end{split}
\eea
\end{theorem}

\begin{proof}
Since $\de<q_0\leq 1-\de$, according to Theorem \ref{theorem-combinedMor-r-weighted-improved:bis},           $\qfc =f_2 \ec_4\qf$     verifies,
 for   any $q_0-1\le q\leq q_0$ and any $s\le k_{small}+29$,
\beaa
E^s_q[\qfc](\tau_2)+   B^s_q[\qfc](\tau_1, \tau_2)      &\les& E^s_q [\qfc](\tau_1)     +    E_{q+1}^{s+1}[\qf](\tau_1)  + I^{s+2}_{q+2}[\Ng](\tau_1,\tau_2).
\eeaa
According to  the definition of  our decay norms above  we have,
\bea
\bsplit
  I^{s+2}_{q+2}[\Ng](\tau_1,\tau_2) &\les  (1+\tau_1)^{q-q_0} \II^{s+2}_{q+2,  q_0-q}[\Ng].
  \end{split}
\eea
Also, according to the definition   \ref{definition:normsdecay-psi} for the decay norms for  $\qf$       we  also   have
\beaa
E_{q+1}^{s+1}[\qf](\tau_1)&\les (1+\tau_1)^{q-q_0} \EE^{s+2}_{q+1, q_0-q}[\qf].
\eeaa

We deduce\footnote{Note that it is important  in what follows that     the $r^q$ weighted estimates  hold  also  for negative values of $q$.}, for  all $q_0-1 \le q\le  q_0$,
   \bea
    \label{eq:Daf-Rodn-estim2-psic'1}
    E^s_q[\qfc](\tau_2)+   \int_{\tau_1}^{\tau_2}M^s_q[\qfc](\tau)    &\les 
           E^s_q[\qfc](\tau_1)     + (1+\tau_1)^{q-q_0} \EEt^{s}_{q, q_0-q} 
            \eea
     where,
\bea       
\EEt^{s}_{q, q_0-q} :=  \EE^{s+1}_{q+1,q_0-q}[\qf]+   \II^{s+2}_{q+2, q_0-q}[\Ng].
\eea
In particular,
\bea\lab{eq:iamunabletofindgoodnamesforlabelastherearetoomanyofthemqfc}
 E^s_{q_0}[\qfc](\tau_2)+\int_{\tau_1}^{\tau_2}    M^s_{q_0-1}[\qfc](\tau) d\tau&\les &E^s_{q_0}[\qfc](\tau_1) +   \EEt_{q_0, 0}^{s}.  
\eea
By the mean value theorem we deduce    that there exists $\tau_0\in[\tau_1,\tau_2]$ such that,
\beaa
M^s_{q_0-1}[\qfc](\tau_0)&\les & \frac{1}{\tau_2-\tau_1} \left(E^s_{q_0}[\qfc](\tau_1) +  \EEt_{q_0, 0}^{s}    \right)\les  \frac{1}{\tau_2-\tau_1} \left( \EE^s_{q_0,0}[\qfc] +  \EEt_{q_0, 0}^{s}    \right).
\eeaa
Thus also,
\bea
\label{equat:Daf-Rodn-enhanceddecay1}
 E^s_{q_0-1}[\qfc](\tau_0)&\les & \frac{1}{\tau_2-\tau_1} \left( \EE^s_{q_0, 0}[\qfc]+  \EEt_{q_0, 0}^{s}  \right).
  \eea
We now make use of \eqref{eq:Daf-Rodn-estim2-psic'1}   to compare    the  quantities  $ E_q[\qfc]$  for  negative weights   ($q=q_0-1$)  at different values of $\tau$.
\beaa
 E^s_{q_0-1}[\qfc](\tau_2)&\les &E^s_{q_0-1}[\qfc](\tau_0) +(1+\tau_0)^{-1} \EEt^s_{q_0-1, 1 }. 
\eeaa
 Combining this with \eqref{equat:Daf-Rodn-enhanceddecay1}  we deduce,
\beaa
E^{s}_{q_0-1}[\qfc](\tau_2)&\les& \frac{1}{\tau_2-\tau_1} \left( \EE^s_{q_0, 0}[\qfc] + \EEt_{q_0, 0}^{s}  \right)+(1+\tau_0)^{-1} \EEt_{q_0-1, 1}^s.
\eeaa
Applying this inequality for $\tau_2=\tau \leq\tau_*$, $\tau_1=\frac 1 2  \tau$,     $\tau_0\in[\tau_1,\tau_2]$  we deduce,
 \bea
    \label{eq:Daf-Rodn-estim2-psic-qnegative-s2}
E^s_{q_0-1}[\qfc](\tau)&\les(1+\tau)^{-1}\left( \EE_{q_0, 0}^s[\qfc]  +    \EEt_{q_0, 0}^{s} +   \EEt_{q_0-1, 1}^s  \right).
   \eea
We now interpolate this  last inequality with the following immediate consequence of \eqref{eq:iamunabletofindgoodnamesforlabelastherearetoomanyofthemqfc}
\beaa
 E^s_{q_0}[\qfc](\tau)&\les & \EE^s_{q_0, 0}[\qfc]+   \EEt_{q_0, 0}^{s}  
\eeaa
to deduce, for all $q_0-1\le q\le  q_0$,
\beaa
E^s_q[\qfc](\tau)&\les & (1+\tau)^{q-q_0}\left( \EE_{q_0, 0}^s[\qfc]  +    \EEt_{q_0, 0}^{s} +   \EEt_{q_0-1, 1}^s  \right)
\eeaa
i.e.,
\beaa
\EE^s_{q, q_0-q}[\qfc]&\les&  \EE_{q_0, 0}^s[\qfc]  +    \EEt_{q_0, 0}^{s} +   \EEt_{q_0-1, 1}^s. 
\eeaa
In view of the definition of $\EEt_{q, q_0-q}^{s}$, this yields for  all $q_0-1\le q\le  q_0$,
\beaa
\EE^s_{q, q_0-q}[\qfc]&\les&  \EE_{q_0, 0}^s[\qfc]  + \EE^{s+1}_{q_0+1,0}[\qf] + \EE^{s+1}_{q_0,1}[\qf]+   \II^{s+2}_{q_0+2, 0}[\Ng]+   \II^{s+2}_{q_0+1, 1}[\Ng].
\eeaa
On the other hand, we have in view of Theorem \ref{theorem-combinedMor-r-weighted-improved:bis},          
\beaa
\EE^s_{q_0,0}[\qfc]     &\les& \EE^s_{q_0}[\qfc](0)     +    \EE_{q_0+1,0}^{s+1}[\qf]  + \II^{s+2}_{q_0+2,0}[\Ng]
\eeaa
and hence
\beaa
\EE^s_{q, q_0-q}[\qfc]&\les& \EE^s_{q_0}[\qfc](0) + \EE^{s+1}_{q_0+1,0}[\qf] + \EE^{s+1}_{q_0,1}[\qf]+   \II^{s+2}_{q_0+2, 0}[\Ng]+   \II^{s+2}_{q_0+1, 1}[\Ng].
\eeaa

Now, since $\de<q_0\leq 1-\de$, we have $\de<q_0< q_0+1\leq 2-\de$ and thus, we may apply Theorem \ref{theorem:decay-psi-general} to obtain for all $q_0-1\leq q\leq q_0$
\bea\lab{eq:iamunabletofindgoodnamesforlabelastherearetoomanyofthemqfc:00}
 \EE^{s+1}_{q+1,q_0-q}[\qf] &\les& \EE^{s+2}_{2-\de}[\qf](0)  +\II_{2-\de, 0}^{s+3}[\Ng]+\II_{\de, 2-2\de}^{s+3}[\Ng].
\eea
We thus infer 
\beaa
\EE^s_{q, q_0-q}[\qfc]&\les& \EE^s_{q_0}[\qfc](0) + \EE^{s+2}_{2-\de}[\qf](0)+   \II^{s+3}_{q_0+2, 0}[\Ng]+   \II^{s+3}_{q_0+1, 1}[\Ng] +\II_{2-\de, 0}^{s+3}[\Ng]+\II_{\de, 2-2\de}^{s+3}[\Ng]
\eeaa
and hence, for  all $q_0-1\le q\le  q_0$,
\bea\lab{eq:iamunabletofindgoodnamesforlabelastherearetoomanyofthemqfc:1}
\EE^s_{q, q_0-q}[\qfc]&\les& \EE^s_{q_0}[\qfc](0) + \EE^{s+2}_{2-\de}[\qf](0)+   \II^{s+3}_{q_0+2, 0}[\Ng]+\II_{\de, 2-2\de}^{s+3}[\Ng].
\eea

Finally, making use of Theorem \ref{theorem-combinedMor-r-weighted-improved:bis} between $\tau$ and $\tau_*$, we have in particular 
\beaa  
B^s_q[\qfc](\tau, \tau_*)    &\les & E^s_q[\qfc](\tau)+E^{s+1}_{q+1}[\qf](\tau)+   I^{s+2}_{q+2}[\Ng]( \tau,\tau_*)\\
&\les&  (1+\tau)^{-(q_0-q)}\Big(\EE^s_{q,q_0-q}[ \qfc]+\EE^{s+1}_{q+1,q_0-q}[\qf]+\II_{q+2, q_0-q}^{s+2}[\Ng]\Big)\\
&\les&  (1+\tau)^{-(q_0-q)}\Big(\EE^s_{q,q_0-q}[ \qfc]+\EE^{s+2}_{2-\de}[\qf](0)+ \II^{s+3}_{q_0+2, 0}[\Ng]+\II_{\de, 2-2\de}^{s+3}[\Ng]\Big)
\eeaa
where we used \eqref{eq:iamunabletofindgoodnamesforlabelastherearetoomanyofthemqfc:00} in the last inequality. Hence, we infer  for all $s\le k_{small}+28$ and for all  $q_0-1\leq q\leq q_0$
\beaa  
\BB^s_{q,q_0-q}[\qfc]   &\les&  \EE^s_{q,q_0-q}[ \qfc]+\EE^{s+2}_{2-\de}[\qf](0)+ \II^{s+3}_{q_0+2, 0}[\Ng]+\II_{\de, 2-2\de}^{s+3}[\Ng].
\eeaa
Together with \eqref{eq:iamunabletofindgoodnamesforlabelastherearetoomanyofthemqfc:1}, this concludes the proof of Theorem \ref{theorem:gen-decay-psic}.
\end{proof}


\subsection{Proof of Theorem \ref{theorem:improveddecay-estimates-psi}}\lab{sec:proofoftheorem:improveddecay-estimates-psi}


In this section, we prove Theorem  \ref{theorem:improveddecay-estimates-psi} by making use of Theorem \ref{theorem:decay-psi-general} and Theorem \ref{theorem:gen-decay-psic}. We start with the main estimate of Theorem \ref{theorem:gen-decay-psic} with $q=-\de$ which we write in the form,
\beaa
E^s_{-\de}[\qfc]&\les(1+\tau)^{-q_0-\de}  C_{q_0}^s
\eeaa
where,
\beaa
 C_{q_0}^s&:=& \EE_{q_0}^s[\qfc] (0)+\EE_{2-\de}^{s+2} [\qf](0) +   \II^{s+3}_{q_0, 0}[\Ng]+  \II^{s+3}_{\de, q_0+2-\de}[\Ng].
\eeaa

In view of the definition \eqref{notation:Eps} of   $E^s_{-\de}[\qfc]$ and   since   $\qfc=f_2\ec_4\qf$,
\beaa
\int_{\Si_{\ge 4m_0}(\tau)} r^{-\de }\left(|\ec_4\qfc|^2+ r^{-2}|\qfc|^2\right) &\les(1+\tau)^{-q_0-\de }  C_{q_0}^s.
\eeaa
Hence,
\bea
\label{eq:improveddecay-psi1}
\Ed^s_{2-\de, 4m_0}[\qf]= \int_{\Si_{\ge 4m_0}(\tau)} r^{2-\de}|\ec_4\qf|^2 &\les(1+\tau)^{-q_0-\de}  C_{q_0}^s.
\eea
In view of the decay estimates \eqref{equation:decaypd-psi} for $\qf$    established in Theorem \ref{theorem:decay-psi-general}
we have,
\beaa
E^{s}(\tau)&\les&(1+\tau)^{-2+2\de}B^{2+s}_{2-\de},\\
B^{2+s}_{2-\de}:&=&\EE^{s+2}_{2-\de}[\qf](0) +  \II^{s+3}_{2-\de, 0}[\Ng]+  \II^{s+3}_{\de, 2-2\de}[\Ng]. 
\eeaa
Thus, the quantity
\beaa
E^s_{2-\de}=E_{2-\de}^s[\qf](\tau) =\Ed^s_{2-\de , 4m_0}[\qf]+E^s[\qf]
\eeaa
verifies,
\bea
\label{eq:improveddecay-psi11}
E^s_{2-\de}&\les&(1+\tau)^{-q_0-\de} \left( C_{q_0}^s+  B^{2+s}_{2-\de}\right).
\eea
On the other hand, $E^s_{2-\de}$ verifies  \eqref{eq:Daf-Rdn-1-s}   for $p=2-\de$, i.e.
\beaa
 E^s_{2-\de} (\tau_2)+   \int_{\tau_1}^{\tau_2}      M_{1-\de} ^s (\tau) d\tau    &\les E^s_{2-\de} (\tau_1)+   I^{s+1}_{2-\de}[\Ng](\tau_1,\tau_2).
   \eeaa
Since
\beaa
 I^{s+1}_{2-\de}[\Ng](\tau_1,\tau_2)&\les& (1+\tau_1)^{-q_0-\de}\II^{s+1}_{2-\de, q_0+\de}[\Ng],
\eeaa
we infer
\bea
\nn E^s_{2-\de} (\tau)+   \int_{\tau/2}^{\tau}      M_{1-\de} ^s (\tau') d\tau'   &\les& E^s_{2-\de} (\tau/2)+   I^{s+1}_{2-\de}[\Ng](\tau/2,\tau)\\
 &\les& (1+\tau)^{-q_0-\de} \left( C_{q_0}^s+B^{2+s}_{2-\de}+\II^{s+1}_{2-\de, q_0+\de}[\Ng]\right).
\eea
Following the same arguments as in  the proof of Theorem \ref{theorem:decay-psi-general} we deduce, for  a $\tau_0\in[\tau/2, \tau]$,
\beaa
E^{s-1}_{1-\de} (\tau_0)&\les& (1+\tau)^{-q_0-1-\de}\left( C_{q_0}^s+B^{2+s}_{2-\de}+\II^{s+1}_{2-\de, q_0+\de}[\Ng]\right)
\eeaa
and since,
\beaa
 E^s_{1-\de} (\tau)  &\les& E^s_{1-\de} (\tau_0)+   I^{s+1}_{1-\de}(\tau_0,\tau)[\Ng],
\eeaa
we infer that,
\begin{equation}\label{eq:improveddecay-psi3}
E^s_{1-\de} (\tau)   \les (1+\tau)^{-q_0-1-\de}\left( C_{q_0}^{s+1}+B^{3+s}_{2-\de}+\II^{s+2}_{2-\de, q_0+\de}[\Ng] +   \II^{s+1}_{1-\de, 1+q_0+\de}[\Ng]    \right).
   \end{equation}
  Interpolating with \eqref{eq:improveddecay-psi11}, i.e.
  \beaa
E^s_{2-\de}&\les&(1+\tau)^{-q_0-\de} \left( C_{q_0}^s+  B^{2+s}_{2-\de}\right)
  \eeaa
   we deduce,
   \beaa
   E_1^s&\les&(E^s_{1-\de})^{1-\de } (E^s_{2-\de})^\de\les  (1+\tau)^{-q_0-1}\left( C_{q_0}^{s+1}+B^{3+s}_{2-\de}+\II^{s+2}_{2-\de, q_0+\de}[\Ng] +   \II^{s+1}_{1-\de, 1+q_0+\de}[\Ng]    \right).
   \eeaa
 Hence,
 \bea
 \label{eq:improveddecay-psi4}
 E_1^s&\les&(1+\tau)^{-q_0-1}\left( C_{q_0}^{s+1}+B^{3+s}_{2-\de}+\II^{s+2}_{2-\de, q_0+\de} +   \II^{s+1}_{1-\de, 1+q_0+\de}    \right).
 \eea
    As in the proof of Theorem  \ref{theorem:decay-psi-general} we 
        repeat the procedure       starting  with  the inequality   \eqref{eq:Daf-Rdn-1-s} for  $p=1$,
\beaa
&& E^{s}_{1}  (\tau )   +   \int_{\tau/2}^{\tau}      M_{0 } ^{s}  (\la ) d\la   \\   
&\les&    E^{s}_{1}  (\tau/2)+  I^{s+1}_{1}[\Ng](\tau/2 ,\tau )  \\
   &\les   & (1+\tau)^{-q_0-1}\left( C_{q_0}^{s+1}+B^{3+s}_{2-\de}+\II^{s+2}_{2-\de, q_0+\de}[\Ng] +   \II^{s+1}_{1-\de, 1+q_0+\de}[\Ng]    \right) +  (1+\tau)^{-1-q_0}    \II^{s+1}_{     1, 1+q_0}[\Ng] \\
    &\les& (1+\tau)^{-q_0-1}\left( C_{q_0}^{s+1}+B^{3+s}_{2-\de}+\II^{s+s}_{2-\de, q_0+\de}[\Ng] +   \II^{s+1}_{1-\de, 1+q_0+\de}[\Ng]  +  \II^{s+1}_{     1, 1+q_0}[\Ng]   \right)
\eeaa
   from which we infer that, for a $\tau_0\in[\tau/2, \tau]$,
   \bea
   \label{eq:improveddecay-psi5}
   && E^{s} (\tau_0 )\\
   \nn &\les& (1+\tau)^{-q_0-2}\left( C_{q_0}^{s+2}+B^{s+4}_{2-\de}+\II^{s+3}_{2-\de, q_0+\de}[\Ng] +   \II^{s+2}_{1-\de, 1+q_0+\de}[\Ng]  +  \II^{s+2}_{     1, 1+q_0}[\Ng]   \right).
   \eea

 Interpolating  \eqref{eq:improveddecay-psi3} and  \eqref{eq:improveddecay-psi5} we deduce, for $\de>0$ sufficiently small
 \beaa
 E^s_\de(\tau_0) &\les&\big( E^s (\tau_0)\big )^{\frac{1-2\de}{1-\de}}  \big(E^s_{1-\de}(\tau_0 )\big)^{\frac{\de}{1-\de}}\\
 &\les& (1+\tau)^{-2-q_0 +\de}\left( C_{q_0}^{s+2}+B^{s+4}_{2-\de}+\II^{s+3}_{2-\de, q_0+\de}[\Ng] +   \II^{s+2}_{1-\de, 1+q_0+\de}[\Ng]  +  \II^{s+2}_{     1, 1+q_0}[\Ng]   \right). 
 \eeaa
 Thus, since we have, as in \eqref{eq:enhanceddecay3},
 \beaa
 E^s_\de(\tau)\les E^s_\de(\tau_0)+I_\de^{s+1}[\Ng](\tau_0,\tau),
 \eeaa
 we deduce
\beaa
E^s_\de(\tau) &\les & (1+\tau)^{-2-q_0 +\de}\left( C_{q_0}^{s+2}+B^{s+4}_{2-\de}+\II^{s+3}_{2-\de, q_0+\de}[\Ng] +   \II^{s+2}_{1-\de, 1+q_0+\de}[\Ng] +   \II^{s+2}_{1, 1+q_0}[\Ng]   \right)\\
&+& (1+\tau)^{-2-q_0 +\de}\II^{s+1}_{\de, 2+q_0-\de }[\Ng]
\eeaa
i.e.,
\beaa
E^s_\de(\tau)&\les& (1+\tau)^{-2-q_0 +\de}\Big( C_{q_0}^{s+2}+B^{s+4}_{2-\de}+\II^{s+3}_{2-\de, q_0+\de}[\Ng] +   \II^{s+2}_{1-\de, 1+q_0+\de}[\Ng] \\
&& +  \II^{s+2}_{     1, 1+q_0}[\Ng] +\II^{s+1}_{\de, 2+q_0-\de }[\Ng]  \Big).
\eeaa
By interpolating the middle terms 
we write,
\beaa
E^s_\de(\tau)&\les& (1+\tau)^{-2-q_0 +\de}\left( C_{q_0}^{s+2}+B^{s+4}_{2-\de}+\II^{s+3}_{2-\de, q_0+\de}[\Ng] +\II^{s+3}_{\de, 2+q_0-\de }[\Ng] \right).
\eeaa
We now  recall,
\beaa
 C_{q_0}^s&:=& \EE_{q_0}^s[\qfc] (0)+\EE_{2-\de}^{s+2} [\qf](0)   +  \II^{s+3}_{q_0+2, 0}[\Ng]+ \II^{s+3}_{\de, q_0+2-\de}[\Ng]\\
  B^{2+s}_{2-\de}:&=&\EE^{s+2}_{2-\de}[\qf](0) +  \II^{s+3}_{2-\de, 0}[\Ng]+ \II^{s+3}_{\de, 2-2\de}[\Ng]. 
  \eeaa
  Hence,
  \beaa
 &&  C_{q_0}^{s+2}+B^{s+4}_{2-\de}+\II^{s+3}_{2-\de, q_0+\de}[\Ng] +\II^{s+3}_{\de, 2+q_0-\de }[\Ng] \\
&= & \EE_{q_0}^{s+2}[\qfc] (0)+\EE_{2-\de}^{s+4} [\qf](0) +  \II^{s+5}_{q_0+2, 0}[\Ng]+ \II^{s+5}_{\de, q_0+2-\de}[\Ng]\\
&+&\EE^{s+4}_{2-\de}[\qf](0) +   \II^{s+5}_{2-\de, 0}[\Ng]+ \II^{s+5}_{\de, 2-2\de}[\Ng] +\II^{s+3}_{2-\de, q_0+\de}[\Ng] +\II^{s+3}_{\de, 2+q_0-\de }[\Ng].
  \eeaa
We deduce,
 \bea\lab{eq:almosttherebutIwillneedtointerpolatebetweenthefollowing3estimatestoconludetheproofofdecay}
\EE^s_{\de, 2+q_0-\de}[\qf] &\les  \EE_{q_0}^{s+2}[\qfc] (0)+\EE_{2-\de}^{s+4} [\qf](0) +  \II^{s+5}_{q_0+2, 0}[\Ng] +\II^{s+5}_{\de, 2+q_0-\de }[\Ng].
\eea
 
We can also simplify the right hand side of \eqref{eq:improveddecay-psi4},
\beaa
&&C_{q_0}^{s+1}+B^{3+s}_{2-\de}+\II^{s+2}_{2-\de, q_0+\de}[\Ng] +   \II^{s+1}_{1-\de, 1+q_0+\de}[\Ng]\\
&&\les \EE_{q_0}^{s+2}[\qfc] (0)+\EE_{2-\de}^{s+4} [\qf](0)  +  \II^{s+5}_{q_0+2, 0}[\Ng] +\II^{s+5}_{\de, 2+q_0-\de }[\Ng].
\eeaa
Thus \eqref{eq:improveddecay-psi3}  becomes,
\bea\lab{eq:almosttherebutIwillneedtointerpolatebetweenthefollowing3estimatestoconludetheproofofdecay:1}
 \EE^s_{1-\de, 1+q_0+\de}&\les& \EE_{q_0}^{s+2}[\qfc] (0)+\EE_{2-\de}^{s+4} [\qf](0)  +  \II^{s+5}_{q_0+2, 0}[\Ng] +\II^{s+5}_{\de, 2+q_0-\de }[\Ng].
 \eea
 Similarly, \eqref{eq:improveddecay-psi11} yields
\bea\lab{eq:almosttherebutIwillneedtointerpolatebetweenthefollowing3estimatestoconludetheproofofdecay:2}
 \EE^s_{2-\de, q_0-\de}&\les& \EE_{q_0}^{s+2}[\qfc] (0)+\EE_{2-\de}^{s+4} [\qf](0)  +  \II^{s+5}_{q_0+2, 0}[\Ng] +\II^{s+5}_{\de, 2+q_0-\de }[\Ng].
 \eea 
Interpolating \eqref{eq:almosttherebutIwillneedtointerpolatebetweenthefollowing3estimatestoconludetheproofofdecay} and \eqref{eq:almosttherebutIwillneedtointerpolatebetweenthefollowing3estimatestoconludetheproofofdecay:1}, as well as \eqref{eq:almosttherebutIwillneedtointerpolatebetweenthefollowing3estimatestoconludetheproofofdecay:1} and \eqref{eq:almosttherebutIwillneedtointerpolatebetweenthefollowing3estimatestoconludetheproofofdecay:2},  we infer for all $s\leq k_{small}+25$ and for all $\de\leq p\leq 2-\de$
\bea\lab{eq:arggggggggggggggggggggggghhhh} 
\EE^s_{p, 2+q_0-p}[\qf] &\les&  \EE_{q_0}^{s+2}[\qfc] (0)+\EE_{2-\de}^{s+4} [\qf](0) +   \II^{s+5}_{q_0+2, 0}[\Ng]  + \II^{s+5}_{\de, 2+q_0 -\de}[\Ng].
\eea

Finally, making use of Theorem \ref{theorem-combinedMor-r-weighted-improved} between $\tau$ and $\tau_*$, we have in particular
\beaa
 B ^s_{p}[\qf](\tau, \tau_*)+F ^s_{p}[\qf](\tau, \tau_*) &\les&  E\,^s _{p}[\qf](\tau)+   I ^{s+1}_p[ N_g]( \tau,\tau_*)\\
 &\les& (1+\tau)^{-(2+q_0-p)}\Big(\EE^s_{p, 2+q_0-p}[\qf]+I ^{s+1}_{p,2+q_0-p}[ N_g]\Big)
\eeaa
and hence, we infer for  all $s\leq k_{small}+25$ and for all $\de\leq p\leq 2-\de$
\beaa
\BB^s_{p, 2+q_0-p}[\qf]+\FF^s_{p, 2+q_0-p}[\qf]    \les  \EE^s_{p, 2+q_0-p}[\qf] +   \II^{s+5}_{q_0+2, 0}[\Ng]  + \II^{s+5}_{\de, 2+q_0 -\de}[\Ng].
\eeaa 
Together with \eqref{eq:arggggggggggggggggggggggghhhh}, this concludes the proof of Theorem \ref{theorem:improveddecay-estimates-psi}.


\subsection{Proof of Proposition \ref{proposition:supnormdecay-psi}}\label{subs:pointwisedecay-psi}

  
Let $\chi$  be  a smooth cut-off function   vanishing for $r\le 4m_0$  and  equal to   $1$ for $ r\ge 6m_0$.  To prove   estimate    \eqref{eq:pointwisedecayS-1} we  consider the identity,
\beaa
e_4\left(\int_{S_r}\chi(\qf^{(s)})^2\right) &=& \int_{S_r}\Big(e_4(\chi(\qf^{(s)})^2)+\ka\chi(\qf^{(s)})^2\Big)\\
&=& \int_{S_r}\Big(\chi(2  \qf^{(s)} e_4\qf^{(s)} + 2r^{-1}(\qf^{(s)})^2)+ \chi '  (\qf^{(s)})^2 + \chi (\ka-2r^{-1}) |\qf^{(s)}|^2\Big)\\
&=& \int_{S_r}\Big( 2 \chi \qf^{(s)}\ec_4 \qf^{(s)}  + \chi '  (\qf^{(s)})^2+ O(r^{-2})|\qf^{(s)}|^2\Big).
\eeaa 
Integrating between $4m_0$ and $r$  for a fixed   $r\ge 6m_0$, we  deduce, in view of the definitions of $E[\qf^{(s)}](\tau)$ and of $ E_{p}[\qf^{(s)}](\tau)$, 
\beaa
\int_{S_r}|\qf^{(s)}|^2&\les&  \int_{\Si(\tau)_{\ge  4m_0}}|\qf^{(s)}| |\ec_4 \qf^{(s)}|+ E[\qf^{(s)}](\tau)\\
&\les& \left(    \int_{\Si(\tau)_{\ge  4m_0}} r^{1+\de}  |\ec_4 \qf^{(s)}|^2\right)^{1/2} \left(    \int_{\Si(\tau)_{\ge  4m_0}} r^{-1-\de}  |\qf^{(s)}|^2\right)^{1/2}+  E[\qf^{(s)}](\tau)\\
&\les&\big(E_{1+\de}[\qf^{(s)}](\tau)\big)^{1/2}\big( E_{1-\de}[\qf^{(s)}](\tau)\big)^{1/2}.
\eeaa
Clearly, this estimate also holds for $r\leq 6m_0$. Together with the definition \eqref{definition:normsdecay-psi} of $\EE^s_{p,d}[\qf^{(s)}]$, we immediately infer 
\beaa
(1+\tau)^{1+q_0}\int_{S_r}|\qf^{(s)} |^2&\les& \left(\EE^s_{1+\de, 1+q_0-\de}[\qf]\right)^{\frac{1}{2}}\left(\EE^s_{1-\de, 1+q_0+\de}[\qf]\right)^{\frac{1}{2}}
\eeaa
which is the desired estimate \eqref{eq:pointwisedecayS-1}.

To prove \eqref{eq:pointwisedecayS-1:bis}  we start  instead  with the identity,
\beaa
e_4\left(r^{-1}\int_{S_r}\chi(\qf^{(s)})^2\right) &=& \int_{S_r}r^{-1}\Big(e_4(\chi(\qf^{(s)})^2)+\ka\chi(\qf^{(s)})^2\Big)-\frac{e_4(r)}{r^2}\int_{S_r}\chi(\qf^{(s)})^2\\
&=& \int_{S_r}r^{-1}\Big(\chi(2  \qf^{(s)} e_4\qf^{(s)} + r^{-1}(\qf^{(s)})^2)+ \chi '  (\qf^{(s)})^2 + \chi (\ka-2r^{-1}) |\qf^{(s)}|^2\Big)\\
&&-\frac{e_4(r)-1}{r^2}\int_{S_r}\chi(\qf^{(s)})^2\\
&=& \int_{S_r}\Big( 2r^{-1}\chi e_4(\qf^{(s)})\qf^{(s)}         +  r^{-1}\chi '(\qf^{(s)})^2+ O(r^{-2})|\qf^{(s)}|^2\Big).
\eeaa 
Integrating between $4m_0$ and $r$  for a fixed   $r\ge 6m_0$, we  deduce, in view of the definitions of $E[\qf^{(s)}](\tau)$ and of $ E_{p}[\qf^{(s)}](\tau)$, 
\beaa
r^{-1} \int_{S_r}|\psi|^2&\les&  \int_{\Si(\tau)_{\ge  4m_0}}   r^{-1}|\qf^{(s)}||e_4(\qf^{(s)})| +  E[\qf^{(s)}](\tau)\\
&\les&2\left(    \int_{\Si(\tau)_{\ge  4m_0}}|e_4(\qf^{(s)})|^2\right)^{1/2} \left(    \int_{\Si(\tau)_{\ge  4m_0}} r^{-2}  |\qf^{(s)}|^2\right)^{1/2}+  E[\qf^{(s)}](\tau)\\
&\les& E[\qf^{(s)}](\tau)\\
&\les&E_{\de}[\qf^{(s)}](\tau).
\eeaa
Clearly, this estimate also holds for $r\leq 6m_0$. Together with the definition \eqref{definition:normsdecay-psi} of $\EE^s_{p,d}[\qf^{(s)}]$, we immediately infer 
\beaa
r^{-1}(1+\tau)^{2+q_0-\de}\int_{S_r}|\qf^{(s)} |^2&\les& \EE^s_{\de, 2+q_0-\de}[\qf]
\eeaa
which is the desired estimate \eqref{eq:pointwisedecayS-1:bis}. This concludes the proof of Proposition \ref{proposition:supnormdecay-psi}.


\subsection{Proof of Proposition \ref{Prop:integrated decay-lastslice-repeat}}\lab{sec:integrated decay-lastslice}


Recall the following definitions
\beaa
F[\psi](\tau_1, \tau_2) &=& \int_{\AA(\tau_1, \tau_2)}\Big( \deh^{-1}   |e_4 \Psi|^2  + \deh  |e_3\Psi|^2 + |\nabb \Psi|^2 + r^{-2} |\Psi|^2\Big)\\
&&  +\int_{\Sigma_*(\tau_1, \tau_2)}\Big(  |e_4 \Psi|^2  + |e_3\Psi|^2 + |\nabb \Psi|^2 + r^{-2} |\Psi|^2\Big),\\
\Fd_p[\psi](\tau_1, \tau_2) &=& \int_{\Sigma_*(\tau_1, \tau_2)}r^p\Big( |e_4\psi|^2+|\nabb\psi|^2 + r^{-2} |\psi|^2 \Big),\\
F_p[\psi](\tau_1, \tau_2) &=& F[\psi](\tau_1, \tau_2)+\Fd_p[\psi](\tau_1, \tau_2),\\
F^s[\psi](\tau_1, \tau_2) &=& \sum_{k\leq s}F[\dk^k\psi](\tau_1, \tau_2),\\
F^s_p[\psi](\tau_1, \tau_2) &=& \sum_{k\leq s}F_p[\dk^k\psi](\tau_1, \tau_2),\\
\FF^s_{p,d}[\psi] &=& \sup_{0\leq\tau\leq\tau_*}(1+\tau)^dF^s_p[\psi](\tau, \tau_*).
\eeaa
We deduce
\beaa
F^s[\qf](\tau, \tau_*) \leq F^s_\de[\qf](\tau, \tau_*) \leq (1+\tau)^{-2-q_0+\de}\FF^s_{\de ,2+q_0-\de}[\qf]
\eeaa
and hence in particular 
\bea\lab{eq:ffffffffflux}
(1+\tau)^{2+q_0-\de}\int_{\Sigma_*(\tau, \tau_*)}\Big(|e_3\dk^{\le s} \qf|^2+r^{-2} |\dk^{\leq s}\qf|^2 \Big) &\les& \FF^s_{\de, 2+q_0-\de}[\qf]
\eea
which yields the desired estimate \eqref{eq:ffffffffflux:00}.

Next, we focus on the proof of \eqref{eq:ffffffffflux:00:bis}. We start with the following trace estimate 
\beaa
\sup_{\Sigma_*(\tau, \tau_*)}\|e_3\dk^{\le s} \qf\|_{L^2(S)} &\les& \|\nu e_3\dk^{\le s} \qf\|_{L^2(\Sigma_*(\tau, \tau_*))}+\|e_3\dk^{\le s} \qf\|_{L^2(\Sigma_*(\tau, \tau_*))}
\eeaa
where we recall that $\nu$ is tangent to $\Sigma_*$, orthogonal to $e_\th$ and given by
\beaa
\nu=e_3+ae_4, \quad -2\leq a\leq -\frac{1}{2}.
\eeaa
We infer
\beaa
\sup_{\Sigma_*(\tau, \tau_*)}\|e_3\dk^{\le s} \qf\|_{L^2(S)} &\les& \|e_3e_3\dk^{\le s} \qf\|_{L^2(\Sigma_*(\tau, \tau_*))}+ \|e_4e_3\dk^{\le s} \qf\|_{L^2(\Sigma_*(\tau, \tau_*))}\\
&&+\|e_3\dk^{\le s} \qf\|_{L^2(\Sigma_*(\tau, \tau_*))}\\
&\les& \|e_3\dk^{\le s+1} \qf\|_{L^2(\Sigma_*(\tau, \tau_*))}+ \|r^{-1}\dk^{\le s+1} \qf\|_{L^2(\Sigma_*(\tau, \tau_*))}\\
&& +\|[e_4,e_3]\dk^{\le s} \qf\|_{L^2(\Sigma_*(\tau, \tau_*))}\\
&\les& \|e_3\dk^{\le s+1} \qf\|_{L^2(\Sigma_*(\tau, \tau_*))}+ \|r^{-1}\dk^{\le s+1} \qf\|_{L^2(\Sigma_*(\tau, \tau_*))}.
\eeaa
In view of \eqref{eq:ffffffffflux}, we deduce
\bea\lab{eq:ffffffffflux:bis}
\sup_{\Sigma_*}\left\{(1+\tau)^{2+q_0-\de}\|e_3\dk^{\le s} \qf\|_{L^2(S)}^2\right\} &\les& \FF^{s+1}_{\de, 2+q_0-\de}[\qf]. 
\eea

Next, we extend \eqref{eq:ffffffffflux:bis} to $r\geq 4m_0$. In view of \eqref{eq:higherorderderivativesintheidentityformula-wave-rpsi}, we have schematically 
 \beaa
e_3  e_4 (r\dk^k\qf) &=& -\dk^{\leq k}(r\square_2 \qf )+ r\lapp_2(\dk^{\leq k}\qf)+r^{-1}\dk^{\leq k+1}\qf\\
&=& -\dk^{\leq k}(r\square_2 \qf )+r^{-1}\dk^{\leq k+2}\qf. 
\eeaa
Also, we have
\beaa
e_4(re_3(\dk^k\qf)) &=& e_3 e_4(r\dk^k\qf)+[e_4, e_3](r\dk^k\qf)-e_4(e_3(r)\dk^k\qf)
\eeaa
and hence, we infer schematically 
 \beaa
e_4(re_3(\dk^k\qf)) &=& -\dk^{\leq k}(r\square_2 \qf )+r^{-1}\dk^{\leq k+2}\qf. 
\eeaa
Now, recall \eqref{eq:combined-improved10}
\beaa
\nn |\dk^k (r\square_2 \qf )|  &\les& r^{-1}\big |\dk^{\le k+1} \qf \big|+ r \big|\dk^{\le k } N_g|+ r^2\big|\dk^{\le k }e_3(N_g)|.
 \eeaa
We deduce
\beaa
|e_4(re_3(\dk^k\qf))| &\les& r^{-1}\big |\dk^{\le k+2} \qf \big|+ r \big|\dk^{\le k } N_g|+ r^2\big|\dk^{\le k }e_3(N_g)|
\eeaa

Now, we have
\beaa
e_4\left(r^{-2}\int_S(re_3(\dk^{\leq s}\qf))^2\right) &=& r^{-2}\int_S\Big(2e_4(re_3(\dk^{\leq s}\qf))re_3(\dk^{\leq s}\qf)+\ka (re_3(\dk^{\leq s}\qf))^2\Big)\\
&& -\frac{2e_4(r)}{r}r^{-2}\int_S(re_3(\dk^{\leq s}\qf))^2\\
&=& r^{-2}\int_S\Big(2e_4(re_3(\dk^{\leq s}\qf))re_3(\dk^{\leq s}\qf)+(\ka-2r^{-1}) (re_3(\dk^{\leq s}\qf))^2\Big)\\
&& -2\frac{e_4(r)-1}{r}r^{-2}\int_S(re_3(\dk^{\leq s}\qf))^2\\
\eeaa
and hence
\beaa
\left|e_4\left(r^{-2}\int_S(re_3(\dk^{\leq s}\qf))^2\right)\right| &\les& r^{-2}\int_S\Bigg\{\Big(r^{-1}\big |\dk^{\le s+2} \qf \big|+ r \big|\dk^{\le s } N_g|+ r^2\big|\dk^{\le s }e_3(N_g)|\Big)|re_3(\dk^{\leq s}\qf)|\\
&&+r^{-2}(re_3(\dk^{\leq s}\qf))^2\Bigg\}\\
&\les& r^{-2}\int_S\Big(r^{-\frac{1}{2}}\big |\dk^{\le s+2} \qf \big|^2+ r^2 \big|\dk^{\le s } N_g|^2+ r^4\big|\dk^{\le s }e_3(N_g)|^2\Big)\\
&&+r^{-\frac{7}{2}}\int_S(re_3(\dk^{\leq s}\qf))^2
\eeaa
Together with  \eqref{eq:pointwisedecayS-1:bis}, this yields 
\beaa
\left|e_4\left(r^{-2}\int_S(re_3(\dk^{\leq s}\qf))^2\right)\right| &\les& r^{-2}\int_S\Big( r^{\frac{7}{2}} \big|\dk^{\le s } N_g|^2+ r^{\frac{11}{2}}\big|\dk^{\le s }e_3(N_g)|^2\Big)\\
&&+r^{-4}\int_S(re_3(\dk^{\leq s}\qf))^2+r^{-\frac{3}{2}}(1+\tau)^{-2-q_0+\de}\EE^{s+2}_{\de, 2+q_0-\de}[\qf].
\eeaa

Now, recall from \eqref{eq:usedlaterforthedecayofe3qfthankstoflux} that we have for $s\leq k_{small}+30$
\beaa
\begin{split}
|\dk^s\Ng| &\les \ep^2r^{-3}\tau^{-1-2\dec+2\de_0}\\
|\dk^s\Ng| &\les \ep^2r^{-1}\tau^{-2-2\dec+2\de_0},\\
|\dk^se_3(\Ng)| &\les \ep^2r^{-3}\tau^{-\frac{3}{2}-2\dec+2\de_0},\\
|\dk^se_3(\Ng)| &\les \ep^2r^{-\frac{7}{2}-\frac{\dt}{2}}\tau^{-1-\dec+2\de_0}.
\end{split}
\eeaa
By interpolation, we infer
\beaa
 r^{-2}\int_S\Big( r^{\frac{7}{2}} \big|\dk^{\le s } N_g|^2+ r^{\frac{11}{2}}\big|\dk^{\le s }e_3(N_g)|^2\Big) &\les&  \ep^4r^{-\frac{3}{2}}\tau^{-\frac{5}{2}-4\dec+4\de_0}+ \ep^4r^{-1-\frac{\dt}{2}}\tau^{-\frac{5}{2}-3\dec+4\de_0}\\
 &\les&   \ep_0^2r^{-1-\frac{\dt}{2}}\tau^{-\frac{5}{2}-3\dec+4\de_0}
\eeaa
and hence
\beaa
\left|e_4\left(r^{-2}\int_S(re_3(\dk^{\leq s}\qf))^2\right)\right| &\les& r^{-4}\int_S(re_3(\dk^{\leq s}\qf))^2\\
&&+\ep_0^2r^{-1-\frac{\dt}{2}}\tau^{-\frac{5}{2}-3\dec+4\de_0}+r^{-\frac{3}{2}}(1+\tau)^{-2-q_0+\de}\EE^{s+2}_{\de, 2+q_0-\de}[\qf].
\eeaa
We integrate from $\Sigma_*$. By Gronwall, and in view of \eqref{eq:ffffffffflux:bis}, we deduce for $r\geq 4m_0$
\beaa
(1+\tau)^{2+q_0-\de}\int_{S_r}(e_3\dk^{\le s} \qf)^2 &\les& \ep_0^2+\FF^{s+1}_{\de, 2+q_0-\de}[\qf]+\EE^{s+2}_{\de, 2+q_0-\de}[\qf].
\eeaa
On the other hand, we have by a trace estimate for $r\leq 4m_0$
\beaa
(1+\tau)^{2+q_0-\de}\int_{S_r}(e_3\dk^{\le s} \qf)^2 &\les& \EE^{s+2}_{0, 2+q_0-\de}[\qf].
\eeaa
We finally deduce on $\MM$
\beaa
(1+\tau)^{2+q_0-\de}\int_{S_r}(e_3\dk^{\le s} \qf)^2 &\les& \ep_0^2+\FF^{s+1}_{\de, 2+q_0-\de}[\qf]+\EE^{s+2}_{\de, 2+q_0-\de}[\qf]
\eeaa
which is the desired estimate \eqref{eq:ffffffffflux:00:bis}. This concludes the proof of Proposition \ref{Prop:integrated decay-lastslice-repeat}.


\chapter{DECAY ESTIMATES FOR $\a$ AND $\aa$ (Theorems M2, M3)}\lab{chap:proofoftheoremM2M3}


In this section, we rely on the decay of $\qf$ to prove the decay estimates for $\a$ and $\aa$. More precisely, we rely on the results of Theorem M1 to prove Theorem M2 and M3.


\section{Proof of Theorem M2}



\subsection{A renormalized frame on $\Mext$}


In Theorem M1, decay estimates are derived for $\qf$ defined with respect to the global frame constructed in Proposition \ref{prop:existenceandestimatesfortheglobalframe:bis}. We have the following control for the Ricci coefficients in that frame. 

\begin{lemma}\lab{lemma:identitiesandnonsharpesitmatesforRiccicoefficientsoftherenormalizedframeofMext}
Consider the  global null frame $(e_3, e_4, e_\th)$ constructed in Proposition \ref{prop:existenceandestimatesfortheglobalframe:bis}. Then, the Ricci coefficients satisfy the following estimates 
\beaa
\max_{0\leq k\leq k_{small}+20}\sup_{\Mext}u^{\frac{1}{2}}\Bigg(r^2\left|\dk^k\left(\om+\frac{m}{r^2}, \ka-\frac{2\Up}{r}, \vth, \ze, \eta, \etab\right)\right|+r\left|\dk^k\left(\xib, \omb, \kab+\frac{2}{r}, \vthb\right)\right|, && \\
+\left|\dk^k\left(e_4(r)-\Up, e_3(r)+1\right)\right|\Bigg) &\les&\ep.
\eeaa
\end{lemma}

\begin{proof}
This follows immediately from the stronger estimates of Lemma \ref{le:interpolatedbootstrap} with the choice $k_{loss}=20$.
\end{proof}


\subsection{A transport equation for $\a$}


To recover $\a$ from $\qf$, we derive below a transport equation for $\a$ where $\qf$ is on the RHS. We are careful to avoid terms of the type $e_3(\omb)$ as they are anomalous w.r.t. decay in $r$. Indeed, they only decay linearly in $r^{-1}$ while all comparable term decay like $r^{-2}$ in $r$.  
\begin{lemma}
We have
\beaa
&& \kab^2e_3\left\{e_3\left(\frac{\a}{\kab^2}\right) -\left(8\omb-\frac{2}{\kab}\left(2\ddd_1\xib -\frac{1}{2}\vthb^2\right)\right)\frac{\a}{\kab^2}\right\}\\ 
&=& \frac{\qf}{r^4} +\left\{10\omb +\frac{4}{\kab}\left(-\ddd_1\xib -2(\eta-3\ze)\xib+\frac{1}{4}\vthb^2\right)\right\}e_3\a\\
&& +\Bigg\{ -2\ddd_1\xib +\left( 6\kab -24\omb +\frac{8}{\kab}\left(2\ddd_1\xib +2(\eta-3\ze)\xib-\frac{1}{2}\vthb^2\right)\right)\omb+\frac{1}{2}\vthb^2-\frac{4}{\kab}e_3((\eta-3\ze)\xib)\\
&& +\left( 16 +\frac{48}{\kab}\omb -\frac{24}{\kab^2}\left(2\ddd_1\xib +2(\eta-3\ze)\xib-\frac{1}{2}\vthb^2\right)\right)\ze\xib
\Bigg\}\a.
\eeaa
\end{lemma}

\begin{proof}
We compute
\beaa
e_3e_3\left(\frac{\a}{\kab^2}\right) &=& e_3\left(\frac{e_3\a}{\kab^2}-\frac{2e_3(\kab)\a}{\kab^3}\right)\\
&=& \frac{1}{\kab^2}\left(e_3e_3\a - 4\frac{e_3(\kab)}{\kab}e_3\a -2\kab^2 e_3\left(\frac{e_3(\kab)}{\kab^3}\right)\a\right).
\eeaa
Now, recall the following null structure equation 
\beaa
e_3(\kab)+\frac{1}{2}\kab^2+2\omb\,\kab= 2\ddd_1\xib +2(\eta-3\ze)\xib-\frac{1}{2}\vthb^2.
\eeaa
We infer
\beaa
\frac{e_3(\kab)}{\kab}= -\frac{1}{2}\kab -2\omb +\frac{1}{\kab}\left(2\ddd_1\xib +2(\eta-3\ze)\xib-\frac{1}{2}\vthb^2\right)
\eeaa
and
\beaa
e_3\left(\frac{e_3(\kab)}{\kab^3}\right) &=& e_3\left(-\frac{1}{2\kab} -2\frac{\omb}{\kab^2} +\frac{1}{\kab^3}\left(2\ddd_1\xib +2(\eta-3\ze)\xib-\frac{1}{2}\vthb^2\right)\right)\\
&=& \frac{e_3(\kab)}{2\kab^2}+e_3\left( -2\frac{\omb}{\kab^2} +\frac{1}{\kab^3}\left(2\ddd_1\xib +2(\eta-3\ze)\xib-\frac{1}{2}\vthb^2\right)\right)\\
&=&  -\frac{1}{4} +\frac{1}{2\kab}\left(-2\omb +\frac{1}{\kab}\left(2\ddd_1\xib +2(\eta-3\ze)\xib-\frac{1}{2}\vthb^2\right)\right)\\
&&+e_3\left( -2\frac{\omb}{\kab^2} +\frac{1}{\kab^3}\left(2\ddd_1\xib +2(\eta-3\ze)\xib-\frac{1}{2}\vthb^2\right)\right)\\
\eeaa
and hence
\beaa
\kab^2e_3e_3\left(\frac{\a}{\kab^2}\right) &=& e_3e_3\a - 4\frac{e_3(\kab)}{\kab}e_3\a -2\kab^2 e_3\left(\frac{e_3(\kab)}{\kab^3}\right)\a\\
&=&  e_3e_3\a +2\kab e_3\a  +\frac{1}{2}\kab^2\a +\left(8\omb -\frac{4}{\kab}\left(2\ddd_1\xib +2(\eta-3\ze)\xib-\frac{1}{2}\vthb^2\right)\right)e_3\a\\
&& +\Bigg\{-\kab\left(-2\omb +\frac{1}{\kab}\left(2\ddd_1\xib +2(\eta-3\ze)\xib-\frac{1}{2}\vthb^2\right)\right)\\
&&-2\kab^2e_3\left( -2\frac{\omb}{\kab^2} +\frac{1}{\kab^3}\left(2\ddd_1\xib +2(\eta-3\ze)\xib-\frac{1}{2}\vthb^2\right)\right)\Bigg\}\a.
\eeaa

Next, recall from section \ref{sec:discussionanddefintionofinvariantquantities} that $\qf$ is defined with respect to a general null frame as follows 
\beaa
\qf&=& r^4\left(e_3(e_3(\a))+(2\kab -6\omb)e_3(\a)+\left(-4e_3(\omb)+8\omb^2-8\omb\,\kab+\frac{1}{2}\kab^2\right)\a\right).
\eeaa
We infer
\beaa
\kab^2e_3e_3\left(\frac{\a}{\kab^2}\right) &=& \frac{\qf}{r^4} +\left(14\omb -\frac{4}{\kab}\left(2\ddd_1\xib +2(\eta-3\ze)\xib-\frac{1}{2}\vthb^2\right)\right)e_3\a\\
&& +\Bigg\{4e_3(\omb)-8\omb^2+10\omb\,\kab -\left(2\ddd_1\xib +2(\eta-3\ze)\xib-\frac{1}{2}\vthb^2\right)\\
&&-2\kab^2e_3\left( -2\frac{\omb}{\kab^2} +\frac{1}{\kab^3}\left(2\ddd_1\xib +2(\eta-3\ze)\xib-\frac{1}{2}\vthb^2\right)\right)\Bigg\}\a.
\eeaa

We rewrite the following terms
\beaa
&& \Bigg\{4e_3(\omb)-2\kab^2e_3\left( -2\frac{\omb}{\kab^2} +\frac{1}{\kab^3}\left(2\ddd_1\xib -\frac{1}{2}\vthb^2\right)\right)\Bigg\}\a\\
&=& \kab^2e_3\left\{\left(8\frac{\omb}{\kab^2}-\frac{2}{\kab^3}\left(2\ddd_1\xib -\frac{1}{2}\vthb^2\right)\right)\a\right\} -4\kab^2\omb e_3\left(\frac{\a}{\kab^2}\right) +\frac{2}{\kab}\left(2\ddd_1\xib -\frac{1}{2}\vthb^2\right)e_3(\a)\\
&=& \kab^2e_3\left\{\left(8\frac{\omb}{\kab^2}-\frac{2}{\kab^3}\left(2\ddd_1\xib -\frac{1}{2}\vthb^2\right)\right)\a\right\} \\
&&+\left( -4\omb +\frac{2}{\kab}\left(2\ddd_1\xib -\frac{1}{2}\vthb^2\right)\right)e_3\a  -4\kab^2\omb e_3\left(\frac{1}{\kab^2}\right)\a 
\eeaa
so that we obtain
\beaa
&& \kab^2e_3\left\{e_3\left(\frac{\a}{\kab^2}\right) -\left(8\frac{\omb}{\kab^2}-\frac{2}{\kab^3}\left(2\ddd_1\xib -\frac{1}{2}\vthb^2\right)\right)\a\right\}\\ 
&=& \frac{\qf}{r^4} +\left\{10\omb -\frac{4}{\kab}\left(\ddd_1\xib +2(\eta-3\ze)\xib-\frac{1}{4}\vthb^2\right)\right\}e_3\a +\Bigg\{-8\omb^2+10\omb\,\kab\\
&& -\left(2\ddd_1\xib +2(\eta-3\ze)\xib-\frac{1}{2}\vthb^2\right)-4\kab^2e_3\left( \frac{1}{\kab^3}(\eta-3\ze)\xib\right)-4\kab^2\omb e_3\left(\frac{1}{\kab^2}\right)\Bigg\}\a
\eeaa
which we rewrite as
\beaa
&& \kab^2e_3\left\{e_3\left(\frac{\a}{\kab^2}\right) -\left(8\omb-\frac{2}{\kab}\left(2\ddd_1\xib -\frac{1}{2}\vthb^2\right)\right)\frac{\a}{\kab^2}\right\}\\ 
&=& \frac{\qf}{r^4} +\left\{10\omb -\frac{4}{\kab}\left(\ddd_1\xib +2(\eta-3\ze)\xib-\frac{1}{4}\vthb^2\right)\right\}e_3\a +\Bigg\{-8\omb^2+10\omb\,\kab \\
&&-\left(2\ddd_1\xib +2(\eta-3\ze)\xib-\frac{1}{2}\vthb^2\right)-\frac{4}{\kab}e_3((\eta-3\ze)\xib) +12\frac{e_3(\kab)}{\kab^2}(\eta-3\ze)\xib
+8\frac{e_3(\kab)}{\kab}\omb\Bigg\}\a.
\eeaa
Now, recall from above that we have
\beaa
\frac{e_3(\kab)}{\kab}= -\frac{1}{2}\kab -2\omb +\frac{1}{\kab}\left(2\ddd_1\xib +2(\eta-3\ze)\xib-\frac{1}{2}\vthb^2\right).
\eeaa
We finally deduce
\beaa
&& \kab^2e_3\left\{e_3\left(\frac{\a}{\kab^2}\right) -\left(8\omb-\frac{2}{\kab}\left(2\ddd_1\xib -\frac{1}{2}\vthb^2\right)\right)\frac{\a}{\kab^2}\right\}\\ 
&=& \frac{\qf}{r^4} +\left\{10\omb +\frac{4}{\kab}\left(-\ddd_1\xib -2(\eta-3\ze)\xib+\frac{1}{4}\vthb^2\right)\right\}e_3\a\\
&& +\Bigg\{ -2\ddd_1\xib +\left( 6\kab -24\omb +\frac{8}{\kab}\left(2\ddd_1\xib +2(\eta-3\ze)\xib-\frac{1}{2}\vthb^2\right)\right)\omb+\frac{1}{2}\vthb^2-\frac{4}{\kab}e_3((\eta-3\ze)\xib)\\
&& +\left( 16 +\frac{48}{\kab}\omb -\frac{24}{\kab^2}\left(2\ddd_1\xib +2(\eta-3\ze)\xib-\frac{1}{2}\vthb^2\right)\right)\ze\xib
\Bigg\}\a.
\eeaa
This concludes the proof of the lemma.
\end{proof}


\subsection{Estimates for transport equations in $e_3$}


The following lemma will be useful to integrate the transport equations in $e_3$. 
\begin{lemma}\lab{lemma:usefullupperboundofintegralsfortranportequationsine3TheoremM2}
Let $p\in\Mext$. Let $\ga[p]$ the unique integral curve of $e_3$ starting from a point on $\CC_1$ terminating at $p$. Then, we have for $l\geq 1$
\beaa
\int_{\ga[p]}\frac{1}{{r'}^{2+l}{u'}^{\frac{1}{2}+\dee}+{r'}^{1+l}{u'}^{1+\dee}} &\les& \frac{1}{r^{1+l}(2r+u)^{\frac{1}{2}+\dee}+r^l(2r+u)^{1+\dee}}
\eeaa
and 
\beaa
\int_{\ga[p]}\frac{1}{{r'}^2{u'}^{\frac{1}{2}+\dee}+{r'}{u'}^{1+\dee}} &\les& \frac{1}{r(2r+u)^{\frac{1}{2}+\dee}+\frac{1}{\log(1+u)}(2r+u)^{1+\dee}}
\eeaa
where $(u,r)$ correspond to $p$ and $(r',  u')$ to a point on $\ga[p]$, and where the integration along $\ga[p]$ relies on a parametrization of  $\ga[p]$ normalized with respect to $e_3$.
\end{lemma}

\begin{proof}
Note first from the construction of $\Mext$ that $\ga[p]$ exists for any $p\in\Mext$ (i.e. any point $p$ can be joined to $\CC_1$ by an integral curve of $e_3$), and $\ga[p]$ is included in $\Mext$. 

Next, recall that the integration along $\ga[p]$ relies on a parametrization of  $\ga[p]$ normalized with respect to $e_3$. To parametrize the integration by $u$ or $r$, we will thus have to derive an upper bound for the corresponding Jacobian of the change of variable, i.e. for
\beaa
\frac{1}{|e_3(u)|}, \quad \frac{1}{|e_3(r)|}.
\eeaa
To this end, note that we have on $\Mext$
\beaa
e_3(u) = \frac{2}{\vsi\Up}\geq \frac{2+O(\ep)}{\Up}\geq \frac{1}{\Up}\geq 1
\eeaa
since $\Up\leq 1$ by definition. Also, we have on $\Mext$ in view of Lemma \ref{lemma:identitiesandnonsharpesitmatesforRiccicoefficientsoftherenormalizedframeofMext}
\beaa
|e_3(r)| &\geq & 1-\left|e_3(r)+1\right|\\
&=& 1+O(\ep)\\
&\geq & \frac{1}{2}. 
\eeaa
Hence, we have obtained on $\Mext$ 
\bea\lab{eq:uuperboundjacobianchangeofvariableur}
\frac{1}{|e_3(u)|}\leq \frac{1}{2}, \qquad \frac{1}{|e_3(r)|}\leq 1.
\eea

Next, since $e_3(u)>0$ and $e_3(r)<0$ in $\Mext$, we have $r'\geq r$ and $1\leq u'\leq u$. We start with the proof of the first inequality. We consider two cases
\begin{itemize}
\item If $r\geq u$, we have 
\beaa
\int_{\ga[p]}\frac{1}{{r'}^{2+l}{u'}^{\frac{1}{2}+\dee}+{r'}^{1+l}{u'}^{1+\dee}} &\leq& \frac{1}{r^{2+l}}\int_0^u\frac{1}{|e_3(u')|}\frac{du'}{{u'}^{\frac{1}{2}+\dee}}\\
&\les& \frac{u^{\frac{1}{2}-\dee}}{r^{2+l}}\\
&\les&  \frac{1}{r^{1+l}(2r+u)^{\frac{1}{2}+\dee}+r^l(2r+u)^{1+\dee}},
\eeaa
where we used \eqref{eq:uuperboundjacobianchangeofvariableur}.

\item If $r\leq u$, we separate the integral in $r'\geq u$, which coincides with $1\leq u'\leq u$, and $r\leq r'\leq u$ and compute
\beaa
&&\int_{\ga[p]}\frac{1}{{r'}^{2+l}{u'}^{\frac{1}{2}+\dee}+{r'}^{1+l}{u'}^{1+\dee}}\\
&=& \int_0^u\frac{1}{{r'}^{2+l}{u'}^{\frac{1}{2}+\dee}+{r'}^{1+l}{u'}^{1+\dee}}\frac{du'}{|e_3(u')|}\\
&&+\int_r^u\frac{1}{{r'}^{2+l}{u'}^{\frac{1}{2}+\dee}+{r'}^{1+l}{u'}^{1+\dee}}\frac{dr'}{|e_3(r')|}\\
 &\les& \frac{1}{u^{2+l}}\int_0^u\frac{1}{|e_3(u')|}\frac{du'}{{u'}^{\frac{1}{2}+\dee}}\\
 &&+\min\left(\frac{1}{u^{\frac{1}{2}+\dee}}\int_r^u\frac{1}{|e_3(r')|}\frac{dr'}{{r'}^{2+l}}, \frac{1}{u^{1+\dee}}\int_r^u\frac{1}{|e_3(r')|}\frac{dr'}{{r'}^{1+l}}\right)\\
&\les& \frac{1}{u^{\frac{5}{2}+\dee}}+\min\left(\frac{1}{u^{\frac{1}{2}+\dee}}\frac{1}{r^{1+l}}, \frac{1}{u^{1+\dee}}\frac{1}{r^l}\right)\\
&\les& \frac{1}{r^{1+l}(2r+u)^{\frac{1}{2}+\dee}+r^l(2r+u)^{1+\dee}},
\eeaa
where we used \eqref{eq:uuperboundjacobianchangeofvariableur}. 
\end{itemize}
This proves the first inequality. 

The second inequality is obtained similarly as follows
\begin{itemize}
\item If $r\geq u$, we have
\beaa
\int_{\ga[p]}\frac{1}{{r'}^2{u'}^{\frac{1}{2}+\dee}+r'{u'}^{1+\dee}} &\leq& \frac{1}{r^2}\int_0^u\frac{1}{|e_3(u')|}\frac{du'}{{u'}^{\frac{1}{2}+\dee}}\\
&\les& \frac{u^{\frac{1}{2}-\dee}}{r^2}\\
&\les&  \frac{1}{r(2r+u)^{\frac{1}{2}+\dee}+(2r+u)^{1+\dee}},
\eeaa
where we used \eqref{eq:uuperboundjacobianchangeofvariableur}. 

\item If $r\leq u$, we separate the integral in $r'\geq u$, which coincides with $1\leq u'\leq u$, and $r\leq r'\leq u$ and compute
\beaa
&&\int_{\ga[p]}\frac{1}{{r'}^2{u'}^{\frac{1}{2}+\dee}+{r'}{u'}^{1+\dee}}\\
&=&\int_0^u\frac{1}{{r'}^2{u'}^{\frac{1}{2}+\dee}+{r'}{u'}^{1+\dee}}\frac{du'}{|e_3(u')|}+\int_r^u\frac{1}{{r'}^2{u'}^{\frac{1}{2}+\dee}+{r'}{u'}^{1+\dee}}\frac{dr'}{|e_3(r')|}\\
&\les& \frac{1}{u^3}\int_0^u\frac{1}{|e_3(u')|}\frac{du'}{{u'}^{\frac{1}{2}+\dee}}\\
 &&+\min\left(\frac{1}{u^{\frac{1}{2}+\dee}}\int_r^u\frac{1}{|e_3(r')|}\frac{dr'}{{r'}^2}, \frac{1}{u^{1+\dee}}\int_r^u\frac{1}{|e_3(r')|}\frac{dr'}{{r'}}\right)\\
&\les& \frac{1}{u^{\frac{5}{2}+\dee}}+\min\left(\frac{1}{u^{\frac{1}{2}+\dee}}\frac{1}{r}, \frac{1}{u^{1+\dee}}\int_{\frac{r}{u}}^1\frac{dr''}{r''}\right)\\
&\les& \frac{1}{r(2r+u)^{\frac{1}{2}+\dee}+\frac{(2r+u)^{1+\dee}}{\log(1+u)}},
\eeaa
where we used \eqref{eq:uuperboundjacobianchangeofvariableur}. 
\end{itemize}
This concludes the proof of the lemma.
\end{proof}

\begin{corollary}\lab{cor:usefullcontroloftranportequationsine3dierctionTheoremM2}
Let $\psi$ a solution of the following transport equation
\beaa
e_3(\psi) = h\textrm{ on }\Mext. 
\eeaa
Let also $0<u_1\leq u_*$. Then
\begin{itemize}
\item If $h$ and $\psi$ satisfy for $l\geq 1$
\beaa
|h| \les \frac{\ep_0}{r^{2+l}u^{\frac{1}{2}+\dee}+r^{1+l}u^{1+\dee}}  \textrm{ on }\Mext(u\leq u_1)\textrm{ and }|\psi| \les \frac{\ep_0}{r^{l+\frac{3}{2}+\dee}}\textrm{ on }\CC_1,
\eeaa
we have 
\beaa
\sup_{\Mext(u\leq u_1)}\Big(r^{1+l}(2r+u)^{\frac{1}{2}+\dee}+r^l(2r+u)^{1+\dee}\Big)|\psi| &\les& \ep_0.
\eeaa

\item   If $h$ and $\psi$ satisfy  
\beaa
|h| \les \frac{\ep_0}{r^2u^{\frac{1}{2}+\dee}+ru^{1+\dee}}  \textrm{ on }\Mext(u\leq u_1)\textrm{ and }|\psi| \les \frac{\ep_0}{r^{\frac{3}{2}+\dee}}\textrm{ on }\CC_1,
\eeaa
we have 
\beaa
\sup_{\Mext(u\leq u_1)}\left(r(2r+u)^{\frac{1}{2}+\dee}+\frac{(2r+u)^{1+\dee}}{\log(1+u)}\right)|\psi| &\les& \ep_0.
\eeaa
\end{itemize}
\end{corollary}

\begin{proof}
This follows immediately from Lemma \ref{lemma:usefullupperboundofintegralsfortranportequationsine3TheoremM2}.
\end{proof}


\subsection{Decay estimates for $\a$}


We start with an estimate for $\a$ on $\CC_1$.
\begin{lemma}\lab{lemma:initalesitmateonCC0foralphaofrenormalizedframeonMext}
We have
\beaa
\max_{0\leq k\leq k_{small}+22}\sup_{\CC_1}r^{\frac{7}{2}+\dee}|\dk^k\a|+\max_{0\leq k\leq k_{small}+21}\sup_{\CC_1}r^{\frac{9}{2}+\dee}|\dk^ke_3\a| &\les&\ep_0.
\eeaa
\end{lemma}

\begin{proof}
Recall that on $\CC_1$, we have obtained in Theorem M0 
\beaa
\max_{0\leq k\leq k_{large}}&&\Bigg\{ \sup_{\CC_1}  \left[r^{\frac{7}{2}  +\de_B}\left( |\dk^k\,{}^{(ext)}\a| + |\dk^k\,{}^{(ext)}\b|\right)+r^{\frac{9}{2}  +\de_B}|\dk^{k-1}e_3(\,{}^{(ext)}\a)|   \right]\\
\nn&&+ \sup_{\CC_1} \left[r^3\ \left|\dk^k\left(\,{}^{(ext)}\rho+\frac{2m_0}{r^3}\right)\right|+r^2|\dk^k\,{}^{(ext)}\bb|+r|\dk^k\,{}^{(ext)}\aa|\right]\Bigg\} \les \ep_0.
\eeaa
Since we have chosen $\dt\geq \dee$, we deduce
\beaa
\max_{0\leq k\leq k_{large}}\sup_{\CC_1}\Big[ r^{\frac{7}{2}  +\dee}|\dk^k\,{}^{(ext)}\a|+r^{\frac{9}{2}  +\dee}|\dk^{k-1}e_3(\,{}^{(ext)}\a)|   \Big]  &\les&\ep_0.
\eeaa

Next, recall  that $\qf$ is defined with respect to the global frame constructed in Proposition \ref{prop:existenceandestimatesfortheglobalframe:bis}. In view of Proposition \ref{prop:existenceandestimatesfortheglobalframe:bis} and Proposition \ref{prop:constructionsecondframeinMext}, and the change of frame formula for $\a$ in Proposition \ref{prop:transformations1}, we have
\bea\lab{eq:thechangeofframeforalphainsecondgloabalfrmaebacktoMext}
\a &=& (\,{}^{(ext)}\Up)^2\left(\,{}^{(ext)}\a+2f\,{}^{(ext)}\b+\frac{3}{2}f^2\,{}^{(ext)}\rho+\lot\right)
\eea
where $f$ satisfies\footnote{Here we use  \eqref{eq:estimateforfinconstructionsecondframeinMext} with $k_{loss}=20$. Note also that the estimates we claim here for $f$ are slightly weaker that those in \eqref{eq:estimateforfinconstructionsecondframeinMext}.}, see \eqref{eq:estimateforfinconstructionsecondframeinMext},
\bea\lab{eq:estimateforfinconstructionsecondframeinMext:usedinproofThM2}
\begin{split}
|\dk^kf| &\les \frac{\ep}{ru^{\frac{1}{2}}+u}, \,\,\,\textrm{ for }k\leq k_{small}+22\textrm{ on }\Mext,\\
 |\dk^{k-1}e_3f| &\les \frac{\ep}{ru}\,\,\,\textrm{ for }k\leq k_{small}+22\textrm{ on }\Mext.
\end{split}
\eea
We easily infer
\beaa
\max_{0\leq k\leq k_{small}+22}\sup_{\CC_1}r^{\frac{7}{2}+\dee}|\dk^k\a|+\max_{0\leq k\leq k_{small}+21}\sup_{\CC_1}r^{\frac{9}{2}+\dee}|\dk^ke_3\a| &\les&\ep_0.
\eeaa
This concludes the proof of the lemma.
\end{proof}

Next, let $0<u_1\leq u_*$. We introduce the following bootstrap assumption for $\a$ on $\Mext(u\leq u_1)$
\begin{equation}\lab{eq:localbootstrapasumptiontorecoveralphainTheoremM2}
\max_{0\leq k\leq k_{small}+20}\sup_{\Mext(u\leq u_1)}\Big(\frac{r^2(2r+u)^{1+\dee}}{\log(1+u)}+r^3(2r+u)^{\frac{1}{2}+\dee}\Big)\Big(|\dk^k\a|+r|\dk^ke_3\a|\Big) \leq \ep.
\end{equation}

The goal of this section will be the following proposition, i.e. the improvement of these bootstrap assumptions.
\begin{proposition}\lab{prop:improvementofbootstrapasumtionsforainTheoremM2}
We have
$$
\max_{0\leq k\leq k_{small}+20}\sup_{\Mext(u\leq u_1)}\Big(\frac{r^2(2r+u)^{1+\dee}}{\log(1+u)}+r^3(2r+u)^{\frac{1}{2}+\dee}\Big)\Big(|\dk^k\a|+r|\dk^ke_3\a|\Big) \les \ep_0.
$$
\end{proposition}

Proposition \ref{prop:improvementofbootstrapasumtionsforainTheoremM2} will be proved at the end of this section. 

Based on the bootstrap assumptions \eqref{eq:localbootstrapasumptiontorecoveralphainTheoremM2}, we estimate the RHS of the transport equation for $\a$.
\begin{lemma}\lab{lemma:estimatefortheRHSoftranportequationforalphawithrespectofrenormalizedframeonMext}
We have
\beaa
e_3\left\{e_3\left(\frac{\a}{\kab^2}\right) -F_1\right\} &=& F_2
\eeaa
where $F_1$ and $F_2$ satisfy
\beaa
\max_{0\leq k\leq k_{small}+20}\sup_{\Mext(u\leq u_1)}\Big(r(2r+u)^{1+\dee}+r^2(2r+u)^{\frac{1}{2}+\dee}\Big)|\dk^k F_1|&&\\
+\max_{0\leq k\leq k_{small}+20}\sup_{\Mext(u\leq u_1)}\Big(r^2u^{1+\dee}+r^3u^{\frac{1}{2}+\dee}\Big)|\dk^k F_2| &\les&  \ep_0.
\eeaa
\end{lemma}

\begin{proof}
Recall that we have
\beaa
&& \kab^2e_3\left\{e_3\left(\frac{\a}{\kab^2}\right) -\left(8\omb-\frac{2}{\kab}\left(2\ddd_1\xib -\frac{1}{2}\vthb^2\right)\right)\frac{\a}{\kab^2}\right\}\\ 
&=& \frac{\qf}{r^4} +\left\{10\omb +\frac{4}{\kab}\left(-\ddd_1\xib -2(\eta-3\ze)\xib+\frac{1}{4}\vthb^2\right)\right\}e_3\a\\
&& +\Bigg\{ -2\ddd_1\xib +\left( 6\kab -24\omb +\frac{8}{\kab}\left(2\ddd_1\xib +2(\eta-3\ze)\xib-\frac{1}{2}\vthb^2\right)\right)\omb+\frac{1}{2}\vthb^2-\frac{4}{\kab}e_3((\eta-3\ze)\xib)\\
&& +\left( 16 +\frac{48}{\kab}\omb -\frac{24}{\kab^2}\left(2\ddd_1\xib +2(\eta-3\ze)\xib-\frac{1}{2}\vthb^2\right)\right)\ze\xib
\Bigg\}\a
\eeaa
which we rewrite as
\beaa
e_3\left\{e_3\left(\frac{\a}{\kab^2}\right) -F_1\right\} &=& F_2
\eeaa
where $F_1$ and $F_2$ are defined by
\beaa
F_1 &:=& \left(8\omb-\frac{2}{\kab}\left(2\ddd_1\xib -\frac{1}{2}\vthb^2\right)\right)\frac{\a}{\kab^2}
\eeaa
and
\beaa
F_2 &:=& \frac{\qf}{r^4\kab^2} +\frac{1}{\kab^2}\left\{10\omb +\frac{4}{\kab}\left(-\ddd_1\xib -2(\eta-3\ze)\xib+\frac{1}{4}\vthb^2\right)\right\}e_3\a\\
&& +\frac{1}{\kab^2}\Bigg\{ -2\ddd_1\xib +\left( 6\kab -24\omb +\frac{8}{\kab}\left(2\ddd_1\xib +2(\eta-3\ze)\xib-\frac{1}{2}\vthb^2\right)\right)\omb+\frac{1}{2}\vthb^2-\frac{4}{\kab}e_3((\eta-3\ze)\xib)\\
&& +\left( 16 +\frac{48}{\kab}\omb -\frac{24}{\kab^2}\left(2\ddd_1\xib +2(\eta-3\ze)\xib-\frac{1}{2}\vthb^2\right)\right)\ze\xib
\Bigg\}\a.
\eeaa
In view of the bootstrap assumptions \eqref{eq:localbootstrapasumptiontorecoveralphainTheoremM2} for $\a$, the estimates of Lemma \ref{lemma:identitiesandnonsharpesitmatesforRiccicoefficientsoftherenormalizedframeofMext} for the Ricci coefficients, and using Theorem M2 to estimate $\qf$, we easily infer
\beaa
&&\max_{0\leq k\leq k_{small}+20}\sup_{\Mext(u\leq u_1)}\Big(r(2r+u)^{1+\dee}+r^2(2r+u)^{\frac{1}{2}+\dee}\Big)|\dk^k F_1|\\
&\les& \ep\max_{0\leq k\leq k_{small}+20}\sup_{\Mext(u\leq u_1)}\left(\frac{r^2(2r+u)^{1+\dee}}{\log(1+u)}+r^3(2r+u)^{\frac{1}{2}+\dee}\right)\Big(|\dk^k\a|+r|\dk^ke_3\a|\Big)\\
&\les& \ep^2\les \ep_0.
\eeaa
and
\beaa
&&\max_{0\leq k\leq k_{small}+20}\sup_{\Mext(u\leq u_1)}\Big(r^2u^{1+\dee}+r^3u^{\frac{1}{2}+\dee}\Big)|\dk^k F_2|\\
&\les& \max_{0\leq k\leq k_{small}+20}\sup_{\Mext(u\leq u_1)}\Big(u^{1+\dee}+ru^{\frac{1}{2}+\dee}\Big)|\dk^k\qf|\\
&& +\ep\max_{0\leq k\leq k_{small}+20}\sup_{\Mext(u\leq u_1)}\left(\frac{r^2(2r+u)^{1+\dee}}{\log(1+u)}+r^3(2r+u)^{\frac{1}{2}+\dee}\right)\Big(|\dk^k\a|+r|\dk^ke_3\a|\Big)\\
&\les& \ep_0+\ep^2\les \ep_0.
\eeaa
This concludes the proof of the lemma.
\end{proof}

\begin{lemma}\lab{lemma:differentiationoftransportequationalphabyangularderivativesforTheoremM2}
For $0\leq k+j\leq k_{small}+20$, we have
\beaa
e_3\left\{e_3\dkb^ke_4^j\left(\frac{\a}{\kab^2}\right)- F_{1,\dkb^k, e_4^j}\right\} &=& F_{2, \dkb^k, e_4^j}
\eeaa
where 
\beaa
\max_{0\leq l\leq k_{small}+20-k}\sup_{\Mext(u\leq u_1)}\Big(r(2r+u)^{1+\dee}+r^2(2r+u)^{\frac{1}{2}+\dee}\Big)|\dk^l F_{1,\dkb^k}|&&\\
+\max_{0\leq l\leq k_{small}+20-k}\sup_{\Mext(u\leq u_1)}\Big(r^2u^{1+\dee}+r^3u^{\frac{1}{2}+\dee}\Big)|\dk^l F_{2,\dkb^k}| &\les&  \ep_0,
\eeaa
and for $j\geq 1$
\beaa
&&\max_{0\leq l\leq k_{small}+20-k-j}\sup_{\Mext(u\leq u_1)}\Big(r^{1+j}(2r+u)^{1+\dee}+r^{2+j}(2r+u)^{\frac{1}{2}+\dee}\Big)|\dk^l F_{1,\dkb^k, e_4^j}|\\
&&+\max_{0\leq l\leq k_{small}+20-k-j}\sup_{\Mext(u\leq u_1)}\Big(r^{2+j}u^{1+\dee}+r^{3+j}u^{\frac{1}{2}+\dee}\Big)|\dk^l F_{2,\dkb^k, e_4^j}|\\
 &\les&  \ep_0+\max_{0\leq j+k\leq k_{small}+20}\sup_{\Mext(u\leq u_1)}\Big(\frac{r^2(2r+u)^{1+\dee}}{\log(1+u)}+r^3(2r+u)^{\frac{1}{2}+\dee}\Big)r\\
 &&\times\Big(|\dkb^k(re_4)^{j-1}e_3\a|+|\dkb^k(re_4)^{j-2}e_3^2\a|\Big).
\eeaa
\end{lemma}

\begin{proof}
Recall from Lemma \ref{lemma:estimatefortheRHSoftranportequationforalphawithrespectofrenormalizedframeonMext} that we have
\beaa
e_3\left\{e_3\left(\frac{\a}{\kab^2}\right) -F_1\right\} &=& F_2
\eeaa
where $F_1$ and $F_2$ satisfy
\beaa
\max_{0\leq k\leq k_{small}+20}\sup_{\Mext}\Big(r(2r+u)^{1+\dee}+r^2(2r+u)^{\frac{1}{2}+\dee}\Big)|\dk^k F_1|&&\\
+\max_{0\leq k\leq k_{small}+20}\sup_{\Mext}\Big(r^2u^{1+\dee}+r^3u^{\frac{1}{2}+\dee}\Big)|\dk^k F_2| &\les&  \ep_0.
\eeaa
Differentiating with $\dkb^k$, this yields
\beaa
e_3\left\{e_3\dkb^k\left(\frac{\a}{\kab^2}\right)+[\dkb^k, e_3]\left(\frac{\a}{\kab^2}\right) -\dkb^kF_1\right\} &=& \dkb^kF_2 - [\dkb^k,e_3]\left\{e_3\left(\frac{\a}{\kab^2}\right) -F_1\right\} 
\eeaa
and hence
\beaa
e_3\left\{e_3\dkb^k\left(\frac{\a}{\kab^2}\right)- F_{1,\dkb^k}\right\} &=& F_{2, \dkb^k}
\eeaa
where
\beaa
F_{1,\dkb^k}:=\dkb^kF_1-[\dkb^k, e_3]\left(\frac{\a}{\kab^2}\right),\quad  F_{2, \dkb^k}:=\dkb^kF_2 - [\dkb^k,e_3]\left\{e_3\left(\frac{\a}{\kab^2}\right) -F_1\right\}. 
\eeaa

In view of Lemma \ref{Le:comme3e4-outgeodesic}, we have schematically
\beaa
[\dkb, e_4] &=& \Gac_g\dk+\Gac_g+r\b,\\
{[\dkb, e_3]} &=& \Gac_b\dk+ \Gac_b+r\bb.
\eeaa
Together with the estimates of Lemma \ref{lemma:identitiesandnonsharpesitmatesforRiccicoefficientsoftherenormalizedframeofMext} for the Ricci coefficients and curvature components as well as the bootstrap assumptions \eqref{eq:localbootstrapasumptiontorecoveralphainTheoremM2} for $\a$ on $\Mext$, we infer
\beaa
\max_{0\leq j\leq k_{small}+20-k}\sup_{\Mext}\Big(r(2r+u)^{1+\dee}+r^2(2r+u)^{\frac{1}{2}+\dee}\Big)|\dk^j F_{1,\dkb^k}|&&\\
+\max_{0\leq j\leq k_{small}+20-k}\sup_{\Mext}\Big(r^2u^{1+\dee}+r^3u^{\frac{1}{2}+\dee}\Big)|\dk^j F_{2,\dkb^k}| &\les&  \ep_0.
\eeaa

Next, we consider the case $j\geq 1$. We have the commutator 
\beaa
[e_4, e_3] &=& 2\om e_3 - 2\omb e_4 -4\ze e_\th.
\eeaa
In view of the estimates of Lemma \ref{lemma:identitiesandnonsharpesitmatesforRiccicoefficientsoftherenormalizedframeofMext} for the Ricci coefficients, and in view of the bootstrap assumptions \eqref{eq:localbootstrapasumptiontorecoveralphainTheoremM2} for $\a$, we infer after commutation by $e_4^j$ for $0\leq k+j\leq k_{small}+20$
\beaa
e_3\left\{e_3\dkb^ke_4^j\left(\frac{\a}{\kab^2}\right)- F_{1,\dkb^k, e_4^j}\right\} &=& F_{2, \dkb^k, e_4^j}
\eeaa
where 
\beaa
&&\max_{0\leq l\leq k_{small}+20-k-j}\sup_{\Mext(u\leq u_1)}\Big(r^{1+j}(2r+u)^{1+\dee}+r^{2+j}(2r+u)^{\frac{1}{2}+\dee}\Big)|\dk^l F_{1,\dkb^k, e_4^j}|\\
&&+\max_{0\leq l\leq k_{small}+20-k-j}\sup_{\Mext(u\leq u_1)}\Big(r^{2+j}u^{1+\dee}+r^{3+j}u^{\frac{1}{2}+\dee}\Big)|\dk^l F_{2,\dkb^k, e_4^j}|\\
 &\les&  \ep_0+\max_{0\leq j+k\leq k_{small}+20}\sup_{\Mext(u\leq u_1)}\Big(\frac{r^2(2r+u)^{1+\dee}}{\log(1+u)}+r^3(2r+u)^{\frac{1}{2}+\dee}\Big)r\\
 &&\times\Big(|\dkb^k(re_4)^{j-1}e_3\a|+|\dkb^k(re_4)^{j-2}e_3^2\a|\Big) \les \ep_0.
\eeaa
This concludes the proof of the lemma.
\end{proof}

We are now ready to prove Proposition \ref{prop:improvementofbootstrapasumtionsforainTheoremM2}. 

{\bf Step 1.} For $0\leq k\leq k_{small}+20$, recall from the above lemma with $j=0$ that we have
\beaa
e_3\left\{e_3\dkb^k\left(\frac{\a}{\kab^2}\right)- F_{1,\dkb^k}\right\} &=& F_{2, \dkb^k}
\eeaa
where 
\beaa
\max_{0\leq j\leq k_{small}+20-k}\sup_{\Mext(u\leq u_1)}\Big(r^2u^{1+\dee}+r^3u^{\frac{1}{2}+\dee}\Big)|\dk^j F_{2,\dkb^k}| &\les&  \ep_0.
\eeaa
Also, we have in view of Lemma \ref{lemma:initalesitmateonCC0foralphaofrenormalizedframeonMext}
\beaa
\max_{0\leq k\leq k_{large}-4}\sup_{\CC_1}r^{\frac{5}{2}+\dee}\left|e_3\dkb^k\left(\frac{\a}{\kab^2}\right)- F_{1,\dkb^k}\right| &\les&\ep_0.
\eeaa
In view of Corollary \ref{cor:usefullcontroloftranportequationsine3dierctionTheoremM2}, we immediately infer for any $0\leq k\leq k_{small}+20$
\beaa
\max_{0\leq k\leq k_{small}+20}\sup_{\Mext(u\leq u_1)}\Big(r^2(2r+u)^{\frac{1}{2}+\dee}+r(2r+u)^{1+\dee}\Big)\left|e_3\dkb^k\left(\frac{\a}{\kab^2}\right)- F_{1,\dkb^k}\right| &\les& \ep_0.
\eeaa
Since we have from the above lemma that 
\beaa
\max_{0\leq j\leq k_{small}+20-k}\sup_{\Mext(u\leq u_1)}\Big(r(2r+u)^{1+\dee}+r^2(2r+u)^{\frac{1}{2}+\dee}\Big)|\dk^j F_{1,\dkb^k}| &\les&  \ep_0,
\eeaa
we deduce that we have  for any $0\leq k\leq k_{small}+20$
\begin{equation}\lab{eq:transportequationine3fore3alphaaoverkabsquare}
\max_{0\leq k\leq k_{small}+20}\sup_{\Mext(u\leq u_1)}\Big(r^2(2r+u)^{\frac{1}{2}+\dee}+r(2r+u)^{1+\dee}\Big)\left|e_3\dkb^k\left(\frac{\a}{\kab^2}\right)\right| \les \ep_0.
\end{equation}

{\bf Step 2.} Next, note that we have in view of Lemma \ref{lemma:initalesitmateonCC0foralphaofrenormalizedframeonMext}
\beaa
\max_{0\leq k\leq k_{large}-3}\sup_{\CC_1}r^{\frac{3}{2}+\dee}\left|\dkb^k\left(\frac{\a}{\kab^2}\right)\right| &\les&\ep_0.
\eeaa
Together with the transport equation \eqref{eq:transportequationine3fore3alphaaoverkabsquare}, and in view of Corollary  \ref{cor:usefullcontroloftranportequationsine3dierctionTheoremM2}, we infer
\beaa
\max_{0\leq k\leq k_{small}+20}\sup_{\Mext(u\leq u_1)}\left(r(2r+u)^{\frac{1}{2}+\dee}+\frac{(2r+u)^{1+\dee}}{\log(1+u)}\right)\left|\dkb^k\left(\frac{\a}{\kab^2}\right)\right| &\les& \ep_0.
\eeaa
In view of the control of $\kab$ provided by Lemma \ref{lemma:identitiesandnonsharpesitmatesforRiccicoefficientsoftherenormalizedframeofMext}, we easily deduce
\beaa
\max_{0\leq k\leq k_{small}+20}\sup_{\Mext(u\leq u_1)}\left(r^3(2r+u)^{\frac{1}{2}+\dee}+\frac{r^2(2r+u)^{1+\dee}}{\log(1+u)}\right)\left|\dkb^k\a\right| &\les& \ep_0.
\eeaa
Together with \eqref{eq:transportequationine3fore3alphaaoverkabsquare}, we infer
\beaa
\max_{0\leq k\leq k_{small}+20}\sup_{\Mext(u\leq u_1)}\left(r^3(2r+u)^{\frac{1}{2}+\dee}+\frac{r^2(2r+u)^{1+\dee}}{\log(1+u)}\right)\Big(|\dkb^k\a|+r|\dkb^ke_3\a|\Big) &\les& \ep_0.
\eeaa

{\bf Step 3.} Next, recall from section \ref{sec:discussionanddefintionofinvariantquantities} that $\qf$ is defined with respect to a general null frame as follows 
\beaa
\qf&=& r^4\left(e_3(e_3(\a))+(2\kab -6\omb)e_3(\a)+\left(-4e_3(\omb)+8\omb^2-8\omb\,\kab+\frac{1}{2}\kab^2\right)\a\right).
\eeaa
We infer
\beaa
e_3(e_3(\a)) &=& \frac{\qf}{r^4} -(2\kab -6\omb)e_3(\a)-\left(-4e_3(\omb)+8\omb^2-8\omb\,\kab+\frac{1}{2}\kab^2\right)\a.
\eeaa
Together with the above estimate for $\a$ and $e_3\a$, we infer by iteration
\beaa
\max_{0\leq k\leq k_{small}+20}\sup_{\Mext(u\leq u_1)}\Big(\frac{r^2(2r+u)^{1+\dee}}{\log(1+u)}+r^3(2r+u)^{\frac{1}{2}+\dee}\Big)&&\\
\times\Big(|(\dkb, e_3)^k\a|+r|(\dkb, e_3)^ke_3\a|\Big) &\les& \ep_0.
\eeaa

{\bf Step 4.} Arguing as for Step 1, but with $j\geq 1$, we infer the following analog of \eqref{eq:transportequationine3fore3alphaaoverkabsquare}
\beaa
&&\max_{0\leq j+k\leq k_{small}+20}\sup_{\Mext(u\leq u_1)}\Big(r^2(2r+u)^{\frac{1}{2}+\dee}+r(2r+u)^{1+\dee}\Big)\left|e_3\dkb^k(re_4)^j\left(\frac{\a}{\kab^2}\right)\right|\\
& \les& \ep_0+\max_{0\leq j+k\leq k_{small}+20}\sup_{\Mext(u\leq u_1)}\Big(\frac{r^2(2r+u)^{1+\dee}}{\log(1+u)}+r^3(2r+u)^{\frac{1}{2}+\dee}\Big)r\\
 &&\times\Big(|\dkb^k(re_4)^{j-1}e_3\a|+|\dkb^k(re_4)^{j-2}e_3^2\a|\Big).
\eeaa

{\bf Step 5.} Arguing as for Step 2, but with $j\geq 1$, we infer the following analog of the last estimate of Step 2
\beaa
&&\max_{0\leq j\leq k_{small}+20}\sup_{\Mext(u\leq u_1)}\left(r^3(2r+u)^{\frac{1}{2}+\dee}+\frac{r^2(2r+u)^{1+\dee}}{\log(1+u)}\right)\Big(|\dkb^k(re_4)^j\a|+r|\dkb^k(re_4)^je_3\a|\Big)\\ 
&\les& \ep_0+\max_{0\leq j+k\leq k_{small}+20}\sup_{\Mext(u\leq u_1)}\Big(\frac{r^2(2r+u)^{1+\dee}}{\log(1+u)}+r^3(2r+u)^{\frac{1}{2}+\dee}\Big)r\\
 &&\times\Big(|\dkb^k(re_4)^{j-1}e_3\a|+|\dkb^k(re_4)^{j-2}e_3^2\a|\Big).
\eeaa

{\bf Step 6.} Arguing as for Step 3, but with $j\geq 1$, we infer the following analog of the last estimate of Step 3
\beaa
&&\max_{0\leq j+k\leq k_{small}+20}\sup_{\Mext(u\leq u_1)}\Big(\frac{r^2(2r+u)^{1+\dee}}{\log(1+u)}+r^3(2r+u)^{\frac{1}{2}+\dee}\Big)\\
&&\times\Big(|(\dkb, e_3)^k(re_4)^j\a|+r|(\dkb, e_3)^k(re_4)^je_3\a|\Big) \\
&\les& \ep_0+\max_{0\leq j+k\leq k_{small}+20}\sup_{\Mext(u\leq u_1)}\Big(\frac{r^2(2r+u)^{1+\dee}}{\log(1+u)}+r^3(2r+u)^{\frac{1}{2}+\dee}\Big)r\\
&&\times\Big(|(\dkb, e_3)^k(re_4)^{j-1}e_3\a|+r|(\dkb, e_3)^k(re_4)^{j-2}e_3^2\a|\Big).
\eeaa

{\bf Step 7.} Arguing by iteration on $j$, noticing that the last estimate of Step 3 corresponds to desired estimate for $j=0$, and in view of the estimate derived in Step 6, we finally obtain 
\beaa
\max_{0\leq j+k\leq k_{small}+20}\sup_{\Mext(u\leq u_1)}\Big(\frac{r^2(2r+u)^{1+\dee}}{\log(1+u)}+r^3(2r+u)^{\frac{1}{2}+\dee}\Big)&&\\
\times\Big(|(\dkb, e_3)^k(re_4)^j\a|+r|(\dkb, e_3)^k(re_4)^je_3\a|\Big) &\les& \ep_0
\eeaa
and hence
$$
\max_{0\leq k\leq k_{small}+20}\sup_{\Mext(u\leq u_1)}\Big(\frac{r^2(2r+u)^{1+\dee}}{\log(1+u)}+r^3(2r+u)^{\frac{1}{2}+\dee}\Big)\Big(|\dk^k\a|+r|\dk^ke_3\a|\Big) \les \ep_0.
$$
This concludes the proof of Proposition \ref{prop:improvementofbootstrapasumtionsforainTheoremM2}.


\subsection{End of the proof of Theorem M2}


First, note in view of the estimates for $\a$ on $\CC_1$ provided by Lemma \ref{lemma:initalesitmateonCC0foralphaofrenormalizedframeonMext} that the bootstrap assumptions \eqref{eq:localbootstrapasumptiontorecoveralphainTheoremM2} for $\a$ hold by continuity for some sufficiently small $u_1>0$. Then, we may in view of Proposition \ref{prop:improvementofbootstrapasumtionsforainTheoremM2} choose $u_1=u_*$. We deduce therefore
\beaa
\max_{0\leq k\leq k_{small}+20}\sup_{\Mext}\Big(\frac{r^2(2r+u)^{1+\dee}}{\log(1+u)}+r^3(2r+u)^{\frac{1}{2}+\dee}\Big)\Big(|\dk^k\a|+r|\dk^ke_3\a|\Big) &\les& \ep_0.
\eeaa

Next, recall from \eqref{eq:thechangeofframeforalphainsecondgloabalfrmaebacktoMext} and \eqref{eq:estimateforfinconstructionsecondframeinMext:usedinproofThM2} that we have
\beaa
\a &=& (\,{}^{(ext)}\Up)^2\left(\,{}^{(ext)}\a+2f\,{}^{(ext)}\b+\frac{3}{2}f^2\,{}^{(ext)}\rho+\lot\right)
\eeaa
where $f$ satisfies
\beaa
\begin{split}
|\dk^kf| &\les \frac{\ep}{ru^{\frac{1}{2}}+u}, \,\,\,\textrm{ for }k\leq k_{small}+22\textrm{ on }\Mext,\\
 |\dk^{k-1}e_3f| &\les \frac{\ep}{ru}\,\,\,\textrm{ for }k\leq k_{small}+22\textrm{ on }\Mext.
\end{split}
\eeaa
Together with bootstrap assumptions for ${}^{(ext)}\b$ and ${}^{(ext)}\rho$, we easily infer
\beaa
\max_{0\leq k\leq k_{small}+20}\sup_{\Mext}\Big(\frac{r^2(2r+u)^{1+\dee}}{\log(1+u)}+r^3(2r+u)^{\frac{1}{2}+\dee}\Big)\Big(|\dk^k\,{}^{(ext)}\a|+r|\dk^ke_3\,{}^{(ext)}\a|\Big) &\les& \ep_0.
\eeaa
This concludes the proof of Theorem M2.


\section{Proof of Theorem M3}


Theorem M3 contains decay estimates for $\aa$ in $\Mint$ and on $\Si_*$. We first proceed with the estimate on $\Mint$ before moving to $\Mext$. 


\subsection{Estimate for $\aa$ in $\Mint$}


Recall that $\qf$, controlled in Theorem M1, is defined with respect to the global frame of  Proposition \ref{prop:existenceandestimatesfortheglobalframe:bis}. Recall also that we may choose the global null frame to coincide with the ingoing geodesic null frame of $\Mint$ in $\Mint$ (see property (b) in Proposition \ref{prop:existenceandestimatesfortheglobalframe:bis} together with property (d) ii. in Proposition \ref{prop:existenceandestimatesfortheglobalframe}). Thus, in this section, as we only work on $\Mint$, the null frame $(e_4, e_3, e_\th)$ denotes both the frame of $\Mint$ and the global frame with respect to which $\qf$ is defined. We start with the following definition.
\begin{definition}
In $\Mint$, we define with respect to the ingoing geodesic frame of $\Mint$
\bea\lab{eq:defintionofthevectorfieldwidetildeT}
\widetilde{T}:= e_4-\frac{1}{\overline{\kab}}\Big(\overline{\ka}+A\Big)e_3.
\eea
\end{definition}

The estimate for $\aa$ in $\Mint$ relies on the following proposition. 
\begin{proposition}\lab{prop:parabolicequationsatsfiedbyalphabarwithestimateRHShigherorderderivatives}
Let $0\leq k\leq k_{small}+17$. Then, $\aa$ satisfies in $\Mint$
\beaa
6m\widetilde{T}(\dk^k\aa)+r^4\dds_2\dds_1\ddd_1\ddd_2(\dk^k\aa) &=& F_k
\eeaa
where $F_k$ satisfies 
\beaa
\max_{0\leq k\leq k_{small}+17}\int_{\Mint}\ub^{2+2\dec}|\dk^{\leq 1}F_k|^2 &\les& \ep_0^2.
\eeaa
\end{proposition}

\begin{remark}\lab{remark:basicpropertiesofvectorfieldwidetildeT}
In view of the definition of $\widetilde{T}$, we have
\beaa
\widetilde{T}(r) = e_4(r) -\frac{1}{\overline{\kab}}\Big(\overline{\ka}+A\Big)e_3(r)= 0
\eeaa
so that $\widetilde{T}$ is tangent to the hypersurfaces of constant $r$. In particular, $(\widetilde{T}, e_\th)$ spans the tangent space of hypersurfaces of constant $r$. Therefore, in view of Proposition \ref{prop:parabolicequationsatsfiedbyalphabarwithestimateRHShigherorderderivatives}, $\aa$ and its derivatives satisfy on each hyper surface of contant $r$ in $\Mint$, i.e. on $\{r=r_0\}$ for $2m_0(1-\delta_{\HH})\leq r\leq \rh$, a forward parabolic equation. Furthermore, since we have $\widetilde{T}(\ub)=2/\vsib=2+O(\ep)$, $\ub$ plays the role of time in this forward parabolic equation.
\end{remark}

We also derive estimates for the control of the parabolic equation appearing in the statement of Proposition  \ref{prop:parabolicequationsatsfiedbyalphabarwithestimateRHShigherorderderivatives}. 
\begin{lemma}\lab{lemma:basicparabolicestimateforcontrolofaainMint}
Let $f$ and $h$ reduced 2-scalars such that 
\beaa
\Big(6m\widetilde{T}+r^4\dds_2\dds_1\ddd_1\ddd_2\Big)f &=& h.
\eeaa
 Then, for any real number $n\geq 0$ and any $r_0$ such that $2m_0(1-\de_\HH)\leq r_0\leq \rh $, we have
 \beaa
\sup_{1\leq \ub\leq u_*} \int_{S(r=r_0, \ub)}(1+\ub^n)f^2 &\les_n & \int_{S(r=r_0, 1)}f^2 +\ep^2\int_1^{u_*}\int_{S(r=r_0, \ub)}(1+\ub^{n-2})(\dk f)^2\\
&&+\int_{\Mint}(1+\ub^n)(\dk^{\leq 1}h)^2.
 \eeaa
\end{lemma}

We are now in position to control $\aa$ in $\Mint$. Recall from Proposition \ref{prop:parabolicequationsatsfiedbyalphabarwithestimateRHShigherorderderivatives} that $\aa$ satisfies in $\Mint$ for $0\leq k\leq k_{small}+17$ 
\beaa
6m\widetilde{T}(\dk^k\aa)+r^4\dds_2\dds_1\ddd_1\ddd_2(\dk^k\aa) &=& F_k.
\eeaa
Applying Lemma \ref{lemma:basicparabolicestimateforcontrolofaainMint} with $n=2+2\dec$, $f=\dk^k\aa$ and $h=F_k$, we infer for any $r_0$ such that $2m_0(1-\de_\HH)\leq r_0\leq \rh $
 \beaa
\sup_{1\leq \ub\leq u_*} \int_{S(r=r_0, \ub)}(1+\ub^{2+2\dec})(\dk^k\aa)^2 &\les & \int_{S(r=r_0, 1)}(\dk^k\aa)^2 +\ep^2\int_1^{u_*}\int_{S(r=r_0, \ub)}(1+\ub^{2\dec})(\dk^{k+1}\aa)^2\\
&&+\int_{\Mint}(1+\ub^{2+2\dec})(\dk^{\leq 1}F_k)^2.
\eeaa
Together with the bounds for $\aa$ on $\underline{\CC}_1$ provided by Theorem M0, the bootstrap assumptions on decay and energy for $\aa$ in $\Mint$, and the bound for $F_k$ provided by Proposition \ref{prop:parabolicequationsatsfiedbyalphabarwithestimateRHShigherorderderivatives}, we infer for $0\leq k\leq k_{small}+17$ in $\Mint$
 \beaa
\sup_{1\leq \ub\leq u_*} \int_{S(r=r_0, \ub)}(1+\ub^{2+2\dec})(\dk^k\aa)^2 &\les & \ep_0^2.
\eeaa
In particular, we have obtained
\beaa
\max_{0\leq k\leq k_{small}+17}\sup_{\Mint}\ub^{1+\dec}\|\dk^k\aa\|_{L^2(S)} &\les& \ep_0.
\eeaa
Using the Sobolev embedding on 2-surface and the fact that $r$ is bounded on $\Mint$, we infer
\beaa
\max_{0\leq k\leq k_{small}+15}\sup_{\Mint}\ub^{1+\dec}|\dk^k\aa| &\les& \ep_0
\eeaa
and hence
\bea\lab{eq:conclusionofThM3forMint}
{}^{(int)}\mathfrak{D}_{k_{small}+15}[\aa] &\les& \ep_0
\eea
which is the desired estimate for $\aa$ in $\Mint$. 

The proof of Proposition  \ref{prop:parabolicequationsatsfiedbyalphabarwithestimateRHShigherorderderivatives} will be given in section \ref{sec:proofofprop:parabolicequationsatsfiedbyalphabarwithestimateRHShigherorderderivatives}, and to the proof of Lemma \ref{lemma:basicparabolicestimateforcontrolofaainMint} which will be given in section \ref{sec:proofoflemma:basicparabolicestimateforcontrolofaainMint}. But first, we conclude in the next section the proof of Theorem M3 by controlling $\aa$ on $\Si_*$.


\subsection{Estimate for $\aa$ on $\Si_*$}


Recall that $\qf$, controlled in Theorem M1, is defined with respect to the global frame of  Proposition \ref{prop:existenceandestimatesfortheglobalframe:bis}. We will first control $\aa$ in this frame, before coming back to $\Mext$ at the end of the argument. We start with the following definition.
\begin{definition}
In $\Si_*$, we define, with respect to the  the global frame of  Proposition \ref{prop:existenceandestimatesfortheglobalframe:bis}, 
\bea\lab{eq:defintionofthevectorfieldwidetildenu}
\widetilde{\nu}:= e_3+ae_4,
\eea
where the scalar function $a$ is uniquely defined so that $\widetilde{\nu}$ is tangent to $\Si_*$. 
\end{definition}

The estimate for $\aa$ on $\Si_*$ relies on the following proposition. 
\begin{proposition}\lab{prop:parabolicequationsatsfiedbyalphabarwithestimateRHShigherorderderivatives:bis}
Let $0\leq k\leq k_{small}+18$. Then, $\aa$ satisfies on $\Si_*$
\beaa
6m\widetilde{\nu}(\dk^k\aa)+r^4\dds_2\dds_1\ddd_1\ddd_2(\dk^k\aa) &=& F_k
\eeaa
where $F_k$ satisfies 
\beaa
\max_{0\leq k\leq k_{small}+18}\int_{\Si_*}u^{2+2\dec}|F_k|^2 &\les& \ep_0^2.
\eeaa
\end{proposition}

\begin{remark}\lab{remark:basicpropertiesofvectorfieldwidetildenu}
Since $\widetilde{\nu}$ is tangent to $\Si_*$, and since $(\widetilde{\nu}, e_\th)$ spans the tangent space of $\Si_*$, in view of Proposition \ref{prop:parabolicequationsatsfiedbyalphabarwithestimateRHShigherorderderivatives:bis}, $\aa$ and its derivatives satisfy on $\Si_*$ a forward parabolic equation. Furthermore, since we have $\widetilde{\nu}(u)=2+O(\ep)$, $u$ plays the role of time in this forward parabolic equation.
\end{remark}

We also derive estimates for the control of the parabolic equation appearing in the statement of Proposition  \ref{prop:parabolicequationsatsfiedbyalphabarwithestimateRHShigherorderderivatives}. 
\begin{lemma}\lab{lemma:basicparabolicestimateforcontrolofaainSi*}
Let $f$ and $h$ reduced 2-scalars such that 
\beaa
\Big(6m\widetilde{\nu}+r^4\dds_2\dds_1\ddd_1\ddd_2\Big)f &=& h.
\eeaa
 Then, for any real number $n\geq 0$, we have
 \beaa
\int_{\Si_*}(1+u^n)f^2 &\les_n & \int_{\Si_*\cap\CC_1}f^2 +\ep^2\int_{\Si_*}(1+u^{n-2})(\dk f)^2+\int_{\Si_*}(1+u^n)h^2.
 \eeaa
\end{lemma}

We are now in position to control $\aa$ on $\Si_*$. Recall from Proposition \ref{prop:parabolicequationsatsfiedbyalphabarwithestimateRHShigherorderderivatives:bis} that $\aa$ satisfies in $\Si_*$ for $0\leq k\leq k_{small}+18$ 
\beaa
6m\widetilde{\nu}(\dk^k\aa)+r^4\dds_2\dds_1\ddd_1\ddd_2(\dk^k\aa) &=& F_k.
\eeaa
Applying Lemma \ref{lemma:basicparabolicestimateforcontrolofaainSi*} with $n=2+2\dec$, $f=\dk^k\aa$ and $h=F_k$, we infer 
 \beaa
\int_{\Si_*}(1+u^{2+2\dec})(\dk^k\aa)^2 &\les & \int_{\Si_*\cap\CC_1}(\dk^k\aa)^2 +\ep^2\int_{\Si_*}(1+u^{2\dec})(\dk^{k+1}\aa)^2 \\
&&+\int_{\Si_*}(1+u^{2+2\dec})(F_k)^2.
\eeaa
Together with the bounds for $\aa$ on $\underline{\CC}_1$ provided by Theorem M0,  the bootstrap assumptions on decay and energy for $\aa$ in $\Mext$,  and the bound for $F_k$ provided by Proposition \ref{prop:parabolicequationsatsfiedbyalphabarwithestimateRHShigherorderderivatives:bis}, we infer
 \beaa
\int_{\Si_*}(1+u^{2+2\dec})(\dk^k\aa)^2 &\les & \ep_0^2.
\eeaa
In particular, we have obtained
\beaa
\max_{0\leq k\leq k_{small}+18}\int_{\Si_*}(1+u^{2+2\dec})(\dk^k\aa)^2 &\les& \ep_0.
\eeaa

Now, recall that $\aa$ in the above estimate is defined with respect to the global frame constructed in Proposition \ref{prop:existenceandestimatesfortheglobalframe:bis}. In view of Proposition \ref{prop:existenceandestimatesfortheglobalframe:bis} and Proposition \ref{prop:constructionsecondframeinMext}, and the change of frame formula for $\aa$ in Proposition \ref{prop:transformations1}, we have
\beaa
\aa &=& (\,{}^{(ext)}\Up)^{-2}\,{}^{(ext)}\aa.
\eeaa
Hence, we immediately infer
\beaa
\max_{0\leq k\leq k_{small}+18}\int_{\Si_*}(1+u^{2+2\dec})(\dk^k{}^{(ext)}\aa)^2 &\les& \ep_0.
\eeaa
which is the desired estimate in $\Si_*$. Together with \eqref{eq:conclusionofThM3forMint}, this concludes the proof of Theorem M3.

The proof of Proposition  \ref{prop:parabolicequationsatsfiedbyalphabarwithestimateRHShigherorderderivatives:bis} will be given in section \ref{sec:proofofprop:parabolicequationsatsfiedbyalphabarwithestimateRHShigherorderderivatives:bis}, and to the proof of Lemma \ref{lemma:basicparabolicestimateforcontrolofaainSi*} which will be given in section \ref{sec:proofoflemma:basicparabolicestimateforcontrolofaainSi*}.


\subsection{Proof of Proposition \ref{prop:parabolicequationsatsfiedbyalphabarwithestimateRHShigherorderderivatives}}\lab{sec:proofofprop:parabolicequationsatsfiedbyalphabarwithestimateRHShigherorderderivatives} 


In this section, we infer from the Teukolsky-Starobinski identity, see Proposition \ref{Prop:Teuk-Star-main}, a parabolic equation for $\aa$. 
\begin{corollary}\lab{corollary:parabolicequationsatsfiedbyalphabar}
$\aa$ satisfies in $\Mint$ the following equation
\beaa
&& 6m\widetilde{T}\aa+r^4\dds_2\dds_1\ddd_1\ddd_2\aa \\
  &=& \frac{1}{r^3}\Big(e_3(r^2e_3(r\qf))+2\omb r^2e_3(r\qf)\Big) -r^{-3}\err[TS] -\left\{\frac{3}{2}r^4\left(\rho+\frac{2m}{r^3}\right)\kab  - 3mr\left(\kab+\frac{2}{r}\right) \right\}e_4\aa\\
&&  -\left\{-  \frac{3}{2}r^4\left(\rho+\frac{2m}{r^3}\right)\ka   +\frac{3mr}{\overline{\kab}}\left(\overline{\kab}+\frac{2}{r}\right)\overline{\ka} +3mr\check{\ka}  +\frac{6m}{\overline{\kab}}A\right\}e_3\aa
\eeaa
where the vectorfield $\widetilde{T}$ is defined by \eqref{eq:defintionofthevectorfieldwidetildeT}. 
\end{corollary}

\begin{proof}
Recall from \eqref{eq:Teuk-Star-main(seeApp)} that we have
\beaa
e_3(r^2e_3(r\qf))+2\omb r^2e_3(r\qf) &=& r^7\left\{  \dds_2\dds_1\ddd_1\ddd_2\aa  +\frac{3}{2}\rho\Big(\kab e_4 -\ka e_3\Big)\aa\right\} + \err[TS].
\eeaa
This yields
\beaa
\frac{3}{2}r^4\rho\Big(\kab e_4 -\ka e_3\Big)\aa+r^4\dds_2\dds_1\ddd_1\ddd_2\aa   &=& \frac{1}{r^3}\Big(e_3(r^2e_3(r\qf))+2\omb r^2e_3(r\qf)\Big) -r^{-3}\err[TS].
\eeaa
Now, we have in view of the definition of $\widetilde{T}$
\beaa
&&\frac{3}{2}r^4\rho\Big(\kab e_4 -\ka e_3\Big) -6m\widetilde{T}\\
 &=&  \left\{\frac{3}{2}r^4\left(\rho+\frac{2m}{r^3}\right)\kab  - 3mr\left(\kab+\frac{2}{r}\right) \right\}e_4\\
&&  +\left\{-  \frac{3}{2}r^4\left(\rho+\frac{2m}{r^3}\right)\ka   +\frac{3mr}{\overline{\kab}}\left(\overline{\kab}+\frac{2}{r}\right)\overline{\ka} +3mr\check{\ka}  +\frac{6m}{\overline{\kab}}A\right\}e_3.
\eeaa
We infer
\beaa
&& 6m\widetilde{T}\aa+r^4\dds_2\dds_1\ddd_1\ddd_2\aa \\
  &=& \frac{1}{r^3}\Big(e_3(r^2e_3(r\qf))+2\omb r^2e_3(r\qf)\Big) -r^{-3}\err[TS] -\left\{\frac{3}{2}r^4\left(\rho+\frac{2m}{r^3}\right)\kab  - 3mr\left(\kab+\frac{2}{r}\right) \right\}e_4\aa\\
&&  -\left\{-  \frac{3}{2}r^4\left(\rho+\frac{2m}{r^3}\right)\ka   +\frac{3mr}{\overline{\kab}}\left(\overline{\kab}+\frac{2}{r}\right)\overline{\ka} +3mr\check{\ka}  +\frac{6m}{\overline{\kab}}A\right\}e_3\aa.
\eeaa
This concludes the proof of the corollary.
\end{proof}

\begin{corollary}\lab{corollary:parabolicequationsatsfiedbyalphabarwithestimateRHS}
$\aa$ satisfies in $\Mint$
\beaa
6m\widetilde{T}\aa+r^4\dds_2\dds_1\ddd_1\ddd_2\aa &=& F
\eeaa
where $F$ satisfies 
\beaa
\max_{0\leq k\leq k_{small}+18}\int_{\Mint}\ub^{2+2\dec}|\dk^kF|^2 &\les& \ep_0^2.
\eeaa
\end{corollary}

\begin{proof}
In view of Corollary \ref{corollary:parabolicequationsatsfiedbyalphabar}, $\aa$ satisfies 
\beaa
6m\widetilde{T}\aa+r^4\dds_2\dds_1\ddd_1\ddd_2\aa &=& F
\eeaa
with 
\beaa
F &:=& \frac{1}{r^3}\Big(e_3(r^2e_3(r\qf))+2\omb r^2e_3(r\qf)\Big)+F_1,\\
F_1 &:=&   -r^{-3}\err[TS] -\left\{\frac{3}{2}r^4\left(\rho+\frac{2m}{r^3}\right)\kab  - 3mr\left(\kab+\frac{2}{r}\right) \right\}e_4\aa\\
&&  -\left\{-  \frac{3}{2}r^4\left(\rho+\frac{2m}{r^3}\right)\ka   +\frac{3mr}{\overline{\kab}}\left(\overline{\kab}+\frac{2}{r}\right)\overline{\ka} +3mr\check{\ka}  +\frac{6m}{\overline{\kab}}A\right\}e_3\aa.
\eeaa
Using the bootstrap assumptions in $\Mint$ for decay and energies, and in view of the fact that $F_1$ contains only quadratic or higher order terms, we easily derive
\beaa
\max_{0\leq k\leq k_{small}+18}\sup_{\Mint}\ub^{\frac{3}{2}+\frac{3}{2}\dec}|\dk^kF_1| &\les& \ep^2\les\ep_0.
\eeaa
In view of the definition of $F$, this yields
\beaa
\max_{0\leq k\leq k_{small}+18}\int_{\Mint}\ub^{2+2\dec}|\dk^kF|^2 &\les& \ep_0^2+\max_{0\leq k\leq k_{small}+20}\int_{\Mint}\ub^{2+2\dec}|\dk^k\qf|^2.
\eeaa
Together with Theorem M1, and the fact that $\dee>\dec$, we infer
\beaa
\max_{0\leq k\leq k_{small}+18}\int_{\Mint}\ub^{2+2\dec}|\dk^kF|^2 &\les& \ep_0^2.
\eeaa
This concludes the proof of the corollary.
\end{proof}

We are now ready to prove Proposition \ref{prop:parabolicequationsatsfiedbyalphabarwithestimateRHShigherorderderivatives}. In view of Corollary \ref{corollary:parabolicequationsatsfiedbyalphabarwithestimateRHS}, $\aa$ satisfies
\beaa
6m\widetilde{T}\aa+r^4\dds_2\dds_1\ddd_1\ddd_2\aa &=& F.
\eeaa
Commuting with $\dk^k$, we infer
\beaa
6m\widetilde{T}(\dk^k\aa)+r^4\dds_2\dds_1\ddd_1\ddd_2(\dk^k\aa) &=& F_k
\eeaa
where $F_k$ is defined by
\beaa
F_k &:=& -6m[\dk^k, \widetilde{T}]\aa-6\sum_{j=1}^k\dk^j(m)\dk^{k-j}\widetilde{T}\aa -[\dk^k, r\dds_2] r\dds_1 r\ddd_1 r\ddd_2\aa -r\dds_2[\dk^k, r\dds_1] r\ddd_1 r\ddd_2\aa\\
&& -r\dds_2 r\dds_1 [\dk^k, r\ddd_1] r\ddd_2\aa -r\dds_2 r\dds_1 r\ddd_1 [\dk^k, r\ddd_2]\aa + \dk^kF.
\eeaa

Note that we have schematically
\beaa
[\dk, \dkb] = \Gac\dk,\quad [\widetilde{T}, \dkb]=\Big(\dk\Gac+\Gac\Big)\dk,
\eeaa
as well as 
\beaa
 [\widetilde{T}, re_4] &=& e_4(r)e_4 -\frac{1}{\overline{\kab}}\Big(\overline{\ka}+ A \Big)[e_3, re_4]+re_4\left(\frac{1}{\overline{\kab}}\Big(\overline{\ka}+ A \Big)\right)e_3\\
 &=& \frac{r}{2}\Big(\overline{\ka}+ A \Big)e_4  -\frac{1}{\overline{\kab}}\Big(\overline{\ka}+ A \Big)\left(\frac{r}{2}\overline{\kab}e_4+r\Big(-2\om e_3+4\ze e_\th\Big)\right)\\
 && +re_4\left(\frac{\overline{\ka}}{\overline{\kab}}\right)e_3 +re_4\left(\frac{1}{\overline{\kab}}A\right)e_3\\
 &=&    \left\{\frac{2r}{\overline{\kab}}\Big(\overline{\ka}+ A \Big)\om +re_4\left(\frac{\overline{\ka}}{\overline{\kab}}\right) +re_4\left(\frac{1}{\overline{\kab}} A \right)\right\}e_3\\
 &&  -\frac{4r}{\overline{\kab}}\Big(\overline{\ka}+ A \Big)\ze e_\th\\
 &=&    \Bigg\{-\frac{2m}{r}\left(\frac{\overline{\ka}}{\overline{\kab}}+\Up\right) + \frac{2r}{\overline{\kab}}\overline{\ka}\left(\om +\frac{m}{r}\right) -\frac{2m(e_4(r)-\Up)}{r}+2e_4(m) +re_4\left(\frac{\overline{\ka}}{\overline{\kab}}+\Up\right) \\
 && +   \frac{2r}{\overline{\kab}} A \om+re_4\left(\frac{1}{\overline{\kab}} A \right)\Bigg\}e_3  -\frac{4r}{\overline{\kab}}\Big(\overline{\ka}+ A \Big)\ze e_\th\\
 &=& \Big(\dk\Gac+\Gac\Big)\dk,
 \eeaa
and
\beaa
 [\widetilde{T}, e_3] &=& [e_4, e_3] +e_3\left(\frac{1}{\overline{\kab}}\Big(\overline{\ka}+ A \Big)\right)e_3\\
 &=& \left\{2\om +e_3\left(\frac{\overline{\ka}}{\overline{\kab}}\right) + e_3\left(\frac{1}{\overline{\kab}} A \right) \right\}e_3 -4\ze e_\th\\
 &=& \left\{2\left(\om+\frac{m}{r^2}\right)  -\frac{2m(e_3(r)+1)}{r^2}   +\frac{2e_3(m)}{r}  +e_3\left(\frac{\overline{\ka}}{\overline{\kab}}+\Up\right) + e_3\left(\frac{1}{\overline{\kab}} A \right) \right\}e_3 -4\ze e_\th \\
 &=&  \Big(\dk\Gac+\Gac\Big)\dk.
 \eeaa
Together with the bootstrap assumptions in $\Mint$ for decay and energies, and in view of the fact that $F_1$ contains only quadratic or higher order terms, we easily derive
\beaa
\max_{0\leq k\leq k_{small}+18}\sup_{\Mint}\ub^{\frac{3}{2}+\frac{3}{2}\dec}&&\Bigg|-6m[\dk^k, \widetilde{T}]\aa-6\sum_{j=1}^k\dk^j(m)\dk^{k-j}\widetilde{T}\aa -[\dk^k, r\dds_2] r\dds_1 r\ddd_1 r\ddd_2\aa\\
&& -r\dds_2[\dk^k, r\dds_1] r\ddd_1 r\ddd_2\aa -r\dds_2 r\dds_1 [\dk^k, r\ddd_1] r\ddd_2\aa\\
&& -r\dds_2 r\dds_1 r\ddd_1 [\dk^k, r\ddd_2]\aa\Bigg| \les \ep^2.
\eeaa
In view of the definition of $F_k$, this yields
\beaa
\max_{0\leq k\leq k_{small}+18}\int_{\Mint}\ub^{2+2\dec}|F_k|^2 &\les& \ep^4+\max_{0\leq k\leq k_{small}+18}\int_{\Mint}\ub^{2+2\dec}|\dk^kF|^2.
\eeaa
Together with the estimate for $F$ of Corollary \ref{corollary:parabolicequationsatsfiedbyalphabarwithestimateRHS}, we infer
\beaa
\max_{0\leq k\leq k_{small}+18}\int_{\Mint}\ub^{2+2\dec}|F_k|^2 &\les& \ep^4+\ep_0^2\les \ep_0^2.
\eeaa
This concludes the proof of Proposition \ref{prop:parabolicequationsatsfiedbyalphabarwithestimateRHShigherorderderivatives}.


\subsection{Proof of Lemma \ref{lemma:basicparabolicestimateforcontrolofaainMint}}\lab{sec:proofoflemma:basicparabolicestimateforcontrolofaainMint}


In this section we  prove Lemma \ref{lemma:basicparabolicestimateforcontrolofaainMint}, i.e. we derive estimates for the control of the parabolic equation appearing in the statement of Proposition \ref{prop:parabolicequationsatsfiedbyalphabarwithestimateRHShigherorderderivatives}. To this end, we first start with a Poincar\'e inequality.
\begin{lemma}\lab{lemma:poincareinequalityusedforcontrolofparabolicequation}
We have
\beaa
\int_S f\dds_2\dds_1\ddd_1\ddd_2 f &\geq& 24\int_S(1+O(\ep))K^2f^2.
\eeaa
\end{lemma}

\begin{proof}
We have
\beaa
\dds_2\dds_1\ddd_1\ddd_2 &=& \dds_2(-\lapp_1+K)\ddd_2\\
&=& -\dds_2\lapp_1\ddd_2+K\dds_2\ddd_2+\dds_1(K)\ddd_2\\
&=& -\lapp_2\dds_2\ddd_2+\Big(\lapp_2\dds_2-\dds_2\lapp_1\Big)\ddd_2 +K\dds_2\ddd_2+\dds_1(K)\ddd_2\\
&=& (\dds_2\ddd_2-2K)\dds_2\ddd_2+\Big(3K\dds_2-\dds_1(K)\Big)\ddd_2 +K\dds_2\ddd_2+\dds_1(K)\ddd_2\\
&=& (\dds_2\ddd_2)^2+2K\dds_2\ddd_2.
\eeaa
Recall also the Poincar\'e inequality for $\ddd_2$ which holds for any reduced 2-scalar $f$
\beaa
\int_S|\ddd_2f|^2 \geq 4\int_SKf^2.
\eeaa
Then, we easily infer 
\beaa
\int_S f\dds_2\dds_1\ddd_1\ddd_2 f &=& \int_Sf(\dds_2\ddd_2)^2f+\int_S2Kf\dds_2\ddd_2f\\
&\geq& 4^2\int_S(1+O(\ep))K^2f^2+8\int_S(1+O(\ep))K^2f^2\\
&\geq& 24\int_S(1+O(\ep))K^2f^2
\eeaa
where we also used the estimates for the Gauss curvature 
\beaa
K=\frac{1}{r^2}+O\left(\frac{\ep}{r^2}\right),\quad re_\th(K)=O\left(\frac{\ep}{r^2}\right),
\eeaa
which follow from the bootstrap assumptions. 
\end{proof}

The following identity will be useful.
\begin{lemma}\lab{lemma:identityneededforbasicparabolicestimateforcontrolofaainMint}
We have for any reduced scalar $f$
\beaa
\widetilde{T}\left(\int_Sf^2\right) &=& 2\int_S f\widetilde{T}f  + \int_S\left(\frac{2}{\overline{\kab}} Afe_3(f)+\kac f^2\right) -\frac{\overline{\ka}}{\overline{\kab}}\left(\int_S\check{\kab} f^2\right) \\
&&-\frac{1}{\overline{\kab}}A\left(\int_S(2fe_3(f)+\kab f^2)\right)+\err\left[e_4\left(\int_S f^2\right)\right].
\eeaa
\end{lemma}

\begin{proof}
Recall from the definition of  $\widetilde{T}$ that 
\beaa
\widetilde{T} = e_4-\frac{1}{\overline{\kab}}\Big(\overline{\ka}+ A \Big)e_3.
\eeaa
We infer, in view of the analog of Proposition \ref{prop:outgoinggeod-e3e4averages} for an ingoing geodesic foliation, 
\beaa
&&\widetilde{T}\left(\int_Sf^2\right)\\
 &=& e_4\left(\int_Sf^2\right) - \frac{1}{\overline{\kab}}\Big(\overline{\ka}+ A \Big)e_3\left(\int_Sf^2\right)\\
&=& \int_S(2fe_4(f)+\ka f^2)+\err\left[e_4\left(\int_S f^2\right)\right] - \frac{1}{\overline{\kab}}\Big(\overline{\ka}+ A \Big)\left(\int_S(2fe_3(f)+\kab f^2)\right)\\
&=& \int_S\left(2f\widetilde{T}f+\frac{2}{\overline{\kab}}\Big(\overline{\ka}+ A \Big)fe_3(f)+\ka f^2\right)+\err\left[e_4\left(\int_S f^2\right)\right] \\
&& -\frac{1}{\overline{\kab}}\Big(\overline{\ka}+ A \Big)\left(\int_S(2fe_3(f)+\kab f^2)\right)\\
&=& 2\int_S f\widetilde{T}f  + \int_S\left(\frac{2}{\overline{\kab}} Afe_3(f)+\kac f^2\right) -\frac{\overline{\ka}}{\overline{\kab}}\left(\int_S\check{\kab} f^2\right) -\frac{1}{\overline{\kab}}A\left(\int_S(2fe_3(f)+\kab f^2)\right)\\
&&+\err\left[e_4\left(\int_S f^2\right)\right].
\eeaa
This concludes the proof of the lemma.
\end{proof}

We are now ready to prove Lemma \ref{lemma:basicparabolicestimateforcontrolofaainMint}. Recall from Lemma \ref{lemma:identityneededforbasicparabolicestimateforcontrolofaainMint} that we have
\beaa
\widetilde{T}\left(\int_Sf^2\right) &=& 2\int_S f\widetilde{T}f  + \int_S\left(\frac{2}{\overline{\kab}} Afe_3(f)+\kac f^2\right) -\frac{\overline{\ka}}{\overline{\kab}}\left(\int_S\check{\kab} f^2\right)\\
&& -\frac{1}{\overline{\kab}}A\left(\int_S(2fe_3(f)+\kab f^2)\right)+\err\left[e_4\left(\int_S f^2\right)\right].
\eeaa
In view of the equation satisfied by $f$, we infer
\beaa
\widetilde{T}\left(\int_Sf^2\right) &=&- \frac{1}{3m}\left(\int_Sr^4f\dds_2\dds_1\ddd_1\ddd_2f\right)+ \frac{1}{3m}\left(\int_Shf\right) \\
&&+ \int_S\left(\frac{2}{\overline{\kab}} Afe_3(f)+\kac f^2\right) -\frac{\overline{\ka}}{\overline{\kab}}\left(\int_S\check{\kab} f^2\right) -\frac{1}{\overline{\kab}}A\left(\int_S(2fe_3(f)+\kab f^2)\right)\\
&& +\err\left[e_4\left(\int_S f^2\right)\right].
\eeaa
Now, from the definition of $\widetilde{T}$, we have $\widetilde{T}(\ub)=2/\vsib$. We deduce
\beaa
&& \widetilde{T}\left(\ub^n\int_Sf^2\right) +\frac{\ub^n}{3m}\left(\int_Sr^4f\dds_2\dds_1\ddd_1\ddd_2f\right)\\
 &=&\frac{\ub^n}{3m}\left(\int_Shf\right) + \ub^n\int_S\left(\frac{2}{\overline{\kab}} Afe_3(f)+\kac f^2\right) -\ub^n\frac{\overline{\ka}}{\overline{\kab}}\left(\int_S\check{\kab} f^2\right) \\
 &&-\frac{\ub^n}{\overline{\kab}}A\left(\int_S(2fe_3(f)+\kab f^2)\right) +\ub^n\err\left[e_4\left(\int_S f^2\right)\right] +\frac{2}{\vsib}n\ub^{n-1}\int_Sf^2.
\eeaa
This yields in view of the bootstrap assumptions 
\beaa
 &&\widetilde{T}\left(\ub^n\int_Sf^2\right) +\frac{\ub^n}{3m}\left(\int_Sr^4f\dds_2\dds_1\ddd_1\ddd_2f\right)\\
  &\les& \frac{\ub^n}{3m}\|h\|_{L^2(S)}\|f\|_{L^2(S)}+ \ep\ub^{n-1}\int_S|f||\dk^{\leq 1}f| +n\ub^{n-1}\int_Sf^2.
\eeaa
Next, we rely on the Poincar\'e inequality of Lemma \ref{lemma:poincareinequalityusedforcontrolofparabolicequation} to deduce
\beaa
\widetilde{T}\left(\ub^n\int_Sf^2\right) +\ub^n\int_Sf^2 &\les& \ub^n\int_Sh^2 +\ep^2\ub^{n-2}\int_S(\dk f)^2 +n\ub^{n-1}\int_Sf^2.
\eeaa
Integrating in $\ub$ between $1$ and $u_*$, and recalling that $\widetilde{T}(\ub)=2/\vsib$, we infer for any $r_0$ such that $2m_0(1-\de_\HH)\leq r_0\leq \rh $
 \beaa
 &&\int_{S(r=r_0, \ub)}\ub^nf^2 +\int_1^{\ub}\left(\int_{S(r=r_0, \ub')}{\ub'}^nf^2\right)d\ub'\\
  &\les & \int_{S(r=r_0, 1)}f^2 +\int_1^{\ub}\left(\int_{S(r=r_0, \ub')}{\ub'}^nh^2\right)d\ub' +\ep^2\int_1^{\ub}\left(\int_{S(r=r_0, \ub')}{\ub'}^{n-2}(\dk f)^2\right)d\ub'\\
  &&+n\int_0^{\ub}\left(\int_{S(r=r_0, \ub')}{\ub'}^{n-1}f^2\right)d\ub'.
 \eeaa
 In particular, we have for $n=0$
 \beaa
\sup_{1\leq \ub\leq u_*} \int_{S(r=r_0, \ub)}f^2 +\int_1^{u_*}\left(\int_{S(r=r_0, \ub)}f^2\right)d\ub &\les & \int_{S(r=r_0, 1)}f^2 +\int_1^{u_*}\left(\int_{S(r=r_0, \ub)}h^2\right)d\ub\\
&& +\ep^2\int_1^{u_*}\left(\int_{S(r=r_0, \ub)}{\ub}^{-2}(\dk f)^2\right)d\ub.
 \eeaa
Then, starting from the case $n=0$ and arguing by iteration on the largest integer below $n$, one immediately deduces for any real $n\geq 0$ 
 \beaa
&&\sup_{1\leq \ub\leq u_*} \int_{S(r=r_0, \ub)}(1+\ub^n)f^2 +\int_1^{u_*}\left(\int_{S(r=r_0, \ub)}(1+\ub^n)f^2\right)d\ub\\
 &\les & \int_{S(r=r_0, 1)}f^2 +\int_1^{u_*}\left(\int_{S(r=r_0, \ub)}(1+\ub^n)h^2\right)d\ub +\ep^2\int_1^{u_*}\left(\int_{S(r=r_0, \ub)}(1+{\ub}^{n-2})(\dk f)^2\right)d\ub.
 \eeaa
 Now, a simple trace estimate yields
 \beaa
 \int_{S(r=r_0, \ub)}(1+\ub^n)h^2 &\les&  \int_{\CCb_{\ub}}(1+\ub^n)\Big(|h|^2+|e_3(h)|^2\Big)
 \eeaa
 so that
 \beaa
 \int_1^{u_*}\left(\int_{S(r=r_0, \ub)}(1+\ub^n)h^2\right)d\ub &\les& \int_1^{u_*}\int_{\CCb_{\ub}}(1+\ub^n)\Big(|h|^2+|e_3(h)|^2\Big)d\ub\\
 &\les& \int_{\Mint}(1+\ub^n)(\dk^{\leq 1}h)^2.
 \eeaa
 We deduce
 \beaa
&&\sup_{1\leq \ub\leq u_*} \int_{S(r=r_0, \ub)}(1+\ub^n)f^2 +\int_1^{u_*}\left(\int_{S(r=r_0, \ub)}(1+\ub^n)f^2\right)d\ub\\
 &\les & \int_{S(r=r_0, 1)}f^2 + \int_{\Mint}(1+\ub^n)(\dk^{\leq 1}h)^2+\ep^2\int_1^{u_*}\left(\int_{S(r=r_0, \ub)}(1+{\ub}^{n-2})(\dk f)^2\right)d\ub
 \eeaa
which concludes the proof of Lemma \ref{lemma:basicparabolicestimateforcontrolofaainMint}.


\subsection{Proof of Proposition \ref{prop:parabolicequationsatsfiedbyalphabarwithestimateRHShigherorderderivatives:bis}}\lab{sec:proofofprop:parabolicequationsatsfiedbyalphabarwithestimateRHShigherorderderivatives:bis} 


In this section, we infer from the Teukolsky-Starobinski identity, see Proposition \ref{Prop:Teuk-Star-main}, a parabolic equation for $\aa$. 
\begin{corollary}\lab{corollary:parabolicequationsatsfiedbyalphabar:bis}
$\aa$ satisfies in $\Mint$ the following equation
\beaa
&& 6m\widetilde{\nu}\aa+r^4\dds_2\dds_1\ddd_1\ddd_2\aa \\
  &=& \frac{1}{r^3}\Big(e_3(r^2e_3(r\qf))+2\omb r^2e_3(r\qf)\Big) -r^{-3}\err[TS]   -\left\{\frac{3}{2}r^4\rho\kab  - 6am \right\}e_4\aa\\
  && -\left\{-  \frac{3}{2}r^4\left(\rho+\frac{2m}{r^3}\right)\ka  + 3mr\left(\ka-\frac{2\Up}{r}\right)-\frac{12m}{r}\right\}e_3\aa.
\eeaa
where the vectorfield $\widetilde{\nu}$ is defined by \eqref{eq:defintionofthevectorfieldwidetildenu}. 
\end{corollary}

\begin{proof}
Recall from \eqref{eq:Teuk-Star-main(seeApp)} that we have
\beaa
e_3(r^2e_3(r\qf))+2\omb r^2e_3(r\qf) &=& r^7\left\{  \dds_2\dds_1\ddd_1\ddd_2\aa  +\frac{3}{2}\rho\Big(\kab e_4 -\ka e_3\Big)\aa\right\} + \err[TS].
\eeaa
This yields
\beaa
\frac{3}{2}r^4\rho\Big(\kab e_4 -\ka e_3\Big)\aa+r^4\dds_2\dds_1\ddd_1\ddd_2\aa   &=& \frac{1}{r^3}\Big(e_3(r^2e_3(r\qf))+2\omb r^2e_3(r\qf)\Big) -r^{-3}\err[TS].
\eeaa
Now, we have in view of the definition of $\widetilde{\nu}$
\beaa
&&\frac{3}{2}r^4\rho\Big(\kab e_4 -\ka e_3\Big) -6m\widetilde{\nu}\\
 &=&  \left\{\frac{3}{2}r^4\rho\kab  - 6am \right\}e_4+\left\{-  \frac{3}{2}r^4\left(\rho+\frac{2m}{r^3}\right)\ka  + 3mr\left(\ka-\frac{2\Up}{r}\right)-\frac{12m}{r}\right\}e_3.
\eeaa
We infer
\beaa
&& 6m\widetilde{\nu}\aa+r^4\dds_2\dds_1\ddd_1\ddd_2\aa \\
  &=& \frac{1}{r^3}\Big(e_3(r^2e_3(r\qf))+2\omb r^2e_3(r\qf)\Big) -r^{-3}\err[TS]   -\left\{\frac{3}{2}r^4\rho\kab  - 6am \right\}e_4\aa\\
  && -\left\{-  \frac{3}{2}r^4\left(\rho+\frac{2m}{r^3}\right)\ka  + 3mr\left(\ka-\frac{2\Up}{r}\right)-\frac{12m}{r}\right\}e_3\aa.
\eeaa
This concludes the proof of the corollary.
\end{proof}

\begin{corollary}\lab{corollary:parabolicequationsatsfiedbyalphabarwithestimateRHS:bis}
$\aa$ satisfies in $\Mint$
\beaa
6m\widetilde{\nu}\aa+r^4\dds_2\dds_1\ddd_1\ddd_2\aa &=& F
\eeaa
where $F$ satisfies 
\beaa
\max_{0\leq k\leq k_{small}+18}\int_{\Si_*}\ub^{2+2\dec}|\dk^kF|^2 &\les& \ep_0^2.
\eeaa
\end{corollary}

\begin{proof}
In view of Corollary \ref{corollary:parabolicequationsatsfiedbyalphabar}, $\aa$ satisfies 
\beaa
6m\widetilde{\nu}\aa+r^4\dds_2\dds_1\ddd_1\ddd_2\aa &=& F
\eeaa
with 
\beaa
F &:=& e_3(e_3(\qf))+F_1,\\
F_1 &:=& \frac{1}{r^3}\Big(e_3(r^2e_3(r)\qf)+e_3(r^3)e_3(\qf)+2\omb r^2e_3(r\qf)\Big)  -r^{-3}\err[TS] \\
&& -\left\{\frac{3}{2}r^4\rho\kab  - 6am \right\}e_4\aa -\left\{-  \frac{3}{2}r^4\left(\rho+\frac{2m}{r^3}\right)\ka  + 3mr\left(\ka-\frac{2\Up}{r}\right)-\frac{12m}{r}\right\}e_3\aa.
\eeaa
Recall also that $\err[TS]$ is given schematically by, see Proposition \ref{Prop:Teuk-Star-main(seeApp)},
\beaa
\err[TS]&=&r^4\big( \dkb \Ga_b+  r \Ga_b\c \Ga_b) \c   \aa+r^2\big( \Ga_b e_3(r\qf)+( \dk^{\le 1} \Ga_b) r\qf \Big)\\
&+& r^7 \dk^{\le 2} \big(e_3 \eta \c \b\big)+r^5  \dk^{\le 3 }\big( \Ga_b\c\Ga_g\big).
\eeaa
We infer that $F_1$ is given schematically by
\beaa
F_1 &=& r\big( \dkb \Ga_b+  r \Ga_b\c \Ga_b) \c   \aa+r^{-1}\big( \Ga_b e_3(r\qf)+( \dk^{\le 1} \Ga_b) r\qf \Big)\\
&+& r^4 \dk^{\le 2} \big(e_3 \eta \c \b\big)+r^2  \dk^{\le 3 }\big( \Ga_b\c\Ga_g\big)+r^{-1}\Ga_b\\
&=& r^{-1}\Ga_b +r^2  \dk^{\le 3 }\big( \Ga_b\c\Ga_g\big)+r^4  \dk^{\le 3 }\big( \Ga_g\c\b\big)
\eeaa
where we have used
\begin{itemize}
\item The fact we are working here with the global frame of Proposition \ref{prop:existenceandestimatesfortheglobalframe:bis} which has the property that $\eta\in\Ga_g$.

\item The fact that $\Ga_b$ behave better that $r\Ga_g$.

\item The fact that $\aa$ and $\qf$ behaves at least as good as $\Ga_b$.

\item The fact that $\rho+\frac{2m}{r^3}$ behaves as good as $r^{-1}\Ga_g$.

\item The fact that $e_3(r)+1$ belongs to $r\Ga_b$.
\end{itemize}

Now, recall from Lemma \ref{le:interpolatedbootstrap} that the global frame of  Proposition satisfies in particular\footnote{Here we use  \eqref{eq:estimateforfinconstructionsecondframeinMext} with $k_{loss}=22$.}
\bea\lab{eq:recallfrombehaviorestablishedforsecondglabalfrmaeinTheoremM1}
\max_{0\leq k\leq k_{small}+22}\sup_{\MM}\Big\{r^{\frac{7}{2}+\dec-2\de_0}|\dk^k\b|+r^2u^{\frac{1}{2}+\dec-2\de_0}|\dk^k\Ga_g|+ru^{1+\dec-2\de_0}|\dk^k\Ga_b|\Big\} \les \ep.
\eea
Together with the schematic for of $F_1$ and the behavior \eqref{eq:behaviorofronSigmastar} of $r$ on $\Si_*$, and the fact that $\de_0$ can be chosen to satisfy\footnote{Recall from Lemma \ref{le:interpolatedbootstrap} that we have 
\beaa
\de_0=\frac{k_{loss}}{k_{large}-k_{small}}.
\eeaa
Since we have here $k_{loss}=22$, and since we have $2k_{small}\leq k_{large}+1$ and $k_{large}\dec\gg 1$, we deduce $\de_0\ll \dec$ and we have indeed $8\de_0\leq \dec$.} $8\de_0\leq \dec$, we infer
\beaa
\max_{0\leq k\leq k_{small}+18}\sup_{\Si_*}ru^{\frac{3}{2}+\frac{3}{2}\dec}|\dk^kF_1| &\les& u_*^{\frac{3}{2}+\frac{3}{2}\dec}\sup_{\Si_*}(r^{-\frac{1}{2}})\ep+\ep^2\les\ep_0.
\eeaa
In view of the definition of $F$, this yields
\beaa
\max_{0\leq k\leq k_{small}+18}\int_{\Si_*}u^{2+2\dec}|\dk^kF|^2 &\les& \ep_0^2+\max_{0\leq k\leq k_{small}+19}\int_{\Si_*}u^{2+2\dec}|\dk^k e_3(\qf)|^2.
\eeaa
Together with Theorem M1, and the fact that $\dee>\dec$, we infer
\beaa
\max_{0\leq k\leq k_{small}+18}\int_{\Si_*}u^{2+2\dec}|\dk^kF|^2 &\les& \ep_0^2.
\eeaa
This concludes the proof of the corollary.
\end{proof}

We are now ready to prove Proposition \ref{prop:parabolicequationsatsfiedbyalphabarwithestimateRHShigherorderderivatives:bis}. In view of Corollary \ref{corollary:parabolicequationsatsfiedbyalphabarwithestimateRHS:bis}, $\aa$ satisfies
\beaa
6m\widetilde{\nu}\aa+r^4\dds_2\dds_1\ddd_1\ddd_2\aa &=& F.
\eeaa
Commuting with $\dk^k$, we infer
\beaa
6m\widetilde{\nu}(\dk^k\aa)+r^4\dds_2\dds_1\ddd_1\ddd_2(\dk^k\aa) &=& F_k
\eeaa
where $F_k$ is defined by
\beaa
F_k &:=& -6m[\dk^k, \widetilde{\nu}]\aa-6\sum_{j=1}^k\dk^j(m)\dk^{k-j}\widetilde{\nu}\aa -[\dk^k, r\dds_2] r\dds_1 r\ddd_1 r\ddd_2\aa -r\dds_2[\dk^k, r\dds_1] r\ddd_1 r\ddd_2\aa\\
&& -r\dds_2 r\dds_1 [\dk^k, r\ddd_1] r\ddd_2\aa -r\dds_2 r\dds_1 r\ddd_1 [\dk^k, r\ddd_2]\aa + \dk^kF.
\eeaa

Note that we have schematically
\beaa
[\dk, \dkb] = r\Ga_b\dk,\quad [\widetilde{\nu}, \dkb]=\Big(O(r^{-1})+r\Ga_b\Big)\dk,\quad [\widetilde{\nu}, re_4]=O(r^{-1})\dk, \quad [\widetilde{\nu}, e_3]=O(r^{-1})\dk.
\eeaa
Together with the fact that $\aa$ behaves at least as good as $\Ga_b$, we infer, schematically,
\beaa
F_k &=& \dk^kF+r^{-1}\dk^{\leq k+4}\Ga_b+r\dk^{\leq k+4}(\Ga_b^2).
\eeaa
In view of \eqref{eq:recallfrombehaviorestablishedforsecondglabalfrmaeinTheoremM1} and the behavior \eqref{eq:behaviorofronSigmastar} of $r$ on $\Si_*$, we have
\beaa
\max_{0\leq k\leq k_{small}+18}\sup_{\Si_*}ru^{\frac{3}{2}+\frac{3}{2}\dec}|r^{-1}\dk^{\leq k+4}\Ga_b+r\dk^{\leq k+4}(\Ga_b^2)| &\les&u_*^{\frac{3}{2}+\frac{3}{2}\dec}\sup_{\Si_*}(r^{-\frac{1}{2}})\ep+\ep^2\les\ep_0.
\eeaa
This yields
\beaa
\max_{0\leq k\leq k_{small}+18}\int_{\Si_*}u^{2+2\dec}|F_k|^2 &\les& \ep_0^2+\max_{0\leq k\leq k_{small}+18}\sup_{\Si_*}u^{2+2\dec}|\dk^kF|^2.
\eeaa
Together with the estimate for $F$ of Corollary \ref{corollary:parabolicequationsatsfiedbyalphabarwithestimateRHS:bis}, we infer
\beaa
\max_{0\leq k\leq k_{small}+18}\int_{\Si_*}u^{2+2\dec}|F_k|^2 &\les& \ep^4+\ep_0^2\les \ep_0^2.
\eeaa
This concludes the proof of Proposition \ref{prop:parabolicequationsatsfiedbyalphabarwithestimateRHShigherorderderivatives:bis}.


\subsection{Proof of Lemma \ref{lemma:basicparabolicestimateforcontrolofaainSi*}}\lab{sec:proofoflemma:basicparabolicestimateforcontrolofaainSi*}


In this section we  prove Lemma \ref{lemma:basicparabolicestimateforcontrolofaainSi*}, i.e. we derive estimates for the control of the parabolic equation appearing in the statement of Proposition \ref{prop:parabolicequationsatsfiedbyalphabarwithestimateRHShigherorderderivatives:bis}. The following identity will be useful.
\begin{lemma}\lab{lemma:identityneededforbasicparabolicestimateforcontrolofaainSi*}
We have for any reduced scalar $f$
\beaa
&&\widetilde{\nu}\left(\int_Sf^2\right)\\
&=& 2\int_Sf\widetilde{\nu}(f)+\int_S(-2afe_4(f)+\kab f^2)+a\int_S(2fe_4(f)+\ka f^2)+\err\left[e_3\left(\int_S f^2\right)\right].
\eeaa
\end{lemma}

\begin{proof}
Recall from the definition of  $\widetilde{\nu}$ that 
\beaa
\widetilde{\nu} = e_3+ae_4.
\eeaa
We infer, in view of Proposition \ref{prop:outgoinggeod-e3e4averages}, 
\beaa
&&\widetilde{\nu}\left(\int_Sf^2\right)\\
 &=& e_3\left(\int_Sf^2\right) +ae_4\left(\int_Sf^2\right)\\
&=& \int_S(2fe_3(f)+\kab f^2)+\err\left[e_3\left(\int_S f^2\right)\right]+a\int_S(2fe_4(f)+\ka f^2)\\
&=& 2\int_Sf\widetilde{\nu}(f)+\int_S(-2afe_4(f)+\kab f^2)+a\int_S(2fe_4(f)+\ka f^2)+\err\left[e_3\left(\int_S f^2\right)\right].
\eeaa
This concludes the proof of the lemma.
\end{proof}

We are now ready to prove Lemma \ref{lemma:basicparabolicestimateforcontrolofaainMint}. Recall from Lemma \ref{lemma:identityneededforbasicparabolicestimateforcontrolofaainSi*} that we have
\beaa
&&\widetilde{\nu}\left(\int_Sf^2\right)\\
&=& 2\int_Sf\widetilde{\nu}(f)+\int_S(-2afe_4(f)+\kab f^2)+a\int_S(2fe_4(f)+\ka f^2)+\err\left[e_3\left(\int_S f^2\right)\right].
\eeaa
In view of the equation satisfied by $f$, we infer
\beaa
\widetilde{\nu}\left(\int_Sf^2\right) &=&- \frac{1}{3m}\left(\int_Sr^4f\dds_2\dds_1\ddd_1\ddd_2f\right)+ \frac{1}{3m}\left(\int_Shf\right) \\
&&+\int_S(-2afe_4(f)+\kab f^2)+a\int_S(2fe_4(f)+\ka f^2)+\err\left[e_3\left(\int_S f^2\right)\right].
\eeaa
Now, from the definition of $\widetilde{\nu}$, we have $\widetilde{\nu}(u)=2/\vsi$. We deduce
\beaa
&& \widetilde{\nu}\left(u^n\int_Sf^2\right) +\frac{u^n}{3m}\left(\int_Sr^4f\dds_2\dds_1\ddd_1\ddd_2f\right)\\
 &=&\frac{u^n}{3m}\left(\int_Shf\right) + u^n\int_S(-2afe_4(f)+\kab f^2)+au^n\int_S(2fe_4(f)+\ka f^2)\\
 &&+u^n\err\left[e_3\left(\int_S f^2\right)\right]+\frac{2}{\vsi}nu^{n-1}\int_Sf^2.
\eeaa
This yields in view of the bootstrap assumptions 
\beaa
 &&\widetilde{\nu}\left(u^n\int_Sf^2\right) +\frac{u^n}{3m}\left(\int_Sr^4f\dds_2\dds_1\ddd_1\ddd_2f\right)\\
  &\les& \frac{u^n}{3m}\|h\|_{L^2(S)}\|f\|_{L^2(S)}+ \left(\frac{1}{r}+\ep u^{-1}\right)u^n\int_S|f||\dk^{\leq 1}f| +nu^{n-1}\int_Sf^2\\
    &\les& \frac{u^n}{3m}\|h\|_{L^2(S)}\|f\|_{L^2(S)}+ \ep u^{n-1}\int_S|f||\dk^{\leq 1}f| +nu^{n-1}\int_Sf^2
\eeaa
where we have used in the last inequality the behavior \eqref{eq:behaviorofronSigmastar} of $r$ on $\Si_*$. Next, we rely on the Poincar\'e inequality of Lemma \ref{lemma:poincareinequalityusedforcontrolofparabolicequation} to deduce
\beaa
\widetilde{\nu}\left(u^n\int_Sf^2\right) +u^n\int_Sf^2 &\les& u^n\int_Sh^2 +\ep^2 u^{n-2}\int_S(\dk f)^2+nu^{n-1}\int_Sf^2.
\eeaa
Integrating in $u$ between $1$ and $u_*$, and recalling that $\widetilde{\nu}(u)=2/\vsi$, we infer
 \beaa
\int_{\Si_*}u^nf^2 &\les & \int_{\Si_*\cap\CC_1}f^2 +\int_{\Si_*}u^nh^2+\ep^2\int_{\Si_*}u^{n-2}(\dk f)^2+n\int_{\Si_*}u^{n-1}f^2.
 \eeaa
  In particular, we have for $n=0$
 \beaa
\int_{\Si_*}f^2 &\les & \int_{\Si_*\cap\CC_1}f^2 +\int_{\Si_*}h^2+\ep^2\int_{\Si_*}u^{-2}(\dk f)^2.
 \eeaa
Then, starting from the case $n=0$ and arguing by iteration on the largest integer below $n$, one immediately deduces for any real $n\geq 0$ 
 \beaa
\int_{\Si_*}(1+u^n)f^2 &\les & \int_{\Si_*\cap\CC_1}f^2 +\int_{\Si_*}(1+u^n)h^2 +\ep^2\int_{\Si_*}(1+u^{n-2})(\dk f)^2
 \eeaa
which concludes the proof of Lemma \ref{lemma:basicparabolicestimateforcontrolofaainSi*}.


\chapter{DECAY ESTIMATES (Theorems M4, M5)}\lab{chap:proofoftheoremM4M5}


In this chapter, we rely on the decay of $\qf$, $\a$ and $\aa$ to prove the decay estimates for all the other quantities. More precisely, we rely on the results of Theorem M1, M2 and M3 to prove Theorem M4 and M5.

 
 \section{Preliminaries to the proof of Theorem M4}
 
 
In what follows we give a  detailed proof of Theorem {\bf M4},  which, we recall, provides the main decay estimates in $\Mext$.  The proof makes use of  the  bootstrap assumptions {\bf BA-D}, 
 {\bf BA-E},  the results of Theorems  {\bf M1, M2, M3}  and Lemmas  \ref{lemma:estimatesonceandforallforaverages},  \ref{lemma:estimatesonceandforallforHawkingmass}.  In this section, we start with some preliminaries.

 
 \subsection{Geometric structure of $\Si_*$}
 

The proof of Theorem M4 depends in a fundamental way on the geometric properties of the GCM hypersuface $\Si_*$, the spacelike future boundary of $\Mext$  introduced  in  section \ref{sec:defintioncanonicalspacetime}. For the convenience of the reader, we recall below its  main features.  

 \begin{enumerate}
 \item The affine parameter $s$ is initialized on $\Si_*$ such that  $s=r$.
\item There exists a constant $c{_*}$ such that 
\beaa
\Sigma_*:=\{u+ r =c_{*}\}.
\eeaa
\item 
Let $\nu_*=e_3+ a_*  e_4 $  be the unique vectorfield tangent to the hypersurface $\Si_*$,  perpendicular to  the foliation $S(u)$ induced on $\Si_*$ and normalized by the condition $g(\nu_*, e_4)=-2$.  
 The following normalization condition  holds true  at the  South Pole $SP$  of every sphere $S$,
 \bea
 \lab{GCM-conditionfora*-M4}
 a_*\Big|_{SP}=-1 -\frac{2m^\S}{r^\S}.
 \eea

\item We have
\bea
r_*\ge \ep_0^{-\frac{2}{3}} \, u^4  \quad \textrm{ on }\Sigma_*.
\eea

\item   The following GCM conditions hold on   $ \Sigma_*$
\bea
\label{GCM-Si*1}
\ka=\frac{2}{r}, \qquad \dds_2\dds_1\kab=0,\qquad \dds_2\dds_1\mu=0,
\eea
\bea
\label{GCM-Si*2}
   \int_S \eta  e^\Phi = \int_S\xib e^\Phi=0.
\eea
Moreover  on $S_*=\Sigma_*\cap\CC_*$,
\bea
\lab{GCM-Si*3}
\int_{S_*}\b e^\Phi=0, \quad \int_{S_*}e_\th(\kab) e^\Phi=0.
\eea
\item According to the definition of the Hawking mass, i.e.   $1-\frac{2m}{r}=-\frac{r^2}{4} \ov{\ka \kab}$, 
  and the GCM assumption for $\ka$ we also have,
\bea
\ov{\kab}&=&-\frac r 2 \left(1-\frac{2m}{r} \right).
\eea
Thus on $\Si_*$,
\bea
\lab{eq:err-e4ronSi*-M4}
e_3(r)=\frac r 2 \left( \ov{\kab}+\Ab\right)=-\Up+\frac r 2 \Ab, \qquad  e_4(r)=1.
\eea
\item 
In view of  the definition of $\nu_*$ and and that of  $\vsi$ we 
 we easily deduce\footnote{   Indeed, since $\nu_* $ is tangent  to $\Si_*$ along which $u=-r+c_*$,  using also \eqref{eq:err-e4ronSi*-M4},
 $\frac{2}{\vsi}=e_3(u)=\nu_*(u)=-\nu_*(r)= -e_3(r)- a_* e_4(r)  =  -a_*  +\Up    -\frac{r}{2}\Ab.$} the following relation between $a_*$  and  $ \vsi$ on $\Si_*$.
 \bea
 \lab{formula-a*-M4}
 a_*=- \frac{2}{\vsi} +\Up-\frac r 2 \Ab.
 \eea 
 \item Since  on $\Si_*$ we have $r=s$ we deduce,
\bea
 \lab{formula:Omb-onSi*-M4}
\Omb= e_3(r)=-\Up+\frac r 2  \Ab \quad \text{on } \,\, \Si_*.
\eea
\end{enumerate}


\subsection{Main  Assumptions}


We reformulate below  the main bootstrap assumption\footnote{Based on bootstrap assumptions {\bf BA-D}, {\bf BA-E},  Theorems  {\bf M1, M2, M3}  and   Lemmas  \ref{lemma:estimatesonceandforallforaverages},  \ref{lemma:estimatesonceandforallforHawkingmass}.} in  the form needed in the proof of Theorem M4.

\begin{definition}
\label{definition:norms-ProofThm.M4}
  We make  use of  the following   norms on  $S=S(u,r)\subset\Mext$,
  \bea
  \bsplit
  \| f\|_{\infty} (u,r):&=\| f\|_{L^\infty\big(S(u,r)\big)}, \qquad \quad  \| f\|_{2} (u,r):=\| f\|_{L^2\big(S(u,r)\big)}, \\
  \|f\|_{\infty,k}(u, r)&:= \sum_{i=0}^k \|\dk^i f\|_{\infty }(u, r),  \qquad 
\|f\|_{2,k}(u, r):=\sum_{i=0}^k \|\dk^i f\|_{2}(u, r).
\end{split}
  \eea
  \end{definition}

To simplify the exposition it  also helps  to  introduce the following schematic notation  for the    connection coefficients (recall $\om, \xi=0$ and $\ze=-\etab$),
 \bea
 \label{notation-Gag-Gab}
 \bsplit
 \Ga_g&= \Big\{\kac, \vth, \etab,\ze,  \kabc\Big\} \cup \Big\{ \ov{\ka} -\frac{2}{r}, \ov{\kab}+\frac{2\Up}{r}\Big\}, \\
 \Ga_b&=\Big\{\vthb, \eta, \ombc, \xib,  \Ab,     r^{-1} \vsic, r^{-1} \Ombc,     \Big\}    \cup\Big\{\ov{\omb}
 -\frac{m}{r^2}, r^{-1}(  \ov{\vsi}-1),  r^{-1}(\ov{\Omb}+\Up )  \Big\}.
 \end{split}
 \eea
 \begin{remark}
 It is important to note that  $\eta$ belongs to $\Ga_b$ rather  than $\Ga_g$ as it may have been expected.
 Note also that $\Ab\in \Ga_b$  in view of   Proposition \ref{prop:outgoinggeod-e3e4averages-M4} and the fact that $  (\vsic, \Ombc)\in r\Ga_b $. We also note that the averaged quantities  $ \Big\{ \ov{\ka} -\frac{2}{r}, \ov{\kab}+\frac{2\Up}{r}\Big\}$ and $\Big\{\ov{\omb}
 -\frac{m}{r^2}, r^{-1}(  \ov{\vsi}-1),  r^{-1}(\ov{\Omb}+\Up )  \Big\}$ are actually better behaved in view of 
 Lemmas \ref{lemma:estimatesonceandforallforaverages},  \ref{lemma:estimatesonceandforallforHawkingmass}.
 \end{remark}

{\bf Ref 1.} According to our   bootstrap  assumptions  {\bf BA-D}, and the pointwise estimates of Proposition \ref{prop:pointwiseboundsforhighorderderivatives}, which themselves follow from {\bf BA-E}, as well as  the control of averages in Lemma   \ref{lemma:estimatesonceandforallforaverages}    and the control of the Hawking mass in Lemma       \ref{lemma:estimatesonceandforallforHawkingmass},  we have   on  $\MMext$,
\begin{enumerate}
\item  For $0\le k\le k_{small}$,
\bea
\label{Ref1-smallk}
\bsplit
\| \Ga_g\|_{\infty, k}&\les\ep \min\left\{ r^{-2} u^{-\frac{1}{2}-\dec}, \, r^{-1} u^{-1-\dec}   \right\},\\
 \|e_3  \Ga_g\|_{\infty, k-1}  &\les \ep  r^{-2} u^{-1-\dec}, \\   
\|\Ga_b\|_{\infty, k}&\les\ep  r^{-1} u^{-1-\dec}.
\end{split}
\eea

\item   For $ k\le k_{large}-5$
\bea
\label{Ref1-largek}
\|\Ga_g\|_{\infty, k}&\les&\ep  r^{-2}, \qquad  \| \Ga_b\|_{\infty,k} \les  \ep  r^{-1}.
\eea
\end{enumerate}

{\bf Ref  2.}  The quantity\footnote{Recall (see Remark \ref{def:wherewementionthatwealwaysexpressqfinthesecondglobalframe}) that   the quantity $\qf$ we are working with   is defined relative to the global frame of Proposition \ref{prop:existenceandestimatesfortheglobalframe:bis}.} $\qf$ satisfies on $\MMext$, for all  $0\le k\le k_{small}+20$,
 \bea
\bsplit
\| \qf\|_{\infty, k} &\les\ep_0\min\left\{u^{-1-\dee}, r^{-1} u^{-\frac{1}{2}-\dee}\right\},\\
\| e_3 \qf\|_{\infty, k-1} &\les\ep_0 r^{-1} u^{-1-\dee}.
\end{split}
\eea
In addition, on the last slice $\Si_*$, for all $k\le k_{small}+20$,
\bea
\label{improved-qf:onSi*}
\int_{\Sigma_*(\tau, \tau_*)}|e_3\dk^{ k} \qf|^2 +   |e_4\dk^{ k} \qf|^2     + r^{-2} |\qf|^2  &\les \ep^2_0(1+\tau)^{-2-2\dec}.
\eea

 According to Theorem    {\bf M2}   we have on $\MMext$, for all  $0\le k\le k_{small}+20$,
\bea
\begin{split}
\| \a\|_{\infty, k} &\les\ep_0\min\left\{r^{-3} (u+2r)^{-\frac{1}{2}-\dee}, \,  \log(1+u)r^{-2} (u+2r)^{-1-\dee}\right\},\\
\|e_3  \a\|_{\infty, k-1} &\les\ep_0\min\left\{r^{-4} (u+2r)^{-\frac{1}{2}-\dee},\,  \log(1+u)r^{-3} (u+2r)^{-1-\dee}\right\}.
\end{split}
\eea
According to  Theorem   {\bf M3}, the component $\aa$  verifies  the following estimate\footnote{In fact, the corresponding estimate in Theorem M3 holds on $\Mint$, and hence in particular on $\TT$ since $\TT\subset\Mint$.}   holds  on $\TT$,  for $0\le k\le k_{small}+16$,
\bea
\sup_{\TT}u^{1+\dec}|\dk^k\aa| &\les& \ep_0,
\eea
and on the last slice $\Si_*$ for all $k\le k_{small}+16$
\bea
\label{improved-aa:onSi*}
\int_{\Sigma_*(\tau, \tau_*)}| \dk^{ k} \aa |^2 &\les \ep^2_0(1+\tau)^{-2-2\dec}.
\eea

{\bf Ref 3.}  In view of the bootstrap assumptions  {\bf BA-D} and the pointwise estimates of Proposition \ref{prop:pointwiseboundsforhighorderderivatives} for the curvature components, which themselves follow from {\bf BA-E} , we have in $\Mext$,
\begin{itemize}
\item[i.]  For all  $ 0\le k\le  k_{small}$,
\bea
\begin{split}
\| \b\|_{\infty, k} &\les\ep\min\left\{r^{-3} (u+2r)^{-\frac{1}{2}-\dec}, \,  r^{-2} (u+2r)^{-1-\dec}\right\},\\
\|e_3  \b\|_{\infty, k-1} &\les\ep\min\left\{r^{-4} (u+2r)^{-\frac{1}{2}-\dec}, \,  r^{-3} (u+2r)^{-1-\dec}\right\},\\
\left\| \left(\rhoc, \ov{\rho}+\frac{2m}{r^3}\right) \right\|_{\infty, k} &\les \ep\min\left\{r^{-3} u^{-\frac{1}{2}-\dec}, \, r^{-2} u^{-1-\dec}\right\},\\
\left\| e_3\left(  \rhoc, \ov{\rho}+\frac{2m}{r^3}\right)  \right\|_{\infty, k-1} &\les \ep r^{-3} u^{-1-\dec},\\
\left\| \check{\mu}, \ov{\mu}-\frac{2m}{r^3} \right\|_{\infty, k} &\les\ep  r^{-3} u^{-1-\dec},\\
\| \bb  \|_{\infty, k} &\les\ep r^{-2} u^{-1-\dec}.
\end{split}
\eea
Since $K=-\rho-\frac 1 4 \ka\kab +\frac 1 4 \vth \vthb =\frac{1}{r^2} -(\rho -\ov{\rho} )- \frac{1}{4}(\ka\kab- \ov{\ka} \ov{\kab})+\lot$ we also deduce for all  $ 0\le k\le  k_{small}$,
\beaa
\left\|K-\frac{1}{r^2}\right\|_{\infty, k} &\les&\ep\min\left\{r^{-3} u^{-\frac{1}{2}-\dec}, \, r^{-2} u^{-1-\dec}\right\}.
\eeaa
\item[ii.] For all $k\le k_{large}-5$,
\bea\lab{eq:rpweigthedestimatesforcurvaturerestrictedonspheresforTheoremM4}
\bsplit
   r^{\frac{7}{2} +\frac{\de_B}{2}}\Big( \|\a\|_{\infty, k} +    \|\b\|_{\infty , k} \Big)  &\les\ep, \\
   r^3  \|\rhoc\|_{\infty , k}   +r^2\|\bb\|_{\infty, k} +r\|\aa\|_{\infty, k}   &\les\ep. 
   \end{split}
\eea
\end{itemize}

\begin{remark}
In view of the control of averages  Lemma   \ref{lemma:estimatesonceandforallforaverages}  we  have in fact better estimates for   the scalars,
\beaa
\ov{\ka}-\frac 2 r, \quad \ov{\kab}+\frac{2\Up}{r}, \quad \ov{\omb}-\frac{m}{r^2}, \quad \ov{\rho}+\frac{2m}{r^3}.
\eeaa
In particular  they can be estimated by $\ep$ replaced by $\ep_0$ in {\bf Ref 1}.
\end{remark}
\begin{remark}
\lab{extended-definitions-forGagb}
Note that $ r(\rhoc, \ov{\rho}+\frac{2m}{r^3}), r(K-\frac{1}{r^2}) $ behave as  $\Ga_g$. 
For convenience   we shall just simply add  them to  $\Ga_g$.  Similarly $(r \bb, \aa) $ behave  as $\Ga_b$. Thus,  our  extended $\Ga_g, \Ga_b $  are 
\beaa
\Ga_g&=& \Big\{\kac, \vth, \etab,\ze,  \kabc,  r\rhoc \Big\} \cup \left\{ \ov{\ka} -\frac{2}{r}, \ov{\kab}+\frac{2\Up}{r},  r\left(\ov{\rho}+\frac{2m}{r^3}\right)\right\}, \\
 \Ga_b&=&\Big\{\vthb, \eta, \ombc, \xib,  \Ab,     r^{-1} \vsic, r^{-1} \Ombc,  r\bb, \aa     \Big\}    \cup\Big\{\ov{\omb}
 -\frac{m}{r^2}, r^{-1}(  \ov{\vsi}-1),  r^{-1}(\ov{\Omb}+\Up )  \Big\}.
\eeaa
Note also   that we can write $ e_3 (\Ga_g) =  r^{-1}\dk \Ga_b$.
\end{remark}


\subsection{Basic Lemmas}



\subsubsection{Commutation identities}


\begin{lemma}
\label{Cor:comme3e4-outgeodesic-M4}
We have, schematically,
 \bea
\bsplit
\,[\dkb , e_4]\psi&= \Ga_g\dkout  \psi +\lot, \\
\,[\dkb, e_3]\psi&= r \Ga_b   e_3 \psi   +\Ga_b^{\le 1} \dkout \psi      +  \lot 
\end{split}
\eea
\end{lemma}

\begin{proof}
Follows  from    Lemma  \ref{Le:comme3e4-outgeodesic} and the symbolic notation introduced in \eqref{notation-Gag-Gab}, see also Remark \ref{extended-definitions-forGagb}.
\end{proof}


\subsubsection{Product Estimates}


We  estimate quadratic error terms with the help of  the  following,
\begin{lemma}
\lab{lemma:interpolation-decay}
Let $k_{small}< k_{large} $  and   $k_{loss}>0$ three positive integers  verifying the conditions,
\bea
\label{Decay:condition-kloss}
k_{loss}< k_{small},\qquad k_{loss}  \le   \frac{k_{large} -5-k_{small}}{4}. 
\eea
The following product estimates hold true for all $0\le k\le k_{small}+k_{loss} $,
\bea
\bsplit
\|\Ga_g\c \Ga_g\|_{\infty, k}+r\left\|\Big(\check{\rho}, \b, \a\Big)\c \Ga_g\right\|_{\infty, k} &\les\ep_0  r^{-\frac{7}{2}} u^{-1-\dec},\\
\left\|\Ga_g\c \Big(\Ga_b, \aa\Big)\right\|_{\infty, k}+r\|\Ga_g\c \bb\|_{\infty, k}+r\|\check{\rho}\c\Ga_b\|_{\infty, k}\\
+r^{\frac{5}{4}}\left\|\Big(\b, \a\Big)\c\Ga_b\right\|_{\infty, k}&\les\ep_0  r^{-3} u^{-1-\dec},\\
\left\|\Big(\Ga_b, \aa\Big)\c \Ga_b\right\|_{\infty, k} +r\|\bb\c \Ga_b\|_{\infty, k} &\les\ep_0  r^{-2} u^{-1-\dec},\\
\|e_3(\Ga \c \b)\|_{\infty, k} &\les  r^{-5} u^{-1-\dec}. 
 \end{split}
\eea
\end{lemma}

\begin{proof}
All estimates   are easy to prove  in the range $0\le k\le k_{small} $. 
We shall thus assume that $k_{small}\le k\le k_{small}+k_{loss}$.  Since  $k_{loss}< k_{small}$ we have 
$k/2 < k_{small}$ for all $k$ in that range.
We start with the first estimate. Since $r\check{\rho}$ satisfies the same estimates as $\Ga_g$, and as $r\b$ and $r\a$ satisfy even better estimates, it suffices to prove the first estimate for  $\Ga_g\c\Ga_g$. 
For simplicity of notation we write  $L:=k_{large}-5$, \, $S:=k_{small}  $.
By standard interpolation inequalities, for all $S\le k\le L$,
\beaa
\|  \Ga_g \|_{\infty, k}&\les& \|   \Ga_g \|^{\frac{k-S }{ L-S} } _{\infty, L} \|   \Ga_g \|_{\infty, S}^{\frac{L -k }{ L-S } } \les \ep   r^{-2}  \Big[  u^{-\frac 1 2 -\dec}\Big]^ {\frac{L-k }{ L-S }}.
\eeaa
Therefore, for any  $S\le k\le L$
\beaa
\|  \Ga_g \c \Ga_g \|_{\infty, k}&\les& \| \Ga_g\|_{\infty} \|  \Ga_g \|_{\infty, k}\les \| \Ga_g\|_{\infty} \, \ep   r^{-2}  \Big[  u^{-\frac 1 2 -\dec}\Big]^ {\frac{L -k }{ L-S }}\\
&\les& \ep r^{-2} \| \Ga_g\|_{\infty}^\frac{1}{2}    \,  \| \Ga_g\|_{\infty}^\frac{1}{2}\,  u^{-(\frac{1}{2}+\dec)   \frac{L -k }{ L-S}}\\
&\les& \ep^2r^{-2}\, r^{-\frac{1}{2} } u^{-\frac{1}{2}(1+\dec)}  r^{-1 } u^{-\frac{1}{2}(\frac{1}{2}+\dec)} u^{-(\frac{1}{2}+\dec)   \frac{L-k }{ L-S}}\\
&=&\ep^2r^{-\frac{7}{2}} u^{-\frac{3}{4}-\dec-(\frac{1}{2}+\dec)   \frac{L-k }{ L-S }}.
\eeaa
Now, we have
\beaa
-\frac{3}{4}-\frac{1}{2}\frac{L-k }{ L-S } = -1- \frac{1}{2}\left(\frac{1}{2}-\frac{k-S }{ L-S } \right) = -1- \frac{1}{2}\left(\frac{1}{2}-\frac{k_{loss} }{k_{large}-5-k_{small}} \right)\leq -1
\eeaa
where we used the assumption on $k_{loss}$
\beaa
k_{loss}\leq \frac{k_{large}-5-k_{small}}{2}.
\eeaa
We infer
\beaa
\|  \Ga_g \c \Ga_g \|_{\infty, k}&\les& \ep^2r^{-\frac{7}{2}} u^{-1-\dec}.
\eeaa

Next, we consider the second estimate. Since $r\check{\rho}$ satisfies the same estimates as $\Ga_g$, and as $\aa$ and $r\bb$ satisfy the same estimate as $\Ga_b$, it suffices to prove the second estimate for $\Ga_g\c \Ga_b$ and $(\a, \b)\c \Ga_b$.  Starting with,
\beaa
\|  \Ga_b \|_{\infty, k}&\les& \|   \Ga_b \|^{\frac{k-S }{ L-S} } _{\infty, L}     \|   \Ga_b \|_{\infty, S}^{\frac{L -k }{ L-S } } \les   \ep r^{-1}  \Big[  u^{- 1 -\dec}\Big]^ {\frac{L -k }{ L- S  }},
\eeaa
we deduce,
\beaa
\|  \Ga_g \c \Ga_b \|_{\infty, k}&\les& \|  \Ga_g\|_{\infty}  \|\Ga_b\|_{\infty, k}+ \|  \Ga_b\|_{\infty}  \|\Ga_g\|_{\infty, k}\\
&\les& \ep^2r^{-2} u^{-\frac 1 2 -\dec} r^{-1}  \Big[  u^{- 1 -\dec}\Big]^ {\frac{L -k }{ L-S }}+ \ep^2r^{-1} u^{-1-\dec} r^{-2}  \Big[  u^{-\frac 1 2 -\dec}\Big]^ {\frac{L -k }{ L -S  }}\\
&\les& \ep^2r^{-3} u^{-1-\dec}.
\eeaa
Also,
\beaa
\left\| \Big(\a, \b\Big) \c \Ga_b \right\|_{\infty, k}&\les& \| (\a, \b)\|_{\infty}  \|\Ga_b\|_{\infty, k}+ \|  \Ga_b\|_{\infty}  \|(\a, \b)\|_{\infty, k}\\
&\les& \ep^2r^{-\frac{13}{4}} u^{-\frac 1 4 -\dec} r^{-1}  \Big[  u^{- 1 -\dec}\Big]^ {\frac{L -k }{ L-S }}+ \ep^2r^{-1} u^{-1-\dec} r^{-\frac{7}{2}}  \\
&\les& \ep^2r^{-\frac{17}{4}} u^{-1-\dec}.
\eeaa

Finally, the estimates\footnote{Note also that in view of  Remark \ref{extended-definitions-forGagb} we can write $\bb\in r^{-1} \Ga_b,\,  \aa\in \Ga_b$.  } for $\Ga_b \c \Ga_b$,\,  $\aa\c\Ga_b$, \,  $\bb\c\Ga_b$ as well as $e_3(\Ga \c \b)$  are easier and left to the reader.
\end{proof}


 \subsubsection{Elliptic Estimates} 


 We shall often make use of the  results of Proposition \ref{prop:2D-hodge-reduced}  and Lemma \ref{Lemma:poincarefor-dds2}  which we  rewrite as follows with respect to the $L^2$ based  $\hk_k(S)$ spaces introduced in Definition \ref{definitionhkspacesS}.
 
 \begin{lemma} 
\label{prop:2D-hodge-reduced-M4}
Under the assumptions ${\bf Ref1-Ref3}$ the following  elliptic estimates hold true for the Hodge operators
$ \ddd_1, \ddd_2, \dds_1, \dds_2$, for all $k\le k_{small}+20$.
\begin{enumerate}

\item  If   $  f  \in \sk_1(S)$,
\beaa
\| \dkb f\|_{\hk_k (S)} + \|  f\|_{\hk_k  (S)}    \les r   \|\ddd_1   f  \|_{\hk_k(S)}.
\eeaa

\item If $f\in \sk_2(S)$,
\beaa
\| \dkb f\|_{\hk_k (S)} + \|  f\|_{\hk_k  (S)}    \les r   \|\ddd_2   f  \|_{\hk_k(S)}.
\eeaa

\item  If  $f\in \sk_0(S)$,
\beaa
  \|\dkb f\|_{\hk_{k} (S) } \les r  \|\dds_1\, f\|_{\hk_k(S)}.
\eeaa

\item   If  $   f  \in \sk_1(S) $,
\beaa
  \|  f\|_{\hk_{k+1}  (S)}&\les&  r  \|\dds_2\, f\|_{\hk_k(S)}+r^{-2}\left |\int_S e^\Phi f\right|.
\eeaa

\item  If  $   f  \in \sk_1(S) $,
\beaa
\left\| f-  \frac{\int_S f e^\Phi} {\int_S e^{2\Phi} }  e^{\Phi}\right\|_{\hk_{k+1}  (S)}&\les  r  \|\dds_2\, f\|_{\hk_k(S)}.
\eeaa
\end{enumerate}
\end{lemma}


\subsection{Main equations}

 
 The proof  of Theorem {\bf M4}  relies heavily  on   the  null structure  and null Bianchi identities    derived in section \ref{sec:mainequationsforoutgoiggeodesicfoliations}, see Propositions \ref{propos:basiceqts-geod}. We also rely on   Proposition \ref{propos:transportaverages} for equations  verified by the check quantities. We rewrite them below in  a schematic form.

\begin{proposition}[Transport equations for checked quantities]
\label{propos:transportaverages-M4}
We have the following transport equations in the  $e_4$ direction,
\bea
\begin{split}
e_4\check{\ka}+\ov{\ka}\, \check{\ka}&=\Ga_g\c \Ga_g,\\
e_4\check{\kab} +\frac1 2 \ov{\ka} \check{\kab}+\frac 1 2 \check{\ka} \ov{\kab} &=-2 \ddd_1\ze+2\check{\rho}+\Ga_g\c \Ga_b,\\
e_4\check{\omb}&=\check{\rho}+\Ga_g \c \Ga_b,\\
e_4\check{\rho}+\frac  3 2 \ov{\ka} \check{\rho}+\frac 3 2 \ov{\rho}\check{\ka}&=\ddd_1\b+\Ga_b \c \a+ \Ga_g \c \b+\kac\c \rhoc, 
\\
 e_4\check{\mu} +\frac 3 2\ov{ \ka}\check{ \mu} +\frac  3 2 \ov{\mu}\check{\ka}&= r^{-1}\Ga_g\c  \dkb^{\le 1} \Ga_g.
\end{split}
\eea

Also, we have in the $e_3$ direction, 
\bea
\bsplit
e_3 \kabc&= r^{-1}\dkb^{\leq 1}\Ga_b  +\Ga_b\c\dkb^{\le 1 }\Ga_b,\\
e_3\rhoc&= r^{-2}\dkb^{\leq 1}\Ga_b  + r^{-1} \Ga_b\c\dkb^{\le 1 }\Ga_b.
\end{split}
\eea
\end{proposition}
\begin{proof}
The  statements  follow from the  precise formulas  of Proposition \ref{propos:transportaverages}  and the symbolic notation  in \eqref{notation-Gag-Gab}. We also use the convention made in Remark \ref{extended-definitions-forGagb}
 according to which we write  $r  \rhoc, r\muc \in \Ga_g, \,\,  (r\bb,   \aa)\in \Ga_b$ and  $ e_3 (\Ga_g) =  r^{-1} (\dk \Ga_b)$.
 \end{proof}


\subsection{Equations  involving   $\qf$}


Recall that our main quantity  $\qf$  has been introduced  in Definition \ref{definition:materquantity-qf}  with respect to the global frame of Proposition \ref{prop:existenceandestimatesfortheglobalframe:bis} (see Remark \ref{def:wherewementionthatwealwaysexpressqfinthesecondglobalframe}).  The passage from the geodesic frame $(e_3, e_\th, e_4)$  of $\Mext$ to the global frame  $(e_3',  e_\th', e_4')$ is given by 
\bea
 \lab{transformationformulas-global.frame-M4}
e_4' = \Up\left(e_4 +f e_\th  +\frac 1 4 f^2  e_3\right), \qquad 
e_\th' = e_\th +\frac 1 2 f  e_3, \qquad 
e_3'=\Up^{-1}e_3.
\eea
with a reduced scalar  $f$ which was constructed in  Proposition \ref{prop:constructionsecondframeinMext}.  We recall below  the main relevant  statements of Proposition \ref{prop:constructionsecondframeinMext}  in connection to the construction of the global frame.
 \begin{proposition}
 \lab{Proposition-Globalframe-Mext}
 Under  assumptions   {\bf Ref 1-2}   on $\Mext$   there exists  a  frame transformation of the form,
 \eqref{transformationformulas-global.frame-M4}
verifying the following properties\footnote{We denote by primes the Ricci and curvature components w.r.t.  to the primed frame.}:
\begin{enumerate}
\item  Everywhere in $\Mext$  we have $\xi'=0$.
 
 \item The transition  function $f$ verifies, relative to the  background frame  $(e_3, e_\th, e_4), $ the estimates\footnote{In fact, the estimates hold for $k_{small}+k_{loss}$, see Proposition \ref{prop:constructionsecondframeinMext}, and we choose here $k_{loss}=20$.}
 \bea
\begin{split}
|\dk^kf| &\les \frac{\ep}{ru^{\frac{1}{2}+\dec-2\de_0}+u^{1+\dec-2\de_0}}, \,\,\,\textrm{ for }k\leq k_{small}+20\textrm{ on }\Mext,\\
 |\dk^{k-1}e_3'f| &\les \frac{\ep}{ru^{1+\dec-2\de_0}}\,\,\,\textrm{ for }k\leq k_{small}+20\textrm{ on }\Mext.
\end{split}
\eea

 \item  The primed Ricci coefficients and curvature components verify\footnote{Note that $u$ and $r$ here are the outgoing optical function and area radius of  the  foliation of $\Mext$.}
 \beaa
\nn\max_{0\leq k\leq k_{small}+k_{loss}}\sup_{\Mext}&&\Bigg\{\Big(r^2u^{\frac{1}{2}+\dec-2\de_0}+ru^{1+\dec-2\de_0}\Big)|\dk^k\Ga_g'|+ru^{1+\dec-2\de_0}|\dk^k\Ga_b'|\\
\nn&&+r^2u^{1+\dec-2\de_0}\left|\dk^{k-1}e_3'\left(\ka'-\frac{2\Up}{r}, \kab'+\frac{2}{r}, \vth', \ze', \etab', \eta'\right)\right|\\
\nn&&+\Big(r^3(u+2r)^{\frac{1}{2}+\dec-2\de_0}+r^2(u+2r)^{1+\dec-2\de_0}\Big)\Big(|\dk^k\a'|+|\dk^k\b'|\Big)\\
\nn&& +\left(r^3(2r+u)^{1+\dec}+r^4(2r+u)^{\frac{1}{2}+\dec-2\de_0}\right)|\dk^{k-1}e_3'(\a')|\\
\nn&&+\Big(r^3u^{1+\dec}+r^4u^{\frac{1}{2}+\dec-2\de_0}\Big)|\dk^{k-1}e_3'(\b')|\\
\nn&&+\Big(r^3u^{\frac{1}{2}+\dec-2\de_0}+r^2ru^{1+\dec-2\de_0}\Big)|\dk^k\rhoc'|\\
&&+u^{1+\dec-2\de_0}\Big(r^2|\dk^k\bb'|+r|\dk^k\aa'|\Big)\Bigg\} \les \ep.
\eeaa
\end{enumerate}
 \end{proposition}

We have the following analog of Proposition \ref{prop:alternateformulaforqfinvolvingtwoangularderrivativesofrho}.
\begin{proposition}
\lab{prop:alternateformulaforqfinvolvingtwoangularderrivativesofrho:M4}
We have, relative to the  background  frame of $\Mext$,
\bea
\lab{eq:alternateformulaforqfinvolvingtwoangularderrivativesofrho:M4}
r^4\left( \dds_2\dds_1\rho+\frac{3}{4}\kab\rho\vth +\frac{3}{4}\ka\rho\vthb \right) &=&\qf    +\err
\eea
with error term expressed schematically  in the form
\bea
\err&=&   r^{2}  \dkb^{\le 2}(\Ga_b \c  \Ga_g).
\eea
\end{proposition}

\begin{proof}
We make use of Proposition \ref{prop:alternateformulaforqfinvolvingtwoangularderrivativesofrho}. 
Recall (see Remark \ref{def:wherewementionthatwealwaysexpressqfinthesecondglobalframe}) that   the quantity $\qf$ we are working with   is defined relative to the global frame of Proposition \ref{prop:existenceandestimatesfortheglobalframe:bis}. 
We thus write\footnote{The values of $r$ and $r'$ differ only by  lower order terms which do not affect the result.},
\beaa
\qf &=& r^4\left(  (\dds_2)' (\dds_1)'\rho'+\frac{3}{4}\kab'\rho'\vth' +\frac{3}{4}\ka'\rho'\vthb\right) +\err',  \\
\err'&=& r^4 e'_3 \eta' \c \b'+ r^ 2 \dk^{\le 1 }\big( \Ga_b\c\Ga_g),
\eeaa
where the primes refer to the  global frame in which $\qf$ was defined. Since in that  frame $e_3' \eta' \in r^{-1} \dk  \Ga_b$ and $\b'\in r^{-1}\Ga_g $   we    can simplify and write,
\beaa
\err'&=& r^ 2 \dk^{\le 1 }\big( \Ga_b\c\Ga_g).
\eeaa

We also have in view of Proposition \ref{prop:transformations1}
\beaa
\rho' &= &\rho  +f\bb+O(f^2 \aa),\\
\bb' &= &\bb+ \frac{1}{2}f\aa,\\
\aa'&=&\aa,\\
\kab'&=& \kab+f\xib,\\
\ka'&=& \ka+ \ddd_1\,\!'(f)    + f(\ze+\eta)   +O(r^{-1} f^2),\\
\vth' &=& \vth- \dds_2\,\!'(f)   +  f(\ze+\eta)   +      O(r^{-1}  f^2), \\
\vthb' &=& \vthb+ f\xib.
\eeaa
Note that 
\beaa
(\dds_1)' \rho&=& - e_\th' (\rho) = - e_\th \rho-\frac 1 2  f e_3\rho=\dds_1 \rho-\frac 1 2  f e_3 \rho.
\eeaa
We deduce, 
\beaa
 (\dds_2)' (\dds_1)'\rho'&=& (\dds_2)' (\dds_1)'\rho+(\dds_2)' (\dds_1)'(  \Ga_b\c \Ga_g) +\lot\\
 &=&(\dds_2)' \left(\dds_1-\frac 1 2 f e_3\right) \rho + r^{-2}\dkb^{\le 2}(\Ga_b \c \Ga_g)\\
 &=&\dds_2 \left(\dds_1-\frac 1 2 f e_3\right) \rho + r^{-2}\dkb^{\le 2}(\Ga_b \c \Ga_g)\\
 &=& \dds_2\dds_1\rho- \frac 1 2 \dds_2 f e_3 \rho + r^{-2}\dkb^{\le 2}(\Ga_b \c \Ga_g).
\eeaa
Similarly,
\beaa
\kab'\rho'\vth'
&=&\rho  \kab \big(\vth- \dds_2 f \big) + r^{-3}\dkb^{\le 1}(\Ga_g \c \Ga_g),
\\
\ka'\rho'\vthb'&=& \ka \rho \vthb+
 r^{-3}\dkb^{\le 1}(\Ga_b \c \Ga_g).
\eeaa
We deduce,
\beaa
 (\dds_2)' (\dds_1)'\rho'+\frac{3}{4}\kab'\rho'\vth' +\frac{3}{4}\ka'\rho'\vthb'&=&\dds_2 \dds_1 \rho+\frac{3}{4}\kab\rho\vth +\frac{3}{4}\ka\rho\vthb-\frac 1 2  \dds_2 f\left( e_3 \rho+\frac 3 2\kab \rho\right)  \\
 &+&  r^{-2}\dkb^{\le 2}(\Ga_b \c \Ga_g).
\eeaa
Note that,
\beaa
 \dds_2 f\left( e_3 \rho+\frac 3 2\kab \rho\right)  &=&\dds_2  f \left( \ddd_1\bb -\frac 1 2 \vth\aa +\lot\right)= r^{-2} \dkb^{\le 2 }( \Ga_g\c \Ga_b).
\eeaa
Hence
\beaa
(\dds_2)' (\dds_1)'\rho'+\frac{3}{4}\kab'\rho'\vth' +\frac{3}{4}\ka'\rho'\vthb'&=&\dds_2 \dds_1 \rho+\frac{3}{4}\kab\rho\vth +\frac{3}{4}\ka\rho\vthb + r^{-2}\dkb^{\le 2}(\Ga_b \c \Ga_g).
\eeaa
This concludes the proof of  Proposition \ref{prop:alternateformulaforqfinvolvingtwoangularderrivativesofrho:M4}.
\end{proof}

We shall also need the following analogue of  Proposition \ref{Le:Teuk-Star1}.
\begin{proposition}
\lab{Le:Teuk-Star1-M4}
The following identity holds true in $\Mext$, with respect to  its background frame

 \bea
\label{eq:Le-Teuk-Star1-M4}
\bsplit
e_3(r\qf) &= r^5\Bigg\{ \dds_2\dds_1\ddd_1\bb   -\frac{3}{2}\rho\dds_2\dds_1\kab -\frac{3}{2}\kab\rho\dds_2\ze -\frac{3}{2}\ka\rho\aa + \frac{3}{4}(2\rho^2-\ka\kab\rho)\vthb\Bigg\} \\
&+\err[e_3(r\qf)],
\end{split}
 \eea
 where
 \bea
\err[e_3(r\qf)] &=& r^3  \dk^{\le 3 }\big( \Ga_b\c\Ga_g\big).
\eea
\end{proposition}

\begin{proof}
We start with the result of Proposition \ref{Le:Teuk-Star1} which we write in the form,
\beaa
(r')^{-5}e'_3(r'\qf) &=&  (\dds_2\dds_1\ddd_1)' \bb'  -\frac{3}{2}\ka'\rho'\aa'  -\frac{3}{2}\rho' (\dds_2\dds_1)'\kab' -\frac{3}{2}\kab'\rho('\dds_2)'\ze' + \frac{3}{4}(2(\rho')^2-\ka'\kab'\rho')\vthb'\\
&+&(r')^{-5} \err[e'_3(r'\qf)]\\
\err'[e'_3(r'\qf)]&=& r'\Ga_b \qf +r^5  \dk'^{\le 1} \big(e'_3 \eta' \c \b'\big) + r'^3  \dk^{\le 2 }\big( \Ga_b\c\Ga_g\big).
\eeaa
Since $e_3' \eta' \in r^{-1} \Ga_b$  and $\qf\in \Ga_b$, we  deduce,
\beaa
\err'[e'_3(r'\qf)]= r^3  \dk^{\le 2 }\big( \Ga_b\c\Ga_g\big).
\eeaa

Now, in view of Proposition \ref{prop:transformations1},
\beaa
(\dds_2\dds_1\ddd_1)' \bb'&=&(\dds_2\dds_1\ddd_1)'\left(\b +\frac 1 2 f \aa\right)=(\dds_2\dds_1\ddd_1)' \bb+ 
 r^{-2}\dkb^3( \Ga_b\c \Ga_g).
  \eeaa
  Proceeding in the same manner with all other terms we find,
  \beaa
 &&  (\dds_2\dds_1\ddd_1)' \bb'  -\frac{3}{2}\ka'\rho'\aa'  -\frac{3}{2}\rho' (\dds_2\dds_1)'\kab' -\frac{3}{2}\kab'\rho('\dds_2)'\ze' + \frac{3}{4}(2(\rho')^2-\ka'\kab'\rho')\vthb'\\
 &&=\dds_2\dds_1\ddd_1\bb  -\frac{3}{2}\ka\rho\aa  -\frac{3}{2}\rho\dds_2\dds_1\kab -\frac{3}{2}\kab\rho\dds_2\ze + \frac{3}{4}(2\rho^2-\ka\kab\rho)\vthb + r^{-2} \dkb^{\le 3} (\Ga_b\c\Ga_g)
  \eeaa
  from which the result  easily follows.
\end{proof}


\subsection{Additional equations}


The following proposition   is an immediate  corollary  of  Proposition \ref{prop:eqtsfor-ometaxib}.
\begin{proposition}
\lab{cor:eqtsfor-ometaxib-M4}
We have, schematically,
\beaa
 2\dds_1\omb &=& \left(\frac{1}{2}\kab  +2\omb\right)\eta + e_3(\ze) -\bb -\frac{1}{2}\ka\xib+r^{-1}\Ga_g+\Ga_b^2,\\
 2\ddd_2\dds_2\eta&=&\ka\left( -e_3(\ze) +\bb\right) -e_3(e_\th(\ka)) +r^{-2} \dkb^{\le 1 }\Ga_g+r^{-1}\dkb^{\leq 1}(\Ga_b\c \Ga_b),\\
     2\ddd_2\dds_2\xib
&=& \kab\left(e_3(\ze) -\bb\right)   -e_3(e_\th(\kab))+r^{-2} \dkb^{\le 1 }\Ga_g+r^{-1}\dkb^{\leq 1}(\Ga_b\c \Ga_b).
\eeaa
\end{proposition}

\begin{remark}
\lab{remark:cor:eqtsfor-ometaxib-M4}
Note that    in fact $\Ga_g=\{ \kac, \vth, \ze, \kabc, r\rhoc\}$ and $\Ga_b=\{\vthb, \eta, \xib, \ombc, r \bb, \aa\}$ in the derivation of this proposition. 
It is important to note also  that   the terms denoted schematically  by $\dkb(\Ga_b\c \Ga_b)$ do not contain derivatives of $\ombc$.
\end{remark}

The following   corollary   of Proposition \ref{cor:eqtsfor-ometaxib-M4}  which will  be  very useful later on.
\begin{proposition}
\lab{Prop:nu*ofGCM:0}
 The following identities hold true   on $\Si_*$.
   \bea
   \lab{eq:D^5eta-prop:0}
  \bsplit
 2 \dds_2  \dds_1 \ddd_1\ddd_2\dds_2\eta&=
  \ka \Big( e_3 ( \dds_2 \dds_1\mu) +2\dds_2  \dds_1\ddd_1 \bb\Big) -\dds_2 \dds_1 \ddd_1 e_3(e_\th(\ka))
  \\
  &+ r^{-5} \dkb^{ \le 4 }\Ga_g+r^{-4 }\dkb^{\le 4} (\Ga_b\c  \Ga_b)+\lot
  \end{split} 
  \eea  
\bea
\lab{eq:D^5xib-prop:0}
\bsplit
2\dds_2\dds_1\ddd_1\ddd_2\dds_2\xib &= e_3\Big( ( \dds_2\ddd_2+ 2K)       \dds_2 \dds_1 \kab)\Big) - \kab\Big(e_3(\dds_2\dds_1\mu) +2\dds_2\dds_1\ddd_1\bb\Big)\\
&+ r^{-5} \dkb^{ \le 4 }\Ga_g+r^{-4 }\dkb^{\le 4} (\Ga_b\c  \Ga_b)+\lot
\end{split}
\eea
\end{proposition}

\begin{remark}
Here, as in    the remark following            Proposition \ref{cor:eqtsfor-ometaxib-M4},   $\Ga_g=\{ \kac, \vth, \ze, \kabc, r\rhoc\}$ and $\Ga_b=\{\vthb, \eta, \xib, \ombc, r \bb, \aa\}$.
 The quadratic  terms denoted  $\lot$ are  lower order both in terms of  decay in $r, u$ as well in terms of number of derivatives. They also contain only angular derivatives $\dkb$ and not $e_3$ nor $ e_4$.
\end{remark}

\begin{proof}
We make use of Proposition   \ref{cor:eqtsfor-ometaxib-M4} .             We shall also make  use of the conventions  mentioned in  Remark  \ref{extended-definitions-forGagb},  i.e. $\rhoc, \muc \in r^{-1} \Ga_g, \,  \bb\in r^{-1} \Ga_b,\, \aa\in \Ga_b $.

 We start with,
\beaa
 2\ddd_2\dds_2\eta&=&\ka\left( -e_3(\ze) +\bb\right) -e_3(e_\th(\ka)) +r^{-2} \dkb^{\le 1} \Ga_g+r^{-1}\dkb(\Ga_b\c \Ga_b)
\eeaa
We apply $  \dds_1 \ddd_1$ to derive,
\beaa
2  \dds_1 \ddd_1\ddd_2\dds_2\eta&=& \ka\left( -\dds_1\ddd_1 e_3(\ze) + \dds_1\ddd_1\bb\right) - \dds_1 \ddd_1e_3(e_\th(\ka))+  r^{-4} \dkb^{\le 3 }\Ga_g+r^{-3 }\dkb^{3} (\Ga_b\c \Ga_b)\\
&=&  \ka\left( - e_3(\dds_1\ddd_1 (\ze) + \dds_1\ddd_1\bb\right) - \dds_1 \ddd_1e_3(e_\th(\ka))\\
&-& \ka  [\dds_1\ddd_1, e_3]\ze +r^{-4} \dkb^{ \le 3 }\Ga_g+r^{-3 }\dkb^{3} (\Ga_b\c \Ga_b)
\eeaa
Making use of   the commutation  formula,  see Lemma \ref{Cor:comme3e4-outgeodesic-M4}, and the null structure equations for $e_3\ze, e_4 \ze $, 
\beaa
[\ddd_1, e_3]\ze=-\eta e_3 \ze+r^{-2} \dkb \ze + \Ga_b e_4 \ze+ \lot= r^{-1} \Ga_b  \c \Ga_b + r^{-2} \dkb \Ga_g+ \lot
\eeaa
we deduce, schematically,
\beaa
\,[\dds_1\ddd_1, e_3]\ze&=& \dds_1[\ddd_1, e_3]\ze+[\dds_1, e_3]\ddd_1\ze\\
&=& r^{-1} \dkb\left( r^{-1} \Ga_b  \c \Ga_b + r^{-2} \dkb\ze   +\lot\right)+ \Ga_b  e_3 \ddd_1 \ze +r^{-2} \ddd_1 \ze +\lot\\
&=& r^{-2} \dkb(\Ga_b\c \Ga_b)+ r^{-3}\dkb^2 \ze +\Ga_b \left(  \ddd_1  e_3 \ze+ \Ga_b e_3 \ze+ r^{-2} \dkb \ze\right)+\lot
\\
&=& r^{-2} \dkb(\Ga_b\c \Ga_b)+ r^{-2} \Ga_b \dkb( \dk \Ga_b)+ r^{-1} \Ga_b\c \Ga_b \c \Ga_b   + r^{-4} \dkb^2 \Ga_g\\
&=& r^{-2}   \dkb(\Ga_b \dk^{\le 1}  \Ga_b)+ r^{-4} \dkb^2 \Ga_g +\lot\
\eeaa
Hence,
\bea
 \lab{eq:D^5eta-1}
 \bsplit
2  \dds_1 \ddd_1\ddd_2\dds_2\eta&=  \ka\Big( - e_3(\dds_1\ddd_1 \ze) + \dds_1\ddd_1\bb\Big) - \dds_1 \ddd_1e_3(e_\th(\ka))\\
&+r^{-4} \dkb^{\le 3 }\Ga_g+r^{-3 }\dkb^{2} (\Ga_b\c \dkb \Ga_b)
\end{split}
\eea

 Since  $\mu=-\ddd_1\ze -\rho+\frac 1 4 \vth \vthb$, we deduce,
 \beaa
\dds_1 \mu&=&-\dds_2 \ddd_1\ze -\dds_1 \rho+  \frac 1 4\dds_1( \vth \vthb),\\
 e_3 \dds_1 \mu&=&- e_3(\dds_2 \ddd_1\ze) - e_3 \dds_1 \rho+  \frac 1 4 e_3 \dds_1( \vth \vthb)\\
 &=&- e_3(\dds_2 \ddd_1\ze) -  \dds_1 e_3  \rho   -[\dds_1, e_3]\rho+  \frac 1 4  \dds_1 e_3 ( \vth \vthb)+\frac 1 4[ e_3, \dds_1](\vth\c \vthb).
 \eeaa
 Making use  of the equations for $e_3 \rho= \ddd_1\bb  -\frac 3 2 \kab \rho +\Ga_g\c \Ga_b$ and also the  equations for\footnote{This is to avoid the presence of  $e_3, e_4  $ derivatives  in  the error terms.}      $ e_4 \rho, e_3\vth, e_3\vthb, e_4\vth,  e_4\vthb $ (and writing $\ddd_1\b=r^{-1} \dkb \b=r^{-2} \dkb \Ga_b$)
\beaa
\,[ e_3,  \dds_1] \rho&=& \Ga_b  e_3 \rho +\Ga_b e_4 \ze + r^{-2} \dkb  \rho= r^{-2} \Ga_b \dkb \Ga_b + r^{-3} \dkb \Ga_g +\lot, \\
 \, [ e_3,  \dds_1] (\vth\c\vthb)&=&  \Ga_b  e_3 (\vth\c \vthb) + \Ga_b  e_4 (\vth\c \vthb)+ r^{-2} \dkb  (\vth\c\vthb)=r^{-2}  \dkb\big(\Ga_b \c\Ga_g\big)+\lot
 \eeaa
 We deduce, ignoring the lower order terms,
 \beaa
  e_3 \dds_1 \mu&=&- e_3(\dds_2 \ddd_1\ze) -  \dds_1 \big( \ddd_1\bb  -\frac 3 2 \kab \rho +\Ga_g\c \Ga_b\big)+r^{-2}  \Ga_b\dkb  \Ga_b\big)  + r^{-2}  \dkb\big(\Ga_b \Ga_g\big) + r^{-3} \dkb \Ga_g\\
  &=& - e_3(\dds_2 \ddd_1\ze) -\dds_1  \ddd_1\bb  +\frac 3 2 \kab \dds_1 \rho + r^{-3} \dkb \Ga_g  +r^{-2}\dkb^{\le 1}( \Ga_b\c  \Ga_b). 
 \eeaa
 Hence,
 \bea
  \lab{eq:D^5eta-2}
  e_3(\dds_1\ddd_1\ze) 
  &=& -e_3(\dds_1\mu) - \dds_1 \ddd_1 \bb   + r^{-3} \dkb \Ga_g+r^{-2}\dkb^{\le 2}( \Ga_b\c  \Ga_b)+\lot
 \eea
 and thus, back to \eqref{eq:D^5eta-1},
 \bea
  \lab{eq:D^5eta-2'}
 \bsplit
 2  \dds_1 \ddd_1\ddd_2\dds_2\eta&=  \ka \Big( e_3 (\dds_1\mu) +2 \dds_1\ddd_1 \bb\Big) - \dds_1 \ddd_1e_3\big(e_\th(\ka)\big)\\
  &+r^{-4} \dkb^{ \le3 }\Ga_g+r^{-3 }\dkb^{\le 3} (\Ga_b\c  \Ga_b)+\lot
  \end{split}
 \eea
 Applying $\dds_2$  and commuting once more with $e_3$, i.e., 
 \bea
 \lab{eq:D^5eta-3'}
 \bsplit
  2  \dds_2\dds_1 \ddd_1\ddd_2\dds_2\eta&=\ka  \Big( e_3 (\dds_2\dds_1\mu) +2\dds_2 \dds_1\ddd_1 \bb\Big)-\dds_2 \dds_1 \ddd_1e_3\big(e_\th(\ka)\big)\\
  &+\ka[ \dds_2, e_3]\dds_1 \mu+ r^{-1} \dkb \Ga_g\c  \Big( e_3 (\dds_1\mu) +2 \dds_1\ddd_1 \bb\Big)\\
  &+ r^{-5} \dkb^{\le 4} \Ga_g  +r^{-4}\dkb^{\le 4}( \Ga_b\c  \Ga_b). 
  \end{split}
 \eea
 Note that, in view of \eqref{eq:D^5eta-2'} we can write,
 \bea
 \lab{eq:D^5eta-4}
 \bsplit
 e_3 (\dds_1\mu)&= 2 \ka^{-1} \dds_2 \dds_1 \ddd_1e_3\big(e_\th(\ka)\big)    - 2\dds_1\ddd_1 \bb+ 2\ka^{-1} \dds_1 \ddd_1\ddd_2\dds_2\eta \\
 &= r^{-3} \dkb^{\le 4} \Ga_b +\lot
 \end{split}
 \eea
 Hence,
  \beaa
 r^{-1} \dkb   \Ga_g\c  \Big( e_3 (\dds_1\mu) +2 \dds_1\ddd_1 \bb\Big)= r^{-4}\dkb\Ga_g \c \dkb^{\le 4} \Ga_b.
 \eeaa
 Similarly,
 \beaa
  [\dds_2, e_3] \dds_1\mu&=& \Ga_b \c  e_3  \dds_1\mu+   \Ga_b  e_4  \dds_1\mu+  r^{-3} \dkb^2 \mu +\lot\\
  &=&  r^{-3}\Ga_b  \c \dkb^{\le 4} \Ga_b         + \Ga_b \big(  \dds_1e_4  \mu+[e_4, \dds_1] \mu\big) + r^{-4} \dkb^2 \Ga_g+\lot
  \eeaa
  Thus, making use of the equation for $e_4\mu$ and combining with the estimate above,
  \beaa
    \ka[ \dds_2, e_3]\dds_1 \mu+ r^{-1} \dkb\Ga_g\c  \Big( e_3 (\dds_1\mu) +2 \dds_1\ddd_1 \bb\Big)= r^{-4}\Ga_b  \c \dkb^{\le 4} \Ga_b+ r^{-5} \dkb^{\le 2} \Ga_g.
  \eeaa
 Back to \eqref{eq:D^5eta-3'} we deduce,
 \beaa
 \bsplit
  2  \dds_2\dds_1 \ddd_1\ddd_2\dds_2\eta&=\ka  \Big( e_3 (\dds_2\dds_1\mu) +2\dds_2 \dds_1\ddd_1 \bb\Big)-\dds_2 \dds_1 \ddd_1e_3\big(e_\th(\ka)\big)\\
  &+ r^{-5} \dkb^{\le 4} \Ga_g  +r^{-4}\dkb^{\le 4}( \Ga_b\c  \Ga_b) 
  \end{split}
 \eeaa
 as desired.

To prove the second part we start with the formula for $\ddd_2 \dds_2 \,\xib$ in  Corollary
        \ref{cor:eqtsfor-ometaxib-M4}
    \beaa
     2\ddd_2\dds_2\,\xib
&=& \kab\left(e_3(\ze) -\bb\right)   -e_3(e_\th(\kab))+r^{-2} \dkb^{\le 1 }\Ga_g+r^{-1}\dkb(\Ga_b\c \Ga_b).
\eeaa
Applying $\dds_1 \ddd_1 $ and proceeding exactly as before in the derivation of \eqref{eq:D^5eta-1}  we derive,
\bea
  \lab{eq:D^5xib-1}
\bsplit
2\dds_1\ddd_1\ddd_2\dds_2\xib &= - e_3 (\dds_1\ddd_1e_\th(\kab))+ \kab\left( e_3(\dds_1\ddd_1\ze) -\dds_1\ddd_1\bb\right)\\
&+r^{-4} \dkb^{\le 3 }\Ga_g+r^{-3 }\dkb^{2} (\Ga_b\c \dk \Ga_b).
\end{split}
\eea
  Making use of \eqref{eq:D^5eta-2} we deduce, as in \eqref{eq:D^5eta-2'},
  \bea
  \lab{eq:D^5xib-2}
  \bsplit
  2\dds_1\ddd_1\ddd_2\dds_2\xib &= -e_3(\dds_1\ddd_1e_\th(\kab))+ \kab\Big(-e_3(\dds_1\mu) -2\dds_1\ddd_1\bb\Big)
  \\
   &+r^{-4} \dkb^{ 3 }\Ga_g+r^{-3 }\dkb^{\le 2} (\Ga_b\c \dk \Ga_b)+\lot
  \end{split}
  \eea
     Applying  $\dds_2$ and proceeding as in the derivation of \eqref{eq:D^5eta-3'}, by making use  of 
\eqref{eq:D^5xib-1} and       \eqref{eq:D^5eta-4}       we obtain
\beaa
\bsplit
2\dds_2\dds_1\ddd_1\ddd_2\dds_2\xib &= -e_3(\dds_2\dds_1\ddd_1e_\th(\kab)) - \kab\Big(e_3(\dds_2\dds_1\mu) +2\dds_2\dds_1\ddd_1\bb\Big)\\
&+ r^{-5} \dkb^{ \le 4 }\Ga_g+r^{-4 }\dkb^{\le 4} (\Ga_b\c \Ga_b)+\lot
\end{split}
\eeaa
The identity $\dds_1\ddd_1=\ddd_2\dds_2+2K$ yields, together with the bootstrap assumptions, 
\beaa
2\dds_2\dds_1\ddd_1\ddd_2\dds_2\xib &=& -e_3((\dds_2\ddd_2+2K)\dds_2e_\th(\kab)) - \kab\Big(e_3(\dds_2\dds_1\mu) +2\dds_2\dds_1\ddd_1\bb\Big)\\
&+& r^{-5} \dkb^{ \le 4 }\Ga_g+r^{-4 }\dkb^{\le 4} (\Ga_b\c  \Ga_b)+\lot\\
&=&e_3((\dds_2\ddd_2+2K)\dds_2\dds_1(\kab)) - \kab\Big(e_3(\dds_2\dds_1\mu) +2\dds_2\dds_1\ddd_1\bb\Big)\\
&+ &r^{-5} \dkb^{ \le 4 }\Ga_g+r^{-4 }\dkb^{\le 4} (\Ga_b\c  \Ga_b)+\lot
\eeaa
as desired.
\end{proof}


\section{Structure of the proof of Theorem M4}


We rephrase  the statement of Theorem M4 as follows. 
\begin{theorem}
\lab{theoremM4:statement}
Let $\MM=\Mint\cup \Mext$ be a GCM admissible spacetime\footnote{ In particular  the  conditions \eqref{GCM-conditionfora*-M4}--\eqref{GCM-Si*3} hold on  the spacelike  boundary $\Si_*$.}.
Under the basic bootstrap assumptions  and the results of Theorems M1-M4 (all encoded in {\bf Ref1}--{\bf Ref4})  the following estimates\footnote{ See Remark \ref{extended-definitions-forGagb} for the definition of $\Ga_g, \Ga_b$ used here.}
 hold true, for all $k\le k_{small}+8$, everywhere on $\Mext$,
\bea
\bsplit
\| \Ga_g\|_{\infty, k}&\les\ep_0 \min\left\{ r^{-2} u^{-\frac{1}{2}-\dec}, r^{-1} u^{-1-\dec}   \right\},\\
 \|e_3  \Ga_g\|_{\infty, k-1}  &\les \ep_0  r^{-2} u^{-1-\dec}, \\   
\|\Ga_b\|_{\infty, k}&\les\ep_0  r^{-1} u^{-1-\dec},
\end{split}
\eea
and,
\bea
\begin{split}
\| \b\|_{\infty, k} &\les\ep_0\min\left\{  r^{-2} (u+2r)^{-1-\dec}, r^{-3} (u+2r)^{-\frac{1}{2}-\dec}\right\},\\
\|e_3  \b\|_{\infty, k-1} &\les\ep_0  r^{-3} (u+2r)^{-1-\dec}, \\
\| \rhoc \|_{\infty, k} &\les \ep_0\min\left\{ r^{-2} u^{-1-\dec}, r^{-3} u^{-\frac{1}{2}-\dec}\right\},\\
\|  e_3 \rhoc \|_{\infty, k} &\les \ep_0  r^{-3} u^{-1-\dec},\\
\| \check{\mu} \|_{\infty, k} &\les\ep_0  r^{-3} u^{-1-\dec},\\
\| \bb  \|_{\infty, k} &\les\ep_0 r^{-2} u^{-1-\dec}.
\end{split}
\eea
Moreover, everywhere in $\Mext$,
\bea
\| \aa\|_{\infty, k} &\les\ep_0 r^{-1} u^{-1-\dec}.
\eea
\end{theorem}

Here is a short sketch of the proof of the theorem.
\begin{enumerate}
\item   \textit{Estimates on $\Si_*$}. 
 To start with we  only  have good\footnote{ i.e estimates in terms of $\ep_0$.} estimates for $\qf$, $\a$ and $\aa$, according to  {\bf Ref2}. To proceed we make use in an essential way of all 
  the GCM conditions  \eqref{GCM-Si*1}--\eqref{GCM-Si*3} on  the spacelike boundary $\Si_*$  to estimate  all the Ricci and curvature coefficients along $\Si_*$. The main result is stated in Proposition \ref{Lemma: :Estimates-onSi*-1}. The proof is   divided in the following intermediary steps.
\begin{enumerate}
\item In Proposition \ref{Prop.Flux-bb-vthb-eta-xib} we derive   flux type estimates  along $\Si_*$ for the 
quantities  $\bb, \thb, \eta, \xib$. These estimates take advantage in an essential way of the improved flux estimates  for $\qf$  in {\bf Ref 2}, equation  \eqref{improved-qf:onSi*}. This   step also makes  use of 
Proposition \ref{Le:Teuk-Star1-M4}   and  the identities of   Proposition \ref{Prop:nu*ofGCM} for $\eta, \xib$.
\item  We next estimate the $\ell=1$ modes of   the Ricci and curvature coefficients in Proposition \ref{Lemma:Bad-modes-onSi}. Besides the information provided by the estimates for $\qf, \a, \aa$  and the  GCM conditions,  an important   ingredient in the proof is  the vanishing of the $\ell=1$ mode of $ e_\th(K)$, i.e. $\int e_\th(K) e^\Phi=0$.  The flux estimates  derived in Proposition \ref{Prop.Flux-bb-vthb-eta-xib}   play an essential role in deriving the  desired estimate for  the $\ell=1$ mode of $\b$.

\item  We make use of the previous steps   to  complete the proof   all the desired estimates on $\Si_*$ in  Proposition \ref{Lemma: :Estimates-onSi*-1}. This step also  uses, in addition  to the GCM conditions,  Proposition \ref{prop:alternateformulaforqfinvolvingtwoangularderrivativesofrho:M4}    relating  $\qf$ to $\dds_2\dds_1\rho$, the Codazzi equations  and  elliptic estimates on   $2$ surfaces.

\end{enumerate}
\item \textit{First Estimates in $\Mext$.}
 We make use of  the propagation equations in $e_4$  and the estimates on  $\Si_*$  to derive some of   the desired estimates of  Theorem \ref{theoremM4:statement}, more precisely the better estimates in  powers of $r$ for the $\Ga_g$ quantities.  Note that these estimates  decay only like $u^{-1/2-\dec}$  in powers  of $u$.
\begin{enumerate}
\item  We first prove the desired estimates for   $\kac, \muc$  by  simply  integrating the  corresponding  $e_4$ equations. Note that these estimates are also well behaved in terms of powers of $u$. This is done in subsection \ref{subsection-kacmuc}.

\item    We derive  spacetime estimates for all  the   $\ell=1$ modes  in Lemma \ref{Lemma:Bad-modes-onMext}.
This is done by propagating  them from the last slice  in the $e_4$ direction, combined  with  Codazzi  equations 
and the vanishing of the $\ell=1$ mode of $e_\th(K)$.   
 
\item  We  provide  all the optimal estimates  in terms of  powers\footnote{These estimates also provide  weak   decay in $u$, i.e. $u^{-\frac{1}{2}-\dec}$ decay.} of $r$  for the quantities  $ \vth, \ze, \eta,  \kabc,   \b, \rhoc$. This is achieved in Proposition \ref{Prop:top-r-estimates} with the help of the estimates on the last slice, the propagation equation for these quantities and the estimates for the $\ell=1$ modes derived in the previous step.
\end{enumerate}  
\item \textit{Optimal $u$-decay estimates in $\Mext$}.   We  derive all the remaining estimates of Theorem   \ref{theoremM4:statement} for all  but  the quantities 
$\xib, \ombc, \Ombc,\vsic  $. The main  remaining  difficulty is to get the top decay in powers of $u$ for  $ \vth, \ze, \eta, \kabc, \b, \rhoc, \bb$.  The  result is stated in  Proposition \ref{proposition:maindecay-Mext1}.  We proceed as follows.
 \begin{enumerate}

\item One would  like to start with $\vth$  by using the  equation  $  e_4\vth+\ka\vth   =-2\a$.   This unfortunately  cannot work    by  integration\footnote{  It would work however if instead we  would integrate   from  the interior, but  we don't possess  information about  optimal $u$ decay in the interior, for example on the timelike boundary $\TT $ of $\Mext$.}   starting from the last slice $\Si_*$. Similar problems occur for   $\ze$, $\b$, $ \rhoc$.   On the other hand the quantities  $\kabc$ and $\vthb $   could in principle be propagated using their corresponding $e_4$ equations from $\Si_*$, but unfortunately  they are strongly coupled with  the other  quantities for which we don't yet have information. For  example we have,
\beaa
  e_4\check{\kab} +\frac1 2 \overline{\ka} \check{\kab}+\frac 1 2 \check{\ka} \overline{\kab} &=-2 \ddd_1\ze+2\check{\rho}+\Gac_g \c \Gac_b,
  \eeaa
  and therefore we cannot derive the estimate for $\kabc$, by integration,  before  estimating  $\ddd_1\ze$ and $\rhoc$.
  To circumvent  this difficulty we proceed  by an indirect  method as follows.
 
\item We can   derive optimal  decay    information on various  mixed quantity. For example making use of  the equation
\beaa
e_3 \a+\left(\frac 1 2 \kab -4\omb\right)\a&=&-\dds_2\b  -\frac 3 2  \vth \rho  +5\ze \b,
\eeaa
we infer   the desired decay in $u$ for the  quantity $\dds_2\b  -\frac 3 2  \vth \rho$. Other such informations can be derived from the Codazzi equations for $\vth, \vthb$, the Bianchi identity   for $\bb$ and  the identity
\eqref{eq:alternateformulaforqfinvolvingtwoangularderrivativesofrho:M4} of Lemma 
\ref{prop:alternateformulaforqfinvolvingtwoangularderrivativesofrho:M4}.

\item We combine  the control we have for $\a, \kac, \muc$ with the  control for the mixed quantities mentioned above with a propagation equation for  an intermediary  quantity,
 \beaa
 \Xi:=r^2\left(e_\th(\kab)+4r\dds_1\ddd_1\ze-2r^2\dds_1\ddd_1\b\right).
 \eeaa
 We show  in the crucial  Lemma \ref{Lemma:EstimateXi}  that $\Xi$ is a also a good mixed quantity, i.e. it has optimal decay in $u$. It is important to note that this estimate does not depend  linearly on $\aa$ for which we only have information on the last slice and $\TT$.

\item    We can combine the control of $\Xi$     with all other  available information mentioned above,      
  to derive  good estimates, simultaneously,   for  $\dds_2\dds_1 \kab,  \dds_2\ze $ and $\dds_2 \b$.  
  This is achieved  in   a sequence of crucial Lemma in subsection \ref{subsection:CrucialLemma}. 
  Unfortunately this  step is heavily dependent on the  estimate  of {\bf Ref 2} for $\aa$ and therefore the estimates we derive  are only useful  on $\TT$.

  \item  We   also show that  we  have good estimates  for  $ \dds_2 \Big (\ze,    \dds_1\kabc,  \b,   \bb,  \dds_1 \rhoc\Big)$.  To estimate $ \kabc, \ze, \b, \bb, \rhoc $ from  $ \dds_2 \Big(\ze,    \dds_1\kabc,  \b,   \bb,  \dds_1 \rhoc\Big) $    we rely on  the elliptic Hodge   Lemma \ref{prop:2D-hodge-reduced-M4} and the control we have for the $\ell=1$ modes from Lemma   \ref{Lemma:Bad-modes-onMext} derived earlier.  We obtain estimates for $\eta, \vth, \vthb$ as well. 
  This  establishes all the estimates  of   Proposition \ref{proposition:maindecay-Mext1}   on $\TT$.
  \item   The estimates mentioned above on $\TT$  can now be propagated  by  integrating  forward the $e_4$ null structure   and null Bianchi equations. This ends the proof     of   Proposition \ref{proposition:maindecay-Mext1}   in $\Mext$.
 \end{enumerate}
 \item In  Proposition \ref{Prop: endofThm4} we  derive improved decay  estimates for $  e_3 ( \b, \vth, \ze, \kabc,  \rhoc)$  and  estimates for $\xib, \ombc, \Ombc, \vsic$ in terms of  $u^{-1-\dec}$ decay.   The estimates for  $\ombc$ and $\xib$  are propagated from the last slice    using  their $e_4 $  propagation  equations. The estimate for  $\ombc$ can be easily  derived by integrating  $e_4(\check{\omb}) = \check{\rho}+\Ga_g\c \Ga_b $ form the last slice $\Si_*$. The estimate   for       $\xib$  follows by  integrating $ e_4(\xib) = -e_3(\ze)+\bb-\kab\ze+\Ga_b\c \Ga_b$  
and making use of the previously derived estimates for $e_3\ze, \b, \ze$. 
 The  estimates  for   $  \Ombc, \vsic $     follow easily   from  the equations  \eqref{geodesic-foliation-M4}.
\end{enumerate}


 \section{Decay estimates on the last slice  $\Sigma_*$}



\subsection{Preliminaries}


We shall make use of the following norms on $\Si_*$.
\bea
\|\psi\|^*_{\infty, k}(u, r):=\sum_{j\le k} \|\dk_* ^j \psi\|_{L^\infty(S(u,r))}, \qquad \dk_*^j=\sum_{j_1+j_2 \le j} \dkb^{j_1} \, ( \nu_*)^{j_2}. 
\eea
 Recall that 
   $\nu_*= \nu\Big|_{\Si_*} =e_3 + a_* e_4$, is the    tangent vector  to $\Si_*$ and (see \eqref{formula-a*-M4} \eqref{formula:Omb-onSi*-M4}), along $\Si_*$,
\beaa
 a_*&=&-\frac{2}{\vsi}+\Up-\frac r 2 \Ab =- \frac{2}{\vsi}-\Omb.
 \eeaa
 Based on on our assumptions {\bf Ref 1-2} we deduce
 \bea
 \lab{estimate:a_*-1-2m/r}
 \Big| a_*+1  +\frac{2m}{r} \Big|&\les & \ep u^{-1-\dec}.
 \eea
 
 As immediate consequence of  the commutation Corollary \ref{Cor:comme3e4-outgeodesic-M4} we derive the following,
  \begin{lemma}
  \lab{Lemma:Comm-nu*-M4}
We have, schematically,
\bea
 \,[\dkb, \nu_*] \psi &=&r\Ga_b \, ( \nu_*  \psi )  +\dk^{\le 1} \Ga_b \c \dk \psi.
 \eea
 \end{lemma}
 
 \begin{proof}
 Indeed,  see Lemma \ref{Cor:comme3e4-outgeodesic-M4},
 \bea
\bsplit
\,[\dkb , e_4]\psi&= \Ga_g\dkout  \psi,  \\
\,[\dkb, e_3]\psi&= r \Ga_b   e_3 \psi   +\Ga_b \dkout \psi      +  \lot 
\end{split}
\eea
 Hence, since $\dkb a_* \in  r\dkb \Ga_b$,
 \beaa
  \,[\dkb, \nu_*] \psi &=& \,[\dkb,  e_3 +a_* e_4] \psi = r \Ga_b   e_3 \psi   +\Ga_b \dkout \psi 
   + a_* \Ga_g\dkout  \psi+ \dkb a_*  e_4\psi\\
   &=&r \Ga_b\left( \nu_*\psi-  a_* e_4 \psi\right)  + a_* \Ga_g\dkout  \psi+ \dkb a_*  e_4\psi\\
   &=& r \Ga_b  \,  \nu_*  \psi -a_*\left(  \Ga_b \dkout \psi + \Ga_g\dkout  \psi\right)+\dkb \Ga_b \c \dk \psi\\
   &=& r \Ga_b  \,  \nu_*  \psi +\dk^{\le 1} \Ga_b \c \dk \psi
 \eeaa
 as desired.
 \end{proof}

To estimate  derivatives of the $\ell=1$ modes  on $\Si_*$ we make use of the following.
\begin{lemma}
\lab{Lemma:nuSof integrals:M4}
For every scalar function $h$ we have  the formula
\bea
 \nu_*\left(\int_{\S}h\right) &=& (\vsi)^{-1}\int_{\S}\vsi\left(\nu_*(h)+(\kab+a_*\ka)h\right).
\eea
In particular
 \bea
 \nu_*(r) &=& \frac{r}{2}(\vsi)^{-1}\ov{\vsi(\kab+a_*\ka)}.
 \eea
\end{lemma}

\begin{proof}
We  consider the  coordinates $u^\S$, $ \th^\S$ along $\Si_0$ with $\nu^\S(\th^\S)=0$. In these coordinates
we have,
\beaa
\nu^\S=\frac{2}{\vsi^\S}\pr_{u^\S}.
\eeaa
The lemma follows easily by expressing the volume element of the surfaces  $\S\subset\Si_0$   with respect to the coordinates $u^\S, \th^\S$ (see also the proof of  Proposition \ref{prop:outgoinggeod-e3e4averages}).
\end{proof}

\begin{lemma}
\label{lemma:comm-badmode}
Given $\psi\in  \sk_1(\MM)$ we have the formula,
\bea
\nu_* \left(\int_S\psi e^\Phi\right) &= \int_S(\nu_* \psi) e^\Phi+ \frac 3  2 (\ov{\kab}+ a_*  \ov{\ka})\int_S \psi e^\Phi +\err[\psi, \nu_*]
\eea
with error term
\beaa
 \err[\psi, \nu_*]&=&- \vsi^{-1}\vsic \int_S\left(      \nu_*\psi  -\frac 4 r   \psi  \right) e^\Phi+
 \vsi^{-1}\int_S\vsic \left(      \nu_*\psi   -\frac 4 r   \psi  \right) e^\Phi+ O(\ep_0 u^{-1-\dec} )\int_S  |\psi|.
 \eeaa
\end{lemma}

\begin{proof}
We have
\beaa
\nu_* \left(\int_S\psi e^\Phi\right) &=& \vsi^{-1}\int_{S}\vsi\left(\nu_*(\psi e^\Phi)+(\kab+a_*\ka)\psi e^\Phi\right)\\
&=& \vsi^{-1}\int_{\S}\vsi\left(  \nu_*\psi e^\Phi +   e^{-\Phi}\nu_*(e^\Phi)  +\kab+a_*\ka\right)\psi e^\Phi.
\eeaa
Recalling  that $e_4(\Phi)=\frac 1 2 (\ka-\vth) $,  $e_3(\Phi)=\frac 1 2 (\kab-\vthb) $ we deduce
\beaa
 e^{-\Phi}\nu_*(e^\Phi)  +\kab+a_*\ka&=& \frac 32 (\kab +a_* \ka ) -\frac 12 (\vthb-a_* \vth).
 \eeaa
 Hence, writing also  $\vsi=\ov{\vsi}+\vsic$ and then $\ka=\ov{\ka}+\kac$, $\kab=\ov{\kab}+\kabc$
 \beaa
\nu_* \left(\int_S\psi e^\Phi\right) &=& \vsi^{-1}\int_S\vsi\left(      \nu_*\psi    + \frac 32 (\kab +a_* \ka )  \psi  \right)e^\Phi -\frac 1 2  \vsi^{-1}\int_S\vsi  (\vthb-a_* \vth)\psi e^\Phi\\
&=& \vsi^{-1}\ov{\vsi}\int_S\left(      \nu_*\psi  + \frac 32 (\kab +a_* \ka )   \psi \right) e^\Phi +\vsi^{-1}\int_S\vsic \left(      \nu_*\psi  + \frac 32 (\kab +a_* \ka )    \right) e^\Phi  \\
 &-&\frac 1 2  \vsi^{-1}\int_S\vsi  (\vthb-a_* \vth)\psi e^\Phi\\
&=&\int_S\left(      \nu_*\psi  + \frac 32 (\kab +a_* \ka )  \psi  \right) e^\Phi +\vsi^{-1}\int_S\vsic \left(      \nu_*\psi  + \frac 32 (\kab +a_* \ka )  \psi  \right) e^\Phi  \\
&+&( \vsi^{-1}\ov{\vsi}-1)\int_S\left(      \nu_*\psi  + \frac 32 (\kab +a_* \ka ) \psi   \right) e^\Phi-\frac 1 2  \vsi^{-1}\int_S\vsi  (\vthb-a_* \vth)\psi e^\Phi \\
&=& \int_S(\nu_* \psi) e^\Phi+ \frac 3  2 (\ov{\kab}+ a_*  \ov{\ka})\int_S \psi e^\Phi +\err[\psi, \nu_*]
\eeaa
where,
\bea
\bsplit
\err[\psi, \nu_*]&=\vsi^{-1}\int_S\vsic \left(      \nu_*\psi  + \frac 32 (\kab +a_* \ka )  \psi  \right) e^\Phi - \vsi^{-1}\vsic \int_S\left(      \nu_*\psi  + \frac 32 (\kab +a_* \ka )  \psi  \right) e^\Phi \\
&+  \frac 32 \int_S\left( \kabc +a_* \kac    \right) \psi e^\Phi -\frac 1 2  \vsi^{-1}\int_S\vsi  (\vthb-a_* \vth)\psi e^\Phi.
\end{split}
\eea
 Using {\bf Ref1-Ref3} and \eqref{estimate:a_*-1-2m/r}
\beaa
\kab +a_* \ka&=& -\frac{2\Up}{r} +  (-1-\frac{2m}{r}) \frac{2}{r}+ O(\ep r^{-1}  u^{-1-\dec}) =-\frac{4}{r}+  O(\ep r^{-1}  u^{-1-\dec}),
\eeaa
we deduce,
 \beaa
 \err[\psi, \nu_*]&=&- \vsi^{-1}\vsic \int_S\left(      \nu_*\psi  -\frac 4 r   \psi  \right) e^\Phi+
 \vsi^{-1}\int_S\vsic \left(      \nu_*\psi   -\frac 4 r   \psi  \right) e^\Phi+  O(\ep  u^{-1-\dec}) \int_S  |\psi|
 \eeaa
 as stated.
\end{proof}


\subsection{Differential identities involving GCM conditions on $\Si_*$} 


Recall our  our GCM conditions  on $\Sigma_*$ 
\bea
\label{GCMconditions-Si*-M4}
\ka=\frac{2}{r}, \quad \dds_2\dds_1\mu=0, \quad \dds_2\dds_1\kab=0, \quad \int_S\eta e^\Phi=0, \quad \int_S\xib e^\Phi=0.
\eea
Also, on  $S_*$, the last cut of  $\Si_*$,
\bea
\int_{S_*}\b e^\Phi=0, \quad \int_{S_*}e_\th(\kab) e^\Phi = 0.
\eea

 The goal of the subsection is to  derive identities involving the GCM conditions which will be used later, see  Lemma \ref{lemma:firstfluxestimatesonSi*-etaxib}.
\begin{proposition}
\lab{Prop:nu*ofGCM}
 The following identities hold true   on $\Si_*$.
   \bea
   \lab{eq:D^5eta-prop}
  \bsplit
 2 \dds_2  \dds_1 \ddd_1\ddd_2\dds_2\eta&=
  \ka \Big( \nu_* ( \dds_2 \dds_1\mu) +2\dds_2  \dds_1\ddd_1 \bb\Big) -\dds_2 \dds_1 \ddd_1\nu_*(e_\th(\ka))
  \\
  &+ r^{-5} \dkb^{ \le 4 }\Ga_g+r^{-4 }\dkb^{\le 4} (\Ga_b\c  \Ga_b)+\lot,
  \end{split} 
  \eea  
\bea
\lab{eq:D^5xib-prop}
\bsplit
2\dds_2\dds_1\ddd_1\ddd_2\dds_2\xib &= \nu_*\Big( ( \dds_2\ddd_2+ 2K)       \dds_2 \dds_1 \kab)\Big) - \kab\Big(\nu_*(\dds_2\dds_1\mu) +2\dds_2\dds_1\ddd_1\bb\Big)\\
&+ r^{-5} \dkb^{ \le 4 }\Ga_g+r^{-4 }\dkb^{\le 4} (\Ga_b\c  \Ga_b)+\lot
\end{split}
\eea
The quadratic  terms denoted  $\lot$ are  lower order both in terms of  decay in $r, u$ as well in terms of number of derivatives.

 In particular, if the GCM conditions \eqref{GCMconditions-Si*-M4} are verified, we   deduce,
  \bea
   \lab{eq:D^5eta-xib-prop}
   \bsplit
  \dds_2  \dds_1 \ddd_1\ddd_2\dds_2\eta&=
  \ka  \dds_2  \dds_1\ddd_1 \bb + r^{-5} \dkb^{ \le 4 }\Ga_g+r^{-4 }\dkb^{\le 4} (\Ga_b\c  \Ga_b)+\lot,\\
  \dds_2\dds_1\ddd_1\ddd_2\dds_2\xib & =-\kab \dds_2\dds_1\ddd_1\bb + r^{-5} \dkb^{ \le 4 }\Ga_g+r^{-4 }\dkb^{\le 4} (\Ga_b\c  \Ga_b)+\lot
  \end{split} 
  \eea  
\end{proposition}

\begin{proof}
The proof is a straightforward application of  Proposition \ref{Prop:nu*ofGCM:0}. Indeed  according to
\eqref{eq:D^5eta-prop:0} we have
\beaa
  \bsplit
 2 \dds_2  \dds_1 \ddd_1\ddd_2\dds_2\eta&=
  \ka \Big( e_3 ( \dds_2 \dds_1\mu) +2\dds_2  \dds_1\ddd_1 \bb\Big) -\dds_2 \dds_1 \ddd_1 e_3(e_\th(\ka))
  \\
  &+ r^{-5} \dkb^{ \le 4 }\Ga_g+r^{-4 }\dkb^{\le 4} (\Ga_b\c  \Ga_b)+\lot
  \end{split} 
  \eeaa  
On the other hand since       $\nu_*= e_3 +a_* e_4$ with   $a_*= \frac{2}{\vsi_*} -\Up+\frac r 2 \Ab $, see \eqref{formula-a*-M4},
   \beaa
   e_3 ( \dds_2 \dds_1\mu)&=& \nu_*\left(  \dds_2 \dds_1\mu\right)- a_* e_4 \left(  \dds_2 \dds_1\mu\right)\\  
   &=& \nu_*\left(  \dds_2 \dds_1\mu\right) - a_* \left(  \dds_2 \dds_1  e_4 \mu +[e_4, \dds_2 \dds_1]\muc \right).
        \eeaa
    Also, in the same fashion\footnote{Note that in view of  \eqref{formula-a*-M4}  we have  $\dkb a_*\in r\Ga_b$.},  
       \beaa
\dds_2 \dds_1 \ddd_1e_3(e_\th(\ka))&=&\dds_2 \dds_1 \ddd_1\left[ \nu_*(e_\th(\ka))- a_* e_4 e_\th \ka\right]\\
&=& \dds_2 \dds_1 \ddd_1\left[( \nu_*(e_\th(\ka)) \right]-a_* \dds_2 \dds_1 \ddd_1( e_4 e_\th \ka)+   r^{-3}\sum_{i+j=2} \dkb^i a_* \dkb^j ( e_4 e_\th \ka) \\
&=&\dds_2 \dds_1 \ddd_1\left[ \nu_*(e_\th(\ka))- a_*  e_\th e_4 \ka-[e_4, e_\th] \ka \right]\\
&=& r^{-2}\sum_{i+j=2} \dkb^i   \Ga_b  \dkb^j  \left(e_\th  (e_4 \ka) +[e_\th, e_4] \ka\right).
\eeaa
      Thus, after using the transport equations for $e_4 \mu, e_4\ka$ and the commutator lemma  applied to $[e_4, e_\th]$ we  easily deduce,
    \beaa
  \bsplit
 2 \dds_2  \dds_1 \ddd_1\ddd_2\dds_2\eta&=
  \ka \Big( \nu_* ( \dds_2 \dds_1\mu) +2\dds_2  \dds_1\ddd_1 \bb\Big) -\dds_2 \dds_1 \ddd_1\nu_*(e_\th(\ka))
  \\
  &+ r^{-5} \dkb^{ \le 4 }\Ga_g+r^{-4 }\dkb^{\le 4} (\Ga_b\c  \Ga_b)+\lot
  \end{split} 
  \eeaa   
       which confirms the first  identity of the proposition.  
       
       The second part of the proposition can be derived in the same manner starting with the identity
       \eqref{eq:D^5xib-prop:0}
       \beaa
\bsplit
2\dds_2\dds_1\ddd_1\ddd_2\dds_2\xib &= e_3\Big( ( \dds_2\ddd_2+ 2K)       \dds_2 \dds_1 \kab)\Big) - \kab\Big(e_3(\dds_2\dds_1\mu) +2\dds_2\dds_1\ddd_1\bb\Big)\\
&+ r^{-5} \dkb^{ \le 4 }\Ga_g+r^{-4 }\dkb^{\le 4} (\Ga_b\c  \Ga_b)+\lot
\end{split}
     \eeaa
     This concludes the proof of the proposition.
\end{proof}


\subsection{Control of the flux of some quantities on $\Sigma_*$}


The goal of this subsection is to establish the following.
\begin{proposition}
\lab{Prop.Flux-bb-vthb-eta-xib}
The following estimate holds true for all  $k\le k_{small}+ 15$
\bea
\lab{Estimate:Flux-bb-vthb-eta-xib}
\int_{\Si_*(u, \ub)}  r^2 \big|\dk_*^ k \bb  \big|^2 + \big|\dk_*^k(\vthb, \eta, \xib)  \big|^2 &\les \ep_0^2 u^{-2-2\dec}.
\eea
We also have the weaker estimates for $k\le k_{small}+13$.
\bea
\lab{Estimate:Flux-bb-vthb-eta-xib-weak}
\bsplit
\|\bb \|^*_{\infty, k} &\les\ep_0  r^{-2} u^{-1-\dec},\\
\|(\vthb, \eta, \xib ) \|^*_{\infty, k} &\les \ep_0 r^{-1} u^{-1-\dec}.
\end{split}
\eea
In the process we also prove the following estimates  for the $\ell=1$ modes of  $\nu_*$- derivatives of $\xib, \eta$  (see \eqref{Estimates:BadModed-eta-xib-onSi_*})
\beaa
\left| \int_S( \nu_*^k \eta) e^\Phi\right|+ \left| \int_S (\nu_*^k \xib) e^\Phi\right|&\les&  \ep_0 r \,  u^{-1-\dec}, \qquad  k\le k_{small}+ 15.
\eeaa
\end{proposition}

\begin{proof} 
We concentrate  our attention on deriving \eqref{Estimate:Flux-bb-vthb-eta-xib}. Following its proof below  one can easily also  check   the  weaker estimates \eqref{Estimate:Flux-bb-vthb-eta-xib-weak} which require in fact much less work. We divide the proof in the following steps.

{\bf Step 1.} We  first   prove the corresponding   estimates for   $\bb$  away from  its   $\ell=1$ mode.  More precisely we prove.
\begin{lemma}
\lab{Le:Flux-dds_2bb}
The following estimates holds true  for all $k\le k_{small}+ 15$
\bea
\lab{Estim:Flux-dds_2bb}
\int_{\Si_*(u, \ub)}  r^4  \big| \dds_2 ( \dk_*^ {k}   \bb  )\big|^2   &\les \ep_0^2 u^{-2-2\dec}.
\eea
\end{lemma}
We postpone the proof of the lemma to  Step 7 in this subsection.

{\bf Step 2.}
We          make use of the result    of Lemma  \ref{Le:Flux-dds_2bb}    first prove the desired  estimate for  $\vthb$  i.e.,
\bea
\int_{\Si_*(u, \ub)}   \big|\dk_*^k \vthb|^2  \les \ep_0^2 u^{-2-2\dec}, \qquad k\le k_{small}+15.
\eea
\begin{proof}
One starts with the Codazzi equation
\beaa 
\ddd_2 \vthb &=& -2\bb   -\dds_1 (  \kab) -\ze \kab   + \Ga_g\c \Ga_b.
\eeaa
Differentiating w.r.t. $ \dds_2 $ and then  taking tangential derivatives $\dk_*^k$ we derive,
\beaa
\dk_* ^k  \dds_2 \ddd_2 \vthb &=& -2\dk_*^k  \dds_2 \bb   - \dk_*^k\dds_2 \dds_1 (  \kab)  - \dk_*^k\left[  r^{-2}\dkb \Ga_g +r^{-1}\dkb\left( \Ga_g\c \Ga_b\right) \right].
\eeaa
Making use of the GCM condition  $\dds_2 \dds_1 \kab=0$ along $\Si_*$  and our assumptions  {\bf Ref1-Ref2} we deduce\footnote{Note that we use \eqref{Ref1-largek}  of  {\bf Ref1}  to estimate the linear term $r^{-2}\dkb \Ga_g $. }, for all $k\le k_{small}+15$,
\beaa
\dk_* ^k  \dds_2 \ddd_2 \vthb &=& -2\dk_*^k  \dds_2 \bb+ r^{-2} \dk^{k+1} \Ga_g + r^{-1} \dk^{k+1} \Ga_g\c \Ga_b\\
&=&  -2\dk_*^k  \dds_2 \bb + \ep r^{-4}  +\lot
\eeaa
or, since   $r\ge\frac{\ep}{\ep_0}  u^{1 +\dec}$, 
\beaa
\dk_* ^k  \dds_2 \ddd_2 \vthb &=& -2\dk_*^k  \dds_2 \bb+ \ep_0 r^{-3}  u^{-1-\dec}. 
\eeaa
Moreover,
\beaa
 \dds_2 \ddd_2 \dk_* ^k \vthb  &=& -2\dk_*^k  \dds_2 \bb+ \ep_0 r^{-3}  u^{-3/2-\dec} +[ \dk_* ^k,   \dds_2 \ddd_2 ] \vthb. 
\eeaa
Using the commutator estimates  of Lemma \ref{Lemma:Comm-nu*-M4}
we derive,
\beaa
 \dds_2 \ddd_2 \dk_* ^k \vthb  &=& -2\dk_*^k  \dds_2 \bb+ \ep_0 r^{-3}  u^{-1-\dec}. 
\eeaa
Integrating and using the previously derived estimate for $\bb$ we deduce,
\beaa
\int_{\Sigma_*(u, u_*)}r^4\left| \dds_2 \ddd_2\dk_*^k \vthb  \right|^2 &\les& \ep_0^2 u^{-2-2\dec}, \qquad k\le k_{small}+15.
\eeaa
In view of the coercivity of $ \dds_2 \ddd_2 $ we  infer that,
\beaa
\int_{\Sigma_*(u, u_*)}\left| \dk_*^k  \vthb  \right|^2 &\les& \ep_0^2 u^{-2-2\dec}, \qquad k\le  k_{small}+15
\eeaa
as desired.
\end{proof}

{\bf Step 3.}  We next derive  a weak\footnote{i.e. of order $ O(\ep)$.}   estimate for the $\ell=1$ mode of $\bb$  with the help of the Codazzi equation for $\vthb$,
\beaa
2\bb&=&-\ddd_2\vthb +   e_\th(  \kab)- \ov{\kab} \ze  + \Ga_g\c \Ga_b\nn\\
&=&-\ddd_2\vthb +   e_\th(  \kabc)+\frac{2\Up}{r}  \ze  + \Ga_g\c \Ga_b.
\eeaa 
Projecting on the $\ell=1$ mode  and using the bootstrap assumptions {\bf Ref1-Ref2} we deduce,
\beaa
\lab{firstestimateforbadmodeof-b}
2\int_S\bb e^\Phi &=& \frac{2\Up}{r}  \int_S\ze e^\Phi + \int_S e_\th (\kabc)  e^\Phi  +\int_S \Ga_g\c \Ga_b  e^\Phi\\
&=& O(\ep)+O(\ep_0u^{-3/2-2\dec}).\nn
\eeaa
The same estimate holds true for  the tangential  derivatives. More precisely  taking tangential derivatives and projecting on the $\ell=1$ mode,
\beaa
2\int_S( \dk_*^k  \bb )e^\Phi &=&\int_S \dk_*^k     \Big[ e_\th(  \kabc)+\frac{2\Up}{r}  \ze  + \Ga_g\c \Ga_b\Big] e^\Phi- \int_S \big[\dk_*^k, \ddd_2\big] \vthb e^\Phi. 
\eeaa
Using the bootstrap assumptions {\bf Ref1-Ref2}  and the commutator Lemma \ref{Lemma:Comm-nu*-M4} we  deduce,
\bea
\lab{Estimate-BadMode-bb}
\int_S( \dk_*^k  \bb )e^\Phi=O(\ep).
\eea

{\bf Step 4.} We combine the result of Lemma
\ref{Le:Flux-dds_2bb} with  \ref{Estimate-BadMode-bb}
to deduce
\bea
\int_{\Si_*(u, \ub)}  r^2   \big|\dk_*^  k \bb  \big|^2   &\les \ep_0^2 u^{-2-2\dec}, \qquad k\le k_{small}+15.
\eea
Indeed, according   the  last  elliptic estimate   of Lemma \ref{prop:2D-hodge-reduced-M4},
 \beaa
  \int_S  r^2  \big|\dk_*^  k \bb  \big|^2&\les& r^4   \int_S   \big|\dds_2(\dk_*^  k \bb)  \big|^2+ r^{-2}\left|\int_S( \dk_*^k  \bb )e^\Phi\right|^2\les  r^4   \int_S   \big|\dds_2(\dk_*^  k \bb)  \big|^2+ \ep^2 r^{-2}.
 \eeaa
 Since $r\ge (\ep \ep_0^{-1}) u^{-3/2 -\dec} $ on $\Si_*$ we deduce,
 \beaa
  \int_S  r^2  \big|\dk_*^  k \bb  \big|^2 \les  r^4   \int_S   \big|\dds_2(\dk_*^  k \bb)  \big|^2+ \ep_0^2 u^{-3-2\dec}. 
  \eeaa
  Thus, integrating and making use of estimate \eqref{Estim:Flux-dds_2bb},
 \beaa
 \int_{\Si(u, u_*)}   r^2  \big|\dk_*^  k \bb  \big|^2&\les &  \int_{\Si(u, u_*)}   r^4  \big| \dds_2(\dk_*^  k \bb)  \big|^2+\ep_0^2 u^{-2-2\dec}\\
 &\les&\ep_0^2 u^{-2-2\dec}
 \eeaa
as stated.  This establishes the estimate for $\bb$ in Proposition \ref{Prop.Flux-bb-vthb-eta-xib}.

{\bf Step 5.}  To finish the proof of Proposition \ref{Prop.Flux-bb-vthb-eta-xib} it remains to establish the estimates for $\eta$ and $\xib$.
As for $\bb$ we  prove  separately   estimates for $\dds_2 (\eta, \xib)$ and   estimates  for  the $\ell=1$ modes of  the tangential derivatives with respect to $\nu_*$ of    $\eta, \xib$ (recall that the $\ell=1$ modes of $ \eta, \xib$ were  set to zero).  We      then  combine them, as in the case  of $\bb$,  to provide the desired estimates.  

\begin{lemma}\lab{lemma:firstfluxestimatesonSi*-etaxib}
We have
\beaa
\int_{\Si_*(u, u_*) } r^2 \Big(  | \dds_2(\dk_*^k \eta)|^2+| \dds_2(\dk_*^k \xib)|^2\Big) &\les& \ep_0^2 u^{-2-2\dec},  \qquad  k\le k_{small}+ 15.
\eeaa
\end{lemma}
We postpone the proof of the lemma to  Step 8 of this section.

{\bf Step 6.} We provide below estimates for the $\ell=1$ modes of $\xib$  and  $\eta$ and use them, in combination with  Lemma \ref{lemma:firstfluxestimatesonSi*-etaxib},  to close the estimates of Proposition \ref{Prop.Flux-bb-vthb-eta-xib}, i.e. we show that,
\bea
\lab{Estimates:Flux-eta-xib}
 \int_{\Si_*(u, u_*) } \Big(  |\dk_*^k \eta|^2+| \dk_*^k \xib|^2\Big) &\les& \ep_0^2 u^{-2-2\dec}, \qquad  k\le k_{small}+ 15.
\eea
In the process we also prove the following estimates  for the $\ell=1$ modes of  $\nu_*$- derivatives of $\xib, \eta$,
\bea
\lab{Estimates:BadModed-eta-xib-onSi_*}
\left| \int_S( \nu_*^k \eta) e^\Phi\right|+ \left| \int_S (\nu_*^k \xib) e^\Phi\right|&\les&  \ep_0 r \,  u^{-1-\dec}, \qquad  k\le k_{small}+ 15.
\eea

\begin{proof}   We prove separately the estimates for $\xib$ and $\eta$.

{\bf Step 6a.}  We prove first the estimate for $\xib$.     
Since the $\ell=1$ mode of $\xib$ vanishes we have, in view  of  the elliptic estimates Lemma \ref{prop:2D-hodge-reduced-M4} and  the estimates of Lemma \ref{lemma:firstfluxestimatesonSi*-etaxib},
\bea
\lab{eq:firstfluxestimatesonSi*-etaxib1}
\int_{\Si(u, u_*)}  \big|\dkb  \xib \big|^2  +\big|\xib \big|^2    &\les&   \int_{\Si(u, u_*)} r^2   \big|\dds_2  \xib \big|^2         \les      \ep_0^2 u^{-2-2\dec}. 
\eea
According to our GCM conditions we have   $\int_S \xib e^\Phi=0$.
In view of   Lemma \ref{lemma:comm-badmode} we  deduce,
\beaa
 \int_S(\nu_* \xib ) e^\Phi&=&\err[\xib, \nu_*]
 \eeaa
 where,
 \beaa
 \err[\xib, \nu_*]&=&- \vsi^{-1}\vsic \int_S\left(      \nu_*\xib  -\frac 4 r   \xib  \right) e^\Phi+
 \vsi^{-1}\int_S\vsic \left(      \nu_*\xib  -\frac 4 r   \xib \right) e^\Phi+ O(\ep_0u^{-1-\dec} )  \int_S  |\xib|.
 \eeaa
 We deduce,
\beaa
\int_S(\nu_* \xib ) e^\Phi&=& \vsi^{-1} \left[ -\vsic \int_S (\nu_*\xib ) e^\Phi    +\int_S \vsic  (\nu_*\xib ) e^\Phi     \right]+O(\ep_0   u^{-1-\dec} )\int_S|\xib|. 
\eeaa
Hence,
\beaa
\left|\int_S(\nu_* \xib ) e^\Phi-  \vsi^{-1} \int_S \vsic  (\nu_*\xib ) e^\Phi  \right|&\les& O(\ep) \left|\int_S (\nu_*\xib ) e^\Phi \right|+O(\ep_0) r    u^{-1-\dec} 
\eeaa
i.e., setting $B_1:= \int_S(\nu_* \xib ) e^\Phi-$ and $I_1= \vsi^{-1} \int_S \vsic  (\nu_*\xib ) e^\Phi$,
\bea
\big| B_1-I_1\big|&\les \ep B_1+\ep_0 r  u^{-1-\dec}. 
\eea

To control   $I_1$  we  decompose $ \nu_*\xib$ into its $\ell=1$ mode $B_1=\int_S(\nu_* \xib ) e^\Phi$ and its orthogonal complement
\beaa
 \nu_*\xib = B_1 \frac{1}{\int_S e^{2\Phi}} e^\Phi+ ( \nu_*\xib)^\perp.
\eeaa
Thus, in view of our elliptic estimates,   
\beaa
\int_S \Big| (\nu_*\xib)^\perp\Big|^2 &\les&  r^2 \int_S \Big|\dds_2(\nu_*\xib)^\perp\Big|^2=
 r^2 \int_S \Big|\dds_2(\nu_*\xib) \Big|^2.
\eeaa
We deduce,
\beaa
I_1&=& \int_S \vsic  (\nu_*\xib ) e^\Phi=  \frac{B_1}{ \int_S  e^{2\Phi} }\int \vsic e^{2\Phi}+ \int_S \vsic  (\nu_*\xib )^\perp e^\Phi\\
&\les& \ep |B_1|+\ep r^2 \left( \int_S\Big | (\nu_*\xib )^\perp \Big|^2\right)^{\frac{1}{2}}\\
&\les&  \ep |B_1|+ \ep r^3 \left( \int_S\Big |\dds_2  (\nu_*\xib ) \Big|^2\right)^{\frac{1}{2}}.
\eeaa
Hence,
\bea
\lab{estimate-forB1-M4}
|B_1|&\les&  \ep r^3 \left( \int_S\Big |\dds_2  (\nu_*\xib ) \Big|^2\right)^{1/2}+ \ep_0 r  u^{-1-\dec}. 
\eea
Finally,
\beaa
\int_S \big|\nu_*\xib \big|^2 &\les&  r^2 \int_S \big|\dds_2 (\nu_*\xib) \big|^2 +r^{-4}     B_1^2\\
&\les&  r^2 \int_S \big|\dds_2 (\nu_*\xib) \big|^2+\ep_0^2  r^{-2}   u^{-2-2\dec} 
\eeaa
or, since $r \ge u^{1/2} $,
\beaa
\int_S \big|\nu_*\xib \big|^2 &\les& r^2 \int_S \big|\dds_2 (\nu_*\xib) \big|^2+\ep_0^2   u^{-3-2\dec}.
\eeaa
Integrating   we derive,
\beaa
\int_{\Si(u, u_*)}  \big|\nu_*\xib \big|^2 &\les&\int_{\Si(u, u_*)} r^2  \big|\dds_2 (\nu_*\xib) \big|^2+\ep_0^2 u^{-2-2\dec}. 
\eeaa
Thus, making use of \eqref{lemma:firstfluxestimatesonSi*-etaxib},
\beaa
\int_{\Si(u, u_*)}  \big|\nu_*\xib \big|^2 &\les& \ep_0^2 u^{-2-2\dec} 
\eeaa
which, together with \eqref{eq:firstfluxestimatesonSi*-etaxib1}, yields
\beaa
\int_{\Si(u, u_*)}  \big|\dk_* \xib \big|^2    + \big| \xib \big|^2    &\les& \ep_0^2 u^{-2-2\dec}. 
\eeaa
Moreover  we 
as desired. Returning to \eqref{estimate-forB1-M4} we also infer that,
\beaa
|B_1|&\les&  \ep r^3 \left( \int_S\Big |\dds_2  (\nu_*\xib ) \Big|^2\right)^{1/2}+ \ep_0 r  u^{-1-\dec} \les  \ep_0 r  u^{-1-\dec}. 
\eeaa
This proves  the estimates \eqref{Estimates:Flux-eta-xib}, \eqref{Estimates:BadModed-eta-xib-onSi_*}  for $\xib$ and $k\le 1$.
The proof for the higher  tangential derivatives   can be  derived in the same manner.
 
{\bf Step 6b.}  To prove the desired estimates for $\eta$ we make use of the GCM condition $\int_S \eta e^\Phi=0$.
We deduce, by elliptic estimates, 
\beaa
\int_S |\eta|^2 +|\dkb \eta |^2 &\les& \int_S r^2 | \dds_2 \eta|^2+r^{-4}\left| \int_S \eta e^\Phi \right| ^2= \int_S r^2 | \dds_2 \eta|^2.
\eeaa
Hence, integrating and using  the estimates of Lemma \ref{lemma:firstfluxestimatesonSi*-etaxib},
\bea
\lab{eq:firstfluxestimatesonSi*-etaxib2}
\int_{\Si(u, u_*)}  |\eta|^2+|\dkb\eta| ^2 &\les& \ep_0^2 u^{-2-2\dec}
\eea
which is the exact analogue of \eqref{eq:firstfluxestimatesonSi*-etaxib1} in Step 6a.

To estimate  $\int_S |\nu_* \eta|^2$  we proceed  exactly as in step  6a  by  making use of the GCM condition 
 $\int_S \xib e^\Phi=0$ and Lemma  \ref{lemma:comm-badmode}  to deduce
\beaa
 \int_S(\nu_* \eta) e^\Phi&=& \err[\eta,  \nu_*]
\\
&=&- \vsi^{-1}\vsic \int_S\left(      \nu_*\eta -\frac 4 r   \eta  \right) e^\Phi+
 \vsi^{-1}\int_S\vsic \left(      \nu_*\eta -\frac 4 r   \eta \right) e^\Phi+O(\ep_0   u^{-1-\dec} )\int_S|\eta|. 
 \eeaa
The term $\err[ \eta, \nu_*]$ can be dealt with precisely as  $\err[ \xib, \nu_*]$ in Step 6a.
Proceeding exactly  as before in Step 6a  we deduce,
\beaa
\int_{\Si(u, u_*)}|\nu_* \eta| ^2 &\les& \ep_0^2 u^{-2-2\dec}.
\eeaa
Thus, combined with  \eqref{eq:firstfluxestimatesonSi*-etaxib2}, we infer that,
\beaa
\int_{\Si(u, u_*)}  |\eta|^2+|\dk_*\eta| ^2 &\les& \ep_0^2 u^{-2-2\dec}
\eeaa
and
\beaa
\left|\int_S(\nu_* \eta) e^\Phi\right| &\les& \ep_0 r u^{-1-\dec} 
\eeaa
which  establishes  the estimates \eqref{Estimates:Flux-eta-xib}, \eqref{Estimates:BadModed-eta-xib-onSi_*}  for $\eta$ and $k\le 1$.
The estimates for  the higher tangential derivatives can be proved in the same manner. This completes the proof of 
  both  inequalities  \eqref{Estimates:Flux-eta-xib} and \eqref{Estimates:BadModed-eta-xib-onSi_*}.
\end{proof}

At this point we have  reduced  all estimates of Proposition  \ref{Prop.Flux-bb-vthb-eta-xib}   to the proofs of 
Lemmas \ref{Le:Flux-dds_2bb}  and \ref{lemma:firstfluxestimatesonSi*-etaxib}.

{\bf Step 7.} We prove  Lemma  \ref{Le:Flux-dds_2bb} i.e. we show that,
\beaa
\int_{\Si_*(u, \ub)}  r^4  \big| \dds_2 ( \dk_*^ {k}   \bb  )\big|^2   &\les \ep_0^2 u^{-2-2\dec}, \qquad k\le k_{small}+15.
\eeaa

We make use of Proposition \ref{Le:Teuk-Star1-M4}  according to which
\beaa
e_3(r\qf) &=& r^5\Bigg\{ \dds_2\dds_1\ddd_1\bb  -\frac{3}{2}\ka\rho\aa  -\frac{3}{2}\rho\dds_2\dds_1\kab -\frac{3}{2}\kab\rho\dds_2\ze + \frac{3}{4}(2\rho^2-\ka\kab\rho)\vthb\Bigg\} \\
&+&\err[e_3(r\qf)]
 \eeaa
 where,
 \beaa
 \err[e_3(r\qf)] &=& r^3  \dk^{\le 3 }\big( \Ga_b\c\Ga_g\big).
 \eeaa
 It is easy to check,  with the help of   the estimates  {\bf Ref1} and {\bf  Ref2} and  the product Lemma \ref{lemma:interpolation-decay}, that  for all $k\le k_{small}+16$,
\bea
\lab{estimate-err[e_3(rqf)]}
\|   \err[e_3(r\qf)] \|_{\infty, k}(u, r) &\les& \ep_0\left(  r^{-1/2} u^{-1 -\dec} + u^{-3/2-\dec}\right).
\eea
We can also check, making use of  the estimates\footnote{which hold for all $k\le k_{large}-2$.} 
 \eqref{Ref1-largek} of {\bf Ref1} for large $r$,
\beaa
\|\rho\dds_2\dds_1\kab,\,\,  \kab\rho\dds_2\ze,\,\,  \ka\kab\rho\vthb, \, \,   \rho^2 \vthb \|_{\infty, k} \les \ep\Big( r^{-7}+ r^{-6} u^{-1-\dec} \Big).
\eeaa
In view of our assumption  for $r$ on $\Si_*$ we have  $r\ge\frac{\ep}{\ep_0}  u^{1+\dec}$, we thus  deduce for all $k\le k_{small}+17$
\beaa
\Big\| e_3(r\qf) -r^5\left(\dds_2\dds_1\ddd_1\bb -\frac{3}{2}\ka\rho\aa \right)\Big\|_{\infty, k}& \les& \ep r^{-2}
 +\ep_0\big( r^{-1/2} u^{-1 -\dec} + u^{-3/2-\dec}\big)\\
&\les &\ep_0  u^{-3/2-\dec}.
\eeaa
We  infer that,
\beaa
\|r^4\dds_2\dds_1\ddd_1\bb\|_{\infty, k}&\les& \| r^{-1} e_3(r\qf) \|_{\infty, k}+ \| \aa\|_{\infty, k}+ \ep_0r^{-1}  u^{-3/2-\dec}.
\eeaa
Thus integrating on the last slice $\Si_*$ and making use of   the assumptions \eqref{improved-qf:onSi*} and \eqref{improved-aa:onSi*},  i.e. 
\beaa
\int_{\Sigma_*(u, u_*)}|e_3\dk^{ k} \qf|^2+ r^{-2} |\qf|^2 +  | \dk^{ k} \aa |^2 &\les \ep^2_0(1+u)^{-2-2\dec},\qquad  k\le k_{small}+14,
\eeaa
 we deduce,
 \beaa
\int_{\Si_*(u, u_*)} r^8 \big| \dk^k (\dds_2\dds_1\ddd_1\bb)\big|^2&\les&\ep_0^2 u^{-2-2\dec}.
\eeaa
 In particular,
 \beaa
\int_{\Si_*(u, u_*)} r^8 \big| \dk_*^k (\dds_2\dds_1\ddd_1\bb)\big|^2&\les&\ep_0^2 u^{-2-2\dec}
\eeaa
 where $\dk_*^k =\nu_*^{k_1} \dkb ^{k_2}$ denote the tangential derivatives to $\Si_*$. Taking into account the commutator Lemma \ref{Lemma:Comm-nu*-M4} we deduce, for $k\le k_{small} +14$,
 \bea
 \int_{\Si_*(u, u_*)} r^8 \big| \dds_2\dds_1\ddd_1 ( \dk_*^k\bb)\big|^2&\les&\ep_0^2 u^{-2-2\dec}.
 \eea 
Since
\beaa
\dds_1\ddd_1=\ddd_2\dds_2+2K,
\eeaa 
we infer that
\beaa
\int_{\Sigma_*(u, u_*)}r^8\left| (\dds_2\ddd_2+2K)\dds_2 (\dk^k_*\bb)\right|^2 &\les& \ep_0^2 u^{-2-2\dec}. 
\eeaa
In view of the  coercivity  of $\dds_2\ddd_2+2K$ we deduce,
\beaa
\int_{\Sigma_*(u, u_*)}r^4\left| \dds_2 (\dk^k_*\bb)\right|^2 &\les& \ep_0^2 u^{-2-2\dec}, \qquad k\le k_{small}+16.
\eeaa
This concludes the proof of Lemma  \ref{Le:Flux-dds_2bb}. In view of the results  derived in the first three steps this also establishes, unconditionally,  the
 estimate,
 \bea
 \lab{Estimate:Flux-bb}
\int_{\Si_*(u, \ub)}  r^2   \big|\dk_*^  k \bb  \big|^2   &\les \ep_0^2 u^{-2-2\dec}, \qquad k\le k_{small}+15
\eea
which will be used in the next step.

{\bf Step 8.}  We  prove Lemma  \ref{lemma:firstfluxestimatesonSi*-etaxib} based on the identities of  Proposition \ref{Prop:nu*ofGCM}.
  To derive the desired flux estimate for  $\eta$ we make use of  the first part of Proposition \ref{Prop:nu*ofGCM}
  according to which we have,
   \beaa
  \bsplit
 2 \dds_2  \dds_1 \ddd_1\ddd_2\dds_2\eta&=
  \ka \Big( \nu_* ( \dds_2 \dds_1\mu) +2\dds_2  \dds_1\ddd_1 \bb\Big) -\dds_2 \dds_1 \ddd_1\nu_*(e_\th(\ka))
  \\
  &+ r^{-5} \dkb^{ \le 4 }\Ga_g+r^{-4 }\dkb^{\le 4} (\Ga_b\c  \Ga_b)+\lot
  \end{split} 
  \eeaa  
  Since,
   $\dds_1\ddd_1=\ddd_2\dds_2+2K$, we deduce,
   \beaa
    \dds_2 (\ddd_2\dds_2+2K ) \ddd_2\dds_2\eta&=&\frac 1 2\Big[  \ka  \nu_* ( \dds_2 \dds_1\mu)  -\dds_2 \dds_1 \ddd_1\nu_*(e_\th(\ka))\Big]+
 \ka    \dds_2 (\ddd_2\dds_2+2K )  \bb\\
    &+& r^{-5} \dkb^{ \le 4 }\Ga_g+r^{-4 }\dkb^{\le 4} (\Ga_b\c  \Ga_b)+\lot
   \eeaa
   or,
   \beaa
    (\ddd_2\dds_2+2K )\left[ \dds_2  \ddd_2\dds_2\eta - \ka \dds_2 \bb\right]&=&\frac 1 2\Big[  \ka  \nu_* ( \dds_2 \dds_1\mu)  -\dds_2 \dds_1 \ddd_1\nu_*(e_\th(\ka))\Big]\\
     &+& r^{-5} \dkb^{ \le 4 }\Ga_g+r^{-4 }\dkb^{\le 4} (\Ga_b\c  \Ga_b)+\lot
   \eeaa  
    Taking higher tangential derivatives and   using our GCM assumptions on $\Si_*$
    \beaa
   \dk_*^k  (\ddd_2\dds_2+2K )\Big[ \dds_2  \ddd_2\dds_2\eta - \ka \dds_2 \bb\Big]&=& \dk_*^k \Big[ r^{-5}  \dkb^{ \le 4 }\Ga_g+r^{-4 }\dkb^{\le 3} (\Ga_b\c \dk \Ga_b)\Big]+\lot
    \eeaa
   Making use of  the  commutation Lemma  \ref{Lemma:Comm-nu*-M4}  we can rewrite,
   \beaa
    r^2  (\ddd_2\dds_2+2K )\Big[ \dds_2  \ddd_2  \dds_2( \dk_*^k\eta) - \ka \dds_2(  \dk_*^k \bb)\Big]&=& \sum_{j\le k}  \dkb^{\le 2}  \Big[ r^{-3} \dkb^{ \le 2  } \dk_*^j\Ga_g+r^{-2 }\dkb^{\le 1} \dk_*^j (\Ga_b\c \dk \Ga_b)\Big].
   \eeaa
   Using the    ellipticity  of the operator  $(\ddd_2\dds_2+2K )$,  assumptions {\bf Ref 2- Ref 4},  interpolation Lemma  \ref{lemma:interpolation-decay} and  condition $r\ge( \ep \ep_0^{-1}) \,  u^{ 3/2 +\dec}$ on $\Si_* $   we derive,
   \beaa
   \dds_2  \ddd_2  \dds_2 (\dk_*^k\eta)&=&\ka  \dds_2( \dk_*^k \bb)+O\Big(\ep  r^{-5} + \ep_0 r^{-4} u^{-2-2\dec} \Big)\\
   &=& \ka  \dds_2( \dk_*^k  \bb)+O\Big(\ep_0 r^{-4} u^{-3/2 -\dec}\Big).
   \eeaa
   Using also the ellipticity of $ \dds_2  \ddd_2 $ and  the assumption  $r\ge \ep (\ep_0^{-1})  u^{ 3/2 +\dec}$ on $\Si_* $,
   \beaa
   \|\dds_2 ( \dk_*^k\eta)\|_{L^2(S)} \les & r \|( \dk_*^k\bb)\|_{L^2(S)} + \ep_0 r^{-1} u^{-3/2-\dec}.
   \eeaa
   Finally, squaring,   integrating on $ \Si_*$ and taking into account the  flux estimate for $\bb$  in  \eqref{Estimate:Flux-bb} we deduce
   \beaa
   \int_{\Si_*(u, u_*) } r^2 \big|   \dds_2(\dk_*^k  \eta)\big |^2 &\les& \int_{\Si_*(u, u_*) }  r^4 \big| (\dk_*^k  \bb)\big|^2 + \ep_0 u^{-2-2\dec} \les \ep_0 u^{-2-2\dec}.
   \eeaa  
   Hence,
        \beaa
   \int_{\Si_*(u, u_*) } r^2 \big|   \dds_2(\dk_*^k  \eta)\big |^2 &\les& \ep_0 u^{-2-2\dec} \les \ep_0 u^{-2-2\dec}
   \eeaa    
  as  stated. This completes the proof of Lemma \ref{lemma:firstfluxestimatesonSi*-etaxib}  for $\eta$. The proof for $\xib$ is very similar and will thus be omitted. This   therefore also completes the proof of Proposition  \ref{Prop.Flux-bb-vthb-eta-xib}.
   \end{proof}


\subsection{Estimates for the $\ell=1$ modes on $\Si_*$}


In the previous subsection we have already derived estimates for the $\ell=1$ modes of $\bb, \eta, \xib$. In what follows we  derive estimates for the remaining $\ell=1$ modes. We  summarize the results in the following proposition.
\begin{proposition}
\lab{Lemma:Bad-modes-onSi}
The following estimates hold true for all  $k\le k_{small}+16$,
\bea
\label{Lemma:Bad-modes-onSi-eq1}
\bsplit
 r^{-1} \left| \int_S(\nu^k_* \eta) e^\Phi \right|  +r^{-1} \left| \int_S(\nu^k_* \xib ) e^\Phi \right|     &\les  \ep_0 u^{-1-\dec}, \\
\left|\int_S( \nu_*^k  \bb )e^\Phi\right|  + r^{-1}     \left| \int_S(\nu^k_* \ze ) e^\Phi \right|  + \left| \int_S(\nu^k_* ( e_\th \kab ) e^\Phi \right|   &\les   \ep_0  u^{-1-\dec},\\
 \left|\int_S \nu_*^k( e_\th\rho) e^\Phi\right| +\left|\int_S \nu_*^k( e_\th\mu) e^\Phi\right|+\left| \int_S  (\nu_*^k \b )  e^\Phi\right|  &\les   \ep_0 r^{-1} u^{-1-\dec},\\
\left| \int_S  \nu_*^k( e_\th \omb)  e^\Phi\right| &\les   \ep_0 r u^{-1 -\dec}.
\end{split}
\eea
\end{proposition}

\begin{proof}
Note that the estimates in the first line of \eqref{Lemma:Bad-modes-onSi-eq1}      have  already been  derived in the previous subsection.  The proof  of the remaining ones  is done in a series of steps starting with the estimate for the $\ell=1$ mode 
of $e_\th(\kab) $ and ending with that for the $\ell=1$ mode of $\b$. To derive the correct estimate for  the latter we need to derive in fact  stronger  intermediary  estimates than those stated in the proposition.

{\bf Step 1.}  We prove the desired estimate for  $\int_S \ze e^\Phi$.
Indeed,
 in view of  the  Codazzi equations and the GCM condition on $\ka$,
\beaa
\ddd_2  \vth  &=& -2\b+ (e_\th(\ka) + \ze\ka )+   \Ga_g\c \Ga_g=-2\b+\frac 2 r \ze + \Ga_g\c \Ga_g.
\eeaa
Integrating,
\beaa
\int_S \ze e^\Phi &=&  - r \int_S \b e^\Phi + r^4  \Ga_g\c \Ga_g.
\eeaa
Thus,    in view of  {\bf Ref1-3 } for $\b$ and $\Ga_g$, 
\beaa
\left| \int_S \ze e^\Phi \right| &\les & \ep r^{1/2-\de_B/2} +\ep_0 u^{-1-2\dec}.
\eeaa
The  higher derivative estimates can be proved in the same manner.
Hence,
\bea
\lab{estimate-Badmode-ze-first}
 \left| \int_S(\nu^k_* \ze ) e^\Phi \right| \les  \ep r^{1/2-\de_B/2}  +\ep_0 u^{-1-2\dec}, \qquad k\le k_{small}+16.
\eea
In particular, for  $ r \ge\Big( \frac{\ep}{\ep_0}\Big)^2  u^{2+2\dec}$,
\beaa
\left| \int_S(\nu^k_* \ze ) e^\Phi \right| \les   r \ep_0 u^{-1-\dec}, \qquad k\le k_{small}+16
\eeaa
 as  stated in the proposition.

{\bf Step 2.} We prove here  the following stronger estimate for the $\ell=1$ mode of $e_\th \kab$.
\bea
\lab{Estimate-badmode-e_thkab}
\left|\int_S( \nu_*^k  e_\th \kab  )e^\Phi\right| &\les & \ep_0 u^{-2-\dec}, \qquad k\le k_{small}+16.
\eea

To  control the $\ell=1$ mode of $e_\th(\kab)$ on $\Sigma_*$   we need  the precise  identity\footnote{Note that the schematic  form of Corollary   \ref{cor:eqtsfor-ometaxib-M4}  is not suitable here.} of Proposition
\ref{prop:eqtsfor-ometaxib}
\beaa
 e_3(e_\th(\kab))&=& - 2\ddd_2\dds_2\xib + \kab\left(e_3(\ze) -\bb\right) +\kab^2\ze-\frac 3 2 \kab e_\th \kab + 6\rho\xib -2\omb e_\th(\kab)+\err[\ddd_2\dds_2\xib],
\\
\err[\ddd_2\dds_2\xib]&=&\left(2\ddd_1\xib+\frac{1}{2}\kab\,\vthb+2\eta\xib-\frac{1}{2}\vthb^2\right)\eta + 2e_\th(\eta\xib)  -\frac{1}{2}e_\th(\vthb^2)\\
&+&\kab\left(\frac{1}{2}\vthb\ze  -\frac{1}{2}\vth\xib\right)-\frac 1 2 \vthb e_\th \kab  -\frac{1}{2}\vth\vthb\xib
-\ze\left(2\ddd_1\xib+2(\eta-3\ze)\xib-\frac{1}{2}\vthb^2\right)\\
&+&\xib\Big(-\vth\vthb -2\ddd_1\ze+2\ze^2\Big)-6\eta\ze\xib -6e_\th(\ze\xib).
\eeaa
The error term can be written schematically as,
\beaa
\err[\ddd_2\dds_2\xib] &=& \left(2\ddd_1\xib+\frac{1}{2}\kab\,\vthb+2\eta\xib-\frac{1}{2}\vthb^2\right)\eta + 2e_\th(\eta\xib)  -\frac{1}{2}e_\th(\vthb^2)\\
&&+r^{-1}\dkb^{\le 1}( \Ga_g\c \Ga_b ) +\Ga_g\c\Ga_b\c \Ga_b.
\eeaa
Note also that we can write, schematically,
\beaa
\kab^2\ze-\frac 3 2 \kab e_\th \kab + 6\rho\xib -2\omb e_\th(\kab) &=& r^{-2}  \ze+ r^{-1} e_\th(\kab) + r^{-3}\xib +r^{-1}\dkb^{\le 1}( \Ga_g\c \Ga_b ) +\Ga_g\c\Ga_b\c \Ga_b.
\eeaa
Also, using the transport equation for $e_4(\kab)$,
\beaa
e_4(e_\th(\kab))&=&  e_\th e_4\kab+ [e_4, e_\th] \kab\\
&=& e_\th\left[-\frac 1 2 \ka\, \kab  -2\ddd_1\ze + 2\rho  +\Ga_g \c \Ga_b \right]+ \frac 1 2  \ka  e_\th\kab +\lot
\\
&=& r^{-1} e_\th \kab +\dds_1\ddd_1\ze  +e_\th(\rho) + r^{-1}\dkb(\Ga_g \c \Ga_b)\\
&=& r^{-1} e_\th \kab + (\ddd_2 \dds_2 + 2K) \ze  +e_\th(\rho) + r^{-1}\dkb(\Ga_g \c \Ga_b).
\eeaa
We can also write,  since $\nu_*=e_3+ a_* e_4$
\beaa
e_3(\ze)=\nu_* (\ze)- a_* e_4 (\ze)=\nu_* (\ze) - a_*\left( - \ka\ze -\b +  \Ga_g\c \Ga_g \right).
\eeaa
Therefore, writing also $\kab=-\frac{2\Up}{r}+\Ga_g$,
\bea
\lab{Estimate-BadModeSi*30}
\bsplit
 \nu_* (e_\th(\kab))&=-2\ddd_2\dds_2\xib- \frac{2\Up}{r} \left(\nu_* (\ze) - \bb\right) +E_1+E_2 + E_3,\\
E_1&= r^{-2}  \ze+ r^{-1} e_\th(\kab) + r^{-3}\xib  +\ddd_2\dds_2\ze   + r^{-1} \b  +e_\th(\rho),\\
E_2&= \left(2\ddd_1\xib+\frac{1}{2}\kab\,\vthb+2\eta\xib-\frac{1}{2}\vthb^2\right)\eta + 2e_\th(\eta\xib), 
 -\frac{1}{2}e_\th(\vthb^2)\\
 E_3&= r^{-1}\dkb^{\le 1}( \Ga_g\c \Ga_b ) +\Ga_g\c\Ga_b\c \Ga_b.
 \end{split}
\eea
Projecting on the $\ell=1$ mode we derive,
\bea
\lab{Estimate-BadModeSi*31}
\int_S \nu_*\Big( e_\th(\kab)   - \kab \ze\Big)   e^\Phi&=&- \frac{2\Up}{r} \int_S  \bb e^\Phi +\int_S (E_1+E_2+E_3) e^\Phi.
\eea
On the other hand, according to Lemma \ref{lemma:comm-badmode}
\beaa
\nu_* \left(\int_S e_\th (\kab ) e^\Phi\right) &= \int_S(\nu_*  e_\th (\kab) ) e^\Phi+ \frac 3  2 (\ov{\kab}+ a_*  \ov{\ka})\int_S  e_\th(\kab) e^\Phi +\err[e_\th(\kab) , \nu_*]
\eeaa
with error term
\beaa
 \err[e_\th(\kab), \nu_*]&=&- \vsi^{-1}\vsic \int_S\left(      \nu_*e_\th(\kab)  -\frac 4 r   e_\th(\kab)  \right) e^\Phi+
 \vsi^{-1}\int_S\vsic \left(      \nu_*e_\th(\kab)   -\frac 4 r   e_\th(\kab)  \right) e^\Phi\\
 &+& O(\ep_0 u^{-1-\dec} )\int_S  |e_\th(\kab)|.
 \eeaa
We easily  deduce
\bea
\lab{esstimate:err-nu*1}
\err[e_\th(\kab) , \nu_*]&=& r  \dkb^{\le 1}  \Ga_b \c \Ga_b +\lot
\eea
i.e.,
\beaa
\nu_* \left(\int_S e_\th (\kab ) e^\Phi\right) &=& \int_S(\nu_*  e_\th (\kab) ) e^\Phi+  O(r^{-1} ) \int_S  e_\th(\kab) e^\Phi + r  \dkb^{\le 1}  \Ga_b \c \Ga_b +\lot\\
&=& \int_S(\nu_*  e_\th (\kab) ) e^\Phi + r\dkb \Ga_g + r  \dkb^{\le 1}  \Ga_b \c \Ga_b +\lot
\eeaa
Similarly,
\beaa
\nu_* \left(\int_S ( \kab \ze )  e^\Phi\right) &= \int_S(\nu_*  \kab  \ze ) e^\Phi   + r\dkb \Ga_g+ r  \dkb^{\le 1}  \Ga_b \c \Ga_b +\lot 
\eeaa

Combining with \eqref{Estimate-BadModeSi*31}
 we deduce,
\beaa
\bsplit
 \nu_* \int_S \Big( e_\th(\kab)   - \kab \ze\Big)   e^\Phi&=- \frac{2\Up}{r} \int_S  \bb e^\Phi     +\int_S (E_1+E_2+E_3) e^\Phi\\ 
&+ r\dkb \Ga_g + r  \dkb^{\le 1}  \Ga_b \c \Ga_b +\lot
\end{split}
\eeaa
Now schematically, in view of  {\bf Ref 1-2},
\beaa
\int_S  E_1 e^\Phi&=&\int_S\Big(r^{-2}  \ze+ r^{-1} e_\th(\kab)     + r^{-1} \b  +e_\th(\rho)\Big) e^\Phi= r \dkb^{\le 1} \Ga_g\\
\int_S  E_3 e^\Phi&=&  r^2 \dkb^{\le 1}( \Ga_g\c \Ga_b ) + r^3 \Ga_g\c\Ga_b\c \Ga_b.
\eeaa
Hence,
\bea
\lab{Estimate-BadModeSi*32}
\bsplit
 \nu_*\int_S \Big( e_\th(\kab)   - \kab \ze\Big)   e^\Phi&=- \frac{2\Up}{r} \int_S  \bb e^\Phi   + r\dkb^{\le 1} \Ga_g +\int_S E_2 e^\Phi \\ 
& + r  \dkb^{\le 1}  \Ga_b \c \Ga_b   + r^3 \Ga_g\c\Ga_b\c \Ga_b +\lot
\end{split}
\eea
In particular, in view of the estimate for the $\ell=1$ mode of $\bb$ derived in the previous subsection (see \eqref{Estimate-BadMode-bb})
 and  our assumptions on $\Ga_g, \Ga_b$, 
\beaa
 \nu_* \left(\int_S\Big (e_\th( \kab)  -\kab \ze\Big) e^\Phi\right)= \int_S E_2 e^\Phi  + O( \ep r^{-1}) 
\eeaa
 or, in view of the condition $r\ge (\ep \ep_0^{-1}) u^{3+\dec}$, 
\beaa
 \nu_* \left(\int_S\Big (e_\th( \kab)  -\kab \ze\Big) e^\Phi\right)=     \int_S E_2 e^\Phi   +      O\Big(\ep_0 u^{-3-\dec}\Big).
\eeaa
Integrating  along $\Sigma_*$ from $S_*$  and making use of the vanishing  of the $\ell=1$ mode   of $e_\th(\kab)$ on $S_*$ and the estimate for  the $\ell=1$ mode of $\ze$  derived earlier  in step 0, see \eqref{estimate-Badmode-ze-first}, 
 we deduce,
\beaa
\int_{S(u)} \Big (e_\th( \kab)  -\kab \ze\Big) e^\Phi&=&\int_{S_*} \Big (e_\th( \kab)  -\kab \ze\Big)    e^\Phi  +O\Big (\ep_0 u^{-2-\dec} \Big)
 +\int_{\Si(u, u_*)}  E_2 e^\Phi\\
&=&\frac{2\Up }{r} \int_{S_*}  \ze   e^\Phi  +O\Big (\ep_0 u^{-2-\dec} \Big)    +\int_{\Si(u, u_*)}  E_2 e^\Phi\\
   &=&   O(\ep r^{-1/2-\frac 1 2 \de_B} )+       O\Big (\ep_0 u^{-2-\dec} \Big) +\int_{\Si(u, u_*)}  E_2 e^\Phi.
\eeaa 
Since  $\min_{\Si_*} r  \ge (\ep \ep_0^{-1})^2  u^{4} $ and $\de_B \ge 2\dec$ 
we infer that
\beaa
\left|\int_{S(u)} \Big (e_\th( \kab)  -\kab \ze\Big) e^\Phi\right| &\les \ep_0 u^{-2-\dec}  +\int_{\Si(u, u_*)}    r|E_2|.
\eeaa

It remains to estimate the  term  $\int_{\Si(u, u_*)} r| E_2|$ where, recall,
\beaa
E_2&= \left(2\ddd_1\xib+\frac{1}{2}\kab\,\vthb+2\eta\xib-\frac{1}{2}\vthb^2\right)\eta + 2e_\th(\eta\xib).  
 -\frac{1}{2}e_\th(\vthb^2).
 \eeaa
Hence, appealing to Proposition \ref{Prop.Flux-bb-vthb-eta-xib}
\beaa
 \int_{\Si(u, u_*)}  r  \big| E_2 \big|&\les& \int_{\Si(u, u_*)} \Big(\big|\dkb^{\le 1} \vthb \big |^2+\big|\dkb^{\le 1} \eta\big |^2+\big|\dkb^{\le 1} \xib\big|^2\Big)\\
 &\les& \ep_0^2 u^{-2-2\dec}.
\eeaa
Thus,
\beaa
\left|\int_{S(u)} |e_\th( \kab) e^\Phi\right| &\les &    \left| \int_{S(u)} \kab \ze e^\Phi \right|+   \ep_0^2 u^{-2-2\dec}\les  r^{-1/2} \ep+ \ep_0^2 u^{-2-2\dec}\\
&\les & \ep_0^2 u^{-2-2\dec}
\eeaa
which establishes the desired estimate in this case. 

To derive a similar estimate for the   first  tangential  derivative we go back to  \eqref{Estimate-BadModeSi*31}
\beaa
\int_S \nu_*\Big( e_\th(\kab)   - \kab \ze\Big)   e^\Phi&=&- \frac{2\Up}{r} \int_S  \bb e^\Phi +\int_S (E_1+E_2+E_3) e^\Phi
\eeaa
 from which we easily derive, as before,
 \beaa
 \left| \int_S \nu_*\Big( e_\th(\kab)   - \kab \ze\Big)   e^\Phi\right|&\les& \ep r^{-1}+ \ep_0 u^{-2-2\dec} \les \ep_0 u^{-2-2\dec}
 \eeaa
 and therefore,
 \beaa
 \left| \int_S \nu_* e_\th(\kab)\right| &\les&\ep r^{-1/2}+  \ep_0 u^{-2-2\dec}\les \ep_0 u^{-2-2\dec}.
\eeaa
To estimate   the second tangential derivative we make use  once more of  Lemma \ref{lemma:comm-badmode}.
Setting $\psi:= \nu_*\Big( e_\th(\kab)   - \kab \ze\Big)  $ we  deduce,
\beaa
\int_S(\nu_* \psi) e^\Phi&=&\nu_* \left(\int_S\psi e^\Phi\right) -\err[\psi, \nu_*]\\
&=&\nu_*\left(- \frac{2\Up}{r} \int_S  \bb e^\Phi +\int_S Ee^\Phi\right) -\err[\psi, \nu_*]
\eeaa
with $E=E_1+E_2+E_3 $ and  error term
 \beaa
\err[\psi, \nu_*]&=&-\vsi^{-1} \check{\vsi} \int_S (e_3 \psi+ \frac 3 2  \kab \psi)e^\Phi +\vsi^{-1}\int_S \check{\vsi}\big(e_3 \psi + \frac 32 \kab \psi\big) e^\Phi )+O(\ep_0 u^{-1-\dec} )\int_S  |\psi |\\
\eeaa
which can be estimated as  in  \ref{esstimate:err-nu*1} and shown to  be lower order.
Note that, using again  Lemma \ref{lemma:comm-badmode},
\beaa
\nu_*\left(- \frac{2\Up}{r} \int_S  \bb e^\Phi +\int_S E e^\Phi\right)&=&
-  \frac{2\Up}{r}     \int_S (\nu_* \bb) e^\Phi +\int_S (\nu_* E) e^\Phi+\lot\\
&=& O(\ep r^{-1})+ O(\ep_0 u^{-2-2\dec} )=  O(\ep_0 u^{-2-2\dec})
\eeaa
and we then proceed as before.  All higher tangential derivatives can be treated in the same manner.

{\bf Step 3.} We establish the estimate for the $\ell=1$ mode of $\b$.
Recall that we had, see \eqref{firstestimateforbadmodeof-b},
\beaa
2\int_S\bb e^\Phi &=& \frac{2\Up}{r}  \int_S\ze e^\Phi + \int_S e_\th (\kabc)  e^\Phi  +\int_S \Ga_g\c \Ga_b  e^\Phi.
\eeaa
Thus, using the estimates already derived in Steps 0 and 1,
\beaa
\left|\int_S\bb e^\Phi \right|\les \ep_0u^{-1-\dec}.
\eeaa
The higher derivative estimates  are derived in the same manner.

{\bf Step 4.}  We establish the estimate 
\bea
\label{Estimate:Badmode-e_thrho--1}
\left|\int_S \nu_*^k   e_\th(\rho) e^\Phi\right| &\les& \ep_0  r^{-1} u^{-2-2\dec}+  \int_S  \big| \nu_*^k \dkb (\vth \vthb) \big|.
\eea
 We start by   differentiating the Gauss equation $ K=-\rho -\frac 1 4 \ka\,\kab + \frac 1 4 \vth \vthb$.  we derive Using the GCM condition for $\ka$ we derive,
\beaa
e_\th(\rho) &=& - e_\th(K)   -\frac{1}{2r} \, e_\th(\kab)  +         \frac 1 4 e_\th(\vth \vthb).  
\eeaa
We  make use of  the vanishing of the $\ell=1$ mode of $e_\th(K)$ (see Lemma \ref{lemma:Gauss-curvatureS}) to derive
\bea
\lab{Estimate:Badmode-e_thrho-0}
\int_S  e_\th(\rho) e^\Phi &=&   -\frac{1}{2r} \int_S  e_\th(\kab)  e^\Phi     +  \frac 1 4 \int_S  e_\th(\vth \vthb)   e^\Phi.
\eea
Using the previously deduced estimate for  the $\ell=1$ mode of $e_\th(\kab)$ we  deduce
\bea
\lab{Estimate:Badmode-e_thrho}
\left|\int_S  e_\th(\rho) e^\Phi\right| &\les& \ep_0  r^{-1} u^{-2-2\dec}+ \int_S  |\vthb | |\dkb \vth|+ |\vth | |\dkb \vthb|.
\eea
Note that in particular we deduce,
\beaa
\left|\int_S  e_\th(\rho) e^\Phi\right| &\les& \ep_0  r^{-1} u^{-3/2-2\dec}
\eeaa
but we shall need  the precise form \eqref{Estimate:Badmode-e_thrho}  in Step  7.

To estimate  $\int_S( \nu_*  e_\th(\rho)) e^\Phi$ we apply $\nu_*$ to  \eqref{Estimate:Badmode-e_thrho}  and make use of  Lemma \ref{lemma:comm-badmode} as in  Step 2.
\beaa
\int_S(\nu_*  e_\th (\rho) ) e^\Phi&=&\nu_* \left(\int_S e_\th (\rho ) e^\Phi\right) -\frac 3  2 (\ov{\kab}+ a_*  \ov{\ka})\int_S  e_\th(\rho) e^\Phi -\err[e_\th(\rho) , \nu_*]\\
&=& \nu_*\left[  -\frac{1}{2r} \int_S  e_\th(\kab)  e^\Phi     +  \frac 1 4 \int_S  e_\th(\vth \vthb)   e^\Phi\right] +O(r^{-1} ) \int_S  e_\th(\rho) e^\Phi - \err[e_\th(\rho) , \nu_*]\\
&=& -\frac{1}{2r}  \nu_*\Big( \int_S  e_\th(\kab)  e^\Phi\Big) +\frac 1 4  \int_S  \nu_*\big[ e_\th(\vth \vthb) \big]  e^\Phi+\lot
\eeaa
Hence, in view of the results derived in the first step, we deduce \eqref{Estimate:Badmode-e_thrho--1} for $k=1$. The case of higher derivatives can be treated in the same manner.

{\bf Step 5.}  To estimate the $\ell=1$ mode of $\mu$ we  differentiate the relation $\mu=-\div \ze -\rho+ \Ga_g\c \Ga_b $,
\beaa
e_\th(\mu)&=&\dds_1\ddd_1 \ze - e_\th(\rho)+ r^{-1} \dkb(\Ga_g\c \Ga_b) \\
&=&(\ddd_2\dds_2 + 2K) \ze  - e_\th(\rho)+ r^{-1} \dkb(\Ga_g\c \Ga_b) \\
&=&  \ddd_2\dds_2 \ze +\frac{2}{r^2} \ze -  e_\th(\rho) +r^{-1} \dkb(\Ga_g\c \Ga_b)+\lot 
\eeaa
Hence,
\bea
\lab{Estimate:Badmode-e_thmu-0}
\int_S e_\th(\mu)  e^\Phi&=&2 r^{-2} \int_S  \ze e^\Phi -\int_S e_\th(\rho)  e^\Phi+ r^2 \dkb(\Ga_g\c \Ga_b). 
\eea
Using the previous estimates for the $\ell=1$ modes of $\ze$ and $e_\th \rho$ we deduce,
\beaa
\left|\int_S e_\th(\mu)  e^\Phi\right| &\les&   \ep_0  r^{-1} u^{-1-\dec}.
\eeaa
The higher $\nu_*$  derivatives   can be derived exactly as in the previous steps by differentiating 
\eqref{Estimate:Badmode-e_thmu-0} and applying   Lemma \ref{lemma:comm-badmode}.

{\bf Step 6.} We estimate the $\ell=1$ mode of $\omb$ making use of  the  first identity in Corollary \ref{cor:eqtsfor-ometaxib-M4}
\beaa
 2\dds_1\omb &=& \left(\frac{1}{2}\kab  +2\omb\right)\eta + e_3(\ze) -\bb +\frac{1}{2}\ka\xib+r^{-1}\Ga_g+\Ga_b^2\\
 &=&(-\frac{\Up}{r} + 2\frac{m}{r^2})       \eta+e_3(\ze) -\bb+\frac 1 r \xib +r^{-1}\Ga_g+\Ga_b^2+\lot\\
  &=&(-\frac{\Up}{r} + 2\frac{m}{r^2})       \eta+\nu_*(\ze)  - a_* e_4\ze -\bb+\frac 1 r \xib +r^{-1}\Ga_g+\Ga_b^2+\lot\\
  &=&-\frac  1 2 \eta +\nu_*(\ze) -\bb+\frac 1 r \xib +r^{-1}\Ga_g+\Ga_b^2+\lot
 \eeaa 
 Hence,
 \beaa
 2\int_S \dds_1(\omb ) e^\Phi&=&  -\frac 12 \int_S \eta e^\Phi  +\int_S \nu_*\ze e^\Phi        -\int_S \bb e^\Phi+ r^{-1}\int_S \xib e^\Phi+ r^2 \Ga_g + r^2 \Ga_b\c \Ga_b.
 \eeaa
 Taking into account  the  prescribed  GCM  conditions for  the $\ell=1$ modes of $\eta$  and $\xib$ and the   previous results           for the $\ell=1$ modes of $\nu_*(\ze), \bb$  we deduce,
 \beaa
\left| \int_S  e_\th( \omb)  e^\Phi\right| \les   \ep_0 r u^{-3/2 -\dec}.
\eeaa
The higher $\nu_*$ derivatives can be treated as in the previous steps.

{\bf Step 7.}  It remains to estimate the $\ell=1$ mode of $\b$ and  show
\bea
\lab{Estimate:Badmode-beta}
\left| \int_S  \nu_*^k (\b)  e^\Phi\right| \les  \ep_0 r^{-1} u^{-1-\dec}.
\eea
 We start with the $e_3 \b$  equation  which we write in the form
\beaa
e_3(\b)&=&  e_\th(\rho)+  3 \ov{\rho} \, \eta +   J+   O\big(r^{-1}(\a, \b) \big), \qquad J:=3\eta\rhoc-\vth\bb.
\eeaa
Also,  taking into account the $e_4$ equation for $\b$,
\beaa
\nu_*(\b)&=& e_3(\b)+ a_* e_4 \b= e_3 (\b)+ a^*\left(-2\ka \b+ \ddd_1\a+ \ze\a \right))\\
&=&   e_\th(\rho)+  3 \ov{\rho} \, \eta+ J+O\big( r^{-1} (\b,   \dkb^{\le 1} \a) \big)+\lot
\eeaa
Projecting on the $\ell=1$ mode,
\bea
\lab{Estimate-BadModeSi*71}
\int_S \nu_*(\b) e^\Phi&=&  \int_S e_\th(\rho)e^{\Phi}    +\int_S J e^\Phi+O\big( r^{2} (\b,   \dkb^{\le 1} \a) \big).
\eea
On the other hand, making use of Lemma \ref{lemma:comm-badmode},
\beaa
\nu_* \left(\int_S \b  e^\Phi\right) &=& \int_S(\nu_* \b ) e^\Phi+ \frac 3  2 (\ov{\kab}+ a_*  \ov{\ka})\int_S \b  e^\Phi +\err[\b, \nu_*]\\
&=&\int_S(\nu_* \b ) e^\Phi+O\big( r^{2} \b \big)+\err[\b, \nu_*]
\eeaa
with error term
 \beaa
\err[\b, \nu_*]&=&-\vsi^{-1} \check{\vsi} \int_S (e_3 \b + \frac 3 2  \kab  \b )e^\Phi +\vsi^{-1}\int_S \check{\vsi}\big(e_3 \b+ \frac 32 \kab \b \big) e^\Phi )+\ep_0 u^{-1-\dec}\int_S |\b|.
\eeaa
Note that $ e_3\b =e_\th\rho+\lot =r^{-2} \dkb^{\le 1} \Ga_g$. Hence,
\beaa
\err[\b, \nu_*]&=&r^3 ( r^{-2}  \Ga_b\c  \dkb^{\le 1} \Ga_g)+\lot
\eeaa
Hence, 
\bea
\lab{Estimate-BadModeSi*72}
\nu_* \left(\int_S \b  e^\Phi\right) &= \int_S(\nu_* \b ) e^\Phi +O\big( r^{2} \b \big)  + r \Ga_b \c \dkb^{\le 1} \Ga_g+\lot
\eea
Thus combining to \eqref{Estimate-BadModeSi*71},
\beaa
\nu_* \left(\int_S \b  e^\Phi\right) &=&  \int_S e_\th(\rho) e^\Phi +3\ov{\rho}  \int_S \eta  e^\Phi +\int_S J e^\Phi+O\big( r^{2} (\b,   \dkb^{\le 1} \a) \big)+r \Ga_b \c \dkb^{\le 1} \Ga_g
\eeaa
and, since $\nu_*(r) =O(1) $
\beaa
\nu_* \left( r \int_S \b  e^\Phi\right)&=&r \int_S  e_\th(\rho) e^\Phi  +3r\ov{\rho}  \int_S \eta  e^\Phi+r\int_S J e^\Phi+O\big( r^{3} (\b,   \dkb^{\le 1} \a)\Big)+r^2 \Ga_b \c \dkb^{\le 1} \Ga_g.
\eeaa

Note that, based on {\bf Ref1-Ref2} and condition  $ \min_{\Si_*} r\ge  (\ep \ep_0^{-1})^2 r^{ 4}$ and $\de_B\ge 2 \dec$,
\beaa
 r^{3} \big|(\b,   \dkb^{\le 1} \a) \big|&\les & \ep r^{-1/2-1/2\de_B} \les \ep_0  u^{-2-\dec},\\
 r^2 \big|\Ga_b \c \dkb^{\le 1} \Ga_g\big|&\les& \ep r^{-1} \les \ep_0 u^{-2-\dec}.
\eeaa
Recall also  that the $\ell=1$ mode of   $e_\th(\rho)$ has   been estimated  in
\eqref{Estimate:Badmode-e_thrho} 
\beaa
  \Big| r \int_S  e_\th(\rho) e^\Phi\Big| &\les& \ep_0   u^{-2-2\dec}+r \int_S  |\vthb | |\dkb \vth|+ |\vth | |\dkb \vthb|.
\eeaa
Therefore, using also that the $\ell=1$ mode of $\eta$ vanishes on $\Si_*$ due to our GCM conditions,
\bea
\lab{Estimate-BadModeSi*75}
\Big| \nu_* \left( r \int_S \b  e^\Phi\right)\Big|&\les& \ep_0 u^{-2-2\dec} +r \int_S  |\vthb | |\dkb \vth|+ |\vth | |\dkb \vthb|+ r|J|.
\eea

Hence, integrating on $\Si_*$ starting from $S_*$  and using the GCM condition $\int_{S_*} \b e^\Phi=0$,
we deduce
\beaa
\Big| r  \int_{S(u)} \b  e^\Phi  \Big| &\les&      \ep_0 u^{-1-\dec}+ \int_{\Si(u, u_*)} r \Big( |\vthb | |\dkb \vth|+ |\vth | |\dkb \vthb|+  r |J |\Big).
\eeaa
Using the flux estimates of Proposition \ref{Prop.Flux-bb-vthb-eta-xib} we   infer that
\beaa
\int_{\Si(u, u_*)} r\left( |\vthb | |\dkb \vth|+ |\vth | |\dkb \vthb|\right)&\les&\left( \int_{\Si(u, u_*)}  |\vthb |^2+ |\dkb\vthb |^2\right)^{1/2}\left( \int_{\Si(u, u_*)}  r^2\left( |\vth |^2+ |\dkb\vth |^2\right)\right)^{1/2}\\
&\les& \ep_0  u^{-1-\dec}\left( \int_{\Si(u, u_*)} r^2\left(  |\vth |^2+ |\dkb\vth |^2\right)\right)^{\frac{1}{2}}. 
\eeaa
On the other hand, according to {\bf Ref 1},
\beaa
 \int_{\Si(u, u_*)} r^2\left(  |\vth |^2+ |\dkb\vth |^2\right)\les \ep^2  \int_{\Si(u, u_*)}   r^{-2} u^{-1-2\dec}  \les \ep^2. 
\eeaa
Hence,
\beaa
\int_{\Si(u, u_*)} r\left( |\vthb | |\dkb \vth|+ |\vth | |\dkb \vthb|\right)\les \ep \ep_0     u^{-1-\dec}. 
\eeaa
Similarly, recalling the definition of $J=3\eta\rhoc-\vth\bb$,
\beaa
\int_{\Si(u, u_*)} r^2 |J| &\les&\int_{\Si(u, u_*)} r^2\left( \big |\eta\big| \big| \rhoc\big|   + \big|\vth\big|\big|\bb\big|\right)
\\
&\les&\Big(\int_{\Si(u, u_*)}\left( |\eta|^2 + r^2 |\bb|^2 \right)\Big)^{1/2} \c 
\Big(\int_{\Si(u, u_*)} r^2 \left( |\vth|^2 + r^2 |\rhoc|^2 \right)\Big)^{1/2} \\
&\les& \ep_0 u^{-1-\dec}\c  \Big(\int_{\Si(u, u_*)} r^2 \left( |\vth|^2 + r^2 |\rhoc|^2 \right)\Big)^{1/2}\les   \ep_0 \ep u^{-1-\dec}.
\eeaa
Therefore,
\beaa
\left|   \int_S \b  e^\Phi  \right| &\les&       \ep_0 r^{-1}  u^{-1-\dec}
\eeaa
as desired. To derive  \eqref{Estimate:Badmode-beta} for  $k=1$ we  go back to
\eqref{Estimate-BadModeSi*72} which we write in the form 
\beaa
  \int_S(\nu_* \b ) e^\Phi &=&  r^{-1} \nu_* \left( r \int_S  \b  e^\Phi\right) +O\big( r^{2} \b \big)  + r \Ga_b \c \dkb^{\le 1} \Ga_g+\lot\\
  &=&r^{-1} \nu_* \left( r \int_S  \b  e^\Phi\right) +O(\ep r^{-3/2}). 
\eeaa
Hence,  making use of \eqref{Estimate-BadModeSi*75},  the assumptions {\bf Ref1-2} and the condition on $r$, $r\ge (\ep \ep_0^{-1})^2 u^{2+2\dec}$,
\beaa
\left|  \int_S(\nu_* \b ) e^\Phi  \right|&\les&   \ep_0 u^{-2-2\dec} +r \int_S \big( |\vthb | |\dkb \vth|+ |\vth | |\dkb \vthb|+ r|J|\big) +\ep r^{-3/2} \\
&\les& \ep_0 r^{-1} u^{-1-\dec}.
\eeaa
The higher derivative estimates are derived in the same manner.
This completes the proof of Proposition \ref{Lemma:Bad-modes-onSi}.
\end{proof}


\subsection{Decay of  Ricci and  curvature components on $\Si_*$}


The goal  here is to prove the following proposition. 
\begin{proposition}
 \label{Lemma: :Estimates-onSi*-1}
 The following estimates hold true along $\Si_*$  for all $k\le k_{small}+19$
 \bea
 \bsplit
 \| \vthb, \,\eta, \, \xib,  \,  \ombc,  r\bb \|^*_{\infty, k}  &\les \ep_0 r^{-1} u^{-1-\dec},\\
 \|  \Ombc, \,  \ov{\Omb}+\Up, \,  \vsic, \,  \ov{\vsi}-1  \|^*_{\infty, k} &\les \ep_0  u^{-1-\dec},\\
 \|\kac,  \, \kabc,  \,  \,r\muc \|^*_{\infty, k} &\les \ep_0 r^{-2} u^{-1-\dec},\\
 \| \vth, \,\ze,  \, r \rhoc \|^*_{\infty, k}&\les  \ep_0 r^{-2} u^{-1/2-\dec},\\
\| \b \|^*_{\infty, k}&\les \ep_0r^{-3} (2r+u)^{-1/2-\dec}. 
\end{split}
 \eea
 \end{proposition}

\begin{proof}
Recall that we have already derived the desired   estimates for $\bb,\vthb,  \eta, \xib$, see \eqref{Estimate:Flux-bb-vthb-eta-xib-weak}, in Proposition \ref{Prop.Flux-bb-vthb-eta-xib}. To prove the remaining estimates we proceed in steps as follows.

 {\bf Step 1.}   In this step  we  prove the estimate,
\bea
\lab{EstimatesonSi*-kabc}
\|\kabc\|^*_{\infty, k}&\les & \ep_0r^{-2} u^{-1-\dec}.
\eea
 This makes use of the GCM condition  $\dds_2\dds_1 \kabc=0$,  the control of the $\ell=1$ mode of $\dds_1\kabc$  provided by Lemma \ref{Lemma:Bad-modes-onSi},  the  bootstrap assumptions and   our dominance condition on $r$ along $\Si_*$.     We also make use of the commutation properties of the  tangential derivatives  on $\Si_*$ with
 the GCM condition.

  According to part 4 of the  elliptic Hodge   Lemma  \ref{prop:2D-hodge-reduced-M4} we derive,
  \beaa
  \|  \dds_1 \kabc\|_{\hk_{k+1}(S)}&\les& r   \|  \dds_2 \dds_1 \kabc\|_{\hk_k(S)}+ r^{-2}  \Big |\int_S e^\Phi \dds_1 \kabc\,   \Big |= r^{-2}  \Big |\int_S e^\Phi \dds_1 \kabc\,   \Big |.
  \eeaa
  According to    Proposition  \ref{Lemma:Bad-modes-onSi}   we have, for all   $0\le k\le k_{small}+20$
 \beaa
  \Big |\int_S e^\Phi \dds_1 \kabc\,   \Big | &\les& \ep_0 u^{-1-\dec}.
\eeaa
Therefore,
\beaa
  \|  \dds_1 \kabc\|_{\hk_{k+1}(S)}&\les& \ep_0  r^{-2} u^{-1-\dec}.
\eeaa
Thus,
\beaa
 \|   \kabc\|_{\hk_{k+2}(S)}&\les& \ep_0  r^{-1} u^{-1-\dec}
\eeaa
and therefore, for $k\le k_{small}+20$
\beaa
 \| \dkb^{\le k}   \kabc\|_{L^\infty(S)} &\les&  \ep_0  r^{-2} u^{-1-\dec}.
 \eeaa
 We   next commute   with    $\nu_*$.         Since  $\nu_* $ is tangent to $\Si_*$  and  $r\dds_2\dds_1\kabc=0$, making also use of the commutator Lemma  \ref{Lemma:Comm-nu*-M4},
\beaa
 \dds_2    ( \nu_* \dds_1 \kab) = r^{-1} [ \nu_*, r \dds_2]  \dds_1\kabc =\Ga_b \,  \nu_* ( \dds_1\kab) ) +\lot
\eeaa
Proceeding as before, 
  \beaa
   \int_S\big|   \nu_* ( \dds_1 \kabc) \big |^2 &\les&  r^2  \int_S |\dds_2 (\nu_* \dds_1 \kabc ) |^2+r^{-4} \Big |\int_S e^\Phi \nu_*( \dds_1 \kabc)\,   \Big |^2\\
   &\les& r^2\int_S \Big|\Ga_b \,  \nu_* ( \dds_1\kab)  \Big|^2+ \ep^2_0 r^{-4} u^{-2-2\dec}\\
    &\les& \ep^2\int_S\big|   \nu_* ( \dds_1 \kabc) \big |^2+ \ep^2_0 r^{-4} u^{-2-2\dec}.
  \eeaa
 We can then proceed as before making use of  the estimate for the $\ell=1$ mode 
 of $\nu_* e_\th \kab $ in Proposition   \ref{Lemma:Bad-modes-onSi}, i.e.    for all   $0\le k\le k_{small}+20$. 
 The  higher $\nu_*$ estimates are proved in the same manner.

{\bf Step 2.}  In exactly  the same manner, making use of  the GCM condition $\dds_2\dds_1\mu=0$ and the estimate  for the $\ell=1$ mode  of $e_\th \mu $ in  Lemma \ref{Lemma:Bad-modes-onSi},  we deduce
\bea
\label{EstimatesonSi*-Part2-2}
\|\muc\|^*_{\infty, k}&\les & \ep_0r^{-3} u^{-1-\dec}.
\eea

{\bf Step 3.} At this point we have  established the estimates of Proposition \ref{Lemma: :Estimates-onSi*-1}
for  $\bb, \vthb, \xib, \etab$  and $\kabc, \muc$.  Next we  derive estimates for $\rhoc$   by making use of the 
 quantity  $\qf$.  We make use of Proposition \ref{prop:alternateformulaforqfinvolvingtwoangularderrivativesofrho:M4} according to which
we have, relative to the  background  frame of $\Mext$,
\beaa
r^4\left( \dds_2\dds_1\rho+\frac{3}{4}\kab\rho\vth +\frac{3}{4}\ka\rho\vthb \right) &=&\qf     +r^{2}  \dkb^{\le 2}(\Ga_b \c  \Ga_g) +\lot 
\eeaa
 Using the estimates { \bf Ref1-2} for the  Ricci and curvature coefficients as well as the product Lemma  \ref{lemma:interpolation-decay}  we derive,
\beaa
\|\qf\|^*_{\infty, k}&\les& \ep_0 r^{-1}  u^{-1/2-\dec},\\
\|\vthb\|^*_{\infty, k} &\les& \ep_0 r^{-1} u^{-1-\dec},\\
\|r^{2}  \dkb^{\le 2}(\Ga_b \c  \Ga_g) \|^*_{\infty, k}&\les& \ep_0 r^{-1} u^{-1-\dec}.
\eeaa
Also,  using our condition on $r$ along $\Si_*$
\beaa
\|\vth\|^*_{\infty, k} &\les& \ep r^{-2}\les \ep_0 r^{-1} u^{-1-\dec}.
\eeaa
We deduce,
\beaa
 \|\dds_2\dds_1 \rhoc\|^*_{\infty, k} &\les&\ep_0r^{-5}  u^{-1/2-\dec}.
\eeaa
Also, after  commuting  $\nu_*$ with $r\dds_2$,
\beaa
 \|\dds_2(\nu^i_*\dds_1 \rho) \|^*_{\infty, k-i} &\les& \ep_0r^{-5}  u^{-1/2-\dec}.
\eeaa
Since, according to Lemma    \ref{Lemma:Bad-modes-onSi},    we also control the $\ell=1$ modes of  $ \nu_*^i  e^j_\th (\rho)$  we  can proceed as before\footnote{i.e. passing to $L^2$, using part 4 of  Lemma \ref{prop:2D-hodge-reduced-M4}, and  then going back to $L^\infty$ norms as before}   to deduce the desired estimate
\bea
\label{EstimatesonSi*-Part2-5}
\| \rhoc\|^*_{\infty, k}&\les& \ep_0 r^{-3} u^{-1/2-\dec}, \qquad \forall \, k\le k_{small}+20.
\eea

 {\bf Step 4.} From the definition of $\mu=- \ddd_1 \ze -\rho+\frac 1 4  \vth \vthb$,
we have,
\beaa
 \ddd_1 \ze &=& - \muc -\rho +\Ga_g\c \Ga_b.
\eeaa
We deduce, for all $k\le k_{small }+19$,
 \beaa
\| \dds_1 \ddd_1 \ze\|^*_{\infty, k} &\les& \|\dds_1 \muc\|^*_{k} +\|\dds_1 \rhoc\|^*_{\infty, k}  +\ep_0  r^{-4} u^{-1-\dec}\\
 &\les& \ep_0 r^{-4} u^{-1/2 -\dec}.
 \eeaa
 Since according to Proposition   \ref{Lemma:Bad-modes-onSi}   we control the $\ell=1$ modes of $\ze$ and their $\nu_*$ derivatives  we deduce,  passing  as before  through $L^2$ norms,
\bea
\label{EstimatesonSi*-Part2-7}
\|\ze \|^*_{\infty, k} \les\ep_0 r^{-2} u^{-1/2 -\dec}, \qquad \forall\, k \le k_{small}+19.
\eea

{\bf Step 5.} 
We next use the Codazzi equations, the bootstrap assumption for $\b$ and the above result for $\ze$ to deduce,
\beaa
\ddd_2 \vth &=& -2\b+ (e_\th(\ka) + \ze\ka )+   \Ga_g\c \Ga_g,\\
&=& -2\b+\frac 2 r \ze + \Ga_g\c \Ga_g.
\eeaa
Making use of the  bootstrap assumption for $\b$ and the previous estimate for $\ze$,
\beaa
\big\|\ddd_2 \vth\big\|^*_{\infty, k} &\les& \big\|\b\big \|_{\infty, k} + r^{-1}  \big\|\ze \big \|^*_{\infty, k} +\ep_0 r^{-3} u^{-1-\dec}\\
&\les& \ep r^{-7/2-\dec} + \ep_0 r^{-3} u^{-1/2-\dec}
\eeaa
 from which we derive, using also the condition   $\min_{\Si_*} r  \ge (\ep \ep_0^{-1})^2  u^{1+ 2 \dec} $,
 \bea
 \label{EstimatesonSi*-Part2-9}
 \big\| \vth\big\|^*_{\infty, k} &\les & \ep_0 r^{-2} u^{-1/2 -\dec},\qquad \forall\, k\le k_{small}+19.
 \eea

{\bf Step 6.}
 We now  estimate $\b$ from the equation,
\beaa
e_3 \a+\left(\frac 1 2 \kab -4\omb\right)\a&=&-\dds_2\b  -\frac 3 2  \vth \rho  +(\ze+4\eta) \b.
\eeaa
Hence,
\beaa
\big\|\dds_2 \b \|^*_{\infty, k-1} &\les& \big \|e_3 \a\big\|^*_{\infty, k-1}+ r^{-1} \big\| \a\big\|^*_{\infty, k-1}+ r^{-3} \big\|\vth\|^*_{\infty, k-1}+\ep_0r^{-4} (2r+u)^{-1/2-\dec}\\
&\les&  \ep_0r^{-4} (2r+u)^{-1/2-\dec}.
\eeaa
We deduce, in particular, using our previous estimate for the $\ell=1$ mode of $\b$
\beaa
\| \b\|_{\hk_{k+1} (S) }&\les&  r \big\| \dds_2 \b \|_{\hk_k(S)} + r^{-2} \left| \int_S \b e^\Phi\right|\\
&\les& \ep_0r^{-2} (2r+u)^{-1/2-\dec}  + \ep_0r^{-3}u^{-1-\dec}\\
&\les& \ep_0r^{-2} (2r+u)^{-1/2-\dec}. 
\eeaa
 We deduce, arguing as before,
\bea
\label{EstimatesonSi*-Part2-11}
\big\| \b\|^*_{\infty, k} &\les& \ep_0r^{-3} (2r+u)^{-1/2-\dec}, \qquad \forall\, k\le k_{small}+19.
\eea
\begin{remark} We can in fact  prove a stronger estimate by making use of the full strength of the estimate for $\a$    in  {\bf Ref 2},
\beaa
|\a|&\les&  \log(1+u) r^{-3} (2r +u)^{-1/2 -\dee},\\
| e_3 \a|&\les&  r^{-4} (2r +u)^{-1/2 -\dee}.
\eeaa
Hence,
\beaa
\big\| \dds_2 \b\|^*_{\infty, k} &\les& \ep_0  \log(1+u)  r^{-4} (2r +u)^{-1/2 -\dee} +\ep_0 r^{-5} u^{-1/2-\dec}
\eeaa
and thus
\bea
\big\|  \b\|^*_{\infty, k} &\les& \ep_0  \log(1+u) r^{-3} (2r +u)^{-1/2 -\dee}. 
\eea
\end{remark}

 {\bf Step 7.}  Next, we estimate $\check{\omb}$ by making use of the first   identity\footnote{This is a precise version of the identity in Proposition  \ref{cor:eqtsfor-ometaxib-M4}.} in Proposition \ref{prop:eqtsfor-ometaxib},
\beaa
\dds_1\,\omb  &=& -\frac 14 \ka \,\xib+\frac{1}{2} e_3 \ze - \frac{1}{2}\bb + \frac{1}{2}\kab\ze + \frac{1}{2}\vthb\ze -\frac 1 4 \vth \,\xib.
\eeaa
We deduce, using \eqref{EstimatesonSi*-Part2-7} and the dominance condition for  $r$ on $\Si_*$,
\beaa
\big\|\dds_1 \ombc\big\|_{2, k-1}&\les& r^{-1} \big\|\xib \big\|_{2, k-1}+\big\| e_3\ze\big\|_{2, k-1}  + r^{-1} \big\| \ze\big\|_{2, k-1}     +\big\| \bb\big\|_{2, k-1}+\ep_0 r^{-2} u^{-1-\dec}\\
&\les& \ep_0 r^{-1} u^{-1-\dec}.
\eeaa
Thus,  using the Poincare   and  Sobolev inequalities,
\bea
 \label{EstimatesonSi*-Part2-16}
 \big\| \ombc    \big\|_{\infty, k}&\les& \ep_0 r^{-1} u^{-1-\dec}.
 \eea

{\bf Step 8.} We estimate $\vsic, \ov{\vsi}-1$
Recall that, see \ref{formula-a*-M4},
 \beaa
 a_*= \frac{2}{\vsi} -\Up+\frac r 2 \Ab.
 \eeaa 
Also,   recall that,  see \eqref{formula:Omb-onSi*-M4},    
\beaa
\Omb= e_3(r)=-\Up+\frac r 2  \Ab, \quad \text{on } \quad \Si_*.
\eeaa

{\bf Step 9.} We determine $\vsic, \ombc $  using the equations, see \eqref{geodesic-foliation-M4}, 
\beaa
e_\th\log \vsi&=&\frac 1 2 (\eta-\ze),
\\
e_\th(\Obc) &=& -\xib-(\eta-\ze) \Omb.
\eeaa

Hence, using the estimates  for $\eta, \ze, \xib$  and the bootstrap assumptions  already derived,
\beaa
\|\vsic  \|_{\hk_{k+1}} &\les& \ep_0   u^{-1-\dec}, \\
\|\Ombc  \|_{\hk_{k+1}} &\les& \ep_0  u^{-1-\dec}. 
\eeaa

{\bf Step 10.} We  estimate $\ov{\vsi}-1$ and $\ov{\Omb} +\Up $  on $\Si_*$ using the equations \eqref{formula-a*-M4},  \eqref{formula:Omb-onSi*-M4}, 
\beaa
 a_*&=&- \frac{2}{\vsi} +\Up-\frac r 2 \Ab,\\
\Omb&=& -\Up+\frac r 2  \Ab,
\eeaa
and the GCM condition for $a_*$, see \eqref{GCM-conditionfora*-M4},
   \beaa
 a_*\Big|_{SP}=-1 -\frac{2m^\S}{r^\S}.
 \eeaa
We deduce,
 \beaa
  \frac{2}{\vsi\big|_{SP} }&=& 2 -\frac r 2 \Ab \Big|_{SP} =2 -(\Omb+\Up)\big|_{SP}
 \eeaa
 i.e.,
 \beaa
 2 \left(1-  \frac{1}{\vsi\big|_{SP} }\right)&=&\Omb\big|_{SP}+\Up.
 \eeaa
 Since 
 \beaa
\ov{\vsi}=\vsi\big|_{SP}+\vsic\big|_{SP}, \qquad \ov{\Omb}=\Omb \big|_{SP}+\Ombc\big|_{SP}
\eeaa
we deduce, in view of the above estimates for $\vsic, \Ombc$,
\beaa
\big|\ov{\vsi}-1\big|&\les&\Big|\ov{\Omb}+\Up\Big| + \ep_0 u^{-1-\dec}.
\eeaa
It thus remains to estimate 
\beaa
\ov{\Omb}+\Up&=&\frac  r 2 \ov{\Ab}. 
\eeaa
This can be done with the help of the following  identity,
  \bea
  \lab{eq:formulafortheaverageofAbS-M4}
\ov{\Ab^\S}  &=& \frac{1}{ \ov{\vsi^\S}}\left(\ov{\check{\vsi}^\S\check{\kab}^\S} -\ov{\check{\vsi}^\S\check{\Ab}^\S}\right)
 \eea
 from which we infer that,
 \beaa
 \Big|\ov{\Omb}+\Up\Big|&\les&  \ep_0 u^{-1-\dec}
  \eeaa
  and therefore also
   \beaa
\big|\ov{\vsi}-1\big|&\les&  \ep_0 u^{-1-\dec}.
  \eeaa
 To prove \eqref{eq:formulafortheaverageofAbS-M4}   we apply  Lemma \ref{Lemma:nuSof integrals:M4} to deduce,
   \beaa
 e_3(r)+a_*&=&\nu_*(r) = \frac{r}{2}\vsi^{-1}\ov{\vsi(\kab+a_*\ka)}= \frac{r}{2}\vsi^{-1}\big(\ov{\vsi}\ov{\kab}+\ov{\vsic\kabc}\big)+\vsi^{-1}\ov{\vsi a_*}\\
 &=& -\Up\vsi^{-1}\ov{\vsi}+\frac{r}{2}\vsi^{-1}\ov{\check{\vsi}\check{\kab}}+ \vsi^{-1}\ov{\vsi a_*}.
 \eeaa
 Since   $e_3( r) =-\Up+\frac{r}{2}\Ab$  we infer that
 \beaa
-\Up+\frac{r}{2}\Ab +a_*&=&  -\Up\vsi^{-1}\ov{\vsi}+\frac{r}{2}\vsi^{-1}\ov{\check{\vsi}\check{\kab}}+ \vsi^{-1}\ov{\vsi a_*}\
 \eeaa
 and therefore
 \beaa
 \Ab  &=& \frac{2}{r}\left(\Up -a_* -\Up\vsi^{-1}\ov{\vsi}+\frac{r}{2}\vsi^{-1}\ov{\check{\vsi}\check{\kab}}+ \vsi^{-1}\ov{\vsi a}_*\right).
 \eeaa 
 In particular, multiplying by $\vsi$ and taking the average, we infer that
 \beaa
 \ov{\vsi\Ab}  &=& \ov{\check{\vsi}\check{\kab}},
 \eeaa
 and hence
  \beaa
\ov{\Ab}  &=& \frac{1}{ \ov{\vsi}}\left(\ov{\check{\vsi}\check{\kab}} -\ov{\check{\vsi}\check{\Ab}}\right)
 \eeaa
as stated.
The estimates for all other derivatives of $ \vsic, \Ombc, \ov{\vsic}-1, \ov{\Omb}+\Up$ can be derived as before. This concludes the proof of Proposition \ref{Lemma: :Estimates-onSi*-1}.
\end{proof}


\section{Control  in $\Mext $, Part I}



\subsection{Preliminaries}



\subsubsection{Commutation Lemmas}


  Here and below we write schematically  
\beaa
\dkb =r \ddd, \quad    \dk_{\near}=\{re_4, \dkb\}, \quad   \dk=( e_3, r e_4, r \ddd),\quad  \T=\frac 12 \left(e_3 +\Up e_4\right).
\eeaa
\begin{lemma} 
\label{lemma-Decay:commutation}
 We have, schematically, 
 \bea
\label{comm:dkbTe3e4schematic}
\bsplit
 [\T, e_4]  =r^{-1}\Ga_b\dkout,      \qquad   \,[\dkb, e_4] &= \check{\Ga}_g \dk_{\near}  +\Ga_g.
 \end{split}
\eea
Also,
\beaa
\T(r)=\frac{1}{ 2 r }\Ab = r^{-1}\Ga_b. 
\eeaa
\end{lemma}

\begin{proof}
The identity for $[\dkb, e_4]$ has already been discussed in  Corollary \ref{Cor:comme3e4-outgeodesic-M4}.
According to lemma \ref{lemma:commTwithe3e4}  we have,
\beaa
\begin{split}
\,[\T, e_4]&=\left(\left( \omb-\frac{m}{r^2} \right)     -\frac{m}{2r } \left( \overline{\ka} -\frac{2}{r}\right)+\frac{e_4 (m)}{r}  \right)e_4+(\eta+\ze) e_\th,\\
\end{split}
\eeaa
In view of {\bf Ref 4}   and bootstrap assumptions  {\bf Ref 2}       the  factors   of $e_4$ and $e_\th$, on the right hand side   behave at worst like  $\Ga_b$.   Thus schematically $[\T, e_4] = r^{-1} \Ga_b \dkout$.
 \end{proof}


\subsubsection{Transport Lemmas}


The following lemma will be used repeatedly in what follows.
\begin{lemma}
\label{le:transportrp-f-Decay}
If $f$ verifies the transport equation 
\beaa
e_4 (f) + \frac p 2  \ov{\ka }  f=F,
\eeaa
we have  for fixed $u$ and any $r_0 \le r\le r_* $,
 \bea
\bsplit
  r^{p}  \|   f  \|_\infty(u, r)&\les    r_0  ^{p} \|  f  \|_\infty (u,  r_\HH )+ \int_{r_0 } ^ {r}      \la ^{p} \| F\|_\infty (u, \la) d\la, \\
    r^{p}  \|   f  \|_\infty (u, r)&\les    r_*  ^{p} \|  f  \|_\infty (u,  r_*  )+ \int_r^{r_*}      \la ^{p} \| F\|_\infty(u, \la) d\la,
    \end{split}
\eea
   where $r $  is the area  radius   at fixed $u$.   
\end{lemma}

\begin{proof}
According to  Corollary   \ref{le:transportrp-f} we have $e_4(r^p f)= r^p F$
The desired  estimates follow easily  by integration with respect to the affine parameter $s$, recall that $e_4(s)=1$. 
\end{proof}

\begin{proposition}
\label{Prop:transportrp-f-Decay-knorms}
The following inequalities hold true for all $k\le k_{large}-5$,  $r_0\le r\le r_*$
\bea
\label{eq:integration-knorms}
\bsplit
  r^p\|f\|_{\infty, k}(u, r) &\les r_*^p\|f\|_{\infty, k}(u, r_*) +\int_r^{r_*} \la^p\| F\|_{\infty, k}(u, \la) d\la, 
\\
 r^p\|f\|_{\infty, k}(u, r) &\les r_0 ^p\|f\|_{\infty, k}(u, r_0 ) +\int_{r_0} ^{r} \la^p\| F\|_{\infty, k} (u, \la) d\la. 
\end{split}
\eea
\end{proposition}

\begin{proof}
Commuting  the equation $e_4( r^p f) =r^p F$  with $\dkb$, applying  the commutation Lemma 
\ref{lemma-Decay:commutation}and our bootstrap assumptions on $\Ga_g$         we  derive,
\beaa
e_4 (r^p |\dkb f|) &\les &r^p |\dkb F|+   r^p|\dkb  F|+\ep r^{-2} r^p\big(|\dkb f| + r e_4 (r^p f) \big)\\
&\les&r^p( |\dkb F|+|F|) + \ep r^{-2} r^p( |\dkb f|+|f|).
\eeaa
Similarly, commuting with $\T$,
\beaa
e_4\big( \T ( r^p f)\big) &=&\T(r^p F )- [\T, e_4] (r^p f)= r^p F - p r^{p-1} \T(r) F-r^{-1}\Ga_b\dkout    r^p f.
\eeaa

Hence,
\beaa
e_4\big( r^p   \T  f \big)  &=& r^p\T( F) - p r^{p-1} \T(r) F-r^{-1}\Ga_b\dkout    r^p f- p  e_4\big( r^{p-1}\T(r) f\big)
\eeaa
i.e.,
\beaa
e_4\big( r^p   |\T  f|  \big)&\les&  r^p \big(  |\T F |+ |F|\big) + O(\ep r^{-2} )\big(  r^p |F|+ |\dkb f|  +|f| \big).
\eeaa
Similarly, commuting the equation with $r e_4$ we derive,
\beaa
e_4\big( r^p   | r   f|  \big)&\les&  r^p \big(   | r e_4  F| +|F|\big)+ O(\ep r^{-2} )\big(  r^p |F|+ |\dkb f|  +|f| \big).
\eeaa
Integrating  the inequalities,
\beaa
e_4 (r^p |\dkb f|) &\les&r^p( |\dkb F|+|F|) + \ep r^{-2} r^p( |\dkb f|+|f|)\\
e_4\big( r^p   |\T  f|  \big)&\les&  r^p \big(  |\T F |+ |F|\big) +\ep r^{-2}  r^p \big(   |\dkb f|  +|f| \big)\\
e_4\big( r^p   | r   f|  \big)&\les&  r^p \big(   | r e_4  F| +|F|\big)+\ep r^{-2} \big(  r^p |F|+ |\dkb f|  +|f| \big)
\eeaa
and applying Gronwall  we derive  the desired estimates in \eqref{eq:integration-knorms}  for $k=1$.

Repeating the procedure for $\dtb^k$, any  combination of derivatives of the form $\dtb^k=\T^{k_1} \dkb^{k_2}$ with $k_1+k_2=k$,  estimating the  corresponding  commutators     using  our  assumptions  {\bf Ref 1},   we deduce for all $0\le k\le k_{large}-5$,
\beaa
e_4 ( r^p |\dtb^{\le k} f| )  &\les &  r^p  |\dtb^{\le k } F|+ \ep r^{-2} r^p |\dtb^{\le k} f|
\eeaa
and the desired estimates follow by integration.
\end{proof}


\subsubsection{Transport equations for $\ell=1$ modes}


To estimate  $\ell=1$ modes we make use of the following. 
\begin{lemma}
\lab{transport:e4badmodes-M4}
The following equation holds true for   reduced  scalars  $\psi \in \sk_1(\Mext)$.
 \bea
 \bsplit
 \lab{eq:propagationof-badmodes}
 e_4\left(\int_S\psi e^\Phi\right)&=  \int_S e_4(\psi) e^\Phi+\frac 3 2 \ov{\ka}\int_S \psi e^\Phi  + \int_S     \frac{1}{2} \left(3\kac-\vth)\psi\right)e^\Phi.\end{split}
 \eea
\end{lemma}
\begin{proof}
This is an immediate consequence of  Proposition \eqref{prop:outgoinggeod-e3e4averages}.  Indeed according 
to it and $e_4\Phi=\frac 1 2 (\ka-\vth)$, 
\beaa
 e_4\left(\int_S\psi e^\Phi\right)&=&\int_S(e_4(\psi e^\Phi )+\ka \psi e^\Phi)
 = \int_S\left(e_4(\psi)+\frac{1}{2}(3\ka-\vth)\psi\right)e^\Phi\\
 &=&  \int_S\left(  e_4(\psi)+  \frac   3 2 \ov{\ka}  \psi \right) e ^\Phi  + \int_S     \frac{1}{2} \left(3\kac-\vth)\psi\right)e^\Phi\\
 &=& \int_S e_4(\psi) e^\Phi+\frac 3 2 \ov{\ka}\int_S \psi e^\Phi  + \int_S     \frac{1}{2} \left(3\kac-\vth)\psi\right)e^\Phi
 \eeaa
 as desired.
\end{proof}


\subsection{Proposition \ref{Prop:top-r-estimates}}


In what follows we prove the  stronger estimates  in terms of powers of $r$ for the  quantities $\kac, \muc, \vth, \ze, \kabc, \b, \rhoc$. More precisely we establish the following.
\begin{proposition}
\label{Prop:top-r-estimates}
The following estimates hold true  in $\Mext$ for  all  $ k\le k_{small}+20$
\bea
\label{eq::top-r-estimates1}
\bsplit
\big\|\kac\|_{\infty, k} &\les \ep_0 r^{-2} u^{-1-\dec},  \\
\big\|\muc \|_{\infty, k} &\les \ep_0 r^{-3} u^{-1-\dec}.
\end{split}
\eea
Also,  for  all  $ k\le k_{small}+18$
\bea
\label{eq::top-r-estimates2}
\bsplit
 \big\| \vth, \,  \ze, \, \kabc, r\rhoc  \big\|_{\infty, k}&\les \ep_0 r^{-2} u^{-1/2-\dec},\\
 \big\| \b \big\|_{\infty, k}&\les \ep_0r^{-3} (2r+u)^{-1/2-\dec},\\
 \big\|  e_3 \b \big\|_{\infty, k}&\les\ep_0 r^{-4} u^{-1/2-\dec},\\
 \big\| e_\th K\big\|_{\infty, k} &\les \ep_0 r^{-4} u^{-1/2-\dec}. 
\end{split}
\eea
\end{proposition}

\begin{remark}
\label{remark:stronger-estim-b}
Note that in fact $\b$  admits the stronger estimate, for all $k\le k_{small}+18$,
\bea
\big\|  \b\|_{\infty, k} &\les& \ep_0  \log(1+u) r^{-3} (2r +u)^{-1/2 -\dee}. 
\eea
\end{remark}


\subsection{Estimates for  $\kac, \muc$ in  $ \Mext$}
\lab{subsection-kacmuc}


 {\bf Step 1.} 
 We prove the following estimates for  $\kac$ in $\Mext$.
\bea
\lab{eq:EstimatesinMext1}
\big\|\kac\|_{\infty, k} &\les& \ep_0 r^{-5/2} u^{-1-\dec}, \qquad k\le k_{small}+20.   
\eea

We make use of the equation
\beaa
e_4(\check{\ka}) +\overline{\ka}\,\check{\ka} =F:= -\frac{1}{2}\check{\ka}^2 -\frac{1}{2}\overline{\check{\ka}^2}-\frac{1}{2}\left(\vth^2-\overline{\vth^2}\right).
\eeaa
In view of   our assumptions {\bf Ref1-2} and Lemma \ref{lemma:interpolation-decay}
\beaa
\Big\| F \big\|_{\infty, k}(u, \la)&\les&  \ep_0\la^{-7/2 } u^{-1-\dec}. 
\eeaa
 Applying Proposition \ref{Prop:transportrp-f-Decay-knorms} we deduce,
   \beaa
     r^2  \|   \kac  \|_{\infty, k}  (u, r)&\les&    r_*  ^2   \|  \kac \|_{\infty, k}  (u,  r_* )+ \ep_0 u^{-1-\dec}  \int_{r } ^ {r_*}      \la ^2   \la^{-7/2}  d\la\\
     &\les&  r_*  ^2   \|  \kac \|_\infty (u,  r_* )+\ep_0 r^{-1/2 } u^{-1-\dec}.
   \eeaa
 In view of the   control    on the last slice we infer that, everywhere  in $\Mext$,
\beaa
 \|   \kac  \|_{\infty, k}  (u, r)&\les& \ep_0 r^{-5/2} u^{-1-\dec}.
\eeaa

\noindent{\bf Step 2.}  We prove  the estimate,
\bea
\lab{eq:EstimatesinMext2}
\big\|\muc \|_{\infty, k} &\les& \ep_0 r^{-3} u^{-1-\dec}, \qquad k\le k_{small}+20.   
\eea
Recall that we have
\beaa
e_4(\check{\mu}) +\frac{3}{2}\overline{\ka}\check{\mu} &=& - \frac{3}{2}\overline{\mu}\check{\ka} +F,\\
F:&=& -\frac{3}{2}\check{\mu}\check{\ka}+\frac{1}{2}\overline{\check{\mu}\check{\ka}}+\err[e_4\check{\mu}]-\overline{\err[e_4\check{\mu}]},\\
\err[e_4\check{\mu}] &=& -\frac 18 \kab\vth^2 -\vth \dds_2\ze -\vth\ze^2 +\left(2e_\th(\ka) -2\b    + \frac{3}{2}\ka\ze\right)\ze.
\eeaa
In view of  Lemma \ref{lemma:interpolation-decay}
we  check,
\beaa
\| F\|_{\infty, k}(u, \la) &\les& \ep_0 \la ^{-9/2}u^{-1-\dec}.
\eeaa
 Applying Proposition \ref{Prop:transportrp-f-Decay-knorms}  and  the estimates on the last slice for $\muc$ we deduce
\beaa
 r ^3  \| \dtb^k  \muc \|_{\infty, k} (u,\la) &\les&  r _*^3  \| \dtb^k  \muc \|_{\infty, k} (u, r_*) +\ep_0 u^{-1-\dec}  \int_r ^{r_*}  \la^3\la^{-9/2}\\
 &\les &  r _*^3  \| \dtb^k  \muc \|_{\infty, k} (u, r_*) +\ep_0 u^{-1-\dec} r^{-1/2}\\
 &\les& \ep_0 u^{-1-\dec}  
\eeaa
 from which  the desired estimate \eqref{eq:EstimatesinMext2} follows.


\subsection{Estimates for the $\ell=1$ modes in $ \Mext$}


We extend the validity of Lemma \ref{Lemma:Bad-modes-onSi}  to the entire region $\Mext$.
 \begin{lemma}
\label{Lemma:Bad-modes-onMext}
The following estimates hold true on $\Mext $ or all $k\le k_{small}+19$,
\bea
\bsplit
\left \|\int_S \b   e^\Phi\right\|_{\infty, k} (u, r)&\les\ep_0 r^{-1} u^{-1-\dec},\\
\left \|\int_S \ze e^\Phi\right\|_{\infty, k} (u, r)&\les\ep_0 r u^{-1-\dec},\\
\left \|\int_S e_\th(\rho)  e^\Phi\right\|_{\infty, k} (u, r)&\les\ep_0 r^{-1} u^{-1-\dec},\\
\left \|\int_S e_\th(\kab)  e^\Phi\right\|_{\infty, k} (u, r)&\les\ep_0  u^{-1-\dec},\\
\left \|\int_S \bb  e^\Phi\right\|_{\infty, k} (u, r)&\les\ep_0  u^{-1-\dec}.
\end{split}
\eea
\end{lemma}

 \begin{proof}   We first note that  the estimate for the $\ell=1$ mode of  $\muc$ is an immediate consequence of  the estimate \eqref{eq:EstimatesinMext2}.
 To prove the remaining estimates we  proceed  in steps as follows.

 {\bf Step 1.}  Observe that the  estimates of  Lemma \ref{Lemma:Bad-modes-onSi}   remain valid  when we replaces  the norms $\|\,\|_{\infty, k}^*$  by $\|\,\|_{\infty, k}$. 
 To show this it suffices to  prove estimates for $ r e_4$ of all $\ell=1$ modes. This can easily be achieved with the   help  of Lemma \ref{transport:e4badmodes-M4}
  and our $e_4$  transport equations for $\ze, \rhoc, \muc, \kabc, \bb$.
 

 {\bf Step 2.} We establish the estimate,
 \bea
 \lab{Bad-modes-onMext-b}
 \left\| \int_S \b e^\Phi\right\|_{\infty, k} &\les \ep_0 r^{-1} u^{-1-\dec}, \qquad  k\le k_{small}+20. 
 \eea

 In view of  \eqref{eq:propagationof-badmodes} and  the Bianchi identity for $e_4(\b)$
\beaa
 e_4\left(\int_S \b e^\Phi\right)&=&  \int_S e_4\b  e^\Phi+\frac 3 2 \ov{\ka}\int_S \b e^\Phi  + \int_S     \frac{1}{2} \left(3\kac-\vth)\b\right)e^\Phi \\
 &=& \int_S (-2\ka \b +\dds_2 \a +\ze \a )       e^\Phi+\frac 3 2 \ov{\ka}\int_S \b e^\Phi  + \int_S     \frac{1}{2} \left(3\kac-\vth)\b\right)e^\Phi \\
 &=& -\frac{\overline{\ka}}{2}\int_S\b e^\Phi+\int_S\left(\ze\a+\frac{1}{2}(-\check{\ka}+\vth)\b\right)e^\Phi,
 \eeaa
and hence
\bea
\lab{Bad-modes-onMext-b1}
e_4\left(\int_S\b e^\Phi\right) + \frac{\overline{\ka}}{2}\int_S\b e^\Phi &=& \int_S\left(\ze\a+\frac{1}{2}(-\check{\ka}+\vth)\b\right)e^\Phi.
\eea
Recall that  
\beaa
\big|(\a, \b)\big|&\les&\ep  r^{-3}( 2r+ u) ^{-1/2-\dec}.
\eeaa
We deduce,
\beaa
\Big|e_4\left(r \int_S\b e^\Phi\right)\Big|&\les&r      \ep_0  r^{-2} u^{-1/2-\dec} r^{-3}( 2r+ u) ^{-1/2-\dec}\int_S  | e^\Phi| \\
&\les&\ep_0  r^{-1} u^{-1/2-\dec} ( 2r+ u) ^{-1/2-\dec}\\
&\les& \ep_0 r^{-1-\de} u^{-1-\dec}
\eeaa
i.e., in view of the estimate on $\Si_*$,  everywhere on $\Mext$,
\bea
\left| \int_S\b e^\Phi\right| &\les & \ep  r^{-1} u^{-1-\dec}.
\eea
Commuting with $\T$, $\dkb$ and $re_4 $ we  also easily  deduce,
\beaa
\Big\|  \dtb^{k_2}   (r e_4)^{k_1} \int_S\b e^\Phi\Big\|_\infty &\les&  \ep  r^{-1} u^{-1-\dec}, \qquad \forall \, k_1+k_2\le k_{small}+20
\eeaa
from which \eqref{Bad-modes-onMext-b} follows.

{\bf Step 3.}
We prove the estimate,
\bea
 \lab{Bad-modes-onMext-ze}
 \left\| \int_S \ze  e^\Phi\right\|_{\infty, k} &\les \ep_0  r^{1/2}   u^{-1-\dec}, \qquad k\le k_{small}+19
 \eea
  which is better than the desired estimate in Lemma \ref{Lemma:Bad-modes-onMext}.
This follows, as  for the corresponding estimate on $\Si_*$,  by  projecting the Codazzi  equation for $\vth$ on the $\ell=1$ mode
\beaa
\int_S\ze e^\Phi   &=&  \frac{r}{2\Up}\left(2\int_S\b e^\Phi -\int_Se_\th(\ka)e^\Phi -\int_S\vth\ze e^\Phi - \int_S\left(\ka-\frac{2\Up}{r}\right)\ze e^\Phi\right).
\eeaa
Note that in view of the estimates for  $\kac$  in \eqref{eq:EstimatesinMext1} already established\footnote{Note that  the estimate  for $\kac$ is stronger 
 in powers of $r$ than the  corresponding bootstrap assumption.}   we have,
\beaa
\left\| \int_Se_\th(\ka)e^\Phi \right\|_{\infty, k} &\les& \ep_0 r^{-1/2} u^{-1-\dec}, \qquad k\le k_{small}+19.
\eeaa
Also, making use of  \eqref{Bad-modes-onMext-b},
\beaa
\left\| \int_S \b e^\Phi \right\|_{\infty, k} &\les& \ep_0 r^{-1}  u^{-1-\dec},   \qquad k\le k_{small}+19.
\eeaa
Thus,
since $\Up$ is bounded away from  zero in $\Mext$,
we easily deduce
\beaa
\left\|\int_S\ze e^\Phi \right\|_{\infty , k} &\les& \ep_0 r^{1/2} u^{-1-\dec},   \qquad k\le k_{small}+19.
\eeaa

{\bf Step 4.}
We prove the estimate,
\bea
 \lab{Bad-modes-onMext-rho}
 \left\| \int_S e_\th(\rho)   e^\Phi\right\|_{\infty, k} &\les \ep_0   r^{-1}  u^{-1-\dec}, \qquad k\le k_{small}+19.
 \eea
 We proceed as in Step 4 of the proof of Lemma \ref{Lemma:Bad-modes-onSi}.
In view of the definition of $\mu$  and the identity $\dds_1\ddd_1=\ddd_2\dds_2+2K$we write,
\beaa
\int_Se_\th(\rho)e^\Phi &=& -\int_Se_\th(\mu)e^\Phi +\int_S\dds_1\ddd_1\ze e^\Phi +\frac{1}{4}\int_Se_\th(\vth\vthb)e^\Phi\\
&=& -\int_Se_\th(\mu)e^\Phi +\int_S(\ddd_2\dds_2+2K)\ze e^\Phi +\frac{1}{4}\int_Se_\th(\vth\vthb)e^\Phi\\
&=& -\int_Se_\th(\mu)e^\Phi +\frac{2}{r^2}\int_S\ze e^\Phi +2\int_S\left(K-\frac{1}{r^2}\right)\ze e^\Phi +\frac{1}{4}\int_Se_\th(\vth\vthb)e^\Phi.
\eeaa
Together with the above estimate for the $\ell=1$ mode of   $\ze$, the estimate \eqref{eq:EstimatesinMext2}   for $\check{\mu}$ and the bootstrap assumptions, we infer  that
\beaa
\left\| \int_Se_\th(\rho) e^\Phi\right\|_{\infty, k}&\les& \ep_0 r^{-1}  u^{-1-\dec}.
\eeaa

{\bf Step 5.}
We prove the estimate,
\bea
 \lab{Bad-modes-onMext-kab}
 \left\| \int_S e_\th(\kab)   e^\Phi\right\|_{\infty, k} &\les \ep_0     u^{-1-\dec}, \qquad k\le k_{small}+19.
 \eea
 As in the corresponding estimate on the last slice we make use of the  remarkable identity for the $\ell=1$ mode of $e_\th(K)$, i.e.
 \beaa
\int_Se_\th(\rho)e^\Phi +\frac{1}{4}\int_Se_\th(\ka\kab)e^\Phi -\frac{1}{4}\int_Se_\th(\vth\vthb)e^\Phi &=& 0.
\eeaa
We infer 
 \beaa
\int_Se_\th(\kab)e^\Phi  &=&  -2r\int_Se_\th(\rho)e^\Phi  - \frac{r}{2}\int_S\kab e_\th(\ka)e^\Phi +\frac{r}{2}\int_Se_\th(\vth\vthb)e^\Phi -\frac{r}{2}\int_S\left(\ka-\frac{2}{r}\right) e_\th(\kab)e^\Phi. 
\eeaa
The estimate  \eqref{Bad-modes-onMext-kab}  follows easily  from with the above estimate for the $\ell=1$ mode of $e_\th(\rho)$,  the estimate for $\kac $ in \eqref{eq:EstimatesinMext1}   and the bootstrap assumptions.

{\bf Step 6.}
We prove the estimate,
\bea
 \lab{Bad-modes-onMext-bb}
 \left\| \int_S \bb   e^\Phi\right\|_{\infty, k} &\les \ep_0    u^{-1-\dec}, \qquad k\le k_{small}+19.
 \eea
 
Projecting the Codazzi for $\vthb$ on the $\ell=1$ mode, we have
\beaa
-2\int_S\bb e^\Phi +\int_Se_\th(\kab)e^\Phi -\int_S\kab\ze e^\Phi +\int_S\vthb\ze e^\Phi &=& 0
\eeaa
and hence
\beaa
\int_S\bb e^\Phi &=& \frac{1}{2}\int_Se_\th(\kab)e^\Phi +\frac{\Up}{r}\int_S\ze e^\Phi -\frac{1}{2}\int_S\left(\kab+\frac{2\Up}{r}\right)\ze e^\Phi+\frac{1}{2}\int_S\vthb\ze e^\Phi.
\eeaa
The desired estimate follows easily  in view  of  the above estimates for the $\ell=1$ mode of $e_\th(\kab)$,  the $\ell=1$ mode of $\ze  $ and the bootstrap assumptions.
 \end{proof}

 
 \subsection{Completion of  the Proof of Proposition \ref{Prop:top-r-estimates}}
 
 
 We  prove the  second part of Proposition \ref{Prop:top-r-estimates}, i.e. we prove   for all  $k\le k_{small}+18$
\bea
\label{eq::top-r-estimates2'}
\bsplit
 \big\| \vth, \,  \ze, \, \kabc, r\rhoc  \big\|_{\infty, k}&\les \ep_0 r^{-2} u^{-1/2-\dec},\\
 \big\| \b \big\|_{\infty, k}&\les \ep_0r^{-3} (2r+u)^{-1/2-\dec},\\
 \big\|  e_3 \b \big\|_{k, \infty}&\les\ep_0 r^{-4} u^{-1/2-\dec},\\
 \big\| e_\th K\big\|_{\infty, k} &\les \ep_0 r^{-4} u^{-1/2-\dec}. 
\end{split}
\eea
We also prove the stronger estimate for $\b$ (see Remark \ref{remark:stronger-estim-b})
\bea
\big\|  \b\|_{\infty, k} &\les& \ep_0  \log(1+u) r^{-3} (2r +u)^{-1/2 -\dee}. 
\eea

\begin{proof} We proceed in steps  as follows.

 {\bf Step 1.}  We  derive the estimate,
 \bea
 \lab{eq:Estimate-inMext-vth}
 \big\| \vth\big\|_{\infty, k}&\les \ep_0 r^{-2} u^{-1/2-\dec}, \qquad \forall \, k\le k_{small}+19,
 \eea
 with the help of  the   equation $e_4\vth+\ov{\ka}\vth   =F:=-2\a-\kac \vth$ and the corresponding estimate on the last slice.
  
 Note  that,
 \beaa
\big\|\a\big\|_{\infty, k} &\les&\ep_0  r^{-3-\de }( 2r+ u) ^{-1/2-\dec}
\eeaa
where $\de >0$ is a small constant, $\de<\dee-\dec$. 
 Thus, using also  the  product estimates of Lemma  \ref{lemma:interpolation-decay},  we easily check that,
 \beaa
 \|F\|_{\infty, k} &\les & \ep_0r^{-3-\de} u^{-1/2-\dec} + \ep_0 r^{-7/2}  u^{-1-\dec}, \qquad k\le k_{small}+20.
 \eeaa
 Making use of Proposition  \ref{Prop:transportrp-f-Decay-knorms}  we deduce, for all $k\le k_{small}+19$,
 \beaa
r^2  \| \dk^k \vth\|_{\infty, k  }(u, r) &\les & r_*^2   \| \dk^k \vth\|_{\infty, k }(u, r_*) +\ep_0 u^{-1/2-\dec}  \int_r^{r_*} \la^{-1-\de} d\la.
 \eeaa
 Thus, in view of the results on the last slice $\Si_*$, we deduce,
 \beaa
  \| \dk^k \vth\|_{\infty }(u, r) &\les & r^{-2} u^{-1/2-\dec}.
 \eeaa

 {\bf Step 2.} We derive the estimate,
  \bea
   \lab{eq:Estimate-inMext-b}
   \bsplit
 \big\| \b \big\|_{\infty, k}&\les\ep_0r^{-3} (2r+u)^{-1/2-\dec}, \qquad \forall \, k\le k_{small}+19. 
  \end{split}
 \eea
We  proceed exactly as in the estimates  for $\b$ on the last slice $\Si_*$ by making use 
of the Bianchi identity  $e_3 \a+\left(\frac 1 2 \kab -4\omb\right)\a=-\dds_2\b  -\frac 3 2  \vth \rho  +5\ze \b$,
 from which we deduce,
 \beaa
 \|\dds_2 \b \|_{\infty, k-1}&\les&  \big \|e_3 \a\big\|_{\infty, k-1}+ r^{-1} \big\| \a\big\|_{\infty, k-1}+ r^{-3} \big\|\vth\|_{\infty, k-1}+\ep_0 r^{-5} u^{-1-\dec}.
 \eeaa
 Thus, in view of the above estimate for $\vth$ and {\bf Ref 2} for $\a$,
 \beaa
  \|\dds_2 \b \|_{\infty, k-1}&\les&  \ep_0r^{-4} (2r+u)^{-1/2-\dec} +\ep_0 r^{-5} u^{-1/2-\dec}.
 \eeaa
 On the other hand we have, according to \eqref{Bad-modes-onMext-b}, 
  \beaa
 \left\| \int_S \b e^\Phi\right\|_{\infty, k} &\les \ep_0 r^{-1} u^{-1-\dec}, \qquad  k\le k_{small}+20. 
 \eeaa 
  Estimate \eqref{eq:Estimate-inMext-b}  follows then easily, according to  the  part 4 of the elliptic Hodge Lemma \ref{prop:2D-hodge-reduced-M4}.

  \medskip
  
  As mentioned in Remark \ref{remark:stronger-estim-b} we can prove a stronger estimate for $\b$.
   Indeed  we have, in view of {\bf Ref 2.}
\beaa
|\a|&\les&  \log(1+u) r^{-3} (2r +u)^{-1/2 -\dee},\\
| e_3 \a|&\les&  r^{-4} (2r +u)^{-1/2 -\dee}.
\eeaa
Hence, using the equation 
$e_3 \a+\left(\frac 1 2 \kab -4\omb\right)\a=-\dds_2\b  -\frac 3 2  \vth \rho  +5\ze \b$,
\beaa
\big\| \dds_2 \b\|_{\infty, k} &\les& \ep_0  \log(1+u)  r^{-4} (2r +u)^{-1/2 -\dee} +\ep_0 r^{-5} u^{-1/2-\dec}.
\eeaa
According to Lemma \ref{prop:2D-hodge-reduced-M4}
\beaa
  \|  \b \|_{\hk_{k+1}  (S)}&\les&  r  \|\dds_2\, \b \|_{\hk_k(S)}+r^{-2}\left |\int_S e^\Phi \b \right|
\eeaa
and thus, in view of  the  estimate \eqref{Bad-modes-onMext-b} for the $\ell=1$ mode of $\b$,
\beaa
 \|  \b \|_{\hk_{k+1}  (S)}&\les& \ep_0  \log(1+u)  r^{-2} (2r +u)^{-1/2 -\dee}+\ep_0 r^{-3} u^{-1-\dec}\\
 &\les& \ep_0  \log(1+u)  r^{-2} (2r +u)^{-1/2 -\dee}.
\eeaa
The estimates for  the $T$ and $e_4$ derivatives  are  derived in the same manner.
and hence,
\bea
\big\|  \b\|_{\infty, k} &\les& \ep_0  \log(1+u) r^{-3} (2r +u)^{-1/2 -\dee}, \qquad \forall \, k\le k_{small}+19.  
\eea
This improvement is needed  in the next step.

 {\bf Step 3.} We derive the estimate
   \bea
   \lab{eq:Estimate-inMext-ze}
 \big\| \ze  \big\|_{\infty, k}&\les \ep_0 r^{-2} u^{-1/2-\dec}, \qquad \forall \, k\le k_{small}+19. 
 \eea
 For this we make use of the transport equation for $\ze$, 
 \beaa
  e_4 \ze+\ov{\ka} \ze = F :=-\b +\Ga_g  \c \Ga_g
  \eeaa
    and the improved estimate  for $\b$  in  the previous  step.
 Thus,  making use of the  product Lemma \ref{lemma:interpolation-decay},
 \beaa
 \big\| F\big\|_{\infty, k} &\les& \big\| \b\big\|_{\infty, k}+ \ep_0 r^{-7/2} u^{-1-\dec}\\
 &\les&   \ep_0  \log(1+u) r^{-3} (2r +u)^{-1/2 -\dee}  + \ep_0 r^{-7/2} u^{-1-\dec}\\
 &\les& \ep_0 r^{-3-\de} u^{-1/2-\dec}  + \ep_0 r^{-7/2} u^{-1-\dec}. 
 \eeaa
  Making use of Proposition  \ref{Prop:transportrp-f-Decay-knorms}  we deduce
 \beaa
r^2  \| \dk^k \ze \|_{\infty, k }(u, r) &\les & r_*^2   \| \dk^k \ze \|_{\infty, k }(u, r_*) +\ep_0 u^{-1/2-\dec}  \int_r^{r_*} \la^{-1-\de} d\la.
 \eeaa
  Thus, in view of  the estimates on the last slice,
  \beaa
  r^2  \| \dk^k \ze \|_{\infty }(u, r) &\les &\ep_0  u^{-1/2-\dec}, \qquad k\le k_{small}+19
  \eeaa
  as desired.
  
 {\bf Step 4.} We derive the estimate
   \bea
   \lab{eq:Estimate-inMext-rho}
 \big\| \rhoc   \big\|_{\infty, k}&\les \ep_0 r^{-3} u^{-1/2-\dec}, \qquad \forall \, k\le k_{small}+18. 
 \eea
 We make use of the definition of $\mu$ from which we infer that,
\beaa
 \muc &=& - \ddd_1 \ze -\rhoc +\Ga_g\c \Ga_b.
\eeaa
Hence, in view of the  product Lemma  and the estimates already derived, for all $k\le k_{small}+18$,
\beaa
\big\|\rhoc\big\|_{\infty, k}&\les&r^{-1} \big \|\ze\big \|_{\infty, k+1} +\big\|\muc\big \|_{\infty, k} + \ep_0 r^{-3} u^{-1-\dec}\\
&\les&\ep_0  r^{-3}  u^{-1/2-\dec}
\eeaa
as desired.

 {\bf Step 5.} We derive the estimate
   \bea
   \lab{eq:Estimate-inMext-kab}
 \big\| \kabc   \big\|_{\infty, k}&\les \ep_0 r^{-2} u^{-1/2-\dec}, \qquad \forall \, k\le k_{small}+18. 
 \eea
 We make use of the equation
 \beaa
 e_4\check{\kab} +\frac1 2 \overline{\ka} \check{\kab} =F:=-2 \ddd_1\ze -\frac 1 2 \check{\ka} \overline{\kab} +2\check{\rho}+\Ga_g \c \Ga_b.
 \eeaa
 In view of the previously derived estimates,
 \beaa
 \big\| F\big\|_{\infty,k} &\les& \ep_0 r^{-3} u^{-1/2-\dec} , \qquad k\le k_{small}+18.
 \eeaa
   Making use of Proposition  \ref{Prop:transportrp-f-Decay-knorms}  we deduce, for all $k\le k_{small}+18$,
 \beaa
r  \| \dk^k \kabc  \|_{\infty, k }(u, r) &\les & r_*    \| \dk^k \kabc  \|_{\infty, k }(u, r_*) +\ep_0 u^{-1/2-\dec}  \int_r^{r_*} \la^{-2} d\la\\
&\les&  r_*    \| \dk^k \kabc  \|_{\infty, k }(u, r_*) +\ep_0  r^{-1} u^{-1/2-\dec}.
 \eeaa
 Thus, in view of the estimates on the last slice,
 \beaa
 r  \| \dk^k \kabc  \|_{\infty }(u, r) &\les &\ep_0 ( r_*)^{-1} u^{-1/2-\dec}  +\ep_0  r^{-1} u^{-1/2-\dec}
 \eeaa
 from which the desired estimate easily follows.
  \end{proof}
 
 {\bf Step 6.}  We derive the estimate
 \bea
 \big\| e_3  \b \big\|_{\infty, k}&\les\ep_0r^{-4}  u ^{-1/2-\dec} , \qquad \forall \, k\le k_{small}+18.
 \eea
 making use of the equation $e_3 \b +(\kab-2\omb) \b=-\dds_1 \rho +3\ze \rho +\Ga_g \bb +\Ga_b \a$ and the estimates derived above for $\b$, $\dds_1 \rho$, $\ze$.
 Hence,
 \beaa
 \big\| e_3  \b \big\|_{\infty, k}&\les&  r^{-1} \big\|  \b \big\|_{\infty, k}+ \big\| \dds_1\rho  \big\|_{\infty, k}+ r^{-3}  \big\| \ze  \big\|_{\infty, k}+ \ep_0 r^{-4} u^{-1-\dec}\\
 &\les& \ep_0 r^{-4} u^{-1/2 -\dec}. 
 \eeaa

 {\bf Step 7.} 
  As a corollary of the above estimates (see also {\bf Ref 4}) we also derive, in $\Mext$, 
 \bea
 \lab{eq:Estimate-inMext-K2}
 \bsplit
 \big\| K- \ov{K} \big\|_{\infty, k-1}&\les \ep_0 r^{-3} u^{-1/2-\dec}, \qquad k\le k_{small}+19,\\
 \big\| K- \frac{1}{r^2}\big\|_{\infty, k-1}&\les \ep_0 r^{-3} u^{-1/2-\dec}, \qquad k\le k_{small}+19.
 \end{split}
 \eea
  In view of the definition of $K$ we have,
 \beaa
 e_\th (K)&=& -e_\th\left( \rhoc -\frac 1 4 \ka \kabc-\frac 1 4 \kab \kac +\frac 1 4 \vth \vthb \right).
 \eeaa
 Thus, in view of the above estimates, 
 \beaa
 \big\| e_\th K\big\|_{\infty, k} &\les& \ep_0 r^{-4} u^{-1/2-\dec} 
 \eeaa
from which the desired estimate easily follows.

  
 \section{Control in $\Mext$, Part II}


  We  derive the crucial  decay estimates  which imply, in particular,   decay of order $u^{-1-\dec}$       for  all   quantities in $\Ga$ and $\Rc$ (except  $\xib, \ombc, \Ombc$ which will be treated separately)    in the interior.
  More precisely we prove the following,
 \begin{proposition}
 \label{proposition:maindecay-Mext1}
 The following estimates hold true in $\Mext$, for all $k\le k_{small}+8$.
 \bea
 \bsplit
 \big\|\vth,\,  \ze,\, \eta, \,  \kabc, \, \vthb, \, r\b,\,  r \rhoc, \, r   \bb, \aa  \|_{\infty, k}&\les \ep_0 r^{-1} u^{-1-\dec}.
 \end{split}
 \eea
 \end{proposition}
To prove the proposition we make use of  the fact that we  already have good      decay  estimates  in  terms of powers of $u$ for  $\kac, \muc$.  We also  derive below  decay estimates  for various renormalized quantities.


\subsection{Estimate for $\eta$}


We start with  the following  simple estimate for $\eta$ in terms of $\ze$.
\bea
\lab{estimate:eta-from-ze}
\| \eta \|_{\infty, k}&\les & \| \ze \|_{\infty, k}   +     \ep_0  r^{-1} u^{-1-\dec}, \qquad \quad k\le k_{small}+17.
\eea
This     can be derived by propagation  from the last slice with the help of the equation,
\beaa
e_4(\eta-\ze)+ \frac 1 2 \ka (\eta-\ze) &=&-\frac 1 2\vth(\eta-\ze) =\Ga_g\c \Ga_b. 
\eeaa
Note that
\beaa
 \|\Ga_g\c \Ga_b \|_{\infty, k}&\les &\ep_0  r^{-3} u^{-1-\dec}.
\eeaa
 Thus making use of Proposition  \ref{Prop:transportrp-f-Decay-knorms}  we deduce
 \beaa
r  \|  \eta-\ze\|_{\infty, k }(u, r) &\les & r_*  \| \eta-\ze \|_{\infty, k }(u, r_*) +\int_r^{r_*} \la \|\Ga_g\c \Ga_b \|_{\infty, k}(u, \la)\\
 &\les & r_*  \| \eta-\ze \|_{\infty, k }(u, r_*)  +  \ep_0 u^{-1-\dec} 
\eeaa
with $r_*$ the  value of $r$ on $C(u)\cap \Si_*$.
On the last slice we have derived the estimates, recorded in  Proposition \ref{Prop.Flux-bb-vthb-eta-xib} and
Proposition \ref{Prop:top-r-estimates}
\beaa
  \| \eta \|^*_{\infty, k }&\les& \ep_0 r^{-1}u^{-1-\dec},\\
    \| \ze \|^*_{\infty, k }&\les& \ep_0r^{-2} u^{-1/2-\dec}. 
\eeaa
In view of the  dominance condition on  $r$ on $\Si_*$ we  deduce,
\beaa
 \| \eta-\ze  \|^*_{\infty, k }(u, r)&\les&\ep_0  r^{-1}u^{-1-\dec}
\eeaa
and therefore also,
\beaa
 r_*  \| \eta-\ze \|_{\infty, k }(u, r_*) &\les&\ep_0  u^{-1-\dec}.
\eeaa
Therefore,
\beaa
r  \|  \eta\|_{\infty, k }(u, r) &\les& r  \|\ze\|_{\infty, k }(u, r) +\ep_0  u^{-1-\dec}
\eeaa
as desired.


  \subsection{Crucial  Lemmas}
  \lab{subsection:CrucialLemma}

  
 We start with the following lemma.
 \begin{lemma}
\label{Lemma:EstimateXi}
The $\sk_1(\MM)$  reduced tensor  
\bea
\Xi:&=& r^2\left(e_\th(\kab)+4r\dds_1\ddd_1\ze-2r^2\dds_1\ddd_1\b\right)
\eea
verifies in $\Mext$ the estimate,
\bea
\big\|\Xi\big\|_{\infty, k}\les \ep_0  u^{-1-\dec},\qquad \forall  \, k\le k_{small}+13.
\eea

\end{lemma}
 \begin{proof}
 To  calculate $e_4\Xi$ we make use of the equations,
\beaa
e_4(\kab)+\frac 1 2 \ka\kab  &=& -2\ddd_1\ze + 2\rho -\frac 1 2  \vth\vthb +2\ze^2,\\
e_4 \ze +\ka\ze &=& -\b   -  \vth\ze,\\
 e_4\b +2\ka \b &=&\ddd_2\a  +\ze\a.
\eeaa
Since we already have  an estimate for $\muc$ we  re-express  $\rho=-\mu-\ddd_1 \ze +\frac 1 4  \vth \vthb$ and derive,
 \beaa
 e_4(\kab)+\frac 1 2 \ka\kab  &=&-2\mu -4 \ddd_1\ze +2\ze^2.
 \eeaa
  Commuting with $\dds_1$  and making use of $[\dds_1, e_4]=\frac 1 2( \ka +\vth) \dds_1  $  we derive,
 \bea
 e_4 (\dds_1\kab)+\ka \dds_1\kab +\frac 1 2 \kab \dds_1\ka &=& -\dds_1 \muc - 4 \dds_1 \ddd_1 \ze +2\dds_1 (\ze^2 )+\vth \dds_1 \kab. 
 \eea
 Hence, since $e_4 (r)=\frac r 2 \ov{\ka}$, 
 \beaa
  e_4 (r^2 \dds_1\kab)&=& r^2 (\ka-\ov{\ka} )\dds_1 \kab    -\frac 1 2 r^2  \kab \dds_1\kac- 4 r^2 \dds_1 \ddd_1 \ze -  r^2 \dds_1 \muc +r^2( \dds_1 (\ze^2 )+\vth \dds_1 \kab)\\
  &=& -\frac 1 2 r^2  \kab \dds_1\kac- 4 r^2 \dds_1 \ddd_1 \ze +\err_1
 \eeaa
 where,
 \beaa
 \err_1 :&=&  -\frac 1 2 r^2  \kab \dds_1\kac -  r^2 \dds_1 \muc+  r^2 \left( \kac \dds_1\kab +\dds_1 (\ze^2 )+\vth \dds_1 \kab\right).
 \eeaa
 In view of the estimate already established for $, \kac, \muc$ and  the product  Lemma \ref{lemma:interpolation-decay} we check,
 \beaa
 \big\| \err_1\|_{\infty, k} &\les& \ep_0 r^{-2}  u^{-1-\dec}, \qquad k\le k_{small}+19.
 \eeaa
 To simplify notation we  introduce the following.
 \begin{definition}
 We  say that a quantity  $\psi\in \sk_k(\MM)$   is $ r^{-p}  Good_a  $ provided that it verifies the estimate, everywhere in $\Mext$,
 \bea
 \big\| \psi\big\|_{\infty, k} &\les \ep_0 r^{-p} u^{-1-\dec}, \qquad \forall  k \le k_{small}+a. 
 \eea
 \end{definition}
 
 Using this notation we write,
 \bea
 \lab{eq:Xi1}
  e_4 (r^2 \dds_1\kab)&=&- 4 r^2 \dds_1 \ddd_1 \ze+ r^{-2} Good_{19}.
 \eea
 
 Using the same notation the transport equation for $\ze$ can be written in the form,  
 \beaa
 e_4 \ze +\ov{\ka}\ze = -\b   -  \vth\ze -\kac \ze =-\b +  r^{-7/2} Good _{20}. 
 \eeaa
  Commuting with   $(r\dds_1) (r\ddd_1) $ (making us of Lemma \ref{lemma-Decay:commutation}) we derive
 \beaa
 e_4( r^2\dds_1  \ddd_1 \ze) +\ov{\ka} (r^2 \dds_1 \ddd_1 \ze) =-r^2\dds_1  \ddd_1 \b +  r^{-7/2}Good_{18}. 
 \eeaa
 Since  $e_4(r)=\frac 1 2 r \ov{\ka}$ we deduce,
 \bea
  \lab{eq:Xi2}
 e_4(r^3 \dds_1  \ddd_1  \ze)&=&- \frac 1 2\ov{ \ka} r^3 \dds_1 \ddd_1  \ze  - r^3\dds_1  \ddd_1 \b  + r^{-5/2} Good_{18}. 
 \eea
 Similarly the transport equation for $\b$  takes the form
 \beaa
 e_4\b +2\ov{\ka} \b &=&\ddd_2\a  +\ze\a- 2 \kac \b =\ddd_2 \a + r^{-9/2} Good_{20} 
 \eeaa
 and,
 \beaa
 e_4 ( r^2 \dds_1\ddd_1 \b )+2\ov{\ka}  r^2 \dds_1 \ddd_1 \b= r^2 \dds_1 \ddd_1 \ddd_2 \a + r^{-9/2} Good_{18}.
 \eeaa
 As before,  since  $e_4(r)=\frac 1 2 r \ov{\ka}$,  we deduce,
 \bea
  \lab{eq:Xi3}
 e_4 ( r^4 \dds_1\ddd_1 \b )=    -\ov{\ka}  r^4 \dds_1 \ddd_1 \b+     r^4 \dds_1 \ddd_1 \ddd_2 \a + r^{-5/2} Good_{18}.
 \eea
 Combining \eqref{eq:Xi1}--\eqref{eq:Xi3} we deduce,
 \beaa
 e_4 \Xi&=& e_4\Big[ r^2\left(- \dds_1 \kab +4r\dds_1\ddd_1\ze-2r^2\dds_1\ddd_1\b\right)\Big]\\
&=&4 r^2 \dds_1 \ddd_1 \ze +4 \left(   - \frac 1 2\ov{ \ka} r^3 \dds_1 \ddd_1  \ze  - r^3\dds_1  \ddd_1 \b  \right) -2 \left(   -\ov{\ka}  r^4 \dds_1 \ddd_1 \b+     r^4 \dds_1 \ddd_1 \ddd_2 \a\right)\\
&+& r^{-2}  Good_{18}\\
&=&-2 \left(\ov{\ka}-\frac 2 r \right) r^3 \dds_1 \ddd_1 \ze+ 2 r^4 \left(\ov{\ka}-\frac 2 r \right) \dds_1 \ddd_1 \b        -2  r^4 \dds_1 \ddd_1\ddd_2 \a+ r^{-2}  Good_{18}.
 \eeaa
 Making use of    {\bf Ref 4} estimates for  $\ov{\ka}-\frac 2 r $ and the estimates for $\a$ in  {\bf Ref 2} , i.e.,
 \beaa
r^4  |\dds_1 \ddd_1\ddd_2 \a|&\les &\ep_0  r^{-1} (2 r+u)^ {-1-\dee} \les \ep_0 r^{-1-\de} u^{-1-\dee +\de}, \qquad 0<\de < \dee
\eeaa
i.e.,
\beaa
 r^4 \dds_1 \ddd_1 \ddd_2 \a&=& r^{-1-\de} Good_{13}
\eeaa
  we thus deduce,
 \beaa
 e_4 \Xi &=&   r^{-1-\de} Good_{13}.
 \eeaa
 We deduce,
 \beaa
 \| \Xi\|_{\infty, k}(u, r) &\les& \| \Xi\|_{\infty, k}(u, r_*)+\ep_0 u^{-1-\dec}  \int_r^{r_*} \la^{-1-\de}  d\la, \qquad \forall  k\le k_{small}+13. 
 \eeaa
 In view of the  estimates on the last slice it is easy to check that 
 \beaa
  \| \Xi\|_{\infty, k}(u, r_*)&\les& \ep_0 u^{-1-\dec} , \qquad \forall  k\le k_{small}+13. 
 \eeaa
 Indeed,  on the last slice,
 \beaa
\| \dds_1 \kab\|_{\infty, k}&\les& \ep_0 r^{-3} u^{-1/2-\dec},\\
\| \dds_1 \ddd_1 \ze \|_{\infty, k}&\les& \ep_0 r^{-4} u^{-1/2-\dec},\\
\| \dds_1 \ddd_1 \b  \|_{\infty, k}&\les& \ep_0 r^{-5} u^{-1/2-\dec}.
 \eeaa
 Hence, since   $r\gg u$ on $\Si_*$, 
 \beaa
  \| \Xi\|_{\infty, k}(u, r_*)&\les&\ep_0  r^{-1} u^{-1/2-\dec}\les \ep_0 u^{-1-\dec}.
 \eeaa
 Thus everywhere on $\Mext$,
  \bea
  \| \Xi\|_{\infty, k}&\les& \ep_0 u^{-1-\dec} , \qquad \forall  k\le k_{small}+13 
 \eea
 as desired.
\end{proof}

In the following lemma, we make use of the control we have already established for $\qf, \a, \aa, \kac, \muc$  in $\Mext $ to derive  two nontrivial  relations between 
 angular derivatives of $\ze, \kabc $ and $\b$.
 
 \begin{remark}      
 According to  Theorem M3  we only  have good estimates for $\aa$   along   $\TT$    and on  the last slice $\Si_*$.  To keep track of this  fact we  denote  by  $ r^{-p}Good_{a}(\aa)$   those 
$ r^{-p}Good_{a}$   terms   which depend linearly  on $\aa$  and their derivatives.
\end{remark}

\begin{lemma}
\label{Lemma:remarkable-AB}
Let  $A, B $ be the  operators  $A:=\dds_2 \ddd_2 -3 \ov{\rho}$, $B=\dds_2 \ddd_2+2 K$. The following identities  hold true,
\bea
\bsplit
AB \dds_2\ze -\frac 3 4 \ov{\ka}\, \ov{\rho} \dds_2 e_\th(\kab)& \in  r^{-6}    Good_{14}(\aa)\\
A^2 B\dds_2 \b +\frac 9 8\big( \ov{\ka} \, \ov{\rho} \big)^2  \dds_2 e_\th (\kab) &\in   r^{-9} Good_{9}(\aa)
\end{split}
 \eea
\end{lemma}
\begin{proof}
 In view of the improved control for $\a$ in Theorem M2, $\aa$ in Theorem M3, and $\qf$ in Theorem M1, the bootstrap assumptions and  product lemma, and the  control  we have already  derived  for $\check{\ka}$ and $\check{\mu}$ in $\Mext$, we obtain
\bea
\label{Eq:remarkable-AB1}
\bsplit
\ddd_2\vth  + 2\b-\ov{\ka}\ze&\in   r^{-3} Good_{20},  \qquad \qquad  \mbox{Codazzi and control of } \kac,\\
\ddd_2\vthb  +2\bb-e_\th(\kab)+\ov{\kab}\ze &\in  r^{-3}Good_{20}, \qquad  \qquad  \mbox{Codazzi},\\
\dds_2\b +\frac{3}{2}\ov{\rho}\vth &\in r^{-3} Good_{15}, \qquad  \qquad \mbox{Bianchi and control of } \a, \\
\dds_2\bb +\frac{3}{2}\ov{\rho}\vthb &\in r^{-2}  Good_{15}(\aa), \qquad \,\, \, \mbox{Bianchi  and control of }\aa,\\
\dds_2\dds_1\rho  +\frac{3}{4}\ov{\kab}\, \ov{\rho} \vth+\frac{3}{4}\ov{\ka}\, \,\,\ov{\rho}\vthb &\in r^{-4}   Good_{18}, \qquad  \quad\,\,\,\,\,\,  \eqref{eq:alternateformulaforqfinvolvingtwoangularderrivativesofrho:M4}\mbox{ \,\,and control of } \qf,
\end{split}
\eea
where we used Codazzi for the two first inequalities, Bianchi for the third and fourth inequalities, the definition of $\mu$ for the fifth one, and the identity  relating $\qf$ and $\dds_2\dds_1\rho$ for the last one.

 Combining  the first statement with the third and  the second with the fourth  we infer that,
\beaa
(\dds_2\ddd_2-3\ov{\rho})\vth - \ov{\ka}\dds_2\ze& \in  & r^{-3}  Good_{14},\\
(\dds_2\ddd_2-3\ov{\rho})\vthb - \dds_2e_\th(\kab) +\ov{\kab}\dds_2\ze & \in  &  r^{-2}Good_{14}(\aa),
\eeaa
or,
setting 
\beaa
A:=\dds_2\ddd_2-3\ov{\rho},
\eeaa
\bea
\label{Eq:Estimates-onTT1}
\bsplit
 A\vth -\ov{ \ka}\dds_2\ze& \in  r^{-3}  Good_{14},\\
 A \vthb - \dds_2e_\th(\kab) +\ov{\kab}\dds_2\ze & \in   r^{-2} Good_{14}(\aa).
 \end{split}
\eea
From the fifth equations we deduce,
\beaa
 A\left(\dds_2\dds_1\rho  +\frac{3}{4}\ov{\kab }\, \ov{\rho} \vth+\frac{3}{4}\ov{\ka}\, \ov{\rho} \vthb\right)\in r^{-6} Good_{18}
\eeaa
i.e.,
\beaa
 A \dds_2\dds_1\rho  +\frac{3}{4}\ov{\kab }\, \ov{\rho}\,  A  \vth+\frac{3}{4}\ov{\ka}\, \ov{\rho} \,  A \vthb\in r^{-6}  Good_{18}.
\eeaa
Making use of \eqref{Eq:Estimates-onTT1} we deduce,
\beaa
 A \dds_2\dds_1\rho  +\frac{3}{4}\ov{\kab }\, \ov{\rho}\, \big(\ov{\ka} \dds_2 \ze  \big)  +\frac{3}{4}\ov{\ka}\, \ov{\rho} \big(\dds_2 e_\th(\kab)-\ov{\kab} \dds_ 2 \ze \big) \in r^{-6}   Good_{14}(\aa).
\eeaa
Hence, simplifying,
\bea
\label{Eq:Estimates-onTT2}
A\dds_2\dds_1\rho +\frac{3}{4} \ov{\ka} \, \ov{\rho} \dds_2e_\th(\kab) \in r^{-6}   Good_{14}(\aa).
\eea
Next,  in view of the identity $\dds_1\ddd_1=\ddd_2\dds_2+2K$,
\beaa
(\dds_2 \ddd_2 +2K)\dds_2 \ze &=& \dds_2 \ddd_2\dds_2 \ze +2 K\dds_2 \ze\\
& =&\dds_2 \left( \ddd_2\dds_2 \ze+2K\ze\right)-2 \dds_2 K \ze \\
&=&\dds_2 \dds_1 \ddd_1 \ze + r^{-9/2} Good_{19}.
\eeaa
Recalling the definition of  $\mu=-\ddd_1\ze -\rho+\frac 1 4  \vth \vthb$  and   the product Lemma we write
\beaa
 \dds_1 \ddd_1 \ze&=&-\dds_1\mu -\dds_1\rho +\frac 1 4 \dds_1( \vth \vthb)= -\dds_1\mu -\dds_1\rho +r^{-4}  Good_{19}.
\eeaa
In view of the estimates for $\muc$  we have already established we deduce,
\beaa
 \dds_1 \ddd_1 \ze=-\dds_1\rho+r^{-4} Good_{19}.
\eeaa
Thus,
\bea
\label{Eq:Estimates-onTT3}
(\dds_2 \ddd_2 +2K)\dds_2 \ze &=&- \dds_2 \dds_1 \rho+ r^{-5} Good_{18}.
\eea
 Therefore, making use of \eqref{Eq:Estimates-onTT2}
\beaa
A(\dds_2 \ddd_2 +2K)\dds_2 \ze &=&- A \dds_2 \dds_1 \rho +r^{-6} Good_{16}(\aa)\\
&=& \frac{3}{4}\ov{\ka}\, \ov{\rho}  \dds_2e_\th(\kab) + r^{-6}  Good_{14}(\aa)
\eeaa
i.e.,
\bea
\label{Eq:Estimates-onTT4}
A(\dds_2 \ddd_2 +2K)\dds_2 \ze -\frac{3}{4}  \ov{  \ka}\, \ov{\rho} \, \dds_2e_\th(\kab) = r^{-6} Good_{14} (\aa)
\eea
as desired.

To prove the second statement of the lemma we write, using 
\eqref{Eq:Estimates-onTT1}
\beaa
(\dds_2\ddd_2+2K) A \vth&= &\ov{ \ka}\,(\dds_2\ddd_2+2K)  \dds_2\ze+ r^{-5}Good_{12}(\aa).
\eeaa
Hence applying $A$  and making use of \eqref{Eq:Estimates-onTT4},
\beaa
A(\dds_2\ddd_2+2K) A \vth&= &   \ov{\ka} A(\dds_2\ddd_2+2K)   \dds_2\ze+ r^{-7}Good_{10}(\aa)\\
&=& \frac 3 4 \ov{\ka} ^2 \, \ov{\rho} \dds_2 e_\th(\kab) +  r^{-7} Good_{10}(\aa).
\eeaa
Finally,  making use of  the relation  $\dds_2\b +\frac{3}{2}\ov{\rho}\, \vth \in  r^{-3} Good_{15},$ we have 
\beaa
A^2(\dds_2\ddd_2+2K)\dds_2\b&=& A (\dds_2\ddd_2+2K)A\dds_2\b+r^{-9} Good_{9}(\aa)\\
&=& - \frac 3 2 \ov{\rho} A (\dds_2\ddd_2+2K) A \vth + r^{-9} Good_{9}(\aa)\\
&=& - \frac 3 2\ov{ \rho}\,  \left( \frac 3 4\ov{ \ka} ^2 \, \ov{\rho} \dds_2 e_\th(\kab) + r^{-8} Good_{10}(\aa)\right)+ r^{-9} Good_{9}\\
&=&-\frac 9 8 \ov{\ka}^2\ov{ \rho}^2  \dds_2 e_\th(\kab)+ r^{-9} Good_{9}(\aa)
\eeaa
as desired. This concludes the proof of the lemma.
\end{proof}

\begin{corollary}
\lab{Corollary:remarkable-AB}
The $\sk_1(\MM)$ tensor $e_\th(\kab)=-\dds_1\kabc$  verifies the following fifth order  elliptic equation in $\Mext$
\bea
\label{Eq:Estimates-onTT5}
A^2  \dds_2 \left( e_\th \kab\right)-\frac{12m}{r^3}  A \dds_2e_\th(\kab)+\frac{36m^2}{r^6}   \dds_2 e_\th(\kab) \in r^{-7}  Good_{8}(\aa).
\eea
\end{corollary}

\begin{proof}
 According to  Lemma \ref{Lemma:remarkable-AB}
 \beaa
\bsplit
AB \dds_2\ze -\frac 3 4 \ov{\ka}\, \ov{\rho} \dds_2 e_\th(\kab)& \in    r^{-6}Good_{14}(\aa),\\
A^2 B\dds_2 \b +\frac 9 8\big( \ov{\ka} \, \ov{\rho} \big)^2  \dds_2 e_\th (\kab) &\in   r^{-9} Good_{9}(\aa),
\end{split} 
\eeaa
we have
\bea
\label{Eq:Estimates-onTT7}
\bsplit
A^2 B\dds_2 \ze &= \frac{3}{4}\ov{\ka} \, \ov{\rho}\,   A \dds_2e_\th(\kab) +r^{-8} Good_{12}(\aa), \\
A^2B\dds_2\b&=-\frac 9 8\big(\ov{ \ka}\, \ov{ \rho} \big)^2  \dds_2 e_\th(\kab)+ r^{-9} Good_{9 }(\aa).
\end{split}
\eea
In view of Lemma \ref{Lemma:EstimateXi} we have on $\Mext$,
\bea
e_\th(\kab)+ 4r\dds_1\ddd_1\ze - 2r^2\dds_1\ddd_1\b\in r^{-2} Good_{13}.
\eea
Thus,
\beaa
A^2  \dds_2 \Big( e_\th(\kab)+ 4r\dds_1\ddd_1\ze - 2r^2\dds_1\ddd_1\b\Big)\in  r^{-7} Good_{8}.
\eeaa
Making use of 
\beaa
\dds_2\dds_1\ddd_1 &=& \dds_2\Big(\ddd_2\dds_2+2K\Big)= (\dds_2\ddd_2+2K)\dds_2 -e_\th(K),
\eeaa
we deduce,
\beaa
A^2  \dds_2 \left( e_\th \kab\right)&=&-4 r A^2 \dds_2     \dds_1\ddd_1\ze+ 2 r^2 A^2 \dds_2 \dds_1\ddd_1\b+r^{-7}Good_{8}\\
&=&- 4 r  A^2 (\dds_2\ddd_2+2K)\dds_2\ze   + 2 r^2 A^2 (\dds_2\ddd_2+2K)\dds_2\b +r^{-7} Good_{8}\\
&=&- 4 r  A^2 B\dds_2 \ze +2 r^2 A^2 B\dds_2 \b  +r^{-7} Good_{8}.
\eeaa
Thus, in view of the lemma,
\beaa
 A^2  \dds_2 \left( e_\th \kab\right)&=&-3 r \left(  \ov{\ka} \, \ov{\rho } A \dds_2e_\th(\kab) +r^{-8}  Good_{12} \right)  -\frac 9 4  r^2\left(  \big(\ov{\ka} \, \ov{\rho } \big)^2  \dds_2 e_\th(\kab)+  r^{-9}  Good_{9}(\aa) \right) \\
 &+&r^{-7} Good_{8}.
\eeaa
We deduce,
\beaa
A^2  \dds_2  e_\th \kab +3 r  \left(  \ov{\ka} \, \ov{\rho } \right) A \dds_2e_\th(\kab) +\frac 9 4  r^2  \left(  \ov{\ka} \, \ov{\rho }\right)^2  \dds_2 e_\th(\kab)\in r^{-7}  Good_{8}(\aa).
\eeaa
Finally,   
\beaa
A^2  \dds_2  e_\th \kab-\frac{12m}{r^3}A \dds_2e_\th(\kab)+\frac{36 m^2}{r^6} \dds_2 e_\th(\kab)\in r^{-7}  Good_{8}(\aa)
\eeaa
as desired.
\end{proof}

\begin{lemma}\lab{lemma:trickypoincareinequalityforthesharpucontrolinMext:thmM4}
We have the following Poincar\'e inequality on $\Mext$ for $ f\in \sk_2(\MM)$ with $A=(\dds_2\ddd_2-3\ov{\rho})$
\beaa
\int_Sf\left( A^2 -\frac{12m}{r^3}A  + \frac{36m^2}{r^6}\right)f &\geq& \frac{1}{4r^2}\int_S(\ddd_2f)^2+ \frac{9}{r^4}\int_Sf^2.
\eeaa
\end{lemma}

\begin{proof}
Recall that we have the following Poincar\'e inequality for $\ddd_2$
\beaa
\int_S(\ddd_2f)^2\geq 4\int_SKf^2.
\eeaa
Since   $\big|K-r^{-2} \big|\les \ep r^{-2} $,
\beaa
\int_S f Af &=&\int_Sf(\dds_2\ddd_2-3\ov{\rho})f \geq \int_S(4K-3\ov{\rho})f^2\\
&=& \left(\frac{4}{r^2}+\frac{6m}{r^3}+O(r^{-2}\ep)\right)\int_Sf^2.
\eeaa
Since $A$ is positive self-adjoint,
\beaa
\int_S f A^2 f&=&\int_S  (A^{1/2} f) A  (A^{1/2}  f) = \left(\frac{4}{r^2}+\frac{6m}{r^3}+O(r^{-2}\ep)\right)\int_S   |A^{1/2} f |^2 \\
&=& \left(\frac{4}{r^2}+\frac{6m}{r^3}+O(r^{-2}\ep)\right)\int_S f A f 
\eeaa

This yields
\beaa
\int_Sf\left(A^2f -\frac{12m}{r^3} Af\right) &=&\left(\frac{4}{r^2}+\frac{6m}{r^3}-\frac{12m}{r^3}+O(r^{-2}\ep)\right) \int_S f A f \\
&=&\left(\frac{4}{r^2}-\frac{6m}{r^3}+O(r^{-2}\ep)\right) \int_S f A f,
\eeaa
and therefore,
\beaa
\int_Sf\left(A^2 -\frac{12m}{r^3} A  + \frac{36m^2}{r^6}\right)&\geq & \left(\frac{4}{r^2}-\frac{6m}{r^3}+O(r^{-2}\ep)\right)\int_Sf A f+ \frac{36m^2}{r^6}\int_Sf^2\\
&=& \left(\frac{4}{r^2}\left(1-\frac{3m}{2r}\right)+O(r^{-2} \ep)\right)\int_Sf A f+ \frac{36m^2}{r^6}\int_Sf^2.
\eeaa

Note that  for $r>2m$ we have,
\beaa
1-\frac{3m}{2r} >\frac 1 4. 
\eeaa
We deduce, for sufficiently small $\ep$, everywhere in $\Mext $,
\beaa
\int_Sf\left(A^2 -\frac{12m}{r^3} A  + \frac{36m^2}{r^6}\right)&> & \frac{1}{r^2}\left( \int_Sf A f+ \frac{36m^2}{r^4}\int_Sf^2\right).
\eeaa
Now,  since $\Big|\ov{\rho} +\frac{2m}{r^3} \Big|\les \ep_0 r^{-3} $
\beaa
\int_Sf A f&=&\int_Sf(\dds_2\ddd_2-3\ov{\rho})f  =\int_S\left(|\ddd_2 f|^2 +\left( \frac{6m}{r^3}  +O(r^{-3} \ep_0)\right)   |f|^2\right).
\eeaa
Hence,
\beaa
\int_Sf A f + \frac{36m^2}{r^4}\int_Sf^2 > \int_S\Big( |\ddd_2 f|^2 +\big( \frac{6m}{r^3}+ \frac{36m^2}{r^6}\big) |f|^2 \Big)> \int_S |\ddd_2 f|^2
\eeaa
or, since  $\int_S(\ddd_2f)^2\geq 4\int_S \frac{ 1}{ r^2} \int_S |f|^2  +O(\ep_0 r^{-3} )\int_S |f|^2 $. We deduce,
\beaa
\int_Sf\left(A^2 -\frac{12m}{r^3} A  + \frac{36m^2}{r^6}\right)&\geq &\frac{1}{4r^2}\int_S(\ddd_2f)^2+ \frac{9}{r^4}\int_Sf^2
\eeaa
as desired. This concludes the proof of the lemma.
\end{proof}

Applying the lemma  to    $ f= \dds_2  e_\th \kab $   in    \eqref{Eq:Estimates-onTT5}, i.e. 
\beaa
A^2  \dds_2 \left( e_\th \kab\right)-\frac{12m}{r^3}  A \dds_2e_\th(\kab)+\frac{36m^2}{r^6}   \dds_2 e_\th(\kab)\in r^{-7}  Good_{8}(\aa)
\eeaa
or, in any region where 
\beaa
\|\aa\|_{2,k} &\les & \ep_0 r^{-1} u^{-1-\dec} , \qquad k\le k_{small}+16,
\eeaa
we have
\beaa
\Big\| A^2  \dds_2 \left( e_\th \kab\right)-\frac{12m}{r^3}  A \dds_2e_\th(\kab)+\frac{36m^2}{r^6}   \dds_2 e_\th(\kab)\Big\|_{2, k}\les \ep_0 r^{-6} u^{-1-\dec}, 
\qquad k\le k_{small}+8.
\eeaa
We deduce, by  $L^2$-elliptic estimates,
\bea
\label{Eq:Estimates-onTT6}
\| \dds_2  e_\th \kabc\|_{2,  k} &\les&\ep_0  r^{-2} u^{-1-\dec}, \qquad k\le k_{small}+12.
\eea
Since we control the $\ell=1$ mode of $ e_\th \kabc$  we infer that,
\beaa
\|   e_\th \kabc\|_{2,  k} &\les&\ep_0  r^{-1} u^{-1-\dec}, \qquad k\le k_{small}+13
\eeaa
i.e.,
\beaa
\|    \kabc\|_{2,  k} &\les&\ep_0   u^{-1-\dec}, \qquad k\le k_{small}+14.
\eeaa
Therefore, using the Sobolev embedding,
\beaa
\|\kabc\|_{\infty, k}  &\les&\ep_0  r^{-1} u^{-1-\dec}  \qquad k\le k_{small}+12.
\eeaa

This proves the following,
\begin{proposition}
\lab{proposition:etimate-kabc-onTT}
In any region of $\Mext$ where,
\beaa
\|\aa\|_{2,k} &\les & \ep_0 r^{-1} u^{-1-\dec} , \qquad k\le k_{small}+16,
\eeaa
 we also have,
\bea
\lab{Eq:Estimates-onTT8}
\|\kabc\|_{\infty, k}  &\les&\ep_0  r^{-1} u^{-1-\dec}  \qquad k\le k_{small}+12.
\eea 
\end{proposition}


\subsection{Proof of Proposition \ref{proposition:maindecay-Mext1}, Part I}


We first prove 
Proposition \ref{proposition:maindecay-Mext1} in the region where the estimate
\bea
\lab{good-estimate:foraa}
\|\aa\|_{2,k} &\les & \ep_0 r^{-1} u^{-1-\dec} , \qquad k\le k_{small}+16,
\eea
holds true.

{\bf Step 1.}  We prove the estimates,
\bea
\label{eq:proposition:maindecay-Mext1-1}
\bsplit
\| \ze \|_{\infty, k}&\les  \ep_0  r^{-1} u^{-1-\dec}, \qquad k\le k_{small}+15,\\
\| \b  \|_{\infty, k}&\les  \ep_0  r^{-2} u^{-1-\dec}, \qquad k\le k_{small}+12.
\end{split}
\eea
According to  \eqref{Eq:Estimates-onTT7}
\beaa
A^2 B\dds_2 \ze &= &\frac{3}{4}\ov{\ka} \, \ov{\rho}\,   A \dds_2e_\th(\kab) +r^{-8}  Good_{12}(\aa), \\
A^2B\dds_2\b&=&-\frac 9 8\big(\ov{ \ka}\, \ov{ \rho} \big)^2  \dds_2 e_\th(\kab)+ r^{-9} Good_{9 }(\aa).
\eeaa
In view of \eqref{Eq:Estimates-onTT6}
 we deduce, in $L^2$ norms, 
 \beaa
 \|A^2B\dds_2 \ze \|_{2, k}&\les & r^{-4}  \|A\dds_2  e_\th \ka \|_{2, k} + \ep_0 r^{-7}      u^{-1-\dec}, \qquad k\le k_{small} +12, \\
  \|A^2B\dds_2 \b  \|_{2, k}&\les & r^{-8} \| \dds_2  e_\th \ka \|_{2, k} + \ep_0 r^{-8} u^{-1-\dec}, \qquad k\le k_{small} +9. 
 \eeaa
 Thus,
 in view of the estimates for  $\kac$ derived above,
\beaa
\|A^2B\dds_2 \ze \|_{2, k}&\les & \ep_0  r^{-7}  u^{-1-\dec}, \qquad k\le k_{small}+12,\\
\|A^2B\dds_2 \b  \|_{2, k}&\les & \ep_0  r^{-8} u^{-1-\dec}, \qquad k\le k_{small}+9.
\eeaa
Thus, by elliptic estimates,
\beaa
\|\dds_2 \ze \|_{2, k}&\les & \ep_0  r^{-1} u^{-1-\dec}, \qquad k\le k_{small}+16,\\
\|\dds_2 \b  \|_{2, k}&\les & \ep_0  r^{-2} u^{-1-\dec}, \qquad k\le k_{small}+13.  
\eeaa
In view of the estimates for the $\ell=1$ modes of  $\ze, \b$  we deduce,
\beaa
\bsplit
\| \ze \|_{2, k}&\les  \ep_0   u^{-1-\dec}, \qquad \quad k\le k_{small}+17,\\
\| \b  \|_{2, k}&\les  \ep_0  r^{-1} u^{-1-\dec}, \qquad k\le k_{small}+14.
\end{split}
\eeaa
Passing to $L^\infty$ norms we derive 
\bea
\lab{estimates:forzeandb-onTT}
\bsplit
\| \ze \|_{\infty, k}&\les  \ep_0  r^{-1}  u^{-1-\dec}, \qquad \quad k\le k_{small}+15,\\
\| \b  \|_{\infty , k}&\les  \ep_0  r^{-2} u^{-1-\dec}, \qquad k\le k_{small}+13.
\end{split}
\eea
{\bf Step 2.}
We prove the estimate
\beaa
\| \eta \|_{\infty, k}&\les &  \ep_0  r^{-1} u^{-1-\dec}, \qquad k\le k_{small}+15.
\eeaa

This follows immediately from the estimate from $\ze$ and   the previously derived estimate \eqref{estimate:eta-from-ze}. Indeed,
\beaa
\| \eta \|_{\infty, k}&\les & \| \ze \|_{\infty, k}   +     \ep_0  r^{-1} u^{-1-\dec}\les  \ep_0  r^{-1} u^{-1-\dec}.
\eeaa

{\bf Step 3.} 
We derive the estimate,
\bea
\lab{estimate:alphaonTT}
\|\vth\|_{\infty, k}&\les& r^{-1} u^{-1-\dec}, \qquad k\le k_{small}+11.
\eea
This follows easily in view  of the equation (see  \eqref{Eq:remarkable-AB1})
\beaa
\ddd_2\vth  + 2\b-\ov{\ka}\ze&\in   r^{-3} Good_{20}
\eeaa
 from which, in view of   Step 1,
 \beaa
\| \ddd_2 \vth\|_{2, k} &\les& \ep_0 r^{-1} u^{-1-\dec},   \qquad k\le k_{small}+12.
\eeaa
The desired estimate follows by elliptic estimates and  Sobolev.

{\bf Step 4.}  We derive the  intermediate  estimate for $\vthb$,
\bea
  \|\vthb\|_{\infty, k}&\les&\ep_0   u^{-1-\dec}, \qquad k\le k_{small}+12. 
\eea
 To show this we  combine  the equations (see  \eqref{Eq:remarkable-AB1})
\beaa
\bsplit
\ddd_2\vthb  +2\bb-e_\th(\kab)+\ov{\kab}\ze &\in  r^{-3}Good_{20},\\
\dds_2\bb +\frac{3}{2}\ov{\rho}\vthb &\in r^{-2}  Good_{15},
\end{split}
\eeaa
to  deduce,
 \beaa
 \dds_2 \ddd_2 \vthb -3\ov{\rho} \vthb &=&\dds_2 e_\th \kab +\ov{\kab} \dds_2 \ze + r^{-2} Good_{15},
 \eeaa
 and hence, 
 \beaa
 \| A\vthb\|_{2, k}&\les&\ep_0  r^{-1} u^{-1-\dec}, \qquad k\le k_{small}+12. 
 \eeaa
 Thus,
 \beaa
 \|\vthb\|_{2, k}&\les&\ep_0  r u^{-1-\dec}, \qquad k\le k_{small}+14 
 \eeaa
 and hence,
 \beaa
  \|\vthb\|_{\infty, k}&\les&\ep_0   u^{-1-\dec}, \qquad k\le k_{small}+12 
 \eeaa
as desired.
 
 {\bf Step 5.}  We derive the estimate,
 \bea
 \lab{estimate:forrhoc-onTT}
  \|\rhoc\|_{\infty, k}&\les \ep_0 r^{-2} u^{-1-\dec}, \qquad  k\le k_{small}+14.
 \eea
 
 From,
 \beaa
 \dds_2\dds_1\rho  +\frac{3}{4}\ov{\kab}\, \ov{\rho} \vth+\frac{3}{4}\ov{\ka}\, \ov{\rho}\vthb &\in r^{-4}  Good_{20}, 
 \eeaa
 we deduce,
 \beaa
 \|  \dds_2\dds_1\rho\|_{2, k} &\les& r^{-4}\left(  \|\th\|_{2, k}+ \|\thb\|_{2, k}\right) +\ep_0 r^{-3} u^{-1-\dec}, \qquad k\le k_{small}+14\\
 &\les&\ep_0 r^{-3} u^{-1-\dec}, \qquad k\le k_{small}+14.
 \eeaa
 Since we control the $\ell=1$ mode of $\dds_1\rho$  (see Lemma \ref{Lemma:Bad-modes-onMext}) we infer that,
 \beaa
 \|\rhoc\|_{2, k}&\les \ep_0 r^{-1} u^{-1-\dec}, \qquad  k\le k_{small}+16
 \eeaa
 i.e.,
 \beaa
  \|\rhoc\|_{\infty, k}&\les \ep_0 r^{-2} u^{-1-\dec}, \qquad  k\le k_{small}+14
 \eeaa
 as desired.
 
 {\bf Step 6.}  We  derive the estimate,
  \bea
  \big\|   \bb\big\|_{\infty, k} &\les& \ep_0 r^{-2} u^{-1-\dec},  \qquad \forall \,  k\le k_{small}+9
  \eea
 with the help of  the identity 
 \beaa
e_3(r\qf) &=& r^5\Bigg\{ \dds_2\dds_1\ddd_1\bb  -\frac{3}{2}\ka\rho\aa  -\frac{3}{2}\rho\dds_2\dds_1\kab -\frac{3}{2}\kab\rho\dds_2\ze + \frac{3}{4}(2\rho^2-\ka\kab\rho)\vthb\Bigg\} +\err[e_3(r\qf)],\\
\err[e_3(r\qf)]&=& r^4 (e_3 \Ga_b) \c  \dkb^{\le 1 }\b+ r\Ga_b\c  \qf + r^{2}\dkb^3 (\Ga_g \c \Ga_b),
 \eeaa
 of Proposition \ref{Le:Teuk-Star1-M4}.
 In view of \eqref{estimate-err[e_3(rqf)]} we have,
\beaa
\|\err[e_3(r\qf)]\|_{\infty, k}(u, r) &\les& \ep_0 u^{-1-\dec}, \qquad  k\le k_{small}+16.
\eeaa
We can now make use of the estimates  for  $\kabc, , \ze, \vth, \vthb$   already derived    and the {\bf Ref 2} estimate for $e_3(\qf)$ and  $\aa$  to deduce, for all $k\le k_{small}+10$,
\beaa
\| \rho\dds_2\dds_1\kab\|_{\infty, k} &\les&\ep_0  r^{-6} u^{-1-\dec},\\
\| \kab\rho\dds_2\ze\|_{\infty, k}   &\les&\ep_0  r^{-6} u^{-1-\dec},\\
\|\rho^2 \vthb\|_{\infty, k} &\les& \ep_0  r^{-6} u^{-1-\dec},\\
\|\ka \kab \vthb\|_{\infty, k} &\les& \ep_0  r^{-5} u^{-1-\dec},\\
\| \ka\rho\aa \|_{\infty, k} &\les& \ep_0  r^{-5} u^{-1-\dec},\\
\|  e_3(r\qf) \|_{\infty, k} &\les& \ep_0   u^{-1-\dec}.
\eeaa
Therefore,
\beaa
\|\dds_2\dds_1\ddd_1\bb\|_{\infty, k} &\les&  \ep_0  r^{-5} u^{-1-\dec}, \qquad k\le k_{small}+10,
\eeaa
i.e.,
\beaa
\|\dds_2\dds_1\ddd_1\bb\|_{2, k} &\les&  \ep_0  r^{-4} u^{-1-\dec}, \qquad k\le k_{small}+10.
\eeaa
Making use of the identity,
\beaa
\dds_1\ddd_1=\ddd_2\dds_2+2K,
\eeaa
we deduce
\beaa
\big\| (\dds_2\ddd_2  +K)  \dds_2 \bb\big\|_{2,  k} &\les& \ep_0 r^{-4} u^{-1-\dec}.
\eeaa
Since $\dds_2\ddd_2  +K$ is coercive we  deduce,
\beaa
\big\|  \dds_2 \bb\big\|_{2,  k} &\les& \ep_0 r^{-2} u^{-1-\dec},  \qquad \forall \,  k\le k_{small}+10.
\eeaa
     Since we control the $\ell=1$ mode of $\bb$ (see Lemma \ref{Lemma:Bad-modes-onMext}\, ) according to Lemma \ref{Lemma:Bad-modes-onSi},
  \beaa
  \big\|   \bb\big\|_{2, k} &\les& \ep_0 r^{-1} u^{-1-\dec},  \qquad \forall \,  k\le k_{small}+11.
  \eeaa
  Hence,
  \bea
  \lab{estimate:forbb-onTT}
  \big\|   \bb\big\|_{\infty, k} &\les& \ep_0 r^{-2} u^{-1-\dec},  \qquad \forall \,  k\le k_{small}+9.
  \eea
  
 {\bf Step 7.} Using the  above estimate for $\bb$ we can improve the estimate for $\vthb$ derived  in Step 4. We show, in the region where the estimate \eqref{good-estimate:foraa}  for $\aa$  holds,
 \bea
\|\vthb\|_{\infty, k}&\les&   \ep_0 r^{-1} u^{-1-\dec}, \qquad k\le k_{small}+9.
\eea
Indeed  in view of  the Codazzi equation 
  \beaa
\ddd_2\vthb  +2\bb-e_\th(\kab)+\ov{\kab}\ze &\in  r^{-3}Good_{20},
  \eeaa
   we infer that,  for all $k\le k_{small}+11$,
  \beaa
  \| \ddd_2 \vthb\|_{2, k}&\les&  \| \bb \|_{2, k}+ r^{-1} \|\kabc\|_{2, k+1}+r^{-1} \|\ze\|_{2, k} + \ep_0 r^{-2} u^{-1-\dec}\\
  &\les&  \ep_0 r^{-2} u^{-1-\dec}.
  \eeaa
  Thus,  for all $k\le k_{small}+12$
   \beaa
  \|  \vthb\|_{2, k}&\les&   \ep_0 r^{-1} u^{-1-\dec}
  \eeaa
  and hence,
  \bea
   \|  \vthb\|_{\infty, k}&\les&   \ep_0 r^{-2} u^{-1-\dec}, \qquad k\le k_{small}+10.
  \eea
 This ends the proof of Proposition \ref{proposition:maindecay-Mext1} in the region for which  the desired estimate \eqref{good-estimate:foraa} for $\aa$ holds true. 
 
 Since \eqref{good-estimate:foraa} for $\aa$ holds true on $\TT$ in view of\footnote{Recall that $r$ is bounded on $\TT$ and that $\TT\subset\Mint$  so that  \eqref{good-estimate:foraa} holds true for ${}^{(int)}\aa$ on $\TT$ in view of Theorem M3. Then, since we have ${}^{(ext)}\aa=({}^{(ext)}\Up)^2\,{}^{(int)}\aa$ on $\TT$, \eqref{good-estimate:foraa} holds indeed true for ${}^{(ext)}\aa$ on $\TT$.} Theorem M3, this ends the proof of Proposition \ref{proposition:maindecay-Mext1} on $\TT$.


 \subsection{Proof of Proposition \ref{proposition:maindecay-Mext1}, Part II}


We  extend the validity of  Proposition \ref{proposition:maindecay-Mext1} to all of $\Mext $ propagating the estimates derived in the first part   on    $\TT$.  We also recall that we have  good decay   estimates for  $\kac$ and $\muc$ everywhere on $\Mext$.

 {\bf Step 1.} We first  derive estimates  for  $\vth$ in $\MM_{ext} $ making use of  the transport  equation
\beaa
e_4(\vth)+\ov{\ka}\vth   &=&-2\a-( \ka-\ov{\ka})\vth= -2\a+     \Ga_g\c \Ga_g. 
\eeaa 
Making use of  Proposition \ref{Prop:transportrp-f-Decay-knorms} we derive, for all $r\ge r_0= r_\TT$,
\beaa
r^2  \|  \vth \|_{\infty, k} (u, r )        &  \les &  r^2_0 \|  \vth  \|_{\infty, k} (u, r_0)    +   \int_{r_0 } ^r  \la^2   \| \a\|_{\infty, k}(u,\la)  d\la+ \ep_0 u^{-1-\dec}.
  \eeaa
We now make use of the estimate,
\beaa
\| \a\| _{\infty, k} (u, r) \les \ep_0 r^{-2}u^{-1-\dec} , \qquad k\le k_{small}+20
\eeaa
and,
\beaa
 \|  \vth  \|_{\infty, k} (u, r_0)   &  \les & \ep_0 u^{-1-\dec}
 \eeaa
 derived   above in \eqref{estimate:alphaonTT},
to derive
 \beaa
r^2  \|  \vth \|_{\infty, k} (u, r )        &  \les & \ep_0 u^{-1-\dec} + \ep_0r u^{-1-\dec}.
\eeaa
Therefore, everywhere on $\Mext$,
\bea
 \|  \vth  \|_{\infty, k} (u, r)   &  \les & \ep_0 r^{-1} u^{-1-\dec}.
\eea

{\bf Step 2.} Next, we estimate $\b$ from the equation,
\beaa
e_4\b +2\ov{\ka } \b &=& \ddd_2 \a -(\ka-\ov{\ka} ) \b + \Gac_g\c\a=  \ddd_2 \a+\Ga_g\c (\a, \b)
\eeaa
to deduce in the same manner
\beaa
r^4  \|  \b \|_{\infty, k} (u, r )        &  \les &  r^2_0 \|  \b  \|_{\infty, k} (u, r_0)    +   \int_{r_0 } ^r  \la^4   \| \ddd_2 \a\|_{\infty, k}(u,\la)  d\la +  \ep_0 r u^{-1-\dec}.
  \eeaa
Thus, in view of the estimates for $\a$ in \eqref{estimates:forzeandb-onTT} and the estimates for $\a$ in {\bf Ref2},
i.e., for $0\le k\le k_{small}+20$, 
\beaa
\| \a\|_{\infty, k} \les\ep_0\min\{  r^{-2} \log(1+u)(u+2r)^{-1-\dee}, r^{-3} (u+2r)^{-\frac{1}{2}-\dee}\}.
\eeaa 
Thus we have with $I(u, r):= \int_{r_0 } ^r  \la^4   \| \ddd_2 \a\|_{\infty, k}(u,\la)  d\la$
\beaa
  I(u, r)&\les
   &\ep_0 \min\Big\{\log(1+u)  \int_{r_0}^r \la (u+2\la )^{-1-\dee} d\la ,\,  \,  \int_{r_0}^r   (u+2\la )^{-1/2-\dee}  d\la\Big\}. 
\eeaa
If $ r\le 2u$ we have,
\beaa
  \int_{r_0}^r \la (u+2\la )^{-1-\dee} d\la\les  r^2  u^{-1-\dee}  \les  r^2(u+2 r) ^{-1-\dee}
\eeaa
and
\beaa
r^{-4}I(u, r)&\les&\ep_0  r^{-2 }  \log(1+u) (u+2 r) ^{-1-\dee}.
\eeaa
If  $r\ge 2u$ we have,
\beaa
\int_{r_0}^r   (u+2\la )^{-1/2-\dee} &\les&(u+2r)^{1/2+\dee}
\eeaa
and
\beaa
r^{-4} I(u, r) &\les r^{-4}(u+2r)^{1/2+\dee}\les r^{-2} (u+2r)  ^{-1-\dee}.
\eeaa
We deduce,
\beaa
\|\b\|_{\infty, k} &\les r^{-4} \|  \b  \|_{\infty, k} (u, r_0) + \ep_0  r^{-2 }  \log(1+u) (u+2 r) ^{-1-\dee}.
\eeaa
 Thus in view of \eqref{estimates:forzeandb-onTT},
\bea
\|\b\|_{\infty, k} &\les  \ep_0  r^{-2 }  \log(1+u) (u+2 r) ^{-1-\dee}.
\eea

{\bf Step 3.} 
We  now estimate $\ze$ using the equation
 \beaa
e_4(\ze)+\ka\ze &=& -\b +\Gac_g\c\Gac_g.
\eeaa
This can be done exactly as in Step 1 making use of the estimates  already derived for $\b$  and the  estimate \eqref{estimates:forzeandb-onTT}  for $\ze$ along $\TT$. We thus derive,
\beaa
\|\ze\|_{\infty, k}&\les& \ep_0 r^{-1} u^{-1-\dec}, \qquad  k\le k_{small+15}.
\eeaa

{\bf Step 4.} We estimate $\check{\rho}$ using   equation
\beaa
\check{\rho} = -\ddd_1\ze -\check{\mu}  +\Gac_g\c\Gac_b,
\eeaa
the previous estimate for $\ze$  and $\muc$ in $\Mext$.  We deduce,
 \bea
 \lab{estimate:forrhoc-onMext}
  \|\rhoc\|_{\infty, k}&\les \ep_0 r^{-2} u^{-1-\dec}, \qquad  k\le k_{small}+14.
 \eea

{\bf Step 5.}
 We estimate $\check{\kab}$ using the equation,
\beaa
e_4\check{\kab} +\frac1 2 \overline{\ka} \check{\kab}+\frac 1 2 \check{\ka} \overline{\kab} &=-2 \ddd_1\ze+2\check{\rho}+\Gac_g \c \Gac_b.
\eeaa
Making use of the estimates  in  $\Mext$ for $\kac$, $\ze$ and $\rhoc $   as well as the estimates for $\kabc$ 
on $\TT$ in Proposition \ref{proposition:etimate-kabc-onTT} we derive, everywhere on $\Mext$,
\bea
  \|\kabc \|_{\infty, k}&\les \ep_0 r^{-1} u^{-1-\dec}, \qquad  k\le k_{small}+12.
\eea

Alternatively we can make use of the estimate  for $\Xi=r^2\left(e_\th(\kab)+4r\dds_1\ddd_1\ze-2r^2\dds_1\ddd_1\b\right)$   in Lemma \ref{Lemma:EstimateXi}, which holds everywhere on $\Mext$, and the above estimates for $ \ze, \b$.

{\bf Step 6.}   We estimate  $\bb$ everywhere on $\Mext$ with the help  of  the equation
\beaa
e_4 \bb +\ov{\ka} \bb =-\ddd_1\rho -3\ze \rho -\vthb \b -(\ka-\ov{\ka}) \bb
\eeaa
together with the estimate \eqref{estimate:forbb-onTT} for  $\bb$ on $\TT$   and the above derived estimates for 
$\rhoc, \ze$   in $ \Mext $ to infer that, 
\bea
  \big\|   \bb\big\|_{\infty, k} &\les& \ep_0 r^{-2} u^{-1-\dec},  \qquad \forall \,  k\le k_{small}+9.
  \eea

{\bf Step 7.} We extend the  for  $\vthb$ everywhere on $\Mext$ by making use of the Codazzi  equation  for $\vthb $  in 
\eqref{Eq:remarkable-AB1},
  \beaa
\ddd_2\vthb  +2\bb-e_\th(\kab)+\ov{\kab}\ze &\in  r^{-3}Good_{20}.
  \eeaa
    Using the estimates already derived above, we infer that,  for all $k\le k_{small}+11$,
  \beaa
  \| \ddd_2 \vthb\|_{2, k}&\les&  \| \bb \|_{2, k}+ r^{-1} \|\kabc\|_{2, k+1}+r^{-1} \|\ze\|_{2, k} + \ep_0 r^{-2} u^{-1-\dec}\\
  &\les&  \ep_0 r^{-2} u^{-1-\dec}.
  \eeaa
Hence, everywhere in $\Mext$,
\beaa
 \|  \vthb\|_{2, k}&\les& \ep_0 r^{-1} u^{-1-\dec}, \qquad \text{for all} \,  k\le k_{small}+12,
\eeaa
and therefore,
\beaa
 \|  \vthb\|_{\infty, k}&\les& \ep_0 r^{-2} u^{-1-\dec}, \qquad \text{for all} \,  k\le k_{small}+10.
\eeaa

{\bf Step 7.}  We  estimate $\aa$ everywhere on $\Mext$ by making use of the  equation
\beaa
e_4 \aa +\frac 12\ov{\ka} \aa =-\dds_2 \bb -\frac 3 2 \vthb  \rho - 5 \ze \bb- \frac 1 2 (\ka-\ov{\ka}) \aa
\eeaa
as well as the estimate \eqref{good-estimate:foraa} for $\aa$ on $\TT$ and the above estimates in all $\Mext$  for 
$\bb$  and $\vthb$. Proceeding as before we derive,
\bea
\|\aa\|_{\infty, k}&\les& \ep_0 r^{-1} u^{-1-\dec} for all k\le k_{small}+8.
\eea
This concludes the proof of  Proposition \ref{proposition:maindecay-Mext1}.

  
 \section{Conclusion of the Proof of Theorem M4}
 
 
 So far we have established  the following estimates, for all $k\le k_{small}+8$
 \bea
 \bsplit
 \big\| \kac, r\muc\|_{\infty, k} &\les \ep_0 r^{-2} u^{-1-\dec},\\
   \big\| \vth, \,  \ze, \, \kabc, r\rhoc  \big\|_{\infty, k}&\les \ep_0 r^{-2} u^{-1/2-\dec},\\
    \big\|\vth,\,  \ze,\, \eta,\,   \kabc, \, \vthb, \, r\b,\,  r \rhoc, \, r   \bb, \, \aa \|_{\infty, k}&\les \ep_0 r^{-1} u^{-1-\dec},\\
    \big\| \b,  re_3 \b \big\|_{\infty, k} &\les\ep_0 r^{-3} (2r+u)^{-1/2-\dec}.
 \end{split}
 \eea
 
It only remains to derive improved decay  estimates for $  e_3 ( \b, \vth, \ze, \kabc,  \rhoc)$  and the estimates for $\xib, \ombc, \vsic,  \Ombc$ as well  as $\ov{\vsi}+1$ and $\ov{\Omb}+\Up$  in terms of  $u^{-1-\dec}$ decay.
More precisely it  remains to prove the following.
\begin{proposition}
\label{Prop: endofThm4}
The following estimates hold true  on $\Mext$ for  all $k\le k_{small}+7$.
\beaa
\big\| e_3( \vth, \, \ze\, ,\kabc), \, r e_3 \b, \,  r e_3 \rhoc \big\|_{\infty, k} &\les&  \ep_0  r^{-2} u^{-1-\dec},\\
\big\|\xib, \, \ombc \big\|_{\infty, k} &\les&  \ep_0  r^{-1} u^{-1-\dec},\\
\big\| \vsic,  \Ombc, \ov{\vsi}+1,\,  \ov{\Omb}+\Up \big\|_{\infty, k} &\les&  \ep_0  u^{-1-\dec}.
\eeaa
\end{proposition} 

\begin{proof} 
We proceed in steps as follows.

{\bf Step 1.}  We make use of the equation $e_3\vth =- \frac 12 \kab\, \vth + 2\omb \vth -2\dds_2\eta -\frac 12 \ka \,\vthb+2\eta^2$  and  the previously derived estimates to 
derive,
\bea
\label{eq:Decay-transversal1}
\big\| e_3\vth\big\|_{\infty,k}&\les &\ep_0  r^{-2} u^{-1-\dec}, \qquad k\le k_{small}+ 9.
\eea

{\bf Step 2.}  We  make use of the equation $e_3 \b +(\kab-2\omb) \b=-\dds_1 \rho +3\eta \rho +\Ga_g \bb +\Ga_b \a$ and the previously derived estimates 
for $\b, \rhoc, \bb$
to derive,
\beaa
\big\| e_3\b\big\|_{\infty,k}&\les &\ep_0  r^{-3} u^{-1-\dec}, \qquad k\le  k_{small}+9.
\eeaa

{\bf Step 3.} 
To estimate $e_3 \ze$ in the next step we actually need a stronger estimate for  $e_3\b$ than the one derived above.  
At the same time we derive an improved estimate for $\b$.
  We show in fact, for some $0< \de$,
\bea
\label{eq:Decay-transversal2}
\bsplit
\big\| \b\big\|_{\infty,k}&\les \ep_0  r^{-2-\de} u^{-1-\dec}, \qquad k\le  k_{small}+10,\\
\big\| e_3\b\big\|_{\infty,k-1}&\les \ep_0  r^{-3-\de} u^{-1-\dec}, \qquad k\le  k_{small}+10.
\end{split} 
\eea
This makes use of  the equation
\beaa
e_4\b + 2\ka \b &=&\ddd_2 \a +\Ga_g \c \a= F:=\ddd_2 \a +\Ga_g \c \a -2 \kac \b 
\eeaa
and the  estimates for $\a$ in {\bf Ref 2.} Thus, for some $0<\de< \dee-\dec$, 
\beaa
\| F\|_{\infty, k} &\les & \log(1+u) r^{-3} (2r+u)^{-1-\dee} +  \ep_0r^{-4}u^{-1-\dec}\\
&\les& \ep_0 u^{-1-\dec}   r^{-3-\de}.
\eeaa
Integrating from $\TT$,  where $r=r_\TT=r_0 \les 1$,   we deduce with the help of Proposition   \ref{Prop:transportrp-f-Decay-knorms}
\beaa
r^4 \|\b\|_{\infty, k}(u, r) &\les & {r_0}^4 \|\b\|_{\infty, k}(u, {r_0})+\int_{{r_0}}^ r \la^4 \| F\|_{\infty, k} (u, \la) d\la \\
&\les&  \|\b\|_{\infty, k}(u, {r_0}) +\ep_0  \int_{{r_0}}^ r  \la^{1-\de} d\la.
\eeaa
Based on the previously derived  estimate for $\b$ we have  $ \|\b\|_{\infty, k}(u, r_\HH)\les  \ep_0 u^{-1-\dec}$.
Hence, 
\beaa
\|\b\|_{\infty, k}(u, r) &\les &  \ep_0 r^{-4} u^{-1-\dec}+ \ep_0  r^{-4} r^{2-\de}  u^{-1-\dec} \les \ep_0  r^{-2-\de} u^{-1-\dec}
\eeaa
as desired.

To prove the second estimate in \eqref{eq:Decay-transversal2} we commute  the  transport equation for $\b$ with $\T$ and make use 
of  the corresponding estimate for $\T\a$ (which follows from {\bf Ref 2}.
\beaa
\|\T\a\|_{\infty, k}&\les&  \ep_0\log(1+u) r^{-4} (2r+u)^{-1-\dee} \les  \ep_0 u^{-1-\dec}   r^{-4-\de}
\eeaa
as well as the fact that we control $\T \b$ on $\TT$, i.e. $\|\T\b\|_{\infty, k-1}(u, {r_0})\les  \ep_0 u^{-1-\dec}$.

{\bf Step 4.}  We make use of  the equation  $e_4 \ze+\ov{\ka} \ze =-\b +\Ga_g\c\Ga_g $ to derive,
\bea
\|e_3\ze\|_{k, \infty} &\les \ep_0  r^{-2} u^{-1-\dec}, \qquad k\le k_{small}+9.
\eea
Indeed commuting the equation with $\T$ we derive,
\beaa
e_4\T  \ze+\ov{\ka} \T  \ze =F:=-\T \b +[\T, e_4]\ze  +\ze \T\ov{ \ka }   + \T(\Ga_g\c\Ga_g ).
\eeaa
It is easy to check, in view of  the commutation Lemma \ref{lemma-Decay:commutation},
\beaa
\|F\|_{\infty, k-1}&\les&\| \T\b\|_{\infty, k-1}+\ep_0 r^{-4} u^{-1-\dec}.
\eeaa
Thus, in view of the estimate for $e_3\ze$  derived in Step 3  and  the  estimate for $e_4\ze $ we infer that,
\beaa
\|F\|_{\infty, k-1}&\les& \ep_0 r^{-3-\de} u^{-1-\dec}.
\eeaa
Integrating from $\TT$  and using  the previously derived estimate $\|\ze\|_{k, \infty}\les \ep_0 r^{-1} u^{-1-\dec}$
\beaa
r^2 \|\T \ze\|_{\infty, k-1} &\les& {r_0} ^2 \|\T \ze\|_{\infty, k-1}(u, {r_0})+\ep_0 u^{-1-\dec} \int_{{r_0}}^r  \la^{-1-\de} d\la  \\
&\les& \| \T \ze\|_{\infty, k-1}(u, {r_0})+\ep_0  r ^{-\de} u^{-1-\dec}\les 
\ep_0 u^{-1-\dec}.
\eeaa
Hence 
\beaa
 \|\T \ze\|_{\infty, k-1} &\les&  \ep_0 r^{-2} u^{-1-\dec} 
\eeaa
 from which the desired estimate easily follows.

{\bf Step 5.}            We   make use of the equation    $e_4(\check{\omb}) = \check{\rho}+\Ga_g\c \Ga_b $  and the previously derived estimates for $\rhoc$  as well
 as the estimates of $\ombc$ on the last slice (see Proposition  \ref{Lemma: :Estimates-onSi*-1})
to derive  the estimate 
\bea
\|\ombc\|_{\infty,k}&\les& \ep_0 r^{-1} u^{-1-\dec}, \qquad k\le k_{small}+9. 
\eea
Indeed, 
\beaa
\| e_4\ombc\|_{\infty, k}&\les&\| \rhoc\|_{\infty,k}+ \ep_0 r^{-3} u^{-1-\dec} \les \ep_0 r^{-2} u ^{-1-\dec}. 
\eeaa
Thus, applying  Proposition   \ref{Prop:transportrp-f-Decay-knorms}, integrating   from $\Si_*$ and using the previously  derived  estimate  for  $\ombc$ on $\Si_*$,
\beaa
\|\ombc\|_{\infty, k}(u, r) &\les& \|\ombc\|_{\infty, k}(u, r_*) +\ep_0 u^{-1-\dec}  \int_r^{r_*}  \la^{-2}  d\la \\
&\les& \ep_0 r^{-1} u^{-1-\dec}
\eeaa
as desired.

{\bf Step 6.}  We  derive the estimate,
\bea
\|\xib\|_{\infty, k}&\les& \ep_0 r^{-1} u^{-1-\dec}, \qquad k\le k_{small}+9
\eea
by   making use of the transport equation $ e_4(\xib) =F:= -e_3(\ze)+\bb-\frac{1}{2}\kab(\ze+\eta)+\Ga_b\c \Ga_b$.
In view of the previously derived estimates for $e_3\ze, \bb, \ze, \eta$ we derive,
\beaa
\|F\|_{\infty, k}&\les& \ep_0 r^{-2} u^{-1-\dec}.
\eeaa
Integrating from  $\Si_*$ and making use of   the estimate for $\xib$ on $\Si_*$ (see Proposition \ref{Lemma: :Estimates-onSi*-1})
we derive,
\beaa
\|\xib\|_{\infty, k}(u, r) &\les& \|\xib\|_{\infty, k}(u, r_*) +\ep_0 r^{-1}  u^{-1-\dec}\les \ep_0  r^{-1} u^{-1-\dec}.
\eeaa

{\bf Step 7.}  We derive  the estimate
\bea
\|\Ombc\|_{\infty, k}&\les& \ep_0 u^{-1-\dec},  \qquad k\le k_{small}+8. 
\eea
This   follows immediately from the   the equation
 $ e_\th(\Omb)=-\xib- (\eta-\ze)\Omb$, see \eqref{geodesic-foliation-M4},  and the    previous estimate for $\xib$. Note that $\ov{\Omb}$ has been estimated  in Lemma \ref{lemma:estimatesonceandforallforaverages}.

{\bf Step 8.} 
We derive the estimate 
\bea
\|\vsi- 1 \|_{\infty, k}&\les& \ep_0 u^{-1-\dec},  \qquad k\le k_{small}+8. 
\eea
The estimate  follows from the propagation equation  $  e_4(\vsi) =0$ and  the  estimate for $\vsi -1$ on the last slice  $\Si_*$.

{\bf Step 9.} 
We derive the estimate,
\bea
\|e_3\rhoc\|_{\infty, k}&\les& \ep_0 r^{-3} u^{-1-\dec},  \qquad k\le k_{small}+8  
\eea
with the help of the equation (see Proposition \ref{propos:transportaverages-M4}) 
\beaa
e_3\rhoc&= r^{-2} \dkb^{\leq 1}\Ga_b  + r^{-1} \Ga_b\c\Ga_b
\eeaa
and the previously derived estimates for $\bb, \kabc, \rhoc, \Ombc, \vsic$.

{\bf Step 10.} We derive the estimate,
\bea
\|e_3\kabc \|_{\infty, k}&\les& \ep_0 r^{-2} u^{-1-\dec},  \qquad k\le k_{small}+8  
\eea
using the  equation (see Proposition \ref{propos:transportaverages-M4}) 
\beaa
e_3 \kabc&= r^{-1} \dkb^{\leq 1}\Ga_b  +\Ga_b\c\Ga_b 
\eeaa
and the previously derived estimates for $\kabc, \xib,\ombc,  \Ombc, \vsic$.
This ends the  proof of Proposition \ref{Prop: endofThm4} and Theorem M4.
\end{proof}


\section{Proof of Theorem M5}


Recall from Theorem M3 that we have obtained the following estimate for $\,{}^{(int)}\aa$ in $\Mint$
\bea\lab{eq:controlofalphabarneededforThmM5andcomingfromThmM3}
\max_{0\leq k\leq k_{small}+16}\sup_{\Mint}\ub^{1+\dec}|\dk^k\aa| &\les& \ep_0. 
\eea

{\bf Step 1.} We consider the control of the other curvature components, as well as the Ricci components on $\TT$. Recall that the $(\ub,\sint)$ foliation is initialized on $\TT$ as follows
\begin{itemize}
\item $\ub$ and $\sint$ are defined on $\TT$ by
\beaa
\ub=u\textrm{ and }\sint=\sext\textrm{ on }\TT.
\eeaa
In particular, the 2-spheres $\S(u, \sint)$ coincide on $\TT$ with the 2-sphere $\S(u, \sext)$. 

\item In view of the above initialization, and the fact that $\TT=\{r=\rh\}$, we infer that 
$$\rint=\rext=\rh, \qquad \mint=\mext.$$

\item The null frame $(\,{}^{(int)}e_3, \,{}^{(int)}e_4, \,{}^{(int)}e_\th)$ is defined on $\TT$ by
$$\,{}^{(int)}e_4= \,{}^{(ext)}\la\,{}^{(ext)}e_4,\,\,\,\, \,{}^{(int)}e_3= (\,{}^{(ext)}\la)^{-1}\,{}^{(ext)}e_3,\,\,\,\, \,{}^{(int)}e_\th=\,{}^{(ext)}e_\th\,\,\textrm{ on }\TT$$
where
$$\,{}^{(ext)}\la=1-\frac{2\,{}^{(ext)}m}{\rext}.$$
\end{itemize}

In particular, we deduce the following identities for the curvature components and Ricci coefficients on $\TT$.
\begin{lemma}\lab{lemma:howtogofromfoliationofMexttofoliationofMintonTT}
We have on $\TT$
\beaa
\,{}^{(int)}\vsib &=& -\frac{\ov{\,{}^{(ext)}\kab}+\,{}^{(ext)}\Ab}{\ov{\,{}^{(ext)}\ka}}\la^{-1}\,{}^{(ext)}\vsi,\\
\,{}^{(int)}\Om &=& \la - \la^2\frac{\ov{\,{}^{(ext)}\ka}}{\ov{\,{}^{(ext)}\kab}+\,{}^{(ext)}\Ab}  - \la\frac{\ov{\,{}^{(ext)}\ka}}{\ov{\,{}^{(ext)}\kab}+\,{}^{(ext)}\Ab}\,{}^{(ext)}\Omb.
\eeaa
where
\beaa
\la=\,{}^{(ext)}\la=1-\frac{2\,{}^{(ext)}m}{\rext}.
\eeaa

Moreover, we have on $\TT$
\beaa
\,{}^{(int)}\a=\la^2\,{}^{(ext)}\a,\,\, \,{}^{(int)}\b=\la\,{}^{(ext)}\b,\,\, \,{}^{(int)}\rho=\,{}^{(ext)}\rho,\,\, \,{}^{(int)}\bb=\la^{-1}\,{}^{(ext)}\bb,\,\, \,{}^{(int)}\aa=\la^{-2}\,{}^{(ext)}\aa,
\eeaa
\beaa
&&\,{}^{(int)}\xib =0, \, \,  \,{}^{(int)}\omb=0, \,\, \,{}^{(int)}\ze = \,{}^{(ext)}\ze, \ \ \,{}^{(int)}\etab=-\,{}^{(ext)}\ze,\\ 
&&\,{}^{(int)}\ka = \la\,{}^{(ext)}\ka, \,\,\, \,{}^{(int)}\vth = \la\,{}^{(ext)}\vth, \,\, \,{}^{(int)}\kab = \la^{-1}\,{}^{(ext)}\kab, \,\, \,{}^{(int)}\vthb = \la^{-1}\,{}^{(ext)}\vthb,\\
\eeaa
and 
\beaa
\,{}^{(int)}\xi &=&  \frac{\la^2\ov{\,{}^{(ext)}\ka}}{\ov{\,{}^{(ext)}\kab}+\,{}^{(ext)}\Ab}(\,{}^{(ext)}\ze - \,{}^{(ext)}\eta),\\
 \,{}^{(int)}\om &=&  \frac{\la\ov{\,{}^{(ext)}\ka}}{\ov{\,{}^{(ext)}\kab}+\,{}^{(ext)}\Ab}\,{}^{(ext)}\omb ,\\
 \,{}^{(int)}\etab &=& \,{}^{(ext)}\ze - \frac{\ov{\,{}^{(ext)}\ka}}{\ov{\,{}^{(ext)}\kab}+\,{}^{(ext)}\Ab}\,{}^{(ext)}\xib.
\eeaa
\end{lemma}

\begin{proof}
The following vectorfield is tangent to $\TT$
\beaa
\nu_\TT &:=& \,{}^{(ext)}e_3 -\frac{\ov{\,{}^{(ext)}\kab}+\,{}^{(ext)}\Ab}{\ov{\,{}^{(ext)}\ka}}\,{}^{(ext)}e_4,
\eeaa
which can also be written as 
\beaa
\nu_\TT &=& \la\,{}^{(int)}e_3 -\frac{\ov{\,{}^{(ext)}\kab}+\,{}^{(ext)}\Ab}{\ov{\,{}^{(ext)}\ka}}\la^{-1}\,{}^{(int)}e_4.
\eeaa
Since $\nu_\TT$ is tangent to $\TT$, and in view of the definition of $\ub$ and $\sint$, we immediately infer
\beaa
\nu_\TT(\ub)=\nu_\TT(u)\textrm{ and }\nu_\TT(\sint)=\nu_\TT(\sext)\textrm{ on }\TT
\eeaa
and hence, using the identities
\beaa
\,{}^{(ext)}e_4(u)=\,{}^{(int)}e_3(\ub)=0, \quad \,{}^{(ext)}e_4(\sext)=1, \quad \,{}^{(int)}e_3(\sint)=-1,
\eeaa
we deduce on $\TT$
\beaa
 -\frac{\ov{\,{}^{(ext)}\kab}+\,{}^{(ext)}\Ab}{\ov{\,{}^{(ext)}\ka}}\la^{-1}\,{}^{(int)}e_4(\ub) &=& \,{}^{(ext)}e_3(u),\\
- \la -\frac{\ov{\,{}^{(ext)}\kab}+\,{}^{(ext)}\Ab}{\ov{\,{}^{(ext)}\ka}}\la^{-1}\,{}^{(int)}e_4(\sint) &=& \,{}^{(ext)}e_3(\sext) -\frac{\ov{\,{}^{(ext)}\kab}+\,{}^{(ext)}\Ab}{\ov{\,{}^{(ext)}\ka}}.
\eeaa
In view of the definition of ${}^{(ext)}\vsi$, ${}^{(int)}\vsib$, ${}^{(ext)}\Omb$ and ${}^{(int)}\Om$, this yields
\beaa
\,{}^{(int)}\vsib &=& -\frac{\ov{\,{}^{(ext)}\kab}+\,{}^{(ext)}\Ab}{\ov{\,{}^{(ext)}\ka}}\la^{-1}\,{}^{(ext)}\vsi,\\
\,{}^{(int)}\Om &=& \la - \la^2\frac{\ov{\,{}^{(ext)}\ka}}{\ov{\,{}^{(ext)}\kab}+\,{}^{(ext)}\Ab}  - \la\frac{\ov{\,{}^{(ext)}\ka}}{\ov{\,{}^{(ext)}\kab}+\,{}^{(ext)}\Ab}\,{}^{(ext)}\Omb.
\eeaa

Next, we consider the Ricci coefficients of $\Mint$ on $\TT$. From 
\beaa
\,{}^{(int)}e_4=  \la\,{}^{(ext)}\, e_4,\qquad {}^{(int)}e_3= \la^{-1}\, \,{}^{(ext)}e_3, \qquad {}^{(int)}e_\th=\,{}^{(ext)}e_\th\,\,\textrm{ on }\TT,
\eeaa
the fact that $\la$ is constant on $\TT$, and the fact that $\,{}^{(ext)}e_\th$ is tangent to $\TT$, we infer on $\TT$
\beaa
\,{}^{(int)}\a=\la^2\,{}^{(ext)}\a,\,\, \,{}^{(int)}\b=\la\,{}^{(ext)}\b,\,\, \,{}^{(int)}\rho=\,{}^{(ext)}\rho,\,\, \,{}^{(int)}\bb=\la^{-1}\,{}^{(ext)}\bb,\,\, \,{}^{(int)}\aa=\la^{-2}\,{}^{(ext)}\aa,
\eeaa
and 
\beaa
\,{}^{(int)}\ze = \,{}^{(ext)}\ze, \,\,\,{}^{(int)}\ka = \la\,{}^{(ext)}\ka, \,\,\, \,{}^{(int)}\vth = \la\,{}^{(ext)}\vth, \,\, \,{}^{(int)}\kab = \la^{-1}\,{}^{(ext)}\kab, \,\, \,{}^{(int)}\vthb = \la^{-1}\,{}^{(ext)}\vthb.
\eeaa
Also, since the foliation of $\Mint$ is ingoing geodesic, we have 
\beaa
\,{}^{(int)}\xib =0, \, \,  \,{}^{(int)}\omb=0, \ \ \,{}^{(int)}\etab=-\,{}^{(int)}\ze. 
\eeaa

It remains to find identities for $\,{}^{(int)}\xi$, $\,{}^{(int)}\om$ and $\,{}^{(int)}\eta$. Since $\la$ is constant on $\TT$ and $\nu_\TT$ tangent to $\TT$, we have on $\TT$
\beaa
D_{\nu_T}\,{}^{(int)}\, e_4 = \la D_{\nu_T}\,{}^{(ext)}\, e_4, \quad D_{\nu_T}\,{}^{(int)}\, e_3 = \la^{-1} D_{\nu_T}\,{}^{(ext)}\, e_3
\eeaa
and hence
\beaa
g(D_{\nu_T}\,{}^{(int)}\, e_4, \,{}^{(int)}\, e_\th)  &=& \la g(D_{\nu_T}\,{}^{(ext)}\, e_4, \,{}^{(ext)}\, e_\th),\\
g(D_{\nu_T}\,{}^{(int)}\, e_4, \,{}^{(int)}\, e_3)  &=& g(D_{\nu_T}\,{}^{(ext)}\, e_4, \,{}^{(ext)}\, e_3),\\
g(D_{\nu_T}\,{}^{(int)}\, e_3, \,{}^{(int)}\, e_\th)  &=& \la^{-1} g(D_{\nu_T}\,{}^{(ext)}\, e_3, \,{}^{(ext)}\, e_\th).
\eeaa
We deduce 
\beaa
2\la\,{}^{(int)}\eta -2\frac{\ov{\,{}^{(ext)}\kab}+\,{}^{(ext)}\Ab}{\ov{\,{}^{(ext)}\ka}}\la^{-1}\,{}^{(int)}\xi &=& \la\left(2\,{}^{(ext)}\eta -2\frac{\ov{\,{}^{(ext)}\kab}+\,{}^{(ext)}\Ab}{\ov{\,{}^{(ext)}\ka}}\,{}^{(ext)}\xi\right),\\
-4\la\,{}^{(int)}\omb  -4\frac{\ov{\,{}^{(ext)}\kab}+\,{}^{(ext)}\Ab}{\ov{\,{}^{(ext)}\ka}}\la^{-1}\,{}^{(int)}\om &=& -4\,{}^{(ext)}\omb -4\frac{\ov{\,{}^{(ext)}\kab}+\,{}^{(ext)}\Ab}{\ov{\,{}^{(ext)}\ka}}\,{}^{(ext)}\om,\\
 2\la\,{}^{(int)}\xib -2\frac{\ov{\,{}^{(ext)}\kab}+\,{}^{(ext)}\Ab}{\ov{\,{}^{(ext)}\ka}}\la^{-1}\,{}^{(int)}\etab &=& \la^{-1}\left(2\,{}^{(ext)}\xib -2\frac{\ov{\,{}^{(ext)}\kab}+\,{}^{(ext)}\Ab}{\ov{\,{}^{(ext)}\ka}}\,{}^{(ext)}\etab\right),
\eeaa
and thus
\beaa
\,{}^{(int)}\xi &=&  \frac{\la^2\ov{\,{}^{(ext)}\ka}}{\ov{\,{}^{(ext)}\kab}+\,{}^{(ext)}\Ab}(\,{}^{(ext)}\ze - \,{}^{(ext)}\eta),\\
 \,{}^{(int)}\om &=&  \frac{\la\ov{\,{}^{(ext)}\ka}}{\ov{\,{}^{(ext)}\kab}+\,{}^{(ext)}\Ab}\,{}^{(ext)}\omb ,\\
 \,{}^{(int)}\etab &=& \,{}^{(ext)}\ze - \frac{\ov{\,{}^{(ext)}\ka}}{\ov{\,{}^{(ext)}\kab}+\,{}^{(ext)}\Ab}\,{}^{(ext)}\xib.
\eeaa
This concludes the proof of the lemma.
\end{proof}

\begin{remark}
Since the 2-spheres $\S(u, \sint)$ coincide on $\TT$ with the 2-sphere $\S(u, \sext)$, the above lemma immediately yields 
\beaa
&& \,{}^{(int)}\check{\rho}=\,{}^{(ext)}\check{\rho},\quad \,{}^{(int)}\check{\ka}=\la \,{}^{(ext)}\check{\ka},\quad \,{}^{(int)}\check{\kab}=\la^{-1}\,{}^{(ext)}\check{\kab}\\
&&\,{}^{(int)}\check{\mub}=-\,{}^{(ext)}\check{\mu}-2\,{}^{(ext)}\check{\rho}+\frac{1}{2}\,{}^{(ext)}\vth\,{}^{(ext)}\vthb-\frac{1}{2}\overline{\,{}^{(ext)}\vth\,{}^{(ext)}\vthb},\\
&& \,{}^{(int)}\check{\om} =  \la\ov{\,{}^{(ext)}\ka}\left(\frac{\,{}^{(ext)}\omb}{\ov{\,{}^{(ext)}\kab}+\,{}^{(ext)}\Ab} - \ov{\frac{\,{}^{(ext)}\omb}{\ov{\,{}^{(ext)}\kab}+\,{}^{(ext)}\Ab}}\right),
\eeaa
and
\beaa
\,{}^{(int)}\check{\vsib} &=& -\frac{1}{\la\ov{\,{}^{(ext)}\ka}}\left((\ov{\,{}^{(ext)}\kab}+\,{}^{(ext)}\Ab)\,{}^{(ext)}\vsi -\ov{(\ov{\,{}^{(ext)}\kab}+\,{}^{(ext)}\Ab)\,{}^{(ext)}\vsi}\right),\\
\,{}^{(int)}\Omc &=&  - \la^2\ov{\,{}^{(ext)}\ka}\left(\frac{1}{\ov{\,{}^{(ext)}\kab}+\,{}^{(ext)}\Ab} - \ov{\frac{1}{\ov{\,{}^{(ext)}\kab}+\,{}^{(ext)}\Ab}}\right)\\
&&  - \la\ov{\,{}^{(ext)}\ka}\left(\frac{\,{}^{(ext)}\Omb}{\ov{\,{}^{(ext)}\kab}+\,{}^{(ext)}\Ab} - \ov{\frac{\,{}^{(ext)}\Omb}{\ov{\,{}^{(ext)}\kab}+\,{}^{(ext)}\Ab}}\right).
\eeaa
\end{remark}

Together with the estimates on $\TT$ for the outgoing geodesic foliation of $\Mext$ derived in  Theorem M4, we infer the control of tangential derivatives to $\TT$, i.e. $(e_\th, T_\TT)$ derivatives. Recovering the traversal derivative thanks to the transport equations in the direction $e_3$, we infer for the ingoing geodesic foliation of $\Mint$ on $\TT$
\beaa
\max_{0\leq k\leq k_{small}+8}\sup_{\TT}u^{1+\dec}\Big\|\dk^k\Big(\,{}^{(int)}\a, \,{}^{(int)}\b, \,{}^{(int)}\check{\rho}, \,{}^{(int)}\bb, \,{}^{(int)}\check{\mub}, \,{}^{(int)}\check{\ka}, \,{}^{(int)}\vth, \,{}^{(int)}\ze,\\
\,{}^{(int)}\etab, \,{}^{(int)}\check{\kab}, \,{}^{(int)}\vthb, \,{}^{(int)}\xi, \,{}^{(int)}\check{\om}, \,{}^{(int)}\check{\vsib}, \,{}^{(int)}\check{\Om}\Big)\Big\|_{L^2(S)}&\les& \ep_0.
\eeaa

{\bf Step 2.} Relying on the estimates of the  ingoing geodesic foliation of $\Mint$ on $\TT$ derived in Step 1, we propagate these estimates to $\Mint$ thanks to transport equations in the $e_3$ direction given by the null structure equations and Bianchi identities. Recalling that $\aa$ has already been estimated in Theorem M3, see \eqref{eq:controlofalphabarneededforThmM5andcomingfromThmM3}, quantities are recovered in the following order
\begin{enumerate}
\item We recover $\check{\kab}$, with a control of $k_{small}+8$ derivatives, from
\beaa
e_3\check{\kab}+\overline{\kab}\, \check{\kab}&=&\err[e_3\check{\kab}].
\eeaa

\item We recover $\vthb$, with a control of $k_{small}+8$ derivatives, from
\beaa
e_3(\vthb)+\kab \, \vthb  &=& -2\aa.
\eeaa

\item We recover $\bb$, with a control of $k_{small}+8$ derivatives, from 
\beaa
e_3\bb +2\kab \,\bb &=& \ddd_2\aa -\ze \aa.
\eeaa

\item We recover $\ze$, with a control of $k_{small}+8$ derivatives, from
\beaa
e_3(\ze) +\kab\ze &=& \bb   - \vthb\ze.
\eeaa

\item We recover $\etab$, with a control of $k_{small}+8$ derivatives, from 
\beaa
e_3(\etab+\ze)+\frac{1}{2}\kab(\etab+\ze) &=&   -\frac 1 2 \vthb (\etab+\ze).
\eeaa

\item We recover $\check{\mub}$, with a control of $k_{small}+8$ derivatives, from
\beaa
e_3\check{\mub} +\frac 3 2\overline{ \kab}\check{ \mub} +\frac  3 2 \overline{\mub}\check{\kab} &=& \err[e_3\check{\mub}].
\eeaa

\item We recover $\check{\rho}$, with a control of $k_{small}+7$ derivatives, from
\beaa
e_3\check{\rho}+\frac  3 2 \overline{\kab} \check{\rho}+\frac 3 2 \overline{\rho}\check{\kab}&=& \ddd_1\bb+\err[e_3\check{\rho}].
\eeaa

\item We recover $\check{\ka}$, with a control of $k_{small}+7$ derivatives, from 
\beaa
e_3\check{\ka} +\frac1 2 \overline{\kab} \check{\ka}+\frac 1 2 \check{\kab} \overline{\ka} &=& 2 \ddd_1\ze+2\check{\rho}+\err[e_3\check{\ka}].
\eeaa

\item We recover $\vth$, with a control of $k_{small}+7$ derivatives, from 
\beaa
e_3\vth +\frac 12 \kab\, \vth  &=& -2\dds_2\ze -\frac 12 \ka \,\vthb+2\ze^2.
\eeaa

\item We recover $\b$, with a control of $k_{small}+6$ derivatives, from 
\beaa
e_3 \b+ \kab \b &=& e_\th(\rho)   + 3\ze\rho- \vth \bb.
\eeaa

\item We recover $\a$, with a control of $k_{small}+5$ derivatives, from 
\beaa
e_3 \a+\frac 1 2 \kab \a &=& -\dds_2\b  -\frac 3 2  \vth \rho  +5\ze \b.
\eeaa

\item We recover $\check{\om}$, with a control of $k_{small}+7$ derivatives, from 
\beaa
e_3\check{\om}&=&\check{\rho}+\err[e_3\check{\om}].
\eeaa

\item We recover $\Oc$, with a control of $k_{small}+7$ derivatives, from 
\beaa
e_3(\Oc) &=& -2 \check{\om} + \overline{\check{\kab}\Oc}.
\eeaa

\item We recover $\xi$, with a control of $k_{small}+6$ derivatives, from 
\beaa
e_3(\xi) &=& e_4(\ze)+\b +\frac{1}{2}\ka(\ze-\etab)+\frac{1}{2}\vth(\ze-\etab).
\eeaa

\item We recover $\vsib$, with a control of $k_{small}+8$ derivatives, from
\beaa
e_3(\vsib-1)=0.
\eeaa
\end{enumerate}

As the estimates are significantly simpler to derive\footnote{Note that $r$ is bounded on $\Mint$ and that all quantities behave the same in $\Mint$.} 
and in the same spirit than the corresponding ones in Theorem M4, we leave the details to the reader. This concludes the proof of Theorem M5.


\chapter{INITIALIZATION AND EXTENSION (Theorems M6, M7, M8)}\lab{chap:proofoftheoremM0M7M8}


In this chapter, we prove M6 concerning initialization, Theorem M7 concerning extension, and Theorem M8 concerning the improvement of hight order weighted energies.


\section{Proof of Theorem M6}


{\bf Step 1.} Let $r_0$ such that 
\bea\lab{eq:choicer0forproofThmM6}
r_0 &:=& 2\ep_0^{-\frac{2}{3}},
\eea
and let $\de_0>0$ sufficiently small. Consider the unique sphere $\ovS$ of the initial data layer on $\CC_{(1+\de_0, \LL_0)}$ with area radius $r_0$. Then, denoting $S(u_{\LL_0}, \sext_{\LL_0})$ the spheres of the outgoing portion of the initial data layer, we have 
\beaa
\ovS=S(\ug, \sg), \qquad \ug=1+\de_0, \qquad |\sg-r_0|\les \ep_0.
\eeaa
Relying on the control of the initial data layer given by \eqref{def:initialdatalayerassumptions}, i.e. 
\beaa
\Ik_{k_{large}+5}\leq \ep_0,
\eeaa
we then invoke Theorem GCMS-II  of section \ref{sec:statementofGCMresultsinChapter3} with the choices
\beaa
\dg=\epg=\ep_0, \quad s_{max}=k_{large}+5,
\eeaa
to produce a unique GCM sphere $S_*$, which is a deformation of $\ovS$, satisfying 
\beaa
\ka^{S_*}=\frac{2}{r^{S_*}}, \quad \dds_2^{S_*}\dds_1^{S_*}\kab^{S_*}=\dds_2^{S_*}\dds_1^{S_*}\mu^{S_*}=0,\quad  \int_{S_*}  \b^{S_*}e^\Phi=0, \quad \int_{S_*}  e_\th^{S_*}(\kab^{S_*}) e^\Phi=0 \textrm{ on }S_*.
\eeaa

{\bf Step 2.} Starting from $S_*$ constructed in Step 1, and relying on the control of the initial data layer, we then invoke Theorem GCMH  of section \ref{sec:statementofGCMresultsinChapter3} to produce a smooth spacelike hypersurface $\Si_*$ included in the initial data layer, passing through  the sphere $S_*$, and a  scalar function $u$ defined  on $ \Si_*$ such that
\begin{itemize}
\item The following GCM conditions holds 
\beaa
\ka=\frac{2}{r}, \quad \dds_2\dds_1\kab=\dds_2\dds_1\mu=0, \quad \int_S\eta e^\Phi=\int_S\xib e^\Phi=0\textrm{ on }\Si_*
\eeaa

\item We have, for some constant $c_{\Si_*}$,  
\beaa
u+r=c_{\Si_*}, \qquad \textit{along} \quad \Si_*.
\eeaa
\item The following normalization condition  holds true  at the  South Pole $SP$  of every sphere $S$,
 \beaa
 a\Big|_{SP}=-1 -\frac{2m}{r}
 \eeaa
 where $a$ is such that we have
 \beaa
 \nu=e_3+ae_4,
 \eeaa
 with $\nu$  the unique vectorfield tangent to the hypersurface $\Si_*$, normal to $S$, and normalized by $g(\nu, e_4)=-2$. 
\end{itemize}
Furthermore, we have\footnote{We have in fact 
\beaa
\max_{k\leq k_{large}+6}\sup_{\Si_*}\Big(\|\dk^kf\|_{L^2(S)}+\|\dk^k\fb\|_{L^2(S)}+\|\dk^k\log(\la)\|_{L^2(S)}\Big) &\les& \ep_0, 
\eeaa
and then use the Sobolev embedding on the 2-spheres $S$ foliating $\Sigma_*$ to deduce \eqref{eq:GCMHoutcomeestimate1}.}
\bea\lab{eq:GCMHoutcomeestimate1}
\max_{k\leq k_{large}+4}\sup_{\Si_*}r\Big(|\dk^kf|+|\dk^k\fb|+|\dk^k\log(\la)|\Big) &\les& \ep_0, 
\eea
and 
\bea\lab{eq:GCMHoutcomeestimate2}
\quad \sup_{\Si_*}\Big(|m - m_0| +|r-r_0|\Big) &\les& \ep_0,
\eea
where $(f, \fb, \la)$ are the transition function from the frame of the initial data layer to the frame of $\Si_*$.

{\bf Step 3.} Provided $\delta_0>0$ has been chosen sufficiently small, the spacelike hypersurface $\Si_*$ of Step 2 intersects the curve of the south poles of the spheres foliating  the outgoing cone $\CC_{(1, \LL_0)}$ of the initial data layer. We then call $S_1$ the unique sphere of $\Sigma_*$ such that its south pole coincides with the south pole of a sphere of $\CC_{(1, \LL_0)}$, and we calibrate $u$ such that $u=1$ on $S_1$. We then can compare $\ug=1+\de_0$ to $u(S_*)$ and obtain 
\beaa
|u(S_*)-1-\de_0|\les \ep_0\de_0,
\eeaa
so that
\beaa
1\leq u\leq u(S_*)\quad \textrm{ on }\Si_*\textrm{ where }1<u(S_*)<1+2\de_0.
\eeaa
Together with the estimate \eqref{eq:GCMHoutcomeestimate2}, and in view of the choice \eqref{eq:choicer0forproofThmM6} for $r_0$, we have 
\beaa
\inf_{\Sigma_*}r\geq \frac{3}{2}\ep_0^{-\frac{2}{3}}\geq \ep_0^{-\frac{2}{3}}(u(S_*))^4
\eeaa
so that the dominant condition \eqref{eq:behaviorofronSigmastar} for $r$ is satisfied since $1\leq u\leq 1+2\de_0$ on $\Sigma_*$. 

{\bf Step 4.} In view of Step 1 to Step 3, $\Si_*$ satisfies all the required properties for the future spacelike boundary of a GCM admissible spacetime, see item 3 of definition \ref{definition:canonical-spacetime}. We now control the outgoing geodesic foliation initialized on $\Si_*$ and covering the region we denote by $\Mext$, which is included in the initial data layer. Let $(f,\fb, \la)$ the transition functions from the frame of the outgoing part of the initial  data layer to the frame of $\Mext$. Since both frames are outgoing geodesic, we may apply Corollary \ref{cor:transportequationsforffbandlambda} which yields for  $(\fb, f, \log(\la))$ the following transport equations 
\beaa
\la^{-1}e_4'(rf) &=& E_1'(f, \Ga),\\
\la^{-1}e_4'(\log(\la)) &=&  E_2'(f, \Ga),\\
\la^{-1}e_4'\Big(r\fb-2r^2e_\th'(\log(\la))+rf\Omb\Big) &=& E_3'(f, \fb, \la, \Ga),
\eeaa
where
\beaa
E_1'(f, \Ga) &=& -\frac{r}{2}\kac f -\frac{r}{2}\vth f+\lot,\\
E_2'(f, \Ga) &=&  f\ze- \frac{1}{2}f^2\omb -\etab f - \frac{1}{4}f^2\kab +\lot,\\
E_3'(f, \fb, \la, \Ga)  &=& -\frac{r}{2}\kac\fb+ r^2\left(\kac-\left(\ov{\ka}-\frac{2}{r}\right)\right)e_\th'(\log(\la)) +r^2\Big(\ddd_1'(f)+\la^{-1}\vth'\Big)e_\th'(\log(\la))\\
&&   -\frac{r}{2}\kac\Omb f    +rE_3(f, \fb, \Ga)-2r^2e_\th'(E_2(f, \Ga))+r\Omb E_1(f, \Ga),
\eeaa
and where $E_1$, $E_2$ and $E_3$ are given in Lemma \ref{lemma:transportequationsforffbandlambda}. Integrating these transport equations from $\Si_*$, using the control \eqref{eq:GCMHoutcomeestimate1} of $(f, \fb, \la)$ on $\Si_*$, and together with the assumption \eqref{def:initialdatalayerassumptions}  for the Ricci coefficients of the foliation of the initial data layer, we obtain 
\begin{equation}\lab{eq:controlffblaonMextwidetildeThmM6}
\sup_{\Mext\left(r\geq 2m_0(1+\deh)\right)}r\Big(|\dk^{\leq k_{large}+4} (f,  \log(\la))|+|\dk^{\leq k_{large}+3}\fb|\Big) \les \ep_0.
\end{equation}
Then, let $\TT=\{r= 2m_0(1+\deh)\}$, i.e. we choose $\rh=2m_0(1+\deh)$. We initialize the ingoing geodesic foliation of $\Mint$ on $\TT$ using the outgoing geodesic foliation of $\Mext$ as in item 4 of definition \ref{definition:canonical-spacetime}. Using the control of $(f, \fb, \la)$ induced on $\TT$ by \eqref{eq:controlffblaonMextwidetildeThmM6}, and using the analog of Corollary \ref{cor:transportequationsforffbandlambda} in the $e_3$ direction for ingoing foliations, we obtain similarly, 
\bea\lab{eq:controlffblaonMintwidetildeThmM6}
\sup_{\Mint}\Big(|\dk^{\leq k_{large}+3} (\fb,  \log(\la))|+|\dk^{\leq k_{large}+2}f|\Big) \les \ep_0.
\eea
Then, in view of \eqref{eq:controlffblaonMextwidetildeThmM6} \eqref{eq:controlffblaonMintwidetildeThmM6}, and the assumption \eqref{def:initialdatalayerassumptions}  for the Ricci coefficients and curvature components of the foliation of the initial data layer, and using the transformation formulas of Proposition \ref{prop:transformations1}, we deduce
\beaa
\max_{k\leq k_{large}}&&\Bigg\{\sup_{\Mext}\Big(r^{\frac{7}{2}+\dt}(|\dk^k\a|+|\dk^k\b|)+r^3|\dk^k\rhoc|+r^2|\dk^k\bb|+r|\dk^k\aa|\Big)\\
&&+\sup_{\Mext}r^{2}(|\dk^k\kac|+|\dk^k\vth|+|\dk^k\ze|+|\dk^k\kabc|)\\
&&+\sup_{\Mext}r(|\dk^k\eta|+|\dk^k\vthb|+|\dk^k\ombc|+|\dk^k\xib|)\Big)\Bigg\} \les \ep_0,
\eeaa
and
\beaa
\max_{k\leq k_{large}}\sup_{\Mint}\Big(|\dk^k\Rc|+|\dk^k\Gac|\Big) &\les& \ep_0.
\eeaa
In particular, we infer that 
\beaa
 \Nk^{(En)}_{k_{large}}+ \Nk^{(Dec)}_{k_{small}} &\les&  \ep_0
\eeaa
which concludes the proof of Theorem M6.


\section{Proof of Theorem M7}\lab{sec:proofofTheoremM7}


From the assumptions of Theorem M7 we are given a GCM admissible spacetime $\MM=\MM(u_*) \in\aleph(u_*)$  verifying   the following improved  bounds,  for a universal constant $C>0$,  
\bea
\lab{Fullestimates:onMM}
 \Nk^{(Dec)}_{k_{small}+5}(\MM)\le  C \ep_0 \qquad 
 \eea
  provided by Theorems M1-M5. We then proceed as follows.
  
  {\bf Step 1.} We extend $\MM$ by a local existence argument, to a strictly   larger  spacetime  $\MM^{(extend)}$,  with a naturally  extended foliation and the following slightly increased bounds
\beaa
 \Nk^{(Dec)}_{k_{small}+5}(\MMextend) \le 2C  \ep_0.
\eeaa
but which may  not verify our admissibility criteria. 

{\bf Step 2.} We then invoke Theorem GCMH  of section \ref{sec:statementofGCMresultsinChapter3} to extend $\Si_*$ in $\MMextend\setminus\MM$  as a smooth spacelike hypersurface  $\Si^{(extend)}_*$, together with a scalar function $u^{(extend)}$, satisfying the same GCM conditions than $\Si_*$. 

{\bf Step 3.} We consider the outgoing geodesic foliation $(u^{(extend)}, s^{(extend)})$ initialized on $\Si^{(extend)}_*$ to the future of $\Si^{(extend)}_*$ in $\MMextend$. Note in particular that we have from the definition  of $\Si_*$ and $\Si^{(extend)}_*$ 
\beaa
u^{(extend)}+s^{(extend)}=c_{\Si_*}.
\eeaa
We define the following spacetime region to the future of $\Si^{(extend)}_*$
\beaa
\widetilde{\RR} &:=& \Big\{u_*\leq u^{(extend)}\leq u_*+\de_{ext}, \quad c_{\Sigma_*}\leq u^{(extend)}+s^{(extend)}\leq c_{\Sigma_*}+\Delta_{ext}\Big\},
\eeaa
where 
\beaa
\Delta_{ext} := \frac{4r_*}{u_*}\de_{ext},\qquad r_*:=r(S_*), \qquad S_*:=\Si_*\cap\CC_*,
\eeaa
and $\de_{ext}>0$ is chosen sufficiently small so that $\widetilde{\RR}\subset\MMextend$.  From now on, for convenience, we drop the index $(extend)$ and simply denote $u^{(extend)}$ and $s^{(extend)}$ by $u$ and $s$. 

{\bf Step 4.} Propagating the GCM quantities in the $e_4$ direction from $\Si^{(extend)}_*$, where they all vanish, we obtain for all $k\leq k_{small}+4$
\beaa
\sup_{\widetilde{\RR}}\left(r^2\left|\dk^k\left(\ka-\frac{2}{r}\right)\right|+r^2|\dk^{k-2}(r^2\dds_2\dds_1\kab)|+r^3|\dk^{k-2}(r^2\dds_2\dds_1\mu)|\right) &\les& \frac{\ep_0}{r}\Delta_{ext}.
\eeaa 
Similarly, we have for all $k\leq k_{small}+4$
\beaa
\sup_{\widetilde{\RR}\cap\{u\geq u_*\}}\left(r\left|\dk^k\left(\int_S\b e^\Phi\right)\right|+\left|\dk^{k-1}\left(\int_Se_\th(\kab) e^\Phi\right)\right|\right) &\les& \frac{\ep_0}{u_*}\de_{ext}+\frac{\ep_0}{r}\Delta_{ext}\les \frac{\ep_0}{r}\Delta_{ext},
\eeaa 
where we used the fact that $u(S_*)=u_*$ and 
\beaa
\int_{S_*}\b e^\Phi=\int_{S_*}e_\th(\ka)e^\Phi=0.
\eeaa 
Also, recall that $\nu=e_3+a_*e_4$ denote the unique tangent vectorfield to $\Si_*$ which is orthogonal to $e_\th$ and normalized by $\g(\nu, e_4)=-2$. Then, one has, since $u+r$ is constant on $\Sigma_*$ and $s=r$ on $\Sigma_*$
\beaa
0=\nu(u+s)=e_3(u)+ae_4(u)+e_3(s)+ae_4(s)=\frac{2}{\vsi}+\Omb+a
\eeaa
and hence
\beaa
a &=& -\frac{2}{\vsi}-\Omb\textrm{ on }\Si_*.
\eeaa
Together with the GCM condition on $a$, we infer
\beaa
\frac{2}{\vsi}+\Omb &=& 1+\frac{2m}{r}\textrm{ on }\Si_*.
\eeaa 
As above, propagating forward in $e_4$, we infer
\beaa
\sup_{\widetilde{\RR}}\left|\frac{2}{\vsi}+\Omb -\left(1+\frac{2m}{r}\right)\right| &\les& \frac{\ep_0}{r}\Delta_{ext}.
\eeaa
 
{\bf Step 5.} We fix the following sphere of the $(u^{(extend)}, s^{(extend)})$ foliation in $\widetilde{\RR}\cap\{u\geq u_*\}$
\bea\lab{eq:checkthatthereisindeedalargeruinnewlastGCMS:5}
\ovS:=S(\ug, \sg), \quad \ug:=u_*+\frac{\de_{ext}}{2}, \quad \sg:=r_*+\frac{3r_*}{u_*}\de_{ext}.
\eea
Define
\beaa
\dg:=\frac{\ep_0}{r}\Delta_{ext}=\frac{\ep_0\de_{ext}}{u_*}, \quad \epg:=\frac{\ep_0}{u^{\frac{1}{2}+\dec}}
\eeaa
and the small spacetime neighborhood of $\ovS$
\beaa
 \RR(\epg, \dg) :=\left\{|u-\ug|\leq\de_{\RR} ,\quad |s-\sg|\leq  \de_\RR \right\}, \qquad \de_\RR = \dg \big(\epg\big)^{-\frac{1}{2}}.
\eeaa
Note that $\RR(\epg, \dg)\subset \widetilde{\RR}$. We are in position to apply Theorem GCMS II of section \ref{sec:statementofGCMresultsinChapter3}, with $s_{max}=k_{small}+4$, which yields the existence of a unique sphere $\widetilde{S}_*$, which is a deformation of $\ovS$, is included in $\RR(\epg, \dg)$, and is such that the following GCM conditions hold on it
 \beaa
\widetilde{\dds}_2\widetilde{\dds}_1\,\widetilde{\kab}=\widetilde{\dds}_2\widetilde{\dds}_1\,\widetilde{\mu}=0, \qquad \widetilde{\ka}=\frac{2}{\widetilde{r}}, \quad \int_{\widetilde{S}_*}\widetilde{\b}e^\Phi=\int_{\widetilde{S}_*}\widetilde{e}_\th(\widetilde{\kab})e^\Phi=0,
\eeaa
where the tilde refer to the quantities and tangential operators on $\widetilde{S}_*$. 

{\bf Step 6.} Starting from $\widetilde{S}_*$ constructed in Step 5, we apply Theorem GCMH of section \ref{sec:statementofGCMresultsinChapter3},  with $s_{max}=k_{small}+4$, which yields the existence of a smooth small piece of spacelike $\widetilde{\Si}_*$ starting from $\widetilde{S}_*$ towards the initial data layer, together with a scalar function $\widetilde{u}$ defined on $\widetilde{\Si}_*$, whose level surfaces are topological spheres denoted by $\widetilde{S}$, so that 
\begin{itemize}
\item the following GCM conditions are verified on $\widetilde{\Si}_*$
 \beaa
\widetilde{\dds}_2\widetilde{\dds}_1\,\widetilde{\kab}=\widetilde{\dds}_2\widetilde{\dds}_1\,\widetilde{\mu}=0, \qquad \widetilde{\ka}=\frac{2}{\widetilde{r}}, \quad \int_{\widetilde{S}}\widetilde{\eta}e^\Phi=\int_{\widetilde{S}}\widetilde{\xib}e^\Phi=0,
\eeaa
where the tilde refer to the quantities and tangential operators on $\widetilde{\Si}_*$. 

\item We have, for some constant $c_{\widetilde{\Si}_*}$,  
\beaa
\widetilde{u}+\widetilde{r}=c_{\widetilde{\Si}_*} , \qquad \textit{along} \quad \widetilde{\Si}_*.
\eeaa
\item The following normalization condition  holds true  at the  South Pole $SP$  of every sphere $\widetilde{S}$,
 \beaa
 \widetilde{a}\Big|_{SP}=-1 -\frac{2\widetilde{m}}{\widetilde{r}}
 \eeaa
 where $\widetilde{a}$ is such that we have
 \beaa
 \widetilde{\nu} =\widetilde{e}_3+ \widetilde{a} \widetilde{e}_4,
 \eeaa
 with $\widetilde{\nu}$  the unique vectorfield tangent to the hypersurface $\widetilde{\Si}_*$, normal to $\widetilde{S}$, and normalized by $g(\widetilde{\nu}, \widetilde{e}_4)=-2$. 
 
\item The transition functions $(f, \fb, \la)$ from the frame of $\MMextend$ to the frame of $\widetilde{\Si}_*$
\beaa
\|(f, \fb, \log(\la))\|_{\hk_{k_{small}+5}} &\les& \dg.
\eeaa
\end{itemize}

{\bf Step 7.} The spacelike GCM hypersurface $\widetilde{\Si}_*$ has been constructed in Step 6 in a small neighborhood of $\widetilde{S}_*$. We now focus on proving that it in fact extends all the way to the initial data layer. To this end, we denote by $u_1$ with
\beaa
1\leq u_1<\ug,
\eeaa
the minimal value of $u$ such that:
\begin{itemize}
\item We have
\bea\lab{eq:aprioriestimatecrucialThmM7:0}
\widetilde{\Sigma}_*\cap\CC_u\neq\emptyset\textrm{ for any }u_1\leq u\leq \ug.
\eea

\item There exists a large constant $D\geq 1$ such that we have for any sphere $\widetilde{S}$ of $\widetilde{\Sigma}_*(u\geq u_1)$
\bea\lab{eq:aprioriestimatecrucialThmM7:1}
\|(f, \fb, \log(\la))\|_{\hk_{k_{small}+5}(\widetilde{S})} &\leq& Du_*\dg.
\eea

\item For the same large constant $D\geq 1$ as above, we have along $\widetilde{\Sigma}_*(u\geq u_1)$
\bea\lab{eq:aprioriestimatecrucialThmM7:2}
|\psi(s)| &\leq& Du_*\dg,
\eea
where the function $\psi(s)$ is such that the curve 
\bea\lab{eq:checkthatthereisindeedalargeruinnewlastGCMS:6}
\Big(u=-s+c_{\widetilde{\Sigma}_*}+\psi(s),\, s, \,\th=0\Big)\textrm{ with }\psi(\sg)=0,
\eea
coincides with the south poles of the sphere $\widetilde{S}$ of $\widetilde{\Sigma}_*$ and the constant $c_{\widetilde{\Sigma}_*}$ is fixed by the condition $\psi(\sg)=0$. 
\end{itemize}
The fact that $\psi(\sg)=0$ together with the bounds of Step 6 implies that \eqref{eq:aprioriestimatecrucialThmM7:0} \eqref{eq:aprioriestimatecrucialThmM7:1} \eqref{eq:aprioriestimatecrucialThmM7:2} hold for $u_1<\ug$ with $u_1$ close enough to $\ug$. By a continuity argument based on reapplying Theorem GCMH, it suffices to show that we may improve the bounds \eqref{eq:aprioriestimatecrucialThmM7:1} \eqref{eq:aprioriestimatecrucialThmM7:2} independently of the value of $u_1$. 

{\bf Step 8.} We now focus on improving the bounds \eqref{eq:aprioriestimatecrucialThmM7:1} \eqref{eq:aprioriestimatecrucialThmM7:2}. We first prove that $\widetilde{\Si}_*(u\geq u_1)$ is included in $\widetilde{\RR}$. Indeed, \eqref{eq:aprioriestimatecrucialThmM7:1} \eqref{eq:aprioriestimatecrucialThmM7:2} imply
\beaa
\sup_{\widetilde{\Si}_*(u\geq u_1)}|u+s-c_{\widetilde{\Sigma}_*}| &\les& \sup_{\widetilde{\Si}_*(u\geq u_1)}\Big(|\psi|+r|f|+r|\fb|)\\
&\les& Du_*\dg\\
&\les& \frac{Du_*}{r}\ep_0\Delta_{ext}\\
&\les& \frac{\ep_0^{\frac{2}{3}}D}{u_*^3}\ep_0\Delta_{ext}\\
&\les& \ep_0\Delta_{ext}.
\eeaa
On the other hand, by construction, $\psi(\sg)=0$ and the south pole of $\ovS$ and $\widetilde{S}_*$ coincide, so that we have
\beaa
c_{\widetilde{\Sigma}_*} &=& \ug+\sg=u_*+r_*+\frac{\de_{ext}}{2}+\frac{3r_*}{u_*}\de_{ext}\\
&=& c_{\Sigma_*}+\frac{3}{4}\left(1+\frac{2u_*}{3r_*}\right)\Delta_{ext}
\eeaa
and hence
\beaa
\sup_{\widetilde{\Si}_*(u\geq u_1)}\left|u+s-c_{\Sigma_*}-\frac{3}{4}\Delta_{ext}\right| &\les& \left(\frac{u_*}{2r_*}+\ep_0\right)\Delta_{ext}\\
&\les& \ep_0^{\frac{2}{3}}\Delta_{ext}.
\eeaa
In view of the definition of $\widetilde{\RR}$, we infer
\bea
\widetilde{\Si}_*(u\geq u_1) \subset \widetilde{\RR}
\eea
as claimed.

{\bf Step 9.} Since $\widetilde{\Si}_*(u\geq u_1) \subset \widetilde{\RR}$, the bound of Step 4 apply, and hence we have 
\beaa
\sup_{\widetilde{\RR}}\left|\frac{2}{\vsi}+\Omb -\left(1+\frac{2m}{r}\right)\right| &\les& \frac{\ep_0}{r}\Delta_{ext}\les \dg,
\eeaa
and for all $k\leq k_{small}+4$
\beaa
\sup_{\widetilde{\RR}}\left(r^2\left|\dk^k\left(\ka-\frac{2}{r}\right)\right|+r^2|\dk^{k-2}(r^2\dds_2\dds_1\kab)|+r^3|\dk^{k-2}(r^2\dds_2\dds_1\mu)|\right) &\les& \frac{\ep_0}{r}\Delta_{ext}\les\dg,
\eeaa 
as well as
\beaa
\sup_{\widetilde{\RR}\cap\{u\geq u_*\}}\left(r\left|\dk^k\left(\int_S\b e^\Phi\right)\right|+\left|\dk^{k-1}\left(\int_Se_\th(\kab) e^\Phi\right)\right|\right) &\les& \frac{\ep_0}{r}\Delta_{ext}\les\dg.
\eeaa 
Together with the a priori estimates of Chapter \ref{chap:proofofGCMprocedure} on the GCM construction, this yields
\beaa
|\psi'(s)| &\les& \left|1+\frac{2\widetilde{m}}{\widetilde{r}}+\Omb+\frac{2}{\vsi}\right|+|\la-1|\\
&\les& \left|\frac{\widetilde{m}}{\widetilde{r}}-\frac{m}{r}\right|+|\la-1|+\frac{\ep_0}{r}\Delta_{ext}.
\eeaa
In view of \eqref{eq:aprioriestimatecrucialThmM7:1}, we have
\bea\lab{eq:checkthatthereisindeedalargeruinnewlastGCMS:9}
\left|\widetilde{r}-r\right|+\left|\widetilde{m}-{m}\right| &\les& \sup_{\widetilde{S}}r(|f|+|\fb|)\les Du_*\dg
\eea
and we infer
\beaa
|\psi'(s)| &\les& \frac{Du_*}{r}\dg+\dg\\
&\les& \left(1+\frac{\ep_0^{\frac{2}{3}}}{(u_*)^3D}\right)\dg\\
&\les& \dg.
\eeaa
Integrating from $\sg$ where $\psi(\sg)=0$, we infer
\beaa
|\psi(s)| &\les& |s-\sg|\dg\\
&\les& u_*\dg
\eeaa
which improves \eqref{eq:aprioriestimatecrucialThmM7:2} for $D\geq 1$ large enough. 

Similarly, we obtain 
\beaa
\|(f, \fb, \log(\la))\|_{\hk_{k_{small}+5}(\widetilde{S})} &\les& r^{-2}\left(\left|\int_Sfe^\Phi\right|+\left|\int_S\fb e^\Phi\right|\right)+\dg
\eeaa
and 
\beaa
\left|e_3\left(\int_Sfe^\Phi\right)\right|+\left|e_3\left(\int_S\fb e^\Phi\right)\right| &\les& r^2\dg+\frac{1}{r}\left(\left|\int_Sfe^\Phi\right|+\left|\int_S\fb e^\Phi\right|\right).
\eeaa
In view of \eqref{eq:aprioriestimatecrucialThmM7:1}, we infer
\beaa
\left|e_3\left(\int_Sfe^\Phi\right)\right|+\left|e_3\left(\int_S\fb e^\Phi\right)\right| &\les& r^2\dg+rDu_*\dg
\eeaa
and integrating from $\widetilde{S}_*$, we infer
\beaa
r^{-2}\left(\left|\int_Sfe^\Phi\right|+\left|\int_S\fb e^\Phi\right|\right) &\les& u_*\dg+\frac{D(u_*)^2}{r}\dg\\
&\les& \left(1+\frac{\ep_0^{\frac{2}{3}}D}{(u_*)^2}\right)u_*\dg\\
&\les& u_*\dg.
\eeaa
This yields
\beaa
\|(f, \fb, \log(\la))\|_{\hk_{k_{small}+5}(\widetilde{S})} &\les& u_*\dg
\eeaa
which improves \eqref{eq:aprioriestimatecrucialThmM7:1} for $D\geq 1$ large enough. We thus conclude that $u_1=1$, $\widetilde{\Si}_*$  extends all the way to the initial data layer, $\widetilde{\Si}_*\subset\widetilde{\RR}$, and we have the bounds
\beaa
\|(f, \fb, \log(\la))\|_{\hk_{k_{small}+5}(\widetilde{S})} \les u_*\dg, \qquad |\psi(s)| \les u_*\dg.
\eeaa
In view of the definition of $\dg$, we infer in particular for any sphere $\widetilde{S}$ of $\widetilde{\Si}_*$
\bea\lab{eq:checkthatthereisindeedalargeruinnewlastGCMS:4}
\|(f, \fb, \log(\la))\|_{\hk_{k_{small}+5}(\widetilde{S})} \les \ep_0\de_{ext}, \qquad  |\psi(s)| \les \ep_0\de_{ext}.
\eea

{\bf Step 10.} As $\widetilde{\Si}_*$  extends all the way to the initial data layer, this allows us to calibrate $\tilde{u}$ along $\widetilde{\Si}_*$ by fixing the value $\widetilde{u}=1$ as in \eqref{eq:calibrationofubythechoiceofS1intersectingSPconeIDL}:
\bea
\widetilde{S}_1=\widetilde{\Si}_*\cap\{\widetilde{u}=1\}\textrm{ is such that }\widetilde{S}_1\cap\CC_{(1,\LL_0)}\cap SP\neq \emptyset,  
\eea
i.e. $\widetilde{S}_1$  is the unique sphere of $\widetilde{\Si}_*$ such that its south pole intersects the south pole of one of the sphere of the outgoing null cone $\CC_{(1,\LL_0)}$ of the initial data layer. 

Now that $\widetilde{u}$ is calibrated, we define
\bea
\tilde{u}_*:=\tilde{u}(\widetilde{S}_*).
\eea
For the proof of Theorem M7, we need in particular to prove that $\tilde{u}_*>u_*$. First, note that, since $\tilde{u}+\tilde{r}$ is constant along $\widetilde{\Si}_*$, we have
\bea\lab{eq:checkthatthereisindeedalargeruinnewlastGCMS:8}
\widetilde{\Si}_*=\Big\{\widetilde{u}+\widetilde{r}=1+\widetilde{r}(\widetilde{S}_1)\Big\}.
\eea
Since $\widetilde{S}_*\subset\widetilde{\Si}_*$, and in view of \eqref{eq:checkthatthereisindeedalargeruinnewlastGCMS:8}, \eqref{eq:checkthatthereisindeedalargeruinnewlastGCMS:5}, \eqref{eq:checkthatthereisindeedalargeruinnewlastGCMS:6}, we infer, 
\beaa
\left|\widetilde{u}(\widetilde{S}_*)-\left(u_*+\frac{\de_{ext}}{2}\right)\right| &=& \left|\widetilde{u}(\widetilde{S}_*)-u(\ovS)\right|\\
&=& \left|1+\widetilde{r}(\widetilde{S}_1)-\widetilde{r}(\widetilde{S}_*)-\left(-s(\ovS)+c_{\widetilde{\Si}_*}\right)\right|.
\eeaa
Next, note from
\beaa
s=r\textrm{ on }\Si_*, \qquad e_4(r-s)=\frac{r}{2}\left(\ov{\ka}-\frac{2}{r}\right)
\eeaa
that we have
\bea\lab{eq:checkthatthereisindeedalargeruinnewlastGCMS:3}
\sup_{\widetilde{\RR}}|r-s| &\les& \frac{\ep_0}{r}\Delta_{ext}\les \ep_0\de_{ext}.
\eea
Together with  \eqref{eq:checkthatthereisindeedalargeruinnewlastGCMS:9}, this yields
\beaa
\left|\widetilde{u}(\widetilde{S}_*)-\left(u_*+\frac{\de_{ext}}{2}\right)\right| &\les& \Big|1+\widetilde{r}(\widetilde{S}_1)-c_{\widetilde{\Si}_*}\Big| +\ep_0\de_{ext}.
\eeaa
Since $c_{\widetilde{\Si}_*}$ in \eqref{eq:checkthatthereisindeedalargeruinnewlastGCMS:6} is a constant, we have in particular 
\beaa
c_{\widetilde{\Si}_*}=u(\widetilde{S}_1)+r(\widetilde{S}_1)-\psi(s(\widetilde{S}_1))
\eeaa
and thus
\beaa
\left|\widetilde{u}(\widetilde{S}_*)-\left(u_*+\frac{\de_{ext}}{2}\right)\right| &\les& \Big|1+\widetilde{r}(\widetilde{S}_1)-u(\widetilde{S}_1)-r(\widetilde{S}_1)+\psi(s(\widetilde{S}_1))\Big| +\ep_0\de_{ext}\\
&\les& \Big|1-u(\widetilde{S}_1)\Big|+\Big|\widetilde{r}(\widetilde{S}_1)-r(\widetilde{S}_1)\Big|+\Big|\psi(s(\widetilde{S}_1))\Big| +\ep_0\de_{ext}.
\eeaa
In view of \eqref{eq:checkthatthereisindeedalargeruinnewlastGCMS:4} and \eqref{eq:checkthatthereisindeedalargeruinnewlastGCMS:9}, we infer
\beaa
\left|\widetilde{u}(\widetilde{S}_*)-\left(u_*+\frac{\de_{ext}}{2}\right)\right| &\les& \Big|1-u(\widetilde{S}_1)\Big|+\ep_0\de_{ext}.
\eeaa
Also, since (recall in particular \eqref{eq:calibrationofubythechoiceofS1intersectingSPconeIDL}) 
\beaa
u=1\textrm{ on }S_1\cap SP, \qquad e_4^{\LL_0}(u)=O\left(\frac{\ep_0}{r^2}\right),
\eeaa
and since the south pole of $S_1$ coincides with the one of the corresponding sphere of $\CC_{\LL_0,1}$, we infer
\beaa
\sup_{\widetilde{\RR}\cap\CC_{\LL_0, 1}\cap SP}|u-1| \les  \Delta_{ext}\frac{\ep_0}{r^2}\les\ep_0\de_{ext}.
\eeaa
This yields
\bea\lab{eq:checkthatthereisindeedalargeruinnewlastGCMS:10}
\left|\widetilde{u}(\widetilde{S}_*)-\left(u_*+\frac{\de_{ext}}{2}\right)\right| &\les& \ep_0\de_{ext}.
\eea
In particular, we deduce, for $\ep_0$ small enough, 
\bea\lab{eq:checkthatthereisindeedalargeruinnewlastGCMS:11}
\widetilde{u}(\widetilde{S}_*)>u_*
\eea
as desired.

{\bf Step 11.} We would like to check that the dominant condition \eqref{eq:behaviorofronSigmastar} for $r$ holds on $\widetilde{\Si}_*$, i.e. we need to prove 
\beaa
\widetilde{r}(\widetilde{S}_*) &\geq& \ep_0^{-\frac{2}{3}}(\widetilde{u}(\widetilde{S}_*))^4. 
\eeaa
To this end, note that we have in view of \eqref{eq:checkthatthereisindeedalargeruinnewlastGCMS:9}, \eqref{eq:checkthatthereisindeedalargeruinnewlastGCMS:3} and \eqref{eq:checkthatthereisindeedalargeruinnewlastGCMS:10}
\beaa
\widetilde{r}(\widetilde{S}_*) - \ep_0^{-\frac{2}{3}}(\widetilde{u}(\widetilde{S}_*))^4 &=& s(\ovS) +O\left(\ep_0\de_{ext}\right) - \ep_0^{-\frac{2}{3}}\left(u_*+\frac{\de_{ext}}{2}+O\left(\ep_0\de_{ext}\right)\right)^4\\
&=& s(\ovS) - \ep_0^{-\frac{2}{3}}(u_*)^4 -2\ep_0^{-\frac{2}{3}}(u_*)^3\de_{ext}+\ep_0^{-\frac{2}{3}}(u_*)^3\de_{ext}O\left(\frac{\de_{ext}}{u_*}+\ep_0\right)+O\left(\ep_0\de_{ext}\right).
\eeaa
Together with \eqref{eq:checkthatthereisindeedalargeruinnewlastGCMS:5}, we infer
\beaa
\widetilde{r}(\widetilde{S}_*) - \ep_0^{-\frac{2}{3}}(\widetilde{u}(\widetilde{S}_*))^4 &=& r_*+\frac{3r_*}{u_*}\de_{ext} - \ep_0^{-\frac{2}{3}}(u_*)^4 -2\ep_0^{-\frac{2}{3}}(u_*)^3\de_{ext}\\
&&+\ep_0^{-\frac{2}{3}}(u_*)^3\de_{ext}O\left(\frac{\de_{ext}}{u_*}+\ep_0\right)+O\left(\ep_0\de_{ext}\right)\\
&=& r_*-\ep_0^{-\frac{2}{3}}(u_*)^4 + \left(3r_*-2\ep_0^{-\frac{2}{3}}(u_*)^4\right)\frac{\de_{ext}}{u_*}\\
&&+\ep_0^{-\frac{2}{3}}(u_*)^3\de_{ext}O\left(\frac{\de_{ext}}{u_*}+\ep_0\right)+O\left(\ep_0\de_{ext}\right).
\eeaa
Since we have by the control \eqref{eq:behaviorofronSigmastar} of $r$ on $\Si_*$
\beaa
r_*\geq \ep_0^{-\frac{2}{3}}(u_*)^4, 
\eeaa
we deduce
\beaa
\widetilde{r}(\widetilde{S}_*) - \ep_0^{-\frac{2}{3}}(\widetilde{u}(\widetilde{S}_*))^4 &\geq &  \frac{r_*\de_{ext}}{u_*}+\ep_0^{-\frac{2}{3}}(u_*)^3\de_{ext}O\left(\frac{\de_{ext}}{u_*}+\ep_0\right)+O\left(\ep_0\de_{ext}\right)\\
&=& \frac{r_*\de_{ext}}{u_*}(1+O(\ep_0))
\eeaa
so that, for $\ep_0$ small enough, we have
\beaa
\widetilde{r}(\widetilde{S}_*) &\geq& \ep_0^{-\frac{2}{3}}(\widetilde{u}(\widetilde{S}_*))^4 
\eeaa
as desired.

{\bf Step 12.} We summarize the properties of $\widetilde{\Si}_*$ obtained so far:
\begin{itemize}
\item $\widetilde{\Si}_*$ is a spacelike hypersurface included in the spacetime region $\widetilde{\RR}$. 

\item The scalar function $\widetilde{u}$ is defined on $\widetilde{\Si}_*$ and it level sets  are topological 2-spheres denoted by $\widetilde{S}$.

\item The following GCM conditions holds on $\widetilde{\Si}_*$
 \beaa
\widetilde{\dds}_2\widetilde{\dds}_1\,\widetilde{\kab}=\widetilde{\dds}_2\widetilde{\dds}_1\,\widetilde{\mu}=0, \qquad \widetilde{\ka}=\frac{2}{\widetilde{r}}, \quad \int_{\widetilde{S}}\widetilde{\eta}e^\Phi=\int_{\widetilde{S}}\widetilde{\xib}e^\Phi=0.
\eeaa

\item In addition, the following GCM conditions holds on the sphere $\widetilde{S}_*$ of $\widetilde{\Si}_*$
 \beaa
\int_{\widetilde{S}_*}\widetilde{\b}e^\Phi=\int_{\widetilde{S}_*}\widetilde{e}_\th(\widetilde{\kab})e^\Phi=0,
\eeaa

\item We have, for some constant $c_{\widetilde{\Si}_*}$,  
\beaa
\widetilde{u}+\widetilde{r}=c_{\widetilde{\Si}_*} , \qquad \textit{along} \quad \widetilde{\Si}_*.
\eeaa
\item The following normalization condition  holds true  at the  South Pole $SP$  of every sphere $\widetilde{S}$,
 \beaa
 \widetilde{a}\Big|_{SP}=-1 -\frac{2\widetilde{m}}{\widetilde{r}}
 \eeaa
 where $\widetilde{a}$ is such that we have
 \beaa
 \widetilde{\nu} =\widetilde{e}_3+ \widetilde{a} \widetilde{e}_4,
 \eeaa
 with $\widetilde{\nu}$  the unique vectorfield tangent to the hypersurface $\widetilde{\Si}_*$, normal to $\widetilde{S}$, and normalized by $g(\widetilde{\nu}, \widetilde{e}_4)=-2$. 
 
 \item The dominant condition \eqref{eq:behaviorofronSigmastar} for $r$ holds on $\widetilde{\Si}_*$, i.e. we have
\beaa
\widetilde{r}(\widetilde{S}_*) &\geq& \ep_0^{-\frac{2}{3}}(\widetilde{u}(\widetilde{S}_*))^4. 
\eeaa

\item $\tilde{u}$ is calibrated along $\widetilde{\Si}_*$ by fixing the value $\widetilde{u}=1$:
\bea
\widetilde{S}_1=\widetilde{\Si}_*\cap\{\widetilde{u}=1\}\textrm{ is such that }\widetilde{S}_1\cap\CC_{(1,\LL_0)}\cap SP\neq \emptyset,  
\eea
i.e. $\widetilde{S}_1$  is the unique sphere of $\widetilde{\Si}_*$ such that its south pole intersects the south pole of one of the sphere of the outgoing null cone $\CC_{(1,\LL_0)}$ of the initial data layer. 
\end{itemize}
Thus $\widetilde{\Si}_*$ satisfies all the required properties for the future spacelike boundary of a GCM admissible spacetime, see item 3 of definition \ref{definition:canonical-spacetime}. Furthermore, we have on  $\widetilde{\Si}_*$
\bea\lab{eq:conclusionofThmM7u'*>u*isok}
\widetilde{u}(\widetilde{S}_*)>u_*,
\eea
and $(f, \fb, \la)$ satisfy in view of \eqref{eq:checkthatthereisindeedalargeruinnewlastGCMS:4} and Corollary Rigidity III of section \ref{sec:statementofGCMresultsinChapter3}
\beaa
\sup_{\widetilde{\Si}_*}\|\dk^{\leq k_{small}+5} (f, \fb, \log(\la))\|_{\hk_{k_{small}+5}(\widetilde{S})} \les \ep_0\de_{ext}.
\eeaa
Together with the Sobolev embedding on the spheres $\widetilde{S}$, and possibly reducing the size of $\de_{ext}>0$, we deduce
\bea\lab{eq:controlffblafinalonSigma*widetildeforThmM7}
\sup_{\widetilde{\Si}_*}\widetilde{r}\,\widetilde{u}^{\frac{1}{2}+\dec}|\dk^{\leq k_{small}+3} (f, \fb, \log(\la))| \les \ep_0.
\eea

{\bf Step 13.} We now control the outgoing geodesic foliation initialized on $\widetilde{\Si}_*$. We denote by $\,^{(ext)}\widetilde{\MM}$ the region covered by this outgoing geodesic foliation. Let $(e_4, e_3, e_\th)$  of $\Mext$ extended to the   spacetime  $\MM^{(extend)}$, and satisfying, as discussed in Step 1 to Step 3
\bea\lab{eq:extendedspacetimebootstrapassumptionsforThmM7}
 \Nk^{(Dec)}_{k_{small}+5}(\MMextend) &\les&  \ep_0.
\eea
Let $(f,\fb, \la)$ the transition functions from the frame $(e_4, e_3, e_\th)$ to the frame $(\widetilde{e}_4, \widetilde{e}_3, \widetilde{e}_\th)$ of $\,^{(ext)}\widetilde{\MM}$. Since both frames are outgoing geodesic, we may apply Corollary \ref{cor:transportequationsforffbandlambda} which yields for  $(\fb, f, \log(\la))$ the following transport equations 
\beaa
\la^{-1}e_4'(rf) &=& E_1'(f, \Ga),\\
\la^{-1}e_4'(\log(\la)) &=&  E_2'(f, \Ga),\\
\la^{-1}e_4'\Big(r\fb-2r^2e_\th'(\log(\la))+rf\Omb\Big) &=& E_3'(f, \fb, \la, \Ga),
\eeaa
where
\beaa
E_1'(f, \Ga) &=& -\frac{r}{2}\kac f -\frac{r}{2}\vth f+\lot,\\
E_2'(f, \Ga) &=&  f\ze- \frac{1}{2}f^2\omb -\etab f - \frac{1}{4}f^2\kab +\lot,\\
E_3'(f, \fb, \la, \Ga)  &=& -\frac{r}{2}\kac\fb+ r^2\left(\kac-\left(\ov{\ka}-\frac{2}{r}\right)\right)e_\th'(\log(\la)) +r^2\Big(\ddd_1'(f)+\la^{-1}\vth'\Big)e_\th'(\log(\la))\\
&&   -\frac{r}{2}\kac\Omb f    +rE_3(f, \fb, \Ga)-2r^2e_\th'(E_2(f, \Ga))+r\Omb E_1(f, \Ga),
\eeaa
and where $E_1$, $E_2$ and $E_3$ are given in Lemma \ref{lemma:transportequationsforffbandlambda}. Integrating these transport equations from $\widetilde{\Si}_*$, using the control \eqref{eq:controlffblafinalonSigma*widetildeforThmM7} of $(f, \fb, \la)$ on $\widetilde{\Si}_*$, and together with the control \eqref{eq:extendedspacetimebootstrapassumptionsforThmM7} for the Ricci coefficients of the foliation of $\MM^{(extend)}$, we obtain 
\begin{equation}\lab{eq:controlffblaonMextwidetildeThmM7}
\sup_{\,^{(ext)}\widetilde{\MM}\left(\widetilde{r}\geq 2m_0(1+\frac{\deh}{2})\right)}\Big(\widetilde{r}\,\widetilde{u}^{\frac{1}{2}+\dec}+\widetilde{u}^{1+\dec}\Big)\Big(|\dk^{\leq k_{small}+3} (f,  \log(\la))|+|\dk^{\leq k_{small}+2}\fb|\Big) \les \ep_0.
\end{equation}
Then, for any $\rh$ in the interval 
\bea\lab{eq:intervalrTTforThmM7}
2m_0\left(1+\frac{\deh}{2}\right) \leq \rh\leq 2m_0\left(1+\frac{3\deh}{2}\right),
\eea
we initialize the ingoing geodesic foliation of $\,^{(int)}\widetilde{\MM}[\rh]$ on $\widetilde{r}=\rh$ using the outgoing geodesic foliation of $\,^{(ext)}\widetilde{\MM}$ as in item 4 of definition \ref{definition:canonical-spacetime}. Using the control of $(f, \fb, \la)$ induced on $\widetilde{r}=\rh$ by \eqref{eq:controlffblaonMextwidetildeThmM7}, and using the analog of Corollary \ref{cor:transportequationsforffbandlambda} in the $e_3$ direction for ingoing foliations, we obtain similarly, for any $\rh$ in the interval \eqref{eq:intervalrTTforThmM7}, 
\bea\lab{eq:controlffblaonMintwidetildeThmM7}
\sup_{\,^{(int)}\widetilde{\MM}[\rh]}\widetilde{\ub}^{1+\dec}\Big(|\dk^{\leq k_{small}+2} (\fb,  \log(\la))|+|\dk^{\leq k_{small}+1}f|\Big) \les \ep_0.
\eea
Let now, for any $\rh$  in the interval \eqref{eq:intervalrTTforThmM7},
\beaa
\MM[\rh] &:=& \,^{(ext)}\widetilde{\MM}(\widetilde{r}\geq \rh)\cup \,^{(int)}\widetilde{\MM}[\rh].
\eeaa
Then, in view of \eqref{eq:controlffblaonMextwidetildeThmM7} \eqref{eq:controlffblaonMintwidetildeThmM7}, and \eqref{eq:extendedspacetimebootstrapassumptionsforThmM7}, and using the transformation formulas of Proposition \ref{prop:transformations1}, we deduce
\beaa
 \Nk^{(Dec)}_{k_{small}}(\MM[\rh]) &\les&  \ep_0
\eeaa
which concludes the proof of Theorem M7.


\section{Proof of Theorem M8}


So far, we have only improved our bootstrap assumptions on decay estimates. We now improve our bootstrap assumptions on energies and weighted energies for $\check{R}$ and $\check{\Ga}$ relying on an iterative procedure which recovers derivatives one by one\footnote{See also \cite{Holz} for a related strategy to recover higher order derivatives from the control of lower order ones.}.

Let $I_{m_0, \deh}$ the interval of $\mathbb{R}$ defined by
\bea
I_{m_0, \deh} &:=& \left[2m_0\left(1+\frac{\deh}{2}\right), 2m_0\left(1+\frac{3\deh}{2}\right)\right].
\eea

\begin{remark}\lab{remark:whydotheresultsofThemM0-M7holdforanyrhproofThM8}
Recall that the results of Theorems M0--M7 hold for any $\rh\in I_{m_0, \deh}$, see Remark  \ref{rmk:infactimproveddecayofThM7holdsforrTininterval}. More precisely
\begin{itemize}
\item they hold on $\Mext(r\geq 2m_0(1+\frac{\deh}{2}))$, and hence on $\Mext(r\geq \rh)$ for any $\rh\in I_{m_0, \deh}$,

\item they hold on $\Mint[\rh]$ for  any $\rh\in I_{m_0, \deh}$, where $\Mint[\rh]$ is initialized on $\TT=\{r=\rh\}$ using $\Mext(r\geq \rh)$ as in section \ref{sec:defintioncanonicalspacetime}.
\end{itemize}
\end{remark}

It is at this stage that we need to make a specific choice of $\rh$ in the context of a Lebesgue point argument. More precisely, we choose $\rh$ such that we have
\bea\lab{eq:choiceofRTTismadebythisinfimum:bis}
\int_{\{r=\rh\}}|\dk^{\leq k_{large}}\Rc|^2 &=& \inf_{r_0\in I_{m_0, \deh}}\int_{\{r=r_0\}}|\dk^{\leq k_{large}}\Rc|^2.
\eea

\begin{remark}
In case the above infimum is achieved for several values of $r$, we choose $\rh$ to be the largest of such values, so that $\rh$ is uniquely defined. Note also that  the infimum could a priori be infinite, and will only be shown to be finite - and more precisely $O(\ep_0)$ -, at the end of the proof of Theorem M8, see section \ref{sec:endofproofThmM8}. This could be made rigorous in the context of a continuity argument. 
\end{remark}

In view of the definition of $\rh$, and since $\TT=\{r=\rh\}$, we have
\beaa
\int_{\TT}|\dk^{\leq k_{large}}\Rc|^2  &\leq &  \frac{1}{2m_0\deh}\int_{I_{m_0, \deh}}\left(\int_{\{r=r_0\}}|\dk^{\leq k_{large}}\Rc|^2\right)dr_0
\eeaa
and hence\footnote{We use the coarea formula,  $d\MM = \frac{1}{\sqrt{\g(\D r, \D r)}}d\{r=r_0\} dr_0$  
and the fact that, for $r\in I_{m_0, \deh}$, $
\g(\D r, \D r) = -e_3(r)e_4(r) = \Up+O(\ep) \geq \frac{\deh}{2}+O(\ep+\deh^2)\geq \frac{\deh}{4}.$ Note  that $\les$ here  depends on $\de_\HH^{-1}$, see  the convention for $\les$ made at the end of section \ref{sec:discussionofsmallnessconstantforthemaintheorem}.}
\bea\lab{eq:consequenceofthechoiceofrhwhichisuseful:bis}
\int_{\TT}|\dk^{\leq k_{large}}\Rc|^2  &\lesssim & \int_{\Mext\Big(r\in I_{m_0, \deh}\Big)}|\dk^{\leq k_{large}}\Rc|^2. 
\eea

From now on, we may thus assume that the spacetime $\MM$ satisfies 
\begin{itemize}
\item the conclusions of Theorem M0, i.e.
\bea\lab{eq:mainconclusionofThm0forproofThM8}
\max_{0\leq k\leq k_{large}}&&\Bigg\{ \sup_{\CC_1}  \left[r^{\frac{7}{2}  +\de_B}\left( |\dk^k\,{}^{(ext)}\a| + |\dk^k\,{}^{(ext)}\b|\right)+r^{\frac{9}{2}  +\de_B}|\dk^{k-1}e_3(\,{}^{(ext)}\a)|   \right]\\
\nn&&+ \sup_{\CC_1} \left[r^3\ \left|\dk^k\left(\,{}^{(ext)}\rho+\frac{2m_0}{r^3}\right)\right|+r^2|\dk^k\,{}^{(ext)}\bb|+r|\dk^k\,{}^{(ext)}\aa|\right]\Bigg\} \les \ep_0
\eea
and
\bea\lab{eq:mainconclusionofThm0forproofThM8:bis}
\nn\max_{0\leq k\leq k_{large}}\sup_{\CCb_1}\Bigg[ |\dk^k\,{}^{(int)}\a| + |\dk^k\,{}^{(int)}\b|+ \left|\dk^k\left(\,{}^{(int)}\rho+\frac{2m_0}{r^3}\right)\right| &&\\
 +|\dk^k\,{}^{(int)}\bb|+|\dk^k\,{}^{(int)}\aa|\Bigg] &\les& \ep_0,
\eea

\item  the conclusions of  Theorem M7, i.e.
\bea\lab{eq:mainconclusionofThm7forproofThM8}
\Nk^{(Dec)}_{k_{small}} \les\ep_0,
\eea
see section \ref{sec:definitionofconcatenatednorm} for the definition of the combined norm on decay $\Nk^{(Dec)}_k$,

\item the estimate
\bea\lab{eq:consequenceofthechoiceofrhwhichisuseful:ter}
\int_{\TT}|\dk^{\leq k_{large}}\Rc|^2  &\lesssim & \int_{\Mext\Big(r\in I_{m_0, \deh}\Big)}|\dk^{\leq k_{large}}\Rc|^2.
\eea
\end{itemize}
The goal of this section is to prove Theorem M8, i.e. to prove that the following bound holds on $\MM$ for the weighted energies
\beaa
\Nk^{(En)}_{k_{large}} \les\ep_0,
\eeaa
see section \ref{sec:definitionofconcatenatednorm} for the definition of the combined norm on weighted energies $\Nk^{(En)}_k$.


\subsection{Main norms}


We recall below our norms for measuring weighted energies for  curvature components and Ricci coefficients, see sections \ref{section:main-normsextregion} and \ref{section:main-normsintregion}. Let $r_0\geq 4m_0$. Then, we have for $\Mext$
\beaa
\Big(\,{}^{(ext)}\mathfrak{R}_0^{\geq r_0}[\Rc]\Big)^2 &=&     \sup_{0\leq u\leq u_*}   \int_{\CC_u(r\geq 4m_0)} \Big(  r^{4+\dt} \a^2+r^4\b^2    \Big)\\
&+&\int_{\Sigma_*}\Big(r^{4+\dt}(\a^2+\b^2)+r^4(\check{\rho})^2+r^2\bb^2+\aa^2\Big)\\
&+&\int_{\Mext(r\geq 4m_0)}\Big(r^{3+\dt} (\a^2+\b^2) +r^{3-\dt}(\rhoc)^2 +r^{1-\dt}\bb^2 +r^{-1-\dt}\aa^2        \Big),
\eeaa
\beaa
\Big(\,{}^{(ext)}\mathfrak{R}_0^{\leq r_0}[\Rc]\Big)^2 &=&   \int_{\Mext(r\leq 4m_0)}\left(1-\frac{3m}{r}\right)^2|\Rc|^2,
\eeaa
\beaa
\,{}^{(ext)}\mathfrak{R}_0[\Rc] &=& \,{}^{(ext)}\mathfrak{R}_0^{\geq 4m_0}[\Rc] + \,{}^{(ext)}\mathfrak{R}_0^{\leq 4m_0}[\Rc],
\eeaa
\beaa
\Big(\,{}^{(ext)}\mathfrak{R}_k[\Rc]\Big)^2 &=&       \Big(\,{}^{(ext)}\mathfrak{R}_0[\dk^{\leq k}\Rc]\Big)^2 +\int_{\Mext(r\leq 4m_0)}\Big(|\dk^{\leq k-1}\N\Rc|^2+|\dk^{\leq k-1}\Rc|^2\Big), \textrm{ for }k\geq 1,
\eeaa
and
\beaa
\Big(\,{}^{(ext)}\mathfrak{G}_k^{\geq r_0}\left[\Gac\right]\Big)^2 &=& \int_{\Sigma_*}\Bigg[r^2\Big((\dk^{\leq k}\vth)^2+(\dk^{\leq k}\check{\ka})^2+(\dk^{\leq k}\ze)^2+(\dk^{\leq k}\check{\kab})^2\Big)+(\dk^{\leq k}\vthb)^2\\
\nn&+&(\dk^{\leq k}\eta)^2+(\dk^{\leq k}\check{\omb})^2+(\dk^{\leq k}\xib)^2\Bigg]\\
&+& \sup_{\la\geq 4m_0}\Bigg(\int_{\{r=\la\}}\Bigg[\la^2\Big((\dk^{\leq k}\vth)^2+(\dk^{\leq k}\check{\ka})^2+(\dk^{\leq k}\ze)^2\Big)\\
\nn&+&\la^{2-\dt}(\dk^{\leq k}\check{\kab})^2+(\dk^{\leq k}\vthb)^2+(\dk^{\leq k}\eta)^2+(\dk^{\leq k}\check{\omb})^2+\la^{-\dt}(\dk^{\leq k}\xib)^2\Bigg]\Bigg),\\
\Big(\,{}^{(ext)}\mathfrak{G}_k^{\leq r_0}\left[\Gac\right]\Big)^2 &=& \int_{\Mext(\leq 4m_0)}\left|\dk^{\leq k}\left(\Gac\right)\right|^2,\\[2mm]
\,{}^{(ext)}\mathfrak{G}_k\left[\Gac\right] &=& \,{}^{(ext)}\mathfrak{G}_k^{\leq 4m_0}\left[\Gac\right] + \,{}^{(ext)}\mathfrak{G}_k^{\geq 4m_0}\left[\Gac\right].
\eeaa

Also, we have for $\Mint$
\beaa
\Big(\,{}^{(int)}\mathfrak{R}_k[\Rc]\Big)^2 &=&   \int_{\Mint}|\dk^{\leq k}\Rc|^2,
\eeaa
and
\beaa
\Big(\,{}^{(int)}\mathfrak{G}_k[\Gac]\Big)^2 &=& \int_{\Mint}|\dk^{\leq k}\Gac|^2.
\eeaa

Finally, we recall the following Morawetz type norms, see section \ref{sec:mainquantitiesforcontrolofwaveequation}. For $\de>0$, we have 
\beaa
B_\de[\psi](\tau_1,\tau_2)&=& \int_{\Mtrap(\tau_1,\tau_2)}  | R\psi|^2 +r^{-2}  |\psi|^2     +\left(1-\frac{3m}{r}\right)^2\left(|\nabb\psi|^2 +\frac{1}{r^2}| T \psi|^2 \right)    \\
  &+& \int_{\Mntrap(\tau_1,\tau_2)}   r^{\de-3}\left( |\dk \psi|^2 + |\psi|^2\right) 
  \eeaa
where the scalar function $\tau$ and the spacetime region $\Mtrap$ have beed introduced in section \ref{sec:foliationofMMbytau}, and where $\Mntrap$ denotes the complement of $\Mtrap$. Also, we have
\beaa
E_\de[\psi](\tau) &=& \int_{\Si(\tau)}\bigg( \frac 1 2  (N_\Si, e_3)^2  \,    |e_4 \psi|^2  +\frac 1 2 (N_\Si, e_4 )^2\,  |e_3\psi|^2 +|\nabb\psi|^2 + r^{-2}|\psi|^2 \bigg)\\
&&+\int_{\Si_{\ge  4m_0}(\tau)}  r^\de\Big(  |e_4\psi|^2+ r^{-2} |\psi|^2 \Big).
\eeaa
Here $\Si(\tau)$ denotes the level set of $\tau$, see section \ref{sec:foliationofMMbytau}, $N_\Si$ denotes a choice for the normal to $\Si$, and recall that  we have 
\beaa
N_{\Si}=\begin{cases}
 N_{\Si}=e_3 \qquad \mbox{on} \quad \Sint,\\
 N_{\Si}=e_4 \qquad \mbox{on} \quad \Sext,
\end{cases}
\eeaa
with $\Sint$ and $\Sext$ defined in section \ref{sec:foliationofMMbytau}, and 
\beaa
 (N_\Si, e_3)\leq -1\textrm{ and }(N_\Si, e_4)\leq -1\qquad  \mbox{on} \quad \Sitrap.
\eeaa
Moreover, we have
\beaa
\nn F_\de[\psi](\tau_1,\tau_2) &=& \int_{\AA(\tau_1,\tau_2)}\Big( \deh^{-1}   |e_4 \Psi|^2  + \deh  |e_3\Psi|^2 + |\nabb \Psi|^2 + r^{-2} |\Psi|^2\Big)\\
&&  +\int_{\Sigma_*(\tau_1,\tau_2)}\Big(  |e_3\Psi|^2+ r^\de\big( |e_4\psi|^2+|\nabb\psi|^2 + r^{-2} |\psi|^2 \big)\Big)
\eeaa
with $\AA(\tau_1,\tau_2)=\AA\cap\MM(\tau_1,\tau_2)$ and $\Sigma_*(\tau_1,\tau_2)=\Sigma_*\cap\MM(\tau_1,\tau_2)$.


\subsection{Control of the global frame}\lab{sec:presentationoftheglobalframeusedintheproofofThmM8}


Some quantities will be controlled based on the wave equation they satisfy, and will thus need to be defined w.r.t. a global frame, i.e. a smooth frame on $\MM$. To this end, we will rely on the global frame of section \ref{subsection:constructionglobalframe}. We recall below the main properties of that global frame.

From definition \ref{def:cutofffunctionforthematchingregion}, the region where the frame of $\Mint$ and a conformal renormalization of the frame of $\Mext$ are matched is given by
\beaa
\mr:=\left(\Mext\cap\left\{\rint\leq 2m_0\left(1+\frac{3}{2}\deh\right)\right\}\right)\cup\left(\Mint\cap\left\{\rint\geq 2m_0\left(1+\frac{1}{2}\deh\right)\right\}\right),
\eeaa
where $\rint$ denotes the area radius of the ingoing geodesic foliation of $\Mint$ and its extension to $\Mext$. 

The following proposition concerning the global frame is an immediate consequence of Proposition \ref{prop:existenceandestimatesfortheglobalframe} and the decay estimates \eqref{eq:mainconclusionofThm7forproofThM8}.
\begin{proposition}\lab{prop:existenceandestimatesfortheglobalframe:ThmM8}
Assume \eqref{eq:mainconclusionofThm7forproofThM8}. Then, there exists a global null frame defined on $\Mint\cup\Mext$ and denoted by $({}^{(glo)}e_4, {}^{(glo)}e_3, {}^{(glo)}e_\th)$ such that
\begin{itemize}
\item[(a)] In $\Mext\setminus\mr$, we have
\beaa
({}^{(glo)}e_4, {}^{(glo)}e_3, {}^{(glo)}e_\th)= \left({}^{(ext)}\Up\,{}^{(ext)}e_4, {}^{(ext)}\Up^{-1}{}^{(ext)}e_3, {}^{(ext)}e_\th\right).
\eeaa

\item[(b)] In $\Mint\setminus\mr$, we have
\beaa
({}^{(glo)}e_4, {}^{(glo)}e_3, {}^{(glo)}e_\th) = \left({}^{(int)}e_4, {}^{(int)}e_3, {}^{(int)}e_\th\right).
\eeaa

\item[(c)] In the matching region, we have
\beaa
\max_{0\leq k\leq k_{small}-2}\sup_{\mr\cap\Mint}\ub^{1+\dec}\left|\dk^k({}^{(glo)}\Gac, {}^{(glo)}\Rc)\right| &\les&\ep_0,\\
\max_{0\leq k\leq k_{small}-2}\sup_{\mr\cap\Mext}u^{1+\dec}\left|\dk^k({}^{(glo)}\Gac, {}^{(glo)}\Rc)\right| &\les& \ep_0,
\eeaa
where ${}^{(glo)}\Rc$ and ${}^{(glo)}\Gac$ are given by
\beaa
{}^{(glo)}\Rc &=& \left\{\a, \b, \rho+\frac{2m}{r^3}, \bb, \aa\right\},\\
{}^{(glo)}\Gac &=& \left\{\xi, \om+\frac{m}{r^2}, \ka-\frac{2\Up}{r}, \vth, \ze, \eta, \etab, \kab+\frac{2}{r}, \vthb, \omb, \xib\right\}.
\eeaa

\item[(d)] Furthermore, we may also choose the global frame such that, in addition, one of the following  two possibilities hold,
\begin{itemize}
\item[i.] We have on  all $\Mext$ 
\beaa
({}^{(glo)}e_4, {}^{(glo)}e_3, {}^{(glo)}e_\th)= \left({}^{(ext)}\Up\,{}^{(ext)}e_4, {}^{(ext)}\Up^{-1}{}^{(ext)}e_3, {}^{(ext)}e_\th\right).
\eeaa

\item[ii.] We have on all  $\Mint$ 
\beaa
({}^{(glo)}e_4, {}^{(glo)}e_3, {}^{(glo)}e_\th) = \left({}^{(int)}e_4, {}^{(int)}e_3, {}^{(int)}e_\th\right).
\eeaa
\end{itemize}
\end{itemize}
\end{proposition}


\subsection{Iterative procedure}


Recall our norms for measuring energies for  curvature components and Ricci coefficients which are given respectively by $\,{}^{(int)}\mathfrak{R}_k[\Rc]$, $\,{}^{(ext)}\mathfrak{R}_k[\Rc]$ and $\,{}^{(int)}\mathfrak{G}_k[\Gac]$, $\,{}^{(ext)}\mathfrak{G}_k[\Gac]$, see sections \ref{section:main-normsextregion} and \ref{section:main-normsintregion}. Recall also our combined weighted energy norm 
\beaa
\Nk^{(En)}_k & =& \,{}^{(ext)}\Rk_k[\Rc]+\,{}^{(ext)}\Gk_k[\Gac]+ \,{}^{(int)}\Rk_k[\Rc]+\,{}^{(int)}\Gk_k[\Gac].
\eeaa
We also introduce the following norm controlling on the matching region the Ricci coefficients and curvature components of the global frame of Proposition \ref{prop:existenceandestimatesfortheglobalframe:ThmM8}
\bea
\NN^{(match)}_k &:=& \left(\int_{\mr}\left|\dk^{\leq k}({}^{(glo)}\Gac, {}^{(glo)}\Rc)\right|^2\right)^{\frac{1}{2}}.  
\eea
The estimate \eqref{eq:mainconclusionofThm7forproofThM8} and Proposition \ref{prop:existenceandestimatesfortheglobalframe:ThmM8} imply in particular 
\bea\lab{eq:initializationofiterationassumptioninproofThmM8}
\Nk^{(En)}_{k_{small}}+\NN^{(match)}_{k_{small}-2} &\les&\ep_0.
 \eea
 
Next, for $J$ such that $k_{small}-2\leq J\leq k_{large}-1$, consider the iteration assumption 
\bea\lab{eq:iterationassumptiondiscussionThM8:bis}
\Nk^{(En)}_{J} +\NN^{(match)}_{J} &\les& \ep_\BB[J],
\eea
where
\bea\lab{eq:defofepJforiteration:bis}
\ep_\BB[J] &:=& \sum_{j=k_{small}-2}^J (\ep_0)^{\ell(j)}\,\BB^{1-\ell(j)}+\ep_0^{\ell(J)} \BB, \qquad     \ell(j):=2^{k_{small}-2-j},\\
\BB &:=& \left(\int_{\Mext\Big(r\in I_{m_0, \deh}\Big)}|\dk^{\leq k_{large}}\Rc|^2\right)^{\frac{1}{2}}.
\eea

\begin{lemma}
\lab{Le:ep[J]ep[J+1]}
The following  estimate holds true for $\ep_\BB[J]$ as defined above
\bea
\ep_\BB[J] +\BB^{\frac{1}{2}} (\ep_\BB[J])^{\frac{1}{2}} + \ep_0\BB\les \ep_\BB[J+1]. 
\eea
\end{lemma} 

\begin{proof}
We clearly have
\bea
\ep_\BB[J]  + \ep_0 \BB\les \ep_\BB[J+1]. 
\eea

Also, we have, using $\ell(j)=2\ell(j+1)$,
\beaa
\BB \ep_\BB[J] &\les& \sum_{j=k_{small}-2}^J (\ep_0)^{\ell(j)}\,\BB^{2-\ell(j)}+\ep_0^{\ell(J)} \BB^2\\
&\les& \sum_{j=k_{small}-1}^{J+1} (\ep_0)^{2\ell(j)}\,\BB^{2-2\ell(j)}+\ep_0^{2\ell(J+1)} \BB^2\\
&\les& \left(\sum_{j=k_{small}-2}^{J+1} (\ep_0)^{\ell(j)}\,\BB^{1-\ell(j)}+\ep_0^{\ell(J+1)} \BB\right)^2\\
&=& (\ep_\BB[J+1])^2
\eeaa
which concludes the proof of the lemma.
\end{proof}

In view of \eqref{eq:initializationofiterationassumptioninproofThmM8}, \eqref{eq:iterationassumptiondiscussionThM8:bis} holds for $J=k_{small}-2$. The propositions below will allow us to prove Theorem M8 in the next section.

\begin{proposition}\lab{prop:controlofrhotildeiterationassupmtionThM8}
Let $J$ such that $k_{small}-2\leq J\leq k_{large}-1$. Consider the global frame constructed in Proposition \ref{prop:existenceandestimatesfortheglobalframe:ThmM8}. In that frame, let 
\bea
\rhot &:=& r^2\rho+2mr^{-1}.
\eea
Then, under the iteration assumption \eqref{eq:iterationassumptiondiscussionThM8:bis}, we have
  \beaa
 \sup_{\tau\in[1,\tau_*] }   E^J_{\de} [\rhot](\tau)+   B^J_{\de}[\rhot](1,\tau_*)  + F^J_{\de}[\rhot](1,\tau_*) &\les& (\ep_\BB[J])^2+\ep_0^2\Big(\Nk^{(En)}_{J+1}+\NN^{(match)}_{J+1}\Big)^2.
  \eeaa
\end{proposition}

\begin{proposition}\lab{prop:controlofalphaplusUpalphabiterationassupmtionThM8}
Let $J$ such that $k_{small}-2\leq J\leq k_{large}-1$. Consider the global frame constructed in Proposition \ref{prop:existenceandestimatesfortheglobalframe:ThmM8}. In that frame, under the iteration assumption \eqref{eq:iterationassumptiondiscussionThM8:bis}, we have
  \beaa
 &&\sup_{\tau\in[1,\tau_*] }   E^J_{\de} [\a+\Up^2\aa](\tau)+   B^J_{\de}[\a+\Up^2\aa](1,\tau_*)  + F^J_{\de}[\a+\Up^2\aa](1,\tau_*)\\
  &\les& (\ep_\BB[J])^2+\ep_0^2\Big(\Nk^{(En)}_{J+1}+\NN^{(match)}_{J+1}\Big)^2.
  \eeaa
\end{proposition}

\begin{proposition}\lab{prop:controlofMorrallcurvcompiterationassupmtionThM8}
Let $J$ such that $k_{small}-2\leq J\leq k_{large}-1$. Consider the global frame constructed in Proposition \ref{prop:existenceandestimatesfortheglobalframe:ThmM8}. In that frame, under the iteration assumption \eqref{eq:iterationassumptiondiscussionThM8:bis}, we have
  \beaa
  B^J_{-2}\Big[\rhoc, \,\a, \,\aa, \, \b, \,\bb\Big](1,\tau_*)   &\les& (\ep_\BB[J])^2+\ep_0^2\Big(\Nk^{(En)}_{J+1}+\NN^{(match)}_{J+1}\Big)^2.
  \eeaa
\end{proposition}

\begin{proposition}\lab{prop:rpweightedestimatesiterationassupmtionThM8}
Let $J$ such that $k_{small}-2\leq J\leq k_{large}-1$. Under the iteration assumption \eqref{eq:iterationassumptiondiscussionThM8:bis}, we have for $r_0\geq 4m_0$
\beaa
\,{}^{(int)}\mathfrak{R}_{J+1}[\Rc]+\,{}^{(ext)}\mathfrak{R}_{J+1}[\Rc] &\leq& \,{}^{(ext)}\mathfrak{R}^{\geq r_0}_{J+1}[\Rc]+O\left(r_0^{10}\left(\ep_\BB[J]+\ep_0\Big(\Nk^{(En)}_{J+1}+\NN^{(match)}_{J+1}\Big)\right)\right)
\eeaa
and
 \beaa
 \,{}^{(ext)}\mathfrak{R}^{\geq r_0}_{J+1}[\Rc] &\les& r_0^{-\dt}\,{}^{(ext)}\mathfrak{G}^{\geq r_0}_k[\Gac]+r_0^{10}\left(\ep_\BB[J]+\ep_0\left(\Nk^{(En)}_{J+1}+\NN^{(match)}_{J+1}\right)\right).
 \eeaa 
\end{proposition}

\begin{proposition}\lab{prop:controlGaextiterationassupmtionThM8}
Let $J$ such that $k_{small}-2\leq J\leq k_{large}-1$. Under the iteration assumption \eqref{eq:iterationassumptiondiscussionThM8:bis}, we have 
\beaa
 \,{}^{(ext)}\mathfrak{G}_{J+1}[\Gac] + \,{}^{(int)}\mathfrak{R}_{J+1}[\Rc]+ \,{}^{(ext)}\mathfrak{R}_{J+1}[\Rc] &\les&\ep_\BB[J]+\ep_0\Big(\Nk^{(En)}_{J+1}+\NN^{(match)}_{J+1}\Big).
 \eeaa
\end{proposition}

\begin{proposition}\lab{prop:improvementoftheiterationassupmtionThM8}
Let $J$ such that $k_{small}-2\leq J\leq k_{large}-1$. Under the iteration assumption \eqref{eq:iterationassumptiondiscussionThM8:bis}, we have 
\beaa
 \,{}^{(int)}\mathfrak{G}_{J+1}[\Gac] &\les& \ep_\BB[J]+\ep_0\Big(\Nk^{(En)}_{J+1}+\NN^{(match)}_{J+1}\Big)+\left(\int_{\TT}|\dk^{J+1}({}^{(ext)}\Rc)|^2\right)^{\frac{1}{2}}.
 \eeaa
\end{proposition}

\begin{proposition}\lab{prop:controlglobalframeiterationassupmtionThM8}
Let $J$ such that $k_{small}-2\leq J\leq k_{large}-1$. Under the iteration assumption \eqref{eq:iterationassumptiondiscussionThM8:bis}, we have 
\beaa
\NN^{(match)}_{J+1} &\les& \Nk^{(En)}_{J+1}+\left(\int_{\TT}|\dk^{J+1}({}^{(ext)}\Rc)|^2\right)^{\frac{1}{2}}.
 \eeaa
\end{proposition}
   
The proof of Propositions \ref{prop:controlofrhotildeiterationassupmtionThM8}, \ref{prop:controlofalphaplusUpalphabiterationassupmtionThM8}, \ref{prop:controlofMorrallcurvcompiterationassupmtionThM8}, \ref{prop:rpweightedestimatesiterationassupmtionThM8}, \ref{prop:controlGaextiterationassupmtionThM8}, \ref{prop:improvementoftheiterationassupmtionThM8} and \ref{prop:controlglobalframeiterationassupmtionThM8} are postponed respectively to sections \ref{sec:proofprop:controlofrhotildeiterationassupmtionThM8}, \ref{sec:proofprop:controlofalphaplusUpalphabiterationassupmtionThM8}, \ref{sec:proofprop:controlofMorrallcurvcompiterationassupmtionThM8}, \ref{sec:proofprop:rpweightedestimatesiterationassupmtionThM8}, \ref{sec:proofprop:controlGaextiterationassupmtionThM8}, \ref{sec:proofprop:improvementoftheiterationassupmtionThM8} and \ref{sec:proofprop:controlglobalframeiterationassupmtionThM8}.


\subsection{End of the proof of Theorem M8}\lab{sec:endofproofThmM8}


 To prove Theorem M8,  we rely on  Propositions \ref{prop:controlGaextiterationassupmtionThM8}, \ref{prop:improvementoftheiterationassupmtionThM8} and \ref{prop:controlglobalframeiterationassupmtionThM8}. Note that   among them  only the second  two  involve the   dangerous boundary term $ \left(\int_{\TT}|\dk^{J+1}({}^{(ext)}\Rc)|^2\right)^{\frac{1}{2}}$.  We proceed   as follows.

{\bf Step 1.} As mentioned earlier,  the estimate \eqref{eq:initializationofiterationassumptioninproofThmM8} trivially implies the iteration assumption  \eqref{eq:iterationassumptiondiscussionThM8:bis} with $J=k_{small}-2$. We assume that the iteration assumption \eqref{eq:iterationassumptiondiscussionThM8:bis} holds for any fixed $J$ such that $k_{small}-2\leq J\leq k_{large}-2$. In view of Proposition \ref{prop:improvementoftheiterationassupmtionThM8}, we have
\bea\lab{eq:intermediaryestimateonMintforproofofThmM8}
 \,{}^{(int)}\mathfrak{G}_{J+1}[\Gac] \les\ep_\BB[J]+\ep_0\Big(\Nk^{(En)}_{J+1}+\NN^{(match)}_{J+1}\Big)+\left(\int_{\TT}|\dk^{J+1}({}^{(ext)}\Rc)|^2\right)^{\frac{1}{2}}.
 \eea
  
We need to deal with the last term in the RHS of \eqref{eq:intermediaryestimateonMintforproofofThmM8}. Relying on a trace theorem in the spacetime region $\Mext(r\in I_{m_0, \deh})$, and the fact that $J+2\leq k_{large}$, we obtain
\bea\lab{eq:intermediaryestimateonMintforproofofThmM8:bis}
\nn\left(\int_{\TT}|\dk^{J+1}({}^{(ext)}\Rc)|^2\right)^{\frac{1}{2}} &\les& \left(\int_{\Mext\Big(r\in I_{m_0, \deh}\Big)}|\dk^{k_{large}}\Rc|^2\right)^{\frac{1}{4}} (\,{}^{(ext)}\mathfrak{R}_{J+1}[\Rc])^{\frac{1}{2}}\\
&&+\,{}^{(ext)}\mathfrak{R}_{J+1}[\Rc].
\eea

Proposition \ref{prop:controlGaextiterationassupmtionThM8}, \eqref{eq:intermediaryestimateonMintforproofofThmM8} and \eqref{eq:intermediaryestimateonMintforproofofThmM8:bis} yield, for $\ep_0>0$ small enough so that we can absorb  some of the terms to the left,
\beaa
\Nk^{(En)}_{J+1} &\les& \ep_\BB[J]+\left(\int_{\Mext\Big(r\in I_{m_0, \deh}\Big)}|\dk^{k_{large}}\Rc|^2\right)^{\frac{1}{4}} \Big(\ep_\BB[J]+\ep_0\Big(\Nk^{(En)}_{J+1}+\NN^{(match)}_{J+1}\Big)\Big)^{\frac{1}{2}}\\
&&+\ep_0\NN^{(match)}_{J+1},
\eeaa
and using also Proposition \ref{prop:controlglobalframeiterationassupmtionThM8},
 \beaa
\NN^{(match)}_{J+1} &\les& \Nk^{(En)}_{J+1}+\left(\int_{\TT}|\dk^{J+1}({}^{(ext)}\Rc)|^2\right)^{\frac{1}{2}}\\
&\les& \ep_\BB[J]+\left(\int_{\Mext\Big(r\in I_{m_0, \deh}\Big)}|\dk^{k_{large}}\Rc|^2\right)^{\frac{1}{4}} \Big(\ep_\BB[J]+\ep_0\Big(\Nk^{(En)}_{J+1}+\NN^{(match)}_{J+1}\Big)\Big)^{\frac{1}{2}}\\
&&+\ep_0\NN^{(match)}_{J+1}.
 \eeaa
 For $\ep_0>0$ small enough, we infer, by absorbing the appropriate terms to the left,
 \beaa
&&\Nk^{(En)}_{J+1}+\NN^{(match)}_{J+1}\\
&\les& \ep_\BB[J]+\left(\int_{\Mext\Big(r\in I_{m_0, \deh}\Big)}|\dk^{k_{large}}\Rc|^2\right)^{\frac{1}{4}}\Big(\ep_\BB[J]+\ep_0\Big(\Nk^{(En)}_{J+1}+\NN^{(match)}_{J+1}\Big)\Big)^{\frac{1}{2}}\\
&\les& \ep_\BB[J]+\left(\int_{\Mext\Big(r\in I_{m_0, \deh}\Big)}|\dk^{k_{large}}\Rc|^2\right)^{\frac{1}{4}}\Big(\ep_\BB[J]\Big)^{\frac{1}{2}}\\
&&+\left(\int_{\Mext\Big(r\in I_{m_0, \deh}\Big)}|\dk^{k_{large}}\Rc|^2\right)^{\frac{1}{4}}\Big(\ep_0\Big(\Nk^{(En)}_{J+1}+\NN^{(match)}_{J+1}\Big)\Big)^{\frac{1}{2}}
\eeaa
and hence
 \beaa
\Nk^{(En)}_{J+1}+\NN^{(match)}_{J+1} &\les& \ep_\BB[J]+\left(\int_{\Mext\Big(r\in I_{m_0, \deh}\Big)}|\dk^{k_{large}}\Rc|^2\right)^{\frac{1}{4}}\Big(\ep_\BB[J]\Big)^{\frac{1}{2}}\\
&&+\ep_0\left(\int_{\Mext\Big(r\in I_{m_0, \deh}\Big)}|\dk^{k_{large}}\Rc|^2\right)^{\frac{1}{2}}.
\eeaa
In view of Lemma \ref{Le:ep[J]ep[J+1]}, we deduce
\beaa
\Nk^{(En)}_{J+1}+\NN^{(match)}_{J+1} &\les& \ep_\BB[J+1]
\eeaa
 which is \eqref{eq:iterationassumptiondiscussionThM8:bis} for $J+1$ derivatives. We deduce that \eqref{eq:iterationassumptiondiscussionThM8:bis} holds for all $J\leq k_{large}-1$, and hence
\bea\lab{eq:intermediaryestimateonMintforproofofThmM8:ter}
\Nk^{(En)}_{k_{large}-1}+\NN^{(match)}_{k_{large}-1} &\les& \ep_\BB[k_{large}-1].
\eea
  
{\bf Step 2.} Next, Proposition \ref{prop:controlGaextiterationassupmtionThM8}  implies in view of \eqref{eq:intermediaryestimateonMintforproofofThmM8:ter} 
 \bea\lab{eq:intermediaryestimateonMintforproofofThmM8:quatre}
 \,{}^{(ext)}\mathfrak{G}_{k_{large}}[\Gac] + \,{}^{(int)}\mathfrak{R}_{k_{large}}[\Rc]+ \,{}^{(ext)}\mathfrak{R}_{k_{large}}[\Rc] &\les& \ep_\BB[k_{large}-1]\\
 \nn&&+\ep_0\Big(\Nk^{(En)}_{k_{large}}+\NN^{(match)}_{k_{large}}\Big).
 \eea
In particular, we have
\beaa
\left(\int_{\Mext\Big(r\in I_{m_0, \deh}\Big)}|\dk^{\leq k_{large}}\Rc|^2\right)^{\frac{1}{2}} \leq \,{}^{(ext)}\mathfrak{R}_{k_{large}}[\Rc] \les\ep_\BB[k_{large}-1]+\ep_0\Big(\Nk^{(En)}_{k_{large}}+\NN^{(match)}_{k_{large}}\Big).
\eeaa
In view of the definition of $\ep_\BB[k_{large}-1]$, we infer for $\ep_0>0$ small enough
\beaa
\left(\int_{\Mext\Big(r\in I_{m_0, \deh}\Big)}|\dk^{\leq k_{large}}\Rc|^2\right)^{\frac{1}{2}} \les\ep_0+\ep_0\Big(\Nk^{(En)}_{k_{large}}+\NN^{(match)}_{k_{large}}\Big)
\eeaa
and hence 
\beaa
\ep_\BB[k_{large}-1]\les \ep_0+\ep_0\Big(\Nk^{(En)}_{k_{large}}+\NN^{(match)}_{k_{large}}\Big)
\eeaa 
which yields, together with \eqref{eq:intermediaryestimateonMintforproofofThmM8:quatre},
\bea\lab{eq:intermediaryestimateonMintforproofofThmM8:cinq}
 \,{}^{(ext)}\mathfrak{G}_{k_{large}}[\Gac] + \,{}^{(int)}\mathfrak{R}_{k_{large}}[\Rc]+ \,{}^{(ext)}\mathfrak{R}_{k_{large}}[\Rc] \les \ep_0+\ep_0\Big(\Nk^{(En)}_{k_{large}}+\NN^{(match)}_{k_{large}}\Big).
 \eea

{\bf Step 3.} Next, Proposition \ref{prop:improvementoftheiterationassupmtionThM8}  implies in view of \eqref{eq:intermediaryestimateonMintforproofofThmM8:cinq},
 \beaa
 \,{}^{(int)}\mathfrak{G}_{k_{large}}[\Gac] &\les& \ep_0+\ep_0\Big(\Nk^{(En)}_{k_{large}}+\NN^{(match)}_{k_{large}}\Big)+\left(\int_{\TT}|\dk^{k_{large}}({}^{(ext)}\Rc)|^2\right)^{\frac{1}{2}}
 \eeaa
 and hence, for $\ep_0>0$ small enough, using again \eqref{eq:intermediaryestimateonMintforproofofThmM8:cinq}, 
 \beaa
 \Nk^{(En)}_{k_{large}} &\les& \ep_0+\ep_0\NN^{(match)}_{k_{large}} +\left(\int_{\TT}|\dk^{k_{large}}({}^{(ext)}\Rc)|^2\right)^{\frac{1}{2}}.
 \eeaa
 Together with Proposition \ref{prop:controlglobalframeiterationassupmtionThM8}, we infer for $\ep_0>0$ small enough
\beaa
\Nk^{(En)}_{k_{large}}+\NN^{(match)}_{k_{large}} &\les& \ep_0+\left(\int_{\TT}|\dk^{J+1}({}^{(ext)}\Rc)|^2\right)^{\frac{1}{2}}.
\eeaa 
 
{\bf Step 4.} It remains to estimate the last term of the RHS of the previous inequality. Now, in view of \eqref{eq:consequenceofthechoiceofrhwhichisuseful:ter} and \eqref{eq:intermediaryestimateonMintforproofofThmM8:cinq}, we have
  \beaa
\left(\int_{\TT}|\dk^{k_{large}}({}^{(ext)}\Rc)|^2\right)^{\frac{1}{2}}  &\lesssim & \left(\int_{\Mext\Big(r\in I_{m_0, \deh}\Big)}|\dk^{\leq k_{large}}\Rc|^2\right)^{\frac{1}{2}}\\
&\les& \,{}^{(ext)}\mathfrak{R}_{k_{large}}[\Rc]\\
&\les& \ep_0+\ep_0\Nk^{(En)}_{k_{large}}
\eeaa
so that we finally obtain, for $\ep_0>0$ small enough,
\beaa
 \Nk^{(En)}_{k_{large}} &\les& \ep_0.
 \eeaa
This concludes the proof of Theorem M8.


\section{Proof of Proposition \ref{prop:controlofrhotildeiterationassupmtionThM8}}\lab{sec:proofprop:controlofrhotildeiterationassupmtionThM8}



\subsection{A wave equation for $\rhot$}\lab{sec:waveeqationfor-rt}


\begin{proposition} 
\label{prop:waveeqfor-rt}
 The following wave equations hold true.
\begin{enumerate}
\item The curvature component $\rho$ verifies the identity
\beaa
 \square_\g \rho &=&  \kab e_4\rho + \ka e_3\rho +\frac{3}{2}\Big( \kab\, \ka   + 2\rho \Big)\rho +\err[\square_\g\rho],
 \eeaa
 where
 \beaa
\err[\square_\g\rho] &:=& \frac{3}{2}\rho\left( -\frac 1 2  \vthb\, \vth +2(\xib\,  \xi+\eta\, \eta)\right) +\left(\frac 3 2 \kab  -2\omb\right)\left(\frac 1 2 \vthb \, \a -\ze\, \b -2(\etab \,\b+ \xi\,\bb)\right)\\
   &&  - \frac 1 2 \vthb  \dds_2\b + (\ze-\eta) e_3\b - \eta  e_3(\Phi)\b -\xib( e_4\b +e_4(\Phi)\b )- \bb\b \\
    &&  -e_3\left( -\frac 1 2 \vthb \, \a +\ze\, \b +2(\etab \,\b+ \xi\,\bb)\right)\\
    && -\dds_1(\kab)\b   +2\dds_1(\omb) \b    + 3\eta \dds_1(\rho) - \ddd_1\Big(- \vth \bb +\xib \a\Big) -2 \eta e_\th\rho.
\eeaa

\item
The small curvature quantity,
\beaa
\rhot:=r^2\left(\rho+\frac{2m}{r^3} \right)
\eeaa
verifies the wave equation,
\beaa
\square_\g(\rhot) + \frac{8m}{r^3}\rhot &=&  -6m\frac{\square_\g(r)-\left(\frac{2}{r}-\frac{2m}{r^2}\right)}{r^2}-\frac{3m}{r}\left(\ka\kab+\frac{4\Up}{r^2}\right)\\
&&-\frac{3m}{r}\left( A\kab +\Ab \ka\right)+\err[\square_g\rhot],
\eeaa
where
\beaa
\err[\square_g\rhot] &:=& -\frac{6m}{r} A\Ab + \frac{3}{r^2}\rhot^2+\frac{3}{2}\Bigg(\frac{4}{3} A \frac{e_3(r)}{r}+\frac{4}{3} \Ab  \frac{e_4(r)}{r}\Bigg)\rhot\\
&& + \left(\frac{3}{2}\Big(\ka\kab -\frac{8m}{r^3}+\frac{2}{3r^2}\square_\g(r^2)\Big)+\frac{8m}{r^3}\right)\rhot\\
&&-A e_3(\rhot) -\Ab   e_4(\rhot)+\frac{2}{r} A e_3(m)+\frac{2}{r}\Ab e_4(m)\\
&&+4D^a(m)D_a\left(\frac{1}{r}\right)+\frac{2}{r}\square_\g(m) +4r\dds_1(r)\dds_1(\rho)+r^2\err[\square_\g\rho],
\eeaa
and where we recall that,
\beaa
A=\frac{2}{r}e_4(r)-\ka, \qquad \underline{A}=\frac{2}{r}e_3(r)-\kab.
\eeaa
\end{enumerate}
\end{proposition}

\begin{proof} 
See appendix \ref{appendix:Proofprop:waveeqfor-rt}.
\end{proof}


\subsection{Control of $\square_\g(r)$}\lab{sec:identitywaveeqrforThM8}


\begin{lemma}\lab{lemma:identitywaveeqrforThM8}
Let $r$ the function on $\MM$ associated to the global frame constructed in Proposition \ref{prop:existenceandestimatesfortheglobalframe:ThmM8}, see definition \ref{def:globalrandmfor1stglobalframe}. Let $J$ such that $k_{small}-2\leq J\leq k_{large}-1$. Under the iteration assumption \eqref{eq:iterationassumptiondiscussionThM8:bis}, we have
\beaa
\int_{\Mint\cup\Mext(r\leq 4m_0)}\left(\dk^J\left(\square_\g(r)-\left(\frac{2}{r}-\frac{2m}{r^2}\right)\right)\right)^2\\
+\sup_{r_0\geq 4m_0}\int_{\{r=r_0\}}\left(\dk^J\left(\square_\g(r)-\left(\frac{2}{r}-\frac{2m}{r^2}\right)\right)\right)^2 &\les&  (\ep_\BB[J])^2+\ep_0^2\Big(\Nk^{(En)}_{J+1}+\NN^{(match)}_{J+1}\Big)^2
\eeaa
and 
\beaa
\int_{\Mtrap}\left(\dk^Je_4\left(\square_\g(\rext)-\left(\frac{2}{\rext}-\frac{2\mext}{(\rext)^2}\right)\right)\right)^2 \les  (\ep_\BB[J])^2+\ep_0^2\Big(\Nk^{(En)}_{J+1}+\NN^{(match)}_{J+1}\Big)^2.
\eeaa
\end{lemma}

\begin{proof}
Recall that, according to definition \ref{def:globalrandmfor1stglobalframe}, $r$ is defined on $\Mext\cup\Mint$ as follows
\begin{itemize}
\item on $\Mext\setminus\mr$, we have
\beaa
{}^{(glo)}r=\rext,
\eeaa

\item on $\Mint\setminus\mr$, we have
\beaa
{}^{(glo)}r=\rint,
\eeaa

\item on the matching region, we have
\beaa
{}^{(glo)}r &=& (1-\psi_{m_0, \deh}(\rint))\rint+\psi_{m_0, \deh}(\rint)\rext,
\eeaa
\end{itemize}
where the matching region of Proposition \ref{prop:existenceandestimatesfortheglobalframe:ThmM8} is given by
\beaa
\mr:=\left(\Mext\cap\left\{\rint\leq 2m_0\left(1+\frac{3}{2}\deh\right)\right\}\right)\cup\left(\Mint\cap\left\{\rint\geq 2m_0\left(1+\frac{1}{2}\deh\right)\right\}\right),
\eeaa
and where $\psi_{m_0, \deh}$ is given by
\beaa
\psi_{m_0, \deh}(r)=\psi\left(\frac{r - 2m_0\left(1+\frac{1}{2}\deh\right)}{2m_0\deh}\right)\textrm{ on }2m_0\left(1+\frac 1 2 \deh\right)\leq r\leq 2m_0\left(1+\frac 3 2 \deh\right).
\eeaa
with $\psi:\RRR\to\RRR$ a smooth cut-off function such that $0\leq\psi\leq 1$, $\psi=0$ on $(-\infty, 0]$ and $\psi=1$ on $[1, +\infty)$. 

We have on $\Mext$
\beaa
\square_\g(\rext) &=& -e_3 e_4(\rext) +\lapp(\rext)+\left(2\omb -\frac 1 2 \kab\right) e_4(\rext)- \frac 1 2 \ka e_3(\rext)+2 \eta e_\th(\rext).
\eeaa
Here, $(e_4, e_3, e_\th)$ denotes the frame of $\Mext$ and the Ricci coefficients are computed w.r.t. frame, so we have
\beaa
e_4(\rext) = \frac{\rext}{2}\ov{\ka}, \qquad e_3(\rext) = \frac{\rext}{2}(\ov{\kab}+\Ab), \qquad e_\th(\rext)=0
\eeaa
and hence
\beaa
\square_\g(\rext) &=& -e_3\left(\frac{\rext}{2}\ov{\ka}\right) +\left(2\omb -\frac 1 2 \kab\right)\frac{\rext}{2}\ov{\ka} - \frac 1 2 \ka\frac{\rext}{2}(\ov{\kab}+\Ab)\\
&=& -\frac{\rext}{2} e_3(\ov{\ka}) - \frac{e_3(\rext)}{2}\ov{\ka} +\left(2\omb -\frac 1 2 \kab\right)\frac{\rext}{2}\ov{\ka} - \frac{\rext}{4}\ka\ov{\kab}  - \frac{\rext}{4}\ka\Ab\\
&=& -\frac{\rext}{2} e_3(\ov{\ka}) - \frac{1}{2}\ov{\ka}\frac{\rext}{2}(\ov{\kab}+\Ab) +\left(2\omb -\frac 1 2 \kab\right)\frac{\rext}{2}\ov{\ka} - \frac{\rext}{4}\ka\ov{\kab}  - \frac{\rext}{4}\ka\Ab.
\eeaa

Now, we have
\beaa
e_3(\ov{\ka}) &=& \ov{e_3(\ka)}+\err[e_3\ov{\ka}]\\
&=& \ov{-\frac{1}{2}\ka\kab+2\omb\ka+2\rho+2\ddd_1\eta-\frac{1}{2}\vth\vthb+2\eta^2}+\err[e_3\ov{\ka}]\\
&=& \ov{-\frac{1}{2}\ka\kab+2\omb\ka+2\rho-\frac{1}{2}\vth\vthb+2\eta^2}+\err[e_3\ov{\ka}]
\eeaa
and hence
\beaa
\square_\g(\rext) &=& -\frac{\rext}{2}\left(\ov{-\frac{1}{2}\ka\kab+2\omb\ka+2\rho-\frac{1}{2}\vth\vthb+2\eta^2}+\err[e_3\ov{\ka}]\right) \\
&&- \frac{1}{2}\ov{\ka}\frac{\rext}{2}(\ov{\kab}+\Ab) +\left(2\omb -\frac 1 2 \kab\right)\frac{\rext}{2}\ov{\ka} - \frac{\rext}{4}\ka\ov{\kab}  - \frac{\rext}{4}\ka\Ab.
\eeaa
Together with \eqref{eq:mainconclusionofThm0forproofThM8} and the iteration assumption \eqref{eq:iterationassumptiondiscussionThM8:bis}, we easily infer\footnote{Recall in particular that $\ov{\rho}$ is under control in view of Lemma \ref{lemma:estimatesonceandforallforaverages}.}
\bea\lab{eq:estimatewaveeqrextforcontrolrhotildeThmM8}
&&\nn\int_{\Mext(r\leq 4m_0)}\left(\dk^J\left(\square_\g(\rext)-\left(\frac{2}{\rext}-\frac{2\mext}{(\rext)^2}\right)\right)\right)^2\\
&&\nn +\sup_{r_0\geq 4m_0}\int_{\{r=r_0\}}\left(\dk^J\left(\square_\g(\rext)-\left(\frac{2}{\rext}-\frac{2\mext}{(\rext)^2}\right)\right)\right)^2 \\
&\les&  (\ep_\BB[J])^2+\ep_0^2\Big(\Nk^{(En)}_{J+1}+\NN^{(match)}_{J+1}\Big)^2.
\eea
Also, using again \eqref{eq:mainconclusionofThm0forproofThM8} and the iteration assumption \eqref{eq:iterationassumptiondiscussionThM8:bis}, we have
\bea\lab{eq:estimatewaveeqrextforcontrolrhotildeThmM8:bis}
\nn&&\int_{\Mtrap}\left(\dk^Je_4\left(\square_\g(\rext)-\left(\frac{2}{\rext}-\frac{2\mext}{(\rext)^2}\right)\right)\right)^2 \\
&\les&  (\ep_\BB[J])^2+\ep_0^2\Big(\Nk^{(En)}_{J+1}+\NN^{(match)}_{J+1}\Big)^2,
\eea
where we have used the null structure equations for $e_4(\ka)$, $e_4(\kab)$, $e_4(\omb)$, $e_4(\vth)$, $e_4(\vthb)$, $e_4(\eta)$, the equations for  $e_4(\Omb)$,$e_4(\vsi)$, $e_4(r)$, and the Bianchi identity for $e_4(\rho)$. 

\begin{remark}
Note that we have used in the last estimate the following observations to avoid a potential loss of one derivative
\beaa
e_4(\kab) &=& -2\ddd_1\ze +\cdots= 2\left(\rho+\mu-\frac{1}{4}\vth\vthb\right)+\cdots,\\
\ov{e_4(\rho)} &=& \ov{\ddd_1\b}+\cdots=\cdots,\\
e_4( \err[e_3\ov{\ka}]) &=& 2e_4(\vsi^{-1}\ov{\check{\vsi}\ddd_1\eta})+\cdots=2\vsi^{-1}\ov{\check{\vsi}\ddd_1e_4\eta}+\cdots=-2\vsi^{-1}\ov{e_\th(\vsi)e_4\eta}+\cdots
\eeaa
Note also that there is no term involving $\dk^J\rho$ (without average) as such a term appears only in the null structure equations for $e_4(\kab)$, as well as $e_4(\omb)$ and vanishes due to the cancellation
\beaa
e_4\left(2\omb -\frac 1 2 \kab\right) &=& 2e_4(\omb) - \frac{1}{2}e_4(\kab)\\
&=& 2\rho+\cdots-\frac{1}{2}( -2\ddd_1\ze+2\rho)+\cdots\\
&=& 2\mu+\cdots
\eeaa
This is important as such a term would otherwise violate \eqref{eq:estimatewaveeqrextforcontrolrhotildeThmM8:bis} at $r=3m$.
\end{remark}

\begin{remark}
Recall that the global frame constructed in Proposition \ref{prop:existenceandestimatesfortheglobalframe:ThmM8} 
\begin{itemize}
\item co•ncides with the frame of $\Mint$ in $\Mint\setminus\mr$,

\item co•ncides with a conformal renormalization of  the frame of $\Mext$ in $\Mext\setminus\mr$.
\end{itemize}
Thus, $J+1$ derivatives of its Ricci coefficients and curvature components are controlled
\begin{itemize}
\item by $\NN^{(match)}_{J+1}$ in $\mr$,

\item by $\Nk^{(En)}_{J+1}$ in $\MM\setminus\mr$,
\end{itemize}
and hence by $\Nk^{(En)}_{J+1}+\NN^{(match)}_{J+1}$ on $\MM$. This explains the occurrence of $\Nk^{(En)}_{J+1}+\NN^{(match)}_{J+1}$ on the  right-hand side of numerous estimates, see for example \eqref{eq:estimatewaveeqrextforcontrolrhotildeThmM8} \eqref{eq:estimatewaveeqrextforcontrolrhotildeThmM8:bis}.
\end{remark} 

Arguing similarly for $\rint$, we obtain the following analog of \eqref{eq:estimatewaveeqrextforcontrolrhotildeThmM8}
\begin{equation}\lab{eq:estimatewaveeqrintforcontrolrhotildeThmM8}
\int_{\Mint}\left(\dk^J\left(\square_\g(\rint)-\left(\frac{2}{\rint}-\frac{2\mint}{(\rint)^2}\right)\right)\right)^2 \les  (\ep_\BB[J])^2+\ep_0^2\Big(\Nk^{(En)}_{J+1}+\NN^{(match)}_{J+1}\Big)^2.
\end{equation}

Then, since 
\begin{itemize}
\item on $\Mext\setminus\mr$, we have
\beaa
\square_\g(r)=\square_\g(\rext),\qquad m=\mext,
\eeaa
\item on $\Mint\setminus\mr$, we have
\beaa
\square_\g(r)=\square_\g(\rint),\qquad m=\mint,
\eeaa
\end{itemize}
we immediately infer from \eqref{eq:estimatewaveeqrextforcontrolrhotildeThmM8}, \eqref{eq:estimatewaveeqrextforcontrolrhotildeThmM8:bis} and \eqref{eq:estimatewaveeqrintforcontrolrhotildeThmM8}
\beaa
&&\int_{\Big(\Mint\cup\Mext(r\leq 4m_0)\Big)\setminus\mr}\left(\dk^J\left(\square_\g(r)-\left(\frac{2}{r}-\frac{2m}{r^2}\right)\right)\right)^2\\
&&+\sup_{r_0\geq 4m_0}\int_{\{r=r_0\}}\left(\dk^J\left(\square_\g(r)-\left(\frac{2}{r}-\frac{2m}{r^2}\right)\right)\right)^2\\
 &\les&  (\ep_\BB[J])^2+\ep_0^2\Big(\Nk^{(En)}_{J+1}+\NN^{(match)}_{J+1}\Big)^2
\eeaa
and
\beaa
\int_{\Mtrap}\left(\dk^Je_4\left(\square_\g(r)-\left(\frac{2}{r}-\frac{2m}{r^2}\right)\right)\right)^2 &\les&  (\ep_\BB[J])^2+\ep_0^2\Big(\Nk^{(En)}_{J+1}+\NN^{(match)}_{J+1}\Big)^2
\eeaa
which are the desired estimates outside of the matching region. Note that we have used the fact that $\Mtrap\cap\mr=\emptyset$.

It remains to derive the desired estimates in the matching region. To this end, we need to estimate $\rext-\rint$ and $\mint-\mext$ in the matching region. Step 7 or the proof of Lemma \ref{lemma:estimatesfortheglobalframeinthematchingregion} in section \ref{sec:proofoflemma:estimatesfortheglobalframeinthematchingregion} yields\footnote{The proof of Lemma \ref{lemma:estimatesfortheglobalframeinthematchingregion} in section \ref{sec:proofoflemma:estimatesfortheglobalframeinthematchingregion} is done in the particular case $J=k_{large}-1$ but extends immediately to the case $k_{small}-2\leq J\leq k_{large}-1$.}
\beaa
\int_{\Mint}\left(\dk^{J+1}\left(\rext-\rint, \mext-\mint\right)\right)^2 \les  (\Nk^{(En)}_{J})^2+(\NN^{(match)}_{J})^2.
\eeaa
We infer, in view of the  iteration assumption \eqref{eq:iterationassumptiondiscussionThM8:bis},
\bea\lab{eq:estimaterintminusrextforcontrolrhotildeThmM8}
\int_{\Mint}\left(\dk^{J+1}\left(\rext-\rint, \mext-\mint\right)\right)^2 \les  (\ep_\BB[J])^2.
\eea

Then, since we have on the matching region, 
\beaa
r&=& (1-\psi_{m_0, \deh}(\rint))\rint+\psi_{m_0, \deh}(\rint)\rext,\\
m&=& (1-\psi_{m_0, \deh}(\rint))\mint+\psi_{m_0, \deh}(\rint)\mext,\\
\square_\g(r) &=& (1-\psi_{m_0, \deh}(\rint))\square_\g(\rint)+\psi_{m_0, \deh}(\rint)\square_\g(\rext)\\
&&+2\psi'_{m_0, \deh}(\rint)\D^\a(\rint)\D_\a(\rext - \rint)\\
&&+(\rext - \rint)\square_\g(\psi_{m_0, \deh}),
\eeaa
we deduce there
\beaa
&&\square_\g(r) -\left(\frac{2}{r}-\frac{2m}{r^2}\right)\\
&=& (1-\psi_{m_0, \deh}(\rint))\left(\square_\g(\rint) -\left(\frac{2}{\rint}-\frac{2\mint}{(\rint)^2}\right)\right)\\
&&+\psi_{m_0, \deh}(\rint)\left(\square_\g(\rext)-\left(\frac{2}{\rext}-\frac{2\mext}{(\rext)^2}\right)\right)\\
&&+(1-\psi_{m_0, \deh}(\rint))\left(\frac{2}{\rint}-\frac{2}{r}-\frac{2\mint}{(\rint)^2}+\frac{2m}{r^2}\right)\\
&&+\psi_{m_0, \deh}(\rint)\left(\frac{2}{\rext}-\frac{2}{r}-\frac{2\mext}{(\rext)^2}+\frac{2m}{r^2}\right)\\
&&+2\psi'_{m_0, \deh}(\rint)\D^\a(\rint)\D_\a(\rext - \rint)+(\rext-\rint)\square_\g(\psi_{m_0, \deh})
\eeaa
and thus, in view of \eqref{eq:estimatewaveeqrextforcontrolrhotildeThmM8}, \eqref{eq:estimatewaveeqrintforcontrolrhotildeThmM8} and \eqref{eq:estimaterintminusrextforcontrolrhotildeThmM8}, we have on the matching region 
\beaa
\int_{\mr}\left(\dk^J\left(\square_\g(r)-\left(\frac{2}{r}-\frac{2m}{r^2}\right)\right)\right)^2 &\les&  (\ep_\BB[J])^2+\ep_0^2\Big(\Nk^{(En)}_{J+1}+\NN^{(match)}_{J+1}\Big)^2
\eeaa
as desired. This concludes the proof of the lemma.
\end{proof}

\begin{corollary}\lab{cor:identityN0RHSwaveeqrhotforThM8}
Let $N_0$ the RHS of the wave equation for $\rhot$ provided by Proposition \ref{prop:waveeqfor-rt},  i.e.
\beaa
N_0 &=& -6m\frac{\square_\g(r)-\left(\frac{2}{r}-\frac{2m}{r^2}\right)}{r^2}-\frac{3m}{r}\left(\ka\kab+\frac{4\Up}{r^2}\right)-\frac{3m}{r}\left( A\kab +\Ab \ka\right)+\err[\square_g\rhot].
\eeaa
Then, $N_0-\err[\square_g\rhot]$ satisfies
\beaa
\int_{\Mint\cup\Mext(r\leq 4m_0)}\left(\dk^J\left(N_0-\err[\square_g\rhot]\right)\right)^2\\
+\sup_{r_0\geq 4m_0}\int_{\{r=r_0\}}\left(\dk^J\left(N_0-\err[\square_g\rhot]\right)\right)^2 &\les&  (\ep_\BB[J])^2+\ep_0^2\Big(\Nk^{(En)}_{J+1}+\NN^{(match)}_{J+1}\Big)^2
\eeaa
and 
\beaa
\dk^Je_4\Big(N_0-\err[\square_g\rhot]\Big) &=&  -\frac{12m\ka}{r}\,\dk^J\rho +a_J\textrm{ on }\Mtrap
\eeaa
where $a^J$ satisfies
\beaa
\int_{\Mtrap}\left(\dk^Je_4a^J\right)^2 \les  (\ep_\BB[J])^2+\ep_0^2\Big(\Nk^{(En)}_{J+1}+\NN^{(match)}_{J+1}\Big)^2.
\eeaa
\end{corollary}

\begin{proof}
The first estimate is an immediate consequence of Lemma \ref{lemma:identitywaveeqrforThM8}, \eqref{eq:mainconclusionofThm0forproofThM8} and the iteration assumption \eqref{eq:iterationassumptiondiscussionThM8:bis}. 

Concerning the second estimate, note that the term  $\dk^J\rho$ is due to the null structure equations for $e_4(\kab)$, i.e.
\beaa
e_4(\kab) &=&  -2\ddd_1\ze+2\rho +\cdots\\
&=& 4\rho +\cdots
\eeaa
Then, the estimate for $a_J$ follows from Lemma \ref{lemma:identitywaveeqrforThM8}, \eqref{eq:mainconclusionofThm0forproofThM8} and the iteration assumption \eqref{eq:iterationassumptiondiscussionThM8:bis}. 
\end{proof}


\subsection{End of the proof of Proposition \ref{prop:controlofrhotildeiterationassupmtionThM8}}


In view of Proposition \ref{prop:waveeqfor-rt}, $\rhot$ satisfies
\beaa
(\square_0+V_0)\rhot &=& N_0, \qquad V_0=\frac{8m}{r^3},
\eeaa
where
\beaa
N_0 &:=& -6m\frac{\square_\g(r)-\left(\frac{2}{r}-\frac{2m}{r^2}\right)}{r^2}-\frac{3m}{r}\left(\ka\kab+\frac{4\Up}{r^2}\right)-\frac{3m}{r}\left( A\kab +\Ab \ka\right)+\err[\square_g\rhot].
\eeaa
We may thus apply the estimate \eqref{eq:thesecondequationoftheorem-recoveringhigherderivatives} of Theorem \ref{theorem-recoveringhigherderivatives} with $\phi=\rhot$ and $s=J$ to obtain for   any $k_{small}\leq J\leq k_{large}-1$ 
  \beaa
\nn  &&\sup_{\tau\in[1,\tau_*] }   E^J_{\de} [\rhot](\tau)+   B^J_{\de}[\rhot](1,\tau_*)  + F^J_{\de}[\rhot](1,\tau_*)\\
  \nn &\les& 
          E^J_{\de}[\rhot](1) +\sup_{\tau\in[1,\tau_*] }   E^{J-1}_{\de} [\rhot](\tau)+   B^{J-1}_{\de}[\rhot](1,\tau_*)  + F^{J-1}_{\de}[\rhot](1,\tau_*)\\
\nn&&+D_J[\Ga] \left(\sup_{\MM}ru_{trap}^{\frac{1}{2}+\dec}|\dk^{\leq k_{small}}\rhot|\right)^2+\int_{\Sigma(\tau_*)}\frac{(\dk^{\leq J}\rhot)^2}{r^3}\\
&&+\int_{\MM}r^{1+\de}|\dk^{\leq J}N_0|^2+\left|\int_{\Mtrap}T(\dk^J\rhot)\dk^JN_0\right|,
\eeaa
where $D_J[\Ga]$ is defined by
\beaa
D_J[\Ga] &:=& \int_{\Mint\cup\Mext(r\leq 4m_0)}(\dk^{\leq J}\Gac)^2\\
&& +\sup_{r_0\geq 4m_0}\left(r_0\int_{\{r=r_0\}}|\dk^{\leq J}\Ga_g|^2+r_0^{-1}\int_{\{r=r_0\}}|\dk^{\leq J}\Ga_b|^2\right).
\eeaa

Next we use the iteration assumption \eqref{eq:iterationassumptiondiscussionThM8:bis} which yields in particular 
\beaa
D_J[\Ga]  &\les& (\ep_\BB[J])^2.
\eeaa
Also, we have 
\beaa
\rhot &=& r^2\left(\ov{\rho}-\frac{2m}{r^3}\right)+r^2\rhoc
\eeaa
and hence, using again the iteration assumption \eqref{eq:iterationassumptiondiscussionThM8:bis}, as well as the control on averages provided by Lemma \ref{lemma:estimatesonceandforallforaverages}, we infer
\beaa
 &&\sup_{\tau\in[1,\tau_*] }   E^{J-1}_{\de} [\rhot](\tau)+   B^{J-1}_{\de}[\rhot](1,\tau_*)  + F^{J-1}_{\de}[\rhot](1,\tau_*)+\int_{\Sigma(\tau_*)}\frac{(\dk^{\leq J}\rhot)^2}{r^3} \les (\ep_\BB[J])^2.
\eeaa
Together with the control of $\dk^{\leq k_{small}}\rhot$ provided by the decay estimate \eqref{eq:mainconclusionofThm7forproofThM8}, we infer from the above estimates 
  \beaa
\nn  &&\sup_{\tau\in[1,\tau_*] }   E^J_{\de} [\rhot](\tau)+   B^J_{\de}[\rhot](1,\tau_*)  + F^J_{\de}[\rhot](1,\tau_*)\\
  \nn &\les& E^J_{\de}[\rhot](1)+(\ep_\BB[J])^2+\int_{\MM}r^{1+\de}|\dk^{\leq J}N_0|^2+\left|\int_{\Mtrap}T(\dk^J\rhot)\dk^JN_0\right|.
\eeaa

Next, using the form of $N_0$, as well as Corollary \ref{cor:identityN0RHSwaveeqrhotforThM8}, we derive
\beaa
\int_{\MM}r^{1+\de}|\dk^{\leq J}N_0|^2 &\les& (\ep_\BB[J])^2+\ep_0^2\Big(\Nk^{(En)}_{J+1}+\NN^{(match)}_{J+1}\Big)^2.
\eeaa
Also, decomposing $T$ as a combination of $R$ and $e_4$, integrating $e_4$ by parts, using again the form of $N_0$, as well as Corollary  \ref{cor:identityN0RHSwaveeqrhotforThM8}, we have
\beaa
&&\left|\int_{\Mtrap}T(\dk^J\rhot)\dk^JN_0\right| \\
&\les& \left|\int_{\Mtrap}R(\dk^J\rhot)\dk^J(N_0-\err[\square_g\rhot])\right|+\left|\int_{\Mtrap}e_4(N_0-\err[\square_g\rhot])\dk^JN_0\right|\\
&&+ \int_{\Mtrap}|T(\dk^J\rhot)||\dk^J\err[\square_g\rhot]|\\
&\les& \left|\int_{\Mtrap}\dk^J\rhot e_4(\dk^J(N_0-\err[\square_g\rhot]))\right|\\
&&+ \left(\ep_\BB[J]+\ep_0\Big(\Nk^{(En)}_{J+1}+\NN^{(match)}_{J+1}\Big)\right)\left(\sup_{\tau\in[1,\tau_*] }   E^J_{\de} [\rhot](\tau)+B^J_{\de}[\rhot](1,\tau_*)\right)^{\frac{1}{2}}\\
&\les& \int_{\Mtrap}(\dk^J\rhot)^2+\left(\ep_\BB[J]+\ep_0\Big(\Nk^{(En)}_{J+1}+\NN^{(match)}_{J+1}\Big)\right)\left(\sup_{\tau\in[1,\tau_*] }   E^J_{\de} [\rhot](\tau)+B^J_{\de}[\rhot](1,\tau_*)\right)^{\frac{1}{2}}.
\eeaa
In view of the above, we infer
  \beaa
 &&\sup_{\tau\in[1,\tau_*] }   E^J_{\de} [\rhot](\tau)+   B^J_{\de}[\rhot](1,\tau_*)  + F^J_{\de}[\rhot](1,\tau_*)\\
  &\les&  \int_{\Mtrap}(\dk^J\rhot)^2+(\ep_\BB[J])^2+\ep_0^2\Big(\Nk^{(En)}_{J+1}+\NN^{(match)}_{J+1}\Big)^2.
  \eeaa

Next, note that we have on 
\beaa
R(r-3m) = \frac{1}{2}(e_4(r)-\Up e_3(r)) - \frac{3}{2}(e_4(m)-\Up e_3(m)) =  \Up+O(\ep_0)\geq \frac{1}{6}\textrm{ on }\Mtrap, 
\eeaa
and hence, using also integration by parts,
\beaa
\int_{\Mtrap}(\dk^J\rhot)^2 &\les& \int_{\Mtrap}R(r-3m)(\dk^J\rhot)^2\\
&\les& \int_{\Mtrap}\left|1-\frac{3m}{r}\right||\dk^J\rhot||R\dk^J\rho|\\
&\les& \ep_\BB[J]\left(\sup_{\tau\in[1,\tau_*] }   E^J_{\de} [\rhot](\tau)+B^J_{\de}[\rhot](1,\tau_*)\right)^{\frac{1}{2}}.
\eeaa
We deduce
  \beaa
 \sup_{\tau\in[1,\tau_*] }   E^J_{\de} [\rhot](\tau)+   B^J_{\de}[\rhot](1,\tau_*)  + F^J_{\de}[\rhot](1,\tau_*) &\les& (\ep_\BB[J])^2+\ep_0^2\Big(\Nk^{(En)}_{J+1}+\NN^{(match)}_{J+1}\Big)^2
  \eeaa
as desired. This concludes the proof of Proposition \ref{prop:controlofrhotildeiterationassupmtionThM8}.


\section{Proof of Proposition \ref{prop:controlofalphaplusUpalphabiterationassupmtionThM8}}\lab{sec:proofprop:controlofalphaplusUpalphabiterationassupmtionThM8}



\subsection{A wave equations for $\a+\Up^2\aa$}


\begin{lemma}\lab{lemma:firstwaveeqalphaplusUpsquarealphabThmM8}
We have
\beaa
\square_2(\a+\Up^2\aa) &=&  \frac{4}{r}\left(1- \frac{3m}{r}\right)  \Big(e_3(\a)-\Up e_4(\aa)\Big)  +\left(-\frac{2}{r^2}+\frac{16m}{r^3}\right)\a\\
&& -\frac{2\Up}{r^2}\left(1-\frac{2m}{r}-\frac{8m^2}{r^2}\right)\aa+\err\Big[\square_2(\a+\Up^2\aa)\Big]
 \eeaa
 where
\beaa
&& \err\Big[\square_2(\a+\Up^2\aa)\Big]\\ 
&=& \left(\Up^2\Vb +\frac{4m}{r^2}\Up\square_g(r) -\frac{8m}{r^3}\Up\D^\a(r)\D_\a(r)+\frac{8m^2}{r^4}\D^\a(r)\D_\a(r)\right)\aa \\
&& +\left(4 \left(\om+\frac{m}{r^2}\right)+2\left(\ka-\frac{2\Up}{r}\right)\right) e_3(\a)-4\omb e_4(\a)\\
&&-4\Up\left(\Up\left(\om +\frac{m}{r^2}\right) +\frac{m}{r^2}(e_4(r)-1) -\frac{e_4(m)}{r}\right)e_3(\aa)\\
&& +\left(4\Up^2 \omb +2\Up^2\left(\kab+\frac{2}{r}\right)-4m\Up \frac{(e_3(r)+1)}{r^2}+4\Up \frac{e_3(m)}{r}\right)e_4(\aa) \\
&& +\left(\left(-4\rho -\frac{8m}{r^3}\right)    +2\left(\om\, \kab -\frac{2m}{r^3}\right)   +\frac 1 2 \left(\ka\,\kab +\frac{4\Up}{r^2}\right)- 4 e_4(\omb)  -8\om\omb -10 \ka\, \omb\right)\a\\
&&+\left(\frac{8\Up}{r^2}\D^\a(m)\D_\a(r)-\frac{4\Up}{r}\square_g(m)-\frac{8m}{r^3}\D^\a(r)\D_\a(m)+\frac{8m}{r}\D^\a(m)\D_\a\left(\frac{m}{r}\right)\right)\aa\\
&&+\Up^2\left(\left(-4\rho -\frac{8m}{r^3}\right)  - 4 \left(e_3(\om)-\frac{2m}{r^3}\right) -10\left(\kab\, \om-\frac{2m}{r^3}\right)+\frac 1 2\left(\ka\,\kab +\frac{4\Up}{r^2}\right) -8\om\omb  +2\ka\omb\right)\aa\\
&&+\frac{4m}{r^2}\Up\left(\square_g(r)-\left(\frac{2}{r}-\frac{2m}{r^2}\right)\right)\aa -\frac{8m}{r^3}\left(1-\frac{3m}{r}\right)\Big(-e_4(r)e_3(r)-\Up+(e_\th(r))^2\Big)\aa\\
&&+4\Up e_\th(\Up)e_\th(\aa) +\err[\square_\g \a]+\Up^2\err[\square_\g \aa].
\eeaa
\end{lemma}

\begin{proof}
Recall from Proposition \ref{proposition:wave-a-aa-qf} that the curvature components $\a$ and $\aa$ verify the following Teukolsky equations
\beaa 
\bsplit
 \square_2\a &= -4\omb e_4(\a)+ (4 \om+2\ka) e_3(\a) + V \a +\err[\square_\g \a],\\
V&=-4\rho   - 4 e_4(\omb)  -8\om\omb  +2\om\, \kab    -10 \ka\, \omb+\frac 1 2 \ka\,\kab,
\end{split}
\eeaa
where 
\beaa
\err(\square_\g\a) &=& \frac{1}{2}\vth e_3(\a) + \frac 3 4 \vth^2\rho+ e_\th(\Phi)\vth \b  -\frac{1}{2}\ka(\ze+4\eta)\b  -(\ze+\etab)e_4(\b) - \xi e_3(\b)\\
&&+ e_\th(\Phi)(2\ze+\etab) \a + \b^2 + e_4(\Phi)\etab\b+ e_3(\Phi)\xi\b - (\ze+4\eta) e_4(\b) \\
&&-(e_4(\ze)+4e_4(\eta))\b - 2(\ka+\om)(\ze+4\eta)\b +2e_\th(\ka+\om) \b  -e_\th((2\ze+\etab)\a)\\
&& -3\xi e_\th(\rho)+2\etab e_\th(\a) + \frac 3 2 \vth  \ddd_1 \b  + 3\rho (\etab+\eta+2\ze)\xi    +  \ddd_1\etab \a +\frac{1}{4}\kab\vth\a -2\omb\vth\a\\
&&  -  \frac{1}{2}\vth\vthb\a +\xi\xib\a+\etab^2\a  + \frac 3 2 \vth\ze\b+3\vth(\etab\b+\xi\bb) -\frac{1}{2}\vth(\ze+4\eta)\b,
\eeaa
and 
\beaa
\bsplit
 \square_2\aa&= -4\om e_3(\aa)+ (4 \omb+2\kab) e_4(\aa) +\Vb \aa+\err[\square_\g \aa],\\
\Vb&=-4\rho   - 4 e_3(\om)  -8\om\omb  +2\omb\ka    -10 \kab\, \om+\frac 1 2 \ka\,\kab,
\end{split}
\eeaa
where 
\beaa
\err(\square_\g\aa) &=& \frac{1}{2}\vthb e_4(\aa) + \frac 3 4 \vthb^2\rho+ e_\th(\Phi)\vthb\, \bb  -\frac{1}{2}\ka(-\ze+4\etab)\bb  -(-\ze+\eta)e_3(\bb) - \xib e_4(\bb)\\
&&+ e_\th(\Phi)(-2\ze+\eta) \aa + \bb^2 + e_3(\Phi)\eta\bb+ e_4(\Phi)\xib\,\bb - (-\ze+4\etab) e_3(\bb) \\
&&-(-e_3(\ze)+4e_3(\etab))\bb - 2(\kab+\omb)(-\ze+4\etab)\bb +2e_\th(\kab+\omb) \bb  -e_\th((-2\ze+\eta)\aa)\\
&& -3\xib e_\th(\rho)+2\eta e_\th(\aa) + \frac 3 2 \vthb  \ddd_1 \bb  + 3\rho (\eta+\etab-2\ze)\xib    +  \ddd_1\eta \aa +\frac{1}{4}\ka\vthb\,\aa -2\om\vthb\,\aa\\
&&  -  \frac{1}{2}\vthb\vth\aa +\xi\xib\aa+\eta^2\aa  -\frac 3 2 \vthb\ze\bb+3\vthb(\eta\bb+\xib\b) -\frac{1}{2}\vthb(-\ze+4\etab)\bb.
\eeaa

We infer from the above wave equations
\beaa
\square_2(\a+\Up^2\aa) &=& \square_2(\a)+\Up^2\square_2(\aa)+2\D^\mu(\Up^2)\D_\mu(\aa)+\square_0(\Up^2)\aa\\
&=& -4\omb e_4(\a)+ (4 \om+2\ka) e_3(\a)  \\
&&+\Up^2\Big(-4\om e_3(\aa)+ (4 \omb+2\kab) e_4(\aa) \Big) -2\Up e_3(\Up)e_4(\aa)-2\Up e_4(\Up)e_3(\aa)\\
&& + V \a +\Big(\Up^2\Vb +\square_0(\Up^2)\Big)\aa +4\Up e_\th(\Up)e_\th(\aa) +\err[\square_\g \a]+\Up^2\err[\square_\g \aa]
\eeaa
and hence
\beaa
\square_2(\a+\Up^2\aa) &=&  \frac{4}{r}\left(1- \frac{3m}{r}\right)  \Big(e_3(\a)-\Up e_4(\aa)\Big)\\
&& + V \a +\left(\Up^2\Vb +\frac{4m}{r^2}\Up\square_g(r) -\frac{8m}{r^3}\Up\D^\a(r)\D_\a(r)+\frac{8m^2}{r^4}\D^\a(r)\D_\a(r)\right)\aa \\
&& +\left(4 \left(\om+\frac{m}{r^2}\right)+2\left(\ka-\frac{2\Up}{r}\right)\right) e_3(\a)-4\omb e_4(\a)\\
&&-4\Up\left(\Up\left(\om +\frac{m}{r^2}\right) +\frac{m}{r^2}(e_4(r)-1) -\frac{e_4(m)}{r}\right)e_3(\aa)\\
&& +\left(4\Up^2 \omb +2\Up^2\left(\kab+\frac{2}{r}\right)-4m\Up \frac{(e_3(r)+1)}{r^2}+4\Up \frac{e_3(m)}{r}\right)e_4(\aa) \\
&&+\left(\frac{8\Up}{r^2}\D^\a(m)\D_\a(r)-\frac{4\Up}{r}\square_g(m)-\frac{8m}{r^3}\D^\a(r)\D_\a(m)+\frac{8m}{r}\D^\a(m)\D_\a\left(\frac{m}{r}\right)\right)\aa\\
&&+4\Up e_\th(\Up)e_\th(\aa) +\err[\square_\g \a]+\Up^2\err[\square_\g \aa].
\eeaa

Next, we have in view of the formula for $V$
\beaa
V&=& -4\rho   - 4 e_4(\omb)  -8\om\omb  +2\om\, \kab    -10 \ka\, \omb+\frac 1 2 \ka\,\kab\\
&=&   -\frac{2}{r^2}+\frac{16m}{r^3}     +\left(-4\rho -\frac{8m}{r^3}\right)    +2\left(\om\, \kab -\frac{2m}{r^3}\right)   +\frac 1 2 \left(\ka\,\kab +\frac{4\Up}{r^2}\right)\\
&&- 4 e_4(\omb)  -8\om\omb -10 \ka\, \omb.
\eeaa
Also, we have in view of the formula for $\Vb$
\beaa
\Vb&=& -4\rho   - 4 e_3(\om)  -8\om\omb  +2\omb\ka    -10 \kab\, \om+\frac 1 2 \ka\,\kab\\
&=& -\frac{2}{r^2}+\left(-4\rho -\frac{8m}{r^3}\right)  - 4 \left(e_3(\om)+\frac{2m}{r^3}\right) -10\left(\kab\, \om-\frac{2m}{r^3}\right)\\
&&+\frac 1 2\left(\ka\,\kab +\frac{4\Up}{r^2}\right) -8\om\omb  +2\ka\omb.    
\eeaa
Moreover, we have
\beaa
&&\frac{4m}{r^2}\Up\square_g(r) -\frac{8m}{r^3}\Up\D^\a(r)\D_\a(r)+\frac{8m^2}{r^4}\D^\a(r)\D_\a(r)\\
&=& \frac{4m\Up}{r^2}\left(\frac{2}{r}-\frac{2m}{r^2}\right)+\frac{4m}{r^2}\Up\left(\square_g(r)-\left(\frac{2}{r}-\frac{2m}{r^2}\right)\right)\\
&& -\frac{8m}{r^3}\left(1-\frac{3m}{r}\right)(-e_4(r)e_3(r)+(e_\th(r))^2)\\
&=& \frac{16m^2\Up}{r^4}+\frac{4m}{r^2}\Up\left(\square_g(r)-\left(\frac{2}{r}-\frac{2m}{r^2}\right)\right) -\frac{8m}{r^3}\left(1-\frac{3m}{r}\right)\Big(-e_4(r)e_3(r)-\Up+(e_\th(r))^2\Big)
\eeaa
and hence
\beaa
&&\Up^2\Vb +\frac{4m}{r^2}\Up\square_g(r) -\frac{8m}{r^3}\Up\D^\a(r)\D_\a(r)+\frac{8m^2}{r^4}\D^\a(r)\D_\a(r)\\
&=& -\frac{2\Up}{r^2}\left(1-\frac{2m}{r}-\frac{8m^2}{r^2}\right)\\
&&+\Up^2\left(\left(-4\rho -\frac{8m}{r^3}\right)  - 4 \left(e_3(\om)+\frac{2m}{r^3}\right) -10\left(\kab\, \om-\frac{2m}{r^3}\right)+\frac 1 2\left(\ka\,\kab +\frac{4\Up}{r^2}\right) -8\om\omb  +2\ka\omb\right)\\
&&+\frac{4m}{r^2}\Up\left(\square_g(r)-\left(\frac{2}{r}-\frac{2m}{r^2}\right)\right) -\frac{8m}{r^3}\left(1-\frac{3m}{r}\right)\Big(-e_4(r)e_3(r)-\Up+(e_\th(r))^2\Big).
\eeaa
We deduce
\beaa
\square_2(\a+\Up^2\aa) &=&  \frac{4}{r}\left(1- \frac{3m}{r}\right)  \Big(e_3(\a)-\Up e_4(\aa)\Big)  +\left(-\frac{2}{r^2}+\frac{16m}{r^3}\right)\a\\
&& -\frac{2\Up}{r^2}\left(1-\frac{2m}{r}-\frac{8m^2}{r^2}\right)\aa+\err\Big[\square_2(\a+\Up^2\aa)\Big]
 \eeaa
 where
\beaa
&& \err\Big[\square_2(\a+\Up^2\aa)\Big]\\ 
&=& \left(\Up^2\Vb +\frac{4m}{r^2}\Up\square_g(r) -\frac{8m}{r^3}\Up\D^\a(r)\D_\a(r)+\frac{8m^2}{r^4}\D^\a(r)\D_\a(r)\right)\aa \\
&& +\left(4 \left(\om+\frac{m}{r^2}\right)+2\left(\ka-\frac{2\Up}{r}\right)\right) e_3(\a)-4\omb e_4(\a)\\
&&-4\Up\left(\Up\left(\om +\frac{m}{r^2}\right) +\frac{m}{r^2}(e_4(r)-1) -\frac{e_4(m)}{r}\right)e_3(\aa)\\
&& +\left(4\Up^2 \omb +2\Up^2\left(\kab+\frac{2}{r}\right)-4m\Up \frac{(e_3(r)+1)}{r^2}+4\Up \frac{e_3(m)}{r}\right)e_4(\aa) \\
&& +\left(\left(-4\rho -\frac{8m}{r^3}\right)    +2\left(\om\, \kab -\frac{2m}{r^3}\right)   +\frac 1 2 \left(\ka\,\kab +\frac{4\Up}{r^2}\right)- 4 e_4(\omb)  -8\om\omb -10 \ka\, \omb\right)\a\\
&&+\left(\frac{8\Up}{r^2}\D^\a(m)\D_\a(r)-\frac{4\Up}{r}\square_g(m)-\frac{8m}{r^3}\D^\a(r)\D_\a(m)+\frac{8m}{r}\D^\a(m)\D_\a\left(\frac{m}{r}\right)\right)\aa\\
&&+\Up^2\left(\left(-4\rho -\frac{8m}{r^3}\right)  - 4 \left(e_3(\om)+\frac{2m}{r^3}\right) -10\left(\kab\, \om-\frac{2m}{r^3}\right)+\frac 1 2\left(\ka\,\kab +\frac{4\Up}{r^2}\right) -8\om\omb  +2\ka\omb\right)\aa\\
&&+\frac{4m}{r^2}\Up\left(\square_g(r)-\left(\frac{2}{r}-\frac{2m}{r^2}\right)\right)\aa -\frac{8m}{r^3}\left(1-\frac{3m}{r}\right)\Big(-e_4(r)e_3(r)-\Up+(e_\th(r))^2\Big)\aa\\
&&+4\Up e_\th(\Up)e_\th(\aa) +\err[\square_\g \a]+\Up^2\err\Big[\square_\g \aa\Big]
\eeaa
as desired. This concludes the proof of the lemma.
\end{proof}

\begin{lemma}\lab{lemma:usefullformulae3alphaande4alphabThmM8}
We have
\beaa
 e_3(\a) &=& -\frac{1}{2}\kab\a-\dds_2\ddd_1^{-1}\Bigg\{e_4\left(\frac{\rhot}{r^2}\right) +\frac{3}{2r^2}\ka\rhot -\frac{3m}{r^3}\left(\ka-\frac{2\Up}{r}\right) +  \frac{6m(e_4(r)-\Up)}{r^4} -  \frac{2e_4(m)}{r^3}\\
   &&+\frac{1}{2}\vthb\a-\ze\b-2(\etab\b+\xi\bb)\Bigg\} +4\omb\a-\frac{3}{2}\vth\rho+(\ze+4\eta)\b.
 \eeaa
and 
 \beaa
 e_4(\aa) &=& -\frac{1}{2}\ka\aa -\dds_2\ddd_1^{-1}\Bigg\{e_3\left(\frac{\rhot}{r^2}\right) +\frac{3}{2r^2}\kab\rhot   -\frac{3m}{r^3}\left(\kab+\frac{2}{r}\right)  +\frac{6m(e_3(r)+1)}{r^4}    -\frac{2e_3(m)}{r^3}\\
  &&+\frac{1}{2}\vth\aa+\ze\bb-2(\eta\bb+\xib\b) \Bigg\} +4\om\aa-\frac{3}{2}\vthb\rho+(-\ze+4\etab)\bb.
\eeaa
\end{lemma}

\begin{proof}
Recall that we have
\beaa
\square_2(\a+\Up^2\aa) &=&  \frac{4}{r}\left(1- \frac{3m}{r}\right)  \Big(e_3(\a)-\Up e_4(\aa)\Big)  +\left(-\frac{2}{r^2}+\frac{16m}{r^3}\right)\a\\
&& -\frac{2\Up}{r^2}\left(1-\frac{2m}{r}-\frac{8m^2}{r^2}\right)\aa+\err\Big[\square_2(\a+\Up^2\aa)\Big].
 \eeaa
 We first express $e_3(\a)-\Up e_4(\aa)$ in terms of $\rhot$, where we recall that $\rhot=r^2\rho+\frac{2m}{r}$. Using Bianchi, we have
 \beaa
 e_3(\a) &=& -\frac{1}{2}\kab\a-\dds_2\b+4\omb\a-\frac{3}{2}\vth\rho+(\ze+4\eta)\b,\\
 \ddd_1\b &=& e_4(\rho)+\frac{3}{2}\ka\rho+\frac{1}{2}\vthb\a-\ze\b-2(\etab\b+\xi\bb)\\
 &=& e_4\left(\frac{\rhot}{r^2}-\frac{2m}{r^3}\right)+\frac{3}{2}\ka\rho+\frac{1}{2}\vthb\a-\ze\b-2(\etab\b+\xi\bb)\\
   &=& e_4\left(\frac{\rhot}{r^2}\right) +\frac{3}{2r^2}\ka\rhot -\frac{3m}{r^3}\left(\ka-\frac{2\Up}{r}\right) +  \frac{6m(e_4(r)-\Up)}{r^4} -  \frac{2e_4(m)}{r^3}\\
   &&+\frac{1}{2}\vthb\a-\ze\b-2(\etab\b+\xi\bb)
 \eeaa
 and hence
 \beaa
 e_3(\a) &=& -\frac{1}{2}\kab\a-\dds_2\ddd_1^{-1}\Bigg\{e_4\left(\frac{\rhot}{r^2}\right) +\frac{3}{2r^2}\ka\rhot -\frac{3m}{r^3}\left(\ka-\frac{2\Up}{r}\right) +  \frac{6m(e_4(r)-\Up)}{r^4} -  \frac{2e_4(m)}{r^3}\\
   &&+\frac{1}{2}\vthb\a-\ze\b-2(\etab\b+\xi\bb)\Bigg\} +4\omb\a-\frac{3}{2}\vth\rho+(\ze+4\eta)\b.
 \eeaa
 
Similarly, we have 
 \beaa
 e_4(\aa) &=& -\frac{1}{2}\ka\aa -\dds_2\bb+4\om\aa-\frac{3}{2}\vthb\rho+(-\ze+4\etab)\bb,\\
 \ddd_1\bb &=& e_3(\rho)+\frac{3}{2}\kab\rho+\frac{1}{2}\vth\aa+\ze\bb-2(\eta\bb+\xib\b)\\
 &=&e_3\left(\frac{\rhot}{r^2}-\frac{2m}{r^3}\right)+\frac{3}{2}\kab\rho+\frac{1}{2}\vth\aa+\ze\bb-2(\eta\bb+\xib\b)\\
  &=&e_3\left(\frac{\rhot}{r^2}\right) +\frac{3}{2r^2}\kab\rhot   -\frac{3m}{r^3}\left(\kab+\frac{2}{r}\right)  +\frac{6m(e_3(r)+1)}{r^4}    -\frac{2e_3(m)}{r^3}\\
  &&+\frac{1}{2}\vth\aa+\ze\bb-2(\eta\bb+\xib\b)
 \eeaa
 and hence
 \beaa
 e_4(\aa) &=& -\frac{1}{2}\ka\aa -\dds_2\ddd_1^{-1}\Bigg\{e_3\left(\frac{\rhot}{r^2}\right) +\frac{3}{2r^2}\kab\rhot   -\frac{3m}{r^3}\left(\kab+\frac{2}{r}\right)  +\frac{6m(e_3(r)+1)}{r^4}    -\frac{2e_3(m)}{r^3}\\
  &&+\frac{1}{2}\vth\aa+\ze\bb-2(\eta\bb+\xib\b) \Bigg\} +4\om\aa-\frac{3}{2}\vthb\rho+(-\ze+4\etab)\bb.
\eeaa
This concludes the proof of the lemma.
\end{proof}

\begin{corollary}\lab{cor:secondwaveeqalphaplusUpsquarealphabThmM8}
We have
\beaa
&&\square_2(\a+\Up^2\aa) - \frac{2}{r^2}\left(1+\frac{2m}{r}\right)(\a+\Up^2\aa)\\
 &=&  -\frac{8}{r}\left(1- \frac{3m}{r}\right)  \dds_2\ddd_1^{-1}R\left(\frac{\rhot}{r^2}\right)  -\frac{6}{r}\left(1- \frac{3m}{r}\right)  (\vth -\Up \vthb)\rho \\
&&-\frac{4}{r}\left(1- \frac{3m}{r}\right)\dds_2\ddd_1^{-1}\left\{\frac{3}{2r^2}\ka\rhot -\frac{3m}{r^3}\left(\ka-\frac{2\Up}{r}\right) +  \frac{6m(e_4(r)-\Up)}{r^4}\right\} \\
   &&  +\frac{4\Up}{r}\left(1- \frac{3m}{r}\right)\dds_2\ddd_1^{-1}\left\{ \frac{3}{2r^2}\kab\rhot   -\frac{3m}{r^3}\left(\kab+\frac{2}{r}\right)  +\frac{6m(e_3(r)+1)}{r^4}     \right\}+\err_1,
   \eeaa
   where 
   \beaa
   \err_1 &:=& \frac{4}{r}\left(1- \frac{3m}{r}\right)\Bigg[4\omb\a+(\ze+4\eta)\b -\Up(-\ze+4\etab)\bb+[\Up,\dds_2\ddd_1^{-1}]e_3\left(\frac{\rhot}{r^2}\right)\\
    &&-\dds_2\ddd_1^{-1}\left\{ -  \frac{2e_4(m)}{r^3}+\frac{1}{2}\vthb\a-\ze\b-2(\etab\b+\xi\bb)\right\} \\  
   &&  +\Up\dds_2\ddd_1^{-1}\left\{-\frac{2e_3(m)}{r^3}+\frac{1}{2}\vth\aa+\ze\bb-2(\eta\bb+\xib\b) \right\}\Bigg]\\
 && -\frac{2}{r}\left(1- \frac{3m}{r}\right)  \left(\kab+\frac{2}{r}\right)\a  +\frac{4\Up}{r}\left(1- \frac{3m}{r}\right)\left(\frac{1}{2}\left(\ka-\frac{2\Up}{r}\right) -4\left(\om+\frac{m}{r^2}\right)\right)\aa  \\
&&+\err\Big[\square_2(\a+\Up^2\aa)\Big].
 \eeaa
\end{corollary}

\begin{proof}
Recall from Lemma \ref{lemma:firstwaveeqalphaplusUpsquarealphabThmM8} that we have 
\beaa
\square_2(\a+\Up^2\aa) &=&  \frac{4}{r}\left(1- \frac{3m}{r}\right)  \Big(e_3(\a)-\Up e_4(\aa)\Big)  +\left(-\frac{2}{r^2}+\frac{16m}{r^3}\right)\a\\
&& -\frac{2\Up}{r^2}\left(1-\frac{2m}{r}-\frac{8m^2}{r^2}\right)\aa+\err\Big[\square_2(\a+\Up^2\aa)\Big].
 \eeaa 
In view of Lemma  \ref{lemma:usefullformulae3alphaande4alphabThmM8}, we have
\beaa
e_3(\a)-\Up e_4(\aa) &=& -2\dds_2\ddd_1^{-1}R\left(\frac{\rhot}{r^2}\right) -\frac{1}{2}\kab\a +\Up\left(\frac{1}{2}\ka -4\om\right)\aa -\frac{3}{2}(\vth -\Up \vthb)\rho \\
&&-\dds_2\ddd_1^{-1}\left\{\frac{3}{2r^2}\ka\rhot -\frac{3m}{r^3}\left(\ka-\frac{2\Up}{r}\right) +  \frac{6m(e_4(r)-\Up)}{r^4}\right\} \\
   &&  +\Up\dds_2\ddd_1^{-1}\left\{ \frac{3}{2r^2}\kab\rhot   -\frac{3m}{r^3}\left(\kab+\frac{2}{r}\right)  +\frac{6m(e_3(r)+1)}{r^4}     \right\}\\
   &&+4\omb\a+(\ze+4\eta)\b -\Up(-\ze+4\etab)\bb+[\Up,\dds_2\ddd_1^{-1}]e_3\left(\frac{\rhot}{r^2}\right)\\
    &&-\dds_2\ddd_1^{-1}\left\{ -  \frac{2e_4(m)}{r^3}+\frac{1}{2}\vthb\a-\ze\b-2(\etab\b+\xi\bb)\right\} \\  
   &&  +\Up\dds_2\ddd_1^{-1}\left\{-\frac{2e_3(m)}{r^3}+\frac{1}{2}\vth\aa+\ze\bb-2(\eta\bb+\xib\b) \right\}.
\eeaa
We infer
\beaa
\square_2(\a+\Up^2\aa) &=&  -\frac{8}{r}\left(1- \frac{3m}{r}\right)  \dds_2\ddd_1^{-1}R\left(\frac{\rhot}{r^2}\right) -\frac{2}{r}\left(1- \frac{3m}{r}\right)  \kab\a +\left(-\frac{2}{r^2}+\frac{16m}{r^3}\right)\a\\
&&+\frac{4\Up}{r}\left(1- \frac{3m}{r}\right)\left(\frac{1}{2}\ka -4\om\right)\aa -\frac{2\Up}{r^2}\left(1-\frac{2m}{r}-\frac{8m^2}{r^2}\right)\aa \\
&&-\frac{6}{r}\left(1- \frac{3m}{r}\right)  (\vth -\Up \vthb)\rho \\
&&-\frac{4}{r}\left(1- \frac{3m}{r}\right)\dds_2\ddd_1^{-1}\left\{\frac{3}{2r^2}\ka\rhot -\frac{3m}{r^3}\left(\ka-\frac{2\Up}{r}\right) +  \frac{6m(e_4(r)-\Up)}{r^4}\right\} \\
   &&  +\frac{4\Up}{r}\left(1- \frac{3m}{r}\right)\dds_2\ddd_1^{-1}\left\{ \frac{3}{2r^2}\kab\rhot   -\frac{3m}{r^3}\left(\kab+\frac{2}{r}\right)  +\frac{6m(e_3(r)+1)}{r^4}     \right\}\\
   &&+\frac{4}{r}\left(1- \frac{3m}{r}\right)\Bigg[4\omb\a+(\ze+4\eta)\b -\Up(-\ze+4\etab)\bb+[\Up,\dds_2\ddd_1^{-1}]e_3\left(\frac{\rhot}{r^2}\right)\\
    &&-\dds_2\ddd_1^{-1}\left\{ -  \frac{2e_4(m)}{r^3}+\frac{1}{2}\vthb\a-\ze\b-2(\etab\b+\xi\bb)\right\} \\  
   &&  +\Up\dds_2\ddd_1^{-1}\left\{-\frac{2e_3(m)}{r^3}+\frac{1}{2}\vth\aa+\ze\bb-2(\eta\bb+\xib\b) \right\}\Bigg]\\
&&+\err\Big[\square_2(\a+\Up^2\aa)\Big].
 \eeaa
Since we have
\beaa
&&-\frac{2}{r}\left(1- \frac{3m}{r}\right)  \kab\a +\frac{4\Up}{r}\left(1- \frac{3m}{r}\right)\left(\frac{1}{2}\ka -4\om\right)\aa \\
&=& \frac{4}{r^2}\left(1- \frac{3m}{r}\right)\a +\frac{4\Up}{r^2}\left(1- \frac{3m}{r}\right)\left(1 +\frac{2m}{r}\right)\aa -\frac{2}{r}\left(1- \frac{3m}{r}\right)  \left(\kab+\frac{2}{r}\right)\a\\
&&  +\frac{4\Up}{r}\left(1- \frac{3m}{r}\right)\left(\frac{1}{2}\left(\ka-\frac{2\Up}{r}\right) -4\left(\om+\frac{m}{r^2}\right)\right)\aa, 
\eeaa 
this yields
\beaa
\square_2(\a+\Up^2\aa) &=&  -\frac{8}{r}\left(1- \frac{3m}{r}\right)  \dds_2\ddd_1^{-1}R\left(\frac{\rhot}{r^2}\right) + \frac{2}{r^2}\left(1+\frac{2m}{r}\right)\a+\frac{2\Up}{r^2}\left(1-\frac{4m^2}{r^2}\right)\aa \\
&&-\frac{6}{r}\left(1- \frac{3m}{r}\right)  (\vth -\Up \vthb)\rho \\
&&-\frac{4}{r}\left(1- \frac{3m}{r}\right)\dds_2\ddd_1^{-1}\left\{\frac{3}{2r^2}\ka\rhot -\frac{3m}{r^3}\left(\ka-\frac{2\Up}{r}\right) +  \frac{6m(e_4(r)-\Up)}{r^4}\right\} \\
   &&  +\frac{4\Up}{r}\left(1- \frac{3m}{r}\right)\dds_2\ddd_1^{-1}\left\{ \frac{3}{2r^2}\kab\rhot   -\frac{3m}{r^3}\left(\kab+\frac{2}{r}\right)  +\frac{6m(e_3(r)+1)}{r^4}     \right\}+\err_1,
   \eeaa
   where
   \beaa
   \err_1 &=& \frac{4}{r}\left(1- \frac{3m}{r}\right)\Bigg[4\omb\a+(\ze+4\eta)\b -\Up(-\ze+4\etab)\bb+[\Up,\dds_2\ddd_1^{-1}]e_3\left(\frac{\rhot}{r^2}\right)\\
    &&-\dds_2\ddd_1^{-1}\left\{ -  \frac{2e_4(m)}{r^3}+\frac{1}{2}\vthb\a-\ze\b-2(\etab\b+\xi\bb)\right\} \\  
   &&  +\Up\dds_2\ddd_1^{-1}\left\{-\frac{2e_3(m)}{r^3}+\frac{1}{2}\vth\aa+\ze\bb-2(\eta\bb+\xib\b) \right\}\Bigg]\\
 && -\frac{2}{r}\left(1- \frac{3m}{r}\right)  \left(\kab+\frac{2}{r}\right)\a  +\frac{4\Up}{r}\left(1- \frac{3m}{r}\right)\left(\frac{1}{2}\left(\ka-\frac{2\Up}{r}\right) -4\left(\om+\frac{m}{r^2}\right)\right)\aa  \\
&&+\err\Big[\square_2(\a+\Up^2\aa)\Big].
 \eeaa
 Now, since we have
 \beaa
 \frac{2}{r^2}\left(1+\frac{2m}{r}\right)\a+\frac{2\Up}{r^2}\left(1-\frac{4m^2}{r^2}\right)\aa &=& \frac{2}{r^2}\left(1+\frac{2m}{r}\right)(\a+\Up^2\aa),
 \eeaa
we infer
\beaa
&&\square_2(\a+\Up^2\aa) - \frac{2}{r^2}\left(1+\frac{2m}{r}\right)(\a+\Up^2\aa)\\
 &=&  -\frac{8}{r}\left(1- \frac{3m}{r}\right)  \dds_2\ddd_1^{-1}R\left(\frac{\rhot}{r^2}\right)  -\frac{6}{r}\left(1- \frac{3m}{r}\right)  (\vth -\Up \vthb)\rho \\
&&-\frac{4}{r}\left(1- \frac{3m}{r}\right)\dds_2\ddd_1^{-1}\left\{\frac{3}{2r^2}\ka\rhot -\frac{3m}{r^3}\left(\ka-\frac{2\Up}{r}\right) +  \frac{6m(e_4(r)-\Up)}{r^4}\right\} \\
   &&  +\frac{4\Up}{r}\left(1- \frac{3m}{r}\right)\dds_2\ddd_1^{-1}\left\{ \frac{3}{2r^2}\kab\rhot   -\frac{3m}{r^3}\left(\kab+\frac{2}{r}\right)  +\frac{6m(e_3(r)+1)}{r^4}     \right\}+\err_1,
   \eeaa
as desired. This concludes the proof of the corollary.
\end{proof}


\subsection{End of the proof of Proposition \ref{prop:controlofalphaplusUpalphabiterationassupmtionThM8}}


In view of Corollary \ref{cor:secondwaveeqalphaplusUpsquarealphabThmM8}, $\a+\Up^2\aa$ satisfies
\beaa
(\square_2+V_2)(\a+\Up^2\aa) &=& N_2, \qquad V_2= - \frac{2}{r^2}\left(1+\frac{2m}{r}\right),
\eeaa
where
\beaa
N_2 &:=&  -\frac{8}{r}\left(1- \frac{3m}{r}\right)  \dds_2\ddd_1^{-1}R\left(\frac{\rhot}{r^2}\right)  -\frac{6}{r}\left(1- \frac{3m}{r}\right)  (\vth -\Up \vthb)\rho \\
&&-\frac{4}{r}\left(1- \frac{3m}{r}\right)\dds_2\ddd_1^{-1}\left\{\frac{3}{2r^2}\ka\rhot -\frac{3m}{r^3}\left(\ka-\frac{2\Up}{r}\right) +  \frac{6m(e_4(r)-\Up)}{r^4}\right\} \\
   &&  +\frac{4\Up}{r}\left(1- \frac{3m}{r}\right)\dds_2\ddd_1^{-1}\left\{ \frac{3}{2r^2}\kab\rhot   -\frac{3m}{r^3}\left(\kab+\frac{2}{r}\right)  +\frac{6m(e_3(r)+1)}{r^4}     \right\}+\err_1.
   \eeaa
We may thus apply the estimate \eqref{eq:thefirstequationoftheorem-recoveringhigherderivatives} of Theorem \ref{theorem-recoveringhigherderivatives} with $\psi=\a+\Up^2\aa$ and $s=J$ to obtain for   any $k_{small}\leq J\leq k_{large}-1$ 
  \beaa
\nn  &&\sup_{\tau\in[1,\tau_*] }   E^J_{\de} [\a+\Up^2\aa](\tau)+   B^J_{\de}[\a+\Up^2\aa](1,\tau_*)  + F^J_{\de}[\a+\Up^2\aa](1,\tau_*)\\
  \nn &\les& 
          E^J_{\de}[\a+\Up^2\aa](1) +\sup_{\tau\in[1,\tau_*] }   E^{J-1}_{\de} [\a+\Up^2\aa](\tau)+   B^{J-1}_{\de}[\a+\Up^2\aa](1,\tau_*)  \\
\nn&&+ F^{J-1}_{\de}[\a+\Up^2\aa](1,\tau_*)+D_J[\Ga] \left(\sup_{\MM}ru_{trap}^{\frac{1}{2}+\dec}|\dk^{\leq k_{small}}(\a+\Up^2\aa)|\right)^2\\
&&+\int_{\MM}r^{1+\de}|\dk^{\leq J}N_2|^2+\left|\int_{\Mtrap}T(\dk^J(\a+\Up^2\aa))\dk^JN_2\right|,
\eeaa
where $D_J[\Ga]$ is defined by
\beaa
D_J[\Ga] &:=& \int_{\Mint\cup\Mext(r\leq 4m_0)}(\dk^{\leq J}\Gac)^2\\
&& +\sup_{r_0\geq 4m_0}\left(r_0\int_{\{r=r_0\}}|\dk^{\leq J}\Ga_g|^2+r_0^{-1}\int_{\{r=r_0\}}|\dk^{\leq J}\Ga_b|^2\right).
\eeaa

Next we use the iteration assumption \eqref{eq:iterationassumptiondiscussionThM8:bis} which yields in particular 
\beaa
D_J[\Ga]  &\les& (\ep_\BB[J])^2
\eeaa
and
\beaa
\sup_{\tau\in[1,\tau_*] }   E^{J-1}_{\de} [\a+\Up^2\aa](\tau)+   B^{J-1}_{\de}[\a+\Up^2\aa](1,\tau_*) + F^{J-1}_{\de}[\a+\Up^2\aa](1,\tau_*)\les (\ep_\BB[J])^2.
\eeaa
Together with the control of $\dk^{\leq k_{small}}(\a+\Up^2\aa)$ provided by the decay estimate \eqref{eq:mainconclusionofThm7forproofThM8}, we infer from the above estimates 
  \beaa
\nn  &&\sup_{\tau\in[1,\tau_*] }   E^J_{\de} [\a+\Up^2\aa](\tau)+   B^J_{\de}[\a+\Up^2\aa](1,\tau_*)  + F^J_{\de}[\a+\Up^2\aa](1,\tau_*)\\
  \nn &\les& E^J_{\de}[\a+\Up^2\aa](1)+(\ep_\BB[J])^2+\int_{\MM}r^{1+\de}|\dk^{\leq J}N_2|^2+\left|\int_{\Mtrap}T(\dk^J(\a+\Up^2\aa))\dk^JN_2\right|.
\eeaa

Next, using the form of $N_2$, as well as the control of $\rhot$ provided by Proposition \ref{prop:controlofrhotildeiterationassupmtionThM8}, we derive
\beaa
\int_{\MM}r^{1+\de}|\dk^{\leq J}N_2|^2 &\les& (\ep_\BB[J])^2+\ep_0^2\Big(\Nk^{(En)}_{J+1}+\NN^{(match)}_{J+1}\Big)^2
\eeaa
and 
\beaa
&&\left|\int_{\Mtrap}T(\dk^J(\a+\Up^2\aa))\dk^JN_2\right| \\
&\les& \int_{\Mtrap}\left|1-\frac{3m}{r}\right||T(\dk^J(\a+\Up^2\aa))|\Big(|R(\dk^J\rhot)|+|\dk^J\rhot|+|\dk^J\Gac|\Big)\\
&& +\int_{\Mtrap}|T(\dk^J(\a+\Up^2\aa))||\err_1|\\
&\les& \left(\ep_\BB[J]+\ep_0\Big(\Nk^{(En)}_{J+1}+\NN^{(match)}_{J+1}\Big)\right)\left(\sup_{\tau\in[1,\tau_*] }   E^J_{\de} [\rhot](\tau)+B^J_{\de}[\rhot](1,\tau_*)\right)^{\frac{1}{2}}.
\eeaa
In view of the above, we infer
  \beaa
 &&\sup_{\tau\in[1,\tau_*] }   E^J_{\de} [\a+\Up^2\aa](\tau)+   B^J_{\de}[\a+\Up^2\aa)](1,\tau_*)  + F^J_{\de}[\a+\Up^2\aa)](1,\tau_*)\\
  &\les&  (\ep_\BB[J])^2+\ep_0^2\Big(\Nk^{(En)}_{J+1}+\NN^{(match)}_{J+1}\Big)^2
  \eeaa
as desired. This concludes the proof of Proposition \ref{prop:controlofalphaplusUpalphabiterationassupmtionThM8}.


\section{Proof of Proposition \ref{prop:controlofMorrallcurvcompiterationassupmtionThM8}}\lab{sec:proofprop:controlofMorrallcurvcompiterationassupmtionThM8}



\subsection{Control of $\a$ and $\Up^2\aa$}


We initiate the proof of Proposition \ref{prop:controlofMorrallcurvcompiterationassupmtionThM8} by deriving a suitable control for $\a$ and $\Up^2\aa$. 
Recall from Lemma \ref{lemma:usefullformulae3alphaande4alphabThmM8} that we have
 \beaa
 e_4(\aa) &=& -\frac{1}{2}\ka\aa -\dds_2\ddd_1^{-1}\Bigg\{e_3\left(\frac{\rhot}{r^2}\right) +\frac{3}{2r^2}\kab\rhot   -\frac{3m}{r^3}\left(\kab+\frac{2}{r}\right)  +\frac{6m(e_3(r)+1)}{r^4}    -\frac{2e_3(m)}{r^3}\\
  &&+\frac{1}{2}\vth\aa+\ze\bb-2(\eta\bb+\xib\b) \Bigg\} +4\om\aa-\frac{3}{2}\vthb\rho+(-\ze+4\etab)\bb.
\eeaa
We infer
\beaa
&&e_4(\a-\Up^2\aa) \\
&=& e_4(\a+\Up^2\aa) -2e_4(\Up^2\aa)\\
&=& e_4(\a+\Up^2\aa) -2\Up^2e_4(\aa) -2e_4(\Up^2)\aa\\
&=& e_4(\a+\Up^2\aa)  +2\Up^2\dds_2\ddd_1^{-1}\Bigg\{e_3\left(\frac{\rhot}{r^2}\right) +\frac{3}{2r^2}\kab\rhot   -\frac{3m}{r^3}\left(\kab+\frac{2}{r}\right)  +\frac{6m(e_3(r)+1)}{r^4}    -\frac{2e_3(m)}{r^3}\\
  &&+\frac{1}{2}\vth\aa+\ze\bb-2(\eta\bb+\xib\b) \Bigg\} +\Up^2\ka\aa -8\Up^2\om\aa +3\Up^2\vthb\rho-2\Up^2(-\ze+4\etab)\bb \\
  &&-\frac{8m\Up e_4(r)}{r^2}\aa  +\frac{8\Up e_4(m)}{r}\aa.
\eeaa
Also, recall from Lemma \ref{lemma:usefullformulae3alphaande4alphabThmM8} that we have
\beaa
 e_3(\a) &=& -\frac{1}{2}\kab\a-\dds_2\ddd_1^{-1}\Bigg\{e_4\left(\frac{\rhot}{r^2}\right) +\frac{3}{2r^2}\ka\rhot -\frac{3m}{r^3}\left(\ka-\frac{2\Up}{r}\right) +  \frac{6m(e_4(r)-\Up)}{r^4} -  \frac{2e_4(m)}{r^3}\\
   &&+\frac{1}{2}\vthb\a-\ze\b-2(\etab\b+\xi\bb)\Bigg\} +4\omb\a-\frac{3}{2}\vth\rho+(\ze+4\eta)\b.
 \eeaa
We infer
\beaa
&&e_3(\a-\Up^2\aa) \\
&=& -e_3(\a+\Up^2\aa) +2e_3(\a)\\
&=&  -e_3(\a+\Up^2\aa) -2\dds_2\ddd_1^{-1}\Bigg\{e_4\left(\frac{\rhot}{r^2}\right) +\frac{3}{2r^2}\ka\rhot -\frac{3m}{r^3}\left(\ka-\frac{2\Up}{r}\right) +  \frac{6m(e_4(r)-\Up)}{r^4} -  \frac{2e_4(m)}{r^3}\\
   &&+\frac{1}{2}\vthb\a-\ze\b-2(\etab\b+\xi\bb)\Bigg\} -\kab\a+8\omb\a-3\vth\rho+2(\ze+4\eta)\b.
\eeaa

In view of the above identities for $e_4(\a-\Up^2\aa)$ and $e_3(\a-\Up^2\aa)$, and using the control for $\rhot$ provided by Proposition \ref{prop:controlofrhotildeiterationassupmtionThM8} as well as the control for $\a+\Up^2\aa$ provided by Proposition \ref{prop:controlofalphaplusUpalphabiterationassupmtionThM8}, and the iteration assumption \eqref{eq:iterationassumptiondiscussionThM8:bis}, we obtain
  \beaa
B^{J-1}_{\de}[e_3(\a-\Up^2\aa)](1,\tau_*)  + B^{J-1}_{\de}[re_4(\a-\Up^2\aa)](1,\tau_*) &\les&  (\ep_\BB[J])^2+\ep_0^2\Big(\Nk^{(En)}_{J+1}+\NN^{(match)}_{J+1}\Big)^2.
  \eeaa

Also, using the Bianchi identity for $\ddd_2\a$ and $\ddd_1\b$, we have
\beaa
\ddd_1\ddd_2\a &=& \ddd_1\Big(e_4\b+2(\ka+\om)\b -(2\ze+\eta)\a-3\xi\rho)\Big)\\
&=& e_4(\ddd_1\b)+[\ddd_1, e_4]\b+\ddd_1\Big(2(\ka+\om)\b -(2\ze+\eta)\a-3\xi\rho)\Big)\\
&=& e_4\left(e_4\rho+\frac{3}{2}\rho+\frac{1}{2}\vthb\a-\ze\b-2(\eta\b+\xi\bb)\right)+[\ddd_1, e_4]\b\\
&&+\ddd_1\Big(2(\ka+\om)\b -(2\ze+\eta)\a-3\xi\rho)\Big)\\
&=& e_4\Bigg[e_4\left(\frac{\rhot}{r^2}\right) +\frac{3}{2r^2}\ka\rhot -\frac{3m}{r^3}\left(\ka-\frac{2\Up}{r}\right) +  \frac{6m(e_4(r)-\Up)}{r^4} -  \frac{2e_4(m)}{r^3}\\
   &&+\frac{1}{2}\vthb\a-\ze\b-2(\etab\b+\xi\bb)\Bigg]+[\ddd_1, e_4]\b+\ddd_1\Big(2(\ka+\om)\b -(2\ze+\eta)\a-3\xi\rho)\Big).
\eeaa
Using the control for $\rhot$ provided by Proposition \ref{prop:controlofrhotildeiterationassupmtionThM8} as well as  the iteration assumption \eqref{eq:iterationassumptiondiscussionThM8:bis}, we obtain
  \beaa
B^{J-2}_{\de}[r^2\ddd_1\ddd_2\a](1,\tau_*)   &\les&  (\ep_\BB[J])^2+\ep_0^2\Big(\Nk^{(En)}_{J+1}+\NN^{(match)}_{J+1}\Big)^2.
  \eeaa
Using  the control for $\a+\Up^2\aa$ provided by Proposition \ref{prop:controlofalphaplusUpalphabiterationassupmtionThM8}, we infer
  \beaa
B^{J-2}_{\de}[r^2\ddd_1\ddd_2(\a-\Up^2\aa)](1,\tau_*) &\les& B^{J-2}_{\de}[r^2\ddd_1\ddd_2\a](1,\tau_*)+B^{J-2}_{\de}[r^2\ddd_1\ddd_2(\a+\Up^2\aa)](1,\tau_*)\\
  &\les&  (\ep_\BB[J])^2+\ep_0^2\Big(\Nk^{(En)}_{J+1}+\NN^{(match)}_{J+1}\Big)^2.
  \eeaa
Using a Poincar\'e  inequality for $\ddd_1$ and for $\ddd_2$, we deduce
  \beaa
B^{J-2}_{\de}[\dkb^2(\a-\Up^2\aa)](1,\tau_*)   &\les&  (\ep_\BB[J])^2+\ep_0^2\Big(\Nk^{(En)}_{J+1}+\NN^{(match)}_{J+1}\Big)^2.
  \eeaa
Together with the above estimate for $e_3(\a-\Up^2\aa)$ and $re_4(\a-\Up^2\aa)$, we deduce
  \beaa
B^J_{\de}[\a-\Up^2\aa](1,\tau_*)   &\les&  (\ep_\BB[J])^2+\ep_0^2\Big(\Nk^{(En)}_{J+1}+\NN^{(match)}_{J+1}\Big)^2.
  \eeaa
Together with the control for $\a+\Up^2\aa$ provided by Proposition \ref{prop:controlofalphaplusUpalphabiterationassupmtionThM8}, we finally obtain
  \bea\lab{eq:controlforalphaandforUpquarealphabarfortheproofofThmM8}
B^J_{\de}[\a](1,\tau_*) + B^J_{\de}[\Up^2\aa](1,\tau_*)   &\les&  (\ep_\BB[J])^2+\ep_0^2\Big(\Nk^{(En)}_{J+1}+\NN^{(match)}_{J+1}\Big)^2.
  \eea


\subsection{Control of $\aa$}


\eqref{eq:controlforalphaandforUpquarealphabarfortheproofofThmM8} provides in particular the control of $\Up^2\aa$. In this section, we infer a suitable control for $\aa$ using the wave equation satisfied by $\aa$ and the redshift vectorfield. 
     
  Let   $Y_{(0)}$ the vectorfield given by  
 \beaa
 Y_{(0)}:= \left(1+\frac{5}{4m}(r-2m)+\Up\right) e_3 +\left(1+\frac{5}{4m}(r-2m)\right)e_4,
 \eeaa 
 where $Y_{(0)}$ has been introduced in Proposition  \ref{prop:red-shift-Mor}  in connection with the redshift vectorfield.
 
\begin{lemma}\lab{lemma:waveeqalphabinMintrelatedtoredshiftforproofThmM8}
We have
\beaa
 \square_2\aa &=& \frac{4m}{r^2\left(1+\frac{5}{4m}(r-2m)+\Up\right)}Y_{(0)}\aa +\widetilde{N}_2
 \eeaa
 where $\widetilde{N}_2$ is given by
 \beaa
 \widetilde{N}_2 &:=&  - \frac{4}{r}\left(1+\frac{m\left(1+\frac{5}{4m}(r-2m)\right)}{r\left(1+\frac{5}{4m}(r-2m)+\Up\right)}\right)\Bigg[-\frac{1}{2}\ka\aa \\
 &&-\dds_2\ddd_1^{-1}\Bigg\{e_3\left(\frac{\rhot}{r^2}\right) +\frac{3}{2r^2}\kab\rhot   -\frac{3m}{r^3}\left(\kab+\frac{2}{r}\right)  +\frac{6m(e_3(r)+1)}{r^4}    -\frac{2e_3(m)}{r^3}\\
  &&+\frac{1}{2}\vth\aa+\ze\bb-2(\eta\bb+\xib\b) \Bigg\} +4\om\aa-\frac{3}{2}\vthb\rho+(-\ze+4\etab)\bb\Bigg] \\
 &&+\Vb \aa -4\left(\om+\frac{m}{r^2}\right) e_3(\aa) + \left(4 \omb+2\left(\kab+\frac{2}{r}\right)\right) e_4(\aa) +\err[\square_\g \aa].
\eeaa
\end{lemma}

\begin{proof}
Recall from Proposition \ref{proposition:wave-a-aa-qf} that $\aa$ verifies the following Teukolsky equation
\beaa
\bsplit
 \square_2\aa&= -4\om e_3(\aa)+ (4 \omb+2\kab) e_4(\aa) +\Vb \aa+\err[\square_\g \aa],\\
\Vb&=-4\rho   - 4 e_3(\om)  -8\om\omb  +2\omb\ka    -10 \kab\, \om+\frac 1 2 \ka\,\kab,
\end{split}
\eeaa
where 
\beaa
\err(\square_\g\aa) &=& \frac{1}{2}\vthb e_4(\aa) + \frac 3 4 \vthb^2\rho+ e_\th(\Phi)\vthb\, \bb  -\frac{1}{2}\ka(-\ze+4\etab)\bb  -(-\ze+\eta)e_3(\bb) - \xib e_4(\bb)\\
&&+ e_\th(\Phi)(-2\ze+\eta) \aa + \bb^2 + e_3(\Phi)\eta\bb+ e_4(\Phi)\xib\,\bb - (-\ze+4\etab) e_3(\bb) \\
&&-(-e_3(\ze)+4e_3(\etab))\bb - 2(\kab+\omb)(-\ze+4\etab)\bb +2e_\th(\kab+\omb) \bb  -e_\th((-2\ze+\eta)\aa)\\
&& -3\xib e_\th(\rho)+2\eta e_\th(\aa) + \frac 3 2 \vthb  \ddd_1 \bb  + 3\rho (\eta+\etab-2\ze)\xib    +  \ddd_1\eta \aa +\frac{1}{4}\ka\vthb\,\aa -2\om\vthb\,\aa\\
&&  -  \frac{1}{2}\vthb\vth\aa +\xi\xib\aa+\eta^2\aa  -\frac 3 2 \vthb\ze\bb+3\vthb(\eta\bb+\xib\b) -\frac{1}{2}\vthb(-\ze+4\etab)\bb.
\eeaa
We deduce 
\beaa
 \square_2\aa &=& \frac{4m}{r^2} e_3(\aa) -\frac{4}{r}e_4(\aa) +\Vb \aa -4\left(\om+\frac{m}{r^2}\right) e_3(\aa) + \left(4 \omb+2\left(\kab+\frac{2}{r}\right)\right) e_4(\aa) \\
 &&+\err[\square_\g \aa].
\eeaa
In view of the definition of  $Y_{(0)}$, we infer
\beaa
 \square_2\aa &=& \frac{4m}{r^2\left(1+\frac{5}{4m}(r-2m)+\Up\right)}Y_{(0)}\aa  - \frac{4}{r}\left(1+\frac{m\left(1+\frac{5}{4m}(r-2m)\right)}{r\left(1+\frac{5}{4m}(r-2m)+\Up\right)}\right)e_4(\aa) \\
 &&+\Vb \aa -4\left(\om+\frac{m}{r^2}\right) e_3(\aa) + \left(4 \omb+2\left(\kab+\frac{2}{r}\right)\right) e_4(\aa) +\err[\square_\g \aa].
\eeaa

Next, recall from Lemma \ref{lemma:usefullformulae3alphaande4alphabThmM8} that we have
 \beaa
 e_4(\aa) &=& -\frac{1}{2}\ka\aa -\dds_2\ddd_1^{-1}\Bigg\{e_3\left(\frac{\rhot}{r^2}\right) +\frac{3}{2r^2}\kab\rhot   -\frac{3m}{r^3}\left(\kab+\frac{2}{r}\right)  +\frac{6m(e_3(r)+1)}{r^4}    -\frac{2e_3(m)}{r^3}\\
  &&+\frac{1}{2}\vth\aa+\ze\bb-2(\eta\bb+\xib\b) \Bigg\} +4\om\aa-\frac{3}{2}\vthb\rho+(-\ze+4\etab)\bb.
\eeaa
We infer
\beaa
 \square_2\aa &=& \frac{4m}{r^2\left(1+\frac{5}{4m}(r-2m)+\Up\right)}Y_{(0)}\aa +\widetilde{N}_2
 \eeaa
 where $\widetilde{N}_2$ is given by
 \beaa
 \widetilde{N}_2 &=&  - \frac{4}{r}\left(1+\frac{m\left(1+\frac{5}{4m}(r-2m)\right)}{r\left(1+\frac{5}{4m}(r-2m)+\Up\right)}\right)\Bigg[-\frac{1}{2}\ka\aa \\
 &&-\dds_2\ddd_1^{-1}\Bigg\{e_3\left(\frac{\rhot}{r^2}\right) +\frac{3}{2r^2}\kab\rhot   -\frac{3m}{r^3}\left(\kab+\frac{2}{r}\right)  +\frac{6m(e_3(r)+1)}{r^4}    -\frac{2e_3(m)}{r^3}\\
  &&+\frac{1}{2}\vth\aa+\ze\bb-2(\eta\bb+\xib\b) \Bigg\} +4\om\aa-\frac{3}{2}\vthb\rho+(-\ze+4\etab)\bb\Bigg] \\
 &&+\Vb \aa -4\left(\om+\frac{m}{r^2}\right) e_3(\aa) + \left(4 \omb+2\left(\kab+\frac{2}{r}\right)\right) e_4(\aa) +\err[\square_\g \aa].
\eeaa
This concludes the proof of the lemma.
\end{proof}

\begin{lemma}\lab{lemma:controlN2widetildewhichisRHSwaveeqaaredshiftforproofThmM8}
$\widetilde{N}_2$, in the RHS of the wave equation for $\aa$ introduced in Lemma \ref{lemma:waveeqalphabinMintrelatedtoredshiftforproofThmM8}, satisfies
\beaa
\int_{\Mint}|\dk^J\widetilde{N}_2|^2 &\les&  (\ep_\BB[J])^2+\ep_0^2\Big(\Nk^{(En)}_{J+1}+\NN^{(match)}_{J+1}\Big)^2.
\eeaa
\end{lemma}

\begin{proof}
The proof of the lemma follows immediately from the form of $\widetilde{N}_2$, see Lemma \ref{lemma:waveeqalphabinMintrelatedtoredshiftforproofThmM8}, as well as the  control for $\rhot$ provided by Proposition \ref{prop:controlofrhotildeiterationassupmtionThM8}, \eqref{eq:mainconclusionofThm7forproofThM8}, and the iteration assumption \eqref{eq:iterationassumptiondiscussionThM8:bis}.
\end{proof}

In view of Lemma \ref{lemma:waveeqalphabinMintrelatedtoredshiftforproofThmM8}, we may apply Proposition \ref{prop:controlonMintbyMextthankstoredshiftneededinThmM8} with 
\beaa
\psi=\aa, \qquad f_2(r,m)=\frac{4m}{r^2\left(1+\frac{5}{4m}(r-2m)+\Up\right)}.
\eeaa
We infer
 \beaa
 \int_{\Mint(1,\tau_*)}(\dk^{J+1}\aa)^2 &\les& E^J_{\de}[\aa](\tau=1)+ \int_{\Mext_{r\leq \frac{5}{2}m_0}(1,\tau_*)}(\dk^{J+1}\aa)^2\\
 && +D_J[\Ga] \left(\sup_{\Mint(1,\tau_*)\cup\Mext_{r\leq \frac{5}{2}m_0}}r|\dk^{\leq k_{small}}\aa|\right)^2\\
 && +\int_{\Mint(1,\tau_*)\cup\Mext_{r\leq \frac{5}{2}m_0}}\Big((\dk^{\leq s}\aa)^2+(\dk^{\leq J+1}\widetilde{N}_2)^2\Big).
 \eeaa

Next we use the iteration assumption \eqref{eq:iterationassumptiondiscussionThM8:bis} which yields in particular 
\beaa
D_J[\Ga]  &\les& (\ep_\BB[J])^2
\eeaa
together with the control of $\dk^{\leq k_{small}}\aa$ provided by the decay estimate \eqref{eq:mainconclusionofThm7forproofThM8}, as well as the iteration assumption and the control for $\widetilde{N}_2$ provided by Lemma \ref{lemma:controlN2widetildewhichisRHSwaveeqaaredshiftforproofThmM8} to deduce
 \beaa
 \int_{\Mint(1,\tau_*)}(\dk^{J+1}\aa)^2 &\les&  (\ep_\BB[J])^2+\ep_0^2\Big(\Nk^{(En)}_{J+1}+\NN^{(match)}_{J+1}\Big)^2 + \int_{\Mext_{r\leq \frac{5}{2}m_0}(1,\tau_*)}(\dk^{J+1}\aa)^2.
 \eeaa
Note that $\Up^2\gtrsim\deh^2>0$ on $\Mext$ and hence
\beaa
 \int_{\Mext_{r\leq \frac{5}{2}m_0}(1,\tau_*)}(\dk^{J+1}\aa)^2 &\les&  \int_{\Mext_{r\leq \frac{5}{2}m_0}(1,\tau_*)}(\dk^{J+1}(\Up^2\aa))^2
\eeaa
which together with the control of $\Up^2\aa$ provided by \eqref{eq:controlforalphaandforUpquarealphabarfortheproofofThmM8} yields
\beaa
 \int_{\Mext_{r\leq \frac{5}{2}m_0}(1,\tau_*)}(\dk^{J+1}\aa)^2 &\les&  (\ep_\BB[J])^2+\ep_0^2\Big(\Nk^{(En)}_{J+1}+\NN^{(match)}_{J+1}\Big)^2
\eeaa
and hence
 \beaa
 \int_{\Mint(1,\tau_*)}(\dk^{J+1}\aa)^2 &\les&  (\ep_\BB[J])^2+\ep_0^2\Big(\Nk^{(En)}_{J+1}+\NN^{(match)}_{J+1}\Big)^2.
 \eeaa
 Since 
 \beaa
 B^J_{\de}[\aa](1,\tau_*) &\les&  \int_{\Mint(1,\tau_*)}(\dk^{\leq J+1}\aa)^2+B^J_{\de}[\Up^2\aa](1,\tau_*),
 \eeaa
 using again \eqref{eq:controlforalphaandforUpquarealphabarfortheproofofThmM8}, we finally obtain
  \bea\lab{eq:controlforalphabarwithoutdegenaracyonMintfortheproofofThmM8}
B^J_{\de}[\aa](1,\tau_*)   &\les&  (\ep_\BB[J])^2+\ep_0^2\Big(\Nk^{(En)}_{J+1}+\NN^{(match)}_{J+1}\Big)^2.
  \eea


\subsection{End of the proof of Proposition \ref{prop:controlofMorrallcurvcompiterationassupmtionThM8}}


We have 
\beaa
\rhoc &=& \frac{\rhot}{r^2} -\left(\ov{\rho}-\frac{2m}{r^3}\right). 
\eeaa
Together with the control for $\rhot$ provided by Proposition \ref{prop:controlofrhotildeiterationassupmtionThM8}, as well as the control on averages provided by Lemma \ref{lemma:estimatesonceandforallforaverages}, we infer
\beaa
B^J_{\de}[\rhoc](1,\tau_*)   &\les&  (\ep_\BB[J])^2+\ep_0^2\Big(\Nk^{(En)}_{J+1}+\NN^{(match)}_{J+1}\Big)^2.
\eeaa
Together with the control for $\a$ provided by \eqref{eq:controlforalphaandforUpquarealphabarfortheproofofThmM8} and the control for $\aa$ provided by \eqref{eq:controlforalphabarwithoutdegenaracyonMintfortheproofofThmM8}, we infer
\beaa
B^J_{\de}[\a, \rhoc, \aa](1,\tau_*)   &\les&  (\ep_\BB[J])^2+\ep_0^2\Big(\Nk^{(En)}_{J+1}+\NN^{(match)}_{J+1}\Big)^2.
\eeaa
Together with the Bianchi identities for $e_4(\b)$, $e_3(\b)$, $\ddd_1\b$, $e_4(\bb)$, $e_3(\bb)$, $\ddd_1\bb$, as well as  the iteration assumption \eqref{eq:iterationassumptiondiscussionThM8:bis}, we infer
\beaa
B^J_{\de-2}[\b, \bb](1,\tau_*)   &\les&  (\ep_\BB[J])^2+\ep_0^2\Big(\Nk^{(En)}_{J+1}+\NN^{(match)}_{J+1}\Big)^2
\eeaa
and hence
\beaa
B^J_{-2}[\a, \b, \rhoc, \bb, \aa](1,\tau_*)   &\les&  (\ep_\BB[J])^2+\ep_0^2\Big(\Nk^{(En)}_{J+1}+\NN^{(match)}_{J+1}\Big)^2
\eeaa
as desired. This concludes the proof of Proposition \ref{prop:controlofMorrallcurvcompiterationassupmtionThM8}.


\section{Proof of Proposition \ref{prop:rpweightedestimatesiterationassupmtionThM8}}\lab{sec:proofprop:rpweightedestimatesiterationassupmtionThM8}


First, note that, by definition of the norms $B^J_{-2}$, $\,{}^{(int)}\mathfrak{R}_{J+1}[\Rc]$ and $\,{}^{(ext)}\mathfrak{R}_{J+1}[\Rc]$,  we have for any $r_0\geq 4m_0$
\beaa
\,{}^{(int)}\mathfrak{R}_{J+1}[\Rc]+\,{}^{(ext)}\mathfrak{R}^{\leq r_0}_{J+1}[\Rc] &\les& r^{10}B^J_{-2}[\a, \b, \rhoc, \bb, \aa](1, \tau_*).
\eeaa
Together with Proposition \ref{prop:controlofMorrallcurvcompiterationassupmtionThM8}, this implies 
\beaa
\,{}^{(int)}\mathfrak{R}_{J+1}[\Rc]+\,{}^{(ext)}\mathfrak{R}^{\leq r_0}_{J+1}[\Rc] &\les& r_0^{10}\Big((\ep_\BB[J])^2+\ep_0^2\Big(\Nk^{(En)}_{J+1}+\NN^{(match)}_{J+1}\Big)^2\Big).
\eeaa
Since we have
\beaa
\,{}^{(int)}\mathfrak{R}_{J+1}[\Rc]+\,{}^{(ext)}\mathfrak{R}_{J+1}[\Rc] &=& \,{}^{(int)}\mathfrak{R}_{J+1}[\Rc]+\,{}^{(ext)}\mathfrak{R}^{\leq r_0}_{J+1}[\Rc]+\,{}^{(ext)}\mathfrak{R}^{\geq r_0}_{J+1}[\Rc],
\eeaa
we deduce for any $r_0\geq 4m_0$
\beaa
\,{}^{(int)}\mathfrak{R}_{J+1}[\Rc]+\,{}^{(ext)}\mathfrak{R}_{J+1}[\Rc] &\leq& \,{}^{(ext)}\mathfrak{R}^{\geq r_0}_{J+1}[\Rc]+O\left(r_0^{10}\left(\ep_\BB[J]+\ep_0\Big(\Nk^{(En)}_{J+1}+\NN^{(match)}_{J+1}\Big)\right)\right).
\eeaa
Thus, to prove Proposition \ref{prop:rpweightedestimatesiterationassupmtionThM8}, it suffices to establish the following inequality
 \beaa
 \,{}^{(ext)}\mathfrak{R}^{\geq r_0}_{J+1}[\Rc] &\les& r_0^{-\dt}\,{}^{(ext)}\mathfrak{G}^{\geq r_0}_k[\Gac]+r_0^{10}\left(\ep_\BB[J]+\ep_0\left(\Nk^{(En)}_{J+1}+\NN^{(match)}_{J+1}\right)\right).
 \eeaa 
This will follow from $r^p$ weighted estimates for the curvature components.


\subsection{$r$-weighted divergence identities for Bianchi pairs}


\begin{lemma}\lab{lemma:basicdivergenceidentitybianchipairrpweightedestimate}
Let $k\geq 1$, let $a_{(1)}$ and $a_{(2)}$ real numbers. We consider the following equations.
\begin{itemize}
\item If $\psi_{(1)}, h_{(1)}\in\mathfrak{s}_k$, $\psi_{(2)}, h_{(2)}\in\mathfrak{s}_{k-1}$,  let $(\psi_{(1)}, \psi_{(2)})$ such that
\bea\lab{eq:modelbainchipairequations1}
\left\{\ba{lll}
e_3(\psi_{(1)})+a_{(1)}\kab\psi_{(1)} &=& -\dds_k\psi_{(2)} +h_{(1)},\\[2mm]
e_4(\psi_{(2)})+a_{(2)}\ka\psi_{(2)} &=& \ddd_k\psi_{(1)} +h_{(2)},
\ea\right.
\eea

\item If $\psi_{(1)}, h_{(1)}\in\mathfrak{s}_{k-1}$, $\psi_{(2)}, h_{(2)}\in\mathfrak{s}_k$,  let $(\psi_{(1)}, \psi_{(2)})$ such that
\bea\lab{eq:modelbainchipairequations2}
\left\{\ba{lll}
e_3(\psi_{(1)})+a_{(1)}\kab\psi_{(1)} &=& \ddd_k\psi_{(2)} +h_{(1)},\\[2mm]
e_4(\psi_{(2)})+a_{(2)}\ka\psi_{(2)} &=& -\dds_k\psi_{(1)} +h_{(2)}.
\ea\right.
\eea
\end{itemize}
Then, the pair $(\psi_{(1)}, \psi_{(2)})$ satisfies for any real number $b$
\bea\lab{eq:basicdivergenceidentitybianchipairrpweightedestimate}
\nn&&\Div\Big(r^b\psi_{(1)}^2e_3\Big) + \Div\Big(r^b\psi_{(2)}^2e_4\Big)-\frac{1}{2}r^b\kab\Big(-4a_{(1)} +b +2\Big)\psi_{(1)}^2    +\frac{1}{2}r^b\ka\Big(4a_{(2)} -b-2\Big)\psi_{(2)}^2\\
\nn&=& 2r^b\ddd_1(\psi_{(1)}\psi_{(2)})   -2r^b\omb\psi_{(1)}^2 -2r^b\om\psi_{(2)}^2 + 2r^b\psi_{(1)} h_{(1)}+2r^b\psi_{(2)}h_{(2)}\\
&& +br^{b-1}\left(e_3(r)-\frac{r}{2}\kab\right)\psi_{(1)}^2+br^{b-1}\left(e_4(r)-\frac{r}{2}\ka\right)\psi_{(2)}^2.
\eea
\end{lemma}

\begin{remark}\lab{remark:howtowritebianchipairinframeworkrpweightedestimates}
Note that the Bianchi identities can be written as systems of equations of the type \eqref{eq:modelbainchipairequations1}  \eqref{eq:modelbainchipairequations2}. In particular
\begin{itemize}
\item the Bianchi pair $(\a, \b)$ satisfies  \eqref{eq:modelbainchipairequations1} with $k=2$, $a_{(1)}=\frac{1}{2}$, $a_{(2)}=2$,

\item the Bianchi pair $(\b, \rho)$ satisfies  \eqref{eq:modelbainchipairequations1} with $k=1$, $a_{(1)}=1$, $a_{(2)}=\frac{3}{2}$,

\item the Bianchi pair $(\rho, \bb)$ satisfies  \eqref{eq:modelbainchipairequations2} with $k=1$, $a_{(1)}=\frac{3}{2}$, $a_{(2)}=1$,

\item the Bianchi pair $(\bb, \aa)$ satisfies  \eqref{eq:modelbainchipairequations2} with $k=2$, $a_{(1)}=2$, $a_{(2)}=\frac{1}{2}$.
\end{itemize}
\end{remark}

\begin{proof}[Proof of Lemma \ref{lemma:basicdivergenceidentitybianchipairrpweightedestimate}]
The proof being identical for \eqref{eq:modelbainchipairequations1} and \eqref{eq:modelbainchipairequations2}, it suffices to prove it in the case where $(\psi_{(1)}, \psi_{(2)})$ satisfies \eqref{eq:modelbainchipairequations1}. 

We compute
\beaa
\D_\ga e_4^\ga &=& -\frac{1}{2}\g(\D_4e_4, e_3)-\frac{1}{2}\g(\D_3e_4, e_4)+\g(D_\th e_4, e_\th)+\g(D_\varphi e_4, e_\varphi)\\
&=& \ka-2\om
\eeaa
and 
\beaa
\D_\ga e_3^\ga &=& -\frac{1}{2}\g(\D_4e_3, e_3)-\frac{1}{2}\g(\D_3e_3, e_4)+\g(D_\th e_3, e_\th)+\g(D_\varphi e_3, e_\varphi)\\
&=& \kab -2\omb.
\eeaa
We infer in view of \eqref{eq:modelbainchipairequations1}
\beaa
&&\D_\ga\Big(r^b\psi_{(1)}^2e_3^\ga\Big)\\
 &=& 2r^b\psi_{(1)} e_3(\psi_{(1)})+br^{b-1}e_3(r)\psi_{(1)}^2+r^b\psi_{(1)}^2\D_\ga e_3^\ga\\
 &=& 2r^b\psi_{(1)}\Big( -a_{(1)}\kab\psi_{(1)} - \dds_k\psi_{(2)} +h_{(1)}\Big) +br^{b-1}e_3(r)\psi_{(1)}^2+r^b\psi_{(1)}^2(\kab -2\omb)\\
 &=&  -2r^b\psi_{(1)}\dds_k\psi_{(2)}  +r^b\Big(-2a_{(1)} +\frac{b}{2} +1\Big) \kab\psi_{(1)}^2 +br^{b-1}\left(e_3(r)-\frac{r}{2}\kab\right)\psi_{(1)}^2 -2\omb r^b\psi_{(1)}^2+ 2r^b\psi_{(1)} h_{(1)}  
 \eeaa
and
\beaa
&&\D_\ga\Big(r^b\psi_{(2)}^2e_4^\ga\Big)\\
 &=& 2r^b\psi_{(2)} e_4(\psi_{(2)})+br^{b-1}e_4(r)\psi_{(2)}^2+r^b\psi_{(2)}^2\D_\ga e_4^\ga\\
&=& 2r^b\psi_{(2)} \Big(-a_{(2)}\ka\psi_{(2)} + \ddd_k\psi_{(1)} +h_{(2)}\Big)+br^{b-1}e_4(r)\psi_{(2)}^2+r^b\psi_{(2)}^2( \ka-2\om)\\
&=& 2r^b\psi_{(2)}\ddd_k\psi_{(1)} +r^b\Big(-2a_{(2)} +\frac{b}{2} +1\Big) \ka\psi_{(2)}^2+br^{b-1}\left(e_4(r)-\frac{r}{2}\ka\right)\psi_{(2)}^2 -2r^b\om\psi_{(2)}^2 +2r^b\psi_{(2)}h_{(2)}.
\eeaa
We sum the two identities
\beaa
&&\D_\ga\Big(r^b\psi_{(1)}^2e_3^\ga\Big) + \D_\ga\Big(r^b\psi_{(2)}^2e_4^\ga\Big)\\
&=& -2r^b\psi_{(1)}\dds_k\psi_{(2)} +2r^b\psi_{(2)}\ddd_k\psi_{(1)}  +r^b\Big(-2a_{(1)} +\frac{b}{2} +1\Big) \kab\psi_{(1)}^2    +r^b\Big(-2a_{(2)} +\frac{b}{2} +1\Big) \ka\psi_{(2)}^2
\\
&&+br^{b-1}\left(e_3(r)-\frac{r}{2}\kab\right)\psi_{(1)}^2+br^{b-1}\left(e_4(r)-\frac{r}{2}\ka\right)\psi_{(2)}^2 -2r^b\omb\psi_{(1)}^2 -2r^b\om\psi_{(2)}^2\\ &&+2r^b\psi_{(2)}h_{(2)} + 2r^b\psi_{(1)} h_{(1)}
\eeaa
and hence
\beaa
&&\D_\ga\Big(r^b\psi_{(1)}^2e_3^\ga\Big) + \D_\ga\Big(r^b\psi_{(2)}^2e_4^\ga\Big)-r^b\kab\Big(-2a_{(1)} +\frac{b}{2} +1\Big)\psi_{(1)}^2    +r^b\ka\Big(2a_{(2)} -\frac{b}{2} -1\Big)\psi_{(2)}^2\\
&=& 2r^b\ddd_1(\psi_{(1)}\psi_{(2)})  +br^{b-1}\left(e_3(r)-\frac{r}{2}\kab\right)\psi_{(1)}^2+br^{b-1}\left(e_4(r)-\frac{r}{2}\ka\right)\psi_{(2)}^2 -2r^b\omb\psi_{(1)}^2 -2r^b\om\psi_{(2)}^2 \\
&&+ 2r^b\psi_{(1)} h_{(1)}+2r^b\psi_{(2)}h_{(2)}. 
\eeaa
This concludes the proof of Lemma \ref{lemma:basicdivergenceidentitybianchipairrpweightedestimate}.
\end{proof}

To obtain $r^p$ weighted estimates for higher order derivatives of the curvature components, we will need several lemmas.
\begin{lemma}\lab{lemma:identityfortheangulardivergencepartofrpweightedestimates}
Let $k\geq 1$ and $s\geq 1$ two integers. Let $\psi_{(1)}\in\mathfrak{s}_k$ and $\psi_{(2)}\in\mathfrak{s}_{k-1}$. Then,  we have
\beaa
 -\dkb^s\psi_{(1)}\dkb^s\dds_k\psi_{(2)}+\dkb^s\psi_{(2)}\dkb^s\ddd_k\psi_{(1)} &=& \ddd_1\Big(\dkb^s\psi_{(1)}\dkb^s\psi_{(2)}\Big)+E[\dkb, s, k, \psi_{(1)}, \psi_{(2)}]
\eeaa 
 where 
 \beaa
 |E[s,k, \psi_{(1)}, \psi_{(2)}]| &\les& r|\dkb^s\psi_{(1)}|\sum_{j=0}^{s-1}|\dkb^{s-1-j}(\psi_{(2)})||\dkb^j(K)|\\
 &&+r|\dkb^s\psi_{(2)}|\sum_{j=0}^{s-1}|\dkb^{s-1-j}(\psi_{(1)})||\dkb^j(K)|.
 \eeaa
\end{lemma}

\begin{proof}
Recall our definition $\dkb^s$ for higher angular derivatives. Given $f$ a $k$-reduced  scalar  and    $s$ a positive integer   we define,
\beaa
\dkb^s f=
\begin{cases}
 r^{2p}\lapp_k^p, \qquad\qquad  \mbox{if} \quad  s=2p,\\
 r^{2p+1} \ddd_k  \lapp_k^p ,\qquad \mbox{if} \quad  s=2p+1.
\end{cases}
\eeaa

We start with the case $s=2p$, i.e. $s$ is even. Since $\psi_{(1)}\in\mathfrak{s}_k$ and $\psi_{(2)}\in\mathfrak{s}_{k-1}$, we have 
\beaa
&& -\dkb^s\psi_{(1)}\dkb^s\dds_k\psi_{(2)}+\dkb^s\psi_{(2)}\dkb^s\ddd_k\psi_{(1)}\\
&=& r^{4p}\Big(-\lapp_k^p\psi_{(1)}\lapp_k ^p\dds_k\psi_{(2)}+\lapp_{k-1}^p\psi_{(2)}\lapp_{k-1} ^p\ddd_k\psi_{(1)}\Big). 
\eeaa
Next, recall the commutation formulas
\beaa
-\ddd_k\lapp_k +\lapp_{k-1}\ddd_k &=& K\ddd_k -ke_\th(K),\\
-\dds_k\lapp_{k-1} +\lapp_k\dds_k &=& (2k-1)K\dds_k +(k-1)e_\th(K).
\eeaa
We infer
\beaa
\lapp_{k-1} ^p\ddd_k &=& \ddd_k\lapp_k^p+\sum_{j=1}^p\lapp_{k-1}^{p-j}\Big(\lapp_{k-1}\ddd_k-\ddd_k\lapp_k\Big)\lapp_k^{j-1}\\
&=& \ddd_k\lapp_k^p+\sum_{j=1}^p\lapp_{k-1}^{p-j}\Big( K\ddd_k -ke_\th(K)\Big)\lapp_k^{j-1}
\eeaa
and
\beaa
\lapp_k ^p\dds_k &=& \dds_k\lapp_{k-1}^p+\sum_{j=1}^p\lapp_k^{p-j}\Big(\lapp_k\dds_k-\dds_k\lapp_{k-1}\Big)\lapp_{k-1}^{j-1}\\
&=& \dds_k\lapp_{k-1}^p+\sum_{j=1}^p\lapp_k^{p-j}\Big((2k-1)K\dds_k +(k-1)e_\th(K)\Big)\lapp_{k-1}^{j-1}\\
\eeaa
This yields
\beaa
&& -\dkb^s\psi_{(1)}\dkb^s\dds_k\psi_{(2)}+\dkb^s\psi_{(2)}\dkb^s\ddd_k\psi_{(1)}\\
&=& r^{4p}\Bigg\{-\lapp_k^p\psi_{(1)}\dds_k\lapp_{k-1}^p\psi_{(2)}-\sum_{j=1}^p\lapp_k^p\psi_{(1)}\lapp_k^{p-j}\Big((2k-1)K\dds_k +(k-1)e_\th(K)\Big)\lapp_{k-1}^{j-1}\psi_{(2)}\\
&&+\lapp_{k-1}^p\psi_{(2)}\ddd_k\lapp_k^p\psi_{(1)}+\sum_{j=1}^p\lapp_{k-1}^p\psi_{(2)}\lapp_{k-1}^{p-j}\Big( K\ddd_k -ke_\th(K)\Big)\lapp_k^{j-1}\psi_{(1)}\Bigg\}\\
&=& \ddd_1\Big(\dkb^s\psi_{(1)}\dkb^s\psi_{(2)}\Big)\\
&& -\sum_{j=1}^p\dkb^s\psi_{(1)}\dkb^{s-2j}\Big((2k-1)r^2K\dds_k +(k-1)r^2e_\th(K)\Big)\dkb^{2j-2}\psi_{(2)}\\
&& +\sum_{j=1}^p\dkb^s\psi_{(2)}\dkb^{s-2j}\Big( r^2K\ddd_k -kr^2e_\th(K)\Big)\dkb^{2j-2}\psi_{(1)}.
\eeaa
Hence, we infer 
\beaa
 -\dkb^s\psi_{(1)}\dkb^s\dds_k\psi_{(2)}+\dkb^s\psi_{(2)}\dkb^s\ddd_k\psi_{(1)} &=& \ddd_1\Big(\dkb^s\psi_{(1)}\dkb^s\psi_{(2)}\Big)+E[s, k, \psi_{(1)}, \psi_{(2)}]
\eeaa 
 where 
 \beaa
 |E[s,k, \psi_{(1)}, \psi_{(2)}]| &\les& r^2|\dkb^s\psi_{(1)}|\sum_{j=0}^{s-1}|\dkb^{s-1-j}(\psi_{(2)})||\dkb^j(K)|\\
 &&+r^2|\dkb^s\psi_{(2)}|\sum_{j=0}^{s-1}|\dkb^{s-1-j}(\psi_{(1)})||\dkb^j(K)|.
 \eeaa

Next, we deal with the case $s=2p+1$, i.e. $s$ odd. Since $\psi_{(1)}\in\mathfrak{s}_k$ and $\psi_{(2)}\in\mathfrak{s}_{k-1}$, we have
\beaa
&& -\dkb^s\psi_{(1)}\dkb^s\dds_k\psi_{(2)}+\dkb^s\psi_{(2)}\dkb^s\ddd_k\psi_{(1)}\\
&=& r^{4p}\Big(-\ddd_k\lapp_k^p\psi_{(1)}\ddd_k\lapp_k ^p\dds_k\psi_{(2)}+\ddd_{k-1}\lapp_{k-1}^p\psi_{(2)}\ddd_{k-1}\lapp_{k-1} ^p\ddd_k\psi_{(1)}\Big) 
\eeaa
In view of the case $s=2p$ above, we infer
\beaa
&& -\dkb^s\psi_{(1)}\dkb^s\dds_k\psi_{(2)}+\dkb^s\psi_{(2)}\dkb^s\ddd_k\psi_{(1)}\\
&=& r^{4p+2}\Bigg\{-\ddd_k\lapp_k^p\psi_{(1)}\ddd_k\dds_k\lapp_{k-1}^p\psi_{(2)}-\sum_{j=1}^p\ddd_k\lapp_k^p\psi_{(1)}\ddd_k\lapp_k^{p-j}\Big((2k-1)K\dds_k +(k-1)e_\th(K)\Big)\lapp_{k-1}^{j-1}\psi_{(2)}\\
&&+\ddd_{k-1}\lapp_{k-1}^p\psi_{(2)}\ddd_{k-1}\ddd_k\lapp_k^p\psi_{(1)}+\sum_{j=1}^p\ddd_{k-1}\lapp_{k-1}^p\psi_{(2)}\ddd_{k-1}\lapp_{k-1}^{p-j}\Big( K\ddd_k -ke_\th(K)\Big)\lapp_k^{j-1}\psi_{(1)}\Bigg\}.
\eeaa
Next, recall the commutation formula
\beaa
\ddd_k\dds_k - \dds_{k-1}\ddd_{k-1} &=& -2(k-1)K.
\eeaa
We infer
\beaa
&& -\dkb^s\psi_{(1)}\dkb^s\dds_k\psi_{(2)}+\dkb^s\psi_{(2)}\dkb^s\ddd_k\psi_{(1)}\\
&=& r^{4p+2}\Bigg\{-\ddd_k\lapp_k^p\psi_{(1)}\Big( \dds_{k-1}\ddd_{k-1}  -2(k-1)K\Big)\lapp_{k-1}^p\psi_{(2)}\\
&&-\sum_{j=1}^p\ddd_k\lapp_k^p\psi_{(1)}\ddd_k\lapp_k^{p-j}\Big((2k-1)K\dds_k +(k-1)e_\th(K)\Big)\lapp_{k-1}^{j-1}\psi_{(2)}\\
&&+\ddd_{k-1}\lapp_{k-1}^p\psi_{(2)}\ddd_{k-1}\ddd_k\lapp_k^p\psi_{(1)}+\sum_{j=1}^p\ddd_{k-1}\lapp_{k-1}^p\psi_{(2)}\ddd_{k-1}\lapp_{k-1}^{p-j}\Big( K\ddd_k -ke_\th(K)\Big)\lapp_k^{j-1}\psi_{(1)}\Bigg\}\\
&=& \ddd_1\Big(\dkb^s\psi_{(1)}\dkb^s\psi_{(2)}\Big)\\
&& +2(k-1)r^2K\dkb^s\psi_{(1)}\dkb^{s-1}\psi_{(2)}-\sum_{j=1}^p\dkb^s\psi_{(1)}\dkb^{s-2j}\Big((2k-1)r^2K\dds_k +(k-1)r^2e_\th(K)\Big)\dkb^{2j-2}\psi_{(2)}\\
&&+\sum_{j=1}^p\dkb^s\psi_{(2)}\dkb^{s-2j}\Big( r^2K\ddd_k -kr^2e_\th(K)\Big)\dkb^{2j-2}\psi_{(1)}.
\eeaa
Hence, we obtain 
\beaa
 -\dkb^s\psi_{(1)}\dkb^s\dds_k\psi_{(2)}+\dkb^s\psi_{(2)}\dkb^s\ddd_k\psi_{(1)} &=& \ddd_1\Big(\dkb^s\psi_{(1)}\dkb^s\psi_{(2)}\Big)+E[\dkb,s, k, \psi_{(1)}, \psi_{(2)}]
\eeaa 
 where 
 \beaa
 |E[s,k, \psi_{(1)}, \psi_{(2)}]| &\les& r^2|\dkb^s\psi_{(1)}|\sum_{j=0}^{s-1}|\dkb^{s-1-j}(\psi_{(2)})||\dkb^j(K)|\\
 &&+r^2|\dkb^s\psi_{(2)}|\sum_{j=0}^{s-1}|\dkb^{s-1-j}(\psi_{(1)})||\dkb^j(K)|.
 \eeaa
This concludes the proof of the lemma.
\end{proof}

\begin{corollary}\lab{cor:basicdivergenceidentitybianchipairrpweightedestimate}
Let $k\geq 1$, let $a_{(1)}$ and $a_{(2)}$ real numbers and let $0\leq s\leq k_{large}$. Consider the outgoing geodesic foliation of $\Mext$. We consider the following equations.
\begin{itemize}
\item If $\psi_{(1)}\in\mathfrak{s}_k$, $\psi_{(2)}\in\mathfrak{s}_{k-1}$,  let $(\psi_{(1)}, \psi_{(2,s)})$ such that
\beaa
\left\{\ba{lll}
e_3(\dkb^s\psi_{(1)})+a_{(1)}\kab\dkb^s\psi_{(1)} &=& -\dkb^s\dds_k\psi_{(2)} +h_{(1,s)},\\[2mm]
e_4(\dkb^s\psi_{(2)})+a_{(2)}\ka\dkb^s\psi_{(2)} &=& \dkb^s\ddd_k\psi_{(1)} +h_{(2,s)},
\ea\right.
\eeaa

\item If $\psi_{(1)}\in\mathfrak{s}_{k-1}$, $\psi_{(2)}, h_{(2)}\in\mathfrak{s}_k$,  let $(\psi_{(1)}, \psi_{(2)})$ such that
\beaa
\left\{\ba{lll}
e_3(\dkb^s\psi_{(1)})+a_{(1)}\kab\dkb^s\psi_{(1)} &=& \dkb^s\ddd_k\psi_{(2)} +h_{(1,s)},\\[2mm]
e_4(\dkb^s\psi_{(2)})+a_{(2)}\ka\dkb^s\psi_{(2)} &=& -\dkb^s\dds_k\psi_{(1)} +h_{(2,s)}.
\ea\right.
\eeaa
\end{itemize}
Then, the pair $(\psi_{(1)}, \psi_{(2)})$ satisfies for any real number $b$
\beaa
\nn&&\Div\Big(r^b(\dkb^s\psi_{(1)})^2e_3\Big) + \Div\Big(r^b(\dkb^s\psi_{(2)})^2e_4\Big)\\
\nn&&-\frac{1}{2}r^b\kab\Big(-4a_{(1)} +b +2\Big)(\dkb^s\psi_{(1)})^2    +\frac{1}{2}r^b\ka\Big(4a_{(2)} -b-2\Big)(\dkb^s\psi_{(2)})^2\\
\nn&=& 2r^b\ddd_1\Big(\dkb^s\psi_{(1)}\dkb^s\psi_{(2)}\Big)+2r^bE[\dkb, s, k, \psi_{(1)}, \psi_{(2)}]  -2r^b\omb(\dkb^s\psi_{(1)})^2 \\
&& + 2r^b\dkb^s\psi_{(1)} h_{(1,s)}+2r^b\dkb^s\psi_{(2)}h_{(2,s)}+br^{b-1}\left(e_3(r)-\frac{r}{2}\kab\right)(\dkb^s\psi_{(1)})^2\\
&&+br^{b-1}\left(e_4(r)-\frac{r}{2}\ka\right)(\dkb^s\psi_{(2)})^2.
\eeaa
where $E[\dkb, s, k, \psi_{(1)}, \psi_{(2)}]$ has been introduced in Lemma \ref{lemma:identityfortheangulardivergencepartofrpweightedestimates}. 
\end{corollary}

\begin{proof}
The proof follows immediately from combining Lemma \ref{lemma:basicdivergenceidentitybianchipairrpweightedestimate} and Lemma \ref{lemma:identityfortheangulardivergencepartofrpweightedestimates}.
\end{proof}

\begin{lemma}\lab{lemma:themaindivergencerelationsforhighderivativegeneralpair}
Let $j, k, l$ three integers. Consider a Bianchi $(\psi_{(1)}, \psi_{(2)})$ satisfying \eqref{eq:modelbainchipairequations1} or \eqref{eq:modelbainchipairequations2}. Then,  the pair $(\psi_{(1)}, \psi_{(2)})$ satisfies for any real number $b$
\beaa
\nn&&\Div\Big(r^b(\dkb^j(re_4)^k\T^l\psi_{(1)})^2e_3\Big) + \Div\Big(r^b(\dkb^j(re_4)^k\T^l\psi_{(2)})^2e_4\Big)\\
\nn&&-\frac{1}{2}r^b\kab\Big(-4a_{(1)} +2k+b +2\Big)(\dkb^j(re_4)^k\T^l\psi_{(1)})^2 \\
\nn&&   +\frac{1}{2}r^b\ka\Big(4a_{(2)}-2k -b-2\Big)(\dkb^j(re_4)^k\T^l\psi_{(2)})^2\\
\nn&=& 2r^b\ddd_1\Big(\dkb^j(re_4)^k\T^l\psi_{(1)}\dkb^j(re_4)^k\T^l\psi_{(2)}\Big)+2r^bE[\dkb, j, k, (re_4)^k\T^l\psi_{(1)}, (re_4)^k\T^l\psi_{(2)}]  \\
&&-2r^b\omb(\dkb^j(re_4)^k\T^l\psi_{(1)})^2  + 2r^b\dkb^j(re_4)^k\T^l\psi_{(1)} h_{(1),j,k,l}+2r^b\dkb^j(re_4)^k\T^l\psi_{(2)}h_{(2),j,k,l}\\
&&+br^{b-1}\left(e_3(r)-\frac{r}{2}\kab\right)(\dkb^j(re_4)^k\T^l\psi_{(1)})^2+br^{b-1}\left(e_4(r)-\frac{r}{2}\ka\right)(\dkb^j(re_4)^k\T^l\psi_{(2)})^2.
\eeaa
where $E[\dkb, s, k, (re_4)^k\T^l\psi_{(1)}, (re_4)^k\T^l\psi_{(2)}]$ has been introduced in Lemma \ref{lemma:identityfortheangulardivergencepartofrpweightedestimates}, and where $h_{(1),j,k,l}$ and $h_{(2),j,k,l}$ are given, schematically, by
\beaa
h_{(1),j,k,l} &=& \dkb^{\leq j+k+l}(h_{(1)})+kr^{-1}\dkb^{j+1}(re_4)^{k-1}\T^l\psi_{(2)} \\
&&+r\dk^{j+k+l}\Big(\Ga_g\big(\psi_{(1)}, \psi_{(2)}\big)\Big) +O(r^{-1})\dk^{\leq j+k+l-1}\big(\psi_{(1)}, \psi_{(2)}\big)
\eeaa
and
\beaa
h_{(2),j,k,l} &=&  \dkb^{\leq j+k+l}(h_{(2)})+kr^{-1}\dkb^{j+1}(re_4)^{k-1}\T^l\psi_{(1)} +r\dk^{j+k+l}\Big(\Ga_g\big(\psi_{(1)}, \psi_{(2)}\big)\Big) \\
&&+O(r^{-1})\dk^{\leq j+k+l-1}\big(\psi_{(1)}, \psi_{(2)}\big). 
\eeaa
\end{lemma}

\begin{proof}
We have the following simple schematic consequences of the commutator identities 
\beaa
&& [T, e_4], [T, e_3] = r^{-1}\Ga_b\dk, \quad [T, \ddd_k] = -\eta e_3+\Ga_g\dk,\\
&& [\dkb, e_4] = \Ga_g\dk+\Ga_g, \quad [\dkb, e_3] =  -r\eta e_3+r\Ga_g\dk, \\
&& [re_4,e_4] = -\frac{r}{2}\ka e_4+\Ga_g\dk, \quad [re_4,e_3] = -\frac{r}{2}\kab e_4+\Ga_b\dk, \quad [re_4, \dkb_k] = r^{-1}\dkb+\Ga_g\dk+\Ga_g.
\eeaa
Then, differentiating with $\dkb^j(re_4)^k\T^l$ the equations
\beaa
\left\{\ba{lll}
e_3(\psi_{(1)})+a_{(1)}\kab\psi_{(1)} &=& -\dds_k\psi_{(2)} +h_{(1)},\\[2mm]
e_4(\psi_{(2)})+a_{(2)}\ka\psi_{(2)} &=& \ddd_k\psi_{(1)} +h_{(2)},
\ea\right.
\eeaa
and using the above commutator identities we infer
\beaa
\left\{\ba{lll}
e_3(\dkb^j(re_4)^k\T^l\psi_{(1)})+\left(a_{(1)}-\frac{k}{2}\right)\kab\dkb^j(re_4)^k\T^l\psi_{(1)} &=& -\dkb^j\dds_k((re_4)^k\T^l\psi_{(2)}) +h_{(1),j,k,l},\\[2mm]
e_4(\dkb^j(re_4)^k\T^l\psi_{(2)})+\left(a_{(2)}-\frac{k}{2}\right)\ka \dkb^j(re_4)^k\T^l\psi_{(2)} &=& \dkb^j\ddd_k((re_4)^k\T^l\psi_{(1)}) +h_{(2),j,k,l},
\ea\right.
\eeaa
were 
\beaa
h_{(1),j,k,l} &=& \dkb^j(re_4)^k\T^l(h_{(1)})+kr^{-1}\dkb^{j+1}(re_4)^{k-1}\T^l\psi_{(2)} +jr\eta \dk^{j+k+l-1}e_3\psi_{(1)}\\
&&+r\dk^{j+k+l}\Big(\Ga_g\big(\psi_{(1)}, \psi_{(2)}\big)\Big) +O(r^{-1})\dk^{\leq j+k+l-1}\big(\psi_{(1)}, \psi_{(2)}\big)
\eeaa
and
\beaa
h_{(2),j,k,l} &=& \dkb^j(re_4)^k\T^l(h_{(2)})+kr^{-1}\dkb^{j+1}(re_4)^{k-1}\T^l\psi_{(1)} +r\dk^{j+k+l}\Big(\Ga_g\big(\psi_{(1)}, \psi_{(2)}\big)\Big) \\
&&+O(r^{-1})\dk^{\leq j+k+l-1}\big(\psi_{(1)}, \psi_{(2)}\big). 
\eeaa
Also, using the equation 
\beaa
e_3(\psi_{(1)}) &=& -a_{(1)}\kab\psi_{(1)}  -\dds_k\psi_{(2)} +h_{(1)},
\eeaa
we obtain 
\beaa
jr\eta \dk^{j+k+l-1}e_3\psi_{(1)} &=& r\dk^{j+k+l}\Big(\Ga_g\big(\psi_{(1)}, \psi_{(2)}\big)\Big) +O(r^{-1})\dk^{\leq j+k+l-1}\big(\psi_{(1)}, \psi_{(2)}\big)+r\eta\dk^{k+j+l-1}(h_{(1)})
\eeaa
and hence, 
\beaa
h_{(1),j,k,l} &=& \dkb^{\leq j+k+l}(h_{(1)})+kr^{-1}\dkb^{j+1}(re_4)^{k-1}\T^l\psi_{(2)} \\
&&+r\dk^{j+k+l}\Big(\Ga_g\big(\psi_{(1)}, \psi_{(2)}\big)\Big) +O(r^{-1})\dk^{\leq j+k+l-1}\big(\psi_{(1)}, \psi_{(2)}\big).
\eeaa
We have thus obtained the desired form for $h_{(1),j,k,l}$ and $h_{(2),j,k,l}$.

The divergence identity now follows from the equations 
\beaa
\left\{\ba{lll}
e_3(\dkb^j(re_4)^k\T^l\psi_{(1)})+\left(a_{(1)}-\frac{k}{2}\right)\kab\dkb^j(re_4)^k\T^l\psi_{(1)} &=& -\dkb^j\dds_k((re_4)^k\T^l\psi_{(2)}) +h_{(1),j,k,l},\\[2mm]
e_4(\dkb^j(re_4)^k\T^l\psi_{(2)})+\left(a_{(2)}-\frac{k}{2}\right)\ka \dkb^j(re_4)^k\T^l\psi_{(2)} &=& \dkb^j\ddd_k((re_4)^k\T^l\psi_{(1)}) +h_{(2),j,k,l},
\ea\right.
\eeaa
together with Corollary \ref{cor:basicdivergenceidentitybianchipairrpweightedestimate}. This concludes the proof of the lemma.
\end{proof}

\begin{corollary}\lab{cor:themainrpweightedesimtatesforageneralpairandgeneralderivatives}
Let $r_0\geq 4m_0$ and $1\leq u_0\leq u_*$. We introduce the spacetime region 
\beaa
\RR_{u_0}=\Mext\cap\{r\geq 4m_0\}\cap\{1\leq u\leq u_0\}, \qquad .
\eeaa
Let $j, k, l$ three integers. Assume that the frame of $\Mext$ satisfies 
\beaa
\sup_{\Mext}\left(\left|e_3(r)-\frac{r}{2}\kab\right|+r\left(|\omb|+\left|e_4(r)-\frac{r}{2}\ka\right|\right)\right) &\les& \ep_0.
\eeaa
Consider a pair $(\psi_{(1)}, \psi_{(2)})$ satisfying \eqref{eq:modelbainchipairequations1} or \eqref{eq:modelbainchipairequations2}. Then,  $(\psi_{(1)}, \psi_{(2)})$ satisfies for any real number $b$
\begin{itemize}
\item[(a)] If 
\beaa
-4a_{(1)} +2k+b +2>0\textrm{ and }4a_{(2)}-2k -b-2>0,
\eeaa
then, we have 
\beaa
&&\int_{\CC_{u_0}(r\geq r_0)}r^b(\dkb^j(re_4)^k\T^l\psi_{(1)})^2+\int_{\Sigma_*(\leq u_0)}r^b\Big((\dkb^j(re_4)^k\T^l\psi_{(1)})^2+(\dkb^j(re_4)^k\T^l\psi_{(2)})^2\Big)\\
&&+\int_{\RR_{u_0}(r\geq r_0)}r^{b-1}\Big((\dkb^j(re_4)^k\T^l\psi_{(1)})^2+(\dkb^j(re_4)^k\T^l\psi_{(1)})^2\Big)\\
&\les& \int_{\Mext(\frac{r_0}{2}\leq r\leq r_0)}r^{b-1}\Big((\dkb^j(re_4)^k\T^l\psi_{(1)})^2+(\dkb^j(re_4)^k\T^l\psi_{(1)})^2\Big)\\
&&+\int_{\RR_{u_0}(r\geq r_0)}r^{b+1}\Big((h_{(1),j,k,l})^2+(h_{(2),j,k,l})^2\Big)\\
&& +\int_{\RR_{u_0}(r\geq r_0)}r^bE[\dkb, j, k, (re_4)^k\T^l\psi_{(1)}, (re_4)^k\T^l\psi_{(2)}].
\eeaa

\item[(b)] If 
\beaa
-4a_{(1)} +2k+b +2>0\textrm{ and }4a_{(2)}-2k -b-2=0,
\eeaa
then, we have 
\beaa
&&\int_{\CC_{u_0}(r\geq r_0)}r^b(\dkb^j(re_4)^k\T^l\psi_{(1)})^2+\int_{\Sigma_*(\leq u_0)}r^b\Big((\dkb^j(re_4)^k\T^l\psi_{(1)})^2+(\dkb^j(re_4)^k\T^l\psi_{(2)})^2\Big)\\
&\les& \int_{\Mext(\frac{r_0}{2}\leq r\leq r_0)}r^{b-1}\Big((\dkb^j(re_4)^k\T^l\psi_{(1)})^2+(\dkb^j(re_4)^k\T^l\psi_{(1)})^2\Big)\\
&&+\int_{\RR_{u_0}(r\geq r_0)}r^{b+1-\dt}(h_{(1),j,k,l})^2+\int_{\RR_{u_0}(r\geq r_0)}r^{b+1+\dt}(h_{(2),j,k,l})^2\\
&&+\int_{\RR_{u_0}(r\geq r_0)}r^{b-1+\dt}(\dkb^j(re_4)^k\T^l\psi_{(1)})^2+\int_{\RR_{u_0}(r\geq r_0)}r^{b-1-\dt}(\dkb^j(re_4)^k\T^l\psi_{(2)})^2\\
&& +\int_{\RR_{u_0}(r\geq r_0)}r^bE[\dkb, j, k, (re_4)^k\T^l\psi_{(1)}, (re_4)^k\T^l\psi_{(2)}].
\eeaa

\item[(c)] If 
\beaa
-4a_{(1)} +2k+b +2>0\textrm{ and }4a_{(2)}-2k -b-2<0,
\eeaa
then, we have
\beaa
&&\int_{\CC_{u_0}(r\geq r_0)}r^b(\dkb^j(re_4)^k\T^l\psi_{(1)})^2+\int_{\Sigma_*(\leq u_0)}r^b\Big((\dkb^j(re_4)^k\T^l\psi_{(1)})^2+(\dkb^j(re_4)^k\T^l\psi_{(2)})^2\Big)\\
&&+\int_{\RR_{u_0}(r\geq r_0)}r^{b-1}(\dkb^j(re_4)^k\T^l\psi_{(1)})^2\\
&\les& \int_{\Mext(\frac{r_0}{2}\leq r\leq r_0)}r^{b-1}\Big((\dkb^j(re_4)^k\T^l\psi_{(1)})^2+(\dkb^j(re_4)^k\T^l\psi_{(1)})^2\Big)\\
&&+\int_{\RR_{u_0}(r\geq r_0)}r^{b+1}(h_{(1),j,k,l})^2+\int_{\RR_{u_0}(r\geq r_0)}r^{b+1}(h_{(2),j,k,l})^2\\
&&+\int_{\RR_{u_0}(r\geq r_0)}r^{b-1}((\dkb^j(re_4)^k\T^l\psi_{(2)})^2\\
&& +\int_{\RR_{u_0}(r\geq r_0)}r^bE[\dkb, j, k, (re_4)^k\T^l\psi_{(1)}, (re_4)^k\T^l\psi_{(2)}].
\eeaa
\end{itemize}
\end{corollary}

\begin{proof}
We multiply the pair $(\psi_{(1)}, \psi_{(2)})$ by a smooth cut-off function in $r$ supported in $r\geq \frac{r_0}{2}$ and identically one for $r\geq r_0$. We obtain again a solution to \eqref{eq:modelbainchipairequations1} or \eqref{eq:modelbainchipairequations2} up to error terms that are supported in the region $\frac{r_0}{2}\leq r\leq r_0$. We then integrate the divergence identities of Lemma \ref{lemma:themaindivergencerelationsforhighderivativegeneralpair} on the region $\RR_{u_0}$ and the corollary follows.
\end{proof}


\subsection{End of the proof of Proposition \ref{prop:rpweightedestimatesiterationassupmtionThM8}}


Let $r_0\geq 4m_0$. Recall that, to prove Proposition \ref{prop:rpweightedestimatesiterationassupmtionThM8}, it suffices to establish the following inequality
 \beaa
 \,{}^{(ext)}\mathfrak{R}^{\geq r_0}_{J+1}[\Rc] &\les& r_0^{-\dt}\,{}^{(ext)}\mathfrak{G}^{\geq r_0}_k[\Gac]+r_0^{10}\left(\ep_\BB[J]+\ep_0\left(\Nk^{(En)}_{J+1}+\NN^{(match)}_{J+1}\right)\right).
 \eeaa 
To this end, we will rely on the $r^p$-weighted estimates derived in Corollary \ref{cor:themainrpweightedesimtatesforageneralpairandgeneralderivatives} applied to the Bianchi pairs, where we recall Remark \ref{remark:howtowritebianchipairinframeworkrpweightedestimates}. 

\begin{remark}
For the Bianchi pair $(\b, \rho)$, we replace the Bianchi identities for $e_4(\rho)$ by its analog for $e_4(\rhoc)$, i.e. 
\beaa
e_4\check{\rho}+\frac  3 2 \ov{\ka} \check{\rho} &=& \ddd_1\b -\frac 3 2 \ov{\rho}\check{\ka} +\err[e_4\check{\rho}],
 \eeaa
while for the Bianchi pair $(\rho, \bb)$, we replace the Bianchi identities for $e_3(\rho)$ by its analog for $e_3(\rhoc)$, i.e. 
\beaa
e_3\check{\rho} +\frac 3 2  \ov{\kab} \check{\rho}&=&  \ddd_1\bb - \frac 3 2 \ov{\rho}\check{\kab} -\frac{3}{2}\ov{\kab}\, \ov{\rho}\vsi^{-1} \check{\vsi} +\frac{3}{2}\ov{\ka}\, \ov{\rho}\left(\Obc +\vsi^{-1}\ov{\Omb}\check{\vsi}\right) +\err[e_3\check{\rho}],
\eeaa
see Proposition \ref{propos:transportaverages} for the derivation of these equations.
\end{remark}
 
Let $j, k, l$ three integers such that 
\beaa
j+k+l=J+1.
\eeaa
To derive $r^p$ weighted curvature estimates for $\dkb^j(re_4)^k\T^l$ derivatives in the region $r\geq r_0$, we proceed as follows.

{\bf Step 1.} We start with the case $k=0$, i.e. we derive $r^p$ weighted curvature estimates for $\dkb^j\T^l$ derivatives with $j+l=J+1$. First, we apply Corollary \ref{cor:themainrpweightedesimtatesforageneralpairandgeneralderivatives} 
\begin{itemize}
\item to the Bianchi pair $(\a, \b)$ with the choice $b=4+\dt$, 

\item to the Bianchi pair $(\b, \rho)$ with the choice $b=4-\dt$,

\item to the Bianchi pair $(\rho, \bb)$ with the choice $b=2-\dt$,

\item to the Bianchi pair $(\bb, \aa)$ with the choice $b=-\dt$.
\end{itemize}

All the above choices are such that we have in case (a) of Corollary \ref{cor:themainrpweightedesimtatesforageneralpairandgeneralderivatives}. In particular, we obtain
\beaa
&&\sum_{j+l=J+1}\Bigg\{\sup_{1\leq u\leq u_*}\int_{\CC_u(r\geq r_0)}\Big(r^{4+\dt}(\dkb^j\T^l\a)^2+r^{4-\dt}(\dkb^j\T^l\b)^2+r^{2-\dt}(\dkb^j\T^l\rhoc)^2\\
&&+r^{-\dt}(\dkb^j\T^l\bb)^2\Big)+\int_{\Sigma_*}\Big(r^{4+\dt}\Big((\dkb^j\T^l\a)^2+(\dkb^j\T^l\b)^2\Big)+r^{4-\dt}(\dkb^j\T^l\rhoc)^2\\
&&+r^{2-\dt}(\dkb^j\T^l\bb)^2+r^{-\dt}(\dkb^j\T^l\aa)^2\Big)+\int_{\Mext(r\geq r_0)}\Big(r^{3+\dt}\Big((\dkb^j\T^l\a)^2+(\dkb^j\T^l\b)^2\Big)\\
&&+r^{3-\dt}(\dkb^j\T^l\rhoc)^2+r^{1-\dt}(\dkb^j\T^l\bb)^2+r^{-1-\dt}(\dkb^j\T^l\aa)^2\Big)\Bigg\}\\
&\les& r_0^{8+\dt}\int_{\Mext(\frac{r_0}{2}\leq r\leq r_0)}\frac{(\dk^{J+1}\Rc)^2}{r^5}+\int_{\Mext(r\geq r_0)}\Bigg\{r^{-1+\dt}(\dkb^j\T^l\vth)^2\\
&&+r^{-1-\dt}\Big((\dkb^j\T^l\eta)^2+(\dkb^j\T^l\check{\ka} )^2\Big)+r^{-3-\dt}\Big((\dkb^j\T^l\check{\kab})^2 +(\dkb^j\T^l\ze)^2\Big)\\
&&+r^{-5-\dt}\Big((\dkb^j\T^l\xib)^2+(\dkb^j\T^l\vthb)^2+(\dkb^j\T^l\check{\vsi})^2 +(\dkb^j\T^l\Obc)^2\Big)\Bigg\}+(\ep_\BB[J])^2+\ep_0^2(\Nk^{(En)}_{J+1})^2.
\eeaa
Using Proposition \ref{prop:controlofMorrallcurvcompiterationassupmtionThM8} to bound the first term on the right-hand side, and using also the definition of the norm $\,{}^{(ext)}\mathfrak{G}^{\geq r_0}_k[\Gac]$, we infer that 
\beaa
&&\sum_{j+l=J+1}\Bigg\{\sup_{1\leq u\leq u_*}\int_{\CC_u(r\geq r_0)}\Big(r^{4+\dt}(\dkb^j\T^l\a)^2+r^{4-\dt}(\dkb^j\T^l\b)^2+r^{2-\dt}(\dkb^j\T^l\rhoc)^2\\
&&+r^{-\dt}(\dkb^j\T^l\bb)^2\Big)+\int_{\Sigma_*}\Big(r^{4+\dt}\Big((\dkb^j\T^l\a)^2+(\dkb^j\T^l\b)^2\Big)+r^{4-\dt}(\dkb^j\T^l\rhoc)^2\\
&&+r^{2-\dt}(\dkb^j\T^l\bb)^2+r^{-\dt}(\dkb^j\T^l\aa)^2\Big)+\int_{\Mext(r\geq r_0)}\Big(r^{3+\dt}\Big((\dkb^j\T^l\a)^2+(\dkb^j\T^l\b)^2\Big)\\
&&+r^{3-\dt}(\dkb^j\T^l\rhoc)^2+r^{1-\dt}(\dkb^j\T^l\bb)^2+r^{-1-\dt}(\dkb^j\T^l\aa)^2\Big)\Bigg\}\\
&\les& \left(\int_{r_0}^{+\infty}\frac{dr}{r^{1+\dt}}\right)\,{}^{(ext)}\mathfrak{G}^{\geq r_0}_k[\Gac]+r_0^{10}\Big((\ep_\BB[J])^2+\ep_0^2\Big(\Nk^{(En)}_{J+1}+\NN^{(match)}_{J+1}\Big)^2\Big)
\eeaa
and hence
\bea\lab{eq:firstrpweigthedestmateforproofThmM8}
\nn&&\sum_{j+l=J+1}\Bigg\{\sup_{1\leq u\leq u_*}\int_{\CC_u(r\geq r_0)}\Big(r^{4+\dt}(\dkb^j\T^l\a)^2+r^{4-\dt}(\dkb^j\T^l\b)^2+r^{2-\dt}(\dkb^j\T^l\rhoc)^2\\
\nn&&+r^{-\dt}(\dkb^j\T^l\bb)^2\Big)+\int_{\Sigma_*}\Big(r^{4+\dt}\Big((\dkb^j\T^l\a)^2+(\dkb^j\T^l\b)^2\Big)+r^{4-\dt}(\dkb^j\T^l\rhoc)^2\\
\nn&&+r^{2-\dt}(\dkb^j\T^l\bb)^2+r^{-\dt}(\dkb^j\T^l\aa)^2\Big)+\int_{\Mext(r\geq r_0)}\Big(r^{3+\dt}\Big((\dkb^j\T^l\a)^2+(\dkb^j\T^l\b)^2\Big)\\
\nn&&+r^{3-\dt}(\dkb^j\T^l\rhoc)^2+r^{1-\dt}(\dkb^j\T^l\bb)^2+r^{-1-\dt}(\dkb^j\T^l\aa)^2\Big)\Bigg\}\\
&\les& r_0^{-\dt}\,{}^{(ext)}\mathfrak{G}^{\geq r_0}_k[\Gac]+r_0^{10}\Big((\ep_\BB[J])^2+\ep_0^2\Big(\Nk^{(En)}_{J+1}+\NN^{(match)}_{J+1}\Big)^2\Big).
\eea

{\bf Step 2.} We derive additional $r^p$ weighted curvature estimates for $\dkb^j\T^l$ derivatives with $j+l=J+1$. To this end, we apply Corollary \ref{cor:themainrpweightedesimtatesforageneralpairandgeneralderivatives} 
\begin{itemize}
 \item to the Bianchi pair $(\b, \rho)$ with the choice $b=4$,

\item to the Bianchi pair $(\rho, \bb)$ with the choice $b=2$,

\item to the Bianchi pair $(\bb, \aa)$ with the choice $b=0$.
\end{itemize}

All the above choices are such that we have in case (b) of Corollary \ref{cor:themainrpweightedesimtatesforageneralpairandgeneralderivatives}. In particular, we obtain
\beaa
&&\sum_{j+l=J+1}\Bigg\{\sup_{1\leq u\leq u_*}\int_{\CC_u(r\geq r_0)}\Big(r^{4}(\dkb^j\T^l\b)^2+r^{2}(\dkb^j\T^l\rhoc)^2+(\dkb^j\T^l\bb)^2\Big)\\
&&+\int_{\Sigma_*}\Big(r^{4}\Big((\dkb^j\T^l\b)^2+(\dkb^j\T^l\rhoc)^2\Big)+r^{2}(\dkb^j\T^l\bb)^2+(\dkb^j\T^l\aa)^2\Big)\Bigg\}\\
&\les& r_0^{8}\int_{\Mext(\frac{r_0}{2}\leq r\leq r_0)}\frac{(\dk^{J+1}\Rc)^2}{r^5}\\
&& +\sum_{j+l=J+1}\Bigg\{\int_{\Mext(r\geq r_0)}\Big(r^{3+\dt}(\dkb^j\T^l\b)^2+r^{3-\dt}(\dkb^j\T^l\rhoc)^2+r^{1-\dt}(\dkb^j\T^l\bb)^2\\
&&+r^{-1-\dt}(\dkb^j\T^l\aa)^2\Big)\Bigg\}+\int_{\Mext(r\geq r_0)}\Bigg\{r^{-1-\dt}(\dkb^j\T^l\eta)^2+r^{-1+\dt}(\dkb^j\T^l\check{\ka} )^2\\
&&+r^{-3+\dt}\Big((\dkb^j\T^l\check{\kab})^2 +(\dkb^j\T^l\ze)^2\Big)\\
&&+r^{-5+\dt}\Big((\dkb^j\T^l\xib)^2+(\dkb^j\T^l\vthb)^2+(\dkb^j\T^l\check{\vsi})^2 +(\dkb^j\T^l\Obc)^2\Big)\Bigg\}+(\ep_\BB[J])^2+\ep_0^2(\Nk^{(En)}_{J+1})^2.
\eeaa
Using Proposition \ref{prop:controlofMorrallcurvcompiterationassupmtionThM8} to bound the first term on the right-hand side, and using also the definition of the norm $\,{}^{(ext)}\mathfrak{G}^{\geq r_0}_k[\Gac]$, we infer that 
\beaa
&&\sum_{j+l=J+1}\Bigg\{\sup_{1\leq u\leq u_*}\int_{\CC_u(r\geq r_0)}\Big(r^{4}(\dkb^j\T^l\b)^2+r^{2}(\dkb^j\T^l\rhoc)^2+(\dkb^j\T^l\bb)^2\Big)\\
&&+\int_{\Sigma_*}\Big(r^{4}\Big((\dkb^j\T^l\b)^2+(\dkb^j\T^l\rhoc)^2\Big)+r^{2}(\dkb^j\T^l\bb)^2+(\dkb^j\T^l\aa)^2\Big)\Bigg\}\\
&\les& \left(\int_{r_0}^{+\infty}\frac{dr}{r^{1+\dt}}\right)\,{}^{(ext)}\mathfrak{G}^{\geq r_0}_k[\Gac]+r_0^{10}\Big((\ep_\BB[J])^2+\ep_0^2\Big(\Nk^{(En)}_{J+1}+\NN^{(match)}_{J+1}\Big)^2\Big)\\
&& +\sum_{j+l=J+1}\Bigg\{\int_{\Mext(r\geq r_0)}\Big(r^{3+\dt}(\dkb^j\T^l\b)^2+r^{3-\dt}(\dkb^j\T^l\rhoc)^2+r^{1-\dt}(\dkb^j\T^l\bb)^2\\
&&+r^{-1-\dt}(\dkb^j\T^l\aa)^2\Big)\Bigg\}
\eeaa
and hence
\beaa
&&\sum_{j+l=J+1}\Bigg\{\sup_{1\leq u\leq u_*}\int_{\CC_u(r\geq r_0)}\Big(r^{4}(\dkb^j\T^l\b)^2+r^{2}(\dkb^j\T^l\rhoc)^2+(\dkb^j\T^l\bb)^2\Big)\\
&&+\int_{\Sigma_*}\Big(r^{4}\Big((\dkb^j\T^l\b)^2+(\dkb^j\T^l\rhoc)^2\Big)+r^{2}(\dkb^j\T^l\bb)^2+(\dkb^j\T^l\aa)^2\Big)\Bigg\}\\
&\les& r_0^{-\dt}\,{}^{(ext)}\mathfrak{G}^{\geq r_0}_k[\Gac]+r_0^{10}\Big((\ep_\BB[J])^2+\ep_0^2\Big(\Nk^{(En)}_{J+1}+\NN^{(match)}_{J+1}\Big)^2\Big)\\
&& +\sum_{j+l=J+1}\Bigg\{\int_{\Mext(r\geq r_0)}\Big(r^{3+\dt}(\dkb^j\T^l\b)^2+r^{3-\dt}(\dkb^j\T^l\rhoc)^2+r^{1-\dt}(\dkb^j\T^l\bb)^2\\
&&+r^{-1-\dt}(\dkb^j\T^l\aa)^2\Big)\Bigg\}.
\eeaa
Together with \eqref{eq:firstrpweigthedestmateforproofThmM8}, we deduce
\bea\lab{eq:firstrpweigthedestmateforproofThmM8:1}
\nn&&\sum_{j+l=J+1}\Bigg\{\sup_{1\leq u\leq u_*}\int_{\CC_u(r\geq r_0)}\Big(r^{4+\dt}(\dkb^j\T^l\a)^2+r^4(\dkb^j\T^l\b)^2+r^2(\dkb^j\T^l\rhoc)^2\\
\nn&&+(\dkb^j\T^l\bb)^2\Big)+\int_{\Sigma_*}\Big(r^{4+\dt}\Big((\dkb^j\T^l\a)^2+(\dkb^j\T^l\b)^2\Big)+r^4(\dkb^j\T^l\rhoc)^2\\
\nn&&+r^2(\dkb^j\T^l\bb)^2+(\dkb^j\T^l\aa)^2\Big)+\int_{\Mext(r\geq r_0)}\Big(r^{3+\dt}\Big((\dkb^j\T^l\a)^2+(\dkb^j\T^l\b)^2\Big)\\
\nn&&+r^{3-\dt}(\dkb^j\T^l\rhoc)^2+r^{1-\dt}(\dkb^j\T^l\bb)^2+r^{-1-\dt}(\dkb^j\T^l\aa)^2\Big)\Bigg\}\\
&\les& r_0^{-\dt}\,{}^{(ext)}\mathfrak{G}^{\geq r_0}_k[\Gac]+r_0^{10}\Big((\ep_\BB[J])^2+\ep_0^2\Big(\Nk^{(En)}_{J+1}+\NN^{(match)}_{J+1}\Big)^2\Big).
\eea

{\bf Step 3.} We now argue by iteration on $k$. For $0\leq k\leq J$, we consider the following iteration assumption 
\bea\lab{eq:firstrpweigthedestmateforproofThmM8:2}
\nn&&\sum_{j+l=J+1-k}\Bigg\{\sup_{1\leq u\leq u_*}\int_{\CC_u(r\geq r_0)}\Big(r^{4+\dt}(\dkb^j(re_4)^k\T^l\a)^2+r^4(\dkb^j(re_4)^k\T^l\b)^2+r^2(\dkb^j(re_4)^k\T^l\rhoc)^2\\
\nn&&+(\dkb^j(re_4)^k\T^l\bb)^2\Big)+\int_{\Sigma_*}\Big(r^{4+\dt}\Big((\dkb^j(re_4)^k\T^l\a)^2+(\dkb^j(re_4)^k\T^l\b)^2\Big)+r^4(\dkb^j(re_4)^k\T^l\rhoc)^2\\
\nn&&+r^2(\dkb^j(re_4)^k\T^l\bb)^2+(\dkb^j(re_4)^k\T^l\aa)^2\Big)+\int_{\Mext(r\geq r_0)}\Big(r^{3+\dt}\Big((\dkb^j(re_4)^k\T^l\a)^2+(\dkb^j(re_4)^k\T^l\b)^2\Big)\\
\nn&&+r^{3-\dt}(\dkb^j(re_4)^k\T^l\rhoc)^2+r^{1-\dt}(\dkb^j(re_4)^k\T^l\bb)^2+r^{-1-\dt}(\dkb^j(re_4)^k\T^l\aa)^2\Big)\Bigg\}\\
&\les& r_0^{-\dt}\,{}^{(ext)}\mathfrak{G}^{\geq r_0}_k[\Gac]+r_0^{10}\Big((\ep_\BB[J])^2+\ep_0^2\Big(\Nk^{(En)}_{J+1}+\NN^{(match)}_{J+1}\Big)^2\Big).
\eea
\eqref{eq:firstrpweigthedestmateforproofThmM8:2} holds true for $k=0$ in view of \eqref{eq:firstrpweigthedestmateforproofThmM8:1}. We now assume that \eqref{eq:firstrpweigthedestmateforproofThmM8:2} holds true for $k$ such that  $0\leq k\leq J$, and our goal is to prove that it also holds for $k+1$. 

First, note that the Bianchi identities for $e_4(\b)$, $e_4(\rhoc)$, $e_4(\bb)$ and $e_4(\aa)$, together with \eqref{eq:firstrpweigthedestmateforproofThmM8:2}, yields
\bea\lab{eq:firstrpweigthedestmateforproofThmM8:3}
\nn&&\sum_{j+l=J+1-(k+1)}\Bigg\{\sup_{1\leq u\leq u_*}\int_{\CC_u(r\geq r_0)}\Big(r^4(\dkb^j(re_4)^{k+1}\T^l\b)^2+r^2(\dkb^j(re_4)^{k+1}\T^l\rhoc)^2\\
\nn&&+(\dkb^j(re_4)^{k+1}\T^l\bb)^2\Big)+\int_{\Sigma_*}\Big(r^{4+\dt}(\dkb^j(re_4)^{k+1}\T^l\b)^2+r^4(\dkb^j(re_4)^{k+1}\T^l\rhoc)^2\\
\nn&&+r^2(\dkb^j(re_4)^{k+1}\T^l\bb)^2+(\dkb^j(re_4)^{k+1}\T^l\aa)^2\Big)+\int_{\Mext(r\geq r_0)}\Big(r^{3+\dt}(\dkb^j(re_4)^{k+1}\T^l\b)^2\\
\nn&&+r^{3-\dt}(\dkb^j(re_4)^{k+1}\T^l\rhoc)^2+r^{1-\dt}(\dkb^j(re_4)^{k+1}\T^l\bb)^2+r^{-1-\dt}(\dkb^j(re_4)^{k+1}\T^l\aa)^2\Big)\Bigg\}\\
&\les& r_0^{-\dt}\,{}^{(ext)}\mathfrak{G}^{\geq r_0}_k[\Gac]+r_0^{10}\Big((\ep_\BB[J])^2+\ep_0^2\Big(\Nk^{(En)}_{J+1}+\NN^{(match)}_{J+1}\Big)^2\Big).
\eea

We still need to estimate $\dkb^j(re_4)^{k+1}\T^l\a$. To this end, we apply Corollary \ref{cor:themainrpweightedesimtatesforageneralpairandgeneralderivatives} to the Bianchi pair $(\a, \b)$ with the choice $b=4+\dt$. Since $k+1\geq 1$, we are in case (c) of Corollary \ref{cor:themainrpweightedesimtatesforageneralpairandgeneralderivatives}. In particular, we obtain, arguing similarly as above,
\beaa
&&\sum_{j+l=J+1-(k+1)}\Bigg\{\sup_{1\leq u\leq u_*}\int_{\CC_u(r\geq r_0)}r^{4+\dt}(\dkb^j(re_4)^{k+1}\T^l\a)^2+\int_{\Sigma_*}r^{4+\dt}(\dkb^j(re_4)^{k+1}\T^l\a)^2\\
&&+\int_{\Mext(r\geq r_0)}r^{3+\dt}(\dkb^j(re_4)^{k+1}\T^l\a)^2\Bigg\}\\
&\les& \sum_{j+l=J+1-(k+1)}\left\{\int_{\Mext(r\geq r_0)}r^{3+\dt}(\dkb^j(re_4)^{k+1}\T^l\b)^2\right\}+r_0^{-\dt}\,{}^{(ext)}\mathfrak{G}^{\geq r_0}_k[\Gac]\\
&&+r_0^{10}\Big((\ep_\BB[J])^2+\ep_0^2\Big(\Nk^{(En)}_{J+1}+\NN^{(match)}_{J+1}\Big)^2\Big).
\eeaa
Together with \eqref{eq:firstrpweigthedestmateforproofThmM8:3}, this implies \eqref{eq:firstrpweigthedestmateforproofThmM8:2} for $k+1$. Hence, by iteration, \eqref{eq:firstrpweigthedestmateforproofThmM8:2} holds for any $0\leq k\leq J+1$. This implies
\beaa
\nn&&\sum_{k\leq J+1}\Bigg\{\sup_{1\leq u\leq u_*}\int_{\CC_u(r\geq r_0)}\Big(r^{4+\dt}(\dk^k\a)^2+r^4(\dk^k\b)^2+r^2(\dk^k\rhoc)^2+(\dk^k\bb)^2\Big)\\
&&+\int_{\Sigma_*}\Big(r^{4+\dt}\Big((\dk^k\a)^2+(\dk^k\b)^2\Big)+r^4(\dk^k\rhoc)^2+r^2(\dk^k\bb)^2+(\dk^k\aa)^2\Big)\\
&&+\int_{\Mext(r\geq r_0)}\Big(r^{3+\dt}\Big((\dk^k\a)^2+(\dk^k\b)^2\Big)+r^{3-\dt}(\dk^k\rhoc)^2+r^{1-\dt}(\dk^k\bb)^2+r^{-1-\dt}(\dk^k\aa)^2\Big)\Bigg\}\\
&\les& r_0^{-\dt}\,{}^{(ext)}\mathfrak{G}^{\geq r_0}_k[\Gac]+r_0^{10}\Big((\ep_\BB[J])^2+\ep_0^2\Big(\Nk^{(En)}_{J+1}+\NN^{(match)}_{J+1}\Big)^2\Big).
\eeaa
Hence, we have obtained
 \beaa
 \,{}^{(ext)}\mathfrak{R}^{\geq r_0}_{J+1}[\Rc] &\les& r_0^{-\dt}\,{}^{(ext)}\mathfrak{G}^{\geq r_0}_k[\Gac]+r_0^{10}\left(\ep_\BB[J]+\ep_0\left(\Nk^{(En)}_{J+1}+\NN^{(match)}_{J+1}\right)\right)
 \eeaa 
 which concludes the proof of Proposition \ref{prop:rpweightedestimatesiterationassupmtionThM8}.


\section{Proof of Proposition \ref{prop:controlGaextiterationassupmtionThM8}}\lab{sec:proofprop:controlGaextiterationassupmtionThM8}


To prove Proposition \ref{prop:controlGaextiterationassupmtionThM8}, we rely on the following three propositions.
\begin{proposition}\lab{prop:controlGaextiterationassupmtionThM8:actualresult:Sigma*}
Let $J$ such that $k_{small}-2\leq J\leq k_{large}-1$. Then, we have 
\beaa
 \,{}^{(\Si_*)}\mathfrak{G}_{J+1}[\Gac] +\,{}^{(\Si_*)}\mathfrak{G}_{J+1}'[\Gac]    &\les& \,{}^{(\Si_*)}\mathfrak{R}_{J+1}[\Rc] +\,{}^{(\Si_*)}\mathfrak{G}_{J}[\Gac],
 \eeaa
where we have introduced the notations
\beaa
\,{}^{(\Si_*)}\mathfrak{G}_k[\Gac] &:=& \int_{\Sigma_*}\Bigg[r^2\Big((\dk^{\leq k}\vth)^2+(\dk^{\leq k}\check{\ka})^2+(\dk^{\leq k}\ze)^2+(\dk^{\leq k}\check{\kab})^2\Big)+(\dk^{\leq k}\vthb)^2\\
&&+(\dk^{\leq k}\eta)^2+(\dk^{\leq k}\check{\omb})^2+(\dk^{\leq k}\xib)^2\Bigg],
 \eeaa
 \beaa
\,{}^{(\Si_*)}\mathfrak{G}_k'[\Gac] &:=& \int_{\Sigma_*}\left[r^2\Big((\dk^{k+1}\dkb\check{\ka})^2+(\dk^{k+1}\check{\ka})^2+(\dk^{\leq k+1}\check{\mu})^2+(\dk^{k+1}\check{\kab})^2+(\dk^{k+1}\ze)^2\right],
 \eeaa
 and
 \beaa
 \,{}^{(\Si_*)}\mathfrak{R}_{k}[\Rc] &:=& \int_{\Sigma_*}\Big(r^{4+\dt}\big((\dk^{\leq k}\a)^2+(\dk^{\leq k}\b)^2\big)+r^4(\dk^{\leq k}\check{\rho})^2+r^2(\dk^{\leq k}\bb)^2+(\dk^{\leq k}\aa)^2\Big).
 \eeaa
\end{proposition}

\begin{proposition}\lab{prop:controlGaextiterationassupmtionThM8:actualresult}
Let $J$ such that $k_{small}-2\leq J\leq k_{large}-1$. Then, we have 
\beaa
 \,{}^{(ext)}\mathfrak{G}^{\geq 4m_0}_{J+1}[\Gac] + \,{}^{(ext)}{\mathfrak{G}^{\geq 4m_0}_{J+1}}'[\Gac]  &\les&  \,{}^{(\Si_*)}\mathfrak{G}_{J+1}[\Gac] +\,{}^{(\Si_*)}\mathfrak{G}_{J+1}'[\Gac] +\,{}^{(ext)}\mathfrak{R}_{J+1}[\Rc] +\,{}^{(ext)}\mathfrak{G}_{J}[\Gac],
 \eeaa
 where we have introduced the notation
 \beaa
\,{}^{(ext)}{\mathfrak{G}^{\geq 4m_0}_{k}}'[\Gac] &:=& \sup_{\la\geq 4m_0}\Bigg(\int_{\{r=\la\}}\Bigg[\la^6\left(\dk^k\left(\ddd_1\dds_1\ka-\vth\Big(\ddd_4\dds_3\ddd_2^{-1}+\dds_2\Big)\ddd_1^{-1}\check{\rho}\right)\right)^2\\
&&+\la^2(\dk^{k+1}\kac)^2  + \la^6\left(\dk^k\left(e_\th(\mu) +\vth\ddd_2\dds_2(\dds_1\ddd_1)^{-1}\bb+2\ze\check{\rho}\right)\right)^2\\
&&+\la^4(\dk^{\leq k}\muc)^2+\la^2\left(\dk^k\left(e_\th(\kab) -4\bb\right)\right)^2+\la^2\left(\dk^k\left(e_3(\ze) +\bb\right)\right)^2\Bigg]\Bigg).
\eeaa
\end{proposition}

\begin{proposition}\lab{prop:controlGaextiterationassupmtionThM8:actualresult:bis}
Let $J$ such that $k_{small}-2\leq J\leq k_{large}-1$. Then, we have 
\beaa
 \,{}^{(ext)}\mathfrak{G}^{\leq 4m_0}_{J+1}[\Gac] + \,{}^{(ext)}{\mathfrak{G}^{\leq 4m_0}_{J+1}}'[\Gac]  &\les&   \,{}^{(ext)}\mathfrak{G}^{\geq 4m_0}_{J+1}[\Gac] + \,{}^{(ext)}{\mathfrak{G}^{\geq 4m_0}_{J+1}}'[\Gac] +\,{}^{(ext)}\mathfrak{R}_{J+1}[\Rc] +\,{}^{(ext)}\mathfrak{G}_{J}[\Gac],
 \eeaa
 where we have introduced the notation
 \beaa
 \,{}^{(ext)}{\mathfrak{G}^{\leq 4m_0}_{k}}'[\Gac] &:=& \sup_{\rh\leq\la\leq 4m_0}\Bigg(\int_{\{r=\la\}}\Bigg[\la^6\left(\dk^k\left(\ddd_1\dds_1\ka-\vth\Big(\ddd_4\dds_3\ddd_2^{-1}+\dds_2\Big)\ddd_1^{-1}\check{\rho}\right)\right)^2\\
&&+\la^2(\dk^{k+1}\kac)^2 + \la^6\left(\dk^k\left(e_\th(\mu) +\vth\ddd_2\dds_2(\dds_1\ddd_1)^{-1}\bb+2\ze\check{\rho}\right)\right)^2\\
&&+\la^4(\dk^{\leq k}\muc)^2+\la^2\left(\dk^k\left(e_\th(\kab) -4\bb\right)\right)^2+\la^2\left(\dk^{k-1}\N\left(e_3(\ze) +\bb\right)\right)^2\Bigg]\Bigg).
\eeaa
\end{proposition}

The proof of Proposition \ref{prop:controlGaextiterationassupmtionThM8:actualresult:Sigma*} is postponed to section \ref{sec:proofofprop:controlGaextiterationassupmtionThM8:actualresult:Sigma*}, the proof of Proposition \ref{prop:controlGaextiterationassupmtionThM8:actualresult} is postponed to section \ref{sec:proofofprop:controlGaextiterationassupmtionThM8:actualresult}, and the proof of Proposition \ref{prop:controlGaextiterationassupmtionThM8:actualresult:bis} is postponed to section \ref{sec:proofofprop:controlGaextiterationassupmtionThM8:actualresult:bis}. The proof of the two latter propositions will rely in particular on basic weighted estimates for transport equations along $e_4$ in $\Mext$ derived in section \ref{weightedestimatesfortheproofofThmM8controlGaMext}, as well as  several renormalized identities derived in section \ref{severalidentitiesfortheproofofThmM8controlGaMext}

We now conclude the proof of Proposition \ref{prop:controlGaextiterationassupmtionThM8}. In view of Propositions  \ref{prop:controlGaextiterationassupmtionThM8:actualresult:Sigma*}, \ref{prop:controlGaextiterationassupmtionThM8:actualresult} and \ref{prop:controlGaextiterationassupmtionThM8:actualresult:bis}, we have, for $J$ such that $k_{small}-2\leq J\leq k_{large}-1$,  
\beaa
 \,{}^{(ext)}\mathfrak{G}_{J+1}[\Gac]   &\les&  \,{}^{(ext)}\mathfrak{R}_{J+1}[\Rc] +\,{}^{(ext)}\mathfrak{G}_{J}[\Gac],
 \eeaa 
where we have used the fact that 
\beaa
 \,{}^{(\Si_*)}\mathfrak{R}_{J+1}[\Rc]\leq \,{}^{(ext)}\mathfrak{R}_{J+1}[\Rc], \qquad  \,{}^{(\Si_*)}\mathfrak{G}_{J}[\Gac]\leq  \,{}^{(ext)}\mathfrak{G}_{J}[\Gac].
 \eeaa
In view of the iteration assumption \eqref{eq:iterationassumptiondiscussionThM8:bis}, we infer  
\beaa
 \,{}^{(ext)}\mathfrak{G}_{J+1}[\Gac]   &\les&  \,{}^{(ext)}\mathfrak{R}_{J+1}[\Rc] +\ep_\BB[J].
 \eeaa 
 Since the estimates in Proposition \ref{prop:controlGaextiterationassupmtionThM8:actualresult} are integrated from $\Si_*$, we obtain similarly, for any $r_0\geq 4m_0$, 
\beaa
 \,{}^{(ext)}\mathfrak{G}_{J+1}^{\geq r_0}[\Gac]   &\les&  \,{}^{(ext)}\mathfrak{R}_{J+1}^{\geq r_0}[\Rc] +\ep_\BB[J].
 \eeaa  
On the other hand, we have in view of Proposition \ref{prop:rpweightedestimatesiterationassupmtionThM8}, for any $r_0\geq 4m_0$,
 \beaa
 \,{}^{(ext)}\mathfrak{R}^{\geq r_0}_{J+1}[\Rc] &\les& r_0^{-\dt}\,{}^{(ext)}\mathfrak{G}^{\geq r_0}_k[\Gac]+r_0^{10}\left(\ep_\BB[J]+\ep_0\left(\Nk^{(En)}_{J+1}+\NN^{(match)}_{J+1}\right)\right).
 \eeaa  
 and
 \beaa
\,{}^{(int)}\mathfrak{R}_{J+1}[\Rc]+\,{}^{(ext)}\mathfrak{R}_{J+1}[\Rc] &\leq& \,{}^{(ext)}\mathfrak{R}^{\geq r_0}_{J+1}[\Rc]+O\left(r_0^{10}\left(\ep_\BB[J]+\ep_0\Big(\Nk^{(En)}_{J+1}+\NN^{(match)}_{J+1}\Big)\right)\right).
\eeaa
 Choosing $r_0\geq 4m_0$ large enough, we infer from the above estimates
 \beaa
 \,{}^{(ext)}\mathfrak{G}_{J+1}[\Gac]+\,{}^{(int)}\mathfrak{R}_{J+1}[\Rc]+\,{}^{(ext)}\mathfrak{R}_{J+1}[\Rc]  &\les& \ep_\BB[J]+\ep_0\left(\Nk^{(En)}_{J+1}+\NN^{(match)}_{J+1}\right).
 \eeaa   
 This concludes the proof of Proposition \ref{prop:controlGaextiterationassupmtionThM8}.


\subsection{Proof of Proposition \ref{prop:controlGaextiterationassupmtionThM8:actualresult:Sigma*}}\lab{sec:proofofprop:controlGaextiterationassupmtionThM8:actualresult:Sigma*}


\noindent{\bf Step 1.} We control $\ka$ on $\Si_*$. Recall the GCM conditions $\ka=2/r$ on $\Sigma_*$. Since $\nu_{\Sigma_*}$ and $e_\th$ are tangent, we infer
\beaa
(\dkb, \nu_{\Sigma_*})^k\left(\ka -\frac{2}{r}\right) &=& 0.
\eeaa
Together with Raychadhuri, we infer
\beaa
&&\max_{k\leq J+2}\int_{\Sigma_*}\left(r^2\left(\dk^k\left(\ka-\frac{2}{r}\right)\right)^2+r^4\left(\dk^ke_\th(\ka)\right)^2\right)\\
 &\les& \Big(\,{}^{(\Si_*)}\mathfrak{R}_{J+1}[\Rc] +\,{}^{(\Si_*)}\mathfrak{G}_{J}[\Gac]+\ep_0\,{}^{(\Si_*)}\mathfrak{G}_{J+1}[\Gac]\Big)^2,
\eeaa
where we have used the fact that $e_3$ is in the span of $e_4$ and $\nu_{\Si_*}$. Note that we have used Codazzi for $\vth$ to control the term $\dk^{J+1}e_4(e_\th(\ka))$.

\noindent{\bf Step 2.} We control the $\ell=1$ modes on $\Sigma_*$. In view of the GCM conditions for $\ka$, and projecting the Codazzi for $\vth$ on the $\ell=1$ mode, we infer on $\Sigma_*$
\beaa
\int_S\ze e^\Phi &=& r\int_S\b e^\Phi+\frac{r}{2}\int_S\vth\ze e^\Phi. 
\eeaa
Since the vectorfield $\nu$ is tangent to $\Si_*$, we infer
\beaa
\nu^{J+2}\left(\int_S\ze e^\Phi\right) &=& r\int_S\nu^{J+2}\b e^\Phi+\frac{r}{2}\int_S\nu^{J+2}(\vth\ze) e^\Phi+\lot\\
&=& r\int_S\nu^{J+2}\b e^\Phi+\frac{r}{2}\int_S\ze\nu^{J+2}(\vth) e^\Phi+\frac{r}{2}\int_S\vth\nu^{J+2}(\ze) e^\Phi+\lot 
\eeaa
where $\lot$ denote, here and below, terms that 
\begin{itemize}
\item either are linear and contain at most $J+1$ derivatives of curvature components and $J$ derivatives of Ricci coefficients,

\item or are quadratic and contain at most $J+1$ derivatives of Ricci coefficients and curvature components.
\end{itemize}
Using Bianchi identities and the null structure equations, we deduce
\beaa
&&\nu^{J+2}\left(\int_S\ze e^\Phi\right)\\
 &=& r\int_S\nu^{J+1}(\ddd_2\a, \dds_1\rho-3\rho\eta) e^\Phi+\frac{r}{2}\int_S\ze\nu^{J+1}\dds_2\eta e^\Phi+\frac{r}{2}\int_S\vth\nu^{J+1}\dds_1\omb e^\Phi+\lot\\
&=& r\int_S(\ddd_2\nu^{J+1}\a, \nu^{J}\dds_1\ddd_1(\b,\bb)) e^\Phi -3\ov{\rho}\int_S\nu^{J+1}\eta e^\Phi +\frac{r}{2}\int_S\ze\dds_2\nu^{J+1}\eta e^\Phi\\
&&+\frac{r}{2}\int_S\vth\dds_1\nu^{J+1}\omb e^\Phi+\lot\\
&=& r\int_S(\ddd_2\nu^{J+1}\a, \dds_1\ddd_1\nu^{J}(\b,\bb)) e^\Phi+\frac{r}{2}\int_S\ze\dds_2\nu^{J+1}\eta e^\Phi+\frac{r}{2}\int_S\vth\dds_1\nu^{J+1}\omb e^\Phi+\lot,
\eeaa
where we have used,  in the last equality, a cancellation due to the fact that $\nu$ is tangent to $\Si_*$  and $\int_S\eta e^\Phi=0$ on $\Si_*$. 
Using the identity $\dds_1\ddd_1=\ddd_2\dds_2+2K$, integration by parts for all terms, and the fact that $\dds_2(e^\Phi)=0$ so that the top order linear term vanish, we infer
\beaa
\nu^{J+2}\left(\int_S\ze e^\Phi\right) &=& \lot
\eeaa
with the above convention for the lower order terms. Also, relying on the null equation for $e_4(\ze)$, i.e.
\beaa
e_4(\ze) &=& -\ka\ze-\b -\vth\ze
\eeaa
we obtain, with more ease since this estimate is at one lower level of derivatives 
\beaa
(re_4, \nu)^{J+2}\left(\int_S\ze e^\Phi\right) &=& \lot
\eeaa
We infer
\beaa
\max_{k\leq J+2}\int_1^{u_*}r^{-2}\left(\dk^k\left(\int_S\ze e^\Phi\right)\right)^2 &\les& \Big(\,{}^{(\Si_*)}\mathfrak{R}_{J+1}[\Rc] +\,{}^{(\Si_*)}\mathfrak{G}_{J}[\Gac]+\ep_0\,{}^{(\Si_*)}\mathfrak{G}_{J+1}[\Gac]\Big)^2.
\eeaa

Next, we have in view of the definition of $\mu$ and the identity $\dds_1\ddd_1=\ddd_2\dds_2+2K$
\beaa
\int_Se_\th(\mu)e^\Phi &=& \int_S\dds_1\ddd_1\ze e^\Phi -\int_Se_\th(\rho)e^\Phi+\frac{1}{4}\int_Se_\th(\vth\vthb)e^\Phi\\
&=& 2\int_SK\ze e^\Phi -\int_Se_\th(\rho)e^\Phi+\frac{1}{4}\int_Se_\th(\vth\vthb)e^\Phi\\
&=& \frac{2}{r^2}\int_S\ze e^\Phi -\int_Se_\th(\rho)e^\Phi +\int_S\left(K-\frac{2}{r^2}\right)\ze e^\Phi +\frac{1}{4}\int_Se_\th(\vth\vthb)e^\Phi.
\eeaa
To estimate the RHS, we use in particular
\begin{itemize}
\item for the second term, in view of Bianchi 
\beaa
&&(e_3, re_4)^{J+2}e_\th(\rho)\\
 &=& (e_3, re_4)^{J+1}\dds_1\ddd_1(r\b, \bb)-\frac{3}{2}(e_3, re_4)^{J+1}\dds_1(r\ka\rho, \kab\rho)\\
 && +(e_3, re_4)^{J+1}\dds_1\big(r\vthb\a, (r\ze, \xib)\b, (\ze, \eta)\bb, \vth\aa\big) +\lot\\
&=& (e_3, re_4)^{J+1}\ddd_2\dds_2(r\b, \bb) +\frac{3}{2}\rho(e_3, re_4)^{J+1}(re_\th(\ka), e_\th(\kab))\\
&& +\big(r\vthb(e_3, re_4)^{J+1}\dds_1\a, (r\ze, \xib)(e_3, re_4)^{J+1}\dds_1\b, (\ze, \eta)(e_3, re_4)^{J+1}\dds_1\bb, \vth(e_3, re_4)^{J+1}\dds_1\aa\big)\\
&& +\big(r\a(e_3, re_4)^{J+1}\dds_1\vthb, \b(e_3, re_4)^{J+1}\dds_1(r\ze, \xib), \bb(e_3, re_4)^{J+1}\dds_1(\ze, \eta), \aa(e_3, re_4)^{J+1}\dds_1\vth\big)+\lot\\
&=& \ddd_2(e_3, re_4)^{J+1}\dds_2(r\b, \bb)+[(e_3, re_4)^{J+1}, \ddd_2]\dds_2(r\b, \bb)+\frac{3}{2}\ov{\rho}(e_3, re_4)^{J+1}(re_\th(\ka), e_\th(\kab))\\
&&-\frac{3}{2}\check{\rho}\dds_1(e_3, re_4)^{J+1}(r\check{\ka}, \check{\kab})\\
&& +\big(r\vthb\dds_1(e_3, re_4)^{J+1}\a, (r\ze, \xib)\dds_1(e_3, re_4)^{J+1}\b, (\ze, \eta)\dds_1(e_3, re_4)^{J+1}\bb, \vth\dds_1(e_3, re_4)^{J+1}\aa\big)\\
&& +\big(r\a\dds_1(e_3, re_4)^{J+1}\vthb, \b\dds_1(e_3, re_4)^{J+1}(r\ze, \xib), \bb\dds_1(e_3, re_4)^{J+1}(\ze, \eta), \aa\dds_1(e_3, re_4)^{J+1}\vth\big)+\lot
\eeaa

\item for the third term 
\beaa
&&(e_3, re_4)^{J+2}\Big(\left(K-\frac{2}{r^2}\right)\ze\Big)\\
&=&  \ze(e_3, re_4)^{J+2}\left(K-\frac{2}{r^2}\right)+\left(K-\frac{2}{r^2}\right)(e_3, re_4)^{J+2}\ze+\lot\\
&=&  \ze(e_3, re_4)^{J+1}\Big(\ddd_1(r\b, \bb, \eta, r^{-1}\xib)\Big)+\left(K-\frac{2}{r^2}\right)(e_3, re_4)^{J+1}e_\th(\omb)+\lot\\
&=&  \ze(e_3, re_4)^{J+1}\ddd_1\Big(r\b, \bb, \eta, r^{-1}\xib\Big)+\left(K-\frac{2}{r^2}\right)(e_3, re_4)^{J+1}e_\th(\omb)+\lot\\
&=&  \ze\ddd_1(e_3, re_4)^{J+1}\Big(r\b, \bb, \eta, r^{-1}\xib\Big)+\left(K-\frac{2}{r^2}\right)\dds_1(e_3, re_4)^{J+1}\check{\omb}+\lot,
\eeaa

\item for the fourth term
\beaa
(e_3, re_4)^{J+2}e_\th(\vth\vthb) &=& (e_3, re_4)^{J+1}e_\th(\vth\dds_2(\xib, r\ze))+(e_3, re_4)^{J+1}e_\th(\vthb\dds_2\eta)+\lot\\
&=&  \vth\dds_1\dds_2(e_3, re_4)^{J+1}(\xib, r\ze)+\vthb\dds_1\dds_2(e_3, re_4)^{J+1}\eta+\lot
\eeaa
\end{itemize}

We infer
\beaa
&&(e_3, re_4)^{J+2}\left(\int_Se_\th(\mu)e^\Phi\right)\\ &=& \frac{2}{r^2}(e_3, re_4)^{J+2}\left(\int_S\ze e^\Phi\right) +\int_S\ddd_2(e_3, re_4)^{J+1}\dds_2(r\b, \bb) e^\Phi\\
&& +\int_S[(e_3, re_4)^{J+1}, \ddd_2]\dds_2(r\b, \bb)e^\Phi +\frac{3}{2}\ov{\rho}\int_S(e_3, re_4)^{J+1}(re_\th(\ka), e_\th(\kab))e^\Phi\\
&&-\frac{3}{2}\int_S\check{\rho}\dds_1(e_3, re_4)^{J+1}(r\check{\ka}, \check{\kab})e^\Phi\\
&& +\int_S\big(r\vthb\dds_1(e_3, re_4)^{J+1}\a, (r\ze, \xib)\dds_1(e_3, re_4)^{J+1}\b, (\ze, \eta)\dds_1(e_3, re_4)^{J+1}\bb, \vth\dds_1(e_3, re_4)^{J+1}\aa\big)e^\Phi\\
&& +\int_S\big(r\a\dds_1(e_3, re_4)^{J+1}\vthb, \b\dds_1(e_3, re_4)^{J+1}(r\ze, \xib), \bb\dds_1(e_3, re_4)^{J+1}(\ze, \eta), \aa\dds_1(e_3, re_4)^{J+1}\vth\big)e^\Phi\\
&&+\int_S\ze\ddd_1(e_3, re_4)^{J+1}\Big(r\b, \bb, \ze, r^{-1}\xib\Big)e^\Phi+\int_S\left(K-\frac{2}{r^2}\right)\dds_1(e_3, re_4)^{J+1}\check{\omb}e^\Phi \\
&&+ \int_S\vth\dds_1\dds_2(e_3, re_4)^{J+1}(\xib, r\ze)e^\Phi+\int_S\vthb\dds_1\dds_2(e_3, re_4)^{J+1}\ze e^\Phi+\lot
\eeaa
and after integrations by parts and the fact that 
\beaa
\ddd_k(Fe^\Phi) = \ddd_{k+1}(F)e^\Phi,\quad \dds_k(Fe^\Phi) = \dds_{k-1}(F)e^\Phi,
\eeaa
we obtain
\beaa
&&(e_3, re_4)^{J+2}\left(\int_Se_\th(\mu)e^\Phi\right)\\ 
&=& \frac{2}{r^2}(e_3, re_4)^{J+2}\left(\int_S\ze e^\Phi\right)  +\int_S\dds_1\Big([(e_3, re_4)^{J+1}, \ddd_2]\Big)(r\b, \bb)e^\Phi\\
&&+\frac{3}{2}\ov{\rho}\int_S(e_3, re_4)^{J+1}(re_\th(\ka), e_\th(\kab))e^\Phi+\frac{3}{2}\int_S\ddd_2\rho (e_3, re_4)^{J+1}(r\check{\ka}, \check{\kab})e^\Phi\\
&& +\int_S\big(r\ddd_2\vthb(e_3, re_4)^{J+1}\a, \ddd_2(r\ze, \xib)(e_3, re_4)^{J+1}\b, \ddd_2(\ze, \eta)(e_3, re_4)^{J+1}\bb,  \ddd_2\vth(e_3, re_4)^{J+1}\aa\big)e^\Phi\\
&& +\int_S\big(r \ddd_2\a(e_3, re_4)^{J+1}\vthb,  \ddd_2\b(e_3, re_4)^{J+1}(r\ze, \xib),  \ddd_2\bb(e_3, re_4)^{J+1}(\ze, \eta),  \ddd_2\aa(e_3, re_4)^{J+1}\vth\big)e^\Phi\\
&&+\int_S\ddd_1\ze(e_3, re_4)^{J+1}\Big(r\b, \bb, \ze, r^{-1}\xib\Big)e^\Phi+\int_S\ddd_2\left(K-\frac{2}{r^2}\right)(e_3, re_4)^{J+1}\check{\omb}e^\Phi\\
&& + \int_S\ddd_3\ddd_2\vth(e_3, re_4)^{J+1}(\xib, r\ze)e^\Phi+\int_S\ddd_3\ddd_2\vthb(e_3, re_4)^{J+1}\ze e^\Phi+\lot
\eeaa
Together with the above estimate for the $\ell=1$ mode of $\ze$ and the estimate of Step 1 for $\ka$, we infer
\beaa
\max_{k\leq J+2}\int_1^{u_*}r^2\left(\dk^k\left(\int_Se_\th(\mu) e^\Phi\right)\right)^2 &\les& \Big(\,{}^{(\Si_*)}\mathfrak{R}_{J+1}[\Rc] +\,{}^{(\Si_*)}\mathfrak{G}_{J}[\Gac]+\ep_0\,{}^{(\Si_*)}\mathfrak{G}_{J+1}[\Gac]\Big)^2\\
&&+\max_{k\leq J+1}\int_1^{u_*}r^{-4}\left(\dk^k\left(\int_Se_\th(\kab) e^\Phi\right)\right)^2.
\eeaa
In view of the dominant condition \eqref{eq:behaviorofronSigmastar} for $r$ on $\Sigma_*$, we infer
\beaa
\max_{k\leq J+2}\int_1^{u_*}r^2\left(\dk^k\left(\int_Se_\th(\mu) e^\Phi\right)\right)^2 &\les& \Big(\,{}^{(\Si_*)}\mathfrak{R}_{J+1}[\Rc] +\,{}^{(\Si_*)}\mathfrak{G}_{J}[\Gac]+\ep_0\,{}^{(\Si_*)}\mathfrak{G}_{J+1}[\Gac]\Big)^2\\
&&+\ep_0\max_{k\leq J+1}\int_1^{u_*}\left(\dk^k\left(\int_Se_\th(\kab) e^\Phi\right)\right)^2.
\eeaa

Next, in view of the remarkable identity for the $\ell=1$ mode of $e_\th(K)$, we have
\beaa
-\int_Se_\th(\rho) e^\Phi -\frac{1}{4}\int_Se_\th(\ka\kab)e^\Phi +\frac{1}{4}\int_Se_\th(\vth\vthb)e^\Phi &=& 0
\eeaa
and hence
\beaa
\int_Se_\th(\kab)e^\Phi &=& -2r\int_Se_\th(\rho) e^\Phi -\frac{r}{2}\int_S\left(\ka-\frac{2}{r}\right)e_\th(\kab)e^\Phi -\frac{r}{2}\int_S\kab e_\th(\ka)e^\Phi  +\frac{r}{2}\int_Se_\th(\vth\vthb)e^\Phi.
\eeaa
Arguing as for the estimate of the $\ell=1$ mode of $e_\th(\mu)$, and using the smallness of $\ep_0$, we infer
\beaa
&&\max_{k\leq J+2}\int_1^{u_*}r^2\left(\dk^k\left(\int_Se_\th(\mu) e^\Phi\right)\right)^2+\max_{k\leq J+2}\int_1^{u_*}\left(\dk^k\left(\int_Se_\th(\kab) e^\Phi\right)\right)^2\\
 &\les& \Big(\,{}^{(\Si_*)}\mathfrak{R}_{J+1}[\Rc] +\,{}^{(\Si_*)}\mathfrak{G}_{J}[\Gac]+\ep_0\,{}^{(\Si_*)}\mathfrak{G}_{J+1}[\Gac]\Big)^2.
\eeaa

We have thus obtained
\beaa
&&\max_{k\leq J+2}\int_1^{u_*}\left(r^{-2}\left(\dk^k\left(\int_S\ze e^\Phi\right)\right)^2+r^2\left(\dk^k\left(\int_Se_\th(\mu) e^\Phi\right)\right)^2+\left(\dk^k\left(\int_Se_\th(\kab) e^\Phi\right)\right)^2\right)\\
 &\les& \Big(\,{}^{(\Si_*)}\mathfrak{R}_{J+1}[\Rc] +\,{}^{(\Si_*)}\mathfrak{G}_{J}[\Gac]+\ep_0\,{}^{(\Si_*)}\mathfrak{G}_{J+1}[\Gac]\Big)^2.
\eeaa

\noindent{\bf Step 3.} Recall the GCM conditions $\dds_2\dds_1\kab=\dds_2\dds_1\mu=0$ on $\Sigma_*$. This yields on $\Sigma_*$
\beaa
e_\th(\mu) = \frac{\int_Se_\th(\mu)e^\Phi}{\int_Se^{2\Phi}}e^\Phi, \quad e_\th(\kab) = \frac{\int_Se_\th(\kab)e^\Phi}{\int_Se^{2\Phi}}e^\Phi.
\eeaa
Together with Step 2, we infer
\beaa
&&\max_{k\leq J+2}\int_{\Sigma_*}\Big(r^4\left((\dkb, \nu_{\Sigma_*})^k\check{\mu}\right)^2+r^2\left((\dkb, \nu_{\Sigma_*})^k\check{\kab}\right)^2\Big)\\ &\les& \Big(\,{}^{(\Si_*)}\mathfrak{R}_{J+1}[\Rc] +\,{}^{(\Si_*)}\mathfrak{G}_{J}[\Gac]+\ep_0\,{}^{(\Si_*)}\mathfrak{G}_{J+1}[\Gac]\Big)^2.
\eeaa
Then, in view of the null structure equations for $e_4(\check{\mu})$ and $e_4(\check{\kab})$, 
\beaa
e_4(\check{\mu}) &=& -\frac{3}{2}\overline{\ka}\check{\mu}-\frac{3}{2}\overline{\mu}\check{\ka}+\err[e_4\check{\mu}]\\
e_4(\check{\kab}) &=& -\frac{1}{2}\overline{\ka}\check{\kab}-\frac{1}{2}\overline{\kab}\check{\ka}+2\check{\mu}+4\check{\rho}+\err[e_4\check{\kab}], 
\eeaa
we infer, together with the control of $\check{\ka}$ provided by Step 1,
\beaa
&&\max_{k\leq J+2}\int_{\Sigma_*}\Big(r^4\left(\dk^k\check{\mu}\right)^2+r^2\left(\dk^k\check{\kab}\right)^2\Big)\\
 &\les& \Big(\,{}^{(\Si_*)}\mathfrak{R}_{J+1}[\Rc] +\,{}^{(\Si_*)}\mathfrak{G}_{J}[\Gac]+\ep_0\,{}^{(\Si_*)}\mathfrak{G}_{J+1}[\Gac]\Big)^2.
\eeaa

\noindent{\bf Step 4.} Recall that we have
\beaa
\ddd_1\ze &=& -\check{\mu}-\check{\rho}+\frac{1}{4}\vth\vthb.
\eeaa
Differentiating, and using the Bianchi identities for $e_4(\check{\rho})$ and $e_3(\check{\rho})$, and the null structure equations for $e_4(\vth)$, $e_3(\vth)$, $e_4(\vthb)$ and $e_3(\vthb)$, we infer
\beaa
\ddd_1\dk^k\ze &=& -\dk^k\check{\mu}-\dk^{k-1}\dkb\Big(\check{\rho}, \b, r^{-1}\bb\Big)+\frac{1}{4}\dk^{k-1}\Big(\dkb(\vth\vthb), r^{-1}\vthb\dkb\eta, \vth\dkb\ze, r^{-1}\vth\dkb\xib\Big) +\lot\\
&=& -\dk^k\check{\mu}-\dkb\dk^{k-1}\Big(\check{\rho}, \b, r^{-1}\bb\Big)+\frac{1}{4}\dkb\Big(\vthb\dk^{k-1}(\vth, r^{-1}\eta)\Big)+\frac{1}{4}\dkb\Big(\vth\dk^{k-1}(\vthb, \ze, r^{-1}\xib)\Big) +\lot
\eeaa
We infer, since $\ddd_1$ is invertible in view of the corresponding Poincar\'e inequality, 
\beaa
\dk^k\ze &=& -r\dkb^{-1}\dk^k\check{\mu}-\dk^{k-1}\Big(\check{\rho}, \b, r^{-1}\bb\Big)+\frac{1}{4}\Big(\vthb\dk^{k-1}(\vth, r^{-1}\eta)\Big)+\frac{1}{4}\Big(\vth\dk^{k-1}(\vthb, \ze, r^{-1}\xib)\Big) +\lot\
\eeaa
Together with the estimate for $\check{\mu}$ of Step 3,  this yields
\beaa
\max_{k\leq J+2}\int_{\Sigma_*}r^2(\dk^k\ze)^2 &\les& \Big(\,{}^{(\Si_*)}\mathfrak{R}_{J+1}[\Rc] +\,{}^{(\Si_*)}\mathfrak{G}_{J}[\Gac]+\ep_0\,{}^{(\Si_*)}\mathfrak{G}_{J+1}[\Gac]\Big)^2.
\eeaa

\noindent{\bf Step 5.} Recall from the GCM condition that we have on $\Sigma_*$
\beaa
\int_S\eta e^\Phi=0.
\eeaa
Together with the transport equation
\beaa
e_4(\eta-\ze) &=& -\frac{1}{2}\ka(\eta-\ze)-\frac{1}{2}\vth(\eta-\ze),
\eeaa
we infer in view of the the estimates for $\ze$ of Step 4, 
\beaa
\max_{k\leq J+1}\int_1^{u_*}r^{-4}\left(\dk^k\left(\int_S\eta e^\Phi\right)\right)^2 &\les& \Big(\,{}^{(\Si_*)}\mathfrak{R}_{J+1}[\Rc] +\,{}^{(\Si_*)}\mathfrak{G}_{J}[\Gac]+\ep_0\,{}^{(\Si_*)}\mathfrak{G}_{J+1}[\Gac]\Big)^2.
\eeaa

Next, recall from Proposition \ref{prop:eqtsfor-ometaxib} that $\eta$ verifies
\beaa
 2\ddd_2\dds_2\eta&=& \ka\left( -e_3(\ze) +\bb\right) -e_3(e_\th(\ka)) - \ka\left(\frac{1}{2}\kab\ze  -2\omb\ze \right)+6\rho\eta-\kab e_\th \ka\\
 &-&\frac 1 2 \ka e_\th(\kab) +2\omb e_\th(\ka) + 2e_\th(\rho)+\err[\ddd_2\dds_2\eta],\\
\err[\ddd_2\dds_2\eta]&=& \left(2\ddd_1\eta -\frac{1}{2}\ka\vthb +2\eta^2    \right)\eta+2e_\th(\eta^2)-\ka\left(\frac{1}{2}\vthb\ze  -\frac{1}{2}\vth\xib\right) -\frac{1}{2}\vthb e_\th(\ka)\\
& -&\left(2\ddd_1\eta-\frac{1}{2}\vth\vthb+2\eta^2\right)\ze  -\frac 1 2  e_\th(\vthb\, \vth) -\frac{1}{2}\vth^2\xib-\frac 3 2 \vth \vthb \eta.
 \eeaa
Together with the estimates for $\ka$ of Step 1, the estimates for $\kab$ of Step 3, and the estimates for $\ze$ of Step 4, 
\beaa
&&\max_{k\leq J+1}\int_{\Sigma_*}\left(\dk^k\left( r^2\ddd_2\dds_2\eta - r^2e_\th(\rho) -\frac{r^2}{2}\ddd_2(\eta^2)-r^2e_\th(\eta^2)+r^2\ddd_1(\ze\eta)  +\frac 1 4  r^2e_\th(\vthb\, \vth)\right)\right)^2\\
 &\les& \max_{k\leq J+1}\int_{\Sigma_*}r^{-2}|\dk^k\eta|^2+\Big(\,{}^{(\Si_*)}\mathfrak{R}_{J+1}[\Rc] +\,{}^{(\Si_*)}\mathfrak{G}_{J}[\Gac]+\ep_0\,{}^{(\Si_*)}\mathfrak{G}_{J+1}[\Gac]\Big)^2.
\eeaa
In view of the dominant condition \eqref{eq:behaviorofronSigmastar} for $r$ on $\Sigma_*$, we infer
\beaa
&&\max_{k\leq J+1}\int_{\Sigma_*}\left(\dk^k\left( r^2\ddd_2\dds_2\eta - r^2e_\th(\rho) -\frac{r^2}{2}\ddd_2(\eta^2)-r^2e_\th(\eta^2)+r^2\ddd_1(\ze\eta)  +\frac 1 4  r^2e_\th(\vthb\, \vth)\right)\right)^2\\
 &\les& \ep_0^{\frac{2}{3}}\max_{k\leq J+1}\int_{\Sigma_*}|\dk^k\eta|^2+\Big(\,{}^{(\Si_*)}\mathfrak{R}_{J+1}[\Rc] +\,{}^{(\Si_*)}\mathfrak{G}_{J}[\Gac]+\ep_0\,{}^{(\Si_*)}\mathfrak{G}_{J+1}[\Gac]\Big)^2.
\eeaa
This yields
\beaa
&&\max_{k\leq J+1}\int_{\Sigma_*}\Bigg( r^2\ddd_2\dds_2\dk^k\eta+r\dds_2[\dk^k,r\ddd_2]\eta+r\ddd_2[\dk^k,r\dds_2]\eta\\
&& - r^2e_\th(\dk^k\rho) -\frac{r^2}{2}\ddd_2\dk^k(\eta^2)-r^2e_\th\dk^k(\eta^2)+r^2\ddd_1\dk^k(\ze\eta)  +\frac 1 4  r^2e_\th\dk^k(\vthb\, \vth)\Bigg)^2\\
 &\les& \ep_0^{\frac{2}{3}}\max_{k\leq J+1}\int_{\Sigma_*}|\dk^k\eta|^2+\Big(\,{}^{(\Si_*)}\mathfrak{R}_{J+1}[\Rc] +\,{}^{(\Si_*)}\mathfrak{G}_{J}[\Gac]+\ep_0\,{}^{(\Si_*)}\mathfrak{G}_{J+1}[\Gac]\Big)^2.
\eeaa
We deduce, using a Poincar\'e inequality for $\ddd_2$,
\beaa
\max_{k\leq J+1}\int_{\Sigma_*}\left( r\dds_2\dk^k\eta\right)^2 &\les& \ep_0^{\frac{2}{3}}\max_{k\leq J+1}\int_{\Sigma_*}|\dk^k\eta|^2+\Big(\,{}^{(\Si_*)}\mathfrak{R}_{J+1}[\Rc] +\,{}^{(\Si_*)}\mathfrak{G}_{J}[\Gac]+\ep_0\,{}^{(\Si_*)}\mathfrak{G}_{J+1}[\Gac]\Big)^2.
\eeaa
Together with a Poincar\'e inequality for $r\dds_2$ and the above control of the $\ell=1$ mode of $\eta$, we infer
\beaa
\max_{k\leq J+1}\int_{\Sigma_*}\left(\dk^k\eta\right)^2 &\les& \ep_0^{\frac{2}{3}}\max_{k\leq J+1}\int_{\Sigma_*}|\dk^k\eta|^2+\Big(\,{}^{(\Si_*)}\mathfrak{R}_{J+1}[\Rc] +\,{}^{(\Si_*)}\mathfrak{G}_{J}[\Gac]+\ep_0\,{}^{(\Si_*)}\mathfrak{G}_{J+1}[\Gac]\Big)^2,
\eeaa
and hence, for $\ep_0$ small enough,
\beaa
\max_{k\leq J+1}\int_{\Sigma_*}\left(\dk^k\eta\right)^2 &\les& \Big(\,{}^{(\Si_*)}\mathfrak{R}_{J+1}[\Rc] +\,{}^{(\Si_*)}\mathfrak{G}_{J}[\Gac]+\ep_0\,{}^{(\Si_*)}\mathfrak{G}_{J+1}[\Gac]\Big)^2.
\eeaa

\noindent{\bf Step 6.} Recall from the GCM condition that we have on $\Sigma_*$
\beaa
\int_S\xib e^\Phi=0.
\eeaa
Together with the transport equation
\beaa
e_4(\xib) &=& -e_3(\ze)+\bb-\kab\ze-\ze\vthb,
\eeaa
we infer in view of the the estimates for $\ze$ of Step 4, the estimates for $\bb$, and the bootstrap assumptions
\beaa
\max_{k\leq J+1}\int_1^{u_*}r^{-4}\left(\dk^k\left(\int_S\xib e^\Phi\right)\right)^2 &\les& \Big(\,{}^{(\Si_*)}\mathfrak{R}_{J+1}[\Rc] +\,{}^{(\Si_*)}\mathfrak{G}_{J}[\Gac]+\ep_0\,{}^{(\Si_*)}\mathfrak{G}_{J+1}[\Gac]\Big)^2.
\eeaa

Next, from  Proposition \ref{prop:eqtsfor-ometaxib} that we have
\beaa
2\ddd_2\dds_2\xib&=&-e_3(e_\th(\kab))+ \kab\left(e_3(\ze) -\bb\right) +\kab^2\ze-\frac 3 2 \kab e_\th \kab + 6\rho\xib -2\omb e_\th(\kab)+\err[\ddd_2\dds_2\xib],\\
\err[\ddd_2\dds_2\xib]&=&\left(2\ddd_1\xib+\frac{1}{2}\kab\,\vthb+2\eta\xib-\frac{1}{2}\vthb^2\right)\eta + 2e_\th(\eta\xib)  -\frac{1}{2}e_\th(\vthb^2)\\
&+&\kab\left(\frac{1}{2}\vthb\ze  -\frac{1}{2}\vth\xib\right)-\frac 1 2 \vthb e_\th \kab  -\frac{1}{2}\vth\vthb\xib
-\ze\left(2\ddd_1\xib+2(\eta-3\ze)\xib-\frac{1}{2}\vthb^2\right)\\
&+&\xib\Big(-\vth\vthb -2\ddd_1\ze+2\ze^2\Big)-6\eta\ze\xib -6e_\th(\ze\xib).
\eeaa
Together with the estimates for $\ka$ of Step 1, the estimates for $\kab$ of Step 3, and the estimates for $\ze$ of Step 4, 
\beaa
&&\max_{k\leq J+1}\int_{\Sigma_*}\left(\dk^k\left( r^2\ddd_2\dds_2\xib +\frac{1}{2}e_\th(e_3(\check{\kab})) -\eta\ddd_1\xib - e_\th(\eta\xib)  +\frac{1}{4}e_\th(\vthb^2)+\ddd_2(\ze\xib)  +3e_\th(\ze\xib)\right)\right)^2\\
 &\les& \max_{k\leq J+1}\int_{\Sigma_*}r^{-2}|\dk^k\xib|^2+\Big(\,{}^{(\Si_*)}\mathfrak{R}_{J+1}[\Rc] +\,{}^{(\Si_*)}\mathfrak{G}_{J}[\Gac]+\ep_0\,{}^{(\Si_*)}\mathfrak{G}_{J+1}[\Gac]\Big)^2.
\eeaa
In view of the dominant condition \eqref{eq:behaviorofronSigmastar} for $r$ on $\Sigma_*$, we infer
\beaa
&&\max_{k\leq J+1}\int_{\Sigma_*}\left(\dk^k\left( r^2\ddd_2\dds_2\xib +\frac{1}{2}e_\th(e_3(\check{\kab})) -\eta\ddd_1\xib - e_\th(\eta\xib)  +\frac{1}{4}e_\th(\vthb^2)+\ddd_2(\ze\xib)  +3e_\th(\ze\xib)\right)\right)^2\\
 &\les& \ep_0^{\frac{2}{3}}\max_{k\leq J+1}\int_{\Sigma_*}|\dk^k\xib|^2+\Big(\,{}^{(\Si_*)}\mathfrak{R}_{J+1}[\Rc] +\,{}^{(\Si_*)}\mathfrak{G}_{J}[\Gac]+\ep_0\,{}^{(\Si_*)}\mathfrak{G}_{J+1}[\Gac]\Big)^2.
\eeaa
This yields
\beaa
&&\max_{k\leq J+1}\int_{\Sigma_*}\Bigg(r^2\ddd_2\dds_2\dk^k\xib+r\dds_2[\dk^k,r\ddd_2]\xib+r\ddd_2[\dk^k,r\dds_2]\xib\\
&& +\frac{1}{2}e_\th(\dk^ke_3(\check{\kab}))  -\ddd_1(\eta\dk^k\xib) - e_\th\dk^k(\eta\xib)  +\frac{1}{4}e_\th\dk^k(\vthb^2)+\ddd_2\dk^k(\ze\xib)  +3e_\th\dk^k(\ze\xib)\Bigg)^2\\
 &\les& \ep_0^{\frac{2}{3}}\max_{k\leq J+1}\int_{\Sigma_*}|\dk^k\xib|^2+\Big(\,{}^{(\Si_*)}\mathfrak{R}_{J+1}[\Rc] +\,{}^{(\Si_*)}\mathfrak{G}_{J}[\Gac]+\ep_0\,{}^{(\Si_*)}\mathfrak{G}_{J+1}[\Gac]\Big)^2.
\eeaa
We deduce, using a Poincar\'e inequality for $\ddd_2$ and the estimates for $\kab$ of Step 3,
\beaa
\max_{k\leq J+1}\int_{\Sigma_*}\left( r\dds_2\dk^k\xib\right)^2 &\les& \ep_0^{\frac{2}{3}}\max_{k\leq J+1}\int_{\Sigma_*}|\dk^k\xib|^2+\Big(\,{}^{(\Si_*)}\mathfrak{R}_{J+1}[\Rc] +\,{}^{(\Si_*)}\mathfrak{G}_{J}[\Gac]+\ep_0\,{}^{(\Si_*)}\mathfrak{G}_{J+1}[\Gac]\Big)^2.
\eeaa
Together with a Poincar\'e inequality for $r\dds_2$ and the above control of the $\ell=1$ mode of $\xib$, we infer
\beaa
\max_{k\leq J+1}\int_{\Sigma_*}\left(\dk^k\xib\right)^2 &\les& \ep_0^{\frac{2}{3}}\max_{k\leq J+1}\int_{\Sigma_*}|\dk^k\xib|^2+\Big(\,{}^{(\Si_*)}\mathfrak{R}_{J+1}[\Rc] +\,{}^{(\Si_*)}\mathfrak{G}_{J}[\Gac]+\ep_0\,{}^{(\Si_*)}\mathfrak{G}_{J+1}[\Gac]\Big)^2,
\eeaa
and hence, for $\ep_0$ small enough,
\beaa
\max_{k\leq J+1}\int_{\Sigma_*}\left(\dk^k\xib\right)^2 &\les& \Big(\,{}^{(\Si_*)}\mathfrak{R}_{J+1}[\Rc] +\,{}^{(\Si_*)}\mathfrak{G}_{J}[\Gac]+\ep_0\,{}^{(\Si_*)}\mathfrak{G}_{J+1}[\Gac]\Big)^2.
\eeaa

\noindent{\bf Step 7.} Using the Codazzi for $\vth$ and $\vthb$, the transport equation for $\vth$ and $\vthb$ in the $e_4$ and $e_3$ direction, the control of $\check{\ka}$ of Step 1, the control of $\check{\kab}$ of Step 3, the control of $\ze$ of Step 4, the control of $\eta$ of Step 5, the control of $\xib$ of Step 6, and a Poincar\'e inequality for $\ddd_2$, we infer
\beaa
\max_{k\leq J+1}\int_{\Sigma_*}\Big(r^2(\dk^k\vth)^2+(\dk^k\vthb)^2\Big) &\les& \Big(\,{}^{(\Si_*)}\mathfrak{R}_{J+1}[\Rc] +\,{}^{(\Si_*)}\mathfrak{G}_{J}[\Gac]+\ep_0^{\frac{2}{3}}\,{}^{(\Si_*)}\mathfrak{G}_{J+1}[\Gac]\Big)^2.
\eeaa

\noindent{\bf Step 8.} Recall  form Proposition \ref{prop:eqtsfor-ometaxib} that   $\omb$ verifies
\beaa
  2\dds_1\omb &=& -\frac{1}{2}\ka\xib+\left(\frac{1}{2}\kab  +2\omb +\frac{1}{2}\vthb\right)\eta + e_3(\ze) -\bb \\
&& +\frac{1}{2}\kab\ze  -2\omb\ze +\frac{1}{2}\vthb\ze  -\frac{1}{2}\vth\xib.
 \eeaa
Together with a Poincar\'e inequality for $\dds_1$, the control of $\xib$ from Step 6, the control of $\eta$ from Step 5, and the control of $\ze$ from Step 4, we infer
\beaa
\max_{k\leq J+1}\int_{\Sigma_*}|\dk^k\check{\omb}|^2 &\les& \Big(\,{}^{(\Si_*)}\mathfrak{R}_{J+1}[\Rc] +\,{}^{(\Si_*)}\mathfrak{G}_{J}[\Gac]+\ep_0^{\frac{2}{3}}\,{}^{(\Si_*)}\mathfrak{G}_{J+1}[\Gac]\Big)^2.
\eeaa

Finally, gathering the estimates of Step 1 to Step 8, we infer
\beaa
\,{}^{(\Si_*)}\mathfrak{G}_{J+1}[\Gac]+\,{}^{(\Si_*)}\mathfrak{G}_{J+1}'[\Gac] &\les& \,{}^{(\Si_*)}\mathfrak{R}_{J+1}[\Rc] +\,{}^{(\Si_*)}\mathfrak{G}_{J}[\Gac]+\ep_0^{\frac{2}{3}}\,{}^{(\Si_*)}\mathfrak{G}_{J+1}[\Gac].
\eeaa
and hence, for $\ep_0$ small enough,
\beaa
\,{}^{(\Si_*)}\mathfrak{G}_{J+1}[\Gac]+\,{}^{(\Si_*)}\mathfrak{G}_{J+1}'[\Gac] &\les& \,{}^{(\Si_*)}\mathfrak{R}_{J+1}[\Rc] +\,{}^{(\Si_*)}\mathfrak{G}_{J}[\Gac]
\eeaa
as desired. This concludes the proof of Proposition \ref{prop:controlGaextiterationassupmtionThM8:actualresult:Sigma*}.


\subsection{Weighted estimates for transport equations along $e_4$ in $\Mext$}
\lab{weightedestimatesfortheproofofThmM8controlGaMext}


\begin{lemma}\lab{lemma:MorawetzforRicciinRR1}
Let the following transport equation in $\Mext$
$$e_4(f)+\frac{a}{2}\ka f=h$$
where $a\in\RRR$ is a given constant, and $f$ and $h$ are scalar functions. Also, let  and $\dt>0$. Then, $f$ satisfies
\beaa
\sup_{r_0\geq 4m_0}\left(r_0^{2a-2}\int_{\{r=r_0\}}f^2\right) &\les&  \int_{\Sigma_*}r^{2a-2} f^2 +\int_{\Mext(\geq 4m_0)} r^{2a-1+\dt}h^2.
\eeaa
\end{lemma}

\begin{proof}
Multiply by $f$ to obtain
$$\frac{1}{2}e_4(f^2)+\frac{a}{2}\ka f^2=hf.$$
Next, integrate over $S_{u,r}$ to obtain
\beaa
\frac{1}{2}e_4\left(\int_{S_{u,r}}f^2\right) &=& \int_{S_{u,r}}\frac{1}{2}(e_4(f^2)+\ka f^2)\\
&=& -\int_{S_{u,r}}\frac{a-1}{2}\ka f^2+\int_{S_{u,r}}hf\\
&=& -\frac{a-1}{2}\ov{\ka}\int_{S_{u,r}}f^2- \frac{a-1}{2}\int_{S_{u,r}}\kac f^2+\int_{S_{u,r}}hf
\eeaa
and hence
$$\frac{1}{2}e_4\left(\int_{S_{u,r}}f^2\right)+\frac{a-1}{2}\ov{\ka}\int_{S_{u,r}}f^2=- \frac{a-1}{2}\int_{S_{u,r}}\kac f^2+\int_{S_{u,r}}hf.$$
Also, we multiply by $r^{2a-2}$ which yields
$$\frac{1}{2}e_4\left(r^{2a-2}\int_{S_{u,r}}f^2 \right)= - \frac{a-1}{2}r^{2a-2}\int_{S_{u,r}}\kac f^2+r^{2a-2}\int_{S_{u,r}}hf$$
where we used the fact that $2e_4(r)=r\ov{\ka}$. This yields
$$-e_4\left(r^{2a-2}\int_{S_{u,r}}f^2 \right)\leq r^{2a-3-\dt}\int_{S_{u,r}}f^2 +\frac{1}{4}r^{2a-1+\dt}\int_{S_{u,r}}h^2 $$
and hence
$$-e_4\left(e^{-\dt^{-1}r^{-\dt}}r^{2a-2}\int_{S_{u,r}}f^2 \right)\les e^{-\dt^{-1}r^{-\dt}}r^{2a-1+\dt}\int_{S_{u,r}}h^2 $$
where we used  the fact that $2e_4(r)=r\ov{\ka}=2+O(\ep_0)$. Integrating between $r=r_0$ and $r=r_*(u)$, where $r_*(u)$ is such that $S_{u,r_*(u)}\subset\Sigma_*$, we infer
\bea\lab{eq:estimatebeforermkonvolumesinMextforThmM8}
r_0^{2a-2}\int_{S_{u,r_0}}f^2 \les r_*(u)^{2a-2}\int_{S_{u,r_*(u)}}f^2 +\int_{r_0}^{r_*(u)}r^{2a-1+\dt}\int_{S_{u,r}}h^2.
\eea

\begin{remark}\lab{rmk:volumesinMextforThmM8}
Note that we have the following consequences of the coarea formula
\beaa
d\Sigma_*=\vsi\sqrt{\frac{2}{\vsi}-\Up+\frac{r}{2}\Ab}\,\,\,d\mu_{u, \Si_*}du, \qquad d\{r=r_0\}= \frac{\vsi\sqrt{-\ov{\kab}-\Ab}}{\sqrt{\ov{\ka}}}\,\,\,d\mu_{u,r_0}du,
\eeaa
where we used in particular that $\Si_*=\{u+r=c_{\Si_*}\}$. Also, we have in $\Mext$
\beaa
d\MM=\frac{4\vsi^2}{r^2\ov{\ka}^2}d\mu_{u, r}dudr.
\eeaa
We infer, in $\Mext$, using in particular the dominant condition of $r$ on $\Si_*$, 
\beaa
d\Sigma_*=\left(1+O\left(\ep_0^{\frac{2}{3}}\right)\right)\,d\mu_{u, \Si_*}du, \quad d\{r=r_0\}=\sqrt{1-\frac{2m_0}{r_0}}(1+O(\ep_0))\,\,\,d\mu_{u,r_0}du,
\eeaa
and
\beaa
d\MM=(1+O(\ep_0))d\mu_{u, r}dudr.
\eeaa
\end{remark}

Integrating \eqref{eq:estimatebeforermkonvolumesinMextforThmM8} in $u\in [1, u_*]$, and relying on Remark \ref{rmk:volumesinMextforThmM8} we deduce for $r_0\geq 4m_0$
$$r_0^{2a-2}\int_{\{r=r_0\}}f^2 \les \int_{\Sigma_*}r^{2a-2} f^2 +\int_{\Mext(r\geq 4m_0)} r^{2a-\frac{1}{2}}h^2$$
as desired. This concludes the proof of the lemma.
\end{proof}

\begin{corollary}\lab{cor:MorawetzforRicciinRR1}
Let the following transport equation in $\Mext$
$$e_4(f)+\frac{a}{2}\ka f=h$$
where $a\in\RRR$ is a given constant, and $f$ and $h$ are scalar functions. Also, let  and $\dt>0$. Then, $f$ satisfies for $5\leq k\leq k_{large}+1$
\beaa
&&\sup_{r_0\geq 4m_0}\left(r_0^{2a-2}\int_{\{r=r_0\}}(\dk^kf)^2\right) \\
&\les&  \int_{\Sigma_*}r^{2a-2}(\dk^{\leq k}f)^2 +\sup_{r_0\geq 4m_0}\left(r_0^{2a-2}\int_{\{r=r_0\}}(\dk^{\leq k-1}f)^2\right)+\int_{\Mext(\geq 4m_0)} r^{2a-1+\dt}(\dk^{\leq k}h)^2\\
&&+\left(\sup_{\Mext(\geq 4m_0)}\big(r^{a}|\dk^{\leq k-5}f|\big)\right)^2\Big(\,{}^{(ext)}\mathfrak{G}^{\geq 4m_0}_{k-1}[\Gac]+\,{}^{(ext)}\mathfrak{G}^{\geq 4m_0}_{k}[\kac]\Big)^2.
\eeaa
\end{corollary}

\begin{proof}
We commute first differentiate the equation for $f$ with $(\dkb, \T)^l$ and obtain
\beaa
e_4((\dkb, \T)^lf)+\frac{a}{2}\ka (\dkb, \T)^lf &=& h_l,\\
h_l &:=&(\dkb, \T)^lh -[(\dkb, \T)^l, e_4]f -\frac{a}{2}[(\dkb, \T)^l,\ka]f.
\eeaa
In view of Lemma \ref{lemma:MorawetzforRicciinRR1}, we deduce
\beaa
\sup_{r_0\geq 4m_0}\left(r_0^{2a-2}\int_{\{r=r_0\}}((\dkb, \T)^lf)^2\right) &\les&  \int_{\Sigma_*}r^{2a-2}(\dk^{l}f)^2 +\int_{\Mext(\geq 4m_0)} r^{2a-1+\dt}h_{l}^2.
\eeaa
Now, we have the following schematic commutation formulas
\beaa
[\dkb, e_4] = \Ga_g\dk+\Ga_g,\quad [\T, e_4] = r^{-1}\Ga_b\dk,
\eeaa
Together with the definition of $h_{l}$ and for $5\leq l\leq k_{large}+1$, we deduce
\beaa
\int_{\Mext(\geq 4m_0)} r^{2a-1+\dt}h_{l}^2 &\les& \int_{\Mext(\geq 4m_0)} r^{2a-1+\dt}(\dk^{l}h)^2+\ep_0^2\int_{\Mext} r^{2a-5+\dt}(\dk^{\leq l}f)^2\\
&&+\left(\sup_{\Mext}\big(r^{a}|\dk^{\leq l-5}f|\big)\right)^2\Big(\,{}^{(ext)}\mathfrak{G}_{l-1}[\Gac]+\,{}^{(ext)}\mathfrak{G}_{l}[\kac]\Big)^2
\eeaa
and hence
\beaa
&&\sup_{r_0\geq \rh}\left(r_0^{2a-2}\int_{\{r=r_0\}}((\dkb, \T)^lf)^2\right) \\
&\les&  \int_{\Sigma_*}r^{2a-2}(\dk^{l}f)^2 +\int_{\Mext} r^{2a-1+\dt}(\dk^{l}h)^2+\ep_0^2\int_{\Mext(\geq 4m_0)} r^{2a-5+\dt}(\dk^{\leq l}f)^2\\
&&+\left(\sup_{\Mext(\geq 4m_0)}\big(r^{a}|\dk^{\leq l-5}f|\big)\right)^2\Big(\,{}^{(ext)}\mathfrak{G}^{\geq 4m_0}_{l-1}[\Gac]+\,{}^{(ext)}\mathfrak{G}^{\geq 4m_0}_{l}[\kac]\Big)^2
\eeaa
or,
\beaa
&&\sup_{r_0\geq 4m_0}\left(r_0^{2a-2}\int_{\{r=r_0\}}((\dkb, \T)^lf)^2\right) \\
&\les&  \int_{\Sigma_*}r^{2a-2}(\dk^{l}f)^2 +\int_{\Mext(\geq 4m_0)} r^{2a-1+\dt}(\dk^{l}h)^2+\ep_0^2\sup_{r_0\geq 4m_0}\left(r_0^{2a-2}\int_{\{r=r_0\}}(\dk^{\leq l}f)^2\right)\\
&&+\left(\sup_{\Mext(\geq 4m_0)}\big(r^{a}|\dk^{\leq l-5}f|\big)\right)^2\Big(\,{}^{(ext)}\mathfrak{G}^{\geq 4m_0}_{l-1}[\Gac]+\,{}^{(ext)}\mathfrak{G}^{\geq 4m_0}_{l}[\kac]\Big)^2.
\eeaa
Together with the first equation which yields
\beaa
re_4((\dkb, \T)^lf)+\frac{a}{2}r\ka (\dkb, \T)^lf &=& rh_l,
\eeaa
and hence
\beaa
(re_4)^j((\dkb, \T)^lf)+\frac{a}{2}(re_4)^{j-1}\Big(r\ka (\dkb, \T)^lf\Big) &=& (re_4)^{j-1}(rh_l),
\eeaa
we infer, for $\ep_0>0$ small enough,  and for $5\leq k\leq k_{large}+1$,
\beaa
&&\sup_{r_0\geq 4m_0}\left(r_0^{2a-2}\int_{\{r=r_0\}}(\dk^kf)^2\right) \\
&\les&  \int_{\Sigma_*}r^{2a-2}(\dk^{\leq k}f)^2 +\int_{\Mext(\geq 4m_0)} r^{2a-1+\dt}(\dk^{\leq k}h)^2+\sup_{r_0\geq 4m_0}\left(r_0^{2a-2}\int_{\{r=r_0\}}(\dk^{\leq k-1}h)^2\right)\\
&&+\sup_{r_0\geq 4m_0}\left(r_0^{2a-2}\int_{\{r=r_0\}}(\dk^{\leq k-1}f)^2\right)\\
&&+\left(\sup_{\Mext(\geq 4m_0)}\big(r^{a}|\dk^{\leq k-5}f|\big)\right)^2\Big(\,{}^{(ext)}\mathfrak{G}^{\geq 4m_0}_{k-1}[\Gac]+\,{}^{(ext)}\mathfrak{G}^{\geq 4m_0}_{k}[\kac]\Big)^2.
\eeaa
Using a trace estimate, we infer
\beaa
&&\sup_{r_0\geq 4m_0}\left(r_0^{2a-2}\int_{\{r=r_0\}}(\dk^kf)^2\right) \\
&\les&  \int_{\Sigma_*}r^{2a-2}(\dk^{\leq k}f)^2 +\int_{\Mext(\geq 4m_0)} r^{2a-1+\dt}(\dk^{\leq k}h)^2+\sup_{r_0\geq 4m_0}\left(r_0^{2a-2}\int_{\{r=r_0\}}(\dk^{\leq k-1}f)^2\right)\\
&&+\left(\sup_{\Mext(\geq 4m_0)}\big(r^{a}|\dk^{\leq k-5}f|\big)\right)^2\Big(\,{}^{(ext)}\mathfrak{G}^{\geq 4m_0}_{k-1}[\Gac]+\,{}^{(ext)}\mathfrak{G}^{\geq 4m_0}_{k}[\kac]\Big)^2
\eeaa
as desired. This concludes the proof of the corollary.
\end{proof}

\begin{lemma}\lab{lemma:MorawetzforRicciinRR1bis}
Let the following transport equation in $\Mext$
$$e_4(f)+\frac{a}{2}\ka f=h$$
where $a\in\RRR$ is a given constant, and $f$ and $h$ are scalar functions. Let $b>2a-2$. Then, $f$ satisfies
\beaa
 \sup_{r_0\geq 4m_0}\left(r_0^b\int_{\{r=r_0\}}f^2\right) +\int_{\Mext(\geq 4m_0)}r^{b-1}f^2 & \les& \int_{\Sigma_*}r^bf^2 +\int_{\Mext(\geq 4m_0)}r^{b+1}h^2 .
\eeaa
\end{lemma}

\begin{proof}
Recall from Lemma \ref{lemma:MorawetzforRicciinRR1} the following identity 
$$\frac{1}{2}e_4\left(\int_{S_{u,r}}f^2\right)+\frac{a-1}{2}\ov{\ka}\int_{S_{u,r}}f^2=- \frac{a-1}{2}\int_{S_{u,r}}\kac f^2+\int_{S_{u,r}}hf.$$
We multiply by $r^b$ which yields
\beaa
\frac{1}{2}e_4\left(r^b\int_{S_{u,r}}f^2 \right)+\frac{1}{2}\left(a-1-\frac{b}{2}\right)\ov{\ka}\int_{S_{u,r}}r^{b}f^2 =- \frac{a-1}{2}r^b\int_{S_{u,r}}\kac f^2+r^b\int_{S_{u,r}}hf 
\eeaa
where we used the fact that $2e_4(r)=r\ov{\ka}$. We choose $b>2a-2$ and integrate  between $r=r_0$ and $r=r_*(u)$, where $r_*(u)$ is such that $S_{u,r_*(u)}\subset\Sigma_*$, which yields
\beaa
\int_{S_{u,r_0}}r^bf^2 +\int_{r_0}^{r_*}\int_{S_{u,r}}r^{b-1}f^2  \les \int_{S_{u,r_*}}r^bf^2 +\int_{r_0}^{r_*}\int_{S_{u,r}}r^{b+1}h^2.
\eeaa
Then, integrating in $u$ in $u\in [1, u_*]$, and relying on Remark \ref{rmk:volumesinMextforThmM8} we deduce for $r_0\geq 4m_0$, 
\beaa
 r_0^b\int_{\{r=r_0\}}f^2 +\int_{\Mext\cap\{r\geq r_0\}}r^{b-1}f^2 & \les& \int_{\Sigma_*}r^bf^2 +\int_{\Mext(\geq 4m_0)}r^{b+1}h^2.
\eeaa
This concludes the proof of the lemma.
\end{proof}

\begin{corollary}\lab{cor:MorawetzforRicciinRR1bis}
Let the following transport equation in $\Mext$
$$e_4(f)+\frac{a}{2}\ka f=h$$
where $a\in\RRR$ is a given constant, and $f$ and $h$ are scalar functions. Let $b>2a-2$. Then, $f$ satisfies for $5\leq l\leq k_{large}+1$
\beaa
&&\sup_{r_0\geq 4m_0}\left(r_0^{b}\int_{\{r=r_0\}}(\dk^kf)^2\right)+\int_{\Mext(\geq 4m_0)}r^{b-1}(\dk^kf)^2 \\
&\les&  \int_{\Sigma_*}r^{b}(\dk^{\leq k}f)^2 +\sup_{r_0\geq 4m_0}\left(r_0^{b}\int_{\{r=r_0\}}(\dk^{\leq k-1}f)^2\right)+\int_{\Mext(\geq 4m_0)} r^{b-1}(\dk^{\leq k}h)^2\\
&& +\left(\sup_{\Mext(\geq 4m_0)}\big(r^{b}|\dk^{\leq k-5}f|\big)\right)^2\Big(\,{}^{(ext)}\mathfrak{G}^{\geq 4m_0}_{k-1}[\Gac]+\,{}^{(ext)}\mathfrak{G}^{\geq 4m_0}_{k}[\kac]\Big)^2.
\eeaa
\end{corollary}

\begin{proof}
The proof is based on Lemma \ref{lemma:MorawetzforRicciinRR1bis}. It is similar to the one of Corollary \ref{cor:MorawetzforRicciinRR1} and left to the reader.
\end{proof}

\begin{lemma}\lab{lemma:MorawetzforRicciinRR1forrleq4m0}
Let the following transport equation in $\Mext$
$$e_4(f)+\frac{a}{2}\ka f=h$$
where $a\in\RRR$ is a given constant, and $f$ and $h$ are scalar functions. 
Then, $f$ satisfies
\beaa
\sup_{\rh\leq r_0\leq 4m_0}\int_{\{r=r_0\}}f^2 &\les&  \int_{\{r=4m_0\}}f^2  + \int_{\Mext(\leq 4m_0)} h^2.
\eeaa
\end{lemma}

\begin{proof}
Let $b>2a-2$. Recall from Lemma \ref{lemma:MorawetzforRicciinRR1bis} the following identity
\beaa
\frac{1}{2}e_4\left(r^b\int_{S_{u,r}}f^2 \right)+\frac{1}{2}\left(a-1-\frac{b}{2}\right)\ov{\ka}\int_{S_{u,r}}r^{b}f^2 =- \frac{a-1}{2}r^b\int_{S_{u,r}}\kac f^2+r^b\int_{S_{u,r}}hf. 
\eeaa
Choosing $b=2a$, we obtain 
\beaa
\frac{1}{2}e_4\left(r^b\int_{S_{u,r}}f^2 \right)-\frac{1}{2}\ov{\ka}\int_{S_{u,r}}r^{b}f^2 =- \frac{a-1}{2}r^b\int_{S_{u,r}}\kac f^2+r^b\int_{S_{u,r}}hf. 
\eeaa

Next, let $1\leq u\leq u_*$ and $\rh\leq r_0\leq 4m_0$. We now integrate in $r_0\leq r\leq 4m_0$ and along $\CC_u$ in $\Mext$. Since $r$ is bounded on $\Mext(r\leq 4m_0)$ from above and below, we obtain, for $\ep_0>0$ small enough,
\beaa
\int_{S_{u,r_0}}f^2   &\les& \int_{S_{u,4m_0}}f^2 +\int_{\rh}^{4m_0}\int_{S_{\ub,r}} h^2.
\eeaa
We may now integrate in $u$ to deduce
\beaa
\int_1^{u_*}\int_{S_{u,r_0}}f^2   &\les& \int_1^{u_*}\int_{S_{u,4m_0 }}f^2 +\int_1^{u_*}\int_{\rh}^{4m_0}\int_{S_{u,r}} h^2.
\eeaa
Relying on Remark \ref{rmk:volumesinMextforThmM8} we deduce 
\beaa
\sup_{\rh\leq r_0\leq 4m_0}\int_{\{r=r_0\}}f^2 &\les&  \int_{\{r=4m_0\}}f^2  + \int_{\Mext(\leq 4m_0)} h^2.
\eeaa
as desired. This concludes the proof of the lemma.
\end{proof}

\begin{corollary}\lab{cor:MorawetzforRicciinRR1forrleq4m0}
Let the following transport equation in $\Mext$
$$e_4(f)+\frac{a}{2}\ka f=h$$
where $a\in\RRR$ is a given constant, and $f$ and $h$ are scalar functions. Then, $f$ satisfies for $5\leq l\leq k_{large}+1$
\beaa
&&\sup_{\rh\leq r_0\leq 4m_0}\int_{\{r=r_0\}}(\dk^kf)^2 \\
&\les&  \int_{\{r=4m_0\}}(\dk^{\leq k}f)^2 +\sup_{\rh\leq r_0\leq 4m_0}\int_{\{r=r_0\}}(\dk^{\leq k-1}f)^2+\int_{\Mext(\leq 4m_0)}(\dk^{\leq k}h)^2\\
&&+\left(\sup_{\Mext(r\leq 4m_0)}\big(|\dk^{\leq k-5}f|\big)\right)^2\Big(\,{}^{(ext)}\mathfrak{G}^{\leq 4m_0}_{k-1}[\Gac]+\,{}^{(ext)}\mathfrak{G}^{\leq 4m_0}_{k}[\kac]\Big)^2.
\eeaa
\end{corollary}

\begin{proof}
The proof is based on Lemma \ref{lemma:MorawetzforRicciinRR1forrleq4m0}. It is similar to the one of Corollary \ref{cor:MorawetzforRicciinRR1} and left to the reader.
\end{proof}


\subsection{Several identities}\lab{severalidentitiesfortheproofofThmM8controlGaMext}


The goal of this section is to prove the identities below that will be used to avoid loosing derivatives when controlling the weighted energies of the Ricci coefficients.
\begin{lemma}
We have
\beaa
&& e_4\left(\ddd_1\dds_1\ka-\vth\Big(\ddd_4\dds_3\ddd_2^{-1}+\dds_2\Big)\ddd_1^{-1}\check{\rho}\right)+2\ka\left(\ddd_1\dds_1\ka-\vth\Big(\ddd_4\dds_3\ddd_2^{-1}+\dds_2\Big)\ddd_1^{-1}\check{\rho}\right)\\ 
&=& -\left(-\frac{1}{2}\vth\dds_2+\ze e_4(\Phi) -\b\right)\dds_1\ka-\frac{1}{2}\vth\ddd_1\dds_1\ka +\frac{1}{2}\dds_1(\ka+\vth)\dds_1\ka +(\dds_1\ka)^2\\
&& -2\ka\vth\Big(\ddd_4\dds_3\ddd_2^{-1}+\dds_2\Big)\ddd_1^{-1}\check{\rho} +(\ka\vth+2\a)\Big(\ddd_4\dds_3\ddd_2^{-1}+\dds_2\Big)\ddd_1^{-1}\check{\rho}\\
&&+\vth\left[\Big(\ddd_4\dds_3\ddd_2^{-1}+\dds_2\Big)\ddd_1^{-1},e_4\right]\check{\rho} +\vth\Big(\ddd_4\dds_3\ddd_2^{-1}+\dds_2\Big)\ddd_1^{-1}\left(\frac{3}{2}\ov{\ka}\check{\rho}+\frac{3}{2}\ov{\rho}\check{\ka}-\err[e_4\check{\rho}]\right)\\
&&  -\frac{1}{2}\vth\ddd_4\dds_3\ddd_2^{-1}(-\dds_1\ka+\ka\ze-\vth\ze)-\frac{1}{2}\vth\dds_2(-\dds_1\ka+\ka\ze-\vth\ze)\\
&& +\frac{1}{2}\dds_3\ddd_2^{-1}(-2\b-\dds_1\ka+\ka\ze-\vth\ze)\dds_3\vth +\frac{1}{2}(-2\b-\dds_1\ka+\ka\ze-\vth\ze)\ddd_2\vth,
\eeaa
\beaa
 && e_4\Big(e_\th(\mu) +\vth\ddd_2\dds_2(\dds_1\ddd_1)^{-1}\bb+2\ze\check{\rho}\Big)+ 2\ka\Big(e_\th(\mu)+\vth\ddd_2\dds_2(\dds_1\ddd_1)^{-1}\bb+2\ze\check{\rho}\Big)\\
  &=& -\frac{3}{2}\mu e_\th(\ka)   -\frac{1}{2}\vth e_\th(\mu)  -(\ka\vth+2\a)\ddd_2\dds_2(\dds_1\ddd_1)^{-1}\bb\\
&&-\vth\Big[\ddd_2\dds_2(\dds_1\ddd_1)^{-1}, e_4\Big]\bb-\vth\ddd_2\dds_2(\dds_1\ddd_1)^{-1}\left(\ka\bb+3\rho\ze+\vthb\b\right)\\
  && +\dds_3\vth\dds_2\ddd_1^{-1}\check{\rho}    - e_\th\left(\vth\dds_2\ddd_1^{-1}\left(-\check{\mu}+\frac{1}{4}\vth\vthb\right)\right)   -e_\th\left(\vth\left(\frac{1}{8}\kab\vth+\ze^2\right)\right)\\
  &&-2\ze\ddd_1\dds_1\ka  -2e_\th(\ka)\dds_2\ze    -2(\ka\ze+\b+\vth\ze)\check{\rho} -2\ze\left(\frac{3}{2}\ov{\ka}\check{\rho}+\frac{3}{2}\ov{\rho}\check{\ka}-\err[e_4\check{\rho}]\right)\\
  && +2\b\dds_2\ze  +\frac{3}{2}e_\th\left(\ka\ze^2\right) +2\ka\Big(\vth\ddd_2\dds_2(\dds_1\ddd_1)^{-1}\bb+2\ze\check{\rho}\Big),
 \eeaa
 and
 \beaa
&& e_4(e_\th(\kab)-4\bb)+\ka(e_\th(\kab)-4\bb)\\
 &=& 2e_\th(\mu)+12\rho\ze -\frac{1}{2}\kab e_\th(\ka) +4\vthb\b -\frac{1}{2}\vth e_\th(\kab) -e_\th(\vth\vthb)+2e_\th(\ze^2)\\
  &=& 2\Big(e_\th(\mu)+\vth\ddd_2\dds_2(\dds_1\ddd_1)^{-1}\bb+2\ze\check{\rho}\Big)+12\rho\ze -\frac{1}{2}\kab e_\th(\ka) \\
  &&-2\vth\ddd_2\dds_2(\dds_1\ddd_1)^{-1}\bb+2\ze\check{\rho}  +4\vthb\b -\frac{1}{2}\vth (e_\th(\kab)-4\bb)-2\vth\bb -e_\th(\vth\vthb)+2e_\th(\ze^2).
\eeaa
\end{lemma}

\begin{proof}
Recall Raychadhuri
\beaa
e_4(\ka) +\frac{1}{2}\ka^2 &=& -\frac{1}{2}\vth^2.
\eeaa
We commute with $\ddd_1\dds_1$ which yields 
\beaa
&& e_4(\ddd_1\dds_1\ka)+2\ka\ddd_1\dds_1\ka\\ 
&=& -\left(-\frac{1}{2}\vth\dds_2+\ze e_4(\Phi) -\b\right)\dds_1\ka-\frac{1}{2}\vth\ddd_1\dds_1\ka +\frac{1}{2}\dds_1(\ka+\vth)\dds_1\ka +(\dds_1\ka)^2  -\frac{1}{2}\ddd_1\dds_1(\vth^2).
\eeaa
We have in view of Codazzi for $\vth$ 
\beaa
\frac{1}{2}\ddd_1\dds_1(\vth^2) &=& \ddd_1(\vth\dds_1\vth)\\
&=& \frac{1}{2}\ddd_1(\vth\dds_3\vth -\vth\ddd_2\vth)\\
&=& \frac{1}{2}\ddd_1\Big(\vth\dds_3\ddd_2^{-1}(-2\b-\dds_1\ka+\ka\ze-\vth\ze) -\vth(-2\b-\dds_1\ka+\ka\ze-\vth\ze)\Big)\\
&=& \frac{1}{2}\vth\ddd_4\dds_3\ddd_2^{-1}(-2\b-\dds_1\ka+\ka\ze-\vth\ze)+\frac{1}{2}\vth\dds_2(-2\b-\dds_1\ka+\ka\ze-\vth\ze)\\
&& -\frac{1}{2}\dds_3\ddd_2^{-1}(-2\b-\dds_1\ka+\ka\ze-\vth\ze)\dds_3\vth -\frac{1}{2}(-2\b-\dds_1\ka+\ka\ze-\vth\ze)\ddd_2\vth\\
&=& -\vth\ddd_4\dds_3\ddd_2^{-1}\b -\vth\dds_2\b +\frac{1}{2}\vth\ddd_4\dds_3\ddd_2^{-1}(-\dds_1\ka+\ka\ze-\vth\ze)+\frac{1}{2}\vth\dds_2(-\dds_1\ka+\ka\ze-\vth\ze)\\
&& -\frac{1}{2}\dds_3\ddd_2^{-1}(-2\b-\dds_1\ka+\ka\ze-\vth\ze)\dds_3\vth -\frac{1}{2}(-2\b-\dds_1\ka+\ka\ze-\vth\ze)\ddd_2\vth.
\eeaa
Together with Bianchi for $e_4(\check{\rho})$, we infer
\beaa
\frac{1}{2}\ddd_1\dds_1(\vth^2) &=& -\vth\Big(\ddd_4\dds_3\ddd_2^{-1}+\dds_2\Big)\ddd_1^{-1}\left(e_4(\check{\rho})+\frac{3}{2}\ov{\ka}\check{\rho}+\frac{3}{2}\ov{\rho}\check{\ka}-\err[e_4\check{\rho}]\right)\\
&&  +\frac{1}{2}\vth\ddd_4\dds_3\ddd_2^{-1}(-\dds_1\ka+\ka\ze-\vth\ze)+\frac{1}{2}\vth\dds_2(-\dds_1\ka+\ka\ze-\vth\ze)\\
&& -\frac{1}{2}\dds_3\ddd_2^{-1}(-2\b-\dds_1\ka+\ka\ze-\vth\ze)\dds_3\vth -\frac{1}{2}(-2\b-\dds_1\ka+\ka\ze-\vth\ze)\ddd_2\vth\\
&=& -\vth e_4\left(\Big(\ddd_4\dds_3\ddd_2^{-1}+\dds_2\Big)\ddd_1^{-1}\check{\rho}\right)-\vth\left[\Big(\ddd_4\dds_3\ddd_2^{-1}+\dds_2\Big)\ddd_1^{-1},e_4\right]\check{\rho}\\
&& -\vth\Big(\ddd_4\dds_3\ddd_2^{-1}+\dds_2\Big)\ddd_1^{-1}\left(\frac{3}{2}\ov{\ka}\check{\rho}+\frac{3}{2}\ov{\rho}\check{\ka}-\err[e_4\check{\rho}]\right)\\
&&  +\frac{1}{2}\vth\ddd_4\dds_3\ddd_2^{-1}(-\dds_1\ka+\ka\ze-\vth\ze)+\frac{1}{2}\vth\dds_2(-\dds_1\ka+\ka\ze-\vth\ze)\\
&& -\frac{1}{2}\dds_3\ddd_2^{-1}(-2\b-\dds_1\ka+\ka\ze-\vth\ze)\dds_3\vth -\frac{1}{2}(-2\b-\dds_1\ka+\ka\ze-\vth\ze)\ddd_2\vth.
\eeaa
In view of the null structure equation for $e_4(\vth)$, we infer
\beaa
\frac{1}{2}\ddd_1\dds_1(\vth^2) &=& -e_4\left(\vth\Big(\ddd_4\dds_3\ddd_2^{-1}+\dds_2\Big)\ddd_1^{-1}\check{\rho}\right) -(\ka\vth+2\a)\Big(\ddd_4\dds_3\ddd_2^{-1}+\dds_2\Big)\ddd_1^{-1}\check{\rho}\\
&&-\vth\left[\Big(\ddd_4\dds_3\ddd_2^{-1}+\dds_2\Big)\ddd_1^{-1},e_4\right]\check{\rho} -\vth\Big(\ddd_4\dds_3\ddd_2^{-1}+\dds_2\Big)\ddd_1^{-1}\left(\frac{3}{2}\ov{\ka}\check{\rho}+\frac{3}{2}\ov{\rho}\check{\ka}-\err[e_4\check{\rho}]\right)\\
&&  +\frac{1}{2}\vth\ddd_4\dds_3\ddd_2^{-1}(-\dds_1\ka+\ka\ze-\vth\ze)+\frac{1}{2}\vth\dds_2(-\dds_1\ka+\ka\ze-\vth\ze)\\
&& -\frac{1}{2}\dds_3\ddd_2^{-1}(-2\b-\dds_1\ka+\ka\ze-\vth\ze)\dds_3\vth -\frac{1}{2}(-2\b-\dds_1\ka+\ka\ze-\vth\ze)\ddd_2\vth.
\eeaa
This yields
\beaa
&& e_4\left(\ddd_1\dds_1\ka-\vth\Big(\ddd_4\dds_3\ddd_2^{-1}+\dds_2\Big)\ddd_1^{-1}\check{\rho}\right)+2\ka\left(\ddd_1\dds_1\ka-\vth\Big(\ddd_4\dds_3\ddd_2^{-1}+\dds_2\Big)\ddd_1^{-1}\check{\rho}\right)\\ 
&=& -\left(-\frac{1}{2}\vth\dds_2+\ze e_4(\Phi) -\b\right)\dds_1\ka-\frac{1}{2}\vth\ddd_1\dds_1\ka +\frac{1}{2}\dds_1(\ka+\vth)\dds_1\ka +(\dds_1\ka)^2\\
&& -2\ka\vth\Big(\ddd_4\dds_3\ddd_2^{-1}+\dds_2\Big)\ddd_1^{-1}\check{\rho} +(\ka\vth+2\a)\Big(\ddd_4\dds_3\ddd_2^{-1}+\dds_2\Big)\ddd_1^{-1}\check{\rho}\\
&&+\vth\left[\Big(\ddd_4\dds_3\ddd_2^{-1}+\dds_2\Big)\ddd_1^{-1},e_4\right]\check{\rho} +\vth\Big(\ddd_4\dds_3\ddd_2^{-1}+\dds_2\Big)\ddd_1^{-1}\left(\frac{3}{2}\ov{\ka}\check{\rho}+\frac{3}{2}\ov{\rho}\check{\ka}-\err[e_4\check{\rho}]\right)\\
&&  -\frac{1}{2}\vth\ddd_4\dds_3\ddd_2^{-1}(-\dds_1\ka+\ka\ze-\vth\ze)-\frac{1}{2}\vth\dds_2(-\dds_1\ka+\ka\ze-\vth\ze)\\
&& +\frac{1}{2}\dds_3\ddd_2^{-1}(-2\b-\dds_1\ka+\ka\ze-\vth\ze)\dds_3\vth +\frac{1}{2}(-2\b-\dds_1\ka+\ka\ze-\vth\ze)\ddd_2\vth.
\eeaa

Next, recall that we have
\beaa
 e_4\mu + \frac{3}{2}\ka\mu &=&   - \vth\dds_2\ze-\vth\left(\frac{1}{8}\kab\vth+\ze^2\right)+\left(2e_\th(\ka)-2\b+\frac{3}{2}\ka\ze\right)\ze
 \eeaa
We commute with $e_\th$ which yields 
\beaa
 && e_4(e_\th(\mu)) + 2\ka e_\th(\mu)\\
  &=& -\frac{3}{2}\mu e_\th(\ka)   -\frac{1}{2}\vth e_\th(\mu) - e_\th\left(\vth\dds_2\ddd_1^{-1}\left(-\check{\mu}-\check{\rho}+\frac{1}{4}\vth\vthb\right)\right)-e_\th\left(\vth\left(\frac{1}{8}\kab\vth+\ze^2\right)\right)\\
  &&+e_\th\left(\left(2e_\th(\ka)-2\b+\frac{3}{2}\ka\ze\right)\ze\right)\\
  &=& -\frac{3}{2}\mu e_\th(\ka)   -\frac{1}{2}\vth e_\th(\mu) + \vth\ddd_2\dds_2(\dds_1\ddd_1)^{-1}\dds_1\rho  +\dds_3\vth\dds_2\ddd_1^{-1}\check{\rho} \\
   &&   - e_\th\left(\vth\dds_2\ddd_1^{-1}\left(-\check{\mu}+\frac{1}{4}\vth\vthb\right)\right)   -e_\th\left(\vth\left(\frac{1}{8}\kab\vth+\ze^2\right)\right)\\
  &&-2\ze\ddd_1\dds_1\ka  -2e_\th(\ka)\dds_2\ze   -2\ze\ddd_1\b  +2\b\dds_2\ze  +\frac{3}{2}e_\th\left(\ka\ze^2\right).
 \eeaa
Now, using the Bianchi identities for $e_4(\bb)$ and $e_4(\check{\rho})$, we have
\beaa
\vth\ddd_2\dds_2(\dds_1\ddd_1)^{-1}\dds_1\rho &=& -\vth\ddd_2\dds_2(\dds_1\ddd_1)^{-1}\left(e_4\bb+\ka\bb+3\rho\ze+\vthb\b\right)\\
&=& -e_4\Big(\vth\ddd_2\dds_2(\dds_1\ddd_1)^{-1}\bb\Big)+e_4(\vth)\ddd_2\dds_2(\dds_1\ddd_1)^{-1}\bb-\vth\Big[\ddd_2\dds_2(\dds_1\ddd_1)^{-1}, e_4\Big]\bb\\
&&-\vth\ddd_2\dds_2(\dds_1\ddd_1)^{-1}\left(\ka\bb+3\rho\ze+\vthb\b\right)\\
&=& -e_4\Big(\vth\ddd_2\dds_2(\dds_1\ddd_1)^{-1}\bb\Big)-(\ka\vth+2\a)\ddd_2\dds_2(\dds_1\ddd_1)^{-1}\bb\\
&&-\vth\Big[\ddd_2\dds_2(\dds_1\ddd_1)^{-1}, e_4\Big]\bb-\vth\ddd_2\dds_2(\dds_1\ddd_1)^{-1}\left(\ka\bb+3\rho\ze+\vthb\b\right)
\eeaa
and 
\beaa
\ze\ddd_1\b &=& \ze\left(e_4(\check{\rho})+\frac{3}{2}\ov{\ka}\check{\rho}+\frac{3}{2}\ov{\rho}\check{\ka}-\err[e_4\check{\rho}]\right)\\
&=& e_4(\ze\check{\rho})-e_4(\ze)\check{\rho} +\ze\left(\frac{3}{2}\ov{\ka}\check{\rho}+\frac{3}{2}\ov{\rho}\check{\ka}-\err[e_4\check{\rho}]\right)\\
&=& e_4(\ze\check{\rho})+(\ka\ze+\b+\vth\ze)\check{\rho} +\ze\left(\frac{3}{2}\ov{\ka}\check{\rho}+\frac{3}{2}\ov{\rho}\check{\ka}-\err[e_4\check{\rho}]\right).
\eeaa
We infer
\beaa
 && e_4\Big(e_\th(\mu)) +\vth\ddd_2\dds_2(\dds_1\ddd_1)^{-1}\bb+2\ze\check{\rho}\Big)+ 2\ka e_\th(\mu)\\
  &=& -\frac{3}{2}\mu e_\th(\ka)   -\frac{1}{2}\vth e_\th(\mu)  -(\ka\vth+2\a)\ddd_2\dds_2(\dds_1\ddd_1)^{-1}\bb\\
&&-\vth\Big[\ddd_2\dds_2(\dds_1\ddd_1)^{-1}, e_4\Big]\bb-\vth\ddd_2\dds_2(\dds_1\ddd_1)^{-1}\left(\ka\bb+3\rho\ze+\vthb\b\right)\\
  && +\dds_3\vth\dds_2\ddd_1^{-1}\check{\rho}    - e_\th\left(\vth\dds_2\ddd_1^{-1}\left(-\check{\mu}+\frac{1}{4}\vth\vthb\right)\right)   -e_\th\left(\vth\left(\frac{1}{8}\kab\vth+\ze^2\right)\right)\\
  &&-2\ze\ddd_1\dds_1\ka  -2e_\th(\ka)\dds_2\ze    -2(\ka\ze+\b+\vth\ze)\check{\rho} -2\ze\left(\frac{3}{2}\ov{\ka}\check{\rho}+\frac{3}{2}\ov{\rho}\check{\ka}-\err[e_4\check{\rho}]\right)\\
  && +2\b\dds_2\ze  +\frac{3}{2}e_\th\left(\ka\ze^2\right)
 \eeaa
and hence
\beaa
 && e_4\Big(e_\th(\mu) +\vth\ddd_2\dds_2(\dds_1\ddd_1)^{-1}\bb+2\ze\check{\rho}\Big)+ 2\ka\Big(e_\th(\mu)+\vth\ddd_2\dds_2(\dds_1\ddd_1)^{-1}\bb+2\ze\check{\rho}\Big)\\
  &=& -\frac{3}{2}\mu e_\th(\ka)   -\frac{1}{2}\vth e_\th(\mu)  -(\ka\vth+2\a)\ddd_2\dds_2(\dds_1\ddd_1)^{-1}\bb\\
&&-\vth\Big[\ddd_2\dds_2(\dds_1\ddd_1)^{-1}, e_4\Big]\bb-\vth\ddd_2\dds_2(\dds_1\ddd_1)^{-1}\left(\ka\bb+3\rho\ze+\vthb\b\right)\\
  && +\dds_3\vth\dds_2\ddd_1^{-1}\check{\rho}    - e_\th\left(\vth\dds_2\ddd_1^{-1}\left(-\check{\mu}+\frac{1}{4}\vth\vthb\right)\right)   -e_\th\left(\vth\left(\frac{1}{8}\kab\vth+\ze^2\right)\right)\\
  &&-2\ze\ddd_1\dds_1\ka  -2e_\th(\ka)\dds_2\ze    -2(\ka\ze+\b+\vth\ze)\check{\rho} -2\ze\left(\frac{3}{2}\ov{\ka}\check{\rho}+\frac{3}{2}\ov{\rho}\check{\ka}-\err[e_4\check{\rho}]\right)\\
  && +2\b\dds_2\ze  +\frac{3}{2}e_\th\left(\ka\ze^2\right) +2\ka\Big(\vth\ddd_2\dds_2(\dds_1\ddd_1)^{-1}\bb+2\ze\check{\rho}\Big).
 \eeaa

Finally, recall that we have
\beaa
e_4(\kab)+\frac{1}{2}\ka\kab &=& -2\ddd_1\ze+2\rho -\frac{1}{2}\vth\vthb+2\ze^2\\
&=& 2\mu+4\rho -\vth\vthb+2\ze^2.
\eeaa
We commute with $e_\th$ which yields 
\beaa
e_4(e_\th(\kab))+\ka e_\th(\kab) &=& 2e_\th(\mu)+4e_\th(\rho) -\frac{1}{2}\kab e_\th(\ka) -\frac{1}{2}\vth e_\th(\kab) -e_\th(\vth\vthb)+2e_\th(\ze^2).
\eeaa
Together with Bianchi for $e_4(\bb)$, we infer
\beaa
&& e_4(e_\th(\kab)-4\bb)+\ka(e_\th(\kab)-4\bb)\\
 &=& 2e_\th(\mu)+12\rho\ze -\frac{1}{2}\kab e_\th(\ka) +4\vthb\b -\frac{1}{2}\vth e_\th(\kab) -e_\th(\vth\vthb)+2e_\th(\ze^2)\\
  &=& 2\Big(e_\th(\mu)+\vth\ddd_2\dds_2(\dds_1\ddd_1)^{-1}\bb+2\ze\check{\rho}\Big)+12\rho\ze -\frac{1}{2}\kab e_\th(\ka) \\
  &&-2\vth\ddd_2\dds_2(\dds_1\ddd_1)^{-1}\bb+2\ze\check{\rho}  +4\vthb\b -\frac{1}{2}\vth (e_\th(\kab)-4\bb)-2\vth\bb -e_\th(\vth\vthb)+2e_\th(\ze^2).
\eeaa
This concludes the proof of the lemma.
\end{proof}


\subsection{Proof of Proposition \ref{prop:controlGaextiterationassupmtionThM8:actualresult}}\lab{sec:proofofprop:controlGaextiterationassupmtionThM8:actualresult}


We introduce the following notation which will constantly appear on the RHS of the equalities below
\bea\lab{eq:defintionofnormNgeq4m0JGacRcproofThmM8}
\nn N^{\geq 4m_0}[J, \Gac, \Rc] &:=& \,{}^{(\Si_*)}\mathfrak{G}_{J+1}[\Gac] +\,{}^{(\Si_*)}\mathfrak{G}_{J+1}'[\Gac] +\,{}^{(ext)}\mathfrak{R}_{J+1}[\Rc] +\,{}^{(ext)}\mathfrak{G}_{J}[\Gac]\\
&&+\ep_0\Big( \,{}^{(ext)}\mathfrak{G}^{\geq 4m_0}_{J+1}[\Gac] + \,{}^{(ext)}{\mathfrak{G}^{\geq 4m_0}_{J+1}}'[\Gac]\Big).
 \eea

{\bf Step 1.} Recall that 
\beaa
e_4(\vth) +\ka\vth &=& -2\a.
\eeaa
In view of Corollary \ref{cor:MorawetzforRicciinRR1} with $a=2$, we have for any $r_0\geq 4m_0$
\beaa
\max_{k\leq J+1}\sup_{r_0\geq 4m_0}r_0^2\int_{\{r=r_0\}}(\dk^k\vth)^2 &\les& \Big(N^{\geq 4m_0}[J, \Gac, \Rc] \Big)^2.
 \eeaa

{\bf Step 2.} Next, recall that 
\beaa
e_4(\check{\ka}) +\overline{\ka}\,\check{\ka} &=&  -\frac{1}{4}\vth^2+\frac{1}{4}\overline{\vth}^2-\overline{\check{\ka}^2}.
\eeaa
In view of Corollary \ref{cor:MorawetzforRicciinRR1} with $a=2$, we have for any $r_0\geq 4m_0$
\beaa
 \max_{k\leq J+2}\sup_{r_0\geq 4m_0}r_0^2\int_{\{r=r_0\}}(\dk^k\check{\ka})^2 &\les& \Big(N^{\geq 4m_0}[J, \Gac, \Rc] \Big)^2
\eeaa
 where we have used the null structure equations for $e_4(\vth)$, $e_3(\vth)$ and $\ddd_2\vth$ to avoid a loss of one derivative for the RHS.  

{\bf Step 3.} Next, recall that 
\beaa
e_4(\ze) +\ka\ze &=& -\b -\vth\ze.
\eeaa
In view of Corollary \ref{cor:MorawetzforRicciinRR1} with $a=2$, we have for any $r_0\geq 4m_0$
\beaa
\max_{k\leq J+1}\sup_{r_0\geq 4m_0}r_0^2\int_{\{r=r_0\}}(\dk^k\ze)^2 &\les& \Big(N^{\geq 4m_0}[J, \Gac, \Rc] \Big)^2.
\eeaa

{\bf Step 4.} Next, recall that 
\beaa
 e_4\left(\check{\mu}\right) + \frac{3}{2}\overline{\ka}\check{\mu} &=&   - \frac{3}{2}\overline{\mu}\check{\ka} +\err[e_4\check{\mu}].
 \eeaa
In view of Corollary \ref{cor:MorawetzforRicciinRR1} with $a=3$, commuting with $\dkb$ and $\T$, we have for any $r_0\geq 4m_0$
\beaa
\max_{k\leq J+1}\sup_{r_0\geq 4m_0} r_0^4\int_{\{r=r_0\}}(\dk^k\check{\mu})^2 &\les&  \Big(N^{\geq 4m_0}[J, \Gac, \Rc] \Big)^2
\eeaa
where we used the estimates for $\check{\ka}$ on $\Mext$  derived in Step 2.

{\bf Step 5.} Next, recall that 
\beaa
e_4(\check{\kab}) + \frac{1}{2}\overline{\ka}\check{\kab}  &=& -\frac{1}{2}\overline{\kab}\check{\ka}+2\check{\rho}-2\ddd_1\ze +\err[e_4\check{\kab}].
\eeaa
In view of Corollary \ref{cor:MorawetzforRicciinRR1bis} with $a=1$ and $b=2-\dt$ which satisfy the constraint $b>2a-2$, we have for any $r_0\geq 4m_0$
\beaa
  \max_{k\leq J+1}\sup_{r_0\geq 4m_0}\left(r_0^{2-\dt}\int_{\{r=r_0\}}(\dk^k\check{\kab})^2\right)+\int_{\Mext(r\geq 4m_0)}r^{1-\dt}(\dk^k\check{\kab})^2 & \les& \Big(N^{\geq 4m_0}[J, \Gac, \Rc] \Big)^2
\eeaa
where we used the estimates for $\check{\ka}$ and $\check{\mu}$ on $\Mext$  derived respectively in Step 2 and Step 4.

{\bf Step 6.} Next, recall that 
\beaa
e_4(\vthb) + \frac{1}{2}\ka\vthb &=& 2\dds_2\ze -\frac{1}{2}\kab\vth+2\ze^2\\
&=& 2\dds_2\ddd_1^{-1}\left(-\mu-\rho+\frac{1}{4}\vth\vthb\right)  -\frac{1}{2}\kab\vth+2\ze^2.
\eeaa
In view of Corollary \ref{cor:MorawetzforRicciinRR1} with $a=1$, we have for any $r_0\geq 4m_0$
\beaa
 \max_{k\leq J+1}\sup_{r_0\geq 4m_0}\int_{\{r=r_0\}}(\dk^k\vthb)^2 &\les& \Big(N^{\geq 4m_0}[J, \Gac, \Rc] \Big)^2
\eeaa
where we used the estimates for $\vth$ and $\check{\mu}$ on $\Mext$  derived respectively in Step 1 and Step 4.

{\bf Step 7.} Next, recall that 
\beaa
e_4(\check{\omb}) &=& \check{\rho} + 3\ze^2 - 3\overline{\ze^2} - \overline{\check{\ka}\check{\omb}}.
\eeaa
In view of Corollary \ref{cor:MorawetzforRicciinRR1bis} with $a=0$ and $b=0$ which satisfy the constraint $b>2a-2$, we have for any $r_0\geq 4m_0$
\beaa
  \max_{k\leq J+1}\sup_{r_0\geq 4m_0}\left(\int_{\{r=r_0\}}(\dk^k\check{\omb})^2\right)+\int_{\Mext(\geq 4m_0)}r^{-1}(\dk^k\check{\omb})^2&\les& \Big(N^{\geq 4m_0}[J, \Gac, \Rc] \Big)^2.
\eeaa

{\bf Step 8.} In order to estimate $\xib$ in Step 9, we derive an estimate for $e_3(\ze)+\bb$. Recall that we have
\beaa
e_4(\ze)+\ka\ze &=& -\b-\vth\ze. 
\eeaa
Commuting with $e_3$, we infer
\beaa
e_4(e_3(\ze))+[e_3, e_4]\ze+\ka e_3(\ze)+e_3(\ka)\ze &=& -e_3(\b)-\vth\ze. 
\eeaa
In view of the null structure equation for $e_3(\ka)$, the Bianchi identity for $e_3(\b)$ and the commutator identity for $[e_3, e_4]$, we infer
\beaa
&& e_4(e_3(\ze))+\Big(2\omb e_4+4\ze e_\th\Big)\ze+\ka e_3(\ze)+\left(-\frac{1}{2}\ka\kab+2\omb\ka+2\ddd_1\ze+2\rho-\frac{1}{2}\vth\vthb+2\ze^2\right)\ze\\
 &=& (\kab-2\omb)\b +\dds_1\rho-3\ze\rho+\vth\bb-\xib\a -\vth\ze. 
\eeaa
Together with the null structure equation for $e_4(\ze)$, the Bianchi identity for $e_4(\bb)$ to get rid of the term $\dds_1\rho$, and the definition of $\mu$, we infer
\beaa
&& e_4(e_3(\ze))+2\omb\left(-\ka\ze  -\b-\vth\ze\right) +4\ze\dds_1\ddd_1^{-1}\left(\check{\mu}+\check{\rho}-\frac{1}{4}\vth\vthb+\frac{1}{4}\overline{\vth\vthb}\right)\\
&&+\ka e_3(\ze)+\left(-\frac{1}{2}\ka\kab+2\omb\ka-2\mu+2\ze^2\right)\ze\\
 &=& (\kab-2\omb)\b -e_4(\bb)-\ka\bb -3\ze\rho-\vthb\b  -3\ze\rho+\vth\bb-\xib\a -\vth\ze. 
\eeaa
and hence
\beaa
&& e_4(e_3(\ze)+\bb)+\ka(e_3(\ze)+\bb)\\
 &=& \kab\b  +\left(\frac{1}{2}\ka\kab+2\mu-6\rho\right)\ze\\
 && -\vthb\b  +\vth\bb-\xib\a -\vth\ze  +2\omb\vth\ze -4\ze\dds_1\ddd_1^{-1}\left(\check{\mu}+\check{\rho}-\frac{1}{4}\vth\vthb+\frac{1}{4}\overline{\vth\vthb}\right) -2\ze^3
\eeaa
In view of Corollary \ref{cor:MorawetzforRicciinRR1} with $a=2$, we have for any $r_0\geq 4m_0$
\beaa
\max_{k\leq J+1}\sup_{r_0\geq 4m_0}r_0^2\int_{\{r=r_0\}}(\dk^k(e_3(\ze)+\bb))^2 &\les& \Big(N^{\geq 4m_0}[J, \Gac, \Rc] \Big)^2
\eeaa
where we used the estimates for $\ze$  derived in Step 3.

{\bf Step 9.} Next, recall that we have
\beaa
e_4(\xib) &=& -e_3(\ze)+\bb -\kab\ze-\ze\vthb\\
&=& -(e_3(\ze)+\bb)+2\bb -\kab\ze-\ze\vthb.
\eeaa
In view of Corollary \ref{cor:MorawetzforRicciinRR1bis} with $a=0$ and $b=-\dt$ which satisfy the constraint $b>2a-2$, we have for any $r_0\geq 4m_0$
\beaa
   \max_{k\leq J+1}\sup_{r_0\geq 4m_0}r_0^{-\dt}\left(\int_{\{r=r_0\}}(\dk^k\xib)^2\right)+\int_{\Mext(r\geq 4m_0)}r^{-1-\dt}(\dk^k\xib)^2 &\les& \Big(N^{\geq 4m_0}[J, \Gac, \Rc] \Big)^2
\eeaa
where we used the estimates for $\ze$ and $e_3(\ze)+\bb$ on $\Mext$  derived respectively in Step 3 and Step 8.

{\bf Step 10.} Recall that we have 
\beaa
&& e_4\left(\ddd_1\dds_1\ka-\vth\Big(\ddd_4\dds_3\ddd_2^{-1}+\dds_2\Big)\ddd_1^{-1}\check{\rho}\right)+2\ka\left(\ddd_1\dds_1\ka-\vth\Big(\ddd_4\dds_3\ddd_2^{-1}+\dds_2\Big)\ddd_1^{-1}\check{\rho}\right)\\ 
&=& -\left(-\frac{1}{2}\vth\dds_2+\ze e_4(\Phi) -\b\right)\dds_1\ka-\frac{1}{2}\vth\ddd_1\dds_1\ka +\frac{1}{2}\dds_1(\ka+\vth)\dds_1\ka +(\dds_1\ka)^2\\
&& -2\ka\vth\Big(\ddd_4\dds_3\ddd_2^{-1}+\dds_2\Big)\ddd_1^{-1}\check{\rho} +(\ka\vth+2\a)\Big(\ddd_4\dds_3\ddd_2^{-1}+\dds_2\Big)\ddd_1^{-1}\check{\rho}\\
&&+\vth\left[\Big(\ddd_4\dds_3\ddd_2^{-1}+\dds_2\Big)\ddd_1^{-1},e_4\right]\check{\rho} +\vth\Big(\ddd_4\dds_3\ddd_2^{-1}+\dds_2\Big)\ddd_1^{-1}\left(\frac{3}{2}\ov{\ka}\check{\rho}+\frac{3}{2}\ov{\rho}\check{\ka}-\err[e_4\check{\rho}]\right)\\
&&  -\frac{1}{2}\vth\ddd_4\dds_3\ddd_2^{-1}(-\dds_1\ka+\ka\ze-\vth\ze)-\frac{1}{2}\vth\dds_2(-\dds_1\ka+\ka\ze-\vth\ze)\\
&& +\frac{1}{2}\dds_3\ddd_2^{-1}(-2\b-\dds_1\ka+\ka\ze-\vth\ze)\dds_3\vth +\frac{1}{2}(-2\b-\dds_1\ka+\ka\ze-\vth\ze)\ddd_2\vth.
\eeaa
In view of Corollary \ref{cor:MorawetzforRicciinRR1} with $a=4$, we have 
\beaa
\max_{k\leq J+1}\sup_{r_0\geq 4m_0}\int_{\{r=r_0\}}r_0^6\left(\dk^k\left(\ddd_1\dds_1\ka-\vth\Big(\ddd_4\dds_3\ddd_2^{-1}+\dds_2\Big)\ddd_1^{-1}\check{\rho}\right)\right)^2 &\les&  \Big(N^{\geq 4m_0}[J, \Gac, \Rc] \Big)^2
\eeaa
where we have used 
\begin{itemize}
\item the fact that $\dkb\vth=\dkb\ddd_2^{-1}\ddd_2\vth$ and Codazzi for $\vth$ to estimate the terms of the RHS with one angular derivative of $\vth$, 

\item the estimates of Step 2 to estimate the terms of the RHS with one derivative of $\check{\ka}$, 

\item the fact that $\dkb\ze=\dkb\ddd_1^{-1}\ddd_1\ze$ and the definition of $\mu$ to estimate terms of the RHS with one angular derivative of $\ze$, 

\item the identity
\beaa
\dkb\dds_1\ka &=& \dkb\ddd_1^{-1}\left(\vth\Big(\ddd_4\dds_3\ddd_2^{-1}+\dds_2\Big)\ddd_1^{-1}\check{\rho}\right)\\
&&+\dkb\ddd_1^{-1}\left(\ddd_1\dds_1\ka-\vth\Big(\ddd_4\dds_3\ddd_2^{-1}+\dds_2\Big)\ddd_1^{-1}\check{\rho}\right)
\eeaa
to estimate the terms of the RHS with two angular derivatives of $\check{\ka}$.
\end{itemize}

{\bf Step 11.} Recall that we have
\beaa
 && e_4\Big(e_\th(\mu) +\vth\ddd_2\dds_2(\dds_1\ddd_1)^{-1}\bb+2\ze\check{\rho}\Big)+ 2\ka\Big(e_\th(\mu)+\vth\ddd_2\dds_2(\dds_1\ddd_1)^{-1}\bb+2\ze\check{\rho}\Big)\\
  &=& -\frac{3}{2}\mu e_\th(\ka)   -\frac{1}{2}\vth e_\th(\mu)  -(\ka\vth+2\a)\ddd_2\dds_2(\dds_1\ddd_1)^{-1}\bb\\
&&-\vth\Big[\ddd_2\dds_2(\dds_1\ddd_1)^{-1}, e_4\Big]\bb-\vth\ddd_2\dds_2(\dds_1\ddd_1)^{-1}\left(\ka\bb+3\rho\ze+\vthb\b\right)\\
  && +\dds_3\vth\dds_2\ddd_1^{-1}\check{\rho}    - e_\th\left(\vth\dds_2\ddd_1^{-1}\left(-\check{\mu}+\frac{1}{4}\vth\vthb\right)\right)   -e_\th\left(\vth\left(\frac{1}{8}\kab\vth+\ze^2\right)\right)\\
  &&-2\ze\ddd_1\dds_1\ka  -2e_\th(\ka)\dds_2\ze    -2(\ka\ze+\b+\vth\ze)\check{\rho} -2\ze\left(\frac{3}{2}\ov{\ka}\check{\rho}+\frac{3}{2}\ov{\rho}\check{\ka}-\err[e_4\check{\rho}]\right)\\
  && +2\b\dds_2\ze  +\frac{3}{2}e_\th\left(\ka\ze^2\right) +2\ka\Big(\vth\ddd_2\dds_2(\dds_1\ddd_1)^{-1}\bb+2\ze\check{\rho}\Big).
 \eeaa
 In view of Corollary \ref{cor:MorawetzforRicciinRR1} with $a=4$, we have 
\beaa
&&\max_{k\leq J+1}\sup_{r_0\geq 4m_0}r_0^6\int_{\{r=r_0\}}\left(\dk^k\left(e_\th(\mu) +\vth\ddd_2\dds_2(\dds_1\ddd_1)^{-1}\bb+2\ze\check{\rho}\right)\right)^2\\
 &\les&  \Big(N^{\geq 4m_0}[J, \Gac, \Rc] \Big)^2+\ep^2\int_{\Mext(\geq 4m_0)}\left(\dk^{\leq J+1}\left(e_\th(\kab) -4\bb\right)\right)^2,
\eeaa
where we have used 
\begin{itemize}
\item the fact that $\dkb\vth=\dkb\ddd_2^{-1}\ddd_2\vth$ and Codazzi for $\vth$ to estimate the terms of the RHS with one angular derivative of $\vth$, 

\item the estimates of Step 2 to estimate the terms of the RHS with one derivative of $\check{\ka}$, 

\item the fact that $\dkb\ze=\dkb\ddd_1^{-1}\ddd_1\ze$ and the definition of $\mu$ to estimate terms of the RHS with one angular derivative of $\ze$, 

\item the fact that $e_\th(\kab)=(e_\th(\kab)-4\bb)+4\bb$ to estimate the term with one angular derivative of $\kab$,

\item the identity
\beaa
\dkb\dds_1\ka &=& \dkb\ddd_1^{-1}\left(\vth\Big(\ddd_4\dds_3\ddd_2^{-1}+\dds_2\Big)\ddd_1^{-1}\check{\rho}\right)\\
&&+\dkb\ddd_1^{-1}\left(\ddd_1\dds_1\ka-\vth\Big(\ddd_4\dds_3\ddd_2^{-1}+\dds_2\Big)\ddd_1^{-1}\check{\rho}\right)
\eeaa
and the estimates of Step 10 to estimate the terms of the RHS with two angular derivatives of $\check{\ka}$.
\end{itemize}

{\bf Step 12.} Recall that we have
\beaa
&& e_4(e_\th(\kab)-4\bb)+\ka(e_\th(\kab)-4\bb)\\
 &=& 2e_\th(\mu)+12\rho\ze -\frac{1}{2}\kab e_\th(\ka) +4\vthb\b -\frac{1}{2}\vth e_\th(\kab) -e_\th(\vth\vthb)+2e_\th(\ze^2)\\
  &=& 2\Big(e_\th(\mu)+\vth\ddd_2\dds_2(\dds_1\ddd_1)^{-1}\bb+2\ze\check{\rho}\Big)+12\rho\ze -\frac{1}{2}\kab e_\th(\ka) \\
  &&-2\vth\ddd_2\dds_2(\dds_1\ddd_1)^{-1}\bb+2\ze\check{\rho}  +4\vthb\b -\frac{1}{2}\vth (e_\th(\kab)-4\bb)-2\vth\bb -e_\th(\vth\vthb)+2e_\th(\ze^2).
\eeaa
In view of Corollary \ref{cor:MorawetzforRicciinRR1} with $a=2$, we have 
\beaa
&&\max_{k\leq J+1}\sup_{r_0\geq 4m_0}r_0^2\int_{\{r=r_0\}}\left(\dk^k\left(e_\th(\kab) -4\bb\right)\right)^2\\
 &\les&  \Big(N^{\geq 4m_0}[J, \Gac, \Rc] \Big)^2+\int_{\Mext(\geq 4m_0)}r^4\left(\dk^{\leq J+1}\left(e_\th(\mu) +\vth\ddd_2\dds_2(\dds_1\ddd_1)^{-1}\bb+2\ze\check{\rho}\right)\right)^2,
\eeaa
where we have used 
\begin{itemize}
\item the fact that $\dkb\vth=\dkb\ddd_2^{-1}\ddd_2\vth$ and Codazzi for $\vth$ to estimate the terms of the RHS with one angular derivative of $\vth$,

\item the fact that $\dkb\vthb=\dkb\ddd_2^{-1}\ddd_2\vthb$ and Codazzi for $\vthb$ to estimate the terms of the RHS with one angular derivative of $\vthb$, 

\item the estimates of Step 2 to estimate the terms of the RHS with one derivative of $\check{\ka}$, 

\item the fact that $\dkb\ze=\dkb\ddd_1^{-1}\ddd_1\ze$ and the definition of $\mu$ to estimate terms of the RHS with one angular derivative of $\ze$, 

\item the estimate for $\ze$ of Step 3.
\end{itemize}
Together with the estimate of Step 11, we infer
\beaa
\max_{k\leq J+1}\sup_{r_0\geq 4m_0}r_0^6\int_{\{r=r_0\}}\left(\dk^k\left(e_\th(\mu) +\vth\ddd_2\dds_2(\dds_1\ddd_1)^{-1}\bb+2\ze\check{\rho}\right)\right)^2\\
 +\max_{k\leq J+1}\sup_{r_0\geq 4m_0}r_0^2\int_{\{r=r_0\}}\left(\dk^k\left(e_\th(\kab) -4\bb\right)\right)^2 &\les&  \Big(N^{\geq 4m_0}[J, \Gac, \Rc] \Big)^2.
\eeaa

Finally, we have obtained
\beaa
\nn\max_{k\leq J+1}\sup_{r_0\geq 4m_0}\Bigg(\int_{\{r=r_0\}}\Bigg(r_0^4(\dk^k\check{\mu})^2+r_0^2(\dk^k\vth)^2+r_0^2(\dk^k\ze)^2+r_0^2(\dk^k(e_3(\ze)+\bb))^2\\
+r_0^{2-\dt}(\dk^k\check{\kab})^2+(\dk^k\vthb)^2+(\dk^k\check{\omb})^2+r_0^{-\dt}(\dk^k\xib)^2\Bigg)\Bigg)&\les&  \Big(N^{\geq 4m_0}[J, \Gac, \Rc] \Big)^2,
\eeaa
\beaa
\nn\max_{k\leq J+2}\sup_{r_0\geq 4m_0}\left(r_0^2\int_{\{r=r_0\}}\left(\dk^k\left(\ka-\frac{2}{r}\right)\right)^2\right) &\les&  \Big(N^{\geq 4m_0}[J, \Gac, \Rc] \Big)^2,
\eeaa
and
\beaa
\max_{k\leq J+1}\sup_{\rh\leq r_0\leq 4m_0}\Bigg(\int_{\{r=r_0\}}\Bigg\{r_0^6\left(\dk^k\left(\ddd_1\dds_1\ka-\vth\Big(\ddd_4\dds_3\ddd_2^{-1}+\dds_2\Big)\ddd_1^{-1}\check{\rho}\right)\right)^2\\
 + r_0^6\left(\dk^k\left(e_\th(\mu) +\vth\ddd_2\dds_2(\dds_1\ddd_1)^{-1}\bb+2\ze\check{\rho}\right)\right)^2\\
 +r_0^2\left(\dk^k\left(e_\th(\kab) -4\bb\right)\right)^2\Bigg\}\Bigg) &\les&\Big(N^{\geq 4m_0}[J, \Gac, \Rc] \Big)^2.
\eeaa
In view of the definition \eqref{eq:defintionofnormNgeq4m0JGacRcproofThmM8} of $N^{\geq 4m_0}[J, \Gac, \Rc]$, and of the various norms, we infer
\beaa
  \,{}^{(ext)}\mathfrak{G}^{\geq 4m_0}_{J+1}[\Gac] + \,{}^{(ext)}{\mathfrak{G}^{\geq 4m_0}_{J+1}}'[\Gac] &\les& \,{}^{(\Si_*)}\mathfrak{G}_{J+1}[\Gac] +\,{}^{(\Si_*)}\mathfrak{G}_{J+1}'[\Gac] +\,{}^{(ext)}\mathfrak{R}_{J+1}[\Rc] +\,{}^{(ext)}\mathfrak{G}_{J}[\Gac]\\
&&+\ep_0\Big( \,{}^{(ext)}\mathfrak{G}^{\geq 4m_0}_{J+1}[\Gac] + \,{}^{(ext)}{\mathfrak{G}^{\geq 4m_0}_{J+1}}'[\Gac]\Big)
\eeaa
and hence, for $\ep_0$ small enough, 
\beaa
  \,{}^{(ext)}\mathfrak{G}^{\geq 4m_0}_{J+1}[\Gac] + \,{}^{(ext)}{\mathfrak{G}^{\geq 4m_0}_{J+1}}'[\Gac] &\les& \,{}^{(\Si_*)}\mathfrak{G}_{J+1}[\Gac] +\,{}^{(\Si_*)}\mathfrak{G}_{J+1}'[\Gac] +\,{}^{(ext)}\mathfrak{R}_{J+1}[\Rc] +\,{}^{(ext)}\mathfrak{G}_{J}[\Gac].
\eeaa
This concludes the proof of Proposition \ref{prop:controlGaextiterationassupmtionThM8:actualresult}.


\subsection{Proof of Proposition \ref{prop:controlGaextiterationassupmtionThM8:actualresult:bis}}\lab{sec:proofofprop:controlGaextiterationassupmtionThM8:actualresult:bis}


In the proof below, we will repeatedly use the following estimate 
\bea\lab{eq:indentitytocontrolGammainMextforrleq4m0}
\nn&&\max_{k\leq J+1}\int_{\Mext(r\leq 4m_0)}(\dk^kf)^2\\
 &\les& \max_{k\leq J}\int_{\Mext(r\leq 4m_0)}\Big((\dk^kf)^2+(\dk^k\N f)^2+(\dk^ke_4f)^2+(\dk^k\dkb f)^2\Big)
\eea
which follows from the fact that $\dk=(\dkb, re_4, e_3)$ and $e_3=\Up e_4 -2\N$, where we recall that 
\beaa
\N &=& \frac{1}{2}\Big(\Up e_4 -e_3\Big).
\eeaa
Also, we introduce the following notation which will constantly appear on the RHS of the equalities below
\bea\lab{eq:defintionofnormNleq4m0JGacRcproofThmM8}
\nn N^{\leq 4m_0}[J, \Gac, \Rc] &:=& \,{}^{(ext)}\mathfrak{G}^{\geq 4m_0}_{J+1}[\Gac] + \,{}^{(ext)}{\mathfrak{G}^{\geq 4m_0}_{J+1}}'[\Gac] +\,{}^{(ext)}\mathfrak{R}_{J+1}[\Rc] +\,{}^{(ext)}\mathfrak{G}_{J}[\Gac]\\
&&+\ep_0\Big(  \,{}^{(ext)}\mathfrak{G}^{\leq 4m_0}_{J+1}[\Gac] + \,{}^{(ext)}{\mathfrak{G}^{\leq 4m_0}_{J+1}}'[\Gac] \Big).
 \eea

{\bf Step 1.} Recall that 
\beaa
e_4(\check{\ka})+\ka\check{\ka} &=& \err[e_4\check{\ka}].
\eeaa
In view of Corollary \ref{cor:MorawetzforRicciinRR1forrleq4m0}, we have
\beaa
\max_{k\leq J+2}\sup_{\rh\leq r_0\leq 4m_0}\int_{\{r=r_0\}}(\dk^k\check{\ka})^2 &\les& \Big(N^{\leq 4m_0}[J, \Gac, \Rc] \Big)^2
\eeaa
where we have used the null structure equations for $e_4(\vth)$, $e_3(\vth)$ and $\ddd_2\vth$ to avoid loosing one derivative. 

{\bf Step 2.} Next, recall that 
\beaa
 e_4\left(\check{\mu}\right) + \frac{3}{2}\overline{\ka}\check{\mu} &=&   - \frac{3}{2}\overline{\mu}\check{\ka} +\err[e_4\check{\mu}].
 \eeaa
In view of Corollary \ref{cor:MorawetzforRicciinRR1forrleq4m0}, we have
\beaa
\max_{k\leq J+1}\sup_{\rh\leq r_0\leq 4m_0}\int_{\{r=r_0\}}(\dk^k\check{\mu})^2 &\les& \Big(N^{\leq 4m_0}[J, \Gac, \Rc] \Big)^2
\eeaa
where we have used the estimates for $\check{\ka}$ of Step 1. 

{\bf Step 3.} Next, recall that 
\beaa
e_4(\ze) +\ka\ze &=& -\b -\vth\ze.
\eeaa
In view of Corollary \ref{cor:MorawetzforRicciinRR1forrleq4m0}, we have
\beaa
\max_{k\leq J}\sup_{\rh\leq r_0\leq 4m_0}\int_{\{r=r_0\}}\Big((\dk^ke_4\ze)^2+(\dk^k\ze)^2\Big) &\les& \Big(N^{\leq 4m_0}[J, \Gac, \Rc] \Big)^2.
\eeaa
Also, commuting first with $\N$, and proceeding analogously, we infer
\beaa
\max_{k\leq J}\sup_{\rh\leq r_0\leq 4m_0}\int_{\{r=r_0\}}(\dk^k\N\ze)^2 &\les& \Big(N^{\leq 4m_0}[J, \Gac, \Rc] \Big)^2
\eeaa
Furthermore, in view of the definition of $\mu$ and a Poincar\'e inequality for $\ddd_1$, we have
\beaa
\max_{k\leq J}\sup_{\rh\leq r_0\leq 4m_0}\int_{\{r=r_0\}}(\dk^k\dkb\ze)^2 &\les& \Big(N^{\leq 4m_0}[J, \Gac, \Rc] \Big)^2
\eeaa
where we have used a trace estimate and the estimate for $\check{\mu}$ of Step 2. The above estimates, together with \eqref{eq:indentitytocontrolGammainMextforrleq4m0}, imply
\beaa
\max_{k\leq J+1}\sup_{\rh\leq r_0\leq 4m_0}\int_{\{r=r_0\}}(\dk^k\ze)^2 &\les&  \Big(N^{\leq 4m_0}[J, \Gac, \Rc] \Big)^2.
\eeaa

{\bf Step 4.} Recall that 
\beaa
e_4(\vth) +\ka\vth &=& -2\a.
\eeaa
In view of Corollary \ref{cor:MorawetzforRicciinRR1forrleq4m0}, we have
\beaa
\max_{k\leq J}\sup_{\rh\leq r_0\leq 4m_0}\int_{\{r=r_0\}}\Big((\dk^ke_4\vth)^2+(\dk^k\vth)^2\Big) &\les& \Big(N^{\leq 4m_0}[J, \Gac, \Rc] \Big)^2.
\eeaa
Also, commuting first one time with $\N$, and proceeding analogously, we infer
\beaa
\max_{k\leq J}\sup_{\rh\leq r_0\leq 4m_0}\int_{\{r=r_0\}}(\dk^k\N\vth)^2 &\les& \Big(N^{\leq 4m_0}[J, \Gac, \Rc] \Big)^2.
\eeaa
Furthermore, in view of Codazzi for $\vth$, and a Poincar\'e inequality for $\ddd_2$, we have
\beaa
\max_{k\leq J}\sup_{\rh\leq r_0\leq 4m_0}\int_{\{r=r_0\}}(\dk^k\dkb\vth)^2  &\les& \Big(N^{\leq 4m_0}[J, \Gac, \Rc] \Big)^2
\eeaa
where we have used a trace estimate, and  the estimate for $\check{\ka}$ and $\ze$ respectively in Step 1 and Step 3. The above estimates, together with \eqref{eq:indentitytocontrolGammainMextforrleq4m0}, imply
\beaa
\max_{k\leq J+1}\sup_{\rh\leq r_0\leq 4m_0}\int_{\{r=r_0\}}(\dk^k\vth)^2 &\les&  \Big(N^{\leq 4m_0}[J, \Gac, \Rc] \Big)^2.
\eeaa

{\bf Step 5.} Recall that we have
\beaa
e_4(\check{\kab})+ \frac{1}{2}\ov{\ka}\check{\kab} &=& -\frac{1}{2}\ov{\kab}\check{\ka}-2\ddd_1\ze+2\check{\rho}+\err[e_4\check{\kab}]\\
&=& -\frac{1}{2}\ov{\kab}\check{\ka}+2\check{\mu}+4\check{\rho}-\frac{1}{2}\vth\vthb+\err[e_4\check{\kab}].
\eeaa
In view of Corollary \ref{cor:MorawetzforRicciinRR1forrleq4m0}, we have
\beaa
\max_{k\leq J}\sup_{\rh\leq r_0\leq 4m_0}\int_{\{r=r_0\}}\Big((\dk^ke_4\check{\kab})^2+(\dk^k\check{\kab})^2\Big) &\les& \Big(N^{\leq 4m_0}[J, \Gac, \Rc] \Big)^2
\eeaa
where we have used the estimates for $\check{\ka}$ and $\check{\mu}$ derived respectively in Step 1 and Step 2. Also, commuting first one time with $\N$, and proceeding analogously, we infer
\beaa
\max_{k\leq J}\sup_{\rh\leq r_0\leq 4m_0}\int_{\{r=r_0\}}(\dk^k\N\check{\kab})^2 &\les& \Big(N^{\leq 4m_0}[J, \Gac, \Rc] \Big)^2
\eeaa
where we have used  the estimates for $\check{\ka}$ and $\check{\mu}$ derived respectively in Step 1 and Step 2. Furthermore, commuting the equation for $e_4(\ka)$ once with $e_\th$, we have
\beaa
e_4(e_\th(\kab))+ \ka e_\th(\kab) &=& -\frac{1}{2}\kab e_\th(\ka)+2e_\th(\mu)+4e_\th(\rho) - e_\th(\vth\vthb)+2e_\th(\ze^2)-\frac{1}{2}\vth e_\th(\kab).
\eeaa
Together with the Bianchi identity for $e_4(\bb)$, we infer
\beaa
e_4(e_\th(\kab)-4\bb)+ \ka (e_\th(\kab)-4\bb) &=& -\frac{1}{2}\kab e_\th(\ka)+2e_\th(\mu) +12\rho\ze\\
&&+4\vthb\b - e_\th(\vth\vthb)+2e_\th(\ze^2)-\frac{1}{2}\vth e_\th(\kab).
\eeaa
In view of Corollary \ref{cor:MorawetzforRicciinRR1forrleq4m0}, we have
\beaa
\max_{k\leq J}\sup_{\rh\leq r_0\leq 4m_0}\int_{\{r=r_0\}}\Big((\dk^k(e_4(e_\th(\kab)-4\bb))^2+(\dk^k(e_\th(\kab)-4\bb))^2\Big) &\les& \Big(N^{\leq 4m_0}[J, \Gac, \Rc] \Big)^2
\eeaa
where we have used  the estimates for $\check{\ka}$, $\check{\mu}$ and $\ze$ derived respectively in Step 1,  Step 2 and Step 3. 

The above estimates, together with \eqref{eq:indentitytocontrolGammainMextforrleq4m0}, imply
\beaa
\max_{k\leq J+1}\sup_{\rh\leq r_0\leq 4m_0}\int_{\{r=r_0\}}(\dk^k\check{\kab})^2 &\les&  \Big(N^{\leq 4m_0}[J, \Gac, \Rc] \Big)^2+\max_{k\leq J}\sup_{\rh\leq r_0\leq 4m_0}\int_{\{r=r_0\}}(\dk^k\bb)^2\\
&\les& \Big(N^{\leq 4m_0}[J, \Gac, \Rc] \Big)^2
\eeaa
where we have used a trace estimate on $\{r=r_0\}$ for $\rh\leq r_0\leq 4m_0$. 

{\bf Step 6.} Recall that we have
\beaa
e_4(\check{\omb}) &=& \check{\rho}+\err[e_4\check{\omb}].
\eeaa
In view of Corollary \ref{cor:MorawetzforRicciinRR1forrleq4m0}, we have
\beaa
\max_{k\leq J}\sup_{\rh\leq r_0\leq 4m_0}\int_{\{r=r_0\}}\Big((\dk^ke_4\check{\omb})^2+(\dk^k\check{\omb})^2\Big) &\les& \Big(N^{\leq 4m_0}[J, \Gac, \Rc] \Big)^2.
\eeaa
Also, commuting first one time with $\N$, and proceeding analogously, we infer
\beaa
\max_{k\leq J}\sup_{\rh\leq r_0\leq 4m_0}\int_{\{r=r_0\}}(\dk^k\N\check{\omb})^2 &\les& \Big(N^{\leq 4m_0}[J, \Gac, \Rc] \Big)^2.
\eeaa

{\bf Step 7.} Recall that we have
\beaa
&& e_4(e_3(\ze)+\bb)+\ka(e_3(\ze)+\bb)\\
 &=& \kab\b  +\left(\frac{1}{2}\ka\kab+2\mu-6\rho\right)\ze\\
 && -\vthb\b  +\vth\bb-\xib\a -\vth\ze  +2\omb\vth\ze -4\ze\dds_1\ddd_1^{-1}\left(\check{\mu}+\check{\rho}-\frac{1}{4}\vth\vthb+\frac{1}{4}\overline{\vth\vthb}\right) -2\ze^3
\eeaa
Commuting first one time with $\N$, and in view of Corollary \ref{cor:MorawetzforRicciinRR1forrleq4m0}, we have
\beaa
\max_{k\leq J}\sup_{\rh\leq r_0\leq 4m_0}\int_{\{r=r_0\}}(\dk^k\N(e_3(\ze)+\bb))^2\Big) &\les& \Big(N^{\leq 4m_0}[J, \Gac, \Rc] \Big)^2
\eeaa
where we have used the estimate for $\ze$ in Step 3.

{\bf Step 8.} Recall that we have
\beaa
e_4(\xib) &=& -e_3(\ze)+\bb-\kab\ze-\ze\vthb.
\eeaa
In view of Corollary \ref{cor:MorawetzforRicciinRR1forrleq4m0}, we have
\beaa
\max_{k\leq J}\sup_{\rh\leq r_0\leq 4m_0}\int_{\{r=r_0\}}\Big((\dk^ke_4\xib)^2+(\dk^k\xib)^2\Big) &\les& \Big(N^{\leq 4m_0}[J, \Gac, \Rc] \Big)^2
\eeaa
where we have used the estimates for $\ze$ derived in Step 3. Also, commuting first one time with $\N$, and proceeding analogously, we infer
\beaa
\max_{k\leq J}\sup_{\rh\leq r_0\leq 4m_0}\int_{\{r=r_0\}}(\dk^k\N\xib)^2 &\les& \Big(N^{\leq 4m_0}[J, \Gac, \Rc] \Big)^2
\eeaa
where we have used the estimates for $e_3(\ze)+\bb$ derived in Step 7.

{\bf Step 9.} Recall
\beaa
2\dds_1\omb &=& e_3\ze+\kab\ze -\bb-\frac{1}{2}\ka\xib +\vthb\ze-\frac{1}{2}\vth\xib.
\eeaa
Using a Poincar\'e inequality for $\dds_1$, we infer
\beaa
\max_{k\leq J}\sup_{\rh\leq r_0\leq 4m_0}\int_{\{r=r_0\}}(\dk^k\dkb\check{\omb})^2 &\les& \Big(N^{\leq 4m_0}[J, \Gac, \Rc] \Big)^2
\eeaa
where we have used a trace estimate and the estimate for $\ze$ and $\xib$ respectively in Step 3 and Step 8. The above estimates, together with the estimates for $\check{\omb}$ of Step 6 and  \eqref{eq:indentitytocontrolGammainMextforrleq4m0}, imply
\beaa
\max_{k\leq J+1}\sup_{\rh\leq r_0\leq 4m_0}\int_{\{r=r_0\}}(\dk^k\check{\omb})^2 &\les&  \Big(N^{\leq 4m_0}[J, \Gac, \Rc] \Big)^2.
\eeaa

{\bf Step 10.} Recall that we have
\beaa
e_4(\Obc) &=& -2\check{\omb}+\overline{\check{\ka}\Obc}.
\eeaa
In view of Corollary \ref{cor:MorawetzforRicciinRR1forrleq4m0}, we have
\beaa
\max_{k\leq J+1}\sup_{\rh\leq r_0\leq 4m_0}\int_{\{r=r_0\}}(\dk^k\Obc)^2 &\les& \Big(N^{\leq 4m_0}[J, \Gac, \Rc] \Big)^2
\eeaa
where we have used the estimates for $\check{\omb}$ derived in Step 9. 

{\bf Step 11.} Recall 
\beaa
2\ddd_1\xib &=& e_3(\check{\kab})+\ov{\kab}\,\check{\kab}+2\ov{\omb}\,\check{\kab} +2\ov{\kab}\,\check{\omb}-\left(\frac{1}{2}\overline{\ka\kab}-2\ov{\rho}\right)\Obc-\err[e_3\check{\kab}].
\eeaa
Using a Poincar\'e inequality for $\ddd_1$, we infer
\beaa
\max_{k\leq J}\sup_{\rh\leq r_0\leq 4m_0}\int_{\{r=r_0\}}(\dk^k\dkb\xib)^2 &\les& \Big(N^{\leq 4m_0}[J, \Gac, \Rc] \Big)^2
\eeaa
where we have used the estimates for $\check{\kab}$, $\check{\omb}$ and $\Obc$ respectively in Step 5, Step 9 and Step 10. The above estimates, together with the estimates for $\xib$ of Step 8 and  \eqref{eq:indentitytocontrolGammainMextforrleq4m0}, imply
\beaa
\max_{k\leq J+1}\sup_{\rh\leq r_0\leq 4m_0}\int_{\{r=r_0\}}(\dk^k\xib)^2 &\les&  \Big(N^{\leq 4m_0}[J, \Gac, \Rc] \Big)^2.
\eeaa

{\bf Step 12.} Recall that 
\beaa
e_4(\vthb) +\frac{1}{2}\ka\vthb &=& 2\dds_2\ze -\frac{1}{2}\kab\vth+2\ze^2\\
&=& 2\dds_2\ddd_1^{-1}\left(-\check{\mu}-\check{\rho}+\frac{1}{4}\vth\vthb-\frac{1}{4}\ov{\vth\vthb}\right) -\frac{1}{2}\kab\vth+2\ze^2.
\eeaa
In view of Corollary \ref{cor:MorawetzforRicciinRR1forrleq4m0}, we have
\beaa
\max_{k\leq J}\sup_{\rh\leq r_0\leq 4m_0}\int_{\{r=r_0\}}\Big((\dk^ke_4\vthb)^2+(\dk^k\vthb)^2\Big) &\les& \Big(N^{\leq 4m_0}[J, \Gac, \Rc] \Big)^2
\eeaa
where we have used the estimate for $\check{\mu}$ and $\vth$ respectively in Step 2 and Step 4. Also, commuting first one time with $\N$, and proceeding analogously, we infer
\beaa
\max_{k\leq J}\sup_{\rh\leq r_0\leq 4m_0}\int_{\{r=r_0\}}(\dk^k\N\vthb)^2 &\les& \Big(N^{\leq 4m_0}[J, \Gac, \Rc] \Big)^2
\eeaa
where we have used the estimate for $\check{\mu}$ and $\vth$ respectively in Step 2 and Step 4. Furthermore, in view of Codazzi for $\vthb$, and a Poincar\'e inequality for $\ddd_2$, we have
\beaa
\max_{k\leq J}\sup_{\rh\leq r_0\leq 4m_0}\int_{\{r=r_0\}}(\dk^k\dkb\vthb)^2  &\les& \Big(N^{\leq 4m_0}[J, \Gac, \Rc] \Big)^2
\eeaa
where we have used a trace estimate and the estimate for $\check{\kab}$ and $\ze$ respectively in Step 5 and Step 3. The above estimates, together with \eqref{eq:indentitytocontrolGammainMextforrleq4m0}, imply
\beaa
\max_{k\leq J+1}\sup_{\rh\leq r_0\leq 4m_0}\int_{\{r=r_0\}}(\dk^k\vthb)^2 &\les&  \Big(N^{\leq 4m_0}[J, \Gac, \Rc] \Big)^2.
\eeaa

{\bf Step 13.} Recall that we have
\beaa
&& e_4\left(\ddd_1\dds_1\ka-\vth\Big(\ddd_4\dds_3\ddd_2^{-1}+\dds_2\Big)\ddd_1^{-1}\check{\rho}\right)+2\ka\left(\ddd_1\dds_1\ka-\vth\Big(\ddd_4\dds_3\ddd_2^{-1}+\dds_2\Big)\ddd_1^{-1}\check{\rho}\right)\\ 
&=& -\left(-\frac{1}{2}\vth\dds_2+\ze e_4(\Phi) -\b\right)\dds_1\ka-\frac{1}{2}\vth\ddd_1\dds_1\ka +\frac{1}{2}\dds_1(\ka+\vth)\dds_1\ka +(\dds_1\ka)^2\\
&& -2\ka\vth\Big(\ddd_4\dds_3\ddd_2^{-1}+\dds_2\Big)\ddd_1^{-1}\check{\rho} +(\ka\vth+2\a)\Big(\ddd_4\dds_3\ddd_2^{-1}+\dds_2\Big)\ddd_1^{-1}\check{\rho}\\
&&+\vth\left[\Big(\ddd_4\dds_3\ddd_2^{-1}+\dds_2\Big)\ddd_1^{-1},e_4\right]\check{\rho} +\vth\Big(\ddd_4\dds_3\ddd_2^{-1}+\dds_2\Big)\ddd_1^{-1}\left(\frac{3}{2}\ov{\ka}\check{\rho}+\frac{3}{2}\ov{\rho}\check{\ka}-\err[e_4\check{\rho}]\right)\\
&&  -\frac{1}{2}\vth\ddd_4\dds_3\ddd_2^{-1}(-\dds_1\ka+\ka\ze-\vth\ze)-\frac{1}{2}\vth\dds_2(-\dds_1\ka+\ka\ze-\vth\ze)\\
&& +\frac{1}{2}\dds_3\ddd_2^{-1}(-2\b-\dds_1\ka+\ka\ze-\vth\ze)\dds_3\vth +\frac{1}{2}(-2\b-\dds_1\ka+\ka\ze-\vth\ze)\ddd_2\vth.
\eeaa
In view of Corollary \ref{cor:MorawetzforRicciinRR1forrleq4m0}, we have
\beaa
\max_{k\leq J+1}\sup_{\rh\leq r_0\leq 4m_0}\int_{\{r=r_0\}}\left(\dk^k\left(\ddd_1\dds_1\ka-\vth\Big(\ddd_4\dds_3\ddd_2^{-1}+\dds_2\Big)\ddd_1^{-1}\check{\rho}\right)\right)^2 &\les&  \Big(N^{\leq 4m_0}[J, \Gac, \Rc] \Big)^2
\eeaa
where we have used 
\begin{itemize}
\item the fact that $\dkb\vth=\dkb\ddd_2^{-1}\ddd_2\vth$ and Codazzi for $\vth$ to estimate the terms of the RHS with one angular derivative of $\vth$, 

\item the estimates of Step 1 to estimate the terms of the RHS with one derivative of $\check{\ka}$, 

\item the fact that $\dkb\ze=\dkb\ddd_1^{-1}\ddd_1\ze$ and the definition of $\mu$ to estimate terms of the RHS with one angular derivative of $\ze$, 

\item the identity
\beaa
\dkb\dds_1\ka &=& \dkb\ddd_1^{-1}\left(\vth\Big(\ddd_4\dds_3\ddd_2^{-1}+\dds_2\Big)\ddd_1^{-1}\check{\rho}\right)\\
&&+\dkb\ddd_1^{-1}\left(\ddd_1\dds_1\ka-\vth\Big(\ddd_4\dds_3\ddd_2^{-1}+\dds_2\Big)\ddd_1^{-1}\check{\rho}\right)
\eeaa
to estimate the terms of the RHS with two angular derivatives of $\check{\ka}$.
\end{itemize}

{\bf Step 14.} Recall that we have
\beaa
 && e_4\Big(e_\th(\mu) +\vth\ddd_2\dds_2(\dds_1\ddd_1)^{-1}\bb+2\ze\check{\rho}\Big)+ 2\ka\Big(e_\th(\mu)+\vth\ddd_2\dds_2(\dds_1\ddd_1)^{-1}\bb+2\ze\check{\rho}\Big)\\
  &=& -\frac{3}{2}\mu e_\th(\ka)   -\frac{1}{2}\vth e_\th(\mu)  -(\ka\vth+2\a)\ddd_2\dds_2(\dds_1\ddd_1)^{-1}\bb\\
&&-\vth\Big[\ddd_2\dds_2(\dds_1\ddd_1)^{-1}, e_4\Big]\bb-\vth\ddd_2\dds_2(\dds_1\ddd_1)^{-1}\left(\ka\bb+3\rho\ze+\vthb\b\right)\\
  && +\dds_3\vth\dds_2\ddd_1^{-1}\check{\rho}    - e_\th\left(\vth\dds_2\ddd_1^{-1}\left(-\check{\mu}+\frac{1}{4}\vth\vthb\right)\right)   -e_\th\left(\vth\left(\frac{1}{8}\kab\vth+\ze^2\right)\right)\\
  &&-2\ze\ddd_1\dds_1\ka  -2e_\th(\ka)\dds_2\ze    -2(\ka\ze+\b+\vth\ze)\check{\rho} -2\ze\left(\frac{3}{2}\ov{\ka}\check{\rho}+\frac{3}{2}\ov{\rho}\check{\ka}-\err[e_4\check{\rho}]\right)\\
  && +2\b\dds_2\ze  +\frac{3}{2}e_\th\left(\ka\ze^2\right) +2\ka\Big(\vth\ddd_2\dds_2(\dds_1\ddd_1)^{-1}\bb+2\ze\check{\rho}\Big).
 \eeaa
 In view of Corollary \ref{cor:MorawetzforRicciinRR1forrleq4m0}, we have
\beaa
&&\max_{k\leq J+1}\sup_{\rh\leq r_0\leq 4m_0}\int_{\{r=r_0\}}\left(\dk^k\left(e_\th(\mu) +\vth\ddd_2\dds_2(\dds_1\ddd_1)^{-1}\bb+2\ze\check{\rho}\right)\right)^2\\
 &\les&  \Big(N^{\leq 4m_0}[J, \Gac, \Rc] \Big)^2+\ep^2\int_{\Mext(\leq 4m_0)}\left(\dk^{\leq J+1}\left(e_\th(\kab) -4\bb\right)\right)^2,
\eeaa
where we have used 
\begin{itemize}
\item the fact that $\dkb\vth=\dkb\ddd_2^{-1}\ddd_2\vth$ and Codazzi for $\vth$ to estimate the terms of the RHS with one angular derivative of $\vth$, 

\item the estimates of Step 1 to estimate the terms of the RHS with one derivative of $\check{\ka}$, 

\item the fact that $\dkb\ze=\dkb\ddd_1^{-1}\ddd_1\ze$ and the definition of $\mu$ to estimate terms of the RHS with one angular derivative of $\ze$, 

\item the fact that $e_\th(\kab)=(e_\th(\kab)-4\bb)+4\bb$ to estimate the term with one angular derivative of $\kab$,

\item the identity
\beaa
\dkb\dds_1\ka &=& \dkb\ddd_1^{-1}\left(\vth\Big(\ddd_4\dds_3\ddd_2^{-1}+\dds_2\Big)\ddd_1^{-1}\check{\rho}\right)\\
&&+\dkb\ddd_1^{-1}\left(\ddd_1\dds_1\ka-\vth\Big(\ddd_4\dds_3\ddd_2^{-1}+\dds_2\Big)\ddd_1^{-1}\check{\rho}\right)
\eeaa
and the estimates of Step13 to estimate the terms of the RHS with two angular derivatives of $\check{\ka}$.
\end{itemize}

{\bf Step 15.} Recall that we have
\beaa
&& e_4(e_\th(\kab)-4\bb)+\ka(e_\th(\kab)-4\bb)\\
 &=& 2e_\th(\mu)+12\rho\ze -\frac{1}{2}\kab e_\th(\ka) +4\vthb\b -\frac{1}{2}\vth e_\th(\kab) -e_\th(\vth\vthb)+2e_\th(\ze^2)\\
  &=& 2\Big(e_\th(\mu)+\vth\ddd_2\dds_2(\dds_1\ddd_1)^{-1}\bb+2\ze\check{\rho}\Big)+12\rho\ze -\frac{1}{2}\kab e_\th(\ka) \\
  &&-2\vth\ddd_2\dds_2(\dds_1\ddd_1)^{-1}\bb+2\ze\check{\rho}  +4\vthb\b -\frac{1}{2}\vth (e_\th(\kab)-4\bb)-2\vth\bb -e_\th(\vth\vthb)+2e_\th(\ze^2).
\eeaa
 In view of Corollary \ref{cor:MorawetzforRicciinRR1forrleq4m0}, we have
\beaa
&&\max_{k\leq J+1}\sup_{\rh\leq r_0\leq 4m_0}\int_{\{r=r_0\}}\left(\dk^k\left(e_\th(\kab) -4\bb\right)\right)^2\\
 &\les&  \Big(N^{\leq 4m_0}[J, \Gac, \Rc] \Big)^2+\int_{\Mext(\leq 4m_0)}\left(\dk^{\leq J+1}\left(e_\th(\mu) +\vth\ddd_2\dds_2(\dds_1\ddd_1)^{-1}\bb+2\ze\check{\rho}\right)\right)^2.
\eeaa
where we have used 
\begin{itemize}
\item the fact that $\dkb\vth=\dkb\ddd_2^{-1}\ddd_2\vth$ and Codazzi for $\vth$ to estimate the terms of the RHS with one angular derivative of $\vth$,

\item the fact that $\dkb\vthb=\dkb\ddd_2^{-1}\ddd_2\vthb$ and Codazzi for $\vthb$ to estimate the terms of the RHS with one angular derivative of $\vthb$, 

\item the estimates of Step 1 to estimate the terms of the RHS with one derivative of $\check{\ka}$, 

\item the fact that $\dkb\ze=\dkb\ddd_1^{-1}\ddd_1\ze$ and the definition of $\mu$ to estimate terms of the RHS with one angular derivative of $\ze$, 

\item the estimate for $\ze$ of Step 3.
\end{itemize}
Together with the estimate of Step 14, we infer
\beaa
\max_{k\leq J+1}\sup_{\rh\leq r_0\leq 4m_0}\int_{\{r=r_0\}}\left(\dk^k\left(e_\th(\mu) +\vth\ddd_2\dds_2(\dds_1\ddd_1)^{-1}\bb+2\ze\check{\rho}\right)\right)^2\\
 +\max_{k\leq J+1}\sup_{\rh\leq r_0\leq 4m_0}\int_{\{r=r_0\}}\left(\dk^k\left(e_\th(\kab) -4\bb\right)\right)^2 &\les&  \Big(N^{\leq 4m_0}[J, \Gac, \Rc] \Big)^2.
\eeaa

In view of Step 1 to Step 15, of the definition \eqref{eq:defintionofnormNleq4m0JGacRcproofThmM8} of $N^{\leq 4m_0}[J, \Gac, \Rc]$, and of the various norms, we infer
\beaa
 \,{}^{(ext)}\mathfrak{G}^{\leq 4m_0}_{J+1}[\Gac] + \,{}^{(ext)}{\mathfrak{G}^{\leq 4m_0}_{J+1}}'[\Gac] &\les&  \,{}^{(ext)}\mathfrak{G}^{\geq 4m_0}_{J+1}[\Gac] + \,{}^{(ext)}{\mathfrak{G}^{\geq 4m_0}_{J+1}}'[\Gac] +\,{}^{(ext)}\mathfrak{R}_{J+1}[\Rc] +\,{}^{(ext)}\mathfrak{G}_{J}[\Gac]\\
&&+\ep_0\Big(  \,{}^{(ext)}\mathfrak{G}^{\leq 4m_0}_{J+1}[\Gac] + \,{}^{(ext)}{\mathfrak{G}^{\leq 4m_0}_{J+1}}'[\Gac] \Big).
\eeaa
and hence, for $\ep_0$ small enough, 
\beaa
 \,{}^{(ext)}\mathfrak{G}^{\leq 4m_0}_{J+1}[\Gac] + \,{}^{(ext)}{\mathfrak{G}^{\leq 4m_0}_{J+1}}'[\Gac] &\les&  \,{}^{(ext)}\mathfrak{G}^{\geq 4m_0}_{J+1}[\Gac] + \,{}^{(ext)}{\mathfrak{G}^{\geq 4m_0}_{J+1}}'[\Gac] +\,{}^{(ext)}\mathfrak{R}_{J+1}[\Rc] +\,{}^{(ext)}\mathfrak{G}_{J}[\Gac].
\eeaa
This concludes the proof of Proposition \ref{prop:controlGaextiterationassupmtionThM8:actualresult:bis}.


\section{Proof of Proposition \ref{prop:improvementoftheiterationassupmtionThM8}}\lab{sec:proofprop:improvementoftheiterationassupmtionThM8} 


To prove Proposition \ref{prop:improvementoftheiterationassupmtionThM8}, we rely on the following proposition.
\begin{proposition}\lab{prop:controlGaintiterationassupmtionThM8:actualresult}
Let $J$ such that $k_{small}-2\leq J\leq k_{large}-1$. Then, we have 
\beaa
 \,{}^{(int)}\mathfrak{G}_{J+1}[\Gac] + \,{}^{(int)}\mathfrak{G}_{J+1}'[\Gac]  &\les& \,{}^{(ext)}\mathfrak{G}_{J+1}[\Gac]+\,{}^{(ext)}\mathfrak{G}_{J+1}'[\Gac]+\,{}^{(int)}\mathfrak{R}_{J+1}[\Rc]\\
 && +\left(\int_{\TT}|\dk^{J+1}({}^{(ext)}\Rc)|^2\right)^{\frac{1}{2}},
 \eeaa
 where the notation $\,{}^{(ext)}\mathfrak{G}_{J+1}'[\Gac]$ has been introduced in Proposition \ref{prop:controlGaextiterationassupmtionThM8:actualresult}, and 
 where we have introduced the notation
 \beaa
\,{}^{(int)}\mathfrak{G}_{k}'[\Gac] &:=& \int_{\Mint}\Big[\left(\dk^ke_\th(\kab)\right)^2+(\dk^{\leq k}\check{\mub})^2+\left(\dk^k\left(e_4(\ze) -\b\right)\right)^2\Big].
\eeaa
\end{proposition}

The proof of Proposition \ref{prop:controlGaintiterationassupmtionThM8:actualresult} is postponed to section \ref{sec:proofofprop:controlGaintiterationassupmtionThM8:actualresult}. It will rely in particular on basic weighted estimates for transport equations along $e_3$ in $\Mint$ derived in section \ref{weightedestimatesfortheproofofThmM8controlGaMint}.\\

We now conclude the proof of Proposition \ref{prop:improvementoftheiterationassupmtionThM8}. In view of Proposition \ref{prop:controlGaintiterationassupmtionThM8:actualresult}, we have
\beaa
 \,{}^{(int)}\mathfrak{G}_{J+1}[\Gac]   &\les& \,{}^{(ext)}\mathfrak{G}_{J+1}[\Gac]+\,{}^{(ext)}\mathfrak{G}_{J+1}'[\Gac]+\,{}^{(int)}\mathfrak{R}_{J+1}[\Rc]\\
 && +\left(\int_{\TT}|\dk^{J+1}({}^{(ext)}\Rc)|^2\right)^{\frac{1}{2}}.
 \eeaa
 Also, we have in view of Proposition \ref{prop:controlGaextiterationassupmtionThM8:actualresult:Sigma*}, Proposition \ref{prop:controlGaextiterationassupmtionThM8:actualresult} and  the iteration assumption \eqref{eq:iterationassumptiondiscussionThM8:bis}
\beaa
 \,{}^{(ext)}\mathfrak{G}_{J+1}[\Gac] + \,{}^{(ext)}\mathfrak{G}_{J+1}'[\Gac]  &\les&  \,{}^{(ext)}\mathfrak{R}_{J+1}[\Rc] +\ep_\BB[J].
 \eeaa
 We infer
\beaa
 \,{}^{(int)}\mathfrak{G}_{J+1}[\Gac]   &\les& \,{}^{(int)}\mathfrak{R}_{J+1}[\Rc]+\,{}^{(ext)}\mathfrak{R}_{J+1}[\Rc] +\ep_\BB[J]+\left(\int_{\TT}|\dk^{J+1}({}^{(ext)}\Rc)|^2\right)^{\frac{1}{2}}.
 \eeaa 
 Together with Proposition \ref{prop:controlGaextiterationassupmtionThM8}, we deduce 
\beaa
 \,{}^{(int)}\mathfrak{G}_{J+1}[\Gac]   &\les& \ep_\BB[J]+\ep_0\left(\Nk^{(En)}_{J+1}+\NN^{(match)}_{J+1}\right)+\left(\int_{\TT}|\dk^{J+1}({}^{(ext)}\Rc)|^2\right)^{\frac{1}{2}}
 \eeaa  
which concludes the proof of  Proposition \ref{prop:improvementoftheiterationassupmtionThM8}.\\

 The rest of this section is dedicated to the proof of Proposition \ref{prop:controlGaintiterationassupmtionThM8:actualresult}.


\subsection{Weighted estimates for transport equations along $e_3$ in $\Mint$}
\lab{weightedestimatesfortheproofofThmM8controlGaMint}


\begin{lemma}\lab{lemma:MorawetzforRicciinRR2}
Let the following transport equation in $\Mint$
$$e_3(f)+\frac{a}{2}\ka f=h$$
where $a\in\RRR$ is a given constant, and $f$ and $h$ are scalar functions. 
Then, $f$ satisfies
\beaa
\int_{\Mint}f^2  &\les& \int_{ \TT}f^2 + \int_{\Mint} h^2.
\eeaa
\end{lemma}

\begin{proof}
Multiply by $f$ to obtain
$$\frac{1}{2}e_3(f^2)+\frac{a}{r}f^2=hf.$$
Next, integrate over $S_{\ub,r}$ to obtain
\beaa
\frac{1}{2}e_3\left(\int_{S_{\ub,r}}f^2\right) &=& \int_{S_{\ub,r}}\frac{1}{2}(e_3(f^2)+\kab f^2)\\
&=& -\int_{S_{\ub,r}}\frac{a-1}{2}\kab f^2+\int_{S_{\ub,r}}hf\\
&=& -\frac{a-1}{2}\ov{\kab}\int_{S_{\ub,r}}f^2- \frac{a-1}{2}\int_{S_{\ub,r}}\kabc f^2+\int_{S_{\ub,r}}hf
\eeaa
and hence
$$\frac{1}{2}e_3\left(r^{b}\int_{S_{\ub,r}}f^2 \right) +\frac{1}{2}\left(a+1+\frac{b}{2}\right)\ov{\kab}r^{b}\int_{S_{\ub,r}}f^2= - \frac{a-1}{2}r^{b}\int_{S_{u,r}}\kabc f^2+r^{b}\int_{S_{\ub,r}}hf$$
where we used the fact that $2e_3(r)=r\ov{\kab}$. Also, choosing $b=-2a$, we obtain
$$\frac{1}{2}e_3\left(r^{-2a}\int_{S_{\ub,r}}f^2 \right) +\frac{1}{2}\ov{\kab}r^{-2a}\int_{S_{\ub,r}}f^2= - \frac{a-1}{2}r^{-2a}\int_{S_{u,r}}\kabc f^2+r^{-2a}\int_{S_{\ub,r}}hf.$$

Next, let $1\leq \ub\leq u_*$. We now integrate in $r$ and along $\underline{\CC}_\ub$ in $\Mint$. Since $r$ is bounded on $\Mint$ from above and below, we obtain, for $\ep_0>0$ small enough,
\beaa
\int_{2m_0-2m_0\de_0}^{\rh }\int_{S_{\ub,r}}f^2   &\les& \int_{S_{\ub,\rh }}f^2 +\int_{2m_0-2m_0\de_0}^{\rh }\int_{S_{\ub,r}} h^2.
\eeaa
We may now integrate in $\ub$ to deduce
\bea\lab{eq:estimatebeforermkonvolumesinMintforThmM8}
\int_1^{u_*} \int_{2m_0-2m_0\de_0}^{\rh }\int_{S_{\ub,r}}f^2   &\les& \int_1^{u_*}\int_{S_{\ub,\rh }}f^2 +\int_1^{u_*}\int_{2m_0-2m_0\de_0}^{\rh }\int_{S_{\ub,r}} h^2.
\eea

\begin{remark}\lab{rmk:volumesinMintforThmM8}
Note that we have the following consequence of the coarea formula
\beaa
d\TT= \frac{\vsi\sqrt{\ov{\ka}+A}}{\sqrt{-\ov{\kab}}}\,\,\,d\mu_{\ub \rh}d\ub,
\eeaa
where we used that $\TT=\{r=\rh\}$. Also, we have in $\Mint$
\beaa
d\MM=\frac{4\vsi^2}{r^2\ov{\kab}^2}d\mu_{\ub, r}d\ub dr.
\eeaa
We infer, in $\Mint$,
\beaa
d\TT=\sqrt{1-\frac{2m_0}{\rh}}(1+O(\ep_0))\,\,\,d\mu_{\ub,r_0}d\ub,
\eeaa
and
\beaa
d\MM=(1+O(\ep_0))d\mu_{\ub, r}d\ub dr.
\eeaa
\end{remark}

Relying on Remark \ref{rmk:volumesinMintforThmM8} we deduce from \eqref{eq:estimatebeforermkonvolumesinMintforThmM8}
\beaa
\int_{\Mint}f^2   &\les& \int_{ \TT}f^2 + \int_{\Mint} h^2
\eeaa
as desired. This concludes the proof of the lemma.
\end{proof}

\begin{corollary}\lab{cor:MorawetzforRicciinRR2}
Let the following transport equation in $\Mint$
$$e_3(f)+\frac{a}{2}\kab f=h$$
where $a\in\RRR$ is a given constant, and $f$ and $h$ are scalar functions. Then, $f$ satisfies for $5\leq l\leq k_{large}+1$
\beaa
\int_{\Mint}(\dk^kf)^2  &\les&  \int_{\TT}(\dk^{\leq k}f)^2 +\int_{\Mint}(\dk^{\leq k-1}f)^2\\
&&+\int_{\Mint}(\dk^{\leq k}h)^2+\left(\sup_{\Mint}|\dk^{\leq k-5}f|\right)^2\Big(\,{}^{(int)}\mathfrak{G}_{k-1}[\Gac]+\,{}^{(int)}\mathfrak{G}_{k}[\kabc]\Big)^2.
\eeaa
\end{corollary}

\begin{proof}
The proof is based on Lemma \ref{lemma:MorawetzforRicciinRR2}. It is similar to the one of Corollary \ref{cor:MorawetzforRicciinRR1} and left to the reader.
\end{proof}


\subsection{Proof of Proposition \ref{prop:controlGaintiterationassupmtionThM8:actualresult}}\lab{sec:proofofprop:controlGaintiterationassupmtionThM8:actualresult}


We introduce the following notation which will constantly appear on the RHS of the equalities below
\bea\lab{eq:defintionofnormNMintGacRcproofThmM8}
\nn N^{(int)}[J, \Gac, \Rc] &:=&   \,{}^{(ext)}\mathfrak{G}_{J+1}[\Gac]+\,{}^{(ext)}\mathfrak{G}_{J+1}'[\Gac]+\,{}^{(int)}\mathfrak{R}_{J+1}[\Rc]\\
 && +\left(\int_{\TT}|\dk^{J+1}({}^{(ext)}\Rc)|^2\right)^{\frac{1}{2}}+\ep_0\Big( \,{}^{(int)}\mathfrak{G}_{J+1}[\Gac] + \,{}^{(int)}\mathfrak{G}_{J+1}'[\Gac] \Big).
 \eea

{\bf Step 1.} In view of Lemma \ref{lemma:howtogofromfoliationofMexttofoliationofMintonTT} relating the Ricci coefficients and curvature components of $\Mint$ to the ones of $\Mext$ on the timelike hypersurface $\TT$, we have
\beaa
\int_\TT\big|\dk^{J+1}({}^{(int)}\Gac)\big|^2 &\les& \int_\TT|\dk^{J+1}({}^{(ext)}\Gac)|^2.
\eeaa
Also, using again Lemma \ref{lemma:howtogofromfoliationofMexttofoliationofMintonTT}, we have
\beaa
&&\int_\TT\left|\dk^{J+1}\Big({}^{(int)}e_\th({}^{(int)}\kab), {}^{(int)}\check{\mub}, {}^{(int)}e_4({}^{(int)}\ze-{}^{(int)}\b)\Big)\right|^2 \\
&\les& \int_\TT\left|\dk^{J+1}\Big({}^{(ext)}e_\th({}^{(int)}\kab)-4{}^{(ext)}\bb, {}^{(int)}\check{\mu}, {}^{(int)}e_3({}^{(int)}\ze)+{}^{(int)}\bb\Big)\right|^2+\int_{\TT}|\dk^{J+1}({}^{(ext)}\Rc)|^2
\eeaa
We deduce, using that $\TT=\{r=\rh\}$ and the definitions of the various norms on $\Mext$, 
\beaa
&&\int_\TT\big|\dk^{J+1}({}^{(int)}\Gac)\big|^2+\int_\TT\left|\dk^{J+1}\Big({}^{(int)}e_\th({}^{(int)}\kab), {}^{(int)}\check{\mub}, {}^{(int)}e_4({}^{(int)}\ze-{}^{(int)}\b)\Big)\right|^2\\
 &\les& \,{}^{(ext)}\mathfrak{G}_{J+1}[\Gac]+\,{}^{(ext)}\mathfrak{G}_{J+1}'[\Gac]+\left(\int_{\TT}|\dk^{J+1}({}^{(ext)}\Rc)|^2\right)^{\frac{1}{2}}.
\eeaa
and hence, in view of \eqref{eq:defintionofnormNMintGacRcproofThmM8}, 
\beaa
&&\int_\TT\big|\dk^{J+1}({}^{(int)}\Gac)\big|^2+\int_\TT\left|\dk^{J+1}\Big({}^{(int)}e_\th({}^{(int)}\kab), {}^{(int)}\check{\mub}, {}^{(int)}e_4({}^{(int)}\ze-{}^{(int)}\b)\Big)\right|^2\\
 &\les& \Big(N^{(int)}[J, \Gac, \Rc]\Big)^2.
\eeaa

From now on, we only consider the frame of $\Mint$. The previous estimate can be written as
\beaa
\nn\max_{k\leq J+1}\Bigg(\int_{\TT}\Big((\dk^k\check{\mub})^2+(\dk^k\ze)^2+(\dk^k\check{\ka})^2+(\dk^k\vth)^2+(\dk^k\check{\kab})^2+(\dk^k\vthb)^2\\
+(\dk^k(e_4(\ze)-\b))^2+(\dk^k\xi)^2+(\dk^k\omc)^2+(\dk^k\Oc)^2\Big)\Bigg) &\les&  \Big(N^{(int)}[J, \Gac, \Rc]\Big)^2
\eeaa
and 
\beaa
\nn\max_{k\leq J+1}\int_{\TT}(\dk^ke_\th(\kab))^2 &\les&   \Big(N^{(int)}[J, \Gac, \Rc]\Big)^2.
\eeaa 

{\bf Step 2.} We have obtained all the desired estimates on $\TT$ for the foliation of $\Mint$ in Step 1. We now derive the desired estimates on $\Mint$. To this end, we rely on the transport equations in the $e_3$ directions which we estimate thanks to Corollary \ref{cor:MorawetzforRicciinRR2}. The initial data on $\TT$ is estimated thanks to Step 1. In particular, we proceed in the following order
\begin{itemize}
\item From 
\beaa
e_3(\check{\kab})+\ov{\kab}\,\kabc &=& \err[e_3\kabc]
\eeaa
and the bootstrap assumptions, we infer 
\beaa
\nn\max_{k\leq J+1}\int_{\Mint}(\dk^k\check{\kab})^2 &\les&   \Big(N^{(int)}[J, \Gac, \Rc]\Big)^2.
\eeaa

\item From 
\beaa
e_3(e_\th(\kab))+\frac{3}{2}\kab e_\th(\kab) &=& -\frac{1}{2}\vthb e_\th(\kab) - \frac{1}{2}e_\th(\vthb^2)
\eeaa
and the bootstrap assumptions, we infer
\beaa
\nn\max_{k\leq J+1}\int_{\Mint}(\dk^ke_\th(\kab))^2 &\les&   \Big(N^{(int)}[J, \Gac, \Rc]\Big)^2.
\eeaa

\item From 
\beaa
e_3(\check{\mub}) + \frac{3}{2}\ov{\kab}\,\check{\mub} &=& -\frac{3}{2}\ov{\mub}\,\check{\kab}+\err[e_3\check{\mub}],
\eeaa
the above control of $\check{\kab}$ and $e_\th(\kab)$ (the control of $e_\th(\kab)$ is needed to estimate $\err[e_3\check{\mub}]$), and the bootstrap assumptions, we infer
\beaa
\nn\max_{k\leq J+1}\int_{\Mint}(\dk^k\check{\mub})^2 &\les&   \Big(N^{(int)}[J, \Gac, \Rc]\Big)^2.
\eeaa

\item From
\beaa
e_3(\vthb) +\kab\,\vthb &=& -2\aa
\eeaa
and the control of $\aa$, we infer
\beaa
\nn\max_{k\leq J+1}\int_{\Mint}(\dk^k\vthb)^2 &\les&   \Big(N^{(int)}[J, \Gac, \Rc]\Big)^2.
\eeaa

\item From
\beaa
e_3(\ze) +\kab\ze &=& \bb -\vthb\ze
\eeaa
the control of $\bb$, and the bootstrap assumptions, we infer
\beaa
\nn\max_{k\leq J+1}\int_{\Mint}(\dk^k\ze)^2 &\les&   \Big(N^{(int)}[J, \Gac, \Rc]\Big)^2.
\eeaa

\item From 
\beaa
e_3(\kac)+\frac{1}{2}\ov{\kab}\kac &=& -\frac{1}{2}\ov{\ka}\kabc +2\ddd_1\ze+2\check{\rho}+\err[e_3\kac]\\
&=& -\frac{1}{2}\ov{\ka}\kabc +2\check{\mub}+4\check{\rho}-\frac{1}{2}\vth\vthb+\frac{1}{2}\ov{\vth\vthb}+\err[e_3\kac],
\eeaa
the control of $\check{\rho}$, the above control of $\check{\kab}$ and $\check{\mub}$, and the bootstrap assumptions, we infer
\beaa
\nn\max_{k\leq J+1}\int_{\Mint}(\dk^k\kac)^2 &\les&   \Big(N^{(int)}[J, \Gac, \Rc]\Big)^2.
\eeaa

\item From
\beaa
e_3(\vth) +\frac{1}{2}\kab\vth &=& 2\dds_2\ze -\frac{1}{2}\ka\vthb+2\ze^2\\
&=& 2\dds_2\ddd_1^{-1}\left(\check{\mub}+\check{\rho}-\frac{1}{4}\vth\vthb+\frac{1}{4}\ov{\vth\vthb}\right) -\frac{1}{2}\ka\vthb+2\ze^2,
\eeaa
the control of $\check{\rho}$, the above control of $\vthb$ and $\check{\mub}$, and the bootstrap assumptions, we infer
\beaa
\nn\max_{k\leq J+1}\int_{\Mint}(\dk^k\vth)^2 &\les&   \Big(N^{(int)}[J, \Gac, \Rc]\Big)^2.
\eeaa

\item From 
\beaa
e_3(\omc) &=& \rhoc +\err[e_3\omc],
\eeaa
the control of $\check{\rho}$, and the bootstrap assumptions, we infer
\beaa
\nn\max_{k\leq J+1}\int_{\Mint}(\dk^k\omc)^2 &\les&   \Big(N^{(int)}[J, \Gac, \Rc]\Big)^2.
\eeaa

\item From
\beaa
&& e_3(e_4(\ze)-\b)+\kab(e_4(\ze)-\b)\\
 &=& -\ka\bb  +\left(\frac{1}{2}\ka\kab+2\mub-6\rho\right)\ze\\
 && +\vth\bb  -\vthb\b+\xi\aa -\vthb\ze  +2\om\vthb\ze -4\ze\dds_1\ddd_1^{-1}\left(\check{\mub}+\check{\rho}-\frac{1}{4}\vth\vthb+\frac{1}{4}\overline{\vth\vthb}\right) -2\ze^3,
\eeaa
the control of $\bb$, the above control of $\ze$, and the bootstrap assumptions, we infer
\beaa
\nn\max_{k\leq J+1}\int_{\Mint}(\dk^k(e_4(\ze)-\b))^2 &\les&   \Big(N^{(int)}[J, \Gac, \Rc]\Big)^2.
\eeaa

\item From 
\beaa
e_3(\xi) &=& (e_4(\ze)-\b)+2\b+\ka\ze+\vth\ze,
\eeaa
the control of $\bb$, the above control of $e_4(\ze)-\b$ and $\ze$, and the bootstrap assumptions, we infer
\beaa
\nn\max_{k\leq J+1}\int_{\Mint}(\dk^k\xi)^2 &\les&   \Big(N^{(int)}[J, \Gac, \Rc]\Big)^2.
\eeaa
\end{itemize}

In view of the above estimates, of the definition \eqref{eq:defintionofnormNMintGacRcproofThmM8} of $N^{\leq 4m_0}[J, \Gac, \Rc]$, and of the various norms, we infer
\beaa
 \,{}^{(int)}\mathfrak{G}_{J+1}[\Gac] + \,{}^{(int)}\mathfrak{G}_{J+1}'[\Gac] &\les&  \,{}^{(ext)}\mathfrak{G}_{J+1}[\Gac]+\,{}^{(ext)}\mathfrak{G}_{J+1}'[\Gac]+\,{}^{(int)}\mathfrak{R}_{J+1}[\Rc]\\
 && +\left(\int_{\TT}|\dk^{J+1}({}^{(ext)}\Rc)|^2\right)^{\frac{1}{2}}+\ep_0\Big( \,{}^{(int)}\mathfrak{G}_{J+1}[\Gac] + \,{}^{(int)}\mathfrak{G}_{J+1}'[\Gac] \Big)
\eeaa
and hence, for $\ep_0$ small enough, 
\beaa
 \,{}^{(int)}\mathfrak{G}_{J+1}[\Gac] + \,{}^{(int)}\mathfrak{G}_{J+1}'[\Gac] &\les&  \,{}^{(ext)}\mathfrak{G}_{J+1}[\Gac]+\,{}^{(ext)}\mathfrak{G}_{J+1}'[\Gac]+\,{}^{(int)}\mathfrak{R}_{J+1}[\Rc]\\
 && +\left(\int_{\TT}|\dk^{J+1}({}^{(ext)}\Rc)|^2\right)^{\frac{1}{2}}.
\eeaa
This concludes the proof of Proposition \ref{prop:controlGaintiterationassupmtionThM8:actualresult}.


\section{Proof of Proposition \ref{prop:controlglobalframeiterationassupmtionThM8}}\lab{sec:proofprop:controlglobalframeiterationassupmtionThM8} 


Lemma \ref{lemma:estimatesfortheglobalframeinthematchingregion} corresponds to the particular case $J=k_{large}-1$ of Proposition \ref{prop:controlglobalframeiterationassupmtionThM8}. Its proof in section \ref{sec:proofoflemma:estimatesfortheglobalframeinthematchingregion}  extends immediately to the case $k_{small}-2\leq J\leq k_{large}-1$ which thus yields the proof of Proposition \ref{prop:controlglobalframeiterationassupmtionThM8}.


\chapter{GCM PROCEDURE}\lab{chap:proofofGCMprocedure}



 \section{Preliminaries} 
 

We consider an  axially symmetric  polarized   spacetime regions  $\RR$ 
  foliated by two functions $(u, s)$ such that
\begin{itemize}
\item On $\RR$, $(u, s)$ defines an outgoing  geodesic foliation as in section \ref{sec:mainequationsforoutgoiggeodesicfoliations}.

\item We denote by $(e_3, e_4, e_\th)$ the null frame adapted to the outgoing geodesic foliation $(u, s)$ on $\RR$. 

\item Let 
\bea
\ovS &:=& S(\ug, \sg)
\eea
and  $\rg$ the area radius of $\ovS$, where $S(u, s)$ denote the 2-spheres  of the outgoing geodesic foliation $(u, s)$ on $\RR$.

\item In adapted coordinates $(u,s,\th, \vphi)$ with $b=0$, see   Proposition \ref{prop:outgoinggeodesiccoordinates}, the  spacetime  metric $\g$ in $\RR $  takes the form,  with $\Omb=e_3(s), \, \,  \underline{b}=e_3 (\th)$,
\bea
 \label{reducedmetric-geodesicfoliation-GCM}
  \g=-2\vsi du ds +\vsi^2 \Omb du^2+ \ga\left( d\th- \frac 1 2 \vsi\underline{b} du\right)^2 + e^{2\Phi} d\vphi^2 , \, 
 \eea
 where $\th$ is chosen such that $b=e_4(\th)=0$.
   
\item
 The  spacetime  metric induced on $S(u,s)$ is given by,
 \bea
\gS= \ga d\th^2 + e^{2\Phi}   d\vphi^2. 
 \eea
 \item The relation between the null frame and coordinate system is given by
 \bea
 \label{transf:coords-frame}
 e_4 =\pr_s, \qquad e_3 =\frac{2}{\vsi} \pr_u+\Omb  \pr_s+\underline{b}  \pr_\th,\qquad e_\th=\ga^{-1/2} \pr_\th. 
 \eea 
 \item
We denote the induced metric on $\ovS$ by 
 \beaa
 \ovgS=\ovga d\th^2 + e^{2\Phi} d\vphi^2.
 \eeaa
\end{itemize}

 \begin{definition} 
\label{defintion:regionRRovr}
Let $0<\dg\leq \epg $   two sufficiently   small   constants.
 Let  $(\ug, \sg)$ real numbers so that
\bea\lab{eq:rangeofugandsg}
1  \leq \ug<+\infty,\quad  4m_0\leq\sg < +\infty.
\eea

We define  $\RR=\RR(\dg, \epg)$  to be the region
\bea
\lab{definition:RR(dg,epg)}
\RR:=\left\{|u-\ug|\leq\de_{\RR} ,\quad |s-\sg|\leq  \de_\RR \right\}, \qquad \de_{\RR}:=\dg \big(\epg\big)^{-\frac{1}{2}},
\eea
such that  assumption {\bf A1-A3} below  with constant $\epg$  on  the background foliation of $\RR$,   are verified. The  smaller constant $\dg$          controls the size of    the GCMS quantities  as it will be made precise   below. 
\end{definition}

 In this section we define the  renormalized  Ricci  and  curvature  components,
  \beaa
\bsplit
\Gac :&=\left\{ \check{\ka}, \, \vth,\, \ze, \, \eta,\, \ov{\ka}-\frac 2 r,\,  \ov{\kab}   +\frac{2\Up}{r}, \check{\kab},\, \vthb, \, \xib,\,  \check{\omb}, \,   \ov{ \omb}  -\frac{m}{r^2},  \, \check{\Omb}, \, \big(\ov{\Omb}+\Up\big),\, \big(\ov{\vsi}+1\big)\right \},\\
\Rc:&=\left\{ \a,\,  \b,\, \rhoc,\, \ov{\rho}+\frac{2m}{r^3},\, \bb, \, \aa\right\}.
\end{split}
\eeaa
Since our foliation is outgoing geodesic we also have, 
\bea
\xi=\om=0, \quad \etab+\ze=0.
\eea 
We decompose  $\Gac=\Ga_g\cup \Ga_b$ where,
\bea
\lab{definition:Gag-Gab-GCM}
\bsplit
\Ga_g&=\left\{ \check{\ka}, \, \vth,\, \ze, \, \kabc, \, \ov{\ka}-\frac 2 r,\,  \ov{\kab}   +\frac{2\Up}{r},    \right\},\\
\Ga_b&=\left\{\eta,\,   \vthb, \, \xib,\,  \check{\omb}, \,    \ov{ \omb}  -\frac{m}{r^2},  \, r^{-1}\Ombc,\, r^{-1} \vsic ,  r^{-1}   \big(\ov{\Omb}+\Up\big),\,  r^{-1}\big(\ov{\vsi}+1\big)\right\}.\\
\end{split}
\eea
Given a $p$-reduced scalar $f\in \sk_p(\MM)$, with respect to the given geodesic foliation   on $\RR$,
    we consider   the following   norms on   spheres  $S=S(u,r)\subset\RR$, 
  \bea
  \label{Norms-spacetimefoliation-GSMS}
  \bsplit
  \| f\|_{\infty} (u,r):&=\| f\|_{L^\infty\big(S(u,r)\big)}, \qquad  \| f\|_{2} (u,r):=\| f\|_{L^2\big(S(u,r)\big)}, \\
  \|f\|_{\infty,k}(u, r)&= \sum_{i=0}^k \|\dk^i f\|_{\infty }(u, r),  \qquad 
\|f\|_{2,k}(u, r)=\sum_{i=0}^k \|\dk^i f\|_{2}(u, r).
\end{split}
  \eea
  where, we recall,   that $\dk^i$ stands for any   combination  of length $i$ of operators  of the from 
   $e_3, r e_4, \dkb$. Recall that,
   \bea
\dkb^s f=
\begin{cases}
 r^{2p}\lapp_k ^p, \qquad\qquad  \mbox{if} \quad  s=2p,\\
 r^{2p+1} \ddd_k  \lapp_k^p ,\qquad \mbox{if} \quad  s=2p+1.
\end{cases}
\eea
   On  a given polarized  surface $\S\subset \RR$, not necessarily   a leaf  $S$ of the  given foliation, we define
\bea
\| f\|_{\hk^q_s(\S)}:&=&\sum_{i=0}^s\| \left(\dkb^\S\right)^i f\|_{L^q(\S)}.
\eea
where $\dkb^\S$ is defined as above with respect to the  intrinsic metric on $\S$. In the particular case when $q=2$ we  omit the upper  index i.e., $\hk_s(\S)=\hk_s^2(\S)$.


\subsection{Main assumptions}\label{GCM:Main Assumptions}


Given an integer $s_{max}$, we assume the following\footnote{In applications, $s_{max}=k_{small}+4$  in Theorem  M7, and $s_{max}=k_{large }+5$  in Theorem M0 and Theorem M6.} 

{\bf A1.} For all $k\le s_{max}$, we have on  $\RR$
\bea\lab{eq:assumtionsonthegivenusfoliationforGCMprocedure} 
\bsplit
\| \Ga_g\|_{k, \infty}&\les  \epg  r^{-2},\\
\| \Ga_b\|_{k, \infty}&\les  \epg  r^{-1},
\end{split}
\eea
and,
\bea
\bsplit
\|\a, \b, \rhoc, \muc\|_{k, \infty}&\les \epg r^{-3},\\
\|e_3( \a, \b)\|_{k-1, \infty}&\les \epg r^{-4},\\
\|\bb\|_{k,\infty}&\les \epg r^{-2},\\
\|\aa\|_{k,\infty}&\les \epg r^{-1}.
\end{split}
\eea

{\bf A2.} We have, with $m_0$ denoting the mass of the unperturbed spacetime,
\bea\lab{eq:assumtionsonthegivenusfoliationforGCMprocedure:Hawkingmass} 
\sup_{\RR}\left|\frac{m}{m_0}-1\right| &\les& \epg.
\eea

{\bf A3.}  The metric coefficients are assumed to satisfy the following assumptions  in $\RR$, for all $k\le s_{max}$
\bea
\lab{eq:assumtionsonthegivenusfoliationforGCMprocedure:bis} 
 r \left\|\left ( \frac{\ga}{r^2}-1 ,\,\,  \underline{b},  \, \, \frac{e^\Phi}{r\sin\th}-1\right)\right\|_{\infty, k}+
\left \|   \Omb+\Up\right\|_{\infty, k} +\left \|  \vsi-1\right\|_{\infty, k}\les \epg.
\eea

\begin{remark}
The above assumptions imply in particular the following 
\beaa
|e_4(r)|,\, |e_3(r)|\les 1, \qquad e_4(s)=1+O(\epg), \qquad e_3(u)=2+O(\epg), \qquad e_4(u)=0.
\eeaa
Hence, since $r=\rg$ at $(\ug, \sg)$, we infer 
\beaa
|r-\rg| &\les& |s-\sg|+|u-\ug|,
\eeaa
and thus, in view of the definition \eqref{definition:RR(dg,epg)} of $\RR$,
\bea\lab{eq:comparisionrminusrgforGCMprocedure}
\sup_{\RR}|r-\rg| &\les&  \dg \big(\epg\big)^{-\frac{1}{2}}.
\eea
\end{remark}

We  will  make use of the following lemma, see Lemmas \ref{lemma:clearlyveryusefulforannoyingaxis1}  and \ref{lemma:clearlyveryusefulforannoyingaxis2}.
\begin{lemma}
\lab{lemma:clearlyveryusefulforannoyingaxis1-2}
Under the assumption {\bf A3} for the metric coefficients we have,
\bea
\lab{eq:clearlyveryusefulforannoyingaxis1}
r\big|e_\th(\Phi)\big|&\le \frac{2}{sin\th}, \qquad 
\frac{1}{\sin\th}&\le 2\left( r|e_\th\Phi|+1\right) .
\eea
Moreover,  for any reduced 1-scalar $h$, we have
\bea
\lab{eq:clearlyveryusefulforannoyingaxis2}
\sup_S\frac{|h|}{e^\Phi} &\les r^{-1}\sup_S(|h|+|\dkb h|), \qquad 
\left\|\frac{h}{e^\Phi}\right\|_{L^2(S)}\les r^{-1}\|h\|_{\hk_1(S)}.
\eea
\end{lemma}


\subsection{Elliptic Hodge  lemma}


 We shall often make use of the  results of Proposition \ref{prop:2D-hodge-reduced}  and Lemma \ref{Lemma:poincarefor-dds2}  which we  rewrite as follows.

 \begin{lemma} 
\label{prop:2D-hodge-reduced-GCM}
Under the assumptions ${\bf A1, A3}$ the following  elliptic estimates hold true for the Hodge operators
$ \ddd_1, \ddd_2, \dds_1, \dds_2$, for all $k\le s_{max}$
\begin{enumerate}
  \item  If   $  f  \in \sk_1(S)$
\beaa
\| \dkb f\|_{\hk_k (S)} + \|  f\|_{\hk_k  (S)}    \les r   \|\ddd_1   f  \|_{\hk_k(S)}.
\eeaa
\item If $f\in \sk_2(S)$
\beaa
\| \dkb f\|_{\hk_k (S)} + \|  f\|_{\hk_k  (S)}    \les r   \|\ddd_2   f  \|_{\hk_k(S)}.
\eeaa

\item  If  $f\in \sk_0(S)$
\beaa
  \|\dkb f\|_{\hk_{k} (S) } \les r  \|\dds_1\, f\|_{\hk_k(S)}.
\eeaa

\item   If  $   f  \in \sk_1(S) $
\beaa
  \|  f\|_{\hk_{k+1}  (S)}&\les&  r  \|\dds_2\, f\|_{\hk_k(S)}+r^{-2}\left |\int_S e^\Phi f\right|.
\eeaa

\item  If  $   f  \in \sk_1(S) $
\beaa
\left\| f-  \frac{\int_S f e^\Phi} {\int_S e^{2\Phi} }  e^{\Phi}\right\|_{\hk_{k+1}  (S)}&\les  r  \|\dds_2\, f\|_{\hk_k(S)}.
\eeaa
\end{enumerate}
\end{lemma}

We shall often make use fo the following non-sharp  product  estimate on $\S$, see Proposition \ref{prop:sobolevandproductesitmatesonS}.
\begin{lemma}
\lab{prop:sobolevandproductesitmatesonS-GCMS}
The following estimates hold true  on   a given polarized  surface  $\S\subset \RR$, for any contraction  between two  reduced   scalars $\psi_1, \psi_2$, $k\ge 2 $,
\beaa
\| \psi_1 \c \psi_2\|_{\hk_k(\S)} &\les&  r^{-1} \| \psi_1 \|_{\hk_k(\S)}   \| \psi_1 \|_{\hk_k(\S)}.
\eeaa
\end{lemma}


\section{Deformations  of $S$ surfaces} \label{section:Deformationofsurfaces}  



 \subsection{Deformations}


  Recall that  $\ovS=S(\ovu, \ovs)$  is a fixed    sphere of the  $(u, s)$ outgoing geodesic foliation of  a fixed  spacetime region $\RR=\RR(\epg,\dg)$.
 \begin{definition}
 \label{definition:Deformations}
 We say that    $\S$ is an $O(\epg)$\,  $\Z$-polarized deformation of $ \ovS$ if there exists    a map $\Psi:\ovS\longrightarrow \S $ of the form,
 \bea
 \Psi(\ovu, \ovs, \th, \vphi)=\left( \ovu+ U(\th), \ovs+S(\th), \th, \vphi \right)
 \eea
 where   $U, S$ are smooth  functions   defined on the interval  $[0,\pi]$
 of amplitude at most $\epg$.  
 We denote by $\psi$ the reduce map defined on the interval $ [0,\pi]$,
 \bea
 \psi(\th)= ( \ovu+ U(\th), \ovs+S(\th), \th).
 \eea
 We restrict ourselves to deformations which fix the South Pole, i.e. 
 \bea
 U(0)=S(0)=0.
 \eea
  \end{definition}

 
 \subsection{Pull-back map}  
 
 
 We recall that given a scalar function $f$ on $\S$ 
 one defines its pull-back on $\ovS$ to be  the function,
  \beaa
 f^\#:=  \Psi^\# f =f\circ \Psi.
  \eeaa 
On the other hand, given a vectorfield  $X$ on $\ovS$ one defines its push-forward   $\Psi_\# X$ to be the vectorfield 
on $\S$  defined by,
\beaa
\Psi_\# X(f)=X(\Psi^\# f)= X( f\circ \Psi).
\eeaa
Given a  covariant tensor $U$ on $\S$,  one defines  its pull back  to $\ovS$ to be  the tensor 
\beaa
\Psi^\# U ( X_1, \ldots, X_k)=  U(\Psi_\#  X_1, \ldots,   \Psi_\#X_k).
\eeaa

 \begin{lemma}\label{le:pullbackS}
 Given a $\Z$-invariant deformation  $\Psi:\ovS\longrightarrow \S$, we have,
  \begin{enumerate}
 \item  Let $\gS^\S$ the induced  metric  on $\S$ and  $\gS^{\S, \#}=  \ga^{\S, \#} d\th^2 + e^{2\Phi^\#} d\vphi^2 $ its pull-back to $\ovS$. 
The metric  coefficients 
$\ga^\S$ and $\ga^{\S, \#} $ are  related by,
\bea
   \label{deormation:formula-gaS}
     \ga^{\S, \#} (\th)=\gaS(\psi(\th))=\gaS(\ovu+U(\th), \ovs+S(\th), \th)
\eea
 where $\gaS$  is defined implicitly  by,
  \bea\lab{eq:formulaforgaSintermsofUprimeSprime}
 ( \gaS)^\#=\ga^\# +\big(\vsi^\#\big)^2 \left(\Omb+\frac{1}{4}\underline{b}^2\ga \right)^\#\, ( U' )^2 -2 \vsi^\# U'  S'- (\ga  \vsi\underline{b})^\#  U',
  \eea
 that is,
  \beaa
 \gaS(\psi(\th))&=&\ga(\psi(\th))  +\vsi^2(\psi(\th))\left(\Omb(\psi(\th))+\frac{1}{4}( \underline{b}(\psi(\th)))^2  \ga(\psi(\th)) \right)\, ( U' (\th))^2\\
  &-&  2\vsi(\psi(\th))  U'(\th)  S'(\th)-\ga(\psi(\th)) \vsi(\psi(\th))\underline{b}(\psi(\th))  U'(\th).
\eeaa

 \item  The $\Z$-invariant  vectorfield $ \pr_\th^\S:=     \Psi_\# (\pr_\th)$ is tangent to $\S$ and 
 \bea
  \pr_\th^\S|_{\Psi(p)}&=&\Big[\left( \pr_\th S-\frac{\vsi}{2} \Omb \pr_\th U\right) e_4+\frac{\vsi}{2} \pr_\th U e_3 +\sqrt{\ga}\left(  1 -\frac{\vsi}{2}  \underline{b}\pr_\th U\right) e_\th\Big] \Big|_{\Psi(p)}. \qquad 
 \eea
 
 \item If $f\in\sk_k(\S)$ and $P^\S$  is a geometric operator acting on $ f$   then,
 \bea
 \label{eq:pullback-geometric-operators}
 (P^\S[f])^\#&=& P^{\S, \#} [f^\#]
 \eea
 where, $P^{\S, \#} $ is the corresponding geometric  operator on $\ovS$ with respect to   the metric $ \gS^{ \S, \#}$ and $f^\#=\psi^\# f$.  
   
 \item  The $L^2$ norm of  $f^\#=\psi^\# f$ with respect to the metric $\gS^{\S, \#}$ is the same as 
 as the $L^2$ norm of  $f$ with respect to the metric $\gS^\S$, i.e.,
 \beaa
 \int_{\ovS}    |f^\#|^2    da_{\gS^{\S, \#}}&=& \int_\S |f|^2  da_{\gS^\S}.
 \eeaa
 
 \item If  $f\in \hk_k(\S)$ and $f^\#$ is its pull-back by $\psi$ then,
 \beaa
 \|f^\#\|_{\hk_k(\ovS,\, \gS^{\S,\#})} = \| f\|_{\hk_k(\S)}.
 \eeaa
  \end{enumerate}  
 \end{lemma}

\begin{proof}
If $\pr_\th$ denotes  the coordinate derivative $\pr_\th =\frac{\pr}{\pr \th}$ then, at every point $p\in \ovS$,
 \beaa
 \Psi_\# (\pr_\th )|_{\Psi(p)}&=&\pr_\th U \pr_u|_{\Psi(p)}  +\pr_\th S \pr_s|_{\Psi(p)} +\pr_\th|_{\Psi(p)}, \qquad   \Psi_\# (\pr_\vphi )= \pr_\vphi.
 \eeaa
 In view of  \eqref{transf:coords-frame} we have 
 \beaa
 \pr_s =e_4,\quad \pr_u=\frac{\vsi}{ 2} \left( e_3-\Omb e_4-\underline{b}\ga^{1/2} e_\th\right),\quad \pr_\th=\sqrt{\ga}   e_\th.
 \eeaa 
 Hence, at a point $\Psi(p)$ on $\S$ we have,
  \beaa
 \Psi_\# (\pr_\th )&=&\left( \pr_\th S-\frac{\vsi}{2} \Omb \pr_\th U\right) e_4+\frac{\vsi}{2} \pr_\th U e_3 +\sqrt{\ga}\left(  1 -\frac{\vsi}{2}  \underline{b}\pr_\th U\right) e_\th. 
 \eeaa
    We denote by $\gS^\#=\Psi^\#(\gS^\S ) $  the  pull back to $\ovS$  of the metric   $\gS^\S$  on $\S$, i.e. at any point $p\in \ovS$,
  \beaa
  \gS^{ \#}(\pr_\th,\pr_\th) &=&\gS^\S(\Psi_\# \pr_\th,\Psi_\#  \pr_\th)=\g(\pr_\th U  \pr_u+ \pr_\th S \pr_s+\pr_\th, \pr_\th U  \pr_u+ \pr_\th S \pr_s+\pr_\th)\\
  &=&( \pr_\th U)^2 \g_{uu}+2 \pr_\th U\pr_\th S \g_{us}+ 2  \pr_\th U \g_{u\th}+\g_{\th\th},\\
    \gS^{ \#}(\pr_\th,\pr_\vphi) &=&0,\\
    \gS^{ \#}(\pr_\vphi ,\pr_\vphi) &=& e^{2\Phi^\#},
  \eeaa
  where,
  \beaa
  \g_{uu}=\vsi^2\big(\Omb+\frac{1}{4} \ga  \underline{b}^2\big), \quad \g_{us}=-\vsi,\quad \g_{u\th}=-\frac{\vsi}{2}  \ga \underline{b}, \quad \g_{ss}=\g_{s\th}=0,\quad  \g_{\th\th}=\ga.
  \eeaa
  Hence the pull-back metric $\Psi^\# (\gS^\S)$  on $\ovS$  is given by,
  \beaa
  \ga^{\S, \#} d\th^2 + e^{2\Phi^\#} d\vphi^2 
  \eeaa
 where
  \bea
    \ga^{\S, \#}  =(\gaS)^\#,
  \eea
with $\gaS$  is defined by,
  \bea
 ( \gaS)^\#=\ga^\# +(\vsi^\#)^2 \left(\Omb+\frac{1}{4} \underline{b}^2\ga\right)^\#\, ( U' )^2 -2\vsi ^\# U'  S'- (\ga \vsi  \underline{b})^\#  U'.
  \eea
     Note that the vectorfield,
  \beaa
  e_\th^\S:=  \frac{1}{(\gaS  )^{1/2} }     \psi_\# (\pr_\th)
  \eeaa
  is tangent, $\Z$ invariant and  forms together with $e_\vphi$ an orthonormal frame on $\S$.
  Note that we can also write,
  \beaa
    e_\th^\S:=  \frac{(\ovga  )^{1/2}}{(\gaS  )^{1/2} } \Psi_\#(e_\th)
  \eeaa
   where $\ovga$ is the coefficient in front of $d\th^2$ of  the metric induced by $\g$ on $\ovS$,
  \beaa
  \ovgS=\ovga d\th^2 + e^{2\Phi} d\vphi^2.
  \eeaa
  
    In general, any geometric calculation on $\S$   can be reduced to a geometric calculation on $\ovS$  with respect to the metric $  \gS^{\S, \#}$. Moreover   the   $L^2$ norm on $\S$  with respect to the metric $\gS^\S$ is  the same as the $L^2$ norm of $f^\#=\psi^\# f$ with respect to the norm  $\gS^{\S,\#}$. This concludes the proof of the lemma.
\end{proof}

 
 \subsection{Comparison of norms between deformations}  
 

  \begin{lemma}\lab{lemma:int-gaS-ga}
 Let  $\Psi:\ovS\longrightarrow \S $  a\,  $\Z$-invariant deformation in $\RR(\epg, \dg)$   with $U, V$ verifying the bounds
 \bea
 \label{assumption-UV-dg}
\sup_{0\leq\th\leq\pi} \Big(|U'(\th)|+|S'(\th)|\Big) &\les& \dg,
 \eea
as well as the bound \eqref{eq:assumtionsonthegivenusfoliationforGCMprocedure:bis} for the coordinates system $(u,s, \th, \varphi)$ of $\RR$.  The following hold true
\begin{enumerate}
\item  We have,
 \bea
 \lab{equation:difference-ofgas}
 \big|\ga^{\S, \#} -\ovga\big|&\les \dg \rg.
 \eea 
 \item  For every $f\in\sk_k(\S)$ we have,
  \bea
 \label{eq:lemma:int-gaS-ga3}
 \|f^\#\|_{L^2(\ovS, \gS^{\S, \#})}   &=&   \|f^\#\|_{L^2(\ovS, \ovgS)}  \Big(1+ O(r^{-1} \dg)  \Big).
 \eea
\item  As a corollary  of \eqref{eq:lemma:int-gaS-ga3} (choosing $f=1$) we deduce\footnote{Recall  also from \eqref{eq:comparisionrminusrgforGCMprocedure} that $r-\ovr=O(\dg \epg^{-1/2})$.}, 
  \bea\lab{eq:comparisionbetweenrSandrgforGCMprocedure}
 \frac{r^\S}{\ovr}= 1 + O(\ovr ^{-1}  \dg )
 \eea
 where $r^\S$ is the area radius of $\S$ and $\ovr$ that of $\ovS$.
\end{enumerate}
\end{lemma}

 \begin{proof}
Recall,
\beaa
\ga^{\S, \#}(\ovu, \ovs, \th) &=&\ga(\psi(\th))  +\vsi^2(\psi(\th)) \left(\Omb(\psi(\th))+\frac{1}{4}( \underline{b}(\psi(\th)))^2  \ga(\psi(\th)) \right)\, ( U' (\th))^2\\
& -&2 \vsi(\psi(\th))  U'(\th)  S'(\th)
  -\ga(\psi(\th)) \vsi(\psi(\th)\underline{b}(\psi(\th))  U'(\th).
\eeaa
In view of our assumptions on $U'$ and $S'$ as well as our estimates \eqref{eq:assumtionsonthegivenusfoliationforGCMprocedure:bis}  for $\ga$, $\Omb$ and $\underline{b}$ and $\vsi$, we infer
\beaa
|\ga^{\S, \#}-\ga| &\les& |\ga^\# -\ga|+\rg \epg^{1/2} \dg.
\eeaa
Also, we have
\beaa
 \ga^\#(\ovu, \ovs, \th) -\ga(\ovu, \ovs, \th) &=& \ga(\ovu+U(\th), \ovs+S(\th), \th)-\ga(\ovu, \ovs, \th)\\
 &=&   \int_0^1 \frac{d}{d\la}\left[ \ga (\ovu+\la U(\th), \ovs+\la  S(\th), \th)\right] d\la\\
    &=& U(\th)\int_0^1\pr_ u \ga(\ovu+\la U(\th), \ovs+\la  S(\th), \th)d\la\\
    &&+S(\th)\int_0^1\pr_ s \ga(\ovu+\la U(\th), \ovs+\la  S(\th), \th)d\la.
 \eeaa
In view of  our estimates \eqref{eq:assumtionsonthegivenusfoliationforGCMprocedure:bis} for $\ga$, the assumption \eqref{assumption-UV-dg} on $(U', V')$  and the fact that 
\beaa
\pr_s =e_4, \quad\pr_u=\frac{\vsi}{ 2} \left( e_3-\Omb e_4-\underline{b}\ga^{1/2} e_\th\right),
\eeaa
we infer\footnote{Note that we also use the assumption $U(0)=S(0)=0$ to estimate  $(U, S)$ from $(U', S')$.}
\beaa
 |\ga^\# -\ga| &\les& \rg \dg.
\eeaa
We have finally, obtained 
\beaa
|\ga^{\S, \#}-\ga| \les |\ga^\# -\ga|+\ovr \dg\les \ovr  \dg.
\eeaa
To prove the second part of the lemma  we write,
 \beaa
 \int_{\ovS}    |f^\#|^2    da_{\gS^{\S, \#}} &=&  \int_{\ovS}  
   |f^\#|^2 \frac{\sqrt{\ga^{\S, \#}}}{\sqrt{\ovga}} da_{\ovgS}= \int_{\ovS}     |f^\#|^2  da_{\ovgS}+ \int_{\ovS}     |f^\#|^2\left( \frac{\sqrt{\ga^{\S, \#}}}{\sqrt{\ovga}}-1\right)  da_{\ovgS}
 \eeaa
which yields, in view of  the first part,
 \beaa
 \int_{\ovS}    |f^\#|^2    da_{\gS^{\S, \#}} &=&  \int_{\ovS}     |f^\#|^2  da_{\ovgS} \Big(1+ O(\ovr^{-1}\dg)  \Big).
 \eeaa
This concludes the proof of the lemma.
\end{proof}

\begin{remark}
In view of \eqref{eq:comparisionbetweenrSandrgforGCMprocedure} and \eqref{eq:comparisionrminusrgforGCMprocedure}, $\ovr, r^\S$  and  the value  of $r$ along $\S$  are all comparable.  
\end{remark}

\begin{corollary}
\lab{corr:lemma:int-gaS-ga}
Under the assumptions of Lemma \ref{lemma:int-gaS-ga}  the following estimate\footnote{Recall that $\RR:=\{|u-\ug|\leq \de_\RR ,\quad |s-\sg|\leq \de_\RR \}$, see \eqref{definition:RR(dg,epg)}.} holds true for an arbitrary scalar $f\in \sk_0(\RR)$,
\beaa
\left|\int_\S f -\int_{\ovS} f\right| &\les& \dg  \,  \rg \,\left(    \sup_{\RR}   |\dkout^{\le 1}   f|+          \sup_\RR  r|  e_3  f|\right).
\eeaa
\end{corollary}

\begin{proof}
We have,
\beaa
\int_\S f -\int_{\ovS} f&=&\int_{\ovS} f^\# \frac{\sqrt{\ga^{\S,\#}}}{\sqrt{\ovga}}-\int_{\ovS} f =
 \int_{\ovS} f^\# \left(\frac{\sqrt{\ga^{\S,\#}}}{\sqrt{\ovga}}-1\right)+\int_{\ovS}( f^\#-f ).
\eeaa
Hence,
\beaa
\left|\int_\S f -\int_{\ovS} f\right|&\les&\dg  \rg \sup_\S| f|+ \int_{\ovS} \big|f^\#- f\big|.
\eeaa
Now, proceeding as in the proof of \eqref{equation:difference-ofgas},
\beaa
f(\ovu+U(\th), \ovs +S(\th)) - f(\ovu, \ovs) &\les&  \int_0^1 \frac{d}{d\la}\left[ f (\ovu+\la U(\th), \ovs+\la  S(\th), \th)\right] d\la\\
    &=& U(\th)\int_0^1\pr_ u f(\ovu+\la U(\th), \ovs+\la  S(\th), \th)d\la\\
    &&+S(\th)\int_0^1\pr_ s  f (\ovu+\la U(\th), \ovs+\la  S(\th), \th)d\la.
\eeaa
Therefore ,
\beaa
\left|\int_\S f -\int_{\ovS} f\right|&\les&  \rg \dg\sup_\S| f|    +     \dg \rg\left(   \sup_{\RR}  \rg   |\dkout  f|  +          \sup_\RR  r  | e_3  f|\right)\\
&\les&  \dg\,  \rg  \left(    \sup_{\RR}   |\dkout^{\le 1}   f|+          \sup_\RR  r  | e_3  f|\right)
\eeaa
as stated.
\end{proof}

To compare higher order Sobolev spaces, we will need the following lemma.
 \begin{lemma}\label{lemma:int-gaS-ga:highersobolevregularity}
 Let $\ovS \subset \RR=\RR(\epg,\dg)$ as in Definition \ref{defintion:regionRRovr} verifying the assumptions {\bf A1-A3}.  Let  $\Psi:\ovS\longrightarrow \S $  be  $\Z$-invariant deformation. Assume the bound
 \bea\lab{eq:boundonU'andS'onS0forequivalenceofhigherorderSobolevnorms}
 \|(U', S')\|_{L^\infty(\ovS)}+ \max_{0\leq s\leq s_{max}} (\ovr)^{-1} \|(U', S')\|_{\hk_s(\ovS, \ovgS)} &\les& \dg.
 \eea
 Then, we have for any reduced scalar $h$ defined on $\RR$
 \beaa
 \|h\|_{\hk_s(\S)} \les r \sup_{\RR}|\dk^{\leq s}h|, \qquad \,\,\textrm{ for }0\leq s \leq s_{max}.
 \eeaa
 Also, if $f\in \hk_s(\S)$ and $f^\#$ is its pull-back by $\psi$, we have
 \beaa
 \|f\|_{\hk_s(\S)}= \|f^\#\|_{\hk_s(\ovS,\, \gS^{\S,\#})} = \| f^\#\|_{\hk_s(\ovS, \ovgS)}(1+O(r^{-1} \dg))\,\,\textrm{ for }0\leq s\leq s_{max}-2.
 \eeaa
\end{lemma} 
\begin{proof}
See appendix \ref{appendix:Proofhigherhk-comaprisonlemma}.
\end{proof}

 \begin{corollary}\label{cor:int-gaS-ga:highersobolevregularity}
 Under the same assumptions as  Lemma \ref{lemma:int-gaS-ga:highersobolevregularity}, 
 we have, for all $0\le k\le s_{max}-2$, 
\bea
\bsplit
\| \dk^{\leq 2}\Ga_g\|_{\hk_k(\S)}&\les  \epg  r^{-1},\\
\| \dk^{\leq 2}\Ga_b\|_{\hk_k(\S) }&\les  \epg,  
\end{split}
\eea
\bea
\bsplit
\|\dk^{\leq 2}\left(\a, \b, \rhoc, \muc\right)\|_{\hk_k(\S)}&\les \epg r^{-2},\\
\|\dk^{\leq 2}\bb\|_{\hk_k(\S) }&\les \epg r^{-1},\\
\|\dk^{\leq 2}\aa\|_{\hk_k(\S) }&\les \epg, 
\end{split}
\eea
\bea
\bsplit
  \left\| \dk^{\leq 2}\left (\frac{\ga}{r^2} -1, \,  \underline{b},\,\, \frac{e^{\Phi}}{r\sin\th }-1 \right) \right\|_{\hk_k(\S)} &\les  \epg,  \\
     \left\| \dk^{\leq 2}\left ( \Omb+\Up\right) \right\|_{\hk_k(\S)}   + \left\| \dk^{\leq 2}\left ( \vsi + 1\right) \right\|_{\hk_k(\S)} &\les  \epg   r. 
   \end{split}
\eea
\end{corollary}    
   
\begin{proof}   
In view of Lemma \ref{lemma:int-gaS-ga:highersobolevregularity}  and assumptions {\bf A1-A3} we have, for $0\le s\le s_{max}-2$ ,
\beaa
 \left\|\dk^{\leq 2}\Gac_g \right\|_{\hk_s(\S)} &\les& r \sup_{\RR}\left| \dk^{\leq s+ 2}\Gac_g\right| \les r^{-1}\epg. 
\eeaa
The other estimates are proved in the same manner.
\end{proof}


 \subsection{Adapted frame transformations}
 
  
 We consider    general null transformations introduced in Lemma \ref{lemma:SSMe:general.composite},
   \bea
\label{SSMe:general.composite:repeatforproofGCM}
\begin{split}
e_4'&=\la\left(e_4 + f e_\th +\frac 1 4 f^2  e_3\right),\\
e_\th'&=\left(1+\frac 1 2   f \fb\right) e_\th  + \frac 1 2  \fb e_4+\frac 1 2 f\left(1+ \frac 1 4 f \fb\right) e_3,\\
e_3'&= \la^{-1} \left( \left(1+\frac 12  f \fb +\frac{1}{16} f^2 \fb^2\right)  e_3 +\fb\left(1+\frac 1 4 f\fb\right) e_\th + \frac 1 4 \fb^2  e_4\right).
\end{split}
\eea

\begin{definition}
 Given a deformation $\Psi:\ovS\longrightarrow \S$ we  say that  a new frame   $(e_3', e_4',  e_\th')$, obtained from the standard frame $(e_3, e_4, e_\th)$  via the transformation \eqref{SSMe:general.composite:repeatforproofGCM},  is $\S$-adapted  if  we have,
 \bea
 e_\th'= e_\th^\S=\frac{1}{(\ga^\S)^{1/2} } \psi_\# (\pr_\th).
 \eea
\end{definition}

\begin{proposition}\label{Proposition:compatible-deformations}
Consider a deformation $\Psi:\ovS\longrightarrow \S$  in $\RR=\RR(\epg, \dg)$      verifying  the assumption ${\bf A3}$.
 The following statements hold true.

\begin{enumerate}
\item A   new     frame  $e_3', e_\th', e_4'$ generated by $(f, \fb, \la=e^a)$  according to  \eqref{SSMe:general.composite:repeatforproofGCM}   is   adapted    to   $\S=\S(\ovu+U, \ovs+S) $ provided that,  at all points $\th\in[0,\pi]$,
\bea
\label{eq:DeformS-6}
\bsplit 
  \sqrt{\ga^\#}\left(1-\frac{\vsi}{2} \underline{b}^\# U'\right) &= \left((\gaS)^\#  \right)^{1/2}\left(1+\frac 1 2  (f\fb)^\#\right), \\
   \vsi  U'&= \left((\gaS)^\#  \right)^{1/2}   f^\#\left( 1 +\frac 1 4 (f\fb)^\#\right),\\
   2\left(  S'-\frac{\vsi}{2} \Omb^\#  U'\right)&=   \left((\gaS)^\#  \right)^{1/2}\fb^\#,
\end{split}
\eea
 where,
 \beaa
   (\ga^ \S)^\# &=&\ga^\#+(\vsi^\#)^2 \left(\Omb+\frac{1}{4} \underline{b}^2  \ga  \right)^\#\, ( \pr_\th U )^2 -2 \vsi^\# \pr_\th U \pr_\th  S-(\ga \vsi  \underline{b})^\# \pr_\th  U
  \eeaa
   and  $\#$ denotes the pull back by $\psi$ of the corresponding   reduced scalars, i.e. for example,
   $f^\#(\th)=f(\ug+U(\th), \sg+S(\th), \th) $.

  \item 
  There exists a small enough   constant\footnote{In later applications, we will have
   \beaa
 \sup_{\RR}\,( |f|+|\fb|)\les r^{-1} \dg.
  \eeaa}  $\de_1$ such that for given  $f, \fb$ on $\RR$   satisfying
  \beaa
 \sup_{\RR}\Big( |f|+|\fb|\Big)\leq r^{-1} \de_1,
  \eeaa
   we can uniquely solve the system \eqref{eq:DeformS-6}   for  $U, S$  subject to the initial conditions,
  \beaa
  U(0)= 0, \qquad S(0)=0.
  \eeaa
  Thus, if $(\ug, \sg, 0)$ corresponds to the south pole of $\ovS$ and $f, \fb$ are given   there exists a unique deformation
   $\S\subset \RR$, given by $U, S:[0,\pi] \longrightarrow \RRR$,   adapted  to    frames generated  by\footnote{Note that $a$ is not restricted in this result.} $(f, \fb)$  which passes through the same south pole.    
   Moreover,
   \bea
   \label{eq:DeformS-6-estimate1}
   \sup_{[0, \pi]} |(U', S')| &\les&  \rg    \sup_\S \left(| f|+|\fb| \right) 
   \eea
   and, for $2\leq s\le s_{max}$,
 \bea
 \label{eq:DeformS-6-estimate2}
 \|(U', S')\|_{L^\infty(\ovS)}+  (\ovr)^{-1} \|(U', S')\|_{\hk_s(\ovS, \ovgS)} &\les&
     \|f, \fb \|_{\hk_s(\S)}
  \eea
  with $ \|f, \fb \|_{\hk_s(\S)} =   \| f \|_{\hk_s(\S)}   +\|\fb \|_{\hk_s(\S)}$.
    
  \item As a consequence of \eqref{eq:DeformS-6-estimate2}  the deformation thus obtained  verifies the    conclusions of Lemmas \ref{lemma:int-gaS-ga}-\ref{lemma:int-gaS-ga:highersobolevregularity} and Corollary \ref{cor:int-gaS-ga:highersobolevregularity}. In particular,
  \begin{enumerate}
  
  \item  We have,
  \beaa
  \left|\ga^{\S,\#}-\ovga  \right| &\les& \de_1  \ovr.
  \eeaa
  
  \item We have 
  \beaa
\left| \frac{r^\S}{\rg}-1\right|\les  \ovr^{-1} \de_1. 
 \eeaa
    \end{enumerate}
  \end{enumerate}
\end{proposition}

\begin{proof}
In view of Lemma \ref{le:pullbackS},  
 The $\Z$-invariant  vectorfield $ e_\th^\S:=  \frac{1}{(\gaS  )^{1/2} }     \Psi_\# (\pr_\th)$ can be expressed by the formula,
 \beaa
  e_\th^\S&=&  \frac{1}{(\gaS  )^{1/2} }  \Big[\left( \pr_\th S-\frac{\vsi}{2} \Omb \pr_\th U\right) e_4+\frac{\vsi}{2} \pr_\th U e_3 +\sqrt{\ga}\left(  1 -\frac{\vsi}{2}  \underline{b}\pr_\th U\right) e_\th\Big]. \qquad 
 \eeaa
 where $\psi(p)=(\ug+U(\th), \sg+S(\th), \th) $ and         $U'=\pr_\th U(\th)$, $ S'=\pr_\th S(\th)$.
  On the other hand, according to \eqref{SSMe:general.composite:repeatforproofGCM}, at  $\Psi(p)\in \S$,
  \beaa
 e_\th'=\left(1+\frac 1 2  f\fb\right)  e_\th +\frac 12 f\left( 1 +\frac 1 4 f \fb\right)  e_3+\frac 1 2\fb e_4.
  \eeaa
  We deduce, at every $\th\in[0,\pi]$,
  \beaa
   \sqrt{\ga^\#}\left(1-\frac{\vsi^\#}{2} \underline{b}^\# U'\right) &=& \left((\gaS)^\#  \right)^{1/2}\left(1+\frac 1 2  (f\fb)^\#\right), \\
   \vsi^\#  U'&=& \left((\gaS)^\#  \right)^{1/2}   f^\#\left( 1 +\frac 1 4 (f\fb)^\#\right),\\
   2\left(  S'-\frac{\vsi^\#}{2} \Omb^\#  U'\right)&=&   \left((\gaS)^\#  \right)^{1/2}\fb^\#,
  \eeaa
  as desired.

  To prove the second part of the lemma we first check for the compatibility of the  three   equations in \eqref{eq:DeformS-6}.  Note that, if we denote,
  \beaa
  A=1+\frac 1 2  (f\fb)^\#, \quad B= f^\#\left( 1 +\frac 1 4 (f\fb)^\#\right), \quad C=\fb^\#,
  \eeaa
  we have  $  A^2- BC=1.$
  Hence, squaring the first equation and subtracting  the product of the other two  we derive,
  \bea\lab{eq:formulaforgaSpullbackpsiintermsofUprimeSprimeinproof}
\nn (\ga^\S)^\# &=&  \left( \sqrt{\ga^\#}\left(1-\frac{\vsi^\#}{2} \underline{b}^\# U'\right)\right)^2 -2U' \vsi^\# \left( S'-\frac{\vsi^\#}{2} \Omb^\#  U'\right)\\
 \nn&=& \ga^\#\left(1-(\vsi \underline{b})^\# U'+\frac{1}{4}  \big((\vsi\underline{b})^\# U'\big)^2 \right) - 2\vsi^\#U'S' +(  \vsi^\#)^2\Omb^\#  (U')^2\\
 &=& \ga^\#+(\vsi^\#)^2  \left( \Omb^\#+\frac 1 4\big( \underline{b}^\# \big)^2  \ga^\#  \right)\, ( U' )^2 -2\vsi^\#  U'  S'-\ga^\# \vsi^\#\underline{b}^\#  U'
  \eea
 which  coincides with the formula   \eqref{eq:formulaforgaSintermsofUprimeSprime}. It thus suffices  to only consider the last two equations 
  in \eqref{eq:DeformS-6}  which we write in the form,
  \bea
  \label{SystemforUS-ffb}
\bsplit
   U'&=  (\vsi^\#)^{-1}((\gaS)^\#  )^{1/2}  f^\#\left( 1 +\frac 1 4 (f\fb)^\#\right),\\
     S' &=   \frac 1 2 ((\gaS)^\#  )^{1/2}\fb^\# + \frac 1 2 \Omb^\#  ((\gaS )^\# )^{1/2}  f^\#\left( \fb^\# +\frac 1 4 (f\fb)^\#\right),
     \end{split}
     \eea
     i.e.,
    \beaa
    U'(\th)&=&\left[(\vsi^\#)^{-1}(\gaS)^{1/2}f\left( 1 +\frac 1 4 (f\fb)\right)\right]( \ug+U(\th), \sg+S(\th), \th),\\
    S'(\th) &=&  \left[ \frac 1 2 (\gaS) ^{1/2}\fb+ \frac 1 2 \Omb  (\gaS  )^{1/2}  f\left( 1 +\frac 1 4 f \fb\right)\right]( \ug+U(\th), \sg+S(\th), \th).
    \eeaa
      Thus under the assumption $\sup_{\RR}( |f|+|\fb|)\leq \rg^{-1} \de_1$, with $\de_1$ sufficiently small, making also  use of 
 the expression \eqref{eq:formulaforgaSpullbackpsiintermsofUprimeSprimeinproof} of $\ga^S$, and the estimates \eqref{eq:assumtionsonthegivenusfoliationforGCMprocedure:bis} for $(\ga,\,  \underline{b},\,  \Omb)$,  for $\epg$ sufficiently small,       we can uniquely solve  for  $U, S$  subject to the initial conditions,
  \beaa
  U(0)= 0, \qquad S(0)=0.
  \eeaa
 Moreover the solution verifies,
  \beaa
  \sup_{[0, \pi]} |(U', S')|  \les    \ovr    \sup_\S \left(| f|+|\fb| \right)      
   \eeaa
     according to the Definition \ref{definition:Deformations}.  Estimate \eqref{eq:DeformS-6-estimate2} can be easily derived by taking higher derivatives and using {\bf A1-A3}.
   This concludes the proof of the lemma.    
\end{proof}


\section{Frame transformations}
 

For the convenience of the reader we start by  recalling   the transformation formulas   recorded in Proposition \ref{prop:transformations1}.
\begin{proposition}[Transformation formulas-GCM]
\label{prop:transformations1-GCM}
Under a general transformation of  type \eqref{SSMe:general.composite:repeatforproofGCM}  with $\la= e^a$  the   Ricci coefficients and curvature components  transform as follows:
\bea
\begin{split}
\xi' &= \la^2\left(\xi+\frac{1}{2}\la^{-1}e_4'(f)+\om f + \frac{1}{4}f\ka\right)+\la^2\err(\xi, \xi'),\\
\err(\xi, \xi')&= \frac{1}{4}f\vth+\lot,\\ 
\xib' &= \la^{-2}\left(\xib+\frac{1}{2}\la e_3'(\fb)+\omb\,\fb + \frac{1}{4}\fb\,\kab\right)+\la^{-2}\err(\xib, \xib'),\\
\err(\xib, \xib')&=  -\frac 18\la\fb^2 e_3'(f)+ \frac{1}{4}\fb\,\vthb+\lot,
\end{split}
\eea
\bea
\bsplit
\ze' &= \ze - e_\th'(\log(\la))  +\frac 14 (- f\kab +\fb \ka ) + \fb \om- f  \omb+\err(\ze,\ze'), \\
\err(\ze,\ze')&= \frac{1}{2}\fb e_\th'(f)  +\frac 1 4( - f \vthb +\fb \vth)+\lot,\\
     \eta'&= \eta +\frac{1}{2}\la e_3'(f)     +\frac 1 4 \ka \fb   -f\omb +\err(\eta, \eta'),\\
     \err(\eta, \eta')&= \frac{1}{4}\fb\vth+\lot, \\
     \etab'&= \etab + \frac{1}{2} \la^{-1}e_4'(\fb)     +\frac 1 4 \kab f   -\fb\om +\err(\etab, \etab'),\\
     \err(\etab, \etab')&=   - \frac{1}{8}\fb^2\la^{-1}e_4'(f) +\frac{1}{4}f\vthb+\lot,
\end{split}
\eea
\bea
\bsplit
\ka'&=\la\left( \ka+ \ddd_1\,\!'(f)    \right) +\la\err(\ka,\ka'),\\
   \err(\ka,\ka')&=  f(\ze+\eta)    +\fb\xi       -\frac 1 4 f^2\kab +f\fb\om -f^2\omb+\lot,\\
\kab'&=\la^{-1}\left( \kab+ \ddd_1\,\!'(\fb)    \right) +\la^{-1}\err(\kab,\kab'),\\
   \err(\kab,\kab')&= -\frac{1}{4}\fb^2e_\th'(f) +\fb(-\ze+\etab)    +f\xib       -\frac 1 4 \fb^2\ka +f\fb\omb -\fb^2\om+\lot,
\end{split}
\eea
\bea
\bsplit
\vth' &= \la\left(\vth- \dds_2\,\!'(f)   \right) + \la\err(\vth,\vth'),\\
\err(\vth,\vth') &= f(\ze+\eta)    +\fb\xi            +\frac 1 4  f\fb \ka +f\fb\om -f^2\omb+\lot\\
\vthb' &= \la^{-1}\left(\vthb- \dds_2\,\!'(\fb)   \right) + \la^{-1}\err(\vthb,\vthb'),\\
\err(\vthb,\vthb') &=  -\frac{1}{4}\fb^2 e_\th'(f)+\fb(-\ze+\etab)    +f\xib            +\frac 1 4  f\fb \kab +f\fb\omb -\fb^2\om+\lot, 
\end{split}
\eea 
\bea
\begin{split}
\om' &= \la\left(\om -\frac{1}{2}\la^{-1}e_4'(\log(\la))\right)+\la\err(\om, \om'),\\
\err(\om, \om')&= \frac{1}{4}\fb e_4'(f)  +\frac{1}{2}\om f\fb - \frac{1}{2}f\etab +\frac{1}{2}\fb\xi +\frac{1}{2}f\ze -\frac{1}{8}\kab f^2+\frac{1}{8}f\fb \ka-\frac{1}{4}\omb f^2+\lot,\\
\omb' &= \la^{-1}\left(\omb +\frac{1}{2}\la e_3'(\log(\la))\right)+\la^{-1}\err(\omb, \omb'),\\
\err(\omb, \omb')&=  - \frac{1}{4}\fb e_3'(f)  +\omb f\fb - \frac{1}{2}\fb\eta +\frac{1}{2}f\xib -\frac{1}{2}\fb\ze -\frac{1}{8}\ka\fb^2+\frac{1}{8}f\fb \kab-\frac{1}{4}\om\fb^2+\lot 
\end{split}
\eea 
  The  lower order terms we denote by $\lot$  are linear 
  with respect    to    $\big\{\xi,\xib,\vth, \ka, \eta,\etab,\ze,\kab,  \vthb\big\}$  and quadratic or  higher order in $f,\fb$, and do not contain derivatives of these latter.

Also,
\bea
\bsplit
\a' &= \la^2\a+\la^2\err(\a, \a'),\\
\err(\a, \a') &= 2f\b+\frac{3}{2}f^2\rho+\lot,\\
\b' &= \la\left(\b+\frac{3}{2}\rho f\right)+\la\err(\b,\b'),\\
\err(\b,\b') &= \frac{1}{2}\fb\a+\lot,\\
\rho' &= \rho+\err(\rho,\rho'),\\
\err(\rho,\rho') &= \frac{3}{2}\rho f\fb+\fb\b  +f\bb+\lot,\\
 \bb' &= \la^{-1}\left(\bb+\frac{3}{2}\rho\fb\right)+\la^{-1}\err(\bb,\bb'),\\
\err(\bb,\bb') &= \frac{1}{2}f\aa+\lot,\\
\aa' &= \la^{-2}\aa+\la^{-2}\err(\aa, \aa'),\\
\err(\aa, \aa') &= 2\fb\,\bb+\frac{3}{2}\fb^2\rho+\lot
\end{split}
\eea
    The  lower order terms we denote by $\lot$ are linear   with respect to the curvature quantities    $\a,\b, \rho, \bb, \,\aa$  
    and quadratic  or higher order  in  $f,\fb$, and do not contain derivatives of these latter.
\end{proposition}

In the following lemma we rewrite   a subset of these transformations in a  more useful form,
\begin{lemma}
\label{Lemma:Transformation-special}
Under a general transformation of  type \eqref{SSMe:general.composite:repeatforproofGCM} with $\la= e^a$  we have, in particular,
\bea
\bsplit
\ze'&= \ze   - e'_\th( a )         -f\omb  +\fb \om                -\frac{1}{2}f\chib+\frac 12  \fb \chi    +\err(\ze,\ze'),\\
\err(\ze,\ze')&= \frac{1}{2}\fb\left(1+\frac 1 4 f\fb\right)e_\th'(f) -\frac{1}{16}\fb^2 e_\th'(f^2) +\frac 1 4( - f \vthb +\fb \vth)+\lot  
\end{split}
\eea
\bea
\bsplit
\ka' &= e^{a} \left( \ka+ \ddd_1'  f   \right) + e^a \err(\ka,\ka'),\\
  \err(\ka,\ka')&= \frac 1 2   f \fb e_\th'(f) -\frac{1}{4}\fb e_\th'(f^2)+f(\ze+\eta)    +\fb\xi       -\frac 1 4 f^2\kab +f\fb\om -f^2\omb+\lot  
  \end{split}
  \eea
  \bea
  \bsplit
 \kab' &= e^{-a} \left( \kab+ \ddd_1'  \fb   \right) +e^{-a}  \err(\kab,\kab'),\\ 
   \err(\kab,\kab')&= -\frac{1}{2}\fb e_\th'\left( f \fb +\frac{1}{8} f^2 \fb^2\right) +\left(\frac{3}{4}f\fb+\frac 1 8 (f\fb)^2\right)e_\th'(\fb)  \\
   &+\frac 1 4\left(1+\frac{1}{2}f\fb\right)\fb e_\th'\left( f\fb\right)
-\frac{1}{4}f\left(1+\frac{1}{4}f\fb\right)e_\th'\left(\fb^2\right)  \\
& + \fb(-\ze+\etab)    +f\xib       -\frac 1 4 \fb^2\ka +f\fb\omb -\fb^2\om+\lot
 \end{split}
 \eea
 Also,
 \bea
\bsplit
\vth' &= \la\left(\vth- \dds_2\,\!'(f)   \right) + \la\err(\vth,\vth'),\\
\err(\vth,\vth') &= \frac 1 2   f \fb e_\th'(f) -\frac{1}{4}\fb e_\th'(f^2)+f(\ze+\eta)    +\fb\xi            +\frac 1 4  f\fb \ka +f\fb\om -f^2\omb+\lot\\
\vthb' &= \la^{-1}\left(\vthb- \dds_2\,\!'(\fb)   \right) + \la^{-1}\err(\vthb,\vthb'),\\
\err(\vthb,\vthb') &=  -\frac{1}{2}\fb e_\th'\left( f \fb +\frac{1}{8} f^2 \fb^2\right) +\left(\frac{3}{4}f\fb+\frac 1 8 (f\fb)^2\right)e_\th'(\fb) \\
& +\frac 1 4\left(1+\frac{1}{2}f\fb\right)\fb e_\th'\left( f\fb\right)
 -\frac{1}{4}f\left(1+\frac{1}{4}f\fb\right)e_\th'\left(\fb^2\right)+\fb(-\ze+\etab)   \\
 & +f\xib            +\frac 1 4  f\fb \kab +f\fb\omb -\fb^2\om+\lot, 
\end{split}
\eea 
 The  lower order terms we denote by $\lot$  are cubic or higher order 
  in   the small   quantities    $\xi,\xib,\vth, \eta,\etab,\ze, \vthb$   as well as $ f,\fb$, and do not contain derivatives of these quantities.
 
 We also have,
 \bea
 \bsplit
 \b' &= \la\left(\b+\frac{3}{2}\rho f\right)+\la\err(\b,\b'),\\
\err(\b,\b') &= \frac{1}{2}\fb\a+\lot,\\
\rho' &= \rho+\err(\rho,\rho'),\\
\err(\rho,\rho') &= \frac{3}{2}\rho f\fb+\fb\b  +f\bb+\lot
\end{split}
 \eea
 The lower order  terms   above   denoted  by   $\lot$    are cubic or higher order   in   
  the small   quantities    $\xi,\xib,\vth, \eta,\etab,\ze, \vthb$  as well as $a, f, \fb$.
\end{lemma}

\begin{lemma}
The following transformation formula holds  true
\beaa
 \mu'&=&\mu+(\ddd_1)'\left(-(\dds_1)'  a      +f\omb  -\fb \om    +\frac{1}{4}f\kab -\frac 14   \fb \ka \right)+\err(\mu,\mu'),\\
\err(\mu,\mu')&=& - \ddd\,'_1 \err(\ze,\ze')  - \err(\rho,\rho')+\frac 1 4\big( \vth'\vthb'-
\vth \vthb\big).
\eeaa
The error term  $\err(\mu,\mu')$ is quadratic  or higher order  with respect to $(f, \fb,  a, \Gac, \Rc)$ and  depends  only on   at most   two angular   derivatives $e_\th'$  of $f$ and one angular derivative $e_\th'$ of $a, \fb$.
\end{lemma}

\begin{proof}
Recall that
\beaa
\mu&=&-\ddd_1 \ze-\rho+\frac 1 4 \vth\vthb.
\eeaa
Therefore,
\beaa
\mu'&=&-\ddd\,'_1 \ze'-\rho'+\frac 1 4 \vth'\vthb'\\
&=& -\ddd\,'_1 \left( \ze   - e'_\th( a )         -f\omb  +\fb \om         -\frac{1}{4}f\kab +\frac 14  \fb \ka    +\err(\ze,\ze')\right)- \rho -  \err(\rho,\rho') +\frac 1 4 \vth' \vthb' \\
&=&  -\ddd\, '_1  \ze-\rho +\frac 1 4 \vth\vthb  -\ddd'_1 \left(  (\dds_1)' a         -f\omb  +\fb \om            -\frac{1}{4}f\kab +\frac 14  \fb \ka    \right) \\
&-& \ddd\,'_1 \err(\ze,\ze')  - \err(\rho,\rho')  +\frac 1 4\big( \vth'\vthb'-
\vth \vthb\big). 
\eeaa
Note that,
\beaa
 -\ddd\, '_1  \ze-\rho +\frac 1 4 \vth\vthb&=&-\ddd\, _1  \ze-\rho +\frac 1 4 \vth\vthb +f e_3 \ze+\fb e_ze+\lot\\
 &=&\mu  +f e_3 \ze+\fb e_4\ze+\lot
 \eeaa
Hence,
\beaa
\mu'&=&\mu+\ddd\,'_1 \left( - (\dds_1)' a  +f\omb  -\fb \om         +\frac{1}{4}f\kab-\frac 14  \fb \ka   \right)+\err(\mu,\mu')
\eeaa
where,
\beaa
\err(\mu,\mu')&=& - \ddd\,'_1 \err(\ze,\ze')  - \err(\rho,\rho')+\frac 1 4\big( \vth'\vthb'-
\vth \vthb\big)+f e_3 \ze+\fb e_4\ze+\lot
\eeaa
In view of the transformation formulas for $\th, \thb$ and the  structure of the error terms $\err(\ze,\ze'),\,  \err(\rho,\rho'),\,  \err(\vth,\vth'), \,  \err(\vthb,\vthb') $ in Lemma \ref{Lemma:Transformation-special}  we   easily deduce that  the error term $\err(\mu,\mu')$  depends  only on   at most   two angular   derivatives $e_\th'$  of $f$ and one angular derivative $e_\th'$ of $a, \fb$.
\end{proof}

We  shall also make use of  the following,
\begin{lemma}
\label{lemma:transfethka'-ethka}
We have the transformation equations,
\bea
\bsplit
 e_\th'(\ka' )     &= e_\th \ka +e_\th'  \ddd_1'  f  +\ka e_\th' a -\frac 1 4 \ka ( f\kab +\fb \ka)+\ka ( f\omb- \om \fb)+f \rho\\
 &+\err(e_\th'\ka', e_\th\ka),\\
 e_\th'(\kab' )     &= e_\th \kab +e_\th'  \ddd_1'  \fb  -\kab e_\th' a -\frac 1 4 \kab ( f\kab +\fb \ka)+\kab ( \fb \om- \omb  f )+\fb \rho\\
 &+\err(e_\th'\kab', e_\th\kab),\\
 e_\th' (\mu')
&= e_\th \mu +e_\th'(\ddd_1)'\left(-(\dds_1)'  a      +f\omb  -\fb \om    +\frac{1}{4}f\kab -\frac 14   \fb \ka \right)+\frac 3 4\rho( f\kab+\fb \ka) \\
&+\err(e_\th'\mu', e_\th \mu),
 \end{split}
\eea
where,
\beaa
 \err(e_\th'\ka', e_\th\ka)&=&( e^a-1)\Big(  e_\th \ka  +e_\th'  \ddd\,'_1  f + \frac 1 2  \fb e_4\ka +\frac 1 2 f e_3\ka \Big)\\
 &+&e^a \left[ e_\th' \, \err(\ka,\ka') +e_\th'( a)\Big(\ddd\,'_1 f  +\err(\ka, \ka') \Big) +\frac 1 2   f \fb e_\th \ka +\frac 1 8 f^2 \fb e_3\ka\right]\\
 &+&\frac 1 2 f \left( 2\ddd_1\eta -\frac 1 2  \vthb\, \vth +2(\xi\xib+\eta^2)\right)\\
&+& \frac 1 2   f \fb e_\th \ka +\frac 1 8 f^2 \fb e_3\ka+\frac 1 2 \fb \left(  2\ddd_1\xi -\frac 1 2 \vth^2 +2(\eta+\etab+2\ze)\xi\right),
\eeaa
\beaa
 \err(e_\th'\kab', e_\th\kab)&=&( e^{-a} -1)\Big(  e_\th \kab  +e_\th'  \ddd\,'_1  \fb + \frac 1 2  f  e_3\kab +\frac 1 2 \fb e_4\kab \Big)\\
 &+&e^{-a} \left[ e_\th' \, \err(\kab,\kab') +e_\th'( a)\Big(\ddd\,'_1 \fb   +\err(\kab, \kab') \Big) +\frac 1 2   f \fb e_\th \kab +\frac 1 8 f^2 \fb e_3\kab\right]\\
 &+&\frac 1 2 \fb  \left( 2\ddd_1\etab -\frac 1 2  \vthb\, \vth +2(\xi\xib+\etab^2)\right)\\
&+& \frac 1 2   f \fb e_\th \kab +\frac 1 8 f^2 \fb e_3\kab+\frac 1 2 f  \left(  2\ddd_1\xib -\frac 1 2 \vthb^2 +2(\eta+\etab-2\ze)\xi\right),
\eeaa
and,
\beaa
\err(e'_\th\mu', e_\th \mu)&=&e_\th'\err(\mu,\mu')+ \frac 1 2   f \fb e_\th \mu +\frac 1 8  f^2\fb e_3\mu\\
&-& \frac 12  f \Big( \ddd_1\bb -\frac 1 2 \vth \, \aa - \ze\, \bb +2(\eta \,\bb+ \xib\,\b)\Big)\\
&-& \frac 1 2 \fb \Big(\ddd_1\b -\frac 1 2 \vthb \, \a +\ze\, \b +2(\etab \,\b+ \xi\,\bb)\Big)\\
&+& \frac 1 2 \fb e_4 \left(-\ddd_1 \ze  +\frac 1 4 \vth \vthb\right)+ \frac 1 2 f  e_3 \left(-\ddd_1 \ze  +\frac 1 4 \vth \vthb\right).
\eeaa
\end{lemma}

\begin{proof}
Applying the  vectorfield  $e_\th'$ to 
\beaa
\ka' &=&e^{a} \left( \ka+ \ddd\,'_1  f  +  \err(\ka,\ka')\right)
\eeaa
we deduce,
\beaa
e_\th'(\ka' )     &=&e^{a} \Big( e_\th' \ka+e_\th'  \ddd\,'_1  f  +e_\th' ( \err(\ka,\ka')  \Big)  +e^{a} e_\th'( a)\Big(\ka +\ddd\,'_1 f  +\err(\ka, \ka') \Big). 
\eeaa
Hence,
\beaa
e^{-a} e_\th'(\ka' )&=&  e_\th' \ka+e_\th'  \ddd\,'_1  f  +e_\th' ( \err(\ka,\ka') +e_\th'( a)\Big(\ka +\ddd\,'_1 f  +\err(\ka, \ka') \Big) 
\eeaa
and thus
\beaa
 e_\th'(\ka' )&=&  e_\th \ka +e_\th'( a)\ka  +e_\th'  \ddd\,'_1  f + \frac 1 2  \fb e_4\ka +\frac 1 2 f e_3\ka +\err_1[e_\th(\ka), e_\th'(\ka')]
\eeaa
 with error term,
 \beaa
 \err_1[e_\th(\ka), e_\th'(\ka')]&=&( e^a-1)\Big(  e_\th \ka  +e_\th'  \ddd\,'_1  f + \frac 1 2  \fb e_4\ka +\frac 1 2 f e_3\ka \Big)\\
 &+&e^a \left[ e_\th' ( \err(\ka,\ka') +e_\th'( a)\Big(\ddd\,'_1 f  +\err(\ka, \ka') \Big) +\frac 1 2   f \fb e_\th \ka +\frac 1 8 f^2 \fb e_3\ka\right].
 \eeaa

Now, making use of 
\beaa
e_\th' \ka &=&\left(1+\frac 1 2   f \fb\right) e_\th\ka   + \frac 1 2  \fb e_4\ka +\frac 1 2 f\left(1+ \frac 1 4 f \fb\right) e_3\ka \\
&=& e_\th \ka  + \frac 1 2  \fb e_4\ka +\frac 1 2 f e_3\ka + \frac 1 2   f \fb e_\th \ka +\frac 1 8 f^2 \fb e_3\ka
\eeaa
and the null structure equations,
\beaa
e_3(\ka)+\frac 1 2 \kab\, \ka-2\omb \ka   &=& 2\ddd_1\eta  + 2\rho -\frac 1 2  \vthb\, \vth +2(\xi\xib+\eta^2),\\
e_4 \ka+\frac 1 2 \ka^2 +2\om \ka&=& 2\ddd_1\xi -\frac 1 2 \vth^2 +2(\eta+\etab+2\ze)\xi,
\eeaa
we  deduce,
\beaa
e_\th' \ka &=& e_\th \ka+\frac 1 2 \fb \left(-\frac 1 2 \ka^2 -2\om \ka\right)+ \frac 1 2 f \left(   - \frac 1 2 \kab\, \ka+2\omb \ka    +2\rho \right)\\
&+&  \frac 1 2   f \fb e_\th \ka +\frac 1 8 f^2 \fb e_3\ka+\frac 1 2 \fb \left(  2\ddd_1\xi -\frac 1 2 \vth^2 +2(\eta+\etab+2\ze)\xi\right)\\
&+&\frac 1 2 f \left( 2\ddd_1\eta -\frac 1 2  \vthb\, \vth +2(\xi\xib+\eta^2)\right).
\eeaa
Hence,
\beaa
 e_\th'(\ka' )&=& e_\th \ka +e_\th'( a)\ka +e_\th'  \ddd_1'  f  +\ka e_\th' a -\frac 1 4 \ka ( f\kab +\fb \ka)+\ka ( f\omb- \om \fb)+f \rho+\err(e_\th'\ka', e_\th\ka)
\eeaa
where,
\beaa
\err(e_\th'\ka', e_\th\ka)&=& \err_1(e_\th'\ka', e_\th\ka)+\frac 1 2 f \left( 2\ddd_1\eta -\frac 1 2  \vthb\, \vth +2(\xi\xib+\eta^2)\right)\\
&+& \frac 1 2   f \fb e_\th \ka +\frac 1 8 f^2 \fb e_3\ka+\frac 1 2 \fb \left(  2\ddd_1\xi -\frac 1 2 \vth^2 +2(\eta+\etab+2\ze)\xi\right)
\eeaa
as desired.
 The formula  for $e_\th'(\kab')$ is easily derived by symmetry from the one on $e_\th'(\ka')$. Note however that  $a$ becomes $-a$ in the transformation.

 Applying the operator  $ e_\th ' =\left(1+\frac 1 2   f \fb\right) e_\th  + \frac 1 2  \fb e_4+\frac 1 2 f\left(1+ \frac 1 4 f \fb\right) e_3$   to the transformation formula  for $\mu$,
\beaa
 \mu'&=&\mu+(\ddd_1)'\left(-(\dds_1)'  a      +f\omb  -\fb \om    +\frac{1}{4}f\kab -\frac 14   \fb \ka \right)+\err(\mu,\mu')
\eeaa
we derive,
\beaa
e_\th' (\mu')&=&e_\th'( \mu) +e_\th'(\ddd_1)'\left(-(\dds_1)'  a      +f\omb  -\fb \om    +\frac{1}{4}f\kab -\frac 14   \fb \ka \right)+e_\th'\err(\mu,\mu')\\
&=& e_\th (\mu) +\frac 1 2 \fb e_4\mu+\frac 1 2  f e_3\mu        +e_\th'(\ddd_1)'\left(-(\dds_1)'  a      +f\omb  -\fb \om    +\frac{1}{4}f\kab -\frac 14   \fb \ka \right)\\
&+&e_\th'\err(\mu,\mu')+ \frac 1 2   f \fb e_\th \mu +\frac 1 8  f^2\fb e_3\mu.
\eeaa
Recalling that $\mu= -\ddd_1 \ze -\rho +\frac 1 4 \vth \vthb$ we find,
\beaa
\frac 1 2 \fb e_4\mu+\frac 1 2  f e_3\mu=- \frac 1 2( f e_3 +\fb  e_4 ) \rho+ \frac 1 2 \fb e_4 \left(-\ddd_1 \ze  +\frac 1 4 \vth \vthb\right)+ \frac 1 2 f  e_3 \left(-\ddd_1 \ze  +\frac 1 4 \vth \vthb\right).
\eeaa
Recalling the Bianchi equations for $e_3\rho, e_4 \rho$
\beaa
e_4 \rho+\frac 3 2 \ka \rho&=\ddd_1 \b  -\frac 1 2 \vthb \, \a +\ze\, \b +2(\etab \,\b+ \xi\,\bb),\\
e_3 \rho+\frac 3 2 \kab \rho&=\ddd_1\bb  -\frac 1 2 \vth \, \aa - \ze\, \bb +2(\eta \,\bb+ \xib\,\b),
\eeaa
 we further  deduce,
\beaa
\frac 1 2 \fb e_4\mu+\frac 1 2  f e_3\mu
&=& \frac 3 4 \rho( f\kab+\fb \ka) \\
&-& \frac 12  f \Big( \ddd_1\bb -\frac 1 2 \vth \, \aa - \ze\, \bb +2(\eta \,\bb+ \xib\,\b)\Big)\\
&-& \frac 1 2 \fb \Big(\ddd_1\b -\frac 1 2 \vthb \, \a +\ze\, \b +2(\etab \,\b+ \xi\,\bb)\Big)\\
&+& \frac 1 2 \fb e_4 \left(-\ddd_1 \ze  +\frac 1 4 \vth \vthb\right)+ \frac 1 2 f  e_3 \left(-\ddd_1 \ze  +\frac 1 4 \vth \vthb\right).
\eeaa
Therefore,
\beaa
e_\th' (\mu')&=&  e_\th (\mu) + \frac 3 4 \rho( f\kab+\fb \ka)        +e_\th'(\ddd_1)'\left(-(\dds_1)'  a      +f\omb  -\fb \om    +\frac{1}{4}f\kab -\frac 14   \fb \ka \right)\\
 &+&\err(e_\th\mu, e_\th \mu)
\eeaa
with,
\beaa
\err(e'_\th\mu', e_\th \mu)&=&e_\th'\err(\mu,\mu')+ \frac 1 2   f \fb e_\th \mu +\frac 1 8  f^2\fb e_3\mu\\
&-& \frac 12  f \Big( \ddd_1\bb -\frac 1 2 \vth \, \aa - \ze\, \bb +2(\eta \,\bb+ \xib\,\b)\Big)\\
&-& \frac 1 2 \fb \Big(\ddd_1\b -\frac 1 2 \vthb \, \a +\ze\, \b +2(\etab \,\b+ \xi\,\bb)\Big)\\
&+& \frac 1 2 \fb e_4 \left(-\ddd_1 \ze  +\frac 1 4 \vth \vthb\right)+ \frac 1 2 f  e_3 \left(-\ddd_1 \ze  +\frac 1 4 \vth \vthb\right)
\eeaa
as desired.
 \end{proof}

Finally recalling the definition of the Hodge   operators  $\ddd_1, \dds_1,  (\ddd_1)', (\dds_1)'$  and  noticing that
\beaa
(\dds_1)'(\ka')&=&(\dds_1)'(\check{\ka}'), \qquad (\dds_1)(\ka)=(\dds_1)(\check{\ka}),\\
(\dds_1)'(\kab')&=&(\dds_1)'(\check{\kab}'), \qquad (\dds_1)(\kab)=(\dds_1)(\check{\kab}),\\
(\dds_1)'(\mu')&=&(\dds_1)'(\check{\mu}'), \qquad (\dds_1)(\mu)=(\dds_1)(\check{\mu}),\\
(\dds_1)'(\mub')&=&(\dds_1)'(\check{\mub}'), \qquad (\dds_1)(\mub)=(\dds_1)(\check{\mub}),
\eeaa
 we   recast   the results of Lemma \ref{lemma:transfethka'-ethka} in the following form.
 \begin{lemma}
 \label{Le:GCMS:kakabmumub}
 We have the transformation equations,
 \bea
\label{GCMS:kakabmumub}
\bsplit
(\dds_1)'(\check{\ka}') &= \dds_1(\check{\ka})  +   (\dds_1)' ( \ddd_1)'  f + \ka (\dds_1)' a - \rho f   +\frac 1 4 \ka (f\kab +\fb \ka)\\
& - \ka ( f\omb- \fb \om) - \err_1,\\
(\dds_1)' (\check{\kab}') &= \dds_1(\check{\kab})  +  (\dds_1)' ( \ddd_1)'  \fb -\kab (\dds_1)' a - \rho \fb   +\frac 1 4 \kab (f\kab +\fb \ka)\\
& - \kab ( \fb \om-  f  \omb) - \err_2,
 \\
 (\dds_1)' (\check{\mu}')
&=  \dds_1(\check{\mu})    +       (\dds_1)'(\ddd_1)'\left(-(\dds_1)'  a      +f\omb  -\fb \om    +\frac{1}{4}f\kab -\frac 14   \fb \ka \right)\\
& -\frac 3 4 \rho \left( f \kab +\fb \ka \right)- \err_3,
\end{split}
\eea
where,
\bea
\label{GCMS:Errorterms}
\bsplit
\err_1&=\err(e_\th'\ka', e_\th\ka)=e_\th' \, \err(\ka,\ka') +a e_\th'  \ddd\,'_1  f  + e_\th'( a) \ddd\,'_1 f \\
 &+ a \left( e_\th \ka+ \frac 1 2\left(  \fb e_4\ka + f e_3\ka\right)\right)+  f\ddd_1 \eta+ \fb \ddd_1\xi +\lot,\\
 \err_2&=\err(e_\th'\kab', e_\th\kab)=e_\th' \, \err(\kab,\kab') -a e_\th'  \ddd\,'_1  \fb  - e_\th'( a) \ddd\,'_1 \fb\\
 &- a \left( e_\th \kab+ \frac 1 2\left(  \fb  e_4\kab + f  e_3\kab\right)\right)+  \fb \ddd_1 \etab+ f  \ddd_1\xib +\lot,
 \\
\err_3&=\err(e_\th'\mu', e_\th \mu)=e_\th'\err(\mu,\mu') -\frac 1 2\left( \fb \ddd_1 \b +f\ddd_1\bb \right)  -\frac 1 2( f e_3 +\fb  e_4 )  \ddd_1 \ze+\lot,
\end{split}
\eea
where the terms denoted by $\lot$ are   cubic or higher order in  $a, f, \fb, \Gac, \Rc$  and contain  no  derivatives of $(a, f, \fb)$.
\end{lemma}


\subsection{Main GCM Equations}


Given a deformation $\Psi:\ovS\longrightarrow \S$ and  adapted frame $(e_3', e_4' , e_\th') $ with  $e_\th'=e_\th^\S$  we derive an elliptic system  for the transition  parameters $(a, f, \fb)$.   The system 
will later be used in the construction of GCM surfaces.

   In what follows we denote by  $\dddS_1, \dddS_2, \ddsS_1, \ddsS_2$   the basic  Hodge operators on $\S$.
Noting that the transformation formulae in \eqref{GCMS:kakabmumub}--\eqref{GCMS:Errorterms}  contain only the operators  $( \ddd_1)'=\dddS_1, (\dds_1)'=\ddsS_1 $  applied to $a, f, \fb$  we introduce  the  simplified notation,
\bea
\dddS:=(\ddd_1)', \qquad  \ddsS:=(\dds_1)', \qquad A^\S:=\ddsS \dddS, \qquad \dds:=\dds_1.
\eea
With these notation  \eqref{GCMS:kakabmumub} takes the following form,
\beaa
\bsplit
\ddsS \check{\ka}^\S &= \dds \check{\ka}  +  A^\S f +\ka \ddsS a - \rho f   +\frac 1 4 \ka (f\kab +\fb \ka)- \ka ( f\omb- \fb \om) - \err_1,\\
\ddsS \check{\kab}^\S &= \dds \check{\kab}  + A^\S \fb -\kab \ddsS a - \rho \fb   +\frac 1 4 \kab (f\kab +\fb \ka) - \kab ( \fb \om-  f  \omb) -
 \err_2,
 \\
\ddsS \check{\mu}^\S
&=  \dds\,\,\check{\mu}  +      A^\S\left(- \ddsS  a      +f\omb  -\fb \om    +\frac{1}{4}f\kab -\frac 14   \fb \ka \right) -\frac 3 4 \rho \left( f \kab +\fb \ka \right)- \err_3,
\end{split}
\eeaa
or,
\bea
\label{GCMS:Near1}
\bsplit
A^\S\left( - \ddsS a   +f\omb  -\fb \om                +\frac{1}{4}f\kab -\frac 14   \fb \ka     \right)-\frac 3 4\rho(\ka\fb +\kab  f )&=\ddsS \check{\mu}^\S-\dds\,\,\check{\mu} +\err_3,\\
\\
A^\S f +\ka \ddsS  a -\rho f +\frac 1 4 \ka( f\kab+\fb \ka)-\ka (f\omb-\fb \om)&=\ddsS \check{\ka}^\S -\dds\,\,\check{\ka}+\err_1, \\
A^\S\fb -\kab \ddsS a -\rho \fb +\frac 1 4 \kab( f\kab+\fb \ka)-\kab (\fb \om-f  \omb)&=\ddsS \check{\kab}^\S -\dds\,\,\check{\kab}+\err_2. 
\end{split}
\eea
Since $A^\S$ is invertible\footnote{We have $\int_\S f A^\S f = \int_\S(\dddS f)^2$
which in view of the identity \eqref{eq:hodgeident1} for $\dddS_1$ and the definition of $\dddS$ implies that $A^\S$ is invertible.}  we can write, setting $z:=\ka\fb +\kab f$,
\beaa
\ddsS a&=& f\omb  -\fb \om    +\frac{1}{4}f\kab -\frac 14   \fb \ka   -\frac 3 4( A^\S)^{-1} \big(  \rho z \big)  +(A^\S)^{-1}\big( -\ddsS \check{\mu}^\S +\dds\,\,\check{\mu} -\err_3\big).
\eeaa
We can thus eliminate  $\ddsS a$ from the last two equations,
\beaa
\bsplit
A^\S f+\left(\frac 1 2 \ka \kab -\rho\right) f-\frac 3 4 \ka (A^\S)^{-1}\big(\rho z \big)&=\ddsS \check{\ka}^\S-\dds\,\,\check{\ka}-\ka  (A^\S)^{-1}\big( -\ddsS \check{\mu}^\S +\dds\,\,\check{\mu}\big)\\
&+\err_4,\\
A^\S \fb+\left(\frac 1 2 \ka\kab -\rho\right)\fb+ \frac 3 4\kab  (A^\S)^{-1}\big(\rho z\big)&=\ddsS \check{\kab}^\S-\dds\,\,\check{\kab}+\kab  (A^\S)^{-1}\big( -\ddsS \check{\mu}^\S +\dds\,\,\check{\mu}\big)\\
&+\err_5,
\end{split}
\eeaa
where,
\beaa
\err_4&=&\err_1 +\ka (A^\S)^{-1} \err_3,\qquad 
\err_5=\err_2 -\kab (A^\S)^{-1} \err_3.
\eeaa
Therefore    the system \eqref{GCMS:Near1} is equivalent to the system,
\beaa
\bsplit
A^\S f+(\frac 1 2 \ka \kab -\rho) f-\frac 3 4 \ka (A^\S)^{-1}\big(\rho z \big)&=\ddsS \check{\ka}^\S-\dds\,\,\check{\ka}-\ka  (A^\S)^{-1}\big( -\ddsS \check{\mu}^\S +\dds\,\,\check{\mu}\big)\\
&+\err_4,\\
A^\S \fb+ (\frac 1 2 \ka \kab -\rho)\fb+ \frac 3 4\kab  (A^\S)^{-1}\big(\rho z \big)&=\ddsS \check{\kab}^\S-\dds\,\,\check{\kab}+\kab  (A^\S)^{-1}\big( -\ddsS \check{\mu}^\S +\dds\,\,\check{\mu}\big)\\
&+\err_5,\\
\ddsS a +\frac 3 4( A^\S)^{-1} \big(  \rho z \big) - f\omb  +\fb \om    -\frac{1}{4}f\kab +\frac 14   \fb \ka &= (A^\S)^{-1}\big( -\ddsS \check{\mu}^\S+\dds\,\,\check{\mu}\big) - (A^\S)^{-1}\err_3.
\end{split}
\eeaa
Furthermore, we have
\beaa
A^\S z &=& A^S(\kab f +\ka \fb)= \kab A^Sf+\ka A^S\fb+[A^S,\kab]f+[A^S,\ka]\fb.
\eeaa
In view of the above equations for $A^Sf$ and $A^S\fb$, we infer
\beaa
A^\S z &=& \kab\left\{-\left(\frac 1 2 \ka \kab -\rho\right) f+\frac 3 4 \ka (A^\S)^{-1}\big(\rho z \big)+\ddsS \check{\ka}^\S-\dds\,\,\check{\ka}-\ka  (A^\S)^{-1}\big( -\ddsS \check{\mu}^\S +\dds\,\,\check{\mu}\big)\right\}\\
&& +\ka\left\{-\left(\frac 1 2 \ka\kab -\rho\right)\fb - \frac 3 4\kab  (A^\S)^{-1}\big(\rho z \big) +\ddsS\kabc^\S      -\dds\,\,\check{\kab}+\kab (A^\S)^{-1}\dds\,\big( -\ddsS \check{\mu}^\S +\dds\,\,\check{\mu}\big)\right\}\\
&& +[A^\S,\kab]f+[A^\S,\ka]\fb+\kab \err_4 +\ka \err_5. 
\eeaa
Furthermore
\beaa
A^S z&=& -\left(\frac 1 2 \ka \kab -\rho\right)z +\kab\Big\{ \ddsS \check{\ka}^\S-\dds\,\,\check{\ka}-\ka  (A^\S)^{-1}\big( -\ddsS \check{\mu}^\S +\dds\,\,\check{\mu}\big)\Big\}\\
&& +\ka\Big\{\ddsS\kabc^\S      -\dds\,\,\check{\kab}+\kab (A^\S)^{-1}\,\big( -\ddsS \check{\mu}^\S +\dds\,\,\check{\mu}\big)\Big\} \\
&&+[A^S,\kab]f+[A^S,\ka]\fb +\kab \err_4 +\ka \err_5\\
&=& -\left(\frac 1 2 \ka \kab -\rho\right)z +\kab \ddsS \kac^\S+ \ka \ddsS \kabc^\S  -\kab\dds\,\,\check{\ka} -\ka\dds\,\,\check{\kab}\\
&&+  \kab\err_4 +\ka\err_5 +[A^S,\kab]f+[A^S,\ka]\fb .
\eeaa

We summarize the results of the above calculation in the following lemma.
\begin{lemma}
\lab{Lemma:GCMS:NEAR-again:bis1}
The original system \eqref{GCMS:kakabmumub}   in $(a, f, \fb) $ associated to a  deformation  sphere $\S$  is  equivalent to the following
\bea\lab{GCMS:NEAR-again:bis1}
\bsplit
\left(A^S+V\right)z &= \kab \ddsS \kac^\S+ \ka \ddsS \kabc^\S  -\kab\dds\,\,\check{\ka} -\ka\dds\,\,\check{\kab}\\
&+  \kab\err_4 +\ka\err_5 +[A^S,\kab]f+[A^S,\ka]\fb,\\
\left(A^\S +V\right) f &= \frac 3 4 \ka (A^\S)^{-1}\big(\rho z \big)+ \ddsS \check{\ka}^\S-\dds\,\,\check{\ka}-\ka  (A^\S)^{-1}\big( -\ddsS \check{\mu}^\S +\dds\,\,\check{\mu}\big)\\
&+\err_4,\\
\left(A^\S + V \right)\fb&= - \frac 3 4\kab  (A^\S)^{-1}\big(\rho z \big)+\ddsS \check{\kab}^\S-\dds\,\,\check{\kab}+\kab  (A^\S)^{-1}\big( -\ddsS \check{\mu}^\S +\dds\,\,\check{\mu}\big)\\
&+\err_5,\\
\ddsS a  &= -\frac 3 4( A^\S)^{-1} \big(  \rho z \big) + f\omb  -\fb \om    +\frac{1}{4}f\kab  -\frac 1  4  \fb \ka+ (A^\S)^{-1}\big( -\ddsS \check{\mu}^\S +\dds\,\,\check{\mu}\big)\\
&-(A^\S)^{-1}\err_3,
\end{split}
\eea
where,
\bea
z:=\kab f +\ka \fb,\qquad V:=-\frac 1 2 \ka \kab +\rho.
\eea
The error terms are given by $\err_1,\err_2,\err_3 $, defined in Lemma \ref{Le:GCMS:kakabmumub},  and
\bea
\label{eq:definitionforerrorterms4and5ofGCMequations}
\err_4&=&\err_1 +\ka (A^\S)^{-1} \err_3,\qquad 
\err_5=\err_2 -\kab (A^\S)^{-1} \err_3.
\eea
\end{lemma}

\begin{remark}
\label{remark:GCMS:NEAR-again:bis1}
We note the following remarks concerning the system \eqref{GCMS:NEAR-again:bis1}.
\begin{enumerate}
\item The right hand side of the equations   is linear in the quantities, 
\beaa
 \ddsS \kac^\S, \quad   \ddsS \kabc^\S, \quad \ddsS\muc^\S,
\qquad
\mbox{as well as} \qquad   \dds \kac, \quad   \dds \kabc, \quad \dds\muc.
\eeaa
The first group  is to  be constrained by  our GCM conditions  in the next section while the second group  depends on assumptions regarding the background foliation of $\RR$. 

\item The error terms contain  only \,  $\S$-angular derivatives of $(a, f, \fb)$ of order at most equal to the order of the corresponding operators on the  left hand sides, see Lemma \ref{Lemma:struct-Err1-5} below.   Thus the system is   in a standard quasilinear elliptic system form. 

\item In order to uniquely solve the equations for $z$, and then $f$ and $\fb$, we  need to 
the coercivity of the operator  $A^\S +V$.
One can easily show that the potential $V$ is positive    for  small values of $r$, i.e. $r$ near $r_\HH= 2m_0(1+\de_\HH) $
but negative for large $r$.  In fact $A^\S+V$ has a nontrivial  kernel for  large $r$ as one can easily see from the following calculation.  Since,
 \beaa
 A^\S=\dds_1\ddd_1 = \ddd_2\,\!^\S\dds_2\,\!^\S+2K, \qquad K=-\rho -\frac{1}{4}\ka\kab+\frac{1}{4}\vth\vthb
 \eeaa
   we deduce,
\beaa
A^\S+V= A^S+\frac 1 2 \ka \kab -\rho &=& \ddd_2\,\!^\S\dds_2\,\!^\S -3\rho +\frac{1}{2}\vth\vthb.
\eeaa
Thus for large enough $r$  the operator $A^\S+V$ behaves like $ \dddS_2\,\ddsS_2$ which  has a nontrivial kernel.

\item To be able to  correct for the  lack of coercivity of the system we need to control the $\ell=1$ modes of $f, \fb, \ddsS a $. In subsection  \ref{subsection:Badmodes-ddsSaf}   we derive an  equation for  the  last. The  $\ell=1$ modes of $(f, \fb)$, on the other hand,  have to be prescribed.

\item The equations do not provide information on  the  average of  $a$. For this we will need  yet another equation derived in subsection \ref{subsect:average-a}.
 \end{enumerate}
\end{remark}
 
 \begin{lemma}
  \label{Lemma:struct-Err1-5}
   The error terms $\err_1,\ldots, \err_5$ can be written schematically as follows,
   \bea
    \label{eq:errorterms-nearregion}
   \bsplit
   r^2 \err_1&= (\dkb^\S)^2\left((f, \fb, a)^2 \right)+ \dkb^\S\left( (f, \fb, a)  (r  \Gac_g) \right),\\
    r \err_2&= r^{-1} (\dkb^\S)^2\left((f, \fb, a)^2 \right)+ \dkb^\S\left( (f, \fb, a)  (  \Gac) \right),\\
    r^3 \err_3&=  (\dkb^\S)^3\left((f, \fb, a)^2 \right)+( \dkb^\S)^2\left( (f, \fb, a)  (r  \Gac_g) \right),\\
    \err_4, \err_5 &= \err_1+ (A^\S)^{-1} \err_3,
   \end{split}
   \eea
   where the  lower order  terms  denoted $\lot$ are  cubic  with respect to $a, f, \fb, \Gac, \Rc$
and   may involve  fewer  angular (along $\S$)  derivatives of $a, f, \fb$.
  \end{lemma}
  
  \begin{remark}
Note that $ \err_2$ behaves worse in powers of $r$ than $\err_1$. The reason is the presence of 
the terms $ f e_\th  \xib,  e_\th(f\xib)$  in the formula for $e_\th^\S(\err(\kab', \kab))$.
\end{remark}

  \begin{proof}
   Note that  in the spacetime region $\RR$ of interest   $r$ and $r^\S $  are  comparable.
   Recall, see \eqref{GCMS:Errorterms},
   \beaa
\bsplit
\err_1&=\err(e_\th'\ka', e_\th\ka)=e_\th' \, \err(\ka,\ka') +a e_\th'  \ddd\,'_1  f  + e_\th'( a) \ddd\,'_1 f \\
 &+ a \left( e_\th \ka+ \frac 1 2\left(  \fb e_4\ka + f e_3\ka\right)\right)+  f\ddd_1 \eta+ \fb \ddd_1\xi +\lot,\\
 \err_2&=\err(e_\th'\kab', e_\th\kab)=e_\th' \, \err(\kab,\kab') -a e_\th'  \ddd\,'_1  \fb  - e_\th'( a) \ddd\,'_1 \fb\\
 &- a \left( e_\th \kab+ \frac 1 2\left(  \fb  e_4\kab + f  e_3\kab\right)\right)+  \fb \ddd_1 \etab+ f  \ddd_1\xib +\lot,
 \\
\err_3&=\err(e_\th'\mu', e_\th \mu)=e_\th'\err(\mu,\mu') -\frac 1 2\left( \fb \ddd_1 \b +f\ddd_1\bb \right)  -\frac 1 2( f e_3 +\fb  e_4 )  \ddd_1 \ze+\lot,\\
\end{split}
\eeaa
and\footnote{Recall also  the outgoing  geodesic conditions  i.e. $\xi=0$,  $\ze +\etab=0$, $\ze-\eta=0$, $\om=0$.},
\beaa
  \err(\ka,\ka')&=& \frac 1 2   f \fb e_\th'(f) -\frac{1}{4}\fb e_\th'(f^2)+f(\ze+\eta)    +\fb\xi       -\frac 1 4 f^2\kab +f\fb\om -f^2\omb+\lot,  \\ 
   \err(\kab,\kab')&=& -\frac{1}{2}\fb e_\th'\left( f \fb +\frac{1}{8} f^2 \fb^2\right) +\left(\frac{3}{4}f\fb+\frac 1 8 (f\fb)^2\right)e_\th'(\fb)  \\
   &+&\frac 1 4\left(1+\frac{1}{2}f\fb\right)\fb e_\th'\left( f\fb\right)
-\frac{1}{4}f\left(1+\frac{1}{4}f\fb\right)e_\th'\left(\fb^2\right)  \\
& +& \fb(-\ze+\etab)    +f\xib       -\frac 1 4 \fb^2\ka +f\fb\omb -\fb^2\om+\lot
 \eeaa
Also,
\beaa
\err(\mu,\mu')&=& - e_\th'  \err(\ze,\ze')  - \err(\rho,\rho')+\frac 1 4\big( \vth'\vthb'-\vth \vthb\big),\\
\err(\ze,\ze')&= &\frac{1}{2}\fb\left(1+\frac 1 4 f\fb\right)e_\th'(f) -\frac{1}{16}\fb^2 e_\th'(f^2) +\frac 1 4( - f \vthb +\fb \vth)+\lot,  \\
\err(\rho,\rho') &=& \frac{3}{2}\rho f\fb+\fb\b  +f\bb+\lot
\eeaa

We write schematically\footnote{The last term  $r^{-2}  \dkb^\S( f^2)$ on the right of the  identity below is due
 to the term $e_\th'(f^2\omb)$   in the expression of $e_\th' \, \err(\kab,\kab')$.},
\beaa
\err_1&=& (f, \fb, a)(r^{-2} \dkb^\S)^2(f, \fb, a)+ (r^{-1} \dkb^\S(f, \fb, a))^2+ r^{-1}\dkb^\S\left( (f, \fb, a)  \Gac_g \right)\\
&+& r^{-2}  \dkb^\S( f^2)+ \frac 1 2 a\left(  \fb e_4\ka + f e_3\ka\right)+\lot
\eeaa
Making use of 
\beaa
e_3(\ka)+\frac 1 2 \kab\, \ka-2\omb \ka   &=& 2\ddd_1\eta  + 2\rho -\frac 1 2  \vthb\, \vth +2(\xi\xib+\eta^2),\\
e_4 \ka+\frac 1 2 \ka^2 +2\om \ka&=& 2\ddd_1\xi -\frac 1 2 \vth^2 +2(\eta+\etab+2\ze)\xi,
\eeaa
 and       treating the curvature terms that appear as  $\Gac_g$ we easily derive,
\beaa
r^2 \err_1&=& (\dkb^\S)^2\left((f, \fb, a)^2 \right)+ \dkb^\S\left( (f, \fb, a)  (r  \Gac_g) \right).
\eeaa

We obtain a worse estimate for $\err_2 $ because of the presence  $e_\th' (f\xib)$, since $\xib\in\Gac_b$.
In fact,
\beaa
r \err_2&=& r^{-1} (\dkb^\S)^2\left((f, \fb, a)^2 \right)+ \dkb^\S\left( (f, \fb, a) \Gac \right).
\eeaa

 For  $\err_3$ we write similarly, treating the curvature terms  that appear as $\Gac_g$,
 \beaa
 e_\th(\mu, \mu')&=&r^{-3} ( \dkb^\S)^3 \left((f, \fb, a)^2 \right)+r^{-3}( \dkb^\S)^2 \left((f, \fb, a) \Gac_g \right)+\lot
 \eeaa
 Using the  null structure equations for $\ze$  we  infer that,
 \beaa
 \err_3&=&e_\th(\mu, \mu')  -\frac 1 2\left( \fb \ddd_1 \b +f\ddd_1\bb \right)  -\frac 1 2( f e_3 +\fb  e_4 )  \ddd_1 \ze+\lot\\
 &=&r^{-3} ( \dkb^\S)^3 \left((f, \fb, a)^2 \right)+r^{-3}( \dkb^\S)^2 \left((f, \fb, a) \Gac_g \right)+\lot
 \eeaa
 as stated.
   \end{proof}

  Making use of the above lemma and  the assumptions {\bf A1-A3}   we can derive the following.
  
\begin{lemma}
\label{Lemma-errorestimates-S}
Assume  given  a deformation $\Psi:\ovS\longrightarrow \S$ in $\RR$  and  adapted frame $(e_3', e_4' , e_\th') $ with  $e_\th'=e_\th^\S$  with transition parameters $a, f, \fb$ defined on $\S$. Assume that\footnote{This is   assumption \eqref{eq:boundonU'andS'onS0forequivalenceofhigherorderSobolevnorms}    of Lemma \ref{lemma:int-gaS-ga:highersobolevregularity} with $\dg$ replaced to $\epg$.}
the following  holds true   
\beaa
 \|(U', S')\|_{L^\infty(\ovS)}+ \max_{0\leq s\leq s_{max}} (\ovr)^{-1} \|(U', S')\|_{\hk_s(\ovS, \ovgS)} &\les& \epg.
 \eeaa
Then,  for $ s= s_{max}+1$, with     $s_{max}\ge 4$,
\bea
\bsplit
 \|\err_1, \err_2\|_{\hk_{s-2 } (\S)}&\les&  r^{-2}   \,  \Big\| ( f, \fb, a) \Big\|_{\hk_{s}(\S)} \left(\epg+  \Big\|  f, \fb, a \Big\|_{\hk_{s}(\S)}  \right),  \\
  \left\| \err_3\right\|_{\hk_{s-3}(\S)}&\les&  r^{-3}   \,  \Big\|  (f, \fb, a) \Big\|_{\hk_{s}(\S)} \left(\epg+  \Big\|  f, \fb, a \Big\|_{\hk_{s}(\S)}  \right) ,\\
      \| \err_4, \err_5 \|_{\hk_{s-2}(\S)}&\les&  r^{-2}    \,  \Big\| ( f, \fb, a) \Big\|_{\hk_{s}(\S)} \left(\epg+  \Big\|  f, \fb, a \Big\|_{\hk_{s}(\S)}  \right).
\end{split}
\eea

\end{lemma}
\begin{proof} The proof  follows easily from Lemma \ref{Lemma:struct-Err1-5}, Corollary \ref{cor:int-gaS-ga:highersobolevregularity}, coercivity of $A^\S$  and obvious product estimates on $\S$.
Consider  for example the term 
 \beaa
 \err_2= r^{-2} (\dkb^\S)^2\left((f, \fb, a)^2 \right)+r^{-1}  \dkb^\S\left( (f, \fb, a)  (  \Gac) \right). 
  \eeaa 
We write,
\beaa
(\dkb^\S)^k \err_2 &=&  r^{-2} (\dkb^\S)^{2+k} \left((f, \fb, a)^2 \right)+ r^{-1} (\dkb^\S)^{1+k} \left( (f, \fb, a)  (  \Gac) \right) +\lot
\eeaa
and
\beaa
(\dkb^\S)^{2+k} \left((f, \fb, a)^2 \right)&=&\sum_{i+j= k+2}  \dkb^i  (f, \fb, a) \c   \dkb^j (f, \fb, a).
\eeaa
Thus,  dividing the  sum  into terms with $i\ge [\frac{k+2}{2} ]$ and $i<  [\frac{k+2}{2} ]$  and using Sobolev estimates for the terms involving  fewer derivatives we derive,  for $[\frac{k+2}{2} ] +2 \le k+2 $,
\beaa
\|(\dkb^\S)^{2+k} \left((f, \fb, a)^2 \right)\|_{L^2(\S)} &\les& r^{-1}   \| (a, f, \fb) \|_{\hk^{k+2}(\S)}^2.   
\eeaa
Similarly, making use of our assumptions for $\Gac$,
\beaa
 (\dkb^\S)^{1+k} \left( (f, \fb, a)  (  \Gac) \right)&\les&r^{-1}  \| (a, f, \fb) \|_{\hk^{k+1}(\S)}\| \Gac \|_{\hk^{k+1}(\S)}\\
 &\les& \epg r^{-1}  \| (a, f, \fb) \|_{\hk^{k+1}(\S)}.
\eeaa
Thus, for all $ 2 \le k\le s_{max}-1   $
\beaa
\|(\dkb^\S)^k \err_2 \|_{L^2(\S)} \les r^{-2}   \, \|  (f, \fb, a)\|_{\hk^{k+2}(\S)} \left(\ep_0+  \Big\| ( f, \fb, a )\Big\|_{\hk_{k+2}(\S)}\right)
\eeaa
i.e., for $4\leq s\leq s_{max}+1$,
\beaa
\|\err_2\|_{\hk_{s-2} (\S)}&\les&  r^{-2}   \,  \Big\| ( f, \fb, a) \Big\|_{\hk_{s}(\S)} \left(\epg+  \Big\|  f, \fb, a \Big\|_{\hk_{s}(\S)}  \right).
\eeaa
All other terms  can be treated similarly.
\end{proof}


 \subsection{Equation for the average of $a$}
 \label{subsect:average-a}
 

 In the proof of  existence and uniqueness of GCMS, see  Theorem \ref{Theorem:ExistenceGCMS}
we will need, in addition of  the equations derived so far, an  equation for the average of $a$. To achieve this we make use of the transformation formula for $\ka$ of Lemma \ref{Lemma:Transformation-special}
\beaa
\ka' &=& e^{a} \left( \ka+ \ddd_1'  f   \right) + e^a \err(\ka,\ka'),\\
 \err(\ka,\ka')&=& f(\ze+\eta)  +\fb \xi+\frac{1}{2}\fb e_4f+\frac 1 2 f e_3 f+  \frac 1 4 f \fb \ka +f\fb\om -\omb f^2+\lot
\eeaa
which we  rewrite in the form,
\beaa
\ka^\S &=& e^{a}\left(\frac{2}{r}    +\check{\ka} + \left(\ov{\ka}-\frac{2}{r}\right)       +\dddS f  +\err(\ka, \ka')\right).
\eeaa
We deduce,
\beaa
(e^{a} -1) \frac 2 r &=&\ka^\S -\frac 2 r - e^{a} \left( \check{\ka} + \ov{\ka}-\frac{2}{r}       +\dddS f \right) -e^{a}\err(\ka, \ka')\\
&=&\ka^\S-\frac{2}{r^\S} +\left( \frac{2}{r^\S}-\frac{2}{r}\right)   -   \left( \check{\ka} + \ov{\ka}-\frac{2}{r}       +\dddS f \right)\\
& -&e^{a}\err(\ka, \ka')-  (e^a-1)  \left( \check{\ka} + \ov{\ka}-\frac{2}{r}       +\dddS f \right)
\eeaa
or,
\beaa
a \frac 2 r &=&\ka^\S-\frac{2}{r^\S} +\left( \frac{2}{r^\S}-\frac{2}{r}\right)   -   \left( \check{\ka} + \ov{\ka}-\frac{2}{r}       +\dddS f \right)\\
& -&e^{a}\err(\ka, \ka')-  (e^a-1)  \left( \check{\ka} + \ov{\ka}-\frac{2}{r}       +\dddS f \right)-\left(e^a-1-a \right) \frac 2 r. 
\eeaa
We deduce,
\beaa
a &=&  \frac{ r^\S}{2} \left( \ka^\S-\frac{2}{r^\S}\right)+\left( 1-\frac{r^\S}{r}\right)- \frac{ r^\S}{2} \left( \check{\ka} + \ov{\ka}-\frac{2}{r}       +\dddS f \right)+\err_6\\
\err_6& =&-\frac{r^\S}{2}\left[e^{a}\err(\ka, \ka')-  (e^a-1)  \left( \check{\ka} + \ov{\ka}-\frac{2}{r}       +\dddS f \right)-\left(e^a-1-a \right) \frac 2 r \right]\\
&-& a \left(\frac{r^\S}{r}-1\right).
\eeaa
Taking the average on $\S$  we infer that,
\bea
\ov{a}^\S&=&\frac{r^\S}{2}\left( \ov{{\ka^\S} }^\S-\frac{2}{r^\S}\right)+\ov{\left( 1-\frac{r^\S}{r}\right)}^\S - \frac{ r^\S}{2} \ov{\left( \check{\ka} + \ov{\ka}-\frac{2}{r}      \right)}^\S+\ov{\err_6}^\S
\eea
where  $\ov{h}^{\S}$ denotes the average of $h$ on $\S$.


\subsection{Transversality conditions}


\begin{lemma}
\lab{Lemma:GCMStransversality1} 
Assume given a   deformed sphere $\S\subset\RR$ with  adapted null frame  $e_3^\S, e_4^\S, e_\th^\S$  and   transition functions $(a, f,\fb) $. We can extend $a,f, \fb$, and thus the frame  $e_3^\S, e_4\S, e_\th^\S$,  in a small neighborhood of $\S$   such that  the following hold true
\bea\lab{GCMS:transversality1} 
\bsplit
\xi^\S&=0, \qquad \om^\S= 0, \qquad \, \etab^\S+\ze^\S=0.
\end{split}
\eea
\end{lemma}

\begin{proof}
According to Proposition \ref{prop:transformations1-GCM} we have,
\beaa
 \bsplit
 \xi^\S&=e^{2a}\left(\xi + \frac{1}{2}e^{-a}e^\S_4(f)+\frac{1}{4}f \ka+f\om \right) +e^{2a}\err(\xi, \xi^\S),\\
\xib^\S&=e^{-2a}   \left(\xib + \frac{1}{2} e^a e^\S_3(\fb)+\frac{1}{4}\fb  \,  \kab +\fb\, \omb \right) +e^{-2a} \err(\xib, \xib^\S),\\
\ze^\S&= \ze   - e^\S_\th( a )         -f\omb  +\fb \om                -\frac{1}{4}f\kab+\frac 14  \fb \ka    +\err(\ze,\ze^\S),\\
 \eta^\S&=\eta +\frac 1 2e^ a  e^\S_3 f -  f \omb +\frac 1 4 f\kab +\err(\eta,\eta^\S),\\
 \etab^\S&=\etab+\frac 1 2 e^{-a}  e^\S_4\fb -  \fb  \om +\frac 1 4 \fb \ka +\err(\etab,\etab^\S), \\
 \om^\S&=e^{a}\left(\om-\frac 1 2e^{-a}  e_4a\right)+e^{a}\err(\om,\om^\S),\\
  \omb^\S&=e^{-a}\left(\omb+ \frac 1 2  e^a  e_3 a\right)+e^{-a}\err(\omb,\omb^\S).
\end{split}
\eeaa
Clearly  the conditions $  \xi^\S=0, \, \om^\S=0$ allows us to determine $e_4^\S f$ and $e_4^\S a$ on  $\S$ while the condition  $\etab^\S+\ze^\S=0$ allows us to determine $ e_4^\S \fb$ on $\S$.
\end{proof} 

\begin{remark}
Note that the equations above also allow us to impose, in addition,  vanishing  conditions on $\xib$ and $\om$ along $\S$. Indeed these are determined by $e^\S_3 f$ and $e^\S_3 a $. 
\end{remark}


\subsection{Equation for  the $\ell=1$ mode of $ \ddsS a$}
\label{subsection:Badmodes-ddsSaf}


 Recall the following change of frame formulas
\beaa
 \bsplit
 e_\th'(\ka' )     &= e_\th \ka +e_\th'  \ddd_1\,'  f  +\ka e_\th' a -\frac 1 4 \ka ( f\kab +\fb \ka)+\ka ( f\omb- \om \fb)+f \rho+\err(e_\th'\ka', e_\th\ka).
\end{split}
\eeaa
where $e_\th'=-\ddsS$, $ \ddd_1\,' =\dddS$,  $\ka'=\ka^\S$,  $\b'=\b^\S$.
Making use of
\beaa
\dds\,'_1\ddd\,'_1=\ddd\,'_2\dds\,'_2+2K^S,\quad K=-\rho-\frac{1}{4}\ka \kab+\frac{1}{4}\vth \vthb,
\eeaa
we deduce,
\beaa
 e_\th'(\ka' )     &= &e_\th \ka   -(\ddd_2) '(\dds_2)'f +\ka e_\th'(a) -\ka\left( \frac 1 4 \ka+ \om\right)\fb + \left(-\frac 1 4 \ka\kab +\ka\omb+\rho -2 K \right)f \\
 &+&\err(e_\th'\ka', e_\th\ka)-2(K^\S-K) f \\
 &=& e_\th \ka  -(\ddd_2) '(\dds_2)'f +\ka e_\th'(a) -\ka\left( \frac 1 4 \ka+ \om\right)\fb+ \left( \frac 1 4 \ka\kab +\ka\omb+3\rho\right)f \\
 &+&\err(e_\th'\ka', e_\th\ka) -2(K^\S-K) f -\frac{1}{2}\vth\vthb f,
\eeaa
or, in view of the condition  $\ka'=\ka^\S=\frac{2}{r^\S} $ we have   $ e_\th'(\ka' )  =-\ddsS (\ka^\S)=0$,
\beaa
 e_\th'(\ka' ) &=&e_\th \ka  -(\ddd_2) '(\dds_2)'f +\frac{2}{r'}  e_\th'(a) -\ka\left( \frac 1 4 \ka+ \om\right)\fb+ \left( \frac 1 4 \ka\kab +\ka\omb+3\rho\right)f \\
 &+&\err(e_\th'\ka', e_\th\ka) +(\ka-\frac{2}{r'}  ) e_\th'(a) -2(K^\S-K) f  -\frac{1}{2}\vth\vthb f.
\eeaa
Multiplying by $e^\Phi$ and integrating on $\S$ we  derive,
 \beaa
 \int_\S  e_\th'(\ka' ) e^\Phi       &=& \int_\S e_\th(\ka) e^\Phi   +\frac{2}{r^\S} \int_\S e_\th^\S(a)e^\Phi -\int_{\S}\ka\left( \frac 1 4 \ka+ \om\right)\fb e^\Phi+ \int_\S \left( \frac 1 4 \ka\kab +\ka\omb+3\rho\right)f e^\Phi\\
&+&\int_\S\left(\err(e_\th^\S\ka^\S, e_\th\ka)-2(K^\S-K) f  -(\ka-\frac{2}{r^\S} ) \ddsS a -\frac{1}{2}\vth\vthb f\right)e^\Phi.
\eeaa
Therefore (recall that $\om=0$),
\beaa
\bsplit
\frac{2}{r^\S} \int_\S e_\th^\S(a)e^\Phi&= \int_\S  e_\th^\S (\ka^\S) e^\Phi   -\int_\S e_\th(\ka) e^\Phi + \frac 1 4 \int_{\S}\ka^2 \fb e^\Phi -\int_\S \left( \frac 1 4 \ka\kab +\ka\omb+3\rho\right)f e^\Phi\\
&+\err_7,\\
\err_7 &= \int_\S\left(-\err(e_\th^\S\ka^\S, e_\th\ka)+2(K^\S-K) f  +(\ka-\ka^\S ) \ddsS a        +\frac{1}{2}\vth\vthb f\right)e^\Phi.
\end{split}
\eeaa

We summarize the  result in the following lemma.
\begin{lemma}
\label{GCMS:-BAD MODES}
We have,
\bea
\bsplit
\frac{2}{r^\S} \int_\S e_\th^\S(a)e^\Phi&= \int_\S  e_\th^\S (\ka^\S) e^\Phi   -\int_\S e_\th(\ka) e^\Phi + \frac 1 4 \int_{\S}\ka^2 \fb e^\Phi \\
&-\int_\S \left( \frac 1 4 \ka\kab +\ka\omb+3\rho\right)f e^\Phi +\err_7,
\end{split}
\eea
where,
\beaa
\err_7 &=& \int_\S\left(-\err(e_\th^\S\ka^\S, e_\th\ka)+2(K^\S-K) f  +\left(\ka-\frac{2}{r^\S}\right)  \ddsS a        +\frac{1}{2}\vth\vthb f\right)e^\Phi.
\eeaa
\end{lemma}


\section{Existence of GCM spheres }
\lab{section-ExistenceGCMspheres}


We now impose the  GCM conditions  on the  deformed sphere $\S$
 \bea
 \label{Conditions:GCMS}
\bsplit
\ddsS_2\ddsS_1\kab^\S&=\ddsS_2\ddsS_1\mu^\S=0, \qquad \ka^\S=\frac{2}{r^\S},\\
\int_{\S}  f  e^\Phi&=\La, \qquad  \int_{\S} \fb e^\Phi=\Lab, 
\end{split}
\eea
where $(f, \fb)$ belong to  the triplet $(f, \fb, \la=e^a)$ which denote  the change of frame coefficients from the frame of $\ovS$ to the one of $\S$. We show  that under these  conditions  the   deformation parameters $a, f, \fb$   verify a coercive elliptic system.

We start with the following simple adaptation of Lemmas \ref{GCMS:NEAR-again:bis1}, \ref{GCMS:-BAD MODES}
 and   the result of subsection \ref{subsect:average-a}.

\begin{proposition}
\lab{proposition-GCMS-system}
Assume that the deformed sphere $\S$ verifies  the GCMS conditions \eqref{Conditions:GCMS}.  Then  the deformation parameters  $a, f, \fb$ verify the  system (recall that $z=\kab f+\ka \fb$ and $V=\frac 1 2 \ka\kab -\rho$)
\bea\lab{GCMS:NEAR-again:bis}
\bsplit
\left(A^\S+V\right)z &= \ka \ddsS \kabc^\S  -\kab\dds\,\,\check{\ka} -\ka\dds\,\,\check{\kab}\\
&+  \kab\err_4 +\ka\err_5 +[A^S,\kab]f+[A^S,\ka]\fb,\\
\left(A^\S +V\right) f &= \frac 3 4 \ka (A^\S)^{-1}\big(\rho z \big)-\dds\,\,\check{\ka}-\ka  (A^\S)^{-1}\big( -\ddsS \check{\mu}^\S +\dds\,\,\check{\mu}\big)\\
&+\err_4,\\
\left(A^\S + V \right)\fb&= - \frac 3 4\kab  (A^\S)^{-1}\big(\rho z \big)+\ddsS \check{\kab}^\S+\dds\,\,\check{\kab}-\kab  (A^\S)^{-1}\big( -\ddsS \check{\mu}^\S +\dds\,\,\check{\mu}\big)\\
&+\err_5,\\
\ddsS a  &= -\frac 3 4( A^\S)^{-1} \big(  \rho z \big) + f\omb   +\frac{1}{4}f\kab  -\frac 1  4  \fb \ka+ (A^\S)^{-1}\big( -\ddsS \check{\mu}^\S +\dds\,\,\check{\mu}\big)\\
&-(A^\S)^{-1}\err_3,\\
\ov{a}^\S&=\ov{\left( 1-\frac{r^\S}{r}\right)}^\S - \frac{ r^\S}{2} \ov{\left( \check{\ka} + \ov{\ka}-\frac{2}{r}      \right)}^\S+\ov{\err_6}^\S,
\end{split}
\eea
and,
\bea
\lab{equations-forBadModes}
\bsplit
\frac{2}{r^\S} \int_\S e_\th^\S(a)e^\Phi&= \int_\S  e^\S_\th(\ka^\S ) e^\Phi  -\int_\S e_\th(\ka) e^\Phi +\frac 1 4 \int_{\S}\ka^2  \fb e^\Phi\\
&-\int_\S \left( \frac 1 4 \ka\kab +\ka\omb+3\rho\right)f e^\Phi+\err_7,
\end{split}
\eea
\bea
\lab{equations-forBadModes-ffb}
\int_\S e^\Phi f=\La, \qquad   \int_\S e^\Phi \fb=\Lab.
\eea
The error terms are given by $\err_1,\err_2,\err_3 $, defined in Lemma \ref{Le:GCMS:kakabmumub},   and 
\beaa
\err_4&=&\err_1 +\ka (A^\S)^{-1} \err_3,\qquad 
\err_5=\err_2 -\kab (A^\S)^{-1} \err_3,
\eeaa
\beaa
\err_6& =&-\frac{r^\S}{2}\left[e^{a}\err(\ka, \ka')-  (e^a-1)  \left( \check{\ka} + \ov{\ka}-\frac{2}{r}       +\dddS f \right)-\left(e^a-1-a \right) \frac 2 r \right]\\
&-& a \left(\frac{r^\S}{r}-1\right),\\
\err_7 &=& \int_\S\left(-\err(e_\th^\S\ka^\S, e_\th\ka)+2(K^\S-K) f  +\left(\frac{2}{r} -\frac{2}{r^\S}\right)  \ddsS a        +\frac{1}{2}\vth\vthb f\right)e^\Phi.
\eeaa
\end{proposition}

We are ready to state the first  main result of this chapter. 
\begin{theorem}[Existence of GCM spheres]
\lab{Theorem:ExistenceGCMS}
 Let  $\ovS=S(\ovu, \ovs)$   be  a fixed    sphere of the  $(u, s)$ outgoing geodesic foliation of  a fixed  spacetime region 
 $\RR$.  Assume in addition to {\bf A1}--{\bf A3} that  the following estimates  hold true on $\RR$, for all $k\le s_{max}$,
\bea
 \label{eq:GCM-improved estimate}
 \bsplit
\left|\dk^{k-2}\dds_2\dds_1\kab\right| &\les \dg  r^{-4}, \\
\left|\dk^{k-2}\dds_2\dds_1\mu\right| &\les \dg   r^{-5},  \\
\left|\dk^k\kac\right| &\les \dg r^{-2},\\
\left|\ov{\ka}-\frac{2}{r}\right|+\left|\ov{\kab}+\frac{2\left(1-\frac{2m}{r}\right)}{r}\right| &\les \dg r^{-2}.
\end{split}
\eea

For any fix   $\La, \Lab \in \RRR$  verifying,
\bea
|\La|,\,  |\Lab|  &\les & \dg r^{2}
\eea
 there 
exists a unique  GCM sphere $\S=\S^{(\La, \Lab)}$, which is a deformation of $\ovS$,  such that  the GCMS conditions \eqref{Conditions:GCMS}  are verified.
Moreover the following estimates hold true.
\begin{enumerate}
 \item  We have
  \bea
\left|\frac{r^S}{\rg}-1\right|\les  r^{-1} \dg.
\eea
In particular $r, \ovr$ and $r^\S$ are all comparable in $\RR$.

\item  The unique  functions $(\la, f, \fb)$ on $\S$, which  relate the original frame $e_3, e_4, e_\th$ 
to the new  frame on $e_3^\S, e_4^\S, e_\th^\S$ according to \eqref{SSMe:general.composite:repeatforproofGCM}, verify  the estimates
\bea
\lab{estimates-transtionF-ThmDCM}
\Big\|  f, \fb, \log \la \Big\|_{\hk_k(\S)} &\les \dg, \qquad k\le s_{max}+1. 
\eea

\item  The  parameters $U, S$ of the  deformation, see Definition  \ref{definition:Deformations},  verify  the estimate
 \bea
 \label{eq:ExistenceGCMS-Thm-US}
 \|(U', S')\|_{L^\infty(\ovS)}+ \max_{0\leq s\leq s_{max}}  r^{-1}\|(U', S')\|_{\hk_s(\ovS, \ovgS)} &\les&  \dg.
  \eea
   
   \item The Hawking mass $m^\S$ verifies the estimate,
   \bea
   \lab{eq:ExistenceGCMS-Thm-m^S}
   \big|m^\S-\ovm\big|&\les &\dg. 
   \eea

\item The  curvature components  $(\a^\S, \b^\S, \rho^\S, \bb^\S,  \aa^\S)$, as well as $\mu^\S$ and  the Ricci coefficients\footnote{All other Ricci coefficients involve    the transversal  derivatives $e_3^\S, e_4^\S$ of the frame.} $(\ka^\S, \vth^\S, \ze^\S, \kab^\S, \vthb^\S)$  on $\S$, verify, for all
$k\le s_{max}$,
\bea
 \label{Estimate-GacS-onS}
 \bsplit
 \| \kac^\S, \vth^\S, \ze^\S, \kabc^\S \|_{\hk_k (\S)}&\les \epg r^{-1}, \\
  \|\vthb^{\S} \|_{\hk_k (\S)}&\les \epg,  \\
   \|\a^\S, \b^\S, \rhoc^\S, \mu^\S \|_{\hk_{ k} (\S)}&\les \epg r^{-2}, \\
   \|\bb^\S \|_{\hk_k (\S)}&\les \epg r^{-1},\\
    \|\aa^\S \|_{\hk_{ k} (\S)}&\les \epg. 
 \end{split}
 \eea

 \item The   functions,  $(\la, f, \fb)$ uniquely defined  above,   can be  smoothly extended to a small neighborhood of $\S$ in such a way that  the  corresponding  Ricci coefficients  verify the  following  transversality conditions  
 \bea
\lab{Transversality-Si*}
\xi^\S=0, \qquad \om^\S=0, \qquad \etab^\S+\ze^\S=0.
\eea
In that case, the following estimates hold\footnote{To be more precise one should replace $r$ by $\rg$ in the estimates below.   Of course $r$  and $\rg$ are  comparable in $\RR$, in particular on $\S$.} for all $k\le s_{max}-1$
\bea
\lab{Transversality-Si*-f-fb-la}
\|e_4(f, \fb, \log \la)\|_{\hk_{k+1}(\S)} &\les& r^{-1}\dg+ r^{-3} \left(|\La|+|\Lab|\right),
\eea 
 and, 
 \bea
 \label{Estimate-GacS-onS-e4}
 \bsplit
 \|e^\S_4 (\kac^\S, \vth^\S, \ze^\S, \kabc^\S )\|_{\hk_{k} (\S)}&\les \epg r^{-2}, \\
  \|e^\S_4(\vthb^{\S}) \|_{\hk_k (\S)}&\les \epg r^{-1},  \\
   \|e_4^\S\left(\a^\S, \b^\S, \rhoc^\S, \mu^\S\right) \|_{\hk_{ k} (\S)}&\les \epg r^{-3}, \\
   \|e_4^\S(\bb^\S) \|_{\hk_k (\S)}&\les \epg r^{-2},\\
    \|e_4^\S(\aa^\S) \|_{\hk_{ k} (\S)}&\les \epg r^{-1}.
 \end{split}
 \eea
\end{enumerate}
\end{theorem}

\begin{remark}  In view of Propositions  \ref{Proposition:compatible-deformations} and \ref{proposition-GCMS-system},  to find a GCM sphere amounts to solve the following  coupled system
  \bea
 \label{eq:DeformS-66}
  \bsplit
   \vsi^\#  U'&= \left((\gaS)^\#  \right)^{1/2}   f^\#\left( 1 +\frac 1 4 (f\fb)^\#\right),\\
     S'-\frac{\vsi^\#}{2} \Omb^\#  U'&= \frac 1 2   \left((\gaS)^\#  \right)^{1/2}\fb^\#,
  \\
     (\ga^ \S)^\# &=\ga^\#+(\vsi^\#)^2 \left(\Omb+\frac{1}{4} \underline{b}^2  \ga  \right)^\#\, ( \pr_\th U )^2 -2 \vsi^\# \pr_\th U \pr_\th  S-(\ga \vsi  \underline{b})^\# \pr_\th  U,\\
     U(0)&= S(0)=0,  
     \end{split}
  \eea
where the inputs $(a, f, \fb)$ verifies  \eqref{GCMS:NEAR-again:bis} and \eqref{equations-forBadModes}.
Recall that  for a reduced scalar $h$ defined on $\S$ we write
\beaa
  h^\#(\ug, \sg, \th) &=& h(\ug+U(\th), \sg+S(\th), \th).
\eeaa
We  will  solve  the coupled  system of equations \eqref{eq:DeformS-66},
\eqref{GCMS:NEAR-again:bis},  \eqref{equations-forBadModes}  and \eqref{equations-forBadModes-ffb} by an iteration argument which will be introduced  below.  Before doing this however it pays to observe that  the system  \eqref{GCMS:NEAR-again:bis} and \eqref{equations-forBadModes} can be interpreted as  an elliptic  system
  on a fixed surface $\S$ for  $(a,  f, \fb)$.   In the next subsection we state a result which  establishes  the coercivity of the system.   The full  proof of the theorem is detailed in   the remaining part   of this section, see subsection \ref{subsection:apriori-GCM} --\ref{subsection:endofproof-155-156}.
  \end{remark}


\subsection{A priori  estimates}
\lab{subsection:apriori-GCM}


The following result   plays  the main  role in the proof of Theorem \ref{Theorem:ExistenceGCMS}.
   \begin{proposition}[Apriori Estimates for GCMS]
   \lab{prop.GCMSequations-fixedS}
    Let  a fixed  spacetime region $\RR$   verifying assumptions ${\bf A1-A3}$ and \eqref{eq:GCM-improved estimate}. Assume  $\S$ is a given surface  in $\RR$  such that  the area radius of $\S$ verifies\footnote{Recall  also from \eqref{eq:comparisionrminusrgforGCMprocedure} that $r-\ovr=O(\dg \epg^{-1/2})$.  Thus  $r, \ovr$  and $r^\S$ are all comparable.}
  \bea\lab{eq:assumptionsrminusrSforpropGCMSequations-fixedS}
   \left| \frac{r^\S}{\ovr}-1\right|&\les r^{-1} \dg.
   \eea
     Then,   for every $\La, \Lab$,  $|\La|+|\Lab|\les \dg$,   there exists a unique  solution      $(\la , f, \fb)$  of  the     system
     \eqref{GCMS:NEAR-again:bis}-\eqref{equations-forBadModes}
    verifying the estimate,
\bea
\label{eq:prop.GCMSequations-fixedS1}
\bsplit
\|(\log \la   , f, \fb) \|_{\hk_{s_{max}+1}(\S) } &\les \dg.
\end{split}
\eea
More precisely, 
\bea
\Big\|  f, \fb, \log \la \Big\|_{\hk_{s_{max}+1}(\S) }&\les&  \dg+\left|\La \right|+\left| \Lab \right|.
\eea
\end{proposition}

As an immediate  corollary we  derive the following rigidity result for GCM spheres.
\begin{corollary}[Rigidity I]
\lab{corr:GCM-rigidity1}
  Let  a fixed  spacetime region 
 $\RR$    verifying assumptions ${\bf A1-A3}$ and \eqref{eq:GCM-improved estimate}. 
Assume $\S$ is a sphere in $\RR$ which verifies the the GCM conditions
\bea
\ka^\S=\frac{2}{r^\S}, \quad \dds_2\,^\S\dds_1\,^\S\kab^\S=\dds_2\,^\S\dds_1\,^\S\mu^\S=0.
\eea

Then the transition functions  $(f, \fb, \log\la)$ from the background frame  of $\RR$ to    to that of $\S$ 
 verifies the  estimates
 \beaa
\|(f, \fb, \log(\la))\|_{\hk_{s_{max}+1}(\S) } &\les& \dg+\left|\int_{\S}fe^\Phi\right|+\left|\int_{\S}\fb e^\Phi\right|.
\eeaa
\end{corollary}
The proof  of Proposition     \ref{prop.GCMSequations-fixedS}   will be given in  the next section. The  proof of 
 Corollary \ref{corr:GCM-rigidity1} is an immediate consequence   of the proposition.

  We assume in addition that   $\S$ is a deformation of  $\ovS$    and prove the following corollary of 
   Proposition \ref{prop.GCMSequations-fixedS}.
  \begin{corollary}
     \lab{prop.GCMSequations-fixedS-2}
    Let  a fixed  spacetime region 
 $\RR$    verifying assumptions ${\bf A1-A3}$ and \eqref{eq:GCM-improved estimate}, and let $\Psi:\ovS \longrightarrow \S$  be a  fixed    deformation in $ \RR $ given  by  the  scalars  $U, S$.   We assume that the deformation verifies, see  \eqref{eq:boundonU'andS'onS0forequivalenceofhigherorderSobolevnorms},   
    \bea
    \label{assumptions:U1-S1-2}
  \|(U', S')\|_{L^\infty(\ovS)} &\les& \dg.  
   \eea
    Then $(a, f, \fb)$ verify the estimate
   \bea
\|(a  , f, \fb) \|_{\hk_{s_{max}+1}(\S)} &\les \dg.
\eea 
    \end{corollary}
    
\begin{proof}
In view of Proposition  \ref{prop.GCMSequations-fixedS}, it suffices to prove that  \eqref{eq:assumptionsrminusrSforpropGCMSequations-fixedS}  holds.  \eqref{eq:assumptionsrminusrSforpropGCMSequations-fixedS} actually follows from the comparison  Lemma \ref{lemma:int-gaS-ga}. 
\end{proof}


\subsection{Comparison of the Hawking mass}  
\lab{subsection:estimate-Hawkingmass}


We establish the estimate \eqref{eq:ExistenceGCMS-Thm-m^S} concerning the  Hawking mass $m^\S$.
Recall that,
 \beaa
 m^\S&=&\frac{r^\S}{2} \left(1+ \frac{1}{16\pi}\int_{\S}\ka^{\S }\kab^\S\right),\\
 \ovm&=&\frac{\rg}{2} \left(1+ \frac{1}{16\pi}\int_{\ovS}\ovka \ovkab\right).
 \eeaa
We write
\beaa
 2\Big(\frac{m^\S}{r^\S} -\frac{\ovm}{\ovr}\Big) &=& \frac{1}{16\pi}\left[ \int_{\S}\left(\ka^{\S }\kab^\S    -\ka\kab \right)    +\left(\int_\S  \ka \kab- \int_{\ovS  }\ka \kab\right)   -\int_{\ovS}\left(  \ka\kab-       \ovka\ovkab \right)  \right]\\
&=& I_1+I_2+I_3.
\eeaa
In view of Proposition \ref{Proposition:compatible-deformations} we have $|r^\S-\ovr|\les\dg$ and $\big|\ga^{\S,\#} -\ovga\big|\les \dg \ovr $. Making use of Corollary \ref{corr:lemma:int-gaS-ga}  and the assumptions  {\bf A1-A3} for $\ka, \kab$ we deduce,
\beaa
\big|I_2\big|=\left|\int_\S  \ka \kab- \int_{\ovS  }\ka \kab\right|&\les&\dg r^{-1}.
\eeaa
Similarly, taking into account  the definition of $\RR:=\left\{|u-\ug|\leq \dg ,\quad |s-\sg|\leq \dg \right\}$,
\beaa
\big|I_3\big|&\les& \dg r^{-1}.
\eeaa
Finally, making also  use of   the transformation formula from the original frame $(e_4, e_3, e_\th)$ to the frame $(e_4^{\S}, e_3^{\S}, e^{\S}_\th)$ of $\S$
\beaa
\ka^{\S}\kab^{\S} &=& \left( \ka+ \ddd^{\S}f   + \err(\ka, \ka^{\S})\right)\left( \kab+ \ddd^{\S}\fb  + \err(\kab,\kab^{\S})\right)
\eeaa
and the  estimates for $(f, \fb, a=\log\la )$ we deduce,
\beaa
\left|\ka^{\S }\kab^\S    -\ka\kab \right|&\les& r^{-3} \dg. 
\eeaa
Hence,
\beaa
\Big|I_1\Big|&\les& \dg r^{-1}.
\eeaa
We infer that,
\beaa
\Big|\frac{m^\S}{r^\S} -\frac{\ovm}{\ovr}\Big|&\les& \dg r^{-1}
\eeaa
 from which the desired estimate  \eqref{eq:ExistenceGCMS-Thm-m^S}   easily follows.


\subsection{Iteration procedure for  Theorem  \ref{Theorem:ExistenceGCMS}}
\lab{section:iterationprocedureGCM}


We  solve  the coupled  system of equations \eqref{eq:DeformS-66},
\eqref{GCMS:NEAR-again:bis} and \eqref{equations-forBadModes} by an iteration argument as follows. 

Starting with   the  trivial quintet  
  \beaa
  Q^{(0)}:=(U^{(0)}, S^{(0)}, a^{(0)},  f^{(0)},  \fb^{(0)})=(0,0,0,0,0), 
  \eeaa
  corresponding to the undeformed sphere $\ovS$,  we define iteratively  the quintet 
 \beaa
 Q^{(n+1)}= (U^{(n+1)}, S^{(n+1)}, a^{(n+1)},  f^{(n+1)},  \fb^{(n+1)})
 \eeaa
  from
  \beaa
     Q^{(n)}= (U^{(n)}, S^{(n)}, a^{(n)},  f^{(n)},  \fb^{(n)})
  \eeaa
   as follows.

  \begin{enumerate}
  \item  The pair  $(U^{(n)}, S^{(n)})$ defines the deformation  sphere $\S(n)$ and the corresponding pull back map  $\#_n$ given by the map $\Psi^{(n)} :\ovS\longrightarrow \S(n)$,
  \bea
  \label{definition:Psin-iteration}
  (\ovu, \ovs, \th, \vphi)\longrightarrow (\ovu+U^{(n)}(\th), \ovs+S^{(n)}(\th), \th, \vphi).
  \eea

   By induction we assume that  the following estimates hold true:
  \bea
\label{eq:prop.GCMSequations-fixedS2induction}
\left\|(a^{(n)}, f^{(n)}, \fb^{(n)}) \right\|_{\hk_{s_{max}-2}(\S(n))} &\les& \dg,
\eea
and
   \bea
 \label{eq:ExistenceGCMS-Thm-US-induction}
 \|\pr_\th \left(U^{(n)}, S^{(n)}\right)\|_{L^\infty(\ovS)}+ \max_{0\leq s\leq s_{max}-2}  r^{-1} \|\pr_\th \left(U^{(n)}, S^{(n)}\right)\|_{\hk_s(\ovS, \ovgS)} \les  \dg.
  \eea

  \item
   We then  define the triplet    $(a ^{(n+1)}, f^{(n+1)},  \fb ^{(n+1)}) $ by solving the system on $\S(n)$ consisting 
   of the equations \eqref{system:af-fb-index-n},  \eqref{system:badmodes-index-n} and \eqref{eq:retrieve-a-indexn} below.
  \bea
  \label{system:af-fb-index-n}
  \bsplit
\left(A^{\S(n)}+ V\right)z^{(n+1)} &= E^{(n+1)},\\
E^{(n+1)}:&= \ka \ddsSn \kabc^{\S(n)} -\kab\dds\,\,\check{\ka} -\ka\dds\,\,\check{\kab}\\
&+\kab\err_4^{(n+1)}+\ka\err_5^{(n+1)} +[A^{\S(n)},\kab]f^{(n+1)}+[A^{\S(n)},\ka]\fb^{(n+1)},\\
\\
 \left(A^{\S(n)} +V\right) f^{(n+1)} &=\frac 3 4 \ka (A^{\S(n)})^{-1}(\rho z^{(n+1)}) +E^{(n+1)},\\
  E^{(n+1)}:&= -\dds\,\,\check{\ka}  -\ka  (A^{\S(n)})^{-1}\big( -\ddsSn \check{\mu}^{\S(n)} +\dds\,\,\check{\mu}\big) +\err^{(n+1)}_4,\\
 \\
  \left( A^{\S(n)} +V \right)\fb^{(n+1)} &=-\frac 3 4 \kab (A^{\S(n))})^{-1}(\rho z^{(n+1)}) +\Eb^{(n+1)},\\
   \Eb^{(n+1)}:&=\ddsSn\kabc^{\S(n)}
   -\dds\,\,\check{\kab}+\kab  (A^{\S(n)})^{-1} \left( -\ddsSn\muc^{(\S(n)}  +  \dds\,\,\check{\mu}\right)\\
   &+\err^{(n+1)}_5,\\
   \dds^{\S(n)} a ^{(n+1)} &=-\frac 3 4( A^{\S(n)})^{-1} \big(  \rho z^{(n+1)} \big) + f^{(n+1)}\omb   
    +\frac{1}{4}f^{(n+1)}\kab- \frac 14   \fb^{(n+1)} \ka\\
    & \widetilde{E}^{(n+1)},\\
  \widetilde{E}^{(n+1)}:&=   (A^{\S(n)})^{-1}\big(-\ddsSn \muc^{\S(n)} +\dds\,\,\check{\mu}\big)
   -(A^{\S(n)})^{-1}\big( \err^{(n+1)}_3\big),
  \end{split}
  \eea
  where,
  \bea
    z^{(n+1)}&=\kab f^{(n+1)} +\ka \fb^{(n+1)},
    \eea
  and the error terms,
\bea
\label{errorterms(n+1)1}
\err^{(n+1)}_1, \err^{(n+1)}_2, \err^{(n+1)}_3, \err^{(n+1)}_4, \err^{(n+1)}_5,
\eea 
 are obtained from     the error terms       $\err_1, \err_2, \err_3, \err_4, \err_5$       
   by setting $(a, f, \fb)=(a^{(n+1)}, f^{(n+1)}, \fb^{(n+1)})$ and their    derivatives  by the corresponding ones on   the sphere $\S(n)$.

   We also  set,
\bea
 \label{system:badmodes-index-n}
\bsplit
\frac{2}{r^{\S(n)} } \int_{\S(n)}  e_\th^{\S(n)} (a^{(n+1)})e^\Phi&= \frac 1 4 \int_{\S(n)}\ka^2 \,\fb^{(n+1)}  e^\Phi -\int_{\S(n)}  \left( \frac 1 4 \ka\kab +\ka\omb+3\rho\right)f^{(n+1)} e^\Phi\\
& -\int_{\S(n)}  e_\th(\ka) e^\Phi +\err^{(n+1)}_7,\\
\int_{\S(n)} e^\Phi  f ^{(n+1)}&=\La, \qquad \int_{\S(n)} e^\Phi \fb^{(n+1)}=\Lab,  
\end{split}
\eea
and,
\bea\lab{eq:retrieve-a-indexn}
\ov{a^{(n+1)}}^{\S(n)}  = \ov{\left( 1-\frac{r^{\S(n)}}{r}\right)}^{\S(n)}         - \frac{r^{\S(n)}}{2}\ov{\left( \check{\ka} + \ov{\ka}-\frac{2}{r} \right)}^{\S(n)} +\ov{\err_6^{(n+1)}}^{\S(n)},
\eea
where   $ \err_6^{(n+1)}$, $ \err_7^{(n+1)}$ are    obtained from   
  the error terms $\err_6 $,  $\err_7 $ as above in \eqref{errorterms(n+1)1},   by setting $(a, f, \fb)=(a^{(n+1)}, f^{(n+1)}, \fb^{(n+1)})$ and their    derivatives  by the corresponding ones on   the sphere $\S(n)$.

\item
In view of the induction hypothesis  \eqref{eq:ExistenceGCMS-Thm-US-induction} 
we are thus in a position to apply  Proposition \ref{prop.GCMSequations-fixedS}, more precisely its Corollary 
 \ref{prop.GCMSequations-fixedS-2}, to construct 
 a unique solution  verifying the estimate, uniformly  for $n\in \NNN$,
\bea
\label{eq:prop.GCMSequations-fixedS2}
\left\|(a^{(n+1)}, f^{(n+1)}, \fb^{(n+1)}) \right\|_{\hk_{s_{max}-2}(\S(n))} &\les& \dg.
\eea

\item
  We use  the new pair     $( f^{(n+1)}, \fb^{(n+1)})$   to solve the equations on $\ovS$,
      \bea
      \label{mainiteration-definUS(n+1)}
  \bsplit
 \vsi^{\#_n}   \pr_\th U^{(1+n)} &=  (\ga^{(n)}  )^{1/2} ( f^{(1+n)})^{\#_{n}}\left( 1 +\frac 1 4\Big(f^{(1+n)} \fb^{(1+n)}\Big)^{\#_n}\right),\\
     \pr_\th S^{(1+n)}-\frac 1 2  \vsi^{\#_n}  \Omb^{\#_{n}}  \pr_\th U^{(1+n)}&=   \frac 1 2 (\ga^{(n)}  )^{1/2}(\fb^{(1+n)})^{\#_{n}},
     \\
   \ga^{(n)} & =\ga^{\#_n}+\big(\vsi^{\#_n}\big)^2  \left(\Omb+\frac 1 4\underline{b}^2  \ga  \right)^{\#_n}\, ( \pr_\th U^{(n)} )^2 \\
     &-2  \vsi^{\#_n}   \pr_\th U ^{(n)} 
     \pr_\th  S^{(n)}-\Big(\ga\vsi  \underline{b}\Big)^{\#_n} \pr_\th U^{(n)},\\
     U^{(1+n)}(0)&= S^{(1+n)}(0)=0,
     \end{split}
  \eea

  where, we repeat, the pull back $\#_{n}$ is defined with respect to  the map  
  $$\Psi^{(n)}(\ug,  \sg, \th)= (\ug+U^{(n)}(\th),  \sg+S^{(n)}(\th), \th),$$ 
  and
     $$ \ga^{(n)}: =  \ga^{\S(n), \#_{n}}.$$  
       The equation \eqref{mainiteration-definUS(n+1)} admits a unique solution $U^{(1+n)}, S^{(1+n)} $, according to the proposition below.
        The new  pair  $( U^{(n+1)} ,  S^{(n+1)}  )$ defines  the  new polarized sphere $\S(n+1)$ and we can proceed with the next step  of the iteration. 
\end{enumerate}


\subsection{Boundedness of the iterates}


  \begin{proposition}
  \label{Prop:estimatesUn+1Sn+1}
  The equation \eqref{mainiteration-definUS(n+1)} admits a unique solution $U^{(1+n)}, S^{(1+n)} $   
   verifying the estimate, 
   \beaa
    \|\pr_\th \left(U^{(n+1)}, S^{(n+1)}\right)\|_{L^\infty(\ovS)}+ r^{-1} \|\pr_\th \left(U^{(n+1)}, S^{(n+1)}\right)\|_{\hk_{s_{max}-2}(\ovS, \ovgS)}  &\les& \dg
    \eeaa
      uniformly for all  $n\in\NNN$.
   \end{proposition}
    \begin{proof}    
  The existence and uniqueness part of the proposition  is an immediate consequence of the standard   results for ODE's.
     
      To prove the desired estimate, we use the equations for $(U^{(1+n)}, S^{(1+n)})$ and
       infer,   for $s= s_{max}-2$,
      \beaa
 \|\pr_\th U^{(n+1)}\|_{\hk_{s}(\ovS, \ovgS)}  &\les& \Bigg\|(\ga^{(n)}  )^{1/2} \big(\vsi^{\#_n}\big)^{-1}( f^{(1+n)})^{\#_{n}}\left( 1 +\frac 1 4\Big(f^{(1+n)} \fb^{(1+n)}\Big)^{\#_n}\right)\Bigg\|_{\hk_{s}(\ovS, \ovgS)}.
     \eeaa
      Together with the  non sharp product estimate          on $(\ovS, \ovgS)$, see Lemma \ref{prop:sobolevandproductesitmatesonS-GCMS},  we infer that, for $s= s_{max}-2$,
      \beaa
 \|\pr_\th U^{(n+1)}\|_{\hk_{s}(\ovS, \ovgS)}   &\les&r^{-1} \left\|(f^{(1+n)})^{\#_n}, (\fb^{(1+n)})^{\#_n}\right\|_{\hk_{s_{max}-2}(\ovS, \ovgS)}\left\|\big(\vsi^{\#_n}\big)^{-1}(\ga^{(n)}  )^{1/2}\right\|_{\hk_{s}(\ovS, \ovgS)}\\&&\times\left(1+\left\|(f^{(1+n)})^{\#_n}, (\fb^{(1+n)})^{\#_n}\right\|_{\hk_{s}(\ovS, \ovgS)}^2\right).
 \eeaa
      In view of Lemma \ref{lemma:int-gaS-ga:highersobolevregularity}, Corollary \ref{cor:int-gaS-ga:highersobolevregularity}, and \eqref{eq:prop.GCMSequations-fixedS2}, we deduce
 \beaa
  \|\pr_\th U^{(n+1)}\|_{\hk_{s}(\ovS, \ovgS)}   &\les& \dg r^{-1} \left\| \big(\vsi^{\#_n}\big)^{-1}(\ga^{(n)}  )^{1/2}\right\|_{\hk_{s}(\ovS, \ovgS)}.
 \eeaa  
     We recall that,
     \beaa
     \ga^{(n)} & =\ga^{\#_n}+\big(\vsi^{\#_n}\big)^2  \left(\Omb+\frac 1 4\underline{b}^2  \ga  \right)^{\#_n}\, ( \pr_\th U^{(n)} )^2 -2  \vsi^{\#_n}   \pr_\th U ^{(n)} -\Big(\ga\vsi  \underline{b}\Big)^{\#_n} \pr_\th U^{(n)}.
     \eeaa
     In view of our assumptions  on the Ricci coefficients  and   the non-sharp product estimates  of Lemma \ref{prop:sobolevandproductesitmatesonS-GCMS}
       \beaa
   \left\|\left( \vsi \big(\Omb+\frac 1 4\underline{b}^2  \ga \big) \right)^{\#_n}\right\|_{\hk_{s_{max}-2}(\ovS, \ovgS)} +  \left\|\Big(\ga \underline{b}\Big)^{\#_n}\right\|_{\hk_{s_{max}-2}(\ovS, \ovgS)}      &\les&\epg r 
     \eeaa 
  we deduce, 
\beaa
\left\| \big(\vsi^{\#_n}\big)^{-1} \ga^{(n)}\right\|_{\hk_{s_{max}-2}(\ovS, \ovgS)}  & \les & \left\|\ga^{\#_n}\right\|_{\hk_{s_{max}-2}(\ovS, \ovgS)}+ \dg r^2.
   \eeaa 
  Together with Lemma \ref{lemma:int-gaS-ga:highersobolevregularity} and Corollary \ref{cor:int-gaS-ga:highersobolevregularity}, we deduce 
\beaa
\left\|\big(\vsi^{\#_n}\big)^{-1} \ga^{(n)}\right\|_{\hk_{s_{max}-2}(\ovS, \ovgS)}  & \les &r^2
  \eeaa  
  and therefore,
  \beaa
 \|\pr_\th U^{(n+1)}\|_{\hk_{s_{max}-2}(\ovS, \ovgS)} \les \dg r.
 \eeaa
 Proceeding in the same manner with equation
  $$   \pr_\th S^{(1+n)}-\frac 1 2  \vsi^{\#_n}  \Omb^{\#_{n}}  \pr_\th U^{(1+n)}=   \frac 1 2 (\ga^{(n)}  )^{1/2}(\fb^{(1+n)})^{\#_{n}}$$
 we infer that,
  \beaa
 r^{-1} \|\pr_\th U^{(n+1)}, \pr_\th S^{(n+1)}\|_{\hk_{s_{max}-2}(\ovS, \ovgS)} &\les&\dg.
 \eeaa  
   This, together with the Sobolev inequality,   concludes the proof of Proposition \ref{Prop:estimatesUn+1Sn+1}.
    \end{proof}

 
 \subsection{Convergence of the Iterates}
 \lab{section:convergenceofiterates-GCM}
 
 
  To finish the proof of Theorem  \ref{Theorem:ExistenceGCMS}  it  remains to prove convergence of the iterates.
 
{\bf Step 1.} In order to prove the convergence of the iterative scheme, we introduce the following quintet $P^{(n)}$
 \beaa
 P^{(0)}=(0,0,0,0,0), \qquad P^{(n)}=\Big(U^{(n)}, S^{(n)}, (a^{(n)})^{\#_{n-1}},  (f^{(n)})^{\#_{n-1}},  (\fb^{(n)})^{\#_{n-1}}\Big),\ \ n\geq 1.
\eeaa
Since $(a^{(n)}, f^{(n)}, \fb^{(n)})$ are defined on $\S(n-1)$, their respective pullback by $\Psi^{(n-1)}$ is defined on $\ovS$ so that $P^{(n)}$ is a quintet of functions on $\ovS$ for any $n$ and we may introduce the following norms to compare the elements of the  sequence 
\bea
\label{quintet-norm}
\bsplit
\| P^{(n)}\|_k: &= \|\pr_\th \left(U^{(n)}, S^{(n)} \right)\|_{L^\infty(\ovS)}  +r^{-1} \|\pr_\th \left(U^{(n)}, S^{(n)}\right) \|_{\hk_{k-2}(\ovS)}  \\
&+ \left\|\left((a^{(n)})^{\#_{n-1}},  (f^{(n)})^{\#_{n-1}},  (\fb^{(n)})^{\#_{n-1}}\right)\right\|_{\hk_{k-2}(\ovS)}.
\end{split}
\eea
  Here are the  substeps needed to implement a convergence argument.
\begin{enumerate}
\item   The quintets  $P^{(n)}$ are bounded  with respect to the norm \eqref{quintet-norm} for the choice $k=s_{max}$.
\item   The quintets  $P^{(n)} $  are contractive with respect to the norm \eqref{quintet-norm} for the choice $k=3$.
\end{enumerate}

The precise statements   are given in the following   propositions.
 \begin{proposition}
 \label{Prop:BondsforQn}
 We have, uniformly for all $n\in \NNN$,
 \beaa
 \| P^{(n)}\|_{s_{max}} &\les& \dg.
 \eeaa
 \end{proposition}
 
\begin{proof} 
The proof is an immediate consequence of Propositions \ref{prop.GCMSequations-fixedS}, \ref{Prop:estimatesUn+1Sn+1}
 and  the estimate,
\bea
\left\|( \Psi^{(n-1)})^\# \left ( f^{(n)}, \fb^{(n)}, a^{(n)} \right)\right\|_{\hk_{s_{max}-2}(\ovS)} \les\left\| \left ( f^{(n)}, \fb^{(n)}, a^{(n)} \right)\right\|_{\hk_{s_{max}-2}(\S(n-1))}
\eea
 which is a consequence of Lemma \ref{lemma:int-gaS-ga:highersobolevregularity}.
\end{proof} 
 
\begin{proposition}
\label{Proposition-contraction}
  We have, uniformly for all $n\in\NNN$,   the contraction estimate,
\beaa
\| P^{(n+1)}- P^{(n)} \|_3 \les \dg \,  \| P^{(n)}- P^{(n-1)} \|_3.
\eeaa
\end{proposition}

The proof of Proposition \ref{Proposition-contraction} is postponed to section \ref{sec:proofofProposition-contraction}.

{\bf Step 2.} In view of Proposition \ref{Proposition-contraction}, we have 
\beaa
\| P^{(n+1)}- P^{(n)} \|_3 \les (\dg)^n\, \| P^{(1)}- P^{(0)} \|_3
\eeaa
which in view of Proposition \ref{Prop:BondsforQn} yields
\beaa
\| P^{(n+1)}- P^{(n)} \|_3 \les (\dg)^{n+1}.
\eeaa
Together with a simple interpolation argument on $\ovS$ and Proposition \ref{Prop:BondsforQn}, we infer
\beaa
\| P^{(n+1)}- P^{(n)} \|_k \les( \dg)^{1+\left(\frac{s_{max}-k}{s_{max}-3}\right)n}, \quad 3\leq k\leq s_{max}. 
\eeaa
We infer the existence of a quintet of functions $P^{(\infty)}$ on $\ovS$ such that 
\bea
 \| P^{(\infty)}\|_{s_{max}}&\les& \dg
\eea
and
\bea\lab{eq:PnconvergestoPinftyinhknormasngoestoinfty}
\lim_{n\to +\infty}\|P^{(n)}-P^{(\infty)}\|_{s_{max}-1}=0.
\eea
Also, we have
\beaa
P^{(\infty)}=\Big(U^{(\infty)}, S^{(\infty)}, a^{(\infty)}_0, f^{(\infty)}_0, \fb^{(\infty)}_0\Big), 
\eeaa
where all function are defined on $\ovS$. The functions  $(U^{(\infty)}, S^{(\infty)})$ defines a sphere $\S(\infty)$ and we introduce the map 
\beaa
\Psi^{(\infty)}(\ug, \sg, \th, \vphi) &=& \Big(\ug+U^{(\infty)}(\th), \sg+S^{(\infty)}(\th), \th, \vphi\Big)
\eeaa
so that $\Psi^{(\infty)}$ is a map from $\ovS$ to $\S(\infty)$. Then, let
\beaa
a^{(\infty)}=a^{(\infty)}_0\circ(\Psi^{(\infty)})^{-1},\quad  f^{(\infty)}=f^{(\infty)}_0\circ(\Psi^{(\infty)})^{-1}, \quad \fb^{(\infty)}=\fb^{(\infty)}_0\circ(\Psi^{(\infty)})^{-1}
\eeaa 
so that $a^{(\infty)}, f^{(\infty)}, \fb^{(\infty)}$ are defined on $\S(\infty)$  and 
\beaa
a^{(\infty)}_0= (a^{(\infty)})^{\#_{\infty}}, \quad f^{(\infty)}_0= (f^{(\infty)})^{\#_{\infty}},\quad \fb^{(\infty)}_0= (\fb^{(\infty)})^{\#_{\infty}}.
\eeaa
From these definitions, the above control of $P^{(\infty)}$ and  Lemma \ref{lemma:int-gaS-ga:highersobolevregularity}, we infer
\beaa
r^{-1}\|(U^{(\infty)}, S^{(\infty)})\|_{\hk_{s_{max}-1}(\ovS)}+\|(a^{(\infty)}, f^{(\infty)}, \fb^{(\infty)})\|_{\hk_{s_{max}-2}(\S(\infty))} &\les& \dg.
\eeaa
In particular, using the Sobolev embedding on $\ovS$, we have $\|({U^{(\infty)}}', {S^{(\infty)}}')\|_{L^\infty(\ovS)} \les \dg$, and hence, in view of Corollary \ref{prop.GCMSequations-fixedS-2}, we deduce
 \beaa
\|(a^{(\infty)}, f^{(\infty)}, \fb^{(\infty)})\|_{\hk_{s_{max}+1}(\S(\infty))} &\les& \dg.
\eeaa
This estimate allows to argue as in Proposition \ref{Prop:estimatesUn+1Sn+1} with $(U^{(n+1)}, S^{(n+1)})$ replaced by $(U^{(\infty)}, S^{(\infty)})$ and $s_{max}-2$ replaced by $s_{max}$. We finally obtain 
\bea\lab{eq:controlofUSaffbforthelimitoftheiterationscheme}
r^{-1}\|({U^{(\infty)}}', {S^{(\infty)}}')\|_{\hk_{s_{max}}(\ovS)}+\|(a^{(\infty)}, f^{(\infty)}, \fb^{(\infty)})\|_{\hk_{s_{max}+1}(\S(\infty))} &\les& \dg.
\eea

{\bf Step 3.} We proceed to  control the area radius $r^{\S(\infty)}$ and the Hawking mass $m^{\S(\infty)}$ of the sphere $\S(\infty)$. First, note from \eqref{eq:controlofUSaffbforthelimitoftheiterationscheme} and the Sobolev embedding on $\ovS$ that we have
\bea
\|(U^{(\infty)}, S^{(\infty)})\|_{L^\infty(\ovS)} &\les& \dg.
\eea
Together with Lemma \ref{lemma:int-gaS-ga}, we infer that\footnote{Here, we also use the fact that, on $S^{(\infty)}$, we have
 \beaa
 |r-\rg|\les  \|(U^{(\infty)}, S^{(\infty)})\|_{L^\infty(\ovS)}\les\dg.
 \eeaa}
\bea\lab{eq:controlofarearadiusofthelimitoftheiterationscheme}
\left|\frac{r^{\S(\infty)}}{r}-1\right| &\les& \dg.
\eea
Next, we denote by $\Ga^{\S(\infty)}$ the connection coefficients of $\S(\infty)$. We have in view of the transformation formula from the original frame $(e_4, e_3, e_\th)$ to the frame $(e_4^{\S(\infty)}, e_3^{\S(\infty)}, e^{\S(\infty)}_\th)$ of $\S(\infty)$
\beaa
\ka^{\S(\infty)}\kab^{\S(\infty)} &=& \left( \ka+ \ddd^{\S(\infty)}f^{(\infty)}   + \err(\ka, \ka^{\S(\infty)})\right)\left( \kab+ \ddd^{\S(\infty)}\fb^{(\infty)}   + \err(\kab,\kab^{\S(\infty)})\right).
\eeaa
Together with the estimate \eqref{eq:controlofUSaffbforthelimitoftheiterationscheme} for $f^{(\infty)}$ and $\fb^{(\infty)}$ and the assumptions {\bf A1-A3}  for $\Gac$ corresponding to the original frame
 $(e_4, e_3, e_\th)$, we infer
\beaa
\left|\ka^{\S(\infty)}\kab^{\S(\infty)} -\ka\kab\right| &\les& \dg r^{-3}.
\eeaa  
Recall that (see   \eqref{eq:GCM-improved estimate})  
\beaa
\Big|\ka-\frac 2 r\Big| \les  \dg r^{-2}, \qquad  \Big|\ov{\kab}+
\frac{2\left(1-\frac{2 m}{ r }\right)}{r} \Big| \les \dg. 
\eeaa
Thus, since $\kab=\ov{\kab}+\kabc$,
\beaa
\ka\kab&=&- \frac{4\left(1-\frac{2m}{r}\right)}{r^2}+\frac 2 r \kabc +O(\dg)  r^{-2}.
\eeaa
We deduce,
\beaa
\left|\ka^{\S(\infty)}\kab^{\S(\infty)} +\frac{4\left(1-\frac{2m }{r }\right)}{r^2} -\frac 2 r \kabc\right| &\les& \dg r^{-3}.
\eeaa
Thus, in view of  \eqref{eq:controlofarearadiusofthelimitoftheiterationscheme}, 
\beaa
\int_{\S(\infty)}\ka^{\S(\infty)}\kab^{\S(\infty)}&=& -\int_{\S(\infty)} \frac{4\left(1-\frac{2 m }{r }\right)}{r^2}+O(\dg) r^{-1}. 
\eeaa
Making use of the definition of the Hawking mass  $m^{\S(\infty)} = \frac{r^{\S(\infty)}}{2}\left(1+ \frac{1}{16\pi}\int_{\S(\infty)}\ka^{\S(\infty)}\kab^{\S(\infty)}\right)$ we easily deduce\footnote{See also subsection \ref{subsection:estimate-Hawkingmass}.}
\bea
\lab{eq:controlofHawkingmassofthelimitoftheiterationscheme}
\Big|m^{\S(\infty)}-m\Big|&\les&  \dg. 
\eea

{\bf Step 4.}  We  make use of Lemma \ref{Lemma:GCMStransversality1}  to   extend $(a^\infty, f^\infty, \fb^\infty)$  as well as the frame  $\big (e_3^{\S(\infty)},\, e_4^{\S(\infty)},   \, e_\th^{\S(\infty)} \big)$ in a small neighborhood of $\S(\infty)$ such that  we have,
\bea
\label{transversalconditionsonSinfty}
\bsplit
\xi^{\S(\infty)}&=0, \qquad \om^{\S(\infty)}=0, \qquad \etab^{\S(\infty)}+\ze^{\S(\infty)}=0,
\end{split}
\eea
and then provide   estimates for  the corresponding Ricci coefficients and  curvature components   
$\Gac^{\S(\infty)}$, $\Rc^{\S(\infty)}$.  More precisely we make use of the assumption {\bf A1}, the estimates in \eqref{eq:controlofUSaffbforthelimitoftheiterationscheme} 
 for  $(a^\infty, f^\infty, \fb^\infty)$, and the transformation formulae  to derive the  desired  estimates \eqref{Estimate-GacS-onS} for $s_{max}$ derivative of the Ricci coefficients and  curvature components   of $\S(\infty)$.

{\bf Step 5.} Thanks to \eqref{eq:PnconvergestoPinftyinhknormasngoestoinfty}, we  can  pass to the limit in \eqref{system:af-fb-index-n} \eqref{eq:retrieve-a-indexn}.
 We deduce that   all equations recorded in Proposition  \ref{proposition-GCMS-system} hold true  and  thus that all  the desired GCMS hold true.


\section{Proof of Proposition \ref{prop.GCMSequations-fixedS}}


Since all estimates  below take place on  $\S$,  we  simplify  our  notation  and  denote 
  the norms  $\|\c \|_{\hk_k(\S)}$  simply by $\|\c \|_k$    in what follows\footnote{To remind the reader   about our convention we shall in fact alternately  use both notations.}.   In the particular case $k=0$ we also write  
   $\|\c \|_0= \|\c \|_\S$.  The  sup norms $\|\c\|_{\hk^\infty_k(\S)} $, though rarely needed,  will be denoted  by $ |\c|_k$. Since $r$ and $r^\S$  are comparable  we will freely   choose one or the other   throughout the proof. 
   
    We  introduce  the notation (recall that $a=\log \la$)
   \bea
   \label{eq:proofGCMapriori-ntation}
   \bsplit
   \|F\|_k&=  \| a\|_{k} +\|f\|_k+\|\fb\|_k,\\
   I(f, \fb)&=\left|  \int_\S f e^\Phi \right|+\left|  \int_\S \fb e^\Phi \right|,\\
    I(a, f, \fb)&= I(f, \fb)+r^\S \left|  \int_\S( e_\th a ) e^\Phi \right|.
   \end{split}
   \eea
      We assume that a solution exists 
  verifying  the   auxiliary  bootstrap assumption\footnote{This is not  really needed, we only use it to simplify  the exposition.}
\bea
\label{BA-affb}
\bsplit
\| (a, f, \fb) \|_{s_{max}+1}&\les\sqrt{ \epg}
\end{split}
\eea
based on which we will prove the  stronger estimate,
\bea
\label{Estimate:improved-BA-affb}
\|(a, f,\, \fb) \|_{s_{max}+1} \les \dg.
\eea

{\bf Step 1.} 
 We write the first equation in \eqref{GCMS:NEAR-again:bis} in the form
\beaa
\left(A^S+V\right)z &=& \frac 2 r   \left( \ddsS \kabc^\S - \dds\,\kabc\right)  -\frac{2\Up}{r}  \dds\,\,\kac +\err(z),\\
\err(z)&:=& \kab\err_4 +\ka\err_5 +[A^S,\kab]f+[A^S,\ka]\fb\\
&+&\left(\ka-\frac 2 r \right)  \left( \ddsS \kabc^\S - \dds\,\kabc\right)+\left(\kab+\frac{2\Up}{r}\right)   \dds\,\,\kac.
\eeaa
Thus,
\beaa
\left\|\left(A^S+V\right) z \right\|_{k }&\les& r^{-1} \left\| \ddsS \kabc^\S - \dds\,\kabc\right\|_{k}+ r^{-1}  \left\| \dds\, \kac\right\|_{k}+\left\| \err(z)\right\|_{k}.
\eeaa
  
In view of Lemma \ref{Lemma-errorestimates-S}  and the auxiliary bootstrap assumption \eqref{BA-affb}  we have\footnote{We also  assume tacitly, throughout the  estimates  below,  that $k\ge  s_{max}/2+2$ so that we have   product estimates in $\| \c\|_k$.}, for $k\le s_{max}-1$,
\beaa
 \| \err_4, \err_5 \|_{\hk_{k}(\S)}&\les&  r^{-2}  \,  \Big\|  f, \fb, a \Big\|_{\hk_{k+2}(\S)} \left( \epg +  \Big\|  f, \fb, a \Big\|_{\hk_{k+2}(\S)}  \right) \\
 & \les  & r^{-2}\,\sqrt{ \epg} \,\ \|F\|_{k+2}. 
\eeaa
Also,
\beaa
 \left\|[A^S,\kab]f+[A^S,\ka]\fb\right\|_{\hk_k(\S)}&\les& r^{-3}\sqrt{ \epg}\, \left\| (f, \fb)\right\|_{\hk_{k+1}(\S)}.
 \eeaa 
 In view of  \eqref{eq:GCM-improved estimate} we have 
    $\left\|\kac, \ov{\ka}-\frac 2 r  \right\|_{\hk_k(\S)}\les \dg r^{-1}$
     and    $\|\dds\,\, \kac \|_{\hk_{k-1}(\S)} \les \dg r^{-2}$
  for all $k\le s_{max}$, and thus
 \beaa
\left\| \left(\ka-\frac 2 r \right)  \left( \ddsS \kabc^\S - \dds\,\kabc\right)\right\|_{\hk_k(\S)}&\les& \dg r^{-2} \left\| \ddsS \kabc^\S - \dds\,\kabc\right\|_{\hk_k(\S)},\\
\left\|\left(\kab+\frac{2\Up}{r} \right) \dds\,\,\kac \right\|_{\hk_k(\S)}&\les&\dg \epg r^{-4},
 \eeaa
 i.e.,
 \beaa
 \left\| \err(z)\right\|_{k}&\les&  r^{-3}\,\sqrt{ \epg} \,\ \|F\|_{k+2} + \dg r^{-2} \left\| \ddsS \kabc^\S - \dds\,\kabc\right\|_{\hk_k(\S)}+\dg \epg r^{-4}.
\eeaa
Hence,
 \bea
 \label{eq:ProofGCMapriori1}
 \left\|\left(A^S+V\right)z\right\|_{k}&\les&  r^{-1} \left\| \ddsS \kabc^\S - \dds\,\kabc\right\|_{k}+ r^{-3} \dg+r^{-3}  \sqrt{\epg}\,  \,  \big\|  F \big\|_{k+2}.
 \eea
  To estimate $ \left\| \ddsS \kabc^\S - \dds\,\kabc\right\|_{k}$ we proceed as follows.
  Note, using the GCMS conditions, that
\beaa
\ddsS_2\left( \ddsS_1\kabc^\S-\dds_1 \kabc\right)&=&-\ddsS_2 \dds_1 \kabc=-\dds_2 \dds_1 \kabc -(\ddsS_2 -\dds_2)\dds_1 \kabc.
\eeaa
According to the first inequality in \eqref{eq:GCM-improved estimate}
$\|\dds_2\dds_1\kab\|_{\hk_k(\S)} \les \dg  r^{-3}$. Hence,
\beaa
\left\| \ddsS_2\left( \ddsS_1\kabc^\S-\dds_1 \kabc\right)\right\|_k&\les&  \|\dds_2 \dds_1 \kabc\|_{k} + \|(\ddsS_2 -\dds_2)\dds_1 \kabc\|_{k}\\
&\les&  \dg  r^{-3}+ \|(\ddsS_2 -\dds_2)\dds_1 \kabc\|_{k}.
\eeaa
Recalling the transformation formulas \eqref{SSMe:general.composite:repeatforproofGCM}
we have schematically,
\beaa
\ddsS_2 - \left(1+\frac 1 2   f \fb\right) \dds_2=  \frac 1 2  \fb e_4+\frac 1 2 f\left(1+ \frac 1 4 f \fb\right) e_3+\lot
\eeaa
Therefore,
\beaa
\|(\ddsS_2 -\dds_2)\dds_1 \kabc\|_{k}&\les&r^{-3} \sqrt{ \epg}\,  \| (f, \fb)\|_k
\eeaa
and thus setting
\bea
B:= \ddsS_1\kabc^\S-\dds_1 \kabc,
\eea
we have
\bea
\label{eq:ProofGCMapriori2}
\left\| \ddsS_2 B\right\|_k&\les&r^{-3}\left( \dg  +\sqrt{\epg}\, \| (f, \fb)\|_k\right).
\eea
According  to    the elliptic estimate  given by Lemma \ref{prop:2D-hodge-reduced-GCM} we have,
\beaa
\|B\|_k&\les& r \|\dds_2 B\|_k+ r^{-2} \left| \int_\S B  e^\Phi \right|
\eeaa
i.e.,
\bea
\label{eq:ProofGCMapriori3}
\|B\|_k&\les&r^{-2} \left| \int_\S B  e^\Phi \right|+r^{-2}\left( \dg  +\sqrt{\epg}\,  \| (f, \fb)\|_k\right).
\eea
To   estimate the  $\ell=1$ mode of $B$, i.e. its projection on the kernel of $\dds_2$, we recall  the transformation formula, see Lemma \ref{lemma:transfethka'-ethka},
\beaa
 e_\th'(\kab' )     &= &e_\th \kab +e_\th'  \ddd_1'  \fb  -\kab e_\th' a -\frac 1 4 \kab ( f\kab +\fb \ka)-\kab\, \omb  f +\fb \rho+\err(e_\th'\kab', e_\th\kab)
\eeaa
i.e., 
\beaa
B&=&  \ddsS_1\dddS_1 f +\kab \dddS_1 f  +\left(\frac 1 4 \kab^2  -\kab\, \omb\right ) f +\left(\frac 1 4  \ka\kab-\rho \right)\fb
 -\err(e_\th'\kab', e_\th\kab)\\
 &=&A^\S f +\kab \dddS_1 f  +   c^\S_1  f +c^\S_2 \fb+\err(B)
\eeaa
where,
\beaa
c^\S_1&=&\left(\frac{\Up^\S}{r^\S}\right)^2-\frac{2m^\S}{(r^\S)^3}, \qquad c^\S_2 =-\frac{\Up^\S}{(r^\S)^2}+\frac{2m^\S}{(r^\S)^3}
\eeaa
are constants on $S$ verifying
\beaa
\Big|(c^\S_1, c^\S_2)\Big|&\les& r^{-2}.
\eeaa
We also denote,
\beaa
c_1&=&\ov{\left(\frac 1 4 \kab^2  -\kab\, \omb\right )}=\left(\frac{\Up}{r}\right)^2-\frac{2m}{ r^3}\qquad 
c_2=\ov{\left(\frac 1 4  \ka\kab-\rho \right)}=-\frac{\Up}{r^2}+\frac{2m}{ r^3}.
\eeaa
The error term $\err(B)$ is given by
\beaa
\err(B)&=& -\err(e_\th'\kab', e_\th\kab)+ f\left[\left(\frac 1 4 \kab^2  -\kab\, \omb\right )-  c_1\right]+\fb \left[\left(\frac 1 4  \ka\kab-\rho \right)- c_2 \right].
\eeaa
We deduce,
 \bea
 \label{eq:ProofGCMapriori4}
\left|  \int_\S B e^\Phi\right| &\les& \left|\int_\S A^\S f e^\Phi \right|+ r^{-2}\left( \left|  \int_\S f e^\Phi\right| +\left|  \int_\S \fb e^\Phi\right| \right)+\left|\int_\S \err(B) e^\Phi\right|.
 \eea
 We can easily check, using  the assumptions of the theorem,
 \beaa
\left| \frac 1 4 \kab^2  -\kab\, \omb -\overline{\left(\frac 1 4 \kab^2  -\kab\, \omb\right )}\right|& \les&\epg  r^{-3},\\
\left|  \frac 1 4  \ka\kab-\rho -\overline{\frac 1 4  \ka\kab-\rho} \right| &\les& \epg r^{-3},\\
\left|c_1^\S-c_1\Big| +\Big|c_2^\S-c_2\right|&\les& \epg r^{-2}.
 \eeaa
 For the last inequality we note that
 \beaa
 \Big|\frac{1}{(r^\S)^2}-\frac{1}{r^2}\Big|&\les&( r^\S)^{-2} \Big|\frac{r^\S}{r}-1\Big| \Big|\frac{r^\S}{r}+1\Big|\les r^{-2}\,\epg.
 \eeaa
 Thus,  by the estimate\footnote{Note that $\err(e_\th'\kab', e_\th\kab)=\err_2+\lot$ from the definition of $\err_2$.}  for $\err_2$ in  Lemma  \ref{Lemma-errorestimates-S},
 \beaa
  \left|\int_\S \err(B) e^\Phi\right|&\les& r^2 \|\err(B)\|_\S\les  r^2 \|\err_2\|_\S + \| (f, \fb)\|_\S\les   \sqrt{\epg}\, \| F\|_2. 
 \eeaa
 Consequently,
   \beaa
\left|  \int_\S B e^\Phi\right| \les \left|\int_\S A^\S f e^\Phi \right|+ r^{-2}\left( \left|  \int_\S f e^\Phi\right| +\left|  \int_\S \fb e^\Phi\right| \right)+  \sqrt{\epg}\, \| F\|_2. 
 \eeaa
 We can  evaluate  the integral $\int_\S A^\S f e^\Phi $ by  using the identity $A^\S= \dddS_2 \ddsS_2+K$.
Thus,
\beaa
\int_\S A^\S f e^\Phi &=& \int_\S  \dddS_2 \ddsS_2  f e^\Phi+ \int_\S K f e^\Phi\\
&=& \int_\S   \ddsS_2  f  \ddsS_2(e^\Phi)+ \int_\S K f e^\Phi=\int_\S K f e^\Phi\\
&=&( r^\S)^{-2}\int_\S  f e^\Phi + \int_\S\left(  \left(K-\frac{1}{r^2} \right) +\frac{1}{r^2}-\frac{1}{(r^\S)^2} \right) f e^\Phi.
\eeaa
Hence,
\beaa
\left| \int_\S A^\S f e^\Phi \right| &\les & r^{-2} \left| \int_\S f e^\Phi \right| + \sqrt{\epg}\, \|f\|_0.
\eeaa
Therefore,
\bea
  \label{eq:ProofGCMapriori5}
\left|  \int_\S B e^\Phi\right| \les r^{-2}\left( \left|  \int_\S f e^\Phi\right| +\left|  \int_\S \fb e^\Phi\right| \right)  +\sqrt{\epg}\,  \| F\|_2. 
\eea
 Recalling \eqref{eq:ProofGCMapriori3},
 \beaa
\left\|  B\right\|_{k} &\les&r^{-2} \left| \int_\S B  e^\Phi \right|+r^{-2}\left( \dg  +\sqrt{\epg}\, \| F\|_k\right)\\
&\les& r^{-4}\left( \left|  \int_\S f e^\Phi\right| +\left|  \int_\S \fb e^\Phi\right| \right)+ r^{-2} \left( \dg  +\sqrt{\epg}\, \| F\|_k \right).
\eeaa
Returning  to \eqref{eq:ProofGCMapriori1} and recalling the definition of $B=  \ddsS \kabc^\S - \dds\,\kabc$,
 \beaa
 \left\|\left(A^S+V\right)z\right\|_{k}&\les&  r^{-1} \left\| B\right\|_{k}+ r^{-3} \dg+r^{-3} \, \sqrt{\epg}\,  \,  \Big\|  F \Big\|_{k+2}\\
&\les&  r^{-5}\left( \left|  \int_\S f e^\Phi\right| +\left|  \int_\S \fb e^\Phi\right| \right)+r^{-3}  \left( \dg  +\sqrt{\epg}\, \| F\|_{k+2} \right)
 \eeaa
or, recalling the definition of $I(f, \fb) $,
\bea
 \label{eq:ProofGCMapriori6}
\left\|\left(A^S+V\right)z\right\|_{k}&\les& r^{-5} I(f, \fb)+r^{-3}  \left( \dg  +\sqrt{\epg}\, \| F\|_{k+2} \right).
\eea
To estimate $z$ we recall that,
\beaa
A^\S +V&=& \dddS_2 \ddsS_2 - 3 \rho +\frac 1 2 \vth\vthb.
\eeaa
Thus,
\beaa
\Big((A^\S +V) z , z\Big)_\S&=&\|\ddsS_s z\|_\S^2+ 
\left( \frac{6m}{r^3} + O(r^{-3}\sqrt{\epg}\,) \right)\| z\|_\S.
\eeaa
We deduce,
\beaa
\|\ddsS_s z\|_\S^2&\les& \Big((A^\S +V) z , z\Big)_\S\les \|(A^\S +V) z\|_\S \|z\|_\S.
\eeaa
We make use of the  elliptic estimate, see Lemma \ref{prop:2D-hodge-reduced-GCM},
\beaa
\|z\|_\S^2 &\les& r^2 \|\ddsS_s z\|_\S^2+ r^{-4}\left|\int_\S z e^\Phi \right|^2\\
&\les & r^2 \|(A^\S +V) z\|_\S \|z\|_\S + r^{-4}\left|\int_\S z e^\Phi \right|^2
\eeaa
from which  we infer that
\bea
 \label{eq:ProofGCMapriori7}
\|z\|_\S  &\les&  r^2 \|(A^\S +V) z\|_\S+ r^{-2}\left|\int_\S z e^\Phi \right|.
\eea

Recalling the definition of  $z$,
\beaa
z&=& \ka \fb+\kab f= \frac{2}{r} \fb -\frac{2\Up}{r} f  +   \left(\ka-\frac{2}{r}\right) \fb +\left(\kab+\frac{2\Up}{r} \right) f \\
  &=&\frac{2}{r^\S} \fb -\frac{2\Up^\S}{r^\S} f -\left(\frac{2}{r^\S} -\frac{2}{r} \right)\fb +\left(\frac{2\Up^\S}{r^\S} -
  \frac{2\Up}{r} \right)f+   \left(\ka-\frac{2}{r}\right) \fb +\left(\kab+\frac{2\Up}{r} \right) f,
  \eeaa
  and making use of our assumption  (in particular $\Big|\frac{r^\S}{r}-1\Big|\les  r^{-1} \epg$) we deduce,
  \beaa
\left|  \int_\S z e^\Phi- \frac{2}{r^\S}\int_\S \fb e^\Phi +\frac{2\Up^\S}{r^\S}\int_\S f e^\Phi\right|&\les& \sqrt{\epg}\, r \left(\|f\|_0+\|\fb\|_0 \right). 
  \eeaa
  Hence,
\beaa
\left|\int_\S z e^\Phi\right|&\les&r^{-1} I(f, \fb)+ \sqrt{\epg}\,  r \| F\|_0
\eeaa
and therefore in view of \eqref{eq:ProofGCMapriori7} and   \eqref{eq:ProofGCMapriori6} (for $k=0$)
\beaa
\|z\|_0  &\les&  r^2 \|(A^\S +V) z\|_0+ r^{-2}\left|\int_\S z e^\Phi \right|\\
&\les&  r^2\left( r^{-5} I(f, \fb)+r^{-3}  \left( \dg  +\sqrt{\epg}\, \| F\|_{2} \right) \right)+r^{-2}\left(r^{-1} I(f, \fb)+ \sqrt{\epg}\,  r \| F\|_0\right)
\\
&\les& r^{-3} I(f, \fb)+r^{-1}  \left( \dg  +\sqrt{\epg}\, \| F\|_2 \right).
\eeaa
By standard elliptic  theory,
\beaa
\|z\|_{k+2} &\les&\|z\|_0+r^2 \|(A^\S +V) z\|_k.
\eeaa
 We  infer that, for $k\le s_{max}-1$,
\bea
\label{eq:ProofGCMapriori8}
\|z\|_{k+2} &\les& r^{-3} I(f, \fb)+ r^{-1}  \left( \dg  +\sqrt{\epg}   \| F\|_{2+k} \right).
\eea

{\bf Step 2.} We now proceed in the same manner to derive  an estimate\footnote{Since most estimates are similar as in   Step 1 above   we   skip some of the intermediary steps.}  for $f$. Using the second equation
in  \eqref{GCMS:NEAR-again:bis}
\beaa
\left(A^\S +V\right) f &=& \frac 3 4 \ka (A^\S)^{-1}\big(\rho z \big)-\dds\,\,\check{\ka}-\ka  (A^\S)^{-1}\big( -\ddsS \check{\mu}^\S +\dds\,\,\check{\mu}\big)+\err_4,
\eeaa
  we derive (note that $\| (A^\S)^{-1} h\|_k\les  r^2 \|h\|_{k-2} $),
\bea
 \label{eq:ProofGCMapriori10}
\bsplit
\|\left(A^\S +V\right) f \|_k &\les r \|\rho z\|_{k-2}+\|\dds\,\,\check{\ka}\|_k+ r\| \ddsS \,\muc^\S- \dds\,\muc\|_{k-2}+
\|\err_4\|_k +\sqrt{\epg}\, r^{-2} \|F\|_k\\
&\les r^{-2}  \|\ z\|_{k-2}+\dg r^{-2}+ r \| C\|_{k-2} +\sqrt{\epg}\, r^{-2} \|F\|_{k+2}
\end{split}
\eea
where,
\bea
\label{eq:ProofGCMaprioriC}
C:&=&  \ddsS\, \muc^\S- \dds\, \muc.
\eea
To estimate $C$ we write, proceeding as in the estimate for $B$ is  Step 1. Making use of the GCM condition
$\ddsS_2\ddsS_1 \mu^\S=0$ we derive, as in \eqref{eq:ProofGCMapriori2},
\bea
\label{eq:ProofGCMapriori11}
\left\| \ddsS_2 C\right\|_k&\les&r^{-4}\left(\dg+\sqrt{\epg}\, \| F\|_k\right)
\eea
which implies, as in \eqref{eq:ProofGCMapriori3},
\bea
\label{eq:ProofGCMapriori12}
\|C\|_k&\les&r^{-2} \left| \int_\S C  e^\Phi \right|+r^{-3}\left(\dg+\sqrt{\epg}\, \| F\|_k\right).
\eea
To   estimate the  $\ell=1$ mode of $C$ we recall  the transformation formula, see Lemma \ref{lemma:transfethka'-ethka},
\beaa
 e_\th' (\mu')
&= &e_\th \mu +e_\th'(\ddd_1)'\left(-(\dds_1)'  a      +f\omb  -\fb \om    +\frac{1}{4}f\kab -\frac 14   \fb \ka \right)+\frac 3 4\rho( f\kab+\fb \ka) \\
&+&\err(e_\th'\mu', e_\th \mu),
\eeaa
i.e.,
\beaa
C&=&A^\S \left(-\ddsS_1 a  +f\omb      +\frac{1}{4}f\kab -\frac 14   \fb \ka \right)-\frac 3 4\rho z+\err(e_\th'\mu', e_\th \mu).
\eeaa
Recall that,
\beaa
A^\S&=&\dddS_1\ddsS_1=\dddS_2 \ddsS_2 +K.
\eeaa
Therefore,  for any $h\in\sk_1$,
\beaa
\int_\S  A^\S h  e^\Phi= \int_\S  h K e^\Phi.
\eeaa
We deduce,
\beaa
\int_\S C e^\Phi&=& \int_\S K\left( -\ddsS_1 a  +f\omb     +\frac{1}{4}f\kab -\frac 14   \fb \ka \right) e^\Phi 
-\frac 3 4 \int_\S( \rho z) e^\Phi  + \int_\S \err(e_\th'\mu', e_\th \mu) e^\Phi\\
&=&(r^\S)^{-2}   \int_\S \left( -\ddsS_1 a  +f\omb      +\frac{1}{4}f\kab -\frac 14   \fb \ka \right) e^\Phi -\frac 3 4 \int_\S (\rho z) e^\Phi  +\int_\S \err(e_\th'\mu', e_\th \mu) e^\Phi\\
&+&\int_\S \left((K -r^{-2}) +(r^{-2}- (r^\S)^{-2}) \right)\left( -\ddsS_1 a  +f\omb     +\frac{1}{4}f\kab -\frac 14   \fb \ka \right) e^\Phi.
\eeaa
Also,
\beaa
  \int_\S \left(  f\omb      +\frac{1}{4}f\kab -\frac 14   \fb \ka \right)e^\Phi &=&\frac{m^\S}{  (r^\S)^2}\int_\S f e^\Phi-
  \frac{\Up^\S}{2 r^\S} \int_\S  f e^\Phi- \frac{1}{2 r^\S} \int_\S  \fb e^\Phi+\lot\\
  \int_\S \rho z e^\Phi&=&- \frac{2m^\S}{(r^\S)^3 }   \int_\S  z e^\Phi+ \lot
\eeaa
Note that, since $ \err(e_\th'\mu', e_\th \mu) =\err_3$, we have in view of Lemma \ref{Lemma-errorestimates-S},
\beaa
\left| \int_\S \err(e_\th'\mu', e_\th \mu) e^\Phi\right|\les  r^2\|\err_3\|_0\les r^{-1}\|F\|_3.
\eeaa
We deduce,
\bea
\label{eq:ProofGCMapriori13}
\left|\int_\S C e^\Phi \right| &\les&  r^{-3} I(a, f, \fb)+ r^{-1} \|F\|_{3}. 
\eea
Back to \eqref{eq:ProofGCMapriori12} we infer that,
\bea
\lab{eq:ProofGCMaprioriC-estimate}
\|C\|_k&\les& r^{-5} I(a, f, \fb)+r^{-3}\left( \dg  +\sqrt{\epg} \| F\|_k\right).
\eea
Back to \eqref{eq:ProofGCMapriori10}, 
\beaa
\|\left(A^\S +V\right) f \|_k 
&\les& r^{-2}  \|\ z\|_{k-2}+\dg r^{-2}+ r \| C\|_{k-2} +\sqrt{\epg}\, r^{-2} \|F\|_{k+2} \\
&\les&  r^{-2}  \|\ z\|_{k-2}+\dg r^{-2} +\sqrt{\epg}\, r^{-2} \|F\|_{k+2} + r^{-4} I(a, f, \fb)
\eeaa
i.e., recalling the estimate  \eqref{eq:ProofGCMapriori8} for $  \|\ z\|_k$ in  Step 1,
\bea
\|\left(A^\S +V\right) f \|_k &\les& \dg r^{-2} +\sqrt{\epg}\, r^{-2} \|F\|_{k+2} + r^{-4} I(a, f, \fb).
\eea
To estimate $\|f\|_0=\|f\|_\S$  we proceed as in the proof of  the  estimate   for  $\|z\|_0$ in Step 1 
 and deduce, as in \eqref{eq:ProofGCMapriori7},
 \bea
\|f\|_\S  &\les&  r^2 \|(A^\S +V) f\|_\S+ r^{-2}\left|\int_\S f e^\Phi \right|.
\eea
Hence,
 \beaa
 \|f\|_0  &\les&  \dg  +\sqrt{\epg}\,  \|F\|_{2} + r^{-2} I(a, f, \fb)+ r^{-2}\left|\int_\S f e^\Phi \right|
 \eeaa
 i.e.,
 \beaa
  \|f\|_0  &\les&  \dg  +\sqrt{\epg}\, \|F\|_{2} + r^{-2} I(a, f, \fb).
 \eeaa
 Finally,
 \beaa
\|f\|_{k+2} &\les&\|f\|_0+r^2 \|(A^\S +V) f\|_k \\
&\les&  \dg  +\sqrt{\epg}\,  \|F\|_{2+k} + r^{-2} I(a, f, \fb).
\eeaa
Hence,
\bea
\label{eq:ProofGCMapriori14}
\| f\|_{k+2} &\les&\dg  +\sqrt{\epg}\,  \|F\|_{k+2} + r^{-2} I(a, f, \fb).
\eea 
Since by the definition of $z$ we have,
\beaa
\fb&=&\ka^{-1} \left( z-\kab f\right)
\eeaa
 we also derive,
 \beaa
 \| \fb\|_{k+2}&\les& r^{-1}\|z\|_{k+2} + \|f\|_{k+2}. 
 \eeaa
 Thus, making use of the estimate \eqref{eq:ProofGCMapriori8} for $z$, we also deduce the  estimate for $\fb$,
 \beaa
\| \fb\|_{k+2} &\les&\dg  +\sqrt{\epg}\,  \|F\|_{k+2} + r^{-2} I(a, f, \fb)
\eeaa 
i.e.,
\bea
\label{eq:ProofGCMapriori15}
\|\left(f,  \fb\right)\|_{k+2} &\les&\dg  +\sqrt{\epg}\,  \|F\|_{k+2} + r^{-2} I(a, f, \fb).
\eea

{\bf Step 3.} 
 It remains to estimate  $a$ by making use of the equation (see Proposition \ref{GCMS:NEAR-again:bis} and definition of $C=  \ddsS\, \muc^\S- \dds\, \muc$ in \eqref{eq:ProofGCMaprioriC} )
\beaa
\ddsS a  &=& -\frac 3 4( A^\S)^{-1} \big(  \rho z \big) + f\omb   +\frac{1}{4}f\kab  -\frac 1  4  \fb \ka- (A^\S)^{-1}\ C -(A^\S)^{-1}\err_3.
\eeaa
Proceeding as before we deduce,
\beaa 
\| \ddsS a \|_k&\les& r^{-1}\|z\|_{k-2} + r^{-1}\|(f, \fb)\|_k+ r^2 \|C\|_{k-2} +r^2 \|\err_3\|_{k-2}.
\eeaa
Making use of   \eqref{eq:ProofGCMapriori8} and,\eqref{eq:ProofGCMaprioriC-estimate}
and the estimate of Lemma \ref{Lemma-errorestimates-S} for $\err_3$,
\beaa
\|z\|_{k-2} &\les& r^{-3} I(f, \fb)+ r^{-1}  \left( \dg  +\sqrt{\epg}\,   \| F\|_{k-2} \right),\\
\|C\|_{k-2}&\les& r^{-5} I(a, f, \fb)+r^{-3}\left( \dg  +\sqrt{\epg}\, \| F\|_{k-2}\right),\\
  \| \err_3\|_{k-2} &\les&  r^{-3}  \sqrt{\epg}\,  \,  \|  F \|_{k+1},
  \eeaa
we further deduce,
\bea
\label{eq:ProofGCMapriori16}
 \| \ddsS a \|_k&\les&   r^{-1}  (\dg+\sqrt{\epg}\,\| F\|_{k+1})+  r^{-3} I(a, f, \fb).
\eea
Combining this with \eqref{eq:ProofGCMapriori15},
\bea
\label{eq:ProofGCMapriori17}
 r^\S \| \ddsS a \|_{k+1}+ \|\left(f,  \fb\right)\|_{k+2} &\les&\dg  +\sqrt{\epg}\,  \|F\|_{k+2} + r^{-2} I(a, f, \fb).
\eea

{\bf Step 4.}   We make use of the equations \eqref{equations-forBadModes} to  estimate $I(a, f, \fb)$.
Recall that,
\beaa
    I(a, f, \fb)= I(f, \fb)+r^\S \left|  \int_\S( e^\S_\th a ) e^\Phi \right|.
\eeaa

{\bf Step 4a.} We estimate the integral  $\int_\S( e^\S_\th a ) e^\Phi$   making use of 
the  the first  equation\footnote{ recall that $\om=0$.} in  \eqref{equations-forBadModes}. 
\beaa
\frac{2}{r^\S} \int_\S e_\th^\S(a)e^\Phi&=&  -\int_\S e_\th(\ka) e^\Phi + \frac 1 4\int_{\S}\ka^2 \,\fb e^\Phi -\int_\S \left( \frac 1 4 \ka\kab +\ka\omb+3\rho\right)f e^\Phi+\err_7\\
&=&-J_1+J_2 -J_3+\err_7
\eeaa
where,
\beaa
\err_7 &=& \int_\S\left(-\err(e_\th^\S\ka^\S, e_\th\ka)+2(K^\S-K) f  +\left(\ka-\frac{2}{r^\S}\right)  \ddsS a        +\frac{1}{2}\vth\vthb f\right)e^\Phi.
\eeaa
We deduce,
\beaa
\left| \int_\S e_\th^\S(a)e^\Phi\right| &\les& r \left(| J_1|+|J_2| +|J_3 |   \right)+ r |\err_6|.
\eeaa
We make use of 
\beaa
e_\th^\S&=&\left(1+\frac 1 2   f \fb\right) e_\th  + \frac 1 2  \fb e_4+\frac 1 2 f\left(1+ \frac 1 4 f \fb\right) e_3
\eeaa
and the  assumptions   of Theorem \ref{prop.GCMSequations-fixedS}, in particular, $\| e_\th^\S\ka\|_\S\les \dg r^{-2}$, to deduce,
\beaa
  |J_1| &\les&r \int_\S \left(| e^\S_\th(\ka)| +| e^\S_\th(\ka)-e_\th(\ka)|\right)\les \dg +\sqrt{\epg}\, \| (f, \fb)\|_\S.
\eeaa
Also,
\beaa
|J_2| &\les & \frac{1}{(r^\S)^2} \left|   \int_{\S}\,\fb e^\Phi\right| 
+ r \int_\S  |\fb|\,\left[\Big| \ka^2-\frac{4}{r^2}\Big|    +  \Big| \frac{4}{(r^\S)^2}-\frac{4}{r^2}\Big|    \right]\\
&\les& r^{-2} \left|   \int_{\S}\,\fb e^\Phi\right| +\sqrt{\epg}\,\| \fb\|_\S.
\eeaa
Similarly,
\beaa
|J_3|&\les& r^{-2} \left|   \int_{\S}\, f  e^\Phi\right| +\sqrt{\epg}\,\| f\|_\S.
\eeaa
Also,
\beaa
|\err_7|&\les& r^2\|\err(e_\th' \ka', e_\th \ka)\|_\S+ r\left( \| K^\S-K\|_\S\,\| f\|_\S+\|\ka-\frac{2}{r^\S} \|_\S \|\ddsS a \|_\S +\| \vth\vthb\|_\S \,\|f\|_\S\right).
\eeaa
Thus, proceeding as before,
\beaa
|\err_7| &\les& \sqrt{\epg}\, \|F\|_2. 
\eeaa
We deduce,
\beaa
\left| \int_\S e_\th^\S(a)e^\Phi\right| &\les& r \left(| J_1|+|J_2| +|J_3 |   \right)+ r |\err_6|\\
&\les& r  \dg +\sqrt{\epg}\, r \| (f, \fb)\|_\S + r^{-1}\left[ \left|   \int_{\S}\,\fb e^\Phi\right| + \left|   \int_{\S}\, f  e^\Phi\right| \right] + r\sqrt{\epg}\, \|F\|_2
\eeaa
i.e.,
\bea
\label{eq:ProofGCMapriori20}
\left| \int_\S e_\th^\S(a)e^\Phi\right| &\les&  r  \dg +r^{-1} I(f, \fb)+  r\sqrt{\epg}\, \|F\|_2.
\eea
As a consequence  we deduce,
\bea
\label{eq:ProofGCMapriori21}
  I(a, f, \fb)\les  I(f, \fb)+ r^2\dg. 
\eea

{\bf Step 5.} We combine estimate \eqref{eq:ProofGCMapriori17} with \eqref{eq:ProofGCMapriori21}
with 
\beaa
 r^\S \| \ddsS a \|_{k+1}+ \|\left(f,  \fb\right)\|_{k+2} &\les&\dg  +\sqrt{\epg}\,  \|F\|_{k+2} + r^{-2} I(f, \fb).
\eeaa
Also, taking into account that
\beaa
 \|F\|_{k+2}^2 &=& \|a\|_{0}+  r^\S \| \ddsS a \|_{k+1}+ \|\left(f,  \fb\right)\|_{k+2}  
\eeaa
and the smallness of $\epg$, we deduce,
\bea
\label{eq:ProofGCMapriori22}
 r^\S \| \ddsS a \|_{k+1}+ \|\left(f,  \fb\right)\|_{k+2} &\les&\dg  +\sqrt{\epg}\,  \|a\|_0 + r^{-2} I(f, \fb).
 \eea

{\bf Step 6.} It remains to get an estimate for  $\|a\|_0=\|a\|_{L^2(\S)}$. For this we need to make use of the  last equation in  for the average of $a$ in \eqref{GCMS:NEAR-again:bis} as well as the GCM condition $\ka^\S=\frac{2}{r^\S}$,
\beaa
\ov{a}^\S&=\ov{\left( 1-\frac{r^\S}{r}\right)}^\S - \frac{ r^\S}{2} \ov{\left( \check{\ka} + \ov{\ka}-\frac{2}{r}      \right)}^\S+\ov{\err_6}^\S
\eeaa
where,
\beaa
\err_6& =&-\frac{r^\S}{2}\left[e^{a}\err(\ka, \ka')-  (e^a-1)  \left( \check{\ka} + \ov{\ka}-\frac{2}{r}       +\dddS f \right)-\left(e^a-1-a \right) \frac 2 r \right]\\
&-& a \left(\frac{r^\S}{r}-1\right).
\eeaa
Hence,
\beaa
|\ov{a}^\S| &\les& \left|\ov{\left( 1-\frac{r^\S}{r}\right)}^\S \right|+ r\left(| \check{\ka}| + \left|\ov{\ka}-\frac{2}{r} \right|\right)+r\Big|\err_6\Big|.
\eeaa
Making use of the assumption\footnote{Note that   full strength  of the assumption is only used here, otherwise  the weaker assumption,      with $\dg$ replaced by $\epg$,   suffices in the rest  of the argument.} \ref{eq:assumptionsrminusrSforpropGCMSequations-fixedS}, \eqref{eq:GCM-improved estimate}  and proceeding as before,      it is easy to derive the estimate,
\bea
\label{eq:ProofGCMapriori23}
|\ov{a}^\S| &\les& r^{-1}\dg. 
\eea

{\bf Step 7.}  By the standard Poincare inequality on $\S$ we have,
\beaa
\big\| a-\ov{a}^\S\big\|_\S&\les&r^\S \|\ddsS a \|_\S. 
\eeaa
Combining this with \eqref{eq:ProofGCMapriori22} and \eqref{eq:ProofGCMapriori23} we  deduce,
\bea
 \|\left( a, f,  \fb\right)\|_{k+2} &\les&\dg + r^{-2} I(f, \fb).
\eea
Finally, making use of the condition 
\beaa
\int_\S f e^\Phi=\La, \qquad   \int_\S \fb e^\Phi=\Lab, \qquad  \La, \Lab  =O(\dg r^2)
\eeaa
   we  derive the desired estimate,
   \bea
    \|\left( a, f,  \fb\right)\|_{k+2} &\les&\dg. 
   \eea
   The uniqueness part is obvious and left to the reader. 
   This ends the proof of  Theorem \ref{prop.GCMSequations-fixedS}.


 \section{Proof of Proposition \ref{Proposition-contraction}}\lab{sec:proofofProposition-contraction}
   


  \subsection{Pull-back of the  main equations}  
  
  
  According to Proposition \ref{Prop:BondsforQn} we may assume   valid the uniform bounds for the quintets $P^{(n)}$.
  To establish a contraction estimate  we need to compare    the quantities, 
    \beaa
  \hn :&=&(\Psi^{(n-1)})^\# f^{(n)}, \quad 
  \hbn:=(\Psi^{(n-1)})^\# \fb^{(n)}, \quad 
  \wn:=(\Psi^{(n-1)})^\# z^{(n)},\quad \\
  \en:&=&(\Psi^{(n-1)})^\# a^{(n)},
  \eeaa 
   and,
   \beaa
  \hnn:&=&(\Psi^{(n)})^\# f^{(n+1)},\, 
  \hbnn:=(\Psi^{(n)})^\# \fb^{(n+1)},\,
  \wnn:=(\Psi^{(n)})^\# z^{(n+1)},\\
  \enn:&=&(\Psi^{(n)})^\# a^{(n+1)}.
  \eeaa
  According to Lemma \ref{le:pullbackS} we have,
  \beaa
 (\Psi^{(n)})^\#\left( \ddd^{\S(n)} f^{(n+1)}\right) = \ddd^{(n)} \hnn,\\
 (\Psi^{(n)})^\#\left( A^{\S(n)} f^{(n+1)}\right) = A^{(n)} \hnn,
  \eeaa
  where $\ddd^{(n)}, \dds^{\,\,(n)},  A^{(n)}$   are the corresponding Hodge operators on $\ovS$  defined with respect to the metric 
  $\gS^{(n)}:=(\Psi^{(n)})^\#(\gS^{\S(n)})$ given by,
  \beaa
\gS^{(n)}=  \ga^{(n)}  d\th ^2 + e^{2\Phi^{\#_n}} d\vphi^2.
  \eeaa
  Consequently   the system \eqref{system:af-fb-index-n} takes the form,
  \bea
  \label{system:af-fb-index-n-Pulled}
  \bsplit
  A^{(n)}w^{(n+1)} +V^{(n)}w^{(n+1)} &= (\Psi^{(n)})^\#  E^{(n+1)},\\
  A^{(n)} \hnn+V^{(n)} \hnn&=\frac 3 4 \ka^{(n)} (A^{(n)})^{-1}(\rho^{(n)} w^{(n+1)})+ (\Psi^{(n)})^\# E^{(n+1)},\\
   A^{(n)} \hbnn+V^{(n)} \hbnn&=-\frac 3 4 \ka^{(n)} (A^{(n)})^{-1}(\rho^{(n)} w^{(n+1)})+(\Psi^{(n)})^\# \Eb^{(n+1)},\\
   \dds\,\,^{(n)} \enn&=\frac 3 4  (A^{(n)})^{-1}(\rho^{(n)} w^{(n+1)}) - h^{(n+1)}\omb^{(n)}    
    -\frac{1}{4}h^{(n+1)}\kab^{(n)} +\frac 14   \hbnn \ka^{(n)}\\
    &+(\Psi^{(n)})^\#  \widetilde{E}^{(n+1)},
     \end{split}
  \eea
  where, $ \ka^{(n)},  \kab^{(n)},  \rho^{(n)}, \om^{(n)},  \omb^{(n)}  $ 
  are the pull backs  by        $\Psi^{(n)} $ of  $\ka, \kab, \rho, \om, \omb$.   Also $ V^{(n)} $ is the pull back by 
    $\Psi^{(n)} $ of the potential  $V=\frac 1 2 \ka \kab -\rho$. 
    
    Equation \eqref{system:badmodes-index-n} takes the form
    \bea
    \bsplit
&    \frac{2}{r^{\S(n)} }\int_{\ovS, \gS^{(n)}} \dds^{\,\,(n)} e^{(n+1)} e^{\Phi^{\#_n}} =
  -   \frac 1 4 \int_{\ovS, \gS^{(n)}} (\ka^{(n)})^2 \,\hb^{(n+1)}  e^{\Phi^{\#_n}} \\
&  +\int_{\ovS, \gS^{(n)}}   \left( \frac 1 4 \ka\kab +\ka\omb+3\rho\right)^{(n)\, }h^{(n+1)}e^{\Phi^{\#_n}} \\
  &  -\int_{\ovS, \gS^{(n)}}   \dds^{(n)} (\ka^{(n)} ) e^{\Phi^{\#_n}} da_{ \gS^{(n)}} +(\Psi^{(n)} )^\#\err^{(n+1)}_6,\\
&\int_{\ovS, \gS^{(n)}}   h^{(n+1)} e^{\Phi^{\#_n}} =\La, \quad \int_{\ovS, \gS^{(n)}}  \hb^{(n+1)} e^{\Phi^{\#_n}} =\Lab,  
    \end{split}
    \eea
 Equation \eqref{eq:retrieve-a-indexn} takes  the form,
    \bea\lab{eq:retrieve-a-indexn-dif}
\ov{e^{(n+1)}}^{\ovS, \gS^{(n)}}  = \ov{\left( 1-\frac{r^{\S(n)}}{r}\right)^{(n)}}^{\ovS, \gS^{(n)}}        - \frac{r^{\S(n)}}{2}\ov{\left( \check{\ka} + \ov{\ka}-\frac{2}{r} \right)^{(n)}}^{\ovS, \gS^{(n)}}+\ov{\err_8^{(n+1)}}^{\ovS, \gS^{(n)}}.
\eea    
  Finally       
    the system \eqref{mainiteration-definUS(n+1)} takes the form,
     \bea\lab{eq:someovertherainbow-pullback}
       \bsplit
  \vsi^{\#_n}   \pr_\th U^{(1+n)} &=  (\ga^{(n)}  )^{1/2}  h^{(1+n)}\left(1 +\frac 1 4h ^{(1+n)}\hb^{(1+n)}\right),  \\
     \pr_\th S^{(1+n)}-\frac 1 2  \vsi^{\#_n} \Omb^{\#_{n}}  \pr_\th U^{(1+n)}&=   \frac 1 2 (\ga^{(n)}  )^{1/2}\hb^{(1+n)},
     \\
   \ga^{(n)} & =\ga^{\#_n}+\big(\vsi^{\#_n}\big)^2  \left(\Omb^{\#_n}+\frac 1 4( \underline{b}^{\#_n} )^2  \ga^{\#_n}  \right)\, ( \pr_\th U^{(n)} )^2 \\
     &-2  \vsi^{\#_n} \pr_\th U ^{(n)} \pr_\th  S^{(n)}-\ga^{\#_n} \vsi^{\#_n} \underline{b}^{\#_n} \pr_\th U^{(n)},\\
     U^{(1+n)}(0)&= S^{(1+n)}(0)=0.
     \end{split}
  \eea

   We recall, see \eqref{quintet-norm}, the definition of the norm  for the quintets $P^{(n)}$ in the particular case $k=3$   
     \beaa
\label{quintet-norm'}
\bsplit
\| P^{(n)}\|_3: &= \|\pr_\th \left(U^{(n)}, S^{(n)} \right)\|_{L^\infty(\ovS)}  +r^{-1} \|\pr_\th \left(U^{(n)}, S^{(n)}\right) \|_{\hk_{1}(\ovS)}  \\
&+ \left\|\left((a^{(n)})^{\#_{n-1}},  (f^{(n)})^{\#_{n-1}},  (\fb^{(n)})^{\#_{n-1}}\right)\right\|_{\hk_{1}(\ovS)}.
\end{split}
\eeaa
        To prove the estimate
    \beaa
    \|   P^{(n+1)}-   P^{(n)}\|_3\les   \dg  \|   P^{(n)}-   P^{(n-1)}\|_3
    \eeaa
    we set,
     \beaa
      \de w^{(n+1)}&=&   w^{(n+1)}- w^{(n)}, \quad      \de  h^{(n+1)}=   h^{(n+1)}-h^{(n)},\quad      \de  h^{(n+1)}=   \hb ^{(n+1)}-\hb ^{(n)},\\
       \de  e^{(n+1)}&= &  e^{(n+1)}-e^{(n)},   \quad \de U^{(n+1)} =        U^{(n+1)}  -U^{(n)}, \quad  \de S^{(n+1)} =        S^{(n+1)}  -S^{(n)}, 
      \eeaa
      and,
       \beaa
      \de w^{(n)}&=&   w^{(n)}- w^{(n-1)}, \quad      \de  h^{(n)}=   h^{(n)}-h^{(n-1)},\quad      \de  h^{(n)}=   \hb ^{(n)}-\hb ^{(n-1)},\\
       \de  e^{(n)}&= &  e^{(n)}-e^{(n-1)},   \quad \de U^{(n)} =        U^{(n)}  -U^{(n-1)}, \quad  \de S^{(n)} =        S^{(n)}  -S^{(n-1)}.
      \eeaa
        We will derive below the following estimates 
    \bea
    \label{differences-Step1}
      \|  \de h^{(n+1)},   \de \hb ^{(n+1)},  \de e^{(n+1)})   \| _{\hk_1(\ovS)} &\les& \dg r^{-1}   \|\pr_\th  \big(\de U^{(n)}, \de S^{(n)}\big)  \|_{\hk_1(\ovS)}  
    \eea
   and    
       \bea      
  r^{-1}  \| \pr_\th\big(\de U^{(n+1)},  \de S^{(n+1)}\big) \|_{\hk_1(\ovS)}\les   \| \de h^{(n+1)}, \de \hb ^{(n+1)}  \|_{\hk_1(\ovS)} +  \epg  r^{-1} \|\pr_\th \big(\de U^{(n)}, \de S^{(n)}\big) \|_{\hk_1(\ovS)}.\nn\\
     \label{differences-Step3}
    \eea
     Proposition  \ref{Proposition-contraction} is then an immediate consequence of \eqref{differences-Step1} \eqref{differences-Step3}. Thus, from now on, we focus on the proof of  \eqref{differences-Step1} \eqref{differences-Step3}. To this end, we will rely on the following lemmas.

     
  \subsection{Basic  Lemmas}


  \begin{lemma}
  \label{Lemma:differencesFn}
  Let $F$ be a  reduced  scalar function  defined in a  neighborhood of $\ovS$ in  $\RR$ and define  its  pull back   $F^{(n)}=(\Psi^{(n)})^\#F $ to  $\ovS$,
  i.e.,
  \beaa
  F^{(n)}(\th)&=& F(\ovu+U^{(n)}(\th),  \ovs+S^{(n)} (\th), \th), \\
   F^{(n-1)}(\th)&=& F(\ovu+U^{(n-1)}(\th),  \ovs+S^{(n-1)} (\th), \th).  
  \eeaa 
  Then\footnote{ Recall $\pr_s =e_4, \,  \pr_u=\frac{\vsi}{2}\left( e_3-\Omb e_4-\underline{b}\ga^{1/2} e_\th\right). $},  for all   $1\le p\le \infty$,  with $\de_n U=U^{(n+1)}-U^{(n)}, \, \de_n S=S^{(n+1)} -S^{(n)}$
  \bea
    \label{eq:LemmadifferencesFn-1}
    \|\de_n F\|_{L^p(\ovS)} &\les& \left( \|\de_n U\|_{L^p(\ovS)}+ \|\de_n S\|_{L^p(\ovS)}\right)  \sup_{\RR }\left( \big|\pr_s F\big|+\big| \pr_u  F\big|\right).
  \eea
  
  Also,
    \bea
      \label{eq:LemmadifferencesFn-2}
  \|\de_n F\|_{\hk_1 (\ovS)} &\les&\left( \|\de_n U\|_{\hk_1(\ovS)}+ \|\de_n S\|_{\hk_1(\ovS)}\right) \sup_\RR| \dk^{\le 1}  \dk F|
  \eea
  where $\de_n U=U^{(n+1)}-U^{(n)}, \, \de_n S=S^{(n+1)} -S^{(n)}$.
  \end{lemma}
  
    \begin{proof}
   We write,
\beaa
\de_n F  &:=& F(u_0+ U^{(n)}(\th), s_0+ S^{(n)}(\th), \th)- F(u_0+ U^{(n-1)}(\th), s_0+ S^{(n-1)}(\th), \th)\\
   &=&\int_0^1 \frac{d}{dt}    F\left(u_0+ tU^{(n)}(\th)+(1-t)  U^{(n-1)}(\th),\,  s_0+ tS^{(n)}(\th)+(1-t)S^{(n-1)}(\th)   , \th\right),
   \eeaa
   i.e., denoting $\de_n U=U^{(n)}-U^{(n-1)}$,  $\de_n S=S^{(n)}-S^{(n-1)}$,
   \beaa
   | \de_n F| &\les&\big|\de_n U \big| \int_0^1\left |\pr_u F\left(u_0+ tU^{(n)}(\th)+(1-t)  U^{(n-1)}(\th),\,  s_0+ tS^{(n)}(\th)+(1-t)S^{(n-1)}(\th), \th\right)\right|\\
   &+&\big|\de_n S \big| \int_0^1\left |\pr_s F\left(u_0+ tU^{(n)}(\th)+(1-t)  U^{(n-1)}(\th), \, s_0+ tS^{(n)}(\th)+(1-t)S^{(n-1)}(\th)   , \th\right)\right|
   \eeaa   
   i.e.,
   \beaa
| \de_n F| &\les&\big|U^{(n)}(\th)-U^{(n-1)}(\th)\big|  \sup_{\ovS+\dg \ovS}  |\pr_ u F|+ \big|S^{(n)}(\th)-S^{(n-1)}(\th)\big|  \sup_{\ovS+\ep \ovS}  |\pr_ s F|
  \eeaa
  from which \eqref{eq:LemmadifferencesFn-1}  easily follows.

    Also, in view    of  $\pr_s =e_4, \,  \pr_u=\frac 1 2\left( e_3-\Omb e_4-\underline{b}\ga^{1/2} e_\th\right) $              and our assumptions for $\Omb$ and $\underline{b}$,
  \beaa
  | \de_n F|  
 & \les & \left( |U^{(n)}-U^{(n-1)} |  +   |S^{(n)}-S^{(n-1)}|\right)| \sup_\RR \dk  F|
  \eeaa
and  by integration,
  \beaa
  \|\de_n F\|_{L^p(\ovS)} &\les&\left( \|\de_n U\|_{L^p(\ovS)}+ \|\de_n S\|_{L^p(\ovS)}\right) \sup_\RR| \dk  F|.
  \eeaa
  Similarly,
  \beaa
    \| \dkb \de_n F\|_{L^2 (\ovS)} &\les&\left( \|\de_n U\|_{\hk^1(\ovS)}+ \|\de_n S\|_{\hk^1(\ovS)}\right) \sup_\RR |\dk^{\le 1}\dk    F|.
  \eeaa
  Hence,
   \beaa
  \|\de_n F\|_{\hk_1 (\ovS)} &\les&\left( \|\de_n U\|_{\hk^1(\ovS)}+ \|\de_n S\|_{\hk^1(\ovS)}\right)\sup_\RR |\dk^{\le 1}\dk    F|
  \eeaa
  as desired.
\end{proof}

\begin{lemma}
\label{Lemma:differencesFn-integral}
Let $\psi, h \in\sk_1(\ovS)$,   and  $\de A^{(n)}= A^{(n)}-A^{(n-1)} $. The following  formula holds true.
\beaa
\left|\int_{(\ovS, \gS^{(n)})}\psi \de A^{(n)} h\right|  &\les&r^{-2} \|\pr_\th\left(\Psi^{(n)}-\Psi^{(n-1)}\right)\|_{\hk_1(\ovS)}\left(\|\psi \|_{\hk_1(\ovS)}\|h\|_{\hk_2(\ovS)} +\|\psi \|_{\hk_2(\ovS)}\|h\|_{\hk_1(\ovS)} \right).
\eeaa
\end{lemma}

\begin{proof}
 Recall that the metric $\gS^{(n)}$ is given by 
     \beaa
\gS^{(n)}=  \ga^{(n)}  d\th ^2 + e^{2\Phi^{\#_n}} d\vphi^2
  \eeaa
  so that the operator $A^{(n)}=-\dds\,\,^{(n)}\ddd^{(n)}$, applied to $\sk_1$ tensors $h$  on $\ovS$  is given by
 \beaa
  A^{(n)} h&=&   -\frac{1}{\sqrt{\ga^{(n)}}} \pr_\th \left(  \frac{1}{\sqrt{\ga^{(n)}}}\left(\pr_\th h +\pr_\th(\Phi^{\#_n}) h \right)    \right).
  \eeaa
    This yields
  \beaa
   \de A^{(n)}h &=& -\left(\frac{1}{\sqrt{\ga^{(n)}}}- \frac{1}{\sqrt{\ga^{(n-1)}}}\right) \pr_\th \left(  \frac{1}{\sqrt{\ga^{(n)}}}\left(\pr_\th h +\pr_\th(\Phi^{\#_n}) h \right)    \right)  \\
    && -\frac{1}{\sqrt{\ga^{(n-1)}}} \pr_\th \left(\left(\frac{1}{\sqrt{\ga^{(n)}}}- \frac{1}{\sqrt{\ga^{(n-1)}}}\right)\left(\pr_\th h +\pr_\th(\Phi^{\#_n}) h \right)    \right)\\
 && -\frac{1}{\sqrt{\ga^{(n-1)}}} \pr_\th \left(  \frac{1}{\sqrt{\ga^{(n-1)}}}\pr_\th(\Phi^{\#_n}-\Phi^{\#_{n-1}}) h \right).
  \eeaa
  Using the previous formula to integrate $\psi \de A^{(n)}h$ on $\ovS$ with the volume of $\gS^{(n)}$, and after integration by parts, we infer
  \beaa
  &&\int_{(\ovS, \gS^{(n)})}\psi \de A^{(n)} h \\
  &=& \int_{(\ovS, \gS^{(n)})}\frac{1}{\sqrt{\ga^{(n)}}}\pr_\th\left(\left(1- \frac{\sqrt{\ga^{(n)}}}{\sqrt{\ga^{(n-1)}}}\right)\psi\right)\frac{1}{\sqrt{\ga^{(n)}}}\left(\pr_\th h_2 +\pr_\th(\Phi^{\#_n})h\right)\\
  && +\int_{(\ovS, \gS^{(n)})}\frac{1}{\sqrt{\ga^{(n)}}}\pr_\th\left( \frac{\sqrt{\ga^{(n)}}}{\sqrt{\ga^{(n-1)}}}\psi\right)\left(\frac{1}{\sqrt{\ga^{(n)}}}- \frac{1}{\sqrt{\ga^{(n-1)}}}\right)\left(\pr_\th h_2 +\pr_\th(\Phi^{\#_n})h\right)\\
  && +\int_{(\ovS, \gS^{(n)})}\frac{1}{\sqrt{\ga^{(n)}}}\pr_\th\left( \frac{\sqrt{\ga^{(n)}}}{\sqrt{\ga^{(n-1)}}}\psi \right) \frac{1}{\sqrt{\ga^{(n-1)}}}\pr_\th\left(\Phi^{\#_n}-\Phi^{\#_{n-1}}\right) h.
  \eeaa
  We now make us of the bounds \eqref{eq:assumtionsonthegivenusfoliationforGCMprocedure:bis} for $(\Omb, \underline{b}, \ga)$ involved in the definition of $\ga^{(n-1)}$ and $\ga^{(n)}$,    the uniform bound of $P^{(n)}$ provided by Proposition \ref{Prop:BondsforQn} and the Sobolev inequality to deduce,
   \bea
 \nn\left|\int_{(\ovS, \gS^{(n)})}\psi \de A^{(n)}h\right|  &\les&r^{-4} \|\ga^{(n)}-\ga^{(n-1)}\|_{\hk_1(\ovS)}\left(\|\psi \|_{\hk_1(\ovS)}\|h\|_{\hk_2(\ovS)} +(\|\psi \|_{\hk_2(\ovS)}\|h\|_{\hk_1(\ovS)} \right)\\
 &&+r^{-3}\left\|\pr_\th\left(\Phi^{\#_n}-\Phi^{\#_{n-1}}\right) h\right\|_{L^2(\ovS)}\|\psi\|_{\hk_1(\ovS)}.
  \eea
  To estimate the term $\ga^{(n)}-\ga^{(n-1)}$  we recall that,
    \beaa
    \ga^{(n)} & =\ga^{\#_n}+\big(\vsi^{\#_n}\big)^2  \left(\Omb^{\#_n}+\frac 1 4( \underline{b}^{\#_n} )^2  \ga^{\#_n}  \right)\, ( \pr_\th U^{(n)} )^2 \\
     &-2  \vsi^{\#_n} \pr_\th U ^{(n)} \pr_\th  S^{(n)}-\ga^{\#_n} \vsi^{\#_n} \underline{b}^{\#_n} \pr_\th U^{(n)}\\
       \ga^{(n-1)} & =\ga^{\#_{n-1}}+\big(\vsi^{\#_{n-1}}\big)^2  \left(\Omb^{\#_{n-1}}+\frac 1 4( \underline{b}^{\#_{n-1}} )^2  \ga^{\#_{n-1}}  \right)\, ( \pr_\th U^{(n-1)} )^2 \\
     &-2  \vsi^{\#_{n-1}} \pr_\th U ^{(n-1)} \pr_\th  S^{(n-1)}-\ga^{\#_{n-1}} \vsi^{\#_{n-1}} \underline{b}^{\#_{n-1}} \pr_\th U^{(n-1)}.
  \eeaa 
  The principal term $ \ga^{\#_n}- \ga^{\#_{n-1} }$ can be estimated with the help of Lemma \ref{Lemma:differencesFn},   the uniform bound of $P^{(n)}$ provided by Proposition \ref{Prop:BondsforQn},   and the bounds provided by {\bf A3}. All other  terms   can be estimated in a similar fashion.  We derive,
  \bea
  \|\ga^{(n)}-\ga^{(n-1)}\|_{\hk_1(\ovS)} &\les&   r^2  \|\pr_\th\left( \Psi^{(n)}-\Psi^{(n-1)}\right)\|_{\hk_1(\ovS)}
  \eea
  where,
  \beaa
  \|\pr_\th\left(\Psi^{(n)}-\Psi^{(n-1)}\right)\|_{\hk_1(\ovS)}:=\|\pr_\th ( U^{(n)}- U^{(n-1)} )\|_{\hk_1(\ovS)}    +\|\pr_\th ( S^{(n)}-S^{(n-1)} ) \|_{\hk_1(\ovS)}. 
  \eeaa
  We deduce,
     \bea
 \nn\left|\int_{(\ovS, \gS^{(n)})}\psi \de A^{(n)}h\right|  &\les&r^{-2} \|\pr_\th\left(\Psi^{(n)}-\Psi^{(n-1)}\right)\|_{\hk_1(\ovS)}\left(\|\psi \|_{\hk_1(\ovS)}\|h\|_{\hk_2(\ovS)} +(\|\psi \|_{\hk_2(\ovS)}\|h\|_{\hk_1(\ovS)} \right)\\
 &&+r^{-3}\left\|\pr_\th\left(\Phi^{\#_n}-\Phi^{\#_{n-1}}\right) h\right\|_{L^2(\ovS)}\|\psi\|_{\hk_1(\ovS)}.
  \eea
  
  The proof of   \ref{Lemma:differencesFn-integral}. is now an immediate consequence of the following.
  \begin{lemma}
  \label{Lemma:differencesFn-integral-Main}
  The following estimate holds true for a reduced   scalar $ h\in \sk_1(\ovS)$
  \bea
 \label{eq1:Lemma:differencesFn-integral}
  \left\|\pr_\th\left(\Phi^{\#_n}-\Phi^{\#_{n-1}}\right) h\right\|_{L^2(\ovS)} &\les&r \|\pr_\th\left(\Psi^{(n)}-\Psi^{(n-1)}\right)\|_{\hk_1(\ovS)}\|h\|_{\hk_2(\ovS)}.
  \eea  
  \end{lemma}
  
  \begin{proof}
  We write,
   \beaa
&&  \pr_\th\left(\Phi^{\#_n}-\Phi^{\#_{n-1}}\right) \\
  &&= \left\{\left( \pr_\th S^{(n)}-\frac 1 2 \Omb \pr_\th U^{(n)}\right) e_4\Phi+\frac 1 2 \pr_\th U^{(n)} e_3\Phi +\sqrt{\ga}\left(  1 -\frac 1 2 \underline{b}\pr_\th U^{(n)}\right) e_\th\Phi\right\}^{\#_n}\\
  && -\left\{\left( \pr_\th S^{(n-1)}-\frac 1 2 \Omb \pr_\th U^{(n-1)}\right) e_4\Phi+\frac 1 2 \pr_\th U^{(n-1)} e_3\Phi +\sqrt{\ga}\left(  1 -\frac 1 2 \underline{b}\pr_\th U^{(n-1)}\right) e_\th\Phi\right\}^{\#_{n-1}}\\
   &&= \left( \pr_\th S^{(n)}-\frac 1 2 \Omb^{\#_n} \pr_\th U^{(n)}\right) (e_4\Phi)^{\#_n}+\frac 1 2 \pr_\th U^{(n)} (e_3\Phi)^{\#_n} +\sqrt{\ga}^{\#_n}\left(  1 -\frac 1 2 \underline{b}^{\#_n}\pr_\th U^{(n)}\right) (e_\th\Phi)^{\#_n}\\
  && - \left( \pr_\th S^{(n-1)}-\frac 1 2 \Omb^{\#_{n-1}} \pr_\th U^{(n-1)}\right) (e_4\Phi)^{\#_{n-1}}-\frac 1 2 \pr_\th U^{(n-1)} (e_3\Phi)^{\#_{n-1}}\\
  && -\sqrt{\ga}^{\#_{n-1}}\left(  1 -\frac 1 2 \underline{b}^{\#_{n-1}}\pr_\th U^{(n-1)}\right) (e_\th\Phi)^{\#_{n-1}}
 \eeaa
 i.e., grouping the terms appropriately,
   \beaa
&& \pr_\th\left(\Phi^{\#_n}-\Phi^{\#_{n-1}}\right)=J_1+J_2+J_3, \\
J_1&& = \left( \pr_\th S^{(n)}-\frac 1 2 \Omb^{\#_n} \pr_\th U^{(n)}\right) (e_4\Phi)^{\#_n}-
  \left( \pr_\th S^{(n-1)}-\frac 1 2 \Omb^{\#_n-1} \pr_\th U^{(n-1)}\right) (e_4\Phi)^{\#_{n-1}},\\
 J_2 &&=\frac 1 2 \pr_\th U^{(n)} (e_3\Phi)^{\#_n} -\frac 1 2 \pr_\th U^{(n-1)} (e_3\Phi)^{\#_{n-1}}, \\
J_3&& = \sqrt{\ga}^{\#_n}\left(  1 -\frac 1 2 \underline{b}^{\#_n}\, \pr_\th U^{(n)}\right) (e_\th\Phi)^{\#_n}-\sqrt{\ga}^{\#_{n-1} }\left(  1 -\frac 1 2 \underline{b}^{\#_{n-1} }\pr_\th U^{(n-1)}\right) (e_\th\Phi)^{\#_{n-1}},
  \eeaa
  and,
   \beaa
  J_3&=&J_{31}+J_{32},\\ 
  J_{31}&=&  (e_\th\Phi)^{\#_{n-1}}\left( \sqrt{\ga}^{\#_n}\left(  1 -\frac 1 2 \underline{b}^{\#_n}\pr_\th U^{(n)}\right)- \sqrt{\ga}^{\#_{n-1}}\left(  1 -\frac 1 2 \underline{b}^{\#_{n-1} }\pr_\th U^{(n-1)}\right)\right),\\
   J_{32}&=& \sqrt{\ga}^{\#_n}\left(  1 -\frac 1 2 \underline{b}^{\#_n}\, \pr_\th U^{(n)}\right)\left( (e_\th\Phi)^{\#_n}- (e_\th\Phi)^{\#_{n-1}}\right).
  \eeaa
  The  contribution  to the estimate of  of Lemma \ref{Lemma:differencesFn-integral-Main}  given by $J_1, J_2, J_{31} $  can be easily estimated by  making use of  the uniform bound of $P^{(n)}$ provided by Proposition \ref{Prop:BondsforQn}, the bound  \eqref{eq:assumtionsonthegivenusfoliationforGCMprocedure:bis} for $(\Omb, \underline{b}, \ga)$,  Lemma \ref{lemma:int-gaS-ga:highersobolevregularity} as well as Lemma \ref{Lemma:differencesFn}. We  thus derive,
  \beaa
    \left\| (J_1, J_2, J_{31})h   \right\|_{L^2(\ovS)} &\les& r \|\pr_\th\Psi^{(n)}-\pr_\th\Psi^{(n-1)}\|_{\hk_1(\ovS)}\|h\|_{\hk_2(\ovS)}.
  \eeaa
  It  remains to estimate  the  term $\left\| J_{32} h\right\|_{L^2(\ovS)}$ which presents a  difficulty at the axis of symmetry where $\sin\th=0$. We can write,
  \beaa
  J_3&=&J_{31}+J_{32},\\ 
  J_{31}&=&  (e_\th\Phi)^{\#_{n-1}}\left( \sqrt{\ga}^{\#_n}\left(  1 -\frac 1 2 \underline{b}^{\#_n}\pr_\th U^{(n)}\right)- \sqrt{\ga}^{\#_{n-1}}\left(  1 -\frac 1 2 \underline{b}^{\#_{n-1} }\pr_\th U^{(n-1)}\right)\right),\\
   J_{32}&=& \sqrt{\ga}^{\#_n}\left(  1 -\frac 1 2 \underline{b}^{\#_n}\, \pr_\th U^{(n)}\right)\left( (e_\th\Phi)^{\#_n}- (e_\th\Phi)^{\#_{n-1}}\right).
  \eeaa
 Clearly, $ \left\| J_{32}\, h \right\|_{L^2(\ovS)} \les  \left\| \left( (e_\th\Phi)^{\#_n}- (e_\th\Phi)^{\#_{n-1}}\right)  h \right\|_{L^2(\ovS)} $. 
 We are thus left to estimate the term  $ \left\| \left( (e_\th\Phi)^{\#_n}- (e_\th\Phi)^{\#_{n-1}}\right)  h \right\|_{L^2(\ovS)}$. Proceeding as in the proof of Lemma \ref{Lemma:differencesFn} we write,  for  $F=e_\th \Phi$,
  \beaa
   | \de_n F| &\les&\big|\de_n U \big| \int_0^1\left |\pr_u F\left(u_0+ tU^{(n)}(\th)+(1-t)  U^{(n-1)}(\th),\,  s_0+ tS^{(n)}(\th)+(1-t)S^{(n-1)}(\th), \th\right)\right|\\
   &+&\big|\de_n S \big| \int_0^1\left |\pr_s F\left(u_0+ tU^{(n)}(\th)+(1-t)  U^{(n-1)}(\th), \, s_0+ tS^{(n)}(\th)+(1-t)S^{(n-1)}(\th)   , \th\right)\right|.
   \eeaa  
    We need to pay special attention on the axis\footnote{Indeed the term  $e_\th (e_\th \Phi)$  is quite singular on the axis. }, where $\sin\th=0$,   to the integral   term involving 
    \beaa
    \pr_u (e_\th \Phi)= \frac 1 2\left( e_3-\Omb e_4-\underline{b}\ga^{1/2} e_\th\right) e_\th \Phi.
    \eeaa
   This leads us to consider the  integral,
   \beaa
   \int_0^1\left[\underline{b}e_\th(e_\th(\Phi))\right](\ovu +t  U^{(n)}(\th)+(1-t)U^{(n-1)}(\th) , \ovs, \th)dt 
   \eeaa
   and  the  $L^2 $ norm of its product with $h$   on $\ovS$.
    We recall  (see Lemma \ref{lemma:Gauss-curvatureS1}) that   $\lapp \Phi=-K$. and Therefore,
    $\big|e_\th(e_\th \Phi)\big| \les r^{-2}+ |e_\th\Phi)|^2 $ The contribution due to $K$  does not present any difficulties on the axis therefore we are led to  consider  the  integral
    \beaa
 I(\th):=   \int_0^1\left[\underline{b} (e_\th(\Phi))^2\right](\ovu +t  U^{(n)}(\th)+(1-t)U^{(n-1)}(\th) , \ovs, \th)dt 
    \eeaa
    and  the $L^2$ norm of its product with $h$  on $\ovS$. Making use  of \eqref{eq:clearlyveryusefulforannoyingaxis1} and then the first estimate of  \eqref{eq:clearlyveryusefulforannoyingaxis2}  of  Lemma \ref{lemma:clearlyveryusefulforannoyingaxis1-2} together with our assumption {\bf A3} 
     we    derive the  bound,
     \beaa
     \big|I(\th)h(\th)\big |&\les&   \frac{1}{\sin^2\th}\left(\int_0^1 \left|\underline{b}(\ovu +t  U^{(n)}(\th)+(1-t)U^{(n-1)}(\th) , \ovs, \th)\right|dt \right) \left|h(\th)\right|\\
   &  \les &\Big| \frac{h(\th) }{\sin\th}\Big|\,  \sup_{\RR}\left|\frac{\underline{b}}{\sin\th}\right|
     \les r ^2\Big| \frac{h(\th) }{e^\Phi }\Big|\,  \sup_{\RR}\left|\frac{\underline{b}}{e^\Phi}\right|\les  \epg  r  \Big| \frac{h(\th) }{e^\Phi }\Big|.
     \eeaa
    Making   use of  the second estimate in  \eqref{eq:clearlyveryusefulforannoyingaxis1} we then derive,
    \beaa
    \|I  \c h \|_{L^2(\ovS)} &\les&\epg  r  \left \|\frac{h}{e^\Phi}\right\|_{L^2(S) }\les \epg   \|h\|_{\hk_1(S)}.
    \eeaa
    This shows that the behavior along the axis in \eqref{eq1:Lemma:differencesFn-integral}  is not an issue.
    This ends the proof of     both Lemma \ref{Lemma:differencesFn-integral-Main} and Lemma \ref{Lemma:differencesFn-integral}.
  \end{proof}
\end{proof}

  \begin{lemma}
  \label{Lemma:reduced-iterate-differences}
  Consider equations on $\ovS $,   of the form,
\bea
\label{sytem:two-iterates}
\bsplit
A^{(n)} h ^{(n+1)}+ V^{(n)} h^{(n+1)} &=H^{(n)},\\
A^{(n-1)} h^{(n)}+ V^{(n-1)} h^{(n)} &=H^{(n-1)}.
\end{split}
\eea
Also,
\beaa
  \int_{\ovS, \gS^{(n)}}   h^{(n+1)} e^{\Phi^{\#_n}} &=& B^{(n)},\\
  \int_{\ovS, \gS^{(n-1)}}   h^{(n)} e^{\Phi^{\#_{n-1}}} &=& B^{(n-1)}.
\eeaa
Then we have,
\bea
\bsplit
 \| h^{(n+1)}- h^{(n)}   \|_{\hk_1(\ovS)}&\les  \dg  r^{-1}      \|\pr_\th\left( \Psi^{(n)}-\Psi^{(n-1)}\right)\|_{\hk_1(\ovS)}\\
   &+ r \| H^{(n+1)}-H^{(n)} \|_{L^2(\ovS)}+r^{-2} \Big| B^{(n)} -B^{(n-1)} \Big|.
   \end{split}
 \eea
    \end{lemma}

  \begin{proof}
  Setting
  \beaa
  \de\hnn:&=&\hnn-\hn, \quad \de A^{(n)}:=A^{(n)}-A^{(n-1)}, \quad \de V^{(n)}:=V^{(n)}-V^{(n-1)}, \quad  \\
  \de H^{(n+1)}:&=&H^{(n+1)} -H^{(n)},\quad  \de B^{(n)}:=B^{(n)}-B^{(n-1)},
  \eeaa
   and  subtracting  the equations \eqref{sytem:two-iterates} we  derive,
  \bea
 ( A^{(n)}  +    V^{(n)} )  \de\hnn=-    ( \de A^{(n)}) \hn- \de V^{(n)}  \hn+\de H^{(n+1)}.  
  \eea
  Also writing   $  \int_{\ovS, \gS^{(n)}} f=\int_{\ovS} f d a_{\gS^{(n)}} $
  \bea
  \label{eq:difference-badmodes}
   \int_{\ovS, \gS^{(n)}}   \de h^{(n+1)}  e^{\Phi^{\#_n}} &=&\de B_n -\int_{\ovS} h^{(n)} \left( e^{\Phi^{\#_n}} da _{\gS^{(n)}} -  e^{\Phi^{\#_{n-1} }} da _{\gS^{(n-1)}}\right).
   \eea
  Throughout the proof below we make  systematic use of  the boundedness result of Proposition \ref {Prop:BondsforQn}   and comparison Lemma \ref{lemma:int-gaS-ga}, according to which  the spaces $\hk_k(\ovS, \gS^{(n)} )$ are all  uniformly equivalent to $\hk_k(\ovS)$.

  {\bf Step 1.} Estimates for   $\| \dds_2^{(n)}\big( \de\hnn\big)\|_{\hk_1(\ovS)}$.
   
  As in  Remark \ref{remark:GCMS:NEAR-again:bis1}, see also      the proof of Proposition \ref{prop.GCMSequations-fixedS} in the previous  section,   we write,
  \beaa
    A^{(n)}= \dds_1\,^{(n)}  \ddd_1\,^{(n)}= \ddd_2\,^{(n)}  \dds_2\,^{(n)}-  (\Psi^{(n)})^\# \left(-3\rho+\frac 1 2\vth\vthb\right).
  \eeaa
   Hence,
   \beaa
   \int_{(\ovS, \gS^{(n)})}\de\hnn(A^{(n)}  +V^{(n)}) \de\hnn&=&  \int_{(\ovS, \gS^{(n)})} |  \dds_2^{(n)} ( \de\hnn)|^2\\
   &+&\left( \frac{6m}{r^3}+ O( r^{-3} \dg )\right) \int_{(\ovS, \gS^{(n)}) }| \de\hnn|^2. 
   \eeaa
   Since $\dg$ is small,
   \beaa
 &&  \int_{(\ovS, \gS^{(n)})} |  \dds_2\,^{(n)} ( \de\hnn)|^2\les   \int_{(\ovS, \gS^{(n)})}\de\hnn(A^{(n)}  +V^{(n)}) \de\hnn\\
 &&\les \left|\int_{(\ovS, \gS^{(n)})}\de\hnn ( \de A^{(n)} h^{(n)} ) \right| 
 +   \int_{(\ovS, \gS^{(n)})} \Big|\de\hnn\de V^{(n)}  \hn\Big|+ \int_{(\ovS, \gS^{(n)})}\Big|\de\hnn\de H^{(n+1)}\Big|.
 \eeaa
 We deduce, making use of the comparison Lemma and the boundedness result of Proposition \ref{Prop:BondsforQn}, 
  \bea
   \lab{eq:GCMS-difference-n-1}
  \bsplit
  \| \dds_2^{(n)} \de\hnn\|^2_{\hk_1(\ovS) }&\les \left|\int_{(\ovS, \gS^{(n)})}\de\hnn ( \de A^{(n)} h^{(n)} ) \right|\\
&+\left(\dg r^{-1} \|\de V^{(n)}\|_{L^2(\ovS)}+\|\de H^{(n+1)}\|_{L^2(\ovS)}\right)\|\de \hnn\|_{L^2(\ovS )}.
\end{split}
  \eea

 {\bf Step 1a.} We estimate the term  $ \left|\int_{(\ovS, \gS^{(n)})}\de\hnn ( \de A^{(n)} h^{(n)} ) \right| $
using Lemma \ref{Lemma:differencesFn-integral} with $\psi= \de\hnn$ and $h=\hn$. We deduce, making use 
of the boundedness 
\beaa
&& \left|\int_{(\ovS, \gS^{(n)})}\de\hnn ( \de A^{(n)} h^{(n)} ) \right| \les  r^{-2} \|\pr_\th\left(\Psi^{(n)}-\Psi^{(n-1)}\right)\|_{\hk_1(\ovS)} \|\de \hnn  \|_{\hk_2(\ovS)}\| \hn \|_{\hk_2(\ovS)} \\
\les&& r^{-2}\dg  \|\de \hnn  \|_{\hk_2(\ovS)}  \|\pr_\th\left(\Psi^{(n)}-\Psi^{(n-1)}\right)\|_{\hk_1(\ovS)}.
\eeaa
Back to \eqref{eq:GCMS-difference-n-1} we deduce,
\bea
\label{eq:GCMS-difference-n-2}
\bsplit
 & \| \dds_2^{(n)} \de\hnn\|^2_{\hk_1(\ovS) }\\
 &\les\left(\dg r^{-1} \|\de V^{(n)}\|_{L^2(\ovS)}+r^{-2} \dg  \|\pr_\th(\de \Psi^{(n)})\|_{\hk_1(\ovS)}  +\|\de H^{(n+1)}\|_{L^2(\ovS)}\right)\|\de \hnn\|_{L^2(\ovS )}.
 \end{split}
\eea

 {\bf Step 1b.}  We estimate  $\|\de V^{(n)}\|_{L^2(\ovS)}$.
 Recall that 
  \beaa
\de V^{(n)} &=& \left(\frac 1 2 \ka \kab -\rho\right)^{\#_n} - \left(\frac 1 2 \ka \kab -\rho\right)^{\#_{n-1}}.
\eeaa  
 In view of  Lemma \ref{Lemma:differencesFn} and our assumptions {\bf A1-A3} we derive,
 \beaa
 \|\de V^{(n)}\|_{L^2(\ovS)} &\les& r^{-2}  \|(\de U^{(n)}, \de S^{(n)})\|_{L^2(\ovS)}.
 \eeaa 
 Therefore, back to \eqref{eq:GCMS-difference-n-2}, we deduce
 \bea
  \| \dds_2^{(n)} \de\hnn\|^2_{\hk_1(\ovS) }
 &\les&\left(r^{-2} \dg  \|\pr_\th(\de \Psi^{(n)})\|_{\hk_1(\ovS)}  +\|\de H^{(n+1)}\|_{L^2(\ovS)}\right)\|\de \hnn\|_{L^2(\ovS )}\nn
 \label{eq:GCMS-difference-n-3}.\\
 \eea
 
  {\bf Step 2.} 
  We make use of the equation \eqref{eq:difference-badmodes}
   \beaa
   \int_{\ovS, \gS^{(n)}}   \de h^{(n+1)}  e^{\Phi^{\#_n}} &=&\de B_n -\int_{\ovS} h^{(n)} \left( e^{\Phi^{\#_n}} da _{\gS^{(n)}} -  e^{\Phi^{\#_{n-1} }} da _{\gS^{(n-1)}}\right)
   \eeaa
from which we deduce,
\beaa
\left|  \int_{\ovS, \gS^{(n)}}   \de h^{(n+1)}  e^{\Phi^{\#_n}} \right|&\les& \left|\int_{\ovS} h^{(n)} \left( e^{\Phi^{\#_n}} da _{\gS^{(n)}} -  e^{\Phi^{\#_{n-1} }} da _{\gS^{(n-1)}}\right)\right|+\Big|\de B_n\Big|\\
&\les& \|\hn\|_{L^2(\ovS)}\left\|  \left( e^{\Phi^{\#_n}} \frac{\sqrt{\ga^{(n)}} }{\sqrt{\ga^{(0)}}}  -  e^{\Phi^{\#_{n-1} }}\frac{\sqrt{\ga^{(n-1)}}} {\sqrt{\ga^{(0)}}}\right)\right\|_{L^2(\ovS)}+\Big|\de B_n\Big|.
\eeaa
On the other hand,
\beaa
 \left( e^{\Phi^{\#_n}} \frac{\sqrt{\ga^{(n)}} }{\sqrt{\ga^{(0)}}}  -  e^{\Phi^{\#_{n-1} }}\frac{\sqrt{\ga^{(n-1)}}} {\sqrt{\ga^{(0)}}}\right)&=& \left( e^{\Phi^{\#_n}}  -  e^{\Phi^{\#_{n-1} }} \right)\frac{\sqrt{\ga^{(n)}} }{\sqrt{\ga^{(0)}}}
 + e^{\Phi^{\#_{n-1} }}  \left( \frac{\sqrt{\ga^{(n)}} }{\sqrt{\ga^{(0)}}}  -  \frac{\sqrt{\ga^{(n-1)}}} {\sqrt{\ga^{(0)}}}\right).
 \eeaa
 Hence, proceeding as before,
 \beaa
 \left\|  \left( e^{\Phi^{\#_n}} \frac{\sqrt{\ga^{(n)}} }{\sqrt{\ga^{(0)}}}  -  e^{\Phi^{\#_{n-1} }}\frac{\sqrt{\ga^{(n-1)}}} {\sqrt{\ga^{(0)}}}\right)\right\|_{L^2(\ovS)}&\les&  \left\| e^{\Phi^{\#_n}}  -  e^{\Phi^{\#_{n-1} }} \right\|_{L^2(\ovS)}
 +r^{-1} \left\| \ga^{\#_n}  -  \ga^{\#_{n-1} } \right\|_{L^2(\ovS)}\\
 &\les& r     \|\pr_\th\left( \Psi^{(n)}-\Psi^{(n-1)}\right)\|_{\hk_1(\ovS)}.
 \eeaa
 Therefore,
 \bea
 \label{eq:GCMS-difference-n-4}
\left|  \int_{\ovS, \gS^{(n)}}   \de h^{(n+1)}  e^{\Phi^{\#_n}} \right|&\les&\dg  r     \|\pr_\th\left( \Psi^{(n)}-\Psi^{(n-1)}\right)\|_{\hk_1(\ovS)}+\Big|\de B_n\Big|.
 \eea
 
 {\bf Step 3.} Making use of Lemma \ref{prop:2D-hodge-reduced-GCM}  together with \eqref{eq:GCMS-difference-n-3},  \eqref{eq:GCMS-difference-n-4} and the comparison Lemma and    we deduce,
 \beaa
  \| \de \hnn \|_{L^2 (\ovS, \gS^{(n)} )}&\les&  r   \| \dds_2^{(n)} \de\hnn\|_{\hk_1(\ovS, \gS^{(n)} ) }+ r^{-2} \left|  \int_{\ovS, \gS^{(n)}}   \de h^{(n+1)}  e^{\Phi^{\#_n}} \right|\\
  &\les& r   \| \dds_2^{(n)} \de\hnn\|_{\hk_1(\ovS) }+\dg  r^{-1}      \|\pr_\th\left( \Psi^{(n)}-\Psi^{(n-1)}\right)\|_{\hk_1(\ovS)}+r^{-2} \Big|\de B_n\Big|\\
&\les&  r  \left(r^{-2} \dg  \|\pr_\th(\de \Psi^{(n)})\|_{\hk_1(\ovS)}  +\|\de H^{(n+1)}\|_{L^2(\ovS)}\right)^{1/2}\|\de \hnn\|^{1/2}_{L^2(\ovS )}\\
&+&\dg  r^{-1}      \|\pr_\th\left( \Psi^{(n)}-\Psi^{(n-1)}\right)\|_{\hk_1(\ovS)}+r^{-2} \Big|\de B_n\Big|.
 \eeaa
 We infer that,
 \beaa
   \| \de \hnn \|_{L^2 (\ovS, \gS^{(n)} )}&\les&  \dg  r^{-1}      \|\pr_\th\left( \Psi^{(n)}-\Psi^{(n-1)}\right)\|_{\hk_1(\ovS)}+ r \|\de H^{(n+1)}\|_{L^2(\ovS)}+r^{-2} \Big|\de B_n\Big|
 \eeaa
 which, together with \eqref{eq:GCMS-difference-n-3},  ends the proof of Lemma \ref{Lemma:reduced-iterate-differences}. 
 \end{proof}

 
  \subsection{Proof of the estimates \eqref{differences-Step1},  \eqref{differences-Step3}}
  \lab{subsection:endofproof-155-156}

  
  We are now in position to prove \eqref{differences-Step1} \eqref{differences-Step3}.
  
     {\bf Step 1.} With start by estimating $  \de h^{(n+1)},   \de \hb ^{(n+1)}$. To this end, we need to apply  Lemma \ref{Lemma:reduced-iterate-differences} to the  equations   for  $\de w^{(n+1)}$, $\de h^{(n+1)}$, $\de\hb^{(n+1)}$,  derived from the first three equations in \eqref{system:af-fb-index-n-Pulled},                and  estimate the   corresponding $\de H^{(n+1)}$  on the right-hand side.  This is tedious  but straightforward  and one derives, in the case of  estimates of $\de w^{(n+1)}$
  \beaa
  \|\de H^{(n+1)}\|_{L^2(\ovS)} &\les& \epg\Bigg\{\|U^{(n)}-U^{(n-1)} \|_{\hk_2(\ovS)}  +   \|S^{(n)}-S^{(n-1)} \|_{\hk_2(\ovS)} \\
 &+&  \| \de h^{(n+1)} \|_{\hk_1(\ovS)}+   \|\de \hb^{(n+1)} \|_{\hk_1(\ovS)} + \| \de e^{(n+1)} \|_{\hk_1(\ovS)}\Bigg\}
 \eeaa
  and in the case of $\de h^{(n+1)}$, $\de\hb^{(n+1)}$ 
  \beaa
  \|\de H^{(n+1)}\|_{L^2(\ovS)} &\les& \| \de w^{(n+1)} \|_{\hk_1(\ovS)} + \epg\Bigg\{\|U^{(n)}-U^{(n-1)} \|_{\hk_2(\ovS)}  +   \|S^{(n)}-S^{(n-1)} \|_{\hk_2(\ovS)} \\
 &+&  \| \de h^{(n+1)} \|_{\hk_1(\ovS)}+   \|\de \hb^{(n+1)} \|_{\hk_1(\ovS)} + \| \de e^{(n+1)} \|_{\hk_1(\ovS)}\Bigg\}.
 \eeaa 
 
    \begin{remark}
   Note that the presence of the  inverse operators    $(A^{(n)})^{-1}$ in the right-hand side of the equations for $\de h^{(n+1)}$, $\de\hb^{(n+1)}$     do  not  create any difficulties when taking differences. Indeed we can write,
   \beaa
    (A^{(n)})^{-1}- (A^{(n-1)})^{-1}&=&  (A^{(n)})^{-1}\left(  A^{(n-1)}- A^{(n)}\right) (A^{(n-1)})^{-1}
   \eeaa    
   and estimate  the difference $\de A^{(n)}=  A^{(n)}- A^{(n-1)}$ as in the proof of Lemma \ref{Lemma:reduced-iterate-differences}.
   \end{remark}

 We infer from  Lemma \ref{Lemma:reduced-iterate-differences} and the above estimates 
  \beaa
     \| \de w^{(n+1)}  \|_{\hk_1(\ovS)}  &\les& \epg\Bigg\{\|U^{(n)}-U^{(n-1)} \|_{\hk_2(\ovS)}  +   \|S^{(n)}-S^{(n-1)} \|_{\hk_2(\ovS)} \\
 &+&  \| \de h^{(n+1)} \|_{\hk_1(\ovS)}+   \|\de \hb^{(n+1)} \|_{\hk_1(\ovS)} + \| \de e^{(n+1)} \|_{\hk_1(\ovS)}\Bigg\}  
 \eeaa
and 
  \beaa
      \| \de h^{(n+1)},   \de \hb ^{(n+1)}    \|_{\hk_1(\ovS)} &\les& \| \de w^{(n+1)} \|_{\hk_1(\ovS)}+ \epg\Bigg\{\|U^{(n)}-U^{(n-1)} \|_{\hk_2(\ovS)}  +   \|S^{(n)}-S^{(n-1)} \|_{\hk_2(\ovS)} \\
 &+&  \| \de h^{(n+1)} \|_{\hk_1(\ovS)}+   \|\de \hb^{(n+1)} \|_{\hk_1(\ovS)} + \| \de e^{(n+1)} \|_{\hk_1(\ovS)}\Bigg\}.  
 \eeaa
This thus yields
  \bea
   \label{Step1:estimatesfor-de-dehb} 
\nn    \| \de h^{(n+1)},   \de \hb ^{(n+1)}     \|_{\hk_1(\ovS)}  &\les& \epg    \|U^{(n)}-U^{(n-1)} \|_{\hk_2(\ovS)}    +   \epg\|S^{(n)}-S^{(n-1)} \|_{\hk_2(\ovS)}\\
    && +\epg  \| \de e^{(n+1)} \|_{\hk_1(\ovS)}.
   \eea

  {\bf Step 2.}     Next, we estimate $\dds\,\,\de e^{(n+1)}$. We make use of the last equation  in
   \eqref{system:af-fb-index-n-Pulled} which we write int the form
   \beaa
     \dds\,\,^{(n)}\enn&=&H^{(n+1)}
     \eeaa
     with,
     \beaa
H^{(n+1)}&=&\frac 3 4  (A^{(n)})^{-1}(\rho^{(n)} w^{(n+1)}) - h^{(n+1)}\omb^{(n)}  +\hbnn \om^{(n)}   -\frac{1}{4}h^{(n+1)}\kab^{(n)} +\frac 14   \fb^{(n+1)} \ka^{(n)}\\
   &+& (A^{(n)})^{-1}\big(-\dds\,\,^{(n)} \check{\mu}^{(n)}  + (\Psi^{(n)})^\#\err^{(n+1)}_3\big).
  \eeaa
   Hence,
   \beaa
      \dds\,\,^{(n)}( \de  \enn)+(\dds\,\,^{(n)}- \dds\,\,^{(n-1)} )e^{(n)}=\de H^{(n+1)}
   \eeaa
   which can be written in the form,
   \beaa
 \frac{\sqrt{\gazero}}{\sqrt{\ga^{(n)}}}  \dds\,\, (\de\enn)&=&- (\dds\,\,^{(n)}- \dds\,\,^{(n-1)} )e^{(n)}+\de H^{(n+1)}\\
 &=&-\left(\frac{\sqrt{\gazero}}{\sqrt{\ga^{(n)}}} -\frac{\sqrt{\gazero}}{\sqrt{\ga^{(n-1)}}} \right)\dds\,\, e^{(n)}+\de H^{(n+1)}
   \eeaa
   or,
   \beaa
    \dds\,\,(\de \enn)&=&\left(1 -\frac{\sqrt{\ga^{(n)}}}{\sqrt{\ga^{(n-1)}}} \right)\dds\,\, e^{(n)}+  \frac{\sqrt{\ga^{(n)}}} {\sqrt{\gazero}} \de H^{(n+1)}.
   \eeaa
   This yields
   \beaa
   \| \dds\,\,(\de \enn) \|_{L^2(\ovS)} &\les&  \epg\| \ga^{(n)}-\ga^{(n-1)} \|_{L^2(\ovS)}+ \| \de H^{(n+1)}\|_{L^2(\ovS)}
   \eeaa
where we used  the uniform bound of $P^{(n)}$ provided by Proposition \ref{Prop:BondsforQn} and the bound  \eqref{eq:assumtionsonthegivenusfoliationforGCMprocedure:bis} for $(\Omb, \underline{b}, \ga)$. $\ga^{(n)}-\ga^{(n-1)}$ can be estimated as in the proof of Lemma \ref{Lemma:reduced-iterate-differences}, and hence
 \beaa
   \| \dds\,\,(\de \enn) \|_{L^2(\ovS)} &\les&  \epg\left(  \|U^{(n)}-U^{(n-1)} \|_{\hk_2(\ovS)} + \|S^{(n)}-S^{(n-1)} \|_{\hk_2(\ovS)} \right)   + \| \de H^{(n+1)}\|_{L^2(\ovS)}.
   \eeaa

The differences $H^{(n+1)}- H^{(n)}$ can also be easily estimated and deduce,
\beaa
\|H^{(n)}- H^{(n-1)}\|_{L^2(\ovS)}&\les& \|\de h^{(n+1)}\|_{\hk_1(\ovS)} +\|\de \hb^{(n+1)}\|_{\hk_1(\ovS)}\\
&+&\epg\left(  \|U^{(n)}-U^{(n-1)} \|_{\hk_2(\ovS)} + \|S^{(n)}-S^{(n-1)} \|_{\hk_2(\ovS)} \right).
\eeaa
We obtain
\bea\lab{youpitralalatsouintsoin}
\|\dds\,\,(\de  e^{(n+1)})    \|_{L^2(\ovS)} &\les&  \|\de h^{(n+1)}\|_{\hk_1(\ovS)} +\|\de \hb^{(n+1)}\|_{\hk_1(\ovS)}\\
\nn&+&\epg\left(  \|U^{(n)}-U^{(n-1)} \|_{\hk_2(\ovS)} + \|S^{(n)}-S^{(n-1)} \|_{\hk_2(\ovS)} \right).
\eea

  {\bf Step 3.} Next, we estimate the average of $\de  e^{(n+1)}$. Recall from \eqref{eq:retrieve-a-indexn-dif}         
  \beaa
\ov{e^{(n+1)}}^{\ovS, \gS^{(n)}}  = \ov{\left( 1-\frac{r^{\S(n)}}{r}\right)^{(n)}}^{\ovS, \gS^{(n)}}        - \frac{r^{\S(n)}}{2}\ov{\left( \check{\ka} + \ov{\ka}-\frac{2}{r} \right)^{(n)}}^{\ovS, \gS^{(n)}}+\ov{\err_8^{(n+1)}}^{\ovS, \gS^{(n)}}.
\eeaa
and 
\beaa
\ov{e^{(n)}}^{\ovS, \gS^{(n)}}  = \ov{\left( 1-\frac{r^{\S(n-1)}}{r}\right)^{(n-1)}}^{\ovS, \gS^{(n-1)}}        - \frac{r^{\S(n-1)}}{2}\ov{\left( \check{\ka} + \ov{\ka}-\frac{2}{r} \right)^{(n)}}^{\ovS, \gS^{(n-1)}}+\ov{\err_8^{(n)}}^{\ovS, \gS^{(n-1)}}.
\eeaa
Taking the difference, recalling that we have in the $(\th, \vphi)$ coordinates system
\beaa
dvol\gS^{(n)} &=& \sqrt{\ga^{(n)}}e^{\Phi^{\#_n}}d\th d\vphi,\quad r^{\S(n)}=\int_{\ovS}\sqrt{\ga^{(n)}}e^{\Phi^{\#_n}}d\th d\vphi
\eeaa
and
\beaa
dvol\gS^{(n-1)} &=& \sqrt{\ga^{(n-1)}}e^{\Phi^{\#_{n-1}}}d\th d\vphi,\quad r^{\S(n-1)}=\int_{\ovS}\sqrt{\ga^{(n-1)}}e^{\Phi^{\#_{n-1}}}d\th d\vphi
\eeaa
and using the uniform bound of $P^{(n)}$ provided by Proposition \ref{Prop:BondsforQn} and  the bounds  {\bf A1}   for $\Gac$, we infer
\beaa
\left|\ov{\de e^{(n+1)}}^{\ovS}\right| &\les& \epg\left\{\|\ga^{(n)}-\ga^{(n-1)}\|_{L^2(\ovS)}+\|\Phi^{\#_n}-\Phi^{\#_{n-1}}\|_{L^2(\ovS)}+\|\de \err_6^{(n+1)}\|_{L^2(\ovS)}\right\}.
\eeaa
Arguing as above, we deduce
\beaa
\left|\ov{\de e^{(n+1)}}^{\ovS}\right| &\les&   \epg\Big\{\|\de h^{(n+1)}\|_{\hk_1(\ovS)} +\|\de \hb^{(n+1)}\|_{\hk_1(\ovS)}+\|\de e^{(n+1)}\|_{\hk_1(\ovS)}\\
\nn&&+  \|U^{(n)}-U^{(n-1)} \|_{\hk_2(\ovS)} + \|S^{(n)}-S^{(n-1)} \|_{\hk_2(\ovS)} \Big\}.
\eeaa
Together with \eqref{Step1:estimatesfor-de-dehb} \eqref{youpitralalatsouintsoin}, and an elliptic estimate for $\dds$, we infer
\bea
 \| \de h^{(n+1)},   \de \hb ^{(n+1)},  \de  e^{(n+1)}    \|_{\hk_1(\ovS)} \les  \epg    \|U^{(n)}-U^{(n-1)} \|_{\hk_2(\ovS)}    +   \epg\|S^{(n)}-S^{(n-1)} \|_{\hk_2(\ovS)}
\eea
which concludes the proof of \eqref{differences-Step1}.

  {\bf Step 4.}  Finally, we focus on \eqref{differences-Step3}. Recall  \eqref{eq:someovertherainbow-pullback} 
  \bea
       \bsplit
  \vsi^{\#_n}   \pr_\th U^{(1+n)} &=  (\ga^{(n)}  )^{1/2}  h^{(1+n)}\left(1 +\frac 1 4h ^{(1+n)}\hb^{(1+n)}\right),  \\
     \pr_\th S^{(1+n)}-\frac 1 2  \vsi^{\#_n} \Omb^{\#_{n}}  \pr_\th U^{(1+n)}&=   \frac 1 2 (\ga^{(n)}  )^{1/2}\hb^{(1+n)},
     \\
   \ga^{(n)} & =\ga^{\#_n}+\big(\vsi^{\#_n}\big)^2  \left(\Omb^{\#_n}+\frac 1 4( \underline{b}^{\#_n} )^2  \ga^{\#_n}  \right)\, ( \pr_\th U^{(n)} )^2 \\
     &-2  \vsi^{\#_n} \pr_\th U ^{(n)} \pr_\th  S^{(n)}-\ga^{\#_n} \vsi^{\#_n} \underline{b}^{\#_n} \pr_\th U^{(n)},\\
     U^{(1+n)}(0)&= S^{(1+n)}(0)=0.
     \end{split}
  \eea  
Taking the difference and arguing as above, we derive
\beaa
\|\pr_\th \de U^{(1+n)}\|_{\hk_1(\ovS)} &\les&  \| \de h^{(n+1)}, \de \hb ^{(n+1)} \|_{\hk_1(\ovS)}+\epg\|\ga^{(n)}-\ga^{(n-1)}\|_{\hk_1(\ovS)},\\
\|\pr_\th \de S^{(1+n)}\|_{\hk_1(\ovS)}  &\les& \| \de U^{(1+n)}\|_{\hk_2(\ovS)}+\epg\| \de S^{(1+n)}\|_{\hk_2(\ovS)}+ \| \de h^{(n+1)}, \de \hb ^{(n+1)} \|_{\hk_1(\ovS)}\\
&&+\epg\|\ga^{(n)}-\ga^{(n-1)}\|_{\hk_1(\ovS)}.
\eeaa
Since $\de U^{(1+n)}=\de S^{(1+n)}=0$, we deduce
\beaa
\|\de U^{(1+n)}\|_{\hk_2(\ovS)} &\les&  \| \de h^{(n+1)}, \de \hb ^{(n+1)} \|_{\hk_1(\ovS)}+\epg\|\ga^{(n)}-\ga^{(n-1)}\|_{\hk_1(\ovS)},\\
\|\de S^{(1+n)}\|_{\hk_2(\ovS)}  &\les& \| \de U^{(1+n)}\|_{\hk_2(\ovS)}+\epg\| \de S^{(1+n)}\|_{\hk_2(\ovS)}+ \| \de h^{(n+1)}, \de \hb ^{(n+1)} \|_{\hk_1(\ovS)}\\
&&+\epg\|\ga^{(n)}-\ga^{(n-1)}\|_{\hk_1(\ovS)}
\eeaa
and hence
       \beaa
    \| \de U^{(n+1)},  \de S^{(n+1)} \|_{\hk_2(\ovS)}\les   \| \de h^{(n+1)}, \de \hb ^{(n+1)}  \|_{\hk_1(\ovS)} +  \epg  \|\de U^{(n)}, \de S^{(n)} \|_{\hk_2(\ovS)}+\epg\|\ga^{(n)}-\ga^{(n-1)}\|_{\hk_1(\ovS)}.
    \eeaa
 Estimating $\ga^{(n)}-\ga^{(n-1)}$ as above, we infer
       \beaa
    \| \de U^{(n+1)},  \de S^{(n+1)} \|_{\hk_2(\ovS)}\les   \| \de h^{(n+1)}, \de \hb ^{(n+1)}  \|_{\hk_1(\ovS)} +  \epg  \|\de U^{(n)}, \de S^{(n)} \|_{\hk_2(\ovS)}+\epg \| \de U^{(n)},  \de S^{(n)} \|_{\hk_2(\ovS)}.
    \eeaa
 This is the desired estimate \eqref{differences-Step3} and hence concludes the proof of Proposition \ref{Proposition-contraction}.


\section{A Corollary to  Theorem  \ref{Theorem:ExistenceGCMS}}
\lab{section:Corollary-ExistenceGCMS}


The following result is a  simple Corollary of  Theorem \ref{Theorem:ExistenceGCMS}.
\begin{theorem}[Existence of GCM spheres] 
\lab{Theorem:ExistenceGCMS2}
In addition to  the assumptions  of Theorem \ref{Theorem:ExistenceGCMS} we  assume that, for any background sphere $S$ in $\RR$, 
\bea
\lab{assumptions:Theorem-ExistenceGCMS2}
 r \left|\int_{S} \b e^\Phi\right|+\left|\int_{S}  e_\th (\kab) e^\Phi\right| &\les& \dg.   
\eea
Then there exists a unique GCM sphere  $\S$, which is a deformation of $\ovS$,  such that  the GCMS conditions  hold true
 \bea
 \label{Conditions:GCMS2}
\bsplit
\ddsS_2\ddsS_1\kab^\S&=\ddsS_2\ddsS_1\mu^\S=0, \qquad \ka^\S=\frac{2}{r^\S},\\
\int_{\S}  \b^\S e^\Phi&=0, \qquad 
\int_{\S}  e^\S_\th(\kab^\S) e^\Phi=0.
\end{split}
\eea
Moreover all  other  estimates of Theorem  \ref{Theorem:ExistenceGCMS} hold true.
\end{theorem}

\begin{proof}
The proof of the theorem follows easily in view of  Theorem  \ref{Theorem:ExistenceGCMS}  and the following lemma.
\begin{lemma}
\lab{Lemma:ExistenceGCMS2}
Let $\S$ be  a deformation of $\ovS$ as in Theorem  \ref{Theorem:ExistenceGCMS} with $\La=\int_\S f e^\Phi, \Lab=\int_\S \fb e^\Phi$.
The following  identities hold true.
\bea
\lab{system-La-Lab}
\bsplit
\La&=\frac{r^3}{3m}\left( \int_{\ovS} \b e^\Phi   -\int_\S \b^\S e^\Phi\right) + F_1(\La, \Lab)\lab{eq:Thm-GCMS2-1},\\
 \Lab&=\frac{r^2}{2(1+\frac m r )}\left( \int_{\ovS}( e_\th \kab) e^\Phi -\int_\S( e_\th^\S \kab^\S) e^\Phi  +\Up \int_{\ovS} e_\th(\ka) e^\Phi \right)\\
&+\frac{\Up}{2(1+\frac m r )}\La
+ F_2(\La, \Lab),  
\end{split}
\eea
 where $F_1, F_2$ are   continuous\footnote{In fact smooth.} in $\La, \Lab$,  with $F_1(0, 0)=F_2(0,0)=0$,  verifying the estimates,
 \beaa
 \big|F_1|+\big|F_2\big| &\les& \epg \dg r^2.
 \eeaa
 \end{lemma}

\begin{proof}
 To prove  \eqref{eq:Thm-GCMS2-1}  we start with   the change of frame formula,
\beaa
\b^\S&=& e^a\left(\b+\frac{3}{2}\rho f\right)+ e^a \err(\b,\b^\S),
\\
\err(\b,\b^\S) &=& \frac{1}{2}\fb\a+\lot
\eeaa
We write\footnote{ Here $(r, m)$ represents the area radius  and Hawking of $\ovS$  while  $(r^\S, m^\S)$  represent the  area radius  and Hawking of $\S$. Since $|\frac{r^\S}{r}-1| \les \dg$, $|m^\S-m |\les \dg $ we   can   interchange freely $r^\S$  with $r$ and $m^\S$ with $m$.}
\beaa
\b^\S&=&\b+\frac{3}{2}\rho f+ (e^a-1) \left(\b+\frac{3}{2}\rho f\right)+ e^a \err(\b,\b^\S)\\
&=& \b+\frac{3}{2}\left (-\frac{2 m}{ r ^3}\right)  f+ \frac{3}{2}\left (\rho+\frac{2 m}{r^3}\right) f+(e^a-1) \left(\b+\frac{3}{2}\rho f\right)+ e^a \err(\b,\b^\S)
\eeaa
and deduce,
\beaa
\b^\S+\frac{3 m^\S}{(r^\S)^3} f &=&\b+\err'(\b, \b^\S)\\
\eeaa
with error term $\err'(\b, \b^\S)$,
\beaa
\err'(\b, \b^\S)&=&\left(\frac{3 m^\S}{(r^\S)^3}      - \frac{3 m}{r^3} f \right)   +  \frac{3}{2}\left (\rho+\frac{2 m}{r^3}\right) f+(e^a-1) \left(\b+\frac{3}{2}\rho f\right)+ e^a \err(\b,\b^\S).
\eeaa
Making use of the assumptions {\bf A1-A3} , the estimates of Theorem \ref{Theorem:ExistenceGCMS} for $(f, \fb,  a)$ as well as  the bounds for 
$\ovr -r^\S$, $\ovm-m^\S$  we deduce
\beaa
\Big| \err'(\b, \b^\S)\Big|&\les& r^{-1} \dg \epg. 
\eeaa
Thus,
\beaa
\frac{3 m^\S}{(r^\S)^3} \int_\S f e^\Phi&=&\int_\S \b e^\Phi  -\int_\S \b^\S e^\Phi +\int_\S \err'(\b, \b^\S) e^\Phi\\
&=& \int_{\ovS} \b e^\Phi   -\int_\S \b^\S e^\Phi +\int_\S \err'(\b, \b^\S) e^\Phi+ \Big(\int_\S \b e^\Phi-\int_{\ovS} \b e^\Phi\Big)
\eeaa
or,
\beaa
\frac{3 m}{ r^3} \int_\S f e^\Phi&=& \int_S \b e^\Phi   -\int_\S \b^\S e^\Phi +\int_\S \err'(\b, \b^\S) e^\Phi+ \Big(\int_\S \b e^\Phi-\int_S \b e^\Phi\Big)\\
&+&\Big(\frac{3 m}{ r^3} -\frac{3 m^\S}{(r^\S)^3}\Big) \int_\S f e^\Phi.
\eeaa
Clearly,
\beaa
\left|\int_\S \err'(\b, \b^\S) e^\Phi\right| &\les& r^2 \dg \epg. 
\eeaa
Also, proceeding exactly as in Corollary \ref{corr:lemma:int-gaS-ga} we 
deduce,
\beaa
\left|\int_\S \b e^\Phi  -\int_{\ovS} \b e^\Phi \right| &\les& \dg\left(    \sup_{\RR}  \rg   \big|\dkout^{\le 1}   (\b e^\Phi)\big|+          \sup_\RR \rg^2 \big| e_3 (\b e^\Phi\big)|\right).
\eeaa
Thus, in view of the assumptions  {\bf A1-A3},
\bea
\left|\int_\S \b e^\Phi  -\int_{\ovS} \b e^\Phi \right| &\les& \dg\epg r^2.
\eea
We deduce,
\beaa
\La&=&\frac{r^3}{3m}\Big( \int_{\ovS} \b e^\Phi   -\int_\S \b^\S e^\Phi\Big) + F_1(\La, \Lab)
\eeaa
 where  the error term $F_1(\La, \Lab)$ is a continuous function of $\La, \Lab$ verifying the estimate,
 \beaa
 \Big|F_1(\La, \Lab)\Big| &\les&\epg \dg r^2. 
 \eeaa
We also recall, see Lemma \ref{lemma:transfethka'-ethka}
\beaa
\bsplit
 e_\th^\S(\kab^\S )     &= e_\th \kab -\ddsS_1 \dddS_1  \fb  -\kab e_\th^\S a -\frac 1 4 \kab ( f\kab +\fb \ka)+\kab ( \fb \om- \omb  f )+\fb \rho\\
 &+\err(e_\th^\S\kab^\S, e_\th\kab)
 \end{split}
\eeaa
where,
\beaa
 \err(e_\th^\S \kab^\S,  e_\th\kab)&=&( e^{-a} -1)\Big(  e_\th \kab  -\ddsS \dddS_1  \fb + \frac 1 2  f  e_3\kab +\frac 1 2 \fb e_4\kab \Big)\\
 &+&e^{-a} \left[ e_\th^\S \, \err(\kab,\kab^\S) +e_\th^\S( a)\Big(\dddS_1 \fb   +\err(\kab, \kab^\S) \Big) +\frac 1 2   f \fb e_\th \kab +\frac 1 8 f^2 \fb e_3\kab\right]\\
 &+&\frac 1 2 \fb  \left( 2\ddd_1\etab -\frac 1 2  \vthb\, \vth +2(\xi\xib+\etab^2)\right)\\
&+& \frac 1 2   f \fb e_\th \kab +\frac 1 8 f^2 \fb e_3\kab+\frac 1 2 f  \left(  2\ddd_1\xib -\frac 1 2 \vthb^2 +2(\eta+\etab-2\ze)\xi\right).
\eeaa
Making  use of the  identity $\ddsS_1\dddS_1=\dddS_2\ddsS_2+2K$ we deduce,
\beaa
 e_\th^\S(\kab^\S )   +\left(\frac 1 4 \ka \kab -\rho+ 2K\right)\fb   &=& e_\th \kab- \dddS_2\ddsS_2  \fb + \kab e_\th^\S a -\frac 1 4 \kab^2   f -\kab  \,\omb  f +\err(e_\th^\S\kab^\S, e_\th\kab).
\eeaa
Writing, $k=\frac{2}{r} +(\ka-\frac 2 r)$,  $ \kab=\frac{-2\Up}{r}+(\kab+\frac{2\Up}{r}$, 
$\rho=-\frac{2m}{r^3}+ (\rho+\frac{2m}{r^3})$, $K=\frac{1}{r^2}+ (K-\frac{1}{r^2})$  
\beaa
\frac 1 4 \ka \kab -\rho+ 2K&=&\frac{1}{r^2}+\frac{4m}{r^3} +\frac{1}{2r}\left( \kab+\frac{2\Up}{r} \right)
-\frac{\Up}{2r} \left(\ka-\frac 2 r \right) +\left(\rho+\frac{2m}{r^3} \right)+2\left( K-\frac{1}{r^2}\right)\\
&=&\frac{1}{r^2}+\frac{4m}{r^3}+O(\epg r^{-2}).
\eeaa
Also,
\beaa
\kab&=&-\frac{2\Up}{r} +\left(\kab+\frac{2\Up}{r}\right)=-\frac{2\Up}{r} + O( r^{-2} \epg),\\
\kab\, \omb&=&-\frac{2m\Up}{r^3}+ O(r^{-2} \epg),
\eeaa
and, since $ a, f, \fb=O(r^{-1} \dg)$,
\beaa
 \err(e_\th^\S \kab^\S,  e_\th\kab)&=& O( r^{-3} \dg \epg).
\eeaa
We deduce,
\beaa
 e_\th^\S(\kab^\S )   +\left(\ \frac{1}{r^2}+\frac{4m}{r^3} \right) \fb &=& e_\th \kab- \ddd_2\dds_2  \fb- \frac{2\Up}{r}  e_\th^\S a +\frac{2m\Up}{r^3} f +\err_1
 \eeaa
 with error term
 \beaa
 \Big|\err_1\Big|&\les \epg \dg  r^{-3}.
 \eeaa
Projecting on  $e^\Phi$  and proceeding as before,
 \bea
  \lab{eq:Thm-GCMS2-11}
 \bsplit
 \left( \frac{1}{r^2}+\frac{4m}{r^3} \right)  \int_\S  \fb e^\Phi&= \int_{\ovS}( e_\th \kab) e^\Phi -\int_\S( e_\th^\S \kab^\S) e^\Phi - \frac{2\Up}{r} \int_\S ( e_\th^\S a) e^\Phi +\frac{2m\Up}{r^3}  \int_\S   f  e^\Phi\\
 &+ I_1(\La, \Lab)
 \end{split}
 \eea
where the error term $I_1$ is  continuous in $\La, \Lab$ and verifies the estimate
\beaa
\Big|I_1\Big|\les \epg \dg. 
\eeaa
We now calculate  $\int_\S ( e_\th^\S a) e^\Phi$. In view of  formula \eqref{equations-forBadModes}  we have,
\beaa
\frac{2}{r^\S} \int_\S e_\th^\S(a)e^\Phi&=& -\int_\S e_\th(\ka) e^\Phi + \frac 1 4\int_{\S}  \ka^2  \fb e^\Phi -\int_\S \left( \frac 1 4 \ka\kab +\ka\omb+3\rho\right)f e^\Phi+\err_7\\
&=& -\int_{\ovS} e_\th(\ka) e^\Phi +\frac{1}{r^2} \Lab -\frac{1}{r^2}(1+\frac m r )\La+I_2(\La, \Lab)
\eeaa
where,
\beaa
\Big|I_2(\La, \Lab)\Big|&\les \epg \dg.
\eeaa
Indeed,  using once more Corollary   \ref{corr:lemma:int-gaS-ga}, we   note that 
\beaa
\left|\int_\S e_\th(\ka) e^\Phi-\int_{\ovS} e_\th(\ka) e^\Phi\right|&\les&\epg \dg.
\eeaa
All other error terms are easily estimated.

Back to \eqref{eq:Thm-GCMS2-11} we deduce,
\beaa
 \left( \frac{1}{r^2}+\frac{4m}{r^3} \right)  \Lab &=& \int_{\ovS}( e_\th \kab) e^\Phi -\int_\S( e_\th^\S \kab^\S) e^\Phi - \frac{2\Up}{r} \int_\S ( e_\th^\S a) e^\Phi +\frac{2m\Up}{r^3}  \La + I_1(\La, \Lab)\\
 &=&  \int_{\ovS}( e_\th \kab) e^\Phi -\int_\S( e_\th^\S \kab^\S) e^\Phi  +\frac{2m\Up}{r^3}  \La + I_1(\La, \Lab) \\
 &-&\Up \left(-\int_{\ovS} e_\th(\ka) e^\Phi +\frac{1}{r^2} \Lab -\frac{1}{r^2}(1+\frac m r )\La+I_2(\La, \Lab)\right)\\
 &=&  \int_{\ovS}( e_\th \kab) e^\Phi -\int_\S( e_\th^\S \kab^\S) e^\Phi  +\Up \int_{\ovS} e_\th(\ka) e^\Phi- \frac{\Up}{r^2} \Lab +\frac{\Up}{r^2}(1+\frac{3m}{r})\La\\
 &+& I_1(\La, \Lab)-\Up I_2(\La, \Lab).  
\eeaa
Hence,
\beaa
\frac{2}{r^2}\left( 1+\frac{m}{r}\right) \Lab&=& \int_{\ovS}( e_\th \kab) e^\Phi -\int_\S( e_\th^\S \kab^\S) e^\Phi  +\Up \int_{\ovS} e_\th(\ka) e^\Phi +\frac{\Up}{r^2}(1+\frac{3m}{r})\La\\
 &+& I_1(\La, \Lab)-\Up I_2(\La, \Lab).
\eeaa
Thus,
\beaa
\Lab&=&\frac{r^2}{2(1+\frac m r )}\left( \int_{\ovS}( e_\th \kab) e^\Phi -\int_\S( e_\th^\S \kab^\S) e^\Phi  +\Up \int_{\ovS} e_\th(\ka) e^\Phi \right)
+\frac{\Up}{2(1+\frac m r )}\La+ F_2(\La, \Lab)
\eeaa
with error term $F_2(\La, \Lab)$ continuous in $\La, \Lab$ and verifying the estimate,
\beaa
\Big| F_2(\La, \Lab)\Big|&\les & \epg\dg r^2. 
\eeaa
This ends the proof of the lemma.
\end{proof}

Under the assumptions of the theorem   the  system 
\beaa
\bsplit
\La&=\frac{r^3}{3m} \int_{\ovS} \b e^\Phi + F_1(\La, \Lab),\\
 \Lab&=\frac{r^2}{2(1+\frac m r )}\left( \int_{\ovS}( e_\th^\S \kab) e^\Phi  +\Up \int_{\ovS} e_\th(\ka) e^\Phi \right)
+\frac{\Up}{2(1+\frac m r )}\La
+ F_2(\La, \Lab), 
\end{split}
\eeaa
has a unique solution $\La_0, \Lab_0$  verifying the estimate 
\beaa
|\La_0|+|\Lab_0| &\les& \dg r^2. 
\eeaa
Taking $\La=\La_0, \Lab=\Lab_0$ in \eqref{system-La-Lab} we deduce,
\beaa
\int_{\S}  \b^\S e^\Phi=0, \qquad 
\int_{\S}  e^\S_\th(\kab^\S) e^\Phi=0,
\eeaa
as stated.
\end{proof}

\begin{corollary}[Rigidity II]
\lab{corr:GCM-rigidity2}
Assume that the spacetime region $\RR$ satisfies assumptions ${\bf A1-A3}$ and \eqref{eq:GCM-improved estimate}, as well as \eqref{assumptions:Theorem-ExistenceGCMS2}. 
Assume given a sphere $\S\subset \RR$ endowed with a compatible  frame $e_3^\S, e_\th^\S, e_4^\S $   which verifies  the GCM conditions
\bea
\ka^\S=\frac{2}{r^\S}, \quad \dds_2\,^\S\dds_1\,^\S\kab^\S=\dds_2\,^\S\dds_1\,^\S\mu^\S=0.
\eea
In addition, we have
\bea
 r \left|\int_{\S} \b^\S e^\Phi\right|+\left|\int_{\S}  e^\S_\th (\kab^\S) e^\Phi\right| &\les& \dg.   
\eea
 Then the transition functions $(f, \fb, \log\la) $  from the background frame of $\RR$ to that of $\S$ 
 verifies the estimates
\beaa
\|(f, \fb, \log(\la))\|_{\hk_{s_{max}-2 }(\S)} &\les& \dg.
\eeaa
\end{corollary}
The proof is an immediate consequence of Lemma \ref{Lemma:ExistenceGCMS2} and the rigidity result of  Corollary \ref{corr:GCM-rigidity1}.   Note that     in the corollary,  the sphere   $\S$ is reinterpreted as a deformation sphere from the unique background sphere sharing the same south pole.


\section{Construction of GCM hypersurfaces}
\lab{section:GCM-hypersurface}


We are ready to  state our main result concerning the construction of GCM hypersurfaces.

\begin{theorem}\lab{theorem:contrutionofGCMhypersurface}
Let  a fixed  spacetime region $\RR$   verifying assumptions ${\bf A1-A3}$ and \eqref{eq:GCM-improved estimate}. In addition we assume that,
 \bea
 \lab{assumptions:badmodeof-eta-xib}
\left| \int_{S(u,s)} \eta e^\phi\right| \les r^2 \dg, \qquad   \left| \int_{S(u,s) }\xib  e^\phi\right| \les r^2 \dg,
 \eea
 and, everywhere on $\Si_0$,
 \bea
 \lab{assumptions:badmodeof-a^SatSP}
\left| \left(\frac{2}{\vsi}+\Omb\right)\Big|_{SP}- 1-\frac{2m}{r} \right| &\les \dg
 \eea
 where $SP$  denotes the South pole, i.e. $\th=0$   relative to the  adapted geodesic  coordinates $u, s,  \th$.
 
 Let  $\S_0=\S_0[\ovu, \ovs, \La_0, \Lab_0]$ be a fixed  GCMS  provided   by   Theorem \ref{Theorem:ExistenceGCMS}. 
 Then,  there exists  then a unique,  local\footnote{i.e. in a neighborhood  of  $\S_0$.},    smooth, $\Z$-invariant  spacelike hypersurface $\Si_0$ passing through $\S_0$,      a  scalar function $u^\S$ defined  on $ \Si_0$,  whose level surfaces  are topological spheres denoted by $\S$,  and a smooth collection of   constants $\La^\S, \Lab^\S$ verifying,
\beaa
\La^{\S_0}=\La_0, \qquad \Lab^{\S_0} =\Lab_0,
\eeaa
    such that the following conditions are verified:
\begin{enumerate}
\item   The surfaces    $\S$   of constant $u^\S$   verifies  all the properties   stated in 
Theorem \ref{Theorem:ExistenceGCMS}  for the prescribed constants $\La^\S, \Lab^\S$. 
  In particular  they come endowed  with   null frames  $(e_4^\S,  e_\th^\S, e_3^\S)$ such that
  \begin{itemize}
  \item[i.]
For each $\S$ the    GCM conditions  \eqref{Conditions:GCMS}  with $\La=\La^\S, \Lab=\Lab^\S$,  are verified.

 \item[ii.] The transition  functions $(f, \fb, a=\log\la) $ verify  the estimates \eqref{estimates-transtionF-ThmDCM}.

  \item[iii.] The   transversality conditions \eqref{Transversality-Si*} are verified.

   \item[iv.] The corresponding Ricci and curvature coefficients  verify the  estimates \eqref{Estimate-GacS-onS} and \eqref{Estimate-GacS-onS-e4}.
\end{itemize}

\item  Denoting $r^\S$ to be the area radius of the spheres  $\S$  we have, for some constant $c_*$,  
\bea
u^\S+r^\S=c_*, \qquad \textit{along} \quad \Si_0.
\eea

\item
Let $\nu^\S$  be the unique vectorfield tangent to the hypersurface $\Si_0$, normal to $\S$, and normalized by $g(\nu^\S, e_4^\S)=-2$. There exists a unique scalar function $a^\S$ on $\Si_0$ such that $\nu^\S$ is given by
 \beaa
 \nu^\S =e_3^\S+ a^\S e_4^\S.
 \eeaa
 The following normalization condition  holds true  at the  South Pole $SP$  of every sphere $\S$, i.e. at $\th=0$,
 \bea
 a^\S\Big|_{SP}=-1 -\frac{2m^\S}{r^\S}.
 \eea
 
 \item We extend $u^\S$ and $r^\S$ in a small neighborhood of  $\Si_0$ such that the  following transversality conditions are verified\footnote{  Here  the average of $\ka^\S$ is taken on $\S$. In view of the GCM conditions \eqref{GCM-conditions-Si*1-intr} we deduce $e_4^\S(r^\S)=1$.} on  $\Si_0$,
\bea
\lab{GCMH:additionaltransversality}
e_4^\S(u^\S)=0,\qquad e_4(r^s) = \frac{r^\S}{2}\ov{\ka^\S}= 1.
\eea

\item   In view of \eqref{GCMH:additionaltransversality}       the Ricci coefficients $\eta^\S, \xib^\S$   are well defined    for every $\S\subset \Si_0$ and verify
\bea
\int_\S \eta^\S e^\Phi= \int_\S \xib^S e^\Phi=0.
\eea

\item  The  following  estimates hold true  for all
$k\le s_{max}$,
\bea
  \|\eta^{\S} \|_{\hk_k (\S)}&\les \epg,  \\
   \|\xib ^{\S} \|_{\hk_k (\S)}&\les \epg,\\
      \left\|a  ^{\S} +1+\frac{2m^\S}{r^\S} \right\|_{\hk_k (\S)}&\les \epg.
\eea
The $e_3^\S$ derivatives of $\kac^\S, \vth^\S, \ze^\S, \kabc^\S, \vthb^\S, \a^\S, \b^\S, \rhoc^\S, \mu^\S, \bb^\S $  are well defined on $\Si_0$  and we have, for all  $k\le s_{max}-1$
\bea
 \label{Estimate-GacS-onS-e3}
 \bsplit
 \|e^\S_3 (\kac^\S, \vth^\S, \ze^\S, \kabc^\S )\|_{\hk_k (\S)}&\les \epg r^{-1}, \\
  \|e^\S_3(\vthb^{\S}) \|_{\hk_k (\S)}&\les \epg ,  \\
   \|e_3^\S\left(\a^\S, \b^\S, \rhoc^\S, \mu^\S\right) \|_{\hk_{ k} (\S)}&\les \epg r^{-2}, \\
   \|e_3^\S(\bb^\S) \|_{\hk_k (\S)}&\les \epg r^{-1},\\
    \|e_3^\S(\aa^\S) \|_{\hk_{ k} (\S)}&\les \epg.
 \end{split}
 \eea

 \item  The transition functions from the background foliation to that of $\Si_0$  verify
 \bea
 \lab{EstimatesforF-onSi_*}
 \|\dk^{\le s_{max}+1} (f, \fb, \log \la) \|_{L^2(\S)}\les \dg. 
 \eea
 \end{enumerate}
\end{theorem}

\begin{corollary}[Rigidity III]
\lab{corr:GCM-rigidity3}
   Let  a fixed  spacetime region $\RR$   verifying assumptions ${\bf A1-A3}$ and the small GCM conditions \eqref{eq:GCM-improved estimate}. Assume given a GCM hypersurface $\Si_0\subset \RR$  foliated by surfaces $\S$ such that 
\beaa
\ka^\S=\frac{2}{r^\S}, \quad \dds_2\,^\S\dds_1\,^\S\kab^\S=\dds_2\,^\S\dds_1\,^\S\mu^\S=0, \quad \int_{\S}\eta^\S e^\Phi=0, \quad \int_{\S}\xib^\S e^\Phi=0.
\eeaa
Assume in addition that   for a  specific  sphere  $\S_0$   on $\Sigma_0$, the transition functions $f, \fb $ from the background foliation  to $\S_0$  verify
\bea
\int_{\S_0}f e^\Phi=O(\dg), \quad \int_{\S_0}\fb e^\Phi=O(\dg).
\eea
Then,  for all derivatives of  the transition functions along $\S$,
\beaa
 \|\dk^{\le s_{max}+1} (f, \fb, \log \la) \|_{L^2(\S_0)}\les \dg. 
\eeaa
\end{corollary}

\bigskip

 We   give  below  the  proof of Theorem \ref{theorem:contrutionofGCMhypersurface} and a short  discussion of the proof of the  Corollary.


\subsection{Definition of \, $\Si_0$}


 As stated in the theorem we  assume given a  spacetime region $\RR=\{|u-\ug|\leq \de_\RR ,\, |s-\sg|\leq \de_\RR \}$  (see   definition\eqref{definition:RR(dg,epg)}) endowed with a background foliation  such  that   the condition {\bf A1-A3} hold true.  We also assume given a  deformation sphere 
\beaa
\S_0 &:=& \S[\ug, \sg, \La_0, \Lab_0]
\eeaa
of  a given sphere  $\ovS=S(\ovu, \ovs)$ of the background foliation   which verify the  conclusions of Theorem  \ref{Theorem:ExistenceGCMS}.  We then proceed to construct, in a small neighborhood  of $\S_0$, a spacelike hypersurface  $\Si_0$ initiating at $\S_0$  verifying all the  desired properties mentioned above.
 In what follows we outline the main steps in the construction.

{\bf Step 1.}   According to Theorem  \ref{Theorem:ExistenceGCMS},      for every  value of the parameters $(u, s)$  in
 $\RR$ (i.e. such that the background  spheres  $S(u,s)\subset \RR$) and every  real numbers  $(\La, \Lab) $,  there exists a unique  GCM sphere   $\S[u, s, \La, \Lab]$, as a   $\Z$-polarized deformation of $S(u,s)$. In particular the   following are verified:
 \begin{itemize}
 \item  $\S$ coincides with $S(u,s) $  at their south poles (i.e. for $\th=0$ in the adapted coordinates).   
 
 \item On $\S$,   the following GCMS  conditions hold
\bea
\lab{GCM-conditions-Si*1-intr}
\ka^\S=\frac{2}{r^\S},\qquad \ddsS_2\ddsS_1\kab^\S=0,\qquad  \ddsS_2\ddsS_1\mu^\S=0,
\eea
\bea
\lab{GCM-conditions-Si*2-intr}
 \int_\S f   e^\Phi=\La^\S, \qquad \int_\S\fb e^\Phi =\Lab^\S,
\eea
where $(f, \fb, \la)$ are  the transition parameters of the  frame transformation  from the background  frame
$(e_3, e_\th, e_4)$ to the adapted  frame $(e_3^\S, e_\th^\S, e_4^\S)$. The constants $\La^\S, \Lab^\S$  depend smoothly on the  surfaces $\S$ and
\beaa
\La^{\S_0}=\La_0, \qquad \Lab^{\S_0}=\Lab_0.
\eeaa

\item There is a   map  $\Xi:S(u,s)\longrightarrow \S$  given by
\bea
\lab{eq:parametrized-def-Si_*}
\Xi:(u,s, \th)=\Big( u+U(\th, u,\, s,\, \La, \, \Lab), \, s+S(\th, u,s,\La, \Lab),\, \th\Big)
\eea
with $U, S$ vanishing at  $\th=0$.
\item The transversality conditions  \eqref{Transversality-Si*}  hold,  i.e. $  \xib^\S=\om^\S=\ze^\S+\etab^\S=0$. 
Note that these specify the $e_4^\S$ derivatives of $(f, \fb, \la)$ on $\S$.
\item The Ricci coefficients\footnote{Consequently the Hawking mass   $m^\S$  is also well defined.} $\ka^\S, \kab^\S, \vth^\S, \vthb^\S, \ze^\S$ are well defined on each sphere $\S$ of $\Si_0$, and hence on $\Si_0$.  The same holds true for all curvature coefficients $\a^\S, \b^\S,\rho^\S$, $\bb^\S, \aa^\S$. Taking into account our   transversality condition we  remark  that the only ill defined Ricci coefficients are   $\eta^\S, \xi^\S, \omb^\S$.
\item  Let $\nu^\S$  be the unique vectorfield tangent to the hypersurface $\Si_0$, normal to $\S$, and normalized by $g(\nu^\S, e_4^\S)=-2$. There exists a unique scalar function $a^\S$ on $\Si_0$ such that $\nu^\S$ is given by
 \beaa
 \nu^\S =e_3^\S+ a^\S e_4^\S.
 \eeaa
 We deduce that   the quantities 
\beaa
g(D_{\nu^\S}e_4^\S, e_\th^\S) &=& 2\eta^\S+2a^\S\xi^\S=2\eta^\S,\\
g(D_{\nu^\S}e_3^\S, e_\th^\S) &=& 2\xib^\S+2a^\S\etab^\S=2(\xib^\S-a^\S\ze^\S),\\
g(D_{\nu^\S}e_3^\S, e_4^\S) &=& 4\omb^\S-4a^\S\om^\S=4\omb^\S,
\eeaa
are well defined on $\Si_0$.  Thus the scalar $a^\S$ allows us to  specify  the  remaining Ricci coefficients,
$\eta^\S, \xib^\S, \omb^\S$ along $\Si_0$, which we do below.
\end{itemize}


\subsection{Extrinsic properties of $\Si_0$} 


We analyze the extrinsic properties of the hypersurfaces  $\Si_0$ defined  in  Step 1.

{\bf Step 2.} We define the scalar function $u^\S$ on $\Si_0$ as
\bea
\lab{EQ:Define-uS}
u^\S:=c_0-r^\S,
\eea
where $r^\S$ is the area radius of \, $\S$ and  the contant $c_0$ is such that $u^\S\big|_{ \S_0} = \ovu$, i.e. $c_0=\ovu+r^\S\big|_{ \S_0}$.

{\bf Step 3.}  We extend $u^\S$ and $r^\S$ in a small neighborhood of  $\Si_0$  such that  the following transversality conditions are verified.
\bea\lab{eq:additionaltransversalityconditionsuands}
e_4^\S(u^\S)=0,
\qquad e_4^\S(r^\S)=\frac{r^\S}{2}\ov{\ka^\S},
\eea
where the average of $\ka^\S$ is taken on $\S$. In view of the GCM conditions \eqref{GCM-conditions-Si*1-intr}
we deduce $e_4^\S(r^\S)=1$.

{\bf Step 4.}  Note that  $e_3^\S(u^\S, r^\S) $  remain undetermined. On the other hand,  since
$e_\th^\S(u^\S)= e_\th^\S(r^\S)=0$, we deduce in view of  \eqref{eq:additionaltransversalityconditionsuands}
\beaa
e_\th^\S (e_3^\S (u^\S))&=&[e^\S_\th, e^\S_3] u^\S=\left[\frac 1 2 (\kab^\S+\vthb^\S) e^\S_\th +(\ze^\S-\eta^\S) e_3^\S+\xib^\S e_4^\S\right] u^\S\\
&=&(\ze^\S-\eta^\S) e_3^\S( u^\S),\\
e_\th^\S (e_3^\S (r^\S ))&=&[e^\S_\th, e^\S_3] r^\S=\left[\frac 1 2 (\kab^\S+\vthb^\S) e^\S_\th +(\ze^\S-\eta^\S) e_3^\S+\xib^\S e_4^\S\right] r^\S\\
&=&(\ze^\S-\eta^\S) e_3^\S( r^\S )+ \xib^\S.
\eeaa
Thus introducing the scalars
\bea
\lab{definition-vsiS-Si}
\vsi^\S &:=& \frac{2}{e_3^\S(u^\S)},
\eea
and,
\bea
\lab{definition-AbS-Si}
 \Ab^\S &:=&\frac{2}{r^\S}(e_3^\S(r^\S)+\Up^\S),
\eea
we deduce,
 \bea
  \lab{equation:ddsS1vsiS}
    e^\S_\th (\log \vsi^\S)&=&(\eta^\S-\ze^\S),\\
    \lab{eq:formulaforethetaofAbS}
 e_\th^\S(\Ab^\S)  &=& -\frac{2\Up^\S}{r^\S}(\ze^\S -\eta^\S)  -\frac{2}{r^\S}\xib^\S + (\ze^\S -\eta^\S)\Ab^\S.
    \eea
We infer that  $ e^\S_\th (\log \vsi^\S)$ and  $e_\th^\S(\Ab^\S) $ are determined in terms of $\eta, \xib$.

{\bf Step 5.}  In view of  the definition of $ \nu^\S$ and $\vsiS$ we make use of  \eqref{eq:additionaltransversalityconditionsuands}  to deduce
\beaa
\nu^\S(u^\S) = e_3^\S(u^\S)+a^\S e_4^\S(u^\S)= \frac{2}{\vsiS}.
\eeaa
On the other hand, since $u^\S:=c_0-r^\S$ along $\Si_0$,
\beaa
\nu^\S(u^\S)=-\nu^\S(r^\S)=- e_3^\S(r^\S)- a^\S e_4^\S(r^\S)=\Up^\S -\frac{r^\S}{2} \Ab^\S- a^\S
\eeaa
and therefore,
\bea
\lab{formula:a^S-vsiS}
a^\S&=&- \frac{2}{\vsiS}+\Up^\S -\frac{r^\S}{2} \Ab^\S=- \frac{2}{\vsiS}-\Omb^\S
\eea
where,
\bea
\Omb^\S:=  e_3^\S(r^\S) =-\Up^\S -\frac{r^\S}{2} \Ab^\S.
\eea

{\bf Step 6.}       The following lemma will be used, in particular\footnote{It  will also be used  below to derive equations for $\La, \Lab$.}, to      determine the  $\ov{\Ab^\S}$.  
\begin{lemma}
\lab{Lemma:nuSof integrals}
For every scalar function $h$ we have  the formula
\bea
 \nu^\S\left(\int_{\S}h\right) &=& (\vsi^\S)^{-1}\int_{\S}\vsi^\S\left(\nu^\S(h)+(\kab^\S+a^\S\ka^\S)h\right).
\eea
In particular
 \beaa
 \nu^\S(r^\S) &=& \frac{r^\S}{2}(\vsi^\S)^{-1}\ov{\vsi^\S(\kab^\S+a^\S\ka^\S)}
 \eeaa
 where the average is with respect to $\S$.
\end{lemma}

\begin{proof}
We  consider the  coordinates $u^\S$, $ \th^\S$ along $\Si_0$ with $\nu^\S(\th^\S)=0$. In these coordinates
we have,
\beaa
\nu^\S=\frac{2}{\vsi^\S}\pr_{u^\S}.
\eeaa
The lemma follows easily by expressing the volume element of the surfaces  $\S\subset\Si_0$   with respect to the coordinates $u^\S, \th^\S$ (see also the proof of  Proposition \ref{prop:outgoinggeod-e3e4averages}).
\end{proof}

{\bf Step 7.}  Note that the  GCM condition $\ka^\S=\frac{2}{r^\S}$ together with the definition of the Hawking mass implies that,
\beaa
\ov{\kab^\S}=-\frac{2\Up^\S}{r^\S}, \qquad \Up^\S=1-\frac{2m^\S}{r^\S},
\eeaa
 where the average is taken  with respect to $\S$.
Thus in view of   Lemma   \ref{Lemma:nuSof integrals}  we deduce
 \beaa
  e_3^\S(r^\S)+a^\S&=&\nu^\S(r^\S) = \frac{r^\S}{2}(\vsi^\S)^{-1}\ov{\vsi^\S(\kab^\S+a^\S\ka^\S)}= \frac{r^\S}{2}(\vsi^\S)^{-1}\big(\ov{\vsi^\S}\ov{\kab^\S}+\ov{\vsic^\S\kabc^\S}\big)+(\vsi^\S)^{-1}\ov{\vsi^\S a^\S}\\
 &=& -\Up^\S(\vsi^\S)^{-1}\ov{\vsi^\S}+\frac{r^\S}{2}(\vsi^\S)^{-1}\ov{\check{\vsi}^\S\check{\kab}^\S}+ (\vsi^\S)^{-1}\ov{\vsi^\S a^\S}.
 \eeaa
Since according to \eqref{definition-AbS-Si}  $e_3^\S( r^\S) =-\Up^S+\frac{r^\S}{2}\Ab^\S$, we deduce
 \beaa
 \Ab^\S  &=& \frac{2}{r^\S}\left(\Up^S -a^\S -\Up^\S(\vsi^\S)^{-1}\ov{\vsi^\S}+\frac{r^\S}{2}(\vsi^\S)^{-1}\ov{\check{\vsi}^\S\check{\kab}^\S}+ (\vsi^\S)^{-1}\ov{\vsi^\S a^\S}\right).
 \eeaa 
 In particular, multiplying by $\vsi^\S$ and taking the average, we infer
 \beaa
 \ov{\vsi^\S\Ab^\S}  &=& \ov{\check{\vsi}^\S\check{\kab}^\S},
 \eeaa
 and hence
  \bea\lab{eq:formulafortheaverageofAbS}
\ov{\Ab^\S}  &=& \frac{1}{ \ov{\vsi^\S}}\left(\ov{\check{\vsi}^\S\check{\kab}^\S} -\ov{\check{\vsi}^\S\check{\Ab}^\S}\right).
 \eea

{\bf Step 8.} We summarize the results  in  Steps 1-7 in  the following.
\begin{proposition}
\lab{proposition:sufficientgenerality-Si}
Let $\Si_0$ be a smooth  spacelike hypersurface foliated   by   framed\footnote{i.e. differentiable  spheres $\S$  endowed   with adapted null  frames 
 $(e_4^\S, e_\th^\S, e_3^\S)$.}   spheres $(\S, e_4^\S, e_\th^\S. e_3^\S) $    whose Ricci coefficients  verify  the  GCM condition  $\ka^\S=\frac{2}{r^\S}$ and transversality condition   \eqref{Transversality-Si*}.
  Define $u^\S$      as in   \eqref{EQ:Define-uS}  such that $u^\S+r^\S$ is constant on $\Si_0$ with $r^\S$ the area radius  of the spheres $\S$.   Extend  $u^\S$ and  $r^\S$ in a neighborhood of $\Si_0$  such that   the  transversality conditions \eqref{eq:additionaltransversalityconditionsuands} are verified. Then, defining the scalars $\vsi^\S, \Ab^\S$ as in \eqref{definition-vsiS-Si}, \eqref{definition-AbS-Si} we establish the following relations between $\eta^\S, \xib^\S$
 and $\vsiS, \Ab^\S$ and $a^\S$, where the latter scalar  is defined   in Step 1,
 \bea
 \bsplit
  e^\S_\th (\log \vsi^\S)&=(\eta^\S-\ze^\S),\\
 e_\th^\S(\Ab^\S)  &= -\frac{2\Up^\S}{r^\S}(\ze^\S -\eta^\S)  -\frac{2}{r^\S}\xib^\S + (\ze^\S -\eta^\S)\Ab^\S,\\
 \ov{\Ab^\S}  &= \frac{1}{ \ov{\vsi^\S}}\left(\ov{\check{\vsi}^\S\check{\kab}^\S} -\ov{\check{\vsi}^\S\check{\Ab}^\S}\right),\\
 a^\S&=- \frac{2}{\vsiS}+\Up^\S -\frac{r^\S}{2} \Ab^\S.
 \end{split}
 \eea
\end{proposition}

\begin{remark}
Note that we lack equations for   $\eta^\S, \xib^\S$ and the average  of $a^\S$.   The latter can be fixed by  fixing the value of $a^\S \big|_{SP}$
and observing that 
\bea
 \ov{a}^\S =a^\S \big|_{SP} -\check{a}^\S \big|_{SP}.
\eea
\end{remark}

 In what follows we state  a result which ties $\eta^\S, \xib^\S$  to  the other GCM conditions in  \eqref{GCM-conditions-Si*1-intr}--    \eqref{GCM-conditions-Si*2-intr}.

{\bf Step 9.}    To state  the proposition below  we  split the Ricci coefficients into the following groups.
\beaa
\GaS_g&=&\left\{\kac^\S,\, \vth^\S, \,\ze^\S,\, \kabc^\S,\,  r\rhoc^\S,  \,    \ov{\ka^\S}-\frac{2}{r^\S}, \, \ov{\rho^\S}+\frac{2m^S}{(r^\S)^3}\right\},\\
\GaS_b&=&\left\{\eta^\S,\, \xib^\S, \, \ombcS, \,  \omb^\S-\frac{m^\S}{(r^\S)^2}, \,   \bb^\S, \aa^\S\right\}.
\eeaa

 \begin{proposition}
 \lab{proposition:sufficientgenerality-Si-estimates}
 The following statements hold true\footnote{$r_\S$ here denotes $r^\S$ the area radius of $\S$.},
 \begin{enumerate}
 \item   Under the same assumptions as in Proposition \ref{proposition:sufficientgenerality-Si},
    the   Ricci coefficients   $\eta^\S,\, \xib^\S, \omb^\S$  verify the following identities.
      \bea
      \lab{identities-etaxivomb-Si}
      \bsplit
2 \ddsS_2  \ddsS_1 \dddS_1\dddS_2\ddsS_2\eta^\S&=\ka^\S \Big(C_1+2\ddsS_2  \ddsS_1\dddS_1 \bb^\S\Big)  -r_\S^{-3} \dkb^3 C_2 \\
&+ r_\S^{-5} (\dkbS)^{ \le 4 }\GaS_g+r_\S^{-4 }(\dkbS)^{\le 4} (\GaS_b\c  \GaS_b)+\lot,\\
  2\ddsS_2\ddsS_1\dddS_1\dddS_2\ddsS_2\xib &=  C_3-\kab^\S\Big(  C_1  +2\ddsS_2\ddsS_1\dddS_1\bb^\S\Big)  \\
  & + r_\S^{-5} (\dkbS)^{ \le 4 }\GaS_g+r_\S^{-4 }(\dkbS)^{\le 4} (\GaS_b\c  \GaS_b)+\lot,\\
\ddsS_1\omb^\S &=\left(\frac{1}{4}\kab^\S  +\omb^\S\right)\eta^\S  -(\ka^\S)^{-1} \dddS_2\ddsS_2 \eta^\S\\
&+\frac{1}{4}\ka^\S\xib^\S -\frac 1 2 (\ka^\S)^{-1} C_2 +r_\S^{-1} (\dkbS)^{\le 1 }\GaS_g+\dkbS(\GaS_b\c \GaS_b),
\end{split}
\eea
     where, 
    \bea
    \lab{definition:C1C2C3-Si}
    \bsplit
C_1&= e^\S_3 ( \ddsS_2 \ddsS_1\mu^\S), \\
C_2&= e^\S_3(e^\S_\th \ka^\S),\\
 C_3&=e_3\Big( ( \ddsS_2\dddS_2+ 2K^\S)       \ddsS_2 \ddsS_1 \kab^\S)\Big). 
 \end{split}
\eea
 The quadratic  terms denoted  $\lot$ are  lower order both in terms of  decay in  as well  as in terms of number of derivatives. They also contain only angular derivatives $\dkbS$ and not $e^\S_3$ or $ e^\S_4$.
We also note  that the error terms $r_\S^{-5} (\dkbS)^{ \le 4 }\GaS_g+r_\S^{-4 }(\dkbS)^{\le 4} (\GaS_b\c  \GaS_b)$  does not in fact contain more than  $3$ derivatives of $\ombcS$.

\item If in addition  \eqref{Estimate-GacS-onS}  of Theorem \ref{Theorem:ExistenceGCMS}  hold true
 then, for  $k\le s_{max}-7$,
 \bea
 \lab{estimates:fordds-etaxibombS}
 \bsplit
  \| \ddsS_2\eta^\S    \|_{\hk_{4+k}(\S)}  &\les  r_\S^3 \|C_1\|_{\hk_{k}(\S)} +r \|C_2\|_{\hk_{3+k}(\S) }+\epg r_\S^{-1}\\
  &+
     r_\S^{-1} \|\GaS_g\|_{\hk_{4+k}(\S)}+  \|\GaS_b\c\GaS_b\|_{\hk_{ 4+k}(\S)}+\lot,\\
      \| \ddsS_2\xib^\S    \|_{\hk_{4+k}(\S)}  &\les  r_\S^4 \|C_3\|_{\hk_{k}(\S)} +r_\S^3 \|C_1\|_{\hk_{3+k}(\S) }+\epg r_\S^{-1}\\
      &+
     r_\S^{-1} \|\GaS_g\|_{\hk_{4+k}(\S)}+ \|\GaS_b\c\GaS_b\|_{\hk_{4+k}(\S)}+\lot,\\
      \|\ddsS_1 \omb\|_{\hk_{2+k}(\S)} &\les r_\S^{-1} \|\eta^\S\|_{\hk_{4+k}(\S) }+r_\S^{-1} \|\xib^\S\|_{\hk_{2+k}(\S)}+r \|C_2\|_{\hk_{2+k}(S)}\\
  &+r_\S^{-1}\|\GaS_g\|_{\hk_{3+k}(\S)}+ \|\GaS_b\c \GaS_b\|_{\hk_{3+k}(\S)}+\lot
  \end{split}
\eea

\item If in addition the GCM conditions  \eqref{GCM-conditions-Si*1-intr}  hold true  along $\Si_0$   and  the estimates \eqref{Estimate-GacS-onS-e4} are also verified then,  for  $k\le s_{max}-7$,
\bea
\lab{Estimate:a^SC1C2C3}
\bsplit
 \big\|  C_1\big\|_{\hk_{k-2}(\S)} &\les& \epg r^{-5}\left( \big|\ov{a^\S}+ 1+\frac{2m^\S}{r^\S} \big|+r^{-1} \big\| \check{a}^\S\big\|_{\hk_{k-2}(\S)} \right),\\
   \big\|  C_2\big\|_{\hk_{k-1}(\S)} &\les& \epg r^{-3}\left( \big|\ov{a^\S} + 1+\frac{2m^\S}{r^\S}\big|+r^{-1} \big\| \check{a}^\S\big\|_{\hk_{k-1}(\S)} \right),\\
   \big\|  C_3\big\|_{\hk_{k-4}(\S)} &\les& \epg r^{-5}\left( \big|\ov{a^\S} + 1+\frac{2m^\S}{r^\S} \big|+r^{-1} \big\| \check{a}^\S\big\|_{\hk_{k-4}(\S)} \right),
   \end{split}
\eea
where $a^\S$ was defined in Step 1 and  can be expressed in terms of $\vsi^\S$  and $ \Ab^\S$ by formula \eqref{formula:a^S-vsiS}.
\end{enumerate}
   \end{proposition} 

\begin{proof}
The proof\footnote{The  equations used in the derivation  of these identities   only require the.  transversality conditions   \eqref{Transversality-Si*}. }   of the first  two identities  in \eqref {identities-etaxivomb-Si} were derived in Proposition \ref{Prop:nu*ofGCM} in  connection to the proof\footnote{Strictly speaking  Proposition \ref{Prop:nu*ofGCM}  requires  the $e_3$  Ricci and Bianchi identities  of a geodesic foliation.  It is easy to justify the application  of these equations in our context by using the  transversality conditions to generate a geodesic foliation in a neighborhood of  $\Si_0$. }  of  Theorem M4,  starting with the  
following\footnote{ These identities were    recorded in Proposition  \ref{cor:eqtsfor-ometaxib-M4}     which was itself a corollary Proposition 
\ref{prop:eqtsfor-ometaxib}.).Note also that   $\dkb(\GaS_b\c \GaS_b)$ does not contain derivatives of $\ombc$.} 
\bea
\lab{identities-etaxivomb-Si-first}
\bsplit
 2\ddsS_1\omb^\S &= \left(\frac{1}{2}\kab^\S  +2\omb^\S\right)\eta^\S + e^\S_3(\ze^\S) -\bb^\S +\frac{1}{2}\ka\xib^\S+r^{-1}\GaS_g+\GaS_b\c \GaS_b,\\
 2\dddS_2\ddsS_2\eta^\S&=\ka^\S\left( -e_3(\ze^\S) +\bb^\S\right) -e^\S_3(e^\S_\th(\ka^\S)) +r_\S^{-2}( \dkbS)^{\le 1 }\GaS_g+r_\S^{-1}\dkb(\GaS_b\c \GaS_b),\\
     2\dddS_2\ddsS_2\xib^\S
&= \kab^\S\left(e_3(\ze^\S) -\bb^\S\right)   -e^\S_3(e^\S_\th(\kab^\S))+r_\S^{-2} (\dkbS)^{\le 1 }\GaS_g+r_\S^{-1}\dkbS(\GaS_b\c \GaS_b).
\end{split}
\eea
The last identity in \eqref{identities-etaxivomb-Si}  follows by combining  the first two identities in \eqref{identities-etaxivomb-Si-first}.

To prove the estimates  for $\eta^\S$ in the second part  of the proposition  we make use of the identity   $\ddsS_1\dddS_1=\dddS_2\ddsS_2+2K^\S$ to  deduce,
  \beaa
    \ddsS_2 (\dddS_2\ddsS_2+2K^\S ) \dddS_2\ddsS_2\eta^\S&=&\frac 1 2 \ka^\S C_1+ \ka^\S    \ddsS_2 (\dddS_2\ddsS_2+2K^\S )  \bb^\S   -r_\S^{-3}( \dkbS)^3 C_2 \\
    &+& r_\S^{-5} (\dkbS)^{ \le 4 }\GaS_g+r_\S^{-4 }( \dkbS)^{\le 4} (\GaS_b\c  \GaS_b)+\lot
\eeaa
i.e.,
\beaa
  \ddsS_2 (\dddS_2\ddsS_2+2K^\S )\Big(\dddS_2\ddsS_2\eta^\S- \ka^\S \bb^\S\Big)&=&\frac 1 2 \ka^\S C_1   -\frac 12 r_\S^{-3}( \dkbS)^3 C_2\\
  & + & r_\S^{-5}( \dkbS)^{ \le 4 }\GaS_g+r_\S^{-4 }(\dkbS)^{\le 4} (\GaS_b\c  \GaS_b)+\lot
\eeaa
Similarly for $\xib^\S$
\beaa
  \ddsS_2 (\dddS_2\ddsS_2+2K^\S )\Big(\dddS_2\ddsS_2\xib^\S+\kab^\S \bb^\S\Big)&=& \frac 1 2 C_3 -\frac 1 2 \kab^\S C_1 \\
  &+& r_\S^{-5}( \dkbS)^{ \le 4 }\GaS_g+r_\S^{-4 }(\dkbS)^{\le 4} (\GaS_b\c  \GaS_b)+\lot
\eeaa
The desired estimates for $\eta^\S $ and $\xib^\S$  follow then  by making use of the coercivity of the operator  $\ddsS_2(\dddS_2\ddsS_2+2K^\S) $. and the estimate  for $\bb=\bb^\S$ in \eqref{Estimate-GacS-onS-e4}.  The estimate for  $\ddsS_1 \omb^\S$  is straightforward from  the last identity in \eqref{identities-etaxivomb-Si-first}.
 
 To prove  the last part of the proposition 
 we make use of   the GCM conditions \eqref{GCM-conditions-Si*1-intr}  on $\Si_0$  to  deduce that
 \beaa
  \nuS ( \ddsS_2 \ddsS_1\mu^\S)= 0, \quad \nuS (e^\S_\th\ka^\S)=0 ,\quad \nuS\Big( ( \ddsS_2\dddS_2+ 2K^\S)       \ddsS_2 \ddsS_1 \kab^\S)\Big) =0.
 \eeaa
 Hence, the quantities $ C_1, C_2, C_3$ in \eqref{definition:C1C2C3-Si} can be expressed in the form
 \beaa
 C_1&=&- a^\S e^\S_4\big( \ddsS_2 \ddsS_1\mu^\S\big),\\
 C_2&=&- a^\S  e_4^\S (e^\S_\th\ka^\S),\\
 C_3&=& a^\S  e_4^\S\Big( ( \ddsS_2\dddS_2+ 2K^\S)       \ddsS_2 \ddsS_1 \kab^\S)\Big).
 \eeaa
  Making use of our  commutation formulas  of Lemma \ref{Le:comme3e4-outgeodesic} and  the  
 estimates  \eqref{Estimate-GacS-onS-e4} and \eqref{Estimate-GacS-onS} we easily  deduce,
  \beaa
\|  e^\S_4\big( \ddsS_2 \ddsS_1\mu^\S\big)\|_{\hk_{k-2}(\S)} &\les& \epg r^{-5},\\
\|  e_4^\S (e^\S_\th\ka^\S)\|_{\hk_{k-1}(\S)}  &\les& \epg r^{-3}.
 \eeaa
 Similarly,
 \beaa
\Big \| e_4^\S\Big( ( \ddsS_2\dddS_2+ 2K^\S)       \ddsS_2 \ddsS_1 \kab^\S)\Big)\ \Big\|_{\hk_{k-4}(\S)}&\les \epg r^{-5}.
 \eeaa
 Writing $a^\S=\ov{a^\S}  +\check{a}^\S$  and making use of product estimates we deduce
 \beaa
  \big\|  C_1\big\|_{\hk_{k-2}(\S)} &\les& \epg r^{-5}\left( \big|\ov{a^\S}+ 1+\frac{2m^\S}{r^\S} \big|+r^{-1} \big\| \check{a}^\S\big\|_{\hk_{k-2}(\S)} \right),\\
   \big\|  C_2\big\|_{\hk_{k-1}(\S)} &\les& \epg r^{-3}\left( \big|\ov{a^\S} + 1+\frac{2m^\S}{r^\S}\big|+r^{-1} \big\| \check{a}^\S\big\|_{\hk_{k-1}(\S)} \right),\\
   \big\|  C_3\big\|_{\hk_{k-4}(\S)} &\les& \epg r^{-5}\left( \big|\ov{a^\S} + 1+\frac{2m^\S}{r^\S} \big|+r^{-1} \big\| \check{a}^\S\big\|_{\hk_{k-4}(\S)} \right),
 \eeaa
 as stated.
\end{proof}

{\bf Step 10.}  Propositions  \ref{proposition:sufficientgenerality-Si}  and  \ref{proposition:sufficientgenerality-Si-estimates}  provide   us with potential\footnote{We cannot close the estimates without being also able to estimate  the $\ell=1$ modes of $\eta^\S, \xib^\S, \ov{\omb^\S}$ and the average  $  \ov{a}^\S$.}  estimates for $\ddsS_2\eta^\S$, $\ddsS_2\xib^\S$, $ \ddsS_1\omb^\S, \ddsS_1\vsiS$. To close we also need to control the $\ell=1$ modes of $\eta^\S, \xib^\S$  the average of $\omb^\S$ and  the average\footnote{ The quantity $\check{a}^\S$ can be determined  using Proposition \ref{proposition:sufficientgenerality-Si}. } of $a^\S$.           Note that the average of $\omb^\S$ can  in fact be  derived form the equation,
  \beaa
    e^\S_3(\ka^\S)+\frac 1 2 \kab^\S\, \ka^\S -2\omb^\S \ka^\S &= 2\dddS_1\eta^\S + 2\rho^\S -\frac 1 2  \vthbS\, \vthS +2(\eta^\S)^2
    \eeaa
    in terms of $\ov{\Ab^\S}$ and $\eta^\S$.
  Indeed, making use of the GCM condition $\ka^\S=\frac{2}{r^\S}$, 
   \beaa
   \omb^\S&=&\frac{1}{2\ka^\S}\left[    e^\S_3(\ka^\S)+\frac 1 2 \kab^\S\, \ka^\S - 2\dddS_1\eta^\S - 2\rho^\S +\frac 1 2  \vthbS\, \vthS -2(\eta^\S)^2\right]\\
   &=&- \frac{1}{2} e_3(r^\S)-\frac{\Up^\S}{ 2r^\S}+\frac{r^\S}{4}\left[    - 2\dddS_1\eta^\S - 2\rho^\S +\frac 1 2  \vthbS\, \vthS -2(\eta^\S)^2\right]\\
   &=&-\frac 1 4 \Ab^\S +\frac{r^\S}{4}\left[    - 2\dddS_1\eta^\S - 2\rho^\S +\frac 1 2  \vthbS\, \vthS -2(\eta^\S)^2\right].
   \eeaa
   Thus, recalling the definition of $\mu^\S$,
   \beaa
   \ov{\omb^\S}&=& -\frac 1 4\ov{ \Ab^\S}+\frac{r^\S}{2}\, \ov{ \mu^\S -(\eta^\S)^2}
   \eeaa
  or,
  \bea
   \lab{equation:ov{om}S}
 \ov{\omb^\S} -\frac{m^\S}{(r^\S)^2} &=& -\frac 1 4\ov{ \Ab^\S}+\frac{r^\S}{2} 
 \Bigg( \ov{ \mu^\S- \frac{m^\S}{(r^\S)^3}}\, -\ov{\eta^\S\c \eta^\S}\Bigg).
 \eea

{\bf Step 11.} In view of the above we     can determine $\eta^\S, \xib^\S, \omb^\S, \vsiS, \Ab^\S$  provided that we control  the $\ell=1$ modes of $\eta^\S, \xib^\S$  and  the average of $\vsiS$.
For this reason we introduce\footnote{ Note that to  prove our main  theorem we have to   construct  our hypersurface $\Si_0$ such that  in fact  $B=\Bb=D=0$. },  along $\Si_0$,
\bea
 \lab{Define:BSBbSDS}
B^\S&=&\int_\S \eta^\S e^\Phi, \qquad  \Bb^\S=\int_\S \xib^\S e^\Phi,\qquad D^\S =a^{\S}\Big|_{SP} +1+\frac{2m^\S}{r^\S}.
\eea

We are now ready to prove the  following 
\begin{proposition}
\lab{Proposition:Estmates-etaSxibSombSvsiSabSaS}
Let $\Si_0$ be a smooth  spacelike hypersurface foliated   by   framed spheres $(\S, e_4^\S, e_\th^\S. e_3^\S) $ 
which verify the GCM conditions \eqref{GCM-conditions-Si*1-intr},  transversality condition   \eqref{Transversality-Si*}  and the estimates   \eqref{Estimate-GacS-onS}-- \eqref{Estimate-GacS-onS-e4} of Theorem \ref{Theorem:ExistenceGCMS}.
 Let $u^\S$ as in   \eqref{EQ:Define-uS}  such that $u^\S+r^\S$ is constant on $\Si_0$.   Extend  $u^\S$ and  $r^\S$ in a neighborhood of $\Si_0$  such that   the  transversality conditions \eqref{eq:additionaltransversalityconditionsuands} are verified. As shown above these   allow us to define $\eta^\S, \xib^\S, \omb^\S, \vsiS, \Ab^\S, a^\S$ and the constants $B^\S, \Bb^\S, D^\S$ as in 
 \eqref{Define:BSBbSDS}.    Finally we assume that,
 \bea
\lab{Aux-BbbD}
r^{-2}\big(|B^\S|+|\Bb^\S|\big)+|D^\S|\le \epg^{1/2}.
\eea
Under these assumptions the following  estimates hold true  for   all $k\le s_{max}-7$,
\begin{enumerate}
\item The Ricci coefficients $\eta^\S, \xib^\S, \omb^\S$ verify
\bea
\lab{eq:Proposition-Estmates-etaSxibSombSvsiSabSaS-1}
\bsplit
\big \|\eta^\S\big\|_{\hk_{5+k}(\S)}&\les\epg+ r_\S^{-2} |B^\S|,\\
\big \|\xib^\S\big\|_{\hk_{5+k}(\S)}&\les\epg+ r_\S^{-2} |\Bb^\S|,\\
\big \|\ombc^\S\big\|_{\hk_{3+k}(\S)}&\les\epg+ r_\S^{-2}\left( |\Bb^\S|+|B^\S|\right),\\
\Big| \ov{\omb^\S}-\frac{m^\S}{(r^\S)^2}\Big| &\les\epg+ r_\S^{-2}\left( |\Bb^\S|+|B^\S|\right).
\end{split}
\eea

\item The scalar $a^\S$ verifies,
\bea
 r_\S^{-1}  \big\| \check{a}^\S \big\| _{\hk_{k+1}(\S)}+ \Big|\ov{a}^\S +1+\frac{2m^\S}{r^\S}  \Big| &\les& \epg+ r_\S^{-2} |B^\S|+|D^\S|.
 \eea

\item
We   also have
\bea
 \bsplit
 \big\| \Ab^\S \big\| _{\hk_{k+1}(\S)}&\les \epg+r_\S^{-2}\left( |\Bb^\S|+|B^\S|\right)+|D^\S|,\\
 r_\S^{-1}    \| \vsic^\S \|_{\hk_{k+1}(\S)}  +  \big|\ov{\vsiS} -1\big| &\les \epg+r_\S^{-2}\left( |\Bb^\S|+|B^\S|\right)+|D^\S|.
 \end{split}
 \eea

\item
We also have, for all  $k\le s_{max}- 4  $
\beaa
 \bsplit
 \|e^\S_3 (\kac^\S, \vth^\S, \ze^\S, \kabc^\S )\|_{\hk_k (\S)}&\les \epg r_\S^{-1}, \\
  \|e^\S_3(\vthb^{\S}) \|_{\hk_k (\S)}&\les \epg ,  \\
   \|e_3^\S\left(\a^\S, \b^\S, \rhoc^\S, \mu^\S\right) \|_{\hk_{ k} (\S)}&\les \epg r_\S^{-2}, \\
   \|e_3^\S(\bb^\S) \|_{\hk_k (\S)}&\les \epg r_\S^{-1},\\
    \|e_3^\S(\aa^\S) \|_{\hk_{ k} (\S)}&\les \epg.
 \end{split}
 \eeaa
\end{enumerate}
  \end{proposition}

\begin{proof}
To simplify the exposition below   we make the auxiliary bootstrap  assumptions,
\bea
\lab{auxilliary-bootstrapSi0}
  \big \| \eta^\S   \big \|_{\hk_{5+k}(\S)} +   \big  \| \xib^\S \big  \|_{\hk_{5+k}(\S)} &\les\epg^{1/2}.
\eea
We start with the  following lemma.
\begin{lemma}
The following estimates hold true
\bea
 \lab{estimate:forcheck{a}^Sandov{a}}
 r_\S^{-1}  \big\| \check{a}^\S \big\| _{\hk_{k+1}(\S)}+ \Big|\ov{a}^\S +1+\frac{2m^\S}{r^\S}  \Big| &\les& \big|D^\S\big| +\|\eta^\S\|_{\hk_{k}(\S)}+ \|\xib^\S\|_{\hk_{k}(\S)}+\epg.
 \eea
\end{lemma}

\begin{proof}
Since $a^\S=\ov{a}^\S+\check{a}^\S$  we deduce $a^\S\big|_{SP} =\ov{a}^\S+\check{a}^\S\big|_{SP}$. Hence,
\bea
\lab{eq:ovaS-D^S}
\ov{a}^\S&=&D^\S-1-\frac{2m^\S}{r^\S} -\check{a}^\S\big|_{SP}.
\eea
We also have (see Proposition \ref{proposition:sufficientgenerality-Si})
\beaa
 a^\S=- \frac{2}{\vsiS } +\Up^\S -\frac{r^\S}{2} \Ab^\S.
 \eeaa
 Hence,
 \beaa
 a^\S&=&- \frac{2}{\ov{\vsi}^\S +\vsic^\S }+\Up^\S -\frac{r^\S}{2} \Ab^\S=- \frac{2}{\ov{\vsi}^\S}\left(1-\frac{\vsic^\S}{\ov{\vsi}^\S} + O\left(\frac{\vsic^\S}{\ov{\vsi}^\S}\right)^2\right)+\Up^\S -\frac{r^\S}{2} \Ab^\S.
 \eeaa
 Taking the average on $\S$  we deduce,
 \bea
 \lab{equationfor-ov{aS}}
 \ov{a^\S}&=&- \frac{2}{\ov{\vsi^\S}}+\Up^\S -\frac{r^\S}{2}\ov{ \Ab}^\S + O\left(\frac{\vsic^\S}{\ov{\vsi}^\S}\right)^2.
 \eea
  Also,  using  \eqref{eq:ovaS-D^S},
  \bea
  \lab{estimate:forcheck{a}^S}
  \check{a}^\S=2 \check{\vsi}^\S -\frac{r^\S}{2} \check{\Ab}^\S+\lot
  \eea
   where $\lot $ denotes higher order terms in $\vsic^\S$ and $\ov{\vsiS}-1$.
   Indeed
 \beaa
 \check{a}^\S= a^\S- \ov{ a^\S}&=& - \frac{2}{\vsiS}+\Up^\S -\frac{r^\S}{2} \Ab^\S -\left(- \frac{2}{\ov{\vsi^\S}}+\Up^\S -\frac{r^\S}{2}\ov{ \Ab}^\S\right)\\
 &=&- \frac{2}{\vsiS}+ \frac{2}{\ov{\vsi^\S}}  -\frac{r^\S}{2} \check{\Ab}^\S=\frac{2\check{\vsi}^\S} { \vsi^\S\ov{\vsi^\S}}  -\frac{r^\S}{2} \check{\Ab}^\S=2 \check{\vsi}^\S -\frac{r^\S}{2} \check{\Ab}^\S+\lot
 \eeaa
 Thus to estimate $\check{a}^\S$ and $\ov{a^\S} $ we first need to estimate $\Ab^\S$, $\vsic^\S$ and $\ov{\vsiS}$. 
  Using the equations (see  Proposition \ref{proposition:sufficientgenerality-Si} )
\beaa
 e_\th^\S(\Ab^\S)  &=& -\frac{2\Up^\S}{r^\S}(\ze^\S -\eta^\S)  -\frac{2}{r^\S}\xib^\S + (\ze^\S -\eta^\S)\Ab^\S,\\
 \ov{\Ab^\S}  &=& \frac{1}{ \ov{\vsi^\S}}\left(\ov{\check{\vsi}^\S\check{\kab}^\S} -\ov{\check{\vsi}^\S\check{\Ab}^\S}\right),
\eeaa
and the auxiliary assumption 
we  derive,
\bea
\lab{estimate:forAb^S}
\big\| \Ab^\S \big\| _{\hk_{k+1}(\S)}&\les \|\eta^\S\|_{\hk_{k}(\S)}+ \|\xib^\S\|_{\hk_{k}(\S)}+\epg r^{-1} \left(1+ \| \vsic^\S\|_{\hk_{k}(\S)}  + \big|\ov{\vsiS} -1\big| \right).
\eea
From the equation 
\beaa
 e^\S_\th (\log \vsi^\S)&=(\eta^\S-\ze^\S).
\eeaa
we also derive,
\bea
\lab{estimate: e_th^Svsis}
 r_\S^{-1} \|  \vsic^\S\|_{\hk_{k+1}(\S)} \les    \|\eta^\S\|_{\hk_{k}(\S)}+\epg  +\epg \big|\ov{\vsi^\S}-1\big|. 
\eea
 To estimate $\ov{\vsi^\S}-1$ we derive from \eqref{equationfor-ov{aS}} and \eqref{estimate:forcheck{a}^S},
 \beaa
 \frac{2}{\ov{\vsi^\S}}&=& -\ov{a^\S} +\Up^\S -\frac{r^\S}{2}\ov{ \Ab}^\S=-\left(D^\S-1-\frac{2m^\S}{r^\S} -\check{a}^\S\big|_{SP}\right) +\Up^\S -\frac{r^\S}{2}\ov{ \Ab}^\S+\lot\\
 &=&-D^\S+ 2 +\check{a}^\S\big|_{SP}  -\frac{r^\S}{2}\ov{ \Ab}^\S+\lot\\
 &=& -D^\S+ 2  +2 \check{\vsi}^\S\big|_{SP}  -\frac{r^\S}{2}\left(\ov{ \Ab}^\S  +\check{\Ab}^\S\big|_{SP} \right)+\lot
 \eeaa
  and therefore,
 \beaa
 \frac{2(1-\ov{\vsiS} )}{\ov{\vsiS}} & =&-D^\S +2 \check{\vsi}^\S\big|_{SP} -\frac{r^\S}{2}\left(\ov{ \Ab}^\S  +\check{\Ab}^\S\big|_{SP}\right)+\lot
 \eeaa
 i.e.,
 \beaa
\ov{\vsiS} -1&=&\frac 1 2 D^\S-  \check{\vsi}^\S\big|_{SP}+\frac{r^\S}{4}\left(\ov{ \Ab}^\S  +\check{\Ab}^\S\big|_{SP}\right)+\lot
 \eeaa
 where $\lot $ denote higher order terms   in $\vsic^\S$ and $\ov{\vsiS}-1$.Thus,
 \beaa
 \big|\ov{\vsiS} -1\big| &\les&  |D^\S|+\|\vsic^\S\|_{L^\infty(\S)} + r^\S\|\Ab^\S\|_{L^\infty(\S)}.
 \eeaa

 Hence, back to \eqref{estimate: e_th^Svsis} we derive,
 \beaa
 r_\S^{-1} \| \vsic^\S \|_{\hk_{k+1}(\S)}  +  \big|\ov{\vsiS} -1\big|  \les |D^\S| +r^\S\|\Ab^\S\|_{L^\infty(\S)}+ r  \|\eta^\S\|_{\hk_{k}(\S)}+\epg.
 \eeaa
 Combining with \eqref{estimate:forAb^S} we deduce,
 \bea
 \bsplit
 \big\| \Ab^\S \big\| _{\hk_{k+1}(\S)}&\les \|\eta^\S\|_{\hk_{k}(\S)}+ \|\xib^\S\|_{\hk_{k}(\S)}+\epg,\\
 r_\S^{-1}    \| \vsic^\S \|_{\hk_{k+1}(\S)}  +  \big|\ov{\vsiS} -1\big| &\les  r  \|\eta^\S\|_{\hk_{k}(\S)}+\epg.
 \end{split}
 \eea
 In view of \eqref{estimate:forcheck{a}^S} we also deduce,
 \beaa
r_\S^{-1}  \big\| \check{a}^\S \big\| _{\hk_{k+1}(\S)}&\les&  r_\S^{-1}  \| \vsic^\S \|_{\hk_{k+1}(\S)}+ \big\| \Ab^\S \big\| _{\hk_{k+1}(\S)}\\
&\les &\|\eta^\S\|_{\hk_{k}(\S)}+ \|\xib^\S\|_{\hk_{k}(\S)}+\epg.
 \eeaa
  From \eqref{eq:ovaS-D^S}  we further deduce 
 \beaa
 \Big|\ov{a}^\S +1+\frac{2m^\S}{r^\S}  \Big|    &\les & \big|D^\S\big|  + \|\check{a}^\S\|_{L^\infty(\S)}\les  \big|D^\S\big| +\|\eta^\S\|_{\hk_{k}(\S)}+ \|\xib^\S\|_{\hk_{k}(\S)}+\epg.
 \eeaa
 Hence,
 \bea
 r_\S^{-1}  \big\| \check{a}^\S \big\| _{\hk_{k+1}(\S)}+ \Big|\ov{a}^\S +1+\frac{2m^\S}{r^\S}  \Big| &\les& \big|D^\S\big| +\|\eta^\S\|_{\hk_{k}(\S)}+ \|\xib^\S\|_{\hk_{k}(\S)}+\epg
 \eea
 as stated.
 \end{proof}
 
 In view of the lemma above  and the assumption $ |D^\S| \les \epg^{1/2} $  the estimates \eqref{Estimate:a^SC1C2C3} become, 
\bea
\lab{Estimate:a^SC1C2C3-1}
\bsplit
\big\|  C_1\big\|_{\hk_{k}(\S)} &\les \epg r_\S^{-4 } \left( \|\eta^\S\|_{\hk_{k}(\S)}+ \|\xib^\S\|_{\hk_{k}(\S)}+\epg^{1/2}\right),\\
\big\|  C_2\big\|_{\hk_{k+3}(\S)} &\les \epg r_\S ^{-2 } \left(  \|\eta^\S\|_{\hk_{k+3}(\S)}+ \|\xib^\S\|_{\hk_{k+3}(\S)}+\epg^{1/2}\right),\\
\big\|  C_3\big\|_{\hk_{k}(\S)} &\les \epg r_\S^{-4 } \left( \|\eta^\S\|_{\hk_{k}(\S)}+ \|\xib^\S\|_{\hk_{k}(\S)}+\epg^{1/2}\right).
\end{split}
\eea

To prove the desired   estimate  for $\eta^\S, \xib^\S, \omb^\S$     we make use of    \eqref{estimates:fordds-etaxibombS}  and   the  following lemma.
\begin{lemma}
The error  term
\beaa
E_k&=& r_\S^{-1} \|\GaS_g\|_{\hk_{4+k}(\S)}+  \|\GaS_b\c\GaS_b\|_{\hk_{ 4+k}(\S)}, \qquad k\le s_{max}-7,
 \eeaa
 appearing in   \eqref{estimates:fordds-etaxibombS} verifies the estimate
 \beaa
 E_k&\les& r_\S^{-1} \epg+    r_\S^{-1} \epg^{1/2}\Big( \big\|( \eta^\S, \xib^\S)\|_{\hk_{4+k}(\S)}    +\|  \ombc^\S\|_{\hk_{k+3}(\S)}   \Big).  
 \eeaa
\end{lemma}

\begin{proof}
Since $\GaS_g$ contains only   terms estimated by   \eqref{Estimate-GacS-onS},
\beaa
\|\GaS_g\|_{\hk_{4+k}(\S)}&\les& r_\S^{-1} \epg
\eeaa
$\GaS_b$  contains $\vthb^\S$, which  is estimated  by    \eqref{Estimate-GacS-onS},  as well
 as $\eta^\S, \xib^\S, \ombc^\S, \ov{ \om}^\S-\frac{m^\S}{(r^\S)^2}$. 
  Thus, in view of the auxiliary  estimates  $\big \| \eta^\S, \xib^\S   \big \|_{\hk_{5+k}(\S)} \les\epg^{1/2}$ and the fact that the  quadratic error  terms   contain one less derivative of $\ombc^\S$, we deduce,
 \beaa
  \|\GaS_b\c\GaS_b\|_{\hk_{ 4+k}(\S)} &\les&  r_\S^{-1} \epg^{1/2}\Bigg( \big\| \eta^\S, \xib^\S\|_{\hk_{4+k}(\S)}    +\|  \ombc^\S\|_{\hk_{k+3}(\S)}       + r ^\S \left| \ov{\om}^\S-\frac{m^\S}{(r^\S)^2}\right|\Bigg).
 \eeaa
 In view of equation \eqref{equation:ov{om}S},  $ \ov{\omb^\S} -\frac{m^\S}{(r^\S)^2} = -\frac 1 4\ov{ \Ab^\S}+\frac{r^\S}{2}  \Big( \ov{ \mu^\S- \frac{m^\S}{(r^\S)^3}}\, -\ov{\eta^\S\c \eta^\S}\Big)$,
\beaa
\left| \ov{\omb^\S}-\frac{m^\S}{(r^\S)^2}\right|&\les& \big| \ov{ \Ab^\S}\big|+ r^\S 
 \left|  \ov{\mu} -\frac{m^\S} {(r^\S)^3}          \right| +|\eta^\S|^2 \\
&\les& r_\S^{-1} \epg   |D^\S| +  r^{-1} \epg^{1/2}\left( \|\eta^\S\|_{\hk_{2}(\S)}+ \|\xib^\S\|_{\hk_{2}(\S)}\right)+\epg r^{-2}\\
&\les& r_\S^{-1} \epg^{1/2}\left( \|\eta^\S\|_{\hk_{2}(\S)}+ \|\xib^\S\|_{\hk_{2}(\S)}+\epg \right).
\eeaa
Hence,
\beaa
 \|\GaS_b\c\GaS_b\|_{\hk_{ 4+k}(\S)} &\les&  r_\S^{-1} \epg^{1/2}\Bigg( \big\| \eta^\S, \xib^\S\|_{\hk_{4+k}(\S)}    +\|  \ombc^\S\|_{\hk_{k+3}(\S)}       +\epg\Bigg)
\eeaa
and,
\beaa
E_k&=&r_\S^{-1} \|\GaS_g\|_{\hk_{4+k}(\S)}+  \|\GaS_b\c\GaS_b\|_{\hk_{ 4+k}(\S)}\\
&\les&r_\S^{-1} \epg  + r_\S^{-1} \epg^{1/2}\Bigg( \big\| \eta^\S, \xib^\S\|_{\hk_{4+k}(\S)}    +\|  \ombc^\S\|_{\hk_{k+3}(\S)}    \Bigg)
\eeaa
as stated.
 \end{proof}
 
 In view of the lemma and estimates \eqref{Estimate:a^SC1C2C3-1} for $C_1, C_2, C_3$  the estimates
 \eqref{estimates:fordds-etaxibombS} of Proposition \ref{proposition:sufficientgenerality-Si-estimates}  become,
  \bea
 \lab{estimates:fordds-etaxibombS-simplified}
 \bsplit
  \| \ddsS_2\eta^\S    \|_{\hk_{4+k}(\S)}  &\les r_\S^{-1} \epg+  r_\S^{-1} \epg^{1/2}\Bigg( \big\| \eta^\S, \xib^\S\|_{\hk_{4+k}(\S)}    +\|  \ombc^\S\|_{\hk_{k+3}(\S)}      \Bigg),
\\
      \| \ddsS_2\xib^\S    \|_{\hk_{4+k}(\S)}  &\les r_\S^{-1} \epg+  r_\S^{-1} \epg^{1/2}\Bigg( \big\| \eta^\S, \xib^\S\|_{\hk_{4+k}(\S)}    +\|  \ombc^\S\|_{\hk_{k+3}(\S)}      \Bigg),\\
      \|\ddsS_1 \omb^\S\|_{\hk_{2+k}(\S)} &\les r_\S^{-1} \|\eta^\S\|_{\hk_{4+k}(\S) }+r_\S^{-1} \|\xib^\S\|_{\hk_{2+k}(\S)} \\
      &+ r_\S^{-1} \epg+  r_\S^{-1} \epg^{1/2}\Bigg( \big\| \eta^\S, \xib^\S\|_{\hk_{3+k}(\S)}    +\|  \ombc^\S\|_{\hk_{2+k}(\S)}      \Bigg).
  \end{split}
\eea
From the last equation we derive,
\beaa
 \| \ombc^\S\|_{\hk_{3+k}(\S)} &\les  \|\eta^\S\|_{\hk_{4+k}(\S) }+ \|\xib^\S\|_{\hk_{2+k}(\S)} +\epg.
\eeaa
Thus the first two equations in \eqref{estimates:fordds-etaxibombS-simplified} become
\bea
\lab{eq:Proposition-Estmates-etaSxibSombSvsiSabSaS-2}
\bsplit
 r ^\S \| \ddsS_2\eta^\S    \|_{\hk_{4+k}(\S)}&\les \epg +\epg^{1/2}\big(  \|\eta^\S\|_{\hk_{4+k}(\S) }+ \|\xib^\S\|_{\hk_{4+k}(\S)}\big),\\
 r^\S  \| \ddsS_2\xib^\S    \|_{\hk_{4+k}(\S)}&\les \epg +\epg^{1/2}\big(  \|\eta^\S\|_{\hk_{4+k}(\S) }+ \|\xib^\S\|_{\hk_{4+k}(\S)}\big),
 \end{split}
\eea
from which we deduce,
\beaa
\big \|\eta^\S\big\|_{\hk_{5+k}(\S)}&\les&\epg+ r_\S^{-2} |B^\S|,\\
\big \|\xib^\S\big\|_{\hk_{5+k}(\S)}&\les&\epg+ r_\S^{-2} |\Bb^\S|,\\
\big \|\ombc^\S\big\|_{\hk_{3+k}(\S)}&\les&\epg+ r_\S ^{-2}\left( |\Bb^\S|+|B^\S|\right),
\eeaa
as stated. We can then go back to the preliminary estimates obtained above for $\vsiS$, $\Ab^\S$ and $a^\S$ to derive the remaining statements (1-4)  of   Proposition \ref{Proposition:Estmates-etaSxibSombSvsiSabSaS}. To prove  the last part of the   Proposition we  make use of the corresponding   Ricci and Bianchi equations in the $e^\S_3$ direction.
\end{proof}

\begin{corollary}
\lab{Cor:Proposition-Estmates-etaSxibSombSvsiSabSaS}
Under the same assumptions  as in the proposition above we have the more precise estimates, with $d(\S)= \int_\S    e^{2\Phi}$,
\beaa
\left\|\eta^\S-    \frac{1}{  d(\S)}   B^\S   e^\Phi \right\|_{\hk_{5+k}(\S)} &\les& \epg,\\
\left\|\xib^\S-     \frac{1}{  d(\S)} \Bb^\S e^\Phi \right\|_{\hk_{5+k}(\S)}& \les &\epg.
\eeaa
Note also that,
\beaa
d(\S) &=& (r^\S)^4\left(\frac{8\pi}{3} +O(\epg)\right).
\eeaa
\end{corollary}

\begin{proof}
In view of  \eqref{eq:Proposition-Estmates-etaSxibSombSvsiSabSaS-2}, \eqref{eq:Proposition-Estmates-etaSxibSombSvsiSabSaS-1} and auxiliary assumption \eqref{Aux-BbbD}  we deduce,
\beaa
\left\|\eta^\S- \left( \frac{\int_\S     \eta^\S e^\Phi} { \int_\S e^{2\Phi}  } \right) e^\Phi \right\|_{\hk_{5+k}(\S)} &\les &   r  \| \ddsS_2\eta^\S    \|_{\hk_{4+k}(\S)}\\
&\les&\epg +\epg^{1/2}\big(  \|\eta^\S\|_{\hk_{4+k}(\S) }+ \|\xib^\S\|_{\hk_{4+k}(\S)}\big)\\
&\les&\epg +\epg^{1/2}\Big( \epg+ r^{-2} \left( |\Bb^\S|+|B^\S|\right)\Big)\\
&\les& \epg.
\eeaa
We deduce,
\beaa
\left\|\eta^\S-    B^\S  \frac{1}{   \int_\S    e^{2\Phi}}   e^\Phi \right\|_{\hk_{5+k}(\S)} \les \epg.
\eeaa
Similarly,
\beaa
\left\|\xib^\S-    \Bb^\S  \frac{1}{   \int_\S    e^{2\Phi}}   e^\Phi \right\|_{\hk_{5+k}(\S)} \les \epg
\eeaa
as desired.
\end{proof}


\subsection{Construction of  $\Si_0$}


To  construct the spacelike hypersurface of  Theorem   \ref{theorem:contrutionofGCMhypersurface}  we proceed as follows.

{\bf Step 12.}   Let $\Psi(s), \La(s), \Lab(s)$ real valued functions that will be carefully chosen later. We  look for the  hypersurface   $\Sigma_0$ in the form,
\bea
\label{def:Sigma*1}
\Sigma_0 &=& \bigcup_{s\geq\sg}\S[P(s)]=  \bigcup_{s\geq\sg}\S[\Psi(s), s, \La(s), \Lab(s)]
\eea
where $P(s)$ is a  curve in the parameter space $P$ given by,
\bea
\label{def:Sigma*2}
P(s)=(\Psi(s), s, \La(s), \Lab(s)).
\eea
   In order for  $\Si_0$ to start at   $\S_0=\S[\ug, \sg, \La_0, \Lab_0]$ we impose the conditions
\bea
\label{def:Sigma*0}
\Psi(\ovs)=\ovu, \quad \La(\ovs)=\La_0, \quad \Lab(\ovs)=\Lab_0.
\eea

{\bf Step 13.}  We expect   $\Si_0$ to be a perturbation of   the timelike surface $u+s=c_0$ for some constant $c_0$. We thus  introduce the notation   
  \beaa
  \psi(s):=\Psi(s)+s-c_0, \textrm{ so that }\Psi(s)=-s+c_0+\psi(s)
  \eeaa
  and expect $\psi(s)=O(\dg)$.
 
{\bf Step 14.}
 In view of \eqref{eq:parametrized-def-Si_*} we can  express  the collection of spheres  $\Sigma_0$ in the form
\bea
\label{def:Sigma*3}
\Sigma_0 = \left\{\Xi(s,\th),\quad s\geq\sg, \,\, \th\in[0,\pi]\right\}
\eea
where the map $\Xi(s, \th) =\Xi(\Psi(s), s, \th)$ is defined as
\bea
\label{def:Sigma*4:0}
\Xi(s,\th):=\Big(\Psi(s)+U(\th,P(s)),\,  s+S(\th,P (s)),\,  \th\Big).
\eea
At the South Pole, i.e. $\th=0$, where $U(0, P)=S(0, P)=0$
\bea
\Xi(s,0)&=&\Big(\Psi(s),\,   s ,\,  0\Big).
\eea

Clearly,
\beaa
\pr_s \Xi(s,\th)&=&\Big(\Psi'(s) +\pr_P U(\th,P(s)) P'(s),\,\,  1+\pr_P S(\th,P(s)) P'(s),\,\,  0\Big),\\
\pr_\th  \Xi(s,\th)&=&\Big(\pr_\th U(\th, P(s)), \, \, \pr_\th S(\th, P(s)), \,\,  1 \Big),
\eeaa
where,
\beaa
\pr_P U(\cdot) P'(s)&=&  \Psi'(s) \pr_u U(\cdot)+ \pr_s U(\cdot) +  \La'(s)\pr_\La U(\cdot)+ \Lab'(s) \pr_\Lab U(\cdot),\\
\pr_P S(\cdot) P'(s)&=&  \Psi'(s) \pr_u S(\cdot)+ \pr_s S(\cdot) +  \La'(s)\pr_\La S(\cdot)+ \Lab'(s) \pr_\Lab S(\cdot).
\eeaa

Given $f$  a function on $\Si_*$ we have,
\beaa
\frac{d}{ds}  f\big(\Xi(s, \th)\big)&=& \Big(\Psi'(s) +\pr_P U(\th,P(s)) P'(s)\Big) \pr_u f +\Big(1+\pr_P S(\th,P(s)) P'(s)\Big)\pr_s f\\
&=& X_* f,\\
\frac{d}{ds}  f\big(\Xi(s, \th)\big)&=& \pr_\th U(\th, P(s)) \pr_sf + \pr_\th S(\th, P(s))\pr_s+\pr_\th f\\
&=& Y_* f,
\eeaa
where $X_*, Y_*$ are   the following tangent  vectorfields along $\Si_*$,
\bea
\label{def:Sigma*4}
\bsplit
X_*(s, \th):&=\Big(\Psi'(s) +\pr_P U(\th,P(s)) P'(s)\Big) \pr_u +\Big(1+\pr_P S(\th,P(s)) P'(s)\Big)\pr_s, \\
Y_*(s, \th):&=\pr_\th U(\th, P(s)) \pr_s + \pr_\th S(\th, P(s))\pr_s+\pr_\th, 
\end{split}
\eea
or,
\bea
\label{def:Sigma*5}
\bsplit
X_*(s, \th):&=\Big(\Psi'(s)+\breve{A}(s, \th)\Big) \pr_u +\Big(1+ \breve{B}(s,\th) P'(s)\Big)\pr_s, \\
Y_*(s, \th):&=\breve{C}(s,\th) \pr_u+  \breve{D}(s,\th)\pr_s+\pr_\th,
\end{split}
\eea
where,
\beaa
 \breve{A}(s,\th):&=& \pr_P U(\th,P(s)) P'(s)\\
 &= &\pr_u U(\th,P(s))  \Psi'(s)+\pr_s U(\th,P(s))+\pr_\La U(\th,P(s)) \La'(s)+\pr_\Lab U(\th,P(s)) \Lab'(s),\\
 \breve{B}(s,\th):&=&\pr_P S(\th,P(s)) P'(s)\\
 &=&\pr_u S(\th,P(s))  \Psi'(s)+\pr_s S(\th,P(s))+\pr_\La U(\th,P(s)) \La'(s)+\pr_\Lab S(\th,P(s)) \Lab'(s),\\
\breve{C}(s,\th):&=& \pr_\th U(\th, P(s)),\\
\breve{D}(s,\th):&=& \pr_\th S(\th, P(s)).
\eeaa

{\bf Step 15.}
 Define the vectorfield, along the South Pole of each $\S\subset \Si_0$, 
\bea
X_*\Big|_{SP} h = \frac{d}{ds} h\big( \Xi(s, 0)\big).
\eea

\begin{lemma}
At  the South Pole  we have the relations (recall  $\nu^\S= e_3^\S+ a^\S e_4^\S$)
   \bea
   X_*\big|_{SP} =\frac{1}{2\la}  \vsi \Psi' \,  \nu^\S\Big|_{SP},
   \eea
   \bea
\lab{eq:a^S-psi'}
a^\S\big|_{SP}&=&\frac{2\la^2}{\Psi'(s)\vsi} \big( 1-\frac 1 2 \Psi' (s)\vsi \Omb\big)|_{SP},
\eea
  or, more precisely,
   \beaa
   a^\S(\Psi(s), s, 0)=\frac{1}{\Psi'(s)}\frac{2\la^2}{\vsi}  \left( 1-\frac 1 2\Psi'(s)  \vsi \Omb\right)(\Psi(s), s, 0).
   \eeaa
Here  $ f, \fb, \la$ are the transition functions   and $\vsi, \Omb $   correspond  to the  background foliation.
   \end{lemma}

\begin{proof}
Note  that  
\beaa
\breve{A}(s, 0)=\breve{B}(s, 0)=\breve{C}(s, 0)=\breve{D}(s, 0)=0.
\eeaa
Thus, at the South Pole  SP,
\beaa
X_*(s, 0)=\Psi'(s) \pr_u+ \pr_s.
\eeaa
 Recall  that
 \beaa
 \pr_s =e_4,\quad \pr_u=\frac 1 2\vsi \left( e_3-\Omb e_4-\underline{b}\ga^{1/2} e_\th\right),\quad \pr_\th=\sqrt{\ga}   e_\th,
 \eeaa
or,  since $\underline{b}$ vanishes at the South Pole,
\beaa
X_*(s, 0)&=&\Psi' \frac 1 2\vsi \left( e_3-\Omb e_4\right)+ e_4= (1-\frac 1 2\Psi'(s) \vsi \Omb) e_4+\frac 1 2\Psi'(s)\vsi e_3.
\eeaa
On the other hand, since the transition functions $f, \fb$ vanish at the South Pole,
\beaa
e_4^\S=\la e_4, \qquad e_3^\S= \la^{-1} e_3.
\eeaa
Hence,
\beaa
X_*(s, 0)&=&\la\big( 1-\frac 1 2 \Psi'(s)\vsi \Omb\big) e^\S_4+\frac  12 \la^{-1} \Psi'(s) \vsi e_3^\S\\
&=&\frac  12\la^{-1} \Psi'(s)\vsi \left( e_3^\S+ \frac{2\la^2}{\Psi'(s)\vsi} \big( 1-\frac 1 2\Psi' (s) \vsi \Omb\big) e^\S_4\right).
\eeaa
In view of the definition of $\nu^\S$ we deduce,
\beaa
X_*(s, 0)&=&\frac  12\la^{-1} \Psi'(s)\vsi \,\nu^\S\big|_{SP}
\eeaa
and\footnote{Note that  $a^\S(s, 0)= a^\S(\Xi(s, 0))$.},
\beaa
a^\S(s, 0)&=& \frac{2\la^2}{\Psi'(s)\vsi} \big( 1-\frac 1 2\Psi' (s) \vsi \Omb\big)
\eeaa
as stated.
\end{proof}

{\bf Step  16.}
The transition functions $(f, \fb, \la)$  are uniquely  determined  on $\S$ by the results of Theorem  \ref{Theorem:ExistenceGCMS} in terms of  $\La, \Lab$. The same  holds  true for  all curvature 
    components  and  the Ricci coefficients
    $\ka^\S,\vth^\S,\ze^\S,  \kab^\S, \vthb^\S$.   One can easily see  from the transformation formulas  that  the values of the $e_3^\S$ derivatives of $(f,\fb, \la)$ are   determined by  the  transversal  Ricci coefficients   $\eta^\S, \xib^\S, \etab^\S$ . Indeed, schematically,  from the transformation formulas  for $\eta, \xib, \omb$  in  Proposition
    \ref{prop:transformations1-GCM}, 
    \bea
    \lab{eq:e_3F-etaS,xibS,ombS}
    \bsplit
  e_3^\S f&=  2(\eta^\S -\eta ) -\frac 1 2 \ka \fb +f \omb  +F\c \Ga_b+\lot, \\
  e_3^\S \fb &=2(\xib ^\S-\xib)  -\frac 1 2 \fb\big( \kab+4\omb)+ F\c \Ga_b +\lot,\\
  e_3^\S(\log \la) &= 2( \omb^\S-\omb)   +\Ga_b \c F+\lot,
  \end{split}
    \eea
    where $F=(f, \fb, \log \la )$ and  $\lot$ denotes terms  which are  linear in $\Ga_g, \Ga_b$ and  linear and higher order  in  $F$.   
   Recall also that the  $e_4^\S$ derivatives  of $F$ are fixed by our transversality condition \eqref{Transversality-Si*}     More precisely we have,
   \beaa
   e^\S_4(f)&=&-\frac 1 2  \ka f  +\lot,\\
   e^\S_4(\fb)&=& 2  e^\S_\th( \log \la  )    -         \fb \ka +2(\omb+\frac 1 4 \kab) f+\lot, \\
   e^\S_4( \log \la )&=&\lot 
   \eeaa
      It  follows  that $\eta^\S, \xib^\S, \omb^\S$    can be determined by $\nu^\S(f,\fb, \la)$  and the   scalar $a^\S$.
   More precisely, 
   \bea
   \lab{eqts:nuS-ffb}
   \bsplit
   \nu^\S (f)   &=  2(\eta^\S -\eta )  -\frac 1 2 ( \ka \fb+a^\S \kab f )      + f\omb  +F\c \Ga_b+\lot,\\
   \nu^\S (\fb)  &=2(\xib ^\S-\xib) - \frac 1 2 \big( \kab+4\omb) (\fb- a^\S f)+ a^\S\left(2  e^\S_\th( \log \la  )    -  \fb \ka \right)  +F\c \Ga_b+\lot
   \end{split}
   \eea

{\bf Step 17.} We derive  equations for $\La(s)=\La(\Psi(s), s, 0) ), \Lab(s)=\Lab(\Psi(s), s, 0) $ as follows.
\begin{lemma}
We have the following identities
\bea
 \lab{system:La'-Lab'}
\bsplit
c(s)  \frac{1}{\Psi'(s) }\La'(s) &= \int_\S  \nu^\S (f) e^\Phi  
   -\frac{6}{r^\S}\La(s) +E(s), \\
 c(s)   \frac{1}{\Psi'(s) } \Lab'(s) &=
\int_{\S(s)}  \nu ^\S( \fb )e^\Phi   -\frac{6}{r^\S}\Lab(s) +     \Eb(s),
  \end{split}
 \eea
 where,
 \beaa
 c(s)&=&\Big(  \frac{2\la}{\vsi} \Big)\Big|_{SP}(s)=  \Big( \frac{2\la}{\vsi} \Big)(\Psi(s), s,0)
\eeaa
 and error terms,
 \beaa
 E(s)  &=&\frac 1 2 \int_{\S(s)} \left(3\kabc^\S- \vthb^\S- a^\S \vth^\S+\frac{3}{r^\S} (a^\S+(1+\frac{2m^\S}{r^\S})\right) f e^\Phi\\
  &+& (\vsi^\S)^{-1}\Big|_{SP}\int_{\S(s)}\left( \vsi^\S -  \vsi^\S\Big|_{SP}\right)\left( \nu^\S (f)- \frac{6 }{r^\S}\La(s)\right) e^\Phi+\lot, \\
   \Eb(s)  &=&\frac 1 2 \int_{\S(s)} \left(3\kabc^\S- \vthb^\S- a^\S \vth^\S+\frac{3}{r^\S} (a^\S+(1+\frac{2m^\S}{r^\S})\right) \fb e^\Phi\\
  &+& (\vsi^\S)^{-1}\Big|_{SP}\int_{\S(s)}\left( \vsi^\S -  \vsi^\S\Big|_{SP}\right)\left( \nu^\S (\fb)- \frac{6 }{r^\S}\Lab(s)\right) e^\Phi+\lot 
 \eeaa
  \end{lemma}
 
 \begin{proof}
 According to Lemma   \ref{Lemma:nuSof integrals} we have 
\beaa
 \nu^\S\left(\int_{\S}h\right) &=& (\vsi^\S)^{-1}\int_{\S}\vsi^\S\left(\nu^\S(h)+(\kab^\S+a^\S\ka^\S)h\right).
\eeaa
   Thus, applying the vectorfield $\nu^\S\big|_{SP}=\frac{2\la}{\vsi \Psi'} X_*\big|_{SP}$   to the formulas  \eqref{GCM-conditions-Si*2-intr},
\beaa
&& \frac{1}{\Psi'(s) } \Big( \frac{2\la}{\vsi}\Big)\Big|_{SP} \frac{d}{ds} \La(s)= \nu^\S\Big|_{SP}(\La)= \nu^\S (\La)\big|_{SP}=\nu^\S\Big(\int_\S f e^\Phi\Big)\Big|_{SP}\\
 &&=  (\vsi^\S)^{-1}\Big|_{SP}\int_{\S(s)}\vsi^\S\Big(\nu^\S(f e^\Phi)+(\kab^\S+a^\S\ka^\S) f e^\Phi\Big).
 \\
 \eeaa
 Introducing
 \bea
J(f)=  e^{-\Phi} \nu^\S(f e^\Phi)+(\kab^\S+a^\S\ka^\S) f 
 \eea
 we   deduce,
 \beaa
 c(s)  \frac{1}{\Psi'(s) } &=&  (\vsi^\S)^{-1}\Big|_{SP}\int_{\S(s)}\vsi^\S J(f) e^\Phi\\
 &=&\int_{\S(s)}J(f) e^\Phi+ (\vsi^\S)^{-1}\Big|_{SP}\int_{\S(s)}\left( \vsi^\S -  \vsi^\S\Big|_{SP}\right) J(f).
 \eeaa
 On the other hand, since $e_3\Phi=\frac 1 2 (\kab-\vthb)$,  $e_4\Phi=\frac 1 2 (\ka-\vth)$
 \beaa
 J(f)&=&\nu^\S (f)+ \left( e^\S_3\Phi+ a^\S e_4^\S\Phi   +\kab^\S+a^\S\ka^\S\right) f \\
 &=&\nu^\S (f)+\frac  1 2 \left( 3\kab^\S -\vthb^\S + a^\S( 3\ka^\S - \vthb^\S)\right) f\\
 &=&\nu^\S (f)+\frac  3 2 \left( \kab^\S+ a^\S \ka^\S\right) -\frac 1 2 \left(\vthb^\S + a^\S\vth^\S\right).
 \eeaa
 Since $\ka^\S =\frac{2}{r^\S}$ and  $\kab^\S=\ov{\kab^\S}+\kabc^\S=-\frac{2\Up^\S}{r^\S}+\kabc^\S$ we deduce,
 \beaa
 J(f)&=& \nu^\S (f)+ \frac{3}{r^\S}\left(-\Up^\S +a^\S \right) f+\frac 1 2  \left( 3\kabc^\S- \vthb^\S- a^\S \vth^\S\right) f\\
 &=& \nu^\S (f)+ \frac{3}{r^\S}\left(-\Up^\S-(1+\frac{2m^\S}{r^\S})\right)  f +\frac 1 2 \left(3\kabc^\S- \vthb^\S- a^\S \vth^\S+\frac{3}{r^\S} (a^\S+(1+\frac{2m^\S}{r^\S})\right) f\\
 &=&\nu^\S (f)- \frac{6 }{r^\S}\La(s)  +\frac 1 2 \left(3\kabc^\S- \vthb^\S- a^\S \vth^\S+\frac{3}{r^\S} (a^\S+(1+\frac{2m^\S}{r^\S})\right) f.
 \eeaa
  We deduce,
  \beaa
   c(s)  \frac{1}{\Psi'(s) } &=&\int_\S  \nu^\S (f) e^\Phi  
   -\frac{6}{r^\S}\La(s) +E(s)
  \eeaa
  where,
  \beaa
  E(s)&=&\frac 1 2 \int_{\S(s)} \left(3\kabc^\S- \vthb^\S- a^\S \vth^\S+\frac{3}{r^\S} (a^\S+(1+\frac{2m^\S}{r^\S})\right) f e^\Phi\\
  &+& (\vsi^\S)^{-1}\Big|_{SP}\int_{\S(s)}\left( \vsi^\S -  \vsi^\S\Big|_{SP}\right) J(f)\\
  &=&\frac 1 2 \int_{\S(s)} \left(3\kabc^\S- \vthb^\S- a^\S \vth^\S+\frac{3}{r^\S} (a^\S+(1+\frac{2m^\S}{r^\S})\right) f e^\Phi\\
  &+& (\vsi^\S)^{-1}\Big|_{SP}\int_{\S(s)}\left( \vsi^\S -  \vsi^\S\Big|_{SP}\right)\left( \nu^\S (f)- \frac{6 }{r^\S}\La(s)\right) e^\Phi+\lot 
  \eeaa
  The proof for $\Lab$ is exactly the same. 
 \end{proof}

 {\bf Step 18.} We make use of   the estimates for $F=(f, \fb, \log \la )$ and $e_4^\S( F)$    derived in  Theorem \ref{Theorem:ExistenceGCMS}   as well as  the estimates for $a^\S, \vsi^\S, \eta^\S, \xib^\S, \omb^\S$ derived 
 in Proposition \ref{Proposition:Estmates-etaSxibSombSvsiSabSaS} to  evaluate the right hand sides of \eqref{system:La'-Lab'}. Recall that   in Proposition \ref{Proposition:Estmates-etaSxibSombSvsiSabSaS}
   we have made the auxiliary  assumption   \eqref{Aux-BbbD}
 i.e.
 \beaa
 r_\S^{-2}\big(|B^\S|+|\Bb^\S|\big)+|D^\S|\le \epg^{1/2} .
 \eeaa
 \begin{proposition}
 \lab{Prop: diff-eqtsforLaLab}
 The following equations hold true  for  the functions\footnote{Note also that  $r^{\S(s)}  =r^{\S(s)}= r|_{SP(\S(s)) } = r(s)$.}   
 \beaa
 \La(s)&=&\La(\Psi(s), s, 0) ),\quad  \Lab(s)=\Lab(\Psi(s), s, 0),  \\
  B(s)&=&\La(\Psi(s), s, 0) ), \quad  \Bb(s)=\La(\Psi(s), s, 0) ), \quad   r(s) =r(\Psi(s), s, 0), 
 \eeaa
 \bea
 \lab{system:La'-Lab'-1}
\bsplit
 \frac{1}{-1+\psi'(s)  }\La'(s)   &= B(s)- \frac 1 2 r(s)^{-1}\Lab(s) - \frac 7  2  r(s)^{-1}\La(s) + O(r^{-1}) \La(s)\\&+N(B, \Bb, D, \La, \Lab, \psi)(s),\\
  \frac{1}{-1+\psi'(s)  }\Lab'(s)   &= \Bb(s)-\frac 7  2 r(s)^{-1}\Lab(s) + \frac 1   2  r(s)^{-1}\La(s) +  O(r^{-1}) \big(\La(s)+\Lab(s) \big)\\
 &+\Nb(B, \Bb, D, \La, \Lab, \psi)(s).
  \end{split}
 \eea
The expressions $N, \Nb$  verify the following properties.
\begin{itemize}
\item They depend on  $ B, \Bb, D,  \La, \Lab, \psi, F=(f, \fb, \la-1)$,  the background Ricci coefficients $\Ga_b, \Ga_g$ and curvature  $\Rc=\{\a,\b, \rhoc, \bb, \aa\}$. 

\item  $N, \Nb$ vanish at $(B, \Bb, D,  \La, \Lab, \psi)=(0,0,0,0,0,0)$.  In fact,
\beaa
|N, \Nb |&\les&  r^{2} \dg.
\eeaa

\item  The linear part  in $B, \Bb, D$ has   $O(\epg)$ coefficients, i.e. coefficients which depend  on  the quantities  $\Ga_b, \Ga_g,\Rc,   F$ and $\La, \Lab, \psi$.

\item The linear part  in $\La, \Lab, \psi$ has   $O(\epg)$ coefficients.
\end{itemize}
 \end{proposition}

 \begin{proof}
  To prove the desired result we  make use of \eqref{eqts:nuS-ffb} to   check    the following,
 \bea
 \lab{eq:Prop-diff-eqtsforLaLab1}
 \bsplit
 \int_{\S(s)}  \nu^\S (f) e^\Phi &=  2 B(s)- r^{-1}\Lab(s) - r^{-1}\La(s) + O(r^{-1}) \La(s)+ O(  r^2 \dg),  \\
  \int_{\S(s)}  \nu^\S (\fb) e^\Phi &= 2 \Bb(s)- r^{-1}  \Lab(s)    +r^{-1}  \La(s)    +   O(r^{-1}) \big(\La(s)+\Lab(s) \big)
+O(r^2 \dg).
 \end{split}
 \eea
 Combining this   with  \eqref{system:La'-Lab'},
 \beaa
c(s)  \frac{1}{\Psi'(s) }\La'(s) &=& \int_\S  \nu^\S (f) e^\Phi  
   -\frac{6}{r^\S}\La(s) +E(s), \\
 c(s)   \frac{1}{\Psi'(s) } \Lab'(s) &=&
\int_{\S(s)}  \nu ^\S( \fb )e^\Phi   -\frac{6}{r^\S}\Lab(s) +     \Eb(s),
 \eeaa
 and  the following estimates for  the error terms $E, \Eb$,
 \bea
  \lab{eq:Prop-diff-eqtsforLaLab2}
 |E(s)|+|\Eb(s)| &\les& r^2\dg,
 \eea
 we deduce,
 \beaa
   \frac{1}{\Psi'(s) }\La'(s)     &=& \frac{1}{c(s)}  \left( 2 B(s)- r^{-1}\Lab(s) - 7  r^{-1}\La(s) + O(r^{-1}) \La(s)+ O(  r^2 \dg)\right), \\
     \frac{1}{\Psi'(s) } \Lab'(s) &=&\frac{1}{c(s)}\left(  2 \Bb(s)-  7 r^{-1}  \Lab(s)    +r^{-1}  \La(s)    +   O(r^{-1}) \big(\La(s)+\Lab(s) \big)
+O(r^2 \dg)\right).
\eeaa
According to  our  assumptions $\vsi=1+O(\epg)$.
Also  according to Theorem   \ref{Theorem:ExistenceGCMS} $\la=1+O(r^{-1} \epg)$. Thus,
\beaa
c(s)=\Big(  \frac{2\la}{\vsi} \Big)\Big|_{SP}(s)= \frac{2(1+ Or^{-1} (\epg )}{1+O(\epg)}= 2 +O(\epg). 
\eeaa
Hence,
\beaa
 \frac{1}{\Psi'(s) }\La'(s)   &=&\frac 1 2 \Big(  2 B(s)- r^{-1}\Lab(s) - 7  r^{-1}\La(s) + O(r^{-1}) \La(s)+ O(  r^2 \dg) \Big)\\
 &=& B(s)- \frac 1 2 r^{-1}\Lab(s) - \frac{7}{2}   r^{-1}\La(s) + O(r^{-1}) \La(s)+ O(  r^2 \dg).
\eeaa
Setting $\Psi(s)=-s+\psi(s) + c_0$ and   recalling the structure of the error terms  we have  denoted by $ O(  r^2 \dg)$
\beaa
 \frac{1}{-1+\psi'(s)  }\La'(s)   &=& B(s)- \frac 1 2 r^{-1}\Lab(s) - \frac 7  2  r^{-1}\La(s) + O(r^{-1}) \La(s)+N(B, \Bb, D, \La, \Lab, \psi)(s)
\eeaa
where $N$ verifies  the properties  mentioned in the proposition.
In the same manner we derive
\beaa
 \frac{1}{-1+\psi'(s) ) }\Lab'(s)   &=& \Bb(s)-\frac 7  2 r^{-1}\Lab(s) + \frac 1   2  r^{-1}\La(s) +  O(r^{-1}) \big(\La(s)+\Lab(s) \big)\\
 &+&\Nb(B, \Bb, D, \La, \Lab, \psi)(s)
\eeaa
 as stated in the proposition.

 It remains to check \eqref{eq:Prop-diff-eqtsforLaLab1} and \eqref{eq:Prop-diff-eqtsforLaLab2}
  According to  \eqref{eqts:nuS-ffb} and our assumptions on the Ricci  coefficients $\ka, \kab, \omb$,    we have along  the sphere $\S$ 
 \beaa
  \nu^\S (f)   &= & 2(\eta^\S -\eta )  -\frac 1 2 \left( \frac 2 r  \fb- a^\S\frac{2\Up}{r}  f \right)      + f\frac{m}{r^2} +F\c \Ga_b+\lot\\
  &=& 2(\eta^\S -\eta )  -r^{-1} \fb + r^{-1}\left(\frac{m}{r}+ a^\S (1-\frac{2m}{r} \right) f +F \c \Ga. 
 \eeaa
 According to \eqref{estimate:forcheck{a}^Sandov{a}} and auxiliary  assumption   \eqref{Aux-BbbD}
 \beaa
\left|a^\S+\left(1+\frac{2m^\S}{r^\S} \right) \right| &\les \epg+r^{-2}\left( |\Bb^\S|+|B^\S|\right)+|D^\S|\les \epg^{1/2}.
\eeaa
 Thus,
 \beaa
   \nu^\S (f)   &= & 2(\eta^\S -\eta )  -r^{-1} \fb - r^{-1}\left(1-\frac{m}{r} -\frac{m ^2}{r^2}\right) f +
    r^{-2} O\left(\dg\, \epg^{1/2}\right).  
 \eeaa

 Since $r$ and $r^\S$ are comparable along $\S$, i.e. $|r-r^\S|\le\dg$, we deduce, recalling the definition of $B$,
 \beaa 
 \int_{\S(s)}  \nu^\S (f) e^\Phi &=& 2 B(s)- 2 \int_{\S(s)}\eta e^\Phi- r^{-1}\Lab(s)  - r^{-1}\left(1-\frac{m}{r} -\frac{m ^2}{r^2}\right)\La(s)+  r O\left(\dg\, \epg^{1/2}\right). 
\eeaa
Making use of the assumption \eqref{assumptions:badmodeof-eta-xib}  for $\eta$ as well as Corollary  \ref{corr:lemma:int-gaS-ga} we easily deduce,
\bea
\left| \int_{\S(s)}\eta e^\Phi\right|\les  r^2 \dg.
\eea
Hence, 
\beaa
 \int_{\S(s)}  \nu^\S (f) e^\Phi &=& 2 B(s)- r^{-1}\Lab(s)  - r^{-1}\big(1 +O(r^{-1})   \big)\La(s)+ O(  r^2 \dg) \\
 &=&   2 B(s)- r^{-1}\Lab(s) - r^{-1}\La(s) + O(r^{-1}) \La(s)+ O(  r^2 \dg). 
\eeaa

 Similarly, starting with,
 \beaa
   \nu^\S (\fb)  &=2(\xib ^\S-\xib) - \frac 1 2 \big( \kab+4\omb) (\fb- a^\S f)+ a^\S\left(2  e^\S_\th( \log \la  )    -  \fb \ka \right)  +F\c \Ga_b+\lot
   \eeaa
   we deduce,
\beaa 
 \int_{\S(s)}  \nu^\S (\fb) e^\Phi &=& 2 \Bb(s)- 2 \int_{\S(s)}\xib e^\Phi +r^{-1}\left(1+\frac{8m}{r}\right)\Lab(s) 
 +r^{-1}\left(1-\frac{2m}{r} - \frac{8m^2}{r^2}\right)\La(s)\\
 &-&  2 \left( 1+\frac{2m}{r}\right)  \int_{\S(s)} e_\th^\S ( \log \la ) e^\Phi  +  r O\left(\dg\, \epg^{1/2}\right). 
\eeaa
 Making use of the assumption \eqref{assumptions:badmodeof-eta-xib}  for $\xib$, as well as Corollary  \ref{corr:lemma:int-gaS-ga},
\bea
\left| \int_{\S(s)}\xib e^\Phi\right|\les  r^2 \dg.
\eea
Also, in view of the estimates of Theorem  \ref{Theorem:ExistenceGCMS}, 
 \beaa
  \left| \int_{\S(s)} e_\th^\S ( \log \la ) e^\Phi  \right|\les  r^2 \dg.
   \eeaa
 We deduce,
 \beaa 
 \int_{\S(s)}  \nu^\S (\fb) e^\Phi &=& 2 \Bb(s)- r^{-1}(1+O(r^{-1}) )\Lab(s) 
 +r^{-1}\big(1 +O(r^{-1} ) \big)\La(s) +O(r^2 \dg)
 \eeaa
 as stated. 
The estimates for $E, \Eb$    in \eqref{eq:Prop-diff-eqtsforLaLab2}  can also be  easily checked.
This ends the proof of Proposition \ref{Prop: diff-eqtsforLaLab}.
  \end{proof}

  {\bf Step 19.} We derive an equation for $\psi$. The main result is stated  in the proposition below.
  \begin{proposition}
  \lab{Prop:equationforpsi'}
  The function  $\psi(s) =\Psi(s)+s-c_0$  defined in Step  13  verifies the following equation
  \bea
  \lab{equation:forpsi'}
  \psi'(s)&=-\frac 1 2 D(s) + O(D(s)^2) + M(s)
  \eea
  where $M(s) $ is a function which  depends only on  $\Ga, R$ of    the  background  foliation, $\psi$   and $(f, \fb, \la-1)$ such that,
  \beaa
  \big|M(s)\big| &\les \dg r(s)^{-1}. 
  \eeaa
    \end{proposition}
  
  \begin{proof} 
 In view of 
     \eqref{eq:a^S-psi'} and the definition of $c(s)=\Big(  \frac{2\la}{\vsi} \Big)\Big|_{SP}(s)$ 
    we  have,
   \beaa
   \Psi'(s)&=&\frac{1}{a^\S\big|_{SP}(s) }   \c   \frac{2\la^2}{\vsi}  \left( 1-\frac 1 2 \Psi'\vsi \Omb\right)\Bigg|_{SP}(s)
   \eeaa  
   or
   \bea
\Psi'(s) &=& \left[\frac{2\la^2}{\vsi}\frac{1}{a^\S+\la^2\Omb}\right]\Big|_{SP}.
\eea
Now, we have
\beaa
\frac{2\la^2}{\vsi}\frac{1}{a^\S+\la^2\Omb} &=& \frac{2}{\vsi}\frac{1}{a^\S+\Omb}+O(\la-1)\\
&=& -1+\frac{a^\S+\frac{2}{\vsi}+\Omb}{a^\S+\Omb}+O(\la-1).
\eeaa
Hence,
\beaa
\psi'(s)&=&\Psi'(s)+1= \left[\frac{a^\S+\frac{2}{\vsi}+\Omb}{a^\S+\Omb}\right]\Big|_{SP} +O(\la-1)=
\left[\frac{a^\S+\frac{2}{\vsi}+\Omb}{a^\S+\Omb}\right]\Big|_{SP} +O( r^{-1} \dg).
\eeaa
We have, see  \eqref{Define:BSBbSDS}, 
\beaa
a^\S\big|_{SP} (s)&=& D(s) -1-\frac{2m^\S}{r^\S}.
\eeaa
Hence,
\beaa
\big(a^\S+\Omb\big) \big|_{SP}(s) &=& D(s) -1-\frac{2m^\S}{r^\S}+\Omb \big|_{SP}(s) = D(s) -1-\frac{2m^\S}{r^\S}- (1-\frac{2m}{r})+O(\epg)\\
&=&D(s)-2 +O(\epg).
\eeaa
In view of the  assumption \eqref{assumptions:badmodeof-a^SatSP}
 \beaa
\left| \left(\frac{2}{\vsi}+\Omb\right)\Big|_{SP}- 1-\frac{2m}{r} \right| &\les r^{-1}  \dg
 \eeaa
 we deduce
 \beaa
\left(a^\S+\frac{2}{\vsi}+\Omb\right)\big|_{SP}(s)&= &a^\S\big|_{SP}+1+\frac{2m}{r}+O(r^{-1}\dg)\\
&=& D(s) +\frac{2m}{r}- \frac{2m^\S}{r^\S}+O(r^{-1}\dg)\\
&=&D(s)+ O(r^{-1}\dg).
\eeaa
Hence,
\beaa
\psi'(s) &=&\left[\frac{a^\S+\frac{2}{\vsi}+\Omb}{a^\S+\Omb}\right]\Big|_{SP} +O( r^{-1} \dg)=- \frac 1 2 D(s) +O(D(s)^2)       + O(r^{-1}\dg)
\eeaa  
as stated.
  \end{proof}

  {\bf Step 20.} We combine Propositions \ref{Prop: diff-eqtsforLaLab} and  \ref {Prop:equationforpsi'} to  derive
  the closed system of equations in $\La, \Lab, \psi$,
  \bea
 \lab{system:La'-Lab'-psi'-1}
\bsplit
\frac{1}{-1+\psi'(s)  }\La'(s)   &= B(s)- \frac 1 2 r(s)^{-1}\Lab(s) - \frac 7  2  r(s)^{-1}\La(s) + O(r^{-1}) \La(s)\\&+N(B, \Bb, D, \La, \Lab, \psi)(s),\\
  \frac{1}{-1+\psi'(s)  }\Lab'(s)   &= \Bb(s)-\frac 7  2 r(s)^{-1}\Lab(s) + \frac 1   2  r(s)^{-1}\La(s) +  O(r^{-1}) \big(\La(s)+\Lab(s) \big)\\
 &+\Nb(B, \Bb, D, \La, \Lab, \psi)(s),\\
    \psi'(s)&=- \frac 1 2 D(s) +O(D(s)^2)       +M(s),
  \end{split}
 \eea
 with  initial conditions
\bea
\lab{eq:initializationoftheODEsystemforLaLabpsi}
\psi(\ovs)=0, \qquad \La(\ovs)=\La_0, \qquad \Lab(\ovs)=\Lab_0.
\eea
Recall also that  $r(s)$  is a smooth function of $\psi(s)$.

 The system \eqref{system:La'-Lab'-psi'-1} is verified for all choices of  $(\La,\Lab, \Psi)$. We now make a suitable particular choice for $(\La,\Lab, \Psi)$ as follows.
 
  Consider  in particular  the system obtained from\eqref{system:La'-Lab'-psi'-1} by setting $B, \Bb, D$ to zero 
\bea
\lab{System:La'-Lab'-D'-reduced} 
\bsplit
  \psi'(s)&=M(s),
\\
\frac{1}{-1+\psi'(s) ) }\La'(s)   &=- \frac 1 2 r(s)^{-1}\Lab(s) - \frac 7  2  r(s)^{-1}\La(s) + O(r^{-1}) \La(s)+\widetilde{N}(\La, \Lab, \psi)(s),\\
  \frac{1}{-1+\psi'(s) ) }\Lab'(s)   &=-\frac 7  2 r(s)^{-1}\Lab(s) + \frac 1   2  r(s)^{-1}\La(s) +  O(r^{-1}) \big(\La(s)+\Lab(s) \big)+\widetilde{\Nb}(\La, \Lab, \psi)(s),
  \end{split}
\eea
where,
\beaa
\widetilde{N}(\La, \Lab, \psi)&=&N(0,0,0, \La, \Lab, \psi),\\
\widetilde{\Nb}(\La, \Lab, \psi)&=&\Nb(0,0,0, \La, \Lab, \psi).
\eeaa

We initialize
 the system at $s=\sg$ as in \eqref{eq:initializationoftheODEsystemforLaLabpsi}, i.e.,
\beaa
\La(\sg)=\psi(\sg)=0, \qquad \La_0, \qquad \Lab(\sg)=\Lab_0.
\eeaa
 The system admits a unique solution $\psi(s)$   defined  in a small   neighborhood $ \ovI  $ of   $\ovs$.
  The function $\Psi(s)= -s+\psi(s)+c_0$  defines the desired hypersurface  $\Si_0$.

 {\bf Step 21.} 
 It remain to show that the function $B, \Bb, D $ vanish on the hypersurface   $\Si_0$ defined above.  Since the system  \eqref{system:La'-Lab'-psi'-1} is verified for all functions $\La, \Lab, \psi$  we deduce, along $\Si_0$,
 \beaa
 D&=&0,\\
 B&=& N(B, \Bb, D, \La, \Lab, \psi)(s)-  N(0,0, 0, \La, \Lab, \psi)(s),\\
  \Bb &=&\Nb(B, \Bb, D, \La, \Lab, \psi)(s)- \Nb(0,0, 0, \La, \Lab, \psi)(s).
  \eeaa
 In view of the properties of $N,\Nb$ we deduce,
 \beaa
\Big| N(B, \Bb, D, \La, \Lab, \psi)(s)-  N(0,0, 0, \La, \Lab, \psi)(s)\Big|&\les& \epg\sup_{\ovI} \Big( |B(s)| +|\Bb(s) |\Big),\\
\Big| \Nb(B, \Bb, D, \La, \Lab, \psi)(s)-  \Nb(0,0, 0, \La, \Lab, \psi)(s)\Big|&\les& \epg\sup_{\ovI} \Big( |B(s)| +|\Bb(s)|\Big).
 \eeaa
 Hence,
 \beaa
  \sup_{\ovI}  |B(s)|+ \sup_{\ovI} |\Bb(s)|  &\les& \dg \big(  \sup_{\ovI}  |B(s)|+ \sup_{\ovI} |\Bb(s)|\big).
  \eeaa
 Hence $B, \Bb, D$  vanish identically on  $\Si_0$.

 {\bf Step 22.} We  have,
 \bea
  \Big|\frac{dr}{ds}-1\Big|\les \epg. 
 \eea
 Indeed,  according to Step 15  and Lemma \ref{Lemma:nuSof integrals} we have
  \beaa
  \frac{d}{ds} r(s)&=& X_*\Big|_{SP} r^\S=\frac{1}{2\la}  \vsi \Psi' \,  \nu^\S(r^\S)\Big|_{SP}=(-1+\psi'(s)) \frac{1}{2\la}  \vsi  \nu^\S(r^\S)\Big|_{SP}\\
  &=&(-1+\psi'(s))\left( \frac{1}{2\la}  \vsi   \frac{r^\S}{2}(\vsi^\S)^{-1}\ov{\vsi^\S(\kab^\S+a^\S\ka^\S)}\right)\Big|_{SP}.
  \eeaa
 In view of Proposition \ref{Prop:equationforpsi'}, with $D=0$,  $\big|\psi' \big|\les  r^{-1} \dg $. We deduce,
  \beaa
  \frac{d}{ds} r(s)&=&-\left( \frac{1}{2\la}  \vsi   \frac{r^\S}{2}(\vsi^\S)^{-1}\ov{\vsi^\S(\kab^\S+a^\S\ka^\S)}\right)\Big|_{SP}+O(\dg).
  \eeaa

 {\bf Step 23.}   Therefore the functions $B, \Bb, D$ vanish identically on  the hypersurface $\Si_0$ defined by the function $\Psi(s)=-s+\psi(s)+c_0$ which accomplishes the main task of Theorem
   \ref{theorem:contrutionofGCMhypersurface}. More  precisely  we have produced a   local  hypersurface 
   $\Si_0$,  as defined in  Step 12,  foliated by the function $u^\S$, defined in Step 2 and extended in Step 3,   such that  the   items  2-5 of the theorem are verified. The estimates in items  6-7  are an immediate consequence    of Proposition \ref{Proposition:Estmates-etaSxibSombSvsiSabSaS}. It only remains to prove the smoothness of     the function $\Xi(s, \th)$   in \eqref{def:Sigma*3}, Step 14 and  the estimates  for $F=(f, \fb, \log\la)$ in the last part of the theorem.    To check the differentiability properties   recall  that,
   \beaa
\pr_s \Xi(s,\th)&=&\Big(\Psi'(s) +\pr_P U(\th,P(s)) P'(s),\,\,  1+\pr_P S(\th,P(s)) P'(s),\,\,  0\Big),\\
\pr_\th  \Xi(s,\th)&=&\Big(\pr_\th U(\th, P(s)), \, \, \pr_\th S(\th, P(s)), \,\,  1 \Big),
\eeaa
where,
\beaa
\pr_P U(\cdot) P'(s)&=&  \Psi'(s) \pr_u U(\cdot)+ \pr_s U(\cdot) +  \La'(s)\pr_\La U(\cdot)+ \Lab'(s) \pr_\Lab U(\cdot),\\
\pr_P S(\cdot) P'(s)&=&  \Psi'(s) \pr_u S(\cdot)+ \pr_s S(\cdot) +  \La'(s)\pr_\La S(\cdot)+ \Lab'(s) \pr_\Lab S(\cdot).
\eeaa
Thus to prove the smoothness of $\Xi$ we need to appeal to the smoothness of $ U, S$ with respect to  the parameters $\La, \Lab$ and $u, s$.   Though tedious, this can be easily done,  by appealing back to    the coupled  system of equations,
\eqref{GCMS:NEAR-again:bis},  \eqref{equations-forBadModes}  and \eqref{equations-forBadModes-ffb}  \eqref{eq:DeformS-66}, as in the proof  of Theorem  \ref{Theorem:ExistenceGCMS} and studying its dependence on  these parameters.

   {\bf Step 24.}   It only remains to derive the   estimates \eqref{EstimatesforF-onSi_*}  for  the transition functions $F=(f, \fb, \log\la)$.
   To start with we  have, in view of the construction of $\Si_0$  and  the estimates for  $F=(f, \fb, \log\la)$  of   Theorem \ref{Theorem:ExistenceGCMS},  for every $\S\subset\Si_0$
    \bea
   \lab {eq:rigidityrecoveryofnuderivatives:0}
 \|F \|_{\hk_{s_{\max}+1 }(\S)}\les \dg. 
 \eea
  To derive the   remaining tangential derivatives     of $F$ along   $\Si_0$
  we the commute the GCM  system \eqref{GCMS:NEAR-again:bis} of Proposition 
 \ref{proposition-GCMS-system}  with respect to   $\nu=\nu^\S= e_3^\S + a^\S e_4 ^\S$ and then proceed, as in the  proof of  the apriori 
  estimates of Theorem \ref{prop.GCMSequations-fixedS}  to derive recursively the estimates, for  $K= k_{max}+1$,
  \bea
  \lab{eq:rigidityrecoveryofnuderivatives}
  \bsplit
\|\nu^l( F) \|_{\hk_{K-l}(\S)} &\les \dg+\left|\int_{\S}\nu^l(f)e^\Phi\right|+\left|\int_{\S}\nu^l(\fb) e^\Phi\right|\\
&+\dg \left\|\nu^{\leq l-1}a^\S\right\|_{\hk_{K-l+1}(\S)}+\|\nu^{\le l-1}  F \|_{\hk_{K-l+1}(\S)}.
\end{split}
\eea
  We already have estimates for the $\ell=1$ modes of $F= (f, \fb)$. To estimate the $\ell=1$ modes of  $\nu^l (f, \fb)$, $l\ge 1$,   we  make use of the equations \eqref{eqts:nuS-ffb}
 and the vanishing of the $\ell=1$ modes of $\eta^\S$,  $ \xib^\S$ along $\Si_0$ to derive, recursively, for all $1\le l\le  K$,
 \bea
 \bsplit
 \lab{eq:rigidityrecoveryofnuderivatives:1}
\left|\int_{\S}\nu^l(f)e^\Phi\right|+\left|\int_{\S}\nu^l(\fb) e^\Phi\right| &\les \dg+\dg \left\|\nu^{\leq l-1}\Big(a, \Omb^\S, \vsi^\S, \xib^\S, \eta^\S, \check{\omb}^\S\Big)\right\|_{\hk_{K-l+1}(\S)}\\
&+\|\nu^{\le l-1} (F)\|_{\hk_{K-l+1}(\S)}.
\end{split}
\eea 
We can then proceed as in the proof of Proposition \ref{Proposition:Estmates-etaSxibSombSvsiSabSaS}  
    derive, recursively, the estimates
\bea\lab{eq:rigidityrecoveryofnuderivatives:2}
\left\|\nu^{\leq l-1}\Big(a^\S, \Omb^\S, \vsi^\S, \xib^\S, \eta^\S, \check{\omb}^\S\Big)\right\|_{\hk_{K-l+1}(\S)} &\les& 1+\sum_{j=0}^{l}\|\nu^j(F)\|_{\hk_{K-j}(\S)}. 
\eea
Combining  \eqref{eq:rigidityrecoveryofnuderivatives} \eqref{eq:rigidityrecoveryofnuderivatives:1} \eqref{eq:rigidityrecoveryofnuderivatives:2}, we obtain
\beaa
\|\nu^l(F)\|_{\hk_{K-l}(\S)} &\les& \dg+\sum_{j=0}^{l-1}\|\nu^j( F )\|_{\hk_{K-j}(\S)},
\eeaa
which, together with \eqref{eq:rigidityrecoveryofnuderivatives:0},  yields  the desired estimate for all tangential derivatives
\beaa
\sum_{j=0}^{K}\|\nu^j(F)\|_{\hk_{K-j}(\S)} &\les& \dg.
\eeaa
To complete the desired estimate for all derivatives    we make use of  the equations  for $e^\S_4(F)$,  due to the transversality conditions  \eqref{Transversality-Si*}. The $e^\S_3$ derivatives   can then be derived from  $\nu^\S= e_3^\S + a^\S e_4^\S$ and the estimates for $a^\S$.
 This ends the proof of Theorem \ref{theorem:contrutionofGCMhypersurface}.
 
{\bf Step 25.}  To prove Corollary \ref{corr:GCM-rigidity3} we  start with  the rigidity statement of Corollary \ref{corr:GCM-rigidity1}. Making use of the assumption $ \int_{\S_0}f e^\Phi=O(\dg), \quad \int_{\S_0}\fb e^\Phi=O(\dg)$
 we infer that the estimate \eqref {eq:rigidityrecoveryofnuderivatives:0}  holds true for $\S_0$.  We then proceed 
  exactly as in Step 24 to derive the estimates   \eqref{eq:rigidityrecoveryofnuderivatives} \eqref{eq:rigidityrecoveryofnuderivatives:1} \eqref{eq:rigidityrecoveryofnuderivatives:2} for our  distinguished sphere $\S_0$.  Note that $\S_0$ can be viewed as a deformation of the unique background sphere sharing the same south pole.


\chapter{REGGE-WHEELER TYPE EQUATIONS}\label{chapter:waveeqtionestimates}


The goal of this chapter is to prove Theorem \ref{theorem-combinedMor-r-weighted} and Theorem \ref{theorem:Daf-Rodn-estim2-psic} concerning the weighted estimates for the solution $\psi$ to 
 \beaa
\square_2 \psi+V\psi =N, \qquad V=\ka\kab. 
\eeaa
Recall that these theorems where used in Chapter \ref{chapter:Thm:mainwavetheorem} to prove Theorem M1.
 
The structure of the chapter is as follows.
 \begin{itemize}
 \item In section \ref{section-BasicMorawetz}, we  prove basic Morawetz estimates for $\psi$.
 
\item In section \ref{section-Basic-rp}, we prove $r^p$-weighted estimates in the spirit of Dafermos-Rodnianski \cite{Da-Ro3} for $\psi$. In particular, we obtain as an immediate corollary the proof of Theorem \ref{theorem-combinedMor-r-weighted} in the case $s=0$ (i.e. without commutating the equation of $\psi$ with derivatives).

\item In section \ref{sec:AnArGatypeestimates}, we use a variation of   the method of \cite{AnArGa}  to derive slightly stronger  weighted   estimates and prove Theorem \ref{theorem:Daf-Rodn-estim2-psic} in the case $s=0$ (i.e. without commutating the equation of $\psic$ with derivatives).

\item In section \ref{esc:higherderivativesestimates}, commuting the equation of $\psi$ with derivatives, we complete the proof of Theorem \ref{theorem-combinedMor-r-weighted} by controlling higher order derivatives of $\psi$, i.e. for $s\le k_{small}+30$. Also, commuting the equation of $\psic$ with derivatives, we complete the proof of Theorem \ref{theorem:Daf-Rodn-estim2-psic} by controlling higher order derivatives of $\psic$, i.e. for $s\le k_{small}+29$. 
 \end{itemize}


\section{Basic  Morawetz Estimates}\lab{section-BasicMorawetz} 


Recall 
\begin{itemize}
\item the definitions in section \ref{sec:foliationofMMbytau} of $\Mtrap$, $\Mntrap$, $\tau$, $\Sigma(\tau)$ and $\,^{(trap)}\Si$,

\item the main quantities involved in the energy and Morawetz estimates, e.g. $ E[\psi](\tau)$, $\Mor[\psi](\tau_1,\tau_2)$, $\Morr[\psi](\tau_1,\tau_2)$, $F[\psi](\tau_1,\tau_2)$, $J_{\de}[\psi, N](\tau_1,\tau_2)$ and $\Bdot^s_{p; R}[\psi](\tau_1, \tau_2)$,  introduced in section \ref{sec:mainquantitiesforcontrolofwaveequation}.
\end{itemize}

The following theorem claims basic Morawetz estimates for the solution $\psi$ of the wave equation \eqref{eq:masterwavepsi}.  
 \begin{theorem}[Morawetz] 
\label{Thm:Morawetz-s}  
Let $\psi$ a reduced 2-scalar solution to 
 \beaa
\square_2 \psi+V\psi =N, \qquad V=\ka\kab. 
\eeaa
Let  $\de>0$ be a fixed   small constant   verifying $0<\ep\ll \de$. 
 The   following estimates hold true  in  $\MM(\tau_1,\tau_2)$, $0\le \tau_1 < \tau_2\le \tau_*$, 
     \bea
     \bsplit
        E[\psi](\tau_2)            +\Morr[\psi](\tau_1,\tau_2)+F[\psi](\tau_1,\tau_2)&\les    E[\psi](\tau_1)  +J_{\de}[\psi, N](\tau_1,\tau_2) \\
        & + O(\ep)\Bdot^s_{\de\,; \,4m_0}[\psi](\tau_1, \tau_2).
        \end{split}
  \eea
\end{theorem}

\begin{remark}
Note that the  bulk term $\Bdot^s_{\de\,; \,4m_0}[\psi](\tau_1, \tau_2)$ cannot yet be absorbed on the left hand side of the inequality. To do that we will rely on  the $r^p$  weighted estimates of Theorem \ref{theorem:Daf-Rodn1-psi-s}.
\end{remark}

\begin{remark}
In addition to $\ep$ and $\de$, the proof of Theorem \ref{Thm:Morawetz-s} will involve several smallness constants: $C^{-1}$, $\widehat{\de}$, $\de_1$, $\deh$, $\ep_\HH$, $\La_\HH^{-1}$ and $\La^{-1}$. These smallness constants will be chosen such that
\bea
0<\ep\ll \widehat{\de}, \, \deh, \,\ep_\HH, \, \La_\HH^{-1}, \, \La^{-1}\ll \de_1\ll C^{-1}.
\eea 
In addition, $\widehat{\de}$, $\ep_\HH$, $\La_\HH^{-1}$ and $\La^{-1}$ will in fact be chosen towards the end of the proof as explicit powers of $\deh$, see \eqref{eq:restrictionson-de...}, \eqref{eq:choiceofLambdaforpositivityofboundaryterm} and Proposition \ref{prop:pre-mor4}.
\end{remark}

The goal of this section is to  prove Theorem \ref{Thm:Morawetz-s}. This will be achieved in section \ref{sec:wheremorawetzinfinalformareactuallyproved}.

 
 \subsection{Structure of the proof of Theorem \ref{Thm:Morawetz-s}}
 

To prove Theorem \ref{Thm:Morawetz-s}, we proceed as follows
\begin{itemize}
\item In section \ref{sec:simplifiedsetofassumptionsformorawetz}, we introduce a simplified set of assumptions of the Ricci coefficients which is sufficient in order to prove Theorem \ref{Thm:Morawetz-s}.

\item In section \ref{sec:notationonfunctionsdependingonmandr}, we discuss notations concerning functions depending on $m$ and $r$.

\item In section \ref{sec:compuationvariousdeformationtensorsformorasetz}, we compute the deformation tensor of the vectorfields $R$, $T$, and $X=f(r,m) R$. 

\item In section \ref{section:basicintegralidentities}, we introduce the basic integral  identities for wave equations that will be used repeatedly in the proof of Theorem \ref{Thm:Morawetz-s}.

\item In section \ref{sec:mainmorawetzidentity}, we derive the main  Morawetz    identity.

\item In section \ref{sec:trapingregionmorawetzpsi}, we derive a first estimate. This estimate is insufficient du to
\begin{itemize}
\item a lack of positivity of the bulk in to the region $3m\leq r\leq 4m$, 

\item a log divergence of a suitable choice of vectorfield at $r=2m$,

\item a degeneracy at $r=2m$.
\end{itemize}

\item In section \ref{subs:improvedLB-awayH}, we add a correction and rely on a Poincar\'e inequality to obtain a positive estimate also on the region  $3m\leq r\leq 4m$.

\item In section \ref{subsection:Mor-cutoff}, we perform a cut-off to remove above mentioned log divergence at $r=2m$.

\item In section \ref{subsection:redshift}, we introduce the red shift vectorfield to remove the above mentioned degeneracy at $r=2m$.

\item In section \ref{sec:combinationofMorawetzwithredshift}, we combine the previous estimates with the redshift vectorfield to obtain a bulk term suitable on the whole spacetime $\MM$.

\item In section \ref{sec:lowerboundonQformorawetz}, we prove the positivity of the boundary terms arising from adding a large multiple of the energy estimate to the Morawetz estimate.

\item In section \ref{sec:firstmorawetzestimate}, combining the good properties of the bulk and of the boundary terms established so far, we obtain a first Morawetz estimate providing in particular the control of the the quantity $\Mor[\psi]$.

\item In section \ref{sec:analysisoftheerrortermEEep}, we analyse an error term appearing in the right-hand side of the above mentioned Morawetz estimate.

\item Finally, in section \ref{sec:wheremorawetzinfinalformareactuallyproved}, we add a correction to upgrade the control of $\Mor[\psi]$ to the control of the quantity $\Morr[\psi]$, hence concluding the proof of   Theorem \ref{Thm:Morawetz-s}.
\end{itemize}

 
 \subsection{A simplified set of assumptions}\lab{sec:simplifiedsetofassumptionsformorawetz}
 
 
To prove Theorem \ref{Thm:Morawetz-s}, it suffices to make a simplified set of assumptions. Define  
 \bea
 u_{trap}=
 \begin{cases}
1+\tau  \qquad \,\,\mbox{for}  \quad \          r\in [ \frac{5m_0}{2},  \frac{7m_0}{2}],
 \\
 1 \qquad\qquad\,\,  \mbox{for} \quad     r\notin  [ \frac{5m_0}{2},  \frac{7m_0}{2}].
  \end{cases}
 \eea
 For $ k=0, 1$, we assume the following.

 \begin{enumerate}
 \item[ {\bf Mor1.}]  
 The renormalized Ricci coefficients   $\check{\Ga}^{\le k}$ verify on $\MM=\Mint\cup \Mext$,
\bea
 \label{eq:assumptions-Moraw2}
\bsplit
|\Gac^{\le k}|&\les \ep r^{-1} u_{trap}^{-1-\dec},\\
\Big|\dk^{\le k}\Big(\om+\frac{m}{r^2}, \xi \Big) \Big|&\les   \ep r^{-2} u_{trap}^{-1-\dec}.
\end{split}
\eea

\item[{\bf Mor2.}] The Gauss curvature $K$ of $S$  and   $\rho$     verify,
 \label{eq:assumptions-Moraw3}
\bea
\begin{split}
\Big|\dk^{\le k}\Big(\rho+\frac{2m}{r^3}\Big)\Big|&\les  \ep r^{-2}  \,u_{trap}^{-1-\dec},\\
\Big|\dk^{\le k}\Big(K-\frac{1}{r^2}\Big)\Big|  &\les \ep r^{-2}  \,u_{trap}^{-1-\dec}.
\end{split}
\eea

\item[{\bf Mor3.}]
We also assume
\bea
 \label{eq:assumptions-Moraw4}
\bsplit
|m-m_0&|\les \ep m_0,\\
|\dk^{\le k}( e_3m,\, r^2 e_4 m)|&\les \ep   \,u_{trap}^{-1-\dec}.
\end{split}
\eea
\end{enumerate}

\begin{remark}
 Note  that in the case when the bootstrap constant  $\ep=0$,  i.e.in  Schwarzschild,     the assumptions made above  are  consistent with   
  the behavior  relative  to the regular   frame (near horizon)  
 \beaa
 e_3=\Up^{-1} \pr_t -\pr_r, e_4=\pr_t+\Up \pr_r.
  \eeaa
\end{remark}


\subsection{Functions depending on $m$ and $r$}\lab{sec:notationonfunctionsdependingonmandr}


In order to prove Theorem \ref{Thm:Morawetz-s}, we will adapt the derivation of the Morawetz estimate for the wave equation in Schwarzschild. In particular, we will need to consider various scalar functions, used to define suitable analogs of the vectorfields in Schwarzschild, which depend on $m$ and $r$. Now, $m$ is now a scalar function unlike the Schwarzschild case where it is constant. To take this into account, we will rely on the following lemma.

\begin{lemma}
\label{remark:slowly varyingf}
Let $f=f(r,m)$ a $C^1$ function of $r$ and $m$. Then, we have 
\beaa
&& e_4\big(f(r, m)\big) = \pr_rf(r, m)e_4(r)+O(\ep r^{-2}u_{trap}^{-1-\dec}|\pr_mf|),\\
&& e_3\big(f(r, m)\big) = \pr_rf(r, m)e_3(r)+O(\ep u_{trap}^{-1-\dec}|\pr_mf|),\\
&& e_4\big(e_3\big(f(r, m)\big)\big) = \pr_r^2f(r,m)e_4(r)e_3(r)+\pr_rf(r,m)e_4(e_3(r))\\
&&\qquad\qquad\qquad\qquad\qquad +O(\ep r^{-2}u_{trap}^{-1-\dec}(r|\pr_r\pr_mf|+|\pr^2_mf|)),\\
&& e_3\big(e_4\big(f(r, m)\big)\big) = \pr_r^2f(r,m)e_4(r)e_3(r)+\pr_rf(r,m)e_3(e_4(r))\\
&&\qquad\qquad\qquad\qquad\qquad +O(\ep r^{-2}u_{trap}^{-1-\dec}(r|\pr_r\pr_mf|+|\pr^2_mf|)),\\
&& e_\th\big(f(r, m)\big)=0.
\eeaa
   \end{lemma}

\begin{proof}
Straightforward verification using \eqref{eq:assumptions-Moraw4}.
\end{proof}

\begin{remark}\lab{remark:slowly varyingf:bis}
Note that in the sequel, $\pr_rf$ will not denote a spacetime coordinate vectorfield applied to $f$, but instead the partial derivative with respect to the variable $r$ of the function $f(r,m)$. 
\end{remark}


\subsection{Deformation tensors of the vectorfields $R, T, X$}\lab{sec:compuationvariousdeformationtensorsformorasetz}


Recall the  definition \eqref{def:TandR} of the   regular  vectorfields\footnote{ In  Schwarzschild,  in  standard coordinates,  we have  $T=\pr_t, \, \,  R=\Up \pr_r $ which are regular      near the horizon. },
 \beaa
 T&=&\frac 1 2 \left(e_4+\Up e_3 \right), \qquad   R= \frac 1 2 \left(  e_4 - \Up  e_3\right).
 \eeaa
 Note that,
 \beaa
- g(T,T)=g(R, R) = \Up, \qquad g(T, R) = 0.
 \eeaa
Note  also  that,
\beaa
R(r)= 1 -\frac{2m}{r}+O(\ep u_{trap}^{-1-\dec}), \qquad T(r)=O(\ep u_{trap}^{-1-\dec}).
\eeaa

 \begin{lemma}
 \label{lemma:componentspiR}
 The following hold true.
 \begin{enumerate}
 \item  The components of the deformation tensor of $R= \frac 1 2 \left(  e_4 - \Up  e_3\right)$  are given by,
\beaa
\Big|\piR_{34}+\frac{4m}{r^2}    \Big| &\les&   \ep r^{-1} u_{trap}^{-1-\dec},\\
\Big|\piR(e_A, e_B)-\frac{2}{r} \Up\de_{AB} \Big| &\les&   \ep r^{-1} u_{trap}^{-1-\dec},\\
\Big|\piR_{33}\Big| &\les & \ep r^{-1} u_{trap}^{-1-\dec},\\
\Big|\piR_{3\th}\Big| &\les & \ep r^{-1} u_{trap}^{-1-\dec},\\
\Big|\piR_{4\th}\Big| &\les & \ep r^{-1} u_{trap}^{-1-\dec}.
\eeaa
Moreover,
\beaa
\Big|\piR_{44} \Big| &\les & \ep r^{-2} u_{trap}^{-1-\dec}.
\eeaa
\item   If  $V=-\ka\, \kab $ we have,
\bea
\bsplit
 e_3(V) &=\frac{8}{r^3} \left(1-\frac{3m}{r}\right)+O( \ep) r^{-3} u_{trap}^{-1-\dec}, \\
 e_4(V) &=-\frac{8\Up}{r^3} \left(1-\frac{3m}{r}\right)+O( \ep) r^{-3} u_{trap}^{-1-\dec}, 
 \end{split}
\eea
and,
\beaa
 R(V)&=& -\frac{8\Up }{r^3} \left(1-\frac{3m}{r}\right)+O( \ep) r^{-3} u_{trap}^{-1-\dec},\\
 T(V)&=& O( \ep) r^{-3} u_{trap}^{-1-\dec}.
 \eeaa
\item  All  components of the deformation tensor of    $T=\frac 1 2 \left(e_4+\Up e_3 \right)$    can be bounded by 
               $O(\ep r^{-1} u_{trap}^{-1-\dec})$.   Moreover,
             \beaa
             \Big|\piT_{44} \Big| &\les & \ep r^{-2} u_{trap}^{-1-\dec}. 
             \eeaa
\end{enumerate}
 \end{lemma}
 
        \begin{proof}
We have
\beaa
\piR_{44}&=&   \g(\D_4 ( e_4-\Up e_3), e_4) = 2  e_4(\Up)  + 4\Up \om,\\
\piR_{34}&=&  \frac 1 2 \g(\D_3 ( e_4-\Up e_3), e_4) + \frac 1 2 \g(\D_4 ( e_4-\Up e_3), e_3)\\
&=& e_3(\Up) -2\Up \omb+ 2\om,\\
\piR_{33}&=&  \g(\D_3( e_4-\Up e_3), e_3) =-4\omb,\\
\piR_{AB} &=& \frac 1 2\g(\D_A ( e_4-\Up e_3), e_B)+\frac 1 2\g(\D_B ( e_4-\Up e_3), e_A),\\
&=&\chiS_{AB}-\Up\chibS_{AB} \\
&=& \frac 1 2(\ka-\Up \kab)\de_{AB} + \chihS_{AB}-\Up\chibhS_{AB}.
\eeaa
Note that,
      \beaa
      e_3(\Up)&=& e_3\left(1-\frac{2m}{r} \right)=\frac{2m}{r^2} e_3(r)-\frac{2e_3 m}{r}=\frac{m}{r}(\kab+\Ab)+ O(\ep r^{-1}  u_{trap}^{-1-\dec})\\
      &=&\frac{m}{r} \kab +O(\ep r^{-1} u_{trap}^{-1-\dec})=      -\frac{2m}{r^2}+O(\ep r^{-1} u_{trap}^{-1-\dec}),\\
      e_4 (\Up)&=& e_4\left(1-\frac{2m}{r}\right)=\frac{2m}{r^2} e_4 (r)-\frac{2e_4 m}{r}=\frac{m}{r}(\ka+A)+ O(\ep r^{-2}  u_{trap}^{-1-\dec})\\
      &=&\frac{m}{r} \ka +O(\ep r^{-2} u_{trap}^{-1-\dec})= \frac{2 m}{r^2} \Up+ O(\ep r^{-2}  u_{trap}^{-1-\dec}).
      \eeaa      
Thus
\beaa
\piR_{44} &=&O( \ep r^{-2} u_{trap}^{-1-\dec}),\\
\piR_{33}&=&O( \ep r^{-1} u_{trap}^{-1-\dec}),\\
\piR_{34}&=&- 4\frac{m}{r^2} +O( \ep r^{-1} u_{trap}^{-1-\dec}),\\
\piR_{AB}&=&\frac{2\Up}{r}\de_{AB} +O( \ep r^{-1} u_{trap}^{-1-\dec}).
\eeaa
   Also, in view of,
 \beaa
 \Big|\xi, \xib, \eta, \etab, \ze\Big|\les \ep  r^{-1} u_{trap}^{-1-\dec},
 \eeaa
 we deduce,
 \beaa
\Big| \piR_{3\th}, \piR_{4\th} \Big|\les  \ep  r^{-1} u_{trap}^{-1-\dec}
 \eeaa
as desired.

To prove the second part of the lemma we write,
 \beaa
 e_3(V)&=&- e_3(\ka)\kab -\ka e_3(\kab)\\
 & =& -\left(-\frac 1 2 \ka \kab +2\omb \ka +2 \rho  +O( \ep r^{-2} u_{trap}^{-1-\dec})\right)\kab -\ka \left(-\frac 1 2 \kab^2-2\omb\, \kab+O( \ep r^{-2} u_{trap}^{-1-\dec})\right)\\
 &=&(\ka \kab-2\rho )\kab +O( \ep r^{-3} u_{trap}^{-1-\dec}).
 \eeaa
 On the other hand,
 \beaa
\ka \kab-2\rho &=&-\left( \frac{2\Up}{r}  +O( \ep) r^{-1} u_{trap}^{-1-\dec} \right)\left(     \frac{2}{r}+O( \ep) r^{-1} u_{trap}^{-1-\dec} \right)+\frac{4m}{r^3}+O( \ep r^{-3} u_{trap}^{-1-\dec})\\
&=&-\frac{4}{r^2} \left(1-\frac{3m}{r}\right)+O( \ep r^{-2} u_{trap}^{-1-\dec}). 
 \eeaa
 Hence,
 \beaa
 e_3(V)&=&(\ka \kab-2\rho )\kab +O( \ep r^{-3} u_{trap}^{-1-\dec})\\
 &=&\frac{8}{r^3} \left(1-\frac{3m}{r}\right)+O( \ep r^{-3} u_{trap}^{-1-\dec})
 \eeaa
 and similarly for $e_4(V)$.
 Thus,
 \beaa
 R(V)&=&\frac 1 2 (e_4-\Up e_3)V=-\frac{8\Up }{r^3} \left(1-\frac{3m}{r}\right)+O( \ep r^{-3} u_{trap}^{-1-\dec}),\\
 T(V)&=& \frac 1 2 (e_4+\Up e_3)V= O( \ep) r^{-3} u_{trap}^{-1-\dec},
 \eeaa
 as desired.
 
 To  prove the last part of the lemma we write,
  \beaa
\piT_{44}&=&   \g\left(\D_4( e_4+\Up e_3), e_4\right) =- 2  e_4(\Up)  -4\Up \om,\\
\piT_{34}&=&  \frac 1 2 \g(\D_3 ( e_4+\Up e_3), e_4) + \frac 1 2 \g(\D_4 ( e_4+\Up e_3), e_3)\\
&=& -e_3(\Up) +2\Up \omb+ 2\om,\\
\piT_{33}&=&  \g\left(\D_3 ( e_4+\Up e_3), e_3\right) =-4\omb,\\
\piT_{AB} &=& \frac 1 2\g\left(\D_A( e_4+\Up e_3), e_B\right)+\frac 1 2\g\left(\D_B( e_4+\Up e_3), e_A\right),\\
&=&\chiS_{AB}+\Up\chibS_{AB} \\
&=& \frac 1 2(\ka+\Up \kab)\de_{AB} + \chihS_{AB}+\Up\chibhS_{AB},
\eeaa
and the proof continues as above in view of our assumptions.
 \end{proof}  
 
 Consider now $X=f(r,m) R$  and $\piX$ its deformation tensor. We have the following lemma.
 \begin{lemma}
 \label{le:piX-decomp}
 Let $X=f(r,m) R$  and $\piX$ its deformation tensor. 
 We have,
 \beaa
 \piX=\pidX+\ep \piddX
 \eeaa
 where\footnote{Recall from Remark \ref{remark:slowly varyingf:bis} that $\pr_rf$ does no denote a spacetime coordinate vectorfield applied to $f$, but instead the partial derivative with respect to the variable $r$ of the function $f(r,m)$.} 
 \begin{itemize}
 \item  The  only nonvanishing  components of $\pidX$ are
 \beaa
 \pidX_{33}&=& 2\pr_r f,\\
 \pidX_{44}&=&2 \pr_r f\Up^2, \\
 \pidX_{34}&=&-\frac{4m}{r^2}  f  -2 \pr_r f \Up, \\
 \pidX_{AB} &=& f \frac{2\Up}{r}\de_{AB}.
 \eeaa
 \item
 All  components of $\piddX$ verify,
 \beaa
 \Big|\piddX\Big| &\les&    r^{-1} u_{trap}^{-1-\dec}  ( |  f| +r|\pr_mf|+      r^2|\pr_r f| ). 
 \eeaa
 Moreover,
 \beaa
 \Big|\piddX_{44}\Big| &\les&    r^{-2} u_{trap}^{-1-\dec} (  |  f| +r|\pr_mf|+      r^2|\pr_r f| ).  
 \eeaa
   \end{itemize}
 \end{lemma}
 
 \begin{proof}
 Clearly,
 \beaa
 \piX_{\mu\nu}= f\piR_{\mu\mu}+ e_\mu f R_\nu+ e_\nu f R_\mu.
 \eeaa
 Therefore, since $\g(R, e_3)=-1, \, \g(R, e_4)=\Up$ and, 
 \beaa
 \Big|e_4(r)-\Up,  e_3(r) +1\Big| &\les \ep\,u_{trap}^{-1-\dec},
 \eeaa
and using Lemma \ref{remark:slowly varyingf}, we deduce,
 \beaa
 \piX_{33}&=& f\piR_{33}- 2 e_3(f)= f\piR_{33}-2 \pr_r f e_3(r) -2 \pr_m f e_3(m)\\
 &=&2 \pr_r f  +O\Big(\ep r^{-1} u_{trap}^{-1-\dec} ( |f| +r|\pr_mf|+r^2|\pr_r f|)\Big)\\
 \piX_{44}&=& f\piR_{44}+2\Up e_4(f)= f\piR_{44}+2\Up \pr_r f e_4(r)+2\Up \pr_m f e_4(m)\\
 &=&2 \pr_r f\Up^2 +O\Big((\ep r^{-2} u_{trap}^{-1-\dec} ( |f| +r|\pr_mf|+r^2|\pr_r f|)\Big)
 \\
 \piX_{34} &=& f\piR_{34}+ e_3(f) \Up - e_4(f)\\
 & = &  f\piR_{34}+ (\pr_r f e_3(r)+\pr_mf e_3(m))\Up - (\pr_r f e_4(r)+\pr_mf e_4(m)) \\
 &=& -\frac{4m}{r^2}  f  -2 \pr_r f \Up +O\Big(\ep r^{-1} u_{trap}^{-1-\dec} ( |f|+r|\pr_mf| +r^2|\pr_r f|)\Big).
 \eeaa
 This concludes the proof of the lemma.
 \end{proof}


\subsection{Basic Integral  Identities}\label{section:basicintegralidentities}


We recall, see    section \ref{subsection:invariantwaveequations},         that    wave equations for $\psi\in \sk_2(\MM)$   of the form
\bea
\label{eq:reduced-wavepsi}
\square_2 \psi =V\psi +N[\psi], \qquad V=-\ka\kab,
\eea
 can be lifted to   the spacetime version,
 \bea
 \label{eq:tensor-waveqf}
 \squared\Psi+V\Psi= N[\Psi]
 \eea
 where $\Psi\in \SS_2(\MM) $ and $N[\Psi]\in \SS_2(\MM)$ are defined according to Proposition \ref{prop:spacetimeversioofwaveforqf}. 
 In fact,
 \beaa
\Psi_{\th\th} &=&-\Psi_{\vphi\vphi} =\psi, \qquad \quad   \qquad \Psi_{\th\vphi}=0.\\
N_{\th\th}[\Psi]&=&N_{\vphi \vphi}[\Psi]= N(\psi),\qquad  N[\Psi]_{\th\vphi}=0.
\eeaa
 All estimates  for \eqref{eq:tensor-waveqf} derived in this section can be  easily transferred to estimates for   \eqref{eq:reduced-wavepsi} and vice versa.

Consider  wave equations of the form,
\bea
\squared_\g \Psi=V\Psi +N
\eea
with $\Psi\in \SS_2(\MM)$ and $N$   a given symmetric traceless tensor, i.e. $\NN\in \SS_2(\MM)$.

\begin{proposition}\lab{prop:summaryofallimportantpropertiesofenergymomentumtensorforwaveequationpsi}
Assume $\Psi\in \SS_2(\MM)$ verifies \eqref{eq:tensor-waveqf}. Then, 
\begin{enumerate}
\item  The energy momentum tensor $\QQ=\QQ[\Psi]$  given by,
\beaa
 \QQ_{\mu\nu}:&=&\Db_\mu  \Psi \c \Db _\nu \Psi 
          -\frac 12 \g_{\mu\nu} \left(\Db_\la \Psi\c\Db^\la \Psi + V\Psi \c \Psi\right)\\
           &=&\Db_\mu  \Psi \c \Db _\nu \Psi    -\frac 12 \g_{\mu\nu}  \LL(\Psi)
 \eeaa
verifies,
 \beaa
 \D^\nu\QQ_{\mu\nu}
  &=& \Db_\mu  \Psi \c\NN[\Psi] + \Db^\nu  \Psi ^A\R_{ A   B   \nu\mu}\Psi^B-\frac 1 2 \D_\mu V \Psi\c \Psi.
 \eeaa
 \item
 The null components of $\QQ$ are given by,
 \beaa
\QQ_{33}&=& |e_3\Psi |^2,   \\
  \QQ_{44}&=&    | e_4\Psi |^2,      \\
\QQ_{34}&=&|\nabb \Psi|^2 +  V|\Psi|^2,
\eeaa
and,
\beaa
\g^{\mu\nu}\QQ_{\mu\nu}&=&-\LL(\Psi)- V|\Psi|^2.
\eeaa
Also,
\beaa
|\LL(\Psi)|&\les& |e_3\Psi| \, |e_4\Psi|+ |\nabb \Psi|^2 +V|\Psi|^2
\eeaa
and
\beaa
|\QQ_{AB}| &\le& |e_3\Psi|  |e_4\Psi| +|\nabb\Psi|^2 +|V||\Psi|^2, \\
|\QQ_{A3}| &\le& |e_3\Psi| |\nabb\Psi|,\\
|\QQ_{A4}| &\le& |e_4\Psi| |\nabb\Psi|.
\eeaa

\item  Introducing
\beaa
\QQh_{34}:=\QQ_{34}-V|\Psi|^2=|\nabb \Psi|^2
\eeaa
we have,
\beaa
 - \QQh_{34}+\QQ_{\th\th}+\QQ_{\vphi\vphi}=-\LL(\Psi).
\eeaa

\item
Let $X= a e_3+b e_4$. Then, since    $\R_{AB34}=0$ in an axially symmetric polarized spacetime, 
\beaa
\D^\mu (\QQ_{\mu\nu} X^\nu)&=&     \frac 1 2 \QQ  \c\piX + X( \Psi )\c \NN[\Psi] 
- \frac 1 2 X( V )  \Psi\c \Psi.
\eeaa

\item
Let $X= a e_3+b e_4$  as above,    $w$ a scalar  function and $M$  a one form. Define,
 \beaa
 \PP_\mu &=&\PP_\mu[X, w, M]=\QQ_{\mu\nu} X^\nu +\frac 1 2  w \Psi \Db_\mu \Psi -\frac 1 4|\Psi|^2   \pr_\mu w +\frac 1 4 |\Psi|^2 M_\mu.
  \eeaa
 Then,   
  \bea
  \label{le:divergPP-gen}
  \begin{split}
  \D^\mu  \PP_\mu[X, w, M] &= \frac 1 2 \QQ  \c\piX - \frac 1 2 X( V )  \Psi\c \Psi+\frac 12  w \LL[\Psi] -\frac 1 4|\Psi|^2   \square_\g  w\\
   &+\frac 1 4  \Db^\mu (|\Psi|^2 M_\mu)      +  \left(X( \Psi )+\frac 1 2   w \Psi\right)\c \NN[\Psi].
   \end{split}
 \eea
\end{enumerate}
\end{proposition}

\begin{proof}
See sections \ref{sec:appendixonenergymomentumtensorwaveeqpsi} and \ref{sec:standardcalculationvectorfieldmethodwaveequationpsi} in the appendix.
\end{proof}

 {\bf Notation.} 
 For convenience    we  introduce the notation,
 \bea
 \label{eq:modified-div}
 \EE[X, w, M](\Psi)&:=& \D^\mu  \PP_\mu[X, w, M] -   \left(X( \Psi )+\frac 1 2   w \Psi\right)\c \NN[\Psi].   
  \eea
  Thus equation  \eqref{le:divergPP-gen} becomes,
   \bea
   \label{le:divergPP-gen-EE}
 \nn  \EE[X, w, M](\Psi)&=& \frac 1 2 \QQ  \c\piX - \frac 1 2 X( V )  \Psi\c \Psi+\frac 12  w \LL[\Psi]\\
   && -\frac 1 4|\Psi|^2   \square_\g  w+\frac 1 4  \Db^\mu (|\Psi|^2 M_\mu).  
 \eea
When   $M=0$ we   simply write  $\EE[X, w](\Psi)$.


\subsection{Main  Morawetz    Identity}\lab{sec:mainmorawetzidentity}
 
 
 \begin{lemma}
 \label{le:QQcpidX}
 Let $f(r,m)$ a function of $r$ and $m$, and let $X$ a vectorfield defined by $X=f(r,m)R$. Then, we have\footnote{Recall from Remark \ref{remark:slowly varyingf:bis} that $\pr_rf$ does no denote a spacetime coordinate vectorfield applied to $f$, but instead the partial derivative with respect to the variable $r$ of the function $f(r,m)$.} ,
 \beaa
 \QQ\c \pidX&=& f \left(-\frac{2m}{r^2}+ \frac{2\Up}{r}\right)|\nabb\Psi|^2  +2 \pr_r f  |R\Psi|^2  -\left(  \frac{2\Up}{r} f   +\Up \pr_r f      \right)\LL-\frac{2m}{r^2}f V|\Psi|^2.
 \eeaa
 where $\pidX$ has been defined in Lemma \ref{le:piX-decomp}.
 \end{lemma} 
 
 \begin{proof}
 In view of Lemma \ref{le:piX-decomp}, we have
 \beaa
 \QQ\c\pidX&=&\frac  1 2 \QQ_{34}\pid_{34}+\frac 1 4 \QQ_{44}\pid_{33}+\frac 1 4 \QQ_{33}\pid_{44}+\QQ_{AB}\pid_{AB}\\
 &=&-\frac{2m}{r^2} f\QQ_{34} - \pr_r f \Up \QQ_{34}+\frac 1 2 \QQ_{44}  \pr_r f +\frac 1 2 \QQ_{33} \Up^2 \pr_r f +  \frac{2\Up}{r} f   \de^{AB}    \QQ_{AB}\\
 &=&-\frac{2m}{r^2} f\QQ_{34} +  \frac{2\Up}{r} f   \de^{AB}    \QQ_{AB}+\frac 12 \pr_r f\left(\QQ_{44} -  2\Up \QQ_{34} +\Up^2\QQ_{33} \right)
 \eeaa
 Note that,
 \beaa
 \left(\QQ_{44} -  2\Up \QQ_{34} +\Up^2\QQ_{33} \right)=4\QQ_{RR}
 \eeaa
 and, since  $\g^{\mu\nu}\QQ_{\mu\nu}=-\LL(\Psi)- V|\Psi|^2$,
 \beaa
 \de^{AB}\QQ_{AB}&=&\QQ_{34}-\LL -V|\Psi|^2=\QQh_{34}-\LL.
 \eeaa
 Hence,
 \beaa
  \QQ\c\pidX&=&-\frac{2m}{r^2}f \QQ_{34} +  \frac{2\Up}{r} f  \left(\QQh_{34}-\LL\right)+2 \pr_r f \QQ_{RR}\\
  &=&-\frac{2m}{r^2}f\left( \QQh_{34}+V|\Psi|^2\right) +  \frac{2\Up}{r} f  \left(\QQh_{34}-\LL\right)+2 \pr_r f \QQ_{RR}\\
  &=& f \left(-\frac{2m}{r^2}+ \frac{2\Up}{r}\right)\QQh_{34} +2 \pr_r f \QQ_{RR} - \frac{2\Up}{r} f\LL-\frac{2m}{r^2}f V|\Psi|^2.
 \eeaa
 Finally,
  \beaa
 \QQ_{RR}&=&|R\Psi|^2 - \frac 1 2 \g(R, R)\LL=|R\Psi|^2  - \frac 1 2 \Up\LL.
 \eeaa
 Hence,
 \beaa
  \QQ\c\pidX&=& 2  f \left(-\frac{m}{r^2}+ \frac{\Up}{r}\right)\QQh_{34} +2 \pr_r f\left(|R\Psi|^2  - \frac 1 2 \Up\LL\right)  - \frac{2\Up}{r} f\LL-\frac{2m}{r^2}f V|\Psi|^2\\
  &=&2  f \left(-\frac{m}{r^2}+ \frac{\Up}{r}\right)|\nabb\Psi|^2+ 2 \pr_r f |R\Psi|^2- \left(   \frac{2\Up}{r} f  + \pr_r f\Up \right)\LL -\frac{2m}{r^2}f V|\Psi|^2.
  \eeaa 
  This concludes the proof of the lemma.
 \end{proof}

 We  shall also make use of the following lemma.
 \begin{lemma}
\label{calculation:square-radial}
If  $f=f(r,m)$, then
\beaa
\square_\g(f(r,m)) &=& r^{-2}\pr_r(r^2\Up\pr_rf) + O(\ep r^{-2}\,u_{trap}^{-1-\dec})  \Big[ r^2 |\pr^2_r f(r,m)| +r|\pr_r f(r,m)| \\
&&+r|\pr_r\pr_mf(r,m)|+|\pr_m^2f(r,m)| \Big].
\eeaa
\end{lemma}

\begin{proof}
 Recall from Lemma \ref{le:square(psi)-null-frame} that, for  a general   scalar $f$, 
 \beaa
 \square_\g  f &=& -\frac 1 2  (e_3 e_4 + e_4 e_3)f +\lapp f +\left(\ombS-\frac 1 2 \trchbS \right) e_4 f +\left(\omS-\frac 1 2 \trchS\right)  e_3 f.
 \eeaa
 Recall that,
\beaa
\trchS&=& 2 \chi -\vth, \quad \trchbS= 2 \chib -\vthb, \quad \omS=\om, \quad \ombS=\omb 
\eeaa
and 
\beaa
\lapp f= e_\th e_\th  f + (e_\th \Phi)^2 e_\th f. 
\eeaa
Using Lemma \ref{remark:slowly varyingf}, we  deduce,    for a function $f=f(r,m)$,
 \beaa
 \square_\g  f &=& -\frac 1 2  (e_3 e_4 + e_4 e_3)f +\left(\omb- \frac{1}{2}\kab \right) e_4 f +\left(\om-\frac{1}{2}\ka \right)  e_3 f \\
 &=& -\pr^2_r f(r,m) e_3(r) e_4(r)-\frac 1 2  \pr_r f(r,m)\left(e_3 e_4(r)+ e_4 e_3(r)\right)\\
 &-& \frac{1}{2}\kab \pr_r f(r,m) e_4 r +\left(\om-\frac{1}{2}\ka \right)\pr_r f(r,m) e_3 (r)+  O(\ep r^{-2}\,u_{trap}^{-1-\dec})\Big[ r^2 |\pr^2_r f(r,m)| \\
 &&+r|\pr_r f(r,m)| +r|\pr_r\pr_mf(r,m)|+|\pr_m^2f(r,m)| \Big]\\
 &=&-\pr^2_r f(r,m)\left(-\Up+ O(\ep u_{trap}^{-1-\dec}\right)+ \pr_r f(r,m)\frac {m}{r^2} +\pr_r f(r,m)\frac{\Up}{r} +\frac{r-m}{r^2}\pr_r f(r,m) \\
 &+&  O(\ep r^{-2}\,u_{trap}^{-1-\dec})\Big[ r^2 |\pr^2_r f(r,m)| +r|\pr_r f(r,m)| +r|\pr_r\pr_mf(r,m)|+|\pr_m^2f(r,m)| \Big]\\
 &=&\Up \pr^2_r f(r,m)+ \pr_r f(r,m)\left( \frac{2}{r}-\frac{2m}{r^2}\right)\\
 &+&O(\ep r^{-2}\,u_{trap}^{-1-\dec})\Big[ r^2 |\pr^2_r f(r,m)| +r|\pr_r f(r,m)| +r|\pr_r\pr_mf(r,m)|+|\pr_m^2f(r,m)| \Big]\\
 &=& r^{-2}\pr_r(r^2\Up\pr_rf) \\
&+&  O(\ep r^{-2}\,u_{trap}^{-1-\dec})\Big[ r^2 |\pr^2_r f(r,m)| +r|\pr_r f(r,m)| +r|\pr_r\pr_mf(r,m)|+|\pr_m^2f(r,m)| \Big]
 \eeaa
 as desired.
\end{proof}

  According to   equation  \eqref{le:divergPP-gen-EE} we have, 
  \beaa
  \EE[X, w ](\Psi) &=& \frac 1 2 \QQ  \c\piX - \frac 1 2 X( V ) |\Psi|^2 +\frac 12  w \LL(\Psi) -\frac 1 4|\Psi|^2\square_\g w.
 \eeaa
 In the next proposition we choose $X$ to be of the form   $X =f(r,m) R$  and     make a choice of $w$  as a function of $f$.
 
\begin{proposition}
  \label{prop-ident-Mor1}
 Assume
 \beaa
 X=f(r,m) R\textrm{ and }   w(r,m)= r^{-2}\Up\pr_r(r^2f).  
 \eeaa
 Then,
  \beaa
 \EE[X, w ](\Psi) &=&\EEd[X, w]+\EE_\ep[X, w]
 \eeaa
 where, with  $\QQh_{34}:=\QQ_{34}-V|\Psi|^2=|\nabb\Psi|^2$,
  \bea
  \begin{split}
\EEd[f R, w] (\Psi)&=  \frac 1 r \left(1-\frac{3m}{r}\right) f \QQh_{34}+ \pr_r f |R(\Psi)|^2 -\frac 1 4 r^{-2}\pr_r(r^2\Up \pr_r w)  |\Psi|^2 \\
&+4 \Up\frac{r-4m}{r^4}  f |\Psi|^2, \\
\EE_\ep[f R, w] (\Psi)&=\ep \frac 1 2  \QQ\c\piddX+ O\Big(\ep r^{-3} u_{trap}^{-1-\dec }\big(|f|+r^2|\pr_r w|+r^3|\pr^2_rw|\\
&+r^2|\pr_r\pr_mw|+r|\pr_m^2w|\big)\Big)|\Psi|^2.
\end{split}
 \eea
  \end{proposition}
 
 \begin{proof}
 According to    Lemma \ref{le:piX-decomp}     and  equation  \eqref{le:divergPP-gen-EE} we have, 
  \beaa
  \EE[X, w ](\Psi) &=& \frac 1 2 \QQ\c  (\pidX  +\ep \piddX)       - \frac 1 2 X( V ) |\Psi|^2 +\frac 12  w \LL(\Psi) -\frac 1 4|\Psi|^2   \square_\g  w.
 \eeaa
 Hence, in view of  lemmas         \ref{le:piX-decomp} and \ref{le:QQcpidX},
 \beaa
 \EE[X, w](\Psi) -\ep\frac 1 2 \QQ\c\piddX        &=&\frac 1 2 \QQ\c \pidX   - \frac 1 2 X( V ) |\Psi|^2 +\frac 12  w \LL(\Psi) -\frac 1 4|\Psi|^2   \square_\g  w\\
 &=& f \left(-\frac{m}{r^2}+ \frac{\Up}{r}\right)|\nabb\Psi|^2  + \pr_r f |R\Psi|^2 -\left( \frac{\Up}{r} f    +\frac 1 2 \Up \pr_r f\right)    \LL(\Psi)\\
  &-& \frac{m}{r^2}f V|\Psi|^2 -\frac 1 2 X( V ) |\Psi|^2+\frac 12  w \LL(\Psi)-\frac 1 4|\Psi|^2   \square_\g  w.
 \eeaa
 Thus, assuming $w= r^{-2}\Up\pr_r(r^2f)=\frac{2\Up}{r}f+\pr_rf\Up$,
 \beaa
 \EE[X, w](\Psi) -\ep\frac 1 2 \QQ\c\piddX        &=&  r^{-1} f \left(   1-\frac{3m}{r} \right)|\nabb\Psi|^2  + \pr_r f |R\Psi|^2\\
 &-& \left( \frac{m}{r^2}f V +\frac 1 2 X( V ) +\frac 1 4   \square_\g  w\right)|\Psi|^2.
 \eeaa
Note that, in view of Lemma \ref{lemma:componentspiR},
 \beaa
 X(V)&=&  f R(V)  =- 8\Up f \frac{r-3m}{r^4} + O(\ep r^{-3} u_{trap}^{-1-\dec}|f| )
 \eeaa
 and,
 \beaa
  \frac{m}{r^2}  f   V     +  \frac 1 2 X( V)&=&f\left( \frac{4m}{r^4}    \Up -       4 \Up  \frac{r-3m}{r^4}\right)+ O(\ep r^{-3} u_{trap}^{-1-\dec}|f| )\\
   &=&- 4f\Up\frac{r-4m}{r^4} + O(\ep r^{-3} \,u_{trap}^{-1-\dec}|f|).
  \eeaa
  Note also that, in view of Lemma \ref{calculation:square-radial}
\beaa
\square_\g(w) &=& r^{-2}\pr_r(r^2\Up \pr_r w)  +O(\ep r^{-2}\,u_{trap}^{-1-\dec})  \Big[ r^2 |\pr^2_r w| +r|\pr_rw| +r|\pr_r\pr_mw|+|\pr_m^2w| \Big].
\eeaa
 Thus,
 \beaa
 \frac{m}{r^2}f V +\frac 1 2 X( V ) +\frac 1 4   \square_\g  w &=& -4 \Up\frac{r-4m}{r^4}  f +\frac{1}{4}r^{-2}\pr_r(r^2\Up \pr_r w)\\
 && + O(\ep r^{-3}\,u_{trap}^{-1-\dec})  \Big[ |f|+r^3 |\pr^2_rw| +r^2|\pr_r w|  \\
 &&+r^2|\pr_r\pr_mw|+r|\pr_m^2w|\Big]
 \eeaa
 and hence
 \beaa
 \EE[X, w](\Psi) -\ep\frac 1 2 \QQ\c\piddX        &=&  r^{-1} f \left(   1-\frac{3m}{r} \right)|\nabb\Psi|^2  + \pr_r f |R\Psi|^2 -\frac 1 4  |\Psi|^2 r^{-2}\pr_r(r^2\Up \pr_r w)\\
 &+&4 \Up\frac{r-4m}{r^4}  f |\Psi|^2\\
 &+& O\Big(\ep r^{-3} \,u_{trap}^{-1-\dec}\big(|f|+r^2|\pr_r w|+r^3|\pr^2_rw|\\
 &&+r^2|\pr_r\pr_mw|+r|\pr_m^2w|\big)  \Big)|\Psi|^2
\eeaa
as  desired.
 \end{proof}


 \subsection{A first estimate}\lab{sec:trapingregionmorawetzpsi}
 
 
We concentrate our attention on the principal  term
\beaa
\EEd[f R, w] (\Psi)&=&  \frac 1 r \left(1-\frac{3m}{r}\right) f \QQh_{34}+ \pr_r f |R(\Psi)|^2 -\frac 1 4 r^{-2}\pr_r(r^2\Up \pr_r w) |\Psi|^2+4 \Up\frac{r-4m}{r^4}  f |\Psi|^2 
\eeaa
and  choose  $f=f(r,m)$ such that the  right  hand side is   positive definite.

 Consider  the quadratic forms,
 \bea
 \label{eq:quadr-Mor}
 \begin{split}
\EEd_0(\Psi):&=   A\QQh_{34}+ B     |R\Psi|^2      + r^{-2} W |\Psi|^2,  \\
\EEd(\Psi) :&= \EEd_0(\Psi) + 4 \Up\frac{r-4m}{r^4}  f |\Psi|^2, 
\end{split}
 \eea
 with the coefficients 
 \bea
 \label{eq:quadr-coeffs}
 A := r^{-1}  f\left(1-\frac{3m}{r}\right), \qquad     B:= \pr_r f ,\qquad  W:=  -\frac{1}{4}\pr_r(r^2\Up \pr_r w).
 \eea
 The goal is   to  show that there  exist choices of $f, w$ verifying the   condition
   of Proposition \ref{prop-ident-Mor1}, i.e. $w=r^{-2}\pr_r(r^2f)$,  which makes  $\EEd(\Psi)$            positive definite,   
    for all    smooth   $S$-valued tensorfields  $\Psi$   defined     in the region     $r\ge   2m_0(1-\deh)$,  which decay    reasonable fast    at infinity. 
     We look  first   for choices     of $f, w$ such that   the coefficient $A, B, W$ are  non-negative.    Note in particular that $f$ must be  increasing  as a function of $r$ and $f=0$ on $r=3m$.
      Following  J. Stogin   \cite{St} we choose $w$ first  to ensure that $W$ is non-negative  and then choose $f$,
       compatible with the equation,
        \bea
        \label{definition:tilde{V}-Morawetz}
   \pr_r(r^2f) = \frac{r^2}{\Up}w, \qquad f=0\textrm{ on }r=3m.
 \eea 
   To   ensure that $A=r^{-2}  f (r-3m) $ is positive we need    a  non-negative $w$  which verifies (modulo error terms\footnote{i.e. terms which  vanish  in  Schwarzschild.})    $W=-\frac 1 4 \pr_r(r^2\Up \pr_r w) \ge 0$.   It is more difficult to choose  $w$ such that  $B=  \pr_r f$  is also non-negative.

 Stogin  defines $w$  based on the following  lemma.
 \begin{lemma}
 \label{le:John2'}
 The scalar function $w$ defined by
 \beaa
 w(r,m)=\begin{cases}
 & \frac{1}{4m} , \qquad  \qquad\qquad  \mbox{if}  \quad    r\leq 4m,\\[1mm]
  & \frac{2\Up}{r} ,  \qquad\qquad\qquad\,\mbox{if} \quad    r\geq 4m,
 \end{cases}
 \eeaa
 is               $C^1$,  non-negative    and  such that  $W=-\frac 1 4 \pr_r(r^2\Up \pr_r w)$ verifies,
\bea
W(r,m)= \begin{cases}
 &    0, \qquad  \qquad   \qquad \qquad  \qquad \qquad\mbox{if}   \,\, \quad  r< 4m,\\
  &  \frac{m}{r^2}\left(3 -\frac{8m}{r}\right),\,\qquad \qquad\qquad\quad\,\,  \mbox{if} \quad  \,\,   r> 4m.
 \end{cases}
\eea
\end{lemma}
  
\begin{proof}
For $r\geq 4m$,   we have
\beaa
w(r,m)=\frac{2\Up}{r}, \quad \pr_r w(r,m) = -\frac{2}{r^2}+\frac{8m}{r^3}, \quad \pr^2_rw(r,m) = \frac{4}{r^3}-\frac{24m}{r^4}.
\eeaa
In particular, we have 
\beaa
w=\frac{1}{4m},\qquad \pr_r w =0\,\,\,\textrm{ at }r=4m 
\eeaa
so that $w$ is indeed $C^1$. Furthermore, we also have 
\beaa
r^{-2}\pr_r(r^2\Up \pr_r w) &=& \Up \pr^2_rw(r)+ \pr_r w(r)\left( \frac{2}{r}-\frac{2m}{r^2}\right)\\
 &=& \Up\left(\frac{4}{r^3}-\frac{24m}{r^4}\right) +\left(-\frac{2}{r^2}+\frac{8m}{r^3}\right)\left( \frac{2}{r}-\frac{2m}{r^2}\right)\\
&=& -\frac{4m}{r^4}\left(3-\frac{8m}{r}\right)
\eeaa
so that, for $r\geq 4m$, 
   \beaa
   W= -\frac 1 4 \pr_r(r^2\Up \pr_r w)  &=&  \frac{m}{r^2}\left( 3 -\frac{8m}{r}\right)
   \eeaa
as desired.
\end{proof}

Once $w$ is defined we can        evaluate  $f$ as follows.
\begin{lemma}
Let $w(r,m)$ defined as in Lemma \ref{le:John2'}. Then, the function $f(r,m)$ given  by,
\bea
\label{eq:John3}
r^2f(r,m):=    \begin{cases}
& 2m^2\log\left(\frac{r-2m}{m}\right)+ (r-3m)      \frac{r^2 +6 mr +30  m^2}{12m},      \qquad     \mbox{for}\quad  r \le  4m,    \\ 
&  C_*m^2+r^2-(4m)^2, \qquad\qquad\qquad\qquad\qquad\,  \mbox{for}\quad r\ge 4m, 
\end{cases}
\eea
with  the constant $C_*$ given by\footnote{$C_*$ is chosen so that $f$ is continuous across $r=4m$.} 
\beaa
C_* &:=& 2 \log(2)  +\frac{35}{6}, \qquad C_*\sim 7.22,
\eeaa
is $C^2$ and satisfies \eqref{definition:tilde{V}-Morawetz}, i.e. we have
\beaa
\pr_r(r^2f) = r^2\Up^{-1}w, \qquad f=0\textrm{ on }r=3m.
\eeaa
\end{lemma}

\begin{proof}
By direct check\footnote{Recall from Remark \ref{remark:slowly varyingf:bis} that $\pr_rf$ does no denote a spacetime coordinate vectorfield applied to $f$, but instead the partial derivative with respect to the variable $r$ of the function $f(r,m)$.}, we have for $r\leq 4m$
\beaa
\pr_r(r^2f)(r,m) &=& \frac{2m^2}{(r-2m)}+  \frac{r^2 +6 mr +30  m^2}{12m}+(r-3m)\frac{2r +6 m}{12m}\\
&=& \frac{r^3}{4m(r-2m)}\\
&=& \frac{r^2}{\Up}\frac{1}{4m}
\eeaa
and for $r\geq 4m$
\beaa
\pr_r(r^2f)(r,m) &=& 2r,
\eeaa
as well as $f=0$ on $r=3m$ so that, in view of the definition of $w(r)$ in Lemma \ref{le:John2'}, we infer
\beaa
\pr_r(r^2f) = r^2\Up^{-1}w, \qquad f=0\textrm{ on }r=3m
\eeaa
as desired. Note also that $w$ being $C^1$, $f$ is thus indeed $C^2$.
\end{proof}

  Next, we derive a lower bound on  $\pr_r f$  for   $r\le 4m$.
   \begin{lemma}
\label{le:John1}
We have for all $r$ and $m$
\beaa
r^3\pr_rf\geq 16m^2.
\eeaa
Also, there exists a constant $C>0$ such that  for all $r$ and $m$
\beaa
\left(1-\frac{3m}{r}\right) f\ge   C^{-1}\left(1-\frac{3m}{r}\right)^2.
\eeaa
\end{lemma}

\begin{proof}
We have
\beaa
\pr_r( r^3 \pr_r f) &=& \pr_r( r\pr_r(r^2f)) - 2\pr_r(r^2f).
\eeaa
Using the identity $\pr_r(r^2f) = r^2\Up^{-1}w$, we infer
\beaa
\pr_r( r^3 \pr_r f) &=& \pr_r( r^3\Up^{-1}w) - 2r^2\Up^{-1}w.
\eeaa
For $r\leq 4m$, we have $w=(4m)^{-1}$  and hence 
\beaa
\pr_r( r^3 \pr_r f) &=& \frac{1}{4m}\left(\pr_r( r^3\Up^{-1}) - 2r^2\Up^{-1}\right)\\
&=& \frac{1}{4m}r^3\pr_r(\Up^{-1})\\
&=& -\frac{r}{2\Up^2}
\eeaa
In particular, $r^3\pr_rf$ is decreasing in $r$ on $r\leq 4m$ and hence
\beaa
r^3\pr_rf\geq (4m)^3\pr_rf(r=4m, m)\textrm{ on }r\leq 4m.
\eeaa
On the other hand, we have, in view of the definition \ref{eq:John3} of $f$
\beaa
\pr_r(r^2f)(r=4m,m) &=& (4m)^2\pr_rf(r=4m,m)+8mf(r=4m,m)\\
&=& (4m)^2\pr_rf(r=4m,m)+\frac{m}{2}C_*
\eeaa
and hence
\beaa
(4m)^2\pr_rf(r=4m,m) &=& \left(8-\frac{C_*}{2}\right)m
\eeaa
so that
\beaa
r^3\pr_rf\geq 2(16-C_*)m^2\textrm{ on }r\leq 4m.
\eeaa
Since $C_*\sim 7.22<8$, we deduce
\beaa
r^3\pr_rf\geq 16m^2\textrm{ on }r\leq 4m.
\eeaa

Also, for $r\geq 4m$, we have
\beaa
f=1-\frac{(16-C_*)m^2}{r^2}
\eeaa
so that
\beaa
\pr_rf = \frac{2(16-C_*)m^2}{r^3}.
\eeaa
Since $C_*\sim 7.22<8$, we deduce 
\beaa
r^3\pr_rf\geq 16m^2\textrm{ on }r\geq 4m
\eeaa
which together with the case $r\leq 4m$ above yields for all $r$ and $m$ the desired estimate for $\pr_rf$
\beaa
r^3\pr_rf\geq 16m^2.
\eeaa
In particular, $\pr_rf>0$ and hence is strictly increasing. On the other hand, $f=0$ on $r=3$ and converges to 1 as $r\to +\infty$. We deduce the existence of a constant $C>0$ such that 
\beaa
\left(1-\frac{3m}{r}\right) f\ge   C^{-1}\left(1-\frac{3m}{r}\right)^2
\eeaa
as desired.
\end{proof}

We  summarize the results in the following. 
    \begin{proposition}
   \label{prop:pre-mor1}
   There exist functions $f\in C^2 , w\in C^1 $   verifying the relation    $w = r^{-2} \Up \pr_r(r^2 f) $ and such that,
   \bea
\label{eq:John3'}
r^2f=    \begin{cases}
&2m^2   \log\left(\frac{r-2m}{m}\right)+ (r-3m)      \frac{r^2 +6 mr +30  m^2}{12m},      \qquad   \quad  \mbox{for}\quad  r \le  4m,    \\ 
&  C_*m^2 +r^2-(4m)^2, \qquad \qquad \qquad  \qquad \qquad \quad\,\,\mbox{for}\quad r \ge 4m ,
\end{cases}
\eea
where $C_*$  is a constant satisfying  $7<C_*<8$.
   In particular,
   \bea
\label{eq:John3''}
f=    \begin{cases}
&\frac{ 2m^2}{r^2}    \log\left(\frac{r-2m}{m}\right)+  O( \frac{r-3m}{m} ),      \qquad  \, \,\,\,\, \mbox{for}\quad  r \le  4m,    \\ 
&  1+O(\frac{m^2}{r^2}),  \qquad  \qquad  \qquad \qquad \quad\,\,\,\mbox{for}\quad r \ge 4m, 
\end{cases}
\eea
 and, for some $C>0$ and all $r\ge 2m$
\bea\lab{eq:usefulllowerboundforfandprrf}
\left(1-\frac{3m}{r}\right) f\ge   C^{-1}\left(1-\frac{3m}{r}\right)^2, \qquad   \pr_r f\ge \frac{16m^2}{r^3}.
\eea
Also, $w$ is given by
   \bea
   w=    \begin{cases}
& \frac{1}{4m},   \qquad \quad \qquad \qquad \qquad   \qquad  \quad \,\,\mbox{for}\quad  r \le  4m,    \\[1mm] 
& \frac{2}{r}\left(1-\frac{2m}{r}\right),  \qquad \,\, \qquad  \qquad \qquad \,\,\,\mbox{for}\quad r \ge 4m.
\end{cases}
   \eea
   Moreover
   $W=  -\frac 1 4 \pr_r(r^2\Up \pr_r w)$ verifies,
\bea
\label{definition:W-Morawetz}
W= \begin{cases}
 &    0, \qquad  \qquad   \qquad \qquad \qquad \quad \,\mbox{if}   \quad  r< 4m,\\
  &  \frac{m}{r^2}\left( 3 -\frac{8m}{r}\right), \, \qquad\,\,\qquad \qquad   \mbox{if} \quad  \,\,   r>4m,
 \end{cases}
\eea
and,
\bea
\label{eq:pre-mor1}
\begin{split}
\EEd_0 [fR, w](\Psi)& = \pr_r f  |R(\Psi)|^2 + r^{-2}W |\Psi|^2
+ r^{-1} \left(1-\frac{3m}{r}\right) f \QQh_{34},\\
\EEd[fR, w](\Psi)&=\EE_0 [fR, w](\Psi)+4 \Up\frac{r-4m}{r^4}  f |\Psi|^2.
\end{split}
\eea
Recall also that,
\beaa
 \QQh_{34}&=&|\nabb \Psi|^2.  
\eeaa
   \end{proposition}
   
   \begin{remark}
   The estimates obtained so far  have two major deficiencies
   \begin{enumerate}
   \item  The  quadratic  form $\EEd_0 [fR, w](\Psi)+4 \Up\frac{r-4m}{r^4}  f |\Psi|^2$       fails to be positive definite 
   in the region $3m\le r\le 4m$   because of the potential  term $\Up\frac{r-4m}{r^4}  f |\Psi|^2$.
   \item The function $f$   blows up logarithmically  at $r=2m$ in $\Mint$.
   \end{enumerate} 
   In the next section we deal with the first issue.  We  handle   the second  problem in the following two  sections.
   \end{remark}


\subsection{Improved lower bound    in $\Mext$}\label{subs:improvedLB-awayH}


     Note that    the term  $4f  \Up\frac{r-4m}{r^4} $ is     negative            for     $3m\le r\le 4m$   and positive everywhere else.        
     An improvement can be obtained   by   using  the following  Poincar\'e inequality.
     \begin{lemma}
We have for $\Psi\in \SS_2(\MM)$,
 \bea
     \label{corr:Poincare-Psi}
\int_S |\nabb \Psi|^2  &\ge & 2r^{-2}\Big(1- O(\ep) \Big) \int_S  \Psi ^2 da_S.
\eea
\end{lemma}

\begin{proof}
See Proposition \ref{proposition:Poincaree}.
\end{proof} 

    According to   Proposition \ref{prop:pre-mor1} we  deduce, 
\bea
\label{eq:intS-EEd}
\begin{split}
&\int_S\EEd [fR, w](\Psi)\ge \int_S  \EEd_1- O(\ep r^{-3})     \int_S  \Psi ^2 da_S,  \\
&\EEd_1:= \pr_r f  |R(\Psi)|^2 + r^{-2}W |\Psi|^2+    2r^{-3} \left(1-\frac{3m}{r}\right) f |\Psi|^2  +4 \Up\frac{r-4m}{r^4}  f |\Psi|^2,
\end{split}
\eea
with $W$ defined in \eqref{definition:W-Morawetz}.
It is easy to see however  that $\EEd_1$  still fails to be  positive for $3m < r<4m$.  To achieve   positivity we also need   to modify the original  energy  density  $\EE [fR, w](\Psi)$ 
   by considering instead   the modified energy   density   $  \EE [fR, w, M](\Psi) $   (see \eqref{le:divergPP-gen} and  notation  \eqref{eq:modified-div}) with $M= 2 h R$ for a function $h=h(r,m)$  supported  for $r\ge  3m$ and constant for  $r\ge 4m$. 
 \beaa
  \EE [fR, w, M](\Psi) &=& \EE[fR, w](\Psi)+\frac 1 4  \Db^\mu (|\Psi|^2 M_\mu) = \EE[fR, w](\Psi)+\frac 1 4 ( \D^\mu  M_\mu) |\Psi|^2 +\frac 1 2 \Psi M(\Psi)\\
  &=&\EE[fR, w](\Psi)+\frac 1 2 \D^\mu ( h R_\mu) |\Psi|^2 + h \Psi R(\Psi). 
  \eeaa

To take into account  the  additional terms in the modified  $\EE [fR, w, M](\Psi)$  we     first derive the following.
\begin{lemma}
Let $h(r,m)$ a $C^1$ function of $r$ and $m$. We have,
\bea
 \D^\mu (h R_\mu) =r^{-2} \pr_r(\Up r^2 h)+  O\left( \ep\, r^{-1}u_{trap}^{-1-\dec}\big(r|\pr_rh|+ |h|+r|\pr_mh|\big)\right).
 \eea
\end{lemma}

\begin{proof}
In view of Lemma \ref{lemma:componentspiR}, which computes the components of $\piR$, as well as Lemma \ref{remark:slowly varyingf} to compute $R(h)$, we calculate,
\beaa
 \D^\mu (h R_\mu) &=& R(h) + h( \D^\mu  R_\mu)= \frac{1}{2}(e_4(h)-\Up e_3(h)) +h \frac 1 2 \tr \,(\piR)\\
 &=& \frac{1}{2}(e_4(r)-\Up e_3(r))\pr_rh +O(\ep \,u_{trap}^{-1-\dec}|\pr_mh|)      + \frac 1 2 h\left( -\piR_{34}+\piR_{\th\th}+\piR_{\vphi\vphi}  \right)\\
 &=&\Up \pr_rh +\frac  12  \left( \frac{4m}{r^2}+  4\frac{\Up}{r}\right) h+  O\left( \ep\,u_{trap}^{-1-\dec} \big(|\pr_rh|+ r^{-1} |h|+|\pr_mh|\big)\right)\\
  &=& r^{-2} \pr_r(\Up r^2 h)+  O\left( \ep\,u_{trap}^{-1-\dec}\big(|\pr_rh|+ r^{-1} |h|+|\pr_mh|\big)\right)
 \eeaa
 as desired.
\end{proof}

  In view of the lemma we      write,
  \bea
  \label{definition:EE[X,w,M]}
  \begin{split}
   \EE [fR, w, 2hR](\Psi) &=\EEd[fR, w,2 hR](\Psi)+\EE_\ep [fR, w, 2hR](\Psi),\\
   \EEd[fR, w,2 hR](\Psi):&= \EEd[fR, w](\Psi)+\frac 1 2 r^{-2} \pr_r(\Up r^2 h)   |\Psi|^2               + h \Psi R(\Psi), \\
   \EE_\ep [fR, w, 2hR](\Psi):&=\EE_\ep [fR, w](\Psi)+    O\left( r^{-1}\ep\,u_{trap}^{-1-\dec}\big(r|\pr_rh|+  |h|+r|\pr_mh|\big)\right)|\Psi|^2.
   \end{split}
  \eea
    
  The main result  of this section  is stated below.
  \begin{proposition}
  \label{prop:pre-mor2}
  There exists    a function $h=h(r,m)$    with bounded   derivative $h'$,   supported   in $r\ge 3m$   such that $h= O(r^{-2}) , h'=O(r^{-3})$ for 
  for $r\ge 4m$  such that,
  \bea
  \bsplit
  \EE[fR, w, 2hR](\Psi)&= \EEd [fR, w, 2hR](\Psi)+ \EE_\ep [fR, w, 2hR](\Psi),\\
 \EE_\ep [fR, w, 2hR](\Psi) &=\ep \frac 1 2  \QQ\c\piddX + O\Big(\ep r^{-3} u_{trap}^{-1-\dec}(|f|  +1)\Big)|\Psi|^2,
 \end{split}
  \eea
    and, for   sufficiently large      universal constant $ C>0$,        in the region   $r\ge \frac{5m}{2}$, 
  \begin{equation}\lab{eq:lowerboundonEEdfRw2hR}
   \int_S  \EEd [fR, w, 2hR](\Psi) \ge C^{-1}\int_{S} \left( \frac{m^2}{r^3} |R(\Psi)|^2 + r^{-1}\left(1-\frac{3m}{r}\right)^2 |\nabb \Psi|^2 +  \frac{m}{r^4} |\Psi|^2    \right).
   \end{equation}
  \end{proposition}
  
  \begin{proof}
  We  first  derive the weaker inequality, 
   \beaa
 \int_S  \EEd [fR, w, 2hR](\Psi) \ge C^{-1}\int_{S} \left( \frac{m^2}{r^3} |R(\Psi)|^2 +  \frac{m}{r^4} |\Psi|^2    \right)\,\,\textrm{ on }r\ge \frac{5m}{2} 
  \eeaa
  by making   full use  of the  Poincar\'e inequality above, i.e.,
  \beaa
  \int_S r^{-1}\left(1-\frac{3m}{r}\right) f(r,m) |\nabb \Psi|^2\ge\int_S (2-O(\ep))r^{-3}\left(1-\frac{3m}{r}\right)  f(r,m)| \Psi|^2.
  \eeaa
  The result will the  easily follow by   writing instead, with a sufficiently small  $\mu>0$,
  \beaa
   &&\int_S r^{-1}\left(1-\frac{3m}{r}\right)f(r,m) |\nabb \Psi|^2 \\
     &=&  \mu  \int_S r^{-1}\left(1-\frac{3m}{r}\right)f(r,m) |\nabb \Psi|^2    +(1-\mu) \int_S r^{-1}\left(1-\frac{3m}{r}\right)     f(r,m) |\nabb \Psi|^2 \\
                & \ge&          \mu  \int_S r^{-1}\left(1-\frac{3m}{r}\right)f(r,m) |\nabb \Psi|^2   +(1-\mu)\int_S  2r^{-3}\left(1-\frac{3m}{r}\right) f(r,m) | \Psi|^2
   \eeaa
and  then proceeding exactly  as below.

\medskip

We start with,
\beaa
  \EEd [fR, w, 2hR](\Psi) &=&\EEd[fR, w](\Psi)+\frac 1 2 r^{-2} \pr_r(\Up r^2 h) |\Psi|^2 + h \Psi R(\Psi).
\eeaa
Recalling the definition of  $\EEd_1$ in \eqref{eq:intS-EEd},
\beaa
\EEd_1:= \pr_r f  |R(\Psi)|^2 + r^{-2}W |\Psi|^2+    2r^{-3} \left(1-\frac{3m}{r}\right) f |\Psi|^2  +4 \Up\frac{r-4m}{r^4}  f |\Psi|^2
\eeaa
 and setting,
\bea
\EEd_2 &:=&\EEd_1 +\frac 1 2 r^{-2} (\Up r^2 h)' |\Psi|^2 + h \Psi R(\Psi)\\
&=&\pr_r f  |R(\Psi)|^2+ 2r^{-3} \left(1-\frac{3m}{r}\right) f |\Psi|^2  +4 \Up\frac{r-4m}{r^4}  f |\Psi|^2+ r^{-2}W |\Psi|^2\nn\\
&+&\frac 1 2 r^{-2}  (\Up   r^2 h)' |\Psi|^2 + h \Psi R(\Psi) \nn
\eea
we  deduce, from \eqref{eq:intS-EEd}
\beaa
\int_S   \EEd [fR, w, 2hR](\Psi) \geq \int_S \EEd_2  - O(\ep r^{-3})     \int_S | \Psi| ^2.
\eeaa

We now substitute,
\beaa
h&=&4\Up r^{-4} \widetilde{h}.
\eeaa
Hence,
\beaa
\frac 1 2 r^{-2}  \pr_r(\Up   r^2 h) |\Psi|^2 + h \Psi R(\Psi)  &=&\frac 1 2 r^{-2}\pr_r( 4\Up^2 r^{-2} \widetilde{h}) |\Psi|^2+ 4 \Up  r^{-4}\widetilde{h} \Psi R(\Psi)\\
&=&\frac 1 2 r^{-2}\pr_r( 4\Up^2 r^{-2}) \widetilde{h} |\Psi|^2  +2 r^{-4}\Up^2  \pr_r\widetilde{h} |\Psi|^2+ 4 \Up  r^{-4}\widetilde{h} \Psi R(\Psi)
\eeaa
or, since  $\frac 1 2 r^{-2}\pr_r( 4\Up^2 r^{-2}) =- 4 r^{-2} \Up \frac{r-4m}{r^4}$,
\beaa
\frac 1 2 r^{-2}  \pr_r(\Up   r^2 h) |\Psi|^2 + h \Psi R(\Psi)  &=&- 4 r^{-2} \Up \frac{r-4m}{r^4} \widetilde{h}|\Psi|^2  +2 r^{-4}\Up^2  \pr_r\widetilde{h} |\Psi|^2+ 4 \Up  r^{-4}\widetilde{h} \Psi R(\Psi).
\eeaa

Thus we have,
\beaa
\EEd_2 &=& \pr_r f  |R(\Psi)|^2+ 2r^{-3} \left(1-\frac{3m}{r}\right) f |\Psi|^2  +4 \Up\frac{r-4m}{r^4}  (f- r^{-2}   \widetilde{h} ) |\Psi|^2\\
 &+&2 r^{-4}\Up^2  \pr_r\widetilde{h} |\Psi|^2+ 4 \Up  r^{-4}\widetilde{h} \Psi R(\Psi)+ r^{-2}W |\Psi|^2.
\eeaa
We also express,
\beaa
 4 \Up  r^{-4}\widetilde{h} \Psi R(\Psi)&=&\frac{2\widetilde{h}}{r^3}(R(\Psi)+ \Up r^{-1} \Psi)^2 - \frac{2 \widetilde{h}}{r^3} |R(\Psi)|^2 -
 \frac{2 \widetilde{h}}{r^5}\Up^ 2 |\Psi|^2
 \eeaa
and   therefore,
\beaa
\EEd_2 &=&(\pr_r f-2 r^{-3} \widetilde{h}) |R(\Psi)|^2    +    \frac{2\widetilde{h}}{r^3}(R(\Psi)+ \Up r^{-1} \Psi)^2 + r^{-2}W |\Psi|^2 \\
 &+&\left[ 2r^{-3} \left(1-\frac{3m}{r}\right) f +4 \Up\frac{r-4m}{r^4}  (f- r^{-2}\widetilde{h})  + 2 r^{-4} \Up \pr_r\widetilde{h} - 2 r^{-5}\Up^2 \widetilde{h} \right]|\Psi|^2.
\eeaa

We choose $\widetilde{h}(r,m)$ as the following continuous and piecewise $C^1$ function,
\beaa
  \widetilde{h}=\begin{cases}
\quad   0,\qquad \qquad \qquad \qquad\qquad  r\le \frac{5m}{2},\\
\quad   \delta_{\widetilde{h}}\left(\frac{5m}{2}-r\right),\qquad \qquad \quad  \frac{5m}{2}\le  r\le \frac{11m}{4},\\
\quad   \delta_{\widetilde{h}}(r-3m),\qquad \qquad \quad   \frac{11m}{4}\le r\le 3m,\\
  \quad r^2 f, \qquad\qquad\qquad\qquad\,\,  3m\le r \le 4m,\\
\quad  (4m)^2 f(4m,m),\qquad\qquad\, r\ge 4m. 
\end{cases}
\eeaa
where the constant $\de_{\widetilde{h}}>0$ will be chosen small enough. We consider the following cases:

{\bf Case 1} ($\frac{5m}{2}\le r\le 3m$).     In view of the definition of $\widetilde{h}$ and since $W=0$, we  deduce,
\beaa
\EEd_2 &=& \pr_r f  |R(\Psi)|^2+ \Bigg[2r^{-3} \left(1-\frac{3m}{r}\right) f   +4 \Up\frac{r-4m}{r^4}  (f- r^{-2}   \widetilde{h} )\\
&& +2 \delta_{\widetilde{h}}r^{-4}\Up^2\Big(1_{\frac{11m}{4}\le r\le 3m}-1_{\frac{5m}{2}\le  r\le \frac{11m}{4}}\Big)\Bigg] |\Psi|^2+  \delta_{\widetilde{h}}O(1)\Psi R(\Psi)1_{\frac{5m}{2}\le  r\le 3m}.
\eeaa
In view of \eqref{eq:usefulllowerboundforfandprrf}, we may assume, choosing for $\delta_{\widetilde{h}}>0$ small enough, that 
\bea
f-\tilde{h}\leq -\frac{1}{2}|f|\textrm{ on }r\leq 3m.
\eea
We infer, using also that $f<0$ on $r\leq 3m$, 
\beaa
\EEd_2 &\geq& \pr_r f  |R(\Psi)|^2+ \Bigg[2r^{-3} \left(1-\frac{3m}{r}\right) f   +2\Up\frac{r-4m}{r^4}f  \\
&&+2 \delta_{\widetilde{h}}r^{-4}\Up^2\Big(1_{\frac{11m}{4}\le r\le 3m}-1_{\frac{5m}{2}\le  r\le \frac{11m}{4}}\Big)\Bigg] |\Psi|^2+  \delta_{\widetilde{h}}O(1)\Psi R(\Psi)1_{\frac{5m}{2}\le  r\le 3m}.
\eeaa
Since we have  
\beaa
\pr_rf\gtrsim 1, \quad 2r^{-3} \left(1-\frac{3m}{r}\right)  +4 \Up\frac{r-4m}{r^4}\lesssim -1, \quad f\lesssim -\left|1-\frac{3m}{r}\right|\quad\textrm{ on }r\leq 3m,
\eeaa
where we have used in particular Lemma \ref{le:John1} and  Proposition \ref{prop:pre-mor1}, we infer
\beaa
\EEd_2 &\gtrsim &  |R(\Psi)|^2+ \left(\left|1-\frac{3m}{r}\right|+\delta_{\widetilde{h}}1_{\frac{11m}{4}\le r\le 3m} -O(1) \delta_{\widetilde{h}}1_{\frac{5m}{2}\le  r\le \frac{11m}{4}}\right) |\Psi|^2 \\
&&-  \delta_{\widetilde{h}}O(1)\Psi R(\Psi)1_{\frac{5m}{2}\le  r\le 3m}\\
&\geq& \frac{1}{2}|R(\Psi)|^2+\left(\left|1-\frac{3m}{r}\right|+\delta_{\widetilde{h}}\Big(1-O(1)\delta_{\widetilde{h}}\Big)1_{\frac{11m}{4}\le r\le 3m}-O(1)\delta_{\widetilde{h}}1_{\frac{5m}{2}\le  r\le \frac{11m}{4}}\right) |\Psi|^2.
\eeaa
 Thus, for $\delta_{\widetilde{h}}>0$ small enough, the exists some large $C>0$ such that
\bea\lab{eq:case1forthecontrolofEEdot2}
\EE_2\ge C^{-1}\left[ |R(\Psi)|^2 + |\Psi|^2 \right]\,\,\textrm{ on }\frac{5m}{2}\leq r\leq 3m.
\eea

{\bf Case 2} ($ 3m\le r    \le 4m$).  Since $\widetilde{h}=r^2f$ and $W=0$, using in particular $\widetilde{h}\geq 0$ on $ 3m\le r    \le 4m$, we  deduce,
\beaa
\EEd_2 &\geq&(\pr_r f-2 r^{-3}(r^2 f)) |R(\Psi)|^2   
 +\left[ 2r^{-3} \left(1-\frac{3m}{r}\right) f  + 2 r^{-4} \Up \pr_r(r^2 f) - 2 r^{-5}\Up^2 (r^2 f) \right]|\Psi|^2\\
 &=&(\pr_r f- 2r^{-1} f)  |R(\Psi)|^2 +\left[  2r^{-3} \left(1-\frac{3m}{r}\right) f     + 2 r^{-4}\Up^2 (2r f+r^2\pr_r f) - 2 r^{-3} \Up^2f \right] |\Psi|^2  \\
 &=&         (\pr_r f- 2r^{-1} f)  |R(\Psi)|^2 +\left[  2r^{-3} \left(1-\frac{3m}{r}\right) f+2 r^{-2} \Up^2 \pr_r f+ 2 r^{-3}\Up^2  f\right]|\Psi|^2.
 \eeaa
Note that  the second term is strictly  positive. It  remains to analyze the first term.
\begin{lemma} 
In the interval $[3m, 4m]$ we have,
\beaa
 \pr_r f-2 r^{-1} f>0.
\eeaa
\end{lemma}

\begin{proof}
Recall  from  Proposition \ref{prop:pre-mor1} that $w=r^{-2}\Up \pr_r(r^2 f)=\frac{1}{4m}$ in the interval $[3m, 4m]$. Using also $f=0$ on $r=3m$, we deduce
\beaa
\pr_r(r^2f) &=& \frac{r^2}{\Up}\frac{1}{4m}.
\eeaa
We compute
\beaa
\pr_r\left(r^2f - \frac{(r-3m)r^2}{4m\Up}\right) &=& -\frac{(r-3m)}{4m}\pr_r\left(\frac{r^2}{\Up}\right)\\
&=& -\frac{(r-3m)(r-4m)}{2m\Up^2}\\
&\leq 0& \textrm{ on }3m\leq r\leq 4m,
\eeaa
so that the differentiated quantity decays in $r$ on $[3m, 4m]$. Since it vanishes on $r=3m$, we infer
\beaa
f &\leq& \frac{(r-3m)}{4m\Up} \textrm{ on }3m\leq r\leq 4m.
\eeaa
Thus, we deduce, using again $\pr_r(r^2f) = \frac{r^2}{\Up}\frac{1}{4m}$, 
\beaa
\pr_rf-\frac{2}{r}f &=& r^{-2}\Big(\pr_r(r^2f)-4rf\Big)\\
&=& \frac{1}{4m\Up} -\frac{4}{r}f\\
&\geq& \frac{1}{4m\Up}  -\frac{(r-3m)}{rm\Up}\\
&\geq& \frac{1}{4m\Up}\left(1-4\left(1-\frac{3m}{r}\right)\right)\\
&>& 0\textrm{ on }3m\leq r< 4m.
\eeaa
On the other hand, we have by direct check at $r=4m$, using \eqref{eq:John3'},
\beaa
\left(\pr_rf-\frac{2}{r}f\right)_{r=4m} =\frac{1}{2m} -\frac{1}{m}f_{r=4m}=\frac{1}{2m}\left(1-\frac{C_*}{8}\right)>0 
\eeaa
since $C_*<8$. Hence, we infer
\beaa
 \pr_r f-2 r^{-1} f>0\textrm{ on }3m\leq r\leq 4m
\eeaa
as desired.
\end{proof}

We thus conclude,  for some $C>0$,  in the interval $[3m, 4m]$
\bea\lab{eq:case2forthecontrolofEEdot2}
\EEd_2\ge C^{-1}\left[ |R(\Psi)|^2 + |\Psi|^2 \right].
\eea

{\bf Case 3} ($r\geq 4m$).  Since $\widetilde{h}$ is constant and positive on $r\geq 4m$, we deduce,
\beaa
\EEd_2 &\geq&(\pr_r f-2 r^{-3} \widetilde{h}) |R(\Psi)|^2   \\
&+& \left[ 2r^{-3} \left(1-\frac{3m}{r}\right) f +4 \Up\frac{r-4m}{r^4}  (f- r^{-2}\widetilde{h})   - 2 r^{-5}\Up^2 \widetilde{h} +r^{-2} W\right] |\Psi|^2. 
\eeaa
We examine the first term. In view of the formula for $f$ for $r\geq 4m$, see \eqref{eq:John3'}, 
\beaa
\pr_rf=\frac{2}{r^3}(16-C_*)m^2, \qquad \widetilde{h}=(4m)^2f(4m,m)=C_*m^2
\eeaa
and hence
\beaa
\pr_r f-2 r^{-3} \widetilde{h} &=& \frac{2(16-2C_*)m^2}{r^3}
\eeaa
and hence, since $C_*<8$, we have
\beaa
\pr_r f-2 r^{-3} \widetilde{h} &\gtrsim& \frac{m^2}{r^3}\,\,\,\textrm{ for }r\geq 4m.
\eeaa

It remains to  analyze the  sign of 
 \beaa
&& 2r^{-3} \left(1-\frac{3m}{r}\right) f +4 \Up\frac{r-4m}{r^4}  (f- r^{-2}\widetilde{h})   - 2 r^{-5}\Up^2 \widetilde{h} \\
&&=\left[2r^{-3} \left(1-\frac{3m}{r}\right)  +4 \Up\frac{r-4m}{r^4} \right] (f- r^{-2}\widetilde{h})
+\left[2r^{-3} \left(1-\frac{3m}{r}\right) - 2r^{-3} \Up^2\right] r^{-2} \widetilde{h}.
\eeaa
The first term,  which   can be written in the form,
\beaa
\left[2r^{-3} \left(1-\frac{3m}{r}\right)  +4 \Up\frac{r-4m}{r^4} \right] r^{-2}\Big( r^2f(r,m)- (4m)^2f(4m,m)\Big)
  \eeaa
is  manifestly positive  for $r\geq 4m$. To evaluate the sign of the second term we  calculate,
\beaa
2r^{-3} \left(1-\frac{3m}{r}\right) - 2r^{-3} \Up^2=2m r^{-5}(r-4m).
\eeaa
Thus,  for $r\ge 4m$,
\beaa
2r^{-3} \left(1-\frac{3m}{r}\right) f +4 \Up\frac{r-4m}{r^4}  (f- r^{-2}\widetilde{h})   - 2 r^{-5}\Up^2 \widetilde{h} &\geq& 0.
\eeaa
Also, since $W=\frac{m}{r^2}\left(3-\frac{8m}{r}\right)$, we have
\beaa
r^{-2}W &\gtrsim& \frac{1}{r^4}.
\eeaa
Thus, in view of the above, we have   for some $C>0$ and  for $r\ge 4m$,   
\bea\lab{eq:case3forthecontrolofEEdot2}
\EEd_2&\ge C^{-1}\left[\frac{1}{r^3} |R(\Psi)|^2 +\frac{1}{r^4} |\Psi|^2 \right]. 
\eea

Gathering \eqref{eq:case1forthecontrolofEEdot2}, \eqref{eq:case2forthecontrolofEEdot2} and \eqref{eq:case3forthecontrolofEEdot2}, we infer  for some $C>0$,   
\beaa
\EEd_2&\ge C^{-1}\left[\frac{1}{r^3} |R(\Psi)|^2 +\frac{1}{r^4} |\Psi|^2 \right]\,\,\textrm{ on }r\ge \frac{5m}{2}. 
\eeaa
Recalling
\beaa
\int_S   \EEd [fR, w, 2hR](\Psi) \geq \int_S \EEd_2  - O(\ep r^{-3})     \int_S  |\Psi| ^2,
\eeaa
we infer
\beaa
\int_S   \EEd [fR, w, 2hR](\Psi) \geq C^{-1}\int_S\left[\frac{1}{r^3} |R(\Psi)|^2 +\frac{1}{r^4} |\Psi|^2 \right] - O(\ep r^{-3})     \int_S  |\Psi| ^2
\eeaa
and hence, for $\ep>0$ small enough,
   \beaa
 \int_S  \EEd [fR, w, 2hR](\Psi) \ge \frac{1}{2}C^{-1}\int_{S} \left[ \frac{1}{r^3} |R(\Psi)|^2 +  \frac{1}{r^4} |\Psi|^2    \right]\,\,\textrm{ on }r\ge \frac{5m}{2}
  \eeaa
as desired.

It remains to analyze the error term, 
\beaa
\EE_\ep[fR, w, 2hR](\Psi)&=&\EE_\ep[fR, w](\Psi)+   O\left( r^{-1}\ep\,u_{trap}^{-1-\dec} ( r|\pr_rh|+  |h| +r|\pr_mh|)\right)|\Psi|^2\\
&=&\ep \frac 1 2 \QQ\piddX+ O(\ep r^{-3} u_{trap}^{-1-\dec}(|f|   +r^2 |\pr_r w|+ r^3|\pr^2_rw| \\
&& +r^2|\pr_r\pr_mw|+r|\pr^2_mw|)     )|\Psi|^2 \\
&&+ O\left( r^{-3}\ep\,u_{trap}^{-1-\dec} ( r^3|\pr_rh|+  r^2|h| +r^3|\pr_mh|)\right)|\Psi|^2.
\eeaa
Recall that,
 \beaa
   w=    \begin{cases}
& \frac{1}{4m},   \qquad \quad \qquad \qquad \qquad   \qquad  \, \,\,\,\, \mbox{for}\quad  r \le  4m,    \\ 
& \frac{2}{r} \left(1-\frac{2m}{r}\right),  \qquad \,\, \qquad  \qquad \qquad \mbox{for}\quad r \ge 4m, 
\end{cases}
   \eeaa
   and $h=4\Up r^{-4} \widetilde{h}$, with
  \beaa
  \widetilde{h}=\begin{cases}
\quad   0,\qquad \qquad \qquad \qquad\qquad  r\le \frac{5m}{2},\\
\quad   \delta_{\widetilde{h}}\left(\frac{5m}{2}-r\right),\qquad \qquad \quad  \frac{5m}{2}\le  r\le \frac{11m}{4},\\
\quad   \delta_{\widetilde{h}}(r-3m),\qquad \qquad \quad   \frac{11m}{4}\le r\le 3m,\\
  \quad r^2 f, \qquad\qquad\qquad\qquad\,\,  3m\le r \le 4m,\\
\quad  (4m)^2 f(4m,m),\qquad\qquad\, r\ge 4m. 
\end{cases}
\eeaa
We deduce,
\beaa
\EE_\ep[fR, w, 2hR](\Psi)&=&\ep \QQ\piddX + O\Big(\ep r^{-3} u_{trap}^{-1-\dec}(|f|  +1)\Big)
\eeaa
which concludes the proof of Proposition \ref{prop:pre-mor2}.
\end{proof}

  
\subsection{Cut-off Correction in $\Mint$}\label{subsection:Mor-cutoff}


 So far we have found a     triplet $(X=fR,  w=  r^{-2}\Up\pr_r \left( r^2 f\right),   M=2hR) $    with   $f$    defined    in Proposition \ref{prop:pre-mor1}     
  and $h$ in Proposition \ref{prop:pre-mor2}   allowing for the lower bound \eqref{eq:lowerboundonEEdfRw2hR} on  $\int_S\EEd [fR, w, M](\Psi)$. 
The main problem  which remains to be addressed  is that
\begin{enumerate}  
\item  $f$ blows up     logarithmically  near  $r=2m$.
\item  The lower bound  for $ \int_S\EEd [fR, w, 2hR](\Psi)$ does not control $e_3(\Psi)$ near $r=2m$.
\end{enumerate}

In this section, we deal with the first problem, while the second problem will be treated in section \ref{subsection:redshift}. To correct for the first problem, i.e. the fact that $f$ blows up     logarithmically  near  $r=2m$, we    have to modify  our choice of $f$ and $w$  there.      
 Introducing
 \beaa
 u:=r^2 f,
 \eeaa
 we have, 
 \bea
  f= r^{-2} u , \qquad w= r^{-2} \Up \pr_r u.
 \eea
 
 {\bf Warning. } The auxiliary    function $u$ introduced  here, and used only in this section,  has of course nothing to do with our previously defined   optical  function on $\Mext$.  \\
 
\begin{definition}\lab{def:udefdewdeWde}
For a given ${\widehat{\de}}>0  $ we define the following functions of $(r,m)$
\beaa
&& u_{\widehat{\de}}:= - \frac{m^2}{\widehat{\de}}   F\left(-\frac{\widehat{\de}}{m^2}   u\right), \qquad    f_{\widehat{\de}}:= r^{-2} u_{\widehat{\de}},\\
&& w_{\widehat{\de}}:= r^{-2} \Up \pr_r  u_{\widehat{\de}}, \qquad\qquad\quad  W_{\widehat{\de}}:= -\frac 1 4 \pr_r \left(  r^2 \Up \pr_r  w_{\widehat{\de}}\right),
\eeaa
 where  $F:\RRR\longrightarrow \RRR$ is   is a fixed, increasing,  smooth function   such that 
  \beaa
  F(x)=
  \begin{cases}  
  &x\qquad  \mbox{for}\quad  x\le 1,\\
&2 \qquad  \mbox{for}\quad  x\ge 3.
  \end{cases}
 \eeaa
\end{definition} 

We now derive useful properties satisfied by $f_{\widehat{\de}}$, $w_{\widehat{\de}}$ and $W_{\widehat{\de}}$. 
\begin{lemma}
Let $f_{\widehat{\de}}$, $w_{\widehat{\de}}$ and $W_{\widehat{\de}}$ introduced in definition \ref{def:udefdewdeWde}. Then,  $f_{\widehat{\de}}\in C^2(r>0)$, $w_{\widehat{\de}}\in C^1(r>0)$, and  we have for ${\widehat{\de}}>0$ sufficiently small
\beaa
f_{\widehat{\de}}=f\qquad\qquad  w_{\widehat{\de}}= w,  \qquad\qquad W_{\widehat{\de}}=W\quad\textrm{ for }r\geq \frac{5m}{2}.
\eeaa
Also, we have for all $r>0$
\bea\lab{eq:lowerboundforfdelta}
r^{-1} f_{\widehat{\de}}\left(1-\frac{3m}{r}\right) \ge   C^{-1}  r^{-1} \left(1-\frac{3m}{r}\right)^2
\eea
and
\bea\lab{eq:lowerboundforpartialfdelta}
\pr_r(f_{\widehat{\de}})\ge     \frac{16m^2}{r^3}. 
\eea
\end{lemma}

\begin{proof} 
Note first that
\beaa
w_{\widehat{\de}}= r^{-2} \Up \pr_r  u_{\widehat{\de}}= r^{-2} \Up F' \left(-\frac{\widehat{\de}}{m^2}  u\right) \pr_r u= w F' \left(-\frac{\widehat{\de}}{m^2} u\right).
\eeaa
In view of the definition of $u_{\widehat{\de}}$, $f_{\widehat{\de}}$, $w_{\widehat{\de}}$ and $W_{\widehat{\de}}$, we have
\beaa
&& u_{\widehat{\de}}=  u, \qquad\qquad\,  f_{\widehat{\de}}=f,\qquad\qquad\quad   w_{\widehat{\de}}= w,  \qquad W_{\widehat{\de}}=W       \qquad \,\mbox{ for}\quad     u  \ge  - \frac{m^2}{\widehat{\de}},  \\
&& u_{\widehat{\de}}= -\frac{2m^2}{{\widehat{\de}}}, \qquad f_{\widehat{\de}}=-\frac{2m^2}{{\widehat{\de}}r^2},  \qquad   \quad  w_{\widehat{\de}}=0,  \qquad  W_{\widehat{\de}}=0   \qquad \,\,  \mbox{ for}\quad   u   \le    -\frac{3m^2}{{\widehat{\de}}}.
 \eeaa
Also, according to    \eqref{eq:John3''}  
 \beaa
u=    \begin{cases}
& 2m^2    \log \frac{r-2m}{m}+  O( m(r-3m)  ),      \qquad\,  \mbox{for}\quad  r \le  4m,    \\ 
&  r^2 +O(m^2),  \qquad \qquad\qquad \qquad\qquad  \,\,\mbox{for}\quad r \ge 4m,
\end{cases}
\eeaa
and hence,  for ${\widehat{\de}}>0$   sufficiently small
\beaa
\left\{ r \ge 2m+  e^{-\frac{1}{3{\widehat{\de}}}}\right\}\cup\left\{u  \ge  - \frac{m^2}{\widehat{\de}}\right\}, \qquad \left\{ r \le 2m+  e^{-\frac{2}{{\widehat{\de}}}}\right\}\subset\left\{u   \le    -\frac{3m^2}{{\widehat{\de}}}\right\}.
\eeaa
This yields
\beaa
f_{\widehat{\de}}=f\qquad\qquad  w_{\widehat{\de}}= w,  \qquad\qquad W_{\widehat{\de}}=W\quad\textrm{ for }r\geq \frac{5m}{2}.
\eeaa
Also, we have
\beaa
 f_{\widehat{\de}}=\begin{cases}
 &  -\frac{2m^2}{{\widehat{\de}}r^2}, \qquad \, \,\mbox{for}  \qquad      r \le 2m+  e^{-\frac{2}{{\widehat{\de}}}},\\
 &  f,   \qquad \qquad\, \mbox{for}  \qquad      r\ge 2m+  e^{-\frac{1}{3{\widehat{\de}}}},
 \end{cases}
 \eeaa
 and
 \beaa
 f_{\widehat{\de}}\gtrsim \frac{1}{\widehat{\de}}\quad\textrm{ on }\quad 2m+  e^{-\frac{2}{{\widehat{\de}}}}\leq r\leq 2m+  e^{-\frac{1}{3{\widehat{\de}}}},
 \eeaa
and thus,  there exists $C>0$ such  that, for all $r>0$,
\beaa
r^{-1} f_{\widehat{\de}}\left(1-\frac{3m}{r}\right) \ge   C^{-1}  r^{-1} \left(1-\frac{3m}{r}\right)^2
\eeaa
which is \eqref{eq:lowerboundforfdelta}.

 For $u \le -\frac{3m^2}{{\widehat{\de}}}$,
\beaa
\pr_r(f_{\widehat{\de}}) &=&  \pr_r( r^{-2}  u_{\widehat{\de}})=- 2  r^{-3}  u_{\widehat{\de}}+ r^{-2}  \pr_r(u_{\widehat{\de}}) =  \frac{4m^2}{{\widehat{\de}}}   r^{-3}.
\eeaa
 For  $-\frac{3m^2}{{\widehat{\de}}} \le  u \le  - \frac{m^2}{\widehat{\de}}  $
 \beaa
\pr_r(f_{\widehat{\de}}) &=&  \pr_r( r^{-2}  u_{\widehat{\de}})=- 2  r^{-3}  u_{\widehat{\de}}+ r^{-2}\pr_r(u_{\widehat{\de}})\\
&=& - 2  r^{-3}  u_{\widehat{\de}}+ r^{-2} F'\left(-\frac{\widehat{\de}}{m^2}  u\right)\pr_ru\\
&=& - 2  r^{-3}  u_{\widehat{\de}}+ r^{-2} F'\left(-\frac{\widehat{\de}}{m^2}  u\right)r^2\Up^{-1}w,
\eeaa
and since $w\geq 0$ and $F'\geq 0$, we deduce
\beaa
\pr_r(f_{\widehat{\de}}) &\ge&- 2  r^{-3}  u_{\widehat{\de}}\ge  2       {\widehat{\de}}^{-1} m^2   r^{-3}. 
 \eeaa
For  $ u \ge   - \frac{m^2}{\widehat{\de}}  $, using Lemma \ref{le:John1}, we have
\beaa
\pr_r(f_{\widehat{\de}}) &=& \pr_r f\ge \frac{16m^2}{r^3}.
\eeaa
Hence,   for   all $r\ge 2m$, ${\widehat{\de}}>0$ sufficiently small,
\beaa
\pr_r(f_{\widehat{\de}})\ge     \frac{16m^2}{r^3}
\eeaa
which is \eqref{eq:lowerboundforpartialfdelta}.
\end{proof}

It remains to   evaluate $W_{\widehat{\de}}$. This is done in the following lemma.
\begin{lemma}
\label{lemma:potentialWde}
Let 
\bea
\overline{W}_{\widehat{\de}}(r,m)&:=& 1_{r\le \frac{5m}{2}} |W_{\widehat{\de}}|.
\eea
Then, $\overline{W}_{\widehat{\de}}$ is supported, for $\de>0$ small enough, in the region 
\beaa
2m+e^{-\frac{2}{\widehat{\de}}}\leq r\leq \frac{9m}{4}.
\eeaa
Moreover  its   primitive,
\bea
\widetilde{W_{\widehat{\de}}}(r,m):=\int_{2m}^ r \overline{W}_{\widehat{\de}}(r',m) dr' 
\eea
verifies the pointwise   estimate
\bea
\widetilde{W_{\widehat{\de}}}(r,m) \les  {\widehat{\de}}.
\eea
\end{lemma}

\begin{proof}
Recall that  we have chosen $w=\frac{1}{4m} $ to be constant      in the region  $ r \le  4 m$.    Hence, in that region,
\beaa
w_{\widehat{\de}}&=&  \frac{1}{4m} F'\left(-\frac{\widehat{\de}}{m^2} u\right), \qquad 
\pr_r  w_{\widehat{\de}}= \frac{1}{4m} \pr_r \left( F'\left(-\frac{\widehat{\de}}{m^2}  u\right)\right).
\eeaa
Hence,
\beaa
W_{\widehat{\de}}&=& -\frac 1 4 r^{-2} \pr_r \left(   \frac{1}{4m} r^2 \Up \pr_r \left( F'\left(-\frac{\widehat{\de}}{m^2} u\right)\right)\right)= -\frac{1}{16m} r^{-2}  \pr_r \left(   r^2 \Up \pr_r \left( F'\left(-\frac{\widehat{\de}}{m^2} u\right)\right)\right).
\eeaa
Now, setting $\de_0=\frac{\widehat{\de}}{m^2}$  for convenience below,
\beaa
r^{-2}  \pr_r \left(   r^2 \Up \pr_r \left( F'\left(-\de_0 u\right)\right)\right) &=&  - \de_0 F'' (-\de_0 u) r^{-2} \pr_r \left( r^2 \Up \pr_r  u \right) +\de_0^2F'''(-\de_0 u)  \Up(\pr_r u)^2. 
\eeaa
Note that, since $r^{-2} \Up \pr_r u =w$ and $w=(4m)^{-1}$ is constant  in $r$ in the region of interest
\beaa
r^{-2} \pr_r \left( r^2 \Up \pr_r  u \right) = r^{-2} \pr_r \left( r^4  r^{-2} \Up  \pr_r  u \right) =r^{-2} \pr_r \left( \frac{r^4}{4m}\right) = \frac{r}{m}.
\eeaa
Hence,
\beaa
r^{-2}  \pr_r \left(   r^2 \Up \pr_r \left( F'\left(-\de_0 u\right)\right)\right) &=&  - \de_0 F'' (-\de_0 u) r^{-2} \pr_r \left( r^2 \Up \pr_r  u \right) +\de_0^2F'''(-\de_0 u)  \Up(\pr_r u)^2\\
&=& - \de_0 F'' (-\de_0 u)\frac{r}{m} +\de_0^2F'''(-\de_0 u)  \Up(\pr_r u)^2.
\eeaa
Hence, for $r\le 4m$,  with $\de_0=\frac{\widehat{\de}}{m^2}$,
\beaa
| W_{\widehat{\de}}|   &\les &\de_0^2  |\Up  | | F'''(- \de_0  u)|  (\pr_r u)^2 +\de_0 |F''(-\de_0 u) |
\eeaa
or, since $|\pr_r u|\les    \frac{1}{r-2m} $, in the region of interest,
\beaa
| W_{\widehat{\de}}|   &\les &  \frac{\de_0^2}{|r-2m|}  | F'''(- \de_0  u)| +\de_0 |F''(-\de_0 u) |.
\eeaa
Since  $ F''(-\de_0 u),      F'''(-\de_0 u) $ are supported in the region   $1 \le   -\de_0 u \le 3 $,  i.e.    $-\frac{3}{\de_0} \le u \le -\frac{1}{\de_0}$,  for  ${\widehat{\de}}>0$  sufficiently small
\beaa
e^{-\frac{2}{\widehat{\de}}} \leq r-2m \leq  e^{-\frac{1}{3{\widehat{\de}}}}\leq  \frac{m}{4}.
\eeaa
Hence,
\beaa
\overline{W}_{\widehat{\de}}= 1_{ r\le  \frac{5}{2}m }\,| W_{\widehat{\de}}| \les {\widehat{\de}}\left(  \frac{\de}{ r-2m} +1\right)\ka_{\widehat{\de}}\left(  r-2m \right)
\eeaa
with $\ka_{\widehat{\de}}(x)$ the characteristic function of  the interval $[e^{-\frac{2}{{\widehat{\de}}}} , \, e^{ -\frac{1}{3{\widehat{\de}}}}    ]     $.
Note that the primitive of $\overline{W}_{\widehat{\de}}$, i.e.  
 \beaa
\widetilde{W_{\widehat{\de}}}(r,m)=\int_{2m}^ r \overline{W}_{\widehat{\de}}(r',m) dr', 
\eeaa
is a positive,  increasing    function. Moreover,
\beaa
\widetilde{W_{\widehat{\de}}}(r)\les \int_{2m} ^ {4m} \overline{W} _{\widehat{\de}}(r)  dr \les \de+  \de^2 \int _{e^{-\frac{2}{\widehat{\de}}}}^  {e^{-\frac{1}{3{\widehat{\de}}}}}     \frac{1}{x}  dx\les \de
\eeaa
as desired.
\end{proof}

We now recall  that, see \eqref{eq:quadr-Mor},
\beaa
\begin{split}
\EEd [fR, w](\Psi)&=\EEd_0 [fR, w](\Psi)+4 \Up\frac{r-4m}{r^4}  f |\Psi|^2,\\
\EEd_0 [fR, w](\Psi)&= \pr_r f  |R(\Psi)|^2 + r^{-2}W |\Psi|^2
+ r^{-1} \left(1-\frac{3m}{r}\right) f \QQh_{34}.
\end{split}
\eeaa
Using   the functions $f_{\widehat{\de}}$, $w_{\widehat{\de}}$ and $W_{\widehat{\de}}$  introduced in definition \ref{def:udefdewdeWde}, we have
 \beaa
\EEd_0[ f_{\widehat{\de}} R,  w_{\widehat{\de}}] (\Psi)&= & \frac 1 r \,  f_{\widehat{\de}}   \left(1-\frac{3m}{r}\right)\QQh_{34} + \pr_r(f _{\widehat{\de}}) |R\Psi|^2 + W_{\widehat{\de}}|\Psi|^2.
\eeaa
 Note that in view of the estimates \eqref{eq:lowerboundforfdelta} \eqref{eq:lowerboundforpartialfdelta}, and Lemma \ref{lemma:potentialWde}, we immediately deduce the existence of  a constant  $C>0$ independent  of ${\widehat{\de}}$ such that 
\bea\label{estim:Mor-cutoff}
\EEd[ f_{\widehat{\de}} R,  w_{{\widehat{\de}}}] (\Psi) &\ge& C^{-1}\left[  | R\Psi|^2 +   |\nabb \Psi|^2  +  \Up|\Psi|^2\right] -\overline{W}_{\widehat{\de}}|\Psi|^2\,\,\textrm{ on } r\le \frac{5m}{2}.
\eea
where $\overline{W}_{\widehat{\de}}$ is a  non-negative  potential  supported  in the region $2m+e^{-\frac{2}{\widehat{\de}}} \le r \le  \frac{ 9m}{4}$,     whose primitive $\widetilde{W_{\widehat{\de}}}(r)=\int_{2m}^r  \overline{W}_{\widehat{\de}}(r'm)dr' $  verifies 
$  \widetilde{W_{\widehat{\de}}}\les  {\widehat{\de}}$. 
Combining  this with estimates    of the previous  section we   derive the following.

\begin{proposition}
\label{prop:pre-mor3}
There exists a constant $C>0$, and for any small enough  ${\widehat{\de}}>0$,   there exists   functions  $f_{\widehat{\de}}\in C^2(r>0)  $, $w_{\widehat{\de}}\in C^1(r>0) $   and  $h\in C^2(r>0)$  
 verifying,   for all  $r>0$,
 \beaa
 |f_{\widehat{\de}}(r)|\les \widehat{\de}^{-1}, \qquad w_{\widehat{\de}}\les  r^{-1}, \qquad h\les  r^{-4},
 \eeaa
such that
 \beaa
 \EE[f_{\widehat{\de}}R, w_{\widehat{\de}}, 2 hR] (\Psi)&=&\EEd[f_{\widehat{\de}}R, w_{\widehat{\de}}, 2 hR] +\EE_\ep[f_{\widehat{\de}}R, w_{\widehat{\de}}, 2 hR](\Psi)
 \eeaa
 satisfies 
 \beaa
 \int_S \EEd[f_{\widehat{\de}}R, w_{\widehat{\de}}, 2 hR] &\ge&  C^{-1}\int_{S} \left( \frac{m^2}{r^3} |R(\Psi)|^2 + r^{-1}\left(1-\frac{3m}{r}\right)^2\left( |\nabb \Psi|^2+\frac{m^2}{r^2} |T\Psi|^2 \right) + \frac{m}{r^4} |\Psi|^2      \right) \\
   &-&\int_S\overline{W}_{\widehat{\de}}|\Psi|^2,\\
   \EE_\ep[f_{\widehat{\de}}R, w_{\widehat{\de}}, 2 hR] &=& \ep\frac{1}{2}\QQ\c \, ^{(f_{\widehat{\de}}R)} \pidd+ O(r^{-3} u_{trap}^{-1-\dec} (1+|f_{\widehat{\de}}|) )|\Psi|^2,
 \eeaa
 where $\overline{W}_{\widehat{\de}}$ is  non-negative, supported  in the region $ 2m+e^{-\frac{2}{\widehat{\de}}}\le r\le \frac{9m}{4} $, and such that its primitive $\widetilde{W_{\widehat{\de}}}(r)=\int_{2m}^r  \overline{W}_{\widehat{\de}}$  verifies $\widetilde{W_{\widehat{\de}}}\les {\widehat{\de}}$.
 \end{proposition}

\begin{proof}
We choose $h$ to be the function of $(r,m)$ introduced in Proposition \ref{prop:pre-mor2}, $f_{\widehat{\de}}$ to be the function of $(r,m)$ introduced in definition \ref{def:udefdewdeWde}, and $\overline{W}_{\widehat{\de}}$, introduced in Lemma \ref{lemma:potentialWde}. Also, by an abuse of notation, we denote by $w_{\widehat{\de},0}$ the function denoted by $w_{\widehat{\de}}$ in definition \ref{def:udefdewdeWde}. Then, combining  Proposition \ref{prop:pre-mor2} in the region $r\geq \frac{5m}{2}$ with 
 the estimate  \eqref{estim:Mor-cutoff} in the region $r\le \frac{5m}{2}$, we immediately obtain
 \bea
 \label{eq:interm-Mor10}
\nn \int_S \EEd[f_{\widehat{\de}}R, w_{{\widehat{\de}},0}, 2 hR] &\ge&  C^{-1}\int_{S} \left( \frac{m^2}{r^3} |R(\Psi)|^2 + r^{-1}\left(1-\frac{3m}{r}\right)^2 |\nabb \Psi|^2 + \frac{m}{r^4} |\Psi|^2      \right) \\
   &&-\int_S\overline{W}_{\widehat{\de}}|\Psi|^2.
 \eea

\eqref{eq:interm-Mor10} corresponds to the desired estimate  without the presence of the term $|T\Psi|^2$ on the right   hand side. To get the improved estimate  of Proposition \ref{prop:pre-mor3},  we set 
\bea\lab{eq:theformulaforwasasumof2terms}
 w_{\widehat{\de}}&:=& w_{\widehat{\de},0}- \de_1  w_1,
 \eea
for a small  parameter  $\de_1 >0$  to be chosen later, where $w_{\widehat{\de},0}$  is    our previous choice introduced in definition \ref{def:udefdewdeWde},  and where  
\bea\lab{eq:theformulaforwasasumof2terms:bis}
w_1(r,m):=r^{-1} \frac{m^2}{r^2}     \Up  \left(1-\frac{3m}{r}\right)^2.
\eea
We   evaluate (modulo the same type  of error terms as before which we include in $\EE_\ep$),
 \beaa
\EEd[ f_{\widehat{\de}}R, w_{\widehat{\de}}, 2hR]  (\Psi)&=& \EEd[ X_{\widehat{\de}}, w_{\widehat{\de},0}, 2hR]-\frac 1 2 \de_1   w_1  \LL(\Psi)+\frac{\de_1 }{4}  |\Psi|^2 r^{-2}\pr_r(r^2\Up\pr_rw_1).
\eeaa
Now, since
\beaa
\LL(\Psi) &=& -e_3\Psi\cdot e_4\Psi +  |\nabb\Psi|^2 +  V|\Psi|^2\\
&=&\Up^{-1}\left( -|T\Psi|^2 + | R\Psi|^2\right) +  |\nabb\Psi|^2 +  V|\Psi|^2, 
   \eeaa
    we have,
 \beaa
 && -\frac 1 2 \de_1   w_1  \LL(\Psi)+\frac{\de_1 }{4}  |\Psi|^2 r^{-2}\pr_r(r^2\Up\pr_rw_1) \\
 &=& -\frac 1 2 \de_1   r^{-1} \frac{m^2}{r^2}     \Up  \left(1-\frac{3m}{r}\right)^2  \LL(\Psi)+\frac{\de_1 }{4}  |\Psi|^2 r^{-2}\pr_r\left(r^2\Up\pr_r\left(r^{-1} \frac{m^2}{r^2}     \Up  \left(1-\frac{3m}{r}\right)^2\right)\right)\\
 &=&  \frac 1 2 \de_1   r^{-1}\left(1-\frac{3m}{r}\right)^2\frac{m^2}{r^2}|T\Psi|^2 +O(\de_1 )\left( \frac{m^2}{r^3} |R(\Psi)|^2 + r^{-1}\left(1-\frac{3m}{r}\right)^2 |\nabb \Psi|^2 + \frac{m}{r^4} |\Psi|^2      \right)
 \eeaa
and hence
 \bea\lab{eq:interm-Mor10:bis}
\EEd[ f_{\widehat{\de}}R, w_{\widehat{\de}}, 2hR]  (\Psi)&=& \EEd[ X_{\widehat{\de}}, w_{\widehat{\de},0}, 2hR] +\frac 1 2 \de_1   r^{-1}\left(1-\frac{3m}{r}\right)^2\frac{m^2}{r^2}|T\Psi|^2 \\
\nn&&+O(\de_1 )\left( \frac{m^2}{r^3} |R(\Psi)|^2 + r^{-1}\left(1-\frac{3m}{r}\right)^2 |\nabb \Psi|^2 + \frac{m}{r^4} |\Psi|^2      \right).
\eea
The desired estimate now follows from \eqref{eq:interm-Mor10} and \eqref{eq:interm-Mor10:bis} provided $\de_1 >0$ is chosen small enough compared to the constant $C>0$ of \eqref{eq:interm-Mor10} so that the last term $O(\de_1)$ in the above identity can be absorbed.
\end{proof}


\subsection{The red shift vectorfield}\label{subsection:redshift}
  
  
  Note that the  vectorfields    $T$ and $R$  become both proportional to $e_4$  for $\Up=0$ which means that      the estimate   of Proposition \ref{prop:pre-mor3}
    degenerates  along   $\Up=0$, i.e.  it does not control $e_3(\Psi)$ there.
     In this section we make use of  the       Dafermos-Rodnianski        red shift vectorfield 
   to     compensate for this degeneracy.          The crucial ingredient here is the  favorable sign of $\om$  in a small neighborhood of  $r=2m$.

      \begin{lemma}
    \label{le:pi3-pi4}
Let $\,\pi^{(3)},   \, \,\pi^{(4)}$  denote  the deformation tensors  of   $e_3, e_4$.  
     In the region $r\le 3m$    all   components         are  $O(\ep)$   with the exception of, 
\beaa
\,\pi^{(3)}_{44} &=& - 8 \om =\frac{8m}{r^2} +O(\ep), \qquad\,\,\,  \,\pi^{(3)}_{\th\th}=\kab+\vthb=-\frac{2}{r}+O(\ep),\\
\pi^{(3)}_{\vphi\vphi} &=& \kab-\vthb=-\frac{2}{r}+O(\ep),\\
 \,\pi^{(4)}_{34}&=&4\om=-\frac{4m}{r^2}+O(\ep), \qquad   \,\pi^{(4)}_{\th\th}=\ka+\vth=\frac{2\Up}{r}+O(\ep),\\
 \,\pi^{(4)}_{\vphi\vphi} &=& \ka-\vth=\frac{2\Up}{r}+O(\ep).
 \eeaa
\end{lemma}

\begin{proof}
Immediate verification  in view of our assumptions.
\end{proof}

\begin{lemma}
\label{le:calc-redshift}
Given   the vectorfield,
\bea
Y= a(r,m)  e_3+b(r,m)  e_4, 
\eea
and assuming 
\beaa
\sup_{r\leq 3m}\Big(|a|+|\pr_ra|+|\pr_ma|+|b|+|\pr_rb|+|\pr_mb|\Big) &\les& 1,
\eeaa
we  have,  for $r\le 3m$,
\beaa
\QQ^{\a\b}\piY_{\a\b}&=&\left( \frac{2m}{r^2} a -\Up \pr_ra\right)\QQ_{33}+\pr_rb\QQ_{44} +
 \left(\pr_ra  - \frac{2m}{r^2} b   -\Up \pr_rb  \right)\QQ_{34}\\
 &+& \frac 2 r (b\Up-a)  e_3 \Psi  \c e_4 \Psi +8 \frac{\Up}{r^3}( a -\Up b)|\Psi|^2 \\
 &+& O(\ep)   \big( |\QQ(\Psi)|+ r^{-2} |\Psi|^2\big).
\eeaa
Moreover,  with the notation \eqref{le:divergPP-gen-EE},
\bea
\label{eq:EEYPsi}
\EE[Y, 0](\Psi) = \frac 1 2 \QQ^{\a\b}\piY_{\a\b}+4\frac{r-3m}{r^4}(-a+b \Up)|\Psi|^2+ O(\ep) r^{-2} |\Psi|^2.
\eea
\end{lemma}

\begin{proof}
In view of  
\beaa
\left|e_4(r)-\Up, e_3(r)+1\right| &\les & \ep
\eeaa
Lemma \ref{remark:slowly varyingf}, and the assumptions on the derivatives of $a$ and $b$ w.r.t. $(r,m)$, we have
\beaa
&& e_4(a) = \Up\pr_ra+O(\ep),\quad  e_3(a) = -\pr_ra+O(\ep), \\
&& \, e_4(b) = \Up\pr_rb+O(\ep),\quad \, e_3(b) = -\pr_rb+O(\ep),\quad e_\th(a)=e_\th(b)=0.
\eeaa
We infer, 
\beaa
\QQ^{\a\b}\piY_{\a\b} &=&a \QQ^{\a\b} \,\pi^{(3)}_{\a\b}  - ( \QQ_{33}    e_4  a+  \QQ_{43}   e_3  a)    +b  \QQ^{\a\b} \,\pi^{(4)}_{\a\b}  - ( \QQ_{34}   e_4   b +\QQ_{44}  e_3   b) +O(\ep)|\QQ(\Psi)|\\
&=&a \QQ^{\a\b} \,\pi^{(3)}_{\a\b} + b \QQ^{\a\b} \,\pi^{(4)}_{\a\b} -\QQ_{33}  \Up\pr_ra-\QQ_{34}\left( - \pr_ra +\Up\pr_rb     \right)  +\QQ_{44}\pr_rb +O(\ep)|\QQ(\Psi)|.
\eeaa
Note that,
\bea
\QQ_{\th\th}+\QQ_{\vphi\vphi}= e_3\Psi \c e_4\Psi - V|\Psi|^2=
e_3\Psi \c e_4\Psi - 4\frac{\Up}  {r^2}|\Psi|^2 + O(\ep)r^{-2} |\Psi|^2. 
\eea
Hence, 
 \beaa
  \QQ^{\a\b} \,\pi^{(3)}_{\a\b}&=&\QQ^{44}\,\pi^{(3)}_{44}+ \QQ^{\th\th}\,\pi^{(3)}_{\th\th }+\QQ^{\vphi\vphi }\,\pi^{(3)}_{\vphi\vphi } +O(\ep) |\QQ(\Psi)|\\
  &=& \frac 14 \QQ_{33} \frac{8m}{r^2}-\frac{2}{r} \left(  \QQ_{\th\th}+\QQ_{\vphi\vphi}\right) +O(\ep) |\QQ(\Psi)|\\
  &=&\frac{2m}{r^2}\QQ_{33}-\frac 2 r  e_3\Psi \c  e_4\Psi   +8 \frac{\Up}{r^3} |\Psi|^2+O(\ep)   \big( |\QQ(\Psi)|+ r^{-2} |\Psi|^2\big),      \\
  \\
  \QQ^{\a\b} \,\pi^{(4)}_{\a\b}&=&   2\QQ^{34} \,\pi^{(4)}_{34}+ \QQ^{\th\th}\,\pi^{(4)}_{\th\th }+\QQ^{\vphi\vphi }\,\pi^{(4)}_{\vphi\vphi } +O(\ep) |\QQ(\Psi)|\\
  &=&\frac 1 2 \QQ_{34}  (-4\frac{m}{r^2})+\frac{2\Up}{r}  \left(  \QQ_{\th\th}+\QQ_{\vphi\vphi}\right) +O(\ep) |\QQ(\Psi)|\\
  &=&-\frac{2m}{r^2}\QQ_{34}+\frac{2\Up}{r} e_3 \Psi\c  e_4 \Psi -8 \frac{\Up^2}{r^3} |\Psi|^2+O(\ep)   \big( |\QQ(\Psi)|+ r^{-2} |\Psi|^2\big).   
\eeaa
Therefore,
\beaa
\QQ^{\a\b}\piY_{\a\b}&=&a\left[\frac{2m}{r^2}\QQ_{33}-\frac 2 r  e_3\Psi  e_4\Psi   +8 \frac{\Up}{r^3} |\Psi|^2  \right]+ b\left[-\frac{2m}{r^2}\QQ_{34}+\frac{2\Up}{r} e_3 \Psi\c e_4 \Psi -8 \frac{\Up^2}{r^3} |\Psi|^2 \right]
\\
&-&\QQ_{33}\Up\pr_ra -\QQ_{34}\left( - \pr_ra +\Up\pr_rb     \right)  +\QQ_{44}\pr_rb +O(\ep)   \big( |\QQ(\Psi)|+ r^{-2} |\Psi|^2\big)
\\
&=&
\left( \frac{2m}{r^2} a -\Up\pr_ra\right)\QQ_{33}+\QQ_{44}\pr_rb+\frac 2 r (b\Up-a) e_3 \Psi e_4 \Psi+
 \left(\pr_ra  - \frac{2m}{r^2} b   -\Up\pr_rb  \right)\QQ_{34}\\
 &+&8 \frac{\Up}{r^3}( a -\Up b) |\Psi|^2 +O(\ep)   \big( |\QQ(\Psi)|+ r^{-2} |\Psi|^2\big).
\eeaa
To prove the second part of the  lemma we   recall   (see \eqref{le:divergPP-gen-EE}),
\beaa
\EE[Y,  0](\Psi)&=&\frac  1  2 \QQ^{\a\b}\piY_{\a\b}  -\frac 1 2 Y(V)|\Psi|^2
\eeaa
and, relying on Lemma \ref{remark:slowly varyingf}, we have on $r\leq 3m$
\beaa
Y(V)&=&(-a+ b \Up)\pr_rV+O(\ep)=(-a+ b \Up)\left(-8 \frac{r-3m}{r^4}\right)+ O(\ep)
\eeaa
which concludes the proof of the lemma.
\end{proof}

\begin{corollary}
\label{corollary:QQ*piY}
If we choose,
\beaa
a(2m,m)=1, \qquad   b(2m,m)=0, \qquad \pr_ra(2m,m)\ge \frac{1}{4m},\qquad \pr_rb(2m,m)\ge \frac{5}{4m},
\eeaa
then, at $r=2m$, we have
  \bea
  \label{eq:red-shift1}
   \QQ^{\a\b}\piY_{\a\b} \ge  \frac{1}{4m} (  | e_3\Psi|^2+ |e_4\Psi|^2+\QQ_{34}) +O(\ep)   \big( |\QQ(\Psi)|+ r^{-2} |\Psi|^2\big)
     \eea
    and,
    \begin{equation}
     \label{eq:red-shiftEE1}
    \EE[Y,  0](\Psi)\ge  \frac{1}{8m} \left(  | e_3\Psi|^2+ |e_4\Psi|^2+\QQ_{34}   +\frac{1}{m^2}|\Psi|^2        \right) +O(\ep)   \big( |\QQ(\Psi)|+ r^{-2} |\Psi|^2\big).
    \end{equation}
      Moreover the estimates    remain true if we add        to     $Y$    a multiple  of   $T= \frac 1 2\left( e_4+\Up e_3 \right)$.
\end{corollary}

\begin{proof}
Recall from Lemma \ref{le:calc-redshift} that we  have,  for $r\le 3m$,
\beaa
\QQ^{\a\b}\piY_{\a\b}&=&\left( \frac{2m}{r^2} a -\Up \pr_ra\right)\QQ_{33}+\pr_rb\QQ_{44} +
 \left(\pr_ra  - \frac{2m}{r^2} b   -\Up \pr_rb  \right)\QQ_{34}\\
 &+& \frac 2 r (b\Up-a)  e_3 \Psi  \c e_4 \Psi +8 \frac{\Up}{r^3}( a -\Up b)|\Psi|^2+ O(\ep)   \big( |\QQ(\Psi)|+ r^{-2} |\Psi|^2\big).
\eeaa
Hence, at $r=2m$, using $\Up=0$, $a=1$, $b=0$, $\pr_ra\geq (4m)^{-1}$ and $\pr_rb\geq 5(4m)^{-1}$, we deduce 
\beaa
\QQ^{\a\b}\piY_{\a\b}&=& \frac{1}{2m}\QQ_{33}+\pr_rb\QQ_{44} +\pr_ra\QQ_{34} - \frac{1}{m}  e_3 \Psi  \c e_4 \Psi + O(\ep)   \big( |\QQ(\Psi)|+ r^{-2} |\Psi|^2\big)\\
&\geq& \frac{1}{2m} |e_3(\Psi)|^2+\frac{5}{4m}|e_4(\Psi)|^2 +\frac{1}{4m}\QQ_{34} - \frac{1}{m}  e_3 \Psi  \c e_4 \Psi + O(\ep)   \big( |\QQ(\Psi)|+ r^{-2} |\Psi|^2\big)
\eeaa
from which the desired lower bound        in \eqref{eq:red-shift1}   follows.

Also, at $r=2m$, using \eqref{eq:EEYPsi}, $\Up=0$, $a=1$, and $b=0$, we have 
\beaa
\EE[Y,  0](\Psi)&=&  \frac 1 2 \QQ^{\a\b}\piY_{\a\b}+4\frac{r-3m}{r^4}(-a+b \Up)|\Psi|^2+ O(\ep) r^{-2} |\Psi|^2    \\
&= &\frac  1  2 \QQ^{\a\b}\piY_{\a\b} +\frac{1}{4m^3} |\Psi|^2\\
&\ge&   \frac{1}{8m} \left(  | e_3\Psi|^2+ |e_4\Psi|^2+\QQ_{34}  +  \frac{1}{m^2} |\Psi|^2\right)+O(\ep)   \big( |\QQ(\Psi)|+ r^{-2} |\Psi|^2\big)
\eeaa
which yields \eqref{eq:red-shiftEE1}.
\end{proof}
   
     We are now ready to prove the following  result.
\begin{proposition}
       \label{prop:red-shift-Mor}
Given a small parameter $\deh>0$  there exists a     smooth vectorfield  $ Y_\HH$     supported in the region 
$|\Up|\le 2 \deh^{\frac{1}{10}}$   such  that    the following estimate holds, 
            \beaa
       \EE[Y_\HH, 0](\Psi)&\ge & \frac{1}{16m}  1_{|\Up|\le \deh^{\frac{1}{10}}}   \left( |e_3\Psi|^2 +|e_4 \Psi|^2   + \QQh_{34}+m^{-2}|\Psi|^2\right) \\
       &-&    \frac 1 m \de_\HH^{-\frac{1}{10}}  1_{\deh^{\frac{1}{10}}\le \Up \le 2 \deh^{\frac{1}{10}}}     \left( |e_3\Psi|^2 +|e_4 \Psi|^2   + \QQh_{34} +m^{-2}|\Psi|^2 \right)\\
       &+&O(\ep) 1_{|\Up|\le 2\deh^{\frac{1}{10}}}  \big( |\QQ(\Psi)|+ m^{-2} |\Psi|^2\big).
       \eeaa
       Moreover, for $|\Up|\leq \deh^{\frac{1}{10}}$, we have
       \beaa
       Y_\HH= e_3+e_4 +O(\deh^{\frac{1}{10}}) (e_3+e_4).
       \eeaa
\end{proposition}

\begin{proof}
 We introduce the vectorfield  
 \beaa
 Y_{(0)}:= a e_3 +b e_4  +2T,\quad a(r,m):=1+\frac{5}{4m}(r-2m), \quad b(r,m):= \frac{5}{4m}(r-2m),
 \eeaa 
 with $ T= \frac 1 2 (e_4+\Up e_3)$. Also, we  pick   positive   bump function $\ka=\ka(r) $, supported in the region  in $[-2, 2]$ and   equal to $1$   for $[-1,1]$  and   define, for sufficiently small  $\de_\HH>0$.
     \bea
     Y_\HH &:=&    \ka_\HH      Y_{(0)}, \qquad  \ka_\HH:= \ka\left(\frac{\Up}{\deh^{\frac{1}{10}}}\right).
     \eea
     We have
   \beaa
       \EE[Y_\HH, 0](\Psi)  &=& \QQ \c \, ^{(Y_\HH)}\pi - Y_\HH(V)|\Psi|^2\\
       &=& \ka_\HH\QQ \c\,^{(Y_0)}\pi +\QQ( Y_{(0)},  d \ka_\HH )+ \ka_\HH Y_{(0)}(V)|\Psi|^2\\
       &=&\ka_\HH\EE[Y_{(0)}, 0](\Psi) +O(\deh^{-\frac{1}{10}})1_{\deh^{\frac{1}{10}}\le \Up \le 2 \deh^{\frac{1}{10}}}\left( |e_3\Psi|^2 +|e_4 \Psi|^2   + \QQh_{34} +m^{-2}|\Psi|^2 \right).
       \eeaa  
     Note from the definition of $Y_{(0)}$ and the choice of $a$ and $b$ that  Corollary \ref{corollary:QQ*piY} applies to $Y_{(0)}$. In particular, we deduce from \eqref{eq:red-shiftEE1} for $\deh>0$ small enough, 
         \beaa
       \EE[Y_\HH, 0](\Psi)&\ge & \frac{1}{16m}  1_{|\Up|\le \deh^{\frac{1}{10}}}   \left( |e_3\Psi|^2 +|e_4 \Psi|^2   + \QQh_{34}+m^{-2}|\Psi|^2\right) \\
       &-&    \frac 1 m \de_\HH^{-\frac{1}{10}}  1_{\deh^{\frac{1}{10}}\le \Up \le 2 \deh^{\frac{1}{10}}}     \left( |e_3\Psi|^2 +|e_4 \Psi|^2   + \QQh_{34} +m^{-2}|\Psi|^2 \right)\\
       &+&O(\ep) 1_{|\Up|\le 2\deh^{\frac{1}{10}}}  \big( |\QQ(\Psi)|+ m^{-2} |\Psi|^2\big)
       \eeaa
     as desired.
        \end{proof}

  
   \subsection{Combined Estimate}\lab{sec:combinationofMorawetzwithredshift}
   
   
   We consider the combined Morawetz triplet  
   \bea
   (X, w,  M ):= (X_{\widehat{\de}}, w_{\widehat{\de}},  2hR )+\ep_\HH( Y_\HH, \, 0, 0 ), 
   \eea
    with $\ep_\HH>0$ 
sufficiently small to be determined later.  Here   $(X_{\widehat{\de}}=f_{\widehat{\de}}R, w_{\widehat{\de}}, 2hR )$   is the triplet   given by Proposition \ref{prop:pre-mor3}        and $Y_\HH$ the vectorfield   of Proposition \ref{prop:red-shift-Mor}.

 Recall,    see  Proposition \ref{prop:pre-mor3}, that $\EEd_{\widehat{\de}}:=\EEd[ f_{\widehat{\de}}R,  w_{{\widehat{\de}}}, 2h R] (\Psi)$  verifies,
  \beaa
\int_S \EEd_{\widehat{\de}}
&\ge & C^{-1}\int_S \left( \frac{m^2}{r^3} |R(\Psi)|^2 
+  \left(1-\frac{3m}{r}\right)^2  r^{-1} \left(  \QQh_{34}  +\frac{m^2}{r^2}|T\Psi|^2 \right)          + \Up \frac{m}{r^4}|\Psi|^2  \right) \\
  &-&\int_S\overline{W}_{\widehat{\de}}|\Psi|^2. 
\eeaa
     According  to  Proposition \ref{prop:red-shift-Mor}, we write     for $\EE_\HH=\EE(Y_\HH, 0,0)(\Psi)$,
     \beaa
     \EE_\HH&=& \EEd_\HH+\EE_{\HH,\ep},\\
   \EEd_\HH&\ge & \frac{1}{8m}  1_{|\Up|\le \deh^{\frac{1}{10}}}   \left( |e_3\Psi|^2 +|e_4 \Psi|^2   + \QQh_{34} +m^{-2}|\Psi|^2 \right) \\
       &-&    \frac 1 m \de_\HH^{-\frac{1}{10}}  1_{\deh^{\frac{1}{10}}\le \Up \le 2 \deh^{\frac{1}{10}}}     \left( |e_3\Psi|^2 +|e_4 \Psi|^2   + \QQh_{34} +m^{-2}|\Psi|^2 \right),\\
       \EE_{\HH,\ep} &=&O(\ep)   \big( |\QQ(\Psi)|+ m^{-2} |\Psi|^2\big) 1_{|\Up|\leq 2\deh^{\frac{1}{10}}}.
     \eeaa
     Note that, for $|\Up|\ge \deh^{\frac{1}{10}}$ we have,
     \beaa
     |R\Psi|^2 +|T\Psi|^2 =\frac 1 2( |e_4\Psi|^2 +\Up^2 |e_3\Psi|^2)\ge \frac 1 2  \de_\HH^\frac{1}{5} ( |e_4\Psi|^2 + |e_3\Psi|^2).
     \eeaa
     We now proceed to find a lower bound for    the expression $\EEd_{\widehat{\de}}+\ep_\HH \EEd_\HH$. For brevity   the   $S$ integration is omitted  below.

           {\bf Region $\deh^{\frac{1}{10}}\le |\Up|\le 2 \deh^{\frac{1}{10}}$.}
       \beaa
\EEd_{\widehat{\de}}+\ep_\HH \EEd_\HH&\ge &  m^{-1} C^{-1}\Big[\de_\HH^\frac{1}{5} ( |e_4\Psi|^2 + |e_3\Psi|^2)+  m^{-2} |\Psi|^2   + |\nabb \Psi|^2 \Big]
     - \overline{W}_{\widehat{\de}}|\Psi|^2 \\
     &-&  \ep_\HH  \frac 1 m \de_\HH^{-\frac{1}{10}}      \left( |e_3\Psi|^2 +|e_4 \Psi|^2   + |\nabb\Psi|^2  +m^{-2}|\Psi|^2 \right).
     \eeaa
     Therefore,   choosing  $\ep_\HH \le (2C)^{-1}\de_\HH^\frac{3}{10}$, we deduce,
       \beaa
\EEd_{\widehat{\de}}+\ep_\HH \EEd_\HH
     &\ge & m^{-1}  \de_\HH^\frac{1}{5} (2C)^{-1}\left(  |e_4\Psi|^2 + |e_3\Psi|^2+|\nabb\Psi|^2  + m^{-2} |\Psi|^2\right)  - \overline{W}_{\widehat{\de}}|\Psi|^2.
     \eeaa    
     
     {\bf Region $|\Up|\le \deh^{\frac{1}{10}}$.}  
       \beaa
        \ep_\HH \EEd_\HH+\EEd_{\widehat{\de}}&\ge & \ep_\HH \frac{1}{16m}   \left( |e_3\Psi|^2 +|e_4 \Psi|^2   + \QQh_{34} +m^{-2}|\Psi|^2 \right)  - \overline{W}_{\widehat{\de}}|\Psi|^2.
        \eeaa
        
{\bf  Region $\Up\ge  2\deh^{\frac{1}{10}}$.} In this region  $   \EEd_{\widehat{\de}}+\ep_\HH  \EEd_\HH=\EEd_{\widehat{\de}}$. Hence (ignoring the $S$-integration),
\beaa
   \EEd_{\widehat{\de}}+\ep_\HH  \EEd_\HH&\ge&    C^{-1}\left( \frac{m^2}{r^3} |R(\Psi)|^2 
+  \left(1-\frac{3m}{r}\right)^2  r^{-1} \left(  \QQh_{34}  +\frac{m^2}{r^2}|T\Psi|^2 \right)          +  \frac{m}{r^4}|\Psi|^2  \right) \\
  &-&\overline{W}_{\widehat{\de}}|\Psi|^2.
\eeaa

To combine  these three  cases together  we  modify the  vectorfields $R, T$ near $r=2m$  according to \eqref{eq:Rc-Tc}, i.e. 
      \beaa
       \begin{split}
  \Rbrev&:=  \th  \frac 1 2 ( e_4-e_3) +(1-\th)        \Up^{-1} R=    \frac 1 2 \left[\thbrev e_4- e_3\right], \\
  \Tbrev&:=\th  \frac 1 2 ( e_4+e_3) +(1-\th)        \Up^{-1} T= \frac 1 2 \left[ \thbrev e_4+ e_3\right],
  \end{split}
      \eeaa 
      where     $\th$  a smooth bump  function   equal $1$ on  $  |\Up|\le\deh^{\frac{1}{10}} $  vanishing for  $|\Up|\ge2 \deh^{\frac{1}{10}}$, and where 
        \beaa
      \thbrev=\th+\Up^{-1} (1-\th) =\begin{cases}  1,&\qquad \mbox{for} \quad |\Up|\le  \deh^{\frac{1}{10}},\\
      \Up^{-1},&\qquad \mbox{for} \quad |\Up|\ge2 \deh^{\frac{1}{10}}.
      \end{cases}
      \eeaa
       Note that    
      \beaa
   2(   |\Rbrev\Psi|^2 +|\Tbrev \Psi|^2)= |e_3\Psi|^2 +\thbrev^2 |e_4\Psi|^2.
      \eeaa
    Thus in the region $|\Up|\le \deh^{\frac{1}{10}}$ we have $ |e_3\Psi|^2+|e_4\Psi|^2=      2(   |\Rbrev\Psi|^2 +|\Tbrev \Psi|^2)$ and therefore,    
    \beaa
  \EEd_{\widehat{\de}}+\ep_\HH \EEd_\HH
     &\ge &   \ep_\HH \frac{1}{16m}   \left( |e_3\Psi|^2 +|e_4 \Psi|^2   + \QQh_{34} +m^{-2}|\Psi|^2 \right)  - \overline{W}_{\widehat{\de}}|\Psi|^2\\
     &=&   \ep_\HH \frac{1}{16m}   \left(   |\Rbrev\Psi|^2 +|\Tbrev \Psi|^2  + \QQh_{34} +m^{-2}|\Psi|^2 \right)  - \overline{W}_{\widehat{\de}}|\Psi|^2.
    \eeaa

     In the region $\deh^{\frac{1}{10}} \le |\Up|\le 2 \deh^{\frac{1}{10}}$, we have $  |\Rbrev\Psi|^2 +|\Tbrev \Psi|^2\les  |e_3\Psi|^2+\de_\HH^{-\frac{1}{5}} |e_4\Psi|^2$. Hence, 
      for $\ep_\HH \le (2C)^{-1}\de_\HH^{\frac{3}{10}}$,   we deduce,   
       \beaa
\EEd_{\widehat{\de}}+\ep_\HH \EEd_\HH
     &\ge & m^{-1}  \de_\HH^\frac{1}{5} (2C)^{-1}\left(  |e_4\Psi|^2 + |e_3\Psi|^2+|\nabb\Psi|^2  + m^{-2} |\Psi|^2\right)  - \overline{W}_{\widehat{\de}}|\Psi|^2\\
     &\ge & m^{-1}  \de_\HH^\frac{1}{5} (2C)^{-1}\left( \de_\HH^{\frac{1}{5}}  \left(  |\Rbrev\Psi|^2 +|\Tbrev \Psi|^2 \right)         +|\nabb\Psi|^2  + m^{-2} |\Psi|^2\right)  - \overline{W}_{\widehat{\de}}|\Psi|^2.
     \eeaa    
     Finally, for  $\Up\ge 2 \de_\HH^{\frac{1}{10}}$ we have $\Rbrev=\Up^{-1} R, \, \Tbrev=\Up^{-1} T$. Hence,
     \beaa
     \EEd_{\widehat{\de}}+\ep_\HH \EEd_\HH& \ge&   C^{-1}\left( \frac{m^2}{r^3} |R(\Psi)|^2 
+  \left(1-\frac{3m}{r}\right)^2  r^{-1} \left(  \QQh_{34}  +\frac{m^2}{r^2}|T\Psi|^2 \right)          + \Up \frac{m}{r^4}|\Psi|^2  \right) \\
&-&\overline{W}_{\widehat{\de}}|\Psi|^2\\
&\geq & C^{-1} \left( \de_\HH^\frac{1}{5} \frac{m^2}{r^3} |\Rbrev(\Psi)|^2 
+  \left(1-\frac{3m}{r}\right)^2  r^{-1} \left(  \QQh_{34}  + \de_\HH^\frac{2}{10}\frac{m^2}{r^2}|\Tbrev\Psi|^2 \right)          + \Up \frac{m}{r^4}|\Psi|^2  \right) \\
&-&\overline{W}_{\widehat{\de}}|\Psi|^2.
     \eeaa
     We deduce the following.
       \begin{proposition}
     \label{prop:pre-mor4} 
     Let $C>0$ the constant of Proposition    \ref{prop:pre-mor3}. Consider the  combined Morawetz triplet  
     \bea
     (X, w,  M ):= (X_{\widehat{\de}}, w_{\widehat{\de}},  2hR )+\ep_\HH( Y_\HH, \, 0, 0 ), 
     \eea
      with $C^{-1}\deh^{\frac{2}{5}}\leq\ep_\HH\le (2C)^{-1}\de_{\HH}^\frac{3}{10} $ where, for given fixed $\widehat{\de}>0$,   $(X_{\widehat{\de}}, w_{\widehat{\de}},  2hR )$ is the triplet of Proposition \ref{prop:pre-mor3}
       and $Y_\HH$  the  vectorfield  of Proposition \ref{prop:red-shift-Mor}, supported   in   
$|\Up|\le 2 \de_\HH^{\frac{1}{10}}$ with  $\de_\HH>0$ sufficiently small, independent of $\widehat{\de}$.   Let $\EEd_{\widehat{\de}}, \EEd_\HH$ be the principal parts of 
$\EE[f_{\widehat{\de}}R, w_{\widehat{\de}}, 2 hR](\Psi)$ and   respectively  $\EE_\HH[Y_\HH, 0,0](\Psi)$ and $\EE_{\widehat{\de}, \ep}$, $\EE_{\HH, \ep}$  the
 corresponding error terms, i.e.,
 \beaa
 \EE[f_{\widehat{\de}}R, w_{\widehat{\de}}, 2 hR](\Psi)=\EEd_{\widehat{\de}}+\EE_{\widehat{\de},\ep}, \qquad     \EE_\HH[Y_\HH, 0,0](\Psi)=\EEd_\HH+ \EE_{\HH,\ep}.
 \eeaa 
 Then, provided $ \de_\HH>0$ is sufficiently small, we have
     \begin{enumerate}
     \item  In the region  $-2\de_\HH^{\frac{1}{10}} \le \Up$, \, $r\le \frac{5m}{2}$, we have with a constant  $\La_\HH^{-1} :=C^{-1} \de_\HH^\frac{2}{5}>0$
      \beaa
 \int_S  ( \EEd_{\widehat{\de}}+\ep_\HH \EEd_\HH)& \ge&    m^{-1}\La_\HH^{-1}\int_S\left( |\Rbrev(\Psi)|^2 +|\Tbrev \Psi|^2    +|\nabb\Psi|^2  + m^{-2} |\Psi|^2\right)  - \int_S\overline{W}_{\widehat{\de}}|\Psi|^2.
     \eeaa
     
     \item  In the region $r\ge  \frac{5m}{2} $,  where $ \EEd_{\widehat{\de}}+\ep_\HH \EEd_\HH=\EEd_{\widehat{\de}}$ and $\overline{W}_{\widehat{\de}}=0$,  we have the same estimate as  in Proposition 
        \ref{prop:pre-mor3}, i.e.
    \beaa
 \int_S ( \EEd_{\widehat{\de}}+\ep_\HH \EEd_\HH)&\ge&  C^{-1}\int_{S} \left( \frac{m^2}{r^3} |R(\Psi)|^2 + r^{-1}\left(1-\frac{3m}{r}\right)^2\left( |\nabb \Psi|^2+\frac{m^2}{r^2} |T\Psi|^2 \right) + \frac{m}{r^4} |\Psi|^2      \right).
  \eeaa
     \item
     The  $\ep$-error  terms  verify the upper bound estimate,
     \beaa
     \EE_{\widehat{\de}, \ep} +\ep_\HH \EE_{\HH, \ep}&\les& C\ep   \widehat{\de}^{-1}  u_{trap}^{-1-\dec}   \left[ r^{-2} |e_3\Psi|^2 + r^{-1} (|e_4 \Psi|^2+|\nabb\Psi|^2)
     \right] \\
     &+& C\ep   \widehat{\de}^{-1}  u_{trap}^{-1-\dec}  r^{-1} |e_3\Psi| \left(  |e_4 \Psi|+|\nabb\Psi|  \right) +  C\ep   \widehat{\de}^{-1}  u_{trap}^{-1-\dec}  r^{-3}    |\Psi|^2.
     \eeaa
     \end{enumerate}
     \end{proposition}
     
        \begin{proof}
        It only remains to  check the last part. In view of Proposition \ref{prop:pre-mor2} we have,
        \beaa
          \EE_{\widehat{\de}, \ep} &=&\EE_\ep[ f_{\widehat{\de}}R, w_{\widehat{\de}},  2hR](\Psi)=\ep \frac 1 2  \QQ\c\, ^{(X_{\widehat{\de}})}\pidd + O\big(\ep r^{-3} u_{trap}^{-1-\dec}(|f_{\widehat{\de}}|  +1)\big)|\Psi|^2
        \eeaa
    and     $|f_{\widehat{\de}} |\les   \widehat{\de}^{-1}$. Hence,
    \beaa
      \EE_{\widehat{\de}, \ep} &=&\EE_\ep[ f_{\widehat{\de}}R, w_{\widehat{\de}},  2hR](\Psi)=\ep\left( \frac 1 2  \QQ\c\, ^{X_{\widehat{\de}}}\pidd+O( \widehat{\de}^{-1} r^{-3} u_{trap}^{-1-\dec}) |\Psi|^2 \right).
    \eeaa
    We write  with $\pidd=^{(X_{\widehat{\de}})}\pidd$ for simplicity,
    \beaa
      \QQ\c\, \pidd&=& \frac 1 4\left( \QQ_{33}\pidd_{44}+ 2 \QQ_{34}\pidd_{34}+\QQ_{44}\pid_{33} \right)-\frac 1 2 \left(\QQ_{A3 }\pidd_{A4}+\QQ_{A4 }\pidd_{A3}\right)+\QQ_{AB}\pidd_{AB}.
    \eeaa
    Thus,    recalling part 1 and 2 of Proposition \ref{prop:summaryofallimportantpropertiesofenergymomentumtensorforwaveequationpsi}, and  Lemma \ref{le:piX-decomp},
    \beaa
     \QQ\c\, \pidd&\les& r^{-2}  u_{trap}^{-1-\dec}|e_3\Psi|^2+ r^{-1}u_{trap}^{-1-\dec}\left(|e_4\Psi|^2+|\nabb\Psi|^2 + r^{-2}  |\Psi|^2 \right)\\
     &+& r^{-1} u_{trap}^{-1-\dec}|e_3\Psi| \left(   |e_4 \Psi|+|\nabb\Psi| \right).
     \eeaa
    Finally, since $r\sim 2m$ and $u_{trap}=1$ on $|\Up|\leq 2\deh^{\frac{1}{10}}$, the error terms  generated by  the red shift  vectorfield $Y_\HH$,
    \beaa
    \EE_{\HH,\ep} &=& O(\ep)1_{|\Up|\leq 2\deh^{\frac{1}{10}}}   \big( |\QQ(\Psi)|+ m^{-2} |\Psi|^2\big)
   \eeaa
   can easily be  absorbed  on the right hand side to derive the desired  estimate.
        \end{proof}


\subsubsection{Elimination of  $\overline{W}_{\widehat{\de}}$}
            

             We now proceed  to eliminate the potential $\overline{W}_{\widehat{\de}}$  by a procedure  analogous to that used in section  
             \ref{subs:improvedLB-awayH}. 
              More precisely  we  set,  in view of  \eqref{le:divergPP-gen-EE}, 
              \beaa
               \EE_{\widehat{\de}}=\EE[f_{\widehat{\de}}R, w_{\widehat{\de}}, 2 h R](\Psi), \qquad  \EE_{\widehat{\de}}'= \EE[f_{\widehat{\de}}R, w_{\widehat{\de}}, 2 (hR+h_2 \Rbrev)](\Psi),
              \eeaa
              and,
 \beaa
  \EE_{\widehat{\de}}'=\EE_{\widehat{\de}}+ h_2\Psi\Rbrev \Psi +\frac 1 2     \D^\mu ( h_2 \Rbrev_\mu )|\Psi|^2,
 \eeaa
 where $h_2$ is a smooth, compactly  supported  function  supported\footnote{Recall that  $\overline{W}_{\widehat{\de}}$  is supported  is  supported       in the region $2m<r\le \frac{5m}{2}$.}             in the  region $r\le \frac{9m}{4}$.

   Thus, we have in view of Proposition \ref{prop:pre-mor4}, ignoring   the    integration on $S$,
    \bea
    \label{eq:Hardy-delta1}
 \nn     \EEd'_{\widehat{\de}}+\ep_\HH \EEd_\HH &=& \EEd_{\widehat{\de}}+\ep_\HH \EEd_\HH+ h_2\Psi\Rbrev \Psi +\frac 1 2     \D^\mu ( h_2 \Rbrev_\mu )|\Psi|^2\\
      &\ge& \II(\Psi)+ m^{-1}\La_\HH^{-1}\left(\frac{1}{2}|\Rbrev(\Psi)|^2+ |\Tbrev \Psi|^2    +|\nabb\Psi|^2  + m^{-2} |\Psi|^2\right)
      \eea
     where,
     \beaa
    \II(\Psi):&=& \frac{1}{2}\La_\HH^{-1}  m^{-1}      |\Rbrev(\Psi)|^2 + \Psi       h_2  \Rbrev \Psi +\frac 1 2     \D^\mu ( h_2 \Rbrev_\mu )|\Psi|^2 -\overline{W}_{\widehat{\de}}|\Psi|^2
     \eeaa
     so that we have
     \bea\lab{eq:lowerboundonIIofPsi}
       \II(\Psi) &\geq&  \frac 1  2\left[    \D^\mu ( h_2 \Rbrev_\mu )-2 \overline{W}_{\widehat{\de}}-  m  \La_\HH h_2^2\right] |\Psi|^2.
     \eea
     
      We focus on the coefficient in front of $|\Psi|^2$ on the RHS of \eqref{eq:lowerboundonIIofPsi}.    Ignoring   the   error terms  in $\ep$    (which can easily be  incorporated in   the upper  bound for  $    \EE_{{\widehat{\de}}, \ep} +\ep_\HH \EE_{\HH, \ep} $ of the previous proposition),      we have,
      \beaa
 \Div  \Rbrev&=&\frac 12  \left(   \D^\mu(\thbrev ( e_4)_\mu) -   \D^\mu( ( e_3)_\mu\right)=\frac 1 4\left( \thbrev   \tr \pi^{(4)}- \tr\pi^{(3)} \right) +\frac 1 2  e_4(\thbrev )=O(\deh^{-\frac{1}{10}})
 \eeaa
 and, using in particular Lemma \ref{remark:slowly varyingf},
    \beaa
   \D^\mu ( h_2 \Rbrev_\mu ) &=& \Rbrev  h_2 + h_2  \Div \Rbrev  =\frac 1 2  \pr_rh_2 ( \thbrev   e_4 r- e_3 r ) + h_2  \Div \Rbrev  \\
   &=&\frac 1 2  \pr_rh_2 ( \thbrev  \Up+ 1 )+  h_2 O(\deh^{-\frac{1}{10}} )\\
   &\ge & \frac 12 \pr_rh_2 +  h_2O(\deh^{-\frac{1}{10}} ).
    \eeaa
 Together with \eqref{eq:lowerboundonIIofPsi}, we infer
   \bea\lab{eq:lowerboundonIIofPsi:bis}
       \II(\Psi) &\geq&  \frac 14\Big[      \pr_rh_2  -4 \overline{W}_{\widehat{\de}}-  4m  \La_\HH h_2^2+  h_2O(\deh^{-\frac{1}{10}} )\Big] |\Psi|^2.
     \eea   
     
     We now consider the choice of the function $h_2=h_2(r,m)$.
      Recall (see Lemma \ref{lemma:potentialWde})   that  $\overline{W}_{\widehat{\de}}$  is supported  in the region $2m+e^{-\frac{2}{\widehat{\de}}}\le  r \le  \frac{9m}{4}$  and  that its primitive 
    $\widetilde{W_{\widehat{\de}}}(r):=\int_{2m}^r  \overline{W}_{\widehat{\de}}$    verifies  $  \widetilde{W_{\widehat{\de}}}\les  m^{-2} {\widehat{\de}}$.  
    We  choose
    \bea\lab{eq:defh2:1}
    h_2 &=:& \left\{\begin{array}{l} 4\widetilde{W_{\widehat{\de}}}, \qquad  \mbox{for}   \quad     r\le  \frac{9m}{4} \\[2mm]
    0, \qquad\quad \,\, \mbox{for}   \quad     r\ge  \frac{5m}{2}
    \end{array}\right.
    \eea
    and since $  \widetilde{W_{\widehat{\de}}}\les  m^{-2} \widehat{\de}$, we may extend $h_2$ in $\frac{9m}{4}\leq r\leq \frac{5m}{2}$ such that $h_2$ is $C^1$ and we have for all $r>0$
    \bea\lab{eq:defh2:2}
    |h_2|\les m^{-2} \widehat{\de}, \qquad |\pr_rh_2| \les m^{-3} {\widehat{\de}}.
    \eea
    In view of \eqref{eq:lowerboundonIIofPsi:bis}, this choice of $h_2$ yields
    \beaa
       \II(\Psi) &\geq& - \frac 14 O\Big(m^{-1}{\widehat{\de}}+\La_\HH m^{-1}\widehat{\de}^2+  \widehat{\de}(\de_\HH)^{-\frac{1}{10}} \Big) |\Psi|^2.
     \eeaa  
    Hence, for $ \widehat{\de}\ll     \deh^{\frac{1}{10}}\La_\HH^{-1}$, i.e. $\widehat{\de}\ll     \deh^{\frac{1}{2}}$ (recall that $\La_\HH^{-1}=C^{-1}\deh^\frac{2}{5}$)   and $h_2$ defined as above, we infer
    \beaa
    \II(\Psi)    &\geq& -\frac{1}{2}m^{-1}\La_\HH^{-1}m^{-2} |\Psi|^2
     \eeaa
   which together with \eqref{eq:Hardy-delta1} finally yields
     \beaa
 \nn     \int_S  ( \EEd'_{\widehat{\de}}+\ep_\HH \EEd_\HH)       &\ge& \int_S\II(\Psi)+ m^{-1}\La_\HH^{-1}\int_S\left(\frac{1}{2}|\Rbrev(\Psi)|^2+ |\Tbrev \Psi|^2    +|\nabb\Psi|^2  + m^{-2} |\Psi|^2\right)\\
 &\ge&  m^{-1}\La_\HH^{-1}\int_S\left(\frac{1}{2}|\Rbrev(\Psi)|^2+ |\Tbrev \Psi|^2    +|\nabb\Psi|^2  + \frac{1}{2}m^{-2} |\Psi|^2\right).
      \eeaa

 
 \subsubsection{Summary of  results so far}

      
  We summarize the result in the following,
\begin{proposition}
\label{prop:pre-mor5}
Consider the   combined Morawetz     triplet    
       \bea
       (X, w, M):=  (f _{\widehat{\de}}R,  w_{\widehat{\de}}, 2 h R  )  +\ep_\HH( Y_\HH, 0, 0)  + (0, 0, 2 h_2 \Rc)  
        \eea
         with  $ (f _{\widehat{\de}}R,  w_{\widehat{\de}}, 2 h R  )$ the triplet of Proposition \ref{prop:pre-mor3}, $Y_\HH$  the red shift vectorfield  of Proposition \ref{prop:red-shift-Mor}      (corresponding to the   small parameter $\de_\HH$)           and $h_2 $ the $C^1$ function above satisfying \eqref{eq:defh2:1} \eqref{eq:defh2:2}. Let  $\EEd[X, w, M]$ the principal part of  $\EE[X,w, M]$ (independent of $\ep$) and $\EE_\ep[X, w, M] $ the   error term in $\ep$ such that  $\EE=\EEd+ \EE_\ep$.

           We choose the  small strictly positive  parameters       $\ep_\HH, \, \de_\HH, \de$ such that\footnote{Note that \eqref{eq:restrictionson-de...} verifies all the   restrictions we have encountered so far, i.e. $\deh^{\frac{2}{5}}\ll\ep_\HH\ll\deh^{\frac{3}{10}}$ and $0<\widehat{\de}\ll\deh^{\frac{1}{2}}$.},
         \bea
         \label{eq:restrictionson-de...}
         \ep_\HH = \de_\HH^{\frac{7}{20}}, \qquad  \widehat{\de}= \de_\HH^{\frac{3}{5}}.
         \eea
             Then, there holds\footnote{Note that $\deh^{\frac{1}{2}}\ll \La_\HH^{-1}$ (recall that $\La_\HH^{-1}=C^{-1}\deh^{\frac{2}{5}}$) and $\deh^{-1}\gg \widehat{\de}^{-1}$ in view of \eqref{eq:restrictionson-de...}.}            
          \bea
          \bsplit
     \int_S   \EE[ X,  w,  M] (\Psi)&\ge\deh^{\frac{1}{2}} \int_S \left[   \frac{m^2}{r^3}  | \Rbrev \Psi|^2 +  r^{-1}\left(1-\frac{3m}{r}\right)^2\left( \QQh_{34} +\frac{m^2}{r^2} |\Tbrev \Psi|^2\right)+ \frac{m}{r^4}|\Psi|^2\right],\\
     \\
      \EE_\ep[ X,  w,  M] (\Psi)&\le  \deh^{-1}\ep   u_{trap}^{-1-\dec}   \left[ r^{-2} |e_3\Psi|^2 + r^{-1} (|e_4 \Psi|^2+|\nabb\Psi|^2)
     \right] \\
     &+ \deh^{-1}\ep   u_{trap}^{-1-\dec}  r^{-1} |e_3\Psi| \left(  |e_4 \Psi|+|\nabb\Psi|  \right) +  \deh^{-1}\ep     u_{trap}^{-1-\dec}  r^{-3}    |\Psi|^2.
     \end{split}
       \eea 
\end{proposition}


\subsection{Lower bounds  for $\QQ$}\lab{sec:lowerboundonQformorawetz}


       In this section we prove lower bounds for    for $\QQ(X+2\La T, e_3)$
        and $\QQ(X+2\La T, e_4)$   in the region $r_\HH\le r $, for    $r_\HH$    
          to be determined and  $\La$ sufficiently large.

      \begin{proposition} 
      \label{prop:boundar-Mor}
       Under the assumptions of    Proposition \ref{prop:pre-mor5}, and with the choice
         \bea\lab{eq:choiceofLambdaforpositivityofboundaryterm}
    \La := \frac1 4\deh^{-\frac{13}{20}},
    \eea  
    the following  inequalities hold true for $r\ge 2m_0(1-\deh)$.
                   \begin{enumerate}
      \item  For the  region  such that $r\geq 2m_0(1-\deh)$ and $\Up\le\deh^{\frac{1}{10}}$, we have
      \beaa
      \QQ( X+\La T,  e_3) &\ge &\frac 1 4  \ep_\HH\QQ_{33}+\frac{1}{2}\Lambda\QQ_{34},  \\
       \QQ( X+\La T,  e_4) &\ge & \frac 1 4  \ep_\HH\QQ_{34}+\frac{1}{2}\Lambda\QQ_{44}. 
      \eeaa
      \item  For the region $\de_\HH^{\frac{1}{10}} \le \Up\le  \frac 1  3 $, 
      we have
         \beaa
      \QQ( X+\La T,  e_3) &\ge &  \deh^{-\frac{1}{2}} \left( \QQ_{33}+\QQ_{34}\right),  \\
       \QQ( X+\La T,  e_4) &\ge & \deh^{-\frac{1}{2}} \left( \QQ_{44}+\QQ_{34}\right). 
      \eeaa
      
      \item  For the  region  $r\ge 3m   $, we have
       \beaa
      \QQ( X+\La T,  e_3) &\ge & \frac 1 4  \La \left( \QQ_{33}+\QQ_{34}\right), \\
       \QQ( X+\La T,  e_4) &\ge &  \frac 1 4 \La \left( \QQ_{44}+\QQ_{34}\right). 
      \eeaa
     \item
 The null components of $\QQ$ are given by (recall  Proposition \ref{prop:summaryofallimportantpropertiesofenergymomentumtensorforwaveequationpsi}),
 \beaa
\QQ_{33}&=& |e_3\Psi |^2, \qquad    \QQ_{44}=     |e_4\Psi |^2, \qquad 
\QQ_{34}=   |\nabb \Psi|^2+   \frac{4\Up}{r^2}(1+O(\ep)) |\Psi|^2.
\eeaa
      \end{enumerate}
           \end{proposition}
           
  \begin{proof}
   Since    $X=f_{\widehat{\de}}R+ \ep_\HH Y_\HH$  and $T=\frac 1 2 (e_4+\Up e_3) $, $R=\frac 1 2 (e_4-\Up e_3) $,      we write,
    \beaa
    \QQ(X+2\La T, e_3)&=&\QQ( X, e_3)+ \La\QQ(e_4+\Up e_3, e_3)=\QQ(X, e_3)+
    \La \left(\QQ_{34}+ \Up \QQ_{33}\right)\\
 &=&   \ep_\HH \QQ(Y_\HH, e_3)+ \La \left(\QQ_{34}+ \Up \QQ_{33}\right)
    +\frac 1 2 f_{\widehat{\de}}\left( \QQ_{34}-\Up \QQ_{33} \right).
    \eeaa
    In the region  $2m_0(1-\deh) \le r \le  2m$ we have 
    $Y_\HH=   e_3+e_4 +O(\de_\HH) (e_3+e_4)$, $\Up\geq 0$ and $f_{\widehat{\de}}<0$. 
      Hence, in that region,
      \beaa
        \QQ(X+2\La T, e_3)&\ge & \frac{1}{2}\ep_\HH\left( \QQ_{33}+\QQ_{34}\right) 
        +\left(\La -\frac 1 2 |f_{\widehat{\de}}|\right) \QQ_{34}-|\Up| \left(\La +\frac 1 2 |f_{\widehat{\de}}|\right)\QQ_{33}\\
        &\ge & \left(\frac{1}{2}\ep_\HH -|\Up| \left(\La +\frac 1 2 |f_{\widehat{\de}}|\right)\right)\QQ_{33} +\left(\frac{1}{2}\ep_\HH +\La -\frac 1 2 |f_{\widehat{\de}}|\right)\QQ_{34}.
      \eeaa
      Thus, we need to choose $\La$  such that  
    \beaa
 \frac 1 2 |f_{\widehat{\de}}| \le\La \le \frac1 4\frac{\ep_\HH}{\deh}-  \frac{1}{2}|f_{\widehat{\de}}|
    \eeaa   
    Now, recall \eqref{eq:restrictionson-de...} as well as the fact that $|f_{\widehat{\de}}|$  is of size $O((\widehat{\de})^{-1})$. Thus it suffices to choose $\La$ such that
    \beaa
 O(\de_\HH^{-\frac{3}{5}}) \leq\La \le \frac1 2\deh^{-\frac{13}{20}}-  O(\de_\HH^{-\frac{3}{5}}),
    \eeaa     
    i.e. it suffices to choose, for $\deh>0$ small enough, 
    \beaa
    \La = \frac1 4\deh^{-\frac{13}{20}},
    \eeaa
    to deduce the inequality,
    \beaa
            \QQ(X+2\La T, e_3)&\ge & \frac 1 4  \ep_\HH\QQ_{33}+\frac{1}{2}\Lambda\QQ_{34}. 
    \eeaa
    
    Next, in the region $0\le \Up \le \de_\HH^{\frac{1}{10}}$, the sign of $\Up$ is more favorable and  we have
    \beaa
        \QQ(X+2\La T, e_3)&\ge & \frac{1}{2}\ep_\HH\left( \QQ_{33}+\QQ_{34}\right) 
        +\left(\La -\frac 1 2 |f_{\widehat{\de}}|\right) \QQ_{34}+|\Up| \left(\La +\frac 1 2 |f_{\widehat{\de}}|\right)\QQ_{33}\\
        &\ge & \left(\frac{1}{2}\ep_\HH +|\Up| \left(\La +\frac 1 2 |f_{\widehat{\de}}|\right)\right)\QQ_{33} +\left(\frac{1}{2}\ep_\HH +\La -\frac 1 2 |f_{\widehat{\de}}|\right)\QQ_{34}.
      \eeaa
    In particular, we simply need $\La \gg \widehat{\de}^{-1}$, which is in particular satisfied by \eqref{eq:choiceofLambdaforpositivityofboundaryterm},   to deduce the same inequality,
    \beaa
                \QQ(X+2\La T, e_3)&\ge & \frac 1 4  \ep_\HH\QQ_{33}+\frac{1}{2}\Lambda\QQ_{34}. 
    \eeaa
    
    In the region   $\de_\HH^{\frac{1}{10}}\le \Up\le  \frac 1 3 $,  where $f_{\widehat{\de}}\le 0$, and using the fact that $|f_{\widehat{\de}}|$  is of size $O((\widehat{\de})^{-1})$
      \beaa
    \QQ(X+2\La T, e_3)
 &=&   \ep_\HH \QQ(Y_\HH, e_3)+ \La \left(\QQ_{34}+ \Up \QQ_{33}\right)
    +\frac 1 2 f_{\widehat{\de}}\left( \QQ_{34}-\Up \QQ_{33} \right)\\
    &\ge & \La \left(\QQ_{34}+ \de_\HH^{\frac{1}{10}}  \QQ_{33}\right)- O(\widehat{\de}^{-1})\QQ_{34}.
    \eeaa
Hence, for the choice \eqref{eq:choiceofLambdaforpositivityofboundaryterm}, and in view of \eqref{eq:restrictionson-de...}, we infer
\beaa
  \QQ(X+2\La T, e_3)\ge  \de_\HH^{-\frac{1}{2}} \left( \QQ_{33}+\QQ_{34}\right).
\eeaa
    Finally, for $r\ge 3 m$ where  we  have $0\le f_{\widehat{\de}}\les 1$, $\frac{1}{3}\leq \Up\leq 1$ and $Y_\HH=0$,
     \beaa
    \QQ(X+2\La T, e_3) &=&  \La \left(\QQ_{34}+ \Up \QQ_{33}\right)
    +\frac 1 2 f_{\widehat{\de}}\left( \QQ_{34}-\Up \QQ_{33} \right)\\
    &\geq&    \La \left(\QQ_{34}+\frac{1}{3} \QQ_{33}\right)- O(1)\QQ_{33} 
    \eeaa
    and hence,  \eqref{eq:choiceofLambdaforpositivityofboundaryterm} implies 
     \beaa
    \QQ(X+2\La T, e_3)     &\geq&    \frac{1}{4}\La \left(\QQ_{34}+ \QQ_{33}\right)
    \eeaa 
    as desired. The proof for $\QQ(X+\La T, e_4)$ is    similar.  
    \end{proof}

    
  \subsection{First  Morawetz Estimate}\lab{sec:firstmorawetzestimate}
      
      
 We are now ready to state our  first     Morawetz  estimate which is  simply obtained  by integrating  the    pointwise inequality
             in Proposition \ref{prop:pre-mor4} on our domain $\MM=\Mint \cup \Mext$  described at the beginning    of the section, with $X$ replaced by $X+\La T$ for $\La>0$ sufficiently large.  In view of the choice of $\tau$, note that we have
             \bea
             \label{definition:normal-sitau}
             N_\Si=ae_3+ b e_4, \qquad 0\le a,\, b\le 1,\qquad a+b\geq 1, 
             \eea
     with 
     \beaa
     b=0, \, a= 1\textrm{ on }\Mint, \quad a=0,\, b=1\textrm{ on }\MM_{r\ge 4m_0}, \quad a, \,b\geq \frac{1}{4}\textrm{ on }\Mtrap.      
     \eeaa

    We recall the following  quantities for $\Psi$ in  regions    $\MM(\tau_1, \tau_2)\subset \MM$    in  the past of $\Si(\tau_2)$ and future of $\Si(\tau_1)$.
    \begin{enumerate}
    \item  Morawetz bulk quantity
    \beaa
\Mor[\Psi](\tau_1, \tau_2)=\int_{\MM(\tau_1, \tau_2)} \frac{m^2}{r^3}  |\Rbrev \psi|^2 +\frac{m}{r^4} |\Psi|^2  +\left(1-\frac{3m}{r}\right)^2\frac{1}{r}\left(|\nabb\psi|^2 +\frac{m^2}{r^2}|\Tbrev\psi|^2 \right).
\eeaa
\item Basic     energy quantity
\beaa
E[\Psi](\tau)&=& \int_{\Si(\tau)}\bigg( \frac 1 2  (N_\Si, e_3)^2  \,    |e_4 \Psi|^2  +\frac 1 2 (N_\Si, e_4 )^2\,  |e_3\Psi|^2 +|\nabb\Psi|^2 + r^{-2}|\Psi|^2 \bigg).
\eeaa
\item Flux   through $\AA$ and $\Sigma_*$
\beaa
F[\Psi](\tau_1, \tau_2) &=&  \int_{\AA(\tau_1, \tau_2)}\left( \deh^{-1}   |e_4 \Psi|^2  + \deh  |e_3\Psi|^2 + |\nabb \Psi|^2 + r^{-2} |\Psi|^2\right)\\
&&  +\int_{\Sigma_*(\tau_1, \tau_2)}\big(  |e_4 \Psi|^2  + |e_3\Psi|^2 + |\nabb \Psi|^2 + r^{-2} |\Psi|^2\big),
  \eeaa
  with $\AA(\tau_1, \tau_2)=\AA\cap \MM(\tau_1, \tau_2) $ and $\Sigma_*(\tau_1, \tau_2)=\Sigma_*\cap \MM(\tau_1, \tau_2) $.
    \end{enumerate}
    
    The following theorem is our   first     Morawetz  estimate.       
        \begin{theorem}
        \label{thm:Morawetz1}
        Consider  the equation \eqref{eq:tensor-waveqf},   i.e. $\squared\Psi= V\Psi+  \NN$,  with $V=  -\ka\kab $ and a 
        domain $\MM(\tau_1, \tau_2)\subset \MM$. Then, we have
        \bea
        \label{eq:thm-Morawetz1}
        \bsplit
        E[\Psi](\tau_2)+  \Mor[\Psi](\tau_1, \tau_2)  + F[\Psi](\tau_1, \tau_2)      &\les   \left(   E[\Psi](\tau_1)+ J[N, \Psi] (\tau_1, \tau_2) +\err_\ep(\tau_1, \tau_2)[\Psi]\right), \\
  J[N, \Psi](\tau_1,\tau_2):&=\int_{\MM(\tau_1,\tau_2)} ( |\Rbrev\Psi|   +|\Tbrev \Psi| +r^{-1} |\Psi| ) |N|,  \\
            \err_\ep[\Psi](\tau_1,\tau_2) &=  \int_{\MM(\tau_1, \tau_2) }    \EE_\ep[\Psi],&
           \end{split}
        \eea
   where,  
   \beaa
    \EE_\ep[\Psi]&\les & \ep   u_{trap}^{-1-\dec}   \left[ r^{-2} |e_3\Psi|^2 + r^{-1} (|e_4 \Psi|^2+|\nabb\Psi|^2+r^{-2}  |\Psi|^2+  |e_3\Psi| \left(  |e_4 \Psi|+|\nabb\Psi|  \right)     )
     \right]. 
\eeaa
        \end{theorem}
         
    \begin{proof}
    Recall that, see \eqref{eq:modified-div}
     \beaa
 \EE[X, w, M](\Psi)&:=& \D^\mu  \PP_\mu[X, w, M] -   \left(X( \Psi )+\frac 1 2   w \Psi\right)\c \NN[\Psi]   
  \eeaa
  where,
  \beaa
  \PP_\mu &=&\PP_\mu[X, w, M]=\QQ_{\mu\nu} X^\nu +\frac 1 2  w \Psi \Db_\mu \Psi -\frac 1 4|\Psi|^2   \pr_\mu w +\frac 1 4 |\Psi|^2 M_\mu
  \eeaa
  with  triplet,
  \beaa
   (X, w, M):=  (f _{\widehat{\de}}R,  w_{\widehat{\de}}, 2 h R  )  +\ep_\HH( Y_\HH, 0, 0)  + (0, 0, 2 h_2 \Rbrev)  
  \eeaa
   given   in   Proposition \ref{prop:pre-mor4}.
   Replacing $X$ by $\check{X}=  X+\La T$  in the calculation above we deduce,
    \beaa
  \check{\PP}_\mu &=&\PP_\mu[\check{X},  w, M]=\QQ_{\mu\nu} \check{X} ^\nu +\frac 1 2  w \Psi \Db_\mu \Psi -\frac 1 4|\Psi|^2   \pr_\mu w +\frac 1 4 |\Psi|^2 M_\mu.
  \eeaa

  By the divergence theorem we have,
    \bea
    \label{eq:boundariesMM}
    \bsplit
             & \int_{\AA} \check{\PP} \c N_\AA + \int_{\Si_2}\check{\PP}\c N_\Si +\int_{\MM(\tau_1, \tau_2)} \EE        +\int_{\Sigma_*} \check{\PP} \c N_{\Si_*}  = \int_{\Si_1}\check{\PP}\c N_\Si\\
              -& \int_{\MM(\tau_1, \tau_2)} ( \check{X}(\Psi) +\frac 1 2  w \Psi)  N[ \Psi ]
              \end{split}
              \eea
              where      $\EE=  \EE[ \check{X},  w,  M] (\Psi)$. 
              Now,
              \beaa
               \EE[ \check{X},  w,  M] (\Psi) &=&   \EE[ X,  w,  M] (\Psi)+\frac 1 2 \La       \QQ \c  \piT -\frac{1}{2}T(V)|\Psi|^2.
              \eeaa

              According to Lemma \ref{lemma:componentspiR} $T(V)= O( \ep) r^{-3} u_{trap}^{-1-\dec}$, and all components of $\piT$ are $O(\ep r^{-1} u_{trap}^{-1-\dec})$ except for
              $\piT_{44}$ which is $O(\ep r^{-2} u_{trap}^{-1-\dec})$.   We easily  deduce,
              \beaa
              \La|\QQ\c \piT|+|T(V)||\Psi|^2 &\les &     \La\EE_\ep. 
              \eeaa
               Thus in view of to  Proposition \ref{prop:pre-mor5},   we have\footnote{Recall from \eqref{eq:choiceofLambdaforpositivityofboundaryterm}                           that we have $\La=\frac{1}{4}\deh^{-\frac{13}{20}}\ll \deh^{-1}$.}, 
          \beaa
          \bsplit
     \int_{\MM(\tau_1, \tau_2)}  \EE&\ge  \deh^{\frac{1}{2}} \int_{\MM(\tau_1, \tau_2)} \left[   \frac{m^2}{r^3}  | \Rbrev \Psi|^2 +  r^{-1}\left(1-\frac{3m}{r}\right)^2\left( |\nabb \Psi|^2 +\frac{m^2}{r^2} |\Tbrev \Psi|^2\right)+ \frac{m}{r^4}|\Psi|^2\right]  \\
     &- O\big( \deh^{-1}\big)          \int_{\MM(\tau_1, \tau_2)}   \EE_\ep
     \end{split}
       \eeaa  
       i.e.,
       \bea
       \label{equation:LBforEE}
        \int_{\MM(\tau_1, \tau_2)}  \EE&\ge \deh^{\frac{1}{2}}  \Mor[\Psi]( \tau_1, \tau_2)-O\big(  \deh^{-1}   \big)   \err_\ep(\tau_1,\tau_2).
       \eea

       We now analyze the boundary terms in \eqref{eq:boundariesMM}.

       
           \subsubsection{Boundary term  along  $\AA$}    
           
           
           Along  the spacelike hypersurface $\AA$, i.e. $r=2m_0(1-\deh)$, the unit normal $N_{\AA}$ is given by 
           \beaa
           N_{\AA} &=& \frac{1}{2\sqrt{\frac{e_4(r)}{e_3(r)}}}\left(e_4+\frac{e_4(r)}{e_3(r)}e_3\right)\\
           &=& \frac{1}{2\sqrt{\deh+O(\ep)}}\Big(e_4+(\deh+O(\ep))e_3\Big),
           \eeaa
           and we have $h, h_2=0$ as well as $w=-\de_1 w_1$ where $\de_1>0$ is a small constant and $w_1$ is given by \eqref{eq:theformulaforwasasumof2terms:bis}
 \beaa          
 w_1(r,m)=r^{-1} \frac{m^2}{r^2}     \Up  \left(1-\frac{3m}{r}\right)^2.
\eeaa
 In particular, we have on $\AA$ in view of the formula for $w_1$ and for $N_\AA$
 \beaa
 |w_1|\les\deh, \qquad |N_\AA(w_1)|\les\sqrt{\deh}.
 \eeaa
                       Hence,
       \beaa
      \PP\c   N_\AA&=& \QQ(X+\La T, N_\AA) -\frac{\de_1}{2}w_1\Psi N_{\AA}(\Psi) +\frac{\de_1}{4}|\Psi|^2  N_{\AA}(w_1) \\
      &=& \frac{2}{\sqrt{\deh+O(\ep)}}\QQ(X+\La T, e_4)+2\sqrt{\deh+O(\ep)}\QQ(X+\La T, e_3) \\
&&        -O(\sqrt{\deh})\Psi e_4(\Psi)  -O(\deh^{\frac{3}{2}})\Psi e_3(\Psi) -O(\sqrt{\deh})|\Psi|^2. 
            \eeaa
      Thus, in view of  Proposition  \ref{prop:boundar-Mor},   we infer
     \beaa
    \PP\c   N_\AA&\geq&  \frac{2}{\sqrt{\deh+O(\ep)}}\left(\frac 1 4  \ep_\HH\QQ_{34}+\frac{1}{2}\Lambda\QQ_{44}\right)  +2\sqrt{\deh+O(\ep)}\left(\frac 1 4  \ep_\HH\QQ_{33}+\frac{1}{2}\Lambda\QQ_{34}\right)\\
&&        -O(\sqrt{\deh})\Psi e_4(\Psi)  -O(\deh^{\frac{3}{2}})\Psi e_3(\Psi) -O(\sqrt{\deh})|\Psi|^2
      \eeaa
Using in particular \eqref{eq:restrictionson-de...} and \eqref{eq:choiceofLambdaforpositivityofboundaryterm}, we deduce 
     \beaa
    \PP\c   N_\AA&\geq&  \frac{2}{\sqrt{\deh+O(\ep)}}\left(\frac 1 4  \deh^{\frac{7}{20}}\Big(|\nabb\Psi|^2+O(\deh)|\Psi|^2\Big)+\frac{1}{8}\deh^{-\frac{13}{20}}|e_4\Psi|^2\right)\\
    &&  +2\sqrt{\deh+O(\ep)}\left(\frac 1 4  \deh^{\frac{7}{20}}|e_3\Psi|^2+\frac{1}{8}\deh^{-\frac{13}{20}}\Big(|\nabb\Psi|^2+O(\deh)|\Psi|^2\Big)\right)\\
&&        -O(\sqrt{\deh})\Psi e_4(\Psi)  -O(\deh^{\frac{3}{2}})\Psi e_3(\Psi) -O(\sqrt{\deh})|\Psi|^2\\
&\geq& \frac{1}{2}\deh^{-\frac{3}{20}}|\nabb\Psi|^2 +  \frac{1}{8}\deh^{-\frac{23}{20}}|e_4\Psi|^2 +  \frac{1}{4}\deh^{\frac{17}{20}}|e_3\Psi|^2 \\
&&  -O(\sqrt{\deh})\Psi e_4(\Psi)  -O(\deh^{\frac{3}{2}})\Psi e_3(\Psi) -O(\sqrt{\deh})|\Psi|^2.
      \eeaa 
 Recalling  the Poincar\'e inequality \eqref{corr:Poincare-Psi},
     \beaa
\int_S |\nabb \Psi|^2  &\ge & 2r^{-2}\big(1- O(\ep)\big)  \int_S  \Psi ^2 da_S,
\eeaa
 we deduce, in this region,
\beaa
 \int_{\AA(\tau_1,\tau_2)}   \PP\c N_\AA \ge \frac 1 8 \int_{\AA(\tau_1, \tau_2)}\Big( \deh^{-1}   |e_4 \Psi|^2  + \deh  |e_3\Psi|^2 + |\nabb \Psi|^2 + r^{-2} |\Psi|^2\Big)
\eeaa
as desired in view of the definition of the flux along $\AA$.

      
      \subsubsection{Boundary terms along $\Si(\tau_1), \Si(\tau_2)$}
      
          
        Along   a hypersurface $\Si(\tau)$ with timelike  unit  future normal  $N_{\Sigma(\tau)}= a e_3 + b e_4 $,    we have,       
      \beaa
      \PP\c N_\Si&=&\QQ( X+\La T, N_\Si)+\frac 1 2 w \Psi N_\Si (\Psi) -\frac 1  4  N_\Si (w) |\Psi|^2 + \frac 1 2        N_\Si \c(  h R+ h_2 \Rbrev )        |\Psi|^2
      \eeaa
      and
       \beaa
E[\Psi](\tau)&=& \int_{\Si(\tau)}\Big( 2 b^2  \,    |e_4 \Psi|^2  +2 a^2\,  |e_3\Psi|^2 +|\nabb\Psi|^2 + r^{-2}|\Psi|^2 \Big).
\eeaa 
      
      \begin{enumerate}
        \item       In the  region     $r\geq 2m_0(1-\deh)$,    $\Up \le \de_\HH^{\frac{1}{10}}$        we  have   
          $h=0$,  $h_2=O(\widehat{\de})$ and $N_\Sigma=e_3$ (i.e. $a=1$, $b=0$).  Also, we have  $w=-\de_1 w_1$,    where $\de_1>0$ is a small constant and $w_1$ is given by \eqref{eq:theformulaforwasasumof2terms:bis}
 \beaa          
 w_1(r,m)=r^{-1} \frac{m^2}{r^2}     \Up  \left(1-\frac{3m}{r}\right)^2.
\eeaa
 In particular, we have in the region of interest,  in view of the formula for $w_1$ and for $N_\Sigma$
 \beaa
 |w_1|\les\deh^{\frac{1}{10}}, \qquad |N_\Sigma(w_1)|=|e_3(w_1)|\les 1.
 \eeaa
We infer
 \beaa
        \PP\c N_\Si&=&\QQ( X+\La T, e_3) -\frac{\de_1}{2} w_1 \Psi e_3(\Psi) +\frac{\de_1}{4}  e_3(w_1) |\Psi|^2 + \frac 1 2 h_2    e_3\c \Rbrev         |\Psi|^2 \\
 &=& \QQ( X+\La T, e_3) -O(\deh^{\frac{1}{10}}) w_1 \Psi e_3(\Psi) -O(1) |\Psi|^2.       
\eeaa                  
  where we used the fact that $\Rbrev=\frac{1}{2}(e_4-e_3)$ in the region of interest. Thus,   according to Proposition \ref{prop:boundar-Mor},         
      \beaa
        \PP\c N_\Si&\geq& \frac 1 4  \ep_\HH\QQ_{33}+\frac{1}{2}\Lambda\QQ_{34}  -O(\deh^{\frac{1}{10}}) |\Psi| |e_3(\Psi)| -O(1) |\Psi|^2.
      \eeaa 
      Using in particular \eqref{eq:restrictionson-de...} and \eqref{eq:choiceofLambdaforpositivityofboundaryterm}, we deduce 
      \beaa
        \PP\c N_\Si&\geq& \frac 1 4  \deh^{\frac{7}{20}}|e_3\Psi|^2+\frac{1}{8}\deh^{-\frac{13}{20}}(|\nabb\Psi|^2+O(\ep)|\Psi|^2)  -O(\deh^{\frac{1}{10}}) |\Psi| |e_3\Psi| -O(1) |\Psi|^2
      \eeaa
      Together with   the Poincar\'e inequality \eqref{corr:Poincare-Psi}, we deduce
\beaa
 \int_{\Si_{r\geq 2m_0(1-\deh), \,\,\Up \le \de_\HH^{\frac{1}{10}}}(\tau)}   \PP\c N_\Si &\ge& \frac 1 8  \deh^{\frac{7}{20}}\int_{\Si_{r\geq 2m_0(1-\deh), \,\,\Up \le \de_\HH^{\frac{1}{10}}}(\tau)}\Big(   |e_3\Psi|^2 +|\nabb\Psi|^2 + r^{-2}|\Psi|^2 \Big)\\
 &\geq&  \frac 1 8  \deh^{\frac{7}{20}}   E_{r\geq 2m_0(1-\deh), \,\,\Up \le \de_\HH^{\frac{1}{10}}}[\Psi](\tau).   
\eeaa

\item  In the region  $\Up\geq \deh^{\frac{1}{10}}$, we have $w = O(r^{-1})$, $N_\Si(w)=O(r^{-2})$, $h=O(r^{-4})$ and $h_2=O(r^{-4})$. We infer   
  \beaa
         \PP\c N_{\Si}  =  a\QQ( X+\La T, e_3)+b\QQ( X+\La T, e_4) -O(r^{-1})|\Psi|(a|e_3\Psi|+b|e_4\Psi|) - O(r^{-2}) |\Psi|^2. 
         \eeaa
        Thus, according to Proposition \ref{prop:boundar-Mor}, 
     \beaa
         \PP\c N_\Si &\ge & \deh^{-\frac{1}{2}}        \left(   a  \QQ_{33} + b\QQ_{44}+ (a+b) \QQ_{34} \right) -O(1)(a^2|e_3\Psi|^2+b^2|e_4\Psi|^2) - O(r^{-2}) |\Psi|^2\\
         &=& \deh^{-\frac{1}{2}}       \left( a | e_3\Psi|^2 + b |e_4 \Psi|^3 + (a+b)\left( |\nabb\Psi|^2 + \frac{4\Up}{r^2} |\Psi|^2\right) \right)  \\
    &&      -O(1)\Big(a^2|e_3\Psi|^2+b^2|e_4\Psi|^2+r^{-2} |\Psi|^2\Big)\\
         &\ge&    \deh^{-\frac{1}{2}}       \left( a | e_3\Psi|^2 + b |e_4 \Psi|^3 + (a+b)\left( |\nabb\Psi|^2 + \frac{4\deh^{\frac{1}{10}}}{r^2} |\Psi|^2\right) \right)  \\
    &&      -O(1)\Big(a^2|e_3\Psi|^2+b^2|e_4\Psi|^2+r^{-2} |\Psi|^2\Big).
      \eeaa
        Hence,   for $\deh>0$ sufficiently small, and since $a^2\leq a$, $b^2\leq b$ and $a+b\geq 1$, we infer in this region
     \beaa
         \PP\c N_\Si &\ge & \deh^{-\frac{1}{5}}\int_{\Si_{\Up\geq \deh^{\frac{1}{10}}}(\tau)}\Big( 2 b^2  \,    |e_4 \Psi|^2  +2 a^2\,  |e_3\Psi|^2 +|\nabb\Psi|^2 + r^{-2}|\Psi|^2 \Big)\\
         &=& \deh^{-\frac{1}{5}} E[\Psi]_{\Up\geq \deh^{\frac{1}{10}}}(\tau)  
     \eeaa    
    \end{enumerate}

In view of the above estimates in $r\geq 2m_0(1-\deh)$,    $\Up \le \de_\HH^{\frac{1}{10}}$ and in $\Up\geq \deh^{\frac{1}{10}}$,  we deduce, everywhere,
\bea
\int_{\Si(\tau)}   \PP\c N_\Si \ge  \frac{1}{8} \deh^{\frac{7}{20}}   E[\Psi](\tau). 
\eea


  \subsubsection{Boundary terms along $\Sigma_*$}
  
  
  On $\Si_*$, we have
  \beaa
  N_{\Si_*} &=& T+O\left(\ep+\frac{m}{r}\right)(e_3+e_4),
  \eeaa
  $w = O(r^{-1})$, $N_{\Si_*}(w)=O(\ep r^{-2})$, $h=O(r^{-4})$ and $h_2=0$.
    
    Proceeding as before, we have along $\Sigma_*$,
   \beaa
         \PP\c N_{\Si} & =&  \left(\frac{1}{2}+ O\left(\ep+\frac{m}{r}\right)\right)\QQ( X+\La T, e_3)+\left(\frac{1}{2}+ O\left(\ep+\frac{m}{r}\right)\right)\QQ( X+\La T, e_4) \\
         &&-O(r^{-1})|\Psi|(|e_3\Psi|+|e_4\Psi|) - O(r^{-2}) |\Psi|^2\\
         &\geq& \frac{1}{4}\QQ( X+\La T, e_3)+\frac{1}{4}\QQ( X+\La T, e_4) -O\Big(|e_3\Psi|^2+|e_4\Psi|^2 + r^{-2} |\Psi|^2\Big).
         \eeaa
   Thus, according to Proposition \ref{prop:boundar-Mor}, we have
  \beaa
         \PP\c N_{\Si} &\geq& \frac{1}{16}\La        \left( \QQ_{33} + \QQ_{44}+ 2 \QQ_{34} \right)  -O\Big(|e_3\Psi|^2+|e_4\Psi|^2 + r^{-2} |\Psi|^2\Big)\\
         &=& \frac{1}{16}\La        \left(| e_3\Psi|^2 + | e_4\Psi|^2+ 2\left( |\nabb\Psi|^2 + \frac{4\Up}{r^2} |\Psi|^2\right)  \right)  -O\Big(|e_3\Psi|^2+|e_4\Psi|^2 + r^{-2} |\Psi|^2\Big)\\
         &\geq& \frac{1}{64}\deh^{-\frac{13}{20}}        \Big(| e_3\Psi|^2 + | e_4\Psi|^2+ |\nabb\Psi|^2 + r^{-2} |\Psi|^2 \Big)  -O\Big(|e_3\Psi|^2+|e_4\Psi|^2 + r^{-2} |\Psi|^2\Big)\\
         &\geq& \deh^{-\frac{1}{2}}        \Big(| e_3\Psi|^2 + | e_4\Psi|^2+ |\nabb\Psi|^2 + r^{-2} |\Psi|^2 \Big)
         \eeaa
and hence
\bea
\int_{\Si_*(\tau_1,\tau_2)}   \PP\c N_{\Si_*} \ge  \deh^{-\frac{1}{2}}  \int_{\Si_*(\tau_1,\tau_2)} \Big(| e_3\Psi|^2 + | e_4\Psi|^2+ |\nabb\Psi|^2 + r^{-2} |\Psi|^2 \Big).
\eea

      
      \subsubsection{The  inhomogeneous  term  $ \int_{\MM(\tau_1, \tau_2)} ( \check{X}(\Psi) +\frac 1 2  w \Psi)  N[ \Psi ]$}

      
     Recall that,
       $\check{X}=  X+\La T         =f_{\widehat{\de}} R +Y_\HH +\La T$.
 We easily check, recalling  the   properties   of $f_{\widehat{\de}},w$ and $\La$ and the definition of $J[N, \Psi]$,
 \bea\lab{eq:usefultorecalllatertreatmentofNterminMorawetzestimate}
\nn \bigg|     \int_{\MM(\tau_1, \tau_2)} \left( \check{X}(\Psi) +\frac 1 2  w \Psi\right)  N[ \Psi ]\bigg|            & \le & \deh^{-\frac{3}{4}}\int_{\MM(\tau_1,\tau_2)}\left(  |\Rbrev \Psi|+|\Tbrev\Psi| + r^{-2} |\Psi|^2\right)|N(\Psi)|\\
 &=&\deh^{-\frac{3}{4}} J[N, \Psi](\tau_1,\tau_2).
 \eea

      Going back  to \eqref{eq:boundariesMM} we deduce, 
      \beaa
      E[\Psi](\tau_2)+\int_{\MM(\tau_1, \tau_2)} \EE \, \,+ F[\Psi](\tau_1, \tau_2)&\le& \deh^{-\frac{7}{20}}\left(E[\Psi](\tau_1)+J[N, \Psi](\tau_1, \tau_2)\right).
      \eeaa
           In view of \eqref{equation:LBforEE} we obtain,
       \beaa
      E[\Psi](\tau_2)+ \Mor[\Psi](\tau_1, \tau_2) + F[\Psi](\tau_1, \tau_2)&\leq&  \deh^{-1}\big(E[\Psi](\tau_1)+J[N, \Psi](\tau_1, \tau_2)\big)\\
      && +O\Big(  \deh^{-\frac{3}{2}}  \Big)    \err_\ep (\tau_1,\tau_2).
      \eeaa
     This concludes the proof of Theorem \ref{thm:Morawetz1}.
     \end{proof}


\subsection{Analysis of the error term $\EE_\ep$}\lab{sec:analysisoftheerrortermEEep}


Recall that   $  \err_\ep(\tau_1,\tau_2) =  \int_{\MM(\tau_1, \tau_2) }    \EE_\ep$
   where,  
   \beaa
    \EE_\ep&\les & \ep   u_{trap}^{-1-\dec}   \Big[ r^{-2} |e_3\Psi|^2 + r^{-1} \big(|e_4 \Psi|^2+|\nabb\Psi|^2+r^{-2}  |\Psi|^2+  |e_3\Psi| \left(  |e_4 \Psi|+|\nabb\Psi|  \right)     \big) \Big]. 
\eeaa

\begin{itemize}
\item In the trapping region  $\MM_{trap}$, i.e. $\frac{5m}{2}\le r\le  \frac{7m}{2}$,  where $u_{trap}=1+\tau$ and $\Si(\tau)$ is strictly spacelike, we have   
\beaa
 \int_{\Si_{trap}(\tau)}\EE_\ep&\les & \ep   \tau_{trap}^{-1-\dec} \int_{\Si_{trap}(\tau)}\left( |e_3\Psi|^2+|e_4\Psi|^2+|\nabb \Psi|^2 + |\Psi|^2\right)\\
 &\les & \ep   \tau_{trap}^{-1-\dec} E[\Psi](\tau).
\eeaa
Thus,
\beaa
\int_{\MM_{trap}(\tau_1,\tau_2)} \EE_\ep&  \les  &\ep \int_{\tau_1}^{\tau_2}\tau_{trap}^{-1-\dec} E[\Psi](\tau)\\
&\les&  \ep \left(\int_{\tau_1}^{\tau_2}    (1+\tau)^{-1-\de}  \right)   \sup_{\tau\in[\tau_1,\tau_2]}\EE[\Psi](\tau) \\
&\les&  \ep   \sup_{\tau\in[\tau_1,\tau_2]}\EE[\Psi](\tau) 
\eeaa
 and therefore, for small $\ep>0$,  the integral  $\int_{\MM_{trap}(\tau_1,\tau_2)} \EE_\ep$ can be absorbed on the left hand side of \eqref{eq:thm-Morawetz1}.

 \item  In the non trapping region  $\MM_{\ntrap}$ we write, with a fixed $\de>0$,
 \beaa
 \EE_\ep& \les & \ep r^{-1-\de}  |e_3\Psi|^2 +  r^{-1+\de} \left( |e_4\Psi|^2 +|\nabb\Psi|^2 +r^{-2}|\Psi|^2\right).
 \eeaa
 Hence,
 \beaa
 \int_{\MM_{\ntrap}(\tau_1,\tau_2)} \EE_\ep&  \les&\ep       \int_{\MM_{r\geq 4m_0}(\tau_1,\tau_2)}  r^{-1-\de}  |e_3\Psi|^2 \\
 &+&\ep \int_{\MM_{r\geq 4m_0}(\tau_1,\tau_2)}   r^{-1+\de} \left( |e_4\Psi|^2 +|\nabb\Psi|^2 +r^{-2}|\Psi|^2\right)\\
 &+& \ep\int_{\Mntrap_{r\leq 4m_0}(\tau_1, \tau_2)}\left( |e_3\Psi|^2+|e_4\Psi|^2+|\nabb \Psi|^2 + |\Psi|^2\right).
 \eeaa
Note that for  $\ep>0$ sufficiently small, the last integral, on $\Mntrap_{r\leq 4m_0}$,  can be absorbed by the  left hand side of 
 \eqref{eq:thm-Morawetz1}.
  \end{itemize}
 
 As a consequence  we   deduce the following.
 \begin{corollary}
 \label{corr:Morawetz1}
 The statement of Theorem \ref{thm:Morawetz1} remains true  if we replace $\err_\ep$  in the statement of the theorem with
 \beaa
 \err_{\ep} &=& \int_{\MM_{r\geq 4m_0}(\tau_1, \tau_2)}\EE_\ep,\\[1mm]
 \EE_\ep &\les& \ep  r^{-1-\de}  |e_3\Psi|^2+ \ep  r^{-1+\de} \left( |e_4\Psi|^2 +|\nabb\Psi|^2 +|r^{-2}|\Psi|^2\right),
 \eeaa
  for  a fixed  $\de>0$.
 \end{corollary}
 
\begin{remark}
Note  that the error terms $\err_\ep$ cannot yet  be  absorbed to the let hand side of    \eqref{eq:thm-Morawetz1}.  In fact we need
 additional estimates.    The Morawetz bulk quantity  \eqref{def:Mor-bulk},
  \beaa
\Mor[\Psi](\tau_1, \tau_2):=\int_{\MM(\tau_1, \tau_2)} \frac{m^2}{r^3}  |\Rbrev \Psi|^2 +\frac{m}{r^4} |\Psi|^2  +\left(1-\frac{3m}{r}\right)^2\frac{1}{r}\left(|\nabb\Psi|^2 +\frac{m^2}{r^2}|\Tbrev\Psi|^2 \right)\nn\\
\eeaa
is     quite weak  for  $r$ large        with regard to the terms $ |\Rbrev \Psi|^2$ and       $     |\Tbrev \Psi|^2 $,  
 while, using   the  Poincar\'e inequality,  $\Mor[\Psi]$ controls   the term  $\int_{\MM_{r\geq 4m_0}(\tau_1, \tau_2)} r^{-1}\left(|\nabb\Psi|^2 + r^{-2} |\Psi|^2\right)$.  In the next section we show how we  can     estimate $ \int_{\MM_{\ge R_0}     (\tau_1, \tau_2)}  r^{-1-\de}  |e_3\Psi|^2$
 by   $ \int_{\MM_{\ge R_0}     (\tau_1, \tau_2)}  r^{-1-\de}  |e_4\Psi|^2$   and then, we  provide estimates for the remaining terms. Note also that  the weight $r^{-1-\de}$ is optimal   in estimating $e_3\Psi$ in the   wave zone region.
\end{remark}


 \subsection{Proof of Theorem \ref{Thm:Morawetz-s}}\lab{sec:wheremorawetzinfinalformareactuallyproved} 


  We are now ready to prove Theorem  \ref{Thm:Morawetz-s}.    Note that it suffices to  improve   the previous  Morawetz estimate   of  Theorem \ref{thm:Morawetz1}   by  replacing  the quantity  $\Mor[\Psi](\tau_1, \tau_2)$  with 
\beaa
\Morr[\Psi](\tau_1, \tau_2):=\Mor[\Psi](\tau_1, \tau_2)+\int_{\MM_{far}(\tau_1, \tau_2)} r^{-1-\de}  |e_3(\Psi)|^2. 
\eeaa
 
     In view of the Morawetz  estimate  \eqref{eq:thm-Morawetz1} and corollary \ref{corr:Morawetz1} we have
   \bea
   \begin{split}
        E[\Psi](\tau_2)            +\Mor[\Psi](\tau_1,\tau_2)+F[\Psi](\tau_1,\tau_2)&\les     E[\Psi](\tau_2)  +J[N, \Psi](\tau_1,\tau_2) +\err_\ep(\tau_1, \tau_2), \\
         J[N, \Psi](\tau_1,\tau_2):&=\int_{\MM(\tau_1,\tau_2)} ( |\Rbrev\Psi|   +|\Tbrev \Psi| +r^{-1} |\Psi| ) |N|,    
          \end{split}
         \eea
         with error  term,
         \beaa
          \err_\ep &\les& \ep \int_{\MM_{\ge 4m_0}     (\tau_1, \tau_2)}  r^{-1-\de}  |e_3\Psi|^2 +  r^{-1+\de} \left( |e_4\Psi|^2 +|\nabb\Psi|^2 +r^{-2}|\Psi|^2\right).\nn
          \eeaa
         We divide   $J[N]=J[N, \Psi]$ as follows:
         \beaa
         J[N]=J[N]_{trap}+J[N]_{\ntrap}
         \eeaa
         where,
         \beaa
         J[N]_{trap}:&=&\int_{\MM_{trap}}  ( |\Rbrev\Psi|   +|\Tbrev \Psi| +r^{-1} |\Psi| ) |N|, \\
          J[N]_{\ntrap}:&=& \int_{\Mntrap}  ( |\Rbrev\Psi|   +|\Tbrev \Psi| +r^{-1} |\Psi| ) |N|.
         \eeaa
         
         For the trapping region, where  the hypersurfaces   $\Si(\tau)$   are strictly  spacelike, we write,
         \beaa
         J[N]_{trap}(\tau_1,\tau_2) &=& \int _{\tau_1}^{\tau_2}  d\tau \int_{\Si_{trap}(\tau)  }  ( |\Rbrev\Psi|   +|\Tbrev \Psi| +r^{-1} |\Psi| ) |N| 
         \\
         &\le & \int_{\tau_1}^{\tau_2}  E[\Psi](\tau)^{1/2}\left( \int_{\Si_{trap}(\tau)  }|N|^2\right)^{1/2}\\
         &\le& \sup_{\tau\in[\tau_1, \tau_2]}  E[\Psi] (\tau)^{1/2}  \int_{\tau_1}^{\tau_2}\left(  \int_{\Si_{trap}(\tau) } |N|^2\right)^{1/2}         \\
         &\les& \la  \sup_{\tau\in[\tau_1, \tau_2]}  E[\Psi] (\tau)+\la^{-1}   \bigg( \int_{\tau_1}^{\tau_2} \|N\|_{L^2(\Si_{trap}(\tau))}\bigg)^2.
         \eeaa
         Hence, for $\la>0$ sufficiently small,  we deduce,
           \beaa
        E[\Psi](\tau_2)            +\Mor[\Psi](\tau_1,\tau_2)+F[\Psi](\tau_1,\tau_2)&\les&     E[\Psi](\tau_2)    +\err_\ep(\tau_1, \tau_2) \\
        &+& J_{\ntrap} [N, \Psi](\tau_1,\tau_2) +     \bigg( \int_{\tau_1}^{\tau_2} \|N\|_{L^2(\Si_{trap}(\tau))}\bigg)^2.
         \eeaa
         On the other hand  we have,
         \beaa
            J[N]_{\ntrap}(\tau_1,\tau_2) &=&\int_{\MM_{\ntrap}(\tau_1,\tau_2)}  ( |\Rbrev\Psi|   +|\Tbrev \Psi| +r^{-1} |\Psi| ) |N| \\
            &\le & \la \int_{\MM_{\ntrap}}  r^{-1-\de}  ( |\Rbrev\Psi|^2   +|\Tbrev \Psi|^2 +r^{-2} |\Psi|^2 )+\la^{-1}  \int_{\MM_{\ntrap}}  r^{1+\de} |N|^2. 
         \eeaa
         The first integral on the right can be  divided further into  integrals  for   $r\le 4m_0$ and $r\ge 4m_0$. The first 
         integral can the  be easily absorbed   by  the term $Mor[\Psi](\tau_1,\tau_2)$, if $\la>0$ is sufficiently small.
         We are thus led to  the estimate,
         \beaa
           E[\Psi](\tau_2)            +\Mor[\Psi](\tau_1,\tau_2)+F[\Psi](\tau_1,\tau_2)&\les&     E[\Psi](\tau_2)    +\err_\ep(\tau_1, \tau_2)+\II_{\de}[N](\tau_1,\tau_2)\\
           &+&           \int_{\MM_{r\ge 4m_0} } r^{-1-\de}  ( | e_3\Psi|^2   +|e_4  \Psi|^2 +r^{-2} |\Psi|^2 )   
\eeaa
   where,   
         \beaa
\II_{\de}[N](\tau_1,\tau_2)  :&=& \int_{\MM_{\ntrap}(\tau_1,\tau_2)} r^{1+\de}  |N|^2 +\bigg( \int_{\tau_1}^{\tau_2} d\tau \|N\|_{L^2(\Si_{trap}(\tau))}\bigg)^2.
\eeaa   
 Recalling the definition of $\err_\ep$ in Corollary \ref{corr:Morawetz1}, we deduce,
      \bea
       E[\Psi](\tau_2)            +\Mor[\Psi](\tau_1,\tau_2)+F[\Psi](\tau_1,\tau_2)&\les     E[\Psi](\tau_2)    +\err_\ep(\tau_1, \tau_2)+\II_{\de}[N](\tau_1,\tau_2).\nn\\
              \label{equation:estmateMor2}
                    \eea
        To eliminate the term in $e_3\Psi$ from the  error term 
         we appeal  to  the following  proposition.
\begin{proposition}
\label{prop:improved-Morawetz}
Assume $\square \Psi=V\Psi+N$   and consider  the  vectorfield   $ X= f_{-\de}    T $ with  $f_{-\de}:= r^{-\de} $ for $r\ge 4m_0$ and  compactly supported in $r\ge \frac{7m_0}{2}  $. With the notation of   Proposition \ref{prop:summaryofallimportantpropertiesofenergymomentumtensorforwaveequationpsi},  let
\beaa
\PP_\mu[f_{-\de} T, 0,0] &=&f_{-\de} \QQ_{\a\mu} T^\mu, \qquad \\
 \EE [f_{-\de}T, 0,0] &=& \D^\mu  \PP_\mu[f_{-\de} T, 0,0]   -    f_{-\de}T( \Psi )  N.
\eeaa
Then, 
\begin{enumerate}
\item    We have, for $r\ge 4m_0$
\bea
\label{estim:Daf-Rodn--de}
\bsplit
\EE[f_{-\de}T, 0,0]  &=\frac{\Up^2}{4}   \de r^{-1-\de} |e_3\Psi|^2  - \frac 1 4   \de r^{-1-\de} |e_4\Psi|^2+O\left(\ep  r^{-1-\de}\, \big( |\D\Psi|^2  + r^{-2} |\Psi|^2\big)\right).     \nn\\ 
\end{split}
\eea

\item  We have,
\beaa
\PP[f_{-\de}T, 0,0]  \c e_4 &=& f_{-\de}\QQ(T, e_4)  \ge 0, \, \qquad 
\PP[f_{-\de}T, 0,0]  \c e_3 = f_{-\de} \QQ(T, e_3)\ge 0.
\eeaa
\end{enumerate}
\end{proposition}

We postponed the proof of Proof of Proposition \ref{prop:improved-Morawetz}  and continue the proof of Theorem  \ref{Thm:Morawetz-s}.
    By integration,   the      proposition provides a bound for\footnote{Note that $\Up^2\geq \frac{1}{4}$ in $r\geq 4m_0$.}
 \beaa
 \int_{\MM_{\ge 4m_0}(\tau_1,\tau_2)} r^{-1-\de}    |e_3\Psi|^2 
 \eeaa 
 in  terms of  \, $E[\Psi](\tau_1)    $,  the integrals  $  \int_{\MM_{\ge \frac{7m_0}{2}  }(\tau_1,\tau_2)} r^{-1-\de}  |e_4\Psi|^2           $ 
 and       $   \int_{\MM_{\ge \frac{7m_0}{2}  }(\tau_1,\tau_2)}   r^{-\de}T( \Psi )  N$,
 as well as   the error terms. The   second bulk integral  involving the inhomogeneous term $N$ can be 
  estimates exactly like before. Thus  combining the new estimate with  that  in the  corollary \ref{corr:Morawetz1} 
   we derive the desired estimate hence concluding the proof of Theorem  \ref{Thm:Morawetz-s}.

 
\subsubsection{Proof of Proposition \ref{prop:improved-Morawetz}}


       We consider vectorfields  of  the form $X= f(r) T$ with  $T=\frac 12 (\Up e_3+e_4) $.
       Recall, see Lemma \ref{lemma:componentspiR},  that     all  components of the deformation
        tensor     $\piT$  of   
        $T=\frac 1 2 \left(e_4+\Up e_3 \right)$    can be bounded  
             by  $O(\ep r^{-1})$.
             Since $ f=O(r^{-\de})$, we deduce,              
\beaa
\piX_{\a\b}=f\piT_{\a\b}+\D_\a f T_\b+\D_\b fT_\a=\D_\a f T_\b+\D_\b fT_\a+O(\ep r^{-1-\de}).
\eeaa
Also,   
\beaa
e_3(f)= f' e_3(r)=-f' +    O(\ep r^{-1-\de})  , \qquad  e_4(f)  =f' e_4(r)= \Up f' +     O(\ep r^{-1-\de}).
\eeaa
Thus, modulo  error terms of the form        $ O(\ep)  r^{-1-\de} \big(  |e_3\Psi|^2    +|e_4\Psi|^2+ |\nabb\Psi|^2 + r^{-2} |\Psi|^2\big)$, we have
\beaa
\QQ\c \piX&=& 2 \QQ^{\a\b} T_\a \D_\b f= 2\left( \QQ^{ 3 \b} T_\b  e_3  f+  \QQ^{ 4 \b} T_\b  e_4  f\right)=-\QQ(e_4, T)   e_3  f -\QQ(e_3, T) e_4 f\\
&=&\frac 1 2 \QQ(e_4, e_4+\Up e_3)f' -\frac 1 2 f'\Up \QQ(e_3, e_4+\Up e_3)\\
&=&\frac 12 f' \left( |e_4\Psi|^2  -   \Up^2 |e_3\Psi|^2 \right).
\eeaa

We now apply Proposition \ref{prop:summaryofallimportantpropertiesofenergymomentumtensorforwaveequationpsi}, as well as \eqref{eq:modified-div} 
\eqref{le:divergPP-gen-EE}, with      $ X= f_{-\de}(r) T$, $w=0$, $M=0$  so that 
\beaa
\PP_\mu[f_{-\de} T, 0,0] =f_{-\de} \QQ_{\a\mu} T^\mu, \qquad  \EE [f_{-\de}T, 0,0] := \D^\mu  \PP_\mu[f_{-\de} T, 0,0]   -    f_{-\de} T( \Psi )  N
\eeaa
 and
 \beaa
 \EE[f_{-\de}T, 0,0]&=&\frac 1 2 \QQ  \c\piX-\frac{1}{2}f_{-\de}T(V)|\Psi|^2 \\
 &=&\frac 1 4 f_{-\de}'(r)\left( |e_4\Psi|^2  -   \Up^2 |e_3\Psi|^2 \right) +O\left(\ep  r^{-1-\de}   \big( |\D\Psi|^2  + r^{-2} |\Psi|^2\big)\right) 
 \eeaa
 with $|\D\Psi|^2= |e_3\Psi|^2    +|e_4\Psi|^2+ |\nabb\Psi|^2 + r^{-2} |\Psi|^2$.
 Since  $f_{-\de}(r)= r^{-\de}$ for $r\ge 4m_0$,  we  deduce,  for $r\ge 4m_0$,
\beaa
\EE[f_{-\de}T, 0,0]  &=&\frac{\Up^2}{4}   \de r^{-1-\de} |e_3\Psi|^2        - \frac 1 4 \de r^{-1-\de} |e_4\Psi|^2       +O\left(\ep  r^{-1-\de}   \big( |\D\Psi|^2  + r^{-2} |\Psi|^2\big)\right).
\eeaa
On the other hand,
\beaa
\PP[f_{-\de}T, 0,0]  \c e_4 &=& f_{-\de} \QQ(T, e_4)  \ge 0,\\
\PP[f_{-\de}T, 0,0]  \c e_3 &=& f_{-\de}\QQ(T, e_3)\ge 0,
\eeaa
as desired. This concludes the proof of Proposition \ref{prop:improved-Morawetz}.


\section{Dafermos-Rodnianski  $r^p$- weighted   estimates}\lab{section-Basic-rp}


For convenience, we work in this section with the   renormalized frame $(e_3', e_4')$ defined in \eqref{equation:renormalizedframe} instead of     the        original  frame $(e_3, e_4)$. To simplify the exposition, we still denote it as $(e_3, e_4)$. Recall that the two  are   frames    are     equivalent up to lower   terms in $m/r$.      

In this  section    we   rely on the Morawetz estimates proved in the previous section    to establish 
$r^p$-weighted estimates in the spirit of Dafermos-Rodnianski \cite{Da-Ro3}. The following theorem claims $r^p$-weighted estimates for the solution $\psi$ of the wave equation \eqref{eq:masterwavepsi}.  

\begin{theorem}[$r^p$-weighted estimates]
\label{theorem:Daf-Rodn1-psi-s}
 Consider  a fixed $\de>0$ and let  $R\gg \frac{m_0}{\de} $, $\ep \ll \de$.
  The following estimates hold true  and for all $\de\le p\le 2-\de$,
 \bea  
    \dot{E}_{p\,;\, R}[\psi](\tau_2)+   \Bdot_{p\,;\,R}[\psi](\tau_1, \tau_2)   +  \dot{F}_p[\psi] (\tau_1, \tau_2)  \les 
          E_{p}[\psi](\tau_1)+   J_p[\psi, N]( \tau_1,\tau_2).
                \label{eq-thm:Daf-Rodn-estim4} 
\eea
\end{theorem}

\begin{remark}\lab{remark:infactcombinedtheoremconsequenceofbasicones}
Note that Theorem \ref{Thm:Morawetz-s} on Morawetz estimates and Theorem \ref{theorem:Daf-Rodn1-psi-s} on $r^p$-weighted estimates immediately yield for all $\de\le p\le 2-\de$, 
  \bea
  \sup_{\tau\in[\tau_1,\tau_2] }   E_{p} [\psi](\tau)+   B_{p}[\psi](\tau_1, \tau_2)  + F_{p}[\psi](\tau_1, \tau_2) \les 
          E_{p}[\psi](\tau_1)+   J_p[\psi, N]( \tau_1,\tau_2),
  \eea
which corresponds to Theorem \ref{theorem-combinedMor-r-weighted} in the case $s=0$. 
\end{remark}

Theorem \ref{theorem:Daf-Rodn1-psi-s} will be proved in section \ref{sec:proofoftheorem:Daf-Rodn1-psi-s:0}. 
  We will need in this section stronger  assumptions  in the region $r  \ge 4m_0$,
away from trapping, than those in \eqref{eq:assumptions-Moraw2}--\eqref{eq:assumptions-Moraw4}   of the previous section. 
For convenience  we  express our conditions  with respect to  the  weights\footnote{The assumptions are consistent with the global frame used in Theorem M1, see Lemma \ref{le:interpolatedbootstrap}. In particular, $\de_0>0$ is such that $\dec-2\de_0>0$ which is the only needed property of $\dec-2\de_0$ to derive the $r^p$ weighted estimates.},
\beaa
w_{p, q}(u,r)=r^{-p}(1+\tau)^{-q-\dec+2\de_0}.
\eeaa

\begin{itemize}
\item[{\bf RP0.} ] The assumptions  {\bf Mor1}--{\bf Mor4} made in the previous section hold true.
\item[{\bf RP1.} ]
  The Ricci coefficients   verify,   for $r\ge 4m_0$
 \bea
 \label{eq:assumptions-Moraw1-2}
 \begin{split}
\big| \xib, \vth, \vthb, \eta, \etab, \ze, \omb\big|  &\les  \ep  w_{1,1},      
\\
   \Big|\kab +\frac 2 r \Big|, \,  \Big|\chib +\frac 1  r \Big|, \,  \Big|e_3\Phi-\chib\Big|       &\les \ep  w_{1,1},\\
\Big|\ka -\frac{2\Up }{r}\Big|, \,      \Big|\chi -\frac{\Up }{r}|, \,       \Big|e_4\Phi-\chi \Big|   &\les\ep \min\{  w_{1, 1} ,     w_{2,1/2}  \},   \\
\Big|\om+\frac{m}{r^2} \Big|,\, |\xi|&\les   \ep   \min\{  w_{2, 1} ,     w_{3,1/2}  \}. 
\end{split}
\eea

\item[{\bf RP2.}]     The derivatives  of $r$ verify, for $r\ge 4m_0$,
\bea
 \label{eq:assumptions-Moraw2-2}
 \bsplit
 \big|e_3(r) +1\big| &\les \ep w_{0,1}, \\
 \big| e_4(r)-\Up\big|&\les \ep    \min\{   w_{0,1} ,    w_{1, 1/2}   \},  \\
 \Big| e_3 e_4 (r)+\frac{2m}{r^2}, \, e_4 e_3(r)\Big|&\les \ep w_{1,1}.
 \end{split}
 \eea
 
 \item[{\bf RP3.}]
  For $r\ge 4m_0$,
 \label{eq:assumptions-Moraw3-2}
\bea
\begin{split}
\Big|\rho+\frac{2m}{r^3}\Big|&\les\ep   w_{3, 1},\\
\Big|K-\frac{1}{r^2}  \Big|  &\les \ep r^{-2},\\
            \big|e_\th(\Phi)\big|&\les r^{-1}.            
\end{split}
\eea

\item[{\bf RP4.}] 
We also assume,  for $r\ge 4m_0$,
\bea
 \label{eq:assumptions-Moraw4-2}
\bsplit
|m-m_0&|\les \ep,\\
|e_3m,\, r  e_4 m|&\les \ep  w_{0,1},  \\
|e_3 e_4(m),\, e_4 e_3(m)|&\les \ep  w_{1,1}.
\end{split}
\eea
\end{itemize}

 Since  the  estimates we are establishing  are  restricted to   the far  region      $r> R$    it  is  convenient, in this section,   to work with   the    with   renormalized frame
\bea
\label{equation:renormalizedframe}
e_3'=\Up e_3, \qquad e_4'=\Up^{-1}  e_4, \qquad e_\th'=e_\th.
\eea

Relative to the new frame $(e_3', e_4', e_\th')$ we have,
\beaa
\xi'=\Up^{-2}\xi, \quad \xib'=\Up^2 \xib, \quad \ze'=\ze,\quad  \eta'=\eta, \quad \chi'=\Up^{-1}\chi, \quad \chib'=\Up \chib
\eeaa
and, 
\beaa
\om'&=&\Up^{-1} \left(\om +\frac 1 2  e_4( \log \Up)\right)= \Up^{-1} \left(  \om +\frac 1 2 \Up^{-1} \Up'  e_4(r)\right)\\
&=& \Up^{-1} \left( -\frac{m}{r^2} +O(\ep r^{-2}  (1+|u|)^{-1/2-\dec}         ) +\Up^{-1} \frac{m}{r^2} (\Up +O(\ep r^{-1} (1+|u|)^{-1/2-\dec}   ) \right)\\
&=& O(\ep r^{-2}),\\  
\omb'&=&\Up \left(\omb -\frac 1 2  e_3( \log \Up)\right)=\Up \left(\omb -\frac 1 2\Up^{-1} \Up' e_3(r) \right)\\
&=&\Up\left(  \omb  -\frac 1 2\Up^{-1} \Up' ( -1+O(\ep  (1+|u|)^{-1-\dec} ) \right)\\
&=& \frac{m}{r^2} +O(\ep r^{-1} (1+|u|)^{-1-\dec}).
\eeaa
Thus in the new frame we  have, for $r\ge 4m_0$,
\begin{itemize}
\item[{\bf RP1'.}] 
  The Ricci coefficients  with respect to the null  frame  $(e_3', e_4', e_\th')$    verify,   for $r\ge 4m_0$:
 \bea
 \label{eq:assumptions-Moraw1-2'}
 \begin{split}
\big| \xib', \vth', \vthb, \eta', \etab', \ze'\big|,\, |\omb'-\frac{m}{r^2}|  &\les \ep w_{1,1},
\\
   \Big|\kab' +\frac{2\Up}{ r } \Big|, \,  \Big|\chib' +\frac \Up  r \Big|, \,  \big|e'_3\Phi-\chib' \big|       &\les \ep w_{1,1},    \\
\Big|\ka' -\frac{2 }{r}\Big|, \,      \Big|\chi' -\frac{1 }{r}\Big|, \,       \big|e'_4\Phi-\chi' \big|   &\les\ep \min\{  w_{1,1}, w_{2,1/2}  \},      \\
\big|\om'  \big|,\, |\xi'|&\les  \ep\min\{  w_{2,1}, w_{3,1/2}  \}.  
\end{split}
\eea

\item[{\bf RP2'.}]  The derivatives  of $r$ verify,
\bea
 \label{eq:assumptions-Moraw2-2'}
 \bsplit
 \big|e'_3(r) +\Up \big| &\les \ep  w_{0,1},\\
 \big| e'_4(r)- 1\big|&\les \ep w_{1,1},  \\
 \Big| e'_3 e' _4 (r), \, e'_4 e'_3(r)  +\frac{2m}{r^2}       \Big|&\les \ep  w_{1,1}.
 \end{split}
 \eea

\item[{\bf RP3'.}] 
 The Gauss curvature $K$ of $S$  and $\rho$    verify,
 \label{eq:assumptions-Moraw3-2'}
\bea
\begin{split}
\Big|\rho+\frac{2m}{r^3}\Big|&\les \ep r^{-3},\\
\Big|K-\frac{1}{r^2}  \Big|  &\les \ep r^{-2}. 
\end{split}
\eea

\item[{\bf RP4'.}] 
We also assume
\bea
 \label{eq:assumptions-Moraw4-2'}
\bsplit
|m-m_0&|\les \ep,\\
|e'_3m,\, re'_4 m|&\les \ep   w_{0,1},   \\
|e'_3 e'_4(m),\, e'_4 e'_3(m)|&\les \ep w_{1,1}.
\end{split}
\eea
\end{itemize}

\begin{remark}
\label{rem:equivalent Mor-norms}
In the  far  region $r\ge 4m_0$  all norms we are using  in our estimates are equivalent when expressed relative to 
 the null frame  $(e_3, e_4, e_\th)$ or $(e_3', e_4', e_\th')$. 
\end{remark}

{\bf Convention.} For the remaining of this section we   shall  do all calculations  with respect to  the renormalized frame $(e_3', e_4', e_\th')$. For convenience we shall 
 drop the primes, throughout this section,     since  there is no danger of confusion.  Note however that  the main   results, which include the  interior region $r\le R$, 
 are always expressed with respect the   original frame.


  \subsection{Vectorfield  $X= f(r)e_4$}


  \begin{lemma}
  \label{le:app-X-fL}
  Consider the vectorfield $X= f(r)e_4$.
  \begin{enumerate}
  \item  We have the decomposition,
    \beaa
   \piX=\LaX \g+\pitX, \qquad \LaX=\frac{2}{r}  f 
   \eeaa
   with symmetric   tensor $\pitX $   which  verifies
  \bea
 \bsplit
  \pitX_{43}&=-2 f'  +\frac{4f}{r}  +O(\ep) w_{1,1}      \left( |f|+r|f'|\right)\\
   \pitX_{33}&= 4 f' \Up -  4 \Up' +O(\ep)w_{1,1} (|f|+ r|f'|)\\
   \pitX_{4\th}&=O(\ep) w_{2,1/2}  |f| \\
 \pitX_{AB} &=O(\ep) w_{2,1/2} |f| \\
 \pitX_{3\th}&=O(\ep)w_{1,1} |f|\\
 \end{split}
 \eea
 \item We have,
 \bea
 \label{eq:squareLa-f}
  \square \LaX&=&\frac{2}{r} f'' +O  \left(\frac{m}{r^4}  +\ep  w_{3,1} \right) \left(|f|+ r|f'|+ r^2 |f''|\right)
 \eea
  \end{enumerate}
  \end{lemma}
  
  \begin{proof}
  See Lemma \ref{lemma:vectorfield-ve4-appendix}  in  appendix. 
  \end{proof}

  
 \subsection{Energy densities for   $X= f(r) e_4$}
 
 
We start with the following proposition.
\begin{proposition}\label{prop:QC-general-multiplier1} 
Assume $\Psi$ verifies the equation   $\squared_\g \Psi=V\Psi+N$  and 
  let   $ X= f e_4 $    and $w=\LaX=\frac{2 f}{r}$ and let   $\EE:=\EE[X, w] =\EE [X=fe_4, w=\frac{2f  }{r}] $.
  
  \begin{enumerate}
  \item We have,
  \beaa
\EE&=&\frac 1 2 f' |e_4 \Psi|^2 +\frac 1 2 \left(-f'+\frac{2f}{r} \right)\left(|\nabb\Psi|^2+  V|\Psi|^2\right)-\frac{1}{2r} f'' |\Psi|^2+\err\left(\ep, \frac{m}{r}, f\right)(\Psi)
\eeaa
where,
\beaa
\err\left(\ep, \frac{m}{r}, f\right)(\Psi)&=&  O\left(\frac{m}{r^2}\right)( |f|+ r |f'| )  |e_4 \Psi|^2 +O  \left(\frac{m}{r^4}  +\ep  w_{3,1} \right) \left(|f|+ r|f'|+ r^2 |f''|\right)|\Psi|^2\\
&+& O(\ep)  w_{1,1}( |f| +r|f'|) \left( |e_4\Psi|^2+  |\nabb \Psi|^2  +r^{-2} |\Psi|^2 \right)\\
&+&   O(\ep) w_{2, 1/2} |f| \Big(  |e_3\Psi |(|e_4\Psi|+r^{-1}|\nabb\Psi|)+  |\nabb \Psi|^2  +r^{-2} |\Psi|^2\Big).
\eeaa

\item   The  current,
\beaa
\PP_\mu&=&\PP_\mu[X, w]=\QQ_{\mu\nu} X^\nu +\frac 1 2  w \Psi\cdot \D_\mu \Psi -\frac 1 4|\Psi|^2   \pr_\mu w
\eeaa
verifies,
\beaa
\PP\c   e_4&=& f\left|e_4\Psi+ \frac 1 r   \Psi\right|^2 - \frac 1 2 r^{-2}  e_4( r f |\Psi|^2)+ O(\ep r^{-3})   f \,  |\Psi|^2, 
\\
\PP\c e_3&=&f \QQ_{34} +\frac  12  r^{-2}e_3\big ( r  f   \psi^2)+  r^{-1}  f' \psi^2+O(mr^{-3}+\ep r^{-2} ) |r f'|  \, |\Psi|^2.
\eeaa

\item  Let   $\th=\th(r) $ supported    for   $r\ge R/2$   with     $\th=1$ for $r\ge R$  such that $f_p=\th(r) r^p$.
Let $^{(p)} \PP:=\PP[  f_p e_4, w_p]$.  Then, for all $r\ge R$,
\beaa
^{(p)} \PP\c e_4 +\frac p 2 r^{-2}  e_4( \th r^{p+1} |\Psi|^2) \ge \frac 1   8 r^{p-2}  (p-1)^2  |\Psi|^2.
\eeaa
\end{enumerate}
\end{proposition}

Before proceeding with the proof of Proposition \ref{prop:QC-general-multiplier1}, we first establish the following lemma.
 
\begin{lemma}\lab{lemma:basiscomputationcontractionofQwithpitildeofX}
We have,
\beaa
\QQ\c \pitX&=&\left( f'+O\left(\frac{m}{r^2} \right)( |f|+ r |f'| ) \right) |e_4 \Psi|^2 +\left(-f'+\frac{2f}{r} \right)\left(|\nabb\Psi|^2+  V|\Psi|^2\right) \\
&+& O(\ep)  w_{1,1}( |f| +r|f'|) \Big( |e_4\Psi|^2+  |\nabb \Psi|^2  +r^{-2} |\Psi|^2 \Big)\\
&+&   O(\ep) w_{2, 1/2} |f| \Big(  |e_3\Psi |(|e_4\Psi|+r^{-1}|\nabb\Psi|)+  |\nabb \Psi|^2  +r^{-2} |\Psi|^2\Big).
\eeaa
\end{lemma}

\begin{proof}
Recall from Proposition \ref{prop:summaryofallimportantpropertiesofenergymomentumtensorforwaveequationpsi} that we have
\beaa
\QQ_{33}=|e_3\Psi|^2, \quad \QQ_{44}=|e_4\Psi|^2,\quad \QQ_{34}=|\nabb\Psi|^2 + V|\Psi|^2,
\eeaa
and,
\beaa
|\QQ_{AB}| \le |e_3\Psi|  |e_4\Psi| +|\nabb\Psi|^2 +|V||\Psi|^2, \quad |\QQ_{A3}| \le |e_3\Psi| |\nabb\Psi|,\quad  |\QQ_{A4}| \le |e_4\Psi| |\nabb\Psi|.
\eeaa
Hence, in view of Lemma \ref{lemma:vectorfield-ve4-appendix} for $\pitX$, we have 
\beaa
\QQ\c \pitX&=& \frac 1 4 \QQ_{44}   \pitX_{33}+       \frac 1 2 \QQ_{34}\pitX_{34}       -  \frac 1 2 \QQ_{4A} \pitX_{3A}-\frac 1 2 \QQ_{3A} \pitX_{4A}+ \QQ_{AB}\pitX_{AB}\\
&=&\Big(   f' \Up -   \Up' f  +O(\ep)w_{1,1} (|f|+ r|f'|) \Big)\QQ_{44}\\
&+&\left(-  f'  +\frac{2f}{r}  +O(\ep)   w_{1,1}  \left( |f|+r|f'|\right) \right)\QQ_{34}\\
&+&O(\ep) w_{1,1}|f| \  \QQ_{4A}  + O(\ep) w_{2, 1/2} |f|\left(  \QQ_{3A}+\QQ_{AB}\right) \\
&=&\left( f'+O\left(\frac{m}{r^2} \right)( |f|+ r |f'| ) \right) \QQ_{44} +\left(-  f'  +\frac{2f}{r}  \right)\QQ_{34}\\
&+&O(\ep)(|f|+r|f'|)w_{1,1}\left(\QQ_{44} + \QQ_{4A}\right) + O(\ep) w_{2, 1/2} |f| \left(\QQ_{AB} + \QQ_{34}\right)\\
&&+ O(\ep) w_{3, 1/2} |f| \QQ_{3A}
\eeaa
from which we deduce,
\beaa
\QQ\c \pitX&=&\left( f'+O\left(\frac{m}{r^2} \right)( |f|+ r |f'| ) \right) |e_4 \Psi|^2 + \left(-  f'  +\frac{2f}{r}  \right) \left( |\nabb \Psi|^2+ V|\Psi|^2\right)\\
&+& O(\ep)  w_{1,1}( |f| +r|f'|) \Big( |e_4\Psi|^2+  |\nabb \Psi|^2  +r^{-2} |\Psi|^2 \Big)\\
&+&   O(\ep) w_{2, 1/2} |f| \Big(  |e_3\Psi |(|e_4\Psi|+r^{-1}|\nabb\Psi|)+  |\nabb \Psi|^2  +r^{-2} |\Psi|^2\Big)
\eeaa
as  desired.
\end{proof}

We are now ready to prove Proposition \ref{prop:QC-general-multiplier1}.

\begin{proof}[Proof of Proposition \ref{prop:QC-general-multiplier1}]
If  $\QQ=\QQ[\Psi]$  is   the energy momentum tensor  of  $\Psi$     (recall $\squared \Psi= V\Psi +N$)   and
\beaa
 \piX=\LaX\, \g+\pitX
\eeaa
we deduce,
\beaa
\QQ  \c\piX &=&\LaX \tr \QQ+  \QQ\c  \pitX =  \LaX\, (-\LL(\Psi)-V|\Psi|^2)+\QQ\c  \pitX.
\eeaa
Hence, for  $ X= f e_4 $    and $w=\LaX=\frac{2f}{r}$, 
\beaa
 \frac 1 2 \QQ  \c\piX + \frac 1 2  w \LL[\Psi]&=&-\frac 1 2 \LaX   V |\Psi|^2 +\frac 1 2   \QQ\c  \pitX.
\eeaa
 In view of \eqref{le:divergPP-gen-EE}, we infer
 \beaa
\EE:&=&\EE [X, w=\LaX, M=0] \\
&=&\frac 1 2 \QQ \c  \pitX -\frac 1 4|\Psi|^2   \square_\g\LaX-\frac 1 2(X(V) +\LaX V)|\Psi|^2.
\eeaa
Recall  that $V=-\ka\kab$. Hence, 
\beaa
X(V) +\LaX V&=& f e_4(V) +\frac{2f}{r} V=       -f \left(e_4(\ka\kab)+  \frac{2}{r}\ka\kab\right)=   f  \left( \ka^2\kab -2\ka\rho -\frac{2}{r}\ka\kab  +      O(\ep)  w_{3,1}       \right)\\
&=&  O  \left(\frac{m}{r^4}  +\ep  w_{3,1} \right)f.
\eeaa
Hence, in view of the computation \eqref{eq:squareLa-f} of $\square \LaX$
\beaa
&& -\frac 1 4|\Psi|^2   \square_\g\LaX-\frac 1 2(X(V) +\LaX V)|\Psi|^2\\
&=&-\frac{1}{2r} f'' |\Psi|^2  +O  \left(\frac{m}{r^4}  +\ep  w_{3,1} \right) \left(|f|+ r|f'|+ r^2 |f''|\right)|\Psi|^2.
\eeaa
We deduce,
\beaa
\EE =\frac 1 2 \QQ\c \pitX-\frac{1}{2r} f'' |\Psi|^2 +O  \left(\frac{m}{r^4}  +\ep  w_{3,1} \right) \left(|f|+ r|f'|+ r^2 |f''|\right)|\Psi|^2
\eeaa
 Using Lemma \ref{lemma:basiscomputationcontractionofQwithpitildeofX}, we deduce,
\beaa
\EE&=&\frac 1 2 f' |e_4 \Psi|^2 +\frac 1 2 \left(-f'+\frac{2f}{r} \right)\left(|\nabb\Psi|^2+  V|\Psi|^2\right)-\frac{1}{2r} f'' |\Psi|^2+\err\left(\ep, \frac{m}{r}, f\right)(\Psi)
\eeaa
where,
\beaa
\err\left(\ep, \frac{m}{r}, f\right)(\Psi)&=&  O\left(\frac{m}{r^2} \right)( |f|+ r |f'| )  |e_4 \Psi|^2 +O  \left(\frac{m}{r^4}  +\ep  w_{3,1} \right) \left(|f|+ r|f'|+ r^2 |f''|\right)|\Psi|^2\\
&+& O(\ep)  w_{1,1}( |f| +r|f'|) \Big( |e_4\Psi|^2+  |\nabb \Psi|^2  +r^{-2} |\Psi|^2 \Big)\\
&+&   O(\ep) w_{2, 1/2} |f| \Big(  |e_3\Psi |(|e_4\Psi|+r^{-1}|\nabb\Psi|)+  |\nabb \Psi|^2  +r^{-2} |\Psi|^2\Big)
\eeaa
which is the first part of Proposition \ref{prop:QC-general-multiplier1}.

To prove the  second    part of Proposition \ref{prop:QC-general-multiplier1}, we compute
\beaa
\PP\c   e_4&=& f\QQ_{44}+ \frac 1 r  f \Psi\cdot  e_4\Psi - \frac 1 2 e_4(r^{-1} f) |\Psi|^2\\
&=&f\left(  |e_4\Psi|^2 +\frac 1 r  \Psi \cdot e_4\Psi\right) - \frac 1 2 e_4(r^{-1} f) |\Psi|^2\\
&=&f\left| e_4\Psi+ \frac 1 r   \Psi\right|^2 - \frac 1 r    f  \Psi \cdot e_4\Psi- r^{-2}    f  |\Psi|^2 - \frac 1 2 e_4(r^{-1} f) |\Psi|^2\\
&=&f\left|e_4\Psi+ \frac 1 r   \Psi\right|^2 - \frac 1 2 r^{-2}  e_4( r  f |\Psi|^2)+ \frac 1 2 r^{-2}  e_4( r  f )|\Psi|^2   - r^{-2}    f  |\Psi|^2      - \frac 1 2 e_4(r^{-1} f) |\Psi|^2\\
&=&f\left|e_4\Psi+ \frac 1 r   \Psi\right|^2 - \frac 1 2 r^{-2}  e_4( r  f |\Psi|^2)+r^{-2}   (e_4(r)-1)       f   |\Psi|^2.
\eeaa
Since 
\beaa
e_4(r)=\frac{r}{2}(\ka+A),
\eeaa
we have\footnote{Note that  so far we have only used  the weaker version
 $e_4(r)-1=O(\ep ) $. This is the first time we need the stronger  version of the estimate in this chapter.}  
\beaa
e_4(r)-1 =O(\ep r^{-1}).
\eeaa
Thus, as desired,
\beaa
\PP\c   e_4&=& f\left|e_4\Psi+ \frac 1 r   \Psi\right|^2 - \frac 1 2 r^{-2}  e_4( r  f |\Psi|^2)+ O(\ep r^{-3})   f  |\Psi|^2. 
\eeaa
 Also, 
\beaa
\PP\c e_3 
&=&f \QQ_{34}+r^{-1}f \Psi \cdot e_3 \Psi-\frac 1 2   e_3 ( r^{-1}f ) |\Psi|^2\\
&=& f  \QQ_{34}+  \frac 1 2 r^{-1}   f e_3(|\Psi|^2) -\frac 1  2  e_3  ( r^{-1}   f  ) |\Psi|^2\\
&=&f \QQ_{34} +\frac  12\left[  r^{-2}e_3\big ( r    f |\Psi|^2) -       r^{-2}e_3(r   f)  |\Psi|^2    \right]-\frac 1  2  e_3  ( r^{-1}   f  ) |\Psi|^2\\
&=& f \QQ_{34} +\frac  12  r^{-2}e_3\big ( r  f |\Psi|^2)+ r^{-1} f' \Up |\Psi|^2- r^{-1} f'  (e_3(r)+\Up) |\Psi|^2\\
&=&f \QQ_{34} +\frac  12  r^{-2}e_3\big ( r  f |\Psi|^2)+  r^{-1}    f'     |\Psi|^2 +O\left(mr^{-3}+\ep r^{-2}\right) ( r|f'|)|\Psi|^2
\eeaa
as desired.

It remains to prove the last part of Proposition \ref{prop:QC-general-multiplier1}. 
We have, for $r\ge R$
\beaa
^{(p)} \PP\c   e_4&=& r^p |e_4\Psi|^2 + \ r^{p-1}  \Psi  e_4\Psi - \frac 1 2 e_4(r^{p-1} ) |\Psi|^2
\eeaa
and,
\beaa
^{(p)} \PP\c   e_4       +\frac  p 2  r^{-2} e_4( \th r^{p+1} |\Psi|^2)          &=& r^p |e_4\Psi|^2 + \ r^{p-1}  \Psi \cdot e_4\Psi - \frac 1 2 e_4(r^{p-1} ) |\Psi|^2\\
&+& p  r^{p-1} \Psi \c  e_4\Psi+\frac{p(p+1)}{2}  r^{p-2} e_4 (r) |\Psi|^2 \\
&=&r^p |e_4\Psi|^2+(p+1)\ r^{p-1}  \Psi \cdot e_4\Psi+\frac{p^2+1}{2} e_4(r)  r^{p-2}  |\Psi|^2 \\
&=& r^p\left[ \left| e_4\Psi+ \frac{p+1}{2r} \Psi\right|^2+\frac{(p-1)^2}{4r^2} |\Psi|^2 \right]\\
&+&\frac{p^2+1}{2}( e_4(r)-1)  r^{p-2}  |\Psi|^2\\
&\ge &r^{p-2}\frac{(p-1)^2}{4} |\Psi|^2- O(\ep) \frac{p^2+1}{2}   r^{p-3}  |\Psi|^2.
\eeaa
This concludes the proof of Proposition \ref{prop:QC-general-multiplier1}.
\end{proof}
  
In applications we would like to apply Proposition \ref{prop:QC-general-multiplier1} to $f=r^p,$  $0<p<2$. We note however  
 that the presence of the term $ -   \frac 1 2   r^{-1} f''  |\Psi|^2$  on the right hand side  of the $\EE$ identity
   requires an additional correction if  $p>1$.  This additional correction is taken into account by the following proposition.

\begin{proposition}\label{prop:QC-general-multiplier2} 
Assume $\Psi$ verifies the equation   $\square_\g \Psi=V\Psi+N$  and  let  
 $X= f(r)  e_4$, $w=\LaX=\frac{2f}{r} $ and $M=2r^{-1} f'  e_4$. Then,
 \begin{enumerate}
 \item  We have, with $\ec_4= r^{-1}e_4(r\cdot) $,
\bea
 \label{eq:rp-Daf-Rodn1}
\EE[X, w , M]&=&\frac  1 2  f' |\ec_4(\Psi)|^2 +\frac 1 2 \left(\frac{2f}{r}- f'\right) \QQ_{34}+\err\left(\ep, \frac{m}{r};  f\right)[\Psi] 
\eea
with error term,
\beaa
\err\left(\ep, \frac{m}{r}, f\right)(\Psi)&=&  O\left(\frac{m}{r^2} \right)( |f|+ r |f'| )  |e_4 \Psi|^2 +O  \left(\frac{m}{r^4}  +\ep  w_{3,1} \right) \left(|f|+ r|f'|+ r^2 |f''|\right)|\Psi|^2\\
&+& O(\ep)  w_{1,1}( |f| +r|f'|) \left( |e_4\Psi|^2+  |\nabb \Psi|^2  +r^{-2} |\Psi|^2 \right)\\
&+&   O(\ep) w_{2, 1/2} |f| \Big(  |e_3\Psi |(|e_4\Psi|+r^{-1}|\nabb\Psi|)+  |\nabb \Psi|^2  +r^{-2} |\Psi|^2\Big).
\eeaa

 \item   The  current,
\beaa
\PP_\mu&=&\PP_\mu[X, w, M]=\QQ_{\mu\nu} X^\nu +\frac 1 2  w \Psi \D_\mu \Psi -\frac 1 4|\Psi|^2   \pr_\mu w+\frac 1 4  M_\mu |\Psi|^2
\eeaa
verifies, 
\beaa
\PP\c e_4&=&f (\ec_4\Psi)^2-\frac 1 2 r^{-2} e_4( r f |\Psi|^2)+ O(\ep r^{-1} ) f (|e_4\Psi|^2+r^{-2} |\Psi|^2), \\
\PP\c e_3&=&f \QQ_{34} +\frac  12  r^{-2}e_3\big ( r  f |\Psi|^2)+O(mr^{-3}+\ep r^{-2}) (|f|+r|f'|)|\Psi|^2. 
\eeaa

\item  Let   $\th=\th(r) $ supported    for   $r\ge R/2$   with     $\th=1$ for $r\ge R$  such that $f_p=\th(r) r^p$.
Let $^{(p)} \PP:=\PP[  f_p e_4, w_p, M_p]$.  Then, for all $r\ge R$,
\beaa
^{(p)} \PP\c e_4 +\frac p 2 r^{-2}  e_4( \th r^{p+1} |\Psi|^2) \ge \frac 1   8 r^{p-2}  (p-1)^2  |\Psi|^2.
\eeaa
\end{enumerate}
     \end{proposition}   
 
\begin{proof}    
We start with the first part of Proposition \ref{prop:QC-general-multiplier2}. To this end, we use
\beaa
\piX_{43} =- 2 e_4 f +4 f \om, \quad \piX_{AB}= 2 f \chiS_{AB}
\eeaa
so that
\beaa
\tr\piX &=& -\piX_{43}+\g^{AB}\piX_{AB}\\
&=& 2 e_4 f -4 f \om+2 f\ka,
\eeaa
and we compute
\beaa
\D^\mu M_\mu&=&\D^\mu( 2 r^{-1}  f' e_4)_\mu=\D^\mu\left(   \frac{2f'}{r f}  X_\mu\right)=  \frac{2f'}{r f} \Div X+X\left(  \frac{2f'}{r f}\right)\\
&=&  \frac{f'}{r f}  \tr\piX+X\left(  \frac{2f'}{r f}\right)\\
&=& \frac{f'}{r f} \left(2 e_4 f -4 f \om+2 f\ka\right)+ 2f e_4\left(  \frac{f'}{r f}\right)\\
&=& \frac{f'}{r f} \left(2 e_4(r)f'+\frac{4f}{r}+O\left(\frac{m}{r^3}+\ep w_{2,1}\right)\right)   
+2f \left( \frac{  f'' rf-     f'(    f+ rf')}{r^2 f^2} \right)e_4 (r)  \\
&=&\frac{4 f'}{r^2}+\frac{2(f')^2}{r f}+ 2 \frac{f''}{r}- \frac{2 f'}{r^2}-\frac{2(f')^2}{r f}+O\left(\frac{m}{r^4}+\ep w_{3,1}\right)(|f|+r|f'|+r^2|f''|)\\
&=&\frac{2 f'}{r^2} +\frac{2f''}{r}+O\left(\frac{m}{r^4}+\ep w_{3,1}\right)(|f|+r|f'|+r^2|f''|).
\eeaa
We also have
\beaa
\frac 12\Psi\cdot \D_\mu\Psi M^\mu &=& r^{-1}f'\Psi\cdot \D_4\Psi.
\eeaa
Since we have
\beaa
\EE[X, w , M] &=& \EE[X, w]+\frac 1 4( \D^\mu M_\mu)|\Psi|^2+\frac 12\Psi\cdot \D_\mu\Psi M^\mu,
\eeaa
we infer
\beaa
\EE[X, w , M] &=& \EE[X, w]+\left(\frac{f'}{2r^2} +\frac{f''}{2r}+O\left(\frac{m}{r^4}+\ep w_{3,1}\right)(|f|+r|f'|+r^2|f''|)\right)|\Psi|^2+r^{-1}f'\Psi\cdot \D_4\Psi.
\eeaa
Together with Proposition \ref{prop:QC-general-multiplier1}, this yields
\beaa
\EE[X, w , M] &=& \frac 1 2 f' \Big(|e_4 \Psi|^2 +2r^{-1}\Psi\cdot \D_4\Psi+r^{-2}|\Psi|^2\Big)+\frac 1 2 \left(-f'+\frac{2f}{r} \right)\left(|\nabb\Psi|^2+  V|\Psi|^2\right)\\
&&+\err\left(\ep, \frac{m}{r}, f\right)(\Psi)+O\left(\frac{m}{r^4}+\ep w_{3,1}\right)(|f|+r|f'|+r^2|f''|)|\Psi|^2\\
&=& \frac 1 2 f' |e_4 \Psi+r^{-1}\Psi|^2 +\frac 1 2 \left(-f'+\frac{2f}{r} \right)\left(|\nabb\Psi|^2+  V|\Psi|^2\right)\\
&&+\err\left(\ep, \frac{m}{r}, f\right)(\Psi)+O\left(\frac{m}{r^4}+\ep w_{3,1}\right)(|f|+r|f'|+r^2|f''|)|\Psi|^2\\
&=& \frac 1 2 f' |\ec_4 \Psi+r^{-1}(1-e_4(r))\Psi|^2 +\frac 1 2 \left(-f'+\frac{2f}{r} \right)\left(|\nabb\Psi|^2+  V|\Psi|^2\right)\\
&&+\err\left(\ep, \frac{m}{r}, f\right)(\Psi)+O\left(\frac{m}{r^4}+\ep w_{3,1}\right)(|f|+r|f'|+r^2|f''|)|\Psi|^2
\eeaa
and hence
\beaa
\EE[X, w , M] &=& \frac 1 2 f' |\ec_4 \Psi|^2 +\frac 1 2 \left(-f'+\frac{2f}{r} \right)\left(|\nabb\Psi|^2+  V|\Psi|^2\right)+\err\left(\ep, \frac{m}{r}, f\right)(\Psi)
\eeaa
where
\beaa
\err\left(\ep, \frac{m}{r}, f\right)(\Psi)&=&  O\left(\frac{m}{r^2} \right)( |f|+ r |f'| )  |e_4 \Psi|^2 +O  \left(\frac{m}{r^4}  +\ep  w_{3,1} \right) \left(|f|+ r|f'|+ r^2 |f''|\right)|\Psi|^2\\
&+& O(\ep)  w_{1,1}( |f| +r|f'|) \left( |e_4\Psi|^2+  |\nabb \Psi|^2  +r^{-2} |\Psi|^2 \right)\\
&+&   O(\ep) w_{2, 1/2} |f| \Big(  |e_3\Psi |(|e_4\Psi|+r^{-1}|\nabb\Psi|)+  |\nabb \Psi|^2  +r^{-2} |\Psi|^2\Big).
\eeaa
This is the desired estimate \eqref{eq:rp-Daf-Rodn1}.

Next, we consider the second part of Proposition \ref{prop:QC-general-multiplier2}. 
    \beaa
\PP_\mu[X, w, M]&=&\PP_\mu[X, w]+\frac 1 4  |\Psi|^2  M_\mu =\PP_\mu[X, w]+\frac 1 2  r^{-1} f' |\Psi| ^2 e_4. 
\eeaa
Hence, in view of  the results in part 2 of Proposition \ref{prop:QC-general-multiplier1},
\beaa
\PP_ 4 [X, w, M]&=&\PP_4[X, w]=f \Big|\ec_4\Psi+(1-e_4(r))\Psi\Big|^2-\frac 1 2 r^{-2} e_4( r f |\Psi|^2)+
O(\ep r^{-3} ) f  |\Psi|^2\\
&=& f |\ec_4\Psi|^2-\frac 1 2 r^{-2} e_4( r f |\Psi|^2)+
O(\ep r^{-1} ) f (|e_4\Psi|^2+r^{-2} |\Psi|^2),\\
\PP_ 3 [X, w, M]&=&\PP_3[X, w] - r^{-1} f' |\Psi|^2 =f \QQ_{34} +\frac  12  r^{-2}e_3\big ( r  f |\Psi|^2)+  r^{-1}  f' |\Psi|^2  - r^{-1} f' |\Psi|^2\\
&+&O(\ep r^{-2} )  (|f|+r|f'|)  |\Psi|^2\\
&=&f \QQ_{34} +\frac  12  r^{-2}e_3\big ( r  f |\Psi|^2)+O(\ep r^{-2} ) (|f|+r|f'| )|\Psi|^2
\eeaa
as desired. The last part follows from the third part of Proposition \ref{prop:QC-general-multiplier1}.
\end{proof}

\begin{lemma}\lab{lemma:contractionofmathcalPwithunitnormaSigma*forlaterintegrationbyparts}
On $\Si_*$, we have
\beaa
\mathcal{P}\cdot N_{\Si_*} &=& \frac{1}{2}f \QQ_{34} +\frac{1}{2}f (\ec_4\Psi)^2 +\frac  12\div_{\Si_*}\Big(r^{-1}  f |\Psi|^2\nu_{\Si_*}\Big)  \\
&&+O(mr^{-1}+\ep) (|f|+r|f'|)(|e_4\Psi|^2+|\nabb\Psi|^2+r^{-2} |\Psi|^2)
\eeaa
\end{lemma}

\begin{proof}
Recall that there exists a constant $c_*$ such that $u+r=c_*$ on $\Sigma_*$. In particular, the unit normal $N_{\Si_*}$ is collinear to 
\beaa
-2\g^{\a\b}\pr_\a(u+r)\pr_\b &=& e_4(u+r)e_3 +e_3(u+r)e_4\\
&=& e_4(r)e_3 +(e_3(u)+e_3(r))e_4
\eeaa
and since
\beaa
g\Big(e_4(r)e_3 +(e_3(u)+e_3(r))e_4, e_4(r)e_3 +(e_3(u)+e_3(r))e_4\Big) &=& -4e_4(r)(e_3(u)+e_3(r)),
\eeaa
we infer
\beaa
N_{\Si_*} &=& \frac{\sqrt{e_4(r)}}{2\sqrt{e_3(u)+e_3(r)}}e_3 +\frac{\sqrt{e_3(u)+e_3(r)}}{2\sqrt{e_4(r)}}e_4.
\eeaa
In particular, we have
\beaa
\mathcal{P}\cdot N_{\Si_*} &=& \frac{\sqrt{e_4(r)}}{2\sqrt{e_3(u)+e_3(r)}}\mathcal{P}\cdot e_3+\frac{\sqrt{e_3(u)+e_3(r)}}{2\sqrt{e_4(r)}}\mathcal{P}\cdot e_4.
\eeaa

Now, recall from Proposition \ref{prop:QC-general-multiplier2} that we have
\beaa
\PP\c e_4&=&f (\ec_4\Psi)^2-\frac 1 2 r^{-2} e_4( r f |\Psi|^2)+ O(\ep r^{-1} ) f (|e_4\Psi|^2+r^{-2} |\Psi|^2), \\
\PP\c e_3&=&f \QQ_{34} +\frac  12  r^{-2}e_3\big ( r  f |\Psi|^2)+O(mr^{-3}+\ep r^{-2}) (|f|+r|f'|)|\Psi|^2. 
\eeaa
We deduce
\beaa
\mathcal{P}\cdot N_{\Si_*} &=& \frac{\sqrt{e_4(r)}}{2\sqrt{e_3(u)+e_3(r)}}\left(f \QQ_{34} +\frac  12  r^{-2}e_3\big ( r  f |\Psi|^2)+O(mr^{-3}+\ep r^{-2}) (|f|+r|f'|)|\Psi|^2\right)\\
&&+\frac{\sqrt{e_3(u)+e_3(r)}}{2\sqrt{e_4(r)}}\left(f (\ec_4\Psi)^2-\frac 1 2 r^{-2} e_4( r f |\Psi|^2)+ O(\ep r^{-1} ) f (|e_4\Psi|^2+r^{-2} |\Psi|^2)\right)\\
&=&\frac{\sqrt{e_4(r)}}{2\sqrt{e_3(u)+e_3(r)}}f \QQ_{34} +\frac{\sqrt{e_3(u)+e_3(r)}}{2\sqrt{e_4(r)}}f (\ec_4\Psi)^2\\
&& +\frac{\sqrt{e_4(r)}}{2\sqrt{e_3(u)+e_3(r)}}\frac  12  r^{-2}e_3\big ( r  f |\Psi|^2) -\frac{\sqrt{e_3(u)+e_3(r)}}{2\sqrt{e_4(r)}}\frac 1 2 r^{-2} e_4( r f |\Psi|^2)\\
&&+O(mr^{-3}+\ep r^{-2}) (|f|+r|f'|)|\Psi|^2+O(\ep r^{-1} ) f (|e_4\Psi|^2+r^{-2} |\Psi|^2)\\
&=& \frac{1}{2\sqrt{2-\Up}}(1+O(\ep))f \QQ_{34} +\frac{\sqrt{2-\Up}}{2}(1+O(\ep))f (\ec_4\Psi)^2 +\frac  12  r^{-2}\nu_{\Sigma_*}\big ( r  f |\Psi|^2) \\
&&+O(mr^{-3}+\ep r^{-2}) (|f|+r|f'|)|\Psi|^2+O(\ep r^{-1} ) f (|e_4\Psi|^2+r^{-2} |\Psi|^2)\\
&=& \frac{1}{2}f \QQ_{34} +\frac{1}{2}f (\ec_4\Psi)^2 +\frac  12  r^{-2}\nu_{\Sigma_*}\big ( r  f |\Psi|^2) \\
&&+O(mr^{-1}+\ep) (|f|+r|f'|)(|e_4\Psi|^2+|\nabb\Psi|^2+r^{-2} |\Psi|^2)
\eeaa
where we used
\beaa
e_4(r) = 1+O(\ep), \quad e_3(r)=-\Up+O(\ep), \quad e_3(u)=2+O(\ep),
\eeaa
and where $\nu_{\Sigma_*}$ denotes the vectorfield
\beaa
\nu_{\Sigma_*} &=& \frac{\sqrt{e_4(r)}}{2\sqrt{e_3(u)+e_3(r)}}e_3 -\frac{\sqrt{e_3(u)+e_3(r)}}{2\sqrt{e_4(r)}}e_4.
\eeaa

Next, note from the formula that $\nu_{\Sigma_*}$ is unitary and orthogonal to $N_{\Si_*}$ so that $\nu_{\Sigma_*}$ is a unit vectorfield, tangent to $\Sigma_*$ and normal to $e_\th$. Furthermore, since $(\nu_{\Si_*}, e_\th, e_\vphi)$ is an orthonormal frame of $\Si_*$, we have
\beaa
\div_{\Si_*}(\nu_{\Si_*}) &=& \g(\D_{\nu_{\Si_*}}\nu_{\Si_*}, \nu_{\Si_*})+\g(\D_{e_\th}\nu_{\Si_*}, e_\th)+\g(\D_{e_\vphi}\nu_{\Si_*}, e_\vphi).
\eeaa
Since $\nu_{\Si_*}$ is a unit vector, the first term vanishes, and hence
\beaa
\div_{\Si_*}(\nu_{\Si_*}) &=& \g(\D_{e_\th}\nu_{\Si_*}, e_\th)+\g(\D_{e_\vphi}\nu_{\Si_*}, e_\vphi)\\
&=& \frac{\sqrt{e_4(r)}}{2\sqrt{e_3(u)+e_3(r)}}g(D_\th e_3, e_\th) -\frac{\sqrt{e_3(u)+e_3(r)}}{2\sqrt{e_4(r)}}g(D_\th e_4, e_\th)+\nu_{\Si_*}(\Phi)\\
&=& \frac{\sqrt{e_4(r)}}{2\sqrt{e_3(u)+e_3(r)}}\kab -\frac{\sqrt{e_3(u)+e_3(r)}}{2\sqrt{e_4(r)}}\ka.
\eeaa
In particular, we have
\beaa
\div_{\Si_*}\Big(r^{-1}  f |\Psi|^2\nu_{\Si_*}\Big) &=& r^{-2}\nu_{\Si_*}\big ( r  f |\Psi|^2)+\nu_{\Si_*}(r^{-2})r f |\Psi|^2+\div_{\Si_*}(\nu_{\Si_*})r^{-1}  f |\Psi|^2\\
&=& r^{-2}\nu_{\Si_*}\big ( r  f |\Psi|^2)+\left(\div_{\Si_*}(\nu_{\Si_*})-\frac{2\nu_{\Si_*}(r)}{r}\right)r^{-1}  f |\Psi|^2\\
&=& r^{-2}\nu_{\Si_*}\big ( r  f |\Psi|^2)+\Bigg(\frac{\sqrt{e_4(r)}}{2\sqrt{e_3(u)+e_3(r)}}\left(\kab-\frac{2e_3(r)}{r}\right)\\
&& -\frac{\sqrt{e_3(u)+e_3(r)}}{2\sqrt{e_4(r)}}\left(\ka -\frac{2e_4(r)}{r}\right)\Bigg)r^{-1}  f |\Psi|^2
\eeaa
and hence
\beaa
r^{-2}\nu_{\Si_*}\big ( r  f |\Psi|^2) &=& \div_{\Si_*}\Big(r^{-1}  f |\Psi|^2\nu_{\Si_*}\Big) +O(\ep r^{-2})  f |\Psi|^2
\eeaa
We finally obtain
\beaa
\mathcal{P}\cdot N_{\Si_*} &=& \frac{1}{2}f \QQ_{34} +\frac{1}{2}f (\ec_4\Psi)^2 +\frac  12  r^{-2}\nu_{\Sigma_*}\big ( r  f |\Psi|^2) \\
&&+O(mr^{-1}+\ep) (|f|+r|f'|)(|e_4\Psi|^2+|\nabb\Psi|^2+r^{-2} |\Psi|^2)\\
&=& \frac{1}{2}f \QQ_{34} +\frac{1}{2}f (\ec_4\Psi)^2 +\frac  12\div_{\Si_*}\Big(r^{-1}  f |\Psi|^2\nu_{\Si_*}\Big)  \\
&&+O(mr^{-1}+\ep) (|f|+r|f'|)(|e_4\Psi|^2+|\nabb\Psi|^2+r^{-2} |\Psi|^2)
\eeaa
which concludes the proof of the lemma.
\end{proof}


\subsection{Proof of Theorem \ref{theorem:Daf-Rodn1-psi-s}}\lab{sec:proofoftheorem:Daf-Rodn1-psi-s}


    Consider  the function $f_p=f_{p,R}$  defined  by,
\bea
\label{definition:fp}
f_p=\begin{cases}& r^p, \qquad \mbox{if}\quad r\ge R,\\
& 0 , \qquad\,\, \mbox{if}\quad \,r\le \frac R 2,
\end{cases}
\eea
where   $R$ is a   fixed, sufficiently large   constant which will be chosen in the proof. We also consider
\beaa
X_p=f_pe_4, \quad w_p=\frac{2f_p}{r}, \quad M_p=\frac{2f_p'}{r}e_4.
\eeaa
 The proof relies  on Proposition \ref{prop:QC-general-multiplier2}. 
 
   {\bf Step 0.} (Reduction to the region $r\ge R$)\,\,  In view of the definition of $\EE[X_p, w_p,   M_p ]$, see \eqref{eq:modified-div}, and in view of the choice of $X_p$ and $w_p$, we have
    \beaa
    \D^\mu \PP_\mu[X_p, w_p,   M_p] &=& \EE[X_p, w_p,   M_p ]  +f_p(r)  \ec_4 \Psi\c  N.
     \eeaa
We integrate  this identity        on the domain  $\MM(\tau_1, \tau_2)$ to derive,     
     \beaa
   \int_{\Si (\tau_2)}   \PP\c  N_\Si  +\int_{\Sigma_*(\tau_1,\tau_2)}\PP \c N_{\Si_*} +\int_{\MM(\tau_1, \tau_2)} \EE=\int_{\Si (\tau_1)}   \PP\c   N_\Si -          \int_{\MM(\tau_1,\tau_2)} f_p \ec_4 \Psi\c N.
     \eeaa
    Denoting  the boundary terms,   
     \beaa
     K_{\ge R}(\tau_1, \tau_2):&=&\int_{\Si_{\ge R}  (\tau_2)}   \PP\c  e_4        +\int_{\Sigma_*(\tau_1,\tau_2)}\PP \c N_{\Si_*}-\int_{\Si_{\ge R} (\tau_1)}   \PP\c   e_4,\\
        K_{\le R}(\tau_1, \tau_2):&=&\int_{\Si_{\le R} (\tau_1)}   \PP\c   N_\Si- \int_{\Si_{\le R} (\tau_2)}   \PP\c   N_\Si,
     \eeaa
     we   write,
     \bea
      K_{\ge R} (\tau_1, \tau_2)+\int_{\MM_{\ge R} (\tau_1, \tau_2)}\EE         &=&   K_{\le R}(\tau_1, \tau_2)    -\int_{\MM_{\le R} (\tau_1, \tau_2)}\EE -     \int_{\MM(\tau_1,\tau_2)} f_p \ec_4 \Psi\c N.\nn\\
     \eea
     
   We have the following lemma.  
   \begin{lemma}
  \label{lemma:Daf-Rodnianski1}
  For $p\geq \de$, we have
  \beaa
      K_{\ge R} (\tau_1, \tau_2)+\int_{\MM_{\ge R} (\tau_1, \tau_2)}\EE \les  R^{p+2}\Big(E[\Psi](\tau_1)  +J_p[N, \psi](\tau_1,\tau_2)     
      +O(\ep)\Bdot^s_{\de\,; \,4m_0}[\psi](\tau_1, \tau_2)    \Big).
\eeaa
   \end{lemma}     
     
  \begin{proof}[Proof of Lemma \ref{lemma:Daf-Rodnianski1}]  
     The terms  $\int_{\Si_{\le R} (\tau)}   \PP\c   N_\Si $ and $ \int_{\MM_{\le R} (\tau_1, \tau_2)}\EE$ on the  right   can be estimated  as follows
     \beaa
     \bigg|\int_{\Si_{\le R} (\tau_1)}   \PP\c   N_\Si\bigg|&\les& R^p E[\Psi](\tau_1),\\
      \bigg|\int_{\Si_{\le R} (\tau_2)}   \PP\c   N_\Si\bigg|&\les& R^p E[\Psi](\tau_2),\\
    \bigg|  \int_{\MM_{\le R} (\tau_1, \tau_2)}\EE\bigg|&\les&  R^{p+2} \Mor[\Psi](\tau_1,\tau_2).
     \eeaa
     Hence,
     \beaa
       K_{\le R}(\tau_1, \tau_2)    -\int_{\MM_{\le R} (\tau_1, \tau_2)}\EE \les  R^{p+2}\left(E[\Psi](\tau_1)+E[\Psi](\tau_2)+ \Mor[\Psi](\tau_1,\tau_2)\right).
     \eeaa
     In view of the improved Morawetz  Theorem  \ref{Thm:Morawetz-s} we have,   for   fixed   $\de>0$,
       \beaa
        E[\Psi](\tau_2)            +\Morr[\Psi](\tau_1,\tau_2)+F[\Psi](\tau_1,\tau_2)&\les&     E[\Psi](\tau_1)  +J_\de[N, \psi](\tau_1,\tau_2)   \\
        &&  +O(\ep)\Bdot^s_{\de\,; \,4m_0}[\psi](\tau_1, \tau_2) 
  \eeaa
  which implies
    \beaa
    \begin{split}
      K_{\ge R} (\tau_1, \tau_2)+\int_{\MM_{\ge R} (\tau_1, \tau_2)}\EE & \les  R^{p+2}\Big(E[\Psi](\tau_1)  +J_\de[N, \psi](\tau_1,\tau_2)     
      +O(\ep)\Bdot^s_{\de\,; \,4m_0}[\psi](\tau_1, \tau_2)    \Big)\nn\\
      &+\left| \int_{\MM(\tau_1,\tau_2)} f_p \ec_4 \Psi\c N\right|.
      \end{split}    
\eeaa
 Together with the definition \eqref{def:normI-JnormforN} of $J_p$ and the fact that $p\geq \de$, we infer
     \beaa
    \begin{split}
      K_{\ge R} (\tau_1, \tau_2)+\int_{\MM_{\ge R} (\tau_1, \tau_2)}\EE & \les  R^{p+2}\Big(E[\Psi](\tau_1)  +J_p[N, \psi](\tau_1,\tau_2)     
      +O(\ep)\Bdot^s_{\de\,; \,4m_0}[\psi](\tau_1, \tau_2)    \Big)
      \end{split}    
\eeaa 
which concludes the proof of Lemma \ref{lemma:Daf-Rodnianski1}.
\end{proof}
   
   The proof of  Theorem \ref{theorem:Daf-Rodn1-psi-s}  now proceeds  according to the following steps.
   
   {\bf Step 1.} (Bulk terms for $r\ge R$) \,\,  We prove the following lower bound for $\int_{\MM_{\ge R} (\tau_1, \tau_2)}\EE$.
   \begin{lemma}\lab{lemma:intermadiaryresult:eq:Dafermos-Rodn1}
 Given a fixed $\de>0$       we   have  for all $\de \le p\le 2-\de$  and $R\gg \frac m \de $, $\ep\ll \de$,
 \bea
 \label{eq:Dafermos-Rodn1}
\nn \int_{\MM_{\ge R}(\tau_1,\tau_2)  }\EE&\geq &\frac 1  4  \int_{\MM_{\ge R}(\tau_1,\tau_2)  }r^{p-1}\Big(  p   |\ec_4(\Psi)|^2+ (2-p) (|\nabb \Psi|^2 + r^{-2}|\Psi|^2)\Big)\\
 &-&O(\ep) \Morr[\Psi](\tau_1, \tau_2).
 \eea
\end{lemma}

\begin{proof}[Proof of Lemma \ref{lemma:intermadiaryresult:eq:Dafermos-Rodn1}]
     We  make use of Proposition \ref{prop:QC-general-multiplier2} according to which,
    \beaa
\EE[X, w , M]&=&\frac  1 2  f_p' |\ec_4(\Psi)|^2 +\frac 1 2 \left(\frac{2f_p}{r}- f_p'\right) \QQ_{34}+\err\left(\ep, \frac{m}{r};  f_p\right)[\Psi] \\
&=& r^{p-1}\left[ \frac{p}{2} |\ec_4(\Psi)|^2+\frac 1 2 (2-p) (      |\nabb\Psi|^2 + V|\Psi|^2 )\right]+\err\left(\ep, \frac{m}{r};  f_p\right)[\Psi] \\
&\ge & r^{p-1}\left[  \frac{p}{2}  |\ec_4(\Psi)|^2+\frac 1 2 (2-p) (|\nabb \Psi|^2 + r^{-2}|\Psi|^2) \right]+\err\left(\ep, \frac{m}{r};  f_p\right)[\Psi] 
\eeaa
where, 
\beaa
\err\left(\ep, \frac{m}{r}, f_p\right)(\Psi)&=&  r^p  O\left(\frac{m}{r^2} \right) \left[  |e_4 \Psi|^2+   r^{-2} \Psi|^2\right] +  r^p O(\ep)  w_{1,1} \left(    |e_4\Psi|^2+  |\nabb \Psi|^2  +r^{-2} |\Psi|^2  \right) \\
&+&    r^pO(\ep)  w_{2,1/2}\Big(  |e_3\Psi |(|e_4\Psi|+r^{-1}|\nabb\Psi|)+  |\nabb \Psi|^2  +r^{-2} |\Psi|^2\Big)\\
&\les& \err\left(\frac m r\right) +\err(\ep),\\
\err\left(\frac m r\right) &=& O\left(\frac m r\right) r^{p-1}    \left[  |\ec_4 \Psi|^2+   r^{-2} |\Psi|^2\right], \\
\err(\ep)&=& O(\ep)  r^{p-1} \left(    |\ec_4\Psi|^2+  |\nabb \Psi|^2  +r^{-2} |\Psi|^2       +r^{-2} |e_3\Psi|^2  \right).
\eeaa
Thus, 
\beaa
\int_{\MM_{\ge R}(\tau_1,\tau_2)  }\EE&\ge &   \int_{\MM_{\ge R}(\tau_1,\tau_2)  } r^{p-1}\Big(\frac{p}{2}  |\ec_4(\Psi)|^2+\frac 1 2 (2-p) (|\nabb \Psi|^2 + r^{-2}|\Psi|^2) \Big)\\
&-& O\left(\frac{m}{R} \right) \int_{\MM_{\ge R}(\tau_1,\tau_2)  } r^{p-1}    \left[  |\ec_4 \Psi|^2+   r^{-2} |\Psi|^2\right] \\
&-&O(\ep) \int_{\MM_{\ge R}(\tau_1,\tau_2)  } r^{p-1}\left(  |\ec_4\Psi|^2+  |\nabb \Psi|^2+r^{-2}|\Psi|^2 \right)\\
&-&O(\ep) \int_{\MM_{\ge R}(\tau_1,\tau_2)  } r^{p-3}|e_3\Psi|^2.
\eeaa
For $  \de \le p\le 2-\de$, $ R\gg \frac{m}{\de} $ and  $\ep\ll \de $  we    can absorb                 all  error terms except the last,  i.e.            
\beaa
\int_{\MM_{\ge R}(\tau_1,\tau_2)  }\EE&\ge &\frac 1 4   \int_{\MM_{\ge R}(\tau_1,\tau_2)  } r^{p-1}\Big( p   |\ec_4(\Psi)|^2+ (2-p) (|\nabb \Psi|^2 + r^{-2}|\Psi|^2)\Big)\\
&-&O(\ep) \int_{\MM_{\ge R}(\tau_1,\tau_2)  } r^{p-3}|e_3\Psi|^2.
\eeaa
 Note   also  that    for all  $\de\le p\le  2-\de$ we  have,
\beaa
 \int_{\MM_{\ge R}(\tau_1,\tau_2)  } r^{p-3}  |e_3\Psi|^2\les \Morr(\tau_1,\tau_2).
\eeaa
Hence,  for all $\de \le p\le 2-\de$  and $R\gg \frac m \de $, $\ep\ll \de$,
 \beaa
 \int_{\MM_{\ge R}(\tau_1,\tau_2)  }\EE&\ge &\frac 1  4  \int_{\MM_{\ge R}(\tau_1,\tau_2)  }r^{p-1}\Big(  p   |\ec_4(\Psi)|^2+ (2-p) (|\nabb \Psi|^2 + r^{-2}|\Psi|^2)\Big)\\
 &-&O(\ep) \Morr[\Psi](\tau_1, \tau_2)\nn
 \eeaa
as desired.
\end{proof}

Combining \eqref{eq:Dafermos-Rodn1}  with Lemma \ref{lemma:Daf-Rodnianski1},     we deduce,
\bea
\label{eq:Daf-Rodnianski2}
 K_{\ge R} (\tau_1, \tau_2)+ \Bdot_{p, R}[\Psi](\tau_1, \tau_2)  & \les&  R^{p+2}\Big(E[\Psi](\tau_1)    +J_p[N, \psi](\tau_1,\tau_2)   \\
\nn&&  +O(\ep)\Bdot^s_{\de\,; \,4m_0}[\psi](\tau_1, \tau_2)    \Big). 
\eea

{\bf Step 2.} (Boundary terms  for $r\ge R$.)
Recall  that according to Proposition \ref{prop:QC-general-multiplier2},
 \beaa
\PP\c e_4&=&f_p |\ec_4\Psi|^2-\frac 1 2 r^{-2} e_4( r f_p |\Psi|^2)+ O(\ep r^{-1} ) f_p (|e_4\Psi|^2+r^{-2} |\Psi|^2),  
\eeaa
and according to Lemma \ref{lemma:contractionofmathcalPwithunitnormaSigma*forlaterintegrationbyparts}
\beaa
\mathcal{P}\cdot N_{\Si_*} &=& \frac{1}{2}f \QQ_{34} +\frac{1}{2}f (\ec_4\Psi)^2 +\frac  12\div_{\Si_*}\Big(r^{-1}  f |\Psi|^2\nu_{\Si_*}\Big)  \\
&&+O(mr^{-1}+\ep) (|f|+r|f'|)(|e_4\Psi|^2+|\nabb\Psi|^2+r^{-2} |\Psi|^2)
\eeaa

Recalling the definition of 
$$K_{\ge R} =\int_{\Si_{\ge R}  (\tau_2)}   \PP\c  e_4        +\int_{\Sigma_*(\tau_1,\tau_2)}\PP \c N_{\Si_*} -\int_{\Si_{\ge R} (\tau_1)}   \PP\c   e_4$$ 
we write,
    \beaa
    K_{\ge R}&=&\int_{\Si_{\ge R} (\tau_2)} f_p |\ec_4\Psi|^2 +\frac{1}{2} \int_{\Sigma_*(\tau_1, \tau_2)}  r^p  \Big(\QQ_{34} + (\ec_4\Psi)^2\Big)     -\int_{\Si_{\ge R}(\tau_1)} f_p (\ec_4\Psi)^2\\
    &-&\frac 1 2 \int_{\Si_{\ge R} (\tau_2)}  r^{-2} e_4( r f_p |\Psi|^2)+\frac 1 2  \int_{\Sigma_*(\tau_1, \tau_2)}\div_{\Si_*}\Big(r^{-1}  f |\Psi|^2\nu_{\Si_*}\Big)\\
    &+&\frac 1 2 \int_{\Si_{\ge R} (\tau_1)}  r^{-2} e_4( r f_p |\Psi|^2)+  \int_{\Sigma_*(\tau_1, \tau_2)}O(mr^{-1}+\ep) r^{p-2} (|e_4\Psi|^2+|\nabb\Psi|^2+r^{-2} |\Psi|^2)\\
    &+& O(\ep) \left(\int_{\Si_{\ge R} (\tau_2)} r^{p-1} (|e_4\Psi|^2+r^{-2} |\Psi|^2)  -\int_{\Si_{\ge R} (\tau_1)} r^{p-1}  (|e_4\Psi|^2+r^{-2} |\Psi|^2)\right).
    \eeaa
    Now,   the following integrations by parts hold true
  \beaa
   &&\int_{\Si_{\ge R} (\tau)}  r^{-2} e_4( r f_p |\Psi|^2)\\
    &=& \int_{r\geq R}\left(\int_{S_r}r^{-2} e_4( r f_p |\Psi|^2)\right)\frac{1}{e_4(r)}\\
   &=& \int_{r\geq R}\frac{1}{e_4(r)}e_4\left(\int_{S_r}r^{-1} f_p |\Psi|^2\right) - \int_{\Si_{\ge R} (\tau)} \left(e_4\left(r^{-2}\right)+\ka r^{-2}\right)r f_p |\Psi|^2\\
      &=& \int_{S_*(\tau)}r^{p-1}|\Psi|^2 - \int_{S_R(\tau)}r^{p-1}|\Psi|^2 - \int_{\Si_{\ge R} (\tau)}\left(\ka -\frac{2e_4(r)}{r}\right)r^{-1} f_p |\Psi|^2\\
         &=& \int_{S_*(\tau)}r^{p-1}|\Psi|^2  - \int_{S_R(\tau)}r^{p-1}|\Psi|^2 +O(\ep)\int_{\Si_{\ge R} (\tau)}  r^{p-3}  |\Psi|^2
     \eeaa
  and 
  \beaa
   \int_{\Sigma_*(\tau_1, \tau_2)}\div_{\Si_*}\Big(r^{-1}  f |\Psi|^2\nu_{\Si_*}\Big) &=& \int_{S_*(\tau_2)}r^{p-1}|\Psi|^2 -\int_{S_*(\tau_1)}r^{p-1}|\Psi|^2
  \eeaa
  where $S_*(\tau)$ denotes the 2-sphere $\Si_*\cap\Sigma(\tau)$. Note that the boundary terms cancel, except the one on $r=R$, and hence
   \beaa
    K_{\ge R}&=&\int_{\Si_{\ge R} (\tau_2)} f_p |\ec_4\Psi|^2 +\frac{1}{2} \int_{\Sigma_*(\tau_1, \tau_2)}  r^p  \Big(\QQ_{34} + (\ec_4\Psi)^2\Big)     -\int_{\Si_{\ge R}(\tau_1)} f_p |\ec_4\Psi|^2\\
    &+&  \int_{\Sigma_*(\tau_1, \tau_2)}O(mr^{-1}+\ep) r^{p-2} (|e_4\Psi|^2+|\nabb\Psi|^2+r^{-2} |\Psi|^2)\\
    &+&O(\ep) \left(\int_{\Si_{\ge R} (\tau_2)} r^{p-1} (|e_4\Psi|^2+r^{-2} |\Psi|^2)  -\int_{\Si_{\ge R} (\tau_1)} r^{p-1}  (|e_4\Psi|^2+r^{-2} |\Psi|^2)\right)\\
    &+&  \frac{1}{2}\int_{S_R(\tau_2)}r^{p-1}|\Psi|^2 -\frac{1}{2} \int_{S_R(\tau_1)}r^{p-1}|\Psi|^2.
    \eeaa
    Using 
  \beaa
  Q_{34}+|\ec_4\Psi|^2 &=& |\nabb\Psi|^2+\frac{4\Up}{r^2}|\Psi|^2+\left|e_4\Psi+\frac{1}{r}\Psi\right|^2\\
  &=& |\nabb\Psi|^2+\frac{4\Up}{r^2}|\Psi|^2+(e_4\Psi)^2+\frac{1}{r^2}|\Psi|^2+\frac{2}{r}\Psi\cdot e_4(\Psi)\\
  &\geq &  |\nabb\Psi|^2+\frac{4\Up-3}{r^2}|\Psi|^2+\frac{2}{3}|e_4\Psi|^2
  \eeaa  
    and the fact that $4\Up\geq 3+2/3$ for $r\geq R$ and $R$ large enough, we infer
     \bea\lab{eq:intermadiaryesitmatefortheboundarytermsintheproofoftherpweightedestimate}
\nn      K_{\ge R}&\ge &\frac 1 2\left( \int_{\Si_{\ge R} (\tau_2)}r^p |\ec_4\Psi|^2 -   \int_{\Si_{\ge R} (\tau_1)}r^p |\ec_4\Psi|^2  +\dot{F}_p[\Psi](\tau_1,\tau_2) \right) \\
\nn     &+&O(\ep) \left(\int_{\Si_{\ge R} (\tau_2)} r^{p-3} |\Psi|^2 -\int_{\Si_{\ge R} (\tau_1)} r^{p-3} |\Psi|^2\right)\\
&+&   \frac{1}{2}\int_{S_R(\tau_2)}r^{p-1}|\Psi|^2 -\frac{1}{2} \int_{S_R(\tau_1)}r^{p-1}|\Psi|^2 
      \eea

Next, recall  that according to Proposition \ref{prop:QC-general-multiplier2},  we have    
 \beaa
\PP\c e_4 \ge \frac 1   8 r^{p-2}  (p-1)^2  |\Psi|^2  -\frac p 2 r^{-2}  e_4(  r f_p  |\Psi|^2).
\eeaa
We infer
\beaa
\int_{\Si_{\ge R}  (\tau_2)}   \PP\c  e_4 &\geq & \frac 1   8 \int_{\Si_{\ge R}  (\tau_2)} r^{p-2}  (p-1)^2  |\Psi|^2  -\frac p 2 \int_{\Si_{\ge R}  (\tau_2)}r^{-2}  e_4(  r f_p  |\Psi|^2).
\eeaa
Integrating by parts similarly as before, we infer
\beaa
\int_{\Si_{\ge R}  (\tau_2)}   \PP\c  e_4 &\geq & \frac 1   8 \int_{\Si_{\ge R}  (\tau_2)} r^{p-2}  (p-1)^2  |\Psi|^2  
 -\frac{p}{2}\int_{S_*(\tau)}r^{p-1}|\Psi|^2 \\
 && +\frac{p}{2} \int_{S_R(\tau)}r^{p-1}|\Psi|^2 +O(\ep)\int_{\Si_{\ge R} (\tau)}  r^{p-3}  |\Psi|^2.
\eeaa
Arguing as for the proof of \eqref{eq:intermadiaryesitmatefortheboundarytermsintheproofoftherpweightedestimate} except for the boundary term on $\Si_{\ge R}  (\tau_2)$ for which we use the above estimate, we deduce
    \beaa
\nn      K_{\ge R}&\ge & \frac 1   8 \int_{\Si_{\ge R}  (\tau_2)} r^{p-2}  (p-1)^2  |\Psi|^2  + \frac 1 2\left(  -   \int_{\Si_{\ge R} (\tau_1)} r^p |\ec_4\Psi|^2  +\dot{F}_p[\Psi](\tau_1,\tau_2) \right) \\
\nn     &+&O(\ep) \left(\int_{\Si_{\ge R} (\tau_2)} r^{p-1} (|e_4\Psi|^2+r^{-2} |\Psi|^2)  -\int_{\Si_{\ge R} (\tau_1)} r^{p-1}  (|e_4\Psi|^2+r^{-2} |\Psi|^2)\right)\\
&+&  \frac{1-p}{2}\int_{S_*(\tau_2)}r^{p-1}|\Psi|^2 + \frac{p}{2}\int_{S_R(\tau_2)}r^{p-1}|\Psi|^2 -\frac{1}{2} \int_{S_R(\tau_1)}r^{p-1}|\Psi|^2. 
      \eeaa

We first focus on the case $\de\leq p\leq 1-\de$, in which case the previous estimate yields
    \beaa
\nn      K_{\ge R}&\ge &  \frac{\de^2}{8} \int_{\Si_{\ge R}  (\tau_2)} r^{p-2}|\Psi|^2 + \frac 1 2\left(  -   \int_{\Si_{\ge R} (\tau_1)} r^p |\ec_4\Psi|^2  +\dot{F}_p[\Psi](\tau_1,\tau_2) \right) \\
\nn     &+&O(\ep) \left(\int_{\Si_{\ge R} (\tau_2)} r^{p-1} (|e_4\Psi|^2+r^{-2} |\Psi|^2)  -\int_{\Si_{\ge R} (\tau_1)} r^{p-3} |\Psi|^2\right)\\
&+&   \frac{1}{2}\int_{S_R(\tau_2)}r^{p-1}|\Psi|^2 -\frac{1}{2} \int_{S_R(\tau_1)}r^{p-1}|\Psi|^2. 
      \eeaa
Together with \eqref{eq:intermadiaryesitmatefortheboundarytermsintheproofoftherpweightedestimate} and the fact that $\ep\ll \de^2$ by assumption, we infer in view of the definition of $\dot{E}_{p, R}[\Psi]$ for $\de\leq p\leq 1-\de$,
\beaa
\dot{E}_{p, R}[\Psi](\tau_2)+\dot{F}_p[\Psi](\tau_1,\tau_2) &\les& K_{\ge R}+\dot{E}_{p, R}[\Psi](\tau_1)+\int_{S_R(\tau_2)}r^{p-1}|\Psi|^2.
\eeaa
Together with \eqref{eq:Daf-Rodnianski2}, we deduce for $\de\leq p\leq 1-\de$ 
\beaa
&&\dot{E}_{p, R}[\Psi](\tau_2)+\dot{F}_p[\Psi](\tau_1,\tau_2)+ \Bdot_{p, R}[\Psi](\tau_1, \tau_2)\\
 &\les& \dot{E}_{p, R}[\Psi](\tau_1)+\int_{S_R(\tau_2)}r^{p-1}|\Psi|^2+R^{p+2}\Big(E[\Psi](\tau_1)   +J_p[N, \psi](\tau_1,\tau_2)\\
 &&   +O(\ep)\Bdot^s_{\de\,; \,4m_0}[\psi](\tau_1, \tau_2)       \Big).
\eeaa
In view of the improved Morawetz  Theorem  \ref{Thm:Morawetz-s}, and thanks also to the term $\Bdot_{p, R}[\Psi](\tau_1, \tau_2)$ on the left hand side, we may absorb the term $O(\ep)\Bdot^s_{\de\,; \,4m_0}[\psi](\tau_1, \tau_2)       \Big)$ and obtain 
\bea\lab{eq:rpewightedestimatefordeltalessthanplessthan1minusdeltaintheproof}
\nn&&\dot{E}_{p, R}[\Psi](\tau_2)+\dot{F}_p[\Psi](\tau_1,\tau_2)+ \Bdot_{p, R}[\Psi](\tau_1, \tau_2)\\
 &\les& R^{p+2}\Big(E_p[\Psi](\tau_1)    +J_p[N, \psi](\tau_1,\tau_2)       \Big)
\eea
which is the desired estimate in the case  $\de\leq p\leq 1-\de$.

Finally, we focus on the remaining case, i.e.  $1-\de\leq p\leq 2-\de$. Combining \eqref{eq:intermadiaryesitmatefortheboundarytermsintheproofoftherpweightedestimate} and \eqref{eq:Daf-Rodnianski2}, arguing as in the proof of \eqref{eq:rpewightedestimatefordeltalessthanplessthan1minusdeltaintheproof}, and in view of the definition of $\dot{E}_{p, R}[\Psi]$ for $1-\de\leq p\leq 2-\de$, we obtain 
\beaa
\nn&&\dot{E}_{p, R}[\Psi](\tau_2)+\dot{F}_p[\Psi](\tau_1,\tau_2)+ \Bdot_{p, R}[\Psi](\tau_1, \tau_2)\\
 &\les& R^{p+2}\left(E_p[\Psi](\tau_1)     +J_p[N, \psi](\tau_1,\tau_2)      \right)\\
 && +O(\ep)\int_{\Si_{\ge R} (\tau_2)} r^{p-3} |\Psi|^2+\int_{\Si_{\ge R} (\tau_2)} r^{-1-\de} |\Psi|^2\\
 &\les& R^{p+2}\left(E_p[\Psi](\tau_1)     +J_p[N, \psi](\tau_1,\tau_2)       \right)+E_{1-\de}[\Psi](\tau_2)
\eeaa
where we also used the fact that $p\leq 2-\de$ so that $p-3\leq -1-\de$. Together with the fact that
\beaa
\dot{E}_{p, R}[\Psi](\tau)\geq \dot{E}_{1-\de, R}[\Psi](\tau)\textrm{ for }p\geq 1-\de
\eeaa
and \eqref{eq:rpewightedestimatefordeltalessthanplessthan1minusdeltaintheproof}, we infer
\beaa
\dot{E}_{p, R}[\Psi](\tau_2)+\dot{F}_p[\Psi](\tau_1,\tau_2)+ \Bdot_{p, R}[\Psi](\tau_1, \tau_2)\les R^{p+2}\left(E_p[\Psi](\tau_1)     +J_p[N, \psi](\tau_1,\tau_2)     \right)
\eeaa
for all $\de\leq p\leq 2-\de$ as desired. This concludes the proof of Theorem \ref{theorem:Daf-Rodn1-psi-s}.


\section{Higher  Weighted  Estimates}\lab{sec:AnArGatypeestimates}


We use a variation of   the method of \cite{AnArGa}  to derive slightly stronger  weighted   estimates. This allows us to prove Theorem \ref{theorem:Daf-Rodn-estim2-psic} for $s=0$ in section \ref{sec:proofoftheorem:Daf-Rodn-estim2-psic}. The proof for higher order derivatives $s\le k_{small}+29$ will be provided in section \ref{sec:proofoftheorem:Daf-Rodn-estim2-psic}. 

As in the  previous    section we  rely on the assumptions  \eqref{eq:assumptions-Moraw1-2'}--\eqref{eq:assumptions-Moraw4-2'} to which we add,

{\bf RP5. }  The  assumptions  {\bf RP0}--{\bf RP4} hold true  for one extra derivative with respect to 
  $\dk$.
  
{\bf RP6.} $e_4(m)$ satisfies the following improvement of {\bf RP4}  
\bea
|\dk^{\leq 2}e_4(m)| &\les& \ep w_{2,1}.
\eea


\subsection{Wave equation for $\psic$}


\begin{proposition}
\label{square-psic-modified}
Assume  $\psi$  verifies   $\square_2 \psi= -\ka\kab\psi +N$.
 Then $\psic=f_2 \ec_4\psi$  verifies:
 \begin{enumerate}
 \item In the region $r\ge 6m_0 $,
\beaa
(\square_2+\ka\kab)\psic &=& r^2\left(e_4(N) + \frac{3}{r}N\right)  +\frac{2}{r}\left(1-\frac{3m}{r}\right) e_4\psic   +O(r^{-2})\dk^{\leq 1}\psi\\
&&+r\Ga_b e_4\dk\psi +\dk^{\leq 1}(\Ga_b)\dk^{\leq 1}\psi  +r\dk^{\leq 1}(\Ga_g)e_3\psi +\dk^{\leq 1}(\Ga_g)\dk^2\psi.
\eeaa

\item  In the region $4m_0\leq r\leq 6m_0$,
\beaa
(\square_2+\ka\kab)\psic &=& f_2\left(e_4(N) + \frac{3}{r}N\right)+O(1)\dk^{\leq 2}\psi.
\eeaa
\end{enumerate}

\end{proposition}

The proof of Proposition \ref{square-psic-modified} is postponed to Appendix \ref{appendix:proofof:propsquare-psic-modified}.


  \subsection{The $r^p$ weighted estimates for $\psic$}\lab{sec:proofoftheorem:Daf-Rodn-estim2-psic:s=0}


The goal of this section is to prove Theorem \ref{theorem:Daf-Rodn-estim2-psic} in the case $s=0$. The proof for higher order derivatives $s\le k_{small}+29$ will be provided in section \ref{sec:proofoftheorem:Daf-Rodn-estim2-psic}. 

\begin{proof}[Proof of Theorem \ref{theorem:Daf-Rodn-estim2-psic} in the case $s=0$] We write, in accordance to Proposition \ref{square-psic-modified} 
\beaa
\square\psic -V\psic&=&\check{N}+ f_2 \left(e_4 +\frac 3  r \right) N
\eeaa
where,
\bea
\check N=
\begin{cases}
&  \frac{2}{r}\left(1-\frac{3m}{r}\right) e_4\psic   +O(r^{-2})\dk^{\leq 1}\psi\\
 & +r\Ga_b e_4\dk\psi +\dk^{\leq 1}(\Ga_b)\dk^{\leq 1}\psi  +r\dk^{\leq 1}(\Ga_g)e_3\psi +\dk^{\leq 1}(\Ga_g)\dk^2\psi, \quad  r\ge 6m_0,\\
 \\
& O(1)\dk^{\leq 2}\psi, \qquad\qquad \quad \quad\,\, 4m_0\le r\le  6m_0.
\end{cases}
\eea
We apply the first part of  Proposition \ref{prop:QC-general-multiplier2}  to  $\psi$ replaced  by  $ \psic$.  This yields, using also \eqref{eq:modified-div},
  \beaa
    \Div \PP_q(\psic)&=& \EE_q(\psic)   + f_q  \ec_4 \psic\c   \check{N}+  f_{q}   \ec_4 \psic\c  f_2 \left(e_4 +\frac 3  r \right) N,
    \eeaa
    where, with $f=f_q$,   $ X_q=  f_q e_4, w_q= \frac{2f_q}{r}, M_q=2r^{-1} f_q'  e_4$,
  \beaa
 \EE_q(\psic)&=& \EE[X_q, w_q , M_q]=\frac  1 2  f_q' |\ec_4(\psic)|^2 +\frac 1 2 \left(\frac{2f_q}{r}- f'_q\right) \QQ_{34}(\psic) +\err_q(\psic),\\
 \err_q(\psic):&=&\err\left(\ep, \frac{m}{r};  f_q\right)[\psic] \\
&= & O\left(\frac{m}{r^2} \right)  r^q |e_4 \psic|^2 +O  \left(\frac{m}{r^4}  +\ep  w_{3,1} \right) r^q|\psic|^2+ O(\ep)  w_{1,1} r^q   \left( |e_4\psic|^2+  |\nabb \psic|^2  +r^{-2} |\psic|^2 \right)\\
&+ &  O(\ep) w_{2, 1/2}  r^q  \Big( \left(|e_4\psic|+r^{-1}|\nabb \psic|  \right) |e_3\psic |+  |\nabb \psic|^2  +r^{-2} |\psic|^2\Big),\\
 \PP_k(\psic)&=& \PP[X_q, w_q,   M_q](\psic).
 \eeaa

 We then integrate      on the domain  $\MM(\tau_1, \tau_2)$   to derive, exactly as  in the proof of Theorem \ref{theorem:Daf-Rodn1-psi-s} (see section  \ref{sec:proofoftheorem:Daf-Rodn1-psi-s}),  
     \bea
     \label{identity:enhanced-Daf-Rodn}
\nn  && \int_{\Si (\tau_2)}   \PP_q\c e_4  +\int_{\Sigma_*(\tau_1,\tau_2)}\PP_q \c N_{\Si_*} +\int_{\MM(\tau_1, \tau_2)}\left( \EE_q +f_q \ec_4 \psic \check{N}\right)     \\
   &=&\int_{\Si (\tau_1)}   \PP_q\c   e_4 -  \int_{\MM(\tau_1,\tau_2)}    f_{q}   \ec_4 \psic\c  f_2 \left(e_4 +\frac 3  r \right) N.
     \eea
     All terms can be treated exactly as  in the proof of Theorem \ref{theorem:Daf-Rodn1-psi-s},  except for the bulk term, i.e. we obtain the following analog of \eqref{eq-thm:Daf-Rodn-estim4}
  \beaa 
&&    \dot{E}_{q\,;\, R}[\psic](\tau_2)+   \int_{\MM(\tau_1, \tau_2)} (\EE_q+ r^q \ec_4(\psic) \check{N})   +  \dot{F}_q[\psic] (\tau_1, \tau_2) \\ 
    &\les&   E_{q}[\psic](\tau_1)+   J_q\left[\psic, f_2 \left(e_4 +\frac 3  r \right) N\right]( \tau_1,\tau_2).
\eeaa
Since all terms for $r\leq R$ can be controlled by one derivative of $\psi$, we infer 
  \bea\lab{identity:enhanced-Daf-Rodn:notquiteseebelow}
\nn&&    \dot{E}_{q}[\psic](\tau_2)+\Morr[\psic](\tau_1,\tau_2) +  \int_{\MM_{\ge R}(\tau_1, \tau_2)} (\EE_q+ r^q \ec_4(\psic) \check{N})   +  \dot{F}_q[\psic] (\tau_1, \tau_2) \\ 
    &\les&   E_{q}[\psic](\tau_1)+   \check{J}_q[\psic, N]( \tau_1,\tau_2) +R^{q+3}(E^1[\psi](\tau_2)+\Morr^1[\psi](\tau_1,\tau_2)).
\eea
Also, since $\de\leq \max(q,\de)\leq 1-\de$, we have in view of Theorem \ref{theorem-combinedMor-r-weighted} in the case $s=1$\footnote{The proof of Theorem \ref{theorem-combinedMor-r-weighted} for higher derivatives $s\geq 1$, even though proved later in section \ref{sec:proofoftheorem-combinedMor-r-weighted}, is in fact independent of the proof of Theorem   \ref{theorem:Daf-Rodn-estim2-psic} and can thus be invoked here.}
 \bea\lab{Ijustrecallrpweightedestimatesastheyareusedfortheproof}
\nn  &&\sup_{\tau\in[\tau_1,\tau_2] }   E^1_{\max(q,\de)} [\psi](\tau)+   B^1_{\max(q,\de)}[\psi](\tau_1, \tau_2) \\
    &\les& 
          E^1_{\max(q,\de)}[\psi](\tau_1)+   J^1_{\max(q,\de)}[\psi, N]( \tau_1,\tau_2),
  \eea
In view of \eqref{identity:enhanced-Daf-Rodn:notquiteseebelow} and \eqref{Ijustrecallrpweightedestimatesastheyareusedfortheproof}, it thus only remains to  estimate  the integral
\beaa
\int_{\MM_{\ge R}(\tau_1, \tau_2)} (\EE_q+ r^q \ec_4(\psic) \check{N}),
\eeaa
 i.e. we need to derive the analog of \eqref{eq:Dafermos-Rodn1} used in the proof of Theorem \ref{theorem:Daf-Rodn1-psi-s}. This is achieved in Proposition \ref{proposition:enhanced Dafermos-Rodn} below, which together with \eqref{identity:enhanced-Daf-Rodn:notquiteseebelow} and \eqref{Ijustrecallrpweightedestimatesastheyareusedfortheproof} immediately yields 
the proof of Theorem   \ref{theorem:Daf-Rodn-estim2-psic} in the case $s=0$.
   \end{proof}

     \begin{proposition}\label{proposition:enhanced Dafermos-Rodn}
     The following estimate holds true,
     \bea
     \bsplit
     \int_{\MM_{\ge R}(\tau_1, \tau_2)} (\EE_q+ r^q \ec_4(\psic) \check{N})&\ge  \frac 1 8   \int_{\MM_{\ge R}(\tau_1, \tau_2)} r^{q-1}\left(   (2+q) | \ec_4 \psic |^2 + (2-q)|\nabb\psic|^2+ 2 r^{-2}|\psic|^2\right)\\   
 &- O(\ep)  \sup_{\tau_1\le \tau\le \tau_2} \Edot_{q,R}[\psic](\tau)    \\
     &- O(1)\left(E_{\max(q,\de)}^1[\psi](\tau_1)+J_{\max(q,\de)}^1[\psi,N]\right).
     \end{split}
     \eea
     \end{proposition}

   We now focus on the proof of Proposition \ref{proposition:enhanced Dafermos-Rodn}.   In view of the definition of $\check N$, we have for $r\ge R$,
   \beaa
    \check{N}   &=&  A_0+A_1 +A_2,\\
A_0&=&     \frac 2 r  e_4\psic=  \frac{2}{r} (\ec_4\psic- r^{-1}\psic),   \\
A_1&=&   -\frac{6m}{r^2}e_4\psic+  O(r^{-2})\dk^{\leq 1}\psi,  \\
A_2&=& \err[\square_\g\psic],\\
\err[\square_\g\psic] &=& r^2\Ga_g e_4\dk\psi+r\dk^{\leq 1}(\Ga_g)\dk^{\leq 1}\psi+\dk^{\leq 1}(\Ga_g)\dk^2\psi.
    \eeaa
    Also, recall that we have for $r\geq R$
          \beaa
 \EE_q(\psic)&=& \EE[X_q, w_q , M_q]=\frac  q 2  r^{q-1} |\ec_4(\psic)|^2 +\frac{2-q}{2}r^{q-1} \QQ_{34}(\psic) +\err_q(\psic).
 \eeaa
    Consequently, we write,
     \bea
     \begin{split}
     \label{equation: Daf-Rodn-enhabced1} 
      (\EE_q+ r^q \ec_4(\psic) \check{N})&=I_0+I_1   +I_2,\\
I_0:&=\frac  1 2  r^{q-1}\left(   q| \ec_4 \psic |^2 + (2-q)|\nabb\psic|^2 +4(2-q) r^{-2}\psic^2 \right)
+2  r^{q-1}\ec_4\psic  (  \ec_4\psic- r^{-1} \psic)\\ 
&=\frac  1 2  r^{q-1}\left(   (q+4) | \ec_4 \psic |^2 + (2-q)|\nabb\psic|^2 +4(2-q) r^{-2} \psic^2   -4r^{-1}    \ec_4\psic    \psic    \right),\\
I_1:&= r^{q-2}\ec_4(\psic )      \left[-6 m e_4 \psic +O(1)\dk^{\leq 1}\psi    \right]+O\left(\frac{m}{r}\right)r^{q-3}(\psic)^2,\\
I_2:&=\err_q(\psic)+r^q \ec_4(\psic)A_2.
\end{split}
     \eea

     We will rely on the following two lemmas.
     \begin{lemma}
     \label{lemma: Daf-Rodn-enhabced2} 
     The following lower bound estimate holds true for $q\leq 1-\de$ and $r\ge R$, where $R$ is sufficiently large,
      \bea
     \label{equation: Daf-Rodn-enhanced2} 
     I_0+I_1\ge \frac 1 4 r^{q-1}\left(   (2+q) | \ec_4 \psic |^2 + (2-q)|\nabb\psic|^2+ 2 r^{-2}|\psic|^2\right)-O(1)r^{q-3}(\dk^{\leq 1}\psi)^2.
     \eea
     \end{lemma}

\begin{lemma}
\label{lemma:Daf-Rodn-enhabced22} 
The following estimate holds true for the error  term $I_2$
\beaa
 \int_{\MM_{\ge R}(\tau_1,\tau_2)}|I_2|  &\les& \ep\sup_{\tau_1\le \tau\le \tau_2} \Edot_{q,R}[\psic](\tau)+ \left(\frac{m_0}{R}+\ep\right)\Bdot_{q,R}[\psic](\tau_1, \tau_2)\\
 &&+ \ep\left(\sup_{\tau_1\le \tau\le \tau_2}E^1_q[\psi](\tau)+ B^1_q[\psi](\tau_1, \tau_2)+J_q[\psi, N](\tau_1, \tau_2)\right).
\eeaa
\end{lemma}

We postpone the proof of Lemma \ref{lemma: Daf-Rodn-enhabced2} and Lemma \ref{lemma:Daf-Rodn-enhabced22} to finish the proof of  Proposition \ref{proposition:enhanced Dafermos-Rodn}.

\begin{proof}[Proof of Proposition \ref{proposition:enhanced Dafermos-Rodn}]
In view of Lemma \ref{lemma: Daf-Rodn-enhabced2} and Lemma \ref{lemma:Daf-Rodn-enhabced22}, we have
\beaa
\int_{\MM_{\ge R}(\tau_1, \tau_2)} (\EE_q+ r^q \ec_4(\psic) \check{N})&= &\int_{\MM_{\ge R}(\tau_1, \tau_2)}(I_0+I_1) +\int_{\MM_{\ge R}(\tau_1, \tau_2)} I_2\\
&\ge & \frac 1 4 \int_{\MM_{\ge R}(\tau_1, \tau_2)}  r^{q-1}\left(   (2+q) | \ec_4 \psic |^2 + (2-q)|\nabb\psic|^2+ 2 r^{-2}|\psic|^2\right)\\
&- &O(1) \int_{\MM_{\ge R}(\tau_1, \tau_2)} r^{q-3}(\psi)^2\\
&-& O(\ep)\sup_{\tau_1\le \tau\le \tau_2} \Edot_{q,R}[\psic](\tau)+ O\left(\frac{m_0}{R}+\ep\right)\Bdot_{q,R}[\psic](\tau_1, \tau_2)\\
 &-& O(\ep)\left(\sup_{\tau_1\le \tau\le \tau_2}E^1_q[\psi](\tau)+ B^1_q[\psi](\tau_1, \tau_2)+J_q[\psi, N](\tau_1, \tau_2)\right)
\eeaa
so that, since $1-\de<q\leq 1-\de$, and for $R$ sufficiently large and  small $\ep$,
\beaa
\int_{\MM_{\ge R}(\tau_1, \tau_2)} (\EE_q+ r^q \ec_4(\psic) \check{N})&\ge & \frac{1}{8}  \int_{\MM_{\ge R}(\tau_1, \tau_2)}  r^{q-1}\left(   (2+q) | \ec_4 \psic |^2 + (2-q)|\nabb\psic|^2+ 2 r^{-2}|\psic|^2\right)\\ 
&-& O(\ep)\sup_{\tau_1\le \tau\le \tau_2} \Edot_{q,R}[\psic](\tau)\\
&-& O(\ep)\left(\sup_{\tau_1\le \tau\le \tau_2}E^1_q[\psi](\tau)+ B^1_q[\psi](\tau_1, \tau_2)+J_q[\psi, N](\tau_1, \tau_2)\right).
\eeaa
In view of \eqref{Ijustrecallrpweightedestimatesastheyareusedfortheproof}, we infer
   \beaa
     \bsplit
     \int_{\MM_{\ge R}(\tau_1, \tau_2)} (\EE_q+ r^q \ec_4(\psic) \check{N})&\ge  \frac 1 8   \int_{\MM_{\ge R}(\tau_1, \tau_2)} r^{q-1}\left(   (2+q) | \ec_4 \psic |^2 + (2-q)|\nabb\psic|^2+ 2 r^{-2}|\psic|^2\right)\\   
 &- O(\ep)  \sup_{\tau_1\le \tau\le \tau_2} \Edot_{q,R}[\psic](\tau)    \\
     &- O(1)\left(E_{\max(q,\de)}^1[\psi](\tau_1)+J_{\max(q,\de)}^1[\psi,N]\right)
     \end{split}
     \eeaa
which concludes the proof.
\end{proof}

It finally remains to prove Lemma \ref{lemma: Daf-Rodn-enhabced2} and Lemma \ref{lemma:Daf-Rodn-enhabced22}.

    \begin{proof}[Proof of Lemma \ref{lemma: Daf-Rodn-enhabced2}]
     Note  that,
     \beaa
     &&( q+4)| \ec_4 \psic |^2 -4r^{-1}   ( \ec_4\psic )   \psic+4(2-q) r^{-2} |\psic|^2\\
     &=&(q+2)| \ec_4 \psic |^2+ (6-4q) r^{-2}|\psic|^2 +2\left( \ec_4\psic- r^{-1} \psic\right)^2\\
     &\ge &(q+2)| \ec_4 \psic |^2+ (6-4q) r^{-2}|\psic|^2\\
     &\geq& (q+2)| \ec_4 \psic |^2+ 2 r^{-2}|\psic|^2, 
     \eeaa
     where we used the fact that $q\leq 1-\de$.  Hence,
     \beaa
     I_0&\ge & \frac  1 2  r^{q-1}\left(   (2+q) | \ec_4 \psic |^2 + (2-q)|\nabb\psic|^2+ 2 r^{-2}|\psic|^2\right).
     \eeaa
    We also  have,
     \beaa
     I_1&\leq&O\left(\frac{m}{r}\right) r^{q-1}\left( |\ec_4\psic|^2 + r^{-2}|\psic|^2\right)+O(1)\left(r^{q-1}(\ec_4\psic)^2\right)^{\frac{1}{2}}\left(r^{q-3}(\dk^{\leq 1}\psi)^2\right)^{\frac{1}{2}}.
     \eeaa
     Thus if $m_0/R$ is sufficiently small, and since $q\leq 1-\de$,  we deduce, for $r\ge R$,
     \beaa
     I_0+I_1\ge \frac 1 4 r^{q-1}\left( (2+q) | \ec_4 \psic |^2 + (2-q)|\nabb\psic|^2+ 2 r^{-2}|\psic|^2\right)-O(1)r^{q-3}(\dk^{\leq 1}\psi)^2
     \eeaa
     as desired.
\end{proof}

\begin{proof}[Proof of Lemma \ref{lemma:Daf-Rodn-enhabced22}]
Recall that
\beaa
I_2&=&\err_q(\psic)+r^q \ec_4(\psic)A_2,\\
A_2&=& \err[\square_\g\psic],\\
\err[\square_\g\psic] &=& r\Ga_b e_4\dk\psi +\dk^{\leq 1}(\Ga_b)\dk^{\leq 1}\psi  +r\dk^{\leq 1}(\Ga_g)e_3\psi +\dk^{\leq 1}(\Ga_g)\dk^2\psi,\\
 \err_q(\psic)&= & O\left(\frac{m}{r^2} \right)  r^q |e_4 \psic|^2 +O  \left(\frac{m}{r^4}  +\ep  w_{3,1} \right) r^q|\psic|^2+ O(\ep)  w_{1,1} r^q   \left( |e_4\psic|^2+  |\nabb \psic|^2  +r^{-2} |\psic|^2 \right)\\
&+ &  O(\ep) w_{2, 1/2}  r^q  \Big( \left(|e_4\psic|+r^{-1}|\nabb \psic|  \right) |e_3\psic |+  |\nabb \psic|^2  +r^{-2} |\psic|^2\Big).
\eeaa
Hence,
\beaa
|I_2| &\les& \ep r^q|\ec_4(\psic)|\Big(\tau^{-1-\dec}\big(|\ec_4\dk^{\leq 1}\psi|+|\dk^{\leq 1}\psi|\big)+r^{-1}\tau^{-\frac{1}{2}-\dec}\big(|e_3\psi|+r^{-1}|\dk^2\psi|\big)\Big)\\
&& +\left(\frac{m}{r}+\ep\right)r^{q-1}\big(|\ec_4\psic|^2+|\nabb\psic|^2+r^{-2}|\psic|^2\big)+\ep r^{q-2}\tau^{-\frac{1}{2}-\dec}|e_4\psic| |e_3\psic |\\
&&+\ep r^{q-3}|\nabb \psic||e_3\psic |.
\eeaa
This yields, using $q\leq 1-\de$,
\beaa
 \int_{\MM_{\ge R}(\tau_1,\tau_2)}|I_2| &\les& \ep\left(\sup_{\tau_1\le \tau\le \tau_2} \Edot_{q,R}[\psic](\tau) \right)^\frac{1}{2} \left(\sup_{\tau_1\le \tau\le \tau_2}E^1_q[\psi](\tau)+ B^1_q[\psi](\tau_1, \tau_2)\right)^\frac{1}{2} \\
 && + \left(\frac{m_0}{R}+\ep\right)\Bdot_{q,R}[\psic](\tau_1, \tau_2)\\
 &&+\ep\left(\sup_{\tau_1\le \tau\le \tau_2} \Edot_{q,R}[\psic](\tau)+\Bdot_q[\psic](\tau_1, \tau_2) \right)^\frac{1}{2} \left(\int_{\MM_{\ge R}(\tau_1,\tau_2)}r^{q-4}|e_3\psic|^2\right)^\frac{1}{2}\\
 &\les& \ep\sup_{\tau_1\le \tau\le \tau_2} \Edot_{q,R}[\psic](\tau)+ \left(\frac{m_0}{R}+\ep\right)\Bdot_{q,R}[\psic](\tau_1, \tau_2)\\
 &&+ \ep\left(\sup_{\tau_1\le \tau\le \tau_2}E^1_q[\psi](\tau)+ B^1_q[\psi](\tau_1, \tau_2)\right) +\ep \int_{\MM_{\ge R}(\tau_1,\tau_2)}r^{q-4}|e_3\psic|^2.
\eeaa

Next, we estimate the term involving $e_3\psic$. For this we  need  to appeal to   the formula in  Lemma  \ref{lemma:formula-wave-rpsi}     which we recall below,
\beaa
\square_2\psi&=& -e_3e_4\psi+\lapp_2\psi+\left(2\omb-\frac{1}{2}\kab\right)e_4\psi-\frac{1}{2}\ka e_3\psi+2\eta e_\th\psi
\eeaa
We have for $r\geq 6m_0$
\beaa
e_3\psic &=& e_3(re_4(r\psi)) =re_3(re_4\psi)+e_3(r)e_4(r\psi)=r^2e_3e_4\psi+2re_3(r)e_4\psi+e_3(r)e_4(r)\psi\\
&=& r^2\left(-\square_2\psi+\lapp_2\psi+\left(2\omb-\frac{1}{2}\kab\right)e_4\psi-\frac{1}{2}\ka e_3\psi+2\eta e_\th\psi\right)+2re_3(r)e_4\psi+e_3(r)e_4(r)\psi
\eeaa
so that
\beaa
|e_3\psic| &\les& r^2|N|+r|e_3\psi|+|\dk^{\leq 2}\psi|
\eeaa
and hence
\beaa
\int_{\MM_{\ge R}(\tau_1,\tau_2)}r^{q-4}|e_3\psic|^2 &\les& \int_{\MM_{\ge R}(\tau_1,\tau_2)}r^{q-4}\Big(r^4|N|^2+r^2|e_3\psi|^2+|\dk^{\leq 2}\psi|^2\Big).
\eeaa
Since $q\leq 1-\de$, we infer
\beaa
\int_{\MM_{\ge R}(\tau_1,\tau_2)}r^{q-4}|e_3\psic|^2 &\les& \int_{\Mntrap(\tau_1,\tau_2)}r^{1-\de}|N|^2+B^1_q[\psi](\tau_1, \tau_2)\\
&\les& J_q[\psi, N](\tau_1, \tau_2)+B^1_q[\psi](\tau_1, \tau_2)
\eeaa
and thus
\beaa
 \int_{\MM_{\ge R}(\tau_1,\tau_2)}|I_2| 
 &\les& \ep\sup_{\tau_1\le \tau\le \tau_2} \Edot_{q,R}[\psic](\tau)+ \left(\frac{m_0}{R}+\ep\right)\Bdot_{q,R}[\psic](\tau_1, \tau_2)\\
 &&+ \ep\left(\sup_{\tau_1\le \tau\le \tau_2}E^1_q[\psi](\tau)+ B^1_q[\psi](\tau_1, \tau_2)\right) +\ep \int_{\MM_{\ge R}(\tau_1,\tau_2)}r^{q-4}|e_3\psic|^2\\
 &\les& \ep\sup_{\tau_1\le \tau\le \tau_2} \Edot_{q,R}[\psic](\tau)+ \left(\frac{m_0}{R}+\ep\right)\Bdot_{q,R}[\psic](\tau_1, \tau_2)\\
 &&+ \ep\left(\sup_{\tau_1\le \tau\le \tau_2}E^1_q[\psi](\tau)+ B^1_q[\psi](\tau_1, \tau_2)+J_q[\psi, N](\tau_1, \tau_2)\right).
\eeaa
which concludes the proof of Lemma \ref{lemma:Daf-Rodn-enhabced22}.
\end{proof}


 \section{Higher  Derivative      Estimates}\lab{esc:higherderivativesestimates}


We have proved, respectively in section \ref{section-Basic-rp} and section \ref{sec:proofoftheorem:Daf-Rodn-estim2-psic:s=0},  Theorem \ref{theorem-combinedMor-r-weighted} on basic weighted estimates (see Remark \ref{remark:infactcombinedtheoremconsequenceofbasicones}) and Theorem \ref{theorem:Daf-Rodn-estim2-psic} on higher weighted estimates only in the case $s=0$. In this section, we conclude the proof of these theorems by recovering higher order derivatives\footnote{Respectively $s\leq k_{small}+30$ in the case of Theorem \ref{theorem-combinedMor-r-weighted}, and  $s\leq k_{small}+29$ in the case of Theorem \ref{theorem:Daf-Rodn-estim2-psic}.} one by one.

 
\subsection{Basic assumptions}\lab{sec:necessaryassumptionstorecoveryhigherderivatives} 


Recall that any Ricci coefficient either belongs to  $\Ga_g$ or $\Ga_b$, where $\Ga_g$ and $\Ga_b$ are defined in section  \ref{subsection:Main assumptions-wave}. We   make use  of the following non sharp consequence of the estimates of Lemma \ref{le:interpolatedbootstrap}. 
We assume, concerning the Ricci coefficients
\beaa
 |\dk^k(\Ga_g)| &\les& \frac{\ep}{r^2u_{trap}^{1+\dec-2\de_0}} \quad\textrm{ for }k\leq k_{small}+30,\\
 |\dk^k(\Ga_b)| &\les& \frac{\ep}{ru_{trap}^{1+\dec-2\de_0}}\quad\textrm{ for }k\leq k_{small}+30,\\
 |\dk^k(\a, \b, \rhoc)| &\les& \frac{\ep}{r^3u_{trap}^{1+\dec-2\de_0}} \quad\textrm{ for }k\leq k_{small}+30,\\
 |\dk^k\aa|+r|\dk^k\bb| &\les& \frac{\ep}{ru_{trap}^{1+\dec-2\de_0}}\quad\textrm{ for }k\leq k_{small}+30,
\eeaa
where we recall that $\dec$ and $\de_0$ are such that we have in particular $0<2\de_0<\dec$. 


 \subsection{Strategy for recovering higher order derivatives}\lab{sec:strategyforrecoveringhigherorderderivatives}
 

So far, we have proved Theorem \ref{theorem-combinedMor-r-weighted} in the case $s=0$\footnote{Recall that Theorem \ref{theorem-combinedMor-r-weighted} in the case $s=0$ is obtained as a consequence of Theorem \ref{Thm:Morawetz-s} on Morawetz and energy estimates, and Theorem \ref{theorem:Daf-Rodn1-psi-s} on $r^p$-weighted estimates, see Remark \ref{remark:infactcombinedtheoremconsequenceofbasicones}.} in section  \ref{section-Basic-rp}, and Theorem \ref{theorem:Daf-Rodn-estim2-psic} on higher weighted estimates in the case $s=0$ in section \ref{sec:AnArGatypeestimates}. We now conclude the proof of these theorems by recovering higher order derivatives one by one. Since going from $s=0$ to $s=1$ is analogous to going from $s$ to $s+1$, we will in fact consider only the former. More precisely, we assume the following bounds proved respectively  in section section \ref{section-Basic-rp} and section \ref{sec:AnArGatypeestimates},
 \bea\lab{eq:recoverhigherderivativesforestimatewaveequationpsi:2}
  \sup_{\tau\in[\tau_1,\tau_2] }   E_{p} [\psi](\tau)+   B_{p}[\psi](\tau_1, \tau_2)  + F_{p}[\psi](\tau_1, \tau_2) \les 
          E_{p}[\psi](\tau_1)+   J_p[\psi, N]( \tau_1,\tau_2),
\eea
and 
    \bea\lab{eq:recoverhigherderivativesforestimatewaveequationpsi:3}
   \bsplit
  \sup_{\tau\in[\tau_1,\tau_2] }  E_q[\psic](\tau)+   B_q[\psic](\tau_1, \tau_2)      &\les 
           E_q [\psic](\tau_1)      +  \Jc_q[\psic, N] (\tau_1,\tau_2)  \\
&+    E\, ^1_{\max(q, \de)}[\psi](\tau_1)  + J^{1}_{\max(q, \de)}[\psi, N],
      &               \end{split}
\eea
and our goal is to prove the corresponding estimates for $s=1$. We will proceed as follows
\begin{enumerate}
\item we first commute the wave equation for $\psi$ and $\psic$ with $T$ and derive \eqref{eq:recoverhigherderivativesforestimatewaveequationpsi:2} for $T\psi$ instead of $\psi$, 
and \eqref{eq:recoverhigherderivativesforestimatewaveequationpsi:3} for $T\psic$ instead of $\psic$,

\item we then commute the wave equation for $\psi$ and $\psic$ with $r\ddd_2$ and derive 
\eqref{eq:recoverhigherderivativesforestimatewaveequationpsi:2} for $r\ddd_2\psi$ instead of $\psi$, 
and \eqref{eq:recoverhigherderivativesforestimatewaveequationpsi:3} for $r\ddd_2\psic$ instead of $\psic$,

\item we then use the wave equation satisfied by $\psi$ to derive an estimate for $R^2\psi$ in $r\leq 6m_0$\footnote{Note that any finite region in $r$ strictly containing the trapping region would suffice.} with a degeneracy at $r=3m$,

\item we then commute the wave equation for $\psi$ with $R$ and remove the degeneracy at $r=3m$ for $R^2\psi$,

\item we then commute the wave equation for $\psi$ with the redshift vectorfield $Y_\HH$ and derive   
\eqref{eq:recoverhigherderivativesforestimatewaveequationpsi:2} for $Y_\HH\psi$ instead of $\psi$,

\item we then commute the wave equation for $\psi$ and $\psic$ with $f_1e_4$ and derive   
\eqref{eq:recoverhigherderivativesforestimatewaveequationpsi:2} for $re_4\psi$ instead of $\psi$, 
and \eqref{eq:recoverhigherderivativesforestimatewaveequationpsi:3} for $f_1e_4\psic$ instead of $\psic$, where $f_1=r$ for $r\geq 6m_0$ and $f_1=0$ for $r\leq 4m_0$,

\item we finally gather all estimates and conclude.
\end{enumerate}

We will follow the above strategy  in section \ref{sec:proofoftheorem-combinedMor-r-weighted} to prove Theorem \ref{theorem-combinedMor-r-weighted}, and in section \ref{sec:proofoftheorem:Daf-Rodn-estim2-psic} to prove Theorem \ref{theorem:Daf-Rodn-estim2-psic}. To this end, we first derive several commutator identities and estimates.


 \subsection{Commutation formulas with the wave equation}\lab{sec:commutationformulasforthewaveequationhigherderivatives}
 


 \subsubsection{Commutation with $T$}
 

\begin{lemma}
We have, schematically, the following commutator formulae
\beaa
[T , e_4] = \Ga_g\dk,\quad [T , e_3] = \Ga_b\dk, \quad [T , \ddd_k] = \Ga_b\dk+\Ga_b, \quad [T , \dds_k] = \Ga_b\dk+\Ga_b.
\eeaa
\end{lemma}

\begin{proof}
Recall that we have
\beaa
\,[e_3, e_4]&=& 2 \omb e_4-2\om e_3+2(\eta-\etab) e_\th.
\eeaa
We infer 
\beaa
2[T , e_4] &=& [e_4+\Up e_3, e_4]\\
&=& \Up[e_3, e_4]-e_4(\Up)e_3\\
&=& \Up\left(2 \omb e_4-2\om e_3+2(\eta-\etab) e_\th\right)-e_4(\Up)e_3\\
&=& \Up\left(2 \omb e_4-2\left(\om +\frac{1}{2}\Up^{-1}e_4(\Up)\right)e_3+2(\eta-\etab) e_\th\right)\\
&=& (r^{-1}\Ga_b+\Ga_g)\dk\\
&=& \Ga_g\dk,
\eeaa
and
\beaa
2[T , e_3] &=& [e_4+\Up e_3, e_3]\\
&=& [e_4, e_3]-e_3(\Up)e_3\\
&=& -2 \omb e_4+2\om e_3-2(\eta-\etab) e_\th -e_3(\Up)e_3\\
&=& -2 \omb e_4+2\left(\om -\frac{1}{2}e_3(\Up)\right)e_3 -2(\eta-\etab) e_\th \\
&=& -2 \omb e_4+2\left(\om +\frac{1}{2}\Up^{-1}e_4(\Up) -\frac{1}{2\Up}T (\Up)\right)e_3 -2(\eta-\etab) e_\th \\
&=& (\Ga_g+\Ga_b)\dk\\
&=& \Ga_b\dk.
\eeaa

Next, recall in view of Lemma \ref{Le:comme3e4}, the following commutation formulae for reduced scalars.
\begin{enumerate} 
\item If $f\in \mathfrak{s}_k$,
\beaa
\bsplit
\,[\ddd_k, e_3] f&=\frac 1 2 \kab  \ddd_k f+\comb_k(f),\\
\comb_k(f)&=- \frac 1 2 \vthb  \dds_{k+1} f + (\ze-\eta) e_3 f - k \eta  e_3\Phi f -\xib( e_4  f  +k e_4(\Phi)  f )- k \bb f,\\
\,[\ddd_k, e_4] &=\frac 1 2 \ka  \ddd_k f+\com_k(f),\\
\com_k(f)&=- \frac 1 2 \vth  \dds_{k+1} f - (\ze+\etab) e_4 f - k \etab  e_4\Phi f -\xi( e_3  f  +k e_3(\Phi)  f )- k \b f.
\end{split}
\eeaa
\item If  $f\in \mathfrak{s}_{k-1}$
\beaa
\bsplit
\,[\dds_k, e_3]f &=\frac 1 2 \kab  \dds_k f+\comb^*_k(f),\\
\comb^*_k(f)&=- \frac 1 2 \vthb  \ddd_{k-1} f - (\ze-\eta) e_3 f  -(k-1)  \eta  e_3\Phi f +\xib(e_4  f  -(k-1)  e_4(\Phi)  f )\\
&- (k-1) \bb f,\\
\,[\dds_k, e_4]f &=\frac 1 2 \ka  \ddd_k f+\com^*_k(f),\\
\com^*_k(f)&=- \frac 1 2 \vth  \ddd_{k-1} f + (\ze+\etab) e_4 f - (k-1)  \etab  e_4\Phi f +\xi( e_3  f  -(k-1) e_3(\Phi)  f )\\
&- (k-1)  \b f.
\end{split}
\eeaa
\end{enumerate}
We infer, schematically,
\beaa
2[T , \ddd_k] &=& [e_4+\Up e_3, \ddd_k]\\
&=& [e_4, \ddd_k]+\Up[e_3, \ddd_k] - e_\th(\Up)e_3\\
&=& -\frac{1}{2}(\ka+\Up\kab)\ddd_k +\Ga_b\dk+r^{-1}\Ga_b+2e_\th\left(\frac{m}{r}\right)e_3\\
&=& \Ga_b\dk+\Ga_b.
\eeaa
The estimate for $[T , \dds_k]$ is similar and left to the reader. This concludes the proof of the lemma.
\end{proof}

\begin{lemma}
We have
\beaa
T (\ka) = \dk^{\leq 1}\Ga_g, \quad\quad T \left(2\omb -\frac 1 2 \kab\right) =\dk^{\leq 1}\Ga_b, \quad\quad T(K) = \dk^{\leq 1}\Ga_g.
\eeaa
\end{lemma}

\begin{proof}
We have
\beaa
2T (\ka) &=& (e_4+\Up e_3)\ka\\
&=& -\frac{1}{2}\ka^2-2\om\ka+2\ddd_1\xi+2(\etab+\eta+2\ze)\xi -\frac{1}{2}\vth^2\\
&& +\Up\left(-\frac{1}{2}\ka\kab+2\omb\ka+2\ddd_1\eta+2\rho-\frac{1}{2}\vth\vthb+2(\xi\xib+\eta^2)\right)\\
&=& -\frac{1}{2}\ka^2-2\om\ka -\frac{1}{2}\kab\ka\Up +2\omb\ka\Up+2\Up\rho\\
&&+2\ddd_1\xi+2\Up\ddd_1\eta+2(\etab+\eta+2\ze)\xi -\frac{1}{2}\vth^2 +\Up\left(-\frac{1}{2}\vth\vthb+2(\xi\xib+\eta^2)\right)\\
&=& r^{-1}\dk\Ga_b+r^{-1}\Ga_b\\
&=& \dk^{\leq 1}\Ga_g.
\eeaa

Also, we have
\beaa
T (r)=e_4(r)+\Up e_3(r)=e_4(r)-\Up+\Up(e_3(r)-1)\in r\Ga_b.
\eeaa
We infer
\beaa
T \left(2\omb -\frac 1 2 \kab\right) &=& T \left(\frac{1}{r}\right)+\Ga_b -\frac 1 2 T \left(  \kab+\frac{2}{r}\right)\\
&=& -\frac{T(r)}{r^2}+\dk\Ga_b\\
&=& \dk^{\leq 1}\Ga_b
\eeaa
and 
\beaa
T(K) &=& T\left(\frac{1}{r^2}\right)+T\left(K-\frac{1}{r^2}\right)\\
&=& -\frac{2T(r)}{r^3}+r^{-1}\Ga_b\\
&=&  r^{-1}\dk(\Ga_b)+r^{-1}\Ga_b\\
&=& \dk^{\leq 1}\Ga_g.
\eeaa
This concludes the proof of the lemma.
\end{proof}

\begin{corollary}\lab{cor:keycorollaryforcommutationwavewithT}
We have
\beaa
[T , \square_2]\psi &=& \dk^{\leq 1}(\Ga_g)\dk^{\leq 2}\psi.
\eeaa
\end{corollary}

\begin{proof}
Recall that we have
\beaa
\square_2 \psi&=&-e_3 e_4 \psi +\lapp_2\psi+\left(2\omb -\frac 1 2 \kab\right) e_4\psi- \frac 1 2 \ka e_3\psi+2 \eta e_\th \psi.
\eeaa
We infer
\beaa
[T , \square_2]\psi&=& -[T , e_3]e_4\psi - e_3[T , e_4]\psi +[T , \lapp_2]\psi +\left(2\omb -\frac 1 2 \kab\right) [T , e_4]\psi\\
&&+T \left(2\omb -\frac 1 2 \kab\right) e_4\psi - \frac 1 2 \ka [T , e_3]\psi - \frac 1 2 T (\ka) e_3\psi+2 \eta [T , e_\th]\psi +2 T (\eta) e_\th \psi.
\eeaa
and hence, using also $\lapp_2=-\dds_2\ddd_2+2K$,
\beaa
[T , \square_2]\psi&=& -[T , e_3](r^{-1}\dk\psi) - \dk[T , e_4]\psi -r^{-1}\dk[T , \ddd_2]\psi -[T , \dds_2]r^{-1}\dk\psi +2T(K)\psi\\
&& +\left(2\omb -\frac 1 2 \kab\right) [T , e_4]\psi+T \left(2\omb -\frac 1 2 \kab\right) r^{-1}\dk\psi - \frac 1 2 \ka [T , e_3]\psi - \frac 1 2 T (\ka)\dk\psi\\
&&+2 \eta [T , e_\th]\psi +2 T (\eta)r^{-1}\dk\psi.
\eeaa
In view of 
\beaa
[T , e_4] = \Ga_g\dk,\quad [T , e_3] = \Ga_b\dk, \quad [T , \ddd_k] = \Ga_b\dk+\Ga_b, \quad [T , \dds_k] = \Ga_b\dk+\Ga_b
\eeaa
and 
\beaa
T (\ka) = \dk^{\leq 1}\Ga_g, \quad\quad T \left(2\omb -\frac 1 2 \kab\right) =\dk^{\leq 1}\Ga_b, \quad\quad T(K) = \dk^{\leq 1}\Ga_g,
\eeaa
we deduce, schematically,
\beaa
[T , \square_2]\psi&=& \dk^{\leq 1}(\Ga_g)\dk^{\leq 2}\psi+r^{-1}\dk^{\leq 1}(\Ga_b)\dk^{\leq 2}\psi\\
&=& \dk^{\leq 1}(\Ga_g)\dk^{\leq 2}\psi.
\eeaa
This concludes the proof of the corollary.
\end{proof}


\subsubsection{Commutation with angular derivatives}
 

\begin{lemma}
We have, schematically,
\beaa
&&\,[r\ddd_k, e_4]f, \, \,[r\dds_k, e_4]f=\Ga_g\dk^{\leq 1}f,\, \, [r^2\lapp_k, e_4]f=\dk^{\leq 1}(\Ga_g)\dk^{\leq 2}f\\
&& \,[r\ddd_k, e_3]f=-r\eta e_3(f)+\Ga_b\dk^{\leq 1}f, \, \,[r\dds_k, e_3]f=r\eta e_3(f)+\Ga_b\dk^{\leq 1}f.
\eeaa
\end{lemma}

\begin{proof}
Recall from Lemma \ref{Le:comme3e4-outgeodesic} that the following commutation formulae holds true,
\begin{enumerate}     
 \item If $f\in \mathfrak{s}_k$,
 \beaa
\bsplit
 \,[r\ddd_k, e_4] &=r\left[ \com_k(f)-\frac 1 2 A\ddd_k  f \right],\\
 \,[r\ddd_k, e_3]f&= r\left[ \comb_k(f)-\frac 1 2 \Ab \ddd_k  f \right].
 \end{split}
 \eeaa
 
\item If  $f\in \mathfrak{s}_{k-1}$
\beaa
\bsplit
\,[r\dds_k, e_4]f &=r\left[\com^*_k(f)  -\frac 1 2 A \dds_k f   \right],   \\
\,[r \dds_k, e_3]f &= r\left[\comb^*_k(f)  -\frac 1 2 \Ab  \dds_k f   \right],     
\end{split}
\eeaa
\end{enumerate}
where $A = 2/r e_4(r) -\ka$ and $\Ab=2/re_3(r)-\kab$. Now, we have
\beaa
&&\com_k(f)=r^{-1}\Ga_g\dk^{\leq 1}f,\quad  \com^*_k(f)=r^{-1}\Ga_g\dk^{\leq 1}f, \\
&&\comb_k(f)=-\eta e_3(f)+r^{-1}\Ga_b\dk^{\leq 1}f,  \quad \comb^*_k(f)=\eta e_3(f)+r^{-1}\Ga_b\dk^{\leq 1}f,
\eeaa
which together with the fact that $A\in \Ga_g$ and $\Ab\in \Ga_b$ implies, schematically,
\beaa
&&\,[r\ddd_k, e_4]f, \, \,[r\dds_k, e_4]f=\Ga_g\dk^{\leq 1}f, \\
&& \,[r\ddd_k, e_3]f=-r\eta e_3(f)+\Ga_b\dk^{\leq 1}f, \, \,[r\dds_k, e_3]f=r\eta e_3(f)+\Ga_b\dk^{\leq 1}f.
\eeaa
Since $\lapp_k=-\dds_k\ddd_k+kK$, We infer
\beaa
\,[r^2\lapp_k, e_4] &=& [-r^2\dds_k\ddd_k+kr^2K, e_4]\\
&=& -[r\dds_k, e_4]r\ddd_2-r\dds_k[r\ddd_k, e_4]+
\eeaa
This concludes the proof of the lemma.
\end{proof}

\begin{corollary}\lab{cor:keycorollaryforcommutationwavewithangularderivatives}
We have
\beaa
 r\ddd_2(\square_2\psi) -(\square_1-K)(r\ddd_2\psi) &=&  -r\eta\square_2 \psi +\dk^{\leq 1}(\Ga_g)\dk^{\leq 2}\psi 
\eeaa
and 
\beaa
 r\dds_2(\square_1\phi) -(\square_2-3K)(r\dds_2\phi)  &=& r\eta\square_1\phi +\dk^{\leq 1}(\Ga_g)\dk^{\leq 2}\phi.
\eeaa
\end{corollary}

\begin{proof}
Recall that we have
\beaa
\square_2 \psi&=&-e_3 e_4 \psi +\lapp_2\psi+\left(2\omb -\frac 1 2 \kab\right) e_4\psi- \frac 1 2 \ka e_3\psi+2 \eta e_\th \psi
\eeaa
and 
\beaa
\square_1\phi &=&-e_3 e_4\phi +\lapp_2\phi+\left(2\omb -\frac 1 2 \kab\right) e_4\phi- \frac 1 2 \ka e_3\phi+2 \eta e_\th \phi
\eeaa
We infer
\beaa
r\ddd_2(\square_2\psi) -\square_1(r\ddd_2\psi) &=& -[r\ddd_2, e_3] e_4 \psi -e_3[r\ddd_2, e_4]\psi +r(\ddd_2\lapp_2 -\lapp_1\ddd_2)\psi +[r,\lapp_1]\ddd_2\psi \\
&&+re_\th\left(2\omb -\frac 1 2 \kab\right)e_4\psi +\left(2\omb -\frac 1 2 \kab\right)[r\ddd_2, e_4]\psi - \frac 1 2 re_\th(\ka)e_3\psi\\
&&  - \frac 1 2 \ka [r\ddd_2, e_3]\psi +2r\ddd_2(\eta e_\th \psi) - 2\eta e_\th(r\ddd_2\psi),
\eeaa
and 
\beaa
r\dds_2(\square_1\phi) -\square_2(r\dds_2\phi) &=& -[r\dds_2, e_3] e_4 \phi -e_3[r\dds_2, e_4]\phi +r(\dds_2\lapp_1 -\lapp_2\dds_2)\phi +[r,\lapp_2]\dds_2\phi \\
&& -re_\th\left(2\omb -\frac 1 2 \kab\right)e_4\phi +\left(2\omb -\frac 1 2 \kab\right)[r\dds_2, e_4]\phi + \frac 1 2 re_\th(\ka) e_3\phi\\
&&  - \frac 1 2 \ka [r\dds_2, e_3]\phi +2r\ddd_2(\eta e_\th \phi) - 2\eta e_\th(r\ddd_2\phi),
\eeaa
and hence, using also in particular the following identities from Proposition \ref{prop:DDd--ddd}
\beaa
\ddd_2\lapp_2 -\lapp_1\ddd_2 &=& -K\ddd_2+2e_\th(K),\\
\dds_2\lapp_1 -\lapp_2\dds_2 &=& -3K\dds_2-e_\th(K),
\eeaa
we infer, 
\beaa
r\ddd_2(\square_2\psi) -(\square_1-K)(r\ddd_2\psi) &=& -[r\ddd_2, e_3]e_4\psi -\dk[r\ddd_2, e_4]\psi +2re_\th(K)\psi +[r,\lapp_1](r^{-1}\dk\psi) \\
&&+re_\th\left(2\omb -\frac 1 2 \kab\right)r^{-1}\dk\psi +\left(2\omb -\frac 1 2 \kab\right)[r\ddd_2, e_4]\psi - \frac 1 2 re_\th(\ka)\dk\psi\\
&&  - \frac 1 2 \ka [r\ddd_2, e_3]\psi +2\dk(r^{-1}\eta\dk\psi) - 2r^{-1}\eta\dk(\dk\psi),
\eeaa
and 
\beaa
r\dds_2(\square_1\phi) -(\square_2-3K)(r\dds_2\phi) &=& -[r\dds_2, e_3]e_4\phi -\dk[r\dds_2, e_4]\phi -re_\th(K)\phi +[r,\lapp_2](r^{-1}\dk\phi) \\
&& -re_\th\left(2\omb -\frac 1 2 \kab\right)r^{-1}\dk\phi +\left(2\omb -\frac 1 2 \kab\right)[r\dds_2, e_4]\phi + \frac 1 2 re_\th(\ka)\dk\phi\\
&&  - \frac 1 2 \ka [r\dds_2, e_3]\phi +2\dk(r^{-1}\eta\dk\phi) - 2r^{-1}\eta\dk(\dk\phi).
\eeaa
This yields, schematically,
\beaa
&& r\ddd_2(\square_2\psi) -(\square_1-K)(r\ddd_2\psi)\\
 &=& -[r\ddd_2, e_3]e_4\psi -\dk[r\ddd_2, e_4]\psi  +r^{-1}[r\ddd_2, e_4]\psi - \frac 1 2 \ka [r\ddd_2, e_3]\psi  +\dk^{\leq 1}(\Ga_g)\dk^{\leq 2}\psi 
\eeaa
and 
\beaa
&& r\dds_2(\square_1\phi) -(\square_2-3K)(r\dds_2\phi)\\
 &=& -[r\dds_2, e_3]e_4\phi -\dk[r\dds_2, e_4]\phi   +r^{-1}[r\dds_2, e_4]\phi   - \frac 1 2 \ka [r\dds_2, e_3]\phi  +\dk^{\leq 1}(\Ga_g)\dk^{\leq 2}\phi
\eeaa
where we used the fact that $r{-1}\dk^{\leq 1}\Ga_b$ is at least as good as $\dk^{\leq 1}\Ga_g$ and the fact that $r^{-1}e_\th(r)$ is $\Ga_g$. 

Next, we rely on 
\beaa
&&\,[r\ddd_k, e_4]f, \, \,[r\dds_k, e_4]f=\Ga_g\dk^{\leq 1}f, \\
&& \,[r\ddd_k, e_3]f=-r\eta e_3(f)+\Ga_b\dk^{\leq 1}f, \, \,[r\dds_k, e_3]f=r\eta e_3(f)+\Ga_b\dk^{\leq 1}f
\eeaa
to infer
\beaa
 r\ddd_2(\square_2\psi) -(\square_1-K)(r\ddd_2\psi) &=& r\eta e_3e_4\psi   + \frac 1 2 r\ka\eta e_3\psi  +\dk^{\leq 1}(\Ga_g)\dk^{\leq 2}\psi 
\eeaa
and 
\beaa
 r\dds_2(\square_1\phi) -(\square_2-3K)(r\dds_2\phi) &=& -r\eta e_3e_4\phi     - \frac 1 2 r\ka\eta e_3\phi  +\dk^{\leq 1}(\Ga_g)\dk^{\leq 2}\phi.
\eeaa
This yields
\beaa
 r\ddd_2(\square_2\psi) -(\square_1-K)(r\ddd_2\psi) &=& r\eta\left(-\square_2 \psi+\lapp_2\psi+\left(2\omb -\frac 1 2 \kab\right) e_4\psi+2 \eta e_\th \psi\right) \\
 && +\dk^{\leq 1}(\Ga_g)\dk^{\leq 2}\psi \\
  &=& -r\eta\square_2 \psi +\dk^{\leq 1}(\Ga_g)\dk^{\leq 2}\psi 
\eeaa
and 
\beaa
 r\dds_2(\square_1\phi) -(\square_2-3K)(r\dds_2\phi) &=& -r\eta\left( -\square_1\phi+\lapp_2\phi+\left(2\omb -\frac 1 2 \kab\right) e_4\phi+2 \eta e_\th \phi\right) \\
 && +\dk^{\leq 1}(\Ga_g)\dk^{\leq 2}\phi\\
 &=& r\eta\square_1\phi +\dk^{\leq 1}(\Ga_g)\dk^{\leq 2}\phi.
\eeaa
where we used the fact that $r{-1}\dk^{\leq 1}\Ga_b$ is at least as good as $\dk^{\leq 1}\Ga_g$. This concludes the proof of the corollary.
\end{proof}


\subsubsection{Commutation with $R$ in the region $r\leq r_0$}
 

We derive in the following lemma commutator identities that are non sharp as far as decay in $r$ is concerned. This is sufficient for our needs since we will commute the wave equation with $R$ only in the region $r\leq r_0$ for a fixed $r_0\geq 4m_0$ large enough. We will use in particular the following estimate 
Also, recall that 
\bea\lab{eq:someusefullnonsharpestimateforGagood}
\max_{k\leq k_{small}+30}|\dk(\Ga_g)|\les\frac{\ep}{r^2u_{trap}^{1+\dec-2\de_0}}.
\eea

\begin{lemma}
We have
\beaa
&& [R, e_4] = O\left(\frac{\ep}{u_{trap}^{1+\dec-2\de_0}}\right)\dk, \qquad [R, e_3] = -\frac{2m}{r^2}e_3+O\left(\frac{\ep}{u_{trap}^{1+\dec-2\de_0}}\right)\dk,\\
&&\,[r\ddd_k, R]f, \, \,[r\dds_k, R]f=O\left(\frac{\ep}{u_{trap}^{1+\dec-2\de_0}}\right)\dk^{\leq 1}f,\, \, [r^2\lapp_k, R]f=O\left(\frac{\ep}{u_{trap}^{1+\dec-2\de_0}}\right)\dk^{\leq 2}f.
\eeaa
\end{lemma}

\begin{proof}
Recall that $R$ is defined by
\beaa
R=\frac{1}{2}(e_4-\Up e_3)
\eeaa
and that we have
\beaa
[e_3, e_4] &=& 2\omb e_4 - 2\om e_3+2(\eta-\etab)e_\th.
\eeaa
We infer
\beaa
[R, e_4] = \frac{1}{2}[-\Up e_3, e_4] = -\frac{\Up}{2}[e_3, e_4] +\frac{1}{2}e_4(\Up) e_3=\left(\Up\om e_3+\frac{m}{r^2}e_4(r)\right)e_3+O\left(\frac{\ep}{u_{trap}^{1+\dec-2\de_0}}\right)\dk, 
\eeaa
and
\beaa
[R, e_3] = \frac{1}{2}[e_4-\Up e_3, e_3] = \frac{1}{2}[e_4, e_3]  +\frac{1}{2}e_3(\Up) e_3=\left(\om e_3+\frac{m}{r^2}e_3(r)\right)e_3+O\left(\frac{\ep}{u_{trap}^{1+\dec-2\de_0}}\right)\dk,
\eeaa
and hence, 
\beaa
[R, e_4] = O\left(\frac{\ep}{u_{trap}^{1+\dec-2\de_0}}\right)\dk, \qquad [R, e_3] = -\frac{2m}{r^2}e_3+O\left(\frac{\ep}{u_{trap}^{1+\dec-2\de_0}}\right)\dk.
\eeaa

Also, recall that we have
\beaa
&&\,[r\ddd_k, e_4]f, \, \,[r\dds_k, e_4]f=\Ga_g\dk^{\leq 1}f,\, \, [r^2\lapp_k, e_4]f=\dk^{\leq 1}(\Ga_g)\dk^{\leq 2}f\\
&& \,[r\ddd_k, e_3]f=-r\eta e_3(f)+\Ga_b\dk^{\leq 1}f, \, \,[r\dds_k, e_3]f=r\eta e_3(f)+\Ga_b\dk^{\leq 1}f.
\eeaa
We infer
\beaa
&&\,[r\ddd_k, e_4]f, \, \,[r\dds_k, e_4]f=O\left(\frac{\ep}{u_{trap}^{1+\dec-2\de_0}}\right)\dk^{\leq 1}f,\, \, [r^2\lapp_k, e_4]f=O\left(\frac{\ep}{u_{trap}^{1+\dec-2\de_0}}\right)\dk^{\leq 2}f\\
&& \,[r\ddd_k, e_3]f=O\left(\frac{\ep}{u_{trap}^{1+\dec-2\de_0}}\right)\dk^{\leq 1}f, \, \,[r\dds_k, e_3]f=O\left(\frac{\ep}{u_{trap}^{1+\dec-2\de_0}}\right)\dk^{\leq 1}f.
\eeaa
Together with the definition for $R$, we deduce
\beaa
\,[r\ddd_k, R]f, \, \,[r\dds_k, R]f=O\left(\frac{\ep}{u_{trap}^{1+\dec-2\de_0}}\right)\dk^{\leq 1}f,\, \, [r^2\lapp_k, R]f=O\left(\frac{\ep}{u_{trap}^{1+\dec-2\de_0}}\right)\dk^{\leq 2}f.
\eeaa
This concludes the proof of the lemma.
\end{proof}

\begin{corollary}\lab{cor:waveequationforRpsiinrleq8m0}
We have in the region $r\leq r_0$
\beaa
\square_2(R\psi) &=&  \left(1-\frac{3m}{r}\right)\dk^2\psi +O\left(\frac{\ep}{u_{trap}^{1+\dec-2\de_0}}\right)\dk^2\psi + O(1)\dk^{\leq 1}\psi+O(1)\dk^{\leq 1}N.
\eeaa
\end{corollary}

\begin{proof}
Recall that we have
\beaa
\square_2 \psi &=& -e_4(e_3( \psi))+\lapp_2\psi -\frac{1}{2}\kab e_4\psi +\left(-\frac{1}{2}\ka+2\om\right) e_3\psi+2\etab e_\th\psi.
\eeaa
Multiplying by $r^2$, we infer
\beaa
r^2\square_2 \psi &=& -r^2e_4(e_3( \psi))+r^2\lapp_2\psi -\frac{1}{2}r^2\kab e_4\psi +r^2\left(-\frac{1}{2}\ka+2\om\right) e_3\psi+2r\etab re_\th\psi
\eeaa
and hence
\beaa
R(r^2\square_2 \psi) &=& r^2\square_2(R\psi) -[R, r^2e_4e_3]\psi +[R, r^2\lapp_2]\psi -\frac{1}{2}R(r^2\kab) e_4\psi -\frac{1}{2}r^2\kab [R, e_4]\psi \\
&& +R\left(r^2\left(-\frac{1}{2}\ka+2\om\right)\right) e_3\psi  +r^2\left(-\frac{1}{2}\ka+2\om\right) [R, e_3]\psi +2R(r\etab) re_\th\psi \\
&& +2r\etab [R, re_\th]\psi.
\eeaa
Using the commutation identities of the previous lemma, we infer  in the region $r\leq r_0$
\beaa
R(r^2\square_2 \psi) &=& r^2\square_2(R\psi) -[R, r^2e_4e_3]\psi +O\left(\frac{\ep}{u_{trap}^{1+\dec-2\de_0}}\right)\dk^2\psi + O(1)\dk\psi.
\eeaa
Also, since $\psi$ satisfies $\square_2\psi=V\psi+N$, we infer  in the region $r\leq r_0$
\beaa
r^2\square_2(R\psi) &=&  [R, r^2e_4e_3]\psi +O\left(\frac{\ep}{u_{trap}^{1+\dec-2\de_0}}\right)\dk^2\psi + O(1)\dk^{\leq 1}\psi+O(1)\dk^{\leq 1}N.
\eeaa

Next, recall that we have 
\beaa
[R, e_4] = O\left(\frac{\ep}{u_{trap}^{1+\dec-2\de_0}}\right)\dk, \qquad [R, e_3] = -\frac{2m}{r^2}e_3+O(\ep)\dk.
\eeaa
We infer 
\beaa
[R, r^2e_4e_3]\psi &=& R(r^2)e_4e_3+r^2[R,e_4]e_3+r^2e_4[R,e_3]\\
&=& \frac{1}{2}\Big(e_4(r^2)-\Up e_3(r^2)\Big)e_4e_3\psi +r^2e_4\left( -\frac{2m}{r^2}e_3\psi\right)+O\left(\frac{\ep}{u_{trap}^{1+\dec-2\de_0}}\right)\dk^2\psi\\
&=& 2\Big(r-3m\Big)e_4e_3\psi +O\left(\frac{\ep}{u_{trap}^{1+\dec-2\de_0}}\right)\dk^2\psi+O(1)\dk\psi
\eeaa
and thus,  in the region $r\leq r_0$, 
\beaa
\square_2(R\psi) &=&  \left(1-\frac{3m}{r}\right)\dk^2\psi +O\left(\frac{\ep}{u_{trap}^{1+\dec-2\de_0}}\right)\dk^2\psi + O(1)\dk^{\leq 1}\psi+O(1)\dk^{\leq 1}N
\eeaa
as desired. 
\end{proof}


\subsubsection{Commutation with the redshift vectorfield}
 

Let a   positive   bump function $\ka=\ka(r)$, supported in the region  in $[-2, 2]$ and   equal to $1$   for $[-1,1]$. Recall that the redshift vectorfield is given by
     \beaa
     Y_\HH &=&    \ka_\HH      Y_{(0)}, \qquad  \ka_\HH:= \ka\left(\frac{\Up}{\de_\HH}\right)
     \eeaa
where $ Y_{(0)}$ is defined by     
\beaa
Y_{(0)}= a e_3 +b e_4  +2T, \quad a=1+\frac{5}{4m}(r-2m), \quad b= \frac{5}{4m}(r-2m).
\eeaa 

\begin{lemma}
We have
\beaa
[\square_2, e_3]\psi &=& -2\om e_3(e_3\psi) +\kab e_4(e_3\psi)+\kab\square_2 \psi  +\dk^{\leq 1}(\Ga_g)\dk^2\psi+r^{-2}\dk^{\leq 1}\psi.
\eeaa
\end{lemma}

\begin{proof}
Recall that we have
\beaa
\square_2 \psi &=& -e_4(e_3( \psi))+\lapp_2\psi -\frac{1}{2}\kab e_4\psi +\left(-\frac{1}{2}\ka+2\om\right) e_3\psi+2\etab e_\th\psi.
\eeaa
Since we have
\beaa
&& [e_4,e_3] = 2\om e_3 +r^{-1}\Ga_b\dkb, \quad [\ddd_k, e_3]=\frac{1}{2}\kab\ddd_k+\Ga_b\dk+r^{-1}\Ga_b,\\
&& [\dds_k, e_3]=\frac{1}{2}\kab\dds_k+\Ga_b\dk+r^{-1}\Ga_b 
\eeaa
We infer
\beaa
[\square_2, e_3]\psi &=& -[e_4,e_3](e_3( \psi))+[\lapp_2, e_3] -\frac{1}{2}\kab [e_4,e_3](\psi) \\
&&+\frac{1}{2}e_3(\kab) e_4(\psi)  -e_3\left(-\frac{1}{2}\ka+2\om\right) e_3(\psi)   +2\etab [e_\th, e_3]\psi  -2e_3(\etab) e_\th(\psi)\\
&=& -2\om e_3(e_3\psi)+\kab\lapp_2\psi +\dk^{\leq 1}(\Ga_g)\dk^2\psi+r^{-2}\dk^{\leq 1}\psi.
\eeaa
Using again 
\beaa
\square_2 \psi &=& -e_4(e_3( \psi))+\lapp_2\psi -\frac{1}{2}\kab e_4\psi +\left(-\frac{1}{2}\ka+2\om\right) e_3\psi+2\etab e_\th\psi,
\eeaa
we deduce
\beaa
[\square_2, e_3]\psi &=& -2\om e_3(e_3 \psi)+\kab\Big(\square_2 \psi +e_4(e_3\psi)\Big) +\dk^{\leq 1}(\Ga_g)\dk^2\psi+r^{-2}\dk^{\leq 1}\psi\\
&=& -2\om e_3(e_3\psi) +\kab e_4(e_3\psi)+\kab\square_2 \psi  +\dk^{\leq 1}(\Ga_g)\dk^2\psi+r^{-2}\dk^{\leq 1}\psi.
\eeaa
This concludes the proof of the lemma.
\end{proof}

\begin{lemma}\lab{lemma:keycorollaryforcommutationwavewithredshift}
The exists a scalar function $d_0$ satisfying the bound
\beaa
d_0 &=& \frac{1}{2m_0}+O(\deh)\textrm{ on the support of }\ka_\HH,
\eeaa
such that we have, schematically 
\beaa
[\square_2, Y_\HH]\psi &=& d_0Y_{(0)}(Y_\HH\psi) +1_{\Up\leq 2\deh}\left(\square_2\psi+\dk T\psi+\dk^{\leq 1}(\Ga_g)\dk^2\psi+\frac{1}{\deh^2}\dk^{\leq 1}\psi\right)\\
&&+\frac{1}{\deh}1_{\deh\leq \Up\leq 2\deh}\dk^{\leq 2}\psi.
\eeaa
\end{lemma}

\begin{proof}
We have
\beaa
Y_{(0)}= a e_3 +b e_4  +2T &=&  a e_3 +b(2T-\Up e_3)  +2T\\
&=& (a-\Up b)e_3+2(1+b)T.
\eeaa
Thus, in view of the commutator identities 
\beaa
\,[T , \square_2]\psi &=& \dk^{\leq 1}(\Ga_g)\dk^{\leq 2}\psi,\\
\,[\square_2, e_3]\psi &=& -2\om e_3(e_3\psi) +\kab e_4(e_3\psi)+\kab\square_2 \psi  +\dk^{\leq 1}(\Ga_g)\dk^2\psi+r^{-2}\dk^{\leq 1}\psi,
\eeaa
we deduce, schematically,
\beaa
[\square_2, Y_{(0)}]\psi &=& [\square_2, (a-\Up b)e_3]\psi+[\square_2, 2(1+b)T]\psi\\
&=& (a-\Up b)[\square_2, e_3]\psi+\g^{\a\b}\D_\a(a)\D_\b e_3\psi+2(1+b)[\square_2, T]\psi\\
&&+2\g^{\a\b}\D_\a(b)\D_\b T\psi+\dk^{\leq 1}\psi\\
&=&  (a-\Up b)\Big(-2\om e_3(e_3\psi) +\kab e_4(e_3\psi)+\kab\square_2 \psi  \Big) -\frac{1}{2}e_3(a)e_4(e_3\psi) -\frac{1}{2}e_4(a)e_3(e_3\psi)\\
&& +\dk T\psi+\dk^{\leq 1}(\Ga_g)\dk^2\psi+\dk^{\leq 1}\psi.
\eeaa
Since $e_4=-\Up e_3+2T$, we infer schematically
\beaa
[\square_2, Y_{(0)}]\psi &=&  \left((a-\Up b)(-2\om-\Up\kab)+\frac{\Up}{2}e_3(a) -\frac{1}{2}e_4(a)\right)e_3(e_3\psi)\\
&& +\square_2\psi+\dk T\psi+\dk^{\leq 1}(\Ga_g)\dk^2\psi+\dk^{\leq 1}\psi.
\eeaa
We deduce, 
\beaa
[\square_2, Y_\HH]\psi &=& [\square_2, \ka_\HH Y_{(0)}]\psi\\
&=&  \ka_\HH[\square_2,  Y_{(0)}]\psi +\ka_\HH'\dk^{\leq 2}\psi+\ka_\HH''\dk^{\leq 1}\psi\\
&=& \ka_\HH\left((a-\Up b)(-2\om-\Up\kab)+\frac{\Up}{2}e_3(a) -\frac{1}{2}e_4(a)\right)e_3(e_3\psi)\\
&& +1_{\Up\leq 2\deh}\left(\square_2\psi+\dk T\psi+\dk^{\leq 1}(\Ga_g)\dk^2\psi+\frac{1}{\deh^2}\dk^{\leq 1}\psi\right)+\frac{1}{\deh}1_{\deh\leq \Up\leq 2\deh}\dk^{\leq 2}\psi.
\eeaa
Now, we have
\beaa
\ka_\HH e_3(e_3\psi) &=& \frac{1}{a-\Up b}\ka_\HH Y_{(0)}(e_3\psi)+T\dk \psi\\
&=& \frac{1}{(a-\Up b)^2}\ka_\HH Y_{(0)}(Y_{(0)}\psi)+\dk T\psi+\dk^{\leq 1}\psi\\
&=& \frac{1}{(a-\Up b)^2} Y_{(0)}(Y_\HH\psi)+\dk T\psi+\frac{1}{\deh}\dk^{\leq 1}\psi
\eeaa
and hence
\beaa
[\square_2, Y_\HH]\psi &=& \frac{(a-\Up b)(-2\om-\Up\kab)+\frac{\Up}{2}e_3(a) -\frac{1}{2}e_4(a)}{{(a-\Up b)^2}}Y_{(0)}(Y_\HH\psi)\\
&& +1_{\Up\leq 2\deh}\left(\square_2\psi+\dk T\psi+\dk^{\leq 1}(\Ga_g)\dk^2\psi+\frac{1}{\deh^2}\dk^{\leq 1}\psi\right)+\frac{1}{\deh}1_{\deh\leq \Up\leq 2\deh}\dk^{\leq 2}\psi.
\eeaa
Now, we have in view of the definition of $a$ and $b$,
\beaa
\frac{(a-\Up b)(-2\om-\Up\kab)+\frac{\Up}{2}e_3(a) -\frac{1}{2}e_4(a)}{{(a-\Up b)^2}} &=& \frac{(1+O(\Up))(-2\om+O(\Up))+O(\Up)}{{(1+O(\Up))^2}}\\
&=& \frac{1}{2m}+O(\ep)+O(\Up)\\
&=& \frac{1}{2m_0}+O(\ep)+O(\Up)
\eeaa
where we used also our assumptions on $\om$ and $m$. Thus, we have on the support of $\ka_\HH$
\beaa
\frac{(a-\Up b)(-2\om-\Up\kab)+\frac{\Up}{2}e_3(a) -\frac{1}{2}e_4(a)}{{(a-\Up b)^2}} &=& \frac{1}{2m_0}+O(\ep+\deh)\\
&=& \frac{1}{2m_0}+O(\deh)
\eeaa
where we used the fact that $\ep\ll \deh$ by assumption. Setting
\beaa
d_0 &:=& \frac{(a-\Up b)(-2\om-\Up\kab)+\frac{\Up}{2}e_3(a) -\frac{1}{2}e_4(a)}{{(a-\Up b)^2}}, 
\eeaa
this concludes the proof of the lemma.
\end{proof}


\subsubsection{Commutation with $re_4$}
 

\begin{lemma}\lab{lemma:keycorollaryforcommutationwavewithre4}
We have, schematically,
\beaa
[\square_2, re_4]\psi &=& \frac{\Up}{r}\left(1+\frac{2m}{r\Up^2} \right)\ec_4(re_4\psi) +\square_2 \psi   +\Ga_g\dk^2\psi+\frac{1}{\Up}r^{-2}\dk T\psi+r^{-2}\dkb^2\psi+r^{-2}\dk\psi.
\eeaa
\end{lemma}

\begin{proof}
Recall that we have
\beaa
\square_2 \psi &=& -e_3(e_4( \psi))+\lapp_2\psi +\left(2\omb-\frac{1}{2}\kab\right) e_4\psi -\frac{1}{2}\ka e_3\psi+2\eta e_\th\psi.
\eeaa
Since we have
\beaa
&& [re_4,e_3] = 2r\om e_3-\frac{r}{2}\kab e_4 +\Ga_b\dk, \quad [re_4,e_4]=-\frac{r}{2}\ka e_4+\Ga_g\dk,\\
&& [\ddd_k, re_4]=\frac{1}{2}r\ka\ddd_k+\Ga_g\dk+\Ga_g,\quad  [\dds_k, re_4]=\frac{1}{2}r\ka\dds_k+\Ga_g\dk+\Ga_g, 
\eeaa
we infer, schematically,
\beaa
[\square_2, re_4]\psi&=& -[e_3, re_4]e_4\psi - e_3[e_4, re_4]\psi +[\lapp_2, re_4]\psi +\left(2\omb -\frac 1 2 \kab\right) [e_4, re_4]\psi\\
&& -re_4\left(2\omb -\frac 1 2 \kab\right) e_4\psi - \frac 1 2 \ka [e_3, re_4]\psi + \frac 1 2 re_4(\ka) e_3\psi+2 \eta [e_\th, re_4]\psi -2 re_4(\eta) e_\th \psi\\
&=& \left(2r\om e_3-\frac{r}{2}\kab e_4 \right)e_4\psi -\frac{1}{2}e_3\left(r\ka e_4\psi\right) +r\ka\lapp_2\psi   + \frac 1 2 re_4(\ka) e_3\psi  +\Ga_g\dk^2\psi+r^{-2}\dk\psi\\
&=& \left(2r\om e_3-\frac{r}{2}\kab e_4 \right)e_4\psi -\frac{1}{2}r\ka e_3e_4\psi   - \frac 1 4 r\ka^2 e_3\psi   +\Ga_g\dk^2\psi+r^{-2}\dkb^2\psi+r^{-2}\dk\psi.
\eeaa
Using again 
\beaa
\square_2 \psi &=& -e_3(e_4( \psi))+\lapp_2\psi +\left(2\omb-\frac{1}{2}\kab\right) e_4\psi -\frac{1}{2}\ka e_3\psi+2\eta e_\th\psi.
\eeaa
we have
\beaa
-\frac{1}{2}r\ka e_3e_4\psi   - \frac 1 4 r\ka^2 e_3\psi &=& \frac{1}{2}r\ka\square_2 \psi +r^{-2}\dkb^2\psi+r^{-2}\dk\psi  
\eeaa
and hence
\beaa
[\square_2, re_4]\psi &=& \left(2r\om e_3-\frac{r}{2}\kab e_4 \right)e_4\psi +\square_2 \psi   +\Ga_g\dk^2\psi+r^{-2}\dkb^2\psi+r^{-2}\dk\psi\\
&=& \left(2r\om \frac{1}{\Up}(2T-e_4)-\frac{r}{2}\kab e_4 \right)e_4\psi +\square_2 \psi   +\Ga_g\dk^2\psi+r^{-2}\dkb^2\psi+r^{-2}\dk\psi\\
&=& \left(-2r\om \frac{1}{\Up}-\frac{r}{2}\kab \right)e_4(e_4\psi) +\square_2 \psi   +\Ga_g\dk^2\psi+\frac{1}{\Up}r^{-2}\dk T\psi+r^{-2}\dkb^2\psi+r^{-2}\dk\psi\\
&=& \left(\Up+\frac{2m}{r\Up} \right)e_4(e_4\psi) +\square_2 \psi   +\Ga_g\dk^2\psi+\frac{1}{\Up}r^{-2}\dk T\psi+r^{-2}\dkb^2\psi+r^{-2}\dk\psi\\
&=& \frac{\Up}{r}\left(1+\frac{2m}{r\Up^2} \right)\ec_4(re_4\psi) +\square_2 \psi   +\Ga_g\dk^2\psi+\frac{1}{\Up}r^{-2}\dk T\psi+r^{-2}\dkb^2\psi+r^{-2}\dk\psi.
\eeaa
This concludes the proof of the lemma.
\end{proof}


 \subsection{Some weighted estimates for wave equations}
 

Recall from Corollary \ref{cor:keycorollaryforcommutationwavewithangularderivatives} that we have the following commutator identity
\beaa
 r\ddd_2(\square_2\psi) -(\square_1-K)(r\ddd_2\psi) &=&  -r\eta\square_2 \psi +\dk^{\leq 1}(\Ga_g)\dk^{\leq 2}\psi. 
\eeaa
In particular, to derive weighted estimates for $r\ddd_2$, we need to derive weighted estimates for solutions $\phi$ to wave equations of the type
\beaa
(\square_1-V_1)\phi &=& N,
\eeaa
where $\phi$ is a reduced 1-scalar and the potential $V_1$ is given by $V_1=V+K=-\ka\kab+K$. This is done in the following theorem.

\begin{theorem}\lab{thm:nonsharpweightedesitmatesforreduced1scalarsolutionofwaveequationwithpotential}
Let $\phi$ a reduced 1-scalar solution to 
\beaa
(\square_1-V_1)\phi &=& N,\qquad V_1=-\ka\kab+K.
\eeaa
Then, $\phi$ satisfies for all $\de\leq p\leq 2-\de$,
 \bea\lab{eq:nonsharpweightedesitmatesforreduced1scalarsolutionofwaveequationwithpotential:1}
 \nn && \sup_{\tau\in[\tau_1,\tau_2] }   E_{p} [\phi](\tau)+   B_{p}[\phi](\tau_1, \tau_2)  + F_{p}[\phi](\tau_1, \tau_2)\\
 \nn  & \les&  E_{p}[\phi](\tau_1)+   J_p[\phi, N]( \tau_1,\tau_2)+\int_{\Mtrap(\tau_1,\tau_2)}\left|1-\frac{3m}{r}\right||\phi|(|\phi|+|R\phi|)\\
 && +\int_{\Mntrap(\tau_1,\tau_2)}r^{p-3}|\phi|(|\phi|+|\dk\phi|),
\eea
and $\check{\phi}=f_2\ec_4\phi$ satisfies for all $1-\de<q\leq 1-\de$,
    \bea\lab{eq:nonsharpweightedesitmatesforreduced1scalarsolutionofwaveequationwithpotential:2}
 \nn&& \sup_{\tau\in[\tau_1,\tau_2] }  E_q[\check{\phi}](\tau)+   B_q[\check{\phi}](\tau_1, \tau_2) \\
\nn  &\les&   E_q [\check{\phi}](\tau_1)      +  \Jc_q[\check{\phi}, N] (\tau_1,\tau_2)  +    E\, ^1_{\max(q, \de)}[\phi](\tau_1)  + J^{1}_{\max(q, \de)}[\phi, N]\\
&&+ \int_{\MM(\tau_1,\tau_2)}r^{q-3}|\check{\phi}|(|\check{\phi}|+|\dk\check{\phi}|).
\eea
\end{theorem}

\begin{remark}
Although we will not need it, we expect that the last 2 terms in the right-hand side of \eqref{eq:nonsharpweightedesitmatesforreduced1scalarsolutionofwaveequationwithpotential:1} and the last term in the right-hand side of \eqref{eq:nonsharpweightedesitmatesforreduced1scalarsolutionofwaveequationwithpotential:2} could be removed.   
\end{remark}

\begin{proof}
We start with the following observations.
\begin{itemize}
\item \eqref{eq:nonsharpweightedesitmatesforreduced1scalarsolutionofwaveequationwithpotential:2} is the analog of \eqref{eq:recoverhigherderivativesforestimatewaveequationpsi:2}, i.e. of Theorem \ref{theorem:Daf-Rodn-estim2-psic} in the case $s=0$, with $V$ replaced by $V_1$, and with the reduced 2-scalar $\psi$ replaced by the reduced 1-scalar $\phi$. The proof is in fact significantly easier in view of the presence of the term
\beaa
 \int_{\MM(\tau_1,\tau_2)}r^{q-3}|\check{\phi}|(|\check{\phi}|+|\dk\check{\phi}|)
\eeaa
on the right-hand side of \eqref{eq:nonsharpweightedesitmatesforreduced1scalarsolutionofwaveequationwithpotential:2}. 

\item \eqref{eq:nonsharpweightedesitmatesforreduced1scalarsolutionofwaveequationwithpotential:1} is the analog of \eqref{eq:recoverhigherderivativesforestimatewaveequationpsi:3}, i.e. of Theorem \ref{theorem-combinedMor-r-weighted} in the case $s=0$, with $V$ replaced by $V_1$, and with the reduced 2-scalar $\psi$ replaced by the reduced 1-scalar $\phi$. 
 The proof is in fact significantly easier in view of the presence of the terms
\beaa
\int_{\Mtrap(\tau_1,\tau_2)}\left|1-\frac{3m}{r}\right||\phi|(|\phi|+|R\phi|) +\int_{\Mntrap(\tau_1,\tau_2)}r^{p-3}|\phi|(|\phi|+|\dk\phi|)
\eeaa
on the right-hand side of \eqref{eq:nonsharpweightedesitmatesforreduced1scalarsolutionofwaveequationwithpotential:1}. 

\item The boundary terms can be treated as in the proof of \eqref{eq:recoverhigherderivativesforestimatewaveequationpsi:2} and \eqref{eq:recoverhigherderivativesforestimatewaveequationpsi:3} in view of the fact that $V_1$ is a positive potential\footnote{We have 
\beaa
V_1=-\ka\kab+K=\frac{4\Up+1+O(\ep)}{r^2}
\eeaa
in view of the assumptions so that $V_1$ is indeed a positive potential.}.

\item The only place where there might a potential difficulty concerns the proof of \eqref{eq:nonsharpweightedesitmatesforreduced1scalarsolutionofwaveequationwithpotential:2} in $\Mtrap$ where the second to last term on the right-hand side is required to have a more precise structure.
\end{itemize}

In view of the above observations, and in particular of the last one, we focus on recovering the bulk term leading to \eqref{eq:nonsharpweightedesitmatesforreduced1scalarsolutionofwaveequationwithpotential:2} in $\Mtrap$.  To this end, we choose $f$ ad $w$ as in  Proposition \ref{prop:pre-mor1}. This yields\footnote{Note that Proposition \ref{prop:pre-mor1} does not use the particular form of the potential and the type of the reduced scalar $\phi$ and hence holds in our more general case.}
\beaa
\EEd [fR, w](\Psi)&\ge f'  |R(\Psi)|^2 + r^{-1} \left(1-\frac{3m}{r}\right) f|\nabb \Psi|^2+O\left(\frac{1-\frac{3m}{r}}{r^3}\right)|\Psi|^2.
\eeaa
We infer
 \beaa
  \EEd [fR, w, M=2hR](\Psi)   &\geq&  f'  |R(\Psi)|^2 + r^{-1} \left(1-\frac{3m}{r}\right) f|\nabb \Psi|^2+O\left(\frac{1-\frac{3m}{r}}{r^3}\right)|\Psi|^2\\
  && +\frac 1 2r^{-2} (\Up r^2 h)' |\Psi|^2 + h \Psi R(\Psi).
  \eeaa

We now choose a smooth $h$, compactly supported in $[5/2m_0, 7/2m_0]$, such that $h(3m)=0$ and $h'(3m)=1$\footnote{This differs from the choice of $h$  in the proof of \eqref{eq:recoverhigherderivativesforestimatewaveequationpsi:2} in order to avoid using a Poincar\'e inequality (which depends of the type of the reduced scalar) and the particular form of the potential $V_1$.}. We infer $r^{-2} (\Up r^2 h)'(3m)=1/3>0$ and hence
 \beaa
  \EEd [fR, w, M](\Psi)   &\geq&  f'  |R(\Psi)|^2 + r^{-1} \left(1-\frac{3m}{r}\right) f|\nabb \Psi|^2+\frac{\Up}{r^3}|\Psi|^2\\
  &&+O\left(\frac{1-\frac{3m}{r}}{r^3}\right)|\Psi|(|\Psi|+|R(\Psi)|).
  \eeaa
In view of the choice of $f$ in  Proposition \ref{prop:pre-mor1}, we have
\beaa
f'\gtrsim  \frac{1}{r^3}, \qquad \left(1-\frac{3m}{r}\right)f\geq \left(1-\frac{3m}{r}\right)^2,
\eeaa
and hence, there exists two constants $c_0>0$ and $C_0>0$ such that 
 \beaa
  \EEd [fR, w_0, M](\Psi)   &\geq&  c_0\left(\frac{1}{r^3}|R(\Psi)|^2 + r^{-1} \left(1-\frac{3m}{r}\right)^2|\nabb \Psi|^2+\frac{\Up}{r^3}|\Psi|^2\right)\\
  && -C_0\frac{\left|1-\frac{3m}{r}\right|}{r^3}|\Psi|(|\Psi|+|R(\Psi)|).
  \eeaa
 The last term above is responsible for  the  second to last term on the right-hand side of \eqref{eq:nonsharpweightedesitmatesforreduced1scalarsolutionofwaveequationwithpotential:2}. 
\end{proof}

Next, we have the following consequence of \eqref{eq:recoverhigherderivativesforestimatewaveequationpsi:2} and Theorem \ref{thm:nonsharpweightedesitmatesforreduced1scalarsolutionofwaveequationwithpotential}.

\begin{corollary}\lab{cor:estimatesforwaveequationskscalarswithphi1phi2RHS}
Let $\phi$ be a reduced $k$-scalar for $k=1, 2$ such that $\phi$ satisfies\footnote{Recall that we have $\de_0\ll\dec$ in view of \eqref{eq:deandde0muchsmallerthananyotherdeltaintheproof}, and hence $\dec-2\de_0>0$.} 
\beaa
(\square_k-W)\phi &=& O\left(\frac{\ep}{r^2u_{trap}^{1+\dec-2\de_0}}\right)\dk\phi_1+\phi_2
\eeaa
where $\phi_1$ and $\phi_2$ are given reduced scalars, and where $W=V$ in the case $k=2$ and $W=V_1$ in the case $k=1$. Then, $\phi$ satisfies  for all $\de\leq p\leq 2-\de$,
 \beaa
 \nn && \sup_{\tau\in[\tau_1,\tau_2] }   E_{p} [\phi](\tau)+   B_{p}[\phi](\tau_1, \tau_2)  + F_{p}[\phi](\tau_1, \tau_2)\\
   & \les&  E_{p}[\phi](\tau_1)+  \ep^2\left(\sup_{[\tau_1, \tau_2]}E[\phi_1](\tau)+B_p[\phi_1](\tau_1, \tau_2)\right)+J_p[\phi, \phi_2]( \tau_1,\tau_2)\\
      &&+\int_{\Mtrap(\tau_1,\tau_2)}\left|1-\frac{3m}{r}\right||\phi|(|\phi|+|R\phi|) +\int_{\Mntrap(\tau_1,\tau_2)}r^{p-3}|\phi|(|\phi|+|\dk\phi|).
\eeaa
\end{corollary}

\begin{proof}
The wave equation for $\phi$ satisfies the assumptions of \eqref{eq:recoverhigherderivativesforestimatewaveequationpsi:2} and Theorem \ref{thm:nonsharpweightedesitmatesforreduced1scalarsolutionofwaveequationwithpotential} with 
\beaa
N &=& O\left(\frac{\ep}{r^2u_{trap}^{1+\dec-2\de_0}}\right)\dk\phi_1+\phi_2.
\eeaa 
 We deduce
 \beaa
&&  \sup_{\tau\in[\tau_1,\tau_2] }   E_{p} [\phi](\tau)+   B_{p}[\phi](\tau_1, \tau_2)  + F_{p}[\phi](\tau_1, \tau_2)\\
  & \les&      E_{p}[\phi](\tau_1)+   J_p\left[\phi, O\left(\frac{\ep}{r^2u_{trap}^{1+\dec-2\de_0}}\right)\dk\phi_1+\phi_2\right]( \tau_1,\tau_2)\\
  &&+\int_{\Mtrap(\tau_1,\tau_2)}\left|1-\frac{3m}{r}\right||\phi|(|\phi|+|R\phi|)+\int_{\Mntrap(\tau_1,\tau_2)}r^{p-3}|\phi|(|\phi|+|\dk\phi|).
\eeaa

Now, in view of the definition 
\beaa
\bsplit
J_{p, R} [\psi, N](\tau_1,\tau_2)&= \bigg| \int_{\MM_{\ge R}(\tau_1,\tau_2)  } r^ p \ec_4 \psi N\bigg|,\\
J_p[\psi, N](\tau_1,\tau_2)&=\bigg( \int_{\tau_1}^{\tau_2} d\tau \|N\|_{L^2(\Sitrap(\tau))}\bigg)^2+\int_{\Mntrap(\tau_1,\tau_2)}r^{1+\de}|N|^2\\
& +J_{p, 4m_0} [\psi, N](\tau_1,\tau_2), 
\end{split}
\eeaa
we have
\beaa
&& J_p\left[\phi, O\left(\frac{\ep}{r^2u_{trap}^{1+\dec-2\de_0}}\right)\dk\phi_1+\phi_2\right]( \tau_1,\tau_2)\\
&\les&  J_p\left[\phi, O\left(\frac{\ep}{r^2u_{trap}^{1+\dec-2\de_0}}\right)\dk\phi_1\right]( \tau_1,\tau_2)+J_p\left[\phi, \phi_2\right]( \tau_1,\tau_2)
\eeaa
and, for $\de\leq p\leq 2-\de$, using also $\dec-2\de_0>0$, we have
\beaa
&&J_p\left[\phi, O\left(\frac{\ep}{r^2u_{trap}^{1+\dec-2\de_0}}\right)\dk\phi_1\right]( \tau_1,\tau_2)\\
 &\les& \ep^2\bigg( \int_{\tau_1}^{\tau_2} \|\dk\phi_1\|_{L^2(\Sitrap(\tau))}\frac{d\tau}{\tau^{1+\dec-2\de_0}} \bigg)^2+\ep^2\int_{\Mntrap(\tau_1,\tau_2)}r^{\de-3}|\dk\phi_1|^2\\
 &&+\ep\bigg| \int_{\MM_{\ge 4m_0}(\tau_1,\tau_2)  } r^{p-2} \ec_4(\psi)\dk\phi_1\bigg|\\
  &\les& \ep^2\sup_{[\tau_1, \tau_2]}\|\dk\phi_1\|_{L^2(\Sitrap(\tau))}^2+\ep^2\int_{\Mntrap(\tau_1,\tau_2)}r^{p-3}|\dk^{\leq 2}\psi|^2\\
  &&+\ep\left(\int_{\Mntrap(\tau_1,\tau_2)}r^{p-3}|\dk\phi|^2\right)^{\frac{1}{2}}\left(\int_{\Mntrap(\tau_1,\tau_2)}r^{p-3}|\dk\phi_1|^2\right)^{\frac{1}{2}}\\
  &\les& \ep^2\left(\sup_{[\tau_1, \tau_2]}E^1[\phi_1](\tau)+B^1_p[\phi_1](\tau_1, \tau_2)\right)+\ep\left(B^1_p[\phi_1](\tau_1, \tau_2)\right)^{\frac{1}{2}}\left(B^1_p[\phi](\tau_1, \tau_2)\right)^{\frac{1}{2}}.
\eeaa
We immediately deduce
 \beaa
 \nn && \sup_{\tau\in[\tau_1,\tau_2] }   E_{p} [\phi](\tau)+   B_{p}[\phi](\tau_1, \tau_2)  + F_{p}[\phi](\tau_1, \tau_2)\\
   & \les&  E_{p}[\phi](\tau_1)+  \ep^2\left(\sup_{[\tau_1, \tau_2]}E[\phi_1](\tau)+B_p[\phi_1](\tau_1, \tau_2)\right)+J_p[\phi, \phi_2]( \tau_1,\tau_2)\\
   &&+\int_{\Mtrap(\tau_1,\tau_2)}\left|1-\frac{3m}{r}\right||\phi|(|\phi|+|R\phi|)+\int_{\Mntrap(\tau_1,\tau_2)}r^{p-3}|\phi|(|\phi|+|\dk\phi|).
\eeaa
This concludes the proof of the corollary.
\end{proof}

Finally, we end this section with the following lemma.
\begin{lemma}\lab{lemma:usefulllemmaforabsorbingtermsinhigheorderderivativeestimate}
Let $\phi$ be a reduced $k$-scalar for $k=1, 2$, and let $X$ a vectorfield. We have for all $\de\leq p\leq 2-\de$,
\beaa
&&\int_{\Mtrap(\tau_1,\tau_2)}\left|1-\frac{3m}{r}\right||\dk\phi|(|X\phi|+|R(X\phi)|)+\int_{\Mntrap(\tau_1,\tau_2)}r^{p-3}|\dk\phi|(|X\phi|+|\dk(X\phi)|)\\
&\les& \left(B_p[X\phi](\tau_1, \tau_2)\right)^{\frac{1}{2}}\left(B_p[\phi](\tau_1, \tau_2)\right)^{\frac{1}{2}}.
\eeaa
\end{lemma}

\begin{proof}
The proof follows immediately from the definition of $B_p[\phi](\tau_1, \tau_2)$. 
\end{proof}


 \subsection{Proof of Theorem \ref{theorem-combinedMor-r-weighted}}\lab{sec:proofoftheorem-combinedMor-r-weighted}
 

We now conclude the proof of Theorem \ref{theorem-combinedMor-r-weighted} for all $0\leq s\leq k_{small}+30$ by 
recovering higher derivatives $s\geq 1$ one by one starting from the estimate $s=0$ provided by  \eqref{eq:recoverhigherderivativesforestimatewaveequationpsi:2}. As explained in section \ref{sec:strategyforrecoveringhigherorderderivatives}, it suffices to recover the estimates for $s=1$ from the one for $s=0$ as the procedure to recover the estimate for $s+1$ from the one for $s$ is completely analogous. We now follow the strategy outlined in section \ref{sec:strategyforrecoveringhigherorderderivatives}.


 \subsubsection{Recovering estimates for $T\psi$}
 

Recall that $\psi$ satisfies 
 \beaa
\square_2 \psi+V\psi =N, \qquad V=\ka\kab, 
\eeaa
and recall also from Corollary \ref{cor:keycorollaryforcommutationwavewithT} that we have
\beaa
[T , \square_2]\psi &=& \dk^{\leq 1}(\Ga_g)\dk^{\leq 2}\psi.
\eeaa
We infer
\beaa
\square_2(T\psi) +VT(\psi) &=& T(N)+\dk^{\leq 1}(\Ga_g)\dk^{\leq 2}\psi.
\eeaa
In view of Corollary \ref{cor:estimatesforwaveequationskscalarswithphi1phi2RHS} with $\phi=T(\psi)$, $\phi_1=\dk^{\leq 1}\psi$ and $\phi_2=T(N)$, and in view of \eqref{eq:someusefullnonsharpestimateforGagood}, we deduce
 \beaa
&&  \sup_{\tau\in[\tau_1,\tau_2] }   E_{p} [T\psi](\tau)+   B_{p}[T\psi](\tau_1, \tau_2)  + F_{p}[T\psi](\tau_1, \tau_2)\\
  & \les&      E_{p}[T\psi](\tau_1)+    \ep^2\left(\sup_{[\tau_1, \tau_2]}E[\dk^{\leq 1}\psi](\tau)+B_p[\dk^{\leq 1}\psi](\tau_1, \tau_2)\right)+J_p[T(\psi), T(N)]( \tau_1,\tau_2)\\
       &&+\int_{\Mtrap(\tau_1,\tau_2)}\left|1-\frac{3m}{r}\right||T\phi|(|T\phi|+|R(T\phi)|)\\
       && +\int_{\Mntrap(\tau_1,\tau_2)}r^{p-3}|T\phi|(|T\phi|+|\dk(T\phi)|),
\eeaa 
and hence, using Lemma \ref{lemma:usefulllemmaforabsorbingtermsinhigheorderderivativeestimate} with $X=T$, we infer for any $\de\leq p\leq 2-\de$,
\bea\lab{eq:recoverhigherderivativesforestimatewaveequationpsi:1:T}
&& \sup_{\tau\in[\tau_1,\tau_2] }   E_{p} [T\psi](\tau)+   B_{p}[T\psi](\tau_1, \tau_2)  + F_{p}[T\psi](\tau_1, \tau_2)\\
\nn & \les& 
          E_{p}[T\psi](\tau_1)+   J^1_p[\psi, N]( \tau_1,\tau_2)+\ep^2\left(\sup_{[\tau_1, \tau_2]}E^1[\psi](\tau)+B^1_p[\psi](\tau_1, \tau_2)\right) +B_p[\psi](\tau_1, \tau_2).
\eea


 \subsubsection{Recovering estimates for $r\ddd_2\psi$}
 

Recall that $\psi$ satisfies 
 \beaa
\square_2 \psi+V\psi =N, \qquad V=\ka\kab,
\eeaa
and recall also from Corollary \ref{cor:keycorollaryforcommutationwavewithangularderivatives} that we have
\beaa
 r\ddd_2(\square_2\psi) -(\square_1-K)(r\ddd_2\psi) &=&  -r\eta\square_2 \psi +\dk^{\leq 1}(\Ga_g)\dk^{\leq 2}\psi 
\eeaa
We infer
\beaa
\square_1(r\ddd_2\psi) +(V-K)r\ddd_2\psi &=& r\eta\square_2\psi+r\ddd_2(N)+\dk^{\leq 1}(\Ga_g)\dk^{\leq 2}\psi\\
&=& -r\eta N+r\ddd_2(N)+\dk^{\leq 1}(\Ga_g)\dk^{\leq 2}\psi.
\eeaa
and hence
\beaa
\square_1(r\ddd_2\psi) +(V-K)r\ddd_2\psi &=& -r\eta N+r\ddd_2(N)+\dk^{\leq 1}(\Ga_g)\dk^{\leq 2}\psi.
\eeaa
In view of Corollary \ref{cor:estimatesforwaveequationskscalarswithphi1phi2RHS} with $\phi=r\ddd_2\psi$, $\phi_1=\dk^{\leq 1}\psi$ and $\phi_2=-r\eta N+r\ddd_2(N)$, and in view of \eqref{eq:someusefullnonsharpestimateforGagood}, we deduce
 \beaa
&&  \sup_{\tau\in[\tau_1,\tau_2] }   E_{p} [r\ddd_2\psi](\tau)+   B_{p}[r\ddd_2\psi](\tau_1, \tau_2)  + F_{p}[r\ddd_2\psi](\tau_1, \tau_2)\\
  & \les&      E_{p}[r\ddd_2\psi](\tau_1)+    \ep^2\left(\sup_{[\tau_1, \tau_2]}E[\dk^{\leq 1}\psi](\tau)+B_p[\dk^{\leq 1}\psi](\tau_1, \tau_2)\right)\\
  &&+J_p\Big[r\ddd_2\psi, -r\eta N+r\ddd_2(N)\Big]( \tau_1,\tau_2)+\int_{\Mtrap(\tau_1,\tau_2)}\left|1-\frac{3m}{r}\right||r\ddd_2\phi|(|r\ddd_2\phi|+|R(r\ddd_2\phi)|)\\
       && +\int_{\Mntrap(\tau_1,\tau_2)}r^{p-3}|r\ddd_2\phi|(|r\ddd_2\phi|+|\dk(r\ddd_2\phi)|),
\eeaa 
and hence, using Lemma \ref{lemma:usefulllemmaforabsorbingtermsinhigheorderderivativeestimate} with $X=r\ddd_2$, we infer for any $\de\leq p\leq 2-\de$,
\bea\lab{eq:recoverhigherderivativesforestimatewaveequationpsi:1:dd2}
&& \sup_{\tau\in[\tau_1,\tau_2] }   E_{p} [r\ddd_2\psi](\tau)+   B_{p}[r\ddd_2\psi](\tau_1, \tau_2)  + F_{p}[r\ddd_2\psi](\tau_1, \tau_2)\\
\nn & \les&   E_{p}[r\ddd_2\psi](\tau_1)+   J^1_p[\psi, N]( \tau_1,\tau_2)+\ep^2\left(\sup_{[\tau_1, \tau_2]}E^1[\psi](\tau)+B^1_p[\psi](\tau_1, \tau_2)\right)+B_p[\psi](\tau_1, \tau_2).
\eea


 \subsubsection{Recovering estimates for $R\psi$ in $r\leq r_0$}
 

We start with the following lemma.
\begin{lemma}
Let $\psi$ satisfy 
\beaa
\square_2\psi =V\psi+N, \qquad V=\ka\kab.
\eeaa
Then, $R^2\psi$ satisfies 
\beaa
R^2\psi &=& -\Up N+T^2\psi +O(r^{-2})\dkb^2\psi +O(r^{-1})\dk\psi +O(r^{-2})\psi.
\eeaa
\end{lemma}

\begin{proof}
Recall that we have
\beaa
\square_2 \psi &=& -e_3(e_4( \psi))+\lapp_2\psi -\frac{1}{2}\ka e_3\psi +\left(-\frac{1}{2}\kab+2\omb\right) e_4\psi+2\eta e_\th\psi
\eeaa
and 
\beaa
e_4=T+R, \quad \Up e_3=(T-R).
\eeaa
We infer
\beaa
\Up\square_2 \psi &=& -(T-R)(T+R)\psi+\Up\lapp_2\psi -\frac{\Up}{2}\ka e_3\psi +\Up\left(-\frac{1}{2}\kab+2\omb\right) e_4\psi+2\Up\eta e_\th\psi\\
&=& -T^2\psi+R^2\psi-[T,R]\psi+O(r^{-2})\dkb^2\psi +O(r^{-1})\dk\psi 
\eeaa
and hence
\beaa
R^2\psi &=& -\Up\square_2 \psi+T^2\psi -[T,R]\psi+O(r^{-2})\dkb^2\psi +O(r^{-1})\dk\psi\\
&=& -\Up N+T^2\psi -[T,R]\psi+O(r^{-2})\dkb^2\psi +O(r^{-1})\dk\psi +O(r^{-2})\psi
\eeaa
where we used the fact that $\square_2\psi=V\psi+N$ and $V=\ka\kab=O(r^{-2})$. Also, we have
\beaa
[T,R]\psi &=& \frac{1}{4}[e_4+\Up e_3, e_4-\Up e_3]\psi\\
&=& \frac{1}{2}\Big(-e_4(\Up)e_3 +\Up[ e_3, e_4]\Big)\psi\\
&=& O(r^{-2})\dk\psi
\eeaa
and thus
\beaa
R^2\psi &=& -\Up N+T^2\psi +O(r^{-2})\dkb^2\psi +O(r^{-1})\dk\psi +O(r^{-2})\psi.
\eeaa
This concludes the proof of the lemma.
\end{proof}

We now estimate $R\psi$ in $r\leq r_0$ for a fixed $r_0\geq 4m_0$ that will be chosen large enough. First, in view of the identity of the previous lemma, i.e.
\beaa
R^2\psi &=& -\Up N+T^2\psi +O(r^{-2})\dkb^2\psi +O(r^{-1})\dk\psi +O(r^{-2})\psi,
\eeaa
we infer
\bea\lab{eq:firstestimateforRpsiwithdegeneracyatr=3m}
&&\sup_{[\tau_1, \tau_2]}\int_{\Sigma_{r\leq r_0}(\tau)}|R^2\psi|^2+\int_{\MM_{r\leq r_0}(\tau_1, \tau_2)}\left(1-\frac{3m}{r}\right)^2(R^2\psi)^2\\
\nn &\les& \sup_{[\tau_1, \tau_2]}\left(E[T\psi]+E[r\ddd_2\psi]+E[\psi]+\int_{\Sigma_{r\leq r_0}(\tau)}N^2\right)\\
\nn&&+\int_{\MM_{r\leq r_0}(\tau_1, \tau_2)}N^2+\Morr[T\psi](\tau_1, \tau_2)+\Morr[r\ddd_2\psi](\tau_1, \tau_2)+\Morr[\psi](\tau_1, \tau_2).
\eea

Next, we remove the degeneracy of the above estimate at $r=3m$. Recall from Corollary \ref{cor:waveequationforRpsiinrleq8m0} that we have in the region $r\leq 4m_0$
\beaa
\square_2(R\psi) &=&  \left(1-\frac{3m}{r}\right)\dk^2\psi +O\left(\frac{\ep}{u_{trap}^{1+\dec-2\de_0}}\right)\dk^2\psi + O(1)\dk^{\leq 1}\psi+O(1)\dk^{\leq 1}N.
\eeaa
Then,
\begin{enumerate}
\item multiplying $R\psi$ with a cut-off function equal to one on $[5/2m_0, 7/2m_0]$ and vanishing on $[9/4m_0, 4m_0]$ and inferring the corresponding wave equation from the above one for $R\psi$, 

\item relying on the Morawetz estimate of Proposition \ref{prop-ident-Mor1} with the particular choice $f(r)=r-3m$, 

\item adding a large multiple of the energy estimate,

\item using Proposition \ref{prop:boundar-Mor} for the boundary terms,
\end{enumerate}
we easily infer the following estimate 
\beaa
\int_{\Mtrap(\tau_1, \tau_2)}(R^2\psi)^2 &\les& \int_{\MM_{r\leq4m_0}(\tau_1, \tau_2)}\left(\left(1-\frac{3m}{r}\right)^2(\dk^2\psi)^2+(\dk\psi)^2+(\dk^{\leq 1}N)^2\right)\\
&&+E[R\psi](\tau_1)+\ep\sup_{[\tau_1,\tau_2]}E^1[\psi](\tau).
\eeaa
Together with \eqref{eq:firstestimateforRpsiwithdegeneracyatr=3m}, we infer
 \bea\lab{eq:firstestimateforRpsiwithnodegeneracy}
&&\sup_{[\tau_1, \tau_2]}\int_{\Sigma_{r\leq r_0}(\tau)}|R^2\psi|^2+\int_{\MM_{r\leq r_0}(\tau_1, \tau_2)}|R^2\psi|^2\\
\nn &\les& E[R\psi](\tau_1)+\sup_{[\tau_1, \tau_2]}\Big(\ep E^1[\psi](\tau)+E[T\psi](\tau)+E[r\ddd_2\psi](\tau)+E[\psi](\tau)\Big)\\
\nn&&+J^1_p[\psi, N](\tau_1, \tau_2)+\Morr[T\psi](\tau_1, \tau_2)+\Morr[r\ddd_2\psi](\tau_1, \tau_2)+\Morr[\psi](\tau_1, \tau_2).
\eea


 \subsubsection{Recovering estimates for $Y_\HH\psi$}
 

Recall that $\psi$ satisfies 
 \beaa
\square_2 \psi+V\psi =N, \qquad V=\ka\kab, 
\eeaa
and recall also from Lemma \ref{lemma:keycorollaryforcommutationwavewithredshift}
\beaa
[\square_2, Y_\HH]\psi &=& d_0Y_{(0)}(Y_\HH\psi) +1_{\Up\leq 2\deh}\left(\square_2\psi+\dk T\psi+\dk^{\leq 1}(\Ga_g)\dk^2\psi+\frac{1}{\deh^2}\dk^{\leq 1}\psi\right)\\
&&+\frac{1}{\deh}1_{\deh\leq \Up\leq 2\deh}\dk^{\leq 2}\psi
\eeaa
where the scalar function $d_0$ satisfying the bound
\beaa
d_0 &=& \frac{1}{2m_0}+O(\deh)\textrm{ on the support of }\ka_\HH.
\eeaa
We infer
\beaa
\square_2(Y_\HH\psi) +VY_\HH(\psi) &=& d_0Y_{(0)}(Y_\HH\psi) +1_{\Up\leq 2\deh}\left(N+\dk T\psi+\ep\dk^2\psi+\frac{1}{\deh^2}\dk^{\leq 1}\psi\right)\\
&&+\frac{1}{\deh}1_{\deh\leq \Up\leq 2\deh}\dk^{\leq 2}\psi+Y_\HH(N).
\eeaa
Then, 
\begin{enumerate}
\item we use the redshift vectorfield $Y_\HH$ as a multiplier,

\item we rely on Proposition \ref{prop:red-shift-Mor}, 

\item we use the fact that $d_0\geq 0$,

\item we add a large multiple of the energy,

\item we use Proposition \ref{prop:boundar-Mor} for the boundary terms. 
\end{enumerate}
We easily infer
\bea\lab{eq:firstestimateforYHHpsi}
\nn\sup_{[\tau_1, \tau_2]}E[Y_\HH\psi]+\Morr[Y_\HH\psi](\tau_1, \tau_2) &\les& E[Y_\HH\psi](\tau_1)+J^1_p[\psi, N](\tau_1, \tau_2)+\ep\Morr[\dk\psi](\tau_1, \tau_2)\\
\nn&&+\Morr[R\psi](\tau_1, \tau_2)+\Morr[T\psi](\tau_1, \tau_2)\\
&&+\Morr[r\ddd_2\psi](\tau_1, \tau_2)+\Morr[\psi](\tau_1, \tau_2).
\eea


 \subsubsection{Recovering estimates for $re_4\psi$ in $r\geq r_0$}
 

Recall that $\psi$ satisfies 
 \beaa
\square_2 \psi+V\psi =N, \qquad V=\ka\kab, 
\eeaa
and recall also from Lemma \ref{lemma:keycorollaryforcommutationwavewithre4}
\beaa
[\square_2, re_4]\psi &=& \frac{\Up}{r}\left(1+\frac{2m}{r\Up^2} \right)\ec_4(re_4\psi) +\square_2 \psi   +\Ga_g\dk^2\psi+\frac{1}{\Up}r^{-2}\dk T\psi+r^{-2}\dkb^2\psi+r^{-2}\dk\psi.
\eeaa
We infer
\beaa
\square_2(re_4\psi) +Vre_4(\psi) &=& \frac{\Up}{r}\left(1+\frac{2m}{r\Up^2} \right)\ec_4(re_4\psi) +O\left(\frac{\ep}{r^2}\right)\dk^2\psi+\frac{1}{\Up}r^{-2}\dk T\psi+r^{-2}\dkb^2\psi\\
&&+r^{-2}\dk^{\leq 1}\psi+N+re_4(N).
\eeaa
Then, 
\begin{enumerate}
\item as in section \ref{sec:proofoftheorem:Daf-Rodn1-psi-s}, we use the vectorfield $f_pe_4$ as a multiplier, where $f_p=\th_{r_0}(r)r^pe_4$ and the cut-off $\th_{r_0}(r)$ is equal to one in the region $r\geq r_0$ and vanishes in the region $r\leq r_0/2$, 

\item we rely on Proposition \ref{prop:QC-general-multiplier2} to control the bulk and the boundary terms,

\item we use the fact that the prefactor of the term $\ec_4(re_4\psi)$ on the right-hand side is positive for $r\geq 4m_0$, i.e.
\beaa
\frac{\Up}{r}\left(1+\frac{2m}{r\Up^2} \right)\geq 0\textrm{ for }r\geq 4m_0.
\eeaa
\end{enumerate}
We easily infer
\bea\lab{eq:recoverhigherderivativesforestimatewaveequationpsi:1:re_4psiforrleqr0}
\nn&& \sup_{\tau\in[\tau_1,\tau_2] }   E_{p, r\geq r_0}[re_4\psi](\tau)+   B_{p, r\geq r_0}[re_4\psi](\tau_1, \tau_2)  + F_{p, r\geq r_0}[re_4\psi](\tau_1, \tau_2)\\
\nn & \les& 
          E^1_p[re_4\psi](\tau_1)+   J^1_p[\psi, N]( \tau_1,\tau_2)+B^1_{p, r_0/2\leq r< r_0}[\psi](\tau_1, \tau_2)+\ep B^1_p[\psi](\tau_1, \tau_2) \\
          &&+B_p[T\psi](\tau_1, \tau_2)+B_p[r\ddd_2\psi](\tau_1, \tau_2)+B_p[\psi](\tau_1, \tau_2).
\eea


 \subsubsection{Conclusion of the proof of Theorem \ref{theorem-combinedMor-r-weighted}}
 

Gathering the estimates \eqref{eq:recoverhigherderivativesforestimatewaveequationpsi:1:T}, \eqref{eq:recoverhigherderivativesforestimatewaveequationpsi:1:dd2}, \eqref{eq:firstestimateforRpsiwithnodegeneracy}, \eqref{eq:firstestimateforYHHpsi} and \eqref{eq:recoverhigherderivativesforestimatewaveequationpsi:1:re_4psiforrleqr0}, we infer for any $\de\leq p\leq 2-\de$,
\beaa
&& \sup_{\tau\in[\tau_1,\tau_2] }   E^1_{p} [\psi](\tau)+   B^1_{p}[\psi](\tau_1, \tau_2)  + F^1_{p}[\psi](\tau_1, \tau_2)\\
\nn & \les& 
          E^1_{p}[\psi](\tau_1)+   J^1_p[\psi, N]( \tau_1,\tau_2)+\ep^2\left(\sup_{[\tau_1, \tau_2]}E^1_p[\psi](\tau)+B^1_p[\psi](\tau_1, \tau_2)\right) \\
          &&  +\sup_{[\tau_1, \tau_2]}E_p[\psi](\tau) +B_p[\psi](\tau_1, \tau_2),
\eeaa
and hence
\beaa
&& \sup_{\tau\in[\tau_1,\tau_2] }   E^1_{p} [\psi](\tau)+   B^1_{p}[\psi](\tau_1, \tau_2)  + F^1_{p}[\psi](\tau_1, \tau_2)\\
\nn & \les& 
          E^1_{p}[\psi](\tau_1)+   J^1_p[\psi, N]( \tau_1,\tau_2)  +\sup_{[\tau_1, \tau_2]}E_p[\psi](\tau) +B_p[\psi](\tau_1, \tau_2).
\eeaa
In view of \eqref{eq:recoverhigherderivativesforestimatewaveequationpsi:2}, we deduce
\beaa
\sup_{\tau\in[\tau_1,\tau_2] }   E^1_{p} [\psi](\tau)+   B^1_{p}[\psi](\tau_1, \tau_2)  + F^1_{p}[\psi](\tau_1, \tau_2) & \les& 
          E^1_{p}[\psi](\tau_1)+   J^1_p[\psi, N]( \tau_1,\tau_2)
\eeaa
which is Theorem \ref{theorem-combinedMor-r-weighted} in the case $s=1$. We have thus deduced Theorem \ref{theorem-combinedMor-r-weighted} in the case $s=1$ from the case $s=0$, i.e. \eqref{eq:recoverhigherderivativesforestimatewaveequationpsi:2}. Since going from $s=0$ to $s=1$ is analogous to going from $s$ to $s+1$, higher order derivatives $k\leq k_{small}+30$ are recovered in the same fashion. This concludes the proof of Theorem \ref{theorem-combinedMor-r-weighted}.


 \subsection{Proof of Theorem \ref{theorem:Daf-Rodn-estim2-psic}}\lab{sec:proofoftheorem:Daf-Rodn-estim2-psic}
 

We now conclude the proof of Theorem \ref{theorem:Daf-Rodn-estim2-psic} for all $0\leq s\leq k_{small}+29$ by 
recovering higher derivatives $s\geq 1$ one by one starting from the estimate $s=0$ provided by  \eqref{eq:recoverhigherderivativesforestimatewaveequationpsi:3}.  As explained in section \ref{sec:strategyforrecoveringhigherorderderivatives}, it suffices to recover the estimates for $s=1$ from the one for $s=0$ as the procedure to recover the estimate for $s+1$ from the one for $s$ is completely analogous. We now follow the strategy outlined in section \ref{sec:strategyforrecoveringhigherorderderivatives}.


 \subsubsection{Recovering estimates for $T\psic$}
 

Recall from Proposition \ref{square-psic-modified} that $\psic=f_2 \ec_4\psi$ satisfies  
\beaa
\square\psic -V\psic&=&  \frac{2}{r}\left(1-\frac{3m}{r}\right) e_4\psic+\check{N}+ f_2 \left(e_4 +\frac 3  r \right) N
\eeaa
where,
\beaa
\check N=
\begin{cases}
&     O(r^{-2})\dk^{\leq 1}\psi+r\Ga_b e_4\dk\psi +\dk^{\leq 1}(\Ga_b)\dk^{\leq 1}\psi  +r\dk^{\leq 1}(\Ga_g)e_3\psi +\dk^{\leq 1}(\Ga_g)\dk^2\psi, \quad  r\ge 6m_0,\\
\\
& O(1)\dk^{\leq 2}\psi, \qquad\qquad \quad \quad\,\, 4m_0\le r\le  6m_0,
\end{cases}
\eeaa
and recall also from Corollary \ref{cor:keycorollaryforcommutationwavewithT} that we have
\beaa
[T , \square_2]\psic &=& \dk^{\leq 1}(\Ga_g)\dk^{\leq 2}\psic.
\eeaa
We infer
\beaa
\square(T\psic) -VT(\psic) &=& \frac{2}{r}\left(1-\frac{3m}{r}\right) e_4(T\psic)+N^T+ T\left(f_2 \left(e_4 +\frac 3  r \right) N\right),
\eeaa
where we have, in view of the estimates\footnote{Here, unlike the proof of Theorem \ref{theorem-combinedMor-r-weighted} above, the non sharp estimates of section \ref{sec:necessaryassumptionstorecoveryhigherderivatives} are not enough, and we need instead to rely on the stronger estimates provided by Lemma \ref{le:interpolatedbootstrap}.} 
 of Lemma \ref{le:interpolatedbootstrap} for $\dk^k\Ga_g$ and $\dk^k\Ga_b$ with $k\leq k_{small}+30$,
\beaa
N^T=
\begin{cases}
&  O\left(\frac{1}{\tau^{1+\dec-2\de_0}}\right)\Big(e_4\dk^2\psi +r^{-1}\dk^{\leq 2}\psi\Big)  \\
&+O\left(\frac{1}{r\tau^{\frac{1}{2}+\dec-2\de_0}}\right)e_3\dk^{\leq 1}\psi +O\left(\frac{1}{r^2}\right)\Big(\dk^{\leq 3}\psi+\ep\dk^{\leq 2}\psic\Big), \quad  r\ge 6m_0,\\
\\
&O(1)\dk^{\leq 3}\psi, \qquad\qquad \quad \quad\,\, 4m_0\le r\le  6m_0.
\end{cases}
\eeaa

In view of \eqref{eq:recoverhigherderivativesforestimatewaveequationpsi:3} with $T\psic$ instead of $\psic$ and with 
\beaa
N^T+ T\left(f_2 \left(e_4 +\frac 3  r \right) N\right)
\eeaa
instead of $\check{N}+ f_2 \left(e_4 +\frac 3  r \right) N$, we deduce
    \beaa
&&  \sup_{\tau\in[\tau_1,\tau_2] }  E_q[T\psic](\tau)+   B_q[T\psic](\tau_1, \tau_2)    \\
    &\les& 
           E_q [T\psic](\tau_1)      +  J_q\left[T\psic, N^T+ T\left(f_2 \left(e_4 +\frac 3  r \right) N\right)\right] (\tau_1,\tau_2)  \\
           &&+    E\, ^1_{\max(q, \de)}[T\psi](\tau_1)  + J^{1}_{\max(q, \de)}[T\psi, TN]\\
               &\les& 
           E_q [T\psic](\tau_1)      +  \Jc_q^1\left[\psic, N \right] (\tau_1,\tau_2) +  J_q\left[T\psic, N^T\right] (\tau_1,\tau_2) \\
           &&+    E\, ^2_{\max(q, \de)}[\psi](\tau_1)  + J^2_{\max(q, \de)}[\psi, N],
\eeaa
so that it remains to estimate 
\beaa
J_q[T\psic, N^T] (\tau_1,\tau_2) = J_{q,4m_0}\left[T\psic, N^T\right](\tau_1,\tau_2) = \left|\int_{\MM_{\geq 4m_0}(\tau_1, \tau_2)}r^q(\ec_4(T\psic))N^T\right|.
\eeaa
We have, in view of the definition of $N^T$, 
\beaa
&& J_q[T\psic, N^T] (\tau_1,\tau_2)\\
&\les& \left|\int_{\MM_{\geq 4m_0}(\tau_1, \tau_2)}r^q(\ec_4(T\psic))\frac{1}{\tau^{1+\dec-2\de_0}}\Big(e_4\dk^2\psi +r^{-1}\dk^{\leq 2}\psi\Big)\right|\\
&&+\left|\int_{\MM_{\geq 4m_0}(\tau_1, \tau_2)}r^q(\ec_4(T\psic))\frac{1}{r\tau^{\frac{1}{2}+\dec-2\de_0}}e_3\dk^{\leq 1}\psi\right|\\
&&+\left|\int_{\MM_{\geq 4m_0}(\tau_1, \tau_2)}r^q(\ec_4(T\psic))\frac{1}{r^2}\dk^{\leq 3}\psi\right|+\left|\int_{\MM_{\geq 4m_0}(\tau_1, \tau_2)}r^q(\ec_4(T\psic))\frac{\ep}{r^2}\dk^{\leq 2}\psic\right|\\
&\les& \left(\sup_{[\tau_1, \tau_2]}\int_{\Sigma(\tau)}r^q(\ec_4(T\psic))^2\right)^{\frac{1}{2}}\Bigg\{\left(\sup_{[\tau_1, \tau_2]}\int_{\Sigma(\tau)}r^q\Big((e_4\dk^2\psi)^2 +r^{-2}(\dk^{\leq 2}\psi)^2\Big)\right)^{\frac{1}{2}}\\
&&+\left(\int_{\MM_{r\geq 4m_0}(\tau_1, \tau_2)}r^{q-2}(e_3\dk^{\leq 1}\psi)^2\right)^{\frac{1}{2}}\Bigg\}+\ep\int_{\MM(\tau_1, \tau_2)}r^{q-3}(\dk^{\leq 2}\psic)^2\\
&&+\left(\int_{\MM(\tau_1, \tau_2)}r^{q-1}(\ec_4(T\psic))^2\right)^{\frac{1}{2}}\left(\int_{\MM(\tau_1, \tau_2)}r^{q-3}(\dk^{\leq 3}\psi)^2\right)^{\frac{1}{2}}
\eeaa
which yields, using in particular the fact that $q\leq 1-\de$,
\beaa
&& J_q[T\psic, N^T] (\tau_1,\tau_2)\\
&\les& \left(\sup_{[\tau_1, \tau_2]}E_q[T\psic](\tau)\right)^{\frac{1}{2}}\left\{\sup_{[\tau_1, \tau_2]}E^2_{\max(q,\de)}[\psi](\tau)+B^1_{\max(q,\de)}[\psi](\tau_1, \tau_2)\right\}^{\frac{1}{2}}+\ep B^1_q[\psic](\tau_1, \tau_2)\\
&&+\Big(B_q[T\psic](\tau_1, \tau_2)\Big)^{\frac{1}{2}}\Big(B^2_{\max(q,\de)}[\psi](\tau_1, \tau_2)\Big)^{\frac{1}{2}}.
\eeaa
 We deduce 
    \beaa
  &&\sup_{\tau\in[\tau_1,\tau_2] }  E_q[T\psic](\tau)+   B_q[T\psic](\tau_1, \tau_2)          \\
       &\les& 
           E_q [T\psic](\tau_1)      +  \Jc_q^1\left[\psic, N \right] (\tau_1,\tau_2) + \ep B^1_q[\psic](\tau_1, \tau_2)\\
           &&+\sup_{[\tau_1, \tau_2]}E^2_{\max(q,\de)}[\psi](\tau)+B^2_{\max(q,\de)}[\psi](\tau_1, \tau_2)  + J^2_{\max(q, \de)}[\psi, N].
\eeaa 
Together with Theorem \ref{theorem-combinedMor-r-weighted}, this yields
   \bea\lab{eq:weightedenergyesimtateforTpsics}
 \nn\sup_{\tau\in[\tau_1,\tau_2] }  E_q[T\psic](\tau)+   B_q[T\psic](\tau_1, \tau_2)                   &\les& 
           E_q [T\psic](\tau_1)      +  \Jc_q^1\left[\psic, N \right] (\tau_1,\tau_2)  + \ep B^1_q[\psic](\tau_1, \tau_2)\\
           &&+    E\, ^2_{\max(q, \de)}[\psi](\tau_1)  + J^2_{\max(q, \de)}[\psi, N].
\eea


 \subsubsection{Recovering estimates for $r\ddd_2\psic$}
 

Recall from Proposition \ref{square-psic-modified} that $\psic=f_2 \ec_4\psi$ satisfies  
\beaa
\square\psic -V\psic&=&  \frac{2}{r}\left(1-\frac{3m}{r}\right) e_4\psic+\check{N}+ f_2 \left(e_4 +\frac 3  r \right) N.
\eeaa
Recall also from Corollary \ref{cor:keycorollaryforcommutationwavewithangularderivatives} that we have
\beaa
 r\ddd_2(\square_2\psic) -(\square_1-K)(r\ddd_2\psic) &=&  -r\eta\square_2 \psic +\dk^{\leq 1}(\Ga_g)\dk^{\leq 2}\psic 
\eeaa
We infer
\beaa
\square_1(r\ddd_2\psic) +(V-K)r\ddd_2\psic &=& \frac{2}{r}\left(1-\frac{3m}{r}\right)e_4(r\ddd_2\psic) +N^{r\ddd_2}+ (r\ddd_2 -r\eta)\left(f_2 \left(e_4 +\frac 3  r \right) N\right)
\eeaa
where
\beaa
N^{r\ddd_2} &=& -r\eta \check{N}+r\ddd_2(\check{N})+\dk^{\leq 1}(\Ga_g)\dk^{\leq 2}\psic.
\eeaa

In view of \eqref{eq:nonsharpweightedesitmatesforreduced1scalarsolutionofwaveequationwithpotential:2} with $r\ddd_2\psic$ instead of $\psic$  and with 
\beaa
N^{r\ddd_2}+ (r\ddd_2 -r\eta)\left(f_2 \left(e_4 +\frac 3  r \right) N\right)
\eeaa
instead of $\check{N}+ f_2 \left(e_4 +\frac 3  r \right) N$, we deduce
    \beaa
 \nn&& \sup_{\tau\in[\tau_1,\tau_2] }  E_q[r\ddd_2\psic](\tau)+   B_q[r\ddd_2\psic](\tau_1, \tau_2) \\
\nn  &\les&   E_q [r\ddd_2\psic](\tau_1)      +  \Jc_q\left[r\ddd_2\psic, N^{r\ddd_2}+ (r\ddd_2 -r\eta)\left(f_2 \left(e_4 +\frac 3  r \right) N\right)\right] (\tau_1,\tau_2)\\
&&  +    E\, ^1_{\max(q, \de)}[r\ddd_2\psi](\tau_1)  + J^{1}_{\max(q, \de)}[r\ddd_2\psi, r\ddd_2N]\\
&&+ \int_{\MM(\tau_1,\tau_2)}r^{q-3}|r\ddd_2\psic|(|r\ddd_2\psic|+|\dk(r\ddd_2\psic)|)\\
\nn  &\les&   E_q [r\ddd_2\psic](\tau_1)      +  \Jc_q\left[r\ddd_2\psic, N^{r\ddd_2}\right] (\tau_1,\tau_2)\\
&&  +    E\, ^2_{\max(q, \de)}[\psi](\tau_1)  + J^2_{\max(q, \de)}[\psi, N]+ \Big(B_q[r\ddd_2\psic](\tau_1, \tau_2)\Big)^{\frac{1}{2}} \Big(B_q[\psic](\tau_1, \tau_2)\Big)^{\frac{1}{2}}
\eeaa
so that it remains to estimate 
\beaa
\Jc_q\left[r\ddd_2\psic, N^{r\ddd_2}\right] (\tau_1,\tau_2) = J_{q,4m_0}\left[r\ddd_2\psic, N^{r\ddd_2}\right](\tau_1,\tau_2) = \left|\int_{\MM_{\geq 4m_0}(\tau_1, \tau_2)}r^q(\ec_4(r\ddd_2\psic))N^{r\ddd_2}\right|.
\eeaa
The estimate follows along the same lines as the above one for $J_q[T\psic, N^T] (\tau_1,\tau_2)$ so we leave the details to the reader. In the end, we arrive at the following analog of \eqref{eq:weightedenergyesimtateforTpsics}
   \bea\lab{eq:weightedenergyesimtateforrddd2psics}
 \nn&&\sup_{\tau\in[\tau_1,\tau_2] }  E_q[r\ddd_2\psic](\tau)+   B_q[r\ddd_2\psic](\tau_1, \tau_2)              \\
 \nn     &\les& 
           E_q [r\ddd_2\psic](\tau_1)+B_q[\psic](\tau_1, \tau_2)      +  \Jc_q^1\left[\psic, N \right] (\tau_1,\tau_2)  + \ep B^1_q[\psic](\tau_1, \tau_2)\\
           &&+    E\, ^2_{\max(q, \de)}[\psi](\tau_1)  + J^2_{\max(q, \de)}[\psi, N].
\eea


 \subsubsection{Recovering estimates for $re_4\psic$}
 

Recall from Proposition \ref{square-psic-modified} that $\psic=f_2 \ec_4\psi$ satisfies  
\beaa
\square\psic -V\psic&=&  \frac{2}{r}\left(1-\frac{3m}{r}\right) e_4\psic+\check{N}+ f_2 \left(e_4 +\frac 3  r \right) N.
\eeaa
Recall also from Lemma \ref{lemma:keycorollaryforcommutationwavewithre4} that we have
\beaa
[\square_2, re_4]\psic &=& \frac{\Up}{r}\left(1+\frac{2m}{r\Up^2} \right)\ec_4(re_4\psic) +\square_2 \psic   +\Ga_g\dk^2\psic+\frac{1}{\Up}r^{-2}\dk T\psic+r^{-2}\dkb^2\psic+r^{-2}\dk\psic.
\eeaa
We infer\footnote{Notice that the coefficient in front of the term $e_4(re_4\psic)$ in the RHS of the wave equation for $re_4\psic$ differs from the one in front of the term $e_4\psic$ in the RHS of the wave equation for $\psic$. Nevertheless, we may apply  \eqref{eq:recoverhigherderivativesforestimatewaveequationpsi:3} with $re_4\psic$ instead of $\psic$ since the only property of this coefficient which is used in that it is positive on $r\geq 4m_0$, i.e. 
\beaa
\frac{2}{r}\left(1-\frac{3m}{r}\right)+\frac{\Up}{r}\left(1+\frac{2m}{r\Up^2} \right)\geq 0\textrm{ on }r\geq 4m_0.
\eeaa}
\beaa
\square(re_4\psic) -Vre_4(\psic) &=& \left(\frac{2}{r}\left(1-\frac{3m}{r}\right)+\frac{\Up}{r}\left(1+\frac{2m}{r\Up^2} \right)\right) e_4(re_4\psic)+N^{re_4}\\
&&+ re_4\left(f_2 \left(e_4 +\frac 3  r \right) N\right),
\eeaa
where 
\beaa
N^{re_4} &=& re_4(\check{N})+\check{N}+\Ga_g\dk^2\psic+\frac{1}{\Up}r^{-2}\dk T\psic+r^{-2}\dkb^2\psic+r^{-2}\dk\psic.
\eeaa
The rest follows along the same lines as the estimate for $T\psic$ and we arrive at the following analog of \eqref{eq:weightedenergyesimtateforTpsics}
   \bea\lab{eq:weightedenergyesimtateforre4psics}
 &&\sup_{\tau\in[\tau_1,\tau_2] }  E_q[re_4\psic](\tau)+   B_q[re_4\psic](\tau_1, \tau_2)              \\
 \nn     &\les& 
           E_q [re_4\psic](\tau_1)+B_q[\psic](\tau_1, \tau_2) +B_q[T\psic](\tau_1, \tau_2)+B_q[r\ddd_2\psic](\tau_1, \tau_2) \\
 \nn          && +  \Jc_q^1\left[\psic, N \right] (\tau_1,\tau_2)  + \ep B^1_q[\psic](\tau_1, \tau_2)+    E\, ^2_{\max(q, \de)}[\psi](\tau_1)  + J^2_{\max(q, \de)}[\psi, N].
\eea


 \subsubsection{Conclusion of the proof of Theorem \ref{theorem:Daf-Rodn-estim2-psic}}
 

Gathering the estimates \eqref{eq:weightedenergyesimtateforTpsics}, \eqref{eq:weightedenergyesimtateforrddd2psics}  and \eqref{eq:weightedenergyesimtateforre4psics}, we infer for any $-1+\de<q\leq 1-\de$,
 \beaa
 \sup_{\tau\in[\tau_1,\tau_2] }   E^1_q [\psic](\tau)+   B^1_q[\psic](\tau_1, \tau_2)   & \les& 
          E_q^1 [\psic](\tau_1)      +  \Jc^1_q[\psic, N] (\tau_1,\tau_2) +B_q[\psic](\tau_1, \tau_2)\\
          &&+\ep B^1_q[\psic](\tau_1, \tau_2)   +E\, ^2_{\max(q, \de)}[\psi](\tau_1)  + J^2_{\max(q, \de)}[\psi, N]
\eeaa
and hence
 \beaa
 \sup_{\tau\in[\tau_1,\tau_2] }   E^1_q [\psic](\tau)+   B^1_q[\psic](\tau_1, \tau_2)   & \les& 
          E_q^1 [\psic](\tau_1)      +  \Jc^1_q[\psic, N] (\tau_1,\tau_2) +B_q[\psic](\tau_1, \tau_2) \\
          &&  +E\, ^2_{\max(q, \de)}[\psi](\tau_1)  + J^2_{\max(q, \de)}[\psi, N].
\eeaa
In view of \eqref{eq:recoverhigherderivativesforestimatewaveequationpsi:3}, we deduce
 \beaa
 \sup_{\tau\in[\tau_1,\tau_2] }   E^1_q [\psic](\tau)+   B^1_q[\psic](\tau_1, \tau_2)   & \les& 
          E_q^1 [\psic](\tau_1)      +  \Jc^1_q[\psic, N] (\tau_1,\tau_2)  \\
          &&  +E\, ^2_{\max(q, \de)}[\psi](\tau_1)  + J^2_{\max(q, \de)}[\psi, N]
\eeaa
which is Theorem \ref{theorem:Daf-Rodn-estim2-psic} in the case $s=1$. We have thus deduced Theorem \ref{theorem:Daf-Rodn-estim2-psic} in the case $s=1$ from the case $s=0$, i.e. \eqref{eq:recoverhigherderivativesforestimatewaveequationpsi:3}. Since going from $s=0$ to $s=1$ is analogous to going from $s$ to $s+1$, higher order derivatives $k\leq k_{small}+29$ are recovered in the same fashion. This concludes the proof of Theorem \ref{theorem:Daf-Rodn-estim2-psic}.


\section{More weighted estimates for wave equations}


The goal of this section is to derive Theorem \ref{theorem-recoveringhigherderivatives} and Proposition  \ref{prop:controlonMintbyMextthankstoredshiftneededinThmM8}, see below, which is needed for the proof of Theorem M8 in Chapter \ref{chap:proofoftheoremM0M7M8}. Recall that we have used so far in Chapter \ref{chapter:waveeqtionestimates} the global frame of Proposition \ref{prop:existenceandestimatesfortheglobalframe:bis}. For this last section of Chapter \ref{chapter:waveeqtionestimates}, we rely instead on the global frame used in Theorem M8, i.e.  the one of Proposition \ref{prop:existenceandestimatesfortheglobalframe}, as it is more regular and allows us to derive estimates for up to $k_{large}$ derivatives.

\begin{remark}
Recall that in the frame of Proposition \ref{prop:existenceandestimatesfortheglobalframe}, we only have\footnote{Unlike the frame of Proposition \ref{prop:existenceandestimatesfortheglobalframe:bis} for which $\eta\in\Ga_g$.} $\eta\in\Ga_b$. Note that the assumptions on the frame used in Chapter \ref{chapter:waveeqtionestimates} are all consistent with $\eta\in\Ga_b$, so that all results in this chapter apply for the frame of Proposition \ref{prop:existenceandestimatesfortheglobalframe}.
\end{remark}

 \begin{theorem}
 \label{theorem-recoveringhigherderivatives}
Let $\psi$ a reduced 2-scalar, and $\phi$ a reduced 0-scalar satisfying respectively 
\beaa
(\square_2+V_2)\psi = N_2, \qquad (\square_0+V_0)\phi = N_0, \qquad V_2=- \frac{2}{r^2}\left(1+\frac{2m}{r}\right), \qquad V_0=\frac{8m}{r^3}.
\eeaa
Also, assume that the Ricci coefficients and curvature components associated to the global null frame we are using satisfy the estimates of section \ref{sec:necessaryassumptionstorecoveryhigherderivatives} for $k\leq k_{small}$ derivatives. Then, for   any $1\leq s\leq k_{large}-1$, we have
  \bea\lab{eq:thefirstequationoftheorem-recoveringhigherderivatives}
 \nn &&\sup_{\tau\in[\tau_1,\tau_2] }   E^s_{\de} [\psi](\tau)+   B^s_{\de}[\psi](\tau_1, \tau_2)  + F^s_{\de}[\psi](\tau_1, \tau_2)\\
  \nn &\les& 
          E^s_{\de}[\psi](\tau_1) +\sup_{\tau\in[\tau_1,\tau_2] }   E^{s-1}_{\de} [\psi](\tau)+   B^{s-1}_{\de}[\psi](\tau_1, \tau_2)  + F^{s-1}_{\de}[\psi](\tau_1, \tau_2)\\
 \nn&&+ D_s[\Ga]\left(\sup_{\MM(\tau_1, \tau_2)}ru_{trap}^{\frac{1}{2}+\dec}|\dk^{\leq k_{small}}\psi|\right)^2\\
 &&+\int_{\MM(\tau_1, \tau_2)}r^{1+\de}|\dk^{\leq s}N_2|^2+\left|\int_{\Mtrap(\tau_1, \tau_2)}T(\dk^s\phi)\dk^sN_2\right|
\eea
and
 \bea\lab{eq:thesecondequationoftheorem-recoveringhigherderivatives}
\nn  &&\sup_{\tau\in[\tau_1,\tau_2] }   E^s_{\de} [\phi](\tau)+   B^s_{\de}[\phi](\tau_1, \tau_2)  + F^s_{\de}[\phi](\tau_1, \tau_2)\\
  \nn &\les& 
          E^s_{\de}[\phi](\tau_1) +\sup_{\tau\in[\tau_1,\tau_2] }   E^{s-1}_{\de} [\phi](\tau)+   B^{s-1}_{\de}[\phi](\tau_1, \tau_2)  + F^{s-1}_{\de}[\phi](\tau_1, \tau_2)\\
\nn&&+D_s[\Ga] \left(\sup_{\MM(\tau_1, \tau_2)}ru_{trap}^{\frac{1}{2}+\dec}|\dk^{\leq k_{small}}\phi|\right)^2+\int_{\Sigma(\tau_2)}\frac{(\dk^{\leq s}\phi)^2}{r^3}\\
&&+\int_{\MM(\tau_1, \tau_2)}r^{1+\de}|\dk^{\leq s}N_0|^2+\left|\int_{\Mtrap(\tau_1, \tau_2)}T(\dk^s\phi)\dk^sN_0\right|,
\eea
where $D_s[\Ga]$ is defined by
\beaa
D_s[\Ga] &:=& \int_{\Mint\cup\Mext(r\leq 4m_0)}(\dk^{\leq s}\Gac)^2\\
&& +\sup_{r_0\geq 4m_0}\left(r_0\int_{\{r=r_0\}}|\dk^{\leq s}\Ga_g|^2+r_0^{-1}\int_{\{r=r_0\}}|\dk^{\leq s}\Ga_b|^2\right).
\eeaa
\end{theorem}

The proof of Theorem \ref{theorem-recoveringhigherderivatives} relies on the following theorem.
\begin{theorem}\lab{thm:nonsharpweightedesitmatesforreduced0scalarsolutionofwaveequationwithpotential}
Let $\psi$ a reduced scalar, and $\phi$ a reduced 0-scalar satisfying respectively 
\beaa
(\square_2+V_2)\psi = N_2, \qquad (\square_0+V_0)\phi = N_0, \qquad V_2=- \frac{2}{r^2}\left(1+\frac{2m}{r}\right), \qquad V_0=\frac{8m}{r^3}.
\eeaa
Then, we have
 \bea\lab{eq:nonsharpweightedesitmatesforreduced0scalarsolutionofwaveequationwithpotential:1:00}
 \nn && \sup_{\tau\in[\tau_1,\tau_2] }   E_\de [\psi](\tau)+   B_\de[\psi](\tau_1, \tau_2)  + F_\de[\psi](\tau_1, \tau_2)\\
 \nn  & \les&  E_\de[\psi](\tau_1)+\int_{\Mtrap(\tau_1,\tau_2)}\left|1-\frac{3m}{r}\right||\psi|(|\psi|+|R\psi|) +\int_{\Mntrap(\tau_1,\tau_2)}r^{\de-3}|\psi|(|\psi|+|\dk\psi|)\\
 &&+\int_{\MM(\tau_1, \tau_2)}r^{1+\de}|N_2|^2+\left|\int_{\Mtrap(\tau_1, \tau_2)}T(\psi)N_2\right|,
\eea
and
 \bea\lab{eq:nonsharpweightedesitmatesforreduced0scalarsolutionofwaveequationwithpotential:1}
 \nn && \sup_{\tau\in[\tau_1,\tau_2] }   E_\de [\phi](\tau)+   B_\de[\phi](\tau_1, \tau_2)  + F_\de[\phi](\tau_1, \tau_2)\\
   & \les&  E_\de[\phi](\tau_1)+\int_{\AA(\tau_1, \tau_2)\cup\Sigma(\tau_2)\cup\Sigma_*(\tau_1, \tau_2)}\frac{\phi^2}{r^3}\\
 \nn&&+\int_{\Mtrap(\tau_1,\tau_2)}\left|1-\frac{3m}{r}\right||\phi|(|\phi|+|R\phi|) +\int_{\Mntrap(\tau_1,\tau_2)}r^{\de-3}|\phi|(|\phi|+|\dk\phi|)\\
 \nn&&+\int_{\MM(\tau_1, \tau_2)}r^{1+\de}|N_0|^2+\left|\int_{\Mtrap(\tau_1, \tau_2)}T(\phi)N_0\right|.
\eea 
\end{theorem}

\begin{proof}
The proof of Theorem \ref{thm:nonsharpweightedesitmatesforreduced0scalarsolutionofwaveequationwithpotential} is analogous to the one of Theorem \ref{thm:nonsharpweightedesitmatesforreduced1scalarsolutionofwaveequationwithpotential}. The only differences are
\begin{itemize}
\item The treatment of the right-hand sides $N_0$ and $N_2$ in the spacetime region $\Mtrap$.

\item The boundary term on $\AA(\tau_1, \tau_2)\cup\Sigma(\tau_2)\cup\Sigma_*(\tau_1, \tau_2)$ appearing in the right-hand side of\footnote{This boundary term, as discussed below, is due to the fact that $V_0$ is positive, which explains why no such term is present in  \eqref{eq:nonsharpweightedesitmatesforreduced0scalarsolutionofwaveequationwithpotential:1:00} due to the negativity of the potential $V_2$ for the wave equation satisfied by $\psi$.} \eqref{eq:nonsharpweightedesitmatesforreduced0scalarsolutionofwaveequationwithpotential:1}. 
\end{itemize}

The treatment of $N_0$ and $N_2$ is similar, so we focus on the one of $N_2$. The only estimate in which $N_2$ appear in the trapping region is the Morawetz estimate. More precisely, it appear under the form, see \eqref{eq:usefultorecalllatertreatmentofNterminMorawetzestimate},
 \beaa
 \bigg|     \int_{\Mtrap(\tau_1, \tau_2)} \left( f_{\widehat{\de}} R(\Psi) +\La T(\Psi) +\frac 1 2  w \Psi\right)  N_2\bigg|,
 \eeaa
where we recall that $\La$ is a constant, and $f_{\widehat{\de}}$, $w$ are functions which are in particular bounded on $\Mtrap$. We infer
 \beaa
&& \bigg|     \int_{\Mtrap(\tau_1, \tau_2)} \left( f_{\widehat{\de}} R(\Psi) +\La T(\Psi) +\frac 1 2  w \Psi\right)  N_2\bigg|\\
&\les&   \int_{\Mtrap(\tau_1, \tau_2)} (|R(\Psi)| +|\Psi|)|N_2|+ \bigg|     \int_{\Mtrap(\tau_1, \tau_2)}T(\Psi)N_2\bigg|\\
&\les&   \la B_\de[\psi](\tau_1, \tau_2)+ \la^{-1}\int_{\Mtrap(\tau_1, \tau_2)}|N_2|^2 + \bigg|     \int_{\Mtrap(\tau_1, \tau_2)}T(\Psi)N_2\bigg| 
 \eeaa
which yields the desired control provided $\la>0$ is chosen small enough so that the term $\la B_\de[\psi](\tau_1, \tau_2)$ can be absorbed by the LHS in  \eqref{eq:nonsharpweightedesitmatesforreduced0scalarsolutionofwaveequationwithpotential:1:00}.

Concerning the boundary terms on $\AA(\tau_1, \tau_2)\cup\Sigma(\tau_2)\cup\Sigma_*(\tau_1, \tau_2)$ appearing in the right-hand side of  \eqref{eq:nonsharpweightedesitmatesforreduced0scalarsolutionofwaveequationwithpotential:1}, the potential $V_0$ does not appear in the boundary term of the $r^p$ weighted estimates, but it does appear in the boundary term of the energy estimates\footnote{The boundary term of the $r^p$ weighted estimates involves only $Q_{44}=(e_4\phi)^2$, while the one of the energy estimate involves also $Q_{34}=|\nabb\phi|^2+V_0\phi^2$.}. More precisely, it appears in
\beaa
\int_{\AA(\tau_1, \tau_2)\cup\Sigma(\tau_2)\cup\Sigma_*(\tau_1, \tau_2)}Q_{34}=\int_{\AA(\tau_1, \tau_2)\cup\Sigma(\tau_2)\cup\Sigma_*(\tau_1, \tau_2)}\Big(|\nabb\phi|^2+V_0\phi^2\Big).
\eeaa
Now, we have in view of the definition of $V_0$
\beaa
\int_{\AA(\tau_1, \tau_2)\cup\Sigma(\tau_2)\cup\Sigma_*(\tau_1, \tau_2)}Q_{34} \geq \int_{\AA(\tau_1, \tau_2)\cup\Sigma(\tau_2)\cup\Sigma_*(\tau_1, \tau_2)}|\nabb\phi|^2 -O(1)\int_{\AA(\tau_1, \tau_2)\cup\Sigma(\tau_2)\cup\Sigma_*(\tau_1, \tau_2)}\frac{\phi^2}{r^3}
\eeaa
and the control of the boundary terms follows. This concludes the proof of  \ref{thm:nonsharpweightedesitmatesforreduced0scalarsolutionofwaveequationwithpotential}.
\end{proof}

We are now in position to prove Theorem \ref{theorem-recoveringhigherderivatives}. Note first that we have
 \beaa
\int_{\AA(\tau_1, \tau_2)\cup\Sigma_*(\tau_1, \tau_2)}\frac{(\dk^{\leq s}\phi)^2}{r^3}  &\les& F^{s-1}_{\de}[\phi](\tau_1, \tau_2)
\eeaa
which explains why the term $\int_{\AA(\tau_1, \tau_2)\cup\Sigma_*(\tau_1, \tau_2)}\frac{(\dk^{\leq s}\phi)^2}{r^3}$, that one would a priori would expect in view of \eqref{eq:nonsharpweightedesitmatesforreduced0scalarsolutionofwaveequationwithpotential:1}, is not present on the right-hand side of \eqref{eq:thesecondequationoftheorem-recoveringhigherderivatives}. Also, the estimates for $\psi$ and $\phi$ are similar, so we focus on the estimate for $\psi$. 

\begin{proof}[Proof of Theorem \ref{theorem-recoveringhigherderivatives}] The proof of Theorem \ref{theorem-recoveringhigherderivatives} follows along the same lines as the one of Theorem \ref{theorem-combinedMor-r-weighted}. More precisely,  following the strategy in section \ref{sec:strategyforrecoveringhigherorderderivatives}, we recover derivatives one by one starting from Theorem \ref{thm:nonsharpweightedesitmatesforreduced0scalarsolutionofwaveequationwithpotential} and use it iteratively in conjonction with the commutator estimates of section \ref{sec:commutationformulasforthewaveequationhigherderivatives}. The only difference is the treatment of the derivatives for $s\geq k_{small}+1$ as we assume that the estimates of  section \ref{sec:necessaryassumptionstorecoveryhigherderivatives} for the Ricci coefficients and curvature components only hold for $k\leq k_{small}$ derivatives. Thus, to conclude, we need to consider the terms for which at least  $k_{small}+1$ derivatives fall on the Ricci coefficients and curvature components. Since on the other hand we have $s\leq k_{large}-1$,  in view of the definition  \eqref{eq:choiceksmallmaintheorem} of $k_{small}$ in terms of $k_{large}$, and in view of the commutator estimates of  section \ref{sec:commutationformulasforthewaveequationhigherderivatives}, one easily checks that these terms are bounded in absolute value from above by
\beaa
\Big(|\dk^{\leq s}(\Ga_g)|+r^{-1}|\dk^{\leq s}(\Ga_b)|\Big)|\dk^{\leq k_{small}}\psi|.
\eeaa
We thus need, in view of Theorem \ref{thm:nonsharpweightedesitmatesforreduced0scalarsolutionofwaveequationwithpotential}, to estimate
\beaa
&&\int_{\MM(\tau_1, \tau_2)}r^{1+\de}\Big(|\dk^{\leq s}(\Ga_g)|+r^{-1}|\dk^{\leq s}(\Ga_b)|\Big)^2|\dk^{\leq k_{small}}\psi|^2\\
&&+\int_{\Mtrap(\tau_1, \tau_2)}|T\dk^s\psi||\dk^{\leq s}(\Gac)||\dk^{\leq k_{small}}\psi|\\
&\les& \sup_{\MM(\tau_1, \tau_2)}\Big(r^2|\dk^{\leq k_{small}}\psi|^2\Big)\int_{\MM(\tau_1, \tau_2)}r^{-1+\de}\Big(|\dk^{\leq s}(\Ga_g)|+r^{-1}|\dk^{\leq s}(\Ga_b)|\Big)^2\\
&&+\left(\sup_{\Mtrap(\tau_1, \tau_2)}u^{\frac{1}{2}+\dec}|\dk^{\leq k_{small}}\psi|\right)\left(\sup_{\tau\in[\tau_1,\tau_2] }   E^s_\de [\psi](\tau)\right)^{\frac{1}{2}}\left(\int_{\Mtrap(\tau_1, \tau_2)}|\dk^{\leq s}\Gac|^2\right)^{\frac{1}{2}}\\
&\les& \left(\sup_{\MM(\tau_1, \tau_2)}ru_{trap}^{\frac{1}{2}+\dec}|\dk^{\leq k_{small}}\psi|\right)^2D_s[\Ga]\\
&&+\left(\sup_{\MM(\tau_1, \tau_2)}ru_{trap}^{\frac{1}{2}+\dec}|\dk^{\leq k_{small}}\psi|\right)\left(\sup_{\tau\in[\tau_1,\tau_2] }   E^s_\de [\psi](\tau)\right)^{\frac{1}{2}}\sqrt{D_s[\Ga]}
\eeaa
where we have used the definition of $D_s[\Ga]$. We infer
\beaa
&&\int_{\MM(\tau_1, \tau_2)}r^{1+\de}\Big(|\dk^{\leq s}(\Ga_g)|+r^{-1}|\dk^{\leq s}(\Ga_b)|\Big)^2|\dk^{\leq k_{small}}\psi|^2\\
&&+\int_{\Mtrap(\tau_1, \tau_2)}|T\dk^s\psi||\dk^{\leq s}(\Gac)||\dk^{\leq k_{small}}\psi|\\
&\les& \la^{-1}\left(\sup_{\MM(\tau_1, \tau_2)}ru_{trap}^{\frac{1}{2}+\dec}|\dk^{\leq k_{small}}\psi|\right)^2D_s[\Ga]+\la\sup_{\tau\in[\tau_1,\tau_2] }   E^s_\de [\psi](\tau)
\eeaa
for any $\la>0$ and the last term is then absorbed from the left-hand side of the desired estimate by choosing $\la>0$ small enough which concludes the proof of Theorem \ref{theorem-recoveringhigherderivatives}.
\end{proof}

\begin{proposition}\lab{prop:controlonMintbyMextthankstoredshiftneededinThmM8}
Let $\psi$ a reduced 2-scalar satisfying  
\beaa
\square_2\psi =f_2(r,m)Y_{(0)}\psi+ \widetilde{N}_2, 
\eeaa
where the function $f_2$ is smooth and positive, and where the vectorfield $Y_{(0)}$ has been introduced in Proposition  \ref{prop:red-shift-Mor} in connection with the redshift vectorfield and is given by
 \beaa
 Y_{(0)}:= \left(1+\frac{5}{4m}(r-2m)+\Up\right) e_3 +\left(1+\frac{5}{4m}(r-2m)\right)e_4.
 \eeaa  
Also, assume that the Ricci coefficients and curvature components associated to the global null frame we are using satisfy the estimates of section \ref{sec:necessaryassumptionstorecoveryhigherderivatives} for $k\leq k_{small}$ derivatives. Then, for   any $1\leq s\leq k_{large}-1$, we have
 \beaa
 \int_{\Mint(\tau_1,\tau_2)}(\dk^{s+1}\psi)^2 &\les& E^s_{\de}[\psi](\tau_1)+ \int_{\Mext_{r\leq \frac{5}{2}m_0}(\tau_1,\tau_2)}(\dk^{s+1}\psi)^2\\
 && +D_s[\Ga] \left(\sup_{\Mint(\tau_1,\tau_2)\cup\Mext_{r\leq \frac{5}{2}m_0}}r|\dk^{\leq k_{small}}\psi|\right)^2\\
 && +\int_{\Mint(\tau_1,\tau_2)\cup\Mext_{r\leq \frac{5}{2}m_0}}\Big((\dk^{\leq s}\psi)^2+(\dk^{\leq s+1}\widetilde{N}_2)^2\Big).
 \eeaa
 \end{proposition}

\begin{proof}
Recall from Proposition  \ref{prop:red-shift-Mor} that the redshift vectorfield is given by  
\beaa
  Y_\HH &:=&    \ka_\HH      Y_{(0)}, \qquad  \ka_\HH:= \ka\left(\frac{\Up}{\deh^{\frac{1}{10}}}\right),
     \eeaa
   where $\ka$ is a positive   bump function $\ka=\ka(r)$, supported in the region  in $[-2, 2]$ and   equal to $1$   for $[-1,1]$.

To estimate $\psi$ in $\Mint$, we consider 
\beaa
\widetilde{\psi} &:=& \widetilde{\ka}\left(\frac{r-2m_0(1+2\deh)}{2m_0\deh}\right)
\eeaa
where  $\widetilde{\ka}$ is a positive   bump function $\ka=\ka(r)$, supported in the region  in $(-\infty, 1]$ and   equal to $1$   for $(-\infty, 0]$. Since $\Mint$ is included in $r\leq 2m_0(1+2\deh)$, we infer in particular
\beaa
\widetilde{\psi}=\psi\textrm{ on }\Mint,\qquad \textrm{supp}(\widetilde{\psi})\subset  \Mint(\tau_1,\tau_2)\cup\Mext_{r\leq 2m_0(1+3\deh)}.
\eeaa
Also, we have, in view of the wave equation for $\psi$, 
\beaa
\square_2\widetilde{\psi} =f_2(r,m)Y_{(0)}\widetilde{\psi}+\widetilde{N}_2'
\eeaa
where $\widetilde{N}_2'$ satisfies 
\beaa
\int_{\Mint(\tau_1,\tau_2)\cup\Mext_{r\leq \frac{5}{2}m_0}}(\dk^{\leq s+1}\widetilde{N}_2')^2\Big) &\les& \int_{\Mext_{r\leq \frac{5}{2}m_0}(\tau_1,\tau_2)}(\dk^{\leq s+1}\psi)^2\\
&&+\int_{\Mint(\tau_1,\tau_2)\cup\Mext_{r\leq \frac{5}{2}m_0}}(\dk^{\leq s+1}\widetilde{N}_2)^2.
\eeaa
Since $\widetilde{\psi}=\psi$ on $\Mint$, it thus suffices to prove for $\widetilde{\psi}$ the following estimate 
\beaa
 \int_{\MM(\tau_1,\tau_2)}(\dk^{s+1}\widetilde{\psi})^2 &\les& E^s_{\de}[\widetilde{\psi}](\tau_1) +D_s[\Ga] \left(\sup_{\Mint(\tau_1,\tau_2)\cup\Mext_{r\leq \frac{5}{2}m_0}}r|\dk^{\leq k_{small}}\widetilde{\psi}|\right)^2\\
 && +\int_{\Mint(\tau_1,\tau_2)\cup\Mext_{r\leq \frac{5}{2}m_0}}\Big((\dk^{\leq s}\widetilde{\psi})^2+(\dk^{\leq s+1}\widetilde{N}_2')^2\Big).
 \eeaa
This estimate follows from first deriving the corresponding estimate for $s=0$ by using the redshift as a multiplier, and then by  recover derivatives one by one using commutation with $T$, $\dkb$ and the redshift vectorfield. Note that
\begin{itemize}
\item $\widetilde{\psi}$ is supported on $r\leq 2m_0(1+2\deh)$ and hence is estimated on 
\beaa
2m_0(1-2\deh)\leq r\leq 2m_0(1+2\deh)
\eeaa
so that the redshift vectorfield $Y_\HH$ has good properties, both as a multiplier and as a commutator, on the support of $\widetilde{\psi}$.

\item The term $f_2(r,m)Y_{(0)}$ yields a good sign when using $Y_\HH$ as a multiplier since the function $f_2(r,m)$ is positive, and since $Y_\HH =  \ka_\HH      Y_{(0)}$.
\end{itemize}

This concludes the proof of the proposition.
\end{proof}


\appendix


\chapter{APPENDIX TO CHAPTER \ref{chap:preliminaries}}



\section{Proof of Proposition \ref{prop:outgoinggeod-e3e4averages}}\lab{sec:proofofprop:outgoinggeod-e3e4averages}


In a neighborhood of a given sphere $S$, we consider a $(u, s, \th, \vphi)$ coordinates system, where $\th$ is such that $e_4(\th)=0$. Then, in this coordinates system, we have
\beaa
\pr_s = e_4.
\eeaa
Since we have
\beaa
\pr_s\left(\int_Sf\right) &=& \int_S\Big(\pr_sf+g(\D_{e_\th}\pr_s, e_\th)f+g(\D_{e_\vphi}\pr_s, e_{\vphi})f\Big),
\eeaa
we infer
\beaa
e_4\left(\int_Sf\right) = \int_S(e_4(f)+\ka f).
\eeaa
In particular, choosing $f=1$, we deduce
\beaa
\frac{1}{|S|} e_4(|S|) &=&\ov{ \kab}  
\eeaa
and since $|S|=4\pi r^2 $, 
\beaa
e_4(r)=\frac{r\overline{\ka}}{2}.
\eeaa

Next, let $\pr_u$ the coordinates vectorfield in the $(u, s, \th, \vphi)$ coordinates system. We have
\beaa
\pr_u\left(\int_Sf\right) &=& \int_S\Big(\pr_uf+g(\D_{e_\th}\pr_u, e_\th)f+g(\D_{e_\vphi}\pr_u, e_{\vphi})f\Big)\\
&=& \int_S\Big(\pr_uf+g(\D_{e_\th}\pr_u, e_\th)f -g(\pr_u, \D_{e_\vphi}e_\vphi)f\Big)\\
&=& \int_S\Big(\pr_uf+g(\D_{e_\th}\pr_u, e_\th)f +g(\pr_u, D^a(\Phi)e_a)f\Big)
\eeaa
On the other hand, we have we have, see  \eqref{eq:propoutgoinggeodesiccoordinates2}, 
\beaa
\pr_u&=& \varsigma\left(\frac{1}{2}e_3-\frac{1}{2}\Omb e_4-\frac{1}{2}\sqrt{\ga}\underline{b}  e_\th\right).
\eeaa
We infer
\beaa
g(D_{e_\th}\pr_u, e_\th) +g(\pr_u, D^a(\Phi)e_a) &=& \frac{1}{2}\varsigma\kab -\frac{1}{2}\varsigma\Omb\ka-\frac{1}{2}\ddd_1(\varsigma\sqrt{\ga}\underline{b})
\eeaa
and thus
\beaa
\varsigma\left(\frac{1}{2}e_3-\frac{1}{2}\Omb e_4\right)\left(\int_Sf\right) &=& \int_S\Bigg(\varsigma\left(\frac{1}{2}e_3-\frac{1}{2}\Omb e_4 -\frac{1}{2}\sqrt{\ga}\underline{b}  e_\th\right)f+\frac{1}{2}\varsigma\kab f -\frac{1}{2}\varsigma\Omb\ka f\\
 && -\frac{1}{2}\ddd_1(\sqrt{\ga}\underline{b})f\Bigg).
\eeaa
We deduce
\beaa
e_3\left(\int_Sf\right) &=&  \Omb e_4\left(\int_Sf\right)+\varsigma^{-1}\int_S\Bigg(\varsigma e_3f -\varsigma\Omb e_4f +\varsigma\kab f - \varsigma\Omb\ka f -\ddd_1(\varsigma\sqrt{\ga}\underline{b}f)\Bigg).
\eeaa

Next, we use
\beaa
e_4\left(\int_Sf\right) &=& \int_S(e_4(f)+\ka f)
\eeaa
and 
\beaa
\int_S\ddd_1(\varsigma\sqrt{\ga}\underline{b}f) &=& 0.
\eeaa
We infer
\beaa
e_3\left(\int_Sf\right) &=&  \Omb\int_S(e_4(f)+\ka f)+\varsigma^{-1}\int_S\Bigg(\varsigma e_3f -\varsigma\Omb e_4f +\varsigma\kab f -\varsigma\Omb\ka f \Bigg)\\
&=& \varsigma^{-1}\int_S\varsigma(e_3f+\kab f)+\left(\Obc +\varsigma^{-1}\ov{\Omb}\check{\varsigma}\right)\int_S(e_4f +\ka f)\\
&& -\varsigma^{-1}\ov{\Omb}\int_S\check{\varsigma}(e_4f +\ka f) -\varsigma^{-1}\int_S\Obc\varsigma(e_4f +\ka f).
\eeaa
We further write,
\beaa
 \varsigma^{-1}\int_S\varsigma(e_3f+\kab f)&=& \vsi^{-1}\ov{\vsi} \int_S(e_3f+\kab f)+ \vsi^{-1}  \int_S\check{\vsi}\,(e_3f+\kab f)\\
 &=&\int_S(e_3f+\kab f)+ ( \vsi^{-1}\ov{\vsi} -1)\int_S(e_3f+\kab f)+ \vsi^{-1}  \int_S\check{\vsi}\,(e_3f+\kab f)\\
 &=&\int_S(e_3f+\kab f)- \vsi^{-1} \check{\vsi} \int_S(e_3f+\kab f)+ \vsi^{-1}  \int_S\check{\vsi}\,(e_3f+\kab f).
\eeaa
Hence,
\beaa
e_3\left(\int_Sf\right) &=&\int_S(e_3f+\kab f)+\err\left[e_3\left(\int_S f\right)\right],\\
\err\left[e_3\left(\int_S f\right)\right]&=&- \vsi^{-1} \check{\vsi} \int_S(e_3f+\kab f)+ \vsi^{-1}  \int_S\check{\vsi}\,(e_3f+\kab f)\\
&+&\left(\Obc +\varsigma^{-1}\ov{\Omb}\check{\varsigma}\right)\int_S(e_4f +\ka f)-\varsigma^{-1}\ov{\Omb}\int_S\check{\varsigma}(e_4f +\ka f)\\
& -&\varsigma^{-1}\int_S\Obc\varsigma(e_4f +\ka f)
\eeaa
as desired.

In particular, choosing $f=1$, we infer
\beaa
\frac{1}{|S|} e_3(|S|) &=&\ov{ \kab}  -\vsi^{-1}\, \check{\vsi}\,\ov{ \kab}+\vsi^{-1} \,\ov{\check{\vsi} \kab}+
\left(\Obc +\varsigma^{-1}\ov{\Omb}\check{\varsigma}\right)\ov{\ka} -\varsigma^{-1}\ov{\Omb}\, \ov{\check{\vsi}\,\ka }-\vsi ^{-1}\ov{ \Obc\, \vsi\, \ka}\\
&=&\ov{ \kab}  -\vsi^{-1}\, \check{\vsi}\,\ov{ \kab}+\vsi^{-1} \,\ov{\check{\vsi} \kabc}+
\left(\Obc +\varsigma^{-1}\ov{\Omb}\check{\varsigma}\right)\ov{\ka} -\varsigma^{-1}\ov{\Omb}\, \ov{\check{\vsi}\,\kac }-\vsi ^{-1}\ov{ \Obc\, \vsi\, \ka}.
\eeaa
Hence, since $|S|=4\pi r^2 $, recalling the definition of $\Ab$,
\beaa
\frac{2e_3(r)}{r} 
&=&\ov{ \kab}  -\vsi^{-1}\, \check{\vsi}\,\ov{ \kab}+\vsi^{-1} \,\ov{\check{\vsi} \kabc}+
\left(\Obc +\varsigma^{-1}\ov{\Omb}\check{\varsigma}\right)\ov{\ka} -\varsigma^{-1}\ov{\Omb}\, \ov{\check{\vsi}\,\kac } -\vsi ^{-1}\ov{ \Obc\, \vsi\, \ka}\\
&=&\ov{\kab} + \Ab.
\eeaa
 This  concludes the proof of Proposition \ref{prop:outgoinggeod-e3e4averages}.


\section{Proof of Proposition \ref{prop:derivativesHawkingmass}}\lab{sec:proofofprop:derivativesHawkingmass}


We start with the proof for $e_4(m)$. 
Recall that  the Hawking mass  $m$ is given by the formula  $\frac{2m}{r}=1+\frac{1}{16\pi}\int_S\ka\kab$.  
Differentiating in the $e_4$ direction, we deduce,
\beaa
\frac{2e_4(m)}{r}-\frac{2me_4(r)}{r^2} &=& \frac{1}{16\pi}e_4\left(\int_S\ka\kab\right)= \frac{1}{16\pi}\int_S\Big(e_4(\ka\kab)+\kab\ka^2\Big).
\eeaa
Now, making use of the $e_4$  transport equations  of Proposition \ref{propos:basiceqts-geod},
\beaa
e_4(\ka\kab) &=& \kab\left(-\frac{1}{2}\ka^2-\frac{1}{2}\vth^2\right)+\ka\left(-\frac{1}{2}\ka\kab+2\rho-2\ddd_1\ze -\frac{1}{2}\vth\vthb+2\ze^2\right)\\
&=& -\kab\ka^2 +2\ka\rho-2\ka\ddd_1\ze -\frac{1}{2}\kab\vth^2 -\frac{1}{2}\ka\vth\vthb +2\ka\ze^2.
\eeaa
we infer
\beaa
\frac{2e_4(m)}{r}-\frac{m}{r}\ov{\ka} &=& \frac{1}{16\pi}\int_S\left(2\ka\rho-2\ka\ddd_1\ze -\frac{1}{2}\kab\vth^2 -\frac{1}{2}\ka\vth\vthb +2\ka\ze^2\right)\\
&=& \frac{1}{8\pi}|S|\ov{\ka}\,\ov{\rho} + \frac{1}{16\pi}\int_S\left(2\check{\ka}\check{\rho}+2e_\th(\ka)\ze -\frac{1}{2}\kab\vth^2 -\frac{1}{2}\ka\vth\vthb +2\ka\ze^2\right)\\
&=& \frac{r^2}{2}\ov{\ka}\,\ov{\rho} + \frac{1}{16\pi}\int_S\left(2\check{\ka}\check{\rho}+2e_\th(\ka)\ze -\frac{1}{2}\kab\vth^2 -\frac{1}{2}\ka\vth\vthb +2\ka\ze^2\right)
\eeaa
and hence
\beaa
e_4(m) &=& \frac{r^3}{4}\ov{\ka}\left(\ov{\rho}+\frac{2m}{r^3}\right) + \frac{r}{32\pi}\int_S\left(-\frac{1}{2}\kab\vth^2 -\frac{1}{2}\ka\vth\vthb+2\check{\ka}\check{\rho}+2e_\th(\ka)\ze  +2\ka\ze^2\right).
\eeaa
Using  the identity  $\ov{\rho}  = -\frac{2m}{r^3} + \frac{1}{16\pi r^2}\int_S\vth\vthb$  (see   \eqref{identity:overlinerho}  of Proposition \ref{prop:identityaverage-forwgeodfoliation}),
we  deduce
\beaa
e_4(m) &=&  \frac{r}{32\pi}\int_S\left(-\frac{1}{2}\kab\vth^2  -\frac{1}{2} (\ka -\ov{\ka})    \vth\vthb+2\check{\ka}\check{\rho}+2e_\th(\ka)\ze  +2\ka\ze^2\right)\\
&=& \frac{r}{32\pi}\int_S\left(-\frac{1}{2}\kab\vth^2  -\frac{1}{2}\check{\ka}\vth\vthb+2\check{\ka}\check{\rho}+2e_\th(\ka)\ze  +2\ka\ze^2\right)\\
&=& \frac{r}{32\pi}\int_S\err_1
\eeaa
as desired. 

In the same vein,
\beaa
\frac{2e_3(m)}{r}-\frac{2me_3(r)}{r^2} &=& \frac{1}{16\pi}e_3\left(\int_S\ka\kab\right)= \frac{1}{16\pi}\int_S\Big(e_3(\ka\kab)+\kab^2\ka\Big) + E_1,
\eeaa
with $E_1$ the error term defined in Proposition \ref{prop:outgoinggeod-e3e4averages}
\beaa
E_1=\frac{1}{16\pi} \err\left[e_3\left(\int_\S \ka\kab\right)\right].
\eeaa
We  make use of the $e_3$  transport equations  of Proposition \ref{propos:basiceqts-geod},
\beaa
e_3(\ka\kab) &=& \kab\left(- \frac 1 2 \kab\, \ka +2\omb \ka + 2\ddd_1\eta + 2\rho -\frac 1 2  \vthb\, \vth +2\eta^2\right)\\
&+& \ka \left( -\frac 12 \kab^2 -2 \omb \,\kab +2\ddd_1\xib+2(\eta-3\ze)  \xib -\frac 1 2 \vthb^2 \right)\\
&=&-\ka \kab^2+ 2 \kab\ddd_1 \eta +2\ka \ddd_1 \xib+ 2 \rho \kab +\kab\left(2\eta^2-\frac 1 2 \vth \vthb \right)+2\ka \big(\eta-3\ze\big)\xib
-\frac 1 2 \ka \vthb^2.
\eeaa
Therefore, setting  $E_2=\frac{1}{16\pi} \int_S\Big(\kab\big(2\eta^2-\frac 1 2 \vth \vthb \big)+2\ka \big(\eta-3\ze\big)\xib-\frac 1 2 \ka \vthb^2\Big)$,  
\beaa
\frac{2e_3(m)}{r}-\frac{m}{r}(\ov{\kab}+\Ab)&=&\frac{1}{16\pi}\int_S\Big(  2 \kab\ddd_1 \eta +2\ka \ddd_1 \xib+ 2 \rho \kab \Big)+E_1+E_2\\
&=&\frac{1}{16\pi}\int_S\Big( -  2 e_\th( \kab)  \eta - 2 e_\th(\ka)  \xib+ 2(\ov{ \rho } +\rhoc) (\ov{\kab}+\kabc) \Big)+E_1+E_2\\
&=&\frac 1 2    r^2 \ov{ \rho}\,  \ov{\kab} +\frac{1}{16\pi}\int_S\Big( -  2 e_\th( \kab)  \eta - 2 e_\th(\ka)  \xib+ 2\rhoc\, \kabc \Big)+E_1+E_2\\
&=& \frac 1 2    r^2 \ov{\kab} \Big(-\frac{2m}{r^3}+\frac{1}{16\pi r^2 } \int_S \vth\vthb\Big)\\
&+& \frac{1}{16\pi}\int_S\Big( -  2 e_\th( \kab)  \eta - 2 e_\th(\ka)  \xib+ 2\rhoc\, \kabc  \Big)+E_1+E_2.
\eeaa
We deduce 
\beaa
\frac{2e_3(m)}{r}&=& \frac{1}{16\pi}\int_S\Big( -  2 e_\th( \kab)  \eta - 2 e_\th(\ka)  \xib+ 2\rhoc\, \kabc  +\frac 1 2 \ov{\kab} \vth\vthb\Big)+E_1\\
&+& \frac{1}{16\pi} \int_S\left(\kab\left(2\eta^2-\frac 1 2 \vth \vthb \right)+2\ka \big(\eta-3\ze\big)\xib-\frac 1 2 \ka \vthb^2\right)+\frac{m}{r} \Ab\\
&=& \frac{1}{16\pi}\int_S\Big( -  2 e_\th( \kab)  \eta  +2 \kab \eta^2 - 2 e_\th(\ka)  \xib  +2\ka \eta \xib  -\frac 1 2 \ka\vthb^2\Big)\\
&+&  \frac{1}{16\pi}\int_S\Big( 2\rhoc\, \kabc-6\ka\, \ze\, \xib-\frac 1 2 \kabc \vth \, \vthb\Big)+E_1+\frac{m}{r} \Ab,
\eeaa
i.e.,
\beaa
e_3(m)&=&   \frac{r}{32\pi}\int_S\Big( -  2 e_\th( \kab)  \eta  +2 \kab \eta^2 - 2 e_\th(\ka)  \xib  +2\ka \eta \xib  -\frac 1 2 \ka \vthb^2\Big)\\
&+&  \frac{r}{32\pi}\int_S    \Big( 2\rhoc\, \kabc-6\ka\, \ze\, \xib-\frac 1 2 \kabc \vth \, \vthb\Big) +\frac{r}{2}\left(E_1+\frac{m}{r} \Ab\right).
\eeaa

It remains to calculate $E_1+\frac{m}{r}\Ab$. Using the definitions of $E_1$ and $\Ab$  and  grouping similar   terms  appropriately we find
\beaa
&&E_1+\frac m r \Ab   \\
&&=- \vsi^{-1} \check{\vsi} \left[\frac {1}{16\pi}\int_S(e_3(\ka\kab) +\ka \kab^2) +\frac{m}{r} \ov{\kab} \right]+ \vsi^{-1}\left[ \frac {1}{16\pi} \int_S\check{\vsi}\,(e_3(\ka\kab)+\ka \kab^2)
+\frac m r\ov{\check{\vsi}\kabc}\right]\\
&&+\left(\Obc +\vsi^{-1}\ov{\Omb}\check{\vsi}\right)\left[ \frac {1}{16\pi}\int_S(e_4(\ka\kab)  +\ka^2 \kab )+\frac m r \ov{\ka}\right] -\vsi^{-1}\ov{\Omb}\left[\frac{1}{16\pi}\int_S\check{\vsi}(e_4(\ka\kab) +\ka^2  \kab) +\frac m r  \ov{\check{\vsi}\kac }\right]\\
&&-\vsi^{-1}\left[ \frac{1}{16\pi} \int_S\Obc\vsi(e_4(\ka\kab) +\ka^2 \kab)+ \frac{m}{r}   \ov{\Obc \vsi \ka}   \right].
\eeaa
Now, we have from above calculations
\beaa
e_4(\ka\kab) +\kab\ka^2 &=& 2\ka\rho-2\ka\ddd_1\ze+\err[e_4(\ka\kab)],\\
\err[e_4(\ka\kab)] &=& -\frac{1}{2}\kab\vth^2 -\frac{1}{2}\ka\vth\vthb +2\ka\ze^2,\\
e_3(\ka\kab) +\ka \kab^2&=& 2 \rho \kab+2 \kab\ddd_1 \eta +2\ka \ddd_1 \xib+\err[e_3(\ka\kab)],\\
\err[e_3(\ka\kab)] &=& \kab\left(2\eta^2-\frac 1 2 \vth \vthb \right)+2\ka \big(\eta-3\ze\big)\xib-\frac 1 2 \ka \vthb^2.
\eeaa
We infer
\beaa
E_1+\frac m r \Ab   &=&- \vsi^{-1} \check{\vsi} \left[\frac {1}{16\pi}\int_S(2 \rho \kab+2 \kab\ddd_1 \eta +2\ka \ddd_1 \xib+\err[e_3(\ka\kab)]) +\frac{m}{r} \ov{\kab} \right]\\
&&+ \vsi^{-1}\left[ \frac {1}{16\pi} \int_S\check{\vsi}\,(2 \rho \kab+2 \kab\ddd_1 \eta +2\ka \ddd_1 \xib+\err[e_3(\ka\kab)])
+\frac m r\ov{\check{\vsi}\kabc}\right]\\
&&+\left(\Obc +\vsi^{-1}\ov{\Omb}\check{\vsi}\right)\left[ \frac {1}{16\pi}\int_S(2\ka\rho-2\ka\ddd_1\ze+\err[e_4(\ka\kab)] )+\frac m r \ov{\ka}\right] \\
&&-\vsi^{-1}\ov{\Omb}\left[\frac{1}{16\pi}\int_S\check{\vsi}(2\ka\rho-2\ka\ddd_1\ze+\err[e_4(\ka\kab)]) +\frac m r  \ov{\check{\vsi}\kac }\right]\\
&&-\vsi^{-1}\left[ \frac{1}{16\pi} \int_S\Obc\vsi(2\ka\rho-2\ka\ddd_1\ze+\err[e_4(\ka\kab)])+ \frac{m}{r}   \ov{\Obc \vsi \ka}   \right]
\eeaa
and hence
\beaa
E_1+\frac m r \Ab   &=&- \vsi^{-1} \check{\vsi} \left[\frac {1}{16\pi}\int_S\left(2 \rhoc\kabc-2 e_\th(\kab)\eta -2e_\th(\ka)\xib+\frac{1}{2}\ov{\kab}\vth\vthb+\err[e_3(\ka\kab)]\right)  \right]\\
&&+ \vsi^{-1}\left[ \frac {1}{16\pi} \int_S\check{\vsi}\,(2 \ov{\rho}\kabc+2 \rhoc\ov{\kab}+2\rhoc\kabc+2 \kab\ddd_1 \eta +2\ka \ddd_1 \xib+\err[e_3(\ka\kab)])
+\frac m r\ov{\check{\vsi}\kabc}\right]\\
&&+\left(\Obc +\vsi^{-1}\ov{\Omb}\check{\vsi}\right)\left[ \frac {1}{16\pi}\int_S\left(2\kac\rhoc+2e_\th(\ka)\ze+\frac{1}{2}\ov{\kab}\vth\vthb+\err[e_4(\ka\kab)] \right)\right] \\
&&-\vsi^{-1}\ov{\Omb}\left[\frac{1}{16\pi}\int_S\check{\vsi}(2 \ov{\rho}\kac+2 \rhoc\ov{\ka}+2\rhoc\kac-2\ka\ddd_1\ze+\err[e_4(\ka\kab)]) +\frac m r  \ov{\check{\vsi}\kac }\right]\\
&&-\vsi^{-1}\left[ \frac{1}{16\pi} \int_S\Obc\vsi(2 \ov{\rho}\kac+2 \rhoc\ov{\ka}+2\rhoc\kac-2\ka\ddd_1\ze+\err[e_4(\ka\kab)])+  \frac{m}{r}   \ov{\Obc \vsi \ka}  \right].
\eeaa
We deduce
\beaa
e_3(m)&=&   \left(1- \vsi^{-1} \check{\vsi}\right)\frac {r}{32\pi}\int_S\underline{\err}_1+\left(\Obc +\vsi^{-1}\ov{\Omb}\check{\vsi}\right)\frac {r}{32\pi}\int_S\err_1\\
&&+ \vsi^{-1}\frac {r}{32\pi} \int_S\check{\vsi}\,\left(2 \ov{\rho}\kabc+2 \rhoc\ov{\kab}+2 \kab\ddd_1 \eta +2\ka \ddd_1 \xib+\underline{\err}_2\right)\\
&&-\vsi^{-1}\frac{r}{32\pi} \int_S(\ov{\Omb}\check{\vsi}+\Obc\vsi)\left(2 \ov{\rho}\kac+2 \rhoc\ov{\ka}-2\ka\ddd_1\ze+\err_2\right)\\
&&-\frac m r\vsi^{-1}\left[-\ov{\check{\vsi}\kabc}+  \ov{\Omb}\,\ov{\check{\vsi}\kac } +   \ov{\Obc \vsi \ka}\right],
\eeaa
where we have introduced
\beaa
\err_1 &=&  2\kac\rhoc+2e_\th(\ka)\ze+\frac{1}{2}\ov{\kab}\vth\vthb+\err[e_4(\ka\kab)],\\
\underline{\err}_1 &=& 2 \rhoc\kabc-2 e_\th(\kab)\eta -2e_\th(\ka)\xib+\frac{1}{2}\ov{\kab}\vth\vthb+\err[e_3(\ka\kab)],\\
\err_2 &=& 2\rhoc\kac+\err[e_4(\ka\kab)],\\
\underline{\err}_2 &=& 2\rhoc\kabc+\err[e_3(\ka\kab)].
\eeaa
In view of the definition of $\err[e_4(\ka\kab)]$ and $\err[e_3(\ka\kab)]$, this concludes the proof of Proposition \ref{prop:derivativesHawkingmass}.


\section{Proof of Lemma \ref{lemma:transportequationforoverlinekaintheoutgoinggeodesicfoliation}}\lab{sec:proofoflemma:transportequationforoverlinekaintheoutgoinggeodesicfoliation}


Recall that we have
\beaa
e_4(\ka) &=& -\frac{1}{2}\ka^2 -\frac{1}{4}\vth^2.
\eeaa
We infer
\beaa
e_4(\ov{\ka}) &=& \ov{e_4(\ka)} + \ov{\check{\ka}^2}= \ov{-\frac{1}{2}\ka^2 -\frac{1}{4}\vth^2}  + \ov{\check{\ka}^2}\\
&=& -\frac{1}{2}\ov{\ka}^2 -\frac{1}{4}\ov{\vth^2}  + \frac{1}{2}\ov{\check{\ka}^2}
\eeaa
and hence
\beaa
e_4\left(\ov{\ka}-\frac{2}{r}\right) &=& -\frac{1}{2}\ov{\ka}^2 -\frac{1}{4}\ov{\vth^2}  + \frac{1}{2}\ov{\check{\ka}^2} +\frac{2}{r}\frac{e_4(r)}{r}= -\frac{1}{2}\ov{\ka}^2  +\frac{1}{r}\ov{\ka} -\frac{1}{4}\ov{\vth^2}  + \frac{1}{2}\ov{\check{\ka}^2}\\
&=& -\frac{1}{2}\ov{\ka}\left(\ov{\ka}-\frac{2}{r} \right)   -\frac{1}{4}\ov{\vth^2}  + \frac{1}{2}\ov{\check{\ka}^2}.
\eeaa

Next, using
\beaa
e_4(\omb) &=& \rho +\ze(2\eta+\ze)
\eeaa
we infer that
\beaa
e_4(\ov{\omb}) &=& \ov{e_4(\omb)}+\ov{\check{\ka}\check{\omb}}= \ov{\rho} +\ov{\ze(2\eta+\ze)}+\ov{\check{\ka}\check{\omb}},
\eeaa
and hence
\beaa
e_4\left(\ov{\omb}-\frac{m}{r^2}\right) &=& e_4(\ov{\omb}) +\frac{2me_4(r)}{r^3} -\frac{e_4(m)}{r^2}\\
&=& \ov{\rho}+\frac{2m}{r^3} +\frac{m}{r^2}\left(\ov{\ka}-\frac{2}{r}\right) -\frac{e_4(m)}{r^2} +3\ov{\ze(2\eta+\ze)}+\ov{\check{\ka}\check{\omb}}
\eeaa
as stated.

Next, using
\beaa
e_3(\ka) +\frac{1}{2}\ka\kab -2\omb\ka = 2\ddd_1\eta +2\rho -\frac{1}{2}\vth\vthb +2\eta^2
\eeaa
we deduce
\beaa
\ov{e_3(\ka)} &=& -\frac{1}{2}\ov{\ka\kab} + 2\ov{\omb\ka}  +2\ov{\rho} -\frac{1}{2}\ov{\vth\vthb} +2\ov{\eta^2}\\
&=& -\frac{1}{2}\ov{\ka}\, \ov{\kab} + 2\ov{\omb}\,\ov{\ka}   +2\ov{\rho} + 2\ov{\check{\omb}\, \check{\ka}}  -  \frac 1 2\ov{ \kac\,\kabc}    -\frac{1}{2}\ov{\vth\vthb} +2\ov{\eta^2}.
\eeaa
Making use of Corollary \ref{corr:transportcheckf}
\beaa
e_3\left(\ov{\ka}\right) &=& \ov{e_3(\ka)}+\err[e_3\ov{\ka}]\\
&=& -\frac{1}{2}\ov{\ka}\, \ov{\kab} + 2\ov{\omb}\,\ov{\ka}   +2\ov{\rho} + 2\ov{\check{\omb}\, \check{\ka}}  -  \frac 1 2\ov{ \kac\,\kabc}    -\frac{1}{2}\ov{\vth\vthb} +2\ov{\eta^2}+\err[e_3\ov{\ka}]
\eeaa
and,
\beaa
e_3\left(\ov{\ka}-\frac 2 r \right) &=& e_3\left(\ov{\ka}\right) + \frac{2}{r^2} \frac r 2 \left(\ov{\kab}+\Ab\right) \\
&=& -\frac{1}{2}\ov{\ka}\, \ov{\kab} + 2\ov{\omb}\,\ov{\ka}   +2\ov{\rho} + 2\ov{\check{\omb}\, \check{\ka}}  -  \frac 1 2\ov{ \kac\,\kabc}    -\frac{1}{2}\ov{\vth\vthb} +2\ov{\eta^2}+\frac{1}{r }\ov{\kab}
+\frac 1 r \Ab +\err[e_3\ov{\ka}]\\
&=&-\frac 1 2\ov{ \kab}\left( \ov{\ka}-\frac 2 r  \right) + 2\ov{\omb}\,\ov{\ka}   +2\ov{\rho} + 2\ov{\check{\omb}\, \check{\ka}}  -  \frac 1 2\ov{ \kac\,\kabc}    -\frac{1}{2}\ov{\vth\vthb} +2\ov{\eta^2}
+\frac 1 r \Ab +\err[e_3\ov{\ka}].
\eeaa
Now,
\beaa
2\ov{\omb}\,\ov{\ka}+2\ov{\rho}&=&  2 \ov{\omb}\,\left( \ov{\ka}-\frac 2 r \right)+\frac 4 r\ov{ \omb} +    2    \ov{\rho}\\
&=&2 \ov{\omb}\,\left( \ov{\ka}-\frac 2 r \right)+\frac 4 r\left(\ov{ \omb}-\frac{m}{r^2}\right)+ 2\left(  \ov{\rho} +\frac{2m}{r^3}\right).
\eeaa
Hence,
\beaa
e_3\left(\ov{\ka}-\frac 2 r \right) +\frac 1 2  \ov{ \kab} \left(\ov{\ka}-\frac 2 r \right) &=&  2 \ov{\omb}\,\left( \ov{\ka}-\frac 2 r \right)+\frac 4 r\left(\ov{ \omb}-\frac{m}{r^2}\right)+ 2\left(  \ov{\rho} +\frac{2m}{r^3}\right)\\
&+&2\ov{\eta^2}+ 2\ov{\check{\omb}\, \check{\ka}}  -  \frac 1 2\ov{ \kac\,\kabc}    -\frac{1}{2}\ov{\vth\vthb} 
+\frac 1 r \Ab +\err[e_3\ov{\ka}].
\eeaa

In view of  Corollary \ref{corr:transportcheckf} the error term $\err[e_3( \ov{\ka})]$ is given by 
\beaa
\bsplit
\err[e_3( \ov{\ka})]&=-\vsi^{-1} \check{\vsi} \, \left( \ov{e_3\ka+\kab \ka} -\ov{\kab} \ov{\ka} \right)+ \vsi^{-1} \,\left(  \ov{\check{\vsi}(e_3\ka+\kab \ka)}- \ov{\check{\vsi}\kabc} \, \ov{\ka}\right)\\
&+\left(\Obc +\vsi^{-1}\ov{\Omb}\check{\vsi}\right) \,\left( \ov{e_4\ka +\ka^2}-\ov{\ka}^2\right) -\vsi^{-1}\ov{\Omb}\, \left(\ov{ \check{\vsi}(e_4\ka +\ka^2)}- \ov{\check{\vsi} \kac}     \,\ov{\ka} \right)\\
&-\vsi^{-1}\left(  \ov{\Obc\vsi(e_4\ka +\ka^2)}  -  \ov{\Obc\vsi \,\ka} \,\ov{\ka}  \right) +\ov{\kabc\kac}.
\end{split}
\eeaa
Together with the null structure equations for $e_3(\ka)$ and $e_4(\ka)$, we infer
\beaa
\err[e_3( \ov{\ka})]&=&-\vsi^{-1} \check{\vsi} \, \left( \ov{\frac{1}{2}\ka\kab+2\omb\ka +2\rho +2\ddd_1\eta  -\frac{1}{2}\vth\vthb +2\eta^2} -\ov{\kab}\, \ov{\ka} \right)\\
&&+ \vsi^{-1} \,\left(  \ov{\check{\vsi}\left(\frac{1}{2}\ka\kab+2\omb\ka +2\rho +2\ddd_1\eta  -\frac{1}{2}\vth\vthb +2\eta^2\right)}- \ov{\check{\vsi}\kabc} \, \ov{\ka}\right)\\
&&+\left(\Obc +\vsi^{-1}\ov{\Omb}\check{\vsi}\right) \,\left( \ov{\frac{1}{2}\ka^2 -\frac{1}{4}\vth^2}-\ov{\ka}^2\right) -\vsi^{-1}\ov{\Omb}\, \left(\ov{ \check{\vsi}\left(\frac{1}{2}\ka^2 -\frac{1}{4}\vth^2\right)}- \ov{\check{\vsi} \kac}     \,\ov{\ka} \right)\\
&&-\vsi^{-1}\left(  \ov{\Obc\vsi\left(\frac{1}{2}\ka^2 -\frac{1}{4}\vth^2\right)}  -  \ov{\Obc\vsi \,\ka} \,\ov{\ka}  \right) +\ov{\kabc\kac}.
\eeaa
and hence
\bea\lab{eq:precisecomputatione3ovka}
\nn\err[e_3( \ov{\ka})]&=& -\vsi^{-1}\left( -\frac{1}{2}\ov{\kab}\, \ov{\ka}+2\ov{\omb}\,\ov{\ka}+2\ov{\rho} \right)\check{\vsi} -\frac{1}{2}\ov{\ka}^2\left(\Obc +\vsi^{-1}\ov{\Omb}\check{\vsi}\right)\\
\nn&&-\vsi^{-1} \check{\vsi} \, \left( \ov{\frac{1}{2}\kac\kabc+2\ombc\kac    -\frac{1}{2}\vth\vthb +2\eta^2}  \right)\\
\nn&&+ \vsi^{-1} \,\left(  \ov{\check{\vsi}\left(\frac{1}{2}\ka\kab+2\omb\ka +2\rhoc +2\ddd_1\eta  -\frac{1}{2}\vth\vthb +2\eta^2\right)}- \ov{\check{\vsi}\kabc} \, \ov{\ka}\right)\\
\nn&&+\left(\Obc +\vsi^{-1}\ov{\Omb}\check{\vsi}\right)\left(\ov{\frac{1}{2}\kac^2 -\frac{1}{4}\vth^2}\right) -\vsi^{-1}\ov{\Omb}\, \left(\ov{ \check{\vsi}\left(\frac{1}{2}\ka^2 -\frac{1}{4}\vth^2\right)}- \ov{\check{\vsi} \kac}     \,\ov{\ka} \right)\\
&&-\vsi^{-1}\left(  \ov{\Obc\vsi\left(\frac{1}{2}\ka^2 -\frac{1}{4}\vth^2\right)}  -  \ov{\Obc\vsi \,\ka} \,\ov{\ka}  \right) +\ov{\kabc\kac}
\eea
so that, in view of the definition of $\Ab$, we obtain
\beaa
&& e_3\left(\ov{\ka}-\frac 2 r \right) +\frac 1 2  \ov{ \kab} \left(\ov{\ka}-\frac 2 r \right)\\
 &=&  2 \ov{\omb}\,\left( \ov{\ka}-\frac 2 r \right)+\frac 4 r\left(\ov{ \omb}-\frac{m}{r^2}\right)+ 2\left(  \ov{\rho} +\frac{2m}{r^3}\right) -\vsi^{-1}\left( -\frac{1}{2}\ov{\kab}\, \ov{\ka}+2\ov{\omb}\,\ov{\ka}+2\ov{\rho} \right)\check{\vsi}\\
 && -\frac{1}{2}\ov{\ka}^2\left(\Obc +\vsi^{-1}\ov{\Omb}\check{\vsi}\right) -\frac{1}{r}\vsi^{-1}\ov{\kab}\check{\vsi} +\frac{1}{r}\ov{\ka}\left(\Obc +\vsi^{-1}\ov{\Omb}\check{\vsi}\right)+\err\left[e_3\left(\ov{\ka}-\frac 2 r \right)\right],
 \eeaa
 with
 \beaa
\err\left[e_3\left(\ov{\ka}-\frac 2 r \right)\right] &=&2\ov{\eta^2}+ 2\ov{\check{\omb}\, \check{\ka}}  -  \frac 1 2\ov{ \kac\,\kabc}    -\frac{1}{2}\ov{\vth\vthb} + \frac{1}{r}\vsi^{-1}\ov{\check{\vsi}\kabc}  -\frac{1}{r}\vsi^{-1}\ov{\Omb}\,\ov{\check{\vsi}\kac}  -\frac{1}{r}\vsi^{-1}\ov{\Obc\vsi\ka}  \\
&&-\vsi^{-1} \check{\vsi} \, \left( \ov{\frac{1}{2}\kac\kabc+2\ombc\kac    -\frac{1}{2}\vth\vthb +2\eta^2}  \right)\\
&&+ \vsi^{-1} \,\left(  \ov{\check{\vsi}\left(\frac{1}{2}\ka\kab+2\omb\ka +2\rhoc +2\ddd_1\eta  -\frac{1}{2}\vth\vthb +2\eta^2\right)}- \ov{\check{\vsi}\kabc} \, \ov{\ka}\right)\\
&&+\left(\Obc +\vsi^{-1}\ov{\Omb}\check{\vsi}\right)\left(\ov{\frac{1}{2}\kac^2 -\frac{1}{4}\vth^2}\right) -\vsi^{-1}\ov{\Omb}\, \left(\ov{ \check{\vsi}\left(\frac{1}{2}\ka^2 -\frac{1}{4}\vth^2\right)}- \ov{\check{\vsi} \kac}     \,\ov{\ka} \right)\\
&&-\vsi^{-1}\left(  \ov{\Obc\vsi\left(\frac{1}{2}\ka^2 -\frac{1}{4}\vth^2\right)}  -  \ov{\Obc\vsi \,\ka} \,\ov{\ka}  \right) +\ov{\kabc\kac}.
\eeaa
This concludes the proof of Lemma \ref{lemma:transportequationforoverlinekaintheoutgoinggeodesicfoliation}.


\section{Proof of Proposition \ref{propos:transportaverages}}\lab{sec:proofofpropos:transportaverages}


In view of    Corollary \ref{corr:transportcheckf}    applied to  
\beaa
e_4(\ka)+\frac 12 \ka^2  =  -\frac 1 2 \vth^2,
\eeaa    we deduce,
\beaa
e_4\check{\ka}+\ov{\ka} \check{\ka}=-\frac 1 2 \check{\ka}^2 -\frac 1 2 \ov{\check{\ka}^2} -\frac 1 2 (\vth^2-\ov{\vth^2}).
\eeaa
In view of   Corollary \ref{corr:transportcheckf}   applied to 
\beaa
e_4(\kab)+\frac 1 2 \ka\kab  = -2\ddd_1\ze + 2\rho -\frac 1 2  \vth\vthb +2\ze^2
\eeaa
 we deduce,
\beaa
e_4\check{\kab} +\frac1 2 \ov{\ka} \check{\kab}+\frac 1 2 \check{\ka} \ov{\kab}=-\frac 1 2 \check{\ka}\check{\kab}-\frac 1 2 \ov{\check{\ka}\check{\kab}}+F-\ov{F}
\eeaa
where,
\beaa
F-\ov{F}&=&\left(-2\ddd_1\ze + 2\rho -\frac 1 2  \vth\vthb +2\ze^2\right)        -\ov{\left(-2\ddd_1\ze + 2\rho -\frac 1 2  \vth\vthb +2\ze^2\right) }\\
&=&-2 \ddd_1\ze+2\check{\rho} +\left( -\frac 1 2  \vth\vthb +2\ze^2\right)        -\ov{\left( -\frac 1 2  \vth\vthb +2\ze^2\right)}.
\eeaa
Hence,
\beaa
e_4\check{\kab} +\frac1 2 \ov{\ka} \check{\kab}+\frac 1 2 \check{\ka} \ov{\kab} &=&-2 \ddd_1\ze+2\check{\rho}+\err[e_4\check{\kab}]\\
\err[e_4\check{\kab}]:&=&-\frac 1 2 \check{\ka}\check{\kab}-\frac 1 2 \ov{\check{\ka}\check{\kab}} +\left( -\frac 1 2  \vth\vthb +2\ze^2\right)        -\ov{\left( -\frac 1 2  \vth\vthb +2\ze^2\right)}.
\eeaa
In view of   Corollary \ref{corr:transportcheckf}   applied to  $e_4(\omb)  =\rho    +3\ze^2$    we deduce,
\beaa
e_4\check{\omb}&=&-\ov{\check{\ka} \check{\omb}}+(\rho    +3\ze^2)-
\ov{(\rho    +3\ze^2)}=\check{\rho}  -\ov{\check{\ka} \check{\omb}}+ 3 (\ze^2-\ov{\ze^2}).
\eeaa
In view of    Corollary \ref{corr:transportcheckf}     applied to 
\beaa
e_4( \rho)+ \frac 32\ka\rho =\ddd_1 \b -\frac{1}{2}\vthb\a -\ze\b
\eeaa
 we deduce,
\beaa
e_4\check{\rho}+\frac  3 2 \ov{\ka} \check{\rho}+\frac 3 2 \ov{\rho}\check{\ka}=-\frac 3 2 \check{\ka}\check{\rho} +\frac 12 \ov{ \check{\ka}\check{\rho}}+\ddd_1\b - \left(\frac{1}{2}\vthb\a +\ze\b\right)+\ov{\left(\frac{1}{2}\vthb\a +\ze\b\right)}.
\eeaa
\beaa
e_4\mu +\frac 3 2 \ka \mu  =\err[e_4\mu],
\eeaa
we  deduce
\beaa
 e_4\check{\mu} +\frac 3 2\ov{ \ka}\check{ \mu} +\frac  3 2 \ov{\mu}\check{\ka}&=&-\frac 3 2 \check{\ka}\check{\mu}+\frac 1 2 
 \ov{ \check{\ka}\check{\mu}}+\err[e_4\mu]-\ov{\err[e_4\mu]}.
\eeaa
In view of   Corollary \ref{corr:transportcheckf}    applied to  $ e_4(\Om)=-2\omb$
we deduce,
\beaa
- e_4(\Obc) &=& 2 \check{\omb} - \ov{\check{\ka}\Obc}
\eeaa
as stated.

In view of  Corollary \ref{corr:transportcheckf}  applied  to the equation
\beaa
e_3(\ka) +\frac{1}{2}\kab\ka = 2\ddd_1\eta +2\rho +2\eta^2  +2\omb\ka -\frac{1}{2}\vthb\vth 
\eeaa
to deduce,
\beaa
e_3(\kac) &=&e_3(\ka ) -\ov{e_3(\ka) }-\err[e_3(\ov{\ka)}]\\
&=&- \frac{1}{2}\kab\ka + 2\ddd_1\eta  +2\rho +2\eta^2  +2\omb\ka -\frac{1}{2}\vthb\vth \\
&&+\frac{1}{2}\ov{\kab\ka }-2\ov{\rho}- 2\ov{\eta^2}  -2\ov{\omb\ka} +\frac{1}{2}\ov{\vthb\vth }-\err[e_3\ov{\ka}]\\
&=& 2\ddd_1 \eta +2 \rhoc-\frac 1 2\left (\kab \kac+\ka \kabc\right)+  2 \left(\omb \kac+\ka \ombc\right) \\
&& +2\left( \eta^2-\ov{\eta^2}\right)-\frac 12 \ov{\kac \kabc}+ 2\ov{\ombc\kac}-\frac 12 \left(\vth\vthb- \ov{\vth\vthb} \right)-\err[e_3\ov{\ka}].
\eeaa
Now, recall that we have in view of \eqref{eq:precisecomputatione3ovka}, 
\beaa
\nn\err[e_3( \ov{\ka})]&=& -\vsi^{-1}\left( -\frac{1}{2}\ov{\kab}\, \ov{\ka}+2\ov{\omb}\,\ov{\ka}+2\ov{\rho} \right)\check{\vsi} -\frac{1}{2}\ov{\ka}^2\left(\Obc +\vsi^{-1}\ov{\Omb}\check{\vsi}\right)\\
\nn&&-\vsi^{-1} \check{\vsi} \, \left( \ov{\frac{1}{2}\kac\kabc+2\ombc\kac    -\frac{1}{2}\vth\vthb +2\eta^2}  \right)\\
\nn&&+ \vsi^{-1} \,\left(  \ov{\check{\vsi}\left(\frac{1}{2}\ka\kab+2\omb\ka +2\rhoc +2\ddd_1\eta  -\frac{1}{2}\vth\vthb +2\eta^2\right)}- \ov{\check{\vsi}\kabc} \, \ov{\ka}\right)\\
\nn&&+\left(\Obc +\vsi^{-1}\ov{\Omb}\check{\vsi}\right)\left(\ov{\frac{1}{2}\kac^2 -\frac{1}{4}\vth^2}\right) -\vsi^{-1}\ov{\Omb}\, \left(\ov{ \check{\vsi}\left(\frac{1}{2}\ka^2 -\frac{1}{4}\vth^2\right)}- \ov{\check{\vsi} \kac}     \,\ov{\ka} \right)\\
&&-\vsi^{-1}\left(  \ov{\Obc\vsi\left(\frac{1}{2}\ka^2 -\frac{1}{4}\vth^2\right)}  -  \ov{\Obc\vsi \,\ka} \,\ov{\ka}  \right) +\ov{\kabc\kac}.
\eeaa
We deduce
\beaa
e_3(\kac) &=& 2\ddd_1 \eta +2 \rhoc-\frac 1 2\left (\kab \kac+\ka \kabc\right)+  2 \left(\omb \kac+\ka \ombc\right) \\
&& +\vsi^{-1}\left( -\frac{1}{2}\ov{\kab}\, \ov{\ka}+2\ov{\omb}\,\ov{\ka}+2\ov{\rho} \right)\check{\vsi} +\frac{1}{2}\ov{\ka}^2\left(\Obc +\vsi^{-1}\ov{\Omb}\check{\vsi}\right)\\
&& +2\left( \eta^2-\ov{\eta^2}\right)-\frac 12 \ov{\kac \kabc}+ 2\ov{\ombc\kac}-\frac 12 \left(\vth\vthb- \ov{\vth\vthb} \right)+\vsi^{-1} \check{\vsi} \, \left( \ov{\frac{1}{2}\kac\kabc+2\ombc\kac    -\frac{1}{2}\vth\vthb +2\eta^2}  \right)\\
&&- \vsi^{-1} \,\left(  \ov{\check{\vsi}\left(\frac{1}{2}\ka\kab+2\omb\ka +2\rhoc +2\ddd_1\eta  -\frac{1}{2}\vth\vthb +2\eta^2\right)}- \ov{\check{\vsi}\kabc} \, \ov{\ka}\right)\\
&&-\left(\Obc +\vsi^{-1}\ov{\Omb}\check{\vsi}\right)\left(\ov{\frac{1}{2}\kac^2 -\frac{1}{4}\vth^2}\right) +\vsi^{-1}\ov{\Omb}\, \left(\ov{ \check{\vsi}\left(\frac{1}{2}\ka^2 -\frac{1}{4}\vth^2\right)}- \ov{\check{\vsi} \kac}     \,\ov{\ka} \right)\\
&&+\vsi^{-1}\left(  \ov{\Obc\vsi\left(\frac{1}{2}\ka^2 -\frac{1}{4}\vth^2\right)}  -  \ov{\Obc\vsi \,\ka} \,\ov{\ka}  \right) -\ov{\kabc\kac}
\eeaa
as desired.

In view of  Corollary \ref{corr:transportcheckf}  applied  to the equation
\beaa
e_3(\kab)  +\frac{1}{2}\kab^2=2\ddd_1\xib - 2\omb\,\kab  +2(\eta-3 \ze) \xib -\frac{1}{2}\vthb^2
\eeaa
we deduce,
\beaa
e_3(\kabc) +\kab \, \kabc &=& 2\ddd_1\xib- 2\left(\ombc\,\kab+\omb\, \kabc\right)\\
&&- \frac 1 2 \ov{\kabc^2}-2 \ov{\ombc\, \kabc}+ 2(\eta-3 \ze) \xib-\ov{2(\eta-3 \ze) \xib} -\frac{1}{2}\left(\vthb^2-\ov{\vthb^2}\right)-\err[e_3\ov{\kab}]
\eeaa
where,
\beaa
\bsplit
-\err[e_3\ov{\kab}]&=\vsi^{-1} \check{\vsi} \, \left( \ov{e_3\kab+\kab^2} -\ov{\kab} \ov{\kab} \right)- \vsi^{-1} \,\left(  \ov{\check{\vsi}(e_3\kab+\kab^2)}- \ov{\check{\vsi}\kabc} \, \ov{\kab}\right)\\
&-\left(\Obc +\vsi^{-1}\ov{\Omb}\check{\vsi}\right) \,\left( \ov{e_4\kab +\ka \kab)}-\ov{\ka}\, \ov{\kab}\right) +\vsi^{-1}\ov{\Omb}\, \left(\ov{ \check{\vsi}(e_4\kab +\ka \kab)}- \ov{\check{\vsi} \kac}     \,\ov{\kab} \right)\\
&+\vsi^{-1}\left(  \ov{\Obc\vsi(e_4\kab +\ka \kab)}  -  \ov{\Obc\vsi \,\ka} \,\ov{\kab}  \right) -\ov{\kabc^2}.
\end{split}
\eeaa
In view of the null structure equations for $e_3(\kab)$ and $e_4(\kab)$, we infer
\beaa
\bsplit
-\err[e_3\ov{\kab}]&= \vsi^{-1} \check{\vsi} \, \left( -\frac 12 \ov{\kab}^2 -2 \ov{\omb} \,\ov{\kab} \right)-\left(\Obc +\vsi^{-1}\ov{\Omb}\check{\vsi}\right) \,\left( -\frac 1 2 \ov{\ka}\, \ov{\kab}  + 2\ov{\rho}\right) \\
& +\vsi^{-1} \check{\vsi} \, \left( \ov{\frac 12 \kabc^2 -2 \ombc \,\kabc+2(\eta-3\ze)  \xib -\frac 1 2 \vthb^2}  \right)-\left(\Obc +\vsi^{-1}\ov{\Omb}\check{\vsi}\right) \,\left( \ov{\frac 1 2 \kac\, \kabc   -\frac 1 2  \vth\, \vthb +2\ze^2} \right) \\
&- \vsi^{-1} \,\left(  \ov{\check{\vsi}\left(\frac 12 \kab^2 -2 \omb \,\kab+2\ddd_1\xib+2(\eta-3\ze)  \xib -\frac 1 2 \vthb^2\right)}- \ov{\check{\vsi}\kabc} \, \ov{\kab}\right)\\
&+\vsi^{-1}\ov{\Omb}\, \left(\ov{ \check{\vsi}\left(\frac 1 2 \ka\, \kab -2\ddd_1\ze + 2\rho -\frac 1 2  \vth\, \vthb +2\ze^2\right)}- \ov{\check{\vsi} \kac}     \,\ov{\kab} \right)\\
&+\vsi^{-1}\left(  \ov{\Obc\vsi\left(\frac 1 2 \ka\, \kab -2\ddd_1\ze + 2\rho -\frac 1 2  \vth\, \vthb +2\ze^2\right)}  -  \ov{\Obc\vsi \,\ka} \,\ov{\kab}  \right) -\ov{\kabc^2}
\end{split}
\eeaa
and hence
\beaa
e_3(\kabc) +\kab \, \kabc &=& 2\ddd_1\xib- 2\left(\ombc\,\kab+\omb\, \kabc\right) +\vsi^{-1} \check{\vsi} \, \left( -\frac 12 \ov{\kab}^2 -2 \ov{\omb} \,\ov{\kab} \right)\\
&&-\left(\Obc +\vsi^{-1}\ov{\Omb}\check{\vsi}\right) \,\left( -\frac 1 2 \ov{\ka}\, \ov{\kab}  + 2\ov{\rho}\right) +\err[e_3(\kabc)],\\
\err[e_3(\kabc)] &=& - \frac 1 2 \ov{\kabc^2}-2 \ov{\ombc\, \kabc}+ 2(\eta-3 \ze) \xib-\ov{2(\eta-3 \ze) \xib} -\frac{1}{2}\left(\vthb^2-\ov{\vthb^2}\right)\\
&&- \vsi^{-1} \,\left(  \ov{\check{\vsi}\left(\frac 12 \kab^2 -2 \omb \,\kab+2\ddd_1\xib+2(\eta-3\ze)  \xib -\frac 1 2 \vthb^2\right)}- \ov{\check{\vsi}\kabc} \, \ov{\kab}\right)\\
&&+\vsi^{-1}\ov{\Omb}\, \left(\ov{ \check{\vsi}\left(\frac 1 2 \ka\, \kab -2\ddd_1\ze + 2\rho -\frac 1 2  \vth\, \vthb +2\ze^2\right)}- \ov{\check{\vsi} \kac}     \,\ov{\kab} \right)\\
&&+\vsi^{-1}\left(  \ov{\Obc\vsi\left(\frac 1 2 \ka\, \kab -2\ddd_1\ze + 2\rho -\frac 1 2  \vth\, \vthb +2\ze^2\right)}  -  \ov{\Obc\vsi \,\ka} \,\ov{\kab}  \right) -\ov{\kabc^2}
\eeaa
as desired.

In view  of  Corollary \ref{corr:transportcheckf} applied to equation 
\beaa
e_3(\rho) + \frac{3}{2}\kab\rho  = \ddd_1\bb - \frac{1}{2}\vth\aa - \ze\bb +2\eta \bb +2\xib\b
\eeaa
we deduce,
\beaa
e_3\check{\rho} +\frac 3 2  \ov{\kab} \check{\rho}+ \frac 3 2 \check{\kab}\ov{\rho}&=&  \ddd_1\bb - \left(\frac{1}{2}\vth\aa + \ze\bb -2\eta \bb -2\xib\b\right)+\ov{\left(\frac{1}{2}\vth\aa + \ze\bb -2\eta \bb -2\xib\b\right)}\\
&-&\frac{3}{2}\kabc\rhoc -\err[e_3\ov{\rho}]
\eeaa
where
\beaa
\bsplit
-\err[e_3( \ov{\rho})]&=\vsi^{-1} \check{\vsi} \, \left( \ov{e_3\rho+\kab \rho} -\ov{\kab} \ov{\rho} \right)- \vsi^{-1} \,\left(  \ov{\check{\vsi}(e_3\rho+\kab \rho)}- \ov{\check{\vsi}\kabc} \, \ov{\rho}\right)\\
&-\left(\Obc +\vsi^{-1}\ov{\Omb}\check{\vsi}\right) \,\left( \ov{e_4\rho +\ka \rho)}-\ov{\ka}\, \ov{\rho}\right) +\vsi^{-1}\ov{\Omb}\, \left(\ov{ \check{\vsi}(e_4\rho +\ka \rho)}- \ov{\check{\vsi} \kac}     \,\ov{\rho} \right)\\
&+\vsi^{-1}\left(  \ov{\Obc\vsi(e_4\rho +\ka \rho)}  -  \ov{\Obc\vsi \,\ka}   \right) -\ov{\kabc\rhoc}.
\end{split}
\eeaa
In view of the null structure equations for $e_3(\rho)$ and $e_4(\rho)$, we infer
\beaa
\bsplit
-\err[e_3( \ov{\rho})]&=-\frac{3}{2}\ov{\kab}\, \ov{\rho}\vsi^{-1} \check{\vsi} +\frac{3}{2}\ov{\ka}\, \ov{\rho}\left(\Obc +\vsi^{-1}\ov{\Omb}\check{\vsi}\right)\\
&+\vsi^{-1} \check{\vsi} \, \left( \ov{-\frac 1 2 \kabc \rhoc  -\frac 1 2 \vth \, \aa - \ze\, \bb +2(\eta \,\bb+ \xib\,\b)} \right)\\
&- \vsi^{-1} \,\left(  \ov{\check{\vsi}\left(-\frac 1 2 \kab\rho+ \ddd_1\bb  -\frac 1 2 \vth \, \aa - \ze\, \bb +2(\eta \,\bb+ \xib\,\b)\right)}- \ov{\check{\vsi}\kabc} \, \ov{\rho}\right)\\
& -\left(\Obc +\vsi^{-1}\ov{\Omb}\check{\vsi}\right) \,\left( \ov{-\frac 1 2 \kac \rhoc  -\frac 1 2 \vthb \, \a -\ze \b}\right)\\
& +\vsi^{-1}\ov{\Omb}\, \left(\ov{ \check{\vsi}\left(-\frac 1 2 \ka \rho+\ddd_1 \b  -\frac 1 2 \vthb \, \a -\ze \b\right)}- \ov{\check{\vsi} \kac}     \,\ov{\rho} \right)\\
&+\vsi^{-1}\left(  \ov{\Obc\vsi\left(-\frac 1 2 \ka \rho+\ddd_1 \b  -\frac 1 2 \vthb \, \a -\ze \b\right)}  -  \ov{\Obc\vsi \,\ka}   \right) -\ov{\kabc\rhoc}.
\end{split}
\eeaa
and hence
\beaa
e_3\check{\rho} +\frac 3 2  \ov{\kab} \check{\rho}&=& - \frac 3 2 \ov{\rho}\check{\kab}+ \ddd_1\bb -\frac{3}{2}\ov{\kab}\, \ov{\rho}\vsi^{-1} \check{\vsi} +\frac{3}{2}\ov{\ka}\, \ov{\rho}\left(\Obc +\vsi^{-1}\ov{\Omb}\check{\vsi}\right) +\err[e_3\check{\rho}],\\
\err[e_3\check{\rho}] &=&  - \left(\frac{1}{2}\vth\aa + \ze\bb -2\eta \bb -2\xib\b\right)+\ov{\left(\frac{1}{2}\vth\aa + \ze\bb -2\eta \bb -2\xib\b\right)}-\frac{3}{2}\kabc\rhoc \\
&& +\vsi^{-1} \check{\vsi} \, \left( \ov{-\frac 1 2 \kabc \rhoc  -\frac 1 2 \vth \, \aa - \ze\, \bb +2(\eta \,\bb+ \xib\,\b)} \right)\\
&&- \vsi^{-1} \,\left(  \ov{\check{\vsi}\left(-\frac 1 2 \kab\rho+ \ddd_1\bb  -\frac 1 2 \vth \, \aa - \ze\, \bb +2(\eta \,\bb+ \xib\,\b)\right)}- \ov{\check{\vsi}\kabc} \, \ov{\rho}\right)\\
&& -\left(\Obc +\vsi^{-1}\ov{\Omb}\check{\vsi}\right) \,\left( \ov{-\frac 1 2 \kac \rhoc  -\frac 1 2 \vthb \, \a -\ze \b}\right)\\
&& +\vsi^{-1}\ov{\Omb}\, \left(\ov{ \check{\vsi}\left(-\frac 1 2 \ka \rho+\ddd_1 \b  -\frac 1 2 \vthb \, \a -\ze \b\right)}- \ov{\check{\vsi} \kac}     \,\ov{\rho} \right)\\
&&+\vsi^{-1}\left(  \ov{\Obc\vsi\left(-\frac 1 2 \ka \rho+\ddd_1 \b  -\frac 1 2 \vthb \, \a -\ze \b\right)}  -  \ov{\Obc\vsi \,\ka}   \right) -\ov{\kabc\rhoc},
\eeaa
which ends the proof of Proposition \ref{propos:transportaverages}.


\section{Proof of Proposition \ref{prop:eqtsfor-ometaxib}}\lab{sec:proofofprop:eqtsfor-ometaxib}


In view of the null structure equation for $e_3(\ze)$, we have
\beaa
\frac{1}{2}\ka\xib  +2\dds_1\omb &=& e_3(\ze)+\frac{1}{2}\kab(\ze+\eta) -2\omb(\ze-\eta) -\bb+\frac{1}{2}\vthb(\ze+\eta)-\frac{1}{2}\vth\xib
\eeaa
and hence
\beaa
\frac{1}{2}\ka\xib  +2\dds_1\omb &=& \left(\frac{1}{2}\kab  +2\omb +\frac{1}{2}\vthb\right)\eta + e_3(\ze) -\bb\\
 && +\frac{1}{2}\kab\ze  -2\omb\ze +\frac{1}{2}\vthb\ze  -\frac{1}{2}\vth\xib
 \eeaa
which is the first desired identity.

To prove the second identity we start with
\beaa
e_3(\ka)+\frac 1 2 \kab\, \ka -2\omb \ka &= 2\ddd_1\eta + 2\rho -\frac 1 2  \vthb\, \vth +2\eta^2.
\eeaa
Applying $e_\th$,
\beaa
&& e_3(e_\th(\ka))+[e_\th, e_3]\ka+\frac 1 2 \kab e_\th(\ka) +\frac 1 2 \ka e_\th(\kab) -2\omb e_\th(\ka) -2\ka e_\th(\omb)\\
 &=& 2e_\th(\ddd_1\eta) + 2e_\th(\rho) -\frac 1 2  e_\th(\vthb\, \vth) +2e_\th(\eta^2).
\eeaa
Since $[e_\th, e_3]\ka=\frac 1 2 (\kab+\vthb)e_\th \ka  +(\ze-\eta) e_3\ka  -\xib e_4\ka $ we deduce,
\beaa
 2e_\th(\ddd_1\eta)  +\eta e_3(\ka)+2e_\th(\eta^2)&=& -\xib e_4(\ka) -2\ka e_\th(\omb) +e_3(e_\th(\ka)\\
 &+&\frac{1}{2}(\kab+\vthb)e_\th(\ka)+\ze e_3(\ka) +\frac 1 2 \kab e_\th(\ka) +\frac 1 2 \ka e_\th(\kab)\\
  &-&2\omb e_\th(\ka) - 2e_\th(\rho) +\frac 1 2  e_\th(\vthb\, \vth),
\eeaa
or, making use of the  equations for $e_3\ka$ and  $e_4\ka$  in Proposition \ref{propos:basiceqts-geod},
\beaa
&& 2e_\th(\ddd_1\eta)  +\left(-\frac{1}{2}\ka\kab+2\omb\ka+2\ddd_1\eta+2\rho-\frac{1}{2}\vth\vthb+2\eta^2\right)\eta+2e_\th(\eta^2)\\
&=& -\left(-\frac{1}{2}\ka^2-\frac{1}{2}\vth^2\right)\xib  -2\ka e_\th(\omb) +e_3(e_\th(\ka))\\
&&+\frac{1}{2}(\kab+\vthb)e_\th(\ka)+\left(-\frac{1}{2}\ka\kab+2\omb\ka+2\ddd_1\eta+2\rho-\frac{1}{2}\vth\vthb+2\eta^2\right)\ze \\
&&+\frac 1 2 \kab e_\th(\ka) +\frac 1 2 \ka e_\th(\kab) -2\omb e_\th(\ka) - 2e_\th(\rho) +\frac 1 2  e_\th(\vthb\, \vth).
\eeaa
Since  $e_\th=-\dds_1$,  $\dds_1\ddd_1=\ddd_2\dds_2+2K$ and $K=-\rho-\frac 1 4 \ka\kab +\frac 1 4 \vth\vthb$, we infer that
\beaa
&&   \Big( -2\ddd_2\dds_2 +\frac 1 2 \ka \kab +2\omb\ka+2\ddd_1\eta+6 \rho-\frac 3 2 \vth\vthb+2\eta^2\Big)\eta+2e_\th(\eta^2)\\
&=& \ka\left(\frac{1}{2}\ka\xib  +2\dds_1\omb\right) +e_3(e_\th(\ka))\\
&&+\frac{1}{2}(\kab+\vthb)e_\th(\ka)+\left(-\frac{1}{2}\ka\kab+2\omb\ka+2\ddd_1\eta+2\rho-\frac{1}{2}\vth\vthb+2\eta^2\right)\ze \\
&&+\frac 1 2 \kab e_\th(\ka) +\frac 1 2 \ka e_\th(\kab) -2\omb e_\th(\ka) - 2e_\th(\rho) +\frac 1 2  e_\th(\vthb\, \vth) +\frac{1}{2}\vth^2\xib.
\eeaa
Making use of the previously derived identity,
\beaa
  2\dds_1\omb +\frac{1}{2}\ka\xib&=& \left(\frac{1}{2}\kab  +2\omb +\frac{1}{2}\vthb\right)\eta + e_3(\ze) -\bb\\
&& +\frac{1}{2}\kab\ze  -2\omb\ze +\frac{1}{2}\vthb\ze  -\frac{1}{2}\vth\xib,
 \eeaa
we infer that,
\beaa
&&   \Big( -2\ddd_2\dds_2 +\frac 1 2 \ka \kab +2\omb\ka+2\ddd_1\eta+6 \rho-\frac 3 2 \vth\vthb+2\eta^2\Big)\eta+2e_\th(\eta^2)\\
&=&\ka\left(  \left(\frac{1}{2}\kab  +2\omb +\frac{1}{2}\vthb\right)\eta + e_3(\ze) -\bb\right)\\
&& + \ka\left(\frac{1}{2}\kab\ze  -2\omb\ze +\frac{1}{2}\vthb\ze  -\frac{1}{2}\vth\xib\right)
+e_3(e_\th(\ka))\\
&&+\frac{1}{2}(\kab+\vthb)e_\th(\ka)+\left(-\frac{1}{2}\ka\kab+2\omb\ka+2\ddd_1\eta+2\rho-\frac{1}{2}\vth\vthb+2\eta^2\right)\ze \\
&&+\frac 1 2 \kab e_\th(\ka) +\frac 1 2 \ka e_\th(\kab) -2\omb e_\th(\ka) - 2e_\th(\rho) +\frac 1 2  e_\th(\vthb\, \vth) +\frac{1}{2}\vth^2\xib,
\eeaa
or,
\beaa
&&   \left( -2\ddd_2\dds_2  +6\rho+2\ddd_1\eta -\frac{1}{2}\ka\vthb-\frac{3}{2}\vth\vthb+2\eta^2    \right)\eta+2e_\th(\eta^2)\\
&=& \ka\left( e_3(\ze) -\bb\right) +e_3(e_\th(\ka))\\
 && +\ka\left(\frac{1}{2}\kab\ze  -2\omb\ze +\frac{1}{2}\vthb\ze  -\frac{1}{2}\vth\xib\right) \\
&&+\frac{1}{2}(\kab+\vthb)e_\th(\ka)+\left(-\frac{1}{2}\ka\kab+2\omb\ka+2\ddd_1\eta+2\rho-\frac{1}{2}\vth\vthb+2\eta^2\right)\ze \\
&&+\frac 1 2 \kab e_\th(\ka) +\frac 1 2 \ka e_\th(\kab) -2\omb e_\th(\ka) - 2e_\th(\rho) +\frac 1 2  e_\th(\vthb\, \vth) +\frac{1}{2}\vth^2\xib
\eeaa
and hence
\beaa
&&   \left( 2\ddd_2\dds_2  -2\ddd_1\eta +\frac{1}{2}\ka\vthb -2\eta^2    \right)\eta-2e_\th(\eta^2)\\
&=& \ka\left( -e_3(\ze) +\bb\right) -e_3(e_\th(\ka))\\
 && -\ka\left(\frac{1}{2}\kab\ze  -2\omb\ze +\frac{1}{2}\vthb\ze  -\frac{1}{2}\vth\xib\right) +6\rho\eta -\frac{3}{2}\vth\vthb\eta\\
&&-\frac{1}{2}(\kab+\vthb)e_\th(\ka)-\left(-\frac{1}{2}\ka\kab+2\omb\ka+2\ddd_1\eta+2\rho-\frac{1}{2}\vth\vthb+2\eta^2\right)\ze \\
&&-\frac 1 2 \kab e_\th(\ka) -\frac 1 2 \ka e_\th(\kab) +2\omb e_\th(\ka) + 2e_\th(\rho) -\frac 1 2  e_\th(\vthb\, \vth) -\frac{1}{2}\vth^2\xib
\eeaa
which is the second desired identity.

To prove the third identity we start with,
\beaa
e_3(\kab)+\frac{1}{2}\kab^2+2\omb\,\kab &=& 2\ddd_1\xib+2(\eta-3\ze)\xib-\frac{1}{2}\vthb^2.
\eeaa
 Taking $e_\th=-\dds_1 $ and using  $\dds_1\ddd_1=\ddd_2\dds_2+2K $  as before,
\beaa
&& e_3(e_\th(\kab))+[e_\th, e_3]\kab+\kab e_\th(\kab)+2\omb e_\th(\kab)+2\kab e_\th(\omb)\\
 &=& -2\dds_1\ddd_1\xib+2e_\th\Big((\eta-3\ze)\xib\Big)-\frac{1}{2}e_\th(\vthb^2)\\
&=&  -2(\ddd_2\dds_2+2K)\xib+2e_\th\Big((\eta-3\ze)\xib\Big)-\frac{1}{2}e_\th(\vthb^2).
\eeaa
Thus,  since $[e_\th, e_3] \kab=\frac 1 2 (\kab+\vthb) e_\th \kab +\ze-\eta) e_3 \kab -\xib e_4\kab$,
\beaa
-2(\ddd_2\dds_2+2K)\xib&=&e_3(e_\th(\kab)) +2\kab e_\th(\omb)+\frac 1 2 (\kab+\vthb) e_\th \kab +(\ze-\eta) e_3 \kab -\xib e_4\kab\\
&+&\kab e_\th(\kab)+2\omb e_\th(\kab) -2e_\th\Big((\eta-3\ze)\xib\Big)+\frac{1}{2}e_\th(\vthb^2).
\eeaa
Making use of the equations   for $e_3\kab, e_4\ka$ in Proposition \ref{propos:basiceqts-geod}
\beaa
2(\ddd_2\dds_2+2K)\xib&=& 2\kab  \dds_1 \omb -e_3(e_\th(\kab)-\frac 1 2 (\kab+\vthb) e_\th \kab\\
&-&(\ze-\eta)\left(-\frac 12 \kab^2 -2 \omb \,\kab +2\ddd_1\xib+2(\eta-3\ze)  \xib -\frac 1 2 \vthb^2\right)\\
&+&\xib\left(-\frac 1 2 \ka\, \kab   -2\ddd_1\ze + 2\rho -\frac 1 2  \vth\, \vthb +2\ze^2\right)\\
&-&\kab e_\th(\kab)-2\omb e_\th(\kab) +2e_\th\Big((\eta-3\ze)\xib\Big)-\frac{1}{2}e_\th(\vthb^2).
\eeaa
We deduce,
\beaa
 && 2(\ddd_2\dds_2+K)\xib \\
&=& 2\kab\dds_1\omb  +\eta\left(-\frac{1}{2}\kab^2-2\omb\,\kab+2\ddd_1\xib+2(\eta-3\ze)  \xib-\frac{1}{2}\vthb^2\right) +2e_\th\Big((\eta-3\ze)\xib\Big)\\
&&-e_3(e_\th(\kab))  -\frac{1}{2}e_\th(\vthb^2)
-\frac{1}{2}(\kab+\vthb)e_\th(\kab)-\ze\left(-\frac{1}{2}\kab^2-2\omb\,\kab+2\ddd_1\xib+2(\eta-3\ze)\xib-\frac{1}{2}\vthb^2\right)\\
&&+\xib\left(-\frac 1 2 \ka\, \kab -2K  -2\ddd_1\ze + 2\rho -\frac 1 2  \vth\, \vthb +2\ze^2\right)-\kab e_\th(\kab)-2\omb e_\th(\kab) -6\eta\ze\xib -6e_\th(\ze\xib).
\eeaa
Making use of $K=-\rho-\frac 1 4 \ka\kab +\frac 1 4 \vth \vthb $  and reorganizing  we deduce,
\beaa
 && 2(\ddd_2\dds_2+K)\xib \\
&=& 2\kab\dds_1\omb  +\eta\left(-\frac{1}{2}\kab^2-2\omb\,\kab+2\ddd_1\xib+2\eta\xib-\frac{1}{2}\vthb^2\right) + 2e_\th(\eta\xib)-e_3(e_\th(\kab))  -\frac{1}{2}e_\th(\vthb^2)\\
&&-\frac{1}{2}(\kab+\vthb)e_\th(\kab)-\ze\left(-\frac{1}{2}\kab^2-2\omb\,\kab+2\ddd_1\xib+2(\eta-3\ze)\xib-\frac{1}{2}\vthb^2\right)\\
&&+\xib\Big(4\rho-\vth\vthb -2\ddd_1\ze+2\ze^2\Big)-\kab e_\th(\kab)-2\omb e_\th(\kab) -6\eta\ze\xib -6e_\th(\ze\xib).
\eeaa
 We make use again of  the  identity,
\beaa
  2\dds_1\omb +\frac{1}{2}\ka\xib&=& \left(\frac{1}{2}\kab  +2\omb +\frac{1}{2}\vthb\right)\eta + e_3(\ze) -\bb\\
&& +\frac{1}{2}\kab\ze  -2\omb\ze +\frac{1}{2}\vthb\ze  -\frac{1}{2}\vth\xib,
 \eeaa
 to derive,
 \beaa
 && 2(\ddd_2\dds_2+K)\xib \\
 &=&\kab \left(-\frac{1}{2}\ka\xib+ \left(\frac{1}{2}\kab  +2\omb +\frac{1}{2}\vthb\right)\eta + e_3(\ze) -\bb\right)
 +\kab\left(\frac{1}{2}\kab\ze  -2\omb\ze +\frac{1}{2}\vthb\ze  -\frac{1}{2}\vth\xib\right)\\
 &+&\eta\left(-\frac{1}{2}\kab^2-2\omb\,\kab+2\ddd_1\xib+2\eta\xib-\frac{1}{2}\vthb^2\right) + 2e_\th(\eta\xib)-e_3(e_\th(\kab))  -\frac{1}{2}e_\th(\vthb^2)\\
&&-\frac{1}{2}(\kab+\vthb)e_\th(\kab)-\ze\left(-\frac{1}{2}\kab^2-2\omb\,\kab+2\ddd_1\xib+2(\eta-3\ze)\xib-\frac{1}{2}\vthb^2\right)\\
&&+\xib\Big(4\rho-\vth\vthb -2\ddd_1\ze+2\ze^2\Big)-\kab e_\th(\kab)-2\omb e_\th(\kab) -6\eta\ze\xib -6e_\th(\ze\xib).
 \eeaa
 Grouping terms  and using once more the identity $K=-\rho-\frac 1 4 \ka \kab +\frac 1 4 \vth\vthb $ we deduce,
\beaa
 && 2\ddd_2\dds_2\xib\\
&=& -e_3(e_\th(\kab))+ \kab\left(e_3(\ze) -\bb\right) +\left(2\ddd_1\xib+\frac{1}{2}\kab\,\vthb+2\eta\xib-\frac{1}{2}\vthb^2\right)\eta + 2e_\th(\eta\xib)  -\frac{1}{2}e_\th(\vthb^2)\\
 && +\kab\left(\frac{1}{2}\kab\ze  -2\omb\ze +\frac{1}{2}\vthb\ze  -\frac{1}{2}\vth\xib\right) -\frac{1}{2}(\kab+\vthb)e_\th(\kab)  -\frac{1}{2}\vth\vthb\xib\\
 &&-\ze\left(-\frac{1}{2}\kab^2-2\omb\,\kab+2\ddd_1\xib+2(\eta-3\ze)\xib-\frac{1}{2}\vthb^2\right)\\
&&+\xib\Big(6\rho-\vth\vthb -2\ddd_1\ze+2\ze^2\Big)-\kab e_\th(\kab)-2\omb e_\th(\kab) -6\eta\ze\xib -6e_\th(\ze\xib)
\eeaa
which is the third desired identity. This concludes the proof of Proposition \ref{prop:eqtsfor-ometaxib}.


\section{Proof of Proposition \ref{prop:transformations1}}\lab{sec:proofofprop:transformations1}


The proof follows by straightforward calculations  using the definition  of Ricci coefficients and curvature components 
 with respect to the  two frames.  Recall the  transformation    \eqref{SSMe:general.composite}
 \beaa
\begin{split}
e_4'&=\la\left(e_4 + f e_\th +\frac 1 4 f^2  e_3\right),\\
e_\th'&=\left(1+\frac 1 2   f \fb\right) e_\th  + \frac 1 2  \fb e_4+\frac 1 2 f\left(1+ \frac 1 4 f \fb\right) e_3,\\
e_3'&= \la^{-1} \left( \left(1+\frac 12  f \fb +\frac{1}{16} f^2 \fb^2\right)  e_3 +\fb\left(1+\frac 1 4 f\fb\right) e_\th + \frac 1 4 \fb^2  e_4\right).
\end{split}
\eeaa
We first derive the transformation formulae for $\ka$. We have,  under a transformation
 of type   \eqref{SSMe:general.composite}, 
  \beaa
\chi'&=& g( D_{e_\th'}  e_4', e_{\th'})\\
&=& g\left( D_{e_\th'}\left(\la\left(e_4 + f e_\th +\frac 1 4 f^2  e_3\right)\right), e_\th'\right)\\
&=& \la g\left( D_{e_\th'}\left(e_4 + f e_\th +\frac 1 4 f^2  e_3\right), e_\th'\right)\\
&=& \la g\left( D_{e_\th'}e_4, e_\th'\right) +\la e_\th'(f)g(e_\th, e_\th') +\frac{\la}{4}e_\th'(f^2)g(e_3, e_\th') +\la f g\left( D_{e_\th'}e_\th, e_\th'\right) +\frac{\la}{4}f^2 g( D_{e_\th'}e_3, e_\th')\\
&=& \la g\left( D_{e_\th'}e_4, e_\th'\right) +\la\left(1+\frac 1 2   f \fb\right)e_\th'(f) -\frac{\la}{4}\fb e_\th'(f^2) +\la f g\left( D_{e_\th'}e_\th, e_\th'\right) +\frac{\la}{4}f^2\chib +\lot
\eeaa
We recall
that the   lower order terms we denote by $\lot$, here and throughout the proof,  are linear 
  with respect       $\Ga=\{\xi,\xib,\vth, \ka, \eta,\etab,\ze,\kab,  \vthb\}$  and quadratic or  higher order in $f,\fb$, and do not contain derivatives of these latter. We also recall that $\chi=\frac 1 2 (\ka+\vth)$, $\chib=\frac 12 (\kab+\vthb)$.

 Next, we compute
\beaa
g\left( D_{e_\th'}e_4, e_\th'\right) &=& g\left( D_{e_\th'}e_4, \left(1+\frac 1 2   f \fb\right) e_\th  +\frac 1 2 f e_3\right)+\lot\\
&=& \left(1+\frac 1 2   f \fb\right) g\left(D_{\left(1+\frac 1 2   f \fb\right) e_\th  + \frac 1 2  \fb e_4+\frac 1 2 fe_3}e_4, e_\th\right)\\
&&+\frac 1 2 f g\left( D_{e_\th  + \frac 1 2  \fb e_4+\frac 1 2 f e_3}e_4,  e_3\right)+\lot\\
&=& \left(1+f \fb\right)\chi + \fb\xi+f\eta + f\ze + f\fb\om - f^2\omb+\lot,
\eeaa
and 
\beaa
fg\left( D_{e_\th'}e_\th, e_\th'\right) &=& \frac 1 2  f\fb g\left( D_{e_\th'}e_\th,   e_4\right) +\frac 1 2 f^2 g\left( D_{e_\th'}e_\th,   e_3\right)\\
&=& -\frac 1 2  f\fb\chi -\frac 1 2 f^2\chib+\lot
\eeaa
This yields
  \beaa
\chi' &=& \la g\left( D_{e_\th'}e_4, e_\th'\right) +\la\left(1+\frac 1 2   f \fb\right)e_\th'(f) -\frac{\la}{4}\fb e_\th'(f^2) +\la f g\left( D_{e_\th'}e_\th, e_\th'\right) +\frac{\la}{4}f^2\chib +\lot\\
&=& \la\Bigg(\chi+ \left(1+\frac 1 2   f \fb\right)e_\th'(f) -\frac{1}{4}\fb e_\th'(f^2)  + f(\ze+\eta)    +\fb\xi          -  \frac 1 4 f^2 \chib +\frac 1 2  f\fb \chi +f\fb\om -f^2\omb\\
&&  +\lot\Bigg).
\eeaa
Hence,
\beaa
\ka'&=&\chi'+ e'_4 \Phi =\chi' + \la\left(  e_4+ f e_\th +\frac 1 4 f^2 e_3\right )\Phi\\
&=& \la\Bigg(\ka+ e_\th'(f)+e_\th(\Phi)f +\frac 1 8(\kab-\vthb) f^2+\frac 1 2   f \fb e_\th'(f) -\frac{1}{4}\fb e_\th'(f^2)  + f(\ze+\eta)    +\fb\xi   \\
&&       -  \frac 1 4 f^2 \chib +\frac 1 2  f\fb \chi +f\fb\om -f^2\omb  +\lot\Bigg)\\
&=& \la\left(\ka+ \ddd_1\,\!'(f)   +\frac 1 2   f \fb e_\th'(f) -\frac{1}{4}\fb e_\th'(f^2) -\frac 1 4 f^2\kab + f(\ze+\eta)    +\fb\xi        +f\fb\om -f^2\omb  +\lot\right)
\eeaa
and
\beaa
\vth'&=&\chi'- e'_4 \Phi =\chi' - \la\left(  e_4+ f e_\th +\frac 1 4 f^2 e_3\right )\Phi\\
&=& \la\Bigg(\vth+ e_\th'(f)-e_\th(\Phi)f -\frac 1 8(\kab-\vthb) f^2+\frac 1 2   f \fb e_\th'(f) -\frac{1}{4}\fb e_\th'(f^2)  + f(\ze+\eta)    +\fb\xi   \\
&&       -  \frac 1 4 f^2 \chib +\frac 1 2  f\fb \chi +f\fb\om -f^2\omb  +\lot\Bigg)\\
&=& \la\left(\vth- \dds_2\,\!'(f)   +\frac 1 2   f \fb e_\th'(f) -\frac{1}{4}\fb e_\th'(f^2)  + \frac 1 4 f\fb\ka  + f(\ze+\eta)    +\fb\xi        +f\fb\om -f^2\omb  +\lot\right).
\eeaa
This yields
\beaa
\ka'&=&\la\left( \ka+ \ddd_1\,\!'(f)    \right) +\la\err(\ka,\ka'),\\
   \err(\ka,\ka')&=& \frac 1 2   f \fb e_\th'(f) -\frac{1}{4}\fb e_\th'(f^2)+f(\ze+\eta)    +\fb\xi       -\frac 1 4 f^2\kab +f\fb\om -f^2\omb+\lot\nn\\
   &=& f(\ze+\eta)    +\fb\xi       -\frac 1 4 f^2\kab +f\fb\om -f^2\omb+\lot
   \eeaa
and
\beaa
\vth' &=& \la\left(\vth- \dds_2\,\!'(f)   \right) + \la\err(\vth,\vth'),\\
\err(\vth,\vth') &=& \frac 1 2   f \fb e_\th'(f) -\frac{1}{4}\fb e_\th'(f^2)+f(\ze+\eta)    +\fb\xi            +\frac 1 4  f\fb \ka +f\fb\om -f^2\omb+\lot\\
&=& f(\ze+\eta)    +\fb\xi            +\frac 1 4  f\fb \ka +f\fb\om -f^2\omb+\lot
\eeaa

Next, we derive the transformation formula for $ \kab$ and $\vthb$. We have,  under a transformation
 of type   \eqref{SSMe:general.composite}, 
  \beaa
\chib'&=& g( D_{e_\th'}  e_3', e_{\th'})\\
&=& g\left( D_{e_\th'}\left(\la^{-1} \left( \left(1+\frac 12  f \fb +\frac{1}{16} f^2 \fb^2\right)  e_3 +\fb\left(1+\frac 1 4 f\fb\right) e_\th + \frac 1 4 \fb^2  e_4\right)\right), e_\th'\right)\\
&=& \la^{-1}g\left( D_{e_\th'}\left( \left(1+\frac 12  f \fb +\frac{1}{16} f^2 \fb^2\right)  e_3 +\fb\left(1+\frac 1 4 f\fb\right) e_\th + \frac 1 4 \fb^2  e_4\right), e_\th'\right)\\
&=& \frac{\la^{-1}}{2}e_\th'\left( f \fb +\frac{1}{8} f^2 \fb^2\right)g(e_3, e_\th')+\la^{-1}e_\th'\left(\fb\left(1+\frac 1 4 f\fb\right)\right)g(e_\th,  e_\th')+\frac{\la^{-1}}{4}e_\th'\left(\fb^2\right)g(e_4, e_\th')\\
&& +\la^{-1} \left(1+\frac 12  f \fb +\frac{1}{16} f^2 \fb^2\right)g\left( D_{e_\th'}e_3, e_\th'\right)+\la^{-1}\fb\left(1+\frac 1 4 f\fb\right) g\left( D_{e_\th'}e_\th , e_\th'\right)\\
&&+\frac 1 4 \la^{-1}\fb^2 g\left( D_{e_\th'}e_4, e_\th'\right)\\
&=& -\frac{\la^{-1}}{2}\fb e_\th'\left( f \fb +\frac{1}{8} f^2 \fb^2\right)+\la^{-1}\left(1+\frac{1}{2}f\fb\right)e_\th'\left(\fb\left(1+\frac 1 4 f\fb\right)\right)\\
&&-\frac{\la^{-1}}{4}f\left(1+\frac{1}{4}f\fb\right)e_\th'\left(\fb^2\right) +\la^{-1} \left(1+\frac 12  f \fb \right)g\left( D_{e_\th'}e_3, e_\th'\right)+\la^{-1}\fb g\left( D_{e_\th'}e_\th , e_\th'\right)\\
&&+\frac 1 4 \la^{-1}\fb^2\chi+\lot
\eeaa
Then, we easily derive by symmetry from the formula for $\ka$ and $\vth$
\beaa
\kab'&=&\la^{-1}\left( \kab+ \ddd_1\,\!'(\fb)    \right) +\la^{-1}\err(\kab,\kab'),\\
   \err(\kab,\kab')&=& -\frac{1}{2}\fb e_\th'\left( f \fb +\frac{1}{8} f^2 \fb^2\right) +\left(\frac{3}{4}f\fb+\frac 1 8 (f\fb)^2\right)e_\th'(\fb)  +\frac 1 4\left(1+\frac{1}{2}f\fb\right)\fb e_\th'\left( f\fb\right)\\
&&-\frac{1}{4}f\left(1+\frac{1}{4}f\fb\right)e_\th'\left(\fb^2\right)    +\fb(-\ze+\etab)    +f\xib       -\frac 1 4 \fb^2\ka +f\fb\omb -\fb^2\om+\lot\nn\\
&=& -\frac{1}{4}\fb^2e_\th'(f) +\fb(-\ze+\etab)    +f\xib       -\frac 1 4 \fb^2\ka +f\fb\omb -\fb^2\om+\lot
   \eeaa
and
\beaa
\vthb' &=& \la\left(\vthb- \dds_2\,\!'(\fb)   \right) + \la^{-1}\err(\vthb,\vthb'),\\
\err(\vthb,\vthb') &=&  -\frac{1}{2}\fb e_\th'\left( f \fb +\frac{1}{8} f^2 \fb^2\right) +\left(\frac{3}{4}f\fb+\frac 1 8 (f\fb)^2\right)e_\th'(\fb)  +\frac 1 4\left(1+\frac{1}{2}f\fb\right)\fb e_\th'\left( f\fb\right)\\
&&-\frac{1}{4}f\left(1+\frac{1}{4}f\fb\right)e_\th'\left(\fb^2\right)+\fb(-\ze+\etab)    +f\xib            +\frac 1 4  f\fb \kab +f\fb\omb -\fb^2\om+\lot \\
&=&  -\frac{1}{4}\fb^2 e_\th'(f)+\fb(-\ze+\etab)    +f\xib            +\frac 1 4  f\fb \kab +f\fb\omb -\fb^2\om+\lot
\eeaa

Next, we derive the transformation formula for $ \ze$. We have,  under a transformation
 of type   \eqref{SSMe:general.composite}, 
  \beaa
2\ze'&=& g(D_{e_\th'} e_4', e_3')\\
&=& g\left(D_{e_\th'}\left(\la\left(e_4 + f e_\th +\frac 1 4 f^2  e_3\right)\right), e_3'\right)\\
&=& -2e_\th'(\log(\la))+\la g\left(D_{e_\th'}\left(e_4 + f e_\th +\frac 1 4 f^2  e_3\right), e_3'\right)\\
&=& -2e_\th'(\log(\la)) +\la e_\th'(f)g\left(e_\th, e_3'\right)+\frac 1 4 \la e_\th'(f^2)g\left(e_3, e_3'\right)+\la g\left(D_{e_\th'}e_4, e_3'\right) \\
&&+\la f g\left(D_{e_\th'}e_\th, e_3'\right)+\frac 1 4 \la f^2 g\left(D_{e_\th'}e_3, e_3'\right)\\
&=& -2e_\th'(\log(\la)) +\fb\left(1+\frac 1 4 f\fb\right)e_\th'(f) -\frac 1 8  \fb^2 e_\th'(f^2)+\la g\left(D_{e_\th'}e_4, e_3'\right) +\la f g\left(D_{e_\th'}e_\th, e_3'\right)+\lot
\eeaa
We compute 
\beaa
\la g\left(D_{e_\th'}e_4, e_3'\right) &=& g\left(D_{e_\th'}e_4, e_3 +\fb e_\th\right)+\lot\\
&=& g\left(D_{ e_\th  + \frac 1 2  \fb e_4+\frac 1 2 f e_3}e_4, e_3\right) + \fb g\left(D_{e_\th  + \frac 1 2  \fb e_4+\frac 1 2 f e_3}e_4, e_\th\right)+\lot\\
&=& 2\ze + 2\om\fb -2\omb f + \fb \chi   +\lot
\eeaa
and 
\beaa
\la f g\left(D_{e_\th'}e_\th, e_3'\right) &=&  f g\left(D_{e_\th'}e_\th, e_3\right)+\lot\\
&=& f g\left(D_{e_\th  + \frac 1 2  \fb e_4+\frac 1 2 fe_3}e_\th, e_3\right)+\lot\\
&=& -f\chib  +\lot
\eeaa
This yields
  \beaa
2\ze' &=& -2e_\th'(\log(\la)) +\fb\left(1+\frac 1 4 f\fb\right)e_\th'(f) -\frac 1 8  \fb^2 e_\th'(f^2)+\la g\left(D_{e_\th'}e_4, e_3'\right) +\la f g\left(D_{e_\th'}e_\th, e_3'\right)+\lot\\
&=& 2\ze -2e_\th'(\log(\la)) +\fb\left(1+\frac 1 4 f\fb\right)e_\th'(f) -\frac 1 8  \fb^2 e_\th'(f^2) + 2\om\fb -2\omb f + \fb \chi  -f\chib +\lot\\
\eeaa
and hence
  \beaa
\ze' &=& \ze - e_\th'(\log(\la))  +\frac 14 (- f\kab +\fb \ka ) + \fb \om- f  \omb+\err(\ze,\ze'), \\
\err(\ze,\ze')&=& \frac{1}{2}\fb\left(1+\frac 1 4 f\fb\right)e_\th'(f) -\frac{1}{16}\fb^2 e_\th'(f^2) +\frac 1 4( - f \vthb +\fb \vth)+\lot\\
&=& \frac{1}{2}\fb e_\th'(f)  +\frac 1 4( - f \vthb +\fb \vth)+\lot
\eeaa

Next, we derive the transformation formulae for $\eta$. We have,  under a transformation
 of type   \eqref{SSMe:general.composite}, 
\beaa
  2 \eta'&=& g\left(D_{e_3'} e_4', e_\th'\right)\\
  &=& g\left(D_{e_3'}\left(\la\left(e_4 + f e_\th +\frac 1 4 f^2  e_3\right)\right), e_\th'\right)\\
   &=& \la g\left(D_{e_3'}\left(e_4 + f e_\th +\frac 1 4 f^2  e_3\right), e_\th'\right)\\
   &=& \la g\left(D_{e_3'}e_4, e_\th'\right) + \la e_3'(f)g\left(e_\th, e_\th'\right)  + \la f g\left(D_{e_3'}e_\th, e_\th'\right)+\frac 1 4 \la e_3'(f^2)g\left(e_3, e_\th'\right) +\frac{1}{4}\la f^2g(D_{e_3'}e_3, e_\th')  \\
    &=&  \la\left(1+\frac 1 2   f \fb\right) e_3'(f) -\frac 1 4\la\fb e_3'(f^2)+ \la g\left(D_{e_3'}e_4, e_\th'\right)+ \la f g\left(D_{e_3'}e_\th, e_\th'\right) +\lot
\eeaa
We compute 
\beaa
\la g\left(D_{e_3'}e_4, e_\th'\right) &=& \la g\left(D_{e_3'}e_4,  e_\th +\frac 1 2 fe_3\right)+\lot\\
&=& g\left(D_{e_3 +\fb e_\th }e_4,  e_\th\right)+\frac 1 2f g\left(D_{ e_3}e_4,  e_3\right)+\lot\\
&=& 2\eta+\fb\chi -2\omb f +\lot
\eeaa
and
\beaa
\la f g\left(D_{e_3'}e_\th, e_\th'\right) &=& \lot
\eeaa
This yields
\beaa
  2 \eta'   &=&  \la\left(1+\frac 1 2   f \fb\right) e_3'(f) -\frac 1 4\la\fb e_3'(f^2)+ \la g\left(D_{e_3'}e_4, e_\th'\right)+ \la f g\left(D_{e_3'}e_\th, e_\th'\right) +\lot\\
  &=& \la\left(1+\frac 1 2   f \fb\right) e_3'(f) -\frac 1 4\la\fb e_3'(f^2)+ 2\eta+\fb\chi -2\omb f  +\lot
\eeaa
and hence
     \beaa
     \eta'&=& \eta +\frac{1}{2}\la e_3'(f)     +\frac 1 4 \ka \fb   -f\omb +\err(\eta, \eta'),\\
     \err(\eta, \eta')&=& \frac{1}{4}\la f\fb e_3'(f) -\frac 1 8\la\fb e_3'(f^2) +\frac{1}{4}\fb\vth+\lot \\
     &=&  \frac{1}{4}\fb\vth+\lot
     \eeaa

Next, we derive the transformation formulae for $\etab$. We have,  under a transformation
 of type   \eqref{SSMe:general.composite}, 
\beaa
  2 \etab'&=& g\left(D_{e_4'} e_3', e_\th'\right)\\
  &=& g\left(D_{e_4'}\left(\la^{-1} \left( \left(1+\frac 12  f \fb +\frac{1}{16} f^2 \fb^2\right)  e_3 +\fb\left(1+\frac 1 4 f\fb\right) e_\th + \frac 1 4 \fb^2  e_4\right)\right), e_\th'\right)\\
   &=& \la^{-1}g\left(D_{e_4'}\left( \left(1+\frac 12  f \fb +\frac{1}{16} f^2 \fb^2\right)  e_3 +\fb\left(1+\frac 1 4 f\fb\right) e_\th + \frac 1 4 \fb^2  e_4\right), e_\th'\right)\\
&=& \frac 12\la^{-1}e_4'\left(f \fb +\frac{1}{8} f^2 \fb^2\right)g\left(  e_3, e_\th'\right)+\la^{-1}\left(1+\frac 12  f \fb +\frac{1}{16} f^2 \fb^2\right)g\left(D_{e_4'}e_3, e_\th'\right)\\
&& +  \la^{-1}e_4'\left( \fb\left(1+\frac 1 4 f\fb\right)\right) g\left( e_\th, e_\th'\right)  +  \la^{-1}\fb\left(1+\frac 1 4 f\fb\right)g\left(D_{e_4'}e_\th, e_\th'\right)\\
&&   +     \frac 1 4 \la^{-1}e_4'(\fb^2)g\left(e_4, e_\th'\right) +   \frac 1 4 \la^{-1}\fb^2 g\left(D_{e_4'}e_4, e_\th'\right)\\
&=& -\frac 12\la^{-1}\fb e_4'\left(f \fb +\frac{1}{8} f^2 \fb^2\right) +  \la^{-1}\left(1+\frac 1 2   f \fb\right)e_4'\left( \fb\left(1+\frac 1 4 f\fb\right)\right)\\
&& -     \frac 1 4 \la^{-1}f\left(1+ \frac 1 4 f \fb\right) e_4'(\fb^2)+\la^{-1}g\left(D_{e_4'}e_3, e_\th'\right)  +  \la^{-1}\fb g\left(D_{e_4'}e_\th, e_\th'\right)+\lot
\eeaa
We compute
\beaa
\la^{-1}g\left(D_{e_4'}e_3, e_\th'\right) &=& \la^{-1}g\left(D_{e_4'}e_3, e_\th  + \frac 1 2  \fb e_4\right) +\lot\\
&=& 2\etab +f \chib -2\fb\om +\lot
\eeaa
and
\beaa
 \la^{-1}\fb g\left(D_{e_4'}e_\th, e_\th'\right) &=& \lot
\eeaa
This yields
\beaa
  2 \etab' &=& -\frac 12\la^{-1}\fb e_4'\left(f \fb +\frac{1}{8} f^2 \fb^2\right) +  \la^{-1}\left(1+\frac 1 2   f \fb\right)e_4'\left( \fb\left(1+\frac 1 4 f\fb\right)\right)\\
&& -     \frac 1 4 \la^{-1}f\left(1+ \frac 1 4 f \fb\right) e_4'(\fb^2)+\la^{-1}g\left(D_{e_4'}e_3, e_\th'\right)  +  \la^{-1}\fb g\left(D_{e_4'}e_\th, e_\th'\right)+\lot\\
&=&  -\frac 12\la^{-1}\fb e_4'\left(f \fb +\frac{1}{8} f^2 \fb^2\right) +  \la^{-1}\left(1+\frac 1 2   f \fb\right)e_4'\left( \fb\left(1+\frac 1 4 f\fb\right)\right)\\
&& -     \frac 1 4 \la^{-1}f\left(1+ \frac 1 4 f \fb\right) e_4'(\fb^2) +2\etab +f \chib -2\fb\om +\lot
\eeaa
and hence
    \beaa
     \etab'&=& \etab + \frac{1}{2} \la^{-1}e_4'(\fb)     +\frac 1 4 \kab f   -\fb\om +\err(\etab, \etab'),\\
     \err(\etab, \etab')&=&   \frac{1}{4} \la^{-1}f \fb e_4'(\fb)   -\frac 14\la^{-1}\fb e_4'\left(f \fb +\frac{1}{8} f^2 \fb^2\right)   +  \la^{-1}\frac{1}{8}\left(1+\frac 1 2   f \fb\right)e_4'\left(f\fb^2\right)\\
&& -     \frac 1 8 \la^{-1}f\left(1+ \frac 1 4 f \fb\right) e_4'(\fb^2) +\frac{1}{4}f\vthb+\lot \\
&=&  - \frac{1}{8}\fb^2\la^{-1}e_4'(f) +\frac{1}{4}f\vthb+\lot
     \eeaa

Next, we derive the transformation formulae for $\xi$. We have,  under a transformation
 of type   \eqref{SSMe:general.composite}, 
\beaa
  2 \xi'&=& g\left(D_{e_4'} e_4', e_\th'\right)\\
  &=& g\left(D_{e_4'}\left(\la\left(e_4 + f e_\th +\frac 1 4 f^2  e_3\right)\right), e_\th'\right)\\
   &=& \la g\left(D_{e_4'}\left(e_4 + f e_\th +\frac 1 4 f^2  e_3\right), e_\th'\right)\\
     &=& \la g\left(D_{e_4'}e_4, e_\th'\right) +\la e_4'(f) g\left(e_\th, e_\th'\right) +\la f g\left(D_{e_4'}e_\th, e_\th'\right) +\frac 1 4\la e_4'(f^2) g\left(  e_3, e_\th'\right)  +\frac 1 4\la f^2g\left(D_{e_4'}e_3, e_\th'\right)\\
        &=&  \la \left(1+\frac 1 2   f \fb\right)e_4'(f)  -\frac 1 4\la  \fb e_4'(f^2)   +\la g\left(D_{e_4'}e_4, e_\th'\right) +\la f g\left(D_{e_4'}e_\th, e_\th'\right) +\lot
  \eeaa
    We compute
  \beaa
  \la g\left(D_{e_4'}e_4, e_\th'\right) &=& \la g\left(D_{e_4'}e_4, e_\th +\frac 1 2 f e_3\right)+\lot\\
  &=& \la^2g\left(D_{e_4 + f e_\th}e_4, e_\th \right)+\frac 1 2\la^2f g\left(D_{e_4}e_4,  e_3\right)+\lot\\
  &=& 2\la^2\xi  +\la^2f\chi  +2\la^2f\om+\lot
  \eeaa
  and 
  \beaa
  \la f g\left(D_{e_4'}e_\th, e_\th'\right) &=& \lot
  \eeaa
  This yields
\beaa
  2 \xi'  &=&  \la \left(1+\frac 1 2   f \fb\right)e_4'(f)  -\frac 1 4\la  \fb e_4'(f^2)   +\la g\left(D_{e_4'}e_4, e_\th'\right) +\la f g\left(D_{e_4'}e_\th, e_\th'\right) +\lot\\
  &=& \la \left(1+\frac 1 2   f \fb\right)e_4'(f)  -\frac 1 4\la  \fb e_4'(f^2)   +2\la^2\xi  +\la^2f\chi  +2\la^2f\om +\lot
  \eeaa
and hence
 \beaa
\xi' &=& \la^2\left(\xi+\frac{1}{2}\la^{-1}e_4'(f)+\om f + \frac{1}{4}f\ka\right)+\la^2\err(\xi, \xi'),\\
\err(\xi, \xi')&=& \frac 1 4  \la^{-1} f \fb e_4'(f)  -\frac 1 8\la^{-1}  \fb e_4'(f^2)+ \frac{1}{4}f\vth+\lot \\
&=& \frac{1}{4}f\vth+\lot
\eeaa  
In the particular case when $\la=1, \fb=0$, see Remark \ref{remark:lotxixi'},  the error term  takes the form,
\beaa
 \err(\xi, \xi')&=& \frac{1}{4}f\vth +\frac 1 4  f^2\left( \eta+2\ze-\etab \right)-\frac 1 4 f^3 \left(\omb +\frac 1 2 \chib \right)-\frac{3}{32} f^4 \xib.
\eeaa

Next, we derive the transformation formulae for $\xib$. We have,  under a transformation
 of type   \eqref{SSMe:general.composite}, 
\beaa
  2 \xib'&=& g\left(D_{e_3'} e_3', e_\th'\right)\\
  &=& g\left(D_{e_3'}\left(\la^{-1} \left( \left(1+\frac 12  f \fb +\frac{1}{16} f^2 \fb^2\right)  e_3 +\fb\left(1+\frac 1 4 f\fb\right) e_\th + \frac 1 4 \fb^2  e_4\right)\right), e_\th'\right)\\
  &=& \la^{-1}g\left(D_{e_3'}\left( \left(1+\frac 12  f \fb +\frac{1}{16} f^2 \fb^2\right)  e_3 +\fb\left(1+\frac 1 4 f\fb\right) e_\th + \frac 1 4 \fb^2  e_4\right), e_\th'\right)\\
    &=& \frac 12\la^{-1}e_3' \left(  f \fb +\frac{1}{8} f^2 \fb^2\right) g\left(e_3, e_\th'\right)  +\la^{-1}g\left(D_{e_3'}e_3, e_\th'\right)\\
&&  +\la^{-1}e_3'\left( \fb\left(1+\frac 1 4 f\fb\right)\right) g\left(e_\th, e_\th'\right) +\la^{-1}\fb g\left(D_{e_3'}e_\th, e_\th'\right)   +\frac 1 4\la^{-1}e_3'(\fb^2)g\left( e_4, e_\th'\right)   +\lot\\
&=& -\frac 12\la^{-1}\fb e_3' \left(  f \fb +\frac{1}{8} f^2 \fb^2\right)   +\la^{-1}\left(1+\frac 1 2   f \fb\right)e_3'\left( \fb\left(1+\frac 1 4 f\fb\right)\right)\\
&&-\frac 1 4\la^{-1} f\left(1+ \frac 1 4 f \fb\right)e_3'(\fb^2)  +\la^{-1}g\left(D_{e_3'}e_3, e_\th'\right)   +\la^{-1}\fb g\left(D_{e_3'}e_\th, e_\th'\right)     +\lot
\eeaa
   We compute
  \beaa
  \la^{-1}g\left(D_{e_3'}e_3, e_\th'\right) &=& \la^{-1}g\left(D_{e_3'}e_3, e_\th  + \frac 1 2  \fb e_4\right)+\lot\\
  &=& \la^{-2}g\left(D_{   e_3 +\fb e_\th }e_3, e_\th \right) +\frac 1 2  \la^{-2}\fb g\left(D_{e_3}e_3,   e_4\right) +\lot\\
  &=& 2\la^{-2}\xib  +\la^{-2}\fb\,\chib  +2  \la^{-2}\fb\,\omb +\lot\\
  \eeaa
  and 
  \beaa
  \la^{-1}\fb g\left(D_{e_3'}e_\th, e_\th'\right) &=& \lot
  \eeaa
  This yields
\beaa
  2 \xib' &=& -\frac 12\la^{-1}\fb e_3' \left(  f \fb +\frac{1}{8} f^2 \fb^2\right)   +\la^{-1}\left(1+\frac 1 2   f \fb\right)e_3'\left( \fb\left(1+\frac 1 4 f\fb\right)\right)\\
&&-\frac 1 4\la^{-1} f\left(1+ \frac 1 4 f \fb\right)e_3'(\fb^2)  +\la^{-1}g\left(D_{e_3'}e_3, e_\th'\right)   +\la^{-1}\fb g\left(D_{e_3'}e_\th, e_\th'\right)     +\lot\\
&=&  -\frac 12\la^{-1}\fb e_3' \left(  f \fb +\frac{1}{8} f^2 \fb^2\right)   +\la^{-1}\left(1+\frac 1 2   f \fb\right)e_3'\left( \fb\left(1+\frac 1 4 f\fb\right)\right)\\
&&-\frac 1 4\la^{-1} f\left(1+ \frac 1 4 f \fb\right)e_3'(\fb^2)  +2\la^{-2}\xib  +\la^{-2}\fb\,\chib  +2  \la^{-2}\fb\,\omb    +\lot
\eeaa
and hence
 \beaa
\xib' &=& \la^{-2}\left(\xib+\frac{1}{2}\la e_3'(\fb)+\omb\,\fb + \frac{1}{4}\fb\,\kab\right)+\la^{-2}\err(\xib, \xib'),\\
\err(\xib, \xib')&=&  -\frac 14\la\fb e_3' \left(  f \fb +\frac{1}{8} f^2 \fb^2\right)   +\frac 1 4 \la  f \fb e_3'(\fb) +\frac 1 8\la\left(1+\frac 1 2   f \fb\right)e_3'\left(f\fb^2\right)\\
&&-\frac 1 8\la f\left(1+ \frac 1 4 f \fb\right)e_3'(\fb^2)+ \frac{1}{4}\fb\,\vthb+\lot \\
&=& -\frac 18\la\fb^2 e_3'(f)+ \frac{1}{4}\fb\,\vthb+\lot
\eeaa  

Next, we derive the transformation formulae for $\om$. We have,  under a transformation
 of type   \eqref{SSMe:general.composite}, 
\beaa
  4 \om'&=& g\left(D_{e_4'} e_4', e_3'\right)\\
  &=& g\left(D_{e_4'}\left(\la\left(e_4 + f e_\th +\frac 1 4 f^2  e_3\right)\right), e_3'\right)\\
   &=& -2e_4'(\log\la)   +\la g\left(D_{e_4'}\left(e_4 + f e_\th +\frac 1 4 f^2  e_3\right), e_3'\right)\\
     &=& -2e_4'(\log\la)   +\la g\left(D_{e_4'}e_4, e_3'\right)  +\la e_4'(f) g\left( e_\th, e_3'\right) +\la fg\left(D_{e_4'}e_\th, e_3'\right) \\
     && +\frac 1 4\la e_4'(f^2)g\left(e_3, e_3'\right)  +\frac 1 4\la f^2g\left(D_{e_4'}e_3, e_3'\right)\\
         &=& -2e_4'(\log\la)  + \fb\left(1+\frac 1 4 f\fb\right) e_4'(f) -\frac 1 8\fb^2 e_4'(f^2)+\la g\left(D_{e_4'}e_4, e_3'\right)   +\la fg\left(D_{e_4'}e_\th, e_3'\right)   +\lot
\eeaa
  We compute
  \beaa
 \la g\left(D_{e_4'}e_4, e_3'\right) &=&  g\left(D_{e_4'}e_4,  \left(1+\frac 12  f \fb \right)  e_3 +\fb e_\th \right)+\lot\\
 &=& \la\left(1+\frac 12  f \fb \right) g\left(D_{e_4 + f e_\th+\frac{1}{4}f^2e_3}e_4,    e_3  \right) + \la\fb g\left(D_{e_4 + f e_\th}e_4,  e_\th \right)+\lot\\
  &=& 4\la\left(1+\frac 12  f \fb \right)\om +2\la f\ze -\la f^2\omb + 2\la\fb\xi   + \la f\fb\chi +\lot
  \eeaa
  and 
  \beaa
  \la fg\left(D_{e_4'}e_\th, e_3'\right) &=&  fg\left(D_{e_4'}e_\th,    e_3 \right)+\lot\\
  &=& \la fg\left(D_{e_4 + f e_\th }e_\th,    e_3 \right)+\lot\\
  &=& -2\la f\etab -   \la f^2\chib +\lot
  \eeaa
  This yields
\beaa
  4 \om'     &=& -2e_4'(\log\la)  + \fb\left(1+\frac 1 4 f\fb\right) e_4'(f) -\frac 1 8\fb^2 e_4'(f^2)+\la g\left(D_{e_4'}e_4, e_3'\right)   +\la fg\left(D_{e_4'}e_\th, e_3'\right)   +\lot\\
  &=& -2e_4'(\log\la)  + \fb\left(1+\frac 1 4 f\fb\right) e_4'(f) -\frac 1 8\fb^2 e_4'(f^2)+4\la\left(1+\frac 12  f \fb \right)\om +2\la f\ze -\la f^2\omb\\
  && + 2\la\fb\xi   + \la f\fb\chi   -2\la f\etab -   \la f^2\chib  +\lot
\eeaa
and hence
 \beaa
\om' &=& \la\left(\om -\frac{1}{2}\la^{-1}e_4'(\log(\la))\right)+\la\err(\om, \om'),\\
\err(\om, \om')&=& \frac{1}{4}\fb\left(1+\frac 1 4 f\fb\right) e_4'(f) -\frac{1}{32}\fb^2 e_4'(f^2) +\frac{1}{2}\om f\fb - \frac{1}{2}f\etab +\frac{1}{2}\fb\xi +\frac{1}{2}f\ze\\
&& -\frac{1}{8}\kab f^2+\frac{1}{8}f\fb \ka-\frac{1}{4}\omb f^2+\lot \\
&=&  \frac{1}{4}\fb e_4'(f)  +\frac{1}{2}\om f\fb - \frac{1}{2}f\etab +\frac{1}{2}\fb\xi +\frac{1}{2}f\ze -\frac{1}{8}\kab f^2+\frac{1}{8}f\fb \ka-\frac{1}{4}\omb f^2+\lot
\eeaa
In the particular case,  see Remark \ref{remark:lotxixi'},   when $\la=1, \fb=0$ we  have the more  precise formula,
\beaa
\om'&=&\om+\frac 1 2 f(\ze -\etab) -\frac 1 8 f^2\left(2\omb+\kab+\vthb  + f  \xib\right)
\eeaa

Next, we derive the transformation formulae for $\omb$. We have,  under a transformation
 of type   \eqref{SSMe:general.composite}, 
\beaa
  4 \omb'&=& g\left(D_{e_3'} e_3', e_4'\right)\\
  &=& g\left(D_{e_3'}\left(\la^{-1} \left( \left(1+\frac 12  f \fb +\frac{1}{16} f^2 \fb^2\right)  e_3 +\fb\left(1+\frac 1 4 f\fb\right) e_\th + \frac 1 4 \fb^2  e_4\right)\right), e_4'\right)\\
   &=& 2e_3'(\log(\la))\\
     && +\la^{-1}g\left(D_{e_3'}\left( \left(1+\frac 12  f \fb +\frac{1}{16} f^2 \fb^2\right)  e_3 +\fb\left(1+\frac 1 4 f\fb\right) e_\th + \frac 1 4 \fb^2  e_4\right), e_4'\right)\\
     &=& 2e_3'(\log(\la)) +\frac 12\la^{-1}e_3'\left(f \fb +\frac{1}{8} f^2 \fb^2\right)g\left(  e_3, e_4'\right) +\la^{-1}\left(1+\frac 12  f \fb\right) g\left(D_{e_3'}e_3, e_4'\right)\\ 
    && +\la^{-1}e_3'\left(\fb\left(1+\frac 1 4 f\fb\right)\right)g\left( e_\th, e_4'\right) +\la^{-1}\fb g\left(D_{e_3'}e_\th, e_4'\right) + \frac 1 4\la^{-1}e_3'(\fb^2)g\left(e_4, e_4'\right)+\lot\\
     &=& 2e_3'(\log(\la)) - e_3'\left(f \fb +\frac{1}{8} f^2 \fb^2\right)+fe_3'\left(\fb\left(1+\frac 1 4 f\fb\right)\right) - \frac{1}{8}f^2e_3'(\fb^2)\\ 
    &&  +\la^{-1}\left(1+\frac 12  f \fb\right) g\left(D_{e_3'}e_3, e_4'\right) +\la^{-1}\fb g\left(D_{e_3'}e_\th, e_4'\right) +\lot
\eeaa
  We compute
  \beaa
 \la^{-1}\left(1+\frac 12  f \fb\right) g\left(D_{e_3'}e_3, e_4'\right) &=& \left(1+\frac 12  f \fb\right) g\left(D_{e_3'}e_3, e_4+fe_\th\right)\\
 &=& \la^{-1}\left(1+\frac 12  f \fb\right) g\left(D_{\left(1+\frac 12  f \fb\right)  e_3 +\fb e_\th + \frac 1 4 \fb^2  e_4}e_3, e_4+fe_\th\right)+\lot\\
  &=& 4\la^{-1}\left(1+\frac 12  f \fb\right)\omb -2\la^{-1}\fb\ze-\la^{-1}\fb^2\om +2\la^{-1}f\fb\omb \\
  &&+ 2\la^{-1}f\xib + \la^{-1}f\fb\,\chib +\lot
 \eeaa
  and 
  \beaa
  \la^{-1}\fb g\left(D_{e_3'}e_\th, e_4'\right) &=& \fb g\left(D_{e_3'}e_\th, e_4\right)+\lot\\
  &=& \la^{-1}\fb g\left(D_{e_3 +\fb e_\th}e_\th, e_4\right)+\lot\\
  &=& -2\la^{-1}\fb\eta -  \la^{-1}\fb^2\chi+\lot
  \eeaa
  This yields
\beaa
  4 \omb'     &=& 2e_3'(\log(\la)) - e_3'\left(f \fb +\frac{1}{8} f^2 \fb^2\right)+fe_3'\left(\fb\left(1+\frac 1 4 f\fb\right)\right) - \frac{1}{8}f^2e_3'(\fb^2)\\ 
    &&  +4\la^{-1}\left(1+\frac 12  f \fb\right)\omb -2\la^{-1}\fb\ze-\la^{-1}\fb^2\om +2\la^{-1}f\fb\omb \\
  &&+ 2\la^{-1}f\xib + \la^{-1}f\fb\,\chib -2\la^{-1}\fb\eta -  \la^{-1}\fb^2\chi +\lot
\eeaa
and hence
 \beaa
\omb' &=& \la^{-1}\left(\omb +\frac{1}{2}\la e_3'(\log(\la))\right)+\la^{-1}\err(\omb, \omb'),\\
\err(\omb, \omb')&=& - \frac{1}{4}e_3'\left(f \fb +\frac{1}{8} f^2 \fb^2\right)+\frac{1}{4}fe_3'\left(\fb\left(1+\frac 1 4 f\fb\right)\right) - \frac{1}{32}f^2e_3'(\fb^2)\\
&& +\omb f\fb - \frac{1}{2}\fb\eta +\frac{1}{2}f\xib -\frac{1}{2}\fb\ze -\frac{1}{8}\ka\fb^2+\frac{1}{8}f\fb \kab-\frac{1}{4}\om\fb^2+\lot \\
&=&  - \frac{1}{4}\fb e_3'(f)  +\omb f\fb - \frac{1}{2}\fb\eta +\frac{1}{2}f\xib -\frac{1}{2}\fb\ze -\frac{1}{8}\ka\fb^2+\frac{1}{8}f\fb \kab-\frac{1}{4}\om\fb^2+\lot 
\eeaa

Next we derive the formula for $\a$. We have
\beaa
\a' &=& R(e_4', e_4')=\la^2R\left(e_4 + f e_\th +\frac 1 4 f^2  e_3, e_4 + f e_\th +\frac 1 4 f^2  e_3\right)\\
&=& \la^2\left(R_{44}+2fR_{4\th}+f^2R_{\th\th}+\frac{1}{2}f^2R_{34}\right)+\lot\\
&=& \la^2\left(\a+2f\b+\frac{3}{2}f^2\rho\right)+\lot
\eeaa 
and hence
\beaa
\a' &=& \la^2\a+\la^2\err(\a, \a'),\\
\err(\a, \a') &=& 2f\b+\frac{3}{2}f^2\rho+\lot
\eeaa 
  The formula  for $\aa$ is easily derived by symmetry from the one on $\a$.

Next we derive the formula for $\b$. We have
\beaa 
\b' &=& R(e_4', e_\th') = \la R\left(e_4 + f e_\th +\frac 1 4 f^2  e_3, \left(1+\frac 1 2   f \fb\right) e_\th  + \frac 1 2(   \fb e_4+ f e_3)\right)+\lot\\
&=& \la\left(R_{4\th}+fR_{\th\th}+\frac{1}{2}\fb R_{44}+\frac{1}{2}fR_{43}\right)+\lot\\\
&=& \la\left(\b+\frac{3}{2}f\rho+\frac{1}{2}\fb\a\right)+\lot
\eeaa
and hence 
\beaa 
\b' &=& \la\left(\b+\frac{3}{2}f\rho\right)+\la\err(\b,\b'),\\
\err(\b,\b') &=& \frac{1}{2}\fb\a+\lot
\eeaa 
 The formula  for $\bb$ is easily derived by symmetry from the one on $\b$. 
 
 Finally, we derive the formula for $\rho$. We have
 \beaa
 \rho' &=& R(e_4', e_3')= R\left(e_4 + f e_\th +\frac 1 4 f^2  e_3, \left(1+\frac 12  f \fb \right) e_3+ \fb e_\th +\frac 1 4  \fb^2 e_4\right)+\lot\\
 &=& R_{43}+\frac{1}{2}f\fb R_{43}+\fb R_{4\th}  +fR_{\th 3}+f\fb R_{\th\th}+\lot\\
 &=& \rho+\frac{3}{2}\rho f\fb+\fb\b  +f\bb+\lot
 \eeaa
 and hence 
\beaa 
\rho' &=& \rho+\err(\rho,\rho'),\\
\err(\rho,\rho') &=& \frac{3}{2}\rho f\fb+\fb\b  +f\bb+\lot
\eeaa 
 This concludes the proof of Proposition \ref{prop:transformations1}.


\section{Proof of Lemma \ref{lemma:transportequationsforffbandlambda}}\lab{sec:proofoflemma:transportequationsforffbandlambda}


For $\xi'$ and $\om'$, we need more precise transformation formula than the ones of Proposition \ref{prop:transformations1}. 
We have
\beaa
2\xi' &=& g(D_{e_4'}e_4', e_\th')\\
&=& \la^2g(D_{\la^{-1}e_4'}(\la^{-1}e_4'), e_\th')\\
&=& \la^2g\left(D_{\la^{-1}e_4'}(\la^{-1}e_4'),  e_\th  +\frac 1 2 fe_3\right)+\la^2\frac{\fb}{2}g\left(D_{\la^{-1}e_4'}(\la^{-1}e_4'), e_4+ fe_\th  + \frac 1 4 f^2 e_3\right)\\
&=& \la^2g\left(D_{\la^{-1}e_4'}(\la^{-1}e_4'),  e_\th  +\frac 1 2 fe_3\right)+\la^2\frac{\fb}{2}g\left(D_{\la^{-1}e_4'}(\la^{-1}e_4'), \la^{-1}e_4'\right)\\
&=& \la^2g\left(D_{\la^{-1}e_4'}(\la^{-1}e_4'),  e_\th  +\frac 1 2 fe_3\right).
\eeaa

Also, we have
\beaa
4\om' &=&  g(D_{e_4'}e_4', e_3')\\
&=& -2e_4'(\log(\la))+\la g(D_{\la^{-1}e_4'}(\la^{-1}e_4'), \la e_3')\\
&=& -2e_4'(\log(\la))+\la g(D_{\la^{-1}e_4'}(\la^{-1}e_4'),  e_3)+\la\fb g\left(D_{\la^{-1}e_4'}(\la^{-1}e_4'), e_\th+\frac 12  f  e_3\right)\\
&&+\frac{1}{4}\la\fb^2 g\left(D_{\la^{-1}e_4'}(\la^{-1}e_4'), e_4+ f e_\th+\frac{1}{4} f^2  e_3\right)\\
&=& -2e_4'(\log(\la))+\la g(D_{\la^{-1}e_4'}(\la^{-1}e_4'),  e_3)+\la\fb g\left(D_{\la^{-1}e_4'}(\la^{-1}e_4'), e_\th+\frac 12  f  e_3\right)\\
&&+\frac{1}{4}\la\fb^2 g\left(D_{\la^{-1}e_4'}(\la^{-1}e_4'), \la^{-1}e_4'\right)\\
&=& -2e_4'(\log(\la))+\la g(D_{\la^{-1}e_4'}(\la^{-1}e_4'),  e_3)+\la\fb g\left(D_{\la^{-1}e_4'}(\la^{-1}e_4'), e_\th+\frac 12  f  e_3\right).
\eeaa
In view of the change of frame formula for $\xi'$, we infer
\beaa
4\om' &=& -2e_4'(\log(\la))+\la g(D_{\la^{-1}e_4'}(\la^{-1}e_4'),  e_3)+\la^{-1}\fb \xi'.
\eeaa

Next, we compute 
\beaa
g\left(D_{\la^{-1}e_4'}(\la^{-1}e_4'),  e_\th\right) &=& g\left(D_{\la^{-1}e_4'}\left(e_4 + f e_\th +\frac 1 4 f^2  e_3\right),  e_\th\right)\\
&=& g\left(D_{\la^{-1}e_4'}e_4,  e_\th\right) +\la^{-1}e_4'(f) +\frac 1 4 f^2g\left(D_{\la^{-1}e_4'}e_3,  e_\th\right)\\
&=& g\left(D_{e_4 + f e_\th +\frac 1 4 f^2  e_3}e_4,  e_\th\right) +\left(e_4 + f e_\th +\frac 1 4 f^2  e_3\right)f \\
&&+\frac 1 4 f^2g\left(D_{e_4 + f e_\th +\frac 1 4 f^2  e_3}e_3,  e_\th\right)\\
&=& 2\xi+f\chi+\frac{1}{2}f^2\eta+\left(e_4 + f e_\th +\frac 1 4 f^2  e_3\right)f+\frac 1 2 f^2\etab+\frac 1 4 f^3\chib\\
&&+\frac{1}{8}f^4\xib.
\eeaa

Also, we have
\beaa
g\left(D_{\la^{-1}e_4'}(\la^{-1}e_4'),  e_3\right) &=& g\left(D_{\la^{-1}e_4'}\left(e_4 + f e_\th +\frac 1 4 f^2  e_3\right),  e_3\right)\\
 &=& g\left(D_{\la^{-1}e_4'}e_4,  e_3\right) +fg\left(D_{\la^{-1}e_4'}e_\th, e_3\right)\\
 &=& g\left(D_{e_4 + f e_\th +\frac 1 4 f^2  e_3}e_4,  e_3\right) +fg\left(D_{e_4 + f e_\th +\frac 1 4 f^2  e_3}e_\th, e_3\right)\\
 &=& 4\om+2f\ze- f^2\omb -2\etab f -f^2\chib-\frac{1}{2}f^3\xib.
 \eeaa

We deduce
\beaa
2\xi' &=& \la^2g\left(D_{\la^{-1}e_4'}(\la^{-1}e_4'),  e_\th  +\frac 1 2 fe_3\right)\\
&=& \la^2\Bigg\{2\xi+f\chi+\frac{1}{2}f^2\eta+\left(e_4 + f e_\th +\frac 1 4 f^2  e_3\right)f+\frac 1 2 f^2\etab+\frac 1 4 f^3\chib\\
&&+\frac{1}{8}f^4\xib + \frac{1}{2}f\left(4\om+2f\ze- f^2\omb -2\etab f -f^2\chib-\frac{1}{2}f^3\xib\right)\Bigg\}\\
&=& \la^2\Bigg\{2\xi+\left(e_4 + f e_\th +\frac 1 4 f^2  e_3\right)f+f\chi+2f\om +\frac{1}{2}f^2\eta-\frac 1 2 f^2\etab+f^2\ze-\frac 1 4 f^3\chib\\
&&-\frac{1}{2}f^3\omb-\frac{1}{8}f^4\xib\Bigg\}
\eeaa
and
\beaa
4\om' &=& -2e_4'(\log(\la))+\la g(D_{\la^{-1}e_4'}(\la^{-1}e_4'),  e_3)+\la^{-1}\fb \xi'\\
&=& \la\Bigg\{4\om-2\left(e_4 + f e_\th +\frac 1 4 f^2  e_3\right)\log(\la)+2f\ze- f^2\omb -2\etab f -f^2\chib-\frac{1}{2}f^3\xib\Bigg\}\\
&&+\la^{-1}\fb \xi'.
\eeaa

If $\xi'=0$, we infer
\beaa
2\xi+\left(e_4 + f e_\th +\frac 1 4 f^2  e_3\right)f+f\chi+2f\om +\frac{1}{2}f^2\eta-\frac 1 2 f^2\etab+f^2\ze-\frac 1 4 f^3\chib\\
-\frac{1}{2}f^3\omb-\frac{1}{8}f^4\xib &=& 0
\eeaa
and hence
\beaa
\la^{-1}e_4'(f)+\left(\frac{\ka}{2}+2\om\right) f &=& -2\xi -\frac{1}{2}\vth f -\frac{1}{2}f^2\eta+\frac 1 2 f^2\etab-f^2\ze+\frac 1 8 f^3\kab +\frac{1}{2}f^3\omb\\ 
&&+\frac 1 8 f^3\vthb+\frac{1}{8}f^4\xib
\eeaa
which yields the desired transport equation for $f$
\beaa
\la^{-1}e_4'(f)+\left(\frac{\ka}{2}+2\om\right) f &=& -2\xi +E_1(f,\Ga),\\
E_1(f, \Ga) &=& -\frac{1}{2}\vth f -\frac{1}{2}f^2\eta+\frac 1 2 f^2\etab-f^2\ze+\frac 1 8 f^3\kab+\frac{1}{2}f^3\omb+\frac 1 8 f^3\vthb+\frac{1}{8}f^4\xib.
\eeaa

Also, if $\xi'=0$ and $\om'=0$, we infer
\beaa
0 &=& 4\om-2\left(e_4 + f e_\th +\frac 1 4 f^2  e_3\right)\log(\la)+2f\ze- f^2\omb -2\etab f -f^2\chib-\frac{1}{2}f^3\xib
\eeaa
and hence
\beaa
\la^{-1}e_4'(\log(\la)) &=& 2\om+f\ze- \frac{1}{2}f^2\omb -\etab f - \frac{1}{4}f^2\kab - \frac{1}{4}f^2\vthb -\frac{1}{4}f^3\xib
\eeaa
which yields the desired transport equation for $\log(\la)$
\beaa
\la^{-1}e_4'(\log(\la)) &=& 2\om+E_2(f,\Ga),\\
E_2(f,\Ga) &=& f\ze- \frac{1}{2}f^2\omb -\etab f - \frac{1}{4}f^2\kab - \frac{1}{4}f^2\vthb -\frac{1}{4}f^3\xib.
\eeaa

Finally, we derive the transport equation for $\fb$. In view of the transformation formulas of Proposition \ref{prop:transformations1} for $\ze'$ and $\etab'$, and the fact that we assume $\ze'+\etab'=0$, we have
\beaa
 \frac{1}{2} \la^{-1}e_4'(\fb) &=& -(\ze+ \etab)  + e_\th'(\log(\la))  -\frac 14\fb\ka    + f  \omb - \frac{1}{2}\fb e_\th'(f)  +\frac{1}{8}\fb^2\la^{-1}e_4'(f)\\
 && -\frac{1}{4}\fb \vth      +\lot 
\eeaa
 Together with the above identity for $ \la^{-1}e_4'(f)$, we infer
\beaa
 \la^{-1}e_4'(\fb) + \frac{\ka}{2}\fb &=& -2(\ze+ \etab)  + 2e_\th'(\log(\la))      + 2f  \omb      +E_3(f, \fb, \Ga), \\
E_3(f, \fb, \Ga) &=& -\fb e_\th'(f)  -\frac{1}{2}\fb \vth+\lot,
\eeaa
which yields the third identity of the statement. This concludes the proof of Lemma \ref{lemma:transportequationsforffbandlambda}.


\section{Proof of Corollary \ref{cor:transportequationsforffbandlambda}}\lab{sec:proofofcor:transportequationsforffbandlambda}


In view of Lemma \ref{lemma:transportequationsforffbandlambda} and the fact that $(e_3, e_4, e_\th)$ emanates from an outgoing geodesic foliation and hence
\beaa
\xi=0,\quad \om=0,\quad \ze+\etab=0,
\eeaa
we have 
\beaa
\la^{-1}e_4'(f)  +\frac{\ka}{2} f &=& E_1(f, \Ga),\\
\la^{-1}e_4'(\log(\la)) &=&  E_2(f, \Ga),\\
\la^{-1}e_4'(\fb) + \frac{\ka}{2}\fb &=& 2e_\th'(\log(\la))      + 2f  \omb+E_3(f, \fb, \Ga).
\eeaa
The second equation is the desired identity for $\log(\la)$. 

We still need to derive the first and the third identities. We start with the first one. We have
\beaa
\la^{-1}e_4'(rf) &=& r\left(-\frac{\ka}{2} f + E_1(f, \Ga)\right) +\la^{-1}e_4'(r)f\\
&=& -\frac{r}{2}\left(\ka -\frac{2\la^{-1}e_4'(r)}{r}\right)f+rE_1(f, \Ga).
\eeaa
\beaa
\la^{-1}e_4' &=& e_4+fe_\th+\frac{f^2}{4}e_3,
\eeaa
we infer
\beaa
\la^{-1}e_4'(r) &=& \frac{r}{2}\ov{\ka}+\frac{f^2}{4}e_3(r)
\eeaa
and hence
\beaa
\la^{-1}e_4'(rf) &=& -\frac{r}{2}\left(\kac -\frac{e_3(r)}{2r}f^2\right)f+rE_1(f, \Ga)
\eeaa
as desired.

Next, we have
\beaa
&&\la^{-1}e_4'\Big(r\fb-2r^2e_\th'(\log(\la))+rf\Omb\Big)\\
 &=& r\left(-\frac{\ka}{2}\fb + 2e_\th'(\log(\la))      + 2f  \omb+E_3(f, \fb, \Ga)\right) -2r^2e_\th'(E_2(f, \Ga))+r\Omb\left(-\frac{\ka}{2} f + E_1(f, \Ga)\right)\\
&& +\la^{-1}e_4'(r)\fb -2r^2[\la^{-1}e_4', e_\th']\log(\la)-4r\la^{-1}e_4'(r)e_\th'(\log(\la))+r\la^{-1}e_4'(\Omb)f+\la^{-1}e_4'(r)f\Omb\\
 &=& -\frac{r}{2}\left(\ka -\frac{2\la^{-1}e_4'(r)}{r}\right)\fb+ 2r\Big(1-2\la^{-1}e_4'(r)\Big)e_\th'(\log(\la)) -2r^2\la^{-1}[e_4', e_\th']\log(\la)  \\
 &&  +r\Big(\la^{-1}e_4'(\Omb)+ 2\omb\Big)f -\frac{r}{2}\left(\ka -\frac{2\la^{-1}e_4'(r)}{r}\right)\Omb f    -2r^2e_\th'(\log(\la))\la^{-1}e_4'(\log(\la)) \\
&& +rE_3(f, \fb, \Ga)-2r^2e_\th'(E_2(f, \Ga))+r\Omb E_1(f, \Ga).
\eeaa
Since we have
\beaa
\la^{-1}e_4' &=& e_4+fe_\th+\frac{f^2}{4}e_3,
\eeaa
we infer
\beaa
\la^{-1}e_4'(r) &=& \frac{r}{2}\ov{\ka}+\frac{f^2}{4}e_3(r),\\
\la^{-1}e_4'(\Omb) &=& e_4(\Omb) +fe_\th(\Omb)+\frac{f^2}{4}e_3(\Omb)\\
&=& -2\omb+fe_\th(\Omb)+\frac{f^2}{4}e_3(\Omb).
\eeaa
Together with the transport equation for $\log(\la)$ and the commutator identity for $[e_4', e_\th']$, we infer
\beaa
&&\la^{-1}e_4'\Big(r\fb-2r^2e_\th'(\log(\la))+rf\Omb\Big)\\
  &=& -\frac{r}{2}\left(\kac -\frac{e_3(r)}{2r}f^2\right)\fb+ 2r\left(1-r\ov{\ka} -\frac{e_3(r)}{2}f^2\right)e_\th'(\log(\la)) +r^2\la^{-1}(\ka'+\vth')e_\th'(\log(\la))\\
 &&  +r\left(e_\th(\Omb)+\frac{f}{4}e_3(\Omb)\right)f^2 -\frac{r}{2}\left(\kac  -\frac{e_3(r)}{2r}f^2\right)\Omb f   \\
 && -2r^2e_\th'(\log(\la))E_2(f, \Ga) +rE_3(f, \fb, \Ga)-2r^2e_\th'(E_2(f, \Ga))+r\Omb E_1(f, \Ga).
\eeaa
Now, recall the following transformation formulas
\beaa
\la^{-1}\ka' &=& \ka+\ddd_1'(f)+\err(\ka,\ka'),\\
\err(\ka,\ka') &=& f(\ze+\eta)-\frac{1}{4}f^2\kab-f^2\omb+\lot
\eeaa
We infer
\beaa
&&\la^{-1}e_4'\Big(r\fb-2r^2e_\th'(\log(\la))+rf\Omb\Big)\\
  &=& -\frac{r}{2}\left(\kac -\frac{e_3(r)}{2r}f^2\right)\fb+ r^2\left(\kac-\left(\ov{\ka}-\frac{2}{r}\right) -\frac{e_3(r)}{r}f^2\right)e_\th'(\log(\la)) \\
  &&+r^2\Big(\ddd_1'(f)+\err(\ka,\ka')+\la^{-1}\vth'\Big)e_\th'(\log(\la))\\
 &&  +r\left(e_\th(\Omb)+\frac{f}{4}e_3(\Omb)\right)f^2 -\frac{r}{2}\left(\kac  -\frac{e_3(r)}{2r}f^2\right)\Omb f   \\
 && -2r^2e_\th'(\log(\la))E_2(f, \Ga) +rE_3(f, \fb, \Ga)-2r^2e_\th'(E_2(f, \Ga))+r\Omb E_1(f, \Ga).
\eeaa
This concludes the proof of Corollary \ref{cor:transportequationsforffbandlambda}.


\section{Proof of Lemma \ref{remark:lotxixi'}}\lab{sec:remark:lotxixi'}


Recall that we have obtained in section \ref{sec:proofoflemma:transportequationsforffbandlambda}
\beaa
2\xi' &=& \la^2\Bigg\{2\xi+\left(e_4 + f e_\th +\frac 1 4 f^2  e_3\right)f+\left(\frac{1}{2}\ka+2\om\right)f+\frac{1}{2}\vth f +\frac{1}{2}f^2\eta-\frac 1 2 f^2\etab+f^2\ze-\frac 1 4 f^3\chib\\
&&-\frac{1}{2}f^3\omb-\frac{1}{8}f^4\xib\Bigg\},\\
4\om' &=& \la\Bigg\{4\om-2\left(e_4 + f e_\th +\frac 1 4 f^2  e_3\right)\log(\la)+2f\ze- f^2\omb -2\etab f -f^2\chib-\frac{1}{2}f^3\xib\Bigg\}\\
&&+\la^{-1}\fb \xi'.
\eeaa
In the case where $\la=1$ and $\fb=0$, we immediately infer
\beaa
2\xi' &=& 2\xi+\left(e_4 + f e_\th +\frac 1 4 f^2  e_3\right)f+\left(\frac{1}{2}\ka+2\om\right)f+\frac{1}{2}\vth f +\frac{1}{2}f^2\eta-\frac 1 2 f^2\etab+f^2\ze-\frac 1 4 f^3\chib\\
&&-\frac{1}{2}f^3\omb-\frac{1}{8}f^4\xib,\\
4\om' &=& 4\om+2f\ze- f^2\omb -2\etab f -f^2\chib-\frac{1}{2}f^3\xib,
\eeaa
and hence
\beaa
\xi' &=& \xi+\frac{1}{2}e_4'(f)+\left(\frac{1}{4}\ka+\om\right)f+\frac{1}{4}\vth f +\frac{1}{4}f^2\eta-\frac 1 4 f^2\etab+\frac{1}{2}f^2\ze-\frac 1 8 f^3\chib\\
&&-\frac{1}{4}f^3\omb-\frac{1}{16}f^4\xib,\\
\om' &=& \om+\frac{1}{2}f\ze- \frac{1}{4}f^2\omb -\frac{1}{2}\etab f -\frac{1}{8}f^2\kab-\frac{1}{8}f^2\vthb-\frac{1}{8}f^3\xib.
\eeaa

Finally, we compute the change of frame formula for $\ze'$ and $\eta'$ when $\la=1, \fb=0$. We have in this case
\beaa
e_4' &=& e_4+fe_\th+\frac{1}{4}f^2e_3,\\
e_\th' &=& e_\th+\frac{1}{2}fe_3,\\
e_3' &=& e_3,
\eeaa
and hence
\beaa
2\ze' &=&  g\left(D_{e_\th'} e_4', e_3'\right)\\
&=& g\left(D_{e_\th'}\left(e_4 + f e_\th +\frac 1 4 f^2  e_3\right), e_3\right)\\
&=& g\left(D_{e_\th'}e_4, e_3\right)+fg\left(D_{e_\th'}e_\th, e_3\right)\\
&=& g\left(D_{e_\th+\frac{1}{2}fe_3}e_4, e_3\right)+fg\left(D_{e_\th+\frac{1}{2}fe_3}e_\th, e_3\right)\\
&=& 2\ze -2\omb f -\chib f -\xib f^2\\
&=& 2\ze  -\left(\frac{1}{2}\kab+2\omb\right) f - f\left(\frac{1}{2}\vthb +f\xib\right)
\eeaa
and
\beaa
  2 \eta'&=& g\left(D_{e_3'} e_4', e_\th'\right)\\
  &=& g\left(D_{e_3'}\left(e_4 + f e_\th +\frac 1 4 f^2  e_3\right), e_\th'\right)\\
   &=& g\left(D_{e_3'}e_4, e_\th'\right) + e_3'(f)g\left(e_\th, e_\th'\right)  +  f g\left(D_{e_3'}e_\th, e_\th'\right)+\frac 1 4  e_3'(f^2)g\left(e_3, e_\th'\right) +\frac{1}{4} f^2g(D_{e_3'}e_3, e_\th')  \\
    &=&  e_3'(f)+  g\left(D_{e_3}e_4, e_\th+\frac{1}{2}fe_3\right)  + f g\left(D_{e_3}e_\th, \frac{1}{2}fe_3\right) +\frac{1}{4}f^2g(D_{e_3}e_3, e_\th)\\
       &=&  2\eta+ e_3'(f) -2f\omb -\frac{1}{2}f^2\xib
\eeaa
which yields the desired change of frame formula for $\ze'$ and $\eta'$. This concludes the proof of Lemma \ref{remark:lotxixi'}.


\section{Proof of Proposition \ref{prop:alternateformulaforqfinvolvingtwoangularderrivativesofrho}}\lab{sec:proofofprop:alternateformulaforqfinvolvingtwoangularderrivativesofrho}


Recall that we have
\beaa
\qf= r^4\left(e_3(e_3(\a))+(2\kab -6\omb)e_3(\a)+\left(-4e_3(\omb)+8\omb^2-8\omb\,\kab+\frac{1}{2}\kab^2\right)\a\right),
\eeaa
which we write in the form $  \qf= r^4J$ where,
\beaa
J&=& e_3(e_3(\a))+(2\kab -6\omb)e_3(\a)+ V\a, \\
V&=& -4e_3(\omb)+8\omb^2-8\omb\,\kab+\frac{1}{2}\kab^2.
\eeaa
We  make use of the general\footnote{In an arbitrary $\Z$-invariant frame.}  Bianchi equations, see Proposition \ref{prop:reduced-Bianchi}
\beaa
\begin{split}
e_3 \a+\frac 1 2 \kab \a&=-\dds_2\,\b+4\omb \a -\frac 3 2  \vth \rho  +\err[e_3(\a)],\\
e_3 \b+ \kab \b &=-\dds_1\rho +2\omb \b   + 3\eta \rho +\err[e_3(\b)], \\
e_3 \rho+\frac 3 2 \kab \rho&=\ddd_1\bb    +\err[e_3(\rho)] 
\end{split}
\eeaa
as well as  the null structure  equations (see Proposition \ref{prop:null.structure-general})
\beaa
e_3\vth +\frac 12 \kab\, \vth - 2\omb \vth &=&-2\dds_2\,\eta -\frac 12 \ka \,\vthb+\err[e_3(\vth)] \\
e_3(\kab)+\frac 12 \kab^2 +2 \omb \,\kab &=&2\ddd_1\xib +\err[e_3(\kab)]
\eeaa
where $\err[e_3(\a)],  \err[e_3(\b)], +\err[e_3(\rho)], \err[e_3(\vth)], \err[e_3(\kab)] $ denote  the corresponding quadratic terms in each equation.
We also make use of the  commutation  formula (see Lemma  \ref{Le:comme3e4})
\beaa
\, [ e_3, \dds_2 \b]&=&-\frac 12 \kab \dds_2 \b -\comb^*_2(\b)\\
\comb^*_2(\b)&=&- \frac 1 2 \vthb  \ddd_1  \b - (\ze-\eta) e_3 \b   -  \eta  e_3\Phi \b  +\xib(e_4  \b   -  e_4(\Phi)  \b)-\bb\c \b 
\eeaa
Thus,
\beaa
J&=& e_3\left(-\frac{1}{2}\kab\a-\dds_2\b+4\omb\a-\frac{3}{2}\vth\rho+\err[e_3(\a)]\right)+(2\kab -6\omb)e_3(\a)+ V\a\\
&=&(\frac 32 \kab-2\omb) e_3 \a +\big(-\frac 1 2 e_3 \kab  +4 e_3 (\omb)+V\big)\a-\dds_2 e_3\bb-[e_3, \dds_2]\b-\frac 32 \vth e_3 \rho-\frac 32 \rho e_3 \vth\\
&+& e_3 \err[e_3(\a)]\\
&=&(\frac 32 \kab-2\omb) \left(-\frac{1}{2}\kab\a-\dds_2\b+4\omb\a-\frac{3}{2}\vth\rho+\err[e_3(\a)]\right)
+\big(-\frac 1 2 e_3 \kab  +4 e_3 (\omb)+V\big)\a\\
&-&\dds_2 e_3\b+\frac 1 2 \kab \dds_2 \b -\frac 32 \vth e_3 \rho-\frac 32 \rho e_3 \vth  +e_3 \err[e_3(\a)] +\comb^*_2(\b)\\
&=&-\dds_2 e_3\b+(-\kab+2\omb)\dds_2 \bb+ \Big(-\frac 1 2 e_3 \kab  +4 e_3 (\omb)+V+(\frac 32 \kab-2\omb)(-\frac 12 \kab+  4\omb)\Big)\a\\
&-&\frac 32\left(  \vth e_3 \rho+ \rho e_3 \vth+\vth \rho(\frac 32 \kab-2\omb) \right) +e_3 \err[e_3(\a)] +(\frac 32 \kab-2\omb)\err[e_3(\a)]+\comb^*_2(\b)
\eeaa
Hence,
\bea
\lab{eq:alternateformulaforqfinvolvingtwoangularderrivativesofrho-1}
\bsplit
J&=-\dds_2 e_3\b+(-\kab+2\omb)\dds_2 \bb-\frac 32\left(  \vth e_3 \rho+ \rho e_3 \vth+\vth \rho(\frac 32 \kab-2\omb) \right)+W \a + E\\
W:&=-\frac 1 2 e_3 \kab  +4 e_3 (\omb)+V+(\frac 32 \kab-2\omb)(-\frac 12 \kab+  4\omb)\\
E:&=e_3 \err[e_3(\a)] +(\frac 32 \kab-2\omb)\err[e_3(\a)]+\comb^*_2(\b)
\end{split}
\eea

Now, ignoring cubic and higher order terms,
\beaa
-\dds_2 e_3\b+(-\kab+2\omb)\dds_2 \bb&=&-\dds_2\left(-  \kab \b -\dds_1\rho +2\omb \b   + 3\eta \rho+\err[e_3(\b)] \right)+(-\kab+2\omb)\dds_2 \bb\\
&=&\dds_2\dds_1\rho- 3  \rho\dds_2 \eta+ \b\dds_1(\kab-2\omb)- 3 \eta \dds_1 \rho-\dds_2 \err[e_3(\b)]
\eeaa
Also, 
\beaa
&&\vth e_3 \rho+ \rho e_3 \vth+\vth \rho(\frac 32 \kab-2\omb)=\vth\left(-\frac 3 2 \kab \rho+\ddd_1 \bb \right)+\vth \rho(\frac 32 \kab-2\omb)\\
&&+
\rho\left( -\frac 12 \kab\, \vth +2\omb \vth -2\dds_2\,\eta -\frac 12 \ka \,\vthb+\err[e_3(\vth)]  \right)\\
&&=-\frac 12 \kab \rho\vth-\frac 1 2 \ka \rho \vthb - 2 \rho \dds_2 \eta+\vth \ddd_1\bb +\rho \err[e_3(\rho)]
\eeaa
and,
\beaa
W&=&-\frac 1 2 e_3 \kab  +4 e_3 (\omb)+ \left( -4e_3(\omb)+8\omb^2-8\omb\,\kab+\frac{1}{2}\kab^2\right)+(\frac 32 \kab-2\omb)(-\frac 12 \kab+  4\omb)\\
&=&-\frac 1 2 e_3 \kab +\left(8\omb^2-8\omb\,\kab+\frac{1}{2}\kab^2 \right)+\left(-\frac 3 4 \kab^2-8\omb^2+7\omb \kab \right)\\
&=&-\frac 1 2 e_3 \kab-\frac 1 4 \kab^2 -\omb \kab\\
&=& -\frac 1 2\left(-  \frac 12 \kab^2 -2 \omb \,\kab +2\ddd_1\xib +\err[e_3(\kab)]\right) -\frac 1 4 \kab^2 -\omb \kab\\
&=&-\ddd_1\xib-\frac 1 2 \err[e_3(\kab)]
\eeaa
Thus, back to \eqref{eq:alternateformulaforqfinvolvingtwoangularderrivativesofrho-1},
\beaa
J&=&\dds_2\dds_1\rho- 3  \rho\dds_2 \eta + \b\dds_1(\kab-2\omb)- 3 \eta \dds_1 \rho-\dds_2 \err[e_3(\b)] \\
& -&\frac 3 2 \big(-\frac 12 \kab \rho\vth-\frac 1 2 \ka \rho \vthb - 2 \rho \dds_2 \eta+\vth \ddd_1\bb +\rho \err[e_3(\rho)]\Big)-\ddd_1\xib \a+E\\
 &=&\dds_2\dds_1\rho+\frac 3 4\rho(\kab \vth+\ka \vthb) + \b\dds_1(\kab-2\omb)- 3 \eta \dds_1 \rho-\frac 32 \vth \ddd_1\bb -\ddd_1\xib \a-\frac  32  \rho \err[e_3(\rho)]+E
\eeaa
In other words,
\beaa
J&=& \dds_2\dds_1\rho+\frac 3 4\rho(\kab \vth+\ka \vthb) + \err\\
\err:&=&  \b\dds_1(\kab-2\omb)- 3 \eta \dds_1 \rho-\frac 32 \vth \ddd_1\bb -\ddd_1\xib \a\\
&-&\frac  32  \rho \err[e_3(\rho)]+ e_3 \err[e_3(\a)] +(\frac 32 \kab-2\omb)\err[e_3(\a)]+\comb^*_2(\b)+\lot
\eeaa
It remains to analyze the lower order terms according to our convention in  Definition \ref{definition-errortermsforsquareqf}
Note that we can write the first line in  the expression of $\err$
\beaa
 \err_1&=& r^{-1}\Ga_b \c \b + r^{-2} \Ga_g\c \dkb \Ga_g+ r^{-2}\Ga_g \dkb\Ga_b+ r^{-1}\dkb  \Ga_b\c \a \\
 &=&  r^{-1} \dkb^{\le 1} \Ga_b \c \b  + r^{-2}\Ga_g \dkb\Ga_b+\lot
 \eeaa
On the other hand,
\beaa
\err[e_3(\rho)]&=& -\frac 1 2 \vth \, \aa - \ze\, \bb +2(\eta \,\bb+ \xib\,\b)=\Ga_g\c \Ga_b+\Ga_b \c \b\\
\err[e_3(\a)]&=&(\ze+4\eta) \b=\Ga_g\c  \b \\
 e_3 \err[e_3(\a)]&=& e_3(\ze+4\eta )\b + (\ze+4\eta) e_3(\b)\\
 &=&e_3(\ze+4\eta) \c \b + (\ze+4\eta) (-\kab \b -\dds_1\rho +2\omb \b   + 3\eta \rho)\\
 &=&e_3(\ze+4\eta) \c \b+ r^{-1} \Ga_g \b + r^{-2} \Ga_g \dkb \Ga_g + r^{-3} \Ga_g\c \Ga_g
 \eeaa
 \beaa
\comb^*_2(\b)&=&- \frac 1 2 \vthb  \ddd_1  \b - (\ze-\eta) e_3 \b   -  \eta  e_3\Phi \b  +\xib(e_4  \b   -  e_4(\Phi)  \b)-\bb\c \b\\
&=&- \frac 1 2 \vthb  \ddd_1  \b - (\ze-\eta)\left(-\kab \b - \dds_1\rho + 2\omb\, \b +3\eta \rho\right)-\eta e_3(\Phi)\b\\
&+&\xib\left( - 2\ka \b  +\dds_2 \a -2\om \b   \right)- \xib e_4\Phi   \b-\b \c \bb+\lot\\
&=& r^{-1}\Ga_b \c \dkb^{\le 1 }\b + r^{-2}\Ga_g \c\dkb\Ga_b+\lot
\eeaa
Therefore, schematically,
\beaa
\err &=&e_3(\ze+4\eta) \c \b+ r^{-1}\Ga_b \dkb^{\le 1 }\b + r^{-2}\Ga_g \dkb\Ga_b+\lot
\eeaa
and therefore,
\beaa
\err[\qf] = r^4\err=r^4\left(e_3(\ze+4\eta) \c \b+ r^{-1}\Ga_b\c  \dkb^{\le 1 }\b + r^{-2}\Ga_g\c \dkb\Ga_b\right)+\lot
\eeaa
Since $e_3\ze\in r^{-1}\dk \Ga_b$    and     $\b\in r^{-1}\Ga_g$       we rewrite in the form,
\beaa
\err[\qf]&=&r^4 e_3 \eta \c \b+ r^ 2 \dk^{\le 1 }\big( \Ga_b\c\Ga_g).
\eeaa
This concludes the proof of Proposition \ref{prop:alternateformulaforqfinvolvingtwoangularderrivativesofrho}.


\section{Proof of Proposition \ref{Le:Teuk-Star1}}\lab{sec:appendix:proofpropLe:Teuk-Star1}


We start with the formula \eqref{eq:alternateformulaforqfinvolvingtwoangularderrivativesofrho}
 \beaa
r\qf&=& r^5\left(\dds_2\dds_1\rho+\frac{3}{4}\kab\rho\vth +\frac{3}{4}\ka\rho\vthb\right) +r\err[\qf].
\eeaa
with $\err[\qf]$  given by \eqref{eq:alternateformulaforqfinvolvingtwoangularderrivativesofrho-err}.
Taking  the  $e_3$ derivative we deduce,
\bea
\label{eqt:Teuk-Star1}
\bsplit
e_3(r\qf) &= r^5 L + 5e_3(r)\, \qf +  e_3( r\err[\qf]) - 5  e_3(r)  \err[\qf],\\
L:&=e_3\left\{\dds_2\dds_1\rho+\frac{3}{4}e_3(\kab\rho\vth) +\frac{3}{4}e_3(\ka\rho\vthb)\right\}
\end{split}
\eea
We calculate $L$ as follows,
\beaa
L&=&e_3\dds_2\dds_1\rho+\frac{3}{4}e_3(\kab\rho\vth) +\frac{3}{4}e_3(\ka\rho\vthb)\\
&=&\dds_2\dds_1e_3(\rho)+[e_3, \dds_2\dds_1]\rho +\frac{3}{4}e_3(\kab\rho\vth) +\frac{3}{4}e_3(\ka\rho\vthb).
\eeaa
Ignoring cubic and higher order terms
 \beaa
e_3(\kab\rho\vth) &=& \kab\rho e_3(\vth)+e_3(\kab)\rho\vth+\kab e_3(\rho)\vth\\
&=& \kab\rho \left(-\frac 12 \kab\, \vth + 2\omb \vth -2\dds_2\,\eta -\frac 12 \ka \,\vthb +\err[e_3\vth] \right)\\
&&+\left(-\frac 12 \kab^2 -2 \omb \,\kab +2\ddd_1\xib+ \err[e_3\kab]\right)\rho\vth\\
&& +\kab\left(-\frac 3 2 \kab \rho+\ddd_1\bb  -\frac 1 2 \vth \, \aa+ \err[e_3(\rho)]\right)\vth\\
&=& \kab\rho \left(-\frac 52 \kab\, \vth  -2\dds_2\,\eta -\frac 12 \ka \,\vthb\right) +\kab\rho \err[e_3\vth] +2\ddd_1\xib \,\rho\vth+\kab( \ddd_1\bb)  \vth
\eeaa
We deduce,
\bea
\bsplit
e_3(\kab\rho\vth) &= \kab\rho \left(-\frac 52 \kab\, \vth  -2\dds_2\,\eta -\frac 12 \ka \,\vthb\right) +E_1\\
E_1&= 2\ddd_1\xib \,\rho\vth+\kab( \ddd_1\bb)  \vth+\kab \rho \err[e_3(\vth)]
\end{split}
\eea
where  $ \err[e_3(\vth)]$,  $\err[e_3(\kab)],  \err[e_3(\rho)]$ denote the quadratic error terms in the
 corresponding equations.
Also,
\beaa
e_3(\ka\rho\vthb) &=& \ka\rho e_3(\vthb)+e_3(\ka)\rho\vthb+\ka e_3(\rho)\vthb\\
&=& \ka\rho\Big(-\kab \, \vthb  - 2\omb\,  \vthb  -2\aa-2\dds_2\,\xib  +\err[e_3(\vthb)]\Big)\\
&&+\left(-\frac 1 2 \kab\, \ka +2\omb \ka + 2\ddd_1\eta + 2\rho +\err[e_3(\ka)]\right)\rho\vthb\\
&&+\left(-\frac 3 2 \kab \rho+\ddd_1\bb +\err[e_3(\rho)]\right)\ka \vthb
\eeaa
Hence, ignoring the higher order terms,
\bea
\bsplit
e_3(\ka\rho\vthb) &= \ka\rho\Big(-3\kab \, \vthb   -2\aa-2\dds_2\,\xib\Big)+ 2\rho^2\vthb+E_2\\
E_2:&=2 \rho\ddd_1\eta  \vthb+ \ka \ddd_1 \b \vthb+ \ka \rho\,  \err[e_3(\vthb)]
\end{split}
\eea
Also, we have
\beaa
\dds_2\dds_1e_3(\rho) &=& \dds_2\dds_1\left(-\frac 3 2 \kab \rho+\ddd_1\bb +\err[e_3(\rho)]\right)\\
&=&\dds_2\dds_1\ddd_1\bb   -\frac{3}{2}\rho\dds_2\dds_1\kab -\frac{3}{2}\kab\dds_2\dds_1\rho+E_3\\
E_3&=&\dds_2\dds_1  \err[e_3(\rho)]   -3 \dds_1 \kab  \c \dds_1 \rho
\eeaa
i.e.,
\bea
\label{eqt:Teuk-Star11}
\bsplit
e_3(\kab\rho\vth) &= \kab\rho \left(-\frac 52 \kab\, \vth  -2\dds_2\,\eta -\frac 12 \ka \,\vthb\right) +  E_1,\\
e_3(\ka\rho\vthb) &= \ka\rho\Big(-3\kab \, \vthb   -2\aa-2\dds_2\,\xib\Big)+ 2\rho^2\vthb+  E_2\\
\dds_2\dds_1e_3(\rho)&=\dds_2\dds_1\ddd_1\bb   -\frac{3}{2}\rho\dds_2\dds_1\kab -\frac{3}{2}\kab\dds_2\dds_1\rho+E_3
\end{split}
\eea

Now,  in view of Lemma \ref{Le:comme3e4} we have (for $f=\dds_1\rho \in \sk_{2-1}$),
\beaa
\, [e_3, \dds_2]\dds_1\rho&=&-\frac 1 2 \kab  \dds_2\dds_1\rho- \comb^*_2(\dds_1 \rho) 
\eeaa
and
\beaa
 \, [e_3, \dds_1]\rho&=&-[e_3, e_\th ]\rho=- \frac 1 2 \kab  \dds_1 \rho-\frac 1 2 \vthb e_\th \rho  + (\ze-\eta) e_3\rho-     \xib e_4 \rho\\
 &=&- \frac 1 2 \kab  \dds_1 \rho -\frac 1 2 \vthb e_\th \rho +(\ze-\eta) \big( -\frac 3 2\kab \rho+\ddd_1\bb+\err[e_3(\rho)]\big)\\
 &-&\xib \big(-\frac 3 2 \ka \rho+\ddd_1\b +\err[e_4(\rho)]\big)\\
 &=& - \frac 1 2 \kab  \dds_1 \rho -\frac 3 2 \rho\left[ (\ze-\eta)\kab -\xib \ka \right]+ E_{41}\\
 E_{41}&=& (\ze-\eta) \ddd_1\bb-\xib\ddd_1 \b+(\ze-\eta) e_3[(\rho)]    -\frac 1 2 \vthb e_\th \rho   - \xib \err[e_4(\rho)]
 \eeaa
We deduce
\beaa
 \dds_2  [e_3, \dds_1]\rho&=&\dds_2\Big( - \frac 1 2 \kab  \dds_1 \rho -\frac 3 2 \rho\left[\kab  (\ze-\eta)- \ka \xib\right]+ E_{41}\Big)\\
 &=& -\frac 1 2 \kab  \dds_2\dds_1\rho-\frac{3}{2} \rho \Big( \kab  \dds_2(\ze-\eta)    -\ka \dds_2\xib   \Big)+E_4\\
 E_4&=&\dds_2 E_{41}-\frac 1 2 \dds_1 \kab \c \dds_1 \rho-\frac 3 2 (\ze-\eta) \dds_1(\kab \rho)+\frac 32 \xib \dds_1(\ka \rho)
\eeaa
Hence, since $\,[e_3, \dds_2\dds_1] \rho=[e_3, \dds_2]\dds_1\rho+\dds_2 [e_3, \dds_1]\rho$,
\bea
\,[e_3, \dds_2\dds_1] \rho &=& - \kab  \dds_2\dds_1\rho -\frac{3}{2}\Big(  \dds_2(\ze-\eta)\kab\rho    -\dds_2\xib\ka\rho   \Big) - \comb^*_2(\dds_1 \rho) +E_4.
\eea

We deduce, recalling  \eqref{eqt:Teuk-Star11},
\beaa
L&=& \dds_2\dds_1e_3(\rho)+[e_3, \dds_2\dds_1]\rho +\frac{3}{4}e_3(\kab\rho\vth) +\frac{3}{4}e_3(\ka\rho\vthb)\\
&=&\dds_2\dds_1\ddd_1\bb   -\frac{3}{2}\rho\dds_2\dds_1\kab -\frac{3}{2}\kab\dds_2\dds_1\rho+ E_3\\
 &-& \kab  \dds_2\dds_1\rho -\frac{3}{2}\Big(  \dds_2(\ze-\eta)\kab\rho    -\dds_2\xib\ka\rho   \Big) - \comb^*_2(\dds_1 \rho) +E_4\\
 &+&\frac 3 4 \left( \kab\rho \left(-\frac 52 \kab\, \vth  -2\dds_2\,\eta -\frac 12 \ka \,\vthb\right)  \right)+E_1\\
 &+&\frac 3 4 \left( \ka\rho\Big(-3\kab \, \vthb   -2\aa-2\dds_2\,\xib\Big)+ 2\rho^2\vthb+  \err_1 \right)\\
 &=&\dds_2\dds_1\ddd_1\bb  - \frac{5}{2}\kab  \dds_2\dds_1\rho  -\frac{3}{2}\rho\dds_2\dds_1\kab -\frac{3}{2}\ \rho   \kab \dds_2 \ze \\
 &+& \frac 3 4  \kab\rho \left(-\frac 52 \kab\, \vth  -\frac 12 \ka \,\vthb\right) +\frac 3 4  \ka\rho\Big(-3\kab \, \vthb   -2\aa \Big)+ \frac 3 2 \rho^2\vthb \\
 &+&E_1+E_2+E_3 +E_4-\comb^*_2(\dds_1 \rho)
\eeaa
i.e.,
\bea
\label{eqt:Teuk-Star2}
\bsplit
L&=\dds_2\dds_1\ddd_1\bb  - \frac{5}{2}\kab  \dds_2\dds_1\rho  -\frac{3}{2}\rho\dds_2\dds_1\kab -\frac{3}{2}\ \rho   \kab \dds_2 \ze \\
&- \frac 3 4  \kab\rho \left(\frac 52 \kab\, \vth  +\frac 12 \ka \,\vthb\right) -\frac 3 4  \ka\rho\left (3\kab \, \vthb   +2\aa \right)+\frac 3 2 \rho^2\vthb+E \\
E&= E_1+E_2+E_3 +E_4-\comb^*_2(\dds_1 \rho)
 \end{split}
\eea

On the other hand, in view of  \eqref{eq:alternateformulaforqfinvolvingtwoangularderrivativesofrho},  writing $e_3 r=\frac r 2\left( \kab+ \Ab\right)$,
\beaa
5  e_3(r)\qf &=& r  \frac{5}{2}\kab\qf +5   r  \Ab \qf\\
&=&  \frac{5}{2}r^5\kab\left\{ \dds_2\dds_1\rho+\frac{3}{4}\kab\rho\vth +\frac{3}{4}\ka\rho\vthb+\err\right\} +5   r  \Ab \qf
\eeaa
Hence, in view of \eqref{eqt:Teuk-Star1} and  \eqref{eqt:Teuk-Star2},
\beaa
e_3(r\qf)&=& r^5 L + 5 e_3(r) \qf +e_3( r\err) - 5  e_3(r)  \err\\
&=& r^5\Bigg\{\left(\dds_2\dds_1\ddd_1\bb  - \frac{5}{2}\kab  \dds_2\dds_1\rho  -\frac{3}{2}\rho\dds_2\dds_1\kab -\frac{3}{2}\ \rho   \kab \dds_2 \ze \right)\\
&-& \frac 3 4  \kab\rho \left(\frac 52 \kab\, \vth  +\frac 12 \ka \,\vthb\right) -\frac 3 4  \ka\rho\left (3\kab \, \vthb   +2\aa \right)+\frac 3 2 \rho^2\vthb\Bigg\} \\
&+& \frac{5}{2}r^5\kab\left\{ \dds_2\dds_1\rho+\frac{3}{4}\kab\rho\vth +\frac{3}{4}\ka\rho\vthb\right\} +r^5 E+e_3( r^5\err) - 5 \frac{e_3r }{r}  r^5 \err+5   r  \Ab \qf \\
 &=&r^5\left( \dds_2\dds_1\ddd_1\bb     -\frac{3}{2}\rho\dds_2\dds_1\kab   -\frac{3}{2}\kab\rho\dds_2\ze -\frac{3}{2}\ka\rho\aa + \frac{3}{4}(2\rho^2-\ka\kab\rho)\vthb \right)\\
 &+&\err[e_3(r\qf)].
\eeaa
where,
\bea
\lab{expression:err[e_3gf]}
\err[e_3(r\qf)]&=&e_3( r\err[\qf]) - 5 e_3r   \err[\qf]+5   r  \Ab \qf+r^5 E\\
&=&  r e_3 (\err[\qf])     +  \err[\qf]      +5   r  \Ab \qf+r^5 E
\eea
and
\beaa
E&=& E_1+E_2+E_3 +E_4-\comb^*_2(\dds_1 \rho)
\eeaa
with,
\beaa
\bsplit
E_1&= 2\ddd_1\xib \,\rho\vth+\kab( \ddd_1\bb)  \vth+\kab \rho \err[e_3(\kab)]\\
E_2&=2 \rho\ddd_1\eta  \vthb+ \ka \ddd_1 \b \vthb+ \ka \rho\,  \err[e_3(\vthb)]\\
E_3&=\dds_2\dds_1  \err[e_3(\rho)]   -3 \dds_1 \kab  \c \dds_1 \rho\\
E_4&=\dds_2 E_{41}-\frac 1 2 \dds_1 \kab \c \dds_1 \rho-\frac 3 2 (\ze-\eta) \dds_1(\kab \rho)+\frac 32 \xib \dds_1(\ka \rho)\\
E_{41}&= (\ze-\eta) \ddd_1\bb-\xib\ddd_1 \b   -\frac 1 2 \vthb e_\th \rho    \\
\comb^*_2(\dds_1 \rho)&=- \frac 1 2 \vthb  \ddd_1\dds_1  \rho  - (\ze-\eta) e_3 \dds_1 \rho    -   \eta  e_3\Phi \dds_1 \rho +\xib(e_4  \dds_1 \rho   - e_4(\Phi)  \dds_1 \rho )-  \bb \dds_1 \rho.
\end{split}
\eeaa
Note also that,
\beaa
(\ze-\eta) e_3 \dds_1 \rho= (\ze-\eta)  \c \dds_1  e_3 \rho+ (\ze-\eta)\big( - \frac 1 2 \kab  \dds_1 \rho-\frac 1 2 \vthb e_\th \rho+(\ze-\eta) e_3\rho-     \xib e_4 \rhoc \big)
\eeaa
Using our schematic notation
\beaa
\err[e_3(\kab)]&=&\Ga_b\c \Ga_b+\lot\\
\err[e_3(\vthb)]&=&\Ga_b\c \Ga_b+\lot \\
\err[e_3(\rho)]&=&\Ga_g \c \aa+\lot =\Ga_g \c \Ga_b+\lot
\eeaa
and
\beaa
E_1&=& r^{-4}\Ga_b\c \dkb^{\le 1} \Ga_b+\lot\\
E_2&=& r^{-4}\Ga_b\c \dkb^{\le 1} \Ga_b+ r^{-2} \Ga_b\c  \b+\lot\\
E_3&=& r^{-2} \dkb^2( \Ga_g \c \Ga_b)+ r^{-3} (\dkb \Ga_g) \c  (\dkb \Ga_g)\\
E_{41}&=&  r^{-2} \Ga_g \c (\dkb \Ga_b)+ r^{-1}\Ga_b\c \dkb \b+ r^{-2} \Ga_b \dkb\c  \Ga_g\\
&=& r^{-2}\dkb (\Ga_g \c \Ga_b)+\lot\\
E_4&=& r^{-3}\dkb^2 (\Ga_g \c \Ga_b)+\lot\\
\comb^*_2(\dds_1 \rho)&=& r^{-3} \Ga_b\c \dk^{\le 2} \Ga_g+ r^{-2} \dkb \Ga_b \c \Ga_g+ \lot
\eeaa
and, since $r^{-1} \Ga_b$   can be replaced by $\Ga_g$ and $\dkb \b$ can be replaced by $r^{-1} \Ga_g$,
\beaa
E&=& r^{-3}\dkb^2 (\Ga_g \c \Ga_b)+\lot\
\eeaa
Taking into account the expression of       $\err[\qf]$  in Proposition \ref{prop:alternateformulaforqfinvolvingtwoangularderrivativesofrho}  we  write
\beaa
 r e_3 (\err[\qf]) + \err[\qf]&=& r e_3\Big[ r^4 e_3 \eta \c \b+ r^ 2 \dk^{\le 1 }\big( \Ga_b\c\Ga_g)\Big]+r^4 e_3 \eta \c \b+ r^ 2 \dk^{\le 1 }\big( \Ga_b\c\Ga_g)\\
&=&  r^5  \dk^{\le 1} \big(e_3 \eta \c \b\big)+ r^3  \dk^{\le 2 }\big( \Ga_b\c\Ga_g\big)
\eeaa
and therefore, back to \eqref{expression:err[e_3gf]},
\beaa
\err[e_3(r\qf)]&=& e_3 ( r\err[\qf])     +   \err[\qf]      +  r  \Ab \qf+r^5 E\\
&=& r^5  \dk^{\le 1} \big(e_3 \eta \c \b\big)+ r^3  \dk^{\le 2 }\big( \Ga_b\c\Ga_g\big)
  + r\Ga_b \qf + r^2 \dkb^2 (\Ga_g \c \Ga_b)+\lot\\
  &=& r\Ga_b \qf +r^5  \dk^{\le 1} \big(e_3 \eta \c \b\big) + r^3  \dk^{\le 2 }\big( \Ga_b\c\Ga_g\big).
\eeaa
This concludes the proof of Proposition \ref{Le:Teuk-Star1}.


\section{Proof of the Teukolsky-Starobinski identity}\lab{appendix:Teuk-Star}


According to Proposition \ref{Le:Teuk-Star1} we have
\beaa
e_3(r\qf) &=& r^5\Bigg\{ \dds_2\dds_1\ddd_1\bb   -\frac{3}{2}\rho\dds_2\dds_1\kab -\frac{3}{2}\kab\rho\dds_2\ze -\frac{3}{2}\ka\rho\aa + \frac{3}{4}(2\rho^2-\ka\kab\rho)\vthb\Bigg\} +\err[e_3\qf]. 
 \eeaa
  
We infer that
\bea
\lab{equation:Starobinski1}
\bsplit
e_3(r^2e_3(r\qf)) &= r^7\Bigg\{ e_3\dds_2\dds_1\ddd_1\bb   -\frac{3}{2}e_3(\rho\dds_2\dds_1\kab) -\frac{3}{2}e_3(\kab\rho\dds_2\ze) -\frac{3}{2}e_3(\ka\rho\aa)\\
& + \frac{3}{4}e_3\Big((2\rho^2-\ka\kab\rho)\vthb\Big)\Bigg\}+7re_3(r)e_3(r\qf)\\
&+   r^2  e_3 \big( \err[e_3\qf]\big) + r \err[e_3\qf]+\lot  
\end{split}
 \eea
We  first compute
\beaa
e_3\dds_2\dds_1\ddd_1\bb &=& \dds_2\dds_1\ddd_1(e_3\bb) + [e_3,\dds_2]\dds_1\ddd_1\bb + \dds_2[e_3,\dds_1]\ddd_1\bb + \dds_2\dds_1[e_3,\ddd_1]\bb\\
&=&   \dds_2\dds_1\ddd_1(e_3\bb)  + \left(-\frac{1}{2}\kab\dds_2+\frac{1}{2}\vthb\ddd_1 +(\ze-\eta)e_3+\eta e_3\Phi -\xib( e_4 -e_4(\Phi)) +\bb\right)\dds_1\ddd_1\bb \\
&&+ \dds_2\left(\left(-\frac{1}{2}\kab\dds_1-\frac{1}{2}\vthb\dds_1  +(\ze-\eta)e_3 -\xib e_4\right)\ddd_1\bb\right) \\
&&+ \dds_2\dds_1\left(\left(-\frac{1}{2}\kab\ddd_1+\frac{1}{2}\vthb\dds_2 -(\ze-\eta)e_3+\eta e_3\Phi +\xib( e_4 +e_4(\Phi)) +\bb\right)\bb\right)
\eeaa
In view of our general  commutation formulas in  Lemma \ref{Le:comme3e4}      and our notation convention for error terms\footnote{In particular we write $\bb\in  r^{-1}\Ga_b$.}  we have\footnote{We also commute once more $e_3 $ and $e_4$  with $\dds_1, \ddd_1, \dds_2, \ddd_2$ and use Bianchi.} ,
\beaa
 [e_3,\dds_2]\dds_1\ddd_1\bb&=&  \left(-\frac{1}{2}\kab\dds_2+\frac{1}{2}\vthb\ddd_1 +(\ze-\eta)e_3+\eta e_3\Phi -\xib( e_4 -e_4(\Phi)) +\bb\right)\dds_1\ddd_1\bb\\
 &=& -\frac{1}{2}\kab\dds_2\dds_1\ddd_1\bb +r^{-4} \Ga_b \c \dkb^{\le 3} \Ga_b +\lot
 \eeaa
 \beaa
  \dds_2[e_3,\dds_1]\ddd_1\bb &=& \dds_2\left(\left(-\frac{1}{2}\kab\dds_1-\frac{1}{2}\vthb\dds_1  +(\ze-\eta)e_3 -\xib e_4\right)\ddd_1\bb\right) \\
  &=& -\frac{1}{2}\kab\dds_2\dds_1\ddd_1\bb+ r^{-4}\left( \Ga_b \c \dkb^{\le 3} \Ga_b+ \Ga_b^{\le 1}  \c \dkb^{\le 2} \Ga_b \right)+\lot\\
  \dds_2\dds_1[e_3,\ddd_1]\bb&=& \dds_2\dds_1\left(\left(-\frac{1}{2}\kab\ddd_1+\frac{1}{2}\vthb\dds_2 -(\ze-\eta)e_3+\eta e_3\Phi +\xib( e_4 +e_4(\Phi)) +\bb\right)\bb\right)\\
  &=& -\frac{1}{2}\kab\dds_2\dds_1\ddd_1\bb+ r^{-4}\left( \Ga_b \c \dkb^{\le 3} \Ga_b+ \Ga_b^{\le 1}  \c \dkb^{\le 2} \Ga_b \right)+\lot\
\eeaa
Hence, schematically,
\beaa
e_3\dds_2\dds_1\ddd_1\bb &=& \dds_2\dds_1\ddd_1(e_3\bb) -\frac{3}{2}\kab\dds_2\dds_1\ddd_1\bb +r^{-4}  \dkb^{\le 3}( \Ga_b \c \Ga_b)
\eeaa
Using the Bianchi identity  $e_3\bb  = \ddd_2\aa - 2(\kab +\omb)\, \bb+ (-2\ze+\eta) \aa+ 3\xib \rho$  we further deduce,
\beaa
\dds_2\dds_1\ddd_1(e_3\bb) &=& \dds_2\dds_1\ddd_1\ddd_2\aa -2(\kab+\omb)\dds_2\dds_1\ddd_1\bb+3\rho\dds_2\dds_1\ddd_1\xib + r^{-4}  \dkb^{\le 3}( \Ga_b \c \Ga_b)
\eeaa
i.e.,
\bea
\lab{eqt:Teuk-Star4}
\bsplit
e_3\dds_2\dds_1\ddd_1\bb &= \dds_2\dds_1\ddd_1\ddd_2\aa -2(\kab+\omb)\dds_2\dds_1\ddd_1\bb+3\rho\dds_2\dds_1\ddd_1\xib -\frac{3}{2}\kab\dds_2\dds_1\ddd_1\bb\\
& +r^{-4}  \dkb^{\le 3}( \Ga_b \c \Ga_b)
\end{split}
\eea
We next  calculate  the  second  term   $e_3(\rho\dds_2\dds_1\kab)$         on the right hand side of  \eqref{equation:Starobinski1}
\beaa
e_3(\rho\dds_2\dds_1\kab) &=& \rho\dds_2\dds_1e_3\kab+\rho[e_3,\dds_2]\dds_1\kab+\rho\dds_2[e_3,\dds_1]\kab+e_3(\rho)\dds_2\dds_1\kab.
\eeaa

Using the  equation for $e_3\kab $ in Proposition \ref{prop:null.structure-general} we derive,
\beaa
 \rho\dds_2\dds_1e_3\kab&=& \rho\dds_2\dds_1\left(-\frac 12 \kab^2 -2 \omb \,\kab +2\ddd_1\xib+\Ga_b\c \Ga_b\right),\\
 &=&-\rho\left( \kab +2\omb \right)\dds_2\dds_1\kab   -2  \rho \kab \dds_2\dds_1\omb       + 2\rho   \dds_2\dds_1\ddd_1\xib+ r^{-5} \dkb^{\le 2} \Ga_b\c \Ga_b
 \eeaa
 Also,
 
 \beaa
 [e_3,\dds_2]\dds_1\kab&=&\left(-\frac{1}{2}\kab\dds_2+\frac{1}{2}\vthb\ddd_1 +(\ze-\eta)e_3+\eta e_3\Phi -\xib( e_4 -e_4(\Phi)) +\bb\right)\dds_1\kab\\
 &=&-\frac 1 2  \kab\dds_2\dds_1\kab+ r^{-2}\Ga_b \c \dkb^{\le 2} \Ga_g,\\
 \dds_2[e_3,\dds_1]\kab&=&\dds_2\left(\left(-\frac{1}{2}\kab\dds_1-\frac{1}{2}\vthb\dds_1  +(\ze-\eta)e_3 -\xib e_4\right)\kab\right),\\
 &=& -\frac{1}{2}\kab\dds_2 \dds_1\kab+ \dds_2(\ze-\eta) e_3\kab-  \dds_2 \xib e_4\kab  + r^{-2} \dkb^{\le 2}(\Ga_b \c  \Ga_g)
 \eeaa
 Using also. the Bianchi  equation  
 \beaa
 e_3 \rho&=&- \frac 3 2 \kab \rho+\ddd_1\bb  -\frac 1 2 \vth \, \aa - \ze\, \bb +2(\eta \,\bb+ \xib\,\b)
 \eeaa
 We deduce,
 \beaa
 e_3(\rho\dds_2\dds_1\kab) &=& -\rho\left( \kab +2\omb \right)\dds_2\dds_1\kab   -2  \rho \kab \dds_2\dds_1\omb  + 2\rho   \dds_2\dds_1\ddd_1\xib\\
 &- & \rho \kab\dds_2\dds_1\kab+ \rho\left(\dds_2(\ze-\eta)e_3\kab -\dds_2(\xib) e_4\kab\right)-\frac 3 2 \rho \kab\dds_2\dds_1\kab+r^{-5} \dkb^{\le 2}(\Ga_b \c  \Ga_g)
 \eeaa
 i.e.,
 \bea
 \lab{eqt:Teuk-Star4'}
 \bsplit
 e_3(\rho\dds_2\dds_1\kab)&=-\frac{7}{2}\rho\kab\dds_2\dds_1(\kab) -2\rho\omb\dds_2\dds_1\kab -2\rho\kab\dds_2\dds_1\omb  +2\rho\dds_2\dds_1\ddd_1\xib\\
& +\rho\left(\dds_2(\ze-\eta)e_3\kab -\dds_2(\xib) e_4\kab\right)+r^{-5} \dkb^{\le 2}(\Ga_b \c  \Ga_g).
\end{split}
 \eea
 Now,
 \beaa
&&\dds_2(\ze-\eta)e_3\kab -\dds_2(\xib) e_4\kab \\
&&= \dds_2(\ze-\eta)\left(-\frac 12 \kab^2 -2 \omb \,\kab +2\ddd_1\xib+\Ga_b\c \Ga_b\right)\\
&& -\dds_2\xib\left(-\frac{1}{2}\ka\kab+2\om\kab+2\ddd_1\etab+2\rho+\Ga_g\c \Ga_b\right)\\
&& =\dds_2(\ze-\eta)\left(-\frac 12 \kab^2 -2 \omb \,\kab\right) -\dds_2\xib\left(-\frac{1}{2}\ka\kab+2\om\kab+2\rho\right)+ r^{-2} \dkb^{\le 1 }\Ga_b\c   \dkb^{\le 1 }\Ga_b.
\eeaa
Therefore, back to \eqref{eqt:Teuk-Star4'},
\bea
\label{eqt:Teuk-Star5}
\bsplit
e_3(\rho\dds_2\dds_1\kab) &=  -\frac{7}{2}\rho\kab\dds_2\dds_1(\kab) -2\rho\omb\dds_2\dds_1\kab -2\rho\kab\dds_2\dds_1\omb  +2\rho\dds_2\dds_1\ddd_1\xib\\
& +\rho\dds_2(\ze-\eta)\left(-\frac 12 \kab^2 -2 \omb \,\kab\right) -\rho\dds_2\xib\left(-\frac{1}{2}\ka\kab+2\om\kab+2\rho\right)\\
&+r^{-5} \dkb^{\le 2}(\Ga_b \c  \Ga_g)..
\end{split}
\eea

We next estimate the  third   term  $e_3(\kab\rho\dds_2\ze) $ on the right hand side of \eqref{equation:Starobinski1},
\beaa
e_3(\kab\rho\dds_2\ze) &=& \kab\rho\dds_2(e_3\ze) +\kab\rho[e_3, \dds_2]\ze + e_3(\kab)\rho\dds_2\ze + \kab e_3(\rho)\dds_2\ze
\eeaa
Using  again the equations  
\beaa
e_3(\kab)&=&-\frac 12 \kab^2 -2 \omb \,\kab +2\ddd_1\xib+\Ga_b\c \Ga_b\\
e_3\rho&=&-\frac 3 2 \kab \rho+\ddd_1\bb  +\Ga_g\c \Ga_b
\eeaa
i.e.,
\beaa
e_3(\kab)\rho\dds_2\ze&=&(-\frac 12 \kab^2 -2 \omb \,\kab) \rho\dds_2\ze+ r^{-5}\dkb \Ga_b\c \dkb \Ga_g \\
  \kab e_3\rho\dds_2\ze&=&-\frac 3 2 \kab^2 \rho+r^{-4}\dkb \Ga_b \c \dkb \Ga_g 
\eeaa
the equation,
\beaa
e_3\ze&=&-\frac{1}{2}\kab(\ze+\eta)+2\omb(\ze-\eta)+\bb+2\dds_1\omb+2\om\xib+\frac{1}{2}\ka\xib +\Ga_b\c \Ga_b\\
\eeaa
and the commutator formula,
\beaa
[e_3, \dds_2]\ze&=&\left(-\frac{1}{2}\kab\dds_2+\frac{1}{2}\vthb\ddd_1 +(\ze-\eta)e_3+\eta e_3\Phi -\xib( e_4 -e_4(\Phi)) +\bb\right)\ze\\
&=&-\frac 1 2 \kab\dds_2\ze +r^{-1}\Ga_b\c \dkb ^{\le 1} \Ga_g
\eeaa
we deduce
\beaa
e_3(\kab\rho\dds_2\ze)&=& \kab\rho\dds_2\left(-\frac{1}{2}\kab(\ze+\eta)+2\omb(\ze-\eta)+\bb+2\dds_1\omb+2\om\xib+\frac{1}{2}\ka\xib+\Ga_b\c\Ga_b\right)\\
&-&\frac{1}{2}\kab^2\rho \dds_2\ze +r^{-5}\Ga_b\c \dkb ^{\le 1} \Ga_g\\
&-&\frac 12 \kab^2 \rho\dds_2\ze  -2\omb \, \kab \rho  \dds_2\ze    + r^{-5}\dkb^{\le1 }\Ga_b\c \dkb^{\le 1}\Ga_g\\
&-&\frac 3 2 \kab ^2\rho \dds_2\ze + r^{-4} \dkb \Ga_b \c \dkb \Ga_g   \\
&=& \kab\rho\left(-3\kab\dds_2\ze-\frac{1}{2}\kab\dds_2\eta -2\omb\dds_2\eta+\dds_2\bb+2\dds_2\dds_1\omb+2\om\dds_2\xib+\frac{1}{2}\ka\dds_2\xib\right)\\
&+& r^{-4}\dkb^{\le 1 } \Ga_b\c \dkb^{\le 1 } \Ga_g+r^{-5}\dkb^{\le1 }\Ga_b\c \dkb^{\le 1}\Ga_b
\eeaa
i.e.,
\bea
\label{eqt:Teuk-Star6}
\bsplit
e_3(\kab\rho\dds_2\ze) &=\kab\rho\left(-3\kab\dds_2\ze-\frac{1}{2}\kab\dds_2\eta -2\omb\dds_2\eta+\dds_2\bb+2\dds_2\dds_1\omb+2\om\dds_2\xib+\frac{1}{2}\ka\dds_2\xib\right)\\
&+ r^{-4}\dkb^{\le 1 } \Ga_b\c \dkb^{\le 1 } \Ga_g+r^{-5}\dkb^{\le1 }\Ga_b\c \dkb^{\le 1}\Ga_b
\end{split}
\eea
For the fourth term on the right hand side of \eqref{equation:Starobinski1} we have
\beaa
e_3(\ka\rho\aa) &=& \ka\rho e_3(\aa)+e_3(\ka)\rho\aa+\ka e_3(\rho)\aa\\
&=& \ka\rho e_3(\aa)+\left(-\frac 1 2 \kab\, \ka +2\omb \ka + 2\ddd_1\eta + 2\rho -\frac 1 2  \vthb\, \vth +2(\xib\,  \xi+\eta\, \eta)\right)\rho\aa\\
&&+\ka\left(-\frac 3 2 \kab \rho+\ddd_1\bb  -\frac 1 2 \vth \, \aa - \ze\, \bb +2(\eta \,\bb+ \xib\,\b)\right)\aa\\
&=& \ka\rho e_3(\aa)+\left(-2\ka\kab +2\omb \ka  + 2\rho \right)\rho\aa +\big( r^{-3} \dkb \Ga_b+  r^{-2} \Ga_b\c \Ga_b) \c   \aa
\eeaa
i.e.,
\bea
\label{eqt:Teuk-Star7}
\bsplit
e_3(\ka\rho\aa) &=& \ka\rho e_3(\aa)+\left(-2\ka\kab +2\omb \ka  + 2\rho \right)\rho\aa +\big( r^{-3} \dkb \Ga_b+  r^{-2} \Ga_b\c \Ga_b) \c   \aa.
\end{split}
\eea
Finally, for the fifth term on the right hand side of  \eqref{equation:Starobinski1}, using the $ e_3$ equations for $\vthb,\rho, \kab, \ka$,
\beaa
e_3\Big((2\rho^2-\ka\kab\rho)\vthb\Big) &=& (2\rho^2-\ka\kab\rho)e_3\vthb+4\rho e_3(\rho)\vthb-e_3(\ka)\kab\rho\vthb-\ka e_3(\kab)\rho\vthb-\ka\kab e_3(\rho)\vthb\\
&=& (2\rho^2-\ka\kab\rho)\Big(-\kab \, \vthb  - 2\omb\,  \vthb -2\aa-2\dds_2\,\xib  +\Ga_b\c \Ga_b\Big)\\
&& +4\rho\left(-\frac 3 2 \kab \rho+\ddd_1\bb  +\Ga_g\c \Ga_b\right)\vthb\\
&&-\left(-\frac 1 2 \kab\, \ka +2\omb \ka + 2\rho + 2\ddd_1\eta +\Ga_b\c \Ga_b\right)\kab\rho\vthb\\
&& -\ka \left(-\frac 12 \kab^2 -2 \omb \,\kab +2\ddd_1\xib+\Ga_b\c \Ga_b\right)\rho\vthb\\
&&-\ka\kab\left(-\frac 3 2 \kab \rho+\ddd_1\bb+\Ga_g\c \Ga_b\right)\vthb
\eeaa
i.e.,
\beaa
e_3\Big((2\rho^2-\ka\kab\rho)\vthb\Big) &=&  (2\rho^2-\ka\kab\rho)\Big(-\kab \, \vthb  - 2\omb\,  \vthb -2\aa-2\dds_2\,\xib \Big)- 6 \kab \rho^2\vthb\\
&-&\left(-\frac 1 2 \kab\, \ka +2\omb \ka + 2\rho\right)\kab\rho\vthb+ \ka \left(\frac 12 \kab^2 +2 \omb \,\kab \right)\rho\vthb+\frac 3 2 \kab^2 \ka \rho\\
&+& r^{-5}\dkb^{\le 1 } \Ga_b \c \Ga_b
\eeaa
from which,
\bea
e_3\Big((2\rho^2-\ka\kab\rho)\vthb\Big) &=&(2\rho^2-\ka\kab\rho)\Big( -2\aa-2\dds_2\,\xib\Big)+\left(\frac{7}{2}\ka\kab^2\rho-10\kab\rho^2+2\ka\kab\rho\omb-4\omb\rho^2\right)\vthb\nn \\
&+& r^{-5}\dkb^{\le 1 } \Ga_b \c \Ga_b. \label{eqt:Teuk-Star8}
\eea
Recalling \eqref{equation:Starobinski1}
\beaa
r^{-7}e_3(r^2e_3(r\qf)) &=&  e_3\dds_2\dds_1\ddd_1\bb   -\frac{3}{2}e_3(\rho\dds_2\dds_1\kab) -\frac{3}{2}e_3(\kab\rho\dds_2\ze) -\frac{3}{2}e_3(\ka\rho\aa)\\
&& + \frac{3}{4}e_3\Big((2\rho^2-\ka\kab\rho)\vthb\Big)+7r^{-6}e_3(r)e_3(r\qf) \\
&+&   r^2  e_3 \big( \err[e_3\qf]\big) + r \err[e_3\qf]+\lot 
 \eeaa
and making use of \eqref{eqt:Teuk-Star5}--\eqref{eqt:Teuk-Star8} we deduce,
\beaa
 r^{-7}e_3(r^2e_3(r\qf)) &=& \dds_2\dds_1\ddd_1\ddd_2\aa -2(\kab+\omb)\dds_2\dds_1\ddd_1\bb+3\rho\dds_2\dds_1\ddd_1\xib -\frac{3}{2}\kab\dds_2\dds_1\ddd_1\bb\\
 &-&\frac 3 2 \Bigg\{  -\frac{7}{2}\rho\kab\dds_2\dds_1(\kab) -2\rho\omb\dds_2\dds_1\kab -2\rho\kab\dds_2\dds_1\omb  +2\rho\dds_2\dds_1\ddd_1\xib\\
&&\,\, +\rho\dds_2(\ze-\eta)\left(-\frac 12 \kab^2 -2 \omb \,\kab\right) -\rho\dds_2\xib\left(-\frac{1}{2}\ka\kab+2\om\kab+2\rho\right)\Bigg\}\\
&-&\frac 3 2 \Bigg\{ \kab\rho\left(-3\kab\dds_2\ze-\frac{1}{2}\kab\dds_2\eta -2\omb\dds_2\eta+\dds_2\bb+2\dds_2\dds_1\omb+2\om\dds_2\xib+\frac{1}{2}\ka\dds_2\xib\right)\Bigg\}\\
&-&\frac 3 2 \Bigg\{ \ka\rho e_3(\aa)+\left(-2\ka\kab +2\omb \ka  + 2\rho \right)\rho\aa+ \big( r^{-3} \dkb \Ga_b+  r^{-2} \Ga_b\c \Ga_b) \c   \aa\Bigg\}\\
&+&\frac 3 4 \Bigg\{(2\rho^2-\ka\kab\rho)\Big( -2\aa-2\dds_2\,\xib\Big)+\left(\frac{7}{2}\ka\kab^2\rho-10\kab\rho^2+2\ka\kab\rho\omb-4\omb\rho^2\right)\vthb\Bigg\}\\
&+&7r^{-6}e_3(r)e_3(r\qf) + \err+  r^{-7} \Big(r^2  e_3 \big( \err[e_3\qf]\big) + r \err[e_3\qf] \Big)+\lot .
\eeaa
where,
the error term $\err $  is given by 
\bea
\lab{error-Teuk0Star}
\err&=&\big( r^{-3} \dkb \Ga_b+  r^{-2} \Ga_b\c \Ga_b) \c   \aa+  r^{-4}\dkb^{\le 1 } \Ga_b\c \dkb^{\le 1 } \Ga_g+r^{-5}\dkb^{\le1 }\Ga_b\c \dkb^{\le 1}\Ga_b
\eea

Denoting the expression of  left hand side of the identity  \eqref{eq:Teuk-Star-main(seeApp)}  by $I$, i.e.
\beaa
I:=e_3(r^2e_3(r\qf))   +2\omb r^2e_3(r\qf) 
\eeaa
 we deduce,
\beaa
r^{-7}I
&=& \dds_2\dds_1\ddd_1\ddd_2\aa -\left(\frac{7}{2}\kab+2\omb\right)\dds_2\dds_1\ddd_1\bb +\frac{3}{2}\left(\frac{7}{2}\kab +2\omb\right)\rho\dds_2\dds_1\kab  +\frac{3}{2}\left(\frac{7}{2}\kab +2\omb\right)\kab\rho\dds_2\ze\\
&-&\frac{3}{2}\kab\rho\dds_2\bb -\frac{3}{2}\ka\rho e_3(\aa) -\frac{3}{2}\Big(-3\ka\kab +2\omb \ka  + 4\rho \Big)\rho\aa\\
&+&\frac{3}{4}\left(\frac{7}{2}\ka\kab^2\rho-10\kab\rho^2+2\ka\kab\rho\omb-4\omb\rho^2\right)\vthb+r^{-7}\left[7re_3(r)e_3(r\qf)+2\omb r^2e_3(r\qf)\right] +\widetilde{\err}.
 \eeaa
 where the new  error term $\widetilde{\err}$ is given by
 \beaa
 \widetilde{\err}&=& \err+  r^{-7} \Big(r^2  e_3 \big( \err[e_3\qf]\big) + r \err[e_3\qf] \Big)+ 2\omb r^{-5} \err[e_3\qf]
 \eeaa
 
 To calculate the term $J:=7re_3(r)e_3(r\qf)+2\omb r^2e_3(r\qf)$ in the last row
 we make   use once more of  the identity of Lemma \ref{Le:Teuk-Star1} to derive
\beaa
 J&=& r^2\left(\frac{7}{2}\kab+2\omb\right) e_3(r\qf) +7r\left(e_3(r) - \frac{r}{2}\kab\right)e_3(r\qf)\\
 &=& r^2\left(\frac{7}{2}\kab+2\omb\right) e_3(r\qf) + r^2 \Ga_b e_3(r\qf) \\
 &=& r^7\left(\frac{7}{2}\kab+2\omb\right)\left\{  \dds_2\dds_1\ddd_1\bb   -\frac{3}{2}\rho\dds_2\dds_1\kab -\frac{3}{2}\kab\rho\dds_2\ze -\frac{3}{2}\ka\rho\aa + \frac{3}{4}(2\rho^2-\ka\kab\rho)\vthb\right\}\\
 &+& r^2\left(\frac{7}{2}\kab+2\omb\right) \err[e_3(r\qf)] + r^2 \Ga_b e_3(r\qf) 
 \eeaa
 i.e.,
 \beaa
 r^{-7} J&=&\left(\frac{7}{2}\kab+2\omb\right)\left\{  \dds_2\dds_1\ddd_1\bb   -\frac{3}{2}\rho\dds_2\dds_1\kab -\frac{3}{2}\kab\rho\dds_2\ze -\frac{3}{2}\ka\rho\aa + \frac{3}{4}(2\rho^2-\ka\kab\rho)\vthb\right\}\\
 &+& r^{-5}\left(\frac{7}{2}\kab+2\omb\right) \err[e_3(r\qf)] + r^{-5} \Ga_b e_3(r\qf) 
 \eeaa

Combining and simplifying,
\beaa
r^{-7 }I&=&  \dds_2\dds_1\ddd_1\ddd_2\aa  -\frac{3}{2}\kab\rho\dds_2\bb -\frac{3}{2}\ka\rho e_3(\aa) -\frac{3}{2} \left(\frac{1}{2}\ka\kab +4\omb \ka  + 4\rho \right)\rho\aa-\frac{9}{4}\kab\rho^2\vthb+\widetilde{\widetilde{\err}}.
\eeaa
where,
\beaa
\widetilde{\widetilde{\err}}= \widetilde{\err}+r^{-5}\left(\frac{7}{2}\kab+2\omb\right) \err[e_3(r\qf)] + r^{-5} \Ga_b e_3(r\qf) 
\eeaa

Using  Bianchi  to replace $\dds_2\bb$, we deduce
 \beaa
r^{-7}I&=&  \dds_2\dds_1\ddd_1\ddd_2\aa  -\frac{3}{2}\kab\rho\left(-e_4\aa-\frac{1}{2}\ka\aa+4\om\aa-\frac{3}{2}\rho\vthb+ r^{-1}\Ga_g\c \Ga_b \right)   \\
 &-&\frac{3}{2}\ka\rho e_3(\aa) -\frac{3}{2}\left(\frac{1}{2}\ka\kab +4\omb \ka  + 4\rho \right)\rho\aa-\frac{9}{4}\kab\rho^2\vthb+ r^{-5}\Ga_g\c \Ga_b+\lot\\
&=& \dds_2\dds_1\ddd_1\ddd_2\aa  +\frac{3}{2}\kab\rho e_4\aa -\frac{3}{2}\ka\rho e_3(\aa) - 6\Big(\kab\om+\omb \ka  + \rho \Big)\rho\aa+r^{-5}\Ga_g\c \Ga_b.
 \eeaa
 \beaa
 I&=&e_3(r^2e_3(r\qf))   +2\omb r^2e_3(r\qf) \\
 &=& r^7\Bigg\{  \dds_2\dds_1\ddd_1\ddd_2\aa  +\frac{3}{2}\kab\rho e_4\aa -\frac{3}{2}\ka\rho e_3(\aa) \Big\}+\err[ST]
 \eeaa
 where,
 \beaa
 \err[ST]&=& r^7  \widetilde{\err}+r^{2}\left(\frac{7}{2}\kab+2\omb\right) \err[e_3(r\qf)] + r^{2} \Ga_b e_3(r\qf) +r^{2}\Ga_g\c \Ga_b\\
&+& r^7  \err+   \Big(r^2  e_3 \big( \err[e_3\qf]\big) + r \err[e_3\qf] \Big)+ 2\omb r^{2} \err[e_3\qf]  +r^{2} \Ga_b e_3(r\qf) +r^{2}\Ga_g\c \Ga_b
 \eeaa
 Recall that, see 
\eqref{error-Teuk0Star},
\beaa
\err&=&\big( r^{-3} \dkb \Ga_b+  r^{-2} \Ga_b\c \Ga_b) \c   \aa+  r^{-4}\dkb^{\le 1 } \Ga_b\c \dkb^{\le 1 } \Ga_g+r^{-5}\dkb^{\le1 }\Ga_b\c \dkb^{\le 1}\Ga_b
\eeaa
Hence,
\beaa
\err[ST]&=&r^4\big( \dkb \Ga_b+  r \Ga_b\c \Ga_b) \c   \aa+  r^{3}\dkb^{\le 1 } \Ga_b\c \dkb^{\le 1 } \Ga_g+r^{2}\dkb^{\le1 }\Ga_b\c \dkb^{\le 1}\Ga_b+r^{2} \Ga_b e_3(r\qf)\\
&+& r^2  e_3 \big( \err[e_3\qf]\big) + r \err[e_3\qf] + 2\omb r^{2} \err[e_3\qf]
\eeaa
Recall that, see Proposition \ref{Le:Teuk-Star1},
 \beaa
\err[e_3(r\qf)] &=& r\Ga_b \qf +r^5  \dk^{\le 1} \big(e_3 \eta \c \b\big) + r^3  \dk^{\le 2 }\big( \Ga_b\c\Ga_g\big).
\eeaa
Therefore,
\beaa
E&=&r^2  e_3 \big( \err[e_3\qf]\big) + r \err[e_3\qf] + 2\omb r^{2} \err[e_3\qf]\\
&=&r^2\big( \Ga_b e_3(r\qf)+ e_3(\Ga_b) r\qf \Big)+ r^7 \dk^{\le 2} \big(e_3 \eta \c \b\big)+r^5  \dk^{\le 3 }\big( \Ga_b\c\Ga_g\big)\\
&+& r^2\Ga_b \qf +r^6  \dk^{\le 1} \big(e_3 \eta \c \b\big) + r^4  \dk^{\le 2 }\big( \Ga_b\c\Ga_g\big)\\
&=&r^2\big( \Ga_b e_3(r\qf)+( \dk^{\le 1} \Ga_b) r\qf \Big) +r^7 \dk^{\le 2} \big(e_3 \eta \c \b\big)+r^5  \dk^{\le 3 }\big( \Ga_b\c\Ga_g\big)
\eeaa
Thus,
\beaa
\err[TS]&=&r^4\big( \dkb \Ga_b+  r \Ga_b\c \Ga_b) \c   \aa+r^2\big( \Ga_b e_3(r\qf)+( \dk^{\le 1} \Ga_b) r\qf \Big)\\
&+& r^7 \dk^{\le 2} \big(e_3 \eta \c \b\big)+r^5  \dk^{\le 3 }\big( \Ga_b\c\Ga_g\big)
\eeaa
which end the proof of Proposition \ref{Prop:Teuk-Star-main}.


\section{Proof of  Proposition  \ref{proposition:wave-a-aa-qf}}\label{appendix-Prop.lemma:waveeqalphawithquadraticterms}


In this section we give a proof of Proposition \ref{proposition:wave-a-aa-qf}, i.e. we derive 
 the  wave equation for  the extreme curvature component $\a$,
 \bea
\begin{split}
\square_\g\a &=     - 4 \omb e_4(\a)     +\left(2\ka+4\om\right) e_3(\a)+V\a  +\err(\square_\g\a),  \\
V:&= \left(- 4 e_4(\omb) +\frac{1}{2}\ka\kab - 10 \ka\omb +2\kab\om -8\om\omb - 4\rho + 4e_\th(\Phi)^2 \right)\a,
\end{split}
\eea
where
\beaa
\err(\square_\g\a) &=& \frac{1}{2}\vth e_3(\a) + \frac 3 4 \vth^2\rho+ e_\th(\Phi)\vth \b  -\frac{1}{2}\ka(\ze+4\eta)\b  -(\ze+\etab)e_4(\b) - \xi e_3(\b)\\
&&+ e_\th(\Phi)(2\ze+\etab) \a + \b^2 + e_4(\Phi)\etab\b+ e_3(\Phi)\xi\b - (\ze+4\eta) e_4(\b)-(e_4(\ze)+4e_4(\eta))\b \\
&& - 2(\ka+\om)(\ze+4\eta)\b +2e_\th(\ka+\om) \b  -e_\th((2\ze+\etab)\a)-3\xi e_\th(\rho)+2\etab e_\th(\a)\\
&&  + \frac 3 2 \vth (e_\th(\b)+e_\th(\Phi)\b)   + 3\rho (\etab+\eta+2\ze)\xi    +  (e_\th(\etab)+e_\th(\Phi)\etab)\a +\frac{1}{4}\kab\vth\a -2\omb\vth\a\\
&&  -  \frac{1}{2}\vth\vthb\a +\xi\xib\a+\etab^2\a  + \frac 3 2 \vth\ze\b+3\vth(\etab\b+\xi\bb) -\frac{1}{2}\vth(\ze+4\eta)\b.
\eeaa
The equation for $\aa$ can then be easily inferred by symmetry.
 \begin{proof}
 We make use of the  Bianchi identities
\beaa
 e_\th( \b)-  e_\th(\Phi) \b  &=&   e_3 (\a) +\left(\frac{\kab}{2} - 4 \omb\right)\a + \frac 3 2 \vth\,\rho - (\ze+4\eta)\b,\\
   e_4(\b) +2(\ka+\om) \b &=& e_\th(\a)+2e_\th(\Phi)\a +(2\ze+\etab)\a+3\xi\rho.
\eeaa
to  infer that
\beaa
e_4(e_3(\a)) &=&  e_4(e_\th( \b)) -  e_\th(\Phi) e_4(\b)  -e_4(e_\th(\Phi)) \b  -\left(\frac{\kab}{2} - 4 \omb\right)e_4(\a) -\left(\frac{e_4(\kab)}{2} - 4 e_4(\omb)\right)\a\\
&& - \frac 3 2 \vth e_4(\rho) - \frac 3 2 e_4(\vth) \rho + (\ze+4\eta) e_4(\b)+(e_4(\ze)+4e_4(\eta))\b\\
&=&  e_4(e_\th( \b)) -  e_\th(\Phi)\Big(e_\th(\a)+2e_\th(\Phi)\a - 2(\ka+\om) \b +(2\ze+\etab)\a+3\xi\rho\Big)\\
&&  - (D_4D_\th\Phi + D_{D_4e_\th}\Phi)\b  -\left(\frac{\kab}{2} - 4 \omb\right)e_4(\a) -\left(\frac{e_4(\kab)}{2} - 4 e_4(\omb)\right)\a\\
&& - \frac 3 2 \vth e_4(\rho) - \frac 3 2 e_4(\vth) \rho + (\ze+4\eta) e_4(\b)+(e_4(\ze)+4e_4(\eta))\b.
\eeaa
Hence,
\beaa
e_4(e_3(\a)) &=&  e_4(e_\th( \b)) -  e_\th(\Phi)(e_\th(\a)+2e_\th(\Phi)\a) +2e_\th(\Phi)(\ka+\om) \b  -3  e_\th(\Phi)\xi\rho\\
&&  + e_4(\Phi) e_\th(\Phi)\b  -\left(\frac{\kab}{2} - 4 \omb\right)e_4(\a) -\left(\frac{e_4(\kab)}{2} - 4 e_4(\omb)\right)\a\\
&& - \frac 3 2 \vth e_4(\rho) - \frac 3 2 e_4(\vth) \rho - e_\th(\Phi)(2\ze+\etab) \a - \b^2 - e_4(\Phi)\etab\b- e_3(\Phi)\xi\b\\
&& + (\ze+4\eta) e_4(\b)+(e_4(\ze)+4e_4(\eta))\b
\eeaa
and 
\beaa
e_\th(e_\th(\a)) &=&   e_\th(e_4(\b)) +2(\ka+\om) e_\th(\b)  +2e_\th(\ka+\om) \b - 2e_\th(\Phi)e_\th(\a) - 2e_\th(e_\th(\Phi))\a\\
&& -e_\th((2\ze+\etab)\a)-3e_\th(\xi\rho)\\
&=&   e_\th(e_4(\b)) +2(\ka+\om) \Big(e_\th(\Phi)\b + e_3 (\a) +\left(\frac{\kab}{2} - 4 \omb\right)\a + \frac 3 2 \vth\,\rho - (\ze+4\eta)\b\Big)\\
&&   +2e_\th(\ka+\om) \b - 2e_\th(\Phi)e_\th(\a) - 2(D_\th D_\th\Phi + D_{D_\th e_\th}\Phi)\a  -e_\th((2\ze+\etab)\a)-3e_\th(\xi\rho)\\
&=&   e_\th(e_4(\b)) +2(\ka+\om) e_\th(\Phi)\b +2(\ka+\om) e_3 (\a) +2(\ka+\om) \left(\frac{\kab}{2} - 4 \omb\right)\a +3(\ka+\om) \vth\,\rho \\
&&    - 2e_\th(\Phi)e_\th(\a) - 2\left(\rho -e_\th(\Phi)^2 + \frac{1}{2}\chib e_4(\Phi)+\frac{1}{2}\chi e_3(\Phi)\right)\a -3e_\th(\xi)\rho\\
&&  - 2(\ka+\om)(\ze+4\eta)\b +2e_\th(\ka+\om) \b  -e_\th((2\ze+\etab)\a)-3\xi e_\th(\rho).
 \eeaa
In view of Lemma     \ref{le:square(psi)-null-frame}, we have 
\beaa
\square_\g f &=&  -e_4(e_3( f))+e_\th(e_\th( f)) -\frac{1}{2}\kab e_4(f) +\left(-\frac{1}{2}\ka+2\om\right) e_3(f)+e_\th(\Phi)e_\th( f) +2\etab e_\th(f).
\eeaa
We infer
\beaa
\square_\g\a &=&  -e_4(e_3(\a))+e_\th(e_\th(\a)) -\frac{1}{2}\kab e_4(\a) +\left(-\frac{1}{2}\ka+2\om\right) e_3(\a)+e_\th(\Phi)e_\th(\a) +2\etab e_\th(\a)\\
&=&   [e_\th, e_4](\b)    - e_4(\Phi) e_\th(\Phi)\b + \frac 3 2 \vth e_4(\rho) + \frac 3 2 e_4(\vth) \rho  +3(\ka+\om) \vth\,\rho  -3(e_\th(\xi)-e_\th(\Phi)\xi)\rho\\
&&    - 4 \omb e_4(\a)     +\left(\frac{3}{2}\ka+4\om\right) e_3(\a)\\
&&   +\left(\frac{e_4(\kab)}{2} - 4 e_4(\omb) +\ka\kab - 8 \ka\omb +\kab\om -8\om\omb -2\rho + 4e_\th(\Phi)^2 - \chib e_4(\Phi) - \chi e_3(\Phi)\right)\a\\
&&  + e_\th(\Phi)(2\ze+\etab) \a + \b^2 + e_4(\Phi)\etab\b+ e_3(\Phi)\xi\b - (\ze+4\eta) e_4(\b)-(e_4(\ze)+4e_4(\eta))\b \\
&& - 2(\ka+\om)(\ze+4\eta)\b +2e_\th(\ka+\om) \b  -e_\th((2\ze+\etab)\a)-3\xi e_\th(\rho)+2\etab e_\th(\a).
\eeaa

Next, we have 
\beaa
[e_\th, e_4](\b) &=& \chi e_\th(\b) -(\ze+\etab)e_4(\b) - \xi e_3(\b)\\
&=& \chi \left(  e_\th(\Phi) \b  +   e_3 (\a) +\left(\frac{\kab}{2} - 4 \omb\right)\a + \frac 3 2 \vth\,\rho - (\ze+4\eta)\b\right) \\
 &-&(\ze+\etab)e_4(\b) - \xi e_3(\b)
\eeaa 
and hence
\beaa
\square_\g\a &=&   - 4 \omb e_4(\a)     +\left(\frac{3}{2}\ka+\chi+4\om\right) e_3(\a)+V_1\a\\
&+&  \frac 3 2 \vth e_4(\rho) + \frac 3 2 e_4(\vth) \rho  +3(\ka+\om) \vth\,\rho  -3(e_\th(\xi)-e_\th(\Phi)\xi)\rho +   \frac 3 2 \chi\vth\,\rho\\
&+& \err_1
\eeaa
where,
\beaa
V_1&:=&  \frac{e_4(\kab)}{2} - 4 e_4(\omb) +\ka\kab - 8 \ka\omb +\kab\om -8\om\omb -2\rho + 4e_\th(\Phi)^2 - \chib e_4(\Phi) - \chi e_3(\Phi)\\
&+&\chi\frac{\kab}{2} - 4 \chi\omb,\\
\\
\err_1&:=& e_\th(\Phi)\vth \b  -\chi(\ze+4\eta)\b  -(\ze+\etab)e_4(\b) - \xi e_3(\b)\\
&+& e_\th(\Phi)(2\ze+\etab) \a + \b^2 + e_4(\Phi)\etab\b+ e_3(\Phi)\xi\b - (\ze+4\eta) e_4(\b)-(e_4(\ze)+4e_4(\eta))\b \\
& -& 2(\ka+\om)(\ze+4\eta)\b +2e_\th(\ka+\om) \b  -e_\th((2\ze+\etab)\a)-3\xi e_\th(\rho)+2\etab e_\th(\a).
\eeaa
Next, we make use of 
\beaa
e_4(\vth)+\ka\vth +2\om\vth &=& -2\a +2(e_\th(\xi)-e_\th(\Phi)\xi)+2(\etab+\eta+2\ze)\xi,\\
e_4(\rho)+\frac{3}{2}\ka\rho &=& e_\th(\b)+e_\th(\Phi)\b -\frac{1}{2}\vthb\a+\ze\b+2(\etab\b+\xi\bb),
\eeaa
to  calculate the      term
\beaa
I:&=& \frac 3 2 \vth e_4(\rho) + \frac 3 2 e_4(\vth) \rho  +3(\ka+\om) \vth\,\rho  -3(e_\th(\xi)-e_\th(\Phi)\xi)\rho +   \frac 3 2 \chi\vth\,\rho\\
 &=& \frac 3 2 \vth\left(-\frac{3}{2}\ka\rho+\ddd_1\b \right)+ \frac 3 2\rho\left(  -\ka\vth -2\om\vth -2\a+2(e_\th(\xi)-e_\th(\Phi)\xi)  \right)   +3(\ka+\om) \vth\,\rho  \\
 &-&3(e_\th(\xi)-e_\th(\Phi)\xi)\rho +   \frac 3 2 \frac{\ka+\vth}{2} \, \vth\,\rho+\lot\\
 &=&-3 \rho \a + \frac 3 2 \vth \ddd_1\b+\frac 3 4 \vth^2 \rho.
\eeaa
Hence,
\beaa
\square_\g\a &=&   - 4 \omb e_4(\a)     +\left(\frac{3}{2}\ka+\chi+4\om\right) e_3(\a)+\left( V_1 -3 \rho\right)\a\\
&+& \err_1 + \frac 3 2 \vth \ddd_1\b+\frac 3 4 \vth^2 \rho.
\eeaa
Using also,
\beaa
e_4(\kab) +\frac{1}{2}\ka\kab -2\om\kab &=& 2(e_\th(\etab)+e_\th(\Phi)\etab)+2\rho -\frac{1}{2}\vth\vthb +2(\xi\xib+\etab^2)
\eeaa
and the identities,
  $2\chi=\ka+\vth$, as well as $2\chib=\kab+\vthb$, we finally obtain
\beaa
\square_\g\a &=&     - 4 \omb e_4(\a)     +\left(2\ka+4\om\right) e_3(\a) +V\a+\err[\square\a]  
\eeaa
as desired.

We write schematically   the error term,
\beaa
\err[\square \a] && =\frac{1}{2}\vth e_3(\a) + \frac 3 4 \vth^2\rho+ e_\th(\Phi)\vth \b  -\frac{1}{2}\ka(\ze+4\eta)\b  -(\ze+\etab)e_4(\b) - \xi e_3(\b)\\
&&+ e_\th(\Phi)(2\ze+\etab) \a + \b^2 + e_4(\Phi)\etab\b+ e_3(\Phi)\xi\b - (\ze+4\eta) e_4(\b)-(e_4(\ze)+4e_4(\eta))\b \\
&& - 2(\ka+\om)(\ze+4\eta)\b +2e_\th(\ka+\om) \b  -e_\th((2\ze+\etab)\a)-3\xi e_\th(\rho)+2\etab e_\th(\a)\\
&&  + \frac 3 2 \vth (e_\th(\b)+e_\th(\Phi)\b)   + 3\rho (\etab+\eta+2\ze)\xi    +  (e_\th(\etab)+e_\th(\Phi)\etab)\a +\frac{1}{4}\kab\vth\a -2\omb\vth\a\\
&&  -  \frac{1}{2}\vth\vthb\a +\xi\xib\a+\etab^2\a  + \frac 3 2 \vth\ze\b+3\vth(\etab\b+\xi\bb) -\frac{1}{2}\vth(\ze+4\eta)\b
\eeaa
as follows,
\beaa
\err[\square_\g\a] &=& \left(\frac 1 r \Ga_g+\frac 1 r \dkb \Ga_g\right)\a+\Ga_g e_3(\a) +\frac 1 r \Ga_g \dkb \a\\
&+&\left( \b+ \frac 1 r \Ga_g+\frac 1 r \dk \Ga_g\right)\b +\frac 1 r \Ga_g\dk(\b) +\Ga_g e_3(\b) \\
&+&(\Ga_g)^2\rho +\frac 1 r \Ga_g \dkb(\rho) \\
&=&\Ga_g e_3(\a, \b)+r^{-1}   \Ga_g^{\le 1}   \c    \dk^{\le 1} (\a,\b,\rhoc)+\b^2+\Ga_g^2 \rho.
\eeaa
This concludes the proof of  Proposition \ref{proposition:wave-a-aa-qf}.
 \end{proof}


\section{Proof of Theorem \ref{thm:wave-qf} }\label{appendix-Proof-thmwaveqf}


Recall the symbolic  notation used in the statement of the theorem.
\beaa
\Ga_{g} &=& \Big\{  \vth,  \eta, \etab, \ze,  A\Big\}, \qquad \qquad \qquad \qquad \Ga_{b}=\Big\{\vthb , \xib, \underline{A}\Big\},\\
\dk\Ga_g &=& \Big\{  \dk\vth, r e_\th(\ka),        \dk \eta, \, \dk \etab,\dk  \ze, \dk  A\Big\}, \qquad \,  \dk\Ga_{b}=\Big\{\dk \vthb , e_\th(\kab), \dk \xib,\dk \underline{A}\Big\},
\eeaa
where $A=\frac{2}{r}e_4(r)-\ka, \quad \underline{A}=\frac{2}{r}e_3(r)-\kab.$
  We also denote, for $s\ge 2 $,
  \beaa
   \dk^s\Ga_{g}&=&\dk^{s-1}\dk\Ga_g, \qquad \dk^s\Ga_{b}=\dk^{s-1}\dk\Ga_b, 
  \eeaa
 for higher derivatives with respect to $\dk=( e_3, r e_4, \dkb)$ (see  definition \ref{def:angularderivativesonreducedksclars} for the  notation   $\dkb$ and $\dkb^s$). 

We also recall   Remark \ref{remark-notation-qf}.
\begin{remark} 
\label{remark:app-qf1}
According to  the main bootstrap   assumptions   {\bf BA-E}, {\bf BA-D} (see         section \ref{section:bootstrap}.)
  the terms $\Ga_b$ behave worse in powers of $r$ than  the terms in $\Ga_g$. 
  Thus, in the symbolic  expressions below, we  replace   the terms of  the form $\Ga_g+\Ga_b$   by $\Ga_b$.  We also replace $r^{-1}\Ga_b$ by $\Ga_g$.
  We will denote $l.o.t.$  all  cubic and higher error terms in $\Gac, \Rc$.     We also include in $\lot $ terms which decay faster in powers of $r$   that those  taking into account 
 by   the main  quadratic terms.
  \end{remark}

Recall that 
\bea
\qf=  r^4 Q(\a), 
\eea
where $Q $ is the operator
\bea
\bsplit
Q&:=e_3 e_3 +(2\kab -6\omb)e_3 +W, \qquad 
W:= -4e_3(\omb)+8\omb^2-8\omb\,\kab+\frac{1}{2}\kab^2.
\end{split}
\eea

\begin{lemma}
\lab{Le:confinvarianceofqf}
The  quantity $\qf$ is  fully invariant with respect to the conformal  frame transformations 
\beaa
e_3'=\la^{-1} e_3, \qquad  e_4 = \la e_4 , \qquad e_\th'= e_\th.
\eeaa
\end{lemma}

\begin{proof}
The proof is an immediate consequence of  Definition \ref{definition-conformalderivatives} and  Lemmas \ref{Lemma:confonvariance-Qf}, \ref{Lemma:conformalderivativesnab34} below. 
\end{proof}

We  recall that  under   the   above  mentioned  frame transformation we have
\beaa
\a'&=&\la^2 \a,\quad \b'=\la \b,    \quad \rho'=\rho,     \quad \kab'=\la^{-1} \kab, \quad   \ka'=\la\ka,\quad  \eta'=\eta, \quad \etab'=\etab, \\
 \omb'&=& \la^{-1}\left(\omb +\frac{1}{2} e_3(\log \la)\right), \quad \om'= \la\left(\om -\frac{1}{2} e_4(\log \la)\right), \quad
 \ze'= \ze - e_\th (\log \la).
\eeaa

\begin{definition}
\lab{definition-conformalderivatives}
We say that a reduced tensor is conformal invariant of type\footnote{Note that for a given Ricci or curvature coefficient $a$  coincides with the signature of the component.}    $a$, i.e. $a$-conformal invariant, if  under   the conformal change of frames  $e_3'=\la^{-1},  e_4'=\la  e_4 $  it transforms by
\beaa
 f'= \la^a f. 
\eeaa
\end{definition} 

\begin{lemma} 
\lab{Lemma:conformalderivativesnab34}
Let $f$ be an   $a$-conformal invariant   tensor. 
\begin{enumerate}
\item The tensor 
\bea
\nab_3 f:&=& e_3 f - 2 a  \omb f 
\eea
is $a-1$ conformal invariant.
\item   The tensor 
\bea
\nab_4 f:&=& e_4  f + 2 a  \om f 
\eea
is  $a+1$ conformal invariant.
\item       The tensor,
\bea
\nabb^{(c)} _ A f  &=& \nabb_A  f  +  \a \ze_A  f 
\eea
is $a$-conformal invariant.
\end{enumerate}
\end{lemma}

\begin{proof}
Immediate verification.
\end{proof}

\begin{lemma}
\lab{Lemma:confonvariance-Qf}
We have
\beaa
Q(\a) &=&\nab_3( \nab_3  \a) +2 \kab \nab_3 \a +\frac 1 2 \kab^2 \a. 
\eeaa
\end{lemma}

\begin{proof}
We have,
\beaa
\nab_3( \nab_3  \a)&=& \nab_3 ( e_3 \a - 4 \omb \a)= e_3  ( e_3 \a - 4 \omb \a)- 2 \omb  ( e_3 \a - 4 \omb \a)\\
&=& e_3 e_3 \a - 4 e_3\omb \a -4\omb e_3 \a - 2 \omb e_3 \a + 8 \omb^2 \a.
\eeaa
Hence,
\beaa
Q(\a) &=&\nab_3( \nab_3  \a) +2 \kab \nab_3 \a +\frac 1 2 \kab \a \\
&=& e_3 e_3 \a - 4 e_3\omb \a -4\omb e_3 \a - 2 \omb e_3 \a + 8 \omb^2 \a+ 2 \kab ( e_3 \a -  4 \omb  \a) +\frac 1 2 \kab^2 \a \\
&=& e_3 e_3 \a+( 2\kab -6\omb ) e_3 \a +\left(-4 e_3\omb+ 8 \omb^2- 8 \kab\, \omb  +\frac 1 2 \kab^2\right)
\eeaa
as stated.
\end{proof}

\begin{remark}
Using the definitions of $\nab_3, \nab_4$ the   null structure equations for $\ka, \kab$ take the form,
\bea
\bsplit
\nab_3\kab+\frac 12 \kab^2  &=2\ddd_1\xib+\Ga_b\c \Ga_b= r^{-1}\dkb+\lot,\\
\nab_4\kab +\frac 1 2 \ka\, \kab &= 2\ddd_1\etab + 2\rho +\Ga_g\c \Ga_b= 2\rho+r^{-1}\dkb\Ga_g,\\
\nab_3\ka +\frac 1 2 \kab\, \ka&= 2\ddd_1\eta+ 2\rho+\Ga_g\c \Ga_b= 2\rho+r^{-1}\dkb\Ga_g,\\
\nab_4 \ka +\frac 1 2 \ka^2 &=2\ddd_1\xi+\Ga_g\c \Ga_g=r^{-1}\dkb\Ga_g.
\end{split}
\eea
Also, since $\rho$ is $0$-conformal
\bea
\bsplit
\nab_3 \rho+\frac 3 2 \kab \rho&=\ddd_1\bb +\Ga_g\c \Ga_b= r^{-1} \dkb\Ga_g.
\end{split}
\eea
\end{remark}

\begin{definition}
 Given $f$ an $a$-conformal $S$-tangent  tensor we   define   its   a-conformal Laplacian to be
 \beaa
 \lappc f=\nabbc_A\nabbc^A   f. 
 \eeaa
\end{definition}

 \begin{lemma}
The following formula holds true for a $2$-conformal tensor $f$
\beaa
\lappc f  &=&\lapp_2  f  + 4 \ze \nabb  f  + 2 \big(\div \ze +2|\ze|^2\big) f.  
\eeaa
In particular we have,
\beaa
\lappc f  &=&\lapp_2 f + r^{-1} \dkb^{\le 1 } (\Ga_g \c   f  ).
\eeaa
 \end{lemma}
 
\begin{proof}
Immediate verification.
\end{proof}

 The goal of this section is  to  prove Theorem \ref{thm:wave-qf} which we recall below for the convenience of the reader.

\begin{theorem}
\lab{thm:wave-qf-appendix}
The invariant  scalar quantity $\qf$ defined in \eqref{def: gen-qf} 
 verifies  the  equation,
\bea
\square_2 \qf  +\ka \kab\, \qf=\err[\square_2\qf]
\eea
where, schematically,
\bea
\err[\square_2\qf]&:= r^2 \dk^{\le 2}(\Ga_g \c (\a, \b) )         + e_3 \Big( r^3 \dk^{\le 2}(\Ga_g \c (\a, \b) )   \Big) +  \dk^{\le 1 } (\Ga_g \c \qf)+\lot
\eea
\end{theorem}

\begin{definition}
\lab{def:Goodterms-Wave}
Given a quadratic or higher order  $E$ we    say the following
\begin{enumerate}
\item  $E\in \Good$  if $r^4 E$ can be expressed in the form \eqref{thmwaveqf:schematicformerrorterm}.
\item $E\in \Good_1$ if  after applying  $r^ 4 e_3$ or  $ r^3$ it can be expressed in the form \eqref{thmwaveqf:schematicformerrorterm}.
\item $E\in \Good_2$ if  after applying  $r^ 4 e_3 e_3, r^4 e_3 $ or $ r^3$ it can be expressed in the form \eqref{thmwaveqf:schematicformerrorterm}.
\end{enumerate}
\end{definition}

In view of the definition we note that,
\beaa
(e_3 + r^{-1})\Good_1=\Good,\qquad   Q \Good_2=\Good.  
\eeaa
To prove the theorem we have to check that $\err[\square_2\qf]= r^4 \Good$.


\subsection{The Teukolsky equation for $\a$}


We recall below Proposition \ref{proposition:wave-a-aa-qf}.
\begin{lemma}\lab{lemma:waveeqalphawithquadraticterms-appendix}
We have
\beaa
 \square_2\a &=& -4\omb e_4(\a)+ (4 \om+2\ka) e_3(\a) + V \a +\err[\square_\g \a],\\
V&=&-4\rho   - 4 e_4(\omb)  -8\om\omb  +2\om\, \kab    -10 \ka\, \omb+\frac 1 2 \ka\,\kab,
\eeaa
where $\err[\square_\g \a]$ is given schematically by
\beaa
\err(\square_\g\a) &:=& \Ga_g e_3(\a) + r^{-1}\dk^{\leq 1}\Big((\eta, \Ga_g)(\a, \b)\Big)    +\xi (e_3(\b), r^{-1}\dk\rhoc).    
\eeaa
\end{lemma}
\begin{remark}
 Since $\xi$ vanishes for $r\ge 4 m_0$,   $\eta\in \Ga_g$ and $e_3\a = r^{-1}\dk \a $  we   deduce,
\beaa
\err(\square_\g\a) \in \Good_2. 
\eeaa
\end{remark}

\begin{lemma}
\lab{Lemma:derivationwaveqf1}
The Teukolsky  equation for $\a$ can be written  in the form,
\bea
\LL(\a) &=& \Good_2
\eea
where $\LL$ is the operator
\bea
\LL\a&=&-\nab_4 \nab_3 \a +\lappc_2 \a -\frac 5 2\ka  \nab_3\a -\frac 1 2 \kab \nab_4 \a  - \left(-4\rho+\frac 1 2 \ka \kab\right) \a. 
\eea
We also note that, for a  $2$-conformal tensor $f$, 
\bea
\square_2  f  &=& -\nab_4 \nab_3 f  +\lappc_2 f  -\frac 1 2\ka  \nab_3 f -\frac 1 2 \kab \nab_4 f+r^{-1} \Ga_g \c\dkb  f.
\eea
\end{lemma}

\begin{proof}
Recall that we have (see Definition \ref{Definition:square-k})
\beaa
\square_2 \a &=& -e_4(e_3( \a))+\lapp_2\a  -\frac{1}{2}\kab e_4(\a) +\left(-\frac{1}{2}\ka+2\om\right) e_3(\a) +2\etab e_\th(\a).
\eeaa
Therefore,
\beaa
\LL(\a)&=&  -e_4(e_3( \a))+\lapp_2\a  -\frac{1}{2}\kab e_4(\a) +\left(-\frac{1}{2}\ka+2\om\right) e_3(\a) +2\etab e_\th(\a)\\
&+&   4\omb e_4(\a)- (4 \om+2\ka) e_3(\a) - V \a\\
&=& -e_4(e_3( \a))+\lapp_2\a- \left( \frac 1 2 \kab - 4 \omb \right) e_4 \a - \left(\frac 5 2 \ka + 2\om\right)  e_3 \a   +2\etab e_\th(\a)- V\a \\
&=&-e_4(e_3( \a))+\lappc_2\a- \left( \frac 1 2 \kab - 4 \omb \right) e_4 \a - \left(\frac 5 2 \ka + 2\om\right)  e_3 \a  - V\a+\Good_2. 
\eeaa
On the other hand,
\beaa
\nab_4(\nab_3( \a))&=& \nab_4 \big( e_3 \a  - 4\omb \a \big) = e_4  \big( e_3\a  - 4\omb \a \big) + 2\om \big( e_3\a  - 4\omb \a \big) \\
&=& e_4 e_3 \a - 4\omb  e_4 \a - 4 e_4\omb \a+ 2\om e_3 \a - 8 \om \omb \a. 
\eeaa
Hence,
\beaa
-\nab_4 \nab_3 \a -\frac 5 2\ka  \nab_3\a -\frac 1 2 \kab \nab_4 \a&=&- e_4 e_3 \a + 4\omb  e_4 \a + 4 e_4\omb \a- 2\om e_3 \a + 8 \om \omb \a \\
&-& \frac 5 2 \ka \big( e_3 \a -4 \omb \a)-\frac 1 2 \kab ( e_4 \a + 4 \om \a)\\
&=&- e_4 e_3 \a-\frac 1 2 (\kab - 4\omb) e_4 \a - \left(\frac 5 2 \ka +2 \om\right)  e_3 \a \\
&+&\big(4  e_4 \omb + 8 \om \omb + 10 \ka \omb - 2\om \kab\big)\a. 
\eeaa
We deduce, with $V'= -4\rho+\frac 1 2 \ka \kab$,
\beaa
-\nab_4 \nab_3 \a -\frac 5 2\ka  \nab_3\a -\frac 1 2 \kab \nab_4 -  V'\a  &=&- e_4 e_3 \a-\frac 1 2 (\kab - 4\omb) e_4 \a - \left(\frac 5 2 \ka +2 \om\right)  e_3 \a \\
&+&\left(4  e_4 \omb + 8 \om \omb + 10 \ka \omb - 2\om \kab + 4\rho-\frac 1 2 \ka \kab \right)\a \\
&=&- e_4 e_3 \a-\frac 1 2 (\kab - 4\omb) e_4 \a - \left(\frac 5 2 \ka +2 \om\right)  e_3 \a -V\a. 
\eeaa
Hence,
\beaa
\LL\a&=&-\nab_4 \nab_3 \a +\lappc_2 \a -\frac 5 2\ka  \nab_3\a -\frac 1 2 \kab \nab_4 \a  - \left(-4\rho+\frac 1 2 \ka \kab\right) \a \in \Good_2 
\eeaa 
as desired. 
The proof of the second part of the lemma   follows in the same manner.
\end{proof}


\subsection{Commutation lemmas}


The goal of the  following  lemmas is to    calculate the commutator of  $Q$  with  $\LL$.
\begin{lemma}
\lab{Lemma:derivationwaveqf2}
Give  $f$ an $a$-conformal  tensor  we have,
\bea
\,[ \nab_3, \nab_4] f&=&  2 a  \rho f     +    r^{-1}\Ga_ g  \dkb^{\le 1 } f.
\eea
\end{lemma}

\begin{proof}
We have 
\beaa
\,[ \nab_3, \nab_4] f&=& \nab_3 \nab_ 4  f - \nab_4 \nab_3  f\\
 &=& \big( e_3   -2 (a+1) \omb \big) \big ( e_4 f +2 a \om f\big)-
 \big( e_4 + 2(a-1) \om \big)( e_3 f - 2 a \omb f \big)\\
&=&  e_3 e_4 f- 2 (a+1) \omb e_4 f + 2 a e_3(\om f )   -  4 a  (a+1)\omb \om  f \\
&-& e_4 e_3 f - 2 (a-1) \om e_3 f  + 2 a e_4(\omb f) + 4 a (a-1) \om \omb \\
&=&[e_3, e_4] f - 2\omb e_4 f + 2\om   e_3( f) +2 a \Big( e_3 \om + e_4 \omb  - 4 \om \omb \Big) f. 
\eeaa
Recall that,
\beaa
[e_3, e_4] &=& -2\om e_3 +2\omb e_4 +2(\etab-\eta)e_\th,
\\
 e_3 \om + e_4 \omb  - 4 \om \omb&=& \rho+\Ga_g\c \Ga_b .
\eeaa
We deduce\footnote{Recall that $\eta\in \Ga_g$ in  the frame we are using.},
\beaa
\,[ \nab_3, \nab_4] f&=&  2 a  \rho    +    r^{-1}\Ga_ g  \dkb^{\le 1 } f
\eeaa
as stated.
\end{proof}

\begin{lemma}
\lab{Lemma:derivationwaveqf3}
 Assume $ f$ a-conformal and $g$ is b- conformal. Then $fg$ is $a+b$-conformal and 
 \beaa
 \nab_3(fg) &=&f \nab_3 g+ g\nab_3 f, \\
  \nab_4(fg) &=&f \nab_4 g+ g\nab_4 f.
 \eeaa
 \end{lemma}

\begin{proof}
Indeed 
\beaa
 \nab_3(fg)=  e_3(fg)- 2(a+b)\omb   fg= f e_3 g      +   g e_3 f  -2(a+b) \omb fg = f\nab_3 g + g \nab_3  f
\eeaa
 as stated.
\end{proof}

\begin{lemma}
\lab{Lemma:derivationwaveqf4}
We have,
\bea
\bsplit
\,[Q,   \nab_3 ] \a &=\kab^2 \nab_3 \a +\frac 12 \kab^3 \a +\Good_1,\\
\,[Q,   \nab_4 ] \a&=  \big( 2\rho+ \ka \kab\big) \nab_3\a +\frac 1 2 \ka \kab^2 \a +\Good_1.
\end{split}
\eea
Also,
\bea
\bsplit
 \,[Q,   \nab_4\nab_3  ] \a&= \big( - 2   \rho+\ka \kab \big)\nab_3   \nab_3 \a+\kab^2 \nab_4 \nab_3 \a+\frac 1 2 \kab^3   \nab_4 \a \\
 &+\left(3 \rho \kab-\frac 1 2  \ka \kab^2    \right)\nab_3 \a+ \frac 3 2 \kab^2\left(-\frac 1 2 \ka \kab + 2\rho\right) \a +\Good.
 \end{split}
\eea
\end{lemma}

\begin{proof}
We have\footnote{Recall that $r^{-1} \Ga_b=\Ga_g$.},
\beaa
\,[Q,   \nab_3 ] \a &=&\left( \nab_3 \nab _3 + 2 \kab \nab_3 +\frac 1 2 \kab^2\right)\nab_3 \a - \nab_3 \left(\left( \nab_3 \nab _3 + 2 \kab \nab_3 +\frac 1 2 \kab^2\right)\a\right)\\
&=& - 2 \nab_3 (\kab) \nab_3 \a -  \kab (\nab_3\kab )\a \\
&=&-2 \left( -\frac 1 2 \kab^2 + r^{-1} \Ga_b \right)\nab_3 \a  -  \kab  \left(- \frac 1 2 \kab^2 + r^{-1} \Ga_b \right) \a\\
&=&\kab^2 \nab_3 \a +\frac 12 \kab^3 \a + r^{-1} \Ga_g \dk^{\le 1} \a 
\eeaa
and,
\beaa
\,[Q,   \nab_4 ] \a&=&\left( \nab_3 \nab _3 + 2 \kab \nab_3 +\frac 1 2 \kab^2\right)\nab_4 \a - \nab_4 \left(\left( \nab_3 \nab _3 + 2 \kab \nab_3 +\frac 1 2 \kab^2\right)\a\right)\\
&=&\big( \nab_3 \nab _3 \nab_4 -\nab_4 \nab_3 \nab_3 \big) \a + 2 \kab \big(\nab_3\nab_4 -\nab_4\nab_3 \big) \a  -2 \nab_4 \kab \nab_3 \a -\kab (\nab_4 \kab) \a \\
&=&\nab_3\Big( \big[\nab_3, \nab_4]\a\Big)+ [\nab_3, \nab_4]\nab_3 \a+ 2 \kab [\nab_3, \nab_4] \a  -2 \nab_4 \kab \nab_3 \a -\kab (\nab_4 \kab) \a.
\eeaa
In view of  Lemma   \ref{Lemma:derivationwaveqf2} we have,
\beaa
\big[\nab_3, \nab_4]\a&=& 4 \rho \a  + r^{-1} \Ga_g\c  \dkb^{\le 1} \a,\\
\big[\nab_3, \nab_4] \nab_3 \a &=& 2 \rho \nab_3 \a + r^{-1} \Ga_g\c \dkb^{\le 1}\nab_3  \a.
\eeaa
Hence,
\beaa
 \,[Q,   \nab_4 ] \a&=&\nab_3\Big(  4 \rho  \a + r^{-1} \Ga_g \dkb^{\le 1}  \a \Big)+ \Big(  2 \rho  + r^{-1} \Ga_g \dkb^{\le 1}  \Big)\nab_3 \a + 2 \kab \Big(  4 \rho  \a + r^{-1} \Ga_g \dkb^{\le 1} \a \Big)\\
 &-&2 \nab_4 \kab \nab_3 \a -\kab (\nab_4 \kab) \a\\
 &=& \big( 6  \rho- 2 \nab_4 \kab \big)\nab_3 \a +\big( 4 \nab_3 \rho+ 8\kab \rho -\kab \nab_4 \kab\big) \a +\Good_1. 
\eeaa
We now note, using the equations  for $\nab_4 \rho$ and $\nab_4\kab$,
\beaa
4 \nab_3 \rho+ 8\kab \rho -\kab \nab_4 \kab&=& 4\left(-\frac 3 2 \kab \rho+ r^{-1}\dkb \Ga_g\right)+ 8\kab \rho -\kab\left(-\frac 1 2 \ka \kab +2\rho+r^{-1}\dkb\Ga_g\right)\\
&=&\frac 1 2 \ka \kab^2+r^{-1}\dkb\Ga_g\\
 6  \rho- 2 \nab_4 \kab&=& 6\rho- 2  \left(-\frac 1 2 \ka \kab  +2 \rho+ r^{-1} \dkb \Ga_g\right) \\
 &=& 2\rho+\ka \kab + r^{-1} \dkb \Ga_g.
\eeaa
 Hence,
 \beaa
  \,[Q,   \nab_4 ] \a&=& \big( 2\rho+ \ka \kab\big) \nab_3\a +\frac 1 2 \ka \kab^2 \a +\Good_1
 \eeaa
 as stated.
 
Also,
\bea
 \,[Q,   \nab_4\nab_3  ] \a&=& [Q, \nab_4 ] \nab_3 \a + \nab_4\Big([ Q, \nab_3] \a\Big).
\eea
We first calculate, as above, for $f=\nab_3 \a $
\beaa
 [Q, \nab_4 ]  f &=&\left( \nab_3 \nab _3 + 2 \kab \nab_3 +\frac 1 2 \kab^2\right)\nab_4  f  - \nab_4 \left(\left( \nab_3 \nab _3 + 2 \kab \nab_3 +\frac 1 2 \kab^2\right) f \right)\\
 &=&\big( \nab_3 \nab _3 \nab_4 -\nab_4 \nab_3 \nab_3 \big)  f  + 2 \kab \big(\nab_3\nab_4 -\nab_4\nab_3 \big) f  -2 \nab_4 \kab \nab_3 \a -\kab (\nab_4 \kab) f  \\
&=&\nab_3\Big( \big[\nab_3, \nab_4] f \Big)+ [\nab_3, \nab_4]\nab_3  f + 2 \kab [\nab_3, \nab_4] f   -2 \nab_4 \kab \nab_3 f  -\kab (\nab_4 \kab) f.
\eeaa
In view of  Lemma   \ref{Lemma:derivationwaveqf2}, since $f=\nab_3 \a$ is  $1$-conformal  and $\nab_3 f $ is $0$-conformal,  we have
\beaa
\big[\nab_3, \nab_4] f &=& 2  \rho f  + r^{-1} \Ga_g \dkb^{\le 1}  f, \\
\big[\nab_3, \nab_4] \nab_3  f  &=&  r^{-1} \Ga_g \dkb^{\le 1}\nab_3  f. 
\eeaa
Hence
\beaa
 \,[Q,   \nab_4 ] f &=&\nab_3\Big(  2  \rho  f  + r^{-1} \Ga_g \dkb^{\le 1}  f  \Big)+ \Big(   r^{-1} \Ga_g \dkb^{\le 1}  \Big)\nab_3 f  + 2 \kab \Big(  2 \rho   f  + r^{-1} \Ga_g \dkb^{\le 1}  f  \Big)\\
 &-&2 \nab_4 \kab \nab_3  f  -\kab (\nab_4 \kab) f \\
 &=& \big( 2   \rho- 2 \nab_4 \kab \big)\nab_3  f  +\big(  2 \nab_3 \rho+4\kab \rho -\kab \nab_4 \kab\big)  f \\
 &+& r^{-1} \Ga_g \dkb^{\le 1}\nab_3   f + r^{-2} \Ga_g  \dkb^{\le 1}   f.
 \eeaa
 Therefore,
 \beaa
  \,[Q,   \nab_4 ] \nab_3 \a &=&\big( 2   \rho- 2 \nab_4 \kab \big)\nab_3   \nab_3 \a +\big(  2 \nab_3 \rho+4\kab \rho -\kab \nab_4 \kab\big)  \nab_3\a 
  + r^{-2} \Ga_g \dk^{\le 2 } \a.
 \eeaa
 As above,
 \beaa
  2   \rho- 2 \nab_4 \kab&=& 2 \rho-2\left(-\frac 1 2 \ka \kab + 2\rho + r^{-1}\Ga_g \right)= -2\rho +\ka \kab + r^{-1}\Ga_g,\\
   2 \nab_3 \rho+4\kab \rho -\kab \nab_4 \kab&=& 2\left(-\frac 3 2 \kab \rho + r^{-1}\Ga_g\right)+4\kab \rho -\kab \left(-\frac 1 2 \ka \kab + 2\rho+ r^{-1} \Ga_g \right)\\
   &=&\frac 1 2 \ka \kab^2  -\rho\kab   + r^{-1}\Ga_g. 
 \eeaa
 Hence, since $ r^{-1} \Ga_g (\nab_3\nab_3 \a, \nab_3 \a ) = r^{-2} \Ga_g \c \dk^{\le 2 } \a =\Good$,
 \bea
  \,[Q,   \nab_4 ] \nab_3 \a &=&\big( -2\rho +\ka\kab \big)\nab_3\nab_3 \a +\left(\frac 1 2 \ka \kab^2-\rho\kab \right) \nab_3 \a + \Good.
 \eea
 We deduce,
 \beaa
 \,[Q,   \nab_4\nab_3  ] \a&=& [Q, \nab_4 ] \nab_3 \a + \nab_4\Big([ Q, \nab_3] \a\Big)\\
 &=&\big( -2   \rho+\ka \kab \big)\nab_3   \nab_3 \a +\left(\frac 1 2 \ka\kab^2 -\rho\kab \right) \nab_3\a \\
 &+&\nab_4 \left(\kab^2 \nab_3 \a +\frac 12 \kab^3 \a +\Good_1 \right)+\Good\\
 &=&\big( - 2   \rho+\ka \kab \big)\nab_3   \nab_3 \a+\kab^2 \nab_4 \nab_3 \a+\frac 1 2 \kab^3   \nab_4 \a \\
 &+&\left( \nab_4(  \kab^2)+\frac 12 \ka \kab^2 -\rho\kab  \right)\nab_3 \a +\frac 3 2 \kab^2 \nab_4 \kab \a +\Good.
\eeaa
Note that
\beaa
 \nab_4(  \kab^2)+\frac 12 \ka \kab^2 -\rho\kab &=& 2\kab \left(-\frac 1 2 \ka \kab + 2\rho + r^{-1} \dkb \Ga_g \right)+\frac 1 2  \ka \kab^2 -\rho\kab\\
 &=& 3 \rho \kab-\frac 1 2  \ka \kab^2+ r^{-2} \dkb \Ga_g, \\
 \frac 3 2 \kab^2 \nab_4 \kab &=& \frac 3 2 \kab^2\left(-\frac 1 2 \ka \kab + 2\rho+ r^{-1} \dkb\Ga_g\right).
\eeaa
Hence,
 \beaa
 \,[Q,   \nab_4\nab_3  ] \a&=& \big( - 2   \rho+\ka \kab \big)\nab_3   \nab_3 \a+\kab^2 \nab_4 \nab_3 \a+\frac 1 2 \kab^3   \nab_4 \a \\
 &+&\left(3 \rho \kab-\frac 1 2  \ka \kab^2    \right)\nab_3 \a+ \frac 3 2 \kab^2\left(-\frac 1 2 \ka \kab + 2\rho\right) \a +\Good
\eeaa 
as stated.
\end{proof}

\begin{lemma}
Given $f$ a $2$-conformal  tensor in $\sk_2$  we have
\beaa
\,[\nab_3, \lappc]  f  =-\kab \lappc  f +r^{-1} \dk^{\le 2 }( \Ga_b \c f).
\eeaa
\end{lemma}

\begin{proof}
Recall that  for a  $2$-conformal    spacetime   tensor $f$ we have
\beaa
\lappc f  &=&\lapp f + r^{-1} \dkb^{\le 1 } (\Ga_g \c   f  ).
\eeaa
Hence,
\beaa
\,[\nab_3, \lappc]  f &=&\,[\nab_3, \lapp ]  f+\nab_3 \big( r^{-1} \dkb^{\le 1 } (\Ga_g \c   f  )\big)+  r^{-1} \dkb^{\le 1 } (\Ga_g \c \nab_3   f  ).
\eeaa
On the other hand, since $\nabb\omb = r^{-1}\dkb  \Ga_b, \nabb^2 \omb= r^{-2} \dkb^2 \Ga_b$,
\beaa
\,[\nab_3, \lapp ]  f= [e_3- 4\omb, \lapp  ]f = [ e_3, \lapp] f + r^{-2} \dkb^{\le 2 } (\Ga_b \c  f ).
\eeaa
We deduce,
\beaa
\,[\nab_3, \lappc]  f &=&[ e_3, \lapp] f+ r^{-2} \dk^{\le 2 } (\Ga_b \c  f )+e _3 \big( r^{-1} \dkb^{\le 1 } (\Ga_g \c   f  )\big).
\eeaa
In the reduced form, for an $\sk_2$  tensor $f$,
\beaa
\,[\nab_3, \lappc]  f &=&[ e_3, \lapp_2 ] f+ r^{-2} \dk^{\le 2 } (\Ga_b \c  f )+e _3 \big( r^{-1} \dkb^{\le 1 } (\Ga_g \c   f  )\big).
\eeaa
We now recall that $\lapp_2=-\dds_2\ddd_2+2K$. Hence, applying the commutation  Lemma\footnote{Recall that we have
\beaa
\comb_2(f)&=&- \frac 1 2 \vthb  \dds_{3} f + (\ze-\eta) e_3 f - 2\eta  e_3\Phi f -\xib( e_4  f  +k e_4(\Phi)  f )- 2 \bb f,\\
\comb^*_2(f)&=&- \frac 1 2 \vthb  \ddd_{1} f - (\ze-\eta) e_3 f  -  \eta  e_3\Phi f +\xib(e_4  f  -  e_4(\Phi)  f )
-  \bb f.
\eeaa
}  \ref{Le:comme3e4},
\beaa
[\lapp_2, e_3]f &=& [-\dds_2\ddd_2+2K, e_3] f= -\dds_2[\ddd_2, e_3] f  - [\dds_2, e_3]\ddd_2f  -2e_3(K) f\\
&=& -\dds_2\left(\frac{1}{2}\kab\ddd_2+\underline{Com}_2(f)\right)-\left(\frac{1}{2}\kab\dds_2+\underline{Com}^*_2(\ddd_2 f)\right)-2e_3(K)\\
&=& -\kab\dds_2\ddd_2 f -2e_3(K) f  +e_\th(\kab)\ddd_2 f -\dds_2\left(\underline{Com}_2(f)\right)-\underline{Com}^*_2(\ddd_2 f)\\
&=& -\kab\dds_2\ddd_2 f  -2e_3(K) f  +r^{-2} \dk^{\le 2 }( \Ga_b \c f)+ r^{-1}  \dk^{\le 1 }( \Ga_g \c e_3  f) \\
&=&\kab \lapp_2 f-2 (e_3 K +\kab K) f  +r^{-2} \dk^{\le 2 }( \Ga_b \c f)+ r^{-1}  \dk^{\le 1 }( \Ga_g \c e_3  f).
\eeaa
Note  that, ignoring    the quadratic terms, 
\beaa
e_3 K+\kab K&=& -e_3 \left(\rho+ \frac 1 4 \ka \kab\right)         -\kab \left( \rho+\frac 1 4 \ka \kab \right)\\
&=&-e_3 \rho-\kab \rho - \frac 1 4\Big(  e_ 3(\ka \kab) +\ka \kab^2\Big)\\
&=&\frac 12 \kab \rho -\ddd_1\bb-\frac 1 4\left\{ \ka \left(-\frac 1 2 \kab^2 -2\omb\,  \kab  \right)+ \kab\left( -\frac 1 2 \ka\kab +2\omb \ka + 2\ddd_1 \eta+ 2\rho\right) +\ka \kab^2 \right\}\\
&=& -\ddd_1\bb-\frac 1 2 \kab \ddd_1 \eta. 
\eeaa
We deduce,
\beaa
\, [e_3, \lapp_2]=- [\lapp_2, e_3]f &=&-\kab \lapp_2  f +r^{-1} \dk^{\le 2 }( \Ga_b \c f).
\eeaa
  Consequently,
  \beaa
\,[\nab_3, \lappc]  f  =-\kab \lappc  f +r^{-1} \dk^{\le 2 }( \Ga_b \c f)
\eeaa
as stated.
\end{proof} 

\begin{lemma}
\lab{Lemma:derivationwaveqf-comm4}
We have,
\bea
\,[Q, \lappc]  \a&=& -2 \kab \nab_3 \lappc \a-  \frac 5 2 \kab^2  \lappc\a +\Good.
\eea
\end{lemma}

\begin{proof}
We have
\beaa
\,[Q, \lappc_2]  \a&=&  \left[\nab_3\nab_3 +   2 \kab \nab_3 +\frac 1 2 \kab^2  \right]\lappc \a  -\lappc  \left[\nab_3\nab _3 \a+ 2\kab \nab_3 \a+\frac 1 2 \kab^2\a \right]\\
&=& \nab _3  [\nab_3, \lappc]  \a+ [\nab_3, \lappc]  e_3  \a +  [  2 \kab \nab_3, \lappc]  \a+ \left[\frac 1 2 \kab^2, \lappc_2\right] \a.
\eeaa
Note that
\beaa
\, [  2 \kab \nab _3, \lappc]  \a&=&2 \kab [   \nab_3, \lappc]  \a+\Good, \\
\,\left[\frac 1 2 \kab^2, \lappc_2\right] \a&=&\Good.
\eeaa
Hence, using the previous commutation Lemma,
\beaa
\,[Q, \lappc_2]  \a&=& \nab _3  [\nab_3, \lappc]  \a+ [\nab_3, \lappc]  e_3  \a + 2 \kab  [  \nab_3, \lappc]  \a+\Good\\
&=& \nab _3\Big(  - \kab  \lappc \a  +r^{-1} \dk^{\le 2 }( \Ga_b \c \a)\Big)+\Big(  - \kab  \lappc\nab_3  \a  +r^{-1} \dk^{\le 2 }( \Ga_b \c \nab_3 \a)\Big)\\
&+& 2\kab \Big(  - \kab  \lappc \a  +r^{-1} \dk^{\le 2 }( \Ga_b \c \a)\Big)+\Good\\
&=&  - \kab\big( \nab_3  \lappc \a+\lappc \nab_3 \a \big) -\big (\nab_3 \kab +2\kab^2 \big)\lappc\a +\Good\\
&=&- \kab \big( 2  \nab_3  \lappc \a-[\nab_3, \lappc]\a\big)  -\big (\nab_3 \kab +2\kab^2 \big)\lappc\a +\Good\\
&=&- 2 \kab  \nab_3  \lappc \a  -\big (\nab_3 \kab +3\kab^2 \big)\lappc\a +\Good.
\eeaa
Note that
\beaa
(\nab_3 \kab +3\kab^2)\lappc\a &=&\left( \frac  5  2  \kab^2 + r^{-1}\dkb \Ga_b  \right)\lappc\a=  \frac  5  2  \kab^2+ r^{-2}\dkb \Ga_g\c  \dkb^{\le 2} \a. 
\eeaa
Hence,
\beaa
\,[Q, \lappc_2]  \a&=&- 2 \kab  \nab_3  \lappc \a - \frac  5  2  \kab^2\lappc\a+\Good
\eeaa
as stated.
\end{proof}

\begin{lemma}
\lab{Lemma:derivationwaveqf-comm6}
We have
\beaa
Q(f g)= &Q(f ) g+ f Q(g) +   2 \nab _3 f \nab _3 g-  \frac 1 2 \kab^2  fg. 
\eeaa
Also,
\beaa
\, [Q, fe_4] g&=&Q( f) \nab_4 g+ f[Q, e_4] g + 2 \nab_3 f  \nab_3 \nab_4 g - \frac 1 2 \kab^2   f \nab_4 g,\\
\, [Q, f\nab_3] g&=&Q( f) \nab_3  g+ f[Q, \nab_3] g + 2 \nab_3 f  \nab_3 \nab_3 g - \frac 1 2 \kab^2  f \nab_3 g.
\eeaa
\end{lemma}

\begin{proof}
Recall that,
\beaa
Q&=& \nab_3\nab_3+2\kab \nab_3 +\frac 1 2 \kab^2.
\eeaa
Hence,
\beaa
Q(fg)&=& \left[\nab_3\nab_3+2\kab\nab_3 +\frac 1 2 \kab^2  \right](f g)\\
&=& (\nab_3 \nab_3 f) g+ f( \nab_3 \nab_3 g)+  2 \nab_3 f \nab_3 g + 2\kab( \nab_3 f g + f \nab_3 g) + \frac 1 2 \kab^2  fg\\
&=&\Big( \nab_3 \nab_3 f + 2\kab \nab_3 f \Big) g +   2 \nab_3 f \nab_3 g+  f Q(g)\\
&=&Q(f ) g+ f Q(g) +   2 \nab_3 f \nab_3 g- \frac 1 2 \kab^2  fg. 
\eeaa
Also,
\beaa
\, [Q, f\nab_4] g&=& Q( f \nab_4 g)- f \nab_4 Q (g)=Q( f) \nab_4 g + f Q \nab_4 (g)+ 2 \nab_3 f  \nab_3 \nab_4 g - \frac 1 2 \kab^2  f \nab_4 g\\
& - &f \nab_4 Q (g)\\
&=&\left(Q( f)-\frac 1 2 \kab f\right) \nab_4 g+ f[Q, \nab_4] g + 2 \nab_3 f  \nab_3 \nab_4 g. 
\eeaa
Similarly,
\beaa
\, [Q, f\nab_3] g&=&\left(Q( f) - \frac 1 2 \kab^2   f \right) \nab _3 g+ f[Q, \nab_3] g + 2 \nab_3 f  \nab_3 \nab_3 g
\eeaa
as stated.
\end{proof}


\subsection{Main commutation}


\begin{proposition}
\lab{prop:derivation-squareqf-main}
The following identity holds true.
\bea
\bsplit
\, [Q, \LL]\a &=-2 \kab \nab_4 Q(\a)+  C_Q Q(\a)  +\Good, 
\end{split}
\eea
where,
\beaa
C_Q&=& -8 \rho-\frac  72 \ka\kab.
\eeaa
\end{proposition}

\begin{proof}
In view of Lemma \ref{Lemma:derivationwaveqf1}, we have
\beaa
\LL\a&=&-\nab_4 \nab_3 \a +\lappc_2 \a -\frac 5 2\ka  \nab_3\a -\frac 1 2 \kab \nab_4 \a  - \left(-4\rho+\frac 1 2 \ka \kab\right) \a =Good_2. 
\eeaa
Hence, we infer
\bea
\lab{eq:derivation-squareqf-main0}
\nn\, [ Q, \LL] \a&=& - [Q,  \nab_4 \nab_3 ] \a +[ Q, \lapp_2] \a -\frac 1 2[Q, \kab   \nab_4]  \a - \frac 5 2  [Q, \ka   \nab_3]  \a
+\left[Q,  4\rho -\frac 1 2 \ka\kab \right] \a \\
&=& I+J+K +L +M 
\eea
with $I, J, K, L, M$ denoting each of the commutators on the left of \eqref{eq:derivation-squareqf-main0}.


\subsubsection{Expression for $I$}


In view of Lemma \ref{Lemma:derivationwaveqf4} we have, for $I=- [Q,  \nab_4 \nab_3 ] \a$,
\bea
\lab{eq:derivation-squareqf-main1}
\bsplit
I&=( 2   \rho-\ka \kab \big)\nab_3   \nab_3 \a   -\kab^2 \nab_4 \nab_3 \a-\frac 1 2 \kab^3   \nab_4 \a-\left(-\frac 1 2 \ka \kab^2 + 3 \rho\kab\right)\nab_3 \a \\
&- \frac 3 2  \kab^2\left(-\frac 1 2 \ka \kab + 2\rho \right) \a +\Good.
\end{split}
\eea


\subsubsection{Expression for $J$}


Using Lemma \ref{Lemma:derivationwaveqf-comm4},
\beaa
J&=&[ Q, \lappc] \a =-2 \kab \nab_3 \lappc \a- \frac 5 2 \kab^2  \lappc\a.
\eeaa 
Recalling the definition of $\LL$  and the fact that $\LL\a =\Good_1$ we  write,
\beaa
\lapp_2\a &=&  \nab_4 \nab_3 \a+ \frac 5 2\ka  \nab_3\a+\frac 1 2 \kab \nab_4 \a  + \left(-4\rho+\frac 1 2 \ka \kab\right) \a  + \Good_1. 
\eeaa
Hence,
\beaa
J&=&-2\kab\nab_3\left( \nab_4 \nab_3 \a+ \frac 5 2\ka  \nab_3\a+\frac 1 2 \kab \nab_4 \a  + \left(-4\rho+\frac 1 2 \ka \kab\right) \a\right)\\
&-&\frac 5 2 \kab^2 \left( \nab_4 \nab_3 \a+ \frac 5 2\ka  \nab_3\a+\frac 1 2 \kab \nab_4 \a  + \left(-4\rho+\frac 1 2 \ka \kab\right) \a\right)\\
&=&-2\kab\nab_3 \nab_4 \nab_3 \a- 5\ka \kab\nab_3\nab_3 \a -\kab^2  \nab_3\nab_4 \a -2\kab \left(-4\rho+\frac 1 2 \ka \kab\right)\nab_3  \a\\
&-&2\kab\left(\frac 5 2\nab_3 \ka  \nab_3\a+\frac 1 2 \nab_3 \kab \nab_4 \a  +\nab_3  \left(-4\rho+\frac 1 2 \ka \kab\right) \a\right)\\
&-&\frac 5 2 \kab^2 \left( \nab_4 \nab_3 \a+ \frac 5 2\ka  \nab_3\a+\frac 1 2 \kab \nab_4 \a  + \left(-4\rho+\frac 1 2 \ka \kab\right) \a\right).
\eeaa
According to Lemma \ref{Lemma:derivationwaveqf2}
\beaa
\nab_3 \nab_4 \nab_3 \a&=& \nab_4 \nab_3\nab_3\a  +[\nab_3,  \nab_4] \nab_3 \a= \nab_4 \nab_3\nab_3\a + 2   \rho \nab_3 \a      +    r^{-1}\Ga_ g  \dkb^{\le 1 } \nab_3 \a \\
&=&\nab_4 \nab_3\nab_3\a +2   \rho \nab_3 \a+\Good,\\
\nab_3 \nab_4 \a&=&\nab_4\nab_3\a+  4\rho\a +\Good_1.
\eeaa
We deduce, modulo $\Good$ error terms,
\beaa
J&=& -2\kab\big( \nab_4 \nab_3 \nab_3 \a+2\rho\nab_3\a\big) - 5\ka \kab\nab_3\nab_3 \a -\kab^2 \big( \nab_4\nab_3 \a+ 4 \rho \a \big) -2\kab \left(-4\rho+\frac 1 2 \ka \kab\right)\nab_3  \a\\
&-& 5 \kab \nab_3 \ka\nab_3 \a  -\kab\nab_3\kab\nab_4 \a -2 \kab \nab_3 \left(-4\rho+\frac 1 2 \ka \kab\right)\a\\
&-&\frac 5 2 \kab^2 \left( \nab_4 \nab_3 \a+ \frac 5 2\ka  \nab_3\a+\frac 1 2 \kab \nab_4 \a  + \left(-4\rho+\frac 1 2 \ka \kab\right) \a\right).
\eeaa
Grouping terms we rewrite in the form,
\beaa
J&= -2\kab\nab_4 \nab_3 \nab_3 \a  - 5\ka \kab\nab_3\nab_3 \a+J_{43}\nab_4 \nab_3\a  + J_4\nab_4 \a +J_3 \nab_3 \a + J_0 \a.
\eeaa
We calculate the coefficients $ J_{43}, J_4, J_3, J_0$ as follows.
\beaa
J_{43} &=&-\kab^2-\frac 5 2 \kab^2  =-\frac 7 2 \kab^2, 
\\
J_4&=&-\kab\nab_3\kab -\frac  5 4  \kab^3 =-\kab\left(-\frac 1 2 \kab^2 + r^{-1} \dkb\Ga_b\right)-\frac 54 \kab^3 =-\frac 3 4 \kab^3 + r^{-2}\dkb \Ga_b,
\\
J_3&=&- 4\kab\rho-2\kab \left(-4\rho+\frac 1 2 \ka \kab\right)- 5 \kab \nab_3 \ka-\frac{25}{4} \kab^3 \\
&=&  4 \kab\rho-\frac{29}{4} \kab^3 - 5\kab\left(-\frac 1 2 \ka \kab + 2\rho+ r^{-1}\dkb\Ga_g\right)\\
&=&-6\kab \rho-\frac{19}{4} \kab^3 + r^{-2} \dkb\Ga_g,\\
J_0&=&- 4\rho\kab^2  -2 \kab \nab_3 \left(-4\rho+\frac 1 2 \ka \kab\right) -\frac 5 2 \kab^2\left(-4\rho+\frac 1 2 \ka \kab\right) \\
&=&6\rho\kab^2 -\frac 5 4\ka  \kab^3+ 8 \kab \nab_3 \rho-\kab \left(\kab \nab_3\ka+\ka \nab_3 \kab \right)\\
&=&6\rho\kab^2 -\frac 5 4\ka  \kab^3+ 8 \kab  \left( -\frac 3 2 \kab \rho +r^{-1}\dkb \Ga_g \right)\\
&-&\kab \left(\kab\left(-\frac 1 2 \ka \kab + 2\rho +r^{-1}\Ga_g  \right) +\ka \left(-\frac 1 2 \kab^2 +r^{-1}\dkb\Ga_b\right)\right)\\
&=&- 8\kab^2 \rho-\frac 54 \ka \kab^3 +\ka \kab^3  +r^{-3}\dkb \Ga_b+ r^{-2} \Ga_g. 
\eeaa
Hence 
\beaa
J_4\nab_4 \a &=&-\frac 3 4 \kab^3 +\Good, \\
J_3\nab_3\a &=&\left(-6\kab \rho-\frac{19}{4} \kab^3 \right)\nab_3\a+\Good, \\ 
  J_0\a&=&- 8\kab^2 \rho-\frac 14 \ka \kab^3+\Good.
  \eeaa
We finally derive,
\bea
\lab{eq:derivation-squareqf-main2}
\bsplit
J&= -2\kab\nab_4 \nab_3 \nab_3 \a  - 5\ka \kab\nab_3\nab_3 \a-\frac 7 2 \kab^2\nab_4\nab_3 \a\\
&-\frac 3 4 \kab^3 \nab_4 \a -\left( 6\kab \rho+\frac{19}{4}\ka \kab^2\right)\nab_3 \a - \left(8 \kab^2 \rho+\frac 1 4 \ka \kab^3\right) \a+\Good.
\end{split}
\eea


\subsubsection{Expression for $K$}


Also, using Lemma \ref{Lemma:derivationwaveqf-comm6} and Lemma \ref{Lemma:derivationwaveqf4}  (according to which we have the identity 
 $\,[Q,   \nab_4 ] \a=  \big( 2\rho+ \ka \kab\big) \nab_3\a +\frac 1 2 \ka \kab^2 \a +\Good_1$)
\beaa
K&=&-\frac 1 2   \Big [Q, \kab   \nab_4 \Big ]  \a=-\frac 1 2 \left( Q(\kab) \nab_4 \a +\kab [Q, \nab_4] \a + 2\nab_3  \kab \nab_3 \nab_4 \a -\frac 1 2 \kab ^3 \nab_4 \a\right)\\
&=& -\frac 1 2 \left( Q(\kab)-\frac 1  2 \kab^3 \right) \nab_4 \a
 - \frac 1 2 \kab \left( \left( 2\rho+ \ka \kab\right) \nab_3\a +\frac 1 2 \ka \kab^2 \a \right) \\
 &-&\nab_3  \kab \nab_3 \nab_4 \a+\Good.
\eeaa 
Hence,
\beaa
K&=&-\nab_3  \kab \nab_3 \nab_4 \a-\frac 1 2 \left( Q(\kab)-\frac 1  2 \kab^3 \right) \nab_4 \a 
 - \frac 1 2 \kab  \big( 2  \rho+\ka\kab  \big)\nab_3 \a-\frac 1 4 \ka \kab^3 \a+\Good.
 \eeaa
 We calculate the expression,
 \beaa
  Q(\kab)-\frac 1  2 \kab^3&=& \nab_3 \nab_3 \kab +2 \kab\nab_3 \kab =\nab_3\left(-\frac 1 2 \kab ^2 + r^{-1} \dkb \Ga_b\right) + 2 \kab\left(-\frac 1 2 \kab ^2 + r^{-1} \dkb \Ga_b\right)\\
  &=& -\kab \left(\nab_3 \kab+\kab^2 \big)+ \nab_3 \big( r^{-1} \dkb \Ga_b\right)+ r^{-2}  \dkb \Ga_b\\
  &=&-\frac 1 2 \kab^3 + \nab_3 \big( r^{-1} \dkb \Ga_b\big)+ r^{-2}  \dkb \Ga_b.
  \eeaa
Hence,
\beaa
K&=&-\nab_3  \kab \nab_3 \nab_4 \a+\frac 1 4 \kab^3 \nab_4 \a 
 - \frac 1 2 \kab  \big( 2  \rho+\ka\kab  \big)\nab_3 \a-\frac 1 4 \ka \kab^3 \a+ \nab_3 \big( r^{-1} \dkb \Ga_b\big)  \nab_4 \a +\Good.
\eeaa
We note that,
\beaa
 \nab_3 \big( r^{-1} \dkb \Ga_b\big)  \nab_4 \a&=&  \nab_3 \big( r^{-1} \dkb \Ga_b  \nab_4 \a\big) -  r^{-1} \dkb \Ga_b\nab_3 \nab_4 \a \\
 &=& \nab_3 \big( r^{-1} \dkb \Ga_g   \dk^{\le 1} \a \big)- r^{-2}  \dkb \Ga_g\dk^{\le 2 }\a =\Good.
\eeaa
We deduce,
\beaa
K&=&-\nab_3  \kab \nab_3 \nab_4 \a+\frac 1 4 \kab^3  \nab_4 \a 
 - \frac 1 2 \kab  \big( 2  \rho+\ka\kab  \big)\nab_3 \a-\frac 1 4 \ka \kab^3 \a+\Good\\
 &=&-\left(-\frac 1 2 \kab^2 + r^{-1}\dkb\Ga_b \right)\nab_3 \nab_4 \a+\frac 1 4 \kab^3  \nab_4 \a 
 - \frac 1 2 \kab  \big( 2  \rho+\ka\kab  \big)\nab_3 \a-\frac 1 4 \ka \kab^3 \a+\Good\\
&=&\frac 1 2 \kab^2 \nab_3\nab_4\a +\frac 1 4 \kab^3  \nab_4 \a 
 - \frac 1 2 \kab  \big( 2  \rho+\ka\kab  \big)\nab_3 \a-\frac 1 4 \ka \kab^3 \a+\Good.
\eeaa
In view of Lemma \ref{Lemma:derivationwaveqf2}  $\,[ \nab_3, \nab_4] \a=  4   \rho \a     +    r^{-1}\Ga_ g  \dkb^{\le 1 } \a $. Hence
\beaa
K&=&\frac 1 2 \kab^2 \big( \nab_4\nab_3 \a+ 4   \rho \a     +    r^{-1}\Ga_ g  \dkb^{\le 1 } \a\big) +\frac 1 4 \kab^3  \nab_4 \a 
 - \frac 1 2 \kab  \big( 2  \rho+\ka\kab  \big)\nab_3 \a-\frac 1 4 \ka \kab^3 \a+\Good\\
 &=&\frac 1 2 \kab^2  \nab_4\nab_3 \a +\frac 1 4 \kab^3  \nab_4 \a - \frac 1 2 \kab  \left( 2  \rho+\ka\kab  \right)\nab_3 \a+\kab^2\left( 2\rho-\frac 1 4 \ka \kab\right)\a +\Good.
\eeaa
We have thus derived
\bea
\lab{eq:derivation-squareqf-main3}
\bsplit
K&=\frac 1 2 \kab^2  \nab_4\nab_3 \a +\frac 1 4 \kab^3  \nab_4 \a - \frac 1 2 \kab  \big( 2  \rho+\ka\kab  \big)\nab_3 \a+\kab^2\left( 2\rho-\frac 1 4 \ka \kab\right)\a +\Good.
 \end{split}
\eea


\subsubsection{Expression for $L$}


According to Lemma \ref{Lemma:derivationwaveqf-comm6}  and  Lemma \ref{Lemma:derivationwaveqf4}  (according to which we have the identity $ [Q, \nab_3] \a=\kab^2 \nab_3 \a +\frac 12 \kab^3 \a +\Good_1$)  
\beaa
L&=& -\frac  5  2    \Big [Q, \ka   e_3 \Big ]  \a=-\frac 5 2 \left( Q(\ka) \nab_3 \a +\ka [Q, \nab_3] \a + 2\nab_3  \ka \nab_3 \nab_3 \a -\frac 1 2\ka  \kab ^2 \nab_3 \a\right)\\
&=&-\frac 5 2 \left( Q(\ka) \nab_3 \a +\ka\left(\kab^2 \nab_3 \a +\frac 12 \kab^3 \a \right) + 2\nab_3  \ka \nab_3 \nab_3 \a -\frac 1 2\ka  \kab ^2 \nab_3 \a\right)+\Good\\
&=&- 5\nab_3 \ka  \nab_3\nab_3 \a -\frac 5 2\left( Q(\ka)+\frac 1 2 \ka \kab\right)\nab_3 \a-\frac 5 4 \ka \kab^3 \a +\Good.
\eeaa
Note that 
\beaa
Q(\ka)&=& \nab_3 \nab_3 \ka + 2 \kab \nab_3 \ka+ \frac 1 2 \ka \kab^2\\
&=& \nab_3\left(-\frac 1 2 \ka \kab + 2\rho+ r^{-1} \dkb \Ga_g \right) + 2 \kab \left(-\frac 1 2 \ka \kab + 2\rho+ r^{-1} \dkb \Ga_g \right)+ \frac 1 2 \ka \kab^2\\
&=&-\frac 1 2 \big(\ka \nab_3\kab +\kab \nab_3 \ka \big)+ 2 \left(-\frac 3 2 \kab \rho + r^{-1} \dkb \Ga_g \right)-\ka \kab^2 + 4\rho\kab+ \frac 1 2 \ka \kab^2 \\
&+& e_3 \big(  r^{-1} \dkb\Ga_g \big)+ r^{-2} \dkb \Ga_g \\
&=&-\frac 1 2 \big(\ka \nab_3\kab +\kab \nab_3 \ka \big)+\rho \kab-\frac 1 2 \ka \kab.
\eeaa
Therefore,
\beaa
Q(\ka)&=&-\frac 1 2 \ka \left(-\frac 1 2 \kab^2 + r^{-1} \dkb \Ga_b\right)-\frac 1 2 \kab\left(-\frac 1 2 \ka \kab + 2\rho+ r^{-1} \dkb \Ga_g \right)+\kab \rho -\frac 1 2 \ka \kab^2\\
&+& r^{-1}\dk^{\le 2} \Ga_g= r^{-1}\dk^{\le 2} \Ga_g.
\eeaa
We deduce,
\beaa
L&=&- 5\nab_3 \ka  \nab_3\nab_3 \a -\frac 5 4  \ka \kab^2\nab_3 \a +\frac 5 4 \ka \kab^3 \a +\Good\\
&=&- 5\left( -\frac 1 2 \ka \kab +2\rho\right) \nab_3\nab_3 \a  -\frac 5 4  \ka \kab^2\nab_3 \a +\frac 5 4 \ka \kab^3 \a +\Good.
\eeaa
Therefore,
\bea
\lab{eq:derivation-squareqf-main4}
L&=&- 5\left( -\frac 1 2 \ka \kab +2\rho\right) \nab_3\nab_3 \a  -\frac{5}{4}\ka \kab^2 \nab_3 \a+\frac 5 4 \ka \kab^3 \a +\Good.
\eea


\subsubsection{Expression for $M$}


Similarly, according to Lemma \ref{Lemma:derivationwaveqf-comm6},
\beaa
M&=& \left[Q,  4\rho -\frac 1 2 \ka\kab \right] \a   = Q\left(  4\rho -\frac 1 2 \ka\kab   \right)  \a +2 \nab_3  \left( 4\rho -\frac 1 2 \ka\kab\right)  \nab_3\a -  \frac 1 2 \kab^2  \left(  4\rho -\frac 1 2 \ka\kab   \right) \a 
\eeaa
i.e.,
\beaa
M&=& Q\left(  4\rho -\frac 1 2 \ka\kab   \right)  \a +2 \nab_3  \left( 4\rho -\frac 1 2 \ka\kab\right)  \nab_3\a -  \frac 1 2 \kab^2  \left(  4\rho -\frac 1 2 \ka\kab   \right) \a. 
\eeaa
We calculate,
\beaa
 \nab_3  \left( 4\rho -\frac 1 2 \ka\kab\right) &=& 4 \nab_3\rho- \frac 1 2 \ka \nab_3 \kab-\frac 1 2 \kab \nab_3 \ka \\
 &=& 4\left(-\frac 3 2 \kab \rho+ r^{-1} \dkb \Ga_g\right)-\frac 1 2 \ka\left(-\frac 1 2 \kab^2+ r^{-1} \dkb \Ga_b \right)\\
 &-&\frac 1 2 \kab\left(-\frac 1 2 \ka \kab +2 \rho+ r^{-1} \dkb\Ga_g \right)\\
 &=&- 7 \kab \rho+\frac 1 2 \ka \kab^2 + r^{-1} \dkb \Ga_g.
\eeaa
We deduce,
\beaa
M&=&\left(Q\left(  4\rho -\frac 1 2 \ka\kab\right) -  \frac 1 2 \kab^2  \left(  4\rho -\frac 1 2 \ka\kab   \right)  \right)  \a+\left( - 7 \kab \rho+\frac 1 2 \ka \kab^2\right) \nab_3 \a +\Good.
\eeaa
It remains to calculate
\beaa
M_0&=&Q\left(  4\rho -\frac 1 2 \ka\kab\right) -  \frac 1 2 \kab^2  \left(  4\rho -\frac 1 2 \ka\kab   \right) =\nab_3\nab_3 \left(  4\rho -\frac 1 2 \ka\kab\right) +2 \kab\nab_3 \left(  4\rho -\frac 1 2 \ka\kab\right)\\
&=&\nab_3\left(- 7 \kab \rho+\frac 1 2 \ka \kab^2 + r^{-1} \dkb \Ga_g\right)+ 2 \kab\left( - 7 \kab \rho+\frac 1 2 \ka \kab^2 + r^{-1} \dkb \Ga_g\right)\\
&=&- 7 \rho\nab_3 \kab - 7 \kab \nab_3 \rho+\frac  12 \kab^2 \nab_3 \ka +\ka \kab \nab_3\kab + 2 \kab\left( - 7 \kab \rho+\frac 1 2 \ka \kab^2\right)\\
&+&\nab_3 (r^{-1} \dkb \Ga_g )+ r^{-2} \dkb\Ga_g.
\eeaa
Hence,
\beaa
M_0&=&- 7\rho\left( -\frac 1 2 \kab^2 +r^{-1} \dkb \Ga_b \right)- 7\kab \left(-\frac 3 2 \kab \rho+ r^{-1} \dkb \Ga_g \right)+\frac 1 2 \kab^2\left(-\frac 1 2 \ka \kab+ 2\rho+ r^{-1} \dkb \Ga_g\right)\\
&+&\ka \kab \left(-\frac 1 2 \kab^2 + r^{-1}\dkb \Ga_b\right) + 2 \kab\left( - 7 \kab \rho+\frac 1 2 \ka \kab^2\right)+\nab_3 (r^{-1} \dkb \Ga_g )+ r^{-2} \dkb\Ga_g\\
&=&\kab^2 \rho+\frac 14 \ka \kab^3 +\nab_3 (r^{-1} \dkb \Ga_g )+ r^{-2} \dkb\Ga_g.
\eeaa
We conclude,
\bea
\lab{eq:derivation-squareqf-main5}
M&=&\left(\kab^2 \rho+\frac 14 \ka \kab^3\right)\a +2\left( - 7 \kab \rho+\frac 1 2 \ka \kab^2\right) \nab_3 \a +\Good.
\eea
Indeed note that
\beaa
\nab_3 (r^{-1} \dkb \Ga_g ) \a &=& \nab_3 \big( r^{-1} \dkb \Ga_g \a \big) - r^{-1} \dkb \Ga_g \nab_3 \a =\Good. 
\eeaa


\subsubsection{End of the proof of Proposition \ref{prop:derivation-squareqf-main}}


Using   the equations \eqref{eq:derivation-squareqf-main1}--\eqref{eq:derivation-squareqf-main5} we  deduce, back to \eqref{eq:derivation-squareqf-main0},
\beaa
\bsplit
\, [ Q, \LL] \a&=I+J+K +L +M \\
&=\big( 2   \rho-\ka \kab \big)\nab_3   \nab_3 \a   -\kab^2 \nab_4 \nab_3 \a-\frac 1 2 \kab^3   \nab_4 \a-\left(-\frac 1 2 \ka \kab^2 + 3 \rho\kab\right)\nab_3 \a \\
&- \frac 3 2  \kab^2\left(-\frac 1 2 \ka \kab + 2\rho \right) \a -2\kab\nab_4 \nab_3 \nab_3 \a  - 5\ka \kab\nab_3\nab_3 \a-\frac 7 2 \kab^2\nab_4\nab_3 \a\\
&-\frac 3 4 \kab^3 \nab_4 \a -\left( 6\kab \rho+\frac{19}{4}\ka \kab^2\right)\nab_3 \a - \left(8 \kab^2 \rho+\frac 1 4 \ka \kab^3\right) \a \\
&+\frac 1 2 \kab^2  \nab_4\nab_3 \a +\frac 1 4 \kab^3  \nab_4 \a - \frac 1 2 \kab  ( 2  \rho+\ka\kab  \big)\nab_3 \a+\kab^2\left( 2\rho-\frac 1 4 \ka \kab\right)\a\\
&- 5\left( -\frac 1 2 \ka \kab +2\rho\right) \nab_3\nab_3 \a  -\frac{5}{4} \ka \kab^2 \nab_3 \a-\frac 5 4 \ka \kab^3 \a\\
&+\left(\kab^2 \rho+\frac 14 \ka \kab^3\right)\a +2\left( - 7 \kab \rho+\frac 1 2 \ka \kab^2\right) \nab_3 \a +\Good.
\end{split}
\eeaa
We deduce,
\beaa
\, [ Q, \LL] \a&=& -2\kab\nab_4 \nab_3 \nab_3 \a+ C_{33}' \nab_3\nab_3 \a +C'_{43} \nab_4\nab_3 \a +C'_4\nab_4 \a+ C_3'\nab_3\a + C'_0 \a 
\eeaa
with,
\beaa
C_{33}'&=&\big( 2   \rho-\ka \kab \big) -5 \ka \kab - 5\left( -\frac 1 2 \ka \kab +2\rho\right)= -8 \rho-\frac  72 \ka\kab,\\
C_{43}'&=&-\kab^2-\frac 7 2 \kab^2+\frac 1 2 \kab^2 =- 4 \kab^2,\\
C'_4&=&-\frac 1 2\kab^3  -\frac  3 4\kab^3 +\frac 1 4 \kab^3=-\kab^3,\\
C_3'&=&\frac 1 2 \ka \kab^2 - 3 \rho\kab -\left( 6\kab \rho+\frac{19}{4}\ka \kab^2\right) - \frac 1 2 \kab  \big( 2  \rho+\ka\kab  \big)  -\frac{5}{4} \ka \kab^2 
+2\left( - 7 \kab \rho+\frac 1 2 \ka \kab^2\right)\\
&=&-24 \kab \rho-5 \ka\kab^2,\\
C_0'&=&- \frac 3 2  \kab^2\left(-\frac 1 2 \ka \kab + 2\rho \right) - \left(8 \kab^2 \rho+\frac 1 4 \ka \kab^3\right) +\kab^2\left( 2\rho-\frac 1 4 \ka \kab\right)-\frac 5 4 \ka \kab^3 
+\left(\kab^2 \rho+\frac 14 \ka \kab^3\right)\\
&=& -8 \kab^2 \rho-\frac{3}{4} \ka\kab^3.
\eeaa
Finally we write, recalling the definition of $Q=\nab_3\nab_3+ 2\kab \nab_3 +\frac 12 \kab^2 $,
\beaa
\nab_3 \nab_3\a &=&Q(\a) - 2\kab \nab_3 \a -\frac 1 2 \kab^2 \a
\eeaa
and,
\beaa
\nab_4 \nab_3 \nab_3 \a&=&\nab_4 Q(\a) - 2\kab \nab_4 \nab_3 \a -\frac 1 2 \kab^2\nab_4  \a- 2 \nab_4\kab  \nab_3 \a - \kab \nab_4\kab \a. 
\eeaa
Hence,
\beaa
 -2\kab\nab_4 \nab_3 \nab_3 \a+ C_{33}' \nab_3\nab_3 \a&=&-2 \kab \nab_4 Q(\a) + 4 \kab^2   \nab_4 \nab_3 \a+\kab^3 \nab_4 \a +4 \kab \nab_4 \kab \nab_3 \a\\
 &+& 2\kab^2 \nab_4\kab\a+ C'_{33} \left(Q(\a)  - 2\kab \nab_3 \a -\frac 1 2 \kab^2 \a\right).
\eeaa
We deduce,
\beaa
\, [ Q, \LL] \a&=&-2 \kab \nab_4 Q(\a)+  C'_{33} Q(\a)+ 4 \kab^2   \nab_4 \nab_3 \a+\kab^3 \nab_4 \a +\big( 4\kab\nab_4\kab- 2\kab C'_{33}\big)\nab_3 \a \\
&+&\left(2 \kab^2 \nab_4\kab -\frac 1 2 \kab^2 C'_{33}\right)\a +C'_{43} \nab_4\nab_3 \a +C'_4\nab_4 \a+ C_3'\nab_3\a + C'_0 \a. 
\eeaa
Thus, setting $C_Q=C_{33}'$, we deduce,
\beaa
\, [ Q, \LL] \a&=&-2 \kab \nab_4 Q(\a)+  C_Q Q(\a) +C_{43} \nab_4\nab_3 \a +C_4\nab_4 \a+ C_3\nab_3\a + C_0 \a +\Good
\eeaa
where,
\beaa
C_Q&=&C'_{33}= -8 \rho-\frac  72 \ka\kab,\\
C_{43}&=&4 \kab^2+C'_{43} =4 \kab^2- 4\kab^2=0,\\
C_4&=&\kab^3 +C'_4= \kab^3 -\kab^3 =0.
\eeaa
Also,
\beaa
C_3&=& 2\kab \big( 2 \nab_4\kab-  C'_{33}\big)  +C_3'\\
&=& 2\kab \left(-\ka\kab +4 \rho + r^{-1} \dkb\Ga_g  + 8 \rho+\frac  72 \ka\kab\right)+C'_3\\
&=&2\kab\left(12\rho+ \frac 5 2\ka \kab\right) +\big(-24 \kab \rho-5 \ka\kab^2\big)+ r^{-2}\dkb\Ga_g \\
&=& r^{-2} \dkb \Ga_g, 
\\
C_0&=&2\kab^2 \nab_4\kab -\frac 1 2 \kab^2 \ C'_{33}+ C_0'\\
&=&2 \kab^2\left(-\frac 1 2 \ka\kab + 2\rho+r^{-1}\dkb \Ga_g\right)+\frac 12 \kab^2\left(8 \rho+\frac  72 \ka\kab\right) -8 \kab^2 \rho-\frac 3 4  \ka\kab^3\\
&=& 8 \kab^2\rho+\frac 3 4 \ka\kab^3 -8 \kab^2 \rho  -\frac 3 4  \ka\kab^3 + r^{-3} \dkb \Ga_g \\
&=&  r^{-3} \dkb \Ga_g. 
\eeaa
We have therefore checked  that,
\beaa
\, [ Q, \LL] \a&=&-2 \kab \nab_4 Q(\a)+  C_Q Q(\a)  +\Good, \qquad C_Q= -8 \rho-\frac  72 \ka\kab,
\eeaa
as stated  in   Proposition \ref{prop:derivation-squareqf-main}.
\end{proof}


\subsection{Proof of Theorem \ref{thm:wave-qf}}


We start with the following,
\begin{lemma}
\lab{Lemma:derivationwaveqf-comm7}
We have,
\beaa
\square_2 (f r^4)&=&r^4 \square_2 f- 2 r^4 \big(\kab e_4 f +\ka e_3 f \big) +r^4 \big(-  5   \ka \kab - 4 \rho \big)  f  + O( r^4 \dk ^{\le 1} \Ga_g\c f ).
\eeaa
\end{lemma}

We postpone the proof of the lemma to the end of the section and continue below the proof of the theorem. 
According to Lemma \ref{Lemma:derivationwaveqf1}
\beaa
\LL(\a) &=& \Good_2
\eeaa
where $\LL$ is the operator
\beaa
\LL\a&=&-\nab_4 \nab_3 \a +\lappc_2 \a -\frac 5 2\ka  \nab_3\a -\frac 1 2 \kab \nab_4 \a  - \left(-4\rho+\frac 1 2 \ka \kab\right) \a. 
\eeaa
Applying $Q$   and recalling the definition of the error terms $\Good$ we derive,
\beaa
\LL( Q\a ) &=&-[Q,\LL]\a +\Good.
\eeaa
Thus, in view of  Proposition \ref{prop:derivation-squareqf-main},
\beaa
\, [Q, \LL]\a &=-2 \kab \nab_4 Q(\a)+  C_Q Q(\a), \qquad C_Q= -8 \rho-\frac  72 \ka\kab.
\eeaa
We deduce,
\beaa
\LL( Q\a ) &=&2 \kab \nab_4 Q(\a)-  C_Q Q(\a).
\eeaa
Therefore, modulo $\Good$ terms,
\beaa
 2 \kab \nab_4 Q(\a)-  C_Q Q(\a) &=&      -\nab_4 \nab_3 (Q\a) +\lappc_2 (Q\a)  -\frac 5 2\ka  \nab_3 Q(\a) -\frac 1 2 \kab \nab_4 Q( \a) \\
 & -& \left(-4\rho+\frac 1 2 \ka \kab\right) Q(\a).
\eeaa
We deduce
\beaa
 -\nab_4 \nab_3 (Q\a) +\lappc_2 (Q\a)  -\frac 5 2\ka  \nab_3 Q(\a) -\frac 5  2 \kab \nab_4 Q( \a) + \left(C_Q -\left(-4\rho+\frac 1 2 \ka \kab\right)\right) Q(\a)+\Good.
\eeaa
In view of the expression for $\square_2 $ in the   second part of the  Lemma \ref{Lemma:derivationwaveqf1} we  rewrite in the form
\beaa
\square_2 Q(f) -2 \ka  \nab_3 Q(\a) - 2 \kab \nab_4 Q( \a) -\big( 4\rho+ 4\ka \kab  \big) Q(\a)=\Good +r^{-1} \Ga_g \c\dkb  Q(\a).
\eeaa
Finally, making use of  Lemma \ref{Lemma:derivationwaveqf-comm7} and recalling  that $\qf= r^4 Q(\a) $,
\beaa
\square_2 \qf &=&r^4 \square_2 (Q\a) - 2 r^4 \big(\kab e_4 (Q\a)  +\ka e_3 (Q\a)  \big) +r^4 \big(-  5   \ka \kab - 4 \rho \big)  Qf  + O( r^4 \dk ^{\le 1} \Ga_g\c Q( \a)\\
&=&r^4 \Big (2 \ka   \nab_3 Q(\a)+2 \kab  \nab_4 Q( \a) +\big( 4\rho+ 4\ka \kab  \big) Q(\a)+  \Good \Big)\\
&-& 2 r^4 \big(\kab e_4 (Q\a)  +\ka e_3 (Q\a)  \big) +r^4 \big(-  5   \ka \kab - 4 \rho \big)  Qf  + O(  \dk ^{\le 1} \Ga_g\c \qf)\\
&=& - \ka\kab \qf + r^4 \Good.
\eeaa
This ends the proof of Theorem \ref{thm:wave-qf}.


\subsubsection{Proof of Lemma \ref{Lemma:derivationwaveqf-comm7}}


We have,
\beaa
\square_2 (f r^4) &=& \D^\a \D_\a( f r^4)= \D^\a ( \D_\a f  r^4 + f \D_\a r^4)\\
&=& r^4 \square_2 f + 2 \D_\a ( r^4) \D^\a f +  f \square (r)\\
&=&r^4 \square_2 f - \big(e_3(r^4)   e_4 f + e_4 ( r^4) e_3 f\big)+ f\square (r^4)+   r^4 \Ga_g  \dk f \\
&=&r^4 \square_2 f- 4 r^3  \big(  e_3(r)  e_4 f + e_4 (r) e_3 f\big)+ f\square (r^4)+   r^4 \Ga_g \dk  f \\
&=&r^4 \square_2 f-  2 r^4\Big( (\kab+\Ga_b)  e_4 f + (\ka +\Ga_g) e_3 f\Big)+ f\square (r^4)+   r^4 \Ga_g\c \dk   f 
\\
&=&r^4 \square_2 f- 2 r^4 \big(\kab e_4 f +\ka e_3 f \big) + f\square (r^4) +   r^4 \Ga_g\c   f. 
\eeaa
Also,
\beaa
\square (r^4) &=& -e_4(e_3( r^4 ))  -\frac{1}{2}\kab e_4( r^4) +\left(-\frac{1}{2}\ka+2\om\right) e_3(r^4 ) +\lapp ( r^4) +2  \etab e_\th( r^4 )\\
&=&- 4  e_4( r^3 e_3(r) ) - 2 r^3   \kab \frac r 2  (\ka+\Ga_g)  + 4 r^3 \left(-\frac{1}{2}\ka+2\om\right)\frac r 2 ( \kab+\Ga_b) +\lapp ( r^4) +2\etab e_\th(r^4 )\\
&=&-12 r^2  ( e_4 r)( e_3 r)-4 r^3  e_4 e_3 r -  r^4   \kab \ka+ 2  r^4  \left(-\frac{1}{2}\ka+2\om\right)\kab+ O( r^3 \Ga_b)\\
&=& - 3 r^4 (\kab+\Ga_b)( \ka+\Ga_g) - 4 r^3 e_4\left( \frac r 2 ( \kab+\Ga_b) \right)- 2  r^4 \ka \kab + 4 r^4 \om \kab+  O( r^3 \Ga_b).
\eeaa
Hence,
\beaa
\square (r^4) &=& -5  r^4 \ka \kab + 4 r^4 \om \kab - 2  r^3  e_4( r\kab)+  O( r^4 \dk ^{\le 1} \Ga_g).
\eeaa
Note that,
\beaa
e_4( r\kab)&=& r e_4(\kab) + \frac r 2  \kab (\ka+\Ga_g)= r \left(- \frac 1 2 \ka\, \kab +2\om \kab + 2\ddd_1\etab + 2\rho\right) + \frac r 2  \kab (\ka+\Ga_g)\\
&=& 2 r \rho + 2 r \om \kab +O( \dk^{\le 1 }\Ga_g).
\eeaa
Hence,
\beaa
\square (r^4) &=&-5   r^4 \ka \kab + 4 r^4 \om \kab  - 2 r^3  (2r \rho + 2 r \om \kab) + O( r^4 \dk ^{\le 1} \Ga_g)\\
&=& r^4 \big(-5  \ka \kab - 4  \rho \big) + O( r^4 \dk ^{\le 1} \Ga_g).
\eeaa
We conclude
\beaa
\square_2 (f r^4)&=&r^4 \square_2 f- 2 r^4 \big(\kab e_4 f +\ka e_3 f \big) +r^4 \big(-  5   \ka \kab - 4 \rho \big) f  + O( r^4 \dk ^{\le 1} \Ga_g)
\eeaa
 as stated.


\chapter{APPENDIX TO CHAPTER \ref{chap:proofoftheoremM0M7M8}}



\section{Proof of Proposition \ref{prop:waveeqfor-rt}}\label{appendix:Proofprop:waveeqfor-rt}


\begin{proposition} The following wave equations hold true.
\begin{enumerate}
\item The null curvature component $\rho$ verifies the identity
\beaa
 \square_\g \rho &:=&  \kab e_4\rho + \ka e_3\rho +\frac{3}{2}\Big( \kab\, \ka   + 2\rho \Big)\rho +\err[\square_\g\rho],
 \eeaa
 where
 \beaa
\err[\square_\g\rho] &=& \frac{3}{2}\rho\left( -\frac 1 2  \vthb\, \vth +2(\xib\,  \xi+\eta\, \eta)\right) +\left(\frac 3 2 \kab  -2\omb\right)\left(\frac 1 2 \vthb \, \a -\ze\, \b -2(\etab \,\b+ \xi\,\bb)\right)\\
   &&  - \frac 1 2 \vthb  \dds_2\b + (\ze-\eta) e_3\b - \eta  e_3(\Phi)\b -\xib( e_4\b +e_4(\Phi)\b )- \bb\b \\
    &&  -e_3\left( -\frac 1 2 \vthb \, \a +\ze\, \b +2(\etab \,\b+ \xi\,\bb)\right)\\
    && -\dds_1(\kab)\b   +2\dds_1(\omb) \b    + 3\eta \dds_1(\rho) - \ddd_1\Big(- \vth \bb +\xib \a\Big) -2 \eta e_\th\rho.
\eeaa

\item
The small curvature quantity,
\beaa
\rhot:=r^2\left(\rho+\frac{2m}{r^3} \right)
\eeaa
verifies the wave equation,
\beaa
\square_\g(\rhot) + \frac{8m}{r^3}\rhot &=&  -6m\frac{\square_\g(r)-\left(\frac{2}{r}-\frac{2m}{r^2}\right)}{r^2}-\frac{3m}{r}\left(\ka\kab+\frac{4\Up}{r^2}\right)\\
&&-\frac{3m}{r}\left( A\kab +\Ab \ka\right)+\err[\square_g\rhot],
\eeaa
where
\beaa
\err[\square_g\rhot] &:=& -\frac{6m}{r} A\Ab + \frac{3}{r^2}\rhot^2+\frac{3}{2}\Bigg(\frac{4}{3} A \frac{e_3(r)}{r}+\frac{4}{3} \Ab  \frac{e_4(r)}{r}\Bigg)\rhot\\
&& + \left(\frac{3}{2}\Big(\ka\kab -\frac{8m}{r^3}+\frac{2}{3r^2}\square_\g(r^2)\Big)+\frac{8m}{r^3}\right)\rhot\\
&&-A e_3(\rhot) -\Ab   e_4(\rhot)+\frac{2}{r} A e_3(m)+\frac{2}{r}\Ab e_4(m)\\
&&+4D^a(m)D_a\left(\frac{1}{r}\right)+\frac{2}{r}\square_\g(m) +4r\dds_1(r)\dds_1(\rho)+r^2\err[\square_\g\rho].
\eeaa
\end{enumerate}
\end{proposition}
\begin{proof} We prove the result in the following steps.

{\bf Step 1.} We start by  deriving the wave equation for $\rho$.
From Bianchi, $\rho$ satisfies
\beaa
e_4 \rho+\frac 3 2 \ka \rho&=\ddd_1 \b  -\frac 1 2 \vthb \, \a +\ze\, \b +2(\etab \,\b+ \xi\,\bb).
\eeaa
Differentiating with respect to $e_3$, we obtain
\beaa
   e_3(e_4( \rho))+ \frac{3}{2}\ka e_3(\rho) + \frac{3}{2}e_3(\ka)\rho &=& e_3(\ddd_1\b) +e_3\left( -\frac 1 2 \vthb \, \a +\ze\, \b +2(\etab \,\b+ \xi\,\bb)\right).
\eeaa

Also, $\b$ satisfies from Bianchi
\beaa
e_3 \b+ \kab \b &=& -\dds_1\rho +2\omb \b   + 3\eta \rho- \vth \bb +\xib \a.
\eeaa
Differentiating with respect to $\ddd_1$, we infer
\beaa
\ddd_1(e_3 \b)+ \kab \ddd_1\b -\dds_1(\kab)\b  &=& -\ddd_1\dds_1\rho +2\omb \ddd_1\b -2\dds_1(\omb) \b   + 3\rho\ddd_1\eta - 3\eta \dds_1(\rho)\\
&& + \ddd_1\Big(- \vth \bb +\xib \a\Big)
\eeaa
and hence
\beaa
\ddd_1\dds_1\rho  &=& -\ddd_1(e_3 \b)- \kab \ddd_1\b +2\omb \ddd_1\b+ 3\rho\ddd_1\eta\\
&& +\dds_1(\kab)\b   -2\dds_1(\omb) \b    - 3\eta \dds_1(\rho) + \ddd_1\Big(- \vth \bb +\xib \a\Big).
\eeaa

Next, we add the equation for $\ddd_1\dds_1\rho$ from the one for $e_3(e_4( \rho))$. This yields
\beaa
   && e_3(e_4( \rho))+\ddd_1\dds_1\rho+ \frac{3}{2}\ka e_3(\rho) + \frac{3}{2}e_3(\ka)\rho\\
    &=& [e_3, \ddd_1]\b  - \kab \ddd_1\b +2\omb \ddd_1\b+ 3\rho\ddd_1\eta  +e_3\left( -\frac 1 2 \vthb \, \a +\ze\, \b +2(\etab \,\b+ \xi\,\bb)\right)\\
    && +\dds_1(\kab)\b   -2\dds_1(\omb) \b    - 3\eta \dds_1(\rho) + \ddd_1\Big(- \vth \bb +\xib \a\Big).
\eeaa

Next, we recall the following commutator identity 
\beaa
[e_3, \ddd_1]\b &=& -\frac 1 2 \kab  \ddd_1\b + \frac 1 2 \vthb  \dds_2\b - (\ze-\eta) e_3\b + \eta  e_3(\Phi)\b +\xib( e_4\b +e_4(\Phi)\b )+ \bb\b.
\eeaa
We infer
\beaa
   && e_3(e_4( \rho))+\ddd_1\dds_1\rho+ \frac{3}{2}\ka e_3(\rho) + \frac{3}{2}e_3(\ka)\rho +\left(\frac 3 2 \kab  -2\omb\right) \ddd_1\b- 3\rho\ddd_1\eta \\
    &=&    \frac 1 2 \vthb  \dds_2\b - (\ze-\eta) e_3\b + \eta  e_3(\Phi)\b +\xib( e_4\b +e_4(\Phi)\b )+ \bb\b \\
    &&  +e_3\left( -\frac 1 2 \vthb \, \a +\ze\, \b +2(\etab \,\b+ \xi\,\bb)\right)\\
    && +\dds_1(\kab)\b   -2\dds_1(\omb) \b    - 3\eta \dds_1(\rho) + \ddd_1\Big(- \vth \bb +\xib \a\Big).
\eeaa

Next, we make use of the Bianchi identities and the null structure equations to compute
\beaa
&&\frac{3}{2}e_3(\ka)\rho +\left(\frac 3 2 \kab  -2\omb\right) \ddd_1\b- 3\rho\ddd_1\eta\\ 
&=& \frac{3}{2}\rho\left(-\frac 1 2 \kab\, \ka +2\omb \ka + 2\ddd_1\eta + 2\rho -\frac 1 2  \vthb\, \vth +2(\xib\,  \xi+\eta\, \eta)\right)\\
&&+\left(\frac 3 2 \kab  -2\omb\right)\left(e_4 \rho+\frac 3 2 \ka \rho+\frac 1 2 \vthb \, \a -\ze\, \b -2(\etab \,\b+ \xi\,\bb)\right) - 3\rho\ddd_1\eta\\
&=& \left(\frac 3 2 \kab  -2\omb\right)e_4 \rho +\frac{3}{2}\rho\Big( \kab\, \ka   + 2\rho \Big)\\
&& +\frac{3}{2}\rho\left( -\frac 1 2  \vthb\, \vth +2(\xib\,  \xi+\eta\, \eta)\right) +\left(\frac 3 2 \kab  -2\omb\right)\left(\frac 1 2 \vthb \, \a -\ze\, \b -2(\etab \,\b+ \xi\,\bb)\right).
\eeaa
This yields
\beaa
   && e_3(e_4( \rho))-\lapp\rho+ \frac{3}{2}\ka e_3(\rho) + \left(\frac 3 2 \kab  -2\omb\right)e_4 \rho +\frac{3}{2}\rho\Big( \kab\, \ka   + 2\rho \Big)\\
&=&   -\frac{3}{2}\rho\left( -\frac 1 2  \vthb\, \vth +2(\xib\,  \xi+\eta\, \eta)\right) -\left(\frac 3 2 \kab  -2\omb\right)\left(\frac 1 2 \vthb \, \a -\ze\, \b -2(\etab \,\b+ \xi\,\bb)\right)\\
   &&  + \frac 1 2 \vthb  \dds_2\b - (\ze-\eta) e_3\b + \eta  e_3(\Phi)\b +\xib( e_4\b +e_4(\Phi)\b )+ \bb\b \\
    &&  +e_3\left( -\frac 1 2 \vthb \, \a +\ze\, \b +2(\etab \,\b+ \xi\,\bb)\right)\\
    && +\dds_1(\kab)\b   -2\dds_1(\omb) \b    - 3\eta \dds_1(\rho) + \ddd_1\Big(- \vth \bb +\xib \a\Big),
\eeaa
where we used the fact that $\ddd_1\dds_1=-\lapp$.

Next, recall the formula for the wave operator acting on a scalar $\psi$
\beaa
\square_\g \psi&=&-e_3 e_4 \psi +\lapp\psi+\left(2\omb -\frac 1 2 \kab\right) e_4\psi- \frac 1 2 \ka e_3\psi+2 \eta e_\th \psi.
\eeaa
We infer
\beaa
&& e_3(e_4( \rho))-\lapp\rho+ \frac{3}{2}\ka e_3(\rho) + \left(\frac 3 2 \kab  -2\omb\right)e_4 \rho +\frac{3}{2}\rho\Big( \ka\kab + 2\rho \Big)\\
&=& -\square_\g \rho +\left(2\omb -\frac 1 2 \kab\right) e_4\rho - \frac 1 2 \ka e_3\rho+2 \eta e_\th\rho\\
&& + \frac{3}{2}\ka e_3(\rho) + \left(\frac 3 2 \kab  -2\omb\right)e_4 \rho +\frac{3}{2}\rho\Big( \kab\, \ka   + 2\rho \Big)
\eeaa
and hence
\beaa
 \square_\g \rho &=&  \kab e_4\rho + \ka e_3\rho +\frac{3}{2}\Big( \kab\, \ka   + 2\rho \Big)\rho\\
&&  +\frac{3}{2}\rho\left( -\frac 1 2  \vthb\, \vth +2(\xib\,  \xi+\eta\, \eta)\right) +\left(\frac 3 2 \kab  -2\omb\right)\left(\frac 1 2 \vthb \, \a -\ze\, \b -2(\etab \,\b+ \xi\,\bb)\right)\\
   &&  - \frac 1 2 \vthb  \dds_2\b + (\ze-\eta) e_3\b - \eta  e_3(\Phi)\b -\xib( e_4\b +e_4(\Phi)\b )- \bb\b \\
    &&  -e_3\left( -\frac 1 2 \vthb \, \a +\ze\, \b +2(\etab \,\b+ \xi\,\bb)\right)\\
    && -\dds_1(\kab)\b   +2\dds_1(\omb) \b    + 3\eta \dds_1(\rho) - \ddd_1\Big(- \vth \bb +\xib \a\Big) -2 \eta e_\th\rho.
\eeaa

{\bf Step 2.} We  derive the following,
identity
\bea
\label{equation-waveforrt-1}
\bsplit
\square_\g(r^2\rho) &= - A e_3(r^2\rho)-\Ab e_4(r^2\rho)\\
&+\frac{3}{2}\Bigg(\frac{4}{3} A\frac{e_3(r)}{r}+\frac{4}{3}\Ab \frac{e_4(r)}{r}+\ka\kab+2\rho+\frac{2}{3r^2}\square_\g(r^2)\Bigg)r^2\rho \\
&+4r\dds_1(r)\dds_1(\rho)+r^2\err[\square_\g\rho].
\end{split}
\eea

\begin{proof}
$r^2\rho$ satisfies the following wave equation
\beaa
\square_\g(r^2\rho) &=& r^2\square_\g\rho+2D^a(r^2)D_a(\rho)+\rho\square_\g(r^2).
\eeaa
On the other hand, recall that we have
\beaa
 \square_\g \rho &=&  \kab e_4\rho + \ka e_3\rho +\frac{3}{2}\Big( \kab\, \ka   + 2\rho \Big)\rho +\err[\square_\g\rho].
 \eeaa
We deduce
\beaa
\square_\g(r^2\rho) &=& \left(r^2\ka-e_4(r^2)\right)e_3\rho+\left(r^2\kab-e_3(r^2)\right)e_4\rho\\
&&+\frac{3}{2}\left(\ka\kab+2\rho+\frac{2}{3r^2}\square_\g(r^2)\right)r^2\rho+4r\dds_1(r)\dds_1(\rho)+r^2\err[\square_\g\rho]\\
&=& -A e_3(r^2\rho)-\Ab e_4(r^2\rho)\\
&&+\frac{3}{2}\Bigg(\frac{4}{3} A \frac{e_3(r)}{r}+\frac{4}{3}\Ab \frac{e_4(r)}{r}+\ka\kab+2\rho+\frac{2}{3r^2}\square_\g(r^2)\Bigg)r^2\rho +4r\dds_1(r)\dds_1(\rho)\\&&+r^2\err[\square_\g\rho]
\eeaa
as desired.
\end{proof}

{\bf Step 3.} We  now derive the desired formula for $\square_\g\rhot$.
In view of the definition of $\rhot$, we have
\beaa
\square_\g(\rhot) &=& \square_\g(r^2\rho)+\square_\g\left(\frac{2m}{r}\right)\\
&=& \square_\g(r^2\rho)+2m\square_\g\left(\frac{1}{r}\right)+4D^a(m)D_a\left(\frac{1}{r}\right)+\frac{2}{r}\square_\g(m).
\eeaa
Together with \ref{equation-waveforrt-1} we  deduce,
\beaa
\square_\g(\rhot) &=& -A e_3(r^2\rho)-\Ab e_4(r^2\rho)\\
&&+\frac{3}{2}\Bigg(\frac{4}{3} A \frac{e_3(r)}{r}+\frac{4}{3}\Ab \frac{e_4(r)}{r}+\ka\kab+2\rho+\frac{2}{3r^2}\square_\g(r^2)\Bigg)r^2\rho +4r\dds_1(r)\dds_1(\rho)\\&&+2m\square_\g\left(\frac{1}{r}\right)+4D^a(m)D_a\left(\frac{1}{r}\right)+\frac{2}{r}\square_\g(m) +4r\dds_1(r)\dds_1(\rho)+r^2\err[\square_\g\rho].
\eeaa
Next, we use $r^2\rho=\rhot -2mr^{-1}$. This yields
\beaa
&& \square_\g(\rhot) - \frac{3}{2}\Bigg(\ka\kab -\frac{8m}{r^3}+\frac{2}{3r^2}\square_\g(r^2)\Bigg)\rhot\\
&=& 2m\square_\g\left(\frac{1}{r}\right)-\frac{3m}{r}\ka\kab+\frac{12m^2}{r^4}-\frac{2m}{r^3}\square_\g(r^2)\\
&&-6m A \frac{e_3(r)}{r^2} - 6m\Ab\frac{e_4(r)}{r^2}\\
&&+ \frac{3}{r^2}\rhot^2+\frac{3}{2}\Bigg(\frac{4}{3} A \frac{e_3(r)}{r}+\frac{4}{3}\Ab \frac{e_4(r)}{r}\Bigg)\rhot\\
&&- Ae_3(\rhot)-\Ab e_4(\rhot)+\frac{2}{r} Ae_3(m)+\frac{2}{r} \Ab e_4(m)\\
&&+4D^a(m)D_a\left(\frac{1}{r}\right)+\frac{2}{r}\square_\g(m) +4r\dds_1(r)\dds_1(\rho)+r^2\err[\square_\g\rho].
\eeaa
Note that  in Schwarzschild,
\beaa
\frac{3}{2}\left(\ka\kab-\frac{8m}{r^3}+\frac{2}{3r^2}\square_\g(r^2)\right)=-\frac{8m}{r^3}
\eeaa
and hence
\beaa
&& \square_\g(\rhot) + \frac{8m}{r^3}\rhot\\
&=& 2m\square_\g\left(\frac{1}{r}\right)-\frac{3m}{r}\ka\kab+\frac{12m^2}{r^4}-\frac{2m}{r^3}\square_\g(r^2)\\
&&-6m \frac{e_3(r)}{r^2} -6m \Ab \frac{e_4(r)}{r^2}\\
&&+ \frac{3}{r^2}\rhot^2+\frac{3}{2}\Bigg(\frac{4}{3} A\frac{e_3(r)}{r}+\frac{4}{3}\Ab \frac{e_4(r)}{r}\Bigg)\rhot
 + \left(\frac{3}{2}\Bigg(\ka\kab -\frac{8m}{r^3}+\frac{2}{3r^2}\square_\g(r^2)\Bigg)+\frac{8m}{r^3}\right)\rhot\\
&& -A e_3(\rhot)- \Ab e_4(\rhot)+\frac{2}{r} A e_3(m)+\frac{2}{r}\Ab e_4(m)\\
&&+4D^a(m)D_a\left(\frac{1}{r}\right)+\frac{2}{r}\square_\g(m) +4r\dds_1(r)\dds_1(\rho)+r^2\err[\square_\g\rho].
\eeaa

Also, we have
\beaa
\square_\g\left(\frac{1}{r}\right)-\frac{1}{r^3}\square_\g(r^2) &=& -\frac{\square_\g(r)}{r^2}+2\frac{\D^\a(r)\D_\a(r)}{r^3}-2\frac{\square_\g(r)}{r^2}-2\frac{\D^\a(r)\D_\a(r)}{r^3}\\
&=& -3\frac{\square_\g(r)}{r^2}
\eeaa
and hence
\beaa
&& \square_\g(\rhot) + \frac{8m}{r^3}\rhot -6m\frac{\square_\g(r)}{r^2}-\frac{3m}{r}\ka\kab+\frac{12m^2}{r^4}\\
&&-6m \frac{e_3(r)}{r^2} -6m \Ab \frac{e_4(r)}{r^2}+ \frac{3}{r^2}\rhot^2+\frac{3}{2}\Bigg(\frac{4}{3} A\frac{e_3(r)}{r}+\frac{4}{3}\Ab \frac{e_4(r)}{r}\Bigg)\rhot\\
&& + \left(\frac{3}{2}\Bigg(\ka\kab -\frac{8m}{r^3}+\frac{2}{3r^2}\square_\g(r^2)\Bigg)+\frac{8m}{r^3}\right)\rhot\\
&& -Ae_3(\rhot)- \Ab e_4(\rhot)+ \frac{2}{r} Ae_3(m)+\frac{2}{r}\Ab e_4(m)\\
&&+4D^a(m)D_a\left(\frac{1}{r}\right)+\frac{2}{r}\square_\g(m) +4r\dds_1(r)\dds_1(\rho)+r^2\err[\square_\g\rho].
\eeaa

Finally, since
\beaa
-6m \frac{e_3(r)}{r^2} -6m \Ab \frac{e_4(r)}{r^2}&=& -3m A \frac{\kab}{r}- 3m \Ab \frac{\ka}{r}-\frac{6m}{r} A \Ab
\eeaa
and 
\beaa
-6m\frac{\square_\g(r)}{r^2}-\frac{3m}{r}\ka\kab+\frac{12m^2}{r^4} &=&  -6m\frac{\square_\g(r)-\left(\frac{2}{r}-\frac{2m}{r^2}\right)}{r^2}-\frac{3m}{r}\left(\ka\kab+\frac{4\Up}{r^2}\right),
\eeaa
we obtain
\beaa
\square_\g(\rhot) + \frac{8m}{r^3}\rhot &=&  -6m\frac{\square_\g(r)-\left(\frac{2}{r}-\frac{2m}{r^2}\right)}{r^2}-\frac{3m}{r}\left(\ka\kab+\frac{4\Up}{r^2}\right)\\
&& -3m A \frac{\kab}{r}- 3m \Ab \frac{\ka}{r}-\frac{6m}{r} A \Ab\\
&&+ \frac{3}{r^2}\rhot^2+\frac{3}{2}\Bigg(\frac{4}{3} A \frac{e_3(r)}{r}+\frac{4}{3}\Ab \frac{e_4(r)}{r}\Bigg)\rhot\\
&& + \left(\frac{3}{2}\Bigg(\ka\kab -\frac{8m}{r^3}+\frac{2}{3r^2}\square_\g(r^2)\Bigg)+\frac{8m}{r^3}\right)\rhot\\
&& -Ae_3(\rhot)-\Ab e_4(\rhot)+\frac{2}{r} A e_3(m)+\frac{2}{r}\Ab e_4(m)\\
&&+4D^a(m)D_a\left(\frac{1}{r}\right)+\frac{2}{r}\square_\g(m) +4r\dds_1(r)\dds_1(\rho)+r^2\err[\square_\g\rho].
\eeaa
\end{proof}


\chapter{APPENDIX TO CHAPTER \ref{chap:proofofGCMprocedure}}



\section{Proof of Lemma \ref{lemma:int-gaS-ga:highersobolevregularity}}
\label{appendix:Proofhigherhk-comaprisonlemma}


We start with the following
 \begin{lemma}\lab{lemma:formulausefultocomparehknormformetricS0andpullbackmetric}
Let $k\geq 0$ an integer and let $f\in\sk_k(\S)$. Then, we have
   \beaa
  (\ddd_k^\S f)^\#   &=& \frac{\sqrt{\ga}}{\sqrt{\gaS^\#}}\Bigg\{\dddov_k(f^\#) + \Bigg(\frac{k}{2}U\int_0^1\Big(\sqrt{\ga}(e_\th(\ka)-e_\th(\vth))\Big)^{\#_\la}d\la\\
  &&+\frac k 4 S\int_0^1\left(\sqrt{\ga}e_\th \left( \kab-\vthb -\Omb(\ka-\vth) -2\underline{b}\ga^{1/2} e_\th  \Phi \right)\right)^{\#_\la}d\la \\
  &&+\frac k 4\Big( \kab -\vthb -\Omb(\ka-\vth) -2\underline{b}\ga^{1/2} e_\th\Phi\Big)^\# U' +\frac{k}{2}(\ka+\vth)^\# S'\Bigg)f^\#\Bigg\}
\eeaa   
where for $0\leq \la\leq 1$, $\#_\la$ denotes the pull back by 
\beaa
\psi_\la(\ovu, \ovs, \th) &:=& (\ovu+\la U(\th), \ovs+\la  S(\th), \th).
\eeaa
 \end{lemma}
 
 \begin{proof}
 For $p\in\ovS$ and $f$ a $\Z$-invariant scalar function on $S$, we have by definition of the push forward of a vectorfield
\beaa
[\Psi_\#(\pr_\th)f]_{\Psi(p)} &=& [\pr_\th(f\circ\Psi)]_p.
\eeaa
We infer
  \beaa
  (\ddd_k^\S f)^\# &=& \frac{1}{\sqrt{\gaS^\#}}\Big(\pr_\th(f^\#) +k\pr_\th(\Phi^\#) f^\#\Big)
  \eeaa 
  and hence
  \beaa
  (\ddd_k^\S f)^\# &=& \frac{\sqrt{\ga}}{\sqrt{\gaS^\#}}\Big(e_\th(f^\#) +ke_\th(\Phi)f^\#+k(e_\th(\Phi^\#)- e_\th(\Phi))f^\#\Big)\\
  &=& \frac{\sqrt{\ga}}{\sqrt{\gaS^\#}}\Big(\dddov_k(f^\#) +k(e_\th(\Phi^\#)- e_\th(\Phi))f^\#\Big).
  \eeaa   
  
Next, we have
\beaa
e_\th(\Phi^\#)- e_\th(\Phi) &=& \sqrt{\ga}^{-1}\Big(\pr_\th(\Phi^\#) - \pr_\th\Phi\Big)
\eeaa
and
  \beaa
   && \Big(\pr_\th( \Phi^\#)- \pr_\th \Phi\Big)(\ovu, \ovs, \th) \\
    &=&   \pr_\th[   \Phi (\ovu+U(\th), \ovs+ S(\th), \th)] -  \pr_\th \Phi(\ovu, \ovs, \th)\\
    &=&(\pr_\th\Phi) (\ovu+U(\th), \ovs+ S(\th), \th)  -  \pr_\th \Phi(\ovu, \ovs, \th) +\Big[( \pr_u \Phi)^\# U' +( \pr_s \Phi)^\# S'\Big](\ovu, \ovs, \th) \\
    &=&\int_0^1 \frac{d}{d\la}\left[ (\pr_\th\Phi) (\ovu+\la U(\th), \ovs+\la  S(\th), \th)\right] d\la+\Big[( \pr_u \Phi)^\# U' +( \pr_s \Phi)^\# S'\Big](\ovu, \ovs, \th)\\
    &=& U(\th)\int_0^1( \pr_ u \pr_\th \Phi)(\ovu+\la U(\th), \ovs+\la  S(\th), \th)d\la\\
    &&+S(\th)\int_0^1( \pr_ s \pr_\th \Phi)(\ovu+\la U(\th), \ovs+\la  S(\th), \th)d\la +\Big[( \pr_u \Phi)^\# U' +( \pr_s \Phi)^\# S'\Big](\ovu, \ovs, \th)
   \eeaa
   which we rewrite
  \beaa
  \pr_\th( \Phi^\#)- \pr_\th \Phi  &=& U\int_0^1(\pr_ u \pr_\th \Phi)^{\#_\la}d\la+S\int_0^1(\pr_ s \pr_\th \Phi)^{\#_\la}d\la +( \pr_u \Phi)^\# U' +( \pr_s \Phi)^\# S'
   \eeaa   
where $\#_\la$ denotes the pull back by  the map  $\psi_\la(\ovu, \ovs, \th) = (\ovu+\la U(\th), \ovs+\la  S(\th), \th). $

Next,  recall that,
    \beaa
    \pr_s =e_4, \quad \pr_u=\frac 1 2\left( e_3-\Omb e_4-\underline{b}\ga^{1/2} e_\th\right), \quad \pr_\th=\sqrt{\ga}e_\th.
    \eeaa
    Hence,
    \beaa
     \pr_ \th \pr_s \Phi&=&  \sqrt{\ga}e_\th e_4(\Phi) \\
     \pr_ \th \pr_u \Phi&=& \frac 1 2 \sqrt{\ga}e_\th \left( e_3\Phi -\Omb e_4\Phi -\underline{b}\ga^{1/2} e_\th  \Phi \right)
    \eeaa
which yields
 \beaa
  \pr_\th( \Phi^\#)- \pr_\th \Phi  &=& U\int_0^1\Big(\sqrt{\ga}e_\th e_4(\Phi)\Big)^{\#_\la}d\la\\
  &&+S\int_0^1\left(\frac 1 2 \sqrt{\ga}e_\th \left( e_3\Phi -\Omb e_4\Phi -\underline{b}\ga^{1/2} e_\th  \Phi \right)\right)^{\#_\la}d\la \\
  &&+\frac 1 2\Big( e_3\Phi -\Omb e_4\Phi -\underline{b}\ga^{1/2} e_\th\Phi\Big)^\# U' +(e_4\Phi)^\# S'\\
&=& \frac{1}{2}U\int_0^1\Big(\sqrt{\ga}(e_\th(\ka)-e_\th(\vth))\Big)^{\#_\la}d\la\\
  &&+\frac 1 4 S\int_0^1\left(\sqrt{\ga}e_\th \left( \kab-\vthb -\Omb(\ka-\vth) -2\underline{b}\ga^{1/2} e_\th  \Phi \right)\right)^{\#_\la}d\la \\
  &&+\frac 1 4\Big( \kab -\vthb -\Omb(\ka-\vth) -2\underline{b}\ga^{1/2} e_\th\Phi\Big)^{\#} U' +\frac{1}{2}(\ka+\vth)^{\#} S'.
   \eeaa 
 We deduce  
  \beaa
  (\ddd_k^\S f)^\#   &=& \frac{\sqrt{\ga}}{\sqrt{\gaS^\#}}\Bigg\{\dddov_k(f^\#) + \Bigg(\frac{k}{2}U\int_0^1\Big(\sqrt{\ga}(e_\th(\ka)-e_\th(\vth))\Big)^{\#_\la}d\la\\
  &&+\frac k 4 S\int_0^1\left(\sqrt{\ga}e_\th \left( \kab-\vthb -\Omb(\ka-\vth) -2\underline{b}\ga^{1/2} e_\th  \Phi \right)\right)^{\#_\la}d\la \\
  &&+\frac k 4\Big( \kab -\vthb -\Omb(\ka-\vth) -2\underline{b}\ga^{1/2} e_\th\Phi\Big)^\# U' +\frac{k}{2}(\ka+\vth)^\# S'\Bigg)f^\#\Bigg\}.
\eeaa   
This concludes the proof of the lemma.
\end{proof}   

We are ready to prove the  higher derivative  comparison  Lemma \ref{lemma:int-gaS-ga:highersobolevregularity} which we recall below.

 \begin{lemma}
 Let $\ovS \subset \RR=\RR(\epg, \dg)$ as in Definition \ref{defintion:regionRRovr} verifying the assumptions {\bf A1-A3}.  Let  $\Psi:\ovS\longrightarrow \S $  be  $\Z$-invariant deformation. Assume the bound 
 \bea\lab{eq:boundonU'andS'onS0forequivalenceofhigherorderSobolevnorms-app}
 \|(U', S')\|_{L^\infty_1(\ovS)}+\rg^{-1}  \max_{0\leq s\leq s_{max}}\|(U', S')\|_{\hk_s(\ovS, \ovgS)}&\les& \dg.
 \eea
 
 Then, we have for any reduced scalar $h$ defined on $\RR$
 \beaa
 \|h\|_{\hk_s(\S)} \les\sup_{\RR}|\dk^{\leq k}h|\,\,\textrm{ for }0\leq s \leq s_{max}.
 \eeaa
 Also, if $f\in \hk_s(\S)$ and $f^\#$ is its pull-back by $\psi$, we have
 \beaa
 \|f\|_{\hk_s(\S)}= \|f^\#\|_{\hk_s(\ovS,\, \gS^{\S,\#})} = \| f^\#\|_{\hk_s(\ovS, \ovgS)}(1+O(\epg))\,\,\textrm{ for }0\leq s\leq s_{max}.
 \eeaa
\end{lemma} 

\begin{remark} Note that the   estimates of the lemma are  independent of the  size $\rg$  of   the  sphere $\ovS=S(\ug, \sg)\subset \RR$, see Definition \ref{defintion:regionRRovr}.  To  simplify the argument below  we assume $\rg\approx 1$. The general case can be easily deduced by a simple scaling argument or making obvious  adjustments in the inequalities below. 
\end{remark}

\begin{proof}  
We argue by iteration. We consider the following iteration assumptions 
\bea\lab{eq:iterationassupmtionforproofofequivalenceofhigherorderSobolevnorms:0}
\textrm{If }\eqref{eq:boundonU'andS'onS0forequivalenceofhigherorderSobolevnorms}\textrm{ holds, then we have }\|h\|_{\hk_s(\S)} \les \sup_{\RR}|\dk^{\leq s}h|,
\eea
and
\bea\lab{eq:iterationassupmtionforproofofequivalenceofhigherorderSobolevnorms}
\textrm{If }\eqref{eq:boundonU'andS'onS0forequivalenceofhigherorderSobolevnorms}\textrm{ holds, then we have }\|f^\#\|_{\hk_s(\ovS,\, \gS^{\S,\#})} = \| f^\#\|_{\hk_s(\ovS, \ovgS)}(1+O(\dg)).
\eea

First, note that \eqref{eq:iterationassupmtionforproofofequivalenceofhigherorderSobolevnorms:0} holds trivially for $s=0$ and  \eqref{eq:iterationassupmtionforproofofequivalenceofhigherorderSobolevnorms} holds for $s=0$  by Lemma \ref{lemma:int-gaS-ga}. Thus, from now on, we assume that \eqref{eq:iterationassupmtionforproofofequivalenceofhigherorderSobolevnorms:0}  holds for some $s$ with $0\leq s\leq s_{max}-1$ and that \eqref{eq:iterationassupmtionforproofofequivalenceofhigherorderSobolevnorms} holds for some $s$ with $0\leq s\leq s_{max}-3$, and our goal is to prove that it also holds for $s$ replaced by $s+1$.

We start with \eqref{eq:iterationassupmtionforproofofequivalenceofhigherorderSobolevnorms:0}. We have
\beaa
\ddd_k^\S h &=& e_\th^\S h+e_\th^\S(\Phi)h.
\eeaa
Now, recall that we have
\beaa
e_\th^\S = \frac{1}{\sqrt{\ga^\S}} \pr_\th^\S,\qquad  \pr_\th^\S|_{\Psi(p)}= \left( \left(  S'-\frac 1 2 \Omb  U'\right) e_4+\frac 1 2 U' e_3 + \sqrt{\ga}\left(1-\frac 1 2 \underline{b} U'\right) e_\th \right)\Big|_{\Psi(p)}.
 \eeaa
This yields
\beaa
(\ddd_k^\S h)_{|_{\Psi(p)}} &=& \Bigg\{\frac{1}{\sqrt{\ga^\S}}\Bigg(\left(  S'-\frac 1 2 \Omb  U'\right) e_4(h)+\left(  S'-\frac 1 2 \Omb  U'\right) e_4(\Phi)h\\
&&+\frac 1 2 U' e_3(h)+\frac 1 2 U' e_3(\Phi)h + \sqrt{\ga}\left(1-\frac 1 2 \underline{b} U'\right)\ddd_k(h) \Bigg)\Bigg\}_{|_{\Psi(p)}}.
\eeaa 
Together with the iteration assumption \eqref{eq:iterationassupmtionforproofofequivalenceofhigherorderSobolevnorms}, we infer
\beaa
\|\ddd_k^\S h\|_{\hk_s(\S)} &=& \|(\ddd_k^\S h)^\#\|_{\hk_s(\ovS, \ovgS)}(1+O(\dg))\\
&\les& \Bigg\|\frac{1}{\sqrt{\ga^{\S,\#}}}\Bigg(\left(  S'-\frac 1 2 \Omb^\#  U'\right) (e_4(h))^\#+\left(  S'-\frac 1 2 \Omb^\#  U'\right) (e_4(\Phi)h)^\#\\
&&+\frac 1 2 U' (e_3(h))^\#+\frac 1 2 U' (e_3(\Phi)h)^\# + \sqrt{\ga^\#}\left(1-\frac 1 2 \underline{b}^\# U'\right)(\ddd_k(h))^\# \Bigg)\Bigg\|_{\hk_s(\ovS, \ovgS)}
\eeaa
i.e.,
\beaa
 \|\ddd_k^\S h\|_{\hk_s(\S)}&\les& \left\|(\ddd_k(h))^\#\right\|_{\hk_s(\ovS, \ovgS)}+\left\|\left(\frac{\sqrt{\ga^\#}}{\sqrt{\ga^{\S,\#}}}-1\right)(\ddd_k(h))^\#\right\|_{\hk_s(\ovS, \ovgS)}\\
&&+\Bigg\|\frac{1}{\sqrt{\ga^{\S,\#}}}\Bigg(\left(  S'-\frac 1 2 \Omb^\#  U'\right) (e_4(h))^\#+\left(  S'-\frac 1 2 \Omb^\#  U'\right) (e_4(\Phi)h)^\#\\
&&+\frac 1 2 U' (e_3(h))^\#+\frac 1 2 U' (e_3(\Phi)h)^\# -\frac 1 2 \sqrt{\ga^\#}\underline{b}^\# U'(\ddd_k(h))^\#\Bigg)\Bigg\|_{\hk_s(\ovS, \ovgS)}.
\eeaa
Together with a non sharp product rule in $\hk_s(\ovS, \ovgS)$ and the repeated use of the iteration assumptions  \eqref{eq:iterationassupmtionforproofofequivalenceofhigherorderSobolevnorms:0}  \eqref{eq:iterationassupmtionforproofofequivalenceofhigherorderSobolevnorms}, we  can bound the right hand side of the above inequality by  
\beaa
 &\les& \left(1+\left(\left\|\sqrt{\ga}\right\|_{\hk_s(\S)}+\left\|\left(\sqrt{\ga}\right)^\#\right\|_{\hk_1^\infty(\ovS)}\right) \left(\left\|\frac{1}{\sqrt{\ga^\S}}\right\|_{\hk_s(\S)}+\left\|\frac{1}{\sqrt{\ga^{\S,\#}}}\right\|_{\hk_1^\infty(\ovS)}\right)\right)\left\|\ddd_k(h)\right\|_{\hk_s(\S)}\\
&&+\|(U', S')\|_{\hk_s(\ovS, \ovgS)\cap\hk_1^\infty(\ovS)}\left(\left\|\frac{1}{\sqrt{\ga^\S}}\right\|_{\hk_s(\S)}+\left\|\frac{1}{\sqrt{\ga^{\S,\#}}}\right\|_{\hk_1^\infty(\ovS)}\right)\\
&&\times\left(1+\left\|(\Omb, \underline{b}\sqrt{\ga})\right\|_{\hk_s(\S)}+\left\|(\Omb, \underline{b}\sqrt{\ga})^\#\right\|_{\hk_1^\infty(\ovS)}\right)\left\|\Big((e_3, e_4, \ddd_k)h, e_3(\Phi)h, e_4(\Phi)h\Big)\right\|_{\hk_s(\S)}
\eeaa
Therefore $ \|\ddd_k^\S h\|_{\hk_s(\S)}$  can be bounded by
\beaa
&\les&  \left(1+\left(\left\|\sqrt{\ga}\right\|_{\hk_s(\S)}+\left\|\left(\sqrt{\ga}\right)^\#\right\|_{\hk_1^\infty(\ovS)}\right) \left(\left\|\frac{1}{\sqrt{\ga^\S}}\right\|_{\hk_s(\S)}+\left\|\frac{1}{\sqrt{\ga^{\S,\#}}}\right\|_{\hk_1^\infty(\ovS)}\right)\right)\sup_{\RR}\left|\dk^{\leq s}\ddd_kh\right|\\
&&+\left(\left\|\frac{1}{\sqrt{\ga^\S}}\right\|_{\hk_s(\S)}+\left\|\frac{1}{\sqrt{\ga^{\S,\#}}}\right\|_{\hk_1^\infty(\ovS)}\right)\left(1+\left\|(\Omb, \underline{b}\sqrt{\ga})\right\|_{\hk_s(\S)}+\left\|(\Omb, \underline{b}\sqrt{\ga})^\#\right\|_{\hk_1^\infty(\ovS)}\right)\\
&&\times\sup_{\RR}\left|\dk^{\leq s}\left(\dk h, e_3(\Phi)h, e_4(\Phi)h\right)\right|,
\eeaa
where we used in the last inequality the assumption \eqref{eq:boundonU'andS'onS0forequivalenceofhigherorderSobolevnorms} on $(U', S')$. Together with  \eqref{eq:assumtionsonthegivenusfoliationforGCMprocedure} and \eqref{eq:assumtionsonthegivenusfoliationforGCMprocedure:bis}, we infer
\beaa
\|\ddd_k^\S h\|_{\hk_s(\S)} &\les& \Bigg\{ \left(1+\left(1+\left\|\left(\sqrt{\ga}\right)^\#\right\|_{\hk_1^\infty(\ovS)}\right) \left(\left\|\frac{1}{\sqrt{\ga^\S}}\right\|_{\hk_s(\S)}+\left\|\frac{1}{\sqrt{\ga^{\S,\#}}}\right\|_{\hk_1^\infty(\ovS)}\right)\right)\\
&&+\left(1+\left\|\frac{1}{\sqrt{\ga^\S}}\right\|_{\hk_s(\S)}+\left\|\frac{1}{\sqrt{\ga^{\S,\#}}}\right\|_{\hk_1^\infty(\ovS)}\right)\left(1+\left\|(\Omb, \underline{b}\sqrt{\ga})^\#\right\|_{\hk_1^\infty(\ovS)}\right)\Bigg\}\\
&&\times\sup_{\RR}\left|\dk^{\leq s+1}h\right|.
\eeaa
Also, for a reduced scalar $v$ defined on $\RR$, we have in view of  the assumption \eqref{eq:boundonU'andS'onS0forequivalenceofhigherorderSobolevnorms} on $(U', S')$
\bea\lab{eq:estimateforGactocomparehigherordersobolevnormusedbelowintheproof:bis:0}
\nn\|v^\#\|_{\hk_1^\infty(\ovS)} &=& \|v\circ\psi\|_{\hk_1^\infty(\ovS)}\\
\nn&\les& \left(1+\sup_{0\leq\th\leq\pi}|\psi'(\th)|\right)\sup_{\RR}|\dk^{\leq 1}v|\\
\nn&\les& \left(1+\|(U', S')\|_{\hk_1^\infty(\ovS)}\right)\sup_{\RR}|\dk^{\leq 1}v|\\
&\les&(1+ \dg)\sup_{\RR}|\dk^{\leq 1}v|.
\eea
Together with  \eqref{eq:assumtionsonthegivenusfoliationforGCMprocedure} and \eqref{eq:assumtionsonthegivenusfoliationforGCMprocedure:bis}, we infer
\beaa
\|\ddd_k^\S h\|_{\hk_s(\S)} &\les& \Bigg\{ 1+\left\|\frac{1}{\sqrt{\ga^\S}}\right\|_{\hk_s(\S)}+\left\|\frac{1}{\sqrt{\ga^{\S,\#}}}\right\|_{\hk_1^\infty(\ovS)}\Bigg\}\sup_{\RR}\left|\dk^{\leq s+1}h\right|.
\eeaa
Now, recall that 
  \beaa
  \gaS(\psi(\th))&=&\ga(\psi(\th))  +\left(\Omb(\psi(\th))+\frac 1 4( \underline{b}(\psi(\th)))^2  \ga(\psi(\th)) \right)\, ( U' (\th))^2 -2  U'(\th)  S'(\th)\\
  &-&\ga(\psi(\th)) \underline{b}(\psi(\th))  U'(\th).
  \eeaa 
Together with a repeated application of the iteration assumptions and a non sharp product rule in $\hk_s(\ovS, \ovgS)$ and \eqref{eq:estimateforGactocomparehigherordersobolevnormusedbelowintheproof:bis:0}, this yields
\beaa
&&\left\|\ga^\S\right\|_{\hk_s(\S)}+\left\|\ga^{\S,\#}\right\|_{\hk_1^\infty(\ovS)}\\
&\les&  \left(1+\sup_{\RR}|\dk^{\leq 1}(\ga, \Omb, \underline{b}^2\ga, \underline{b}\ga)|+\sup_{\RR}|\dk^{\leq s}(\ga, \Omb, \underline{b}^2\ga, \underline{b}\ga)|\right)\\
&\times& \left(1+\|(U', S')\|_{\hk^\infty_1(\ovS)}+ \|(U', S')\|_{\hk_s(\ovS, \ovgS)}\right)\\
&\les& 1 
\eeaa
where we used in the last estimate  the assumption \eqref{eq:boundonU'andS'onS0forequivalenceofhigherorderSobolevnorms} on $(U', S')$ and \eqref{eq:assumtionsonthegivenusfoliationforGCMprocedure:bis}. We infer
\beaa
\left\|\frac{1}{\sqrt{\ga^\S}}\right\|_{\hk_s(\S)}+\left\|\frac{1}{\sqrt{\ga^{\S,\#}}}\right\|_{\hk_1^\infty(\ovS)} &\les& 1
\eeaa
and hence 
\beaa
\|\ddd_k^\S h\|_{\hk_s(\S)} &\les& \sup_{\RR}\left|\dk^{\leq s+1}h\right|
\eeaa
which corresponds to the first of our iteration assumption \eqref{eq:iterationassupmtionforproofofequivalenceofhigherorderSobolevnorms:0} with $s$ replaced with $s+1$ for $s\leq s_{max}-1$. 

Next, we focus on recovering the second iteration assumption \eqref{eq:iterationassupmtionforproofofequivalenceofhigherorderSobolevnorms} with $s$ replaced with $s+1$ for $s\leq s_{max}-3$. Recall from Lemma \ref{lemma:formulausefultocomparehknormformetricS0andpullbackmetric} that we have for $f\in\sk_k(\S)$
   \beaa
  (\ddd_k^\S f)^\#   &=& \frac{\sqrt{\ga}}{\sqrt{\gaS^\#}}\Bigg\{\dddov_k(f^\#) + \Bigg(\frac{k}{2}U\int_0^1\Big(\sqrt{\ga}(e_\th(\ka)-e_\th(\vth))\Big)^{\#_\la}d\la\\
  &&+\frac k 4 S\int_0^1\left(\sqrt{\ga}e_\th \left( \kab-\vthb -\Omb(\ka-\vth) -2\underline{b}\ga^{1/2} e_\th  \Phi \right)\right)^{\#_\la}d\la \\
  &&+\frac k 4\Big( \kab -\vthb -\Omb(\ka-\vth) -2\underline{b}\ga^{1/2} e_\th\Phi\Big)^\# U' +\frac{k}{2}(\ka+\vth)^\# S'\Bigg)f^\#\Bigg\}
\eeaa   
where for $0\leq \la\leq 1$, $\#_\la$ denotes the pull back by 
\beaa
\psi_\la(\ovu, \ovs, \th) &=& (\ovu+\la U(\th), \ovs+\la  S(\th), \th).
\eeaa
For convenience, we rewrite some of the terms as follows
\beaa
e_\th(\ka)-e_\th(\vth) &=& -\dds_1(\ka) -\frac{1}{2}(\ddd_1\vth -\dds_2\vth),\\
\underline{b}\ga^{1/2} e_\th\Phi &=& \frac{1}{2}\ga^{1/2}(\ddd_1\underline{b} +\dds_2\underline{b}),
\eeaa
and
\beaa
&& e_\th \left( \kab-\vthb -\Omb(\ka-\vth) -2\underline{b}\ga^{1/2} e_\th  \Phi \right) \\
&=& -\dds_1(\kab)-\frac{1}{2}(\ddd_1\vthb -\dds_2\vthb)+\dds_1(\Omb\ka)-\dds_1(\Omb)\vth +\frac{1}{2}\Omb(\ddd_1\vthb -\dds_2\vthb)\\
&&+\dds_1(\ga^{1/2})(\ddd_1\vth +\dds_2\vth)\underline{b}   -2\ga^{1/2} e_\th(\underline{b}e_\th  \Phi)\\ 
&=& -\dds_1(\kab)-\frac{1}{2}(\ddd_1\vthb -\dds_2\vthb)+\dds_1(\Omb\ka)-\dds_1(\Omb)\vth +\frac{1}{2}\Omb(\ddd_1\vthb -\dds_2\vthb)\\
&&+\dds_1(\ga^{1/2})(\ddd_1\vth +\dds_2\vth)\underline{b}   -2\ga^{1/2} (-e_\th(\Phi)\dds_2\underline{b}-K\underline{b})\\
&=& -\dds_1(\kab)-\frac{1}{2}(\ddd_1\vthb -\dds_2\vthb)+\dds_1(\Omb\ka)-\dds_1(\Omb)\vth +\frac{1}{2}\Omb(\ddd_1\vthb -\dds_2\vthb)\\
&&+\dds_1(\ga^{1/2})(\ddd_1\vth +\dds_2\vth)\underline{b}   +\frac{1}{2}\ga^{1/2} (\ddd_2\dds_2\underline{b}+\dds_3\dds_2\underline{b})+2\ga^{1/2}K\underline{b}
\eeaa
where we used the identities
\beaa
e_\th(e_\th(\Phi)) &=& -(e_\th(\Phi))^2-K,\\
2\ga^{1/2} e_\th\Phi\dds_2\underline{b} &=& \frac{1}{2}\ga^{1/2} (\ddd_2\dds_2\underline{b}+\dds_3\dds_2\underline{b}).
\eeaa
This yields
   \beaa
  (\ddd_k^\S f)^\#   &=& \frac{\sqrt{\ga}}{\sqrt{\gaS^\#}}\Bigg\{\dddov_k(f^\#) + \Bigg(\frac{k}{2}U\int_0^1\left(\sqrt{\ga}\left(-\dds_1(\ka) -\frac{1}{2}(\ddd_1\vth -\dds_2\vth)\right)\right)^{\#_\la}d\la\\
  &&+\frac k 4 S\int_0^1\Bigg(\sqrt{\ga}\Bigg(-\dds_1(\kab)-\frac{1}{2}(\ddd_1\vthb -\dds_2\vthb)+\dds_1(\Omb\ka)-\dds_1(\Omb)\vth \\
  &&+\frac{1}{2}\Omb(\ddd_1\vthb -\dds_2\vthb)+\dds_1(\ga^{1/2})(\ddd_1\vth +\dds_2\vth)\underline{b} \\
  &&  +\frac{1}{2}\ga^{1/2} (\ddd_2\dds_2\underline{b}+\dds_3\dds_2\underline{b})+2\ga^{1/2}K\underline{b}\Bigg)\Bigg)^{\#_\la}d\la \\
  &&+\frac k 4\Big( \kab -\vthb -\Omb(\ka-\vth) -\ga^{1/2}(\ddd_1\vth +\dds_2\vth)\underline{b}\Big)^\# U' +\frac{k}{2}(\ka+\vth)^\# S'\Bigg)f^\#\Bigg\}.
\eeaa  

Next, we take the $\hk_s(\ovS,\, \ovgS)$-norm of this identity, and we use the iteration assumption to replace the norm on the left-hand side with the $\hk_s(\ovS,\, \gS^{\S,\#})$-norm. We infer
   \beaa
  &&\|(\ddd_k^\S f)^\#\|_{\hk_s(\ovS,\, \gS^{\S,\#})}(1+O(\dg))  \\
   &=& \Bigg\|\frac{\sqrt{\ga}}{\sqrt{\gaS^\#}}\Bigg\{\dddov_k(f^\#) + \Bigg(\frac{k}{2}U\int_0^1\left(\sqrt{\ga}\left(-\dds_1(\ka) -\frac{1}{2}(\ddd_1\vth -\dds_2\vth)\right)\right)^{\#_\la}d\la\\
  &&+\frac k 4 S\int_0^1\Bigg(\sqrt{\ga}\Bigg(-\dds_1(\kab)-\frac{1}{2}(\ddd_1\vthb -\dds_2\vthb)+\dds_1(\Omb\ka)-\dds_1(\Omb)\vth \\
  &&+\frac{1}{2}\Omb(\ddd_1\vthb -\dds_2\vthb)+\dds_1(\ga^{1/2})(\ddd_1\vth +\dds_2\vth)\underline{b} \\
  &&  +\frac{1}{2}\ga^{1/2} (\ddd_2\dds_2\underline{b}+\dds_3\dds_2\underline{b})+2\ga^{1/2}K\underline{b}\Bigg)\Bigg)^{\#_\la}d\la \\
  &&+\frac k 4\Big( \kab -\vthb -\Omb(\ka-\vth) -\ga^{1/2}(\ddd_1\vth +\dds_2\vth)\underline{b}\Big)^\# U' +\frac{k}{2}(\ka+\vth)^\# S'\Bigg)f^\#\Bigg\}\Bigg\|_{\hk_s(\ovS,\, \ovgS)}.
\eeaa 
Next, we use a non sharp product rule in $\hk_s(\ovS,\, \ovgS)$ to infer
   \beaa
  &&\|(\ddd_k^\S f)^\#\|_{\hk_s(\ovS,\, \gS^{\S,\#})}(1+O(\dg))  \\
   &=& \left(1+O(1)\left\|\frac{\sqrt{\ga}}{\sqrt{\gaS^\#}}-1\right\|_{\hk_s(\ovS,\, \ovgS)\cap\hk^\infty_1(\ovS)}\right)\Bigg\{\left\|\dddov_k(f^\#)\right\|_{\hk_s(\ovS,\, \ovgS)} \\
   && + O(1)\Bigg(\|U\|_{\hk_{s+1}(\ovS,\, \ovgS)}\int_0^1\left\|\left(\sqrt{\ga}\left(-\dds_1(\ka) -\frac{1}{2}(\ddd_1\vth -\dds_2\vth)\right)\right)^{\#_\la}\right\|_{\hk_s(\ovS,\, \ovgS)\cap\hk^\infty_1(\ovS)}d\la\\
  &&+\|S\|_{\hk_{s+1}(\ovS,\, \ovgS)}\int_0^1\Bigg\|\Bigg(\sqrt{\ga}\Bigg(-\dds_1(\kab)-\frac{1}{2}(\ddd_1\vthb -\dds_2\vthb)+\dds_1(\Omb\ka)-\dds_1(\Omb)\vth \\
  &&+\frac{1}{2}\Omb(\ddd_1\vthb -\dds_2\vthb)+\dds_1(\ga^{1/2})(\ddd_1\vth +\dds_2\vth)\underline{b} \\
  &&  +\frac{1}{2}\ga^{1/2} (\ddd_2\dds_2\underline{b}+\dds_3\dds_2\underline{b})+2\ga^{1/2}K\underline{b}\Bigg)\Bigg)^{\#_\la}\Bigg\|_{\hk_s(\ovS,\, \ovgS)\cap\hk^\infty_1(\ovS)}d\la \\
  &&+\left\|\Big( \kab -\vthb -\Omb(\ka-\vth) -\ga^{1/2}(\ddd_1\vth +\dds_2\vth)\underline{b}\Big)^\#\right\|_{\hk_s(\ovS,\, \ovgS)\cap\hk^\infty_1(\ovS)}\|U'\|_{\hk_{s+1}(\ovS,\, \ovgS)} \\
  &&+\left\|(\ka+\vth)^\#\right\|_{\hk_s(\ovS,\, \ovgS)\cap\hk^\infty_1(\ovS)}\|S'\|_{\hk_{s+1}(\ovS,\, \ovgS)}\Bigg)\left\|f^\#\right\|_{\hk_{s+1}(\ovS,\, \ovgS)}\Bigg\}.
\eeaa 
Since $s+1\leq s_{max}-2$, we infer in view of \eqref{eq:boundonU'andS'onS0forequivalenceofhigherorderSobolevnorms}  and the fact that $U(0)=S(0)=0$,
   \beaa
  &&\|(\ddd_k^\S f)^\#\|_{\hk_s(\ovS,\, \gS^{\S,\#})}(1+O(\dg))  \\
   &=& \left(1+O(1)\left\|\frac{\sqrt{\ga}}{\sqrt{\gaS^\#}}-1\right\|_{\hk_s(\ovS,\, \ovgS)\cap\hk^\infty_1(\ovS)}\right)\Bigg\{\left\|\dddov_k(f^\#)\right\|_{\hk_s(\ovS,\, \ovgS)} \\
   && + O(\dg)\Bigg(\int_0^1\left\|\left(\sqrt{\ga}\left(-\dds_1(\ka) -\frac{1}{2}(\ddd_1\vth -\dds_2\vth)\right)\right)^{\#_\la}\right\|_{\hk_s(\ovS,\, \ovgS)\cap\hk^\infty_1(\ovS)}d\la\\
  &&+\int_0^1\Bigg\|\Bigg(\sqrt{\ga}\Bigg(-\dds_1(\kab)-\frac{1}{2}(\ddd_1\vthb -\dds_2\vthb)+\dds_1(\Omb\ka)-\dds_1(\Omb)\vth \\
  &&+\frac{1}{2}\Omb(\ddd_1\vthb -\dds_2\vthb)+\dds_1(\ga^{1/2})(\ddd_1\vth +\dds_2\vth)\underline{b} \\
  &&  +\frac{1}{2}\ga^{1/2} (\ddd_2\dds_2\underline{b}+\dds_3\dds_2\underline{b})+2\ga^{1/2}K\underline{b}\Bigg)\Bigg)^{\#_\la}\Bigg\|_{\hk_s(\ovS,\, \ovgS)\cap\hk^\infty_1(\ovS)}d\la \\
  &&+\left\|\Big( \kab -\vthb -\Omb(\ka-\vth) -\ga^{1/2}(\ddd_1\vth +\dds_2\vth)\underline{b}\Big)^\#\right\|_{\hk_s(\ovS,\, \ovgS)\cap\hk^\infty_1(\ovS)}\\
  &&+\left\|(\ka+\vth)^\#\right\|_{\hk_s(\ovS,\, \ovgS)\cap\hk^\infty_1(\ovS)}\Bigg)\left\|f^\#\right\|_{\hk_{s+1}(\ovS,\, \ovgS)}\Bigg\}.
\eeaa 

Next, we have by the iteration assumption \eqref{eq:iterationassupmtionforproofofequivalenceofhigherorderSobolevnorms}
\beaa
&&\left\|\left(\ddd^{\leq 2}\Big(\Gac, r^{-2}\ga-1, \underline{b}, \Omb+\Up\Big)\right)^\#\right\|_{\hk_s(\ovS,\, \ovgS)}+\sup_{0\leq\la\leq 1}\left\|\left(\ddd^{\leq 2}\Big(\Gac, r^{-2}\ga-1, \underline{b}, \Omb+\Up\Big)\right)^{\#_\la}\right\|_{\hk_s(\ovS,\, \ovgS)}\\ 
&\les& \left\|\left(\ddd^{\leq 2}\Big(\Gac, r^{-2}\ga-1, \underline{b}, \Omb+\Up\Big)\right)^\#\right\|_{\hk_s(\ovS,\, \gS^{\S,\#})}+\sup_{0\leq\la\leq 1}\left\|\left(\ddd^{\leq 2}\Big(\Gac, r^{-2}\ga-1, \underline{b}, \Omb+\Up\Big)\right)^{\#_\la}\right\|_{\hk_s(\ovS,\, \gS^{\S,\#_\la})}\\
&\les& \left\|\ddd^{\leq 2}\Big(\Gac, r^{-2}\ga-1, \underline{b}, \Omb+\Up\Big)\right\|_{\hk_s(S)}+\sup_{0\leq\la\leq 1}\left\|\ddd^{\leq 2}\Big(\Gac, r^{-2}\ga-1, \underline{b}, \Omb+\Up\Big)\right\|_{\hk_s(\S_\la)}
\eeaa
where the surface $\S_\la$ is the image of $\ovS$ by $\psi_\la$. Since $s\leq s_{max}-3$, we infer in view of our iteration assumption \eqref{eq:iterationassupmtionforproofofequivalenceofhigherorderSobolevnorms:0} and our assumptions \eqref{eq:assumtionsonthegivenusfoliationforGCMprocedure} \eqref{eq:assumtionsonthegivenusfoliationforGCMprocedure:bis} on the $(u, s)$-foliation
\bea\lab{eq:estimateforGactocomparehigherordersobolevnormusedbelowintheproof}
\nn &&\left\|\left(\ddd^{\leq 2}\Big(\Gac, r^{-2}\ga-1, \underline{b}, \Omb+\Up\Big)\right)^\#\right\|_{\hk_s(\ovS,\, \ovgS)}+\sup_{0\leq\la\leq 1}\left\|\left(\ddd^{\leq 2}\Big(\Gac, r^{-2}\ga-1, \underline{b}, \Omb+\Up\Big)\right)^{\#_\la}\right\|_{\hk_s(\ovS,\, \ovgS)}\\
\nn &\les& \sup_{\RR}\left|\dk^{\leq s}\ddd^{\leq 2}\Big(\Gac, r^{-2}\ga-1, \underline{b}, \Omb+\Up\Big)\Gac\right|\\
  &\les& \sup_{\RR}\left|\dk^{\leq s+2}\Big(\Gac, r^{-2}\ga-1, \underline{b}, \Omb+\Up\Big)\Gac\right| \les\dg.
\eea
Also, we have
\beaa
\nn&&\left\|\left(\ddd^{\leq 2}\Big(\Gac, r^{-2}\ga-1, \underline{b}, \Omb+\Up\Big)\right)^\#\right\|_{\hk_1^\infty(\ovS)}+\sup_{0\leq\la\leq 1}\left\|\left(\ddd^{\leq 2}\Big(\Gac, r^{-2}\ga-1, \underline{b}, \Omb+\Up\Big)\right)^{\#_\la}\right\|_{\hk_1^\infty(\ovS)}\\
\nn&=& \left\|\left(\ddd^{\leq 2}\Big(\Gac, r^{-2}\ga-1, \underline{b}, \Omb+\Up\Big)\right)\circ\psi\right\|_{\hk_1^\infty(\ovS)}+\sup_{0\leq\la\leq 1}\left\|\left(\ddd^{\leq 2}\Big(\Gac, r^{-2}\ga-1, \underline{b}, \Omb+\Up\Big)\right)\circ\psi_\la\right\|_{\hk_1^\infty(\ovS)} \\
\nn&\les&  \left(\sup_{\RR}\left|\dk^{\leq 3}\Big(\Gac, r^{-2}\ga-1, \underline{b}, \Omb+\Up\Big)\right|\right)\left(1+\sup_{0\leq\th\leq\pi}|\psi'(\th)|\right)\\ 
\nn&\les&  \left(\sup_{\RR}\left|\dk^{\leq 3}\Big(\Gac, r^{-2}\ga-1, \underline{b}, \Omb+\Up\Big)\right|\right)\left(1+\|(U', S')\|_{\hk_1^\infty(\ovS)}\right)\\ 
&\les&\epg
\eeaa
where we used our assumptions \eqref{eq:assumtionsonthegivenusfoliationforGCMprocedure} \eqref{eq:assumtionsonthegivenusfoliationforGCMprocedure:bis} on the $(u, s)$-foliation and our assumption \eqref{eq:boundonU'andS'onS0forequivalenceofhigherorderSobolevnorms} on $(U', S')$.
Therefore,
\bea
\lab{eq:estimateforGactocomparehigherordersobolevnormusedbelowintheproof:bis}
\bsplit
\left\|\left(\ddd^{\leq 2}\Big(\Gac, r^{-2}\ga-1, \underline{b}, \Omb+\Up\Big)\right)^\#\right\|_{\hk_1^\infty(\ovS)}&\les \dg\\
\sup_{0\leq\la\leq 1}\left\|\left(\ddd^{\leq 2}\Big(\Gac, r^{-2}\ga-1, \underline{b}, \Omb+\Up\Big)\right)^{\#_\la}\right\|_{\hk_1^\infty(\ovS)}&\les\dg.
\end{split}
\eea

We deduce
   \beaa
  &&\|(\ddd_k^\S f)^\#\|_{\hk_s(\ovS,\, \gS^{\S,\#})}(1+O(\dg))  \\
   &=& \left(1+O(1)\left\|\frac{\sqrt{\ga}}{\sqrt{\gaS^\#}}-1\right\|_{\hk_s(\ovS,\, \ovgS)\cap\hk^\infty_1(\ovS)}\right)\Bigg\{\left\|\dddov_k(f^\#)\right\|_{\hk_s(\ovS,\, \ovgS)}  + O(\dg)\left\|f^\#\right\|_{\hk_{s+1}(\ovS, \ovgS)}\Bigg\}.
\eeaa 
Next, we estimate the term in the RHS involving $\ga$ and $\gaS^\#$. From the proof of Lemma \ref{lemma:int-gaS-ga}, we have
\beaa
\ga^{\S, \#} -\ga &=& \frac 1 2U\int_0^1\left(\left( e_3-\Omb e_4-\underline{b}\ga^{1/2} e_\th\right)\ga\right)^{\#_\la}d\la+S\int_0^1\left(e_4\ga\right)^{\#_\la}d\la\\
    &&  +\left(\Omb+\frac 1 4 \underline{b}^2\ga \right)^\#\, ( U')^2 -2  U'  S'-\left(\ga \underline{b}\right)^\#  U'.
\eeaa
Using a non sharp product rule, we infer
\beaa
&&\left\|\ga^{\S, \#} -\ga\right\|_{\hk_s(\ovS,\, \ovgS)\cap\hk^\infty_1(\ovS)}\\
&\les& \|U\|_{\hk_s(\ovS,\, \ovgS)\cap\hk^\infty_1(\ovS)}\int_0^1\left\|\left(\left( e_3-\Omb e_4-\underline{b}\ga^{1/2} e_\th\right)\ga\right)^{\#_\la}\right\|_{\hk_s(\ovS,\, \ovgS)\cap\hk^\infty_1(\ovS)}d\la\\
&& \|S\|_{\hk_s(\ovS,\, \ovgS)\cap\hk^\infty_1(\ovS)}\int_0^1\left\|\left(e_4\ga\right)^{\#_\la}\right\|_{\hk_s(\ovS,\, \ovgS)\cap\hk^\infty_1(\ovS)} d\la\\
    &&  +\left\|\left(\Omb+\frac 1 4\underline{b}^2\ga \right)^\#\right\|_{\hk_s(\ovS,\, \ovgS)\cap\hk^\infty_1(\ovS)}\|U'\|_{\hk_s(\ovS,\, \ovgS)\cap\hk^\infty_1(\ovS)}+\|U'\|_{\hk_s(\ovS,\, \ovgS)\cap\hk^\infty_1(\ovS)}\|S'\|_{\hk_s(\ovS,\, \ovgS)\cap\hk^\infty_1(\ovS)}\\
    &&+\left\|\left(\ga \underline{b}\right)^\#\right\|_{\hk_s(\ovS,\, \ovgS)\cap\hk^\infty_1(\ovS)}\|U'\|_{\hk_s(\ovS,\, \ovgS)\cap\hk^\infty_1(\ovS)}\\
        &\les& \dg\int_0^1\left\|\left(\ddd^{\leq 1}\Big(r^{-2}\ga-1, \underline{b}, \Omb+\Up\Big)\right)^{\#_\la}\right\|_{\hk_s(\ovS,\, \ovgS)\cap\hk^\infty_1(\ovS)}d\la\\
        && +\dg\left\|\left(r^{-2}\ga-1, \underline{b}, \Omb+\Up \right)^\#\right\|_{\hk_s(\ovS,\, \ovgS)\cap\hk^\infty_1(\ovS)}+\dg
\eeaa
where we used our assumption \eqref{eq:boundonU'andS'onS0forequivalenceofhigherorderSobolevnorms} on $(U', S')$ and the fact that  $U(0)=S(0)=0$. Using the estimates \eqref{eq:estimateforGactocomparehigherordersobolevnormusedbelowintheproof} \eqref{eq:estimateforGactocomparehigherordersobolevnormusedbelowintheproof:bis} for $(r^{-2}\ga-1, \underline{b}, \Omb+\Up)$, we infer
\beaa
\left\|\ga^{\S, \#} -\ga\right\|_{\hk_s(\ovS,\, \ovgS)\cap\hk^\infty_1(\ovS)} &\les& \dg.
\eeaa
Together with \eqref{eq:assumtionsonthegivenusfoliationforGCMprocedure:bis}  for $\ga$, we infer
\beaa
\left\|\frac{\sqrt{\ga}}{\sqrt{\gaS^\#}}-1\right\|_{\hk_s(\ovS,\, \ovgS)\cap\hk^\infty_1(\ovS)} &\les& \dg
\eeaa
and hence
   \beaa
  \|(\ddd_k^\S f)^\#\|_{\hk_s(\ovS,\, \gS^{\S,\#})}(1+O(\dg))    &=& \left(1+O(\dg)\right)\left\{\left\|\dddov_k(f^\#)\right\|_{\hk_s(\ovS,\, \ovgS)}  + O(\dg)\left\|f^\#\right\|_{\hk_{s+1}(\ovS, \ovgS)}\right\}.
\eeaa 
Now, we have
\beaa
\|f^\#\|_{\hk_{s+1}(\ovS,\, \gS^{\S,\#})} &=& \|f^\#\|_{L^2(\ovS,\, \gS^{\S,\#})}+\|(\ddd_k^\S f)^\#\|_{\hk_s(\ovS,\, \gS^{\S,\#})},\\
\|f^\#\|_{\hk_{s+1}(\ovS,\, \ovgS)} &=& \|f^\#\|_{L^2(\ovS,\, \ovgS)}+\|\dddov_k(f^\#)\|_{\hk_s(\ovS,\, \ovgS)}.
\eeaa
Together with Lemma \ref{lemma:int-gaS-ga}, this yields
\beaa
\|f^\#\|_{\hk_{s+1}(\ovS,\, \gS^{\S,\#})} = \| f^\#\|_{\hk_{s+1}(\ovS, \ovgS)}(1+O(\dg)).
\eeaa

This corresponds to our iteration assumption \eqref{eq:iterationassupmtionforproofofequivalenceofhigherorderSobolevnorms} with $s$ replaced with $s+1$ for $s\leq s_{max}-3$. Thus, we have finally derived both  iteration assumption \eqref{eq:iterationassupmtionforproofofequivalenceofhigherorderSobolevnorms} and \eqref{eq:iterationassupmtionforproofofequivalenceofhigherorderSobolevnorms:0} with $s$ replaced with $s+1$ respectively for $s\leq s_{max}-1$ and $s\leq s_{max}-3$. Hence, we deduce that they hold respectively for $0\leq s\leq s_{max}$ and $0\leq s\leq s_{max}-2$, i.e. 
 \beaa
 \|h\|_{\hk_k(\S)} \les\sup_{\RR}|\dk^{\leq k}h|\textrm{ for }0\leq s\leq s_{max}
 \eeaa
and
 \beaa
  \|f^\#\|_{\hk_s(\ovS,\, \gS^{\S,\#})} = \| f^\#\|_{\hk_s(\ovS, \ovgS)}(1+O(\dg))\textrm{ for }0\leq s\leq s_{max}-2.
 \eeaa
Together with Lemma \ref{le:pullbackS}, we deduce
\beaa
\|f\|_{\hk_s(\S)} = \|f^\#\|_{\hk_{s+1}(\ovS,\, \gS^{\S,\#})} = \| f^\#\|_{\hk_{s+1}(\ovS, \ovgS)}(1+O(\dg))\textrm{ for all }0\leq s\leq s_{max}-2.
\eeaa
This concludes the proof of the lemma.
\end{proof}


\chapter{APPENDIX TO CHAPTER \ref{chapter:waveeqtionestimates}}\label{sect:appendix-sectionMorawetz}



\section{Horizontal $S$-tensors}


 Consider a null pair $e_3, e_4$  on $(\M, \g)$   and, at every point $p\in \M$
 the horizontal  space $S=\{ e_3, e_4\}^\perp$. Let $\ga $ the metric induced on   $S$.  By definition,  for all
     $X, Y\in \T_S  \M$, i.e. vectors in $\M$ tangent to $S$,
 \beaa
 h(X, Y)= \g(X, Y)
 \eeaa
 For any $Y\in T(\M)$ we define its horizontal projection,
 \bea
  Y^\perp = Y+\frac 12 \g( Y, e_3) e_4+\frac 12 \g(Y, e_3) e_4
  \eea
  
  \begin{definition}
  \label{definition:Shorizonthal}
A  $k$-covariant tensor-field $U$ is said to be  $S$-horizontal,  $U\in \T^k_S(\M)$,
if  for any $X_1,\ldots X_k$ we have,
\beaa
U(Y_1,\ldots Y_k)=U(Y^\perp _1,\ldots  Y^\perp_k)
\eeaa
\end{definition}

  We define the projection operator,
  \beaa
  \Pi_\mu^\nu:= \de_\mu^\nu - \frac 1 2 (e_3)_\mu (e_4)^\nu-  \frac 1 2 (e_4)_\mu (e_3)^\nu
  \eeaa
  Clearly $ \Pi_{\a}^{\mu}\Pi_{\mu}^\b=\Pi_{\a}^\b$.  An arbitrary tensor 
$U_{\a_1\ldots\a_m}$ is said to   an $S$- horizontal tensor, or simply $S$-tensor,  if
\beaa
\Pi_{\a_1}^{\b_1}\ldots \Pi_{\a_m}^{\b_m}\, U_{\b_1\ldots\b_m}=U_{\a_1\ldots\a_m}.
\eeaa

\begin{definition}
Given  $X\in \T(\M)$ and $Y \in \T_S(\M)$ we define,
\beaa
\Db_X Y&:=& ( \D_X Y)^\perp
\eeaa
\end{definition} 

\begin{remark}
In the particular case when   $S$ is integrable and   both $X, Y\in \T_S\M$  then
$\Db_X Y$ is the standard  induced covariant differentiation on $S$.
\end{remark}

 \begin{definition}
 \label{definition:S-covariantderivative}
 Given a general, covariant,  $S$- horizontal tensor-field  $U$
 we define its horizontal covariant derivative according to
 the formula,
 \bea
 \Db_X U(Y_1,\ldots Y_k)=X (U(Y_1,\ldots Y_k))&-&U(\Db_X Y_1,\ldots Y_k)-\ldots- U(Y_1,\ldots \Db_XY_k).
 \eea
 where $X\in \T\M$ and $Y_1,\ldots Y_k\in \T_S\M$.
 \end{definition}
 
   \begin{proposition}
 For  all  $X\in\T\M$   and $Y_1, Y_2 \in \T_S\M$,
 \beaa
 X h (Y_1,Y_2)= h (\Db_X Y_1, Y_2)+ h(Y_1, \Db_X Y_2).
 \eeaa
  \end{proposition}
  
  \begin{proof}
  Indeed,
  \beaa
 X h (Y_1,Y_2)&=&X \g (Y_1,Y_2)=\g (\D_X Y_1,Y_2)+\g ( Y_1,\D_XY_2)= \g (\Db_X Y_1,Y_2)+\g ( Y_1,\Db_XY_2)\\
&=&  h (\Db_X Y_1,Y_2)+h ( Y_1,\Db_XY_2)
  \eeaa
   \end{proof}

Given an orthonormal  frame  $e_1, e_2$ on  $S$ we have,
\beaa
\Db_\mu  e_A&=&\sum_{B=1,2}(\La_\mu)_{AB}\, e_B\, \qquad A, B=1,2\
\eeaa
where,
\beaa
(\La_\mu)_{\a\b}:=\g(\D_\mu e_\b, e_\a)
\eeaa


\subsection{Mixed tensors}


We consider tensors  $\T^k \M\otimes   \T_S^l \M  $, i.e. tensors  of the form,
\beaa
U_{\mu_1\ldots \mu_k,  A_1\ldots A_L}
\eeaa
for which we define,
\beaa
\Db_\mu U_{\nu_1\ldots \nu_k,  A_1\ldots A_L}&=& e_\mu U_{\nu_1\ldots \nu_k,  A_1\ldots A_l}-U_{\D_\mu\nu_1\ldots \nu_k,  A_1\ldots A_l}-\ldots- U_{\nu_1\ldots  \D_\mu\nu_k,  A_1\ldots A_l}\\
&-& U_{\nu_1\ldots \nu_k,   \Db_\mu A_1\ldots A_l}-  U_{\nu_1\ldots \nu_k,   A_1 \ldots \Db_\mu A_l}
\eeaa
We are now ready   to prove the following,
\begin{proposition}
We   have  the curvature formula
 \beaa
( \Db _\mu\Db_\nu -\Db_\nu\Db _\mu)\Psi_A=\R_{A}\, ^   B\,_{   \mu\nu}\Psi_B
 \eeaa
 More  generally,
  \beaa
( \Db _\mu\Db_\nu -\Db_\nu\Db _\mu)\Psi_{\la A}=    \R_{\la }\, ^   \si \,_{   \mu\nu}\Psi_{\si A}+    
            \R_{A}\, ^   B\,_{   \mu\nu}\Psi_{\la B}
 \eeaa
 \end{proposition}

\begin{proof}
Straightforward verification.
\end{proof}


\subsection{Invariant Lagrangian}


We introduce,
 \beaa
 \LL&=& g^{\mu\nu} h_{AB}\Db_\mu  \Psi^A \Db_\mu  \Psi^B  +V h_{AB} \Psi^A \Psi^B
 \eeaa
\begin{proposition}
The Euler Lagrange equations are given by:
\beaa
\squared\Psi^A= V \Psi^A
\eeaa
where   $\squared\Psi^A:= \g^{\mu\nu} \Db_\mu\Db_ \nu \Psi^A.$
 \end{proposition}
 
\begin{proof}
The variation of the action is given by,
\beaa
0&=&2\int_\M  h_{AB}\left(  \g^{\mu\nu} \Db_\mu   \Psi^A   \Db_ \nu( \de\Psi)^B +V\Psi^A \de\Psi^B\right) dv_\g\\
&=&2\int_\M   \D_ \nu\left ( \g^{\mu\nu}  h_{AB}\Db_\mu   \Psi^A   ( \de\Psi)^B \right) dv_\g- 2\int_\M  h_{AB} \left ( \g^{\mu\nu}  \Db_ \nu \Db_\mu   \Psi^A   ( \de\Psi)^B  -V\Psi^A \de\Psi^B \right)dv_\g\\
&=&- 2\int_\M  h_{AB} \left ( \g^{\mu\nu}  \Db_ \nu \Db_\mu   \Psi^A   ( \de\Psi)^B  -V\Psi^A \de\Psi^B \right)dv_\g
\eeaa
from which the proposition follows.
\end{proof}


\subsection{Comparison of the Lagrangians}


Let $\Psi\in \SS_2(\MM)$ and $\psi\in\sk_2$ its reduced form.
Note that the Lagrangian of the scalar   equation
\beaa
\square_\g \psi= V\psi     +4 (e_\th\Phi)^2      \psi
\eeaa
is given by,
\beaa
\LL(\psi):&=&\g^{\mu\nu}\pr_\mu \psi \pr_\nu \psi+ (V +4 (e_\th\Phi)^2)\psi^2 
\eeaa
while the Lagrangian for,
\beaa
\squared_\g \Psi= V\Psi    
\eeaa
is given by 
\beaa
\LL(\Psi)&=&\g^{\mu\nu}\Db_\mu \Psi \c \Db_ \nu \Psi+ V \Psi\c\Psi  
\eeaa
\begin{proposition}
We have,
\bea
\LL(\Psi)= 2 \LL(\psi)
\eea
\end{proposition}

\begin{proof}
Observe that,
\beaa
\g^{\mu\nu}\Db_\mu \Psi \Db_ \nu \Psi&=&-\Db_3 \Psi\c  \Db_ 4\Psi +\Db_\th \Psi \c \Db_\th \Psi +\Db_\vphi \Psi \c \Db_\vphi \Psi
\eeaa
Now,   recalling  that,
\beaa
\nabb_\vphi e_\vphi&=&-e_\th \Phi e_\th, \qquad \nabb_\vphi e_\th= e_\th(\Phi) e_\vphi\\
\nabb_\th e_\th &=&0\qquad \qquad\quad\,\,\nabb_\th e_\vphi=0
\eeaa
we  deduce
\beaa
\Db_3 \Psi\c  \Db_ 4\Psi &=& e_3  \Psi  \c e_4\Psi =2 e_3\psi e_4 \psi\\
\Db_\th \Psi \c \Db_\th \Psi &=&\Db_\th \Psi_{\th\th}   \Db_\th \Psi_{\th\th}+2 \Db_\th \Psi_{\th\vphi}   \Db_\th \Psi_{\th\vphi}
+\Db_\th \Psi_{\vphi\vphi }   \Db_\th \Psi_{\vphi\vphi}\\
&=&2 (e_\th \psi)^2  \\
\Db_\vphi \Psi \c \Db_\vphi \Psi&=&\Db_\vphi \Psi_{\th\th}   \Db_\vphi \Psi_{\th\th}+2 \Db_\vphi \Psi_{\th\vphi}   \Db_\vphi \Psi_{\th\vphi}
+\Db_\vphi \Psi_{\vphi\vphi }   \Db_\vphi \Psi_{\vphi\vphi}\\
&=&2 (e_\vphi \psi)^2  +2 (-  \Psi_{\Db_\vphi \th\vphi} -\Psi_{ \th\Db_\vphi\vphi})\c (-  \Psi_{\Db_\vphi \th\vphi} -\Psi_{ \th\Db_\vphi\vphi})\\
&=&2 (e_\vphi \psi)^2  +2( -e_\th(\Phi) \Psi_{\vphi\vphi}+ e_\th (\Phi) \Psi_{\th\th})\c ( -e_\th(\Phi) \Psi_{\vphi\vphi}+ e_\th (\Phi) \Psi_{\th\th})\\
&=&2 (e_\vphi \psi)^2+ 8 (e_\th \Phi)^2 \psi^2
\eeaa
Hence,
\beaa
\g^{\mu\nu}\Db_\mu \Psi \Db_ \nu \Psi&=&-2 e_3\psi e_4 \psi+ 2 (e_\th \psi)^2 + 2 (e_\vphi \psi)^2 +4 (e_\th \Phi)^2 \psi^2
\eeaa
and
\beaa
\LL(\Psi)&=&-2 e_3\Psi e_4 \psi+ 2 (e_\th \psi)^2 + 2 (e_\vphi \psi)^2 +8 (e_\th \Phi)^2 \psi^2+2 V\psi^2 
\eeaa
\end{proof}


\subsection{Energy-Momentum tensor}\lab{sec:appendixonenergymomentumtensorwaveeqpsi}


Consider the   energy-momentum tensor,
 \beaa
 \QQ_{\mu\nu}:=\Db_\mu  \Psi \c \Db _\nu \Psi 
          -\frac 12 \g_{\mu\nu} \left(\Db_\la \Psi\c\Db^\la \Psi + V\Psi \c \Psi\right)
 \eeaa
\begin{lemma}
We have,
 \beaa
 \D^\nu\QQ_{\mu\nu}
  &=& \Db_\mu  \Psi \c\left(\squared \Psi-V\psi\right) + \Db^\nu  \Psi ^A\R_{ A   B   \nu\mu}\Psi^B-\frac 1 2 \D_\mu V \Psi\c \Psi
 \eeaa
\end{lemma}

\begin{proof}
We have,
 \beaa
 \D^\nu\QQ_{\mu\nu}&=&\Db^\nu \Db_\nu  \Psi \c \Db _\mu \Psi
 +  \Db^\nu  \Psi\c  \left( \Db_\nu \Db _\mu -\Db_\mu \Db _\nu \right)\Psi -V\D_\mu \Psi \c \Psi      -\frac 1 2 \D_\mu V \Psi\c\Psi\\
  &=& \Db_\mu  \Psi \c\Db^\nu \Db _\nu \Psi+ \Db^\nu  \Psi ^A\R_{ A   B   \nu\mu}\Psi^B -V\D_\mu \Psi \Psi   -\frac 1 2 \D_\mu V \Psi\c\Psi\\
  &=& \Db_\mu \Psi\left(\squared \Psi- V\Psi\right)+ \Db^\nu  \Psi ^A\R_{ A   B   \nu\mu}\Psi^B -\frac 1 2 \D_\mu V \Psi\c\Psi
 \eeaa
\end{proof}

\begin{lemma} Relative to an arbitrary $\Z$-polarized frame  $e_3, e_4, e_\th, e_\vphi$  we have, 
\beaa
\QQ_{33}&=&|e_3\Psi|^2,\\
\QQ_{44}&=&|e_4\Psi|^2,\\
\QQ_{34}&=&|\nabb \Psi|^2 +V|\Psi|^2.
\eeaa
If $\psi$ is  the reduced  form of $\Psi$,
\beaa
\QQ_{33}&=&2(e_3\psi)^2, \\
  \QQ_{44}&=&2(e_4\psi)^2, \\
\QQ_{34}&=&2 (e_\th\psi)^2+ 2  (e_\vphi \psi)^2+ 2 V|\psi|^2+ 8 ( e_\th \Phi)^2\psi^2.
\eeaa

Also,
\beaa
\g^{\mu\nu}\QQ_{\mu\nu}&=&-\LL(\Psi)- V|\Psi|^2,
\eeaa
\beaa
|\LL(\Psi)|&\les& |e_3\Psi| \, |e_4\Psi|+ |\nabb \Psi|^2 +V|\Psi|^2,
\eeaa
and
\beaa
|\QQ_{AB}| &\le& |e_3\Psi|  |e_4\Psi| +|\nabb\Psi|^2 +|V||\Psi|^2, \\
|\QQ_{A3}| &\le& |e_3\Psi| |\nabb\Psi|,\\
|\QQ_{A4}| &\le& |e_4\Psi| |\nabb\Psi|.
\eeaa
\end{lemma}


\section{Standard Calculation}\lab{sec:standardcalculationvectorfieldmethodwaveequationpsi}


 \begin{proposition}\label{prop-app:stadard-comp-Psi}
 Consider an admissible  spacetime $\MM$  and    $\Psi\in \SS_2(\MM)$ and   $X$  a vectorfield of the form,
 \beaa
    X=  a e_3 +b e_4,
    \eeaa
    \begin{enumerate}
\item 
 The $1$-form   $\PP_\mu=\QQ_{\mu\nu} X^\nu$   verifies,
\beaa
\D^\mu \PP_\mu&=& X^\mu \Db_\mu  \Psi \c\left(\squared  \Psi-V\Psi\right) 
- X( V )  \Psi\c \Psi\
\eeaa
\item
   Let $X$ as above,   $w$ a scalar   and $\M$  a one form. Define,
 \beaa
 \PP_\mu &=&\PP_\mu[X, w, M]=\QQ_{\mu\nu} X^\nu +\frac 1 2  w \Psi \c \Db_\mu \Psi -\frac 1 4|\Psi|^2   \pr_\mu w +\frac 1 4 |\Psi|^2 M_\mu
  \eeaa
 Then,    with $|\Psi|^2:=\Psi\c\Psi$,
  \beaa
  \D^\mu  \PP_\mu[X, w, M] &=& \frac 1 2 \QQ  \c\piX - \frac 1 2 X( V )  \Psi\c \Psi+\frac 12  w \LL[\Psi] -\frac 1 4|\Psi|^2   \square_\g  w\\
   &+&\frac 1 4 |\Psi|^2 \Div M+\frac 1 2 \Psi\c \Db_\mu\Psi \,  M^\mu+  \left(X( \Psi )+\frac 1 2   w \Psi\right)\c \left(\squared  \Psi-V\Psi\right) 
 \eeaa
 \end{enumerate}
\end{proposition}

\begin{proof}
 Let     $\PP_\mu[X, 0, 0]=\QQ_{\mu\nu} X^\nu$,
Then,
\beaa
\D^\mu \PP_\mu&=& X^\mu \Db_\mu  \Psi \c\left(\Db^\nu \Db _\nu \Psi-V\Psi\right) + X^\mu \Db^\nu  \Psi ^A\R_{ A   B   \nu\mu}\Psi^B-\frac 1 2 X^\mu \D_\mu V \Psi\c \Psi\\
&=& X^\mu \Db_\mu  \Psi \c\left(\squared  \Psi-V\Psi\right) -\frac 1 2 X(V)|\Psi|^2
\eeaa
Assume $X= a e_3 +b e_4$. Then, since only the middle components of  $\R$ are relevant,   and recalling that   
$\R_{ A   B   4 3 }=-\rhod \in_{AB}=0$,     we derive,
\beaa
 X^\mu  \Db^\nu  \Psi ^A\R_{ A   B   \nu 3 }\Psi^B=
a  \Db^4  \Psi ^A\R_{ A   B   4 3 }\Psi^B+          b \Db^3  \Psi ^A\R_{ A   B   434  }\Psi^B  =0
\eeaa
To prove the second part of the proposition we write  with $\NN[\Psi]:=\squared  \Psi-V\Psi$,
  \beaa
 \D^\mu  \PP_\mu[X, w, M] &=& \frac 1 2 \QQ  \c\piX + X( \Psi )\c \NN[\Psi]  
-\frac 1 2  X( V )  \Psi\c \Psi +\frac 1 2 \D^\mu w  \,   \Psi\c  \Db_\mu \Psi  \\
 &+& \frac 1 2  w\, \Db^\mu  \Psi \c \Db_\mu \Psi+ \frac 1 2  w \Psi \squared_\g \Psi              -\frac 1 2\Psi \c \Db^\mu \Psi    \pr_\mu w -\frac 1 4|\Psi|^2   \square_\g  w\\
 &+&\frac 1 4 |\Psi|^2 \Div M+\frac 1 2 \Psi\c \Db_\mu\Psi \,  M^\mu\\
 &=& \frac 1 2 \QQ  \c\piX - \frac 1 2 X( V )  \Psi\c \Psi  + \frac 1 2  w\, \Db^\mu  \Psi \c \Db_\mu \Psi+  \frac 1 2  w \Psi\left( V\Psi+\NN[\Psi]\right) \\
  &-& \frac 1 4|\Psi|^2   \square_\g  w+\frac 1 4 |\Psi|^2 \Div M+\frac 1 2 \Psi\c \Db_\mu\Psi \,  M^\mu+ X( \Psi )\c\Psi\c \NN[\Psi] 
 \eeaa
 Hence,
 \beaa
  \D^\mu  \PP_\mu[X, w, M] &=& \frac 1 2 \QQ  \c\piX -\frac 1 2  X( V )  \Psi\c \Psi+\frac 12  w \LL[\Psi] -\frac 1 4|\Psi|^2   \square_\g  w\\
   &+&\frac 1 4 |\Psi|^2 \Div M+\frac 1 2 \Psi\c \Db_\mu\Psi \,  M^\mu+  \left(X( \Psi )+\frac 1 2   w \Psi\right)\c \NN[\Psi] 
 \eeaa
 as  desired.
\end{proof}

\begin{remark}
As consequence  of the proposition above   we  deduce that every time we use vectorfields  of the form $ae_3+b e_4$
 as multipliers,    the equation $\square \Psi-V\Psi=\NN$   is  treated  exactly in the same manner
as the scalar  equation $\square\psi-V\psi=N$.
\end{remark}

\begin{remark}
Note  that  in Schwarzschild our potential $V=\ka \kab=4\Up r^{-2}$ verifies,
\beaa
\frac 1 4 \pr_r V&=& \pr_r\left[  r^{-2}\left(1-\frac{2m}{r}\right)\right]=- 2 r^{-3}\left(1-\frac{2m}{r}\right)+ \frac{2m}{r^4}
\\
&=&- 2\frac{r-3m}{r^4}.
\eeaa
\end{remark}


 \section{Vectorfield  $X_f$}
 
 
 \begin{lemma}
 \label{lemma:vectorfield-ve4-appendix}
 Let  $X_f:= f e_4$.  Then with $\LaX=\frac{2f}{r} $  and  $\pitX=\piX-\LaX\g=\piX-\frac{2f}{r} \g$,
 \begin{itemize}
 \item We have,
 \bea
 \bsplit
\pitX_{44}&=0, \quad \piX_{4\vphi}=0, \quad \piX_{3\vphi}=0,\\
\pitX_{43}&=- 2 e_4 f +4 f \om+ \frac{4f}{r}        =   -2\left( e_4(f) -\frac{2f}{r} \right) +    4 f\om,\\
\pitX_{4\th}&= 2 f \xi,\\
\pitX_{AB}&=2 f \chiS_{AB}-\frac{2f}{r}   \gS_{AB} =  2f\left(  \chiS_{AB}-\frac 1 r \de_{AB}\right),    \\
\pitX_{3\th}&= 2 f(\eta+\ze),\\
\pitX_{33}&=-8 f \omb-4 e_3(f). 
\end{split}
\eea
 
 \item In particular, we have,
  \bea
 \bsplit
  \pitX_{43}&=-2 f'  +\frac{4f}{r}  +O(\ep)\min\{w_{1,1}, w_{2,1/2}\} \left( |f|+r|f'|\right),\\
   \pitX_{4A}&=\ep \min\{w_{2,1}, w_{3,1/2}\}, \\
 \pitX_{AB} &=O(\ep)\min\{w_{1,1}, w_{2,1/2}\} |f|, \\
 \pitX_{3A}&=O(\ep)w_{1,1} |f|,\\
 \pitX_{33}&= 4 f' \Up -  4 \Up' +O(\ep)w_{1,1} (|f|+ r|f'|).
 \end{split}
 \eea
 \item We have,
 \bea
 \label{eq:squareLa-f:app}
  \square \LaX&=&\frac{2}{r} f'' +O  \left(\frac{m}{r^4}  + \ep  w_{3,1} \right) \left(|f|+ r|f'|+ r^2 |f''|\right).
 \eea
  \end{itemize}
 \end{lemma}

\begin{proof}
We calculate  $\piX_{\a\b }= \g( D_{e_\a} X, e_\b)+ \g( \D_{e_\b} X, e_\a)$,
\beaa
\piX_{44}&=&0\\
\piX_{43}&=&- 2 e_4 f +4 f \om\\
\piX_{4\th}&=& 2 f \xi\\
\piX_{AB}&=&2 f \chiS_{AB}\\
\piX_{3\th}&=& 2 f(\eta+\ze)\\
\piX_{33}&=&-8  f \omb   -4 e_3(f) 
\eeaa
We deduce, for $\pitX= \piX-\LaX\g=\piX-\frac{2f}{r} \g$,
\beaa
\pitX_{44}&=&0\\
\pitX_{43}&=&- 2 e_4 f +4 f \om+ \frac{4f}{r}        =   -2\left( e_4(f) -\frac{2f}{r} \right) +    4 f\om\\
\pitX_{4\th}&=& 2 f \xi\\
\pitX_{AB}&=&2 f \chiS_{AB}-\frac{2f}{r}   \gS_{AB} =  2f\left(  \chiS_{AB}-\frac 1 r \de_{AB}\right)    \\
\pitX_{3\th}&=& 2 f(\eta+\ze)\\
\pitX_{33}&=&-8 f \omb-4 e_3(f) 
\eeaa
Under the assumptions   \eqref{eq:assumptions-Moraw1-2'}-- \eqref{eq:assumptions-Moraw2-2'}   on the Ricci coefficients (with respect to the  frame $(e_3', e_4')$),  we
 deduce,
 \beaa
 \pitX_{43}&=&- 2 e_4 f +4 f \om=-2 f'  +\frac{4f}{r}   -2 f'( e_4(r)-1) + 4f (\om-1)\\
 &=&-2 f'  +\frac{4f}{r}  +\ep \min\{w_{1,1}, w_{2,1/2}\} \left( |f|+r|f'|\right)\\
 \pitX_{4A}&=& \ep \min\{w_{2,1}, w_{3,1/2}\}, \\
 \pitX_{AB} &=&\ep \min\{w_{1,1}, w_{2,1/2}\} |f| \\
 \pitX_{3A}&=&\min\{w_{1,1}, w_{2,1/2}\} |f|\\
 \pitX_{33}&=&-8 f \omb-4 e_3(f) =-8f\left(\frac{m}{r^2}+ \ep w_{1,1}\right)- 4f'(-\Up+\ep w_{0,1})\\
 &=& 4 f' \Up -  4 \Up' +\ep w_{1,1} (|f|+ r|f'|)
 \eeaa

 To  prove     formula \eqref{eq:squareLa-f:app}   we make use of the following (see also Lemma \ref{calculation:square-radial}),
 \begin{lemma}
\label{calculation:square-radial'}
If  $h=h(r)$ then
\beaa
\square  h &=&\Up h'' (r)+    \left(  \frac{2}{r}-\frac{2m}{r^2} \right) h' +O(\ep) w_{2,1}\left(|h|+ r|h'|+ r^2 |h'' |\right)
\eeaa
\end{lemma}

\begin{proof}
For a general   scalar $h$, 
 \beaa
 \square  h &=& -\frac 1 2  (e_3 e_4 + e_4 e_3)h +\lapp h +\left(\ombS-\frac 1 2 \trchbS \right) e_4 h +(\omS-\frac 1 2 \trchS)  e_3 h\
 \eeaa
 with     $\lapp h= e_\th e_\th  h + (e_\th \Phi)^2 e_\th h =0$
if $h$ is radial.
 Thus,
 \beaa
  \square  h &=& -\frac 1 2  (e_3 e_4 + e_4 e_3)h +(\ombS-\frac 1 2 \trchbS ) e_4 h +(\omS-\frac 1 2 \trchS)  e_3 h\\
  &=&-  f'' ( e_3 r)(e_4r) -\frac 1 2 h' ( e_3 e_4 + e_4 e_3) r +     h'\left[   (\ombS-\frac 1 2 \trchbS ) e_4 r  +(\omS-\frac 1 2 \trchS) e_3 r \right]\\
  &=&-  h'' (-\Up+O(\ep )w_{0,1})( 1+O(\ep) w_{1,1})+\big(    \frac{m}{r^2} +O(\ep) w_{1,1}\big)h'\\
  &+&  h'\left[  ( \frac{m}{r^2}   +\frac{\Up}{r}     +O(\ep) w_{1,1}      ) (1+O(\ep)  w_{1,1})   + (-   \frac 1  r +O(\ep) w_{1,1} )(-\Up+O(\ep) w_{0,1}   \right]\\
  &=&\Up h'' +    \left(  \frac{2}{r}-\frac{2m}{r^2} \right) h'+O(\ep) w_{2,1}\left(|h|+ r|h'|+ r^2 |h'' |\right)
 \eeaa
 which concludes the proof of Lemma \ref{calculation:square-radial'}.
 \end{proof}
 
 In view of Lemma \ref{calculation:square-radial'},
 \beaa
 \square \LaX&=& \square \left(\frac{2f}{r}\right)= \Up\left(  \frac{2f}{r} \right)''            + \left(\frac{2}{r}-\frac{2m}{r^2}\right)(\frac {2f}{r})'+ O(\ep) w_{3,1}\left(|f|+ r|f'|+ r^2 |f''|\right)
 \eeaa
 Note that,
 \beaa
  \Up\left(  \frac{2f}{r} \right)''            + \left(  \frac{2}{r}-\frac{2m}{r^2} \right)\left(\frac {2f}{r}\right)'&=&\Up\left(\frac{2 f''}{r}-\frac{4f'}{r^2}+ \frac{4 f}{r^3}\right)    + \left(\frac{2}{r}-\frac{2m}{r^2}\right)\left(\frac{2f'}{r}-\frac{2f}{r^2}\right)\\
  &=&\frac{2\Up}{r} f''-(\Up-1)\frac{4f'}{r^2}+(\Up-1)\frac{4f}{r^3} - \frac{2m}{r^2}\left(\frac{2f'}{r}-\frac{2f}{r^2}\right)\\
  &=&\frac{2}{r} +O\left(\frac{m}{r^4}\right) \left(|f|+ r|f'|+ r^2 |f''|\right)
 \eeaa
 Hence,
 \beaa
  \square \LaX&=&\frac{2}{r} f'' +O\left(\frac{m}{r^4}  \ep  w_{3,1}\right) \left(|f|+ r|f'|+ r^2 |f''|\right)
 \eeaa
 as desired. This concludes the proof of Lemma \ref{lemma:vectorfield-ve4-appendix}.
\end{proof}


\section{Proof of Proposition \ref{square-psic-modified}}\lab{appendix:proofof:propsquare-psic-modified}


In view of the following Leibniz rule which holds for any scalar $f$, 
\beaa
-\lapp_2(f\psi) &=& \dds_2\ddd_2(f\psi)+2Kf\psi\\
&=& \dds_2(f\ddd_2\psi+e_\th(f)\psi)+2Kf\psi\\
&=& -f\lapp_2\psi -e_\th(f)\ddd_2\psi +e_\th(f)\dds_3\psi -\lapp_0(f)\psi,
\eeaa
we have the following computation  
\beaa
e_4(\square_2(r\psi)) &=& e_4(r\square_2\psi) -e_4(e_3(r)e_4\psi)-e_4(e_4(r)e_3\psi) -2e_4(e_\th(r)\ddd_2\psi)\\
&&+2e_4(e_\th(r)\dds_3\psi)+e_4(\square_0(r)\psi)\\
&=& e_4(r\square_2\psi) -e_4\left(\frac{r}{2}(\kab+\Ab)e_4\psi\right)-e_4\left(\frac{r}{2}(\ka+A)e_3\psi\right) \\
&&+e_4(\square_0(r)\psi)+r^{-1}\dk^{\leq 1}(\Ga_g)\dk^{\leq 2}\psi\\
&=& e_4(r\square_2\psi) -\frac{1}{2}e_4(r\kab e_4\psi)-\frac{1}{2}e_4\left(r\ka e_3\psi\right) +e_4(\square_0(r)\psi)+r^{-1}\err,
\eeaa
where we have introduced the notation, used throughout the proof of Proposition \ref{square-psic-modified},
\beaa
\err &:=& r^2\Ga_ge_4e_3\psi+r\Ga_b e_4\dk\psi +\dk^{\leq 1}(\Ga_b)\dk^{\leq 1}\psi  +r\dk^{\leq 1}(\Ga_g)e_3\psi +\dk^{\leq 1}(\Ga_g)\dk^2\psi.
\eeaa

Next, recall that we have
\beaa
\square_2\psi &=&-e_4 e_3 \psi +\lapp_2\psi+\left(2\om -\frac 1 2 \ka\right) e_3\psi- \frac 1 2 \kab e_4\psi+2 \etab e_\th \psi.
\eeaa
We infer
\beaa
\square_0(r) &=& -e_4(e_3(r)) +\lapp_2(r)+\left(2\om -\frac 1 2 \ka\right) e_3(r) - \frac 1 2 \kab e_4(r)+2 \etab e_\th(r)\\
&=& -e_4\left(\frac{r}{2}(\kab+\Ab)\right) +\left(2\om -\frac 1 2 \ka\right)\frac{r}{2}(\kab+\Ab) - \frac 1 2 \kab\frac{r}{2}(\ka+A)+r^{-1}\dk^{\leq 1}\Ga_g\\
&=& -\frac{1}{2}e_4(r\kab) +\frac{1}{2}\left(2\om -\frac 1 2 \ka\right)r\kab - \frac 1 4 r\kab\ka+r\dk^{\leq 1}\Ga_g\\
&=& \frac{2}{r}+O(r^{-2})+\dk^{\leq 1}\Ga_b
\eeaa
and hence
\beaa
e_4(\square_2(r\psi)) &=&  e_4(r\square_2\psi) -\frac{1}{2}e_4(r\kab e_4\psi)-\frac{1}{2}e_4\left(r\ka e_3\psi\right) +e_4(\square_0(r)\psi)+\dk^{\leq 1}(\Ga_g)\dk^{\leq 2}\psi\\
&=& e_4(r\square_2\psi) -\frac{1}{2}e_4(r\kab e_4\psi)-\frac{1}{2}e_4\left(r\ka e_3\psi\right)+e_4\left(\frac{2}{r}\psi\right)+O(r^{-3})\dk^{\leq 1}\psi+r^{-1}\err
\eeaa
so that 
\beaa
e_4(r\square_2\psi)  &=& e_4(\square_2(r\psi))+\frac{1}{2}e_4(r\kab e_4\psi)+\frac{1}{2}e_4\left(r\ka e_3\psi\right)-e_4\left(\frac{2}{r}\psi\right)+O(r^{-3})\dk^{\leq 1}\psi+r^{-1}\err.
\eeaa
We infer
\beaa
\square_2(e_4(r\psi))-e_4(r\square_2\psi) &=& [\square_2, e_4](r\psi) -\frac{1}{2}e_4(r\kab e_4\psi)-\frac{1}{2}e_4\left(r\ka e_3\psi\right)+e_4\left(\frac{2}{r}\psi\right)\\
&&+O(r^{-3})\dk^{\leq 1}\psi+r^{-1}\err.
\eeaa

Next, using again 
\beaa
\square_2\psi &=&-e_4 e_3 \psi +\lapp_2\psi+\left(2\om -\frac 1 2 \ka\right) e_3\psi- \frac 1 2 \kab e_4\psi+2 \etab e_\th \psi,
\eeaa
we infer
\beaa
[\square_2,e_4]\psi &=&-e_4 [e_3,e_4] \psi +[\lapp_2, e_4]\psi+\left(2\om -\frac 1 2 \ka\right) [e_3,e_4]\psi -e_4\left(2\om -\frac 1 2 \ka\right) e_3\psi\\
&&+ \frac 1 2 e_4(\kab) e_4\psi+2 \etab [e_\th, e_4]\psi-2 e_4(\etab) e_\th \psi\\
&=& -e_4 [e_3,e_4] \psi +[\lapp_2, e_4]\psi+\left(2\om -\frac 1 2 \ka\right) [e_3,e_4]\psi - \left(2e_4(\om) -\frac 1 2\left(-\frac{1}{2}\ka^2-2\om\ka\right)\right) e_3\psi\\
&&+ \frac 1 2\left(-\frac{1}{2}\ka\kab+2\om\kab+2\rho\right)e_4\psi+2 \etab [e_\th, e_4]\psi+r^{-2}\dk^{\leq 1}(\Ga_g)\dk\psi.
\eeaa
Now, recall 
\beaa
\,[e_3, e_4]&=& 2 \omb e_4-2\om e_3+2(\eta-\etab) e_\th,
\eeaa
and, in view of Lemma \ref{Le:comme3e4}, the following commutation formulae for reduced scalars
\begin{enumerate} 
\item If $f\in \mathfrak{s}_k$,
\beaa
\bsplit
\,[\ddd_k, e_4] &=\frac 1 2 \ka  \ddd_k f+\com_k(f),\\
\com_k(f)&=- \frac 1 2 \vth  \dds_{k+1} f - (\ze+\etab) e_4 f - k \etab  e_4\Phi f -\xi( e_3  f  +k e_3(\Phi)  f )- k \b f.
\end{split}
\eeaa
\item If  $f\in \mathfrak{s}_{k-1}$
\beaa
\bsplit
\,[\dds_k, e_4]f &=\frac 1 2 \ka  \ddd_k f+\com^*_k(f),\\
\com^*_k(f)&=- \frac 1 2 \vth  \ddd_{k-1} f + (\ze+\etab) e_4 f - (k-1)  \etab  e_4\Phi f +\xi( e_3  f  -(k-1) e_3(\Phi)  f )\\
&- (k-1)  \b f.
\end{split}
\eeaa
\end{enumerate}
We infer
\beaa
[\square_2,e_4]\psi &=& -e_4\Big((2 \omb e_4-2\om e_3+2\eta e_\th)\psi\Big) +\ka\lapp_2\psi\\
&&+\left(2\om -\frac 1 2 \ka\right)\Big(2 \omb e_4-2\om e_3+2\eta e_\th\Big)\psi - \left(2e_4(\om) +\frac{1}{4}\ka^2+\om\ka\right) e_3\psi\\
&&+ \frac 1 2\left(-\frac{1}{2}\ka\kab+2\om\kab+2\rho\right)e_4\psi+r^{-2}\dk^{\leq 1}(\Ga_g)\dk^{\leq 2}\psi\\
&=& 2e_4(\om e_3\psi) +\ka\lapp_2\psi-\left(2e_4(\om) +\frac{1}{4}\ka^2\right) e_3\psi-\frac{1}{4}\ka\kab e_4\psi+O(r^{-4})\dk^{\leq 1}\psi+r^{-2}\err\\
&=& 2\om e_4(e_3\psi) +\ka\lapp_2\psi-\frac{1}{4}\ka^2 e_3\psi-\frac{1}{4}\ka\kab e_4\psi+O(r^{-4})\dk^{\leq 1}\psi+r^{-2}\err.
\eeaa
This implies
\beaa
[\square_2,e_4](r\psi) &=& 2\om e_4(e_3(r\psi)) +\ka\lapp_2(r\psi)-\frac{1}{4}\ka^2 e_3(r\psi) -\frac{1}{4}\ka\kab e_4(r\psi) +O(r^{-3})\dk^{\leq 1}\psi+r^{-1}\err\\
&=& 2\om e_3(e_4(r\psi))+2\om [e_4, e_3]r\psi +\ka\lapp_2(r\psi) -\frac{1}{4}\ka^2 e_3(r\psi) -\frac{1}{4}\ka\kab e_4(r\psi)\\
&&+O(r^{-3})\dk^{\leq 1}\psi+r^{-1}\err\\
&=& 2\om e_3(e_4(r\psi)) +\ka\lapp_2(r\psi) -\frac{1}{4}\ka^2 e_3(r\psi) -\frac{1}{4}\ka\kab e_4(r\psi)+O(r^{-3})\dk^{\leq 1}\psi+r^{-1}\err
\eeaa
and hence
\beaa
&&\square_2(e_4(r\psi))-e_4(r\square_2\psi) \\
&=& [\square_2, e_4](r\psi) -\frac{1}{2}e_4(r\kab e_4\psi)-\frac{1}{2}e_4\left(r\ka e_3\psi\right)+e_4\left(\frac{2}{r}\psi\right)+O(r^{-3})\dk^{\leq 1}\psi+r^{-1}\err\\
&=& 2\om e_3(e_4(r\psi)) -\frac{1}{2}e_4(r\kab e_4\psi)-\frac{1}{2}e_4\left(r\ka e_3\psi\right)+\ka\lapp_2(r\psi)- \frac{1}{4}\ka^2 e_3(r\psi)\\
&&-\frac{1}{4}\ka\kab e_4(r\psi) +e_4\left(\frac{2}{r}\psi\right) +O(r^{-3})\dk^{\leq 1}\psi+r^{-1}\err.
\eeaa

Next, we compute
\beaa
&&  -\frac{1}{2}e_4(r\kab e_4\psi)-\frac{1}{2}e_4\left(r\ka e_3\psi\right)+\ka\lapp_2(r\psi)- \frac{1}{4}\ka^2 e_3(r\psi) +e_4\left(\frac{2}{r}\psi\right) \\
&=& -\frac{1}{2}e_4(\kab (e_4(r\psi)-e_4(r)\psi))-\frac{1}{2}e_3(e_4(r\ka\psi))-\frac{1}{2}[e_4,e_3](r\ka\psi)+\frac{1}{2}e_4(e_3(r\ka)\psi)\\
&&+r\ka\lapp_2\psi - \frac{1}{4}r\ka^2 e_3\psi - \frac{1}{4}\ka^2 e_3(r)\psi +\frac{2}{r^2}e_4\left(r\psi\right) +e_4\left(\frac{2}{r^2}\right)r\psi+r^{-1}\dk^{\leq 1}(\Ga_g)\dk^{\leq 2}\psi\\
&=& -\frac{1}{2}e_4(\kab (e_4(r\psi)))+\frac{1}{2}e_4\left(\kab \frac{r}{2}(\ka+A)\psi\right)-\frac{1}{2}e_3(\ka e_4(r\psi))-\frac{1}{2}e_3(e_4(\ka)r\psi)\\
&&-\frac{1}{2}\Big(-2 \omb e_4+2\om e_3-2(\eta-\etab) e_\th\Big)(r\ka\psi)+\frac{1}{2}e_4(e_3(r\ka)\psi)\\
&&+r\ka\lapp_2\psi - \frac{1}{4}r\ka^2 e_3\psi - \frac{1}{4}\ka^2\frac{r}{2}(\kab+\Ab)\psi +\frac{2}{r^2}e_4\left(r\psi\right) -\frac{4e_4(r)}{r^2}\psi +r^{-1}\dk^{\leq 1}(\Ga_g)\dk^{\leq 2}\psi
\eeaa
i.e.
\beaa
&& -\frac{1}{2}e_4(r\kab e_4\psi)-\frac{1}{2}e_4\left(r\ka e_3\psi\right)+\ka\lapp_2(r\psi)- \frac{1}{4}\ka^2 e_3(r\psi) +e_4\left(\frac{2}{r}\psi\right)\\
&=& -\frac{1}{2}e_4(\kab (e_4(r\psi)))+\frac{1}{4}e_4\left(r\kab\ka\psi\right)-\frac{1}{2}e_3(\ka e_4(r\psi))-\frac{1}{2}e_3\left(\left(-\frac{1}{2}\ka^2-2\om\ka\right)r\psi\right)\\
&&-\om e_3(r\ka\psi)+\frac{1}{2}e_4(e_3(r\ka)\psi)+r\ka\lapp_2\psi - \frac{1}{4}r\ka^2 e_3\psi - \frac{1}{8}r\ka^2\kab\psi+\frac{2}{r^2}e_4\left(r\psi\right) -\frac{2\ka}{r}\psi \\
 && +O(r^{-3})\dk^{\leq 1}\psi+r^{-1}\err\\
 &=& -\frac{1}{2}e_4(\kab (e_4(r\psi)))+\frac{1}{4}\kab\ka e_4\left(r\psi\right) +\frac{1}{4}e_4\left(\kab\ka\right)r\psi -\frac{1}{2}e_3(\ka e_4(r\psi))-\frac{1}{2}e_3\left(\left(-\frac{1}{2}\ka^2-2\om\ka\right)r\psi\right)\\
&&-\om e_3(r\ka\psi)+\frac{1}{2}r^{-1}e_3(r\ka)e_4(r\psi)+\frac{1}{2}e_4(r^{-1}e_3(r\ka))r\psi+r\ka\lapp_2\psi - \frac{1}{4}r\ka^2 e_3\psi - \frac{1}{8}r\ka^2\kab\psi\\
&&+\frac{2}{r^2}e_4\left(r\psi\right) -\frac{2\ka}{r}\psi +O(r^{-3})\dk^{\leq 1}\psi+r^{-1}\err.
\eeaa
We infer
\beaa
&&\square_2(e_4(r\psi))-e_4(r\square_2\psi) \\
&=& 2\om e_3(e_4(r\psi)) -\frac{1}{2}e_4(\kab (e_4(r\psi))) +\frac{1}{4}e_4\left(\kab\ka\right)r\psi -\frac{1}{2}e_3(\ka e_4(r\psi))\\
&&-\frac{1}{2}e_3\left(\left(-\frac{1}{2}\ka^2-2\om\ka\right)r\psi\right) -\om e_3(r\ka\psi)+\frac{1}{2}r^{-1}e_3(r\ka)e_4(r\psi)+\frac{1}{2}e_4(r^{-1}e_3(r\ka))r\psi\\
&&+r\ka\lapp_2\psi - \frac{1}{4}r\ka^2 e_3\psi - \frac{1}{8}r\ka^2\kab\psi +\frac{2}{r^2} e_4\left(r\psi\right) -\frac{2\ka}{r}\psi +O(r^{-3})\dk^{\leq 1}\psi+r^{-1}\err.
\eeaa
Since $e_4(r\psi)=r\Up\ec_4\psi$, this may be rewritten as
\beaa
&&\square_2(r\Up\ec_4\psi)-e_4(r\square_2\psi) \\
&=& 2\om e_3(r\Up\ec_4\psi) -\frac{1}{2}e_4(r\Up\kab\ec_4\psi)+\frac{1}{4}e_4(\kab\ka)r\psi -\frac{1}{2}e_3(r\Up\ka\ec_4\psi)\\
&&-\frac{1}{2}e_3\left(\left(-\frac{1}{2}\ka^2-2\om\ka\right)r\psi\right) -\om e_3(r\ka\psi)+\frac{1}{2}e_3(r\ka)\Up\ec_4\psi+\frac{1}{2}e_4(r^{-1}e_3(r\ka))r\psi\\
&&+r\ka\lapp_2\psi - \frac{1}{4}r\ka^2 e_3\psi - \frac{1}{8}r\ka^2\kab\psi +\frac{2}{r}\Up\ec_4\psi  -\frac{2\ka}{r}\psi +O(r^{-3})\dk^{\leq 1}\psi+r^{-1}\err.
\eeaa
Now, since 
\beaa
\square_2 \psi&=&-e_3 e_4 \psi +\lapp_2\psi+\left(2\omb -\frac 1 2 \kab\right) e_4\psi- \frac 1 2 \ka e_3\psi+2 \eta e_\th \psi,
\eeaa
we have
\beaa
r\ka\lapp_2\psi &=& r\ka\square_2 \psi +r\ka e_3 e_4 \psi -r\ka\left(2\omb -\frac 1 2 \kab\right) e_4\psi + \frac 1 2 r\ka^2 e_3\psi +r^{-1}\dk^{\leq 1}(\Ga_b)\dk^{\leq 1}\psi\\
&=& r\ka\square_2 \psi +r\ka e_3(r^{-1}e_4(r\psi))-r\ka e_3(r^{-1}e_4(r) \psi) +\frac 1 2\ka\kab e_4(r\psi)\\
&& -\frac 1 2 \kab\ka e_4(r)\psi + \frac 1 2 r\ka^2 e_3\psi +r^{-1}\dk^{\leq 1}(\Ga_b)\dk^{\leq 1}\psi\\
&=& r\ka\square_2 \psi +r\ka e_3(\Up\ec_4\psi)-\frac{1}{2}r\ka e_3(\ka)\psi +\frac 1 2 \ka\kab r\Up\ec_4\psi -\frac{r}{4}\ka^2\kab\psi  +r^{-1}\dk^{\leq 1}(\Ga_b)\dk^{\leq 1}\psi
\eeaa
and hence
\beaa
&&\square_2(r\Up\ec_4\psi)-e_4(r\square_2\psi) \\
&=& 2\om e_3(r\Up\ec_4\psi) -\frac{1}{2}e_4(r\Up\kab\ec_4\psi) +\frac{1}{4}e_4(\kab\ka)r\psi -\frac{1}{2}e_3(r\Up\ka\ec_4\psi)\\
&&-\frac{1}{2}e_3\left(\left(-\frac{1}{2}\ka^2-2\om\ka\right)r\psi\right) -\om e_3(r\ka\psi)+\frac{1}{2}e_3(r\ka)\Up\ec_4\psi+\frac{1}{2}e_4(r^{-1}e_3(r\ka))r\psi\\
&&+ r\ka\square_2 \psi +r\ka e_3(\Up\ec_4\psi)-\frac{1}{2}r\ka e_3(\ka)\psi  -\frac{r}{4}\ka^2\kab\psi  - \frac{1}{4}r\ka^2 e_3\psi - \frac{1}{8}r\ka^2\kab\psi \\
&& -\frac{2\ka}{r}\psi  +O(r^{-3})\dk^{\leq 1}\psi+r^{-1}\err.
\eeaa

Next, we compute
\beaa
&&\frac{1}{4}e_4(\kab\ka)r\psi -\frac{1}{2}e_3\left(\left(-\frac{1}{2}\ka^2-2\om\ka\right)r\psi\right) -\om e_3(r\ka\psi)+\frac{1}{2}e_4(r^{-1}e_3(r\ka))r\psi\\
&&-\frac{1}{2}r\ka e_3(\ka)\psi  -\frac{r}{4}\ka^2\kab\psi - \frac{1}{4}r\ka^2 e_3\psi - \frac{1}{8}r\ka^2\kab\psi -\frac{2\ka}{r}\psi\\
&=& \frac{r}{2}\ka \rho\psi +r^{-1}\dk^{\leq 1}(\Ga_b)\psi\\
&=& O(r^{-3})\psi+r^{-1}\dk^{\leq 1}(\Ga_b)\psi
\eeaa
so that
\beaa
\square_2(r\Up\ec_4\psi) &=& e_4(r\square_2\psi) + r\ka\square_2 \psi +2\om e_3(r\Up\ec_4\psi) -\frac{1}{2}e_4(r\Up\kab\ec_4\psi)  -\frac{1}{2}e_3(r\Up\ka\ec_4\psi)\\
&&+\frac{1}{2}e_3(r\ka)\Up\ec_4\psi +r\ka e_3(\Up\ec_4\psi) +O(r^{-3})\dk^{\leq 1}\psi+r^{-1}\err.
\eeaa
Since
\beaa
\square_2(r^2\ec_4\psi) &=& r\Up^{-1}\square_2(r\Up\ec_4\psi) -e_3(r\Up^{-1})e_4(r\Up\ec_4\psi) -e_4(r\Up^{-1})e_3(r\Up\ec_4\psi)\\
&&+\square_0(r\Up^{-1})r\Up\ec_4\psi+\dk^{\leq 1}(\Ga_g)\dk^{\leq 2}\psi,
\eeaa
we infer
\beaa
\square_2(r^2\ec_4\psi) &=& r\Up^{-1}e_4(r\square_2\psi) + r^2\Up^{-1}\ka\square_2 \psi +2r\Up^{-1}\om e_3(r\Up\ec_4\psi) -\frac{1}{2}r\Up^{-1}e_4(r\Up\kab\ec_4\psi)\\
&&  -\frac{1}{2}r\Up^{-1}e_3(r\Up\ka\ec_4\psi)+\frac{1}{2}re_3(r\ka)\ec_4\psi +r^2\Up^{-1}\ka e_3(\Up\ec_4\psi) \\
&& -e_3(r\Up^{-1})e_4(r\Up\ec_4\psi) -e_4(r\Up^{-1})e_3(r\Up\ec_4\psi)\\
&&+\square_0(r\Up^{-1})r\Up\ec_4\psi+O(r^{-2})\dk^{\leq 1}\psi+\err.
\eeaa
Now, we have
\beaa
&& 2r\Up^{-1}\om e_3(r\Up\ec_4\psi) -\frac{1}{2}r\Up^{-1}e_4(r\Up\kab\ec_4\psi)\\
&&  -\frac{1}{2}r\Up^{-1}e_3(r\Up\ka\ec_4\psi)+\frac{1}{2}re_3(r\ka)\ec_4\psi +r^2\Up^{-1}\ka e_3(\Up\ec_4\psi) \\
&& -e_3(r\Up^{-1})e_4(r\Up\ec_4\psi) -e_4(r\Up^{-1})e_3(r\Up\ec_4\psi)+\square_0(r\Up^{-1})r\Up\ec_4\psi\\
&=&  2r\frac{1-\frac{3m}{r}}{\Up} e_4(\ec_4\psi)\\
&&+\Bigg\{2r\Up^{-1}\om e_3(r\Up) -\frac{1}{2}r\Up^{-1}e_4(r\Up\kab) -\frac{1}{2}r\Up^{-1}e_3(r\Up\ka)+\frac{1}{2}re_3(r\ka)  +r^2\Up^{-1}\ka e_3(\Up)\\
&&    -e_3(r\Up^{-1})e_4(r\Up)  -e_4(r\Up^{-1})e_3(r\Up) +\square_0(r\Up^{-1})r\Up\Bigg\}\ec_4\psi+\err.
\eeaa
Also, we have
\beaa
&&2r\Up^{-1}\om e_3(r\Up) -\frac{1}{2}r\Up^{-1}e_4(r\Up\kab) -\frac{1}{2}r\Up^{-1}e_3(r\Up\ka)+\frac{1}{2}re_3(r\ka)  +r^2\Up^{-1}\ka e_3(\Up)\\
&&  -e_3(r\Up^{-1})e_4(r\Up)  -e_4(r\Up^{-1})e_3(r\Up) +\square_0(r\Up^{-1})r\Up\\
&=& 4+O(r^{-1})+r\Ga_b\\
&=& -r^2\ka\kab+O(r^{-1})+r\Ga_b.
\eeaa
We infer
\beaa
(\square_2+\ka\kab)(r^2\ec_4\psi) &=& r\Up^{-1}e_4(r\square_2\psi) + r^2\Up^{-1}\ka\square_2 \psi +2r\frac{1-\frac{3m}{r}}{\Up} e_4(\ec_4\psi)\\
&& +O(r^{-2})\dk^{\leq 1}\psi+\err.
\eeaa
In view of the wee equation satisfied by $\psi$, i.e.
\beaa
\square_2\psi +\ka\kab\psi &=& N,
\eeaa
we have
\beaa
&& r\Up^{-1}e_4(r\square_2\psi) + r^2\Up^{-1}\ka\square_2 \psi +2r\frac{1-\frac{3m}{r}}{\Up} e_4(\ec_4\psi)\\
&=& r\Up^{-1}e_4(r(N-\ka\kab\psi)) + r^2\Up^{-1}\ka(N-\ka\kab\psi) +\frac{2}{r}\frac{1-\frac{3m}{r}}{\Up} e_4(r^2\ec_4\psi) -4\frac{1-\frac{3m}{r}}{\Up} e_4(r)\ec_4\psi\\
&=& r\Up^{-1}e_4(rN) + r^2\Up^{-1}\ka N +\frac{2}{r}\frac{1-\frac{3m}{r}}{\Up} e_4(r^2\ec_4\psi) +\frac{4m}{r}\ec_4\psi -2r^2\Up^{-1}\ka\rho\psi  +\dk^{\leq 1}(\Ga_b)\dk^{\leq 1}\psi\\
&=& r^2\left(\Up^{-1}e_4(N) + \frac{3}{r}N\right)  +\frac{2}{r}\frac{1-\frac{3m}{r}}{\Up} e_4(r^2\ec_4\psi) +O(r^{-2})\dk^{\leq 1}\psi  +\dk^{\leq 1}(\Ga_b)\dk^{\leq 1}\psi,
\eeaa
from which we deduce
\beaa
(\square_2+\ka\kab)(r^2\ec_4\psi) &=& r^2\left(\Up^{-1}e_4(N) + \frac{3}{r}N\right)  +\frac{2}{r}\frac{1-\frac{3m}{r}}{\Up} e_4(r^2\ec_4\psi)   +O(r^{-2})\dk^{\leq 1}\psi+\err.
\eeaa

Since 
\beaa
\psic &=& f_2\ec_4\psi=\frac{f_2}{r^2}r^2\ec_4\psi,
\eeaa
we infer
\beaa
(\square_2+\ka\kab)\psic &=& \frac{f_2}{r^2}(\square_2+\ka\kab)(r^2\ec_4\psi) -e_3\left(\frac{f_2}{r^2}\right)e_4(r^2\ec_4\psi) -e_4\left(\frac{f_2}{r^2}\right)e_3(r^2\ec_4\psi)\\
&& +e_\th\left(\frac{f_2}{r^2}\right)\ddd_2(r^2\ec_4\psi) -e_\th\left(\frac{f_2}{r^2}\right)\dds_3(r^2\ec_4\psi)+\square_0\left(\frac{f_2}{r^2}\right)r^2\ec_4\psi
\eeaa
and hence
\beaa
(\square_2+\ka\kab)\psic &=& f_2\left(\Up^{-1}e_4(N) + \frac{3}{r}N\right)  +\frac{f_2}{r^2}\Bigg\{\frac{2}{r}\frac{1-\frac{3m}{r}}{\Up} e_4(r^2\ec_4\psi)  +O(r^{-2})\dk^{\leq 1}\psi+\err\Bigg\}\\
&& -e_3\left(\frac{f_2}{r^2}\right)e_4(r^2\ec_4\psi) -e_4\left(\frac{f_2}{r^2}\right)e_3(r^2\ec_4\psi)\\
&& +e_\th\left(\frac{f_2}{r^2}\right)\ddd_2(r^2\ec_4\psi) -e_\th\left(\frac{f_2}{r^2}\right)\dds_3(r^2\ec_4\psi)+\square_0\left(\frac{f_2}{r^2}\right)r^2\ec_4\psi.
\eeaa
Now, recall that $\err$ is defined by
\beaa
\err &=& r^2\Ga_ge_4e_3\psi+r\Ga_b e_4\dk\psi +\dk^{\leq 1}(\Ga_b)\dk^{\leq 1}\psi  +r\dk^{\leq 1}(\Ga_g)e_3\psi +\dk^{\leq 1}(\Ga_g)\dk^2\psi.
\eeaa
so that 
\beaa
(\square_2+\ka\kab)\psic &=& f_2\left(\Up^{-1}e_4(N) + \frac{3}{r}N\right)  +\frac{f_2}{r^2}\Bigg\{\frac{2}{r}\frac{1-\frac{3m}{r}}{\Up} e_4(r^2\ec_4\psi)  +O(r^{-2})\dk^{\leq 1}\psi+r^2\Ga_ge_4e_3\psi\\
&&+r\Ga_b e_4\dk\psi +\dk^{\leq 1}(\Ga_b)\dk^{\leq 1}\psi  +r\dk^{\leq 1}(\Ga_g)e_3\psi +\dk^{\leq 1}(\Ga_g)\dk^2\psi\Bigg\}\\
&& -e_3\left(\frac{f_2}{r^2}\right)e_4(r^2\ec_4\psi) -e_4\left(\frac{f_2}{r^2}\right)e_3(r^2\ec_4\psi)\\
&& +e_\th\left(\frac{f_2}{r^2}\right)\ddd_2(r^2\ec_4\psi) -e_\th\left(\frac{f_2}{r^2}\right)\dds_3(r^2\ec_4\psi)+\square_0\left(\frac{f_2}{r^2}\right)r^2\ec_4\psi.
\eeaa
In view of 
\beaa
\square_2\psi &=&-e_4 e_3 \psi +\lapp_2\psi+\left(2\om -\frac 1 2 \ka\right) e_3\psi- \frac 1 2 \kab e_4\psi+2 \etab e_\th \psi,
\eeaa
we have
\beaa
r^2\Ga_ge_4e_3\psi &=& r^2\Ga_g\left(-\square_2\psi +\lapp_2\psi+\left(2\om -\frac 1 2 \ka\right) e_3\psi- \frac 1 2 \kab e_4\psi+2 \etab e_\th \psi\right)\\
&=& -r^2\Ga_g N+r\Ga_g e_3\psi +\Ga_g\dk^{\leq 2}\psi
\eeaa
and hence
\beaa
(\square_2+\ka\kab)\psic &=& f_2\left(\Up^{-1}e_4(N) + \frac{3}{r}N\right)  +\frac{f_2}{r^2}\Bigg\{\frac{2}{r}\frac{1-\frac{3m}{r}}{\Up} e_4(r^2\ec_4\psi)  +O(r^{-2})\dk^{\leq 1}\psi\\
&&+r\Ga_b e_4\dk\psi +\dk^{\leq 1}(\Ga_b)\dk^{\leq 1}\psi  +r\dk^{\leq 1}(\Ga_g)e_3\psi +\dk^{\leq 1}(\Ga_g)\dk^2\psi\Bigg\}\\
&& -e_3\left(\frac{f_2}{r^2}\right)e_4(r^2\ec_4\psi) -e_4\left(\frac{f_2}{r^2}\right)e_3(r^2\ec_4\psi)\\
&& +e_\th\left(\frac{f_2}{r^2}\right)\ddd_2(r^2\ec_4\psi) -e_\th\left(\frac{f_2}{r^2}\right)\dds_3(r^2\ec_4\psi)+\square_0\left(\frac{f_2}{r^2}\right)r^2\ec_4\psi.
\eeaa
In particular, we have for $r\geq 6m_0$
\beaa
(\square_2+\ka\kab)\psic &=& r^2\left(\Up^{-1}e_4(N) + \frac{3}{r}N\right)  +\frac{2}{r\Up}\left(1-\frac{3m}{r}\right) e_4\psic  \\
 && +O(r^{-2})\dk^{\leq 1}\psi+r\Ga_b e_4\dk\psi +\dk^{\leq 1}(\Ga_b)\dk^{\leq 1}\psi  +r\dk^{\leq 1}(\Ga_g)e_3\psi +\dk^{\leq 1}(\Ga_g)\dk^2\psi
\eeaa
and for $4m_0\leq r\leq 6m_0$,
\beaa
(\square_2+\ka\kab)\psic &=& f_2\left(\Up^{-1}e_4(N) + \frac{3}{r}N\right)+O(1)\dk^2\psi.
\eeaa
This concludes the proof Proposition \ref{square-psic-modified}.


\end{document}